\documentclass[fleqn,usenatbib]{mnras}

\usepackage{newtxtext,newtxmath}

\usepackage[T1]{fontenc}

\usepackage{graphicx}	

\usepackage{threeparttable}
\usepackage{array}
\usepackage{subcaption}
\usepackage{framed}
\usepackage[utf8]{inputenc}
\DeclareUnicodeCharacter{2212}{-}

\usepackage{tikz}
\definecolor{lime}{HTML}{A6CE39}
\DeclareRobustCommand{\orcidicon}{%
	\begin{tikzpicture}
	\draw[lime, fill=lime] (0,0) 
	circle [radius=0.16] 
	node[white] {{\fontfamily{qag}\selectfont \tiny ID}};
	\draw[white, fill=white] (-0.0625,0.095) 
	circle [radius=0.007];
	\end{tikzpicture}
	\hspace{-2mm}
}
\foreach \x in {A, ..., Z}{%
	\expandafter\xdef\csname orcid\x\endcsname{\noexpand\href{https://orcid.org/\csname orcidauthor\x\endcsname}{\noexpand\orcidicon}}
}

\title[The stellar content of SPT2349$-$56]
{Rapid build-up of the stellar content in the protocluster core SPT2349$-$56 at $z\,{=}\,4.3$}

\author[Hill et al.]
{Ryley Hill,$^{1}$
Scott Chapman,$^{1,2,3}$
Kedar A. Phadke,$^{4}$
Manuel Aravena,$^{5}$
Melanie Archipley,$^{4,6}$ \newauthor
Matthew~L.~N.~Ashby,$^{7}$\orcidA{}
Matthieu B{\'e}thermin,$^{8}$
Rebecca E.~A.~Canning,$^{9}$ 
Anthony Gonzalez,$^{10}$ \newauthor
Thomas R.~Greve,$^{11,12,13}$
Gayathri Gururajan,$^{8}$
Christopher C.~Hayward,$^{14}$ 
Yashar Hezaveh,$^{14,15}$ \newauthor
Sreevani Jarugula,$^{4}$
Duncan MacIntyre,$^{1}$
Daniel P.~Marrone,$^{16}$
Tim Miller,$^{17}$ \newauthor
Douglas Rennehan,$^{18}$
Cassie Reuter,$^{4}$ 
Kaja M.~Rotermund,$^{3}$
Douglas Scott,$^{1}$
Justin Spilker,$^{19,20}$ \newauthor
Joaquin D.~Vieira,$^{4,6}$
George Wang,$^{1}$
Axel Wei{\ss}$^{21}$
\\
$^{1}$Department of Physics and Astronomy, University of British Columbia, 6225 Agricultural Road, Vancouver, V6T 1Z1, Canada\\
$^{2}$National Research Council, Herzberg Astronomy and Astrophysics, 5071 West Saanich Road, Victoria, V9E 2E7, Canada\\
$^{3}$Department of Physics and Atmospheric Science, Dalhousie University, 6310 Coburg Road, B3H 4R2, Halifax, Canada\\
$^{4}$Department of Astronomy, University of Illinois, 1002 West Green Street, Urbana, IL 61801, USA\\
$^{5}$N\'ucleo de Astronom\'{\i}a, Facultad de Ingenier\'{\i}a y Ciencias, Universidad Diego Portales, Av. Ej{\'e}rcito 441, Santiago, 8320000, Chile\\
$^{6}$Center for AstroPhysical Surveys, National Center for Supercomputing Applications, Urbana, IL, 61801, USA\\
$^{7}$Center for Astrophysics $|$ Harvard \& Smithsonian, Optical and Infrared Astronomy Division, 60 Garden St., MS-66, Cambridge, MA 02138, USA\\
$^{8}$Laboratoire d'Astrophysique de Marseille, 38 rue Fr{\'e}d{\'e}ric Joliot-Curie, Marseille, 13013, France\\
$^{9}$Institute of Cosmology and Gravitation, University of
Portsmouth, Dennis Sciama Building, Portsmouth, PO1 3FX, UK\\
$^{10}$Department of Astronomy, University of Florida, 211 Bryant Space Science Center, Gainesville, FL 32611-2055, USA\\
$^{11}$Cosmic Dawn Center\\
$^{12}$DTU Space, Technical University of Denmark, Elektrovej 327, Kgs. Lyngby, DK-2800, Denmark\\
$^{13}$Department of Physics and Astronomy, University College London, Gower Street, London, WC1E 6BT, UK\\
$^{14}$Center for Computational Astrophysics, Flatiron Institute, 162 Fifth Avenue, New York, NY 10010, USA\\
$^{15}$D{\'e}partement de Physique, Universit{\'e} de Montr{\'e}al, 1375 Avenue Th{\'e}r{\`e}se-Lavoie-Roux, Montr{\'e}al, H2V 0B3, Canada\\
$^{16}$Steward Observatory, University of Arizona, 933 North Cherry Avenue, Tucson, AZ 85721, USA\\
$^{17}$Department of Astronomy, Yale University, 52 Hillhouse Avenue, New Haven, CT 06511, USA\\
$^{18}$Department of Physics and Astronomy, University of Victoria, 3800 Finnerty Road, Victoria, V8P 5C2, Canada\\
$^{19}$Department of Astronomy, University of Texas at Austin, 2515 Speedway, Stop C1400, Austin, TX 78712, USA\\
$^{20}$NHFP Hubble Fellow\\
$^{21}$Max-Planck-Institut f{\"u}r Radioastronomie, Auf dem H{\"u}gel 69, Bonn, D-53121, Germany
} 

\date{September 2021}

\pubyear{2021}

\begin{document}
\label{firstpage}
\pagerange{\pageref{firstpage}--\pageref{lastpage}}
\maketitle

\begin{abstract}
\noindent The protocluster SPT2349$-$56 at $z\,{=}\,4.3$ contains one of the most actively star-forming cores known, yet constraints on the total stellar mass of this system are highly uncertain. We have therefore carried out deep optical and infrared observations of this system, probing rest-frame ultraviolet to infrared wavelengths. Using the positions of the spectroscopically-confirmed protocluster members, we identify counterparts and perform detailed source deblending, allowing us to fit spectral energy distributions in order to estimate stellar masses. We show that the galaxies in SPT2349$-$56 have stellar masses proportional to their high star-formation rates, consistent with other protocluster galaxies and field submillimetre galaxies (SMGs) around redshift 4. The galaxies in SPT2349$-$56 have on average lower molecular gas-to-stellar mass fractions and depletion timescales than field SMGs, although with considerable scatter. We construct the stellar-mass function for SPT2349$-$56 and compare it to the stellar-mass function of $z\,{=}\,1$ galaxy clusters, finding consistent shapes between the two. We measure rest-frame galaxy ultraviolet half-light radii from our {\it HST\/}-F160W imaging, finding that on average the galaxies in our sample are similar in size to typical star-forming galaxies at these redshifts. However, the brightest {\it HST\/}-detected galaxy in our sample, found near the luminosity-weighted centre of the protocluster core, remains unresolved at this wavelength. Hydrodynamical simulations predict that the core galaxies will quickly merge into a brightest cluster galaxy, thus our observations provide a direct view of the early formation mechanisms of this class of object.
\end{abstract}

\begin{keywords}
galaxies -- formation: galaxies -- evolution: submm -- galaxies
\end{keywords}

\section{Introduction}
\label{introduction}

In the present day, the large-scale structure of our Universe is made up of filaments, nodes, and voids, a structure that is often described as the `cosmic web'. The nodes of this cosmic web are comprised of galaxy clusters, which are the largest gravitationally bound objects in the Universe. Being such fundamental building blocks, galaxy clusters are well-studied objects; we know that they are seeded by small-amplitude density fluctuations of the sort observed in the cosmic microwave background (CMB), which then grew and collapsed into the massive structures that we see today \citep[e.g.,][]{wright1992,bennett2003,springel2005}. While the CMB is very well understood \citep[e.g.][]{planckI}, and the details of present-day galaxy clusters are well-described \citep[e.g.,][]{biviano1998,giodini2013,bykov2015}, the intermediate phase of evolution, which has become known as the realm of galaxy `protoclusters', still lacks sufficient observations to pin down the models \citep[e.g.][]{overzier2016}.

Traditional galaxy cluster searches have made use of the fact that these objects are virialized, allowing the intergalactic gas to heat up and be detected by X-ray facilities \citep[e.g.,][]{rosati2009,gobat2011,andreon2014,wang2016,mantz2018} or at millimetre wavelengths via the Sunyaev-Zeldovich (SZ) effect \citep[e.g.,][]{planck2014-a36,bleem2015,huang2019}. However, beyond redshifts of around 2, these signatures become too faint for practical detection. Protocluster-detection techniques now include searching large optical and infrared sky maps for overdensities of red galaxies \citep[e.g.,][]{martinache2018,greenslade2018}, Lyman-break galaxies \citep[LBGs; e.g.,][]{steidel2000,dey2016}, or Lyman-$\alpha$ emitters \citep[LAEs; e.g.,][]{shimasaku2003,tamura2009,chiang2015,dey2016,harikane2019}, and searching the area surrounding rare and luminous sources or groups of sources such as radio-loud active galactic nuclei \citep[AGN; e.g.,][]{steidel2005,venemans2007,wylezalek2013,dannerbauer2014,noirot2018} or submillimetre galaxies \citep[SMGs; e.g.,][]{chapman2009,umehata2015,casey2015,hung2016,oteo2018,lacaille2019,long2020}. 

Another protocluster-selection technique that has recently been gaining attention comes from experiments designed to map the CMB. These experiments typically aim to cover huge areas of the sky at submillimetre and millimetre wavelengths with resolution on the order of a few arcminutes, and in the process find some of the brightest and rarest submillimetre and millimetre sources in the sky. After the application of various selection criteria to remove Galactic sources and quasars/blazars, follow-up observations with higher-resolution telescopes have subsequently revealed that many of the remaining sources are gravitational lenses \citep{negrello2010,hezaveh2013,canameras2015,spilker2016}, or genuine overdensities of luminous star-forming galaxies, representing ideal protocluster candidates \citep{planckXXVII,flores-cacho2015,miller2018,kneissl2019,hill2020,wang2020,koyama2021}.

One such source, SPT2349$-$56, was discovered in the South Pole Telescope (SPT)'s extragalactic mm-wave point-source catalogue \citep{vieira2010,mocanu2013,everett2020}, and is now known to contain dozens of spectroscopically-confirmed star-forming galaxies through Atacama Large Millimeter/submillimeter Array \citep[ALMA;][]{wootten2009} observations \citep{miller2018,hill2020}, several LBGs through Gemini Multi-Object Spectrograph \citep[GMOS;][]{hook2004} and near-infrared wide-field imager \citep[FLAMINGOS-2;][]{eikenberry2006} observations \citep{rotermund2021}, and a number of LAEs, alongside a Lyman-$\alpha$ blob (Apostolovski et al.~in prep.), through observations with the Very Large Telescope (VLT)'s Multi Unit Spectroscopic Explorer \citep[MUSE;][]{bacon2010}. This protocluster lies at a redshift of 4.3, and based on observations with the Atacama Pathfinder Experiment (APEX) telescope’s Large APEX BOlometer CAmera \citep[LABOCA;][]{kreysa2003,siringo2009}, has an integrated star-formation rate (SFR) of over 10,000\,M$_{\odot}\,$yr$^{-1}$ within a diameter of about 500 proper kiloparsecs, well above what is seen for a single cluster in a wide variety of cosmological simulations \citep{lim2021}. Hydrodynamical simulations using the known galaxies in SPT2349$-$56 as the initial conditions predict that most of the galaxies in the centre of this object will merge into a single brightest cluster galaxy \citep[BCG;][]{rennehan2019} over a timescale of a few hundred million years. SPT2349$-$56 is believed to be the core of a Mpc-scale protocluster, as evidenced by several infalling subhalos found around it \citep{hill2020}. 

In addition to optical observations of SPT2349$-$56 with GMOS and FLAMINGOS-2, the {\it Spitzer Space Telescope}'s InfraRed Array Camera \citep[IRAC;][]{fazio2004} was used to observe this protocluster core in the infrared \citep{rotermund2021}. These data were used to obtain rest-frame ultraviolet photometry for nine out of the 14 galaxies originally reported by \citet{miller2018}, as well as identifying four LBGs at the same redshift as the structure. While the photometric coverage was sparse, owing to the faintness of the known galaxies in the optical, initial spectral energy distribution (SED) fits suggested that the core has a stellar mass of at least $10^{12}\,$M$_{\odot}$, and the nine detected galaxies appeared to show significant scatter around the $z\,{=}\,4$ galaxy main sequence (MS). A search for an overdensity of LBGs out to about 1\,Mpc found that the overdensity was too low to meet large-field optical survey criteria, meaning that SPT2349$-$56 would not be picked up by traditional optical surveys searching for distant protoclusters.

We have now significantly bolstered our optical and infrared coverage of SPT2349$-$56 using the {\it Hubble Space Telescope} ({\it HST\/}), and by increasing our {\it Spitzer\/}-IRAC integration time by a factor of 10. In addition, we have now identified over 30 galaxy protocluster members. In this paper, we use these new data to analyse the ultraviolet and infrared properties of this much larger sample of protocluster members.

In Section \ref{observations} we describe these new observations in detail. In Section \ref{analysis} we outline our data reduction procedure, including deblending, source matching, flux density extraction, SED fitting, and profile fitting. In Section \ref{results} we show our results, in Section \ref{discussion} we discuss the implications of these observations, and the paper is concluded in Section \ref{conclusion}.

\section{Observations}
\label{observations}

\subsection{\textbf{\textit{HST}}}

{\it HST\/} observed SPT2349$-$56 with the Wide Field Camera 3 (WFC3) instrument in the F110W and F160W filters during Cycle 26 (proposal ID 15701, PI S.~Chapman). Two orbits were carried out for the F110W filter, totalling 1.6 hours on-source, and three orbits were carried out for the F160W filter, totalling 2.4 hours on-source. Since the WFC3 detector pixels undersample the PSF (the plate scale is 0.13\,arcsec\,pixel$^{-1}$, with a full width at half maximum, FWHM, around 1 pixel), the observations utilized a standard sub-pixel dither pattern in order to fully sample the PSF. The field-of-view of the WFC3 instrument is about 4.7\,arcmin$^2$, sufficiently large to image all of the protocluster members in the core, as well as the northern component. \citet{hill2020} found CO(4--3) line emission at $z\,{=}\,4.3$ from three galaxies in a targeted Atacama Large Millimeter/submillimeter Array \citep[ALMA;][]{wootten2009} Band~3 observation of a red {\it Herschel\/}-SPIRE source (named `SPIREc') located about 1.5\,Mpc from the main structure, but these three galaxies are not covered by the {\it HST\/} imaging.

The data were calibrated using {\tt calwf3}, part of the standard {\it HST\/} WFC3 pipeline {\tt wfc3tools} available in {\tt python}.\footnote{\url{https://github.com/spacetelescope/wfc3tools}} The individual exposures were stacked using {\tt astrodrizzle}; the stacking method used was the median, and sky subtraction was performed. We set the final pixel scale to be 0.075\,arcsec\,pixel$^{-1}$, approximately Nyquist sampling (i.e. sampling by a factor of 2) the beam. The final rms reached in these images is 0.50\,nJy for the F110W filter (corresponding to a 5$\sigma$ AB magnitude limit of 30.4), and 0.79\,nJy for the F160W filter (corresponding to a 5$\sigma$ AB magnitude limit of 29.9). 2\,armin\,${\times}$\,2\,arcmin cutouts of the images, smoothed by a 3-pixel FWHM Gaussian, are shown in Fig.~\ref{all_field}.

\subsection{IRAC}

The {\it Spitzer Space Telescope\/} was used to observe SPT2349$-$56 at two wavelengths: 3.6\,$\mu$m and 4.5\,$\mu$m. A total of four observations have now been carried out since 2009, the first two of which were presented in \citet{rotermund2021}, where further details can be found. The two subsequent observations used in this paper were carried out in January 2018 (proposal ID 13224, PI S.~Chapman) and October 2019 (proposal ID 14216, PI S.~Chapman), and uniformly covered 3.6\,$\mu$m and 4.5\,$\mu$m with 324\,${\times}$\,100\,s dithered exposures (162 exposures per observation for each channel). These new data provide a factor of 10 more exposure time, reducing the instrumental rms by a factor of about 3. Given IRAC's very large field-of-view (about 27\,arcmin$^2$), all protocluster members known to date (including the three SPIREc sources) were covered by these observations.

Data from all four observations were combined and set to a pixel scale of 0.6\,arcsec\,pixel$^{-1}$. The final rms levels of the stacked data are 6.9\,nJy at 3.6\,$\mu$m (corresponding to a 5$\sigma$ AB magnitude limit of 27.5), and 6.5\,nJy at 4.5\,$\mu$m (corresponding to a 5$\sigma$ AB magnitude limit of 27.6). Cutouts showing the 2\,armin\,${\times}$\,2\,arcmin region around the main structure are shown in Fig.~\ref{all_field}.

\subsection{Gemini}

The Gemini observations used in this paper were presented in \citet{rotermund2021}, where details of the imaging, calibration, and data reduction can be found. Here we only provide a brief summary. Since the data were taken before many new protocluster members had been confirmed, we also outline which of these new galaxies are covered by the observations.

Data were taken in the $g$, $r$, and $i$ bands using the GMOS instrument \citep{hook2004}, and similarly in the $K_{\rm s}$ band using the FLAMINGOS-2 instrument \citep{eikenberry2004}, both of which are part of the Gemini South Observatory. The rms levels reached in these images are 0.2, 0.3, 0.6, and 15.0\,nJy, respectively. For the GMOS instrument, the $g$-, $r$-, and $i$-band images have a field-of-view of 5.5 arcmin$^2$ and cover all of the core and northern sources, but not the three SPIREc galaxies located 1.5\,Mpc away from the core.

Apostolovski et al. (in prep.) also reports eight LAEs at $z\,{=}\,4.3$, four of which are in the central component of the protocluster, while the remaining four are found in the northern component. These eight galaxies are also covered by the GMOS imaging, and will be included in the analysis below. Finally, \citet{rotermund2021} identified four LBGs around the core of SPT2349$-$56; they identify one of the LBGs (LBG1) with a galaxy from \citet{hill2020} found in the ALMA data (C17), one (LBG4) with an LAE (LAE3), with the remaining two LBGs being unique. However, they note that one of their unique LBGs, LBG2, lies close to galaxy C2 from \citet{hill2020}, and could also be a counterpart. In this paper, we treat LBG2 as the counterpart to C2 as it lies within the 1\,arcsec search radius criteria we outline in Section \ref{matching}, and we include the other unique LBG, LBG3, as a separate source in our analysis below.

\begin{figure*}
\includegraphics[width=0.45\textwidth]{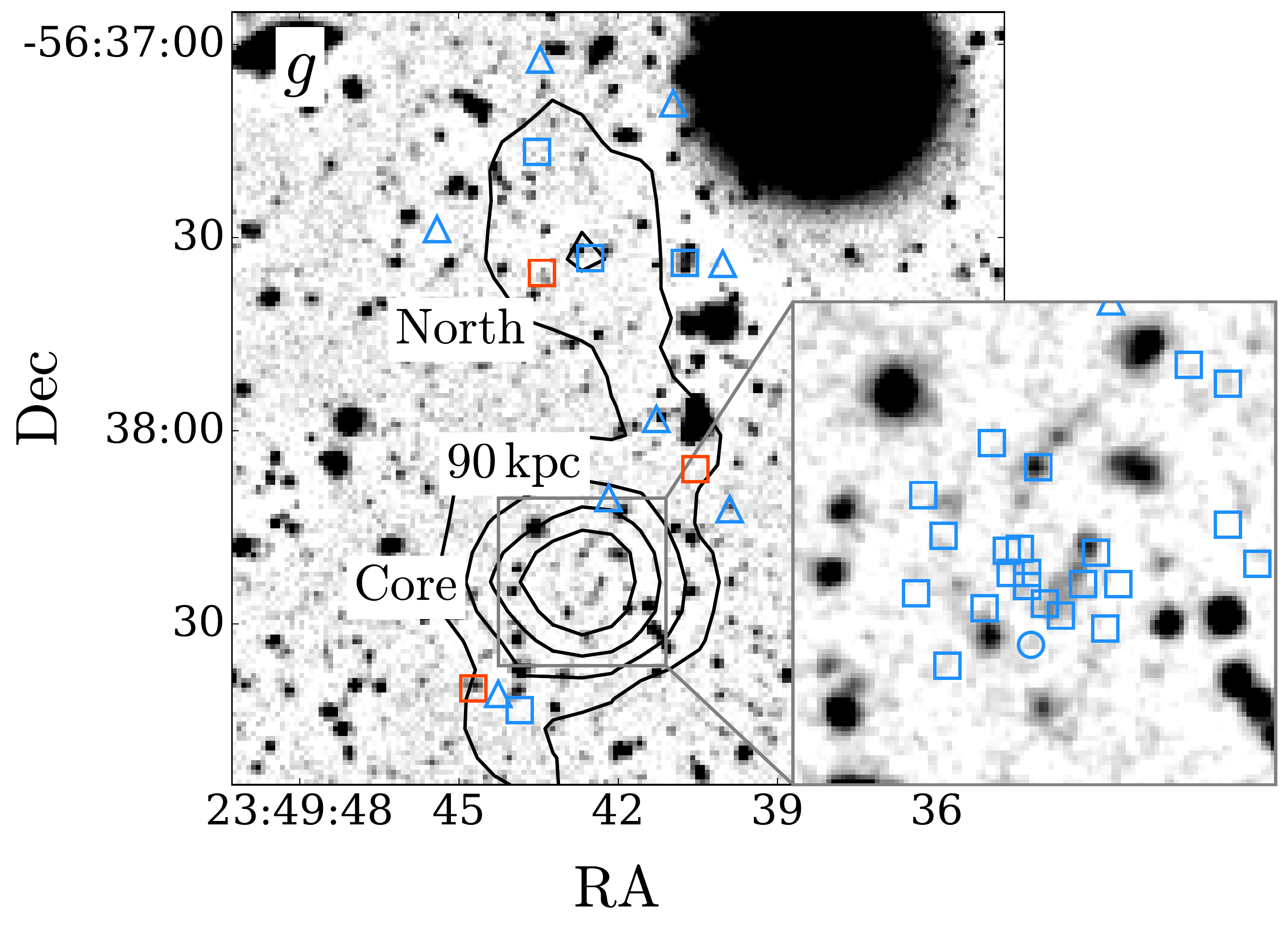}
\includegraphics[width=0.45\textwidth]{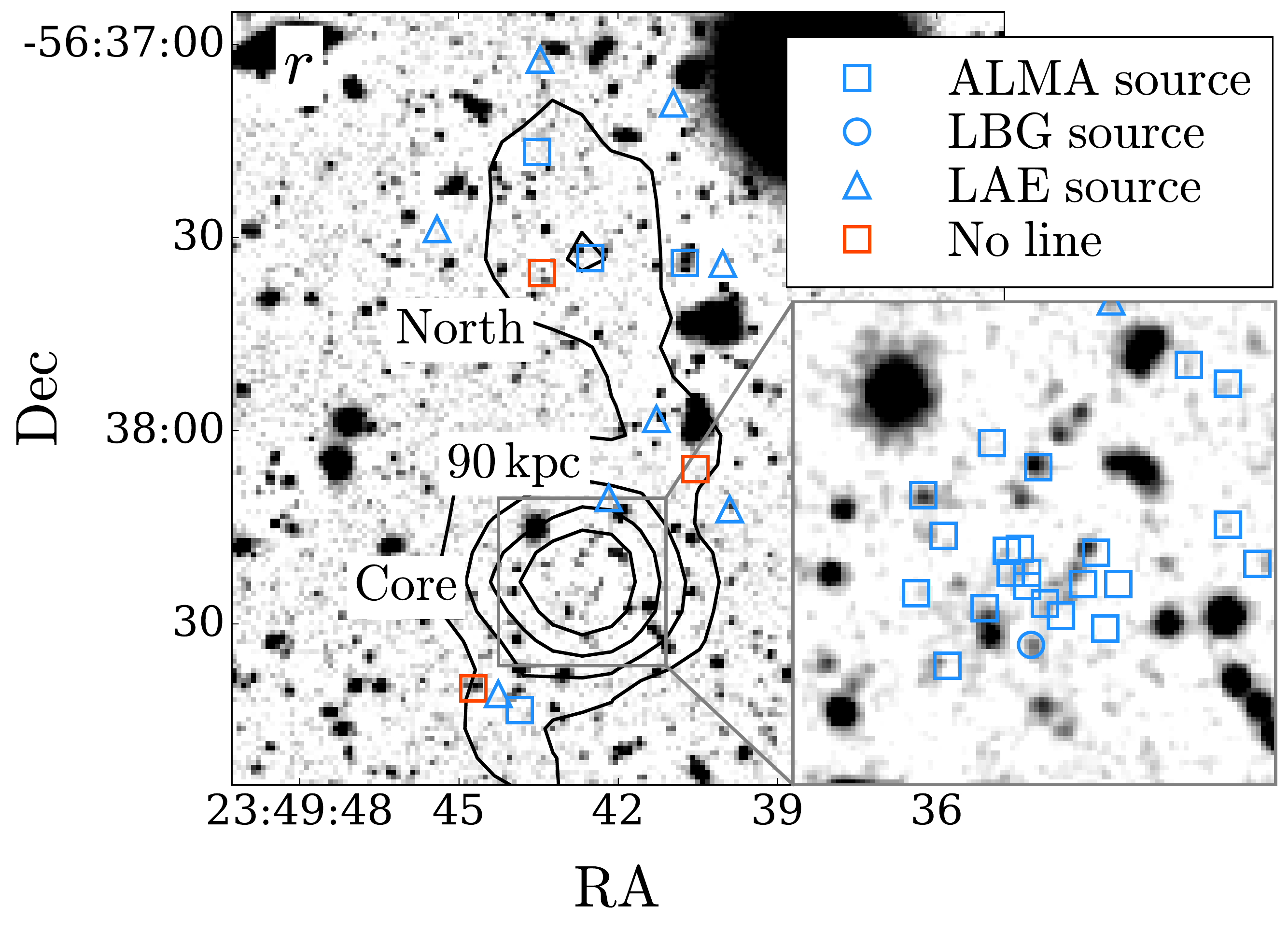}
\includegraphics[width=0.45\textwidth]{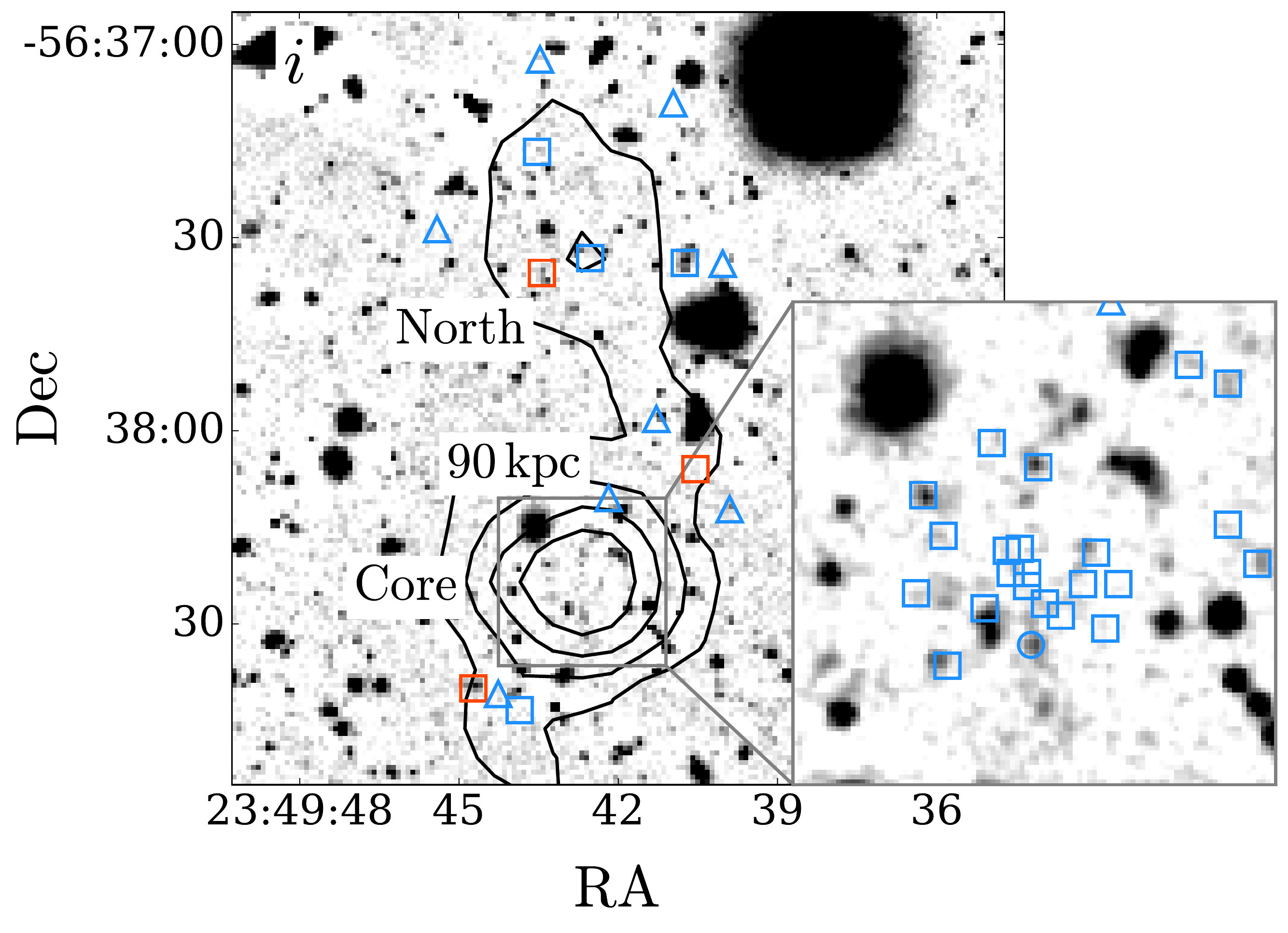}
\includegraphics[width=0.45\textwidth]{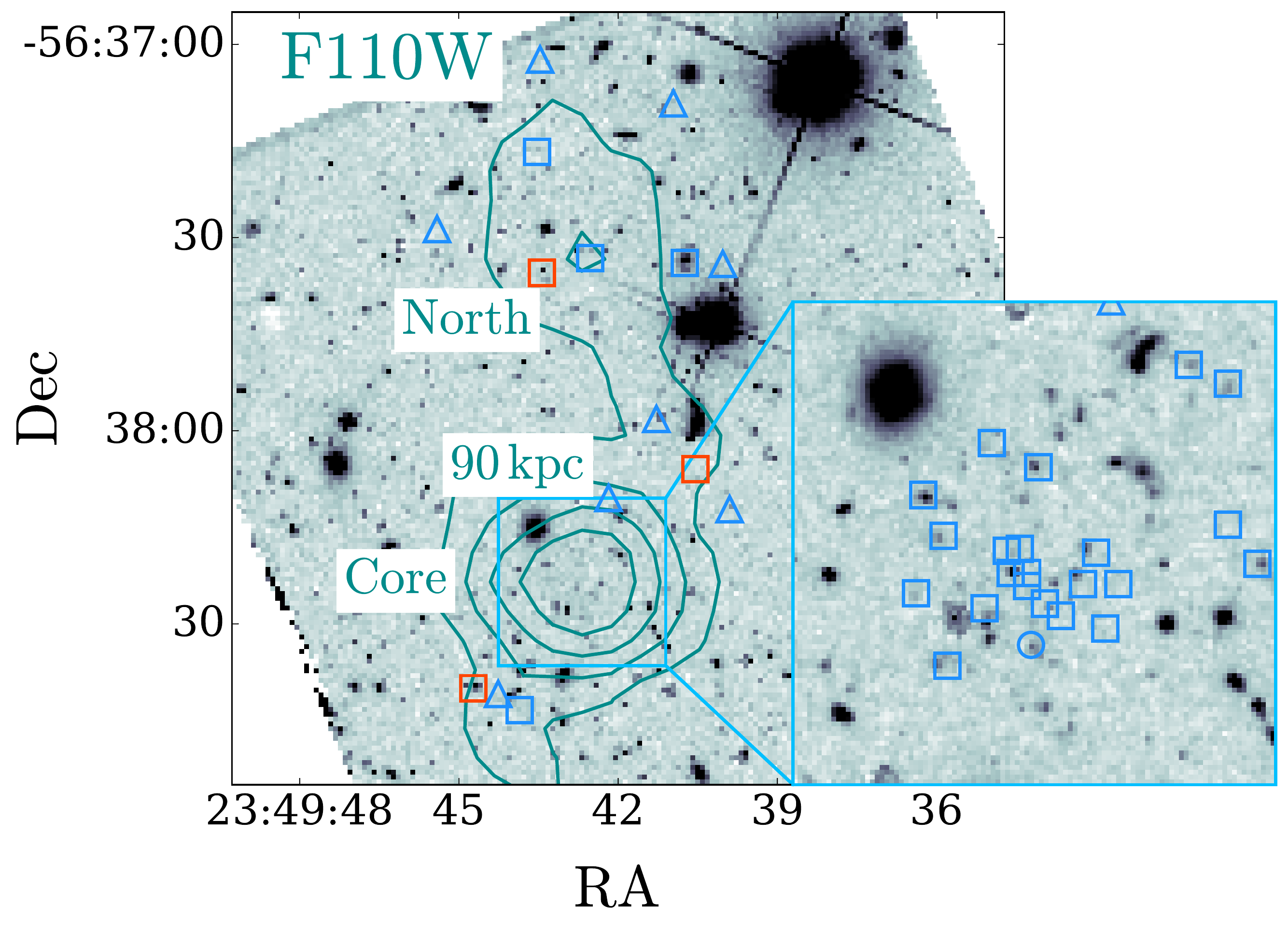}
\includegraphics[width=0.45\textwidth]{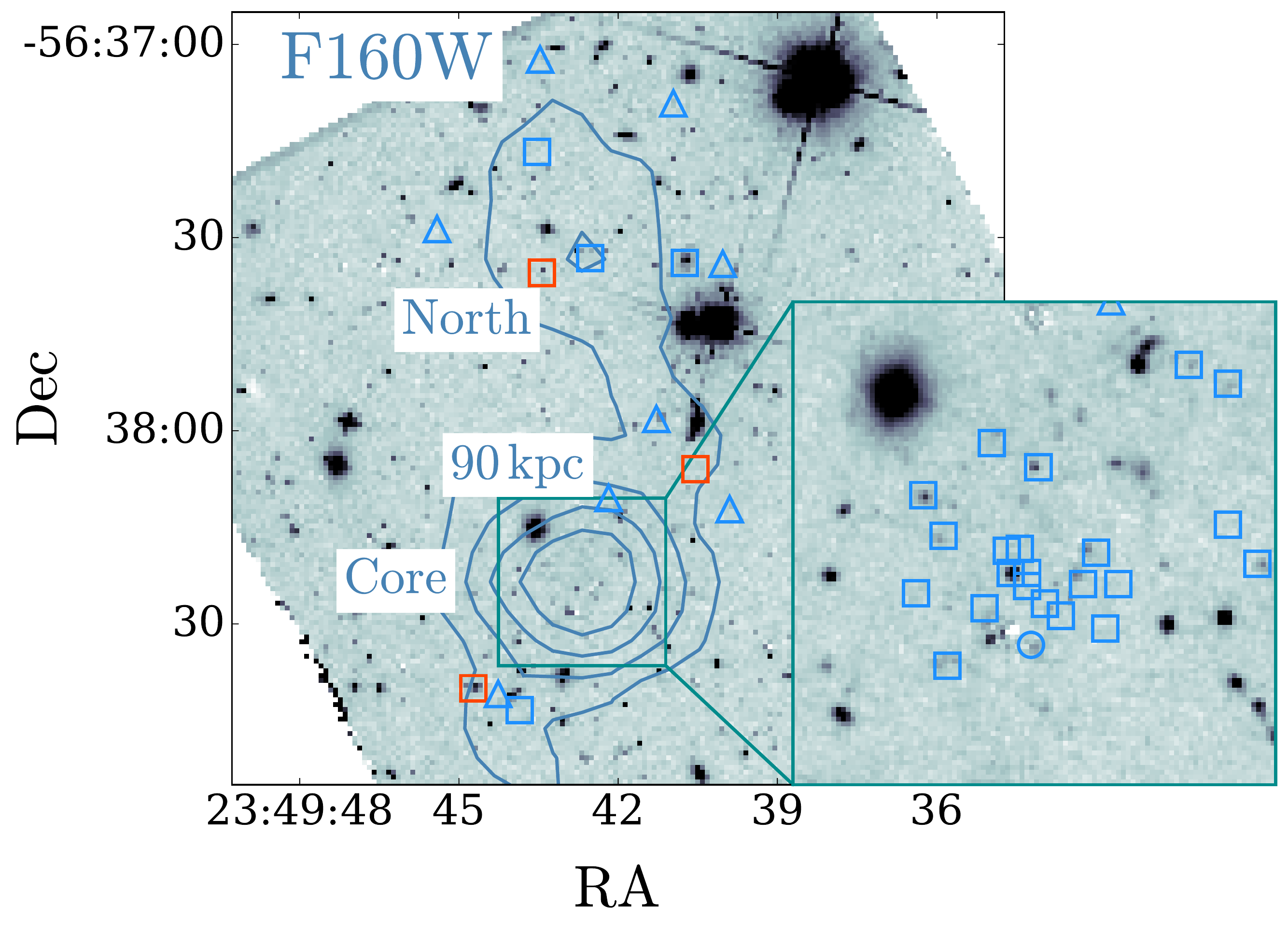}
\includegraphics[width=0.45\textwidth]{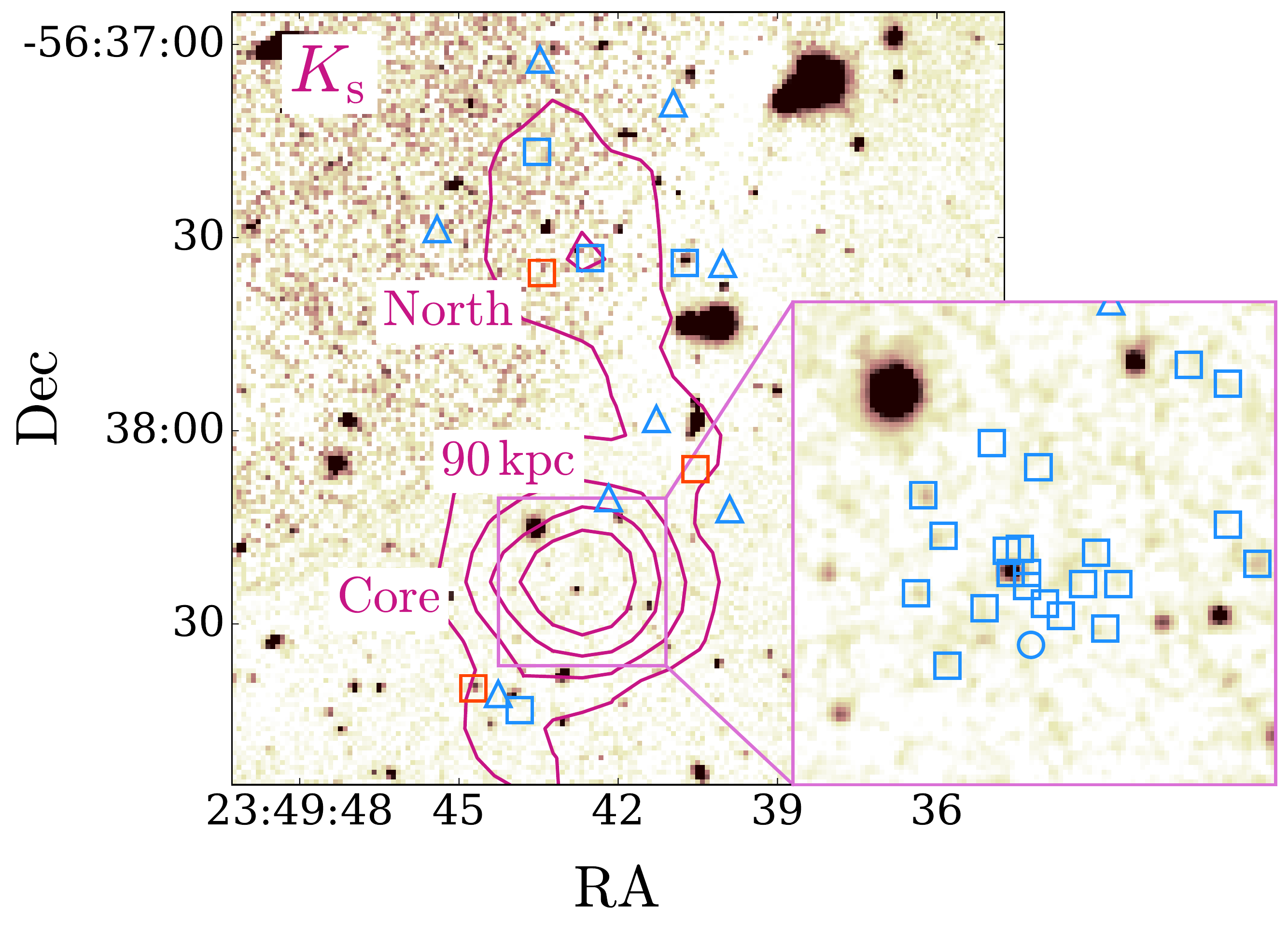}
\includegraphics[width=0.45\textwidth]{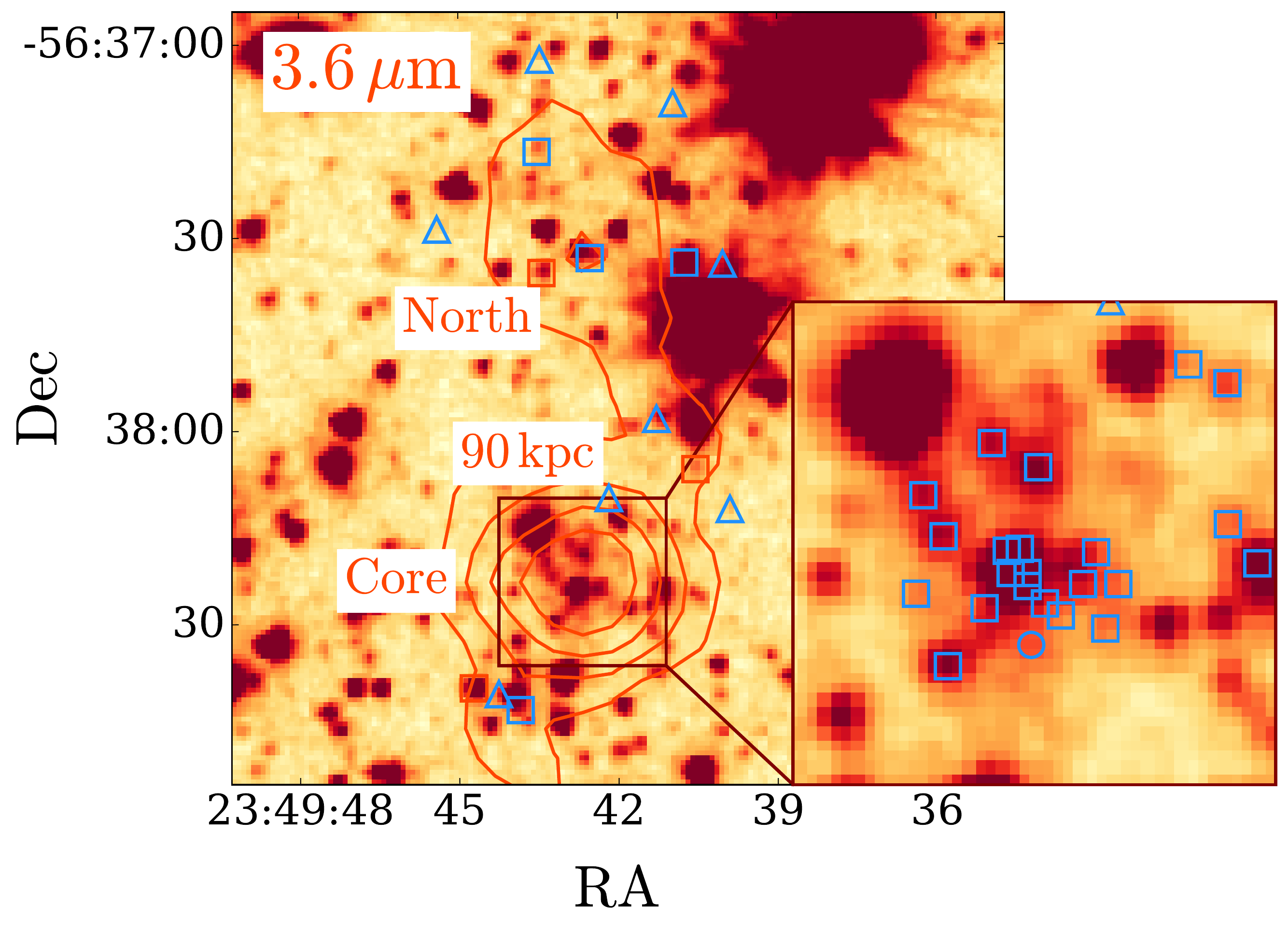}
\includegraphics[width=0.45\textwidth]{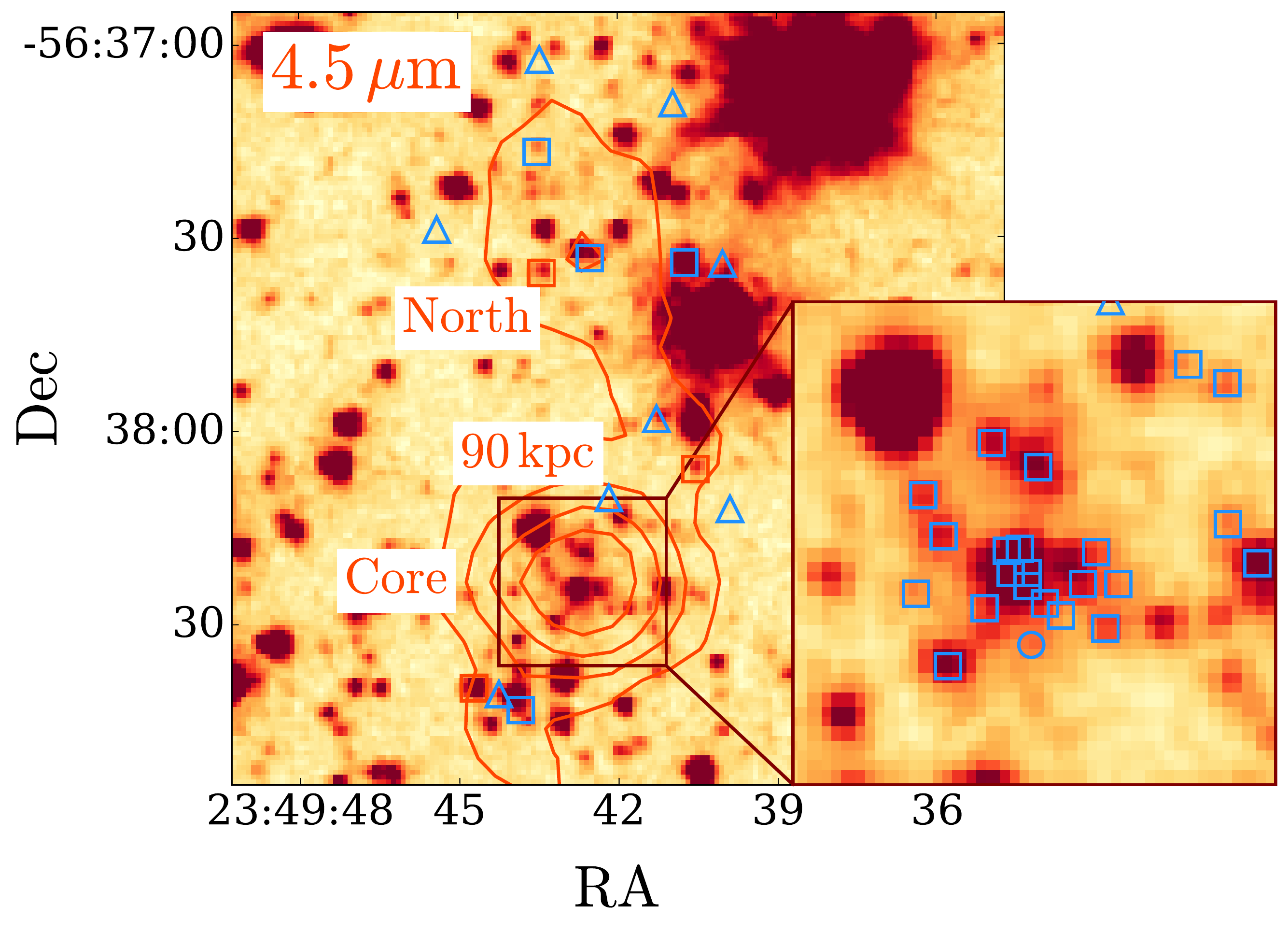}
\caption{Optical and infrared images of SPT2349$-$56. ALMA sources are shown as blue squares, the LBG source as a blue circle, LAE sources as blue triangles, and submm-detected sources with no line emission (i.e. likely foreground/background sources) as red squares. The images are smoothed by a 3-pixel FWHM Gaussian, except for the IRAC images, which have not been smoothed. Single-dish submm imaging of SPT2349$-$56 at 870\,$\mu$m using the LABOCA instrument are shown as the background contours, starting at 4$\sigma$ and increasing in steps of 5$\sigma$, and define the core and northern regions of this protocluster field \citep[see][]{hill2020,wang2020}.}
\label{all_field}
\end{figure*}

For the FLAMINGOS-2 instrument, the field-of-view is circular with a diameter of 6.1\,arcmin. While this field-of-view is large enough to cover all of the core and northern sources described above, due to a lack of nearby guide stars within the field, we were only able to cover the central sources and a few northern sources down to a depth suitable for the detection of the target galaxies in our sample. These include the 23 galaxies from \citet{hill2020} found through their [C{\sc ii}] emission, NL1, NL3, N3, five LAEs, and the LBG. Fig.~\ref{all_field} shows 2\,armin\,${\times}$\,2\,arcmin cutouts for each of these four Gemini fields, smoothed by a 3-pixel FWHM Gaussian.

\section{Data analysis}
\label{analysis}

\subsection{Gemini and \textbf{\textit{HST}} flux density measurements}

We first matched our GEMINI GMOS, FLAMINGOS-2, and {\it HST\/} F110W astrometry to our {\it HST\/}-F160W astrometry using the {\tt python} package {\tt astroalign} \citep{beroiz2020}, which identifies bright stars in source and target images and estimates a transformation matrix that aligns the stars in the target image with the same stars in the source image. We note that we are unable to apply this step to our ALMA data because there are no bright sources easily identifiable in both our {\it HST\/}-F160W imaging and our ALMA imaging; however, as shown below, after matching sources across both images we do not find any significant systematic offsets.

We ran {\tt source-extractor} \citep{bertin1996} on our GEMINI-GMOS and FLAMINGOS-2 images to extract a catalogue of sources and measure their flux densities. The detection images were smoothed with circular Gaussian kernels having FWHMs of three pixels, equivalent to 0.225\,arcsec, as part of the source detection process. Sources were required to consist of at least 3 connected pixels lying 0.8$\sigma$ above the local background. -- in {\tt source-extractor} this means setting {\tt DETECT\_MINAREA} to 3 and {\tt DETECT\_THRESH} to 0.8. These parameters are similar to the ones used by \citet{rotermund2021}, who set {\tt DETECT\_MINAREA} to 3 and varied {\tt DETECT\_THRESH} between 1.1 and 2.5, depending on the band.

The photometry was measured in the unsmoothed images at the locations of the sources identified in the smoothed detection images using the {\tt FLUX\_BEST} option in {\tt source-extractor}, which selects either the flux measured in an adaptively-scaled elliptical aperture ({\tt FLUX\_AUTO}) or sums the pixels found above the chosen threshold, with a correction for missing flux in the wings of the objects ({\tt FLUX\_ISOCOR}). When crowding becomes an issue, the {\tt FLUX\_AUTO} apertures start double counting the flux densities in pixels. For a given source, if 10\,per cent of the {\tt FLUX\_AUTO} measurement comes from other sources, {\tt FLUX\_ISOCOR} is chosen, otherwise {\tt FLUX\_AUTO} is chosen.

{\it HST\/} WFC3 source catalogues were produced using {\tt source-extractor} as well, using the same input parameters described above, except we required nine adjacent pixels to be above 1.2 times the background rms. This setup is similar to the {\tt source-extractor} parameters used in the Cosmic Assembly Near-infrared Deep Extragalactic Legacy Survey \citep[CANDELS;][]{galametz2013} in the `hot' mode, which is optimized to detect small and faint galaxies.

\subsection{IRAC deblending and flux density measurements}

Owing to {\it Spitzer\/}'s poor angular resolution compared to Gemini and {\it HST\/}, source blending can become a serious issue, especially in crowded regions like the centre of SPT2349$-$56. To tackle this issue, we use the publicly-available code {\tt t-phot} \citep{merlin2015,merlin2016}, which uses a high-resolution source catalogue as a prior to deblend a low-resolution, blended image.

To construct an optimal high-resolution catalogue, we stacked (as a weighted mean) both of our F110W and F160W {\it HST\/} images. We then ran {\tt source-extractor} with a higher detection threshold ({\tt DETECT\_MINAREA}\,${=}$\,9 and {\tt DETECT\_THRESH}\,${=}$\,2.0), since having too many faint galaxies used as priors with {\tt t-phot} leads to unrealistic deblending. We also turned off source deblending by setting {\tt DEBLEND\_MINCONT} to 1. This means that within a collection of pixels found above the predefined threshold, local maxima will not be classified as individual sources. This step is necessary because {\tt t-phot} becomes unreliable when given blended priors. We output a segmentation map from {\tt source-extractor}, which is used in conjunction with the catalogue in {\tt t-phot}.

The code {\tt t-phot} requires that the pixel scale of the input prior catalogue is an integer multiple of the low-resolution image, and is also aligned to the same pixel grid. To satisfy these criteria, we used the same {\tt astroalign} code to align our IRAC data to our high-resolution {\it HST\/} data, and then we used the {\tt python} function {\tt reproject\_interp} from the module {\tt reproject} to reproject our IRAC images onto our combined {\it HST\/} image, thus making the IRAC pixels a factor of 8 smaller. 

{\tt t-phot} also requires a convolution kernel that can be used with the high-resolution segmentation image to produce the low-resolution template. Assuming that the galaxies in our {\it HST\/} image are resolved, the appropriate kernel would simply be the IRAC PSF. However, owing to the variation in sensitivity across each IRAC pixel, a more complex point response function (PRF) is more appropriate. A set of 5\,${\times}$\,5 PRFs are available from the NASA/IPAC Infrared Science Archive,\footnote{\url{https://irsa.ipac.caltech.edu/data/SPITZER/docs/irac/calibrationfiles/psfprf/}} where each PRF is the response from a point source illuminating a position on a pixel divided up into a 5\,${\times}$\,5 grid. For the convolution kernel, we selected the PRF corresponding to a point source illuminating the centre of a pixel. The provided PRFs have a pixel scale of 0.24\,arcsec\,pixel$^{-1}$, and we used the same {\tt reproject} function to reproject these pixels to the high-resolution pixel scale of 0.075\,arcsec\,pixel$^{-1}$.

Some sources of interest are not detected in the combined {\it HST\/} image, but are clearly seen in both channels of the IRAC data. For these sources, {\tt t-phot} enables one to provide a second catalogue of unresolved priors as a list of positions that will be treated as delta functions before being convolved with the kernel. Five galaxies from \citet{hill2020} fit this category, and were provided to {\tt t-phot} as unresolved priors, namely C4, C5, C9, C10, and NL1.

With these inputs in hand, we then ran {\tt t-phot} in two passes. The first pass convolves the high-resolution segmentation image with the kernel, and fits an amplitude to each source to best match the low-resolution image. The second pass cross-correlates the model image with the low-resolution image in order to compute small positional shifts for each source, and then fits for the amplitudes a second time.

Since our high-resolution {\it HST\/} imaging did not cover the SPIREc sources, their flux densities were measured independently. We performed aperture photometry at the positions of the three SPIREc sources, using an aperture with a radius of 4 pixels (2.4\,arcsec), and applied aperture corrections of 1.208 and 1.220 for the 3.6\,$\mu$m band and the 4.5\,$\mu$m band, respectively (from the IRAC Instrument Handbook\footnote{\url{https://irsa.ipac.caltech.edu/data/SPITZER/docs/irac/iracinstrumenthandbook/1/}}). SPIREc1 and SPIREc3 are both clearly detected and we were able to measure their flux densities with high signal-to-noise ratio (SNR). SPIREc2 is not a clear detection by eye, and the flux density in the aperture at 3.6\,$\mu$m is consistent with noise, but at 4.5\,$\mu$m we were able to measure a ${>}\,2\sigma$ signal.

\subsection{Source matching}
\label{matching}

In order to perform a multiwavelength analysis of the SEDs of these galaxies, we need to match our sources detected in high-resolution (sub)millimetre imaging to their counterparts in our optical and infrared imaging. This becomes complicated due to the fact that the (sub)millimetre imaging is detecting emission from dust, while the ultraviolet to near-infrared imaging is detecting emission from starlight (including extinction by dust). A galaxy's morphology can be quite complicated in detail, with patches of dust and stars that will not necessarily overlap one another from our line of sight \citep[e.g.][]{goldader2002}. Additionally, although we have tied the all of the optical and infrared astrometry to the {\it HST} F160W frame, we are not able to perform this step to our submm imaging, although the ALMA astrometry relative to that of {\it HST} has been measured to be accurate within 0.1\,arcsec \citep[e.g.,][]{dunlop2017,elbaz2018}.

Theoretically, the 1$\sigma$ uncertainty of our ALMA-derived positions is given by
\begin{equation}
\Delta \mathrm{RA} = \Delta \mathrm{Dec} = \frac{1}{2\sqrt{\ln 2}} \frac{\mathrm{FWHM}}{\mathrm{SNR}},
\end{equation}
\noindent
where FWHM is the size of the synthesized beam, and SNR is the signal-to-noise ratio of the peak pixel \citep{ivison2007}. For our ALMA positions, the synthesized beam is about 0.5\,arcsec \citep{hill2020}, so we expect the 1$\sigma$ angular position uncertainty to be $\Delta \theta \,{\approx}\,$0.3\,arcsec/SNR. Since the probability density of finding a source at an angle $\theta$ from its true position is proportional to $\theta \mathrm{e}^{-\theta^{2}/2\Delta \theta^{2}}$, one must go out to a distance of 3.42$\Delta \theta$ in order to find a correct match with 99.7\,per cent (or 3$\sigma$) certainty. For our ALMA positions, the lowest (spatial) SNR used to measure a position was about 4, meaning a 3$\sigma$ search should go out to about 0.3\,arcsec. However, this becomes more complicated in practice due to the fact that the (sub)millimetre and optical images could have different morphologies, especially for the case of mergers. Previous studies matching ALMA galaxies to {\it HST\/} counterparts have used radial searches around 0.6--1\,arcsec \citep[e.g.,][]{dunlop2017,long2020}, and here we also adopt 1\,arcsec, corresponding to about 6.9 proper kiloparsecs. For comparison, the typical submm size of the galaxies in our sample is about 4 proper kiloparsecs \citep[twice the half-light radius, see][]{hill2020}.

A source is thus considered a match if it lies within 1\,arcsec from the ALMA-derived positions provided in \citet{hill2020}, and we allow the possibility of multiple matches; multiple optical counterparts to a submm source are possible if the submm source is a merging galaxy, for example. For cases where a counterpart is less than 1\,arcsec from two ALMA galaxies, we assign the match to the closest ALMA galaxy. Despite the comparatively large beamsize of the IRAC imaging, we use the same 1\,arcsec matching criteria, since the IRAC positions are nearly identical to the high-resolution {\it HST\/} imaging, which is simply a result of using {\tt t-phot} for the photometry. Also, for reference, 1\,arcsec corresponds to 6.3 pixels in the GMOS images, 13.4 pixels in our WFC3 images, 5.6 pixels in the FLAMINGOS-2 image, and 1.7 pixels in the IRAC images.

Next, we take advantage of our extensive wavelength coverage by imposing the additional constraint that a counterpart is not detected in the $g$ band -- this is simply because at $z\,{=}\,4.3$, the $g$ band probes a galaxy's rest-frame SED at 900\,\AA, where these galaxies should be much fainter than the sensitivity limit of our Gemini data because of neutral hydrogen's efficiency at absorbing light at this wavelength. In other words, if we find a counterpart within 1\,arcsec that is also bright in the $g$ band, we assume that it is a line-of-sight interloper, and remove the match from our sample. This criteria removes the matches to C1, C11, C20, C23, and N3. For the LAEs, this criteria removes LAE1. A galaxy 0.4\,arcsec east of LAE4 is also detected in the $g$ band, but we see that in the $r$ and $i$ bands an extension appears 0.4\,arcsec to the north of this source, which our {\it HST\/} data resolves as a second galaxy. Since the eastern galaxy is detected in the $g$ band but the northern one is not, we only call the northern galaxy a match to LAE4. 

The Gemini-detected source near C1 (bright in the $g$ band and removed from our sample) was confirmed to be a foreground $z\,{=}\,$2.54 galaxy by \citet{rotermund2021} using spectroscopy with the VLT, validating our submm-matching criteria. However, the observed [O{\sc iii}] line strength suggests a stellar mass much smaller than what the IRAC flux densities (2.6 and 2.9\,$\mu$Jy at 3.6 and 4.5\,$\mu$m, respectively) would imply. Quantitatively, \citet{rotermund2021} place an upper limit on the stellar mass of the foreground galaxy of ${<}\,1.6\,{\times}\,10^{9}\,$M$_{\odot}$ based on an [O{\sc iii}] linewidth of 53\,km\,s$^{-1}$. \citet{zhu2010} propose a scaling relation between stellar mass and IRAC continuum flux densities using $g\,{-}\,r$ colours as a proxy for star-formation history, which we use to place an upper limit to the contribution of the measured IRAC flux density from this foreground galaxy. We measure a colour of $g\,{-}\,r\,{=}\,0.10$, corresponding to upper limits of $\nu L_{\nu}\,{<}\,7.2\,{\times}\,10^{8}\,$L$_{\odot}$ and $\nu L_{\nu}\,{<}\,5.0\,{\times}\,10^{8}\,$L$_{\odot}$ at 3.6 and 4.5\,$\mu$m, respectively, or $S_{3.6}\,{<}\,0.06\,\mu$Jy and $S_{4.5}\,{<}\,0.05\,\mu$Jy. Thus, since this foreground galaxy likely contributes less than 2\,per cent to the measured IRAC flux density at the position of C1, we assign all of the measured IRAC flux density to C1.

Lastly, we look by eye for consistent matches across all the data and make sure sources are not double-matched. C3, C12, C13, C16, C22, and C23 are very close to C6, which is clearly the dominant galaxy in the core of the protocluster at the wavelengths covered by {\it HST\/}. In the IRAC imaging, C6 and these six sources are blended within a single beam, yet we expect that nearly all of the measured flux density can be attributed to only C6, as it is ${>}\,$25 times brighter than its surrounding galaxies in the {\it HST\/} imaging. We thus assign all of the flux density within a larger 1.5\,arcsec region to C6, while providing upper limits for the nearby galaxies.

In Appendix \ref{appendix0}, Fig.~\ref{offset}, we show the resulting positional differences between our matching criteria outlined above and the ALMA positions given in \citet{hill2020}, defined as the ALMA position minus the optical/infrared position. We find counterparts in at least one image to 21/29 ALMA-identified galaxies, as well as the single LBG, and 4/6 LAEs. Four galaxies are found to have two counterparts in our F110W image, three have two counterparts in our F160W image, and one has three counterparts in our F110W image. In Fig.~\ref{offset} we show the mean offset found in each band surrounded by a circle whose semi-major and semi-minor axes are equal to the standard deviations of the offsets in each direction. We find a slight offset in the negative $x$ direction of $\Delta $RA$\,{\approx}\,-0.1$\,arcsec (consistent across all wavebands, since the astrometry has been tied to a single frame); this could mean that there is a slight mismatch between the {\it HST\/} astrometry and the ALMA astrometry, but an adjustment of this size would have no effect on our results.

We investigate the purity of our source matches by running our matching algorithm on random locations within each map, allowing us to calculate a false positive rate by taking the ratio of the matches to the number of random locations tested. Using 1000 random locations, we find false positive rates of 0.08, 0.06, 0.23, 0.20, and 0.12 for the $r$, $i$, F110W, F160W, and $K_{\rm s}$ bands, respectively. For reference, our detection rates are 0.32, 0.34, 0.61, 0.47, and 0.25 for the same bands, which are factors of 2--5 higher than the false positive rates. These values are upper limits to the true false positive rates, since we are requiring an optical/infrared-detected galaxy to be less than 1\,arcsec from a submm source (i.e. an ALMA-detected galaxy), which is much less common than a match between two optical/infrared-detected galaxies.

For cases where we found no counterpart match, we estimate upper limits by calculating the mean flux density measurement uncertainty within a given band. We also add the local background to these non-detection upper limits, calculated within a 3 pixel-diametre circular region centred on the source. This takes into account the fact that C1 is blocked by a foreground galaxy in the GMOS and {\it HST\/} imaging \citep[see][]{rotermund2021}, so the only upper limit we can place is that it must be fainter than the interloping foreground galaxy. This is also necessary in the IRAC imaging where confusion and source blending is a reason for many non-detections around galaxy C6. Lastly, we make a 2$\sigma$ cut to the photometry measurements of the matched sources. In Table \ref{table:cont} we provide the resulting flux densities and upper limits for each source, and in Appendix \ref{appendix1} we show 12$\,{\times}\,$12\,arcsec cutouts of each source at each wavelength. We note that there are known bad pixels in our {\it HST\/} data near C21 (in both the F110W and F160W filters as it is a property of the WFC3 detector), but they are outside of the aperture used to measure the flux density of this source.

We compare our resulting optical measurements to the flux densities reported in \citet{rotermund2021}, who used slightly different apertures (a fixed diametre of 1.6\,arcsec for the $g$, $r$, and $i$ images, and {\tt FLUX\_AUTO} for the $K_{\rm s}$ image) and {\tt source-extractor} parameters. We find that we do not detect C2 in the $K_{\rm s}$ band, nor C6 in the $i$ band, but that we detect sources C2 and C14 in the $r$ image, and C2 and C14 in the $i$ image. In terms of the overlapping detections, we find good agreement between the flux density measurements. In the IRAC infrared imaging, we identify counterparts to all the galaxies in \citet{rotermund2021}, except C3 and C13 that are blended with C6, which is expected because the data used here are significantly deeper. The IRAC measurements reported in \citet{rotermund2021} are systematically larger than those reported here, by average factors of 1.7 and 2.1 at 3.6 and 4.5\,$\mu$m, respectively. This is also expected, since \citet{rotermund2021} did not attempt to deblend the IRAC sources while here we did, meaning that the flux inside their fixed apertures should be larger on average due to surrounding source leakage.

\setlength\tabcolsep{2pt}
\setlength\extrarowheight{2pt}
\begin{table*}
\centering
\caption{Gemini-GMOS and FLAMINGOS-2, {\it HST\/}-WFC3, and {\it Spitzer\/}-IRAC flux density measurements for all SPT2349$-$56 galaxies. The names are the same as in \citet{hill2020}, while the names from \citet{miller2018} are given in brackets for reference. Upper limits are 1$\sigma$. Here ellipses indicate where data are not available for a given source.}
\label{table:cont}
\begin{threeparttable}
\begin{tabular}{lcccccccc}
\hline
Name & $S_{0.48}^{\mathrm{a}}$ & $S_{0.63}^{\mathrm{a}}$ & $S_{0.78}^{\mathrm{a}}$ & $S_{1.1}^{\mathrm{b}}$ & $S_{1.5}^{\mathrm{b}}$ & $S_{2.1}^{\mathrm{c}}$ & $S_{3.6}^{\mathrm{d}}$ & $S_{4.5}^{\mathrm{d}}$ \\
& [$\mu$Jy] & [$\mu$Jy] & [$\mu$Jy] & [$\mu$Jy] & [$\mu$Jy] & [$\mu$Jy] & [$\mu$Jy] & [$\mu$Jy] \\
\hline
C1 (A) & $<$0.032 & $<$0.043 & $<$0.045 & $<$0.057 & $<$0.103 & $<$0.19 & 2.6$\pm$0.3 & 2.9$\pm$0.3 \\
C2 (J) & $<$0.010 & 0.071$\pm$0.013 & 0.183$\pm$0.020 & 0.295$\pm$0.009 & 0.381$\pm$0.020 & $<$0.25 & 2.7$\pm$0.4 & 3.2$\pm$0.3 \\
C3 (B) & $<$0.006 & $<$0.006 & $<$0.013 & $<$0.034 & $<$0.055 & $<$0.26 & $<$5.4 & $<$5.6 \\
C4 (D) & $<$0.002 & $<$0.014 & $<$0.019 & $<$0.028 & $<$0.052 & $<$0.17 & 1.1$\pm$0.3 & 1.1$\pm$0.3 \\
C5 (F) & $<$0.003 & $<$0.003 & $<$0.019 & $<$0.031 & $<$0.060 & $<$0.22 & 0.9$\pm$0.3 & 1.0$\pm$0.3 \\
C6 (C) & $<$0.006 & 0.061$\pm$0.011 & $<$0.014 & 0.362$\pm$0.006 & 1.601$\pm$0.012 & 4.78$\pm$0.28 & 9.8$\pm$0.4 & 9.8$\pm$0.3 \\
C7 (K) & $<$0.006 & $<$0.009 & $<$0.012 & 0.082$\pm$0.010 & 0.196$\pm$0.019 & $<$0.10 & 2.1$\pm$0.3 & 2.2$\pm$0.3 \\
C8 (E) & $<$0.006 & 0.024$\pm$0.008 & 0.146$\pm$0.020 & 0.261$\pm$0.014 & 0.242$\pm$0.015 & 1.37$\pm$0.34 & 3.9$\pm$0.4 & 4.2$\pm$0.3 \\
C9 (I) & $<$0.006 & $<$0.006 & $<$0.014 & $<$0.034 & $<$0.054 & $<$0.44 & 0.9$\pm$0.3 & 1.3$\pm$0.3 \\
C10 (H) & $<$0.005 & $<$0.006 & $<$0.022 & 0.037$\pm$0.004 & $<$0.060 & 0.51$\pm$0.18 & 0.7$\pm$0.3 & 0.7$\pm$0.3 \\
C11 (L) & $<$0.013 & $<$0.001 & $<$0.017 & $<$0.030 & $<$0.054 & $<$0.31 & $<$1.2 & $<$1.3 \\
C12 & $<$0.007 & $<$0.019 & $<$0.022 & $<$0.034 & $<$0.064 & $<$0.22 & $<$3.8 & $<$4.0 \\
C13 (G) & $<$0.007 & $<$0.013 & $<$0.024 & 0.028$\pm$0.004 & $<$0.055 & $<$0.57 & $<$4.5 & $<$4.3 \\
C14 (N) & $<$0.012 & 0.037$\pm$0.008 & 0.051$\pm$0.011 & 0.164$\pm$0.013 & 0.167$\pm$0.015 & 0.45$\pm$0.13 & 2.1$\pm$0.3 & 1.8$\pm$0.3 \\
C15 & $<$0.011 & $<$0.015 & 0.076$\pm$0.021 & 0.147$\pm$0.009 & 0.213$\pm$0.018 & $<$0.27 & 1.3$\pm$0.3 & 1.1$\pm$0.3 \\
C16 & $<$0.013 & $<$0.010 & $<$0.021 & $<$0.034 & $<$0.063 & $<$0.21 & $<$3.0 & $<$2.7 \\
C17 (M) & $<$0.009 & 0.120$\pm$0.012 & 0.317$\pm$0.025 & 0.371$\pm$0.009 & 0.551$\pm$0.020 & 0.99$\pm$0.19 & 1.6$\pm$0.5 & 1.2$\pm$0.4 \\
C18 & $<$0.012 & $<$0.017 & $<$0.011 & 0.240$\pm$0.010 & 0.223$\pm$0.017 & $<$0.17 & $<$0.7 & $<$0.8 \\
C19 & $<$0.006 & $<$0.007 & $<$0.022 & $<$0.032 & $<$0.062 & $<$0.16 & $<$0.9 & $<$1.0 \\
C20 & $<$0.016 & $<$0.010 & $<$0.021 & $<$0.026 & $<$0.061 & $<$0.23 & $<$1.7 & $<$1.8 \\
C21 & $<$0.013 & $<$0.027 & 0.351$\pm$0.016 & 0.373$\pm$0.009 & 0.343$\pm$0.014 & $<$0.32 & 1.3$\pm$0.4 & 1.1$\pm$0.4 \\
C22 & $<$0.011 & $<$0.006 & $<$0.012 & $<$0.030 & $<$0.055 & $<$0.37 & $<$1.0 & $<$1.0 \\
C23 & $<$0.013 & $<$0.016 & $<$0.024 & $<$0.034 & $<$0.051 & $<$0.19 & $<$1.4 & $<$1.3 \\
NL1 & $<$0.005 & $<$0.012 & $<$0.021 & $<$0.026 & $<$0.058 & $<$0.30 & 0.9$\pm$0.3 & 1.6$\pm$0.3 \\
NL3 & 0.230$\pm$0.012 & 0.415$\pm$0.019 & 0.542$\pm$0.033 & 1.628$\pm$0.018 & 3.308$\pm$0.029 & 5.15$\pm$0.36 & 8.8$\pm$0.4 & 10.4$\pm$0.3 \\
N1 & $<$0.009 & $<$0.006 & $<$0.015 & 0.115$\pm$0.007 & 0.199$\pm$0.017 & \dots & 2.2$\pm$0.3 & 3.0$\pm$0.3 \\
N2 & $<$0.009 & $<$0.003 & 0.112$\pm$0.019 & 0.094$\pm$0.007 & 0.093$\pm$0.014 & \dots & 1.0$\pm$0.3 & 0.7$\pm$0.3 \\
N3 & 1.176$\pm$0.018 & 1.561$\pm$0.027 & 1.924$\pm$0.039 & 5.644$\pm$0.027 & 8.289$\pm$0.044 & 10.76$\pm$0.46 & 18.8$\pm$0.4 & 22.3$\pm$0.4 \\
NL2 & 0.061$\pm$0.006 & 0.072$\pm$0.009 & 0.180$\pm$0.016 & 0.337$\pm$0.012 & 0.442$\pm$0.018 & \dots & 1.9$\pm$0.4 & 1.7$\pm$0.3 \\
SPIREc1 & \dots & \dots & \dots & \dots & \dots & \dots & 13.0$\pm$0.3 & 19.4$\pm$0.3 \\
SPIREc2 & \dots & \dots & \dots & \dots & \dots & \dots & $<$0.8 & 0.7$\pm$0.3 \\
SPIREc3 & \dots & \dots & \dots & \dots & \dots & \dots & 4.9$\pm$0.3 & 6.8$\pm$0.3 \\
LBG3 & $<$0.011 & 0.077$\pm$0.011 & 0.437$\pm$0.024 & 0.251$\pm$0.014 & 0.329$\pm$0.020 & $<$0.36 & $<$1.1 & $<$1.1 \\
LAE1 & $<$0.031 & $<$0.040 & $<$0.056 & $<$0.051 & $<$0.088 & $<$0.37 & $<$0.8 & $<$0.7 \\
LAE2 & $<$0.007 & $<$0.005 & $<$0.026 & $<$0.033 & $<$0.047 & $<$0.31 & $<$1.0 & $<$0.9 \\
LAE3 & $<$0.005 & 0.054$\pm$0.011 & $<$0.016 & 0.050$\pm$0.005 & $<$0.061 & $<$0.12 & $<$0.3 & $<$0.3 \\
LAE4 & $<$0.021 & $<$0.028 & $<$0.043 & 0.044$\pm$0.006 & 0.066$\pm$0.008 & 0.53$\pm$0.17 & $<$0.5 & $<$0.4 \\
LAE5 & $<$0.010 & 0.107$\pm$0.010 & 0.057$\pm$0.011 & 0.078$\pm$0.006 & 0.089$\pm$0.009 & \dots & $<$0.4 & $<$0.3 \\
LAE6 & $<$0.013 & $<$0.006 & $<$0.010 & 0.030$\pm$0.004 & $<$0.057 & \dots & $<$0.5 & $<$0.5 \\
LAE7 & $<$0.008 & $<$0.006 & $<$0.013 & 0.052$\pm$0.007 & $<$0.052 & \dots & $<$0.5 & $<$0.4 \\
LAE8 & $<$0.003 & 0.082$\pm$0.012 & 0.139$\pm$0.018 & 0.204$\pm$0.010 & 0.052$\pm$0.006 & $<$0.17 & $<$1.6 & $<$1.2 \\
\hline
\end{tabular}
\begin{tablenotes}
\item $^{\rm a}$Gemini-GMOS continuum flux densities at 0.48\,$\mu$m (the $g$ band), 0.63\,$\mu$m (the $r$ band), and 0.78\,$\mu$m (the $i$ band).
\item $^{\rm b}${\it HST\/}-WFC3 continuum flux densities at 1.1\,$\mu$m (the F110W filter) and 1.5\,$\mu$m (the F160W filter).
\item $^{\rm c}$Gemini-FLAMINGOS-2 continuum flux densities at 2.1\,$\mu$m (the $K_{\rm s}$ band).
\item $^{\rm d}${\it Spitzer\/}-IRAC continuum flux densities at 3.6\,$\mu$m and 4.5\,$\mu$m.
\end{tablenotes}
\end{threeparttable}
\end{table*}

\subsection{SED fitting}

To estimate the physical properties of the galaxies belonging to the SPT2349$-$56 protocluster system, we used {\tt CIGALE} \citep{burgarella2005,noll2009,boquien2019} to fit SEDs to the available photometry. This includes all of the photometric data provided in Table~\ref{table:cont}, as well as the 850\,$\mu$m, 1.1\,mm, and 3.2\,mm photometry from \citet{hill2020}. We also input 1$\sigma$ upper limits to the photometry for non-detections, and include {\it Herschel\/}-SPIRE 250, 350, and 500\,$\mu$m constraints from \citet{miller2018}. 

{\tt CIGALE} models a galaxy's rest-frame optical and ultra-violet spectrum using simple stellar population models with variable star-formation histories, including nebular emission lines, and incorporates a flexible dust attenuation curve that allows the slope and strength of the ultraviolet bump to vary. The thermal dust emission is modeled assuming a power-law dust temperature distribution, and energy balance is imposed such that the energy absorbed by dust, primarily at rest-frame ultraviolet and optical wavelengths, is approximately equal to that re-radiated in the infrared, allowing for discrepancies due to non-isotropy from the ultraviolet and optical emission. In our fits we assumed a delayed star-formation history with a single exponential timescale. The SFR as a function of time is parameterized by
\begin{equation}
\mathrm{SFR} \propto \frac{t}{\tau_{\rm SFH}^2} e^{-t/\tau_{\rm SFH}},
\end{equation}
where $\tau_{\rm SFH}$ is the timescale, effectively the time at which the star-formation peaks, and a free parameter of the model. The total current stellar mass, $M_{\ast}$, in this model is found by varying the duration of the star-formation, assuming a Chabrier initial mass function \citep[IMF;][]{chabrier2003} with solar metallicity. The assumed star-formation history can have an effect on the resulting stellar mass (up to a factor of 2, see \citealt{michalowski2012}), although we did not find a large variation in the results after testing several of the available models. Another free parameter is the dust attenuation, given by the amount of extinction present in the $V$ band in magnitudes, $A_V$, which we model using the \citet{calzetti2000} attenuation curve with a variable power-law slope. The dust is modeled following \citet{draine2014}, where the dust is separated into a diffuse component heated by the interstellar radiation field, and a compact component linked to star-forming regions and heated by a variable radiation field. The total dust mass sets the overall normalization.

To obtain posterior distributions for the parameters in the fits, {\tt CIGALE} generates a grid of possible SEDs, and calculates $\chi^2$ for each SED. These $\chi^2$ values are then translated to a global likelihood function, assuming the likelihood is proportional to $e^{-\chi^2/2}$. Marginalized posterior distributions are then calculated for the free parameters, and {\tt CIGALE} returns the mean values and standard deviations of these distributions. The resulting stellar masses are given in Table~\ref{table:cigale}, and the best-fit SEDs are shown in Appendix \ref{appendix2}. Where parameter uncertainties overlap with 0, we provide 1$\sigma$ upper limits. For galaxies C12, C16, C20, C22, C23, LAE1, and LAE2, only upper limits are available for their photometry across all wavelengths (both submm and optical/infrared), so we do not attempt to fit SEDs and derived stellar masses.

The best-fit {\tt CIGALE} ttoal stellar masses range from about 10$^{10}$--10$^{11}\,$M$_{\odot}$. The \citet{rotermund2021} stellar masses are larger than those estimated here in proportion to their reported IRAC flux densities (they differ because we use {\tt t-phot} to improve the IRAC flux density estimates). As a final check, from the fits we calculated the SFR averaged over the past 100\,Myr, and compared these to the SFRs estimated from fitting modified blackbody functions to the far-infrared photometry in \citet{hill2020}; the two estimates are in good agreement considering the simple SFH adopted by our SED models, with a median ratio of far-infrared SFR-to-{\tt CIGALE} SFR of 1.1.

\subsubsection{Galaxy N3}

In Fig.~\ref{cutouts} we see that the centroid of N3 lies right on top of a bright and fully resolved spiral galaxy. N3 was initially identified by \citet{hill2020} using ALMA in Band~3 through the detection of a CO line at 86.502\,GHz, consistent with the frequencies of the CO lines detected in the other protocluster galaxies, and it was assumed that the CO transition was 4--3, placing the redshift at 4.3. However, the coincident position of the CO emission with a resolved spiral galaxy could mean that the CO emission observed is actually from another transition at lower redshift. While an alternative explanation is that there is a genuine protocluster galaxy responsible for the CO emission behind this nearby spiral galaxy, here we investigate the photometric redshift of the spiral galaxy to see if it could in fact emit a CO line at the frequency where a line was observed.

To do this, we used all of our available photometry (Table \ref{table:cont} plus the photometry in \citealt{hill2020}) to estimate a photometric redshift using {\tt CIGALE}. When {\tt CIGALE} is not provided a spectroscopic redshift, it includes a photometric redshift as an additional free parameter, generating an extra dimension of redshifted SEDs before calculating the $\chi^2$ of each model. Photometric redshift uncertainties are computed by converting the $\chi^2$ of each model to a likelihood, and computing the corresponding marginalized probability distribution for the redshift. The resulting photometric redshift is found to be 1.7$\,{\pm}\,$0.3; the only CO transition consistent with this redshift is the 2--1 transition at 230.538\,GHz, which would make the spectroscopic redshift of N3 1.665. Since this redshift matches the photometric redshift quite well, for the remainder of this paper we take this galaxy to be a foreground source at $z\,{=}\,1.665$ and remove it from the subsequent analyses.

\subsubsection{Submm sources with no line detection}

Three galaxies were found in our ALMA observations of SPT2349$-$56 through their continuum only (designated NL in \citealt{hill2020}), making them potential line-of-sight interlopers. In order to verify this hypothesis, we ran {\tt CIGALE} on the complete set of photometry available for these galaxies (Table \ref{table:cont}, and the photometry provided in \citealt{hill2020}) in order to estimate photometric redshifts. For NL1, we found a photometric redshift of 4.3$\,{\pm}\,$0.7; while this is consistent with the redshift of SPT2349$-$56, we do not include it in any subsequent analyses because the uncertainties are still large. Further follow-up will be needed to confirm the membership of this source. For the remaining two galaxies, NL2 has a photometric redshift of 2.7$\,{\pm}\,$0.7, and for NL3 the photometric redshift is 2.2$\,{\pm}\,$0.5, so these are indeed most likely galaxies in the foreground of the protocluster.

\setlength\tabcolsep{10pt}
\setlength\extrarowheight{2pt}
\begin{table*}
\centering
\caption{Best-fit properties of the galaxies in our sample. Here dashes indicate non-detections, while ellipses indicate where data are not available for a given source.}
\label{table:cigale}
\begin{threeparttable}
\begin{tabular}{lccccccc}
\hline
Name & $R_{1/2,\mathrm{UV}}^{\mathrm{a}}$ & $R_{1/2,\mathrm{CII}}^{\mathrm{b}}$ & $M_{\ast}^{\mathrm{c}}$ & $\mu_{\rm gas}^{\mathrm{d}}$ & $\tau_{\mathrm{dep}}^{\mathrm{e}}$ \\
& [kpc] & [kpc] & [10$^{10}$\,M$_{\odot}$] & & [Gyr] \\
\hline
C1 (A) & -- & 2.91$\pm$0.02 & 22.2$\pm$21.3 & 0.34$\pm$0.32 & 0.074$\pm$0.018 \\
C2 (J) & 1.26$\pm$0.76 & 2.57$\pm$0.03 & 4.5$\pm$1.9 & 0.46$\pm$0.21 & 0.100$\pm$0.026 \\
C3 (B) & -- & 1.33$\pm$0.04 & $<$112.6 & -- & 0.046$\pm$0.011 \\
C4 (D) & -- & 1.90$\pm$0.01 & $<$24.0 & -- & 0.053$\pm$0.013 \\
C5 (F) & -- & 2.22$\pm$0.02 & $<$21.8 & -- & 0.024$\pm$0.007 \\
C6 (C) & 0.47$\pm$0.16 & 1.30$\pm$0.04 & 10.9$\pm$3.4 & 0.31$\pm$0.10 & 0.058$\pm$0.014 \\
C7 (K) & -- & 2.22$\pm$0.03 & 3.0$\pm$1.9 & 0.33$\pm$0.22 & 0.119$\pm$0.033 \\
C8 (E) & 1.45$\pm$0.71 & 1.21$\pm$0.03 & 11.1$\pm$4.1 & 0.21$\pm$0.08 & 0.051$\pm$0.013 \\
C9 (I) & -- & 1.42$\pm$0.03 & $<$18.0 & -- & 0.049$\pm$0.014 \\
C10 (H) & -- & 1.14$\pm$0.03 & 1.5$\pm$1.1 & 0.77$\pm$0.57 & 0.060$\pm$0.016 \\
C11 (L) & -- & 1.40$\pm$0.04 & $<$6.7 & -- & 0.084$\pm$0.043 \\
C12 & -- & -- & -- & -- & -- \\
C13 (G) & -- & 1.00$\pm$0.09 & 7.5$\pm$6.8 & 0.11$\pm$0.10 & 0.043$\pm$0.011 \\
C14 (N) & -- & 0.39$\pm$0.10 & 3.4$\pm$1.3 & 0.06$\pm$0.04 & 0.046$\pm$0.030 \\
C15 & -- & -- & 1.2$\pm$0.6 & -- & -- \\
C16 & -- & -- & -- & -- & -- \\
C17 (M) & 1.76$\pm$0.83 & 1.59$\pm$0.06 & 1.2$\pm$0.5 & -- & -- \\
C18 & -- & -- & 0.6$\pm$0.4 & -- & -- \\
C19 & -- & -- & $<$4.8 & -- & -- \\
C20 & -- & -- & -- & -- & -- \\
C21 & -- & -- & 1.0$\pm$0.5 & -- & -- \\
C22 & -- & -- & -- & -- & -- \\
C23 & -- & -- & -- & -- & -- \\
NL1 & -- & -- & -- & -- & -- \\
NL3 & -- & -- & -- & -- & -- \\
N1 & 1.04$\pm$0.87 & \dots & 4.7$\pm$1.7 & 2.56$\pm$0.96 & 0.075$\pm$0.019 \\
N2 & -- & \dots & 1.0$\pm$0.6 & 4.84$\pm$2.68 & 0.066$\pm$0.022 \\
N3 & -- & \dots & -- & -- & -- \\
NL2 & -- & \dots & -- & -- & -- \\
SPIREc1 & \dots & \dots & 39.3$\pm$34.0 & 0.17$\pm$0.15 & -- \\
SPIREc2 & \dots & \dots & $<$4.5 & -- & -- \\
SPIREc3 & \dots & \dots & 12.3$\pm$11.0 & 0.20$\pm$0.18 & -- \\
LBG3 & 1.09$\pm$0.62 & \dots & 0.6$\pm$0.4 & -- & -- \\
LAE1 & -- & \dots & -- & -- & -- \\
LAE2 & -- & \dots & -- & -- & -- \\
LAE3 & -- & \dots & $<$0.1 & -- & -- \\
LAE4 & -- & \dots & 0.5$\pm$0.4 & -- & -- \\
LAE5 & -- & \dots & 0.1$\pm$0.1 & -- & -- \\
LAE6 & -- & \dots & $<$0.8 & -- & -- \\
LAE7 & -- & \dots & 0.4$\pm$0.3 & -- & -- \\
LAE8 & -- & \dots & $<$0.8 & -- & -- \\
\hline
\end{tabular}
\begin{tablenotes}
\item $^{\rm a}$Best-fit rest-frame ultraviolet half-light radius, obtained by fitting S{\'e}rsic profiles to all sources detected in our {\it HST\/} F160W imaging with sufficient SNR (see Section \ref{optical_radius}).
\item $^{\rm b}$Best-fit [C{\sc ii}] profile half-light radius, obtained by fitting S{\'e}rsic profiles to [C{\sc ii}]-detected sources after staking all channels containing line emission (see \citealt{hill2020} and Section \ref{cii_radius}).
\item $^{\rm c}$Best-fit stellar mass from fitting SEDs using the photometry in Table \ref{table:cont} and from \citet{hill2020}, obtained using {\tt CIGALE}.
\item $^{\rm d}$Molecular gas-to-stellar mass fraction, $\mu_{\rm gas}\,{=}\,M_{\rm gas}\,{/}\,M_{\ast}$, with $M_{\rm gas}$ values taken from \citet{hill2020}.
\item $^{\rm e}$Depletion timescale, $\tau_{\rm dep}\,{=}\,M_{\rm gas}\,{/}\,$SFR, with $M_{\rm gas}$ and SFR values taken from \citet{hill2020}.
\end{tablenotes}
\end{threeparttable}
\end{table*}

\subsection{Rest-frame ultraviolet profile fitting}
\label{optical_radius}

We next investigate the morphological profiles of some of the brighter galaxies detected in our F160W image (observed-frame 1.54\,$\mu$m, or 290\,nm in the rest-frame). We are interested in the characteristic sizes of the unobscured stellar emission, compared to their sizes as seen in the submm, where the emission is due to dust and star-formation, thus we choose the longest wavelength covered by our {\it HST\/} data. However, 290\,nm is still in the ultraviolet, where most of the flux density is due to young O and B stars. 

We created 2\,arcsec$\,{\times}\,$2\,arcsec cutouts around each protocluster member galaxy in our sample (except for NL3 and LAE1 where we made 3\,arcsec$\,{\times}\,$3\,arcsec cutouts), and fit elliptical 2D S{\'e}rsic profiles (allowing the S{\'e}rsic index to vary) convolved with the {\it HST\/} beam to the sources with optical counterparts containing pixels greater than 5 times the background rms; the relevant sources are C2, C6, C8, C17, N1, and LBG3. While C21 reaches $>$5 times the background rms as well, we remove it from this analysis due to the nearby bad pixels. We use the {\it HST\/} F160W PSF available from the Space Telescope Science Institute (STScI) instrumentation website,\footnote{\url{https://www.stsci.edu/HST/instrumentation/wfc3/data-analysis/psf}} taking the PSF corresponding to the response from a point source illuminating the centre of a pixel. The available PSF model supersamples the pixel plate scale of 0.13\,arcsec by a factor of 4, and we regrid the PSF model to match the pixel scale of our F160W imaging. This PSF is convolved with each galaxy's profile to produce our models; this is the same modelling technique used in popular profile-fitting packages such as {\tt galfit} \citep{peng2002,peng2010}. The resulting beam-deconvolved half-light radii (the length of the semi-major axes of an ellipse containing half the total flux density) are provided in Table~\ref{table:cigale}, and our models are shown in Appendix \ref{appendix3}. We find half-light radii ranging from 1.1\,kpc to 1.8\,kpc, except for one outlying source, C6, where we find a half-light radius of 0.5\,kpc. For reference, the corresponding half-light radius of the HST F160W beam is about 0.5\,kpc, implying that only galaxy C6 is not resolved. The S{\'e}rsic indices range from 0.3 to 2.3.

Since most of our {\it HST\/} detections are faint and cannot be fit on an individual basis, we performed a stacking analysis in order to evaluate the average rest-frame ultraviolet profile. As we ultimately want to compare this to the average submm profile, we focused on the subset of our sample with corresponding high-resolution ALMA data at 850\,$\mu$m (as detailed in the section below). These include all of the galaxies in the core region of SPT2349$-$56 (C1--C23). We removed C1 from the analysis as the {\it HST\/} detection at its position is a foreground galaxy, and we removed C21 due to the nearby bad pixels. We also excluded C6 since it is clearly an outlier in terms of brightness at this wavelength, and this galaxy will be subject to its own separate analysis. 

The position at which to centre each cutout for the stack is crucial. In order to obtain an unbiased image of the rest-frame ultraviolet light of submm sources, one should centre the rest-frame ultraviolet images on the positions of the submm sources. However, in practice there are physical offsets between rest-frame ultraviolet light and submm light, so this may not be the best choice. Here we choose to centre the cutouts at the position of the peak {\it HST\/} counterpart for cases where one is detected, and otherwise at the position of the peak pixel in each galaxy's average [C{\sc ii}] map, which provides the highest positional accuracy owing to the brightness of the line \citep[see][]{hill2020}. We also masked pixels above 25$\sigma$ (after verifying that this did not mask any sample sources) to remove bright nearby objects not associated with the galaxies in the stack.

The resulting image is shown in Fig.~\ref{stack}. Following the same steps as above, we fit a S{\'e}rsic profile to the stack, and our best-fit model and residual are shown alongside the data. We find a half-light radius of 1.24$\,{\pm}\,$0.29\,kpc and a S{\'e}rsic index of 0.75\,${\pm}\,$0.60, consistent with the sizes found for the individual galaxies and an exponential profile of $n\,{=}\,1$. Similarly, we find an axis ratio of 1.0\,${\pm}\,$0.3, as expected for stacking random orientation angles.

Systematic uncertainties in structural parameters are known to be important for low SNR sources; for example, \citet[][]{vanderwel2012} found that basic size parameterizations can be determined for galaxies detected in the F160W filter down to about 24.5\,mag. For reference, the galaxies in our sample with a peak pixel SNR ${>}\,5$ used to obtain size measurements span a range of 23.4--25.7\,mag, thus we wish to investigate possible systematic uncertainties and biases. To do this, we simulate F160W maps at the SNRs of the sources where we have fit S{\'e}rsic profiles. Our simulation covers a grid of S{\'e}rsic indices from 0.3 to 2.5 and half-light radii from 0.4 to 2\,kpc, and for each S{\'e}rsic index and half-light radius we generate three independent maps. S{\'e}rsic profiles are convolved with the F160W beam, and Gaussian random noise is added to the background such that the peak pixel has the SNR of the given source. For each input half-light radius we calculate the mean and standard deviation of the recovered half-light radii for all values of input S{\'e}rsic indices (effectively marginalizing over this parameter). For the wide range of  S{\'e}rsic indices tested, our algorithm recovers an unbiased estimate of the true half-light radius at all SNRs tested, with scatters of around 0.6\,kpc at SNR\,${=}\,$5 and 0.4\,kpc at SNR\,${=}\,$10. In Table \ref{table:cigale} we include this systematic uncertainty in quadrature with the statistical errors. The uncertainties in our S{\'e}rsic indices are large (often overlapping with 0), and a similar test of the recovered S{\'e}rsic indices from our simulation indicates systematic uncertainties of order ${\pm}\,2$ at a SNR of 5. For the lowest SNR sources our best-fit S{\'e}rsic indices are therefore not likely meaningful, however at the SNRs of our stack and C6 (16 and 112, respectively), the systematic uncertainties are smaller than the statistical uncertainties.

\subsection{[C{\sc ii}] sizes}
\label{cii_radius}

Following the method outlined above and in \citet{hill2020}, we also fit S{\'e}rsic profiles to cutouts of each galaxy's extended [C{\sc ii}] emission. Line emission channels were determined from lower-resolution, deeper ALMA data by fitting Gaussian profiles to the spectra, from which we averaged the high-resolution channels from $-3\sigma$ to $3\sigma$ (where $\sigma$ is the standard deviation of the best-fitting linewidth), or for cases where two Gaussians was a better fit, from $-3\sigma_{\rm L}$ to $+3\sigma_{\rm R}$, where $\sigma_{\rm L}$ and $\sigma_{\rm R}$ are from the left and right Gaussian fits, respectively. 2\,arcsec$\,{\times}\,$2\,arcsec cutouts were made around each source, and models were fit to sources with pixels detected above 5 times the background rms by convolving a S{\'e}rsic profile with the data's synthesized beam. We allowed the position, position angle, ellipticity, half-light radius, and S{\'e}rsic index to vary in our fits. The results are provided alongside our ultraviolet profile fits in Table~\ref{table:cigale}, and the models are shown in Appendix \ref{appendix4}. We find half-light radii in the range of 1.0 to 2.9\,kpc, except for one outlying source, C14, which has a half-light radius of about 0.4\,kpc. For reference, the half-light radius of the ALMA synthesized beam is about 0.7\,kpc, implying that only C14 is not resolved. The S{\'e}rsic indices found range from 0.3 to 2.0.

\begin{figure*}
\includegraphics[width=\textwidth]{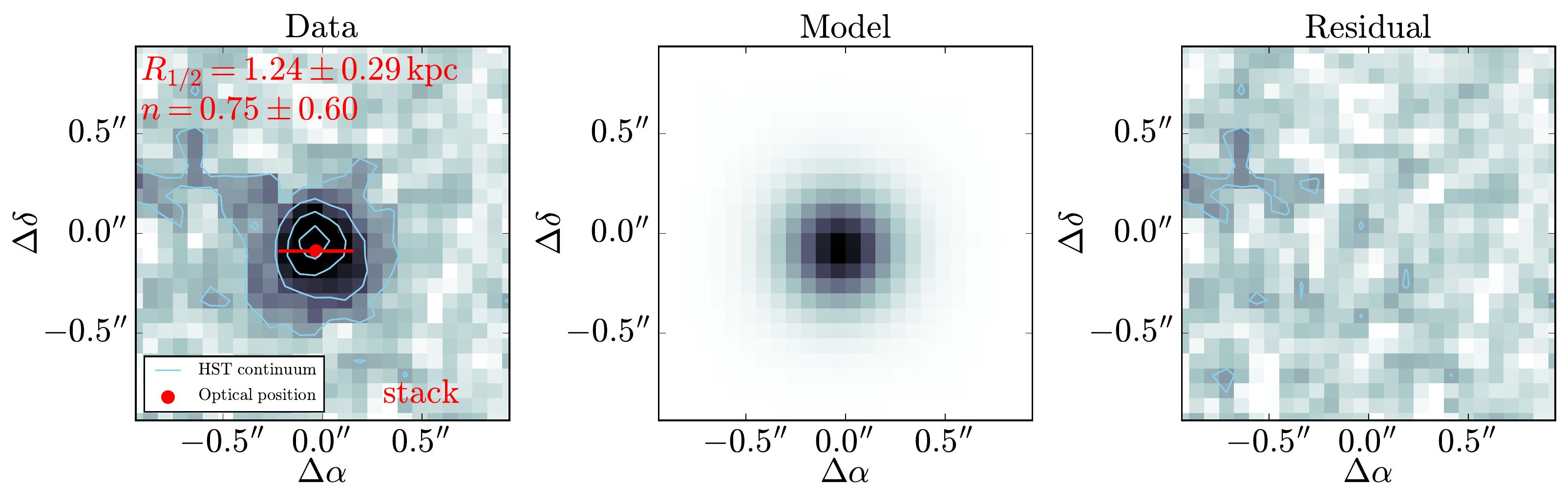}
\includegraphics[width=\textwidth]{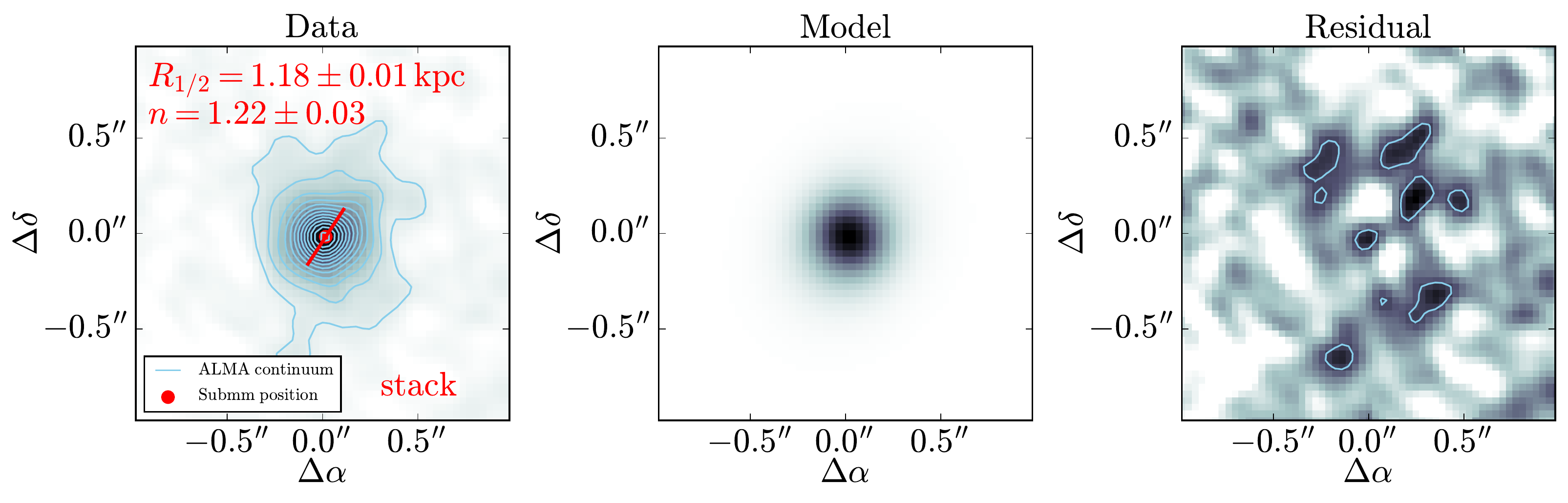}
\caption{{\it Top row:} Stacked F160W image of all the core galaxies in our sample (except C1, C6, and C21; see the text for details). Contours start at 2$\sigma$ and increase in steps of 4$\sigma$. The red dot indicates the position of the best-fit centre, and the red line shows the length and position angle of the best-fit half-light diametre. The middle panel shows the best-fit S{\'e}rsic profile, after convolution with the beam and the pixel window function. The residual map is shown at right, with the same contour levels as the left panel. {\it Bottom row:} Same as the top row, but stacking our high-resolution 850-$\mu$m ALMA continuum images. Contours again start at 2$\sigma$ and increase in steps of 4$\sigma$.}
\label{stack}
\end{figure*}

We next determined the average submm-continuum profile following the same stacking procedure done for our {\it HST\/} data. High resolution continuum maps were already presented in \citet{hill2020}, obtained by stacking the line-free channels. We selected the same galaxies as in the {\it HST\/} analysis (i.e. each core galaxy except for C1, C6, and C20), and centred the cutouts on the peak of the average [C{\sc ii}] map. We then fit a S{\'e}rsic profile to the stack, and found a half-light radius of 1.18$\,{\pm}\,$0.01\,kpc with a S{\'e}rsic index of 1.22\,${\pm}\,$0.03. Similarly, we find an axis ratio of 0.9\,${\pm}\,$0.1, overlapping with 1 as expected for stacking random orientation angles. The stack, along with the fit, is shown in Fig.~\ref{stack}.

\section{Results}
\label{results}

\subsection{The galaxy main sequence}

A primary quantity of interest is the galaxy main-sequence (MS), which is the SFR of a galaxy as a function of its stellar mass \citep[e.g.][]{elbaz2011}, and is commonly used to identify starbursts and quenched galaxies and therefore can place these protocluster galaxies in the context of galaxy evolution. Recently, \citet{rotermund2021} estimated stellar masses for 14 member galaxies of SPT2349$-$56 using SED fits to Gemini and {\it Spitzer\/} photometry, albeit shallower than presented here, and compared these to the $z\,{=}\,4.3$ MS. With our improved optical, infrared, and mm-wavelength coverage, as well as a more detailed IRAC deblending algorithm and an expanded catalogue of protocluster galaxies, we are able to expand on this work in much greater detail.

\begin{figure*}
\includegraphics[width=0.49\textwidth]{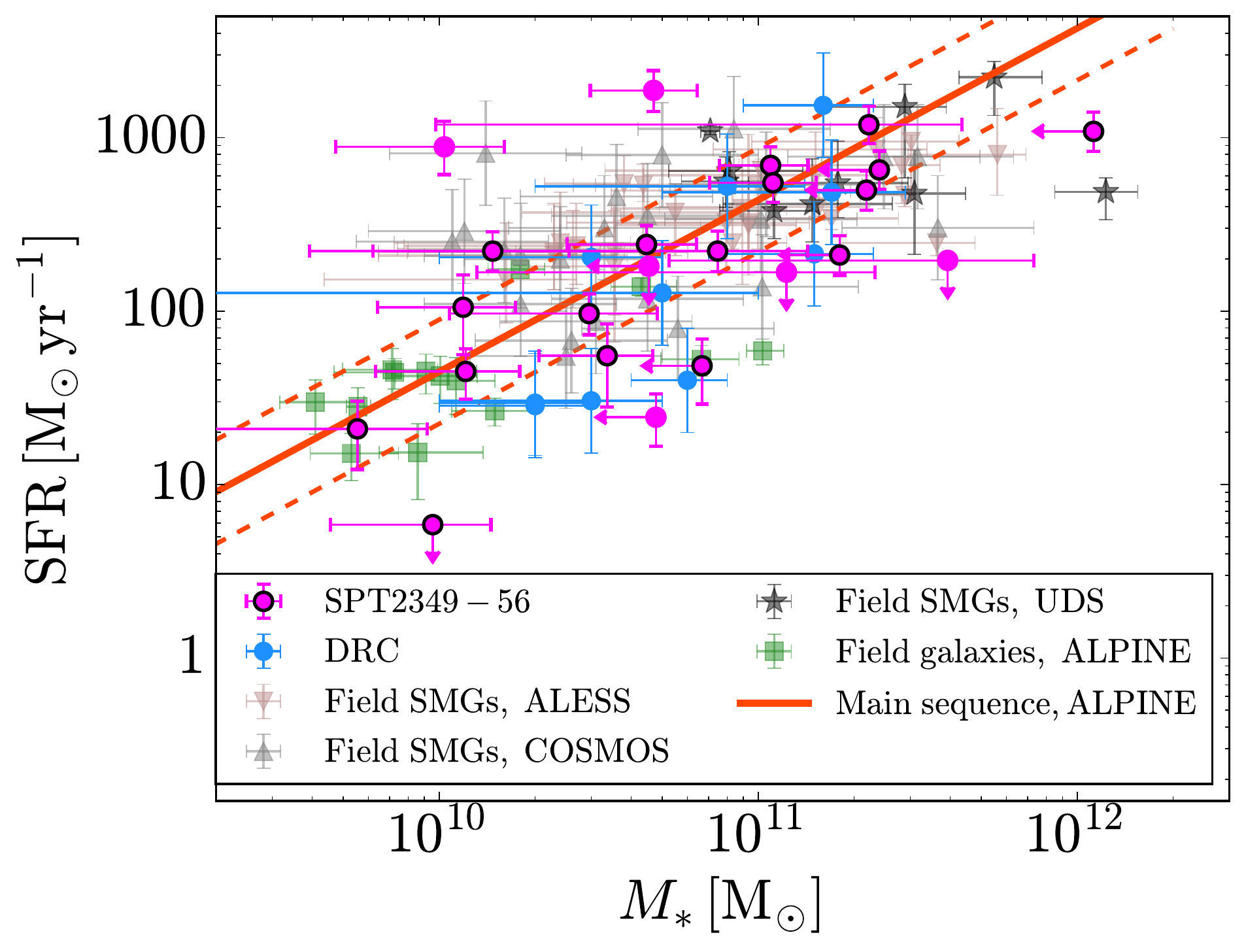}
\includegraphics[width=0.49\textwidth]{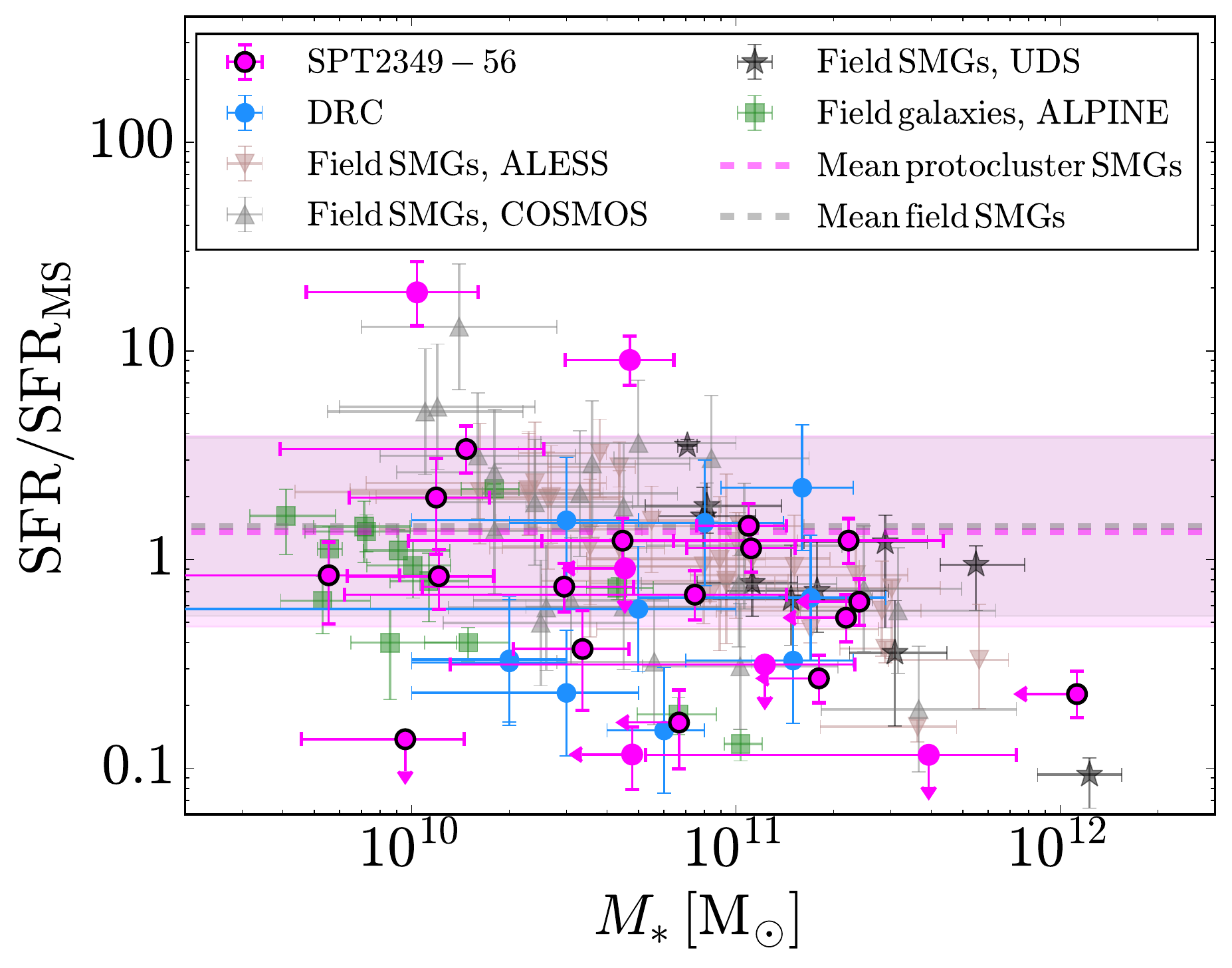}
\includegraphics[width=0.49\textwidth]{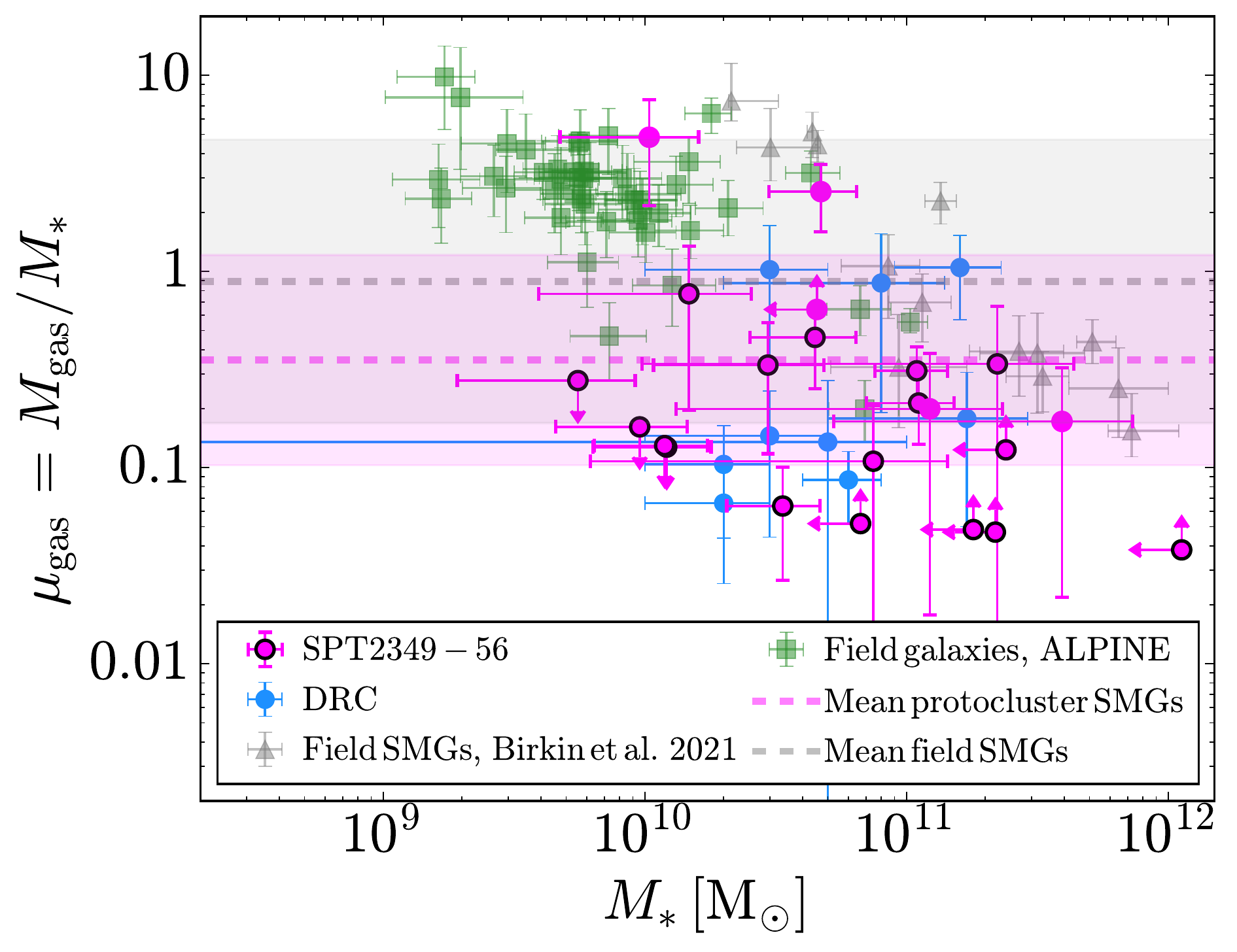}
\includegraphics[width=0.49\textwidth]{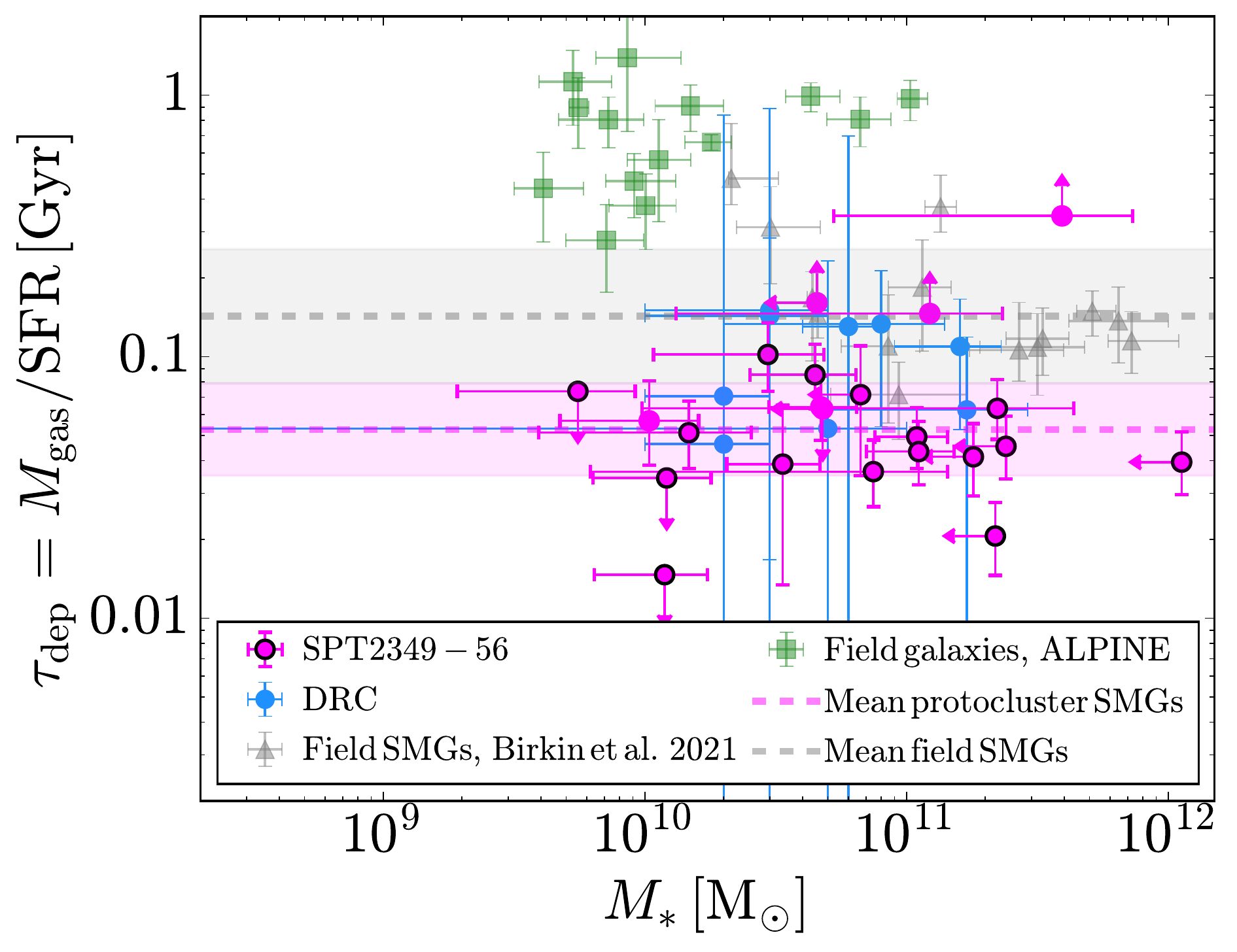}
\caption{{\it Top left:} Stellar mass as a function of SFR (i.e. the galaxy main sequence) for all galaxies in the SPT2349$-$56 protocluster where we could obtain stellar mass estimates from {\tt CIGALE} SED fitting with uncertainties not overlapping with 0. The SFRs shown here were estimated in \citet{hill2020}. Galaxies highlighted in black are part of the core of the structure, defined as those lying within a radius of 90\,kpc of the far-infrared luminosity-weighted centre \citep[see][]{hill2020}. Also shown are $z\,{>}\,3.5$ SMGs from the ALESS survey \citep{dacunha2015}, a sample of SMGs around $z\,{=}\,4.4$ from the COSMOS field \citep{scoville2016}, $z\,{=}\,4$--5 SMGs from the UDS field \citep{dudzeviciute2020}, $z\,{=}\,4.4$--5.9 galaxies from the ALPINE survey \citep{bethermin2020}, and a fit of the MS at $z\,{=}\,4.5$ from the ALPINE survey \citep{khusanova2021}, including an intrinsic scatter of ${\pm}\,0.3\,$dex proposed by \citet{schreiber2015}. Lastly, we show the galaxies from a similar star-forming $z\,{=}\,4$ protocluster known as the DRC \citep{long2020}. {\it Top right:} Measured SFRs divided by the SFR expected for a given stellar mass (SFR$_{\rm MS}$), assuming the MS relation from the ALPINE survey \citep{khusanova2021}, shown for the same galaxies in the top-left panel. Since SFR is distributed log-normally for a given stellar mass, the pink horizontal line and shaded region shows the weighted mean and standard deviation of $\log\,$SFR$\,{/}\,$SFR$_{\rm MS}$ for the protocluster galaxies, respectively (combining our sample with the galaxies from the DRC), while the grey horizontal line and shaded region shows the weighted mean and standard deviation of $\log\,$SFR$\,{/}\,$SFR$_{\rm MS}$ for field SMGs, respectively (combining the galaxies from the ALESS and COSMOS surveys). {\it Bottom left:} Molecular gas-to-stellar mass fraction, $\mu_{\rm gas}$ (Eq.~\ref{gas_mass}), as a function of stellar mass, using molecular gas masses from \citet{hill2020}. Our protocluster sample is compared with the field SMGs from \citet{birkin2021}, where molecular gas masses have been derived from CO observations similar to our sample, and the DRC. We also include the galaxies from the ALPINE survey by converting their published [C{\sc ii}] luminosities to molecular gas masses following the prescription outlined in \citet{dessauges-zavadsky2020}. The weighted mean and standard deviation of $\log\,\mu_{\rm gas}$ for all the protocluster galaxies is shown as the horizontal pink line and shaded region, respectively, and the weighted mean and standard deviation of $\log\,\mu_{\rm gas}$ for the field galaxies is shown as the horizontal grey line. {\it Bottom right:} Depletion timescale, $\tau_{\rm dep}$ (Eq.~\ref{dep_time}), as a function of stellar mass, for the same galaxies shown in the top-left panel, along with the same weighted means and standard deviations of the logarithms of the subsamples as horizontal lines and shaded regions, respectively.}
\label{ms}
\end{figure*}

In Fig.~\ref{ms} (top left panel) we show the stellar masses derived in this paper as a function of SFR. The SFRs shown here were derived in \citet{hill2020} by fitting modified blackbody distributions to the far-infrared photometry observed by ALMA, since the SFRs produced from our SED fitting could be less reliable due to dust obscuration. In the fits, $\beta$ was fixed to 2, and the dust temperature, $T_{\rm d}$, was fixed to 39.6\,K (the temperature consistent with the mean ratio of the measured 850\,$\mu$m flux density to the measured 3.2\,mm flux density of the sample). In \citet{hill2020} the best-fit SEDs were then integrated between 42 and 500\,$\mu$m to obtain far-infrared luminosities, but here we integrate from 8 to 1000\,$\mu$m to obtain more complete infrared luminosities in order to be consistent with comparison samples; this increases the luminosities by a factor of 1.17. These values were converted to SFRs using a factor of 0.95$\,{\times}\,10^{10}\,$M$_{\odot}$\,yr$^{-1}$\,L$_{\odot}^{-1}$ (from \citealt{kennicutt1998}, modified for a Chabrier initial mass function; see \citealt{chabrier2003}). The core galaxies of SPT2349$-$56 (defined as those lying within a 90\,kpc-radius of the far-infrared luminosity-weighted centre, where the primary-beam response of the ALMA observation used to find protocluster members falls to 0.5, see \citealt{hill2020}) are of interest because simulations predict that they will merge into a BCG on a timescale of a few hundred Myr \citep{rennehan2019}. We have highlighted these galaxies in black in Fig.~\ref{ms}. We also show the LAEs and LBGs of this sample as squares in order to distinguish them from the primary submm-selected galaxies making up the core of this protocluster. 

For comparison, we also show in Fig.~\ref{ms} the galaxies found in a similar star-forming protocluster, the Distant Red Core \citep[DRC;][]{oteo2018,long2020}, found at redshift 4. For these protocluster galaxies, the stellar masses were also obtained through optical and infrared SED fitting, while infrared luminosities were derived by scaling the mean template from the ALMA follow-up programme of the LABOCA ECDF-S Submillimetre Survey \citep[ALESS;][]{simpson2014} to match their continuum observations at 2\,mm; we applied the same scale factor used in our work to obtain SFRs.

To see how these protocluster galaxies compare with field galaxies, we focus on samples of high-redshift SMGs, since the SFRs of SMGs typically exceed 100\,M$_{\odot}$, which accounts for the majority of our sample. We show the $z\,{>}\,3.5$ SMGs from the ALESS survey, originally selected as bright 870-$\mu$m point sources in the Extended {\it Chandra} Deep Field South \citep[ECDF-S;][]{simpson2015a}, with stellar masses derived by \citet{dacunha2015} by fitting SEDs to optical and infrared photometry, and for the SFRs we converted the infrared luminosities from \citet{swinbank2014} (obtained by fitting a modified blackbody SED to the available far-infrared photometry) to SFRs using a conversion factor of $0.95\,{\times}\,10^{-10}\,$M$_{\odot}\,$yr$^{-1}\,$L$_{\odot.}^{-1}$. We also show a sample of field SMGs around $z\,{=}\,4.4$ from \citet{scoville2016} that were initially selected in a representative fashion from the Cosmic Evolution Survey (COSMOS) field; for these galaxies, stellar masses were also estimated from SED fits to optical and infrared photometry, and the SFRs were derived from rest-frame ultraviolet and infrared continuum measurements, adopting a factor of 2 uncertainty as recommended in the paper. The final field sample in this comparison comes from a follow-up survey of 707 SMGs detected in the Ultra Deep Survey (UDS) field \citep{dudzeviciute2020}, from which we have taken all galaxies with photometric redshifts between 4 and 5. Here the stellar masses and far-infrared luminosities were obtained by fitting SEDs to photometric data ranging from rest-frame optical to radio wavelengths, and we have converted the far-infrared luminosities to SFRs using the standard conversion factor. While we cannot entirely rule out the possibility that some of these field galaxies are in fact in protocluster environments, none of them are in known protoclusters, nor are any of them located in environments as overdense in the submm as SPT2349$-$56 or the DRC.

Next, we show galaxies from the ALMA Large Program to INvestigatE (ALPINE) survey \citep{lefevre2020,bethermin2020,faisst2020} between $z\,{=}\,4.4$ and $z\,{=}\,4.7$; for the ALPINE galaxies, stellar masses were obtained by fitting SEDs to optical and infrared photometry, and we have taken their infrared luminosities (scaled from 850\,$\mu$m continuum detections assuming a model template of $z\,{\approx}\,4$ MS galaxies) and converted these to SFRs using the same factor of  0.95$\,{\times}\,10^{10}\,$M$_{\odot}$\,yr$^{-1}$\,L$_{\odot}^{-1}$. We then show the best-fit $z\,{=}\,4.5$ MS obtained from the ALPINE survey \citep{khusanova2021}; we note that the parameterization found by the ALPINE survey is consistent with previously-derived MS parameterizations from e.g. \citet{speagle2014} at $z\,{=}\,4.3$. For reference, we show a scatter of a factor of 2 around the MS, the intrinsic scatter proposed by \citet{schreiber2015}.

We see that the galaxies in SPT2349$-$56 follow the $z\,{=}\,4.5$ MS derived by the ALPINE survey, although with considerable scatter, along with the other samples of field SMGs and the star-forming galaxies from the ALPINE survey. To investigate this in detail, in Fig.~\ref{ms} (top right panel) we show the SFRs in SPT2349$-$56 divided by the SFRs predicted by the ALPINE MS for each galaxy's measured stellar mass. We include the field SMGs from the samples described above, as well as the individual star-forming galaxies from the ALPINE survey. 

In order to assess the difference between protocluster galaxies to field galaxies around $z\,{=}\,4$, we combine our sample of protocluster galaxies with those from the DRC, and compare them to the ALESS, COSMOS, and UDS SMGs. Since the SFR is log-normally distributed at a given stellar mass \citep[e.g.][]{speagle2014,schreiber2015}, we compute the weighted mean and weighted standard deviation of the logarithm of the two samples. The weighted mean $\log\,$SFR$\,{/}\,$SFR$_{\rm MS}$ for the protocluster galaxies is 0.13, with a weighted standard deviation of 0.46, while for the field SMGs the weighted mean $\log\,$SFR$\,{/}\,$SFR$_{\rm MS}$ is 0.16, with a weighted standard deviation of 0.43. These values are plotted on Fig.~\ref{ms} for reference. 

We perform an unequal-variance $t$-test on the $\log\,$SFR$\,{/}\,$SFR$_{\rm MS}$ values of each sample. Assuming the two samples in question are drawn from Gaussian distributions, the unequal-variance $t$ score is a statistic used to test the hypothesis that the two distributions have equal means and arbitrary variances. It is worth noting that we are only testing the means of the two samples, meaning that even if they have overlapping scatter, we may still reject the null hypothesis that the means are equal. Using this test, we find a $p$-value of 0.72, which can be interpreted as the probability that the means are the same. From this comparison we cannot reject the null hypothesis that the protocluster galaxies in these samples are different from field galaxies in terms of the MS. However, we emphasize that there could be large systematic errors present, owing from differences in modelling the SEDs used to fit the available photometry to obtain stellar masses; for example, different functional forms for the SFH can be assumed, and there are numerous models of dust extinction available.

\subsection{Molecular gas-to-stellar mass fractions}

Another mass measurement available for the protocluster galaxies in SPT2349$-$56 is the molecular mass, and a useful evolutionary diagnostic is to see if the molecular gas-to-stellar mass fraction scales with the total stellar mass built-up so far. We define the molecular gas-to-stellar mass fraction as
\begin{equation}
\label{gas_mass}
\mu_{\mathrm{gas}} = \frac{M_{\mathrm{gas}}}{M_{\ast}},
\end{equation}
\noindent
where $M_{\rm gas}$ is the molecular gas mass and $M_{\ast}$ is the stellar mass.

We use the molecular gas masses derived in \citet{hill2020}. Briefly, the CO(4--3) transition was observed by ALMA, and line strengths were measured for sources where the line was detected. The CO(4--3) line strength was converted to a CO(1--0) line strength using a factor of 0.60 (the mean line strength ratio of the SPT-SMG sample from \citealt{spilker2014}), and then this was converted to a molecular gas mass using a conversion factor of $\alpha_{\rm CO}\,{=}\,1\,$M$_{\odot}/$(K\,km\,s$^{-1}$\,pc$^2$), similar to other studies of SMGs \citep[see e.g.,][]{aravena2016a,bothwell2017}. The resulting molecular gas-to-stellar mass fractions are given in Table~\ref{table:cigale}; we find values ranging from about 0.04 to 5. Interestingly, since our stellar masses are lower than the values provided by \citet{rotermund2021}, we estimate larger values of $\mu_{\rm gas}$, which is in better agreement with the simulations analyzed by \citet{lim2021}. However, these simulations still predict that $z\,{\approx}\,4$ protocluster galaxies have $\mu_{\rm gas}$ values between 1 and 5, which is only reached by two galaxies (N1 and N2) in our sample.

Figure \ref{ms} (bottom left panel) shows our molecular gas-to-stellar mass fractions as a function of stellar mass (with the core galaxies again highlighted), compared with the galaxies from the DRC \citep{oteo2018,long2020}, where we have converted their CO(6--5) line intensities to molecular gas masses using an $L^{\prime}_{(6-5)}/L^{\prime}_{(1-0)}$ factor of 0.46 (the mean ratio found for the SPT-SMG sample, see \citealt{spilker2014}) and an $\alpha_{\rm CO}$ of 1\,M$_{\odot}/$(K\,km\,s$^{-1}$\,pc$^2$) (the same scale factor used for our sample). 

To compare protocluster galaxies with field galaxies around $z\,{=}\,4$, we require a sample with molecular gas masses estimated with similar CO transition tracers. A recent CO survey of SMGs selected from the COSMOS field, the UDS field, and the ALESS sample spanning $z\,{=}\,1$--5 was carried out by \citet{birkin2021}, and we have selected the galaxies from this survey with $z\,{>}\,3.5$ to use as a field comparison here; this corresponds to four galaxies from the COSMOS field, two galaxies from the UDS field, and nine galaxies from the ALESS sample. All of the galaxies in this sample were originally detected as bright submm sources in large single-dish surveys with SCUBA-2 and LABOCA, and have extensive multiwavelength follow-up observations. The detected lines range from CO(5--4) to CO(2--1), and we use the mean line ratios of the sample to convert these line intensities to the CO(1--0) transition, with the CO(2--1)-to-CO(1--0) ratio fixed to 0.9 (for reference, the mean CO(4--3)-to-CO(1--0) ratio was found to be 0.32, compared to 0.60 used for the SPT2349$-$56 galaxies). We then adopt an $\alpha_{\rm CO}$ of 1\,M$_{\odot}/$(K\,km\,s$^{-1}$\,pc$^2$) for this reference sample. Stellar masses are provided for each source, derived by fitting SEDs to the extensive photometry available in these fields. For comparison with a sample of high-$z$ galaxies that are not SMGs, we include the galaxies from the ALPINE survey here by deriving molecular gas masses from the published [C{\sc ii}] luminosities, following the prescription outlined in \citet{dessauges-zavadsky2020}; however, since the [C{\sc ii}] line is a different tracer, we cannot provide any quantitative comparisons with the samples of CO-derived gas masses.

We again combine our sample of protocluster galaxies with the DRC, and compare this to the sample from \citet{birkin2021}. Looking at Fig. \ref{ms} (bottom left panel) we see that the molecular gas-to-stellar mass also appears log-normally distributed, so we calculate the weighted mean and weighted standard deviation of the logarithm of each sample. The weighted mean $\log\,\mu_{\rm gas}$ of protocluster galaxies is $-$0.45, with a weighted standard deviation of 0.54, while for the field galaxies we find a weighted mean $\log\,\mu_{\rm gas}$ of $-$0.05, with a weighted standard deviation of 0.72 (see Fig.~\ref{ms}). A similar unequal-variance $t$-test on the $\log\,\mu_{\rm gas}$ values results in a $p$-value of 0.04. While these results do provide evidence that the mean molecular gas-to-stellar mass fractions of protocluster galaxies are not the same as those of field galaxies (the null hypothesis can be rejected with ${>}\,95$\,per cent confidence), we nonetheless note that there could still be systematic uncertainties unaccounted for in this analysis.

\subsection{Depletion timescales}

A similar quantity of interest is the gas depletion timescale, which is a measure of the amount of time required to convert all of the available mass in gas into mass in stars if the current star-formation rate were to remain constant. This quantity is defined as
\begin{equation}
\label{dep_time}
\tau_{\mathrm{dep}} = \frac{M_{\mathrm{gas}}}{\mathrm{SFR}},
\end{equation}
\noindent
where this time $M_{\rm gas}$ is divided by the SFR.

In Table \ref{table:cigale} we provide estimates of the depletion timescale for each galaxy with a measurement of molecular gas mass (via CO(4--3) line detection) and SFR (via far-infrared photometry). The galaxies in SPT2349$-$56 have depletion timescales ranging from 0.05 to 0.1\,Gyr.

In Fig.~\ref{ms} (bottom right panel) we show our depletion timescales as a function of stellar mass. In order to compare with the same $z\,{>}\,3.5$ field SMGs from \citet{birkin2021}, we take their provided far-infrared luminosities, obtained by fitting SEDs with the available photometry, and multiply them by the usual factor of 0.95$\,{\times}\,10^{10}\,$M$_{\odot}$\,yr$^{-1}$\,L$_{\odot}^{-1}$. We also show the protocluster galaxies from the DRC, and we include the same ALPINE galaxies for reference to a sample of non-SMGs. 

We find that the depletion timescales appear on average smaller in protocluster galaxies than in field SMGs. We take the same log-normal approach to quantify this difference, finding that the weighted mean $\log\,\left(\tau_{\rm dep}\,{/}\,[\mathrm{Gyr}]\right)$ for the protocluster galaxies (again combining SPT2349$-$56 and the DRC) is $-$1.28, with a weighted standard deviation of 0.18, and the weighted mean $\log\,\left(\tau_{\rm dep}\,{/}\,[\mathrm{Gyr}]\right)$ for the field galaxies from \citet{birkin2021} is $-$0.85, with a weighted standard deviation of 0.26 (see Fig.~\ref{ms}). We therefore find that the depletion timescales for protocluster galaxies are smaller than for field galaxies, as there is very little overlap between the two distributions. The resulting $p$-value from an unequal-variance $t$-test is 5.3$\,{\times}\,10^{-5}$, thus we can reject the null hypothesis that the mean values of the two populations are the same. Lastly, the star-forming galaxies from the ALPINE survey have longer depletion timescales than all of the SMGs in this comparison.

\subsection{The stellar mass function}
\label{counts}

Our large sample of protocluster galaxies with stellar mass estimates allows us to compute the stellar mass function of star-forming galaxies in several different ways. Of most interest is what the stellar mass function of the whole protocluster looks like in comparison to lower-$z$ clusters. However, we know that the core galaxies in SPT2349$-$56 will merge into a single BCG within a few hundred Myr \citep{rennehan2019}, so we would also like to know what the stellar mass function will look like after the merger by summing their masses.

To start, we address the stellar mass completeness of our sample. Most of the galaxies in our sample were selected from line emission surveys in the submm, but others were found through their Ly-$\alpha$ emission or selected as LBGs, making a detailed completeness calculation difficult. Therefore, we begin by simply considering the initial sample of [C{\sc ii}] and CO(4--3)-selected galaxies (C1--C23, N1, and N2), and include LBG3 as this galaxy also shows significant [C{\sc ii}] emission \citep{rotermund2021}. We do not include the three SPIREc sources, since they are at a large projected distance where our submm imaging is not complete.

\begin{figure}
\includegraphics[width=\columnwidth]{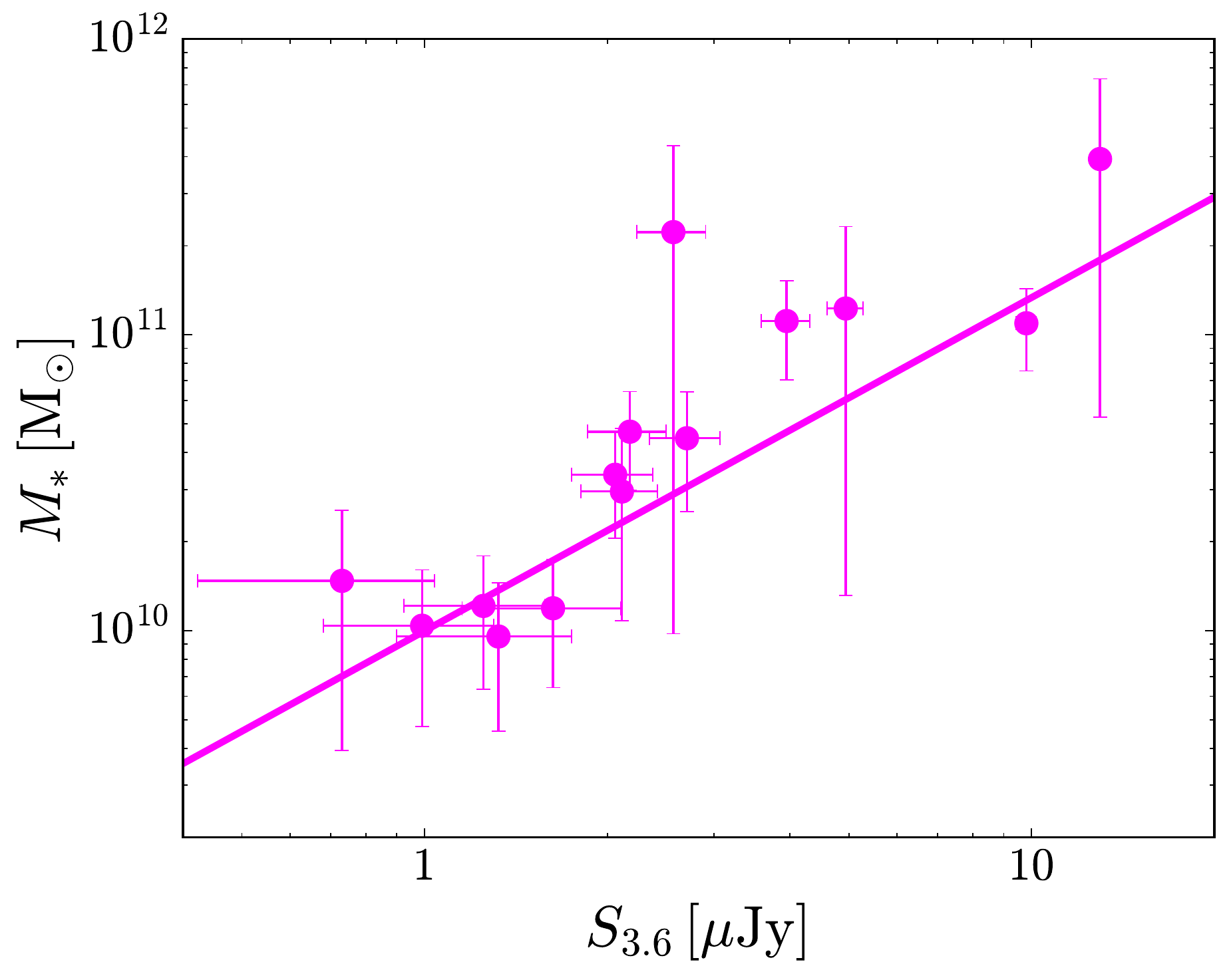}
\caption{The stellar mass as a function of $S_{3.6}$ for all galaxies in our sample with with both measurements available. We find a tight correlation between the two quantities, and fit a power law of the form $M_{\ast}\,{=}\,A \left(S_{3.6}\,{/}\,S_{3.6,0}\right)^{\gamma}$, with $S_{3.6,0}$ fixed to 1\,$\mu$Jy (shown as the solid line). We use this functional form to estimate the completeness of our stellar mass sample, finding 100\,per cent completeness above 10$^{11}\,$M$_{\odot}$, 80\,per cent completeness above 10$^{10}\,$M$_{\odot}$., and 60\,per cent completeness for the whole sample.}
\label{completeness}
\end{figure}

Next, in Table \ref{table:cigale} we see that 15 of the 26 galaxies in this subsample have stellar mass estimates available, thus a lower limit to the completeness of all of our stellar masses is about 60\,per cent. We also have stellar mass upper limits available for most of the undetected sources, however they are of the order ${<}\,10^{11}\,$M$_{\odot}$ and are not very constraining. Instead, we can turn to our IRAC 3.6\,$\mu$m imaging, which is a good tracer of stellar mass, probing rest-frame wavelengths of 680\,nm with good resolution where the large populations of low-mass stars emit. In Fig.~\ref{completeness} we show our derived stellar masses as a function of $S_{3.6}$ for all of the galaxies in our subsample with both quantities available. We see a tight correlation between the two quantities, as expected, and a best-fit power law of the form $M_{\ast}\,{=}\,A \left(S_{3.6}\,{/}\,S_{3.6,0}\right)^{\gamma}$, with $S_{3.6,0}$ fixed to 1\,$\mu$Jy, gives $A\,{=}\,(8.6\,{\pm}\,1.4)\,{\times}\,10^{9}\,$M$_{\odot}$ and $\gamma\,{=}\,1.2\,{\pm}\,0.1$. Using this functional form, we can take the 3.6\,$\mu$m flux density measurements and upper limits for the 11 remaining galaxies with no stellar mass measurements and calculate their expected stellar masses and stellar mass upper limits. We again find that all of the galaxies have upper limits of ${<}\,10^{11}\,$M$_{\odot}$, meaning that our stellar mass subsample is 100\,per cent complete above $10^{11}\,$M$_{\odot}$, while five galaxies have upper limits of ${<}\,10^{10}\,$M$_{\odot}$, so our stellar mass subsample is about 80\,per cent complete above $10^{10}\,$M$_{\odot}$. However, this completeness estimate does not take into account potentially variable dust extinctions between the galaxies, and given the large overall uncertainties in stellar mass estimates, we are not able to provide completeness corrections, so below we present the number counts of our complete subsample and stress the uncertainties.

Figure \ref{counts} (top panels) shows the differential and cumulative number counts of the stellar masses of all the galaxies in our restricted sample. We have normalized our counts by the volume of a sphere of radius 360\,kpc, which is the distance between the core and northern components of SPT2349$-$56 \citep[see][]{hill2020}. The true normalization is highly uncertain, but in this analysis we are only interested in the shape of the mass function. We compare this protocluster star-forming stellar mass function to the stellar mass function of a typical $z\,{\simeq}\,1$ galaxy cluster \citep{vanderburg2013}, obtained by stacking cluster galaxies from a sample of 10 clusters (including their BCGs) within 1\,Mpc of the cluster cores. \citep{vanderburg2013} separate star-forming galaxies from field galaxies, but we have taken their total number counts as we expect our sample to be more representative of the total counts of SPT2349$-$56. We set the normalizing volume to be the volume of 10 spheres of radius 1\,Mpc, and again emphasize that the absolute normalization here is uncertain but has no effect on the shape of the counts. While this reference sample contains galaxies within a larger proper volume than probed by our ALMA data (1\,Mpc versus 360\,kpc), previous studies of $z\,{<}\,1$ galaxy clusters have found that best-fit parameterizations of stellar mass number counts do not vary considerably when calculated within radii ranging from 0.5\,$R_{500}$ to 2\,$R_{500}$ \citep[where $R_{500}$ is around 1\,Mpc, see][]{vanderburg2018}; the slope, $\alpha$, ranges from $-$0.8 to $-$1.0, while the characteristic mass, $M^{\star}$, remains constant within the uncertainties. Therefore, we expect our conclusions would remain unchanged if the $z\,{\simeq}\,1$ comparison sample were limited to galaxies within 360\,kpc, as with our sample. Qualitatively, we see that the shapes of the number counts are in agreement with one another for masses above 10$^{10}\,$M$_{\odot}$.

Next, we compute the differential and cumulative number counts of the molecular gas masses in SPT2349$-$56 using the values provided in \citet{hill2020} for the same galaxies in our restricted sample. These functions are shown alongside the stellar mass function in Fig.~\ref{counts} (top panels). We find that the molecular gas mass function effectively tracks the stellar mass function.

We then assess the state of the number counts after the merger of the BCG galaxies by summing the masses of all the galaxies within 90\,kpc of the far-infrared luminosity-weighted centre (i.e.~the region simulated by \citealt{rennehan2019} where the mergers will take place), and treating this as a single point. These are shown alongside our other number counts in Fig.~\ref{counts} (bottom panels) for comparison. We see that the shapes become linear (in log-log space), and the final mass of the merged galaxies is comparable to the masses of the largest galaxies in the $z\,{=}\,1$ sample. The amplitude of our counts is also larger than the $z\,{=}\,1$ cluster counts at these high-mass bins, although since the normalizations are highly uncertain, we cannot draw any conclusions from this excess.

\begin{figure*}
\includegraphics[width=0.49\textwidth]{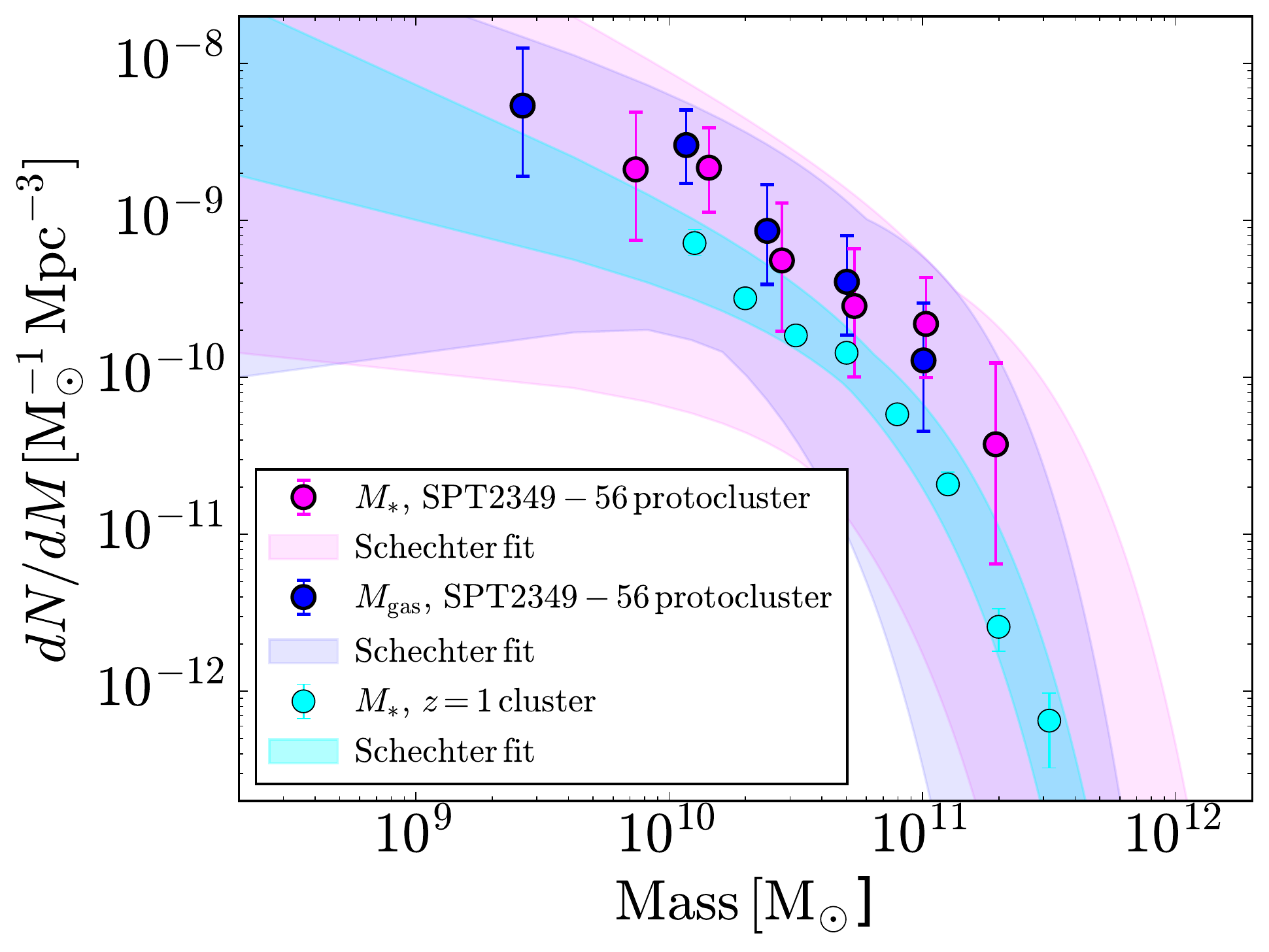}
\includegraphics[width=0.49\textwidth]{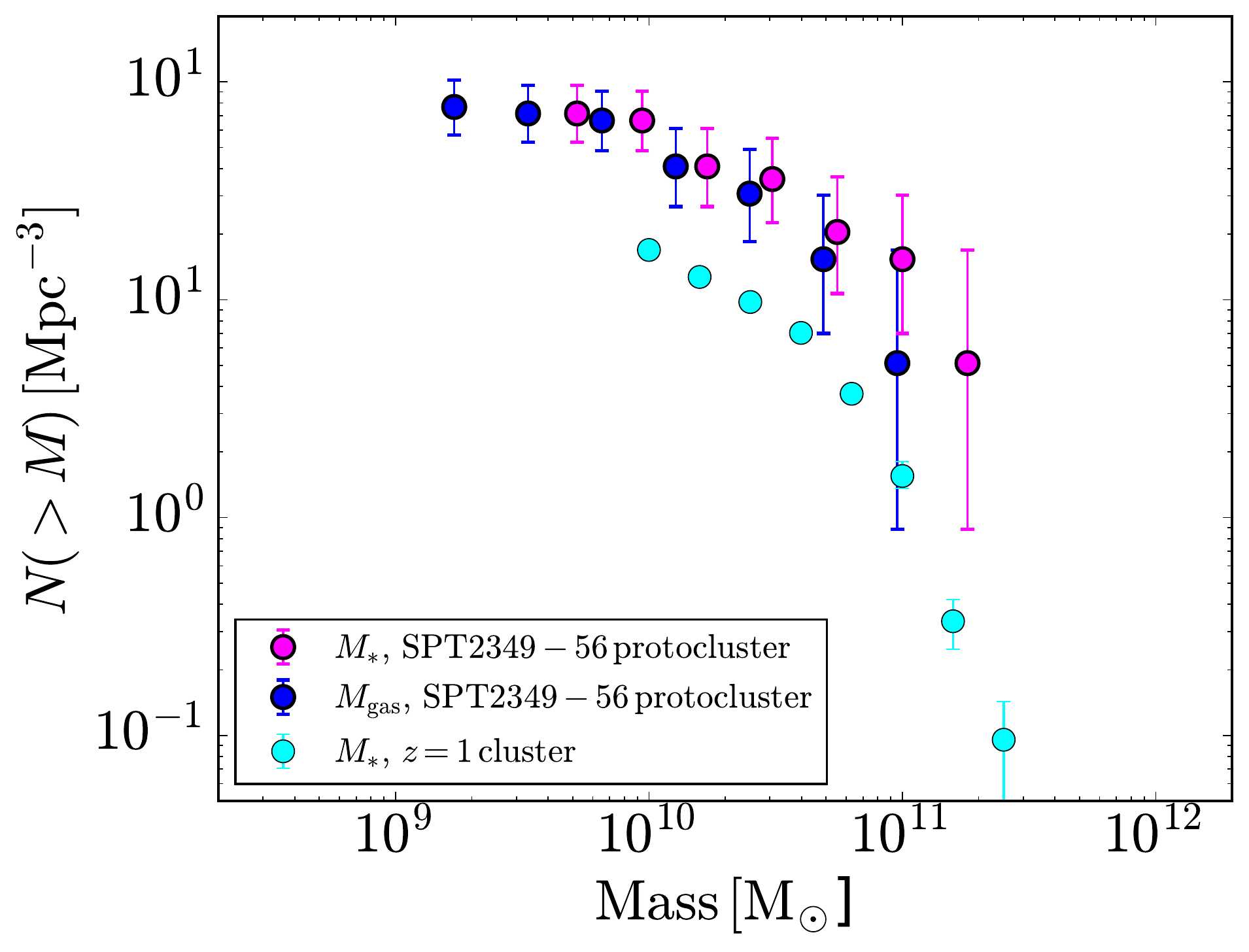}
\includegraphics[width=0.49\textwidth]{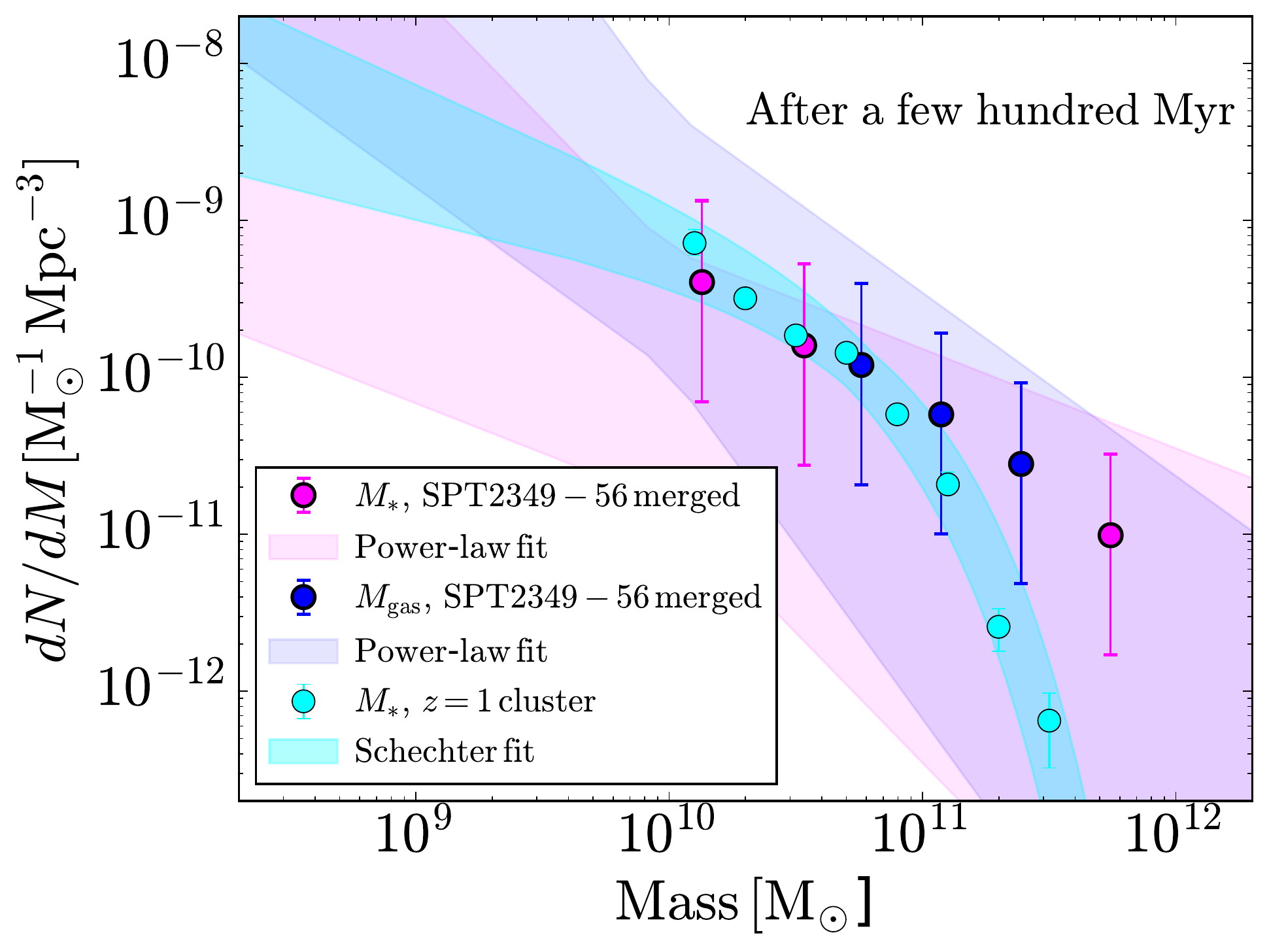}
\includegraphics[width=0.49\textwidth]{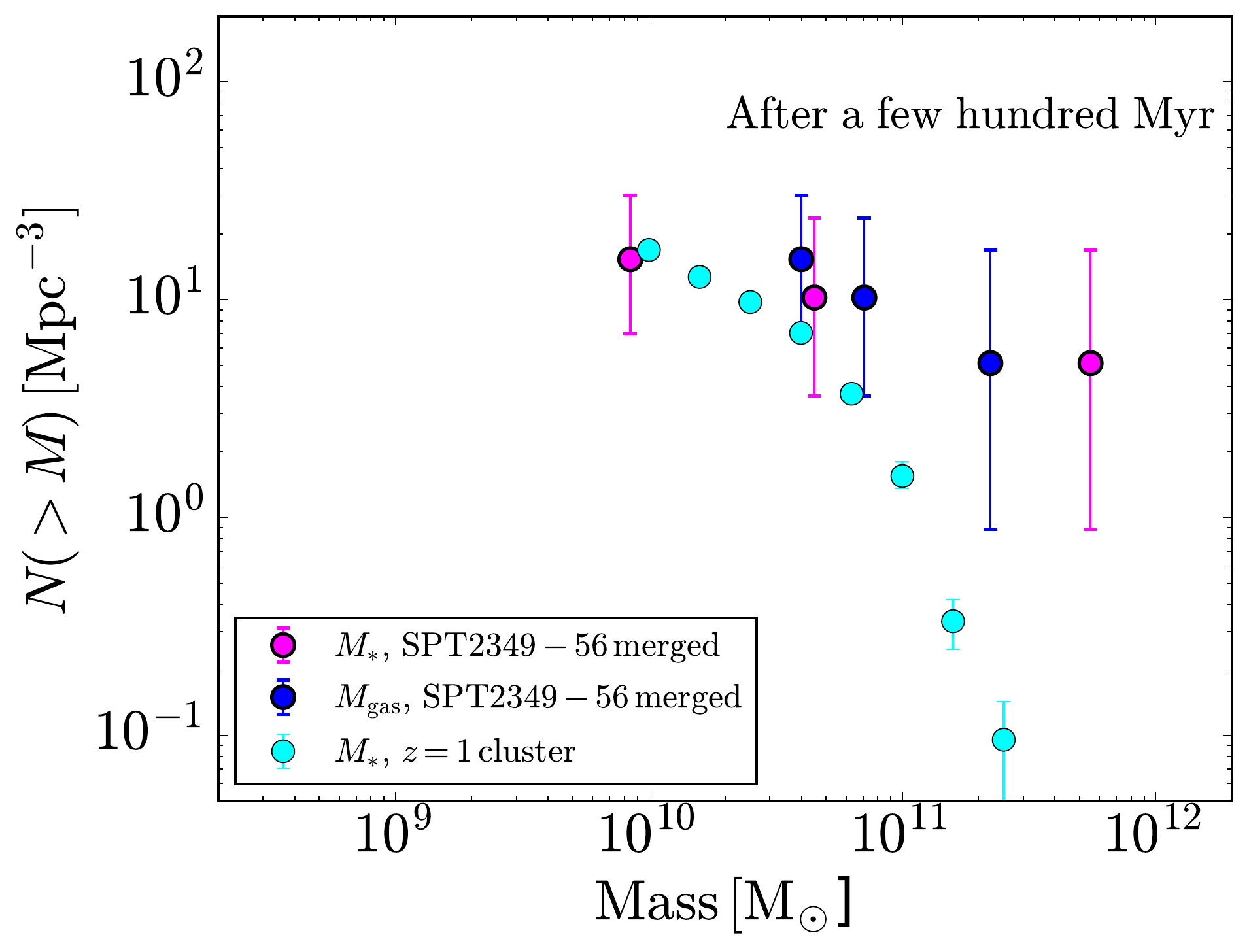}
\caption{{\it Top:} Molecular gas mass and stellar mass differential (left panel) and cumulative (right panel) number counts for SPT2349$-$56. The stellar mass number counts for $z\,{\sim}\,1$ clusters from \citet{vanderburg2013} is shown in cyan for comparison. Best-fit Schechter functions are shown for our differential counts, where the shaded regions are calculated by varying the best-fit parameters within their 68\,per cent confidence intervals and taking the largest difference. The best-fit Schechter function from \citet{vanderburg2013} is shown as well, with the shaded region calculated in the same way. {\it Bottom:} Since the core galaxies will merge into a single BCG in a few hundred Myr \citep{rennehan2019}, we show the molecular gas mass and stellar mass differential (left panel) and cumulative (right panel) number counts after summing the masses of the galaxies within 90\,kpc of the far-infrared luminosity-weighted centre and treating this as a single point. We also show best-fit power-law functions as shaded regions, calculated in the same way as above, and the same best-fit Schechter function for $z\,{\sim}\,1$ clusters from \citet{vanderburg2013}.}
\label{counts}
\end{figure*}

We next fit our protocluster mass functions to functional forms using a maximum-likelihood approach \citep[e.g.,][]{marshall1983,wall2008,hill2020}, where we minimize the negative log-likelihood of our stellar mass measurements assuming that all sources were selected from data of equal depth:
\begin{equation}
S = -2 \ln \mathcal{L} = -2 \sum_{i=1}^N \ln \phi (M_i) + 2 V \int _{M_\mathrm{a}}^{M_\mathrm{b}} \phi (M) dL + C.
\end{equation}
\noindent
In this equation $N$ is the sample size, $\phi(M)$ is the model differential stellar mass number count (in units of M$_{\odot}^{-1}$\,Mpc$^{-3}$), $V$ is the volume of the survey, $M_{\mathrm{a}}$ and $M_{\mathrm{b}}$ are the mass limits of the sample (which we take to be between the smallest and largest masses in the sample), and $C$ is a constant independent of the model. We investigate two models, a single power-law and a Schechter function. The single power-law has two free parameters, a normalization and a power-law index; explicitly,
\begin{equation}
\label{power-law}
\phi(M) = \phi^{\star} \left(\frac{M}{M^{\star}}\right)^{\alpha},
\end{equation}
\noindent
where $\alpha$ is the slope of the power law, $\phi^{\star}$ is the overall normalization, and $M^{\star}$ is fixed to $10^{10}\,$M$_{\odot}$ and is not a free parameter of the model. For the Schechter function, $M^{\star}$ is not fixed but treated as a free parameter, describing the point at which the number counts transition from a power law to an exponential:
\begin{equation}
\label{schechter}
\phi(M) = \phi^{\star} \left(\frac{M}{M^{\star}}\right)^{\alpha} e^{-M/M^{\star}}.
\end{equation}
\noindent

We use a Markov chain Monte Carlo (MCMC) approach to minimize the log-likelihood, and calculate the odds ratio between a single power-law fit and a Schechter fit (simply $\mathcal{L_{\rm Schechter}}/\mathcal{L_{\rm Power-law}}$) to assess which model better-describes the data. We find that a Schechter function is more appropriate for the total protocluster number counts, and that a single power-law function is more appropriate for the protocluster containing a BCG. The resulting fit parameters are provided in Table \ref{table:counts} (where the values are the means of the posterior distributions, and the uncertainties are 68\,per cent confidence intervals), and in Fig.~\ref{counts} we show the best-fit functions, where the shaded regions were calculated by varying the best-fit parameters within their 68\,per cent confidence intervals and taking the largest difference.

We would now like to quantitatively compare our number counts to the stellar mass function of $z\,{=}\,1$ clusters found by \citet{vanderburg2013}. To do this, we note that \citet{vanderburg2013} provide a best-fit Schechter function to the stellar mass number counts measured for a $z\,{=}\,1$ cluster. They find a characteristic stellar mass of $(5.2_{-0.2}^{+1.1})\,{\times}\,10^{10}\,$M$_{\odot}$ and a slope of $-$0.46$_{-0.26}^{+0.08}$ (we ignore the normalization here because it is highly uncertain for our sample). For reference, in Fig.~\ref{counts} we show their best-fit Schechter function as a shaded region encompassing the uncertainties of their best-fit parameters. The characteristic stellar mass found for our total protocluster is $(10.6_{-7.1}^{+2.9})\,{\times}\,10^{10}\,$M$_{\odot}$ and the slope found is $-$0.4$_{-0.4}^{+0.4}$, which is not statistically different from the parameters of \citet{vanderburg2013} given the uncertainties, although we emphasize that the uncertainties are large both due to our small sample size and incompleteness and so we cannot make any further conclusions about the evolution of the star-forming stellar mass function.

\setlength\tabcolsep{4pt}
\setlength\extrarowheight{4pt}
\begin{table*}
\centering
\caption{Best-fit mass function parameters (Eqs.~\ref{power-law} and \ref{schechter}) derived from the stellar mass and molecular gas mass number counts for all of the galaxies in SPT2349$-$56, and SPT2349$-$56 after the central galaxies merge into a BCG, estimated by summing the masses of the central galaxies and treating them as a single source.}
\label{table:counts}
\begin{threeparttable}
\begin{tabular}{lcccc}
\hline
Component & Function & $\phi^{\star}$ & $M^{\star}$ & $\alpha$ \\
& $\phi(M)$ & [10$^{-10}\,$M$_{\odot}^{-1}\,\mathrm{Mpc}^{-3}$] & [10$^{10}\,$M$_{\odot}$] & \\
\hline
Protocluster, stellar mass & $\phi^{\star} \left(\frac{M}{M^{\star}}\right)^{\alpha} e^{-M/M^{\star}}$ & 8$_{-8}^{+1}$ & 10.6$_{-7.1}^{+2.9}$ & -0.4$_{-0.4}^{+0.4}$ \\
Protocluster, gas mass & $\phi^{\star} \left(\frac{M}{M^{\star}}\right)^{\alpha} e^{-M/M^{\star}}$ & 22$_{-18}^{+6}$ & 4.9$_{-3.3}^{+1.1}$ & -0.1$_{-0.5}^{+0.3}$ \\
BCG, stellar mass & $\phi^{\star} \left(\frac{M}{M^{\star}}\right)^{\alpha}$ & 7$_{-6}^{+1}$ & 1$^{\rm a}$ & -1.2$_{-0.5}^{+0.5}$ \\
BCG, gas mass & $\phi^{\star} \left(\frac{M}{M^{\star}}\right)^{\alpha}$ & 39$_{-37}^{+15}$ & 1$^{\rm a}$ & -1.6$_{-0.6}^{+0.4}$ \\
\hline
\end{tabular}
\begin{tablenotes}
\item $^{\rm a}$This parameter was fixed during the fitting.
\end{tablenotes}
\end{threeparttable}
\end{table*}

\subsection{Ultraviolet versus far-infrared sizes}

Many studies have looked at the physical extent of stellar emission compared to dust emission in SMGs \citep[e.g.][]{simpson2015a,lang2019}, finding that the dust emission (and hence the star formation), probed by rest-frame far-infrared observations, is typically more smooth and compact, while the optical stellar emission is clumpy and extended, likely due to patchy dust attenuation \citep[e.g.][]{cochrane2019}. Although the {\it HST} imaging probes a slightly different population of stars in the ultraviolet, we emphasize that {\it HST\/} is still currently the best facility for performing these measurements on objects at such high redshift. The forthcoming {\it James Webb Space Telescope} ({\it JWST}) will operate at the longer wavelengths needed to resolve the average stellar populations in SPT2349$-$56, and we will carry out the measurements when the facility becomes available. On the other hand, our 850-$\mu$m ALMA observations probe the rest-frame at 160\,$\mu$m where the dust is expected to be bright. There are also known correlations between a galaxy sizes and quantities such as SFR, stellar mass, and redshift, the most studied likely being the size-mass relation \citep[e.g.,][]{shen2003,vanderwel2014,mowla2019}. All these correlations can be examined in a high-redshift protocluster environment using the galaxies in SPT2349$-$56.

\begin{figure*}
\includegraphics[width=0.49\textwidth]{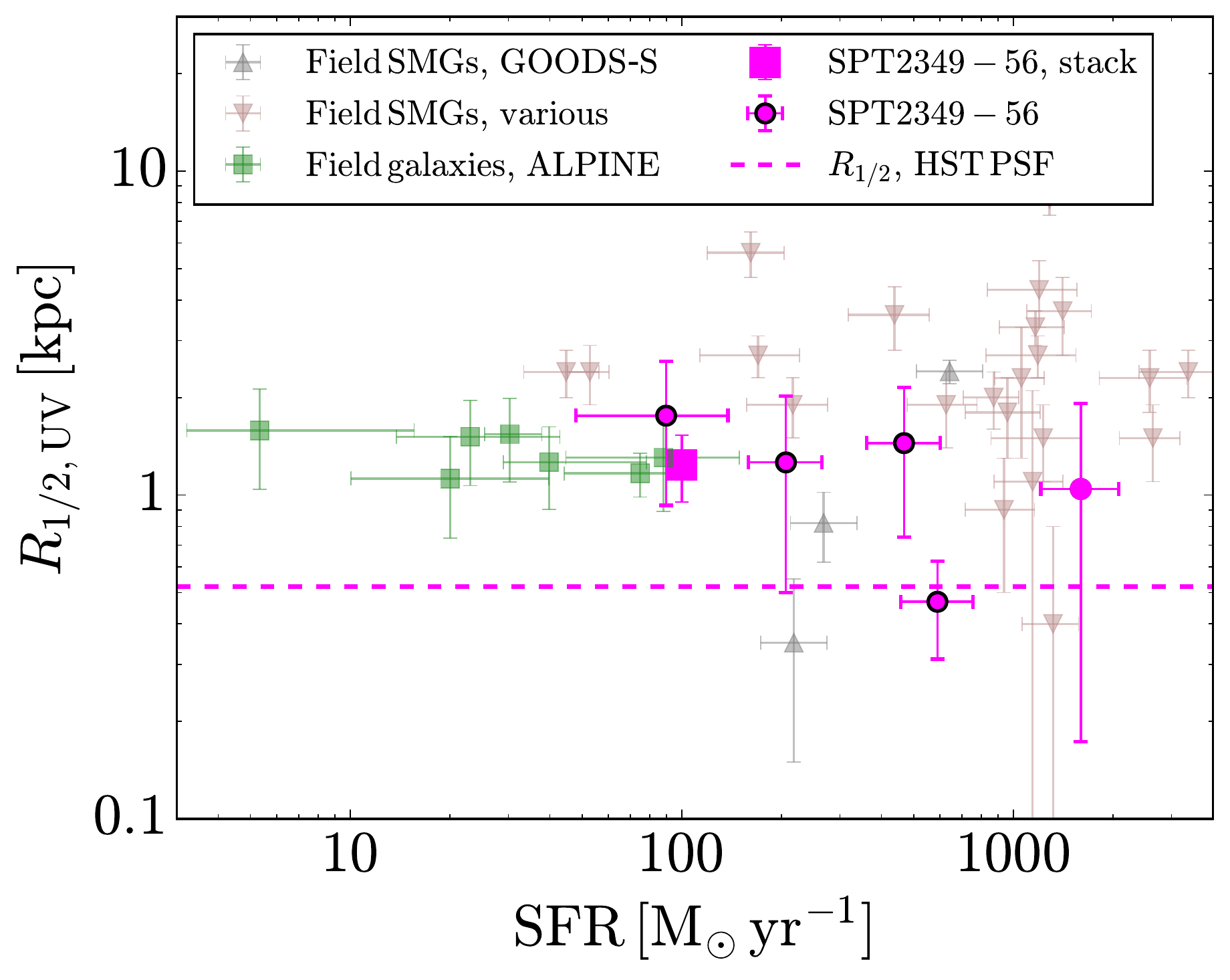}
\includegraphics[width=0.49\textwidth]{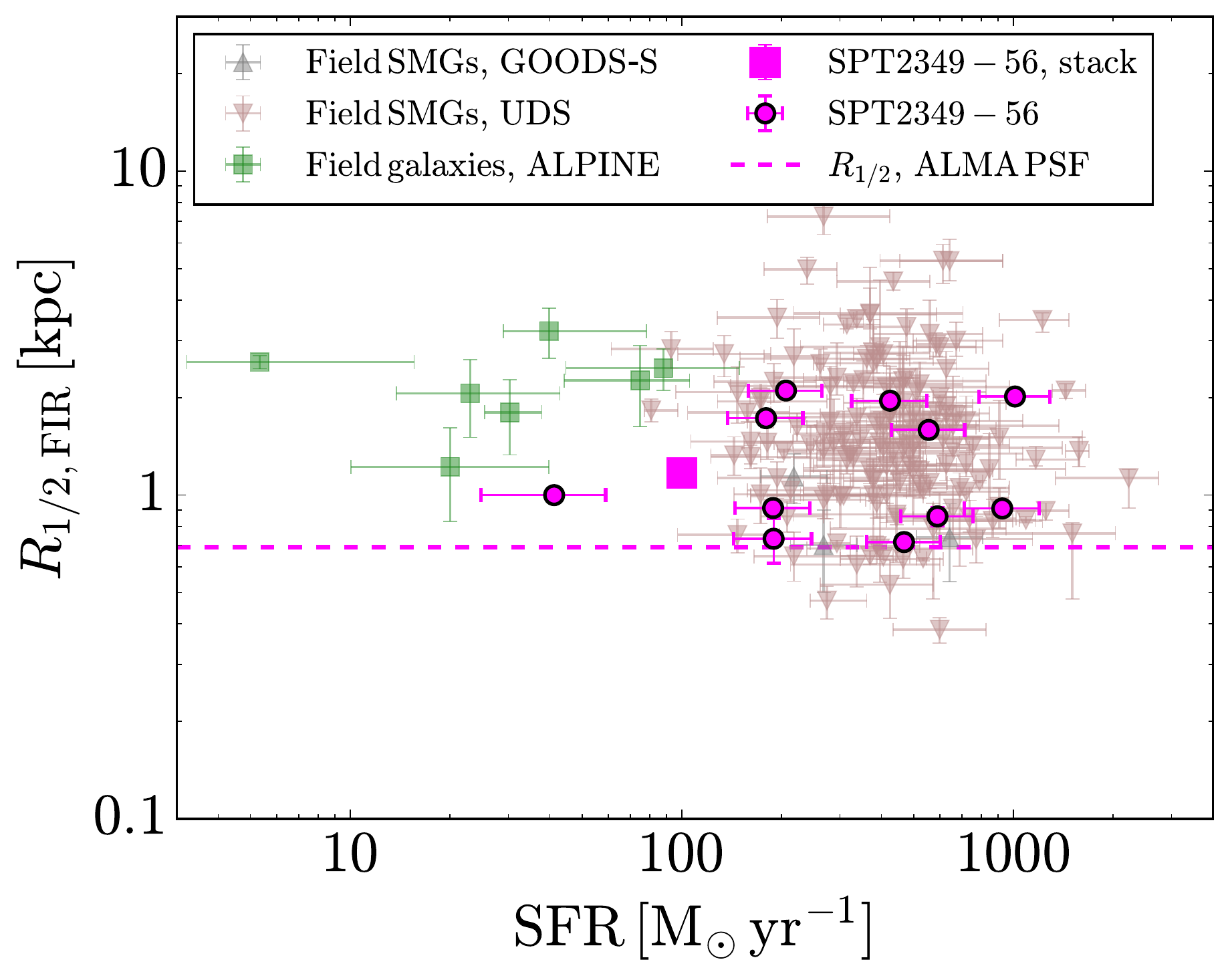}
\caption{{\it Left:} Ultraviolet half-light radius as a function of SFR for all galaxies in the SPT2349$-$56 protocluster detected in the F160W band with pixels above 5 times the local rms. Also shown are results for $z\,{=}\,1$--3 field SMGs from \citet{swinbank2010} and \citet{barro2016}, and $z\,{=}\,4$--5 star-forming galaxies from the ALPINE survey \citep{fujimoto2020}. For reference, the size measured in our {\it HST\/} F160W stack is shown as the pink square, arbitrarily placed at 100\,M$_{\odot}$\,yr$^{-1}$, and the half-light radius of the {\it HST\/} F160W beam is shown as the horizontal dashed line. {\it Right:} Far-infrared half-light radius as a function of SFR from \citet{hill2020}, for all galaxies detected by ALMA at 850\,$\mu$m with pixels above 5 times the local rms, along with our stack shown as the pink square. Also shown are results for the field SMGs from \citet{barro2016}, field SMGs in the UDS field from \citet{gullberg2019}, and [C{\sc ii}] size measurements from the ALPINE survey \citep{fujimoto2020}, converted to far-infrared sizes using the mean $R_{1/2,\mathrm{C[II]}}/R_{1/2,\mathrm{FIR}}$ ratio from our sample (see Fig.~\ref{size}). The half-light radius of the ALMA synthesized beam is shown as the horizontal dashed line.}
\label{size_sfr}
\end{figure*}

Figure \ref{size_sfr} (left panel) shows the derived rest-frame ultraviolet galaxy sizes as a function of SFR, compared to a sample of $z\,{\approx}\,2$ field SMGs from \citet{swinbank2010} (with corresponding SFR estimates from \citealt{chapman2005}), which were observed by the ACS instrument onboard {\it HST\/} in the F775W filter (observed wavelength 770\,nm). These SMGs were selected from SCUBA surveys of various other cosmological fields, and the rest wavelengths probed by the observations range from 170 to 450\,nm, comparable to our coverage of 290\,nm. We next show a sample of three $z\,{\approx}\,2.5$ field dusty star-forming galaxies from \citet{barro2016} that were observed by the ACS F850LP filter (910\,nm in the observed frame, 260\,nm in the rest frame). These galaxies were selected from the CANDELS survey of the Great Observatories Origins Deep Survey-South (GOODS-S) field \citep{grogin2011} for their compact nature and brightness at far-infrared wavelengths, thus follow-up observations found them to be reasonably bright at submm wavelengths (${>}\,1\,$mJy at 870\,$\mu$m) and have large SFRs (${>}\,100\,$M$_{\odot}$\,yr$^{-1}$), so we simply refer to them as SMGs. In these comparison samples, the S{\'e}rsic index was allowed to vary, and the half-light radii are the semi-major axes of an ellipse containing half the total flux density, consistent with our definition of the half-light radius. 

We also include star-forming galaxies in the range $z\,{=}\,4$--5 from the ALPINE survey with reliable fits to imaging in both {\it HST\/}'s F160W filter and in ALMA moment-0 maps of [C{\sc ii}] line emission \citep{fujimoto2020}; for these galaxies, the rest-wavelength observed in the ultraviolet ranges from 230 to 280\,nm. For these measurements, the S{\'e}rsic index was fixed to $n\,{=}\,1$, but the authors note that fixing $n\,{=}\,0.5$ affected their size measurements only at the ${\approx}\,5\,$per cent level. For comparison, the mean S{\'e}rsic index from our fits is 0.84. \citet{fujimoto2020} only provides circularized sizes, defined with respect to our size measurements as $r_{\rm e}\,{=}\,r_{1/2}\sqrt{q}$, where $q$ is the semi minor-to-semi major axis ratio. In order to statistically convert this sample to semi-major axis sizes, we assume that the galaxies are circular discs with finite thickness parameterized by the ratio of the scale height to the disc radius, $q_0$, thus the relationship to the observed semi minor-to-semi major axis ratio $q$ is $\sin^2(i)\,{=}\,(1\,{-}\,q^2)/(1\,{-}q_0^2)$ \citep[e.g.][]{forster2018}, where $i$ is the inclination angle. We take $q_0\,{=}\,0.20$ \citep[e.g.,][]{genzel2008,law2012,vanderwel2014}, and calculate the average expected $\sqrt{q}$ assuming an isotropic distribution of galaxy orientations following the method outlined in Appendix A of \citet{law2009}. We find $\left\langle\sqrt{q}\right\rangle\,{=}\,0.72$, so we divide the ALPINE size measurements by this value. As a check, we calculated circularized sizes for our sample using the best-fit axis ratios from our S{\'e}rsic modelling and compared these directly to the ALPINE circularized size measurements, but did not find any systematic differences. For reference, we show the half-light radius of the {\it HST\/} F160W PSF (approximately half the FWHM of 0.151\,arcsec) as a horizontal dashed line.

\begin{figure*}
\includegraphics[width=0.49\textwidth]{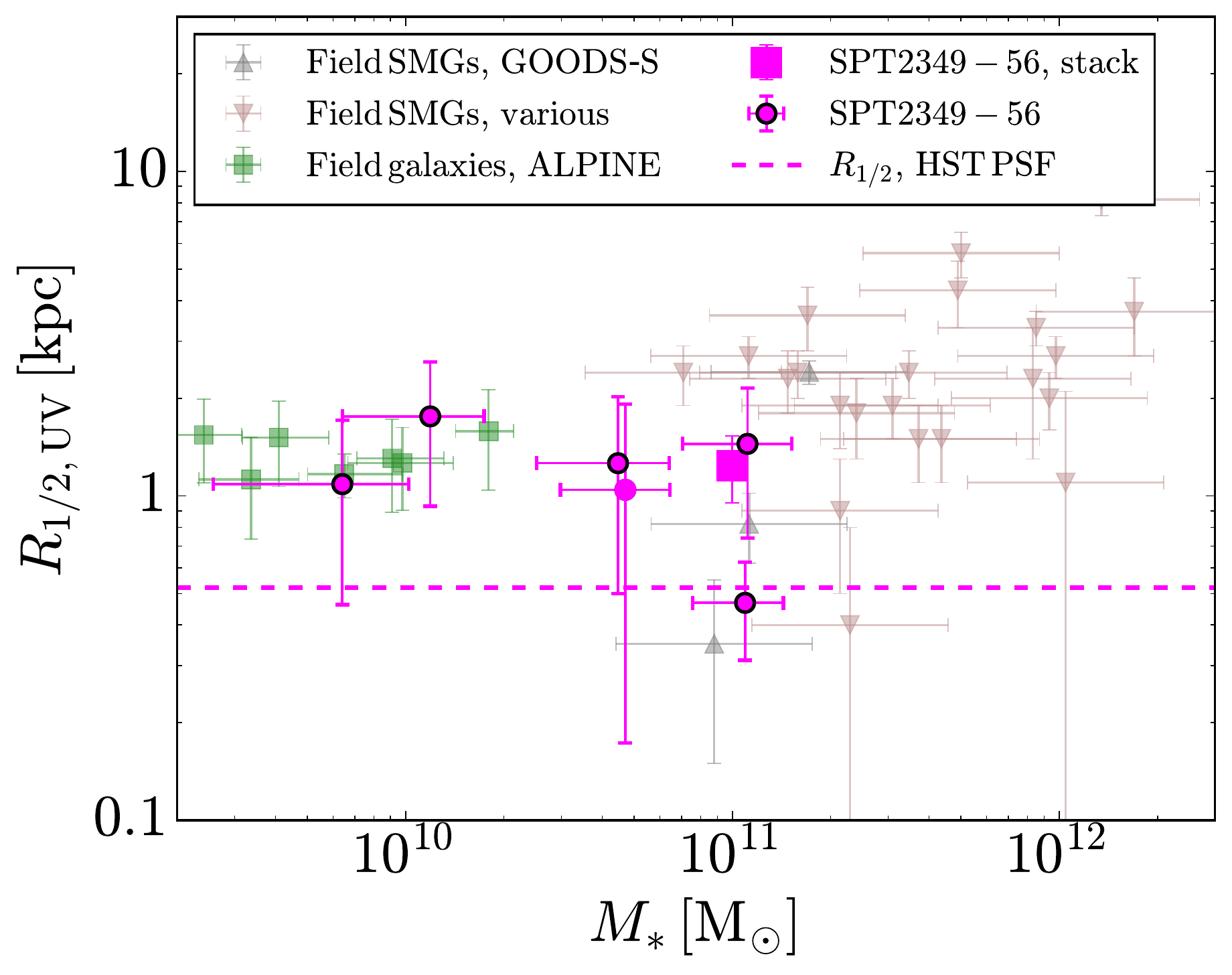}
\includegraphics[width=0.49\textwidth]{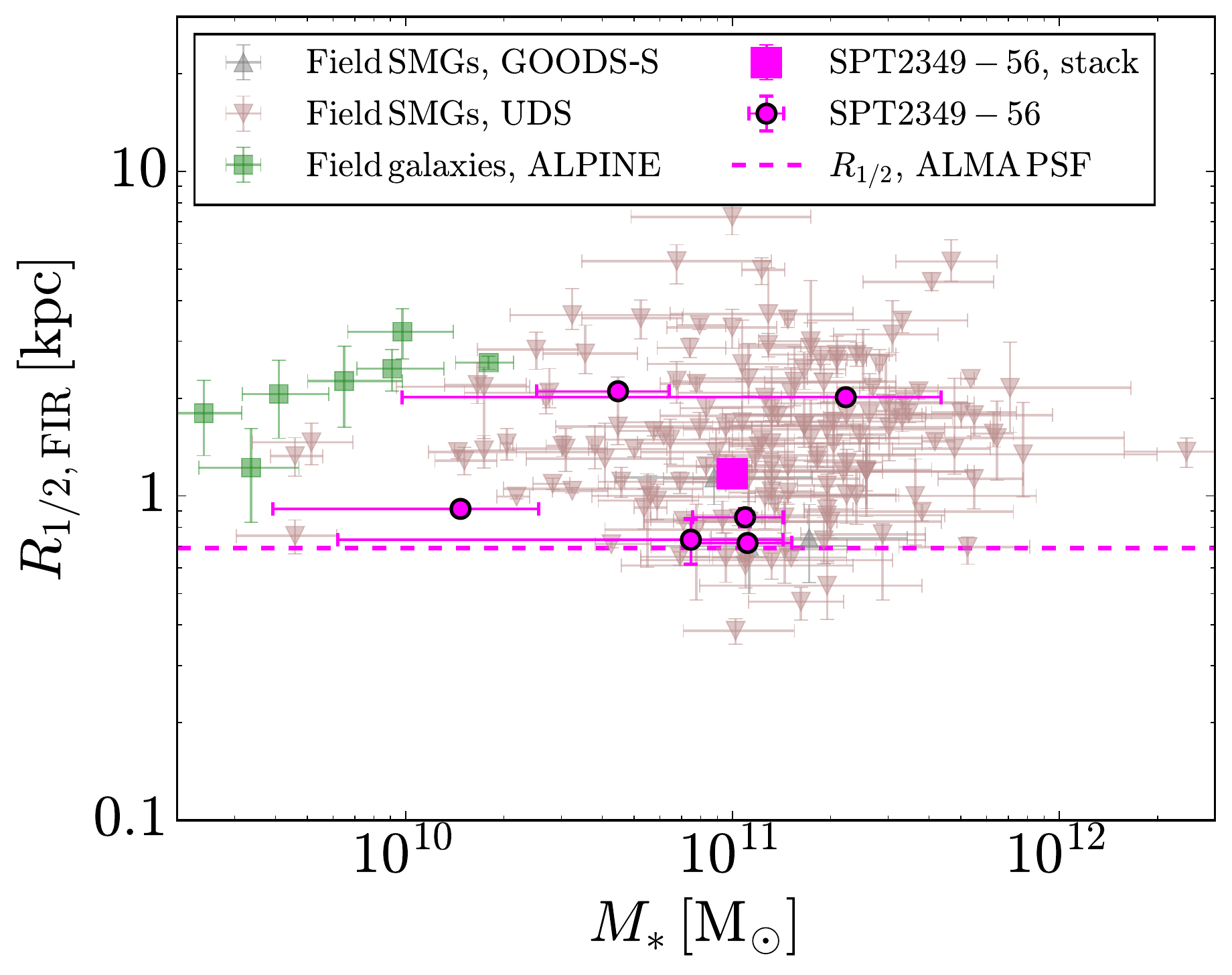}
\caption{{\it Left:} Ultraviolet half-light radius as a function of stellar mass for the same samples shown in Fig.~\ref{size_sfr}. The stacked size measurement (arbitrarily placed at 10$^{11}\,$M$_{\odot}$) and {\it HST} beam size are also shown for comparison. {\it Right:} Far-infrared half-light radius as a function of redshift for the same galaxies shown in Fig.~\ref{size_sfr}, along with the stacked size measurement and the ALMA beam size.}
\label{size_mstar}
\end{figure*}

\begin{figure*}
\includegraphics[width=0.49\textwidth]{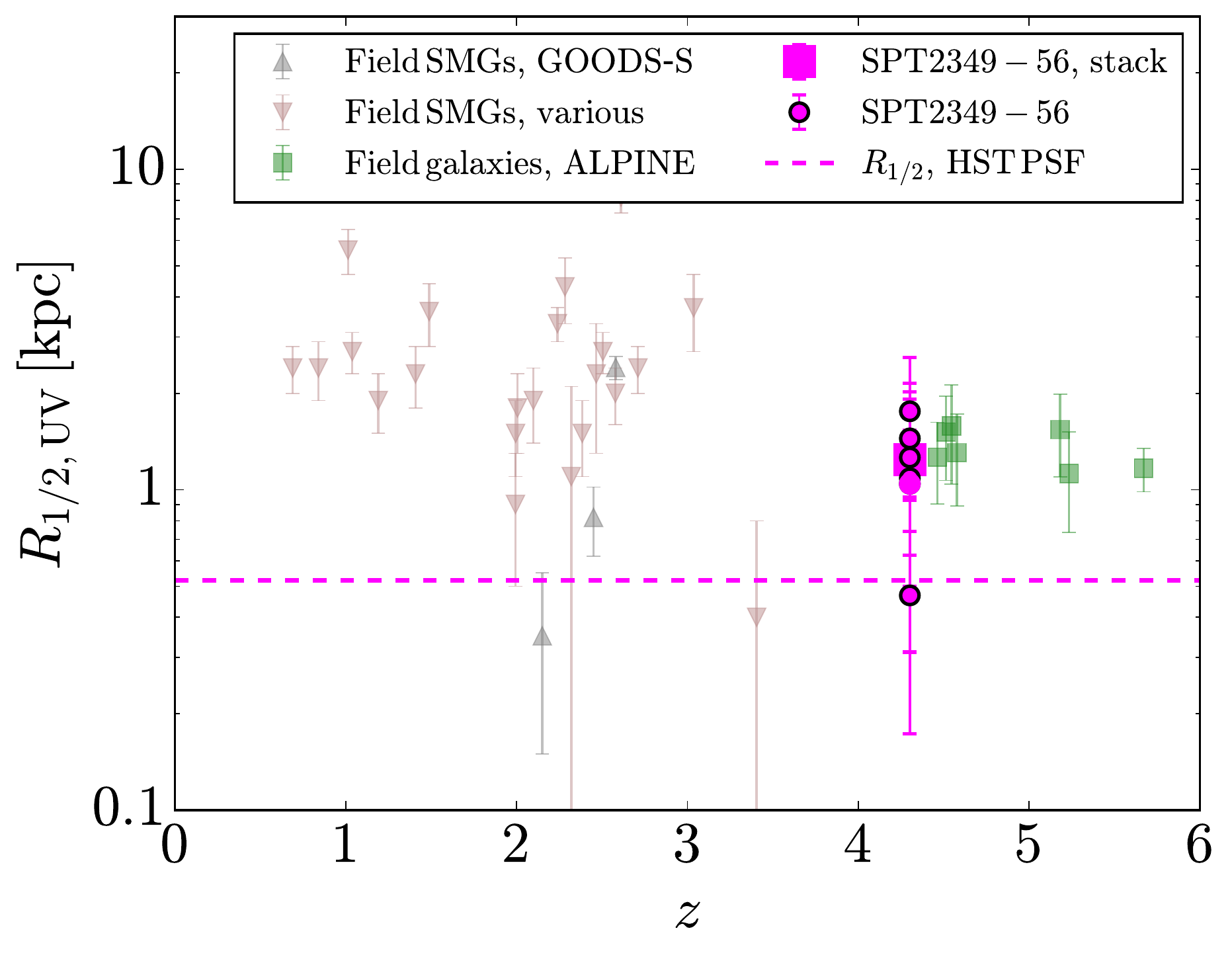}
\includegraphics[width=0.49\textwidth]{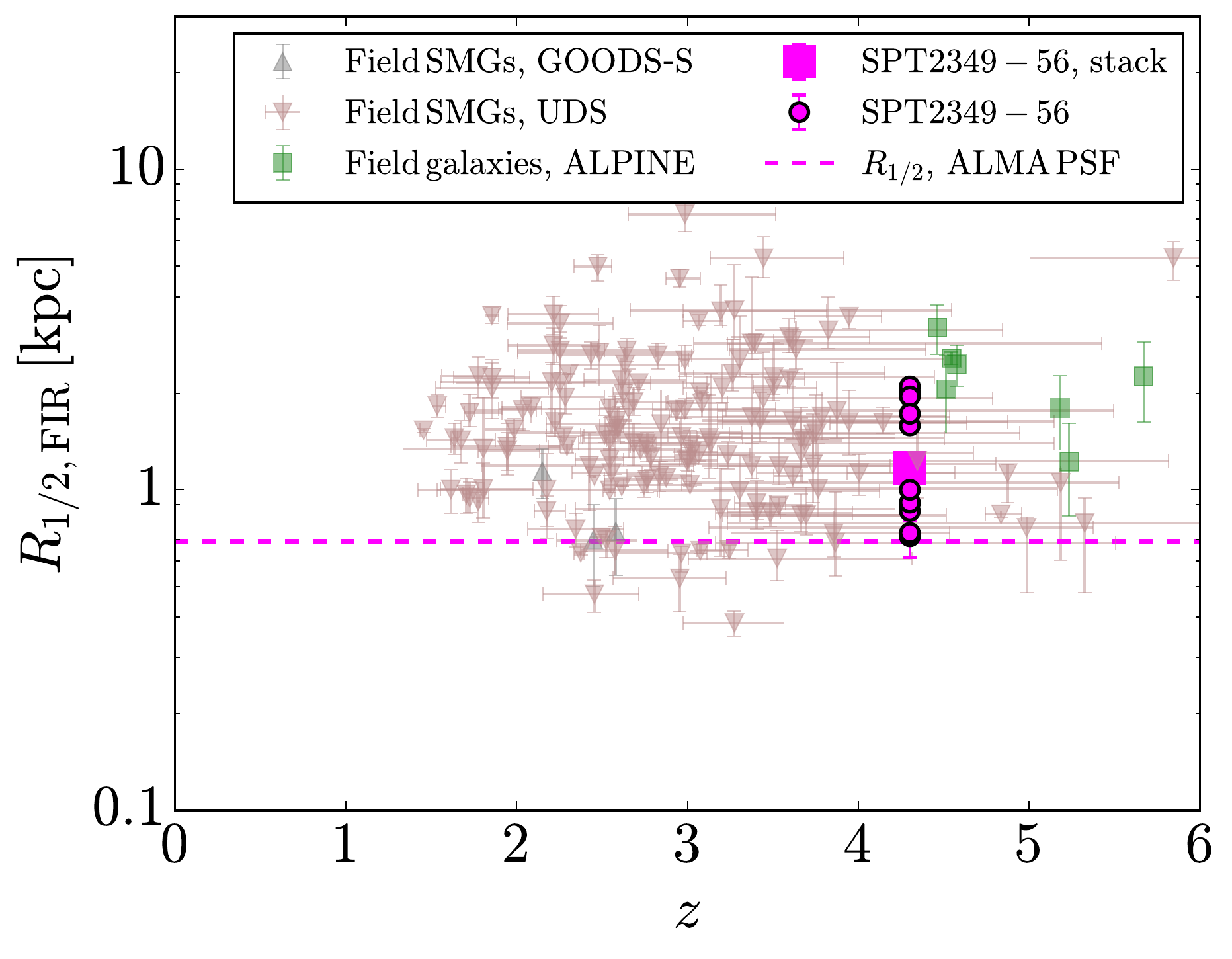}
\caption{{\it Left:} Ultraviolet half-light radius as a function of redshift for the same samples shown in Fig.~\ref{size_sfr}. The stacked size measurement and {\it HST} beam size are also shown for comparison. {\it Right:} Far-infrared half-light radius as a function of redshift for the same galaxies shown in Fig.~\ref{size_sfr}, along with the stacked size measurement and the ALMA beam size.}
\label{size_z}
\end{figure*}

Figure \ref{size_mstar} (left panel) shows the rest-frame ultraviolet sizes as a function of stellar mass. Also shown are the same comparison samples, with stellar masses taken from the same studies except for the $z\,{\approx}\,2$ field SMGs from \citet{swinbank2010}, where we use stellar masses obtained by \citet{michalowski2012}, and we show the size of our stack (arbitrarily placed at 10$^{11}\,$M$_{\odot}$) and the {\it HST\/} F160W PSF. Lastly, in Fig.~\ref{size_z} (left panel) we show rest-frame ultraviolet sizes as a function of spectroscopic redshift. Although no trends in SFR or stellar mass are apparent, the \citet{swinbank2010} SMGs are typically larger than the other comparison samples shown here, although this is to be expected owing to their larger stellar masses (Fig.~\ref{size_mstar}) and the known size-mass relation \citep[e.g.][]{vanderwel2014}.

We next turn to our rest-frame far-infrared sizes of galaxies in SPT2349$-$56, taken from \citet{hill2020}, which were calculated using the same procedure as the [C{\sc ii}] sizes outlined in Section \ref{cii_radius}, only in this case after averaging the line-free ALMA channels. Fig.~\ref{size_sfr} (right panel) shows these size measurements as a function of SFR, and compare them to the same field galaxies from \citet{barro2016} (which measure the rest-frame far-infrared at 250\,$\mu$m). In this comparison we have included the same galaxies from the ALPINE survey with [C{\sc ii}] size measurements from \citet{fujimoto2020} (corrected by the mean squareroot of the axis ratio expected from an isotropic distribution, $\left\langle\sqrt{q}\right\rangle\,{=}\,0.72$), and we have converted these [C{\sc ii}] size measurements to rest-frame far-infrared size measurements using the mean $R_{1/2,\mathrm{C[II]}}/R_{1/2,\mathrm{FIR}}$ ratio from our sample (described below). Again, the fits to the ALPINE galaxies were done with the S{\'e}rsic index fixed to 1, while the mean S{\'e}rsic index from our fits is 0.83, and we checked that a direct comparison between our circularized size measurements the the circularized ALPINE size measurements did not yield any systematic differences. This sample measures the rest-frame far-infrared at 160\,$\mu$m. We then include galaxies between $z\,{=}\,1.5$ and 5.8 with submm size measurements from the UDS field measured by ALMA \citep{gullberg2019}, in this case probing rest wavelengths between 130 and 350\,$\mu$m, converted to the size of the semi-major axis using the axis ratios provided. We also include the size of the stacked image, arbitrarily placed at 100\,M$_{\odot}$\,yr$^{-1}$. For reference, we show the half-light radius of the ALMA synthesized beam ($\sqrt{ab}\,{/}\,2\,{=}\,0.2\,$arcsec, where $a$ and $b$ are the major and minor FWHM, respectively) as a horizontal dashed line. 

In Fig.~\ref{size_mstar} (right panel) we show the same size measurements, this time as a function of stellar mass, including our stack arbitrarily placed at 10$^{11}\,$M$_{\odot}$, and the ALMA PSF. Lastly, Fig.~\ref{size_z} shows the same quantities as a function of redshift. The galaxies in SPT2349$-$56 span a range of sizes comparable to the literature samples, with no discernible trends in SFR, stellar mass, or redshift. This result is in agreement with the idea that not all SMGs are compact starbursts, and instead there is some heterogeneity in the SMG population \citep[e.g.,][]{hayward2012,hayward2013}.

\begin{figure*}
\includegraphics[width=0.49\textwidth]{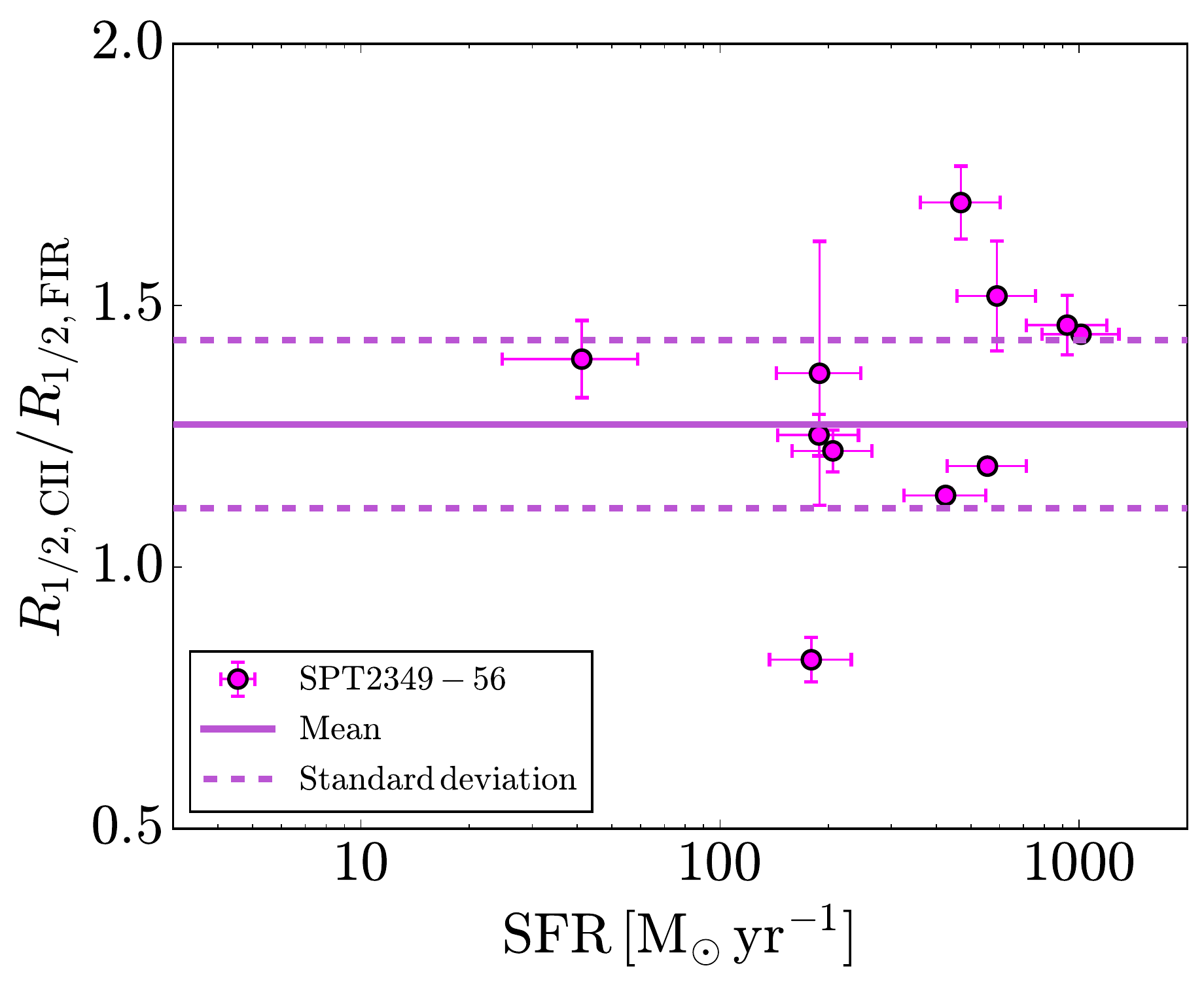}
\includegraphics[width=0.49\textwidth]{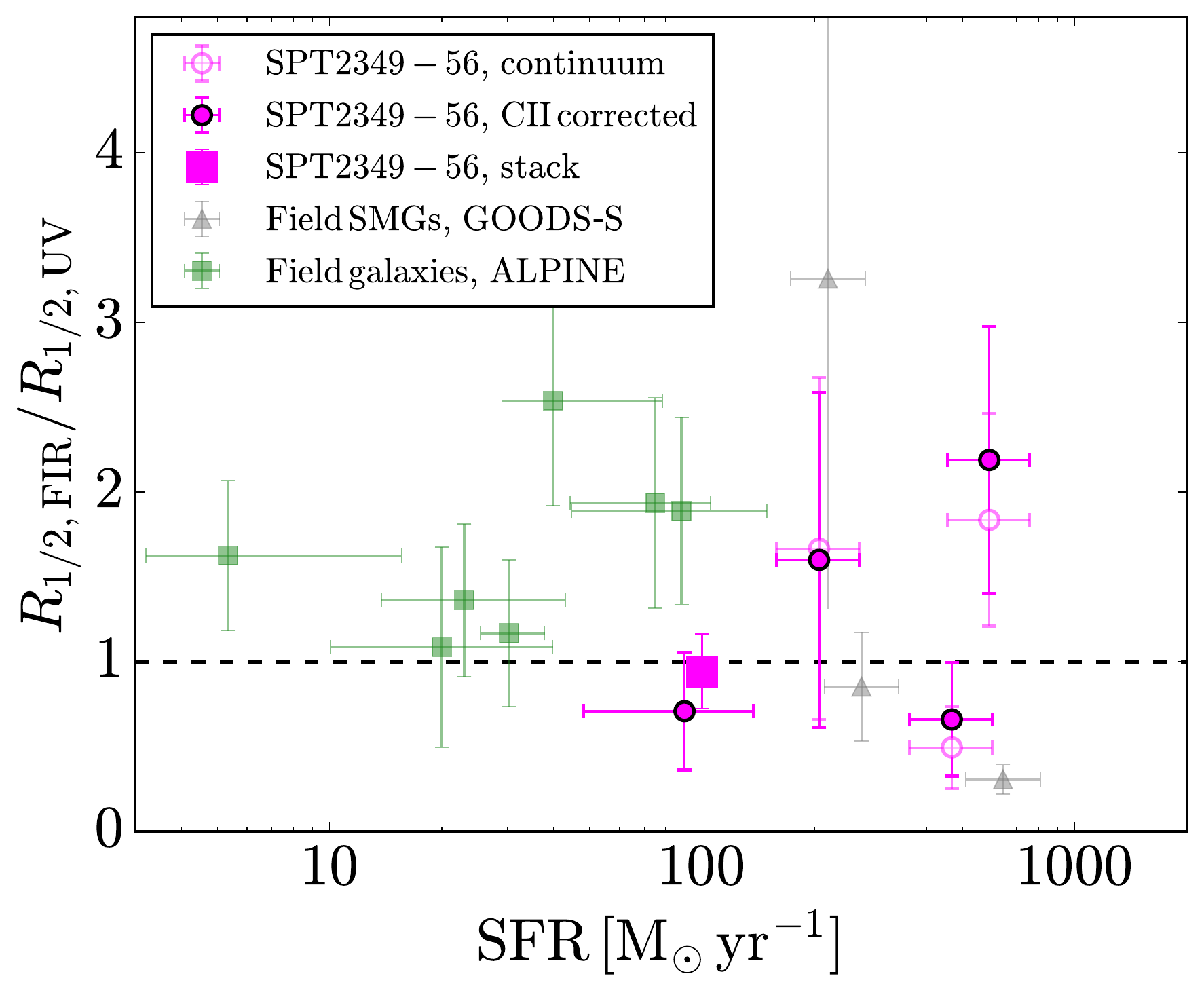}
\caption{{\it Left:} Ratio of [C{\sc ii}] half-light radius to far-infrared half-light radius as a function of SFR for galaxies in SPT2349$-$56. The mean value is 1.3 (solid line) with a standard deviation of 0.2 (dotted line). {\it Right:} Ratio of far-infrared half-light radius to ultraviolet half-light radius; this is shown as the open circles for the four galaxies in SPT2349$-$56 for which both measurements are available. Since there are only four galaxies with both size measurements available, we show the ratio $R_{1/2,\mathrm{C[II]}}/R_{1/2,{\rm UV}}$ as solid circles and correct it using the mean ratio of [C{\sc ii}] to far-infrared size from the left panel in order to estimate $R_{1/2,\mathrm{FIR}}/R_{1/2,{\rm UV}}$. The pink square shows the same ratio for the size measurements of our stacked images. We compare our size ratio measurements to the sample of field SMGs from the GOODS-S field \citep{barro2016} and the star-forming galaxies from the ALPINE survey \citep{fujimoto2020}. A dashed horizontal line is drawn where $R_{1/2,\mathrm{F}}\,{=}\,R_{1/2,{\rm UV}}$ for clarity.}
\label{size}
\end{figure*}

We now turn to the size ratios of our sample, and compare them to the literature. These comparisons should not depend on the choice of circularized size versus semi-major axis length, since we expect that the observed ellipticities of galaxies are equal at ultraviolet and far-infrared lengths. Fig.~\ref{size} (right panel) shows the ratio of the rest-frame far-infrared size to the ultraviolet size, $R_{1/2,\mathrm{FIR}}/R_{1/2,{\rm UV}}$, for the galaxies in SPT2349$-$56, as well as the size ratio for the stacks. We plot these ratios as a function of SFR as the SFRs of our sample are better-constrained than the stellar masses. All of the galaxies here are core galaxies of SPT2349$-$56, since our high-resolution submm imaging only covered this region. The field SMGs from \citet{barro2016} and \citet{fujimoto2020} are shown alongside our sources on this plot, but the field SMGs from \citet{swinbank2010} and \citep{gullberg2019} have not been observed at high resolution at both wavelengths and so are omitted from this comparison. Unfortunately, there are only three sources (C2, C6, and C8) for which we could measure both a far-infrared size and an ultraviolet size. To improve the sample size, we can look at the [C{\sc ii}] sizes for some of our sources, where the emission is much brighter, and use this as a proxy for the dust sizes.

Figure \ref{size} (left panel) plots the ratio $R_{1/2,\mathrm{C[II]}}/R_{1/2,\mathrm{FIR}}$ as a function of SFR for all sources where both measurements are available. We find a relatively consistent ratio across all SFR values, with a mean of 1.3$\,{\pm}\,$0.2. Next, in Fig.~\ref{size} (right panel) we use the ratio $R_{1/2,\mathrm{C[II]}}/R_{1/2,{\rm UV}}$ divided by the mean ratio above to provide an estimate of the underlying continuum size of sources where such a measurement is not available; this adds one more galaxy (C17) to the sample. 

The weighted mean of the size ratio for the galaxies in our sample is 1.4, with a standard deviation of 0.6, consistent with the size ratio of 0.9$\,{\pm}\,$0.2 from our stacked images, and in agreement with simulations of high-$z$ star-forming galaxies \citep[e.g.][]{cochrane2019}. Compared to the literature, given the large uncertainties and small sample sizes, we can only conclude that our galaxies show ratios consistent with the field.

\subsection{Radial surface brightness distributions}

\begin{figure*}
\includegraphics[width=\textwidth]{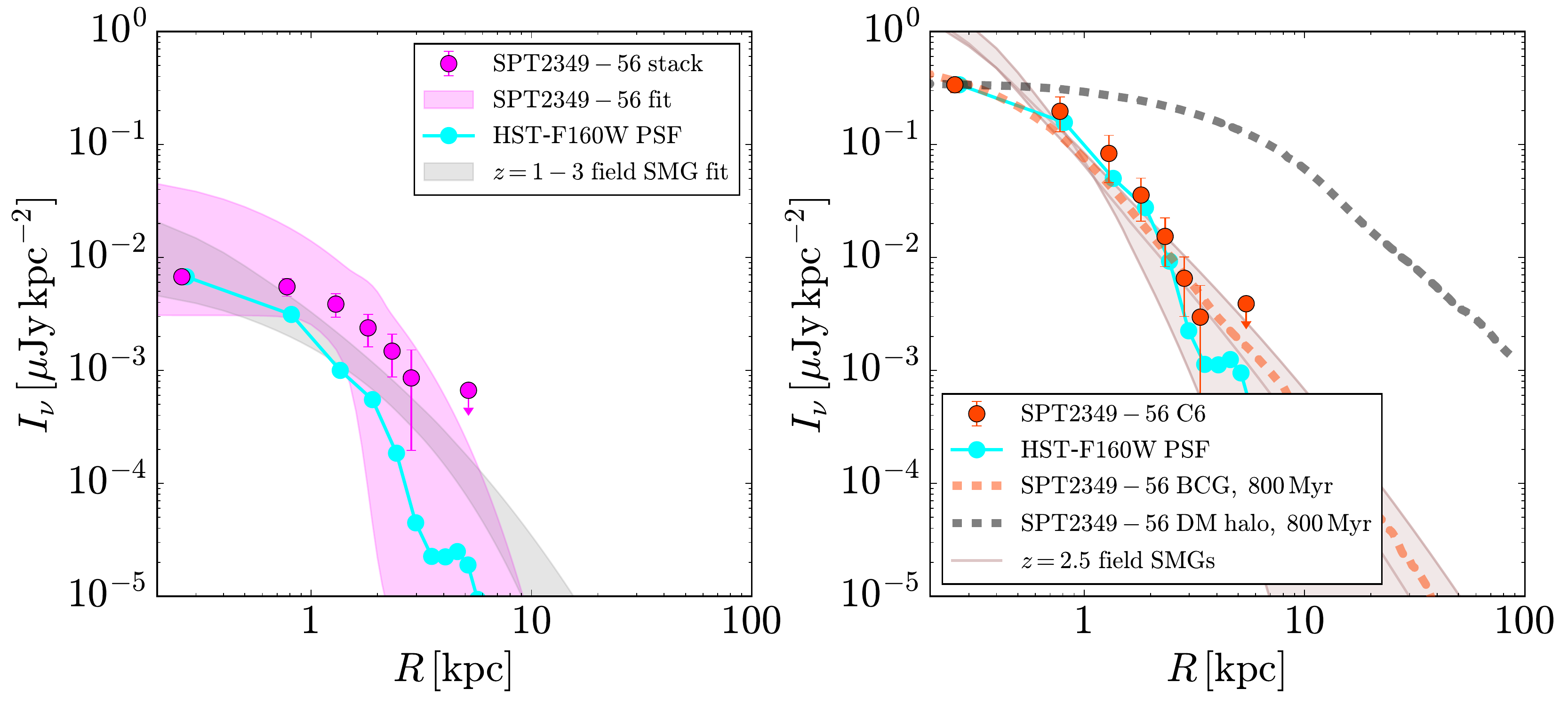}
\caption{Surface brightness as a function of radius for the protocluster galaxies in SPT2349$-$56, determined within elliptical annuli of 1-pixel width. The points and the error bars are the means and standard deviations of the pixel values within each annulus. Where the standard deviation overlaps with zero, we provide 1$\sigma$ upper-limit. {\it Left:} In magenta we show the surface-brightness profile for our stack of {\it HST\/}-detected galaxies (after removing C6, a bright outlier), with the best-fit S{\'e}rsic profile shown as the magenta shaded region, and in cyan we show the {\it HST\/}-F160W PSF. For comparison, we show the best-fit surface-brightness profile from a similar stacking analysis of 25 field SMGs (grey shaded region) observed by {\it HST\/} in the F775W filter by \citet{swinbank2010}, normalized to have the same integrated flux density as our stack. {\it Right:} In red we show the surface brightness profile for galaxy C6, and in cyan we show the same {\it HST\/}-F160W PSF. For comparison, in brown we show a set of three best-fit surface-brightness profiles of SMGs around $z\,{=}\,2.5$ from \citet{barro2016}, normalized by their stellar masses relative to C6. We have then convolved each profile with the {\it HST\/}-F160W PSF, assuming each source is at the same redshift as SPT2349$-$56, in order to provide a direct comparison with our observations. Here we have highlighted their total range for clarity. We show the expected stellar profile (as a dashed red curve) and dark matter profile (as a dashed black curve) of the BCG complex after 800\,Myr, once the central galaxies have merged \citep{rennehan2019}, also normalized by the peak pixel value of C6.}
\label{stack_profile}
\end{figure*}

In Figs.~\ref{size_sfr}--\ref{size_z} we see that C6, the central galaxy of the SPT2349$-$56 protocluster, has a much smaller rest-frame ultraviolet half-light radius compared to the rest of the protocluster galaxies with an available size measurement. However, most of the galaxies in our sample are not well-enough detected to measure ultraviolet sizes, so to perform a more statistical analysis we can turn to our stacked F160W image (Fig.~\ref{stack}). In Fig.~\ref{stack_profile} (left panel) we show the surface brightness of the stack as a function of radius, calculated in elliptical annuli with widths of 1 pixel, where the shape of the annuli was set to the best-fit ellipticity and position angle from the S{\'e}rsic profile fit. In this plot the points and the error bars are the means and standard deviations of the pixel values within each annulus, respectively. Here we have converted our surface brightness measurements to units of $\mu$Jy\,kpc$^{-2}$ by dividing by the number of kpc$^2$ in a pixel of our image. For the last point our measurement is consistent with zero, so we show the 1$\sigma$ upper-limit. We also show the best-fit S{\'e}rsic function as a shaded region, where the width is obtained by varying the best-fit parameters within their 68\,per cent confidence intervals. In Fig.~\ref{stack_profile} (right panel) we show its surface brightness profile of C6, calculated in the same way as with the stack.

In order to assess whether or not we are seeing resolved emission in these surface brightness profiles, we compare them to the {\it HST\/} F160W PSF model described in Section \ref{optical_radius}. Figure \ref{stack_profile} shows the surface profile of the model PSF for reference, normalized to the value of the central pixel of the stacked submm image (left panel) and galaxy C6 (right panel). Beyond about 2 pixels (or about 1\,kpc) our stacked galaxy image shows significantly more emission compared to the stacked star, so we are indeed measuring extended emission, yet galaxy C6 is effectively indistinguishable from an unresolved point source, confirming that C6 is more compact than the average of the other galaxies in SPT2349$-$56.

A similar stacking analysis was performed for the 25 spectroscopically-confirmed $z\,{\approx}$\,2 field SMGs of \citet{swinbank2010} shown in Figs.~\ref{size_sfr}--\ref{size_z}, and any differences in the mean profile of this field population compared to the mean profile of our protocluster galaxies could indicate the presence of interesting environmental effects. Upon rescaling and stacking their detections, it was found that a S{\'e}rsic index of 2.6$\,{\pm}\,$1.0 best described their data, in agreement with our value of 1.72$\,{\pm}\,$0.30. In Fig.~\ref{stack_profile} (left panel) we show their fit as a grey shaded region, where the width corresponds to the uncertainties in their best-fit parameters. We set the scale radius to be 2.7$\,{\pm}\,$0.4\,kpc, corresponding to the median half-light radius of their sample, and scale the amplitude to have the same integrated flux density as our stack, then convolve the 1-D profile with the {\it HST\/} beam in the F160W filter. We see that the surface brightness profile of our stack is consistent out to where our data are sensitive, thus we cannot conclude any differences are seen in the data.

We next turn to comparing galaxy C6 with the literature. The three GOODS-S galaxies at $z\,{\approx}\,2.5$ from \citet{barro2016} observed in the F850LP filter (rest-wavelength 260\,nm) were selected for their small rest-frame optical sizes and high stellar-mass densities, similar to what we are seeing with galaxy C6. In Fig.~\ref{stack_profile} (right panel) we show their resulting best-fit S{\'e}rsic profiles, with each normalization calculated by setting the ratio of the integrated flux density of a given galaxy to the integrated flux density of C6 equal to the ratio of the stellar mass of the same galaxy to the stellar mass of C6. We then convolve each profile with the {\it HST} beam in the F160W filter, converting the units of the beamsize from arcsec to kpc using the redshift of SPT2349$-$56 in order to provide a direct comparison with our observations. We have highlighted the range in surface brightness they span for clarity. We see that the compact SMGs in this sample would all effectively appear unresolved in our {\it HST\/} imaging, similar to C6.

We can also compare the current profile of galaxy C6 to a prediction of its profile from the hydrodynamical simulation presented by \citet{rennehan2019}. Briefly, the simulation evolved the 14 core galaxies initially discovered by \citet{miller2018} for 1\,Gyr in several separate and independent realizations. Each galaxy was initialized with a dark matter halo and a stable gas and stellar disc, with component masses scaled from the available gas mass estimates in \citet{miller2018} and discs modeled following an exponentially decreasing surface density with a scale size related to the angular momentum \citep{robertson2006}. In particular, we select four realizations where the stellar masses were computed assuming a molecular gas-to-stellar mass fraction of 2.3 (for reference, the range measured for these galaxies is 0.1 to 4.8), and the halo masses were scaled from the stellar masses by a factor of 100. In each realization the positions were randomly selected to lie within a 65\,kpc-radius sphere, and the velocities were drawn from a Gaussian distribution matching the measured line-of-sight velocity dispersion. For each realization, we take the median mass profile between 700 and 800\,Myr, then take the mean profile across these realizations. The overall normalization of the profile is highly uncertain in the simulation as it requires radiative transfer models to convert stellar mass into rest-frame ultraviolet flux density, so we simply normalize the profile by the peak pixel in our F160W imaging of C6. The resulting profile is shown in Fig.~\ref{stack_profile}, and we see that after the merger, the BCG will remain quite compact. Nonetheless, we expect the dark matter halo to grow in size after the merger, so in Fig.~\ref{stack_profile} we also show the resulting shape of the dark matter halo, normalized in the same way for easy comparison with the stellar profile. We see that the shape of the dark matter halo after the merger is much more extended than the stellar mass profile.

\section{Discussion}
\label{discussion}

\subsection{The stellar properties of SPT2349$-$56}

The protocluster SPT2349$-$56 is a unique object when observed in the submm. We now know that the total SFR exceeds 10,000\,M$_{\odot}$\,yr$^{-1}$, and its extreme nature in this regard is due to the fact that it was specifically selected as one of the brightest unlensed point sources in the SPT mm-wavelength survey. Now that we have studied this protocluster at optical and infrared wavelengths, we have begun to probe the properties of the constituent stars themselves.

We have found that the stellar masses derived through SED fitting place these galaxies inconspicuously on the MS, as opposed to appearing as outliers above it; given their incredibly high SFRs, this means that they have correspondingly large stellar masses, which are simply hidden by dust. For reference, the total stellar mass of all the known galaxies in SPT2349$-$56 is (1.5$\pm$0.3)$\,{\times}\,10^{12}$M$_{\odot}$, which is comparable to a large BCG at $z\,{<}\,0.1$.

Based on our unequal-variance $t$-test between field SMGs and star-forming protocluster galaxies, assuming that the molecular gas-to-stellar mass fractions and depletion timescales are drawn from Gaussian distributions with arbitrary variances, we are able to reject the null hypothesis that the means of the two distributions are equal, although the scatter between the two populations can overlap. Interestingly, similar studies of CO lines at moderate redshift ($z\,{=}\,1$--3) have found that star-forming galaxies in cluster/protocluster environments at this epoch have systematically higher molecular gas-to-stellar mass fractions and depletion timescales compared to scaling relations derived from the field \citep[e.g.,][]{noble2017,hayashi2018,tadaki2019}, while at $z\,{<}\,1$ galaxy clusters have nearly depleted all of their gas and ceased forming stars \citep[e.g.,][]{young2011,jablonka2013,scott2013,boselli2014,zabel2019}. 

\begin{figure*}
\includegraphics[width=0.75\textwidth]{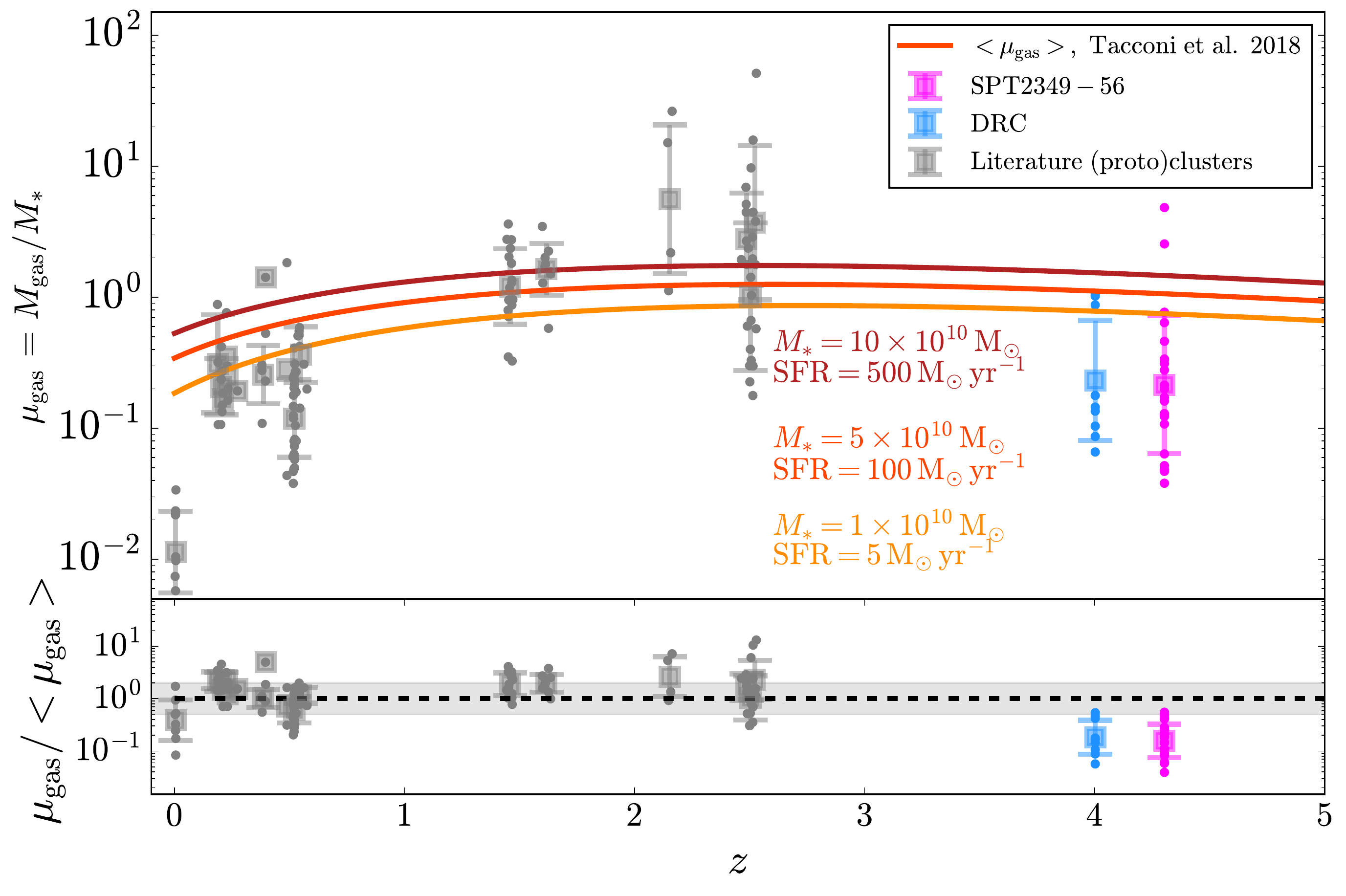}
\includegraphics[width=0.75\textwidth]{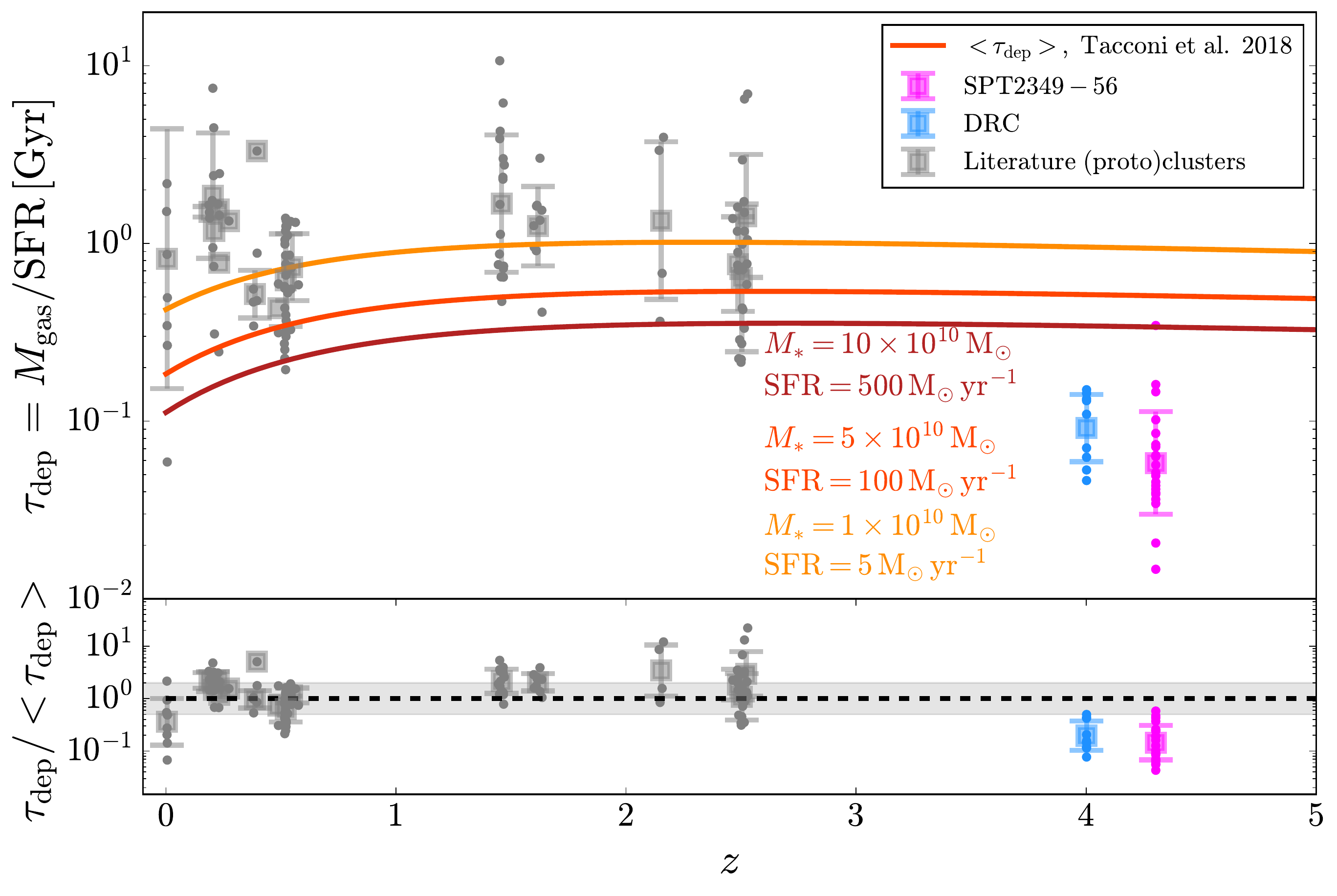}
\caption{{\it Top:} Molecular gas-to-stellar mass fraction ($\mu_{\rm gas}\,{=}\,M_{\rm gas}\,{/}\,M_{\ast}$) as a function of redshift for SPT2349$-$56 (pink), the DRC (blue), and various clusters and protoclusters with similar CO-derived molecular gas mass data available (grey; references are provided in the text). Points indicate values for individual galaxies, while squares show the mean values for each (proto)cluster, and error bars show the standard deviations for (proto)clusters with more than two sources available. Representative models for the mean molecular gas-to-stellar mass fraction of field galaxies ($\left\langle\mu_{\rm gas}\right\rangle$) from \citet{tacconi2018} are shown as the solid curves. The bottom panel shows the ratio $\mu_{\rm gas}\,{/}\,\left\langle\mu_{\rm gas}\right\rangle$, calculated for each galaxy given its redshift, $M_{\ast}$, and SFR, along with the means and standard deviations for each (proto)cluster. The shaded region indicates the expected intrinsic scatter of ${\pm}\,0.3\,$dex \citep{schreiber2015}. {\it Bottom:} Same at the top panel, only showing the depletion timescale ($\tau_{\rm dep}\,{=}\,M_{\rm gas}\,{/}\,$SFR).}
\label{gas_frac_z}
\end{figure*}

To quantify this, in Fig.~\ref{gas_frac_z} we show the molecular gas-to-stellar mass fraction (top panel) and depletion timescale (bottom panel) as a function of redshift for the galaxies in SPT2349$-$56, compared to other clusters and protoclusters with molecular gas mass estimations made through observations of CO, and with sufficient multiwavelength coverage to have stellar mass and SFR estimates. At $z\,{=}\,4.0$ we show the DRC \citep{oteo2018,long2020}, and we have converted CO(6--5) line intensities to molecular gas masses using an $L^{\prime}_{(6-5)}/L^{\prime}_{(1-0)}$ factor of 0.46 (the mean ratio found for the SPT-SMG sample, see \citealt{spilker2014}) and an $\alpha_{\rm CO}$ of 1\,M$_{\odot}/$(K\,km\,s$^{-1}$\,pc$^2$) (the same scale factor used for our sample). At intermediate redshift ($z\,{=}1$--3) we show results from observations of CO(3--2), CO(2--1), and CO(1--0) in members of three {\it Spitzer\/} Adaptation of the Red-sequence Cluster Survey (SpARCS) clusters \citep{noble2017}, XMMXCS J2215.9$-$1738 \citep{hayashi2018}, three Ly$\alpha$-selected protoclusters \citep{tadaki2019}, and two potentially associated overdensities identified in the COSMOS field, CLJ1001 at $z\,{=}\,2.50$ \citep{wang2016,wang2018} and PCL1002 at $z\,{=}\,2.47$ \citep{casey2015,champagne2021}. For the COSMOS structure, we show the measurements of the individual galaxies in CLJ1001 from \citet{wang2018}, and the properties of the single galaxy in PCL1002 with a CO(1--0) detection from \citet{champagne2021}; however, \citet{champagne2021} independently derived the unresolved properties of CLJ1001, finding a 25\,per cent shorter depletion timescale compared to the average reported by \citet{wang2018}. At low redshift ($z\,{<}1$) we take observations of CO(3--2), CO(2--1), and CO(1--0) in members of the Fornax cluster \citep{zabel2019}, Abell 2192 and Abell 963 \citep{cybulski2016}, CL1411.1$−$1148 \citep{sperone2021}, CL0024$+$16 \citep{geach2009,geach2011}, and MACS J0717.5$+$3745, Abell 697, 963, 1763, and 2219 \citep{castignani2020}. These comparison samples consistently used CO conversion factors of $L^{\prime}_{(2-1)}/L^{\prime}_{(1-0)}\,{=}\,0.8$ and $L^{\prime}_{(3-2)}/L^{\prime}_{(1-0)}\,{=}\,0.5$, and $\alpha_{\rm CO}\,{\approx}\,$4\,M$_{\odot}/$(K\,km\,s$^{-1}$\,pc$^2$), with small corrections for metallicity made in some cases. This conversion factor is appropriate for normal star-forming galaxies, which were the targets of these literature studies, and a lower conversion factor around 1\,M$_{\odot}/$(K\,km\,s$^{-1}$\,pc$^2$) is typically used for SMGs, as with SPT2349$-$56 and the DRC. Some galaxies in the sample of \citet{castignani2020} approach the high SFRs in the SMG regime, and they have adopted a smaller conversion factor for those sources. For each cluster in this figure, we show the mean molecular gas-to-stellar mass fraction and depletion timescale as a square symbol, with error bars representing the standard deviation for clusters with more than two galaxies with sufficient data. 

To represent field galaxies in Fig.~\ref{gas_frac_z}, we use the scaling relations for $\left\langle\mu_{\rm gas}\right\rangle$ and $\left\langle\tau_{\rm dep}\right\rangle$ (where $\left\langle \right\rangle$ denotes the average of the field population) estimated by \citet{tacconi2018}, derived from observations of CO lines and continuum flux densities up to millimetre wavelengths in over 1300 field galaxies between $z\,{=}\,0$ and 4 (including the sample from \citealt{scoville2016}). They provide equations for calculating $\left\langle\mu_{\rm gas}\right\rangle$ and $\left\langle\tau_{\rm dep}\right\rangle$ as a function of redshift, $M_{\ast}$, SFR, and the effective radius at rest-frame 500\,nm; however, since we do not have access to size measurements for most of the galaxies in our comparison sample, we use their fits that do not include this parameter. Figure \ref{gas_frac_z} shows several representative curves for $\left\langle\mu_{\rm gas}\right\rangle$ and $\left\langle\tau_{\rm dep}\right\rangle$ given different values of $M_{\ast}$ and SFR. In the bottom panels of Fig.~\ref{gas_frac_z} we show the ratios $\mu_{\rm gas}\,{/}\,\left\langle\mu_{\rm gas}\right\rangle$ and $\tau_{\rm dep}\,{/}\,\left\langle\tau_{\rm dep}\right\rangle$ (including the expected intrinsic scatter of ${\pm}\,0.3\,$dex, see \citealt{schreiber2015}), obtained by dividing each galaxy's molecular gas-to-stellar mass fraction and depletion timescale by the prediction from \citet{tacconi2018}, taking into account each galaxy's redshift, $M_{\ast}$, and SFR. We confirm that at $z\,{=}\,1$--3 most cluster environments are gas-rich, as discussed in previous studies, and we see that the galaxies in SPT2349$-$56 and the DRC continue to fall below the expected molecular gas-to-stellar mass fractions and depletion timescales of field galaxies, although the intrinsic scatters of these populations are still large and often overlap.

In our comparison of the stellar mass function of SPT2349$-$56 with the stellar mass function of $z\,{=}\,1$ galaxy clusters, we found similar shapes well-described by Schechter functions. Since we know that the core galaxies will merge over a timescale of a few hundred Myr, we also computed the stellar mass function of SPT2349$-$56 after summing up the stellar masses of these galaxies and treating them as a single source. In this case we found that the number counts are better-fit by a single power law, indicating that if this structure is to continue along an evolutionary path to become a $z\,{=}\,1$ galaxy cluster, the remaining cluster stellar mass will come from lower-mass galaxies \citep[e.g.][]{naab2009}. However, we have not taken into account the observational biases and incompleteness inherent in our sample, and so it is not clear from the current data whether these galaxies are already within the 1\,Mpc environment of SPT2349$-$56 and have not been detected, or have yet to fall into the galaxy protocluster. It is worth noting that this behaviour is consistent with the notion of `downsizing', where the most massive galaxies formed the earliest times, which has been observed in numerous samples of galaxies \citep[e.g.,][]{cowie1996,magliocchetti2013,miller2015,wilkinson2017}.

Lastly, we have noted that the ratio of far-infrared size to ultraviolet size is comparable to star-forming galaxies found around the same redshift, and field SMGs at lower redshift ($z\,{\approx}\,2.5$). A similar study at $z\,{\approx}\,1$ also found that the stellar emission in cluster galaxies is more compact than in field galaxies \citep{matharu2019}, which is not what we are seeing here, although the sample sizes we are investigating are small, and there could be systematic differences in the size measurements. Nonetheless, we might not expect to see many differences between field galaxies and protocluster galaxies at high redshift as there has not been enough time for the clustering environment to shape the residing galaxies.

\subsection{The properties of a BCG in formation}

In our analysis above, we have paid special attention to separating the central galaxies of SPT2349$-$56 from the wider protocluster. In \citet{rennehan2019}, hydrodynamical simulations using the positions of these central galaxies as the initial conditions predicted a complete merger on timescales of a few hundred Myr, while in \citet{hill2020} it was found that the velocity distribution within this region was consistent with a Gaussian distribution, and that the velocity dispersion predicted a central mass of $(9\,{\pm}\,5)\,{\times}\,10^{12}$\,M$_{\odot}$. Summing up the stellar masses of each of these central galaxies yields a value of $(9\,{\pm}\,2)\,{\times}\,10^{11}$\,M$_{\odot}$, consistent (within the uncertainties) with the mass of $(12\,{\pm}\,3)\,{\times}\,10^{11}$\,M$_{\odot}$ estimated by \citet{rotermund2021}.

Looking at Fig.~\ref{ms}, we see that these central galaxies lie close to the MS found by \citet{khusanova2021}, with no statistically significant offsets given the large uncertainties. Similarly, the two galaxies from the northern component of SPT2349$-$56 are significantly above the MS, but small number statistics again mean that we cannot draw any statistically significant conclusions from this observation. The MS distribution of SPT2349$-$56 overall matches well with what is seen with the DRC, a similar star-forming protocluster from the literature, along with other samples of field SMGs at high redshift.

\begin{figure*}
\includegraphics[width=\textwidth]{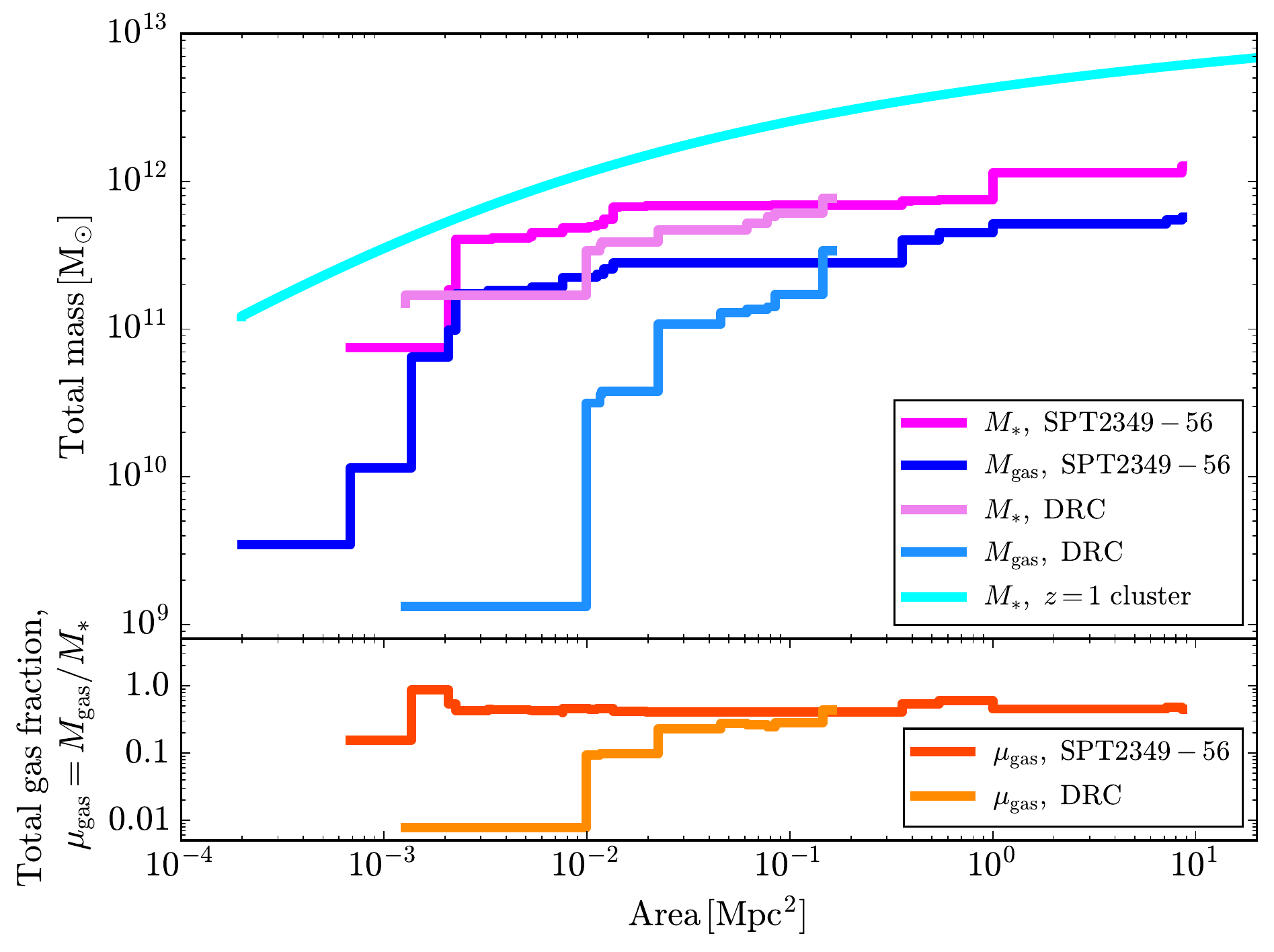}
\caption{{\it Top:} Cumulative mass enclosed within a circular aperture as a function of the area of the aperture, centred on the infrared luminosity-weighted centre of SPT2349$-$56 (see \citealt{hill2020} for details). In blue we show the molecular gas mass, and in magenta we show the stellar mass. The two masses track each other well, implying that the inner region of the protocluster has not had a chance to differentiate from the outer regions. Also shown are the stellar and molecular gas masses of the DRC \citep{oteo2018,long2020}, and the best-fit NFW profile to the stellar mass profile of a stack of $z\,{=}\,1$ clusters from \citet{vanderburg2014}. {\it Bottom:} Total molecular gas-to-stellar mass fraction $\mu_{\rm gas}$ (Eq.~\ref{gas_mass}) enclosed within the same apertures as a function of area.}
\label{growth}
\end{figure*}

In Fig.~\ref{growth} we explore the concept of environmental dependence further by plotting the cumulative mass enclosed within a circular aperture as a function of the area of the aperture, separating out the stellar mass and the molecular gas mass. In this plot the centre of SPT2349$-$56 is the luminosity-weighted centre, as in \citet{hill2020}. We see that the molecular gas mass and stellar mass track one another across all scales probed by our data, from the region of the forming BCG (90\,kpc, or about 0.03\,Mpc$^2$) out to the northern component (about 0.5\,Mpc away, or at 1\,Mpc$^2$). To quantify this, in the bottom panel of Fig.~\ref{growth} we show the total enclosed molecular gas-to-stellar mass fraction, and we can see that it remains roughly constant at a level of about 0.8 (except for near the centre, but these fluctuations suffer from small-number statistics). For comparison, we show the same curve-of-growth for the DRC \citep{oteo2018,long2020}; the behaviour of this protocluster is similar. 

We then compare the stellar mass profiles of these high-$z$ protoclusters to the stellar profile of a typical $z\,{\simeq}\,1$ galaxy cluster from \citet{vanderburg2014}, obtained from the stack of the same sample of 10 clusters discussed in Section \ref{counts}. The authors found a best-fit Navarro-Frenk-White (NFW) profile concentration parameter of 7$^{+1.53}_{-0.99}$, which we use to plot the mass projected within a cylinder of a given area \citep[see e.g.][]{lokas2001}, taking the scale stellar mass to be $M_{200,\ast}\,{=}\,2\,{\times}\,10^{12}\,$M$_{\odot}$, the median of their sample. We see that the shapes of the stellar mass curves between the protoclusters and the $z\,{\simeq}\,1$ clusters are similar, although the slope of the protoclusters becomes shallower than the slope of the $z\,{=}\,1$ clusters at large radii. It is interesting to note that a similar study \citep{alberts2021} investigating stacks of galaxy clusters between $z\,{=}\,0.5$ and 1.6 in the near-infrared (3--8\,$\mu$m in the rest-frame) found similarly-concentrated light profiles, with NFW concentration parameters around 7; this near-infrared light is expected to trace stellar mass. They also investigated stacks in the far-infrared (250--500\,$\mu$m in the rest-frame), tracing dust emission and SFR, and found concentration parameters comparable to the near-infrared light. It was found that 20--30\,per cent of the integrated cluster far-infrared emission comes from high-mass galaxies, while in \citet{hill2020} about 50\,per cent of the low-resolution single-dish far-infrared emission resolved into massive galaxies.

Despite the fact that the galaxies on-track to merge into a BCG are not altogether distinguishable from the rest of the protocluster (or indeed, from most SMGs at these redshifts), galaxy C6 does stand out from our core sample as having the brightest flux density at all optical-through-infrared wavelengths. Our SED modelling found that this galaxy's stellar mass is $(4\,{\pm}\,1)\,{\times}\,10^{11}$\,M$_{\odot}$, or roughly half of the total stellar mass expected to make up the BCG after the mergers are complete, and our brightness profile analysis of this galaxy suggests that it is incredibly compact.

The region around galaxy C6 is clearly at the centre of this BCG-in-formation. \citet{rotermund2021} already discussed a plausible evolutionary track for the growth of the stellar mass of C6. Around $z\,{=}\,0.5$, BCG stellar masses range from 5--10$\,{\times}\,10^{11}\,$M$_{\odot}$ \citep[e.g.,][]{hilton2013}, so C6 is already nearly there, and once the merging is complete, it will be at the upper end of BCG masses known. In Fig.~\ref{stack_profile} we see that C6 is expected to remain compact after the merger, and so if C6 is to continue to grow in stellar mass, we might also expect it to grow by a considerable amount in size. This could happen through dry mergers (mergers between galaxies with little gas, and thus little star formation), which has been found to play an important role in the growth of BCGs \citep[e.g.,][]{liu2009,lin2010,liu2015}, and simulations predict that this process will increase a BCGs size and make it less compact \citep[e.g.,][]{khochfar2006,oogi2012}. This would need to occur after C6 merges with the core galaxies in its current vicinity, as these mergers will be gas-rich. From our simulation we see that the post-merger dark matter halo will have a large and extended profile, and this could become populated by further accretions and dry mergers if C6 grows following an inside-out scenario, where the slope of the outer profile becomes shallower with increasing mass \citep[e.g.,][]{vandokkum2010,bai2014,whitney2019}. While we stress that the precise final mass and size of a BCG depends strongly on its detailed merger history, something we cannot know from system to system, we see that C6 is at least consistent with the picture that it is the progenitor of one of the most massive BCGs seen today, and that it has nearly formed all of its stars at this early epoch of formation.

These observations of a BCG in formation provide a direct measure of the ingredients that built up these massive galaxies in the early Universe. There are many studies of the stellar populations of BCGs that try to piece together their formation histories, which broadly point to a fast and early core formation phase (${>}\,10\,$Gyr ago) followed by a slow and continuous accretion phase (${<}\,10\,$Gyr ago) fuelled by minor mergers respondible for assembling the outer regions \citep[e.g.,][]{delucia2007,collins2009,barbosa2016,cooke2019,edwards2020}. Here we provide a direct observation of this formation in progress, where we have access to information that will likely be lost after the merger of the core galaxies in SPT2349$-$56 into a BCG. In particular, we see that the ingredients of this BCG are numerous galaxies with very high star-formation rates that, once merged, will already have formed most of the stars that make up a typical BCG. Furthermore, stellar population studies are only able to trace BCG histories back to their early star-forming phase, which in the case of SPT2349$-$56, will begin after the merger. But by observing these galaxies before they merge into a BCG, we now have access to information about the stellar history of a BCG back to a much earlier time, as traced by the stellar populations of the pre-merger galaxies.

\section{Conclusion}
\label{conclusion}

SPT2349$-$56 was selected as the brightest protocluster candidate from the SPT-SZ 2500\,deg$^2$ mm-wavelength survey. This object has since been spectroscopically confirmed to be a true protocluster, containing over 30 submm-bright galaxies and dozens more LBGs and LAEs. In this paper we have described our results from an extensive optical-through-infrared follow-up campaign using observations from Gemini-GMOS and FLAMINGOS-2, {\it HST\/}-F110W and F160W, and {\it Spitzer\/}-IRAC.

Owing to the ${\approx}\,2\,$arcsec spatial resolution of the IRAC images, source blending is an issue that needs to be dealt with. We deblended our IRAC data using {\tt t-phot}, which used galaxy positions from our {\it HST\/} imaging as a prior and subsequently convolved their profiles with the IRAC PRF to develop a catalogue of the underlying source distribution as seen by IRAC. Source catalogues for our Gemini and {\it HST\/} imaging were extracted using standard {\tt source-extractor} routines.

We matched the known protocluster galaxies in SPT2349$-$56 discovered by ALMA to our optical and infrared catalogues using a simple radial cut of 1\,arcsec. We found a match in at least one of the eight optical/infrared filters for all but six galaxies. In addition, we searched for ALMA counterparts to a small sample of LBGs and LAEs, and found matches for all but one galaxy.

Taking the photometry measured in these data and combining it with existing mm-wavelength photometry from {\it Herschel\/}-SPIRE and ALMA, we used {\tt CIGALE} to fit SEDs to each galaxy, allowing stellar masses, dust extinctions, ages, star-formation timescales, and star-formation rates to vary.

We found that the galaxies in SPT2349$-$56 follow the galaxy main sequence, consistent with other samples of $z\,{\simeq}\,4$ protocluster galaxies and field SMGs. However, we find small molecular gas-to-stellar mass fractions and short depletion timescales compared to field SMGs at similar redshifts. We perform unequal-variance $t$-tests, rejecting the null hypothesis that the molecular gas-to-stellar mass fractions and depletion timescales of protocluster galaxies and field SMGs have molecular gas-to-stellar mass fractions and depletion timescales drawn from Gaussian distributions with equal means, although the scatter in both populations is large. We find the same result using known scaling relations calibrated from large samples of field galaxies up to redshift 4. This could mean that protocluster galaxies in SPT2349$-$56 are at a late stage in their star-formation phase and have already nearly depleted their gas reservoirs as they build up their stars.

We computed the stellar-mass function and gas-mass function of SPT2349$-$56 in two ways: first, for the entire sample of galaxies; and second, by collapsing the core galaxies into a single source, reflecting the fact that they are expected to merge within a timescale of a few hundred Myr. The stellar- and gas-mass functions track each other well. Comparing the total protocluster stellar mass function to the stellar-mass function of typical $z\,{=}\,1$ galaxy clusters, we find that the samples are consistent with one another and are well-fit by Schechter functions. The best-fit characteristic mass for the $z\,{=}\,1$ galaxy clusters is $(5.2_{-0.2}^{+1.1})\,{\times}\,10^{10}\,$M$_{\odot}$, and the slope is $-$0.46$_{-0.26}^{+0.08}$, compared to SPT2349$-$56, where we find a best-fit characteristic mass of $(6.2_{-3.9}^{+0.8})\,{\times}\,10^{10}\,$M$_{\odot}$ and a slope of $-$0.3$_{-0.3}^{+0.3}$. On the other hand, the protocluster with a merged BCG is better-fit by a single power-law. Thus if SPT2349$-$56 is to follow a similar trajectory as the $z\,{=}\,1$ galaxy clusters, then it must accrete numerous less-massive galaxies, or these less-massive galaxies must already be present but remain undetected in our observations. Due to incompleteness in our sample (about 80\,per cent for $M_{\ast}\,{>}\,10^10$\,M$_{\odot}$), we are unable to distinguish between these two scenarios.

We measured the physical sizes of the galaxies in SPT2349$-$56 in our deep {\it HST\/}-F160W data, which probe rest-frame ultraviolet wavelengths. Upon comparing these measurements with typical star-forming galaxies between redshift 4 and 5, we found that our sample has comparable ultraviolet sizes. We stacked our {\it HST\/} data at the positions of the detected galaxies and compared this with a stack of the same galaxies at submm wavelengths imaged by ALMA, finding a consistent result. Galaxy C6, the brightest protocluster galaxy in our {\it HST\/} data, is the most compact rest-frame ultraviolet source and remains unresolved by our {\it HST\/} imaging. Hydrodynamical simulations predict that this galaxy is at the centre of a major merger, yet after the merger the emission will still remain compact. 

Lastly, we investigated the total projected stellar and molecular gas mass of SPT2349$-$56 as a function of projected area. We found that the molecular gas mass and stellar mass track each other well, with no clear trend in the molecular gas-to-stellar mass fraction as a function of radius (the mean value being about 0.3). The stellar mass distribution of SPT2349$-$56 also does not appear to be markedly different from $z\,{=}\,1$ galaxy clusters, although we note that so far we have only probed the central ${\approx}\,1$\,Mpc of the structure, which is much smaller than the extent of $z\,{=}\,1$ clusters. 

SPT2349$-$56 is a galaxy protocluster in a remarkable phase of its evolution, reaching a total SFR of over 10,000\,M$_{\odot}$\,yr$^{-1}$ within a volume of approximately 0.1\,Mpc$^3$. Having been selected specifically for its star-forming properties, it is interesting that at optical and infrared wavelengths, the galaxies making up SPT2349$-$56 are not very luminous and have fairly typical stellar masses. Galaxy C6 is however an exception; this source is at the centre of a massive merger of over 20 galaxies, and is a likely proto-BCG. The transition of galaxy C6 into a BCG is consistent with the `downsizing' scenario of galaxy formation, and provides a direct observation of the constituents and formation mechanism of a BCG in the early Universe. In this case the high-redshift proto-BCG is undergoing a major merger with dozens of galaxies whose properties are similar to the field, explaining why present-day BCGs contain old cores that seemingly formed very quickly.

\section*{Acknowledgements}

The authors wish to thank Dr.~Adam Muzzin for a useful discussion about the stellar mass function of galaxy clusters around redshift 1. This paper makes use of the following ALMA data: ADS/JAO.ALMA\#2017.1.00273.S; and ADS/JAO.ALMA\#2018.1.00058.S. ALMA is a partnership of ESO (representing its member states), NSF (USA) and NINS (Japan), together with NRC (Canada), MOST and ASIAA (Taiwan), and KASI (Republic of Korea), in cooperation with the Republic of Chile. The Joint ALMA Observatory is operated by ESO, AUI/NRAO, and NAOJ. {\it Herschel\/} is an ESA space observatory with science instruments provided by European-led Principal Investigator consortia and with important participation from NASA. This paper is based on observations made with ESO Telescopes at the La Silla Paranal Observatory under programme ID 299.A-5045(A). The National Radio Astronomy Observatory is a facility of the National Science Foundation operated under cooperative agreement by Associated Universities, Inc. The SPT is supported by the National Science Foundation through grant PLR-1248097, with partial support through PHY-1125897, the Kavli Foundation, and the Gordon and Betty Moore Foundation grant GBMF 947. This work is based in part on observations made with the {\sl Spitzer Space Telescope}, which was operated by the Jet Propulsion Laboratory, California Institute of Technology under a contract with NASA. This work was supported by the Natural Sciences and Engineering Research Council of Canada. The Flatiron Institute is supported by the Simons Foundation. M.A. has been supported by the grant ``CONICYT+PCI+REDES 190194''. D.P.M., J.D.V., K.C.L., and K.P. acknowledge support from the US NSF under grants AST-1715213 and AST-1716127. K.C.L acknowledges support from the US NSF NRAO under grants SOSPA5-001 and SOSPA4-007, respectively. J.D.V. acknowledges support from an A. P. Sloan Foundation Fellowship. M.A. and J.D.V. acknowledge support from the Center for AstroPhysical Surveys at the National Center for Supercomputing Applications in Urbana, IL. S.J. acknowledges support from the US NSF NRAO under grants SOSPA5-001 and SOSPA7-006.

\section*{Data Availability}

All of the data presented in this paper are publicly available. The {\it HST\/} data can be found under the programme 15701 (PI S.~Chapman), the {\it Spitzer\/}-IRAC data can be found under the four programmes 60194 (PI J.~Vieira), 80032 (PI S.~Stanford), 13224 (PI S.~Chapman), and 14216 (PI S.~Chapman), the Gemini-South data can be found under the programme GS-2017B-Q-7 (PI A.~Chapman), and the ALMA data can be found under the two programmes 2017.1.00273.S (PI S.~Chapman) and 2018.1.00058.S (PI S.~Chapman).

\bibliographystyle{mnras}
\bibliography{spt2349_ir}

\appendix

\newpage

\section{Optical counterpart offsets}
\label{appendix0}

\begin{figure*}
\includegraphics[width=0.3\textwidth]{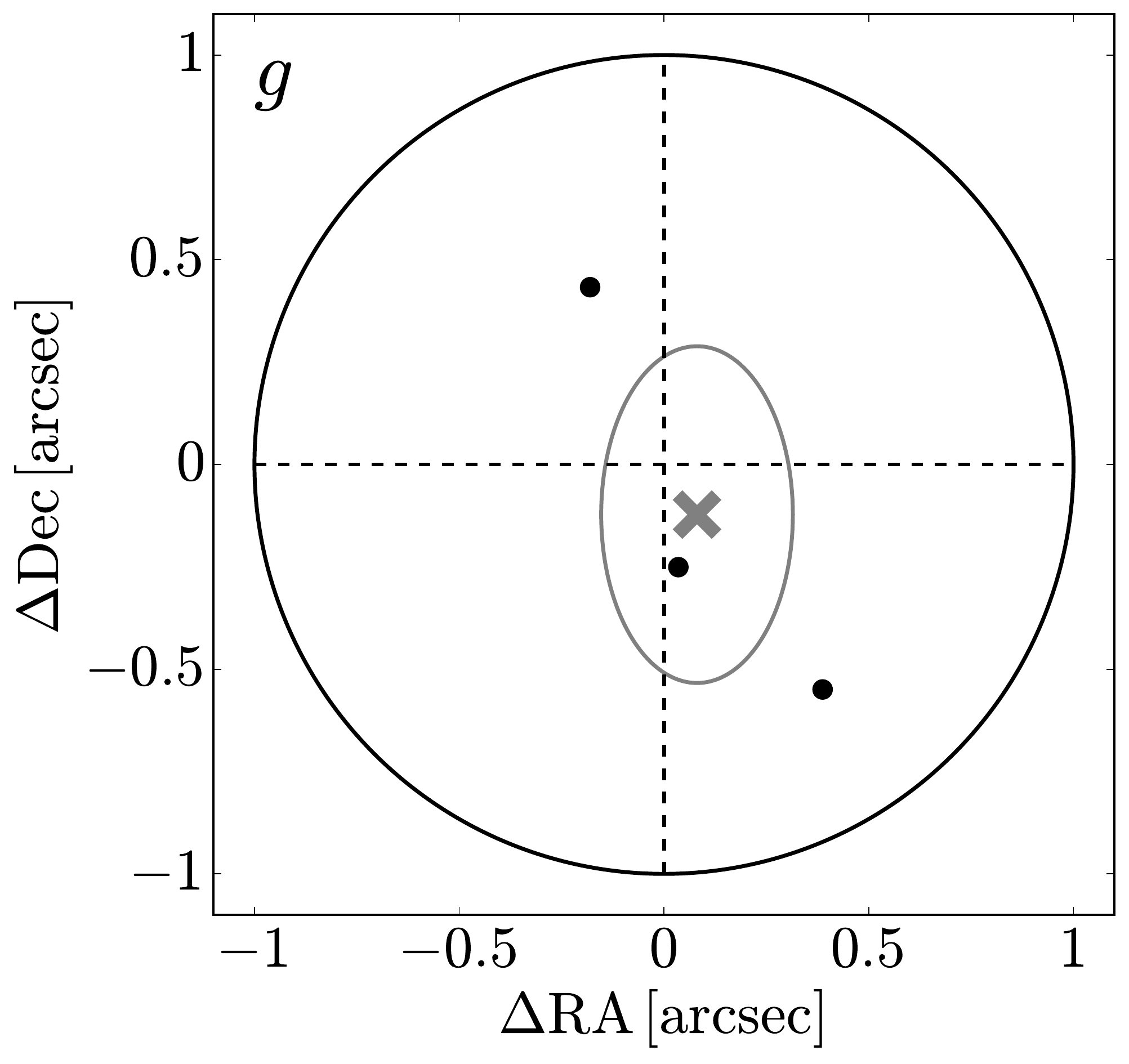}
\includegraphics[width=0.3\textwidth]{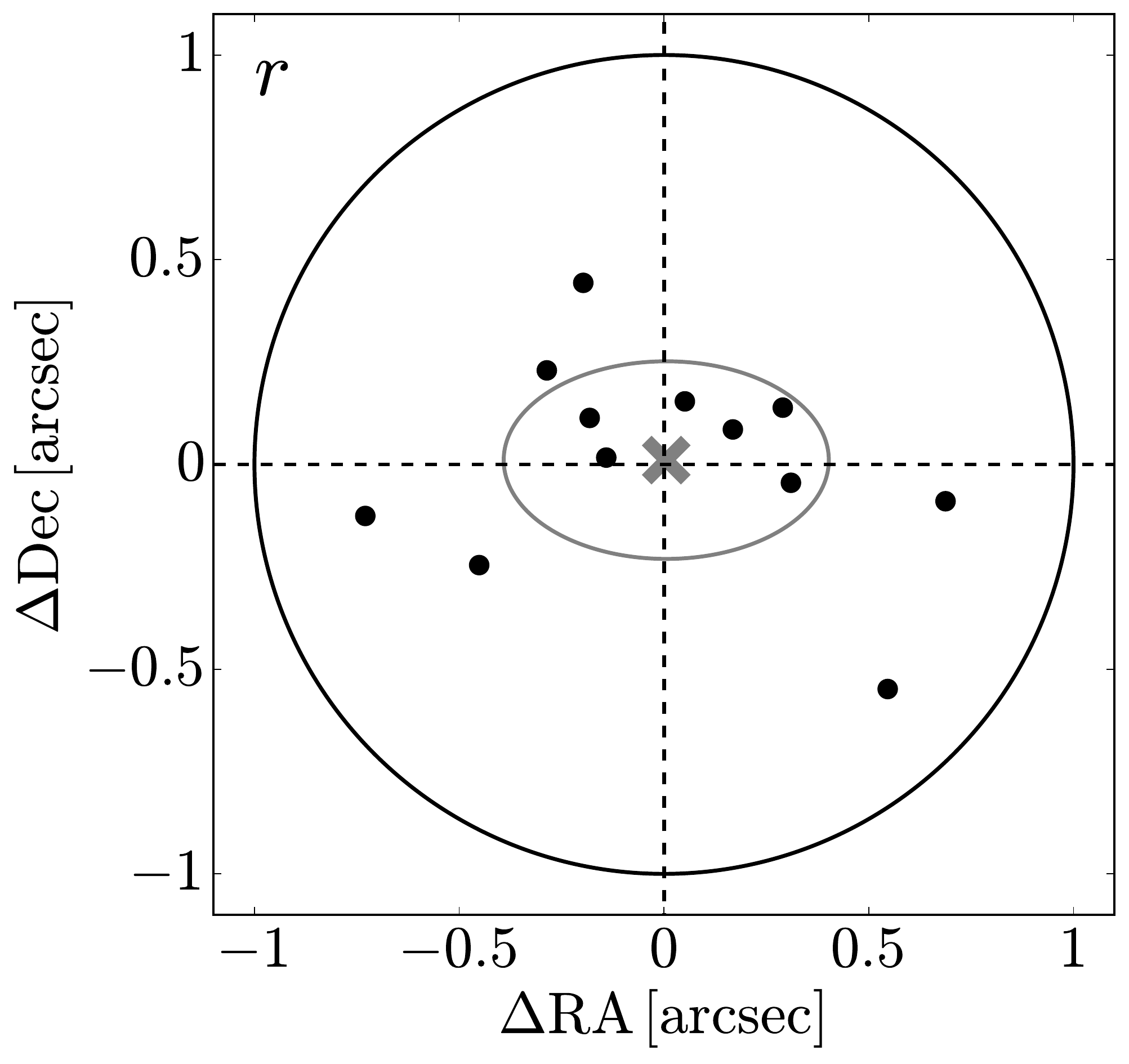}
\includegraphics[width=0.3\textwidth]{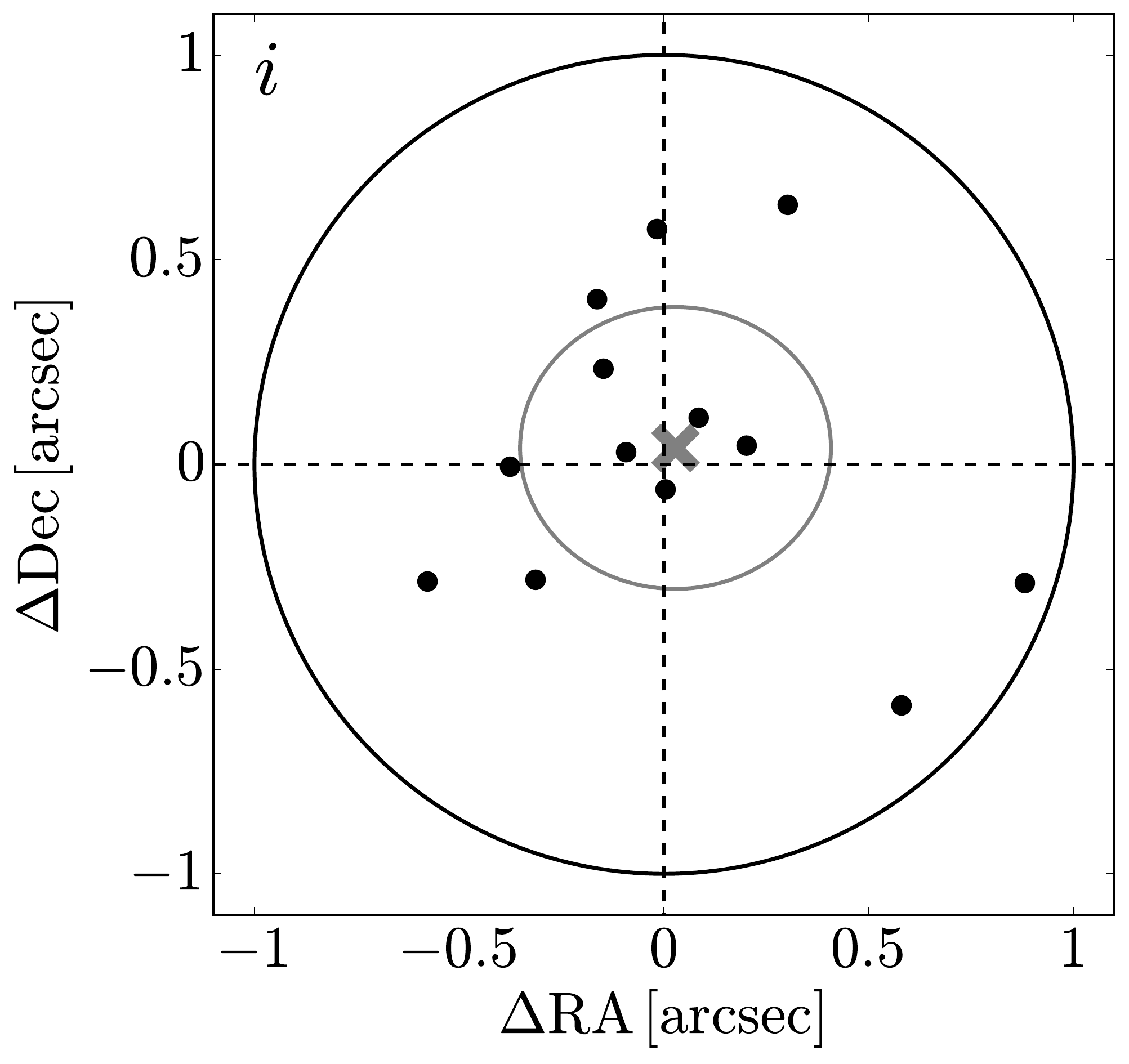}
\includegraphics[width=0.3\textwidth]{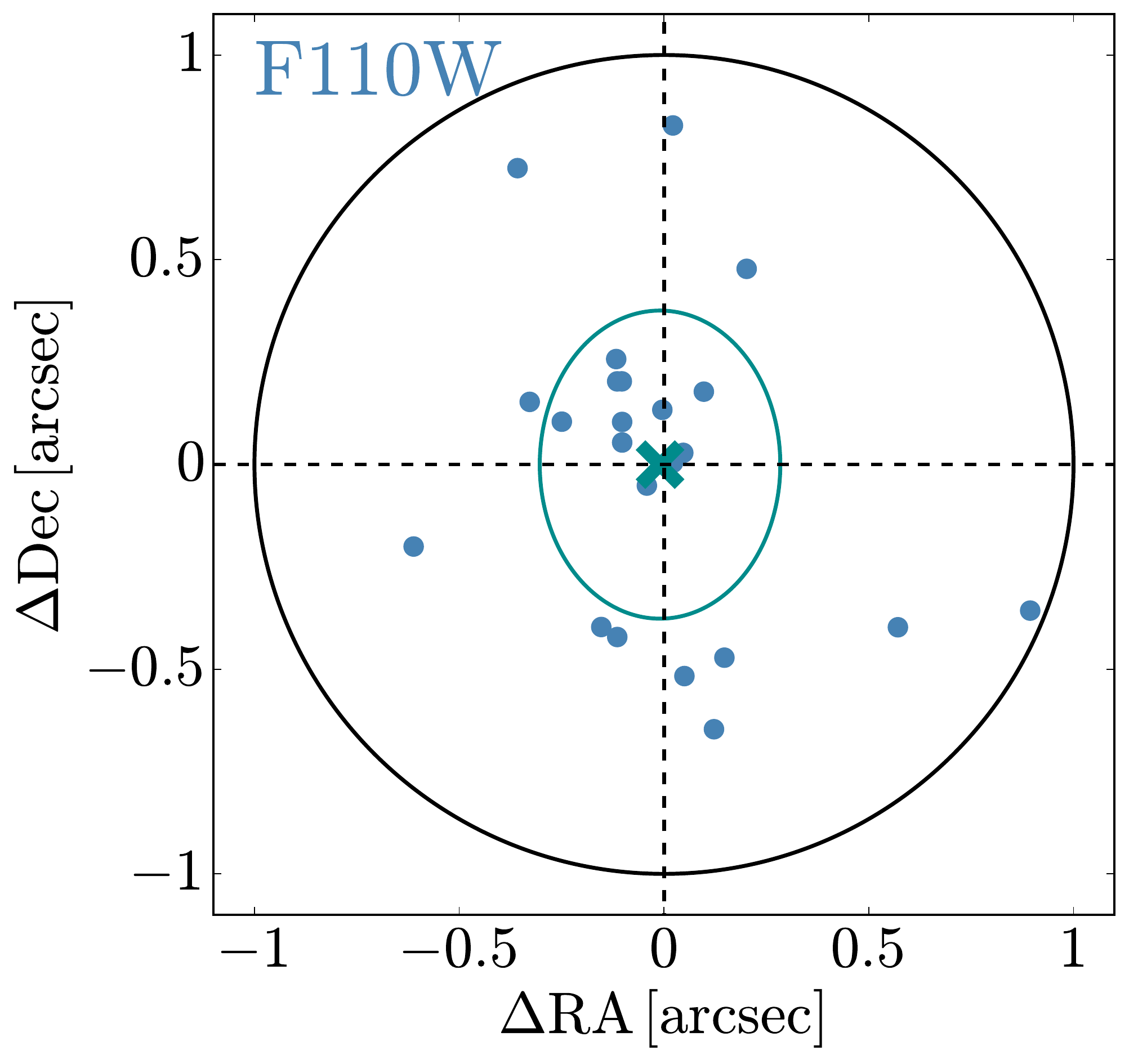}
\includegraphics[width=0.3\textwidth]{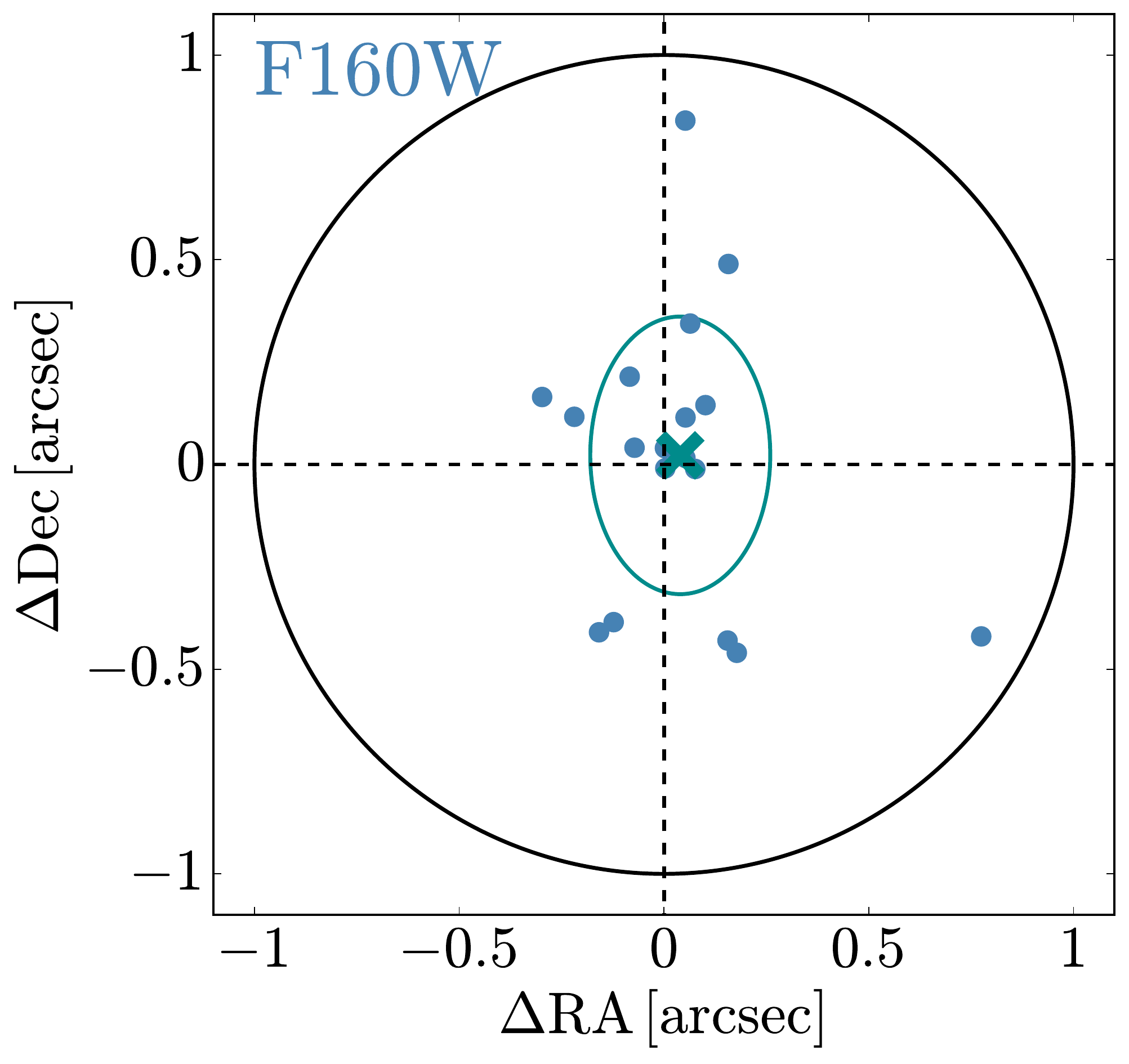}
\includegraphics[width=0.3\textwidth]{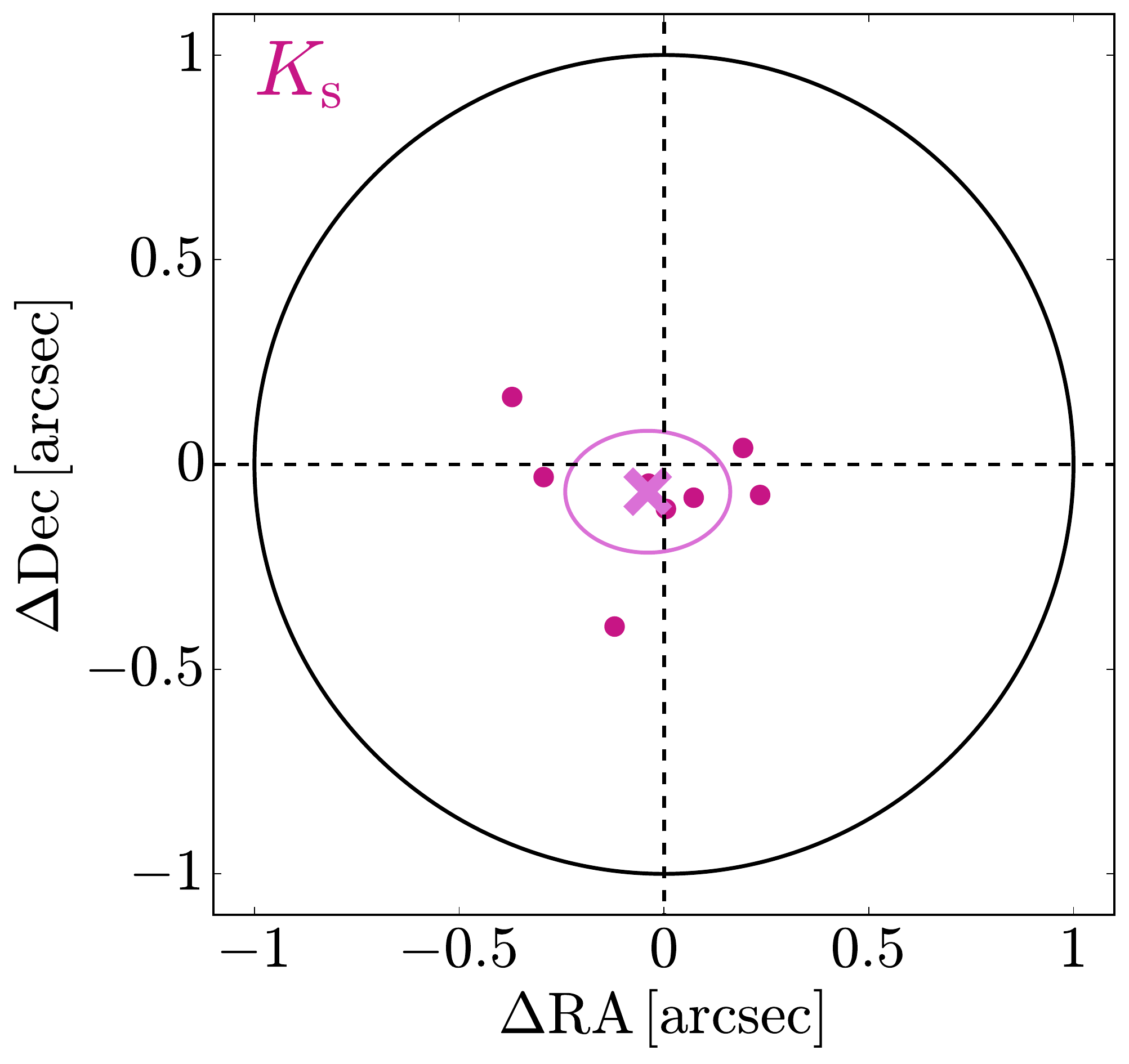}
\includegraphics[width=0.3\textwidth]{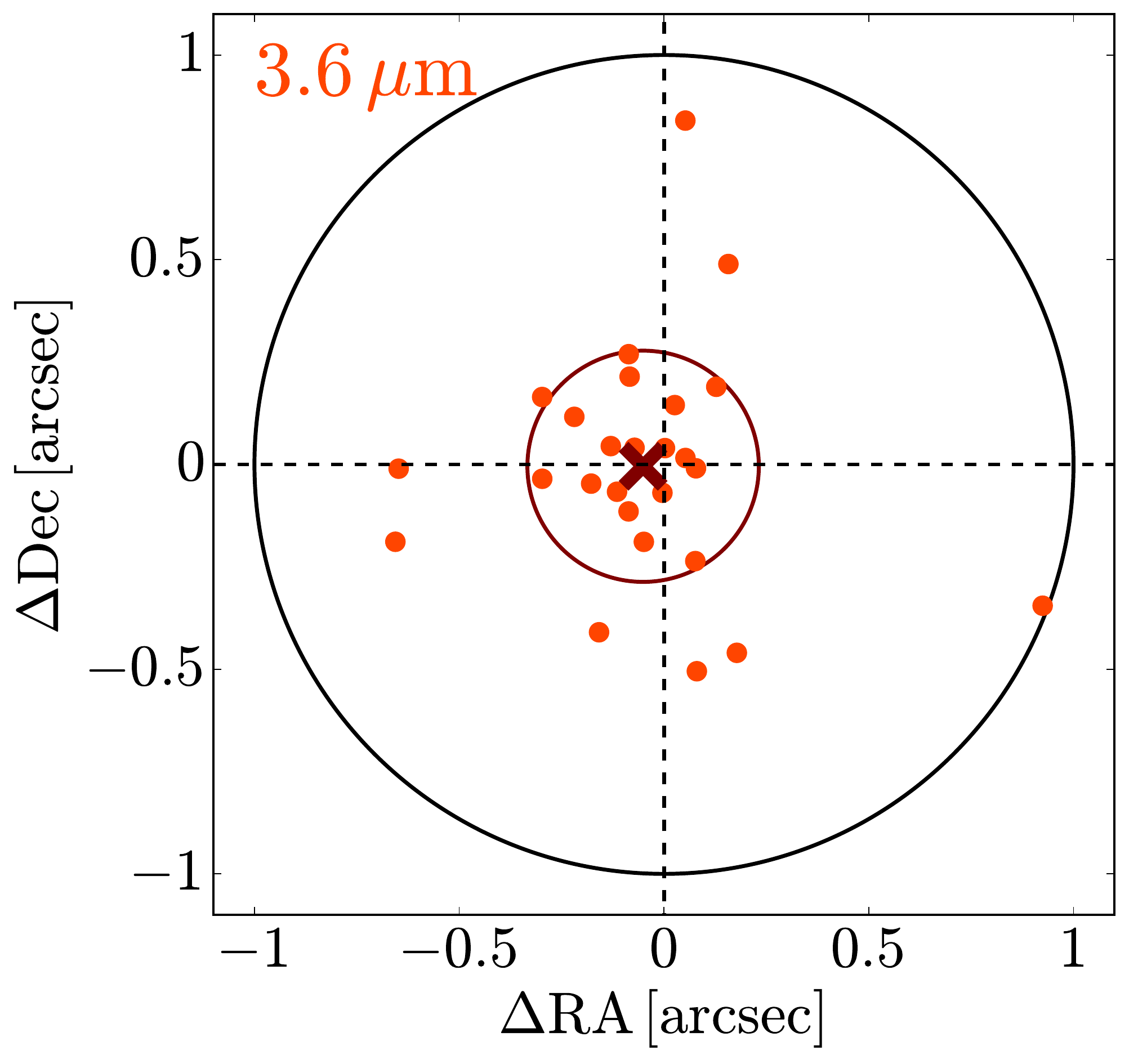}
\includegraphics[width=0.3\textwidth]{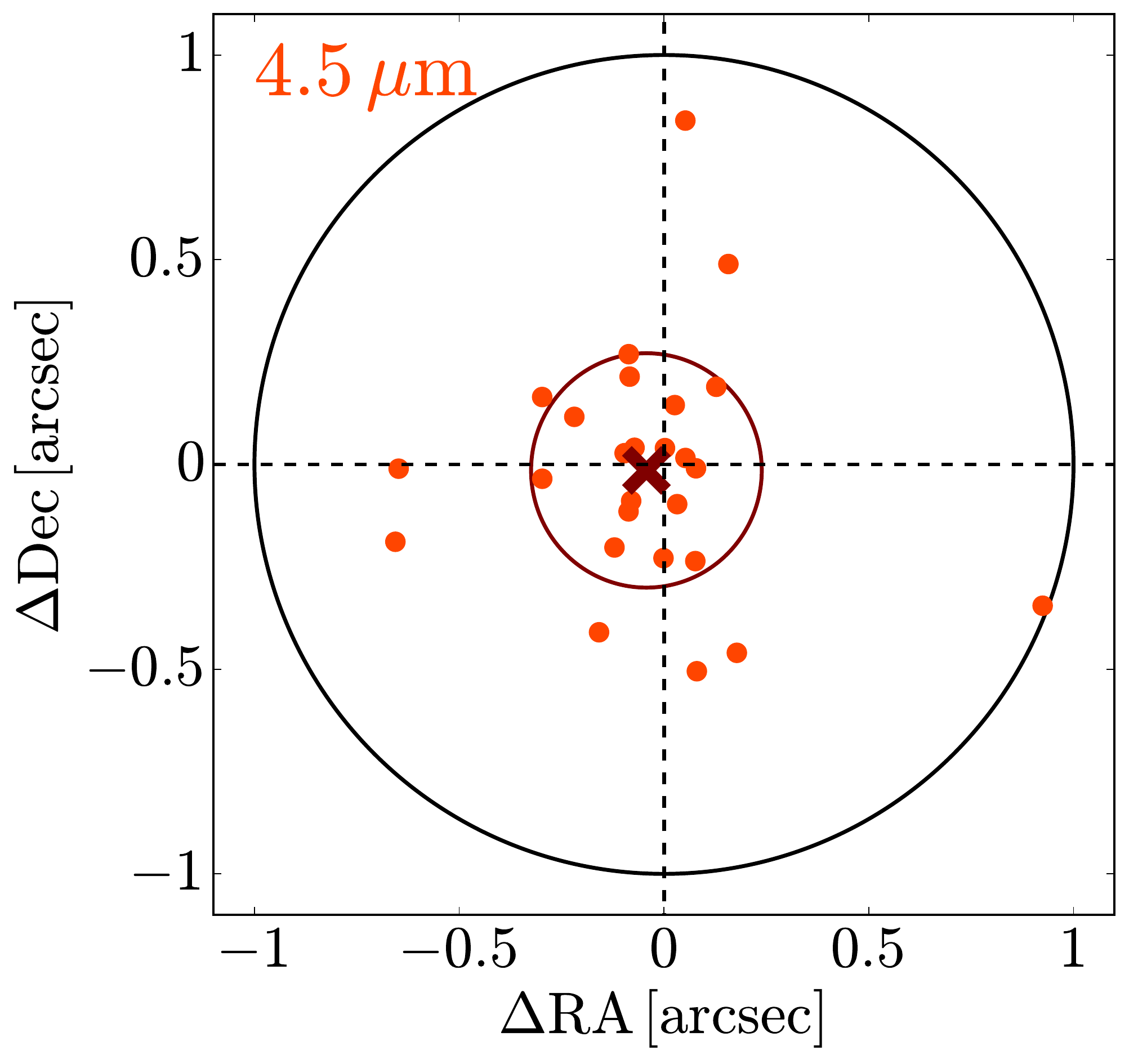}
\caption{Offset in right ascension and declination between the ALMA positions reported in \citet{hill2020} and the counterparts identified in each of the bands used in this work. The mean offset in each band is indicated by a cross, surrounded by an ellipse whose semi-major and semi-minor axes are equal to the standard deviation of the offsets in each direction. In each case we find that the ellipse overlaps with the origin, indicating that there are no systematic offset errors present at a significant level in our data.}
\label{offset}
\end{figure*}

Positional differences between our rest-frame optical and ALMA positions are defined as the ALMA position minus the optical/infrared position. In Fig.~\ref{offset} we show the mean offset found in each band surrounded by a circle whose semi-major and semi-minor axes are equal to the standard deviations of the offsets in each direction. We find a slight offset in the negative $x$ direction of $\Delta $RA$\,{\approx}\,-0.1$\,arcsec (consistent across all wavebands, since the astrometry has been tied to a single frame).

\section{Optical imaging cutouts}
\label{appendix1}

\begin{figure*}
\begin{subfigure}{0.85\textwidth}
\begin{framed}
\includegraphics[width=0.24\textwidth]{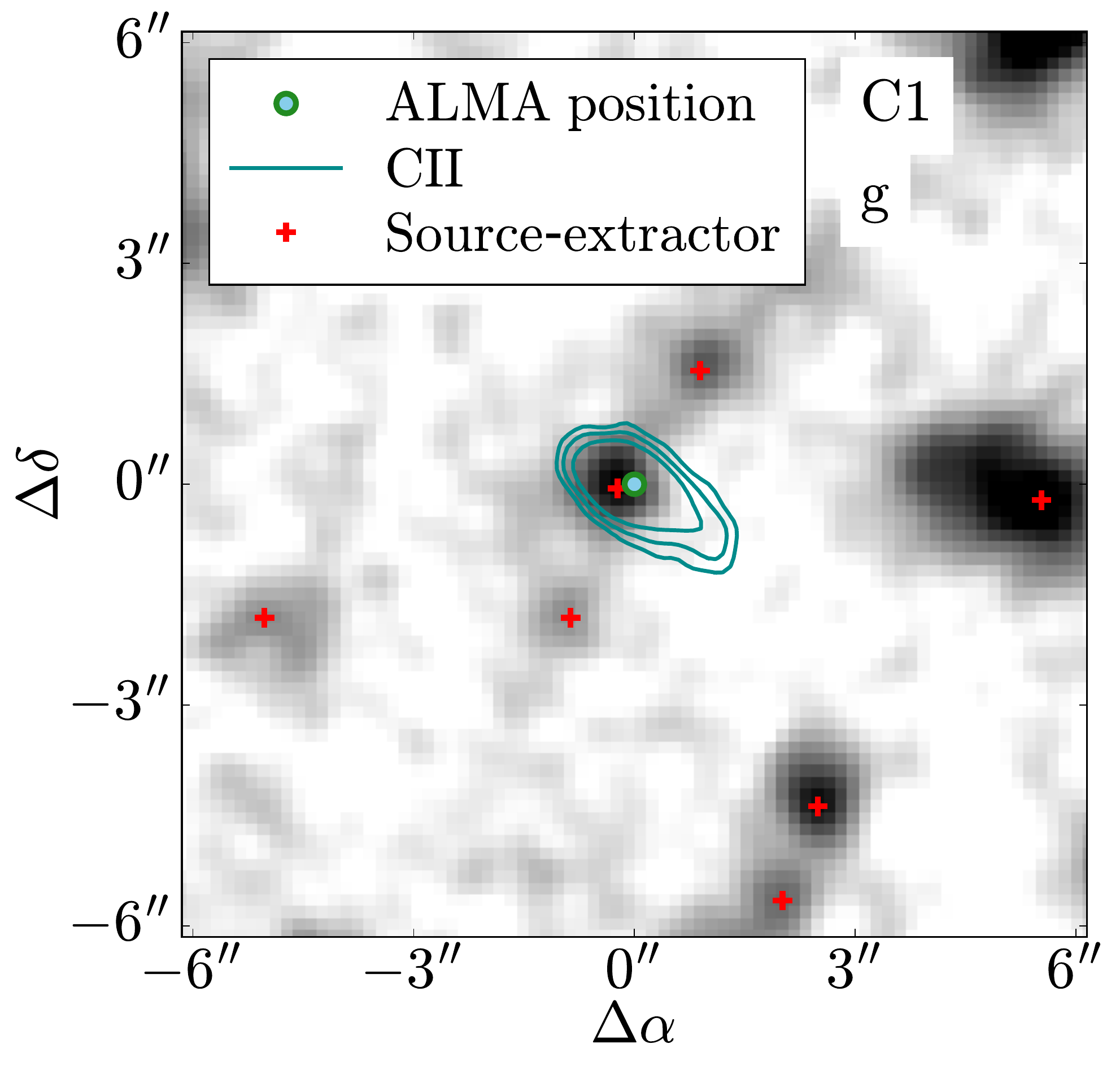}
\includegraphics[width=0.24\textwidth]{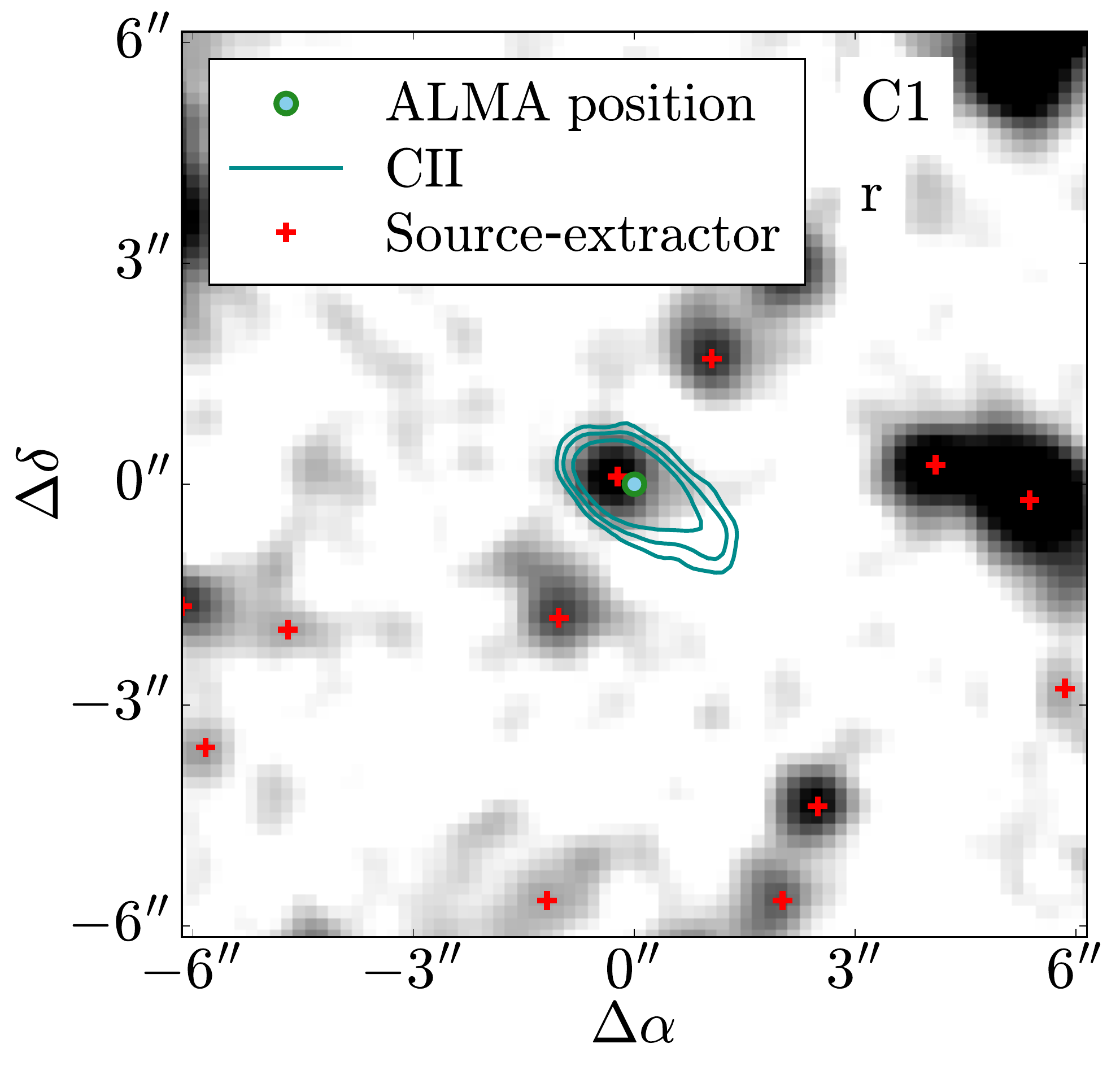}
\includegraphics[width=0.24\textwidth]{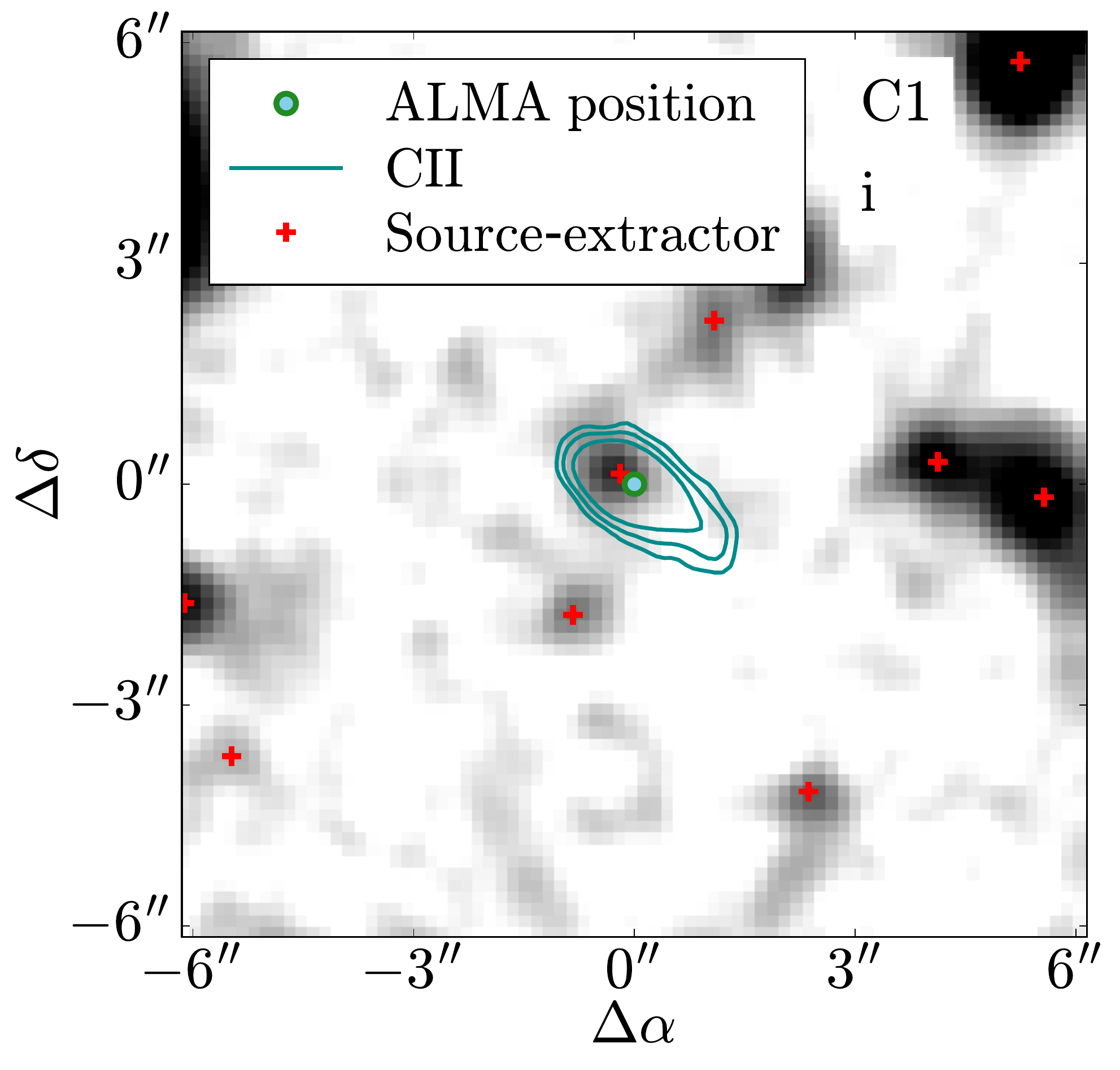}
\includegraphics[width=0.24\textwidth]{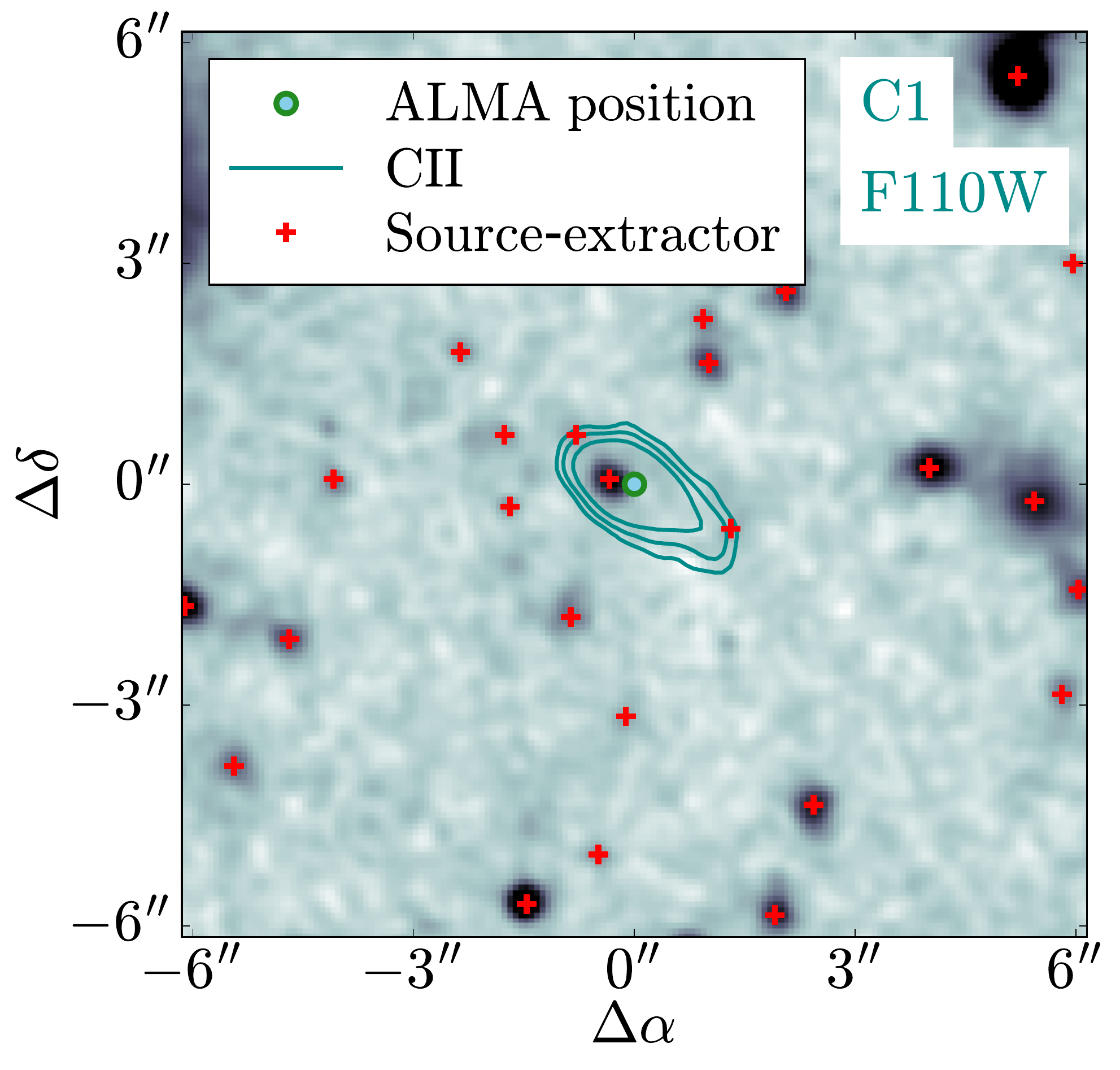}
\includegraphics[width=0.24\textwidth]{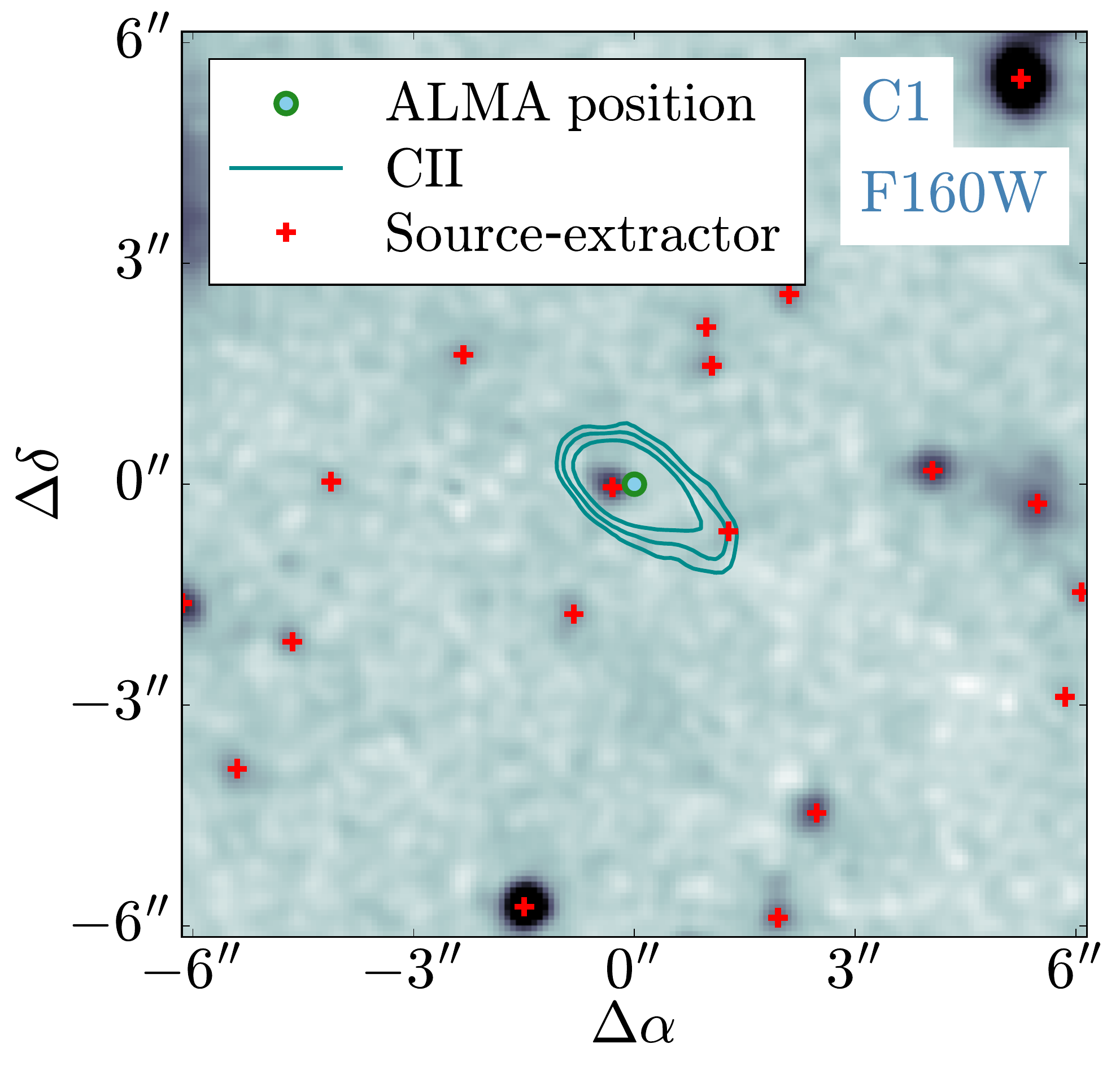}
\includegraphics[width=0.248\textwidth]{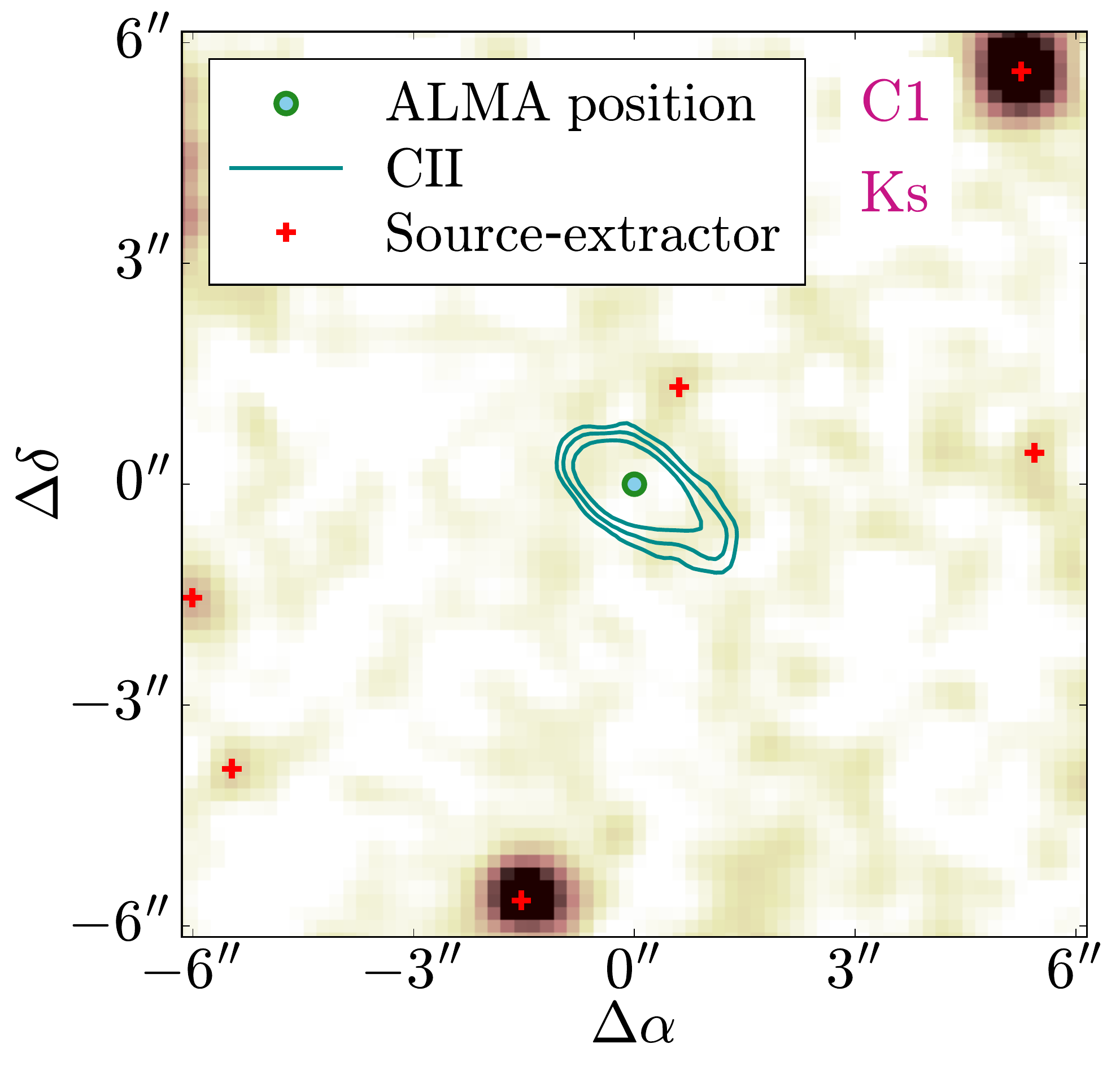}
\includegraphics[width=0.249\textwidth]{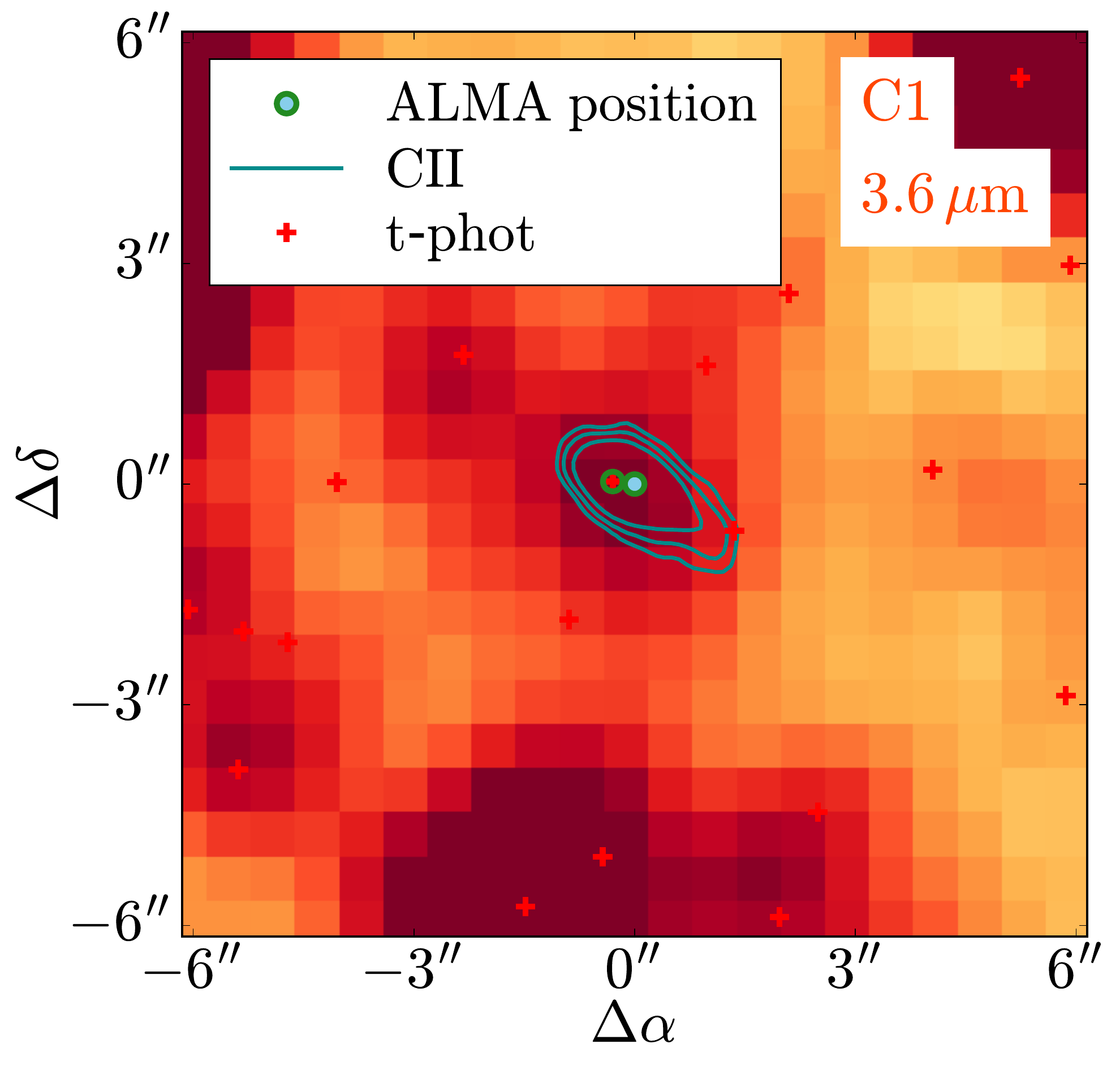}
\includegraphics[width=0.249\textwidth]{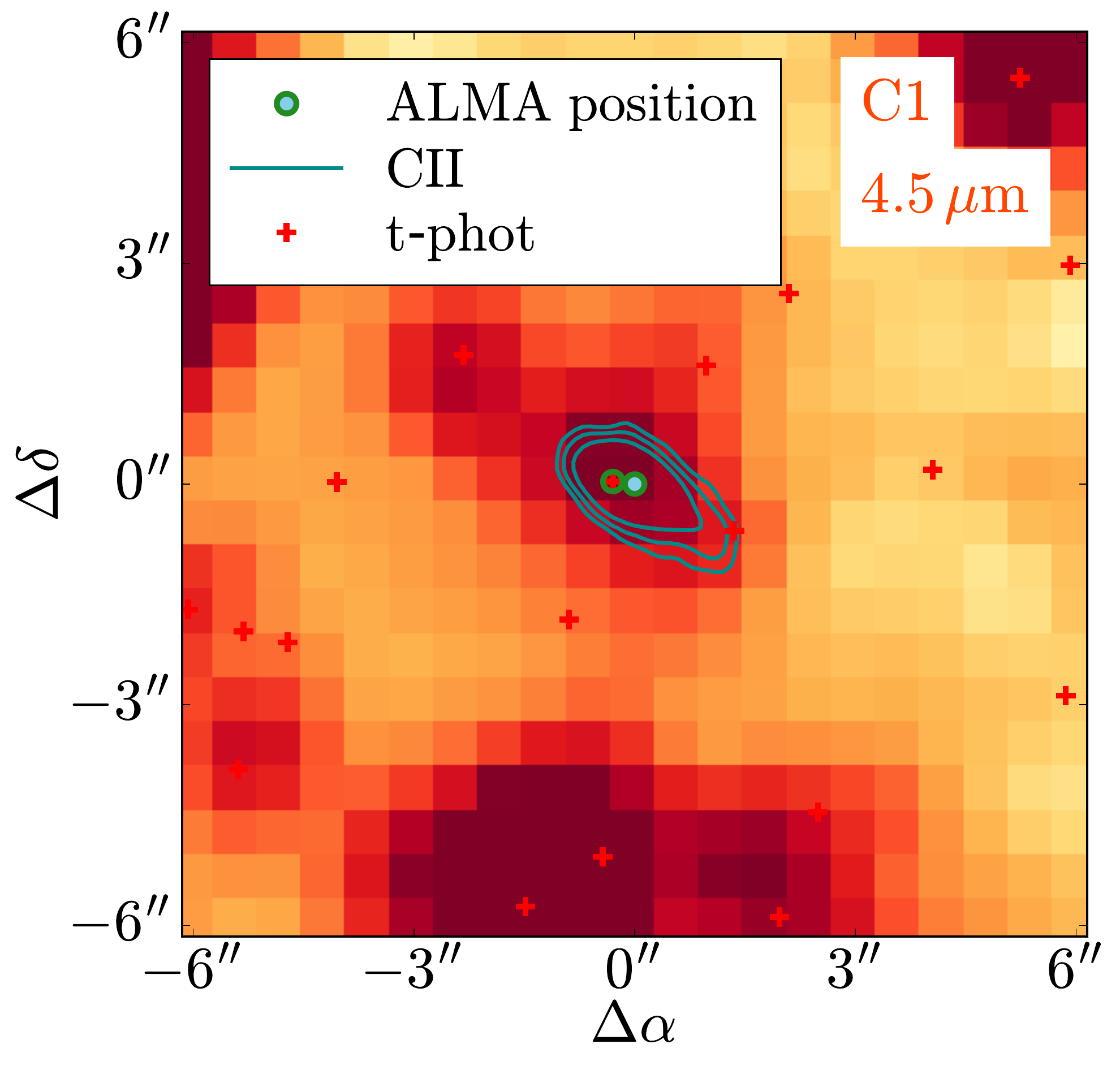}
\end{framed}
\end{subfigure}
\begin{subfigure}{0.85\textwidth}
\begin{framed}
\includegraphics[width=0.24\textwidth]{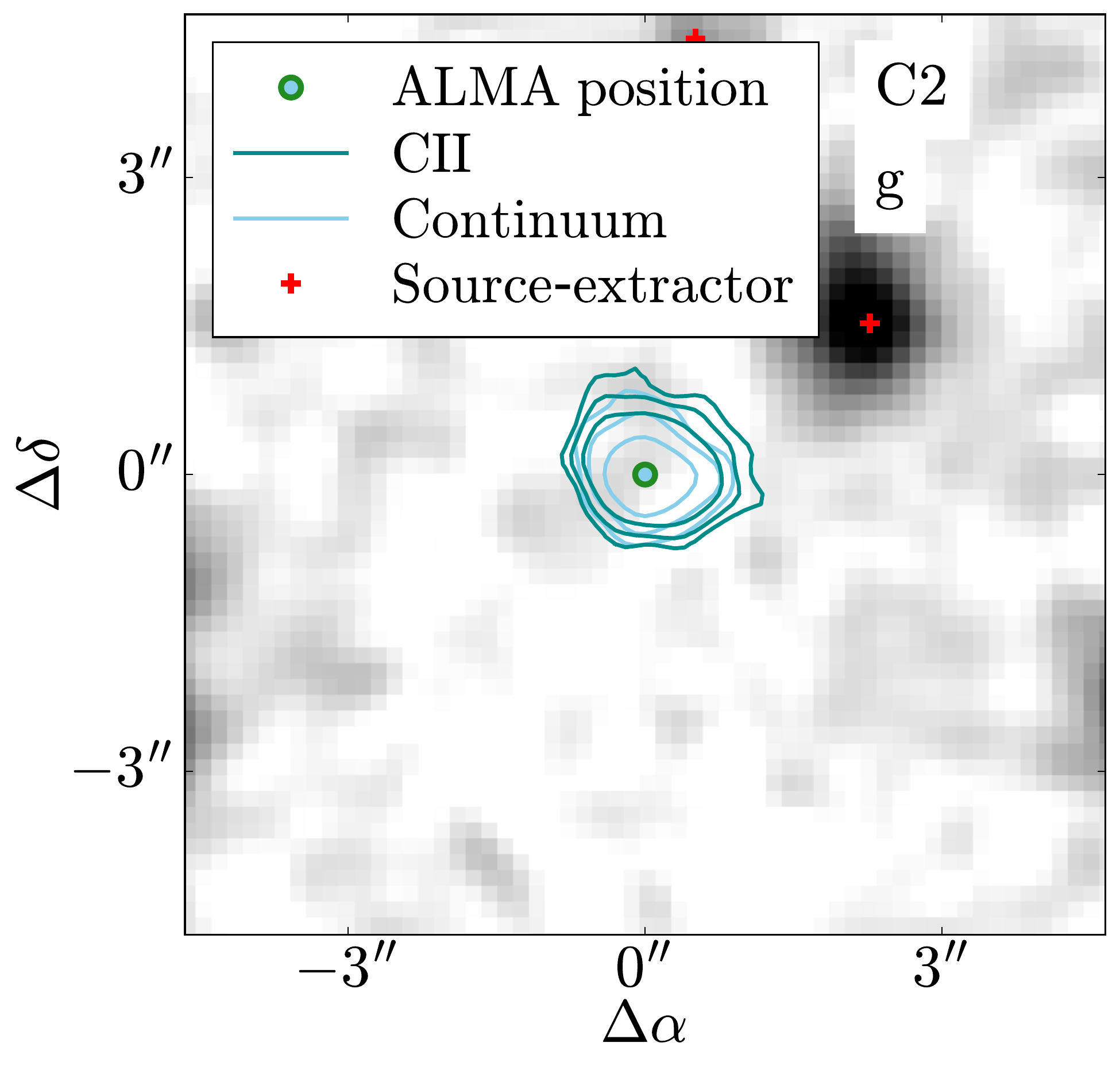}
\includegraphics[width=0.24\textwidth]{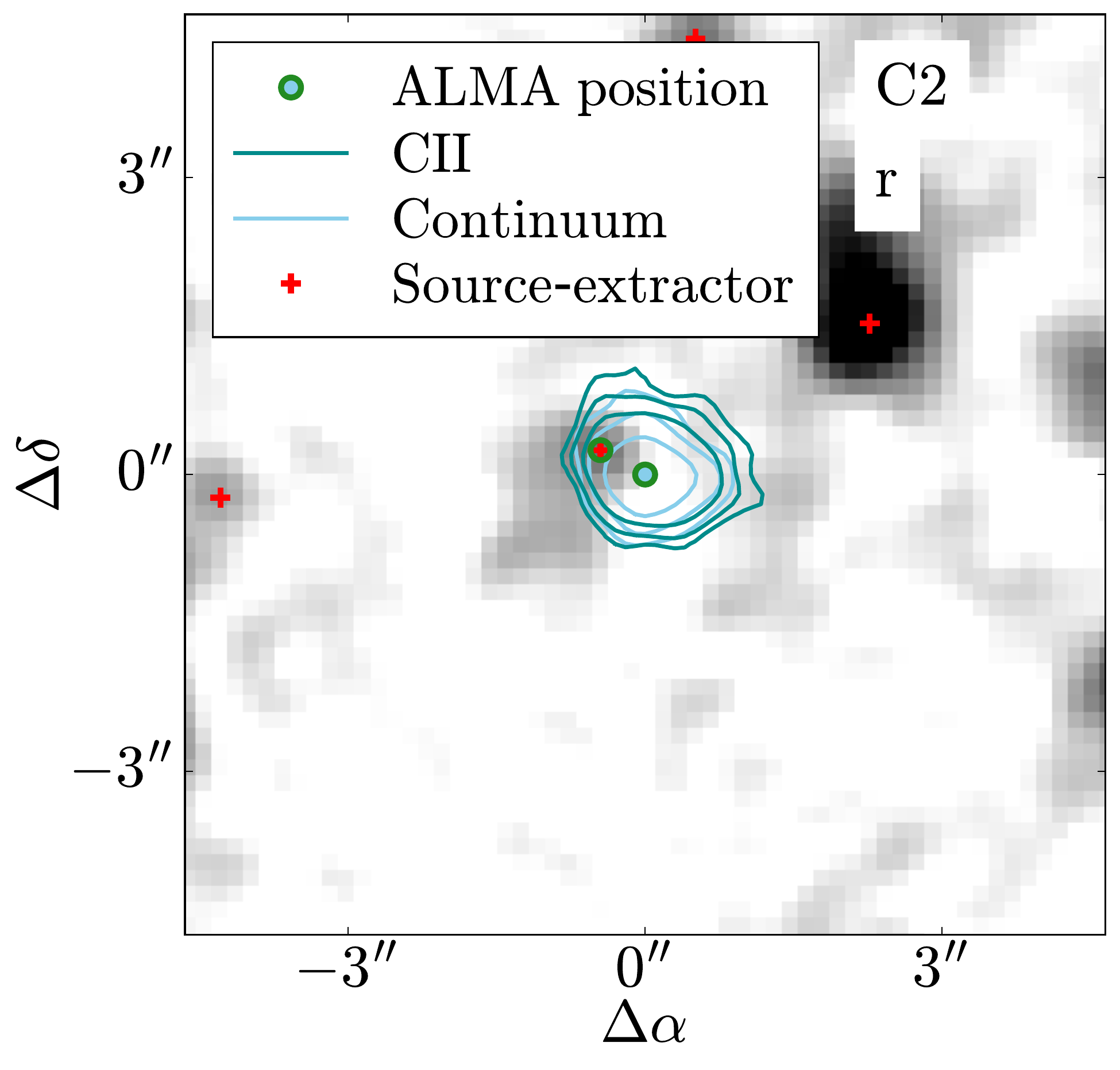}
\includegraphics[width=0.24\textwidth]{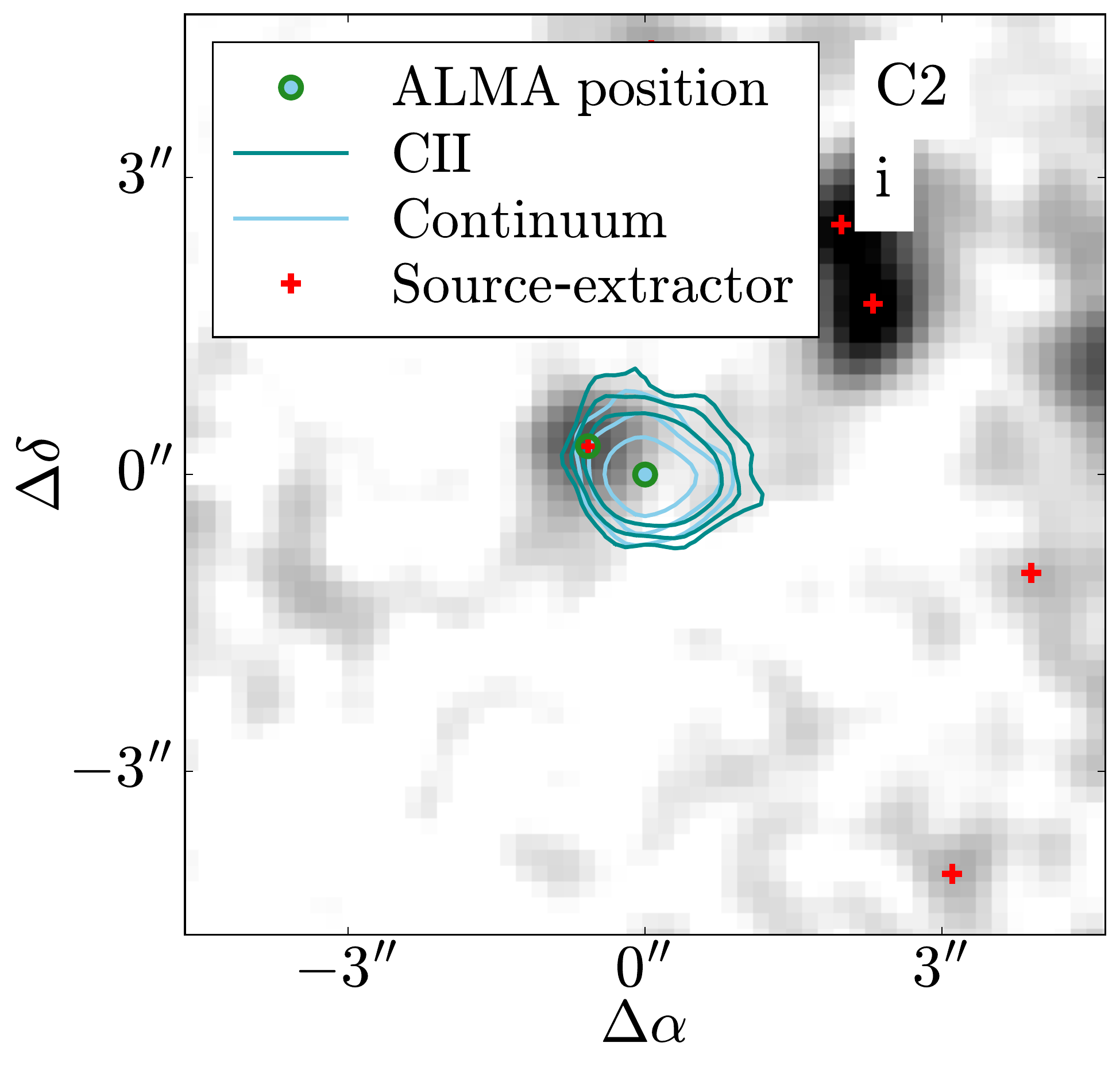}
\includegraphics[width=0.24\textwidth]{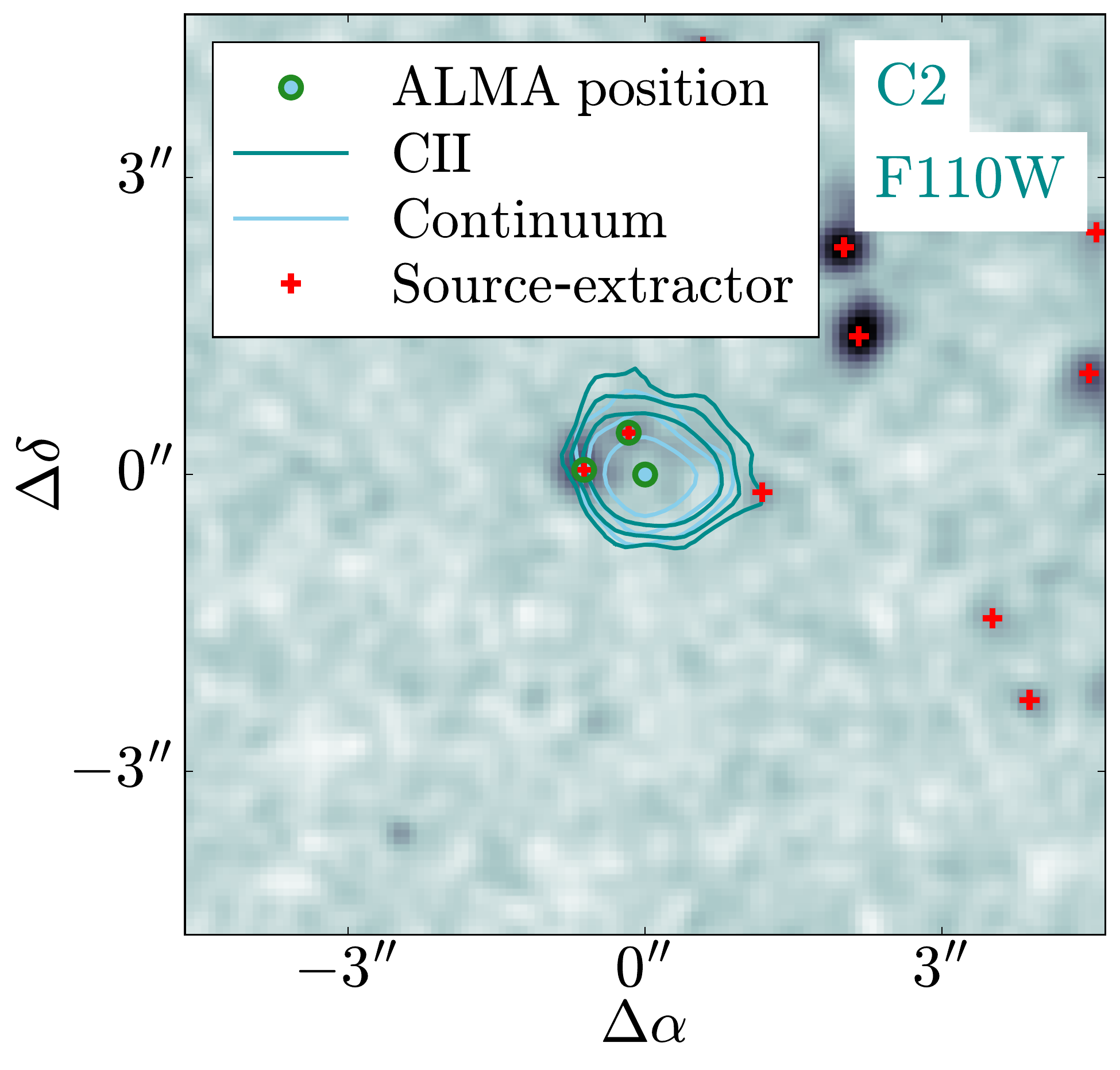}
\includegraphics[width=0.24\textwidth]{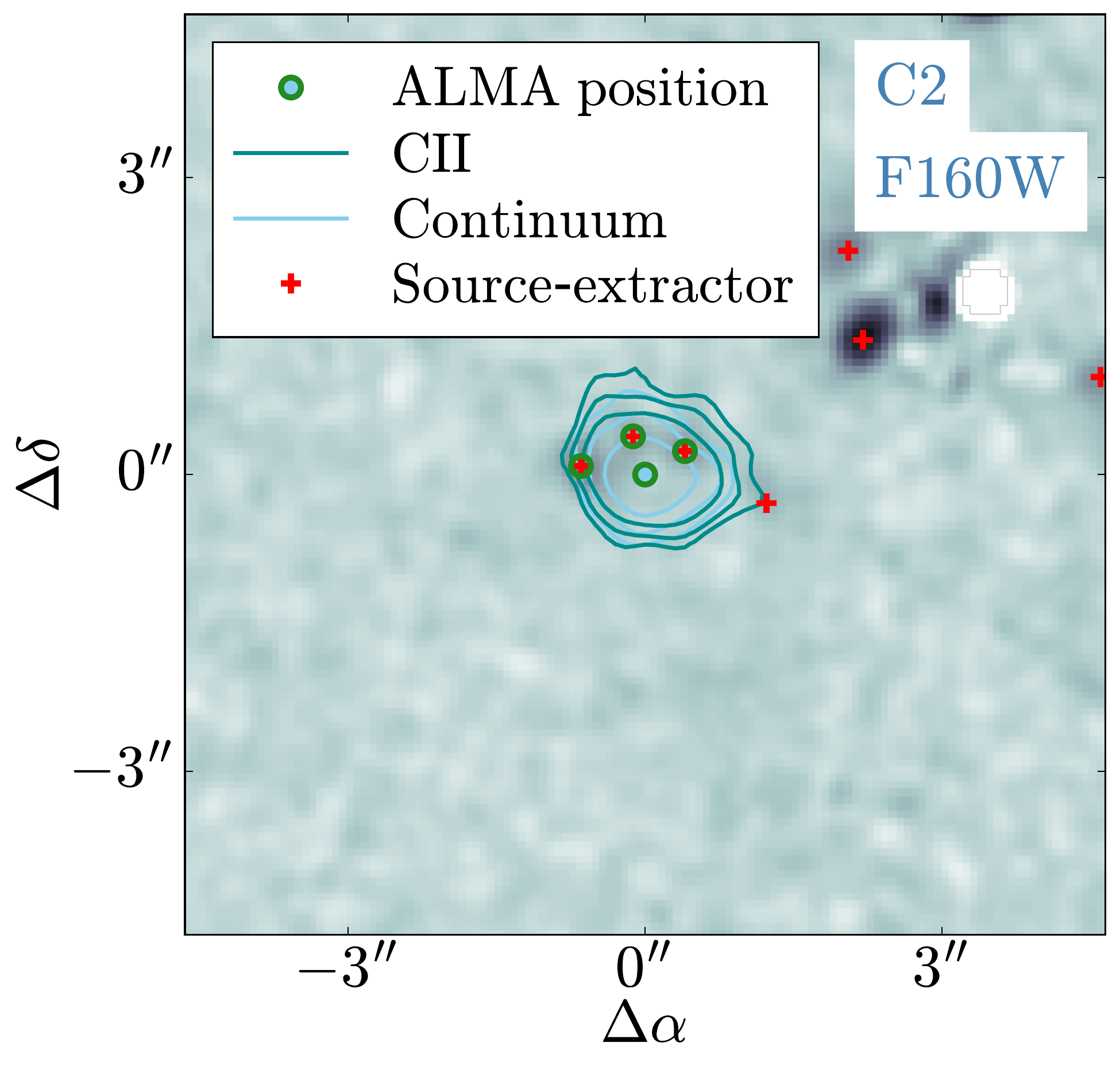}
\includegraphics[width=0.248\textwidth]{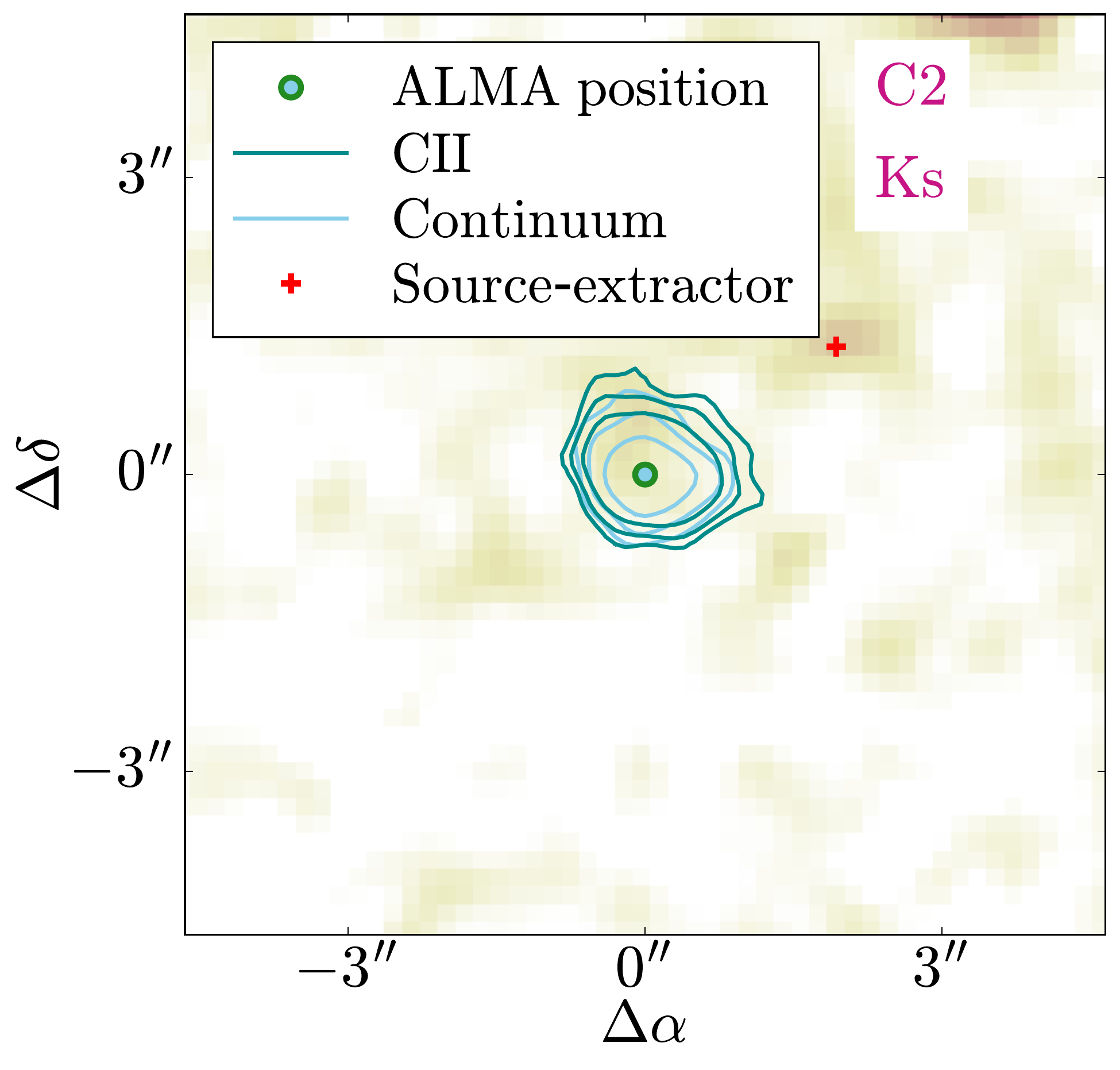}
\includegraphics[width=0.249\textwidth]{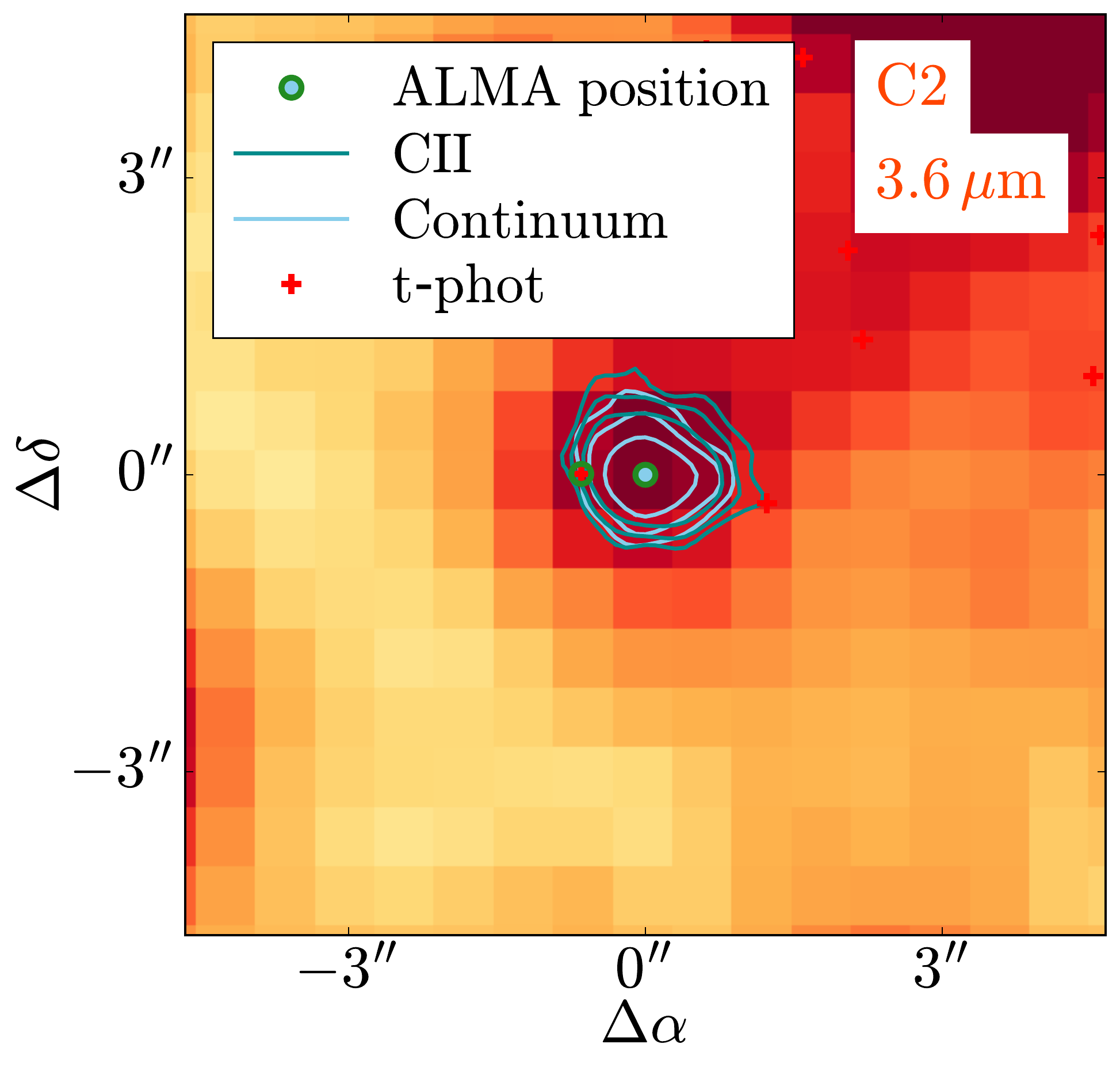}
\includegraphics[width=0.249\textwidth]{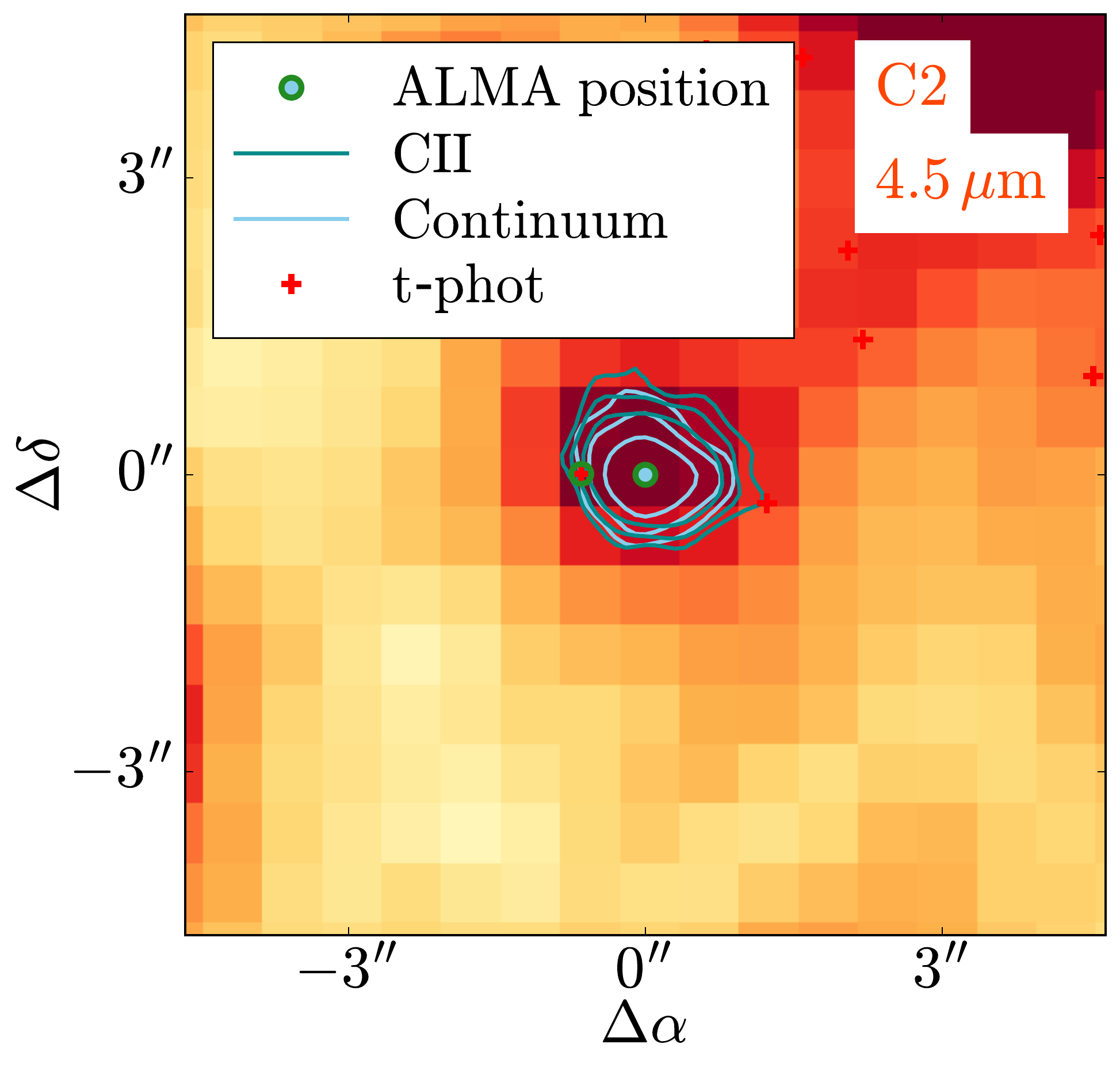}
\end{framed}
\end{subfigure}
\begin{subfigure}{0.85\textwidth}
\begin{framed}
\includegraphics[width=0.24\textwidth]{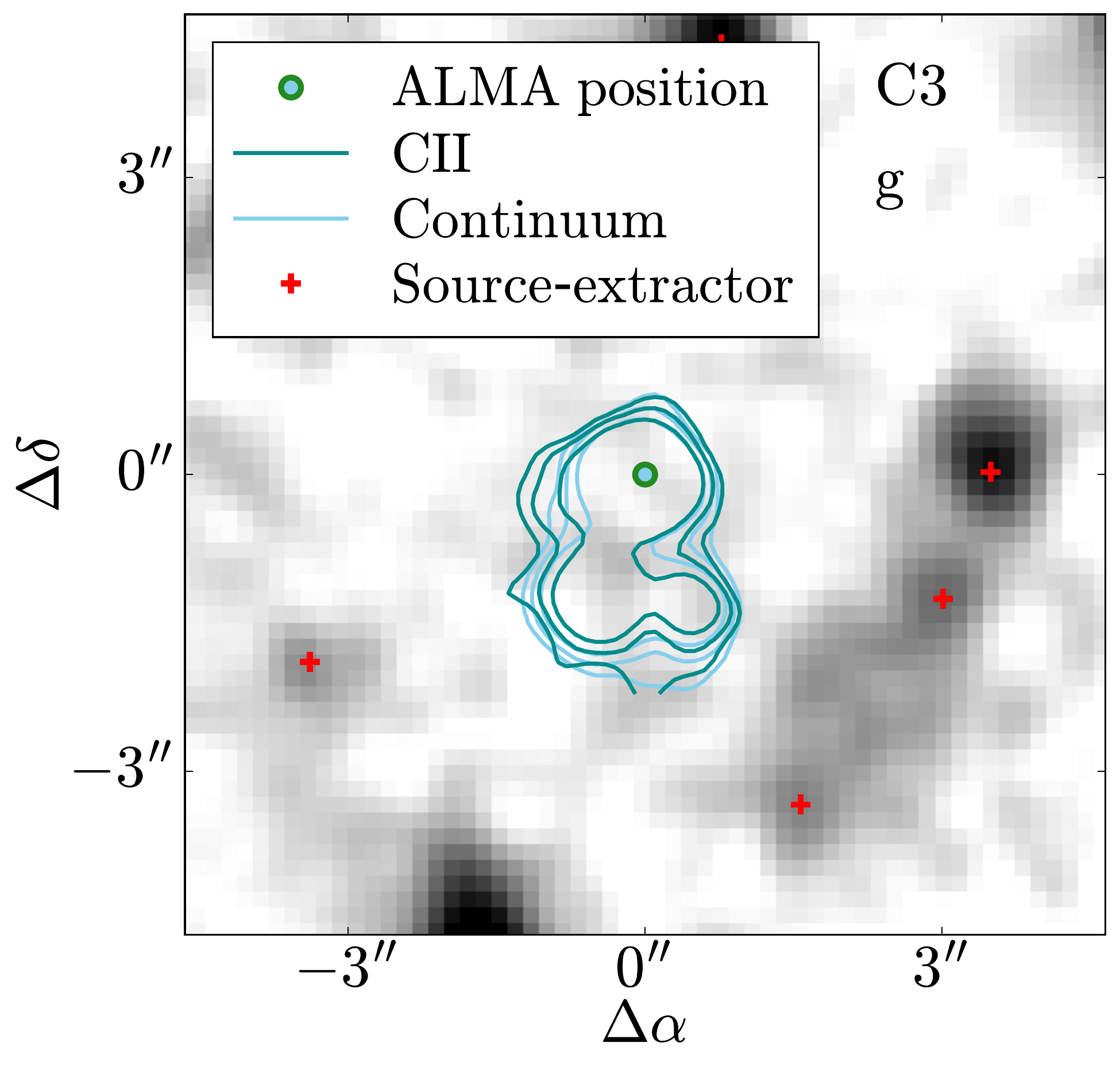}
\includegraphics[width=0.24\textwidth]{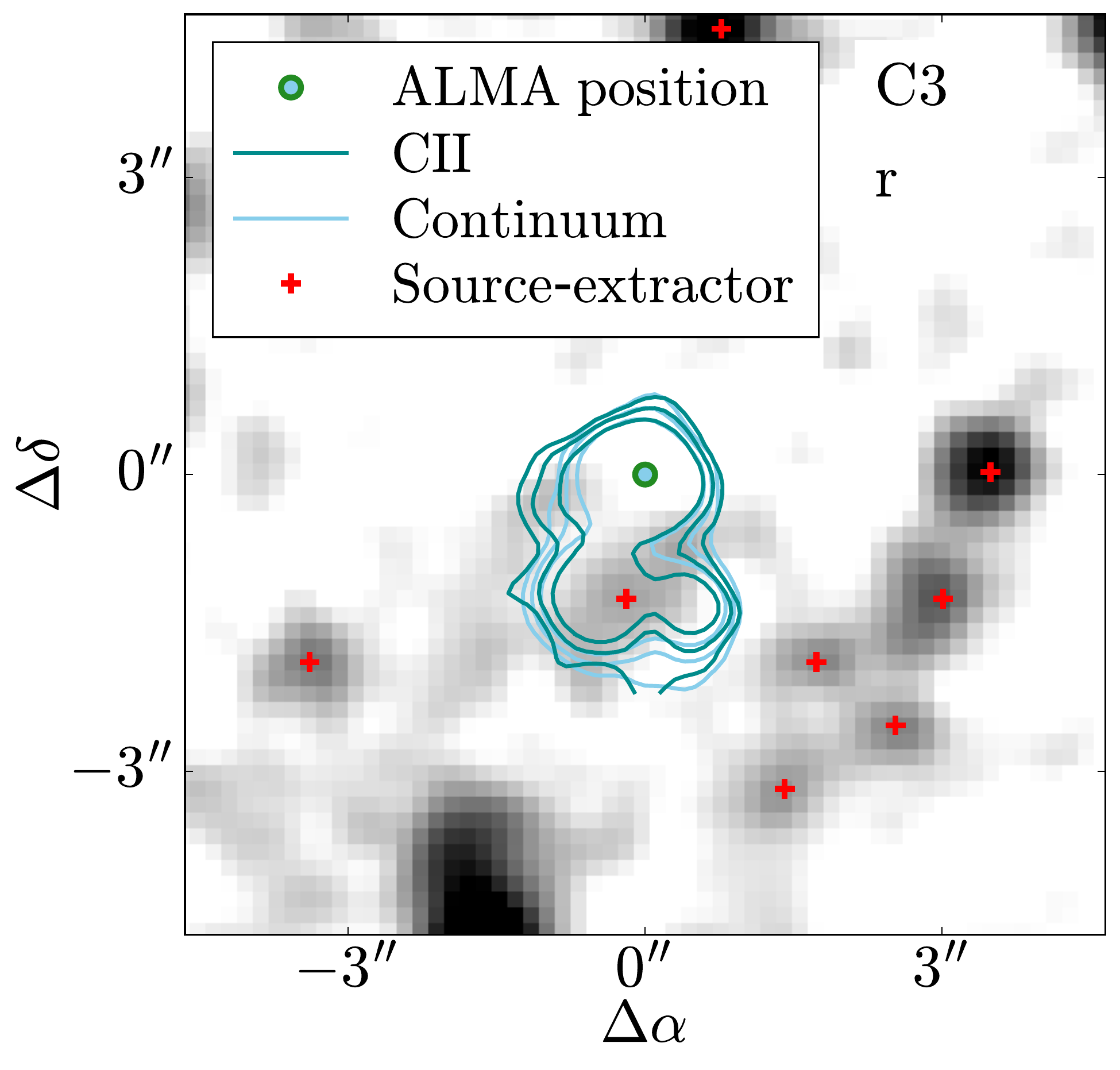}
\includegraphics[width=0.24\textwidth]{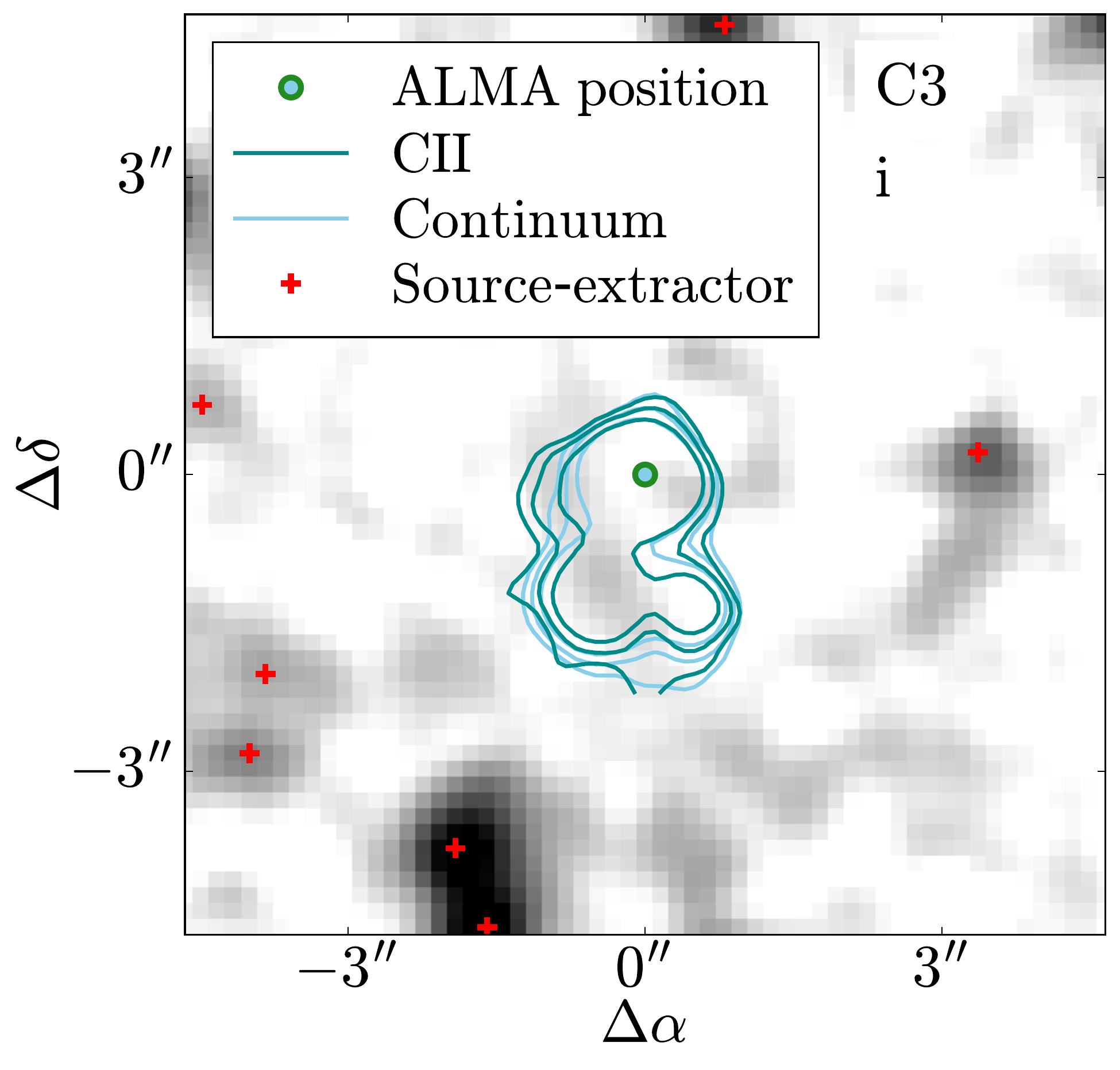}
\includegraphics[width=0.24\textwidth]{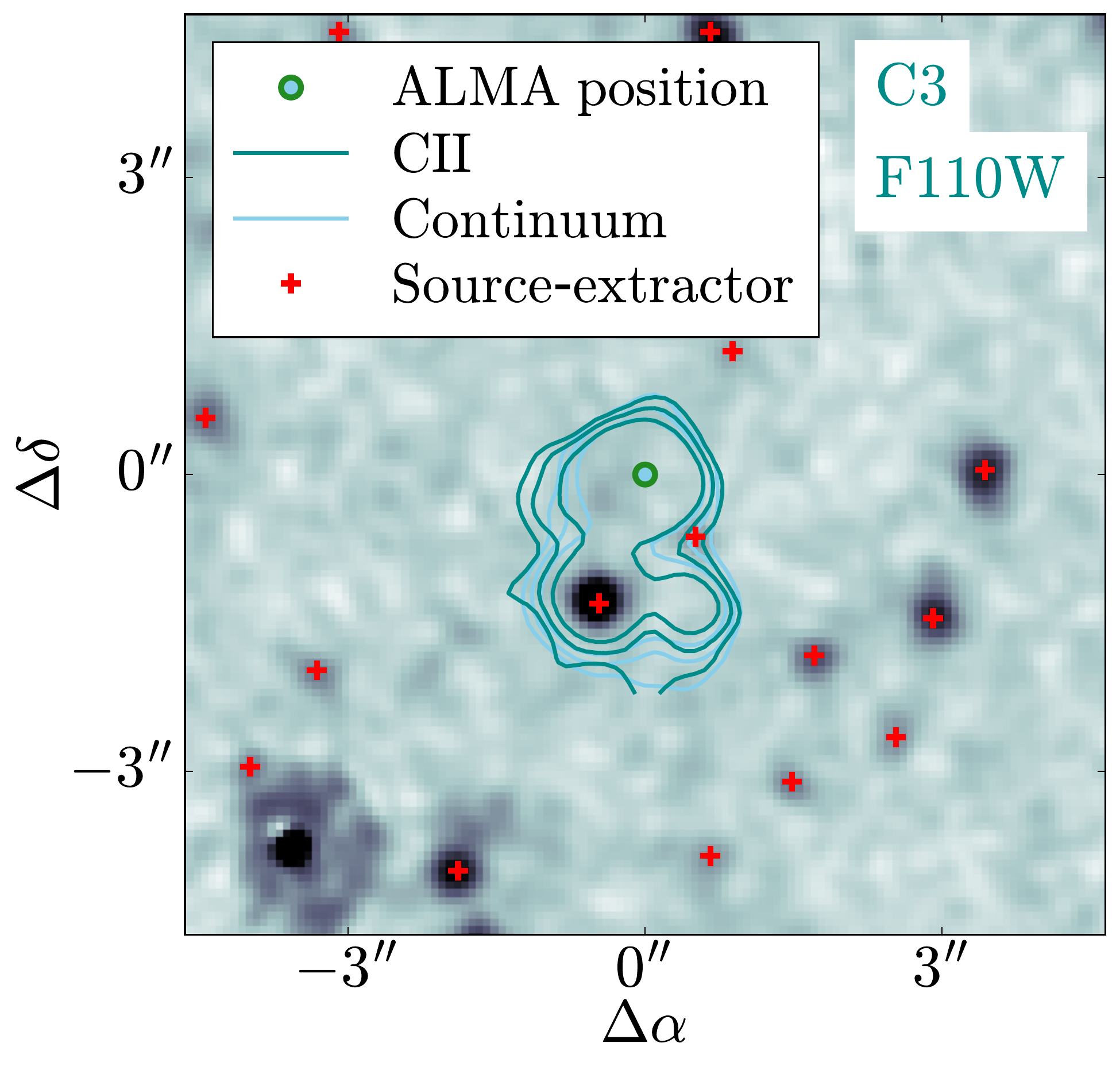}
\includegraphics[width=0.24\textwidth]{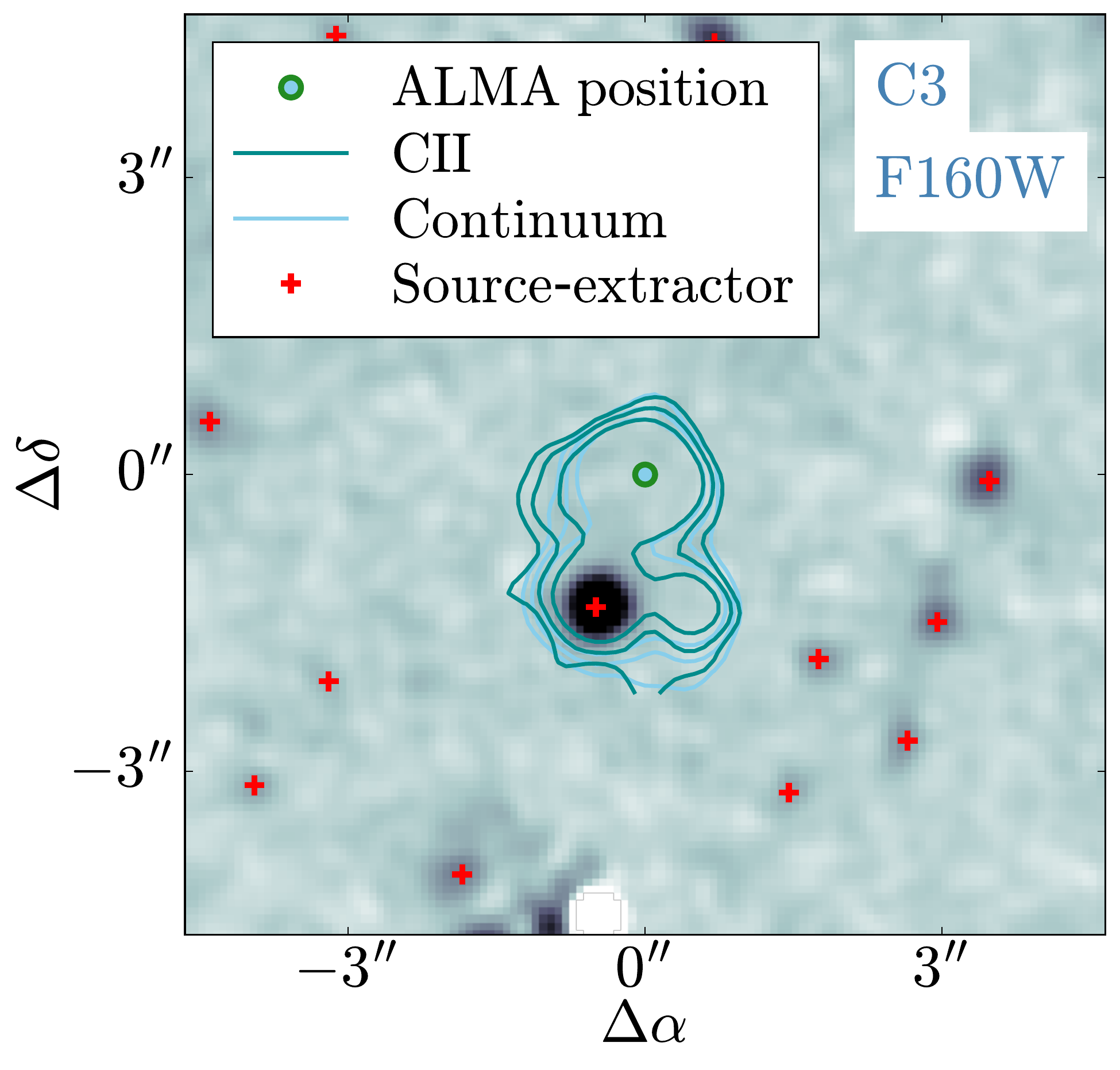}
\includegraphics[width=0.248\textwidth]{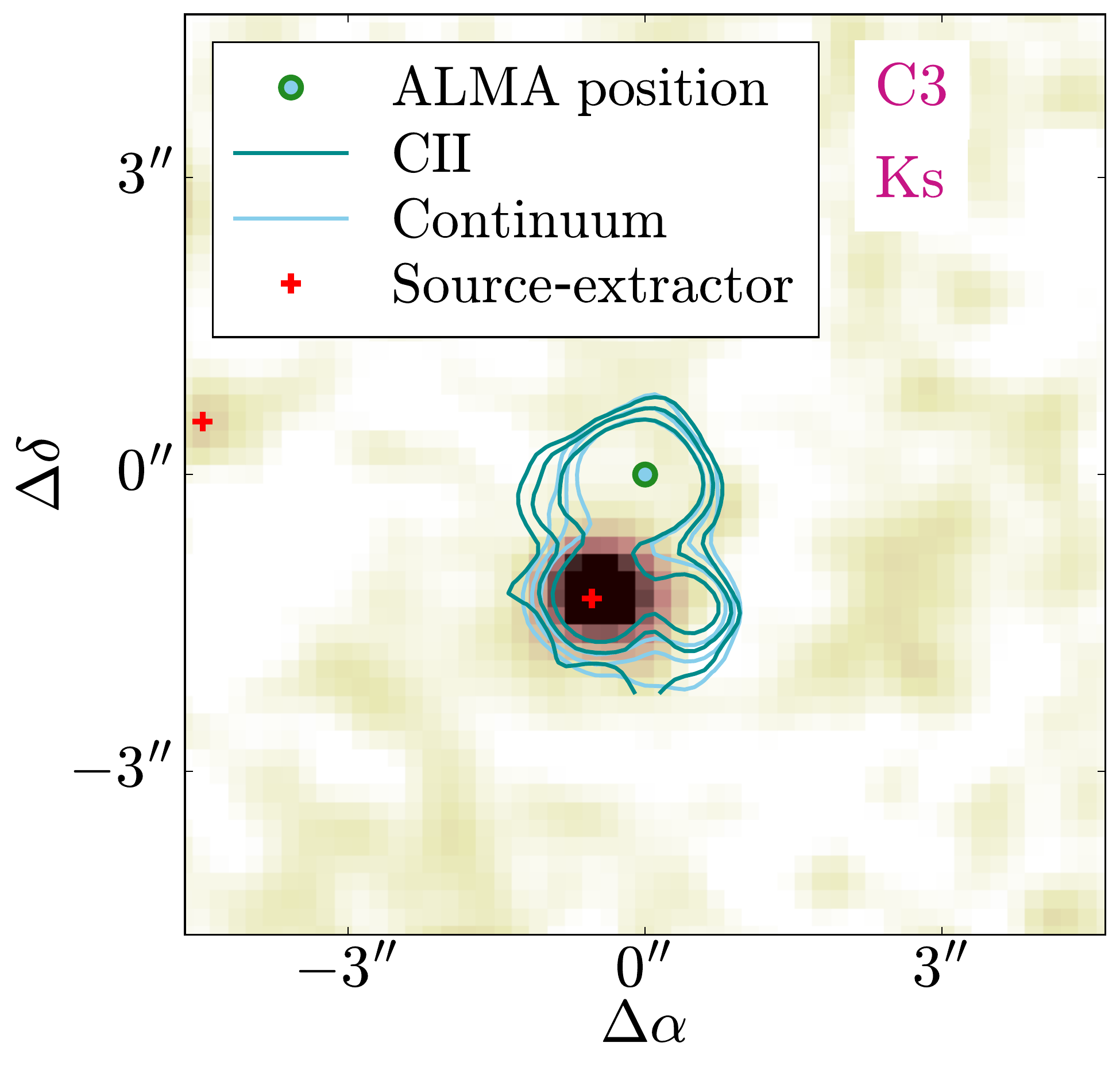}
\includegraphics[width=0.249\textwidth]{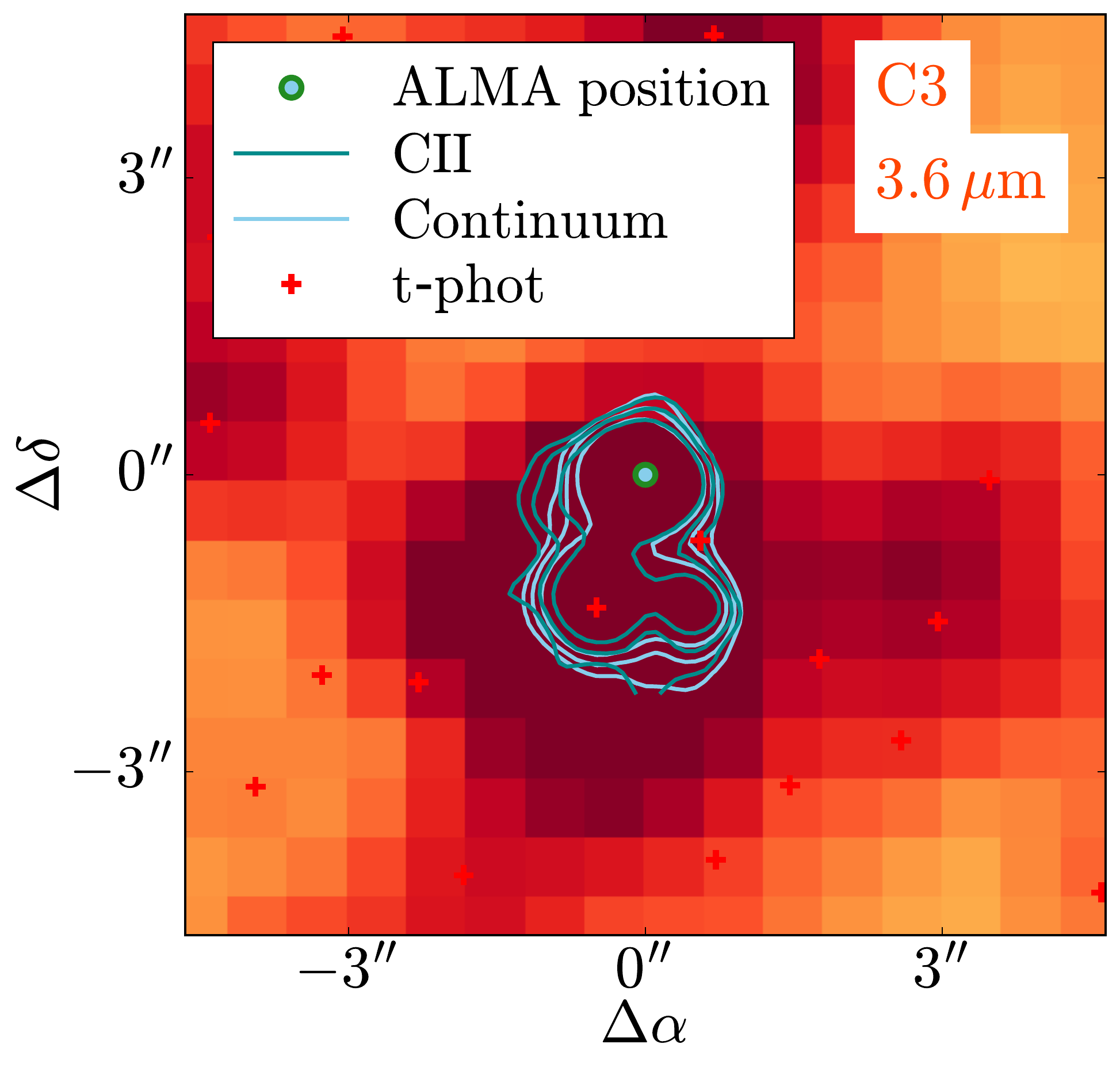}
\includegraphics[width=0.249\textwidth]{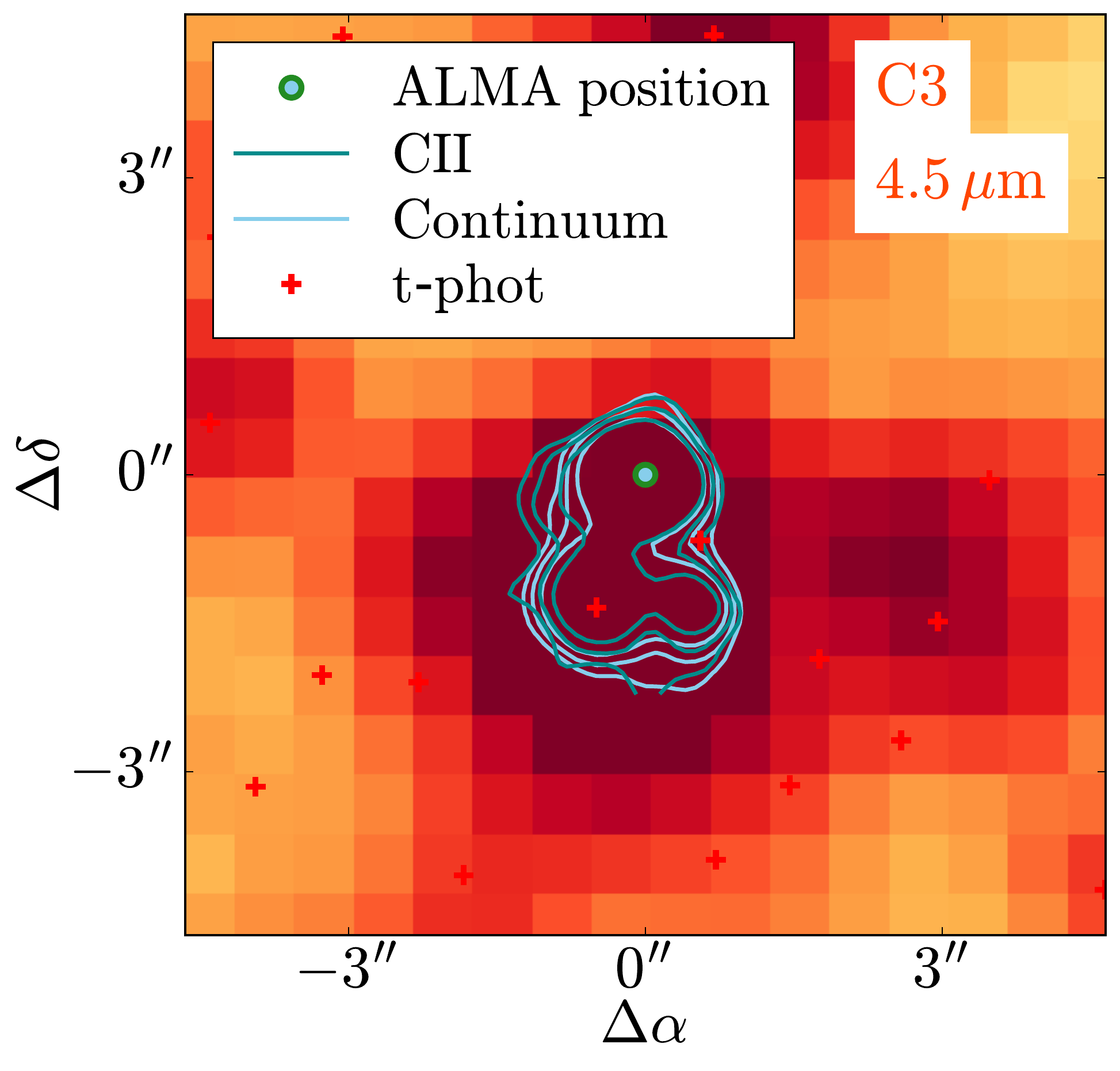}
\end{framed}
\end{subfigure}
\caption{Multiwavelength cutouts for each source in our protocluster sample. In each panel, the ALMA position is shown as the blue point, with ALMA 850-$\mu$m continuum and [C{\sc ii}] line emission contours overlaid. Positions of sources found using {\tt source-extractor} (or {\tt t-phot} for IRAC) are shown as red crosses, with matched ALMA counterparts as circles in green. {\it Clockwise, top-left to bottom-right:} GEMINI-GMOS $g$ band; GEMINI-GMOS $r$ band; GEMINI-GMOS $i$ band; {\it HST\/}-F110W; {\it HST\/}-F160W; GEMINI-FLAMINGOS-2 $K_{\rm s}$ band; {\it Spitzer\/}-IRAC 3.6\,$\mu$m; and {\it Spitzer\/}-IRAC 4.5\,$\mu$m. Panels are blank where we do not have data at a given wavelength for a given source.}
\label{cutouts}
\end{figure*}

In Fig.~\ref{cutouts} we show all 32 12$\,{\times}\,12$\,arcsec source cutouts from our multiwavelength coverage. This includes GEMINI-GMOS in the $g$, $r$, and $i$ bands, GEMINI-FLAMINGOS-2 imaging in the $K_{\rm s}$ band, {\it HST\/} imaging with the F110W and F160W filters, and {\it Spitzer\/}-IRAC imaging at 3.6\,$\mu$m and 4.5\,$\mu$m. In each frame, the ALMA position is shown as the blue point, with ALMA 850-$\mu$m continuum and [C{\sc ii}] line emission contours overlaid. Positions of sources found using {\tt source-extractor} (or {\tt t-phot} for IRAC) are shown as red crosses, with matched ALMA counterparts circled in green. Panels are blank where we do not have data at a given wavelength for a given source.

\renewcommand{\thefigure}{B\arabic{figure} (Cont.)}
\addtocounter{figure}{-1}
\begin{figure*}
\begin{subfigure}{0.85\textwidth}
\begin{framed}
\includegraphics[width=0.24\textwidth]{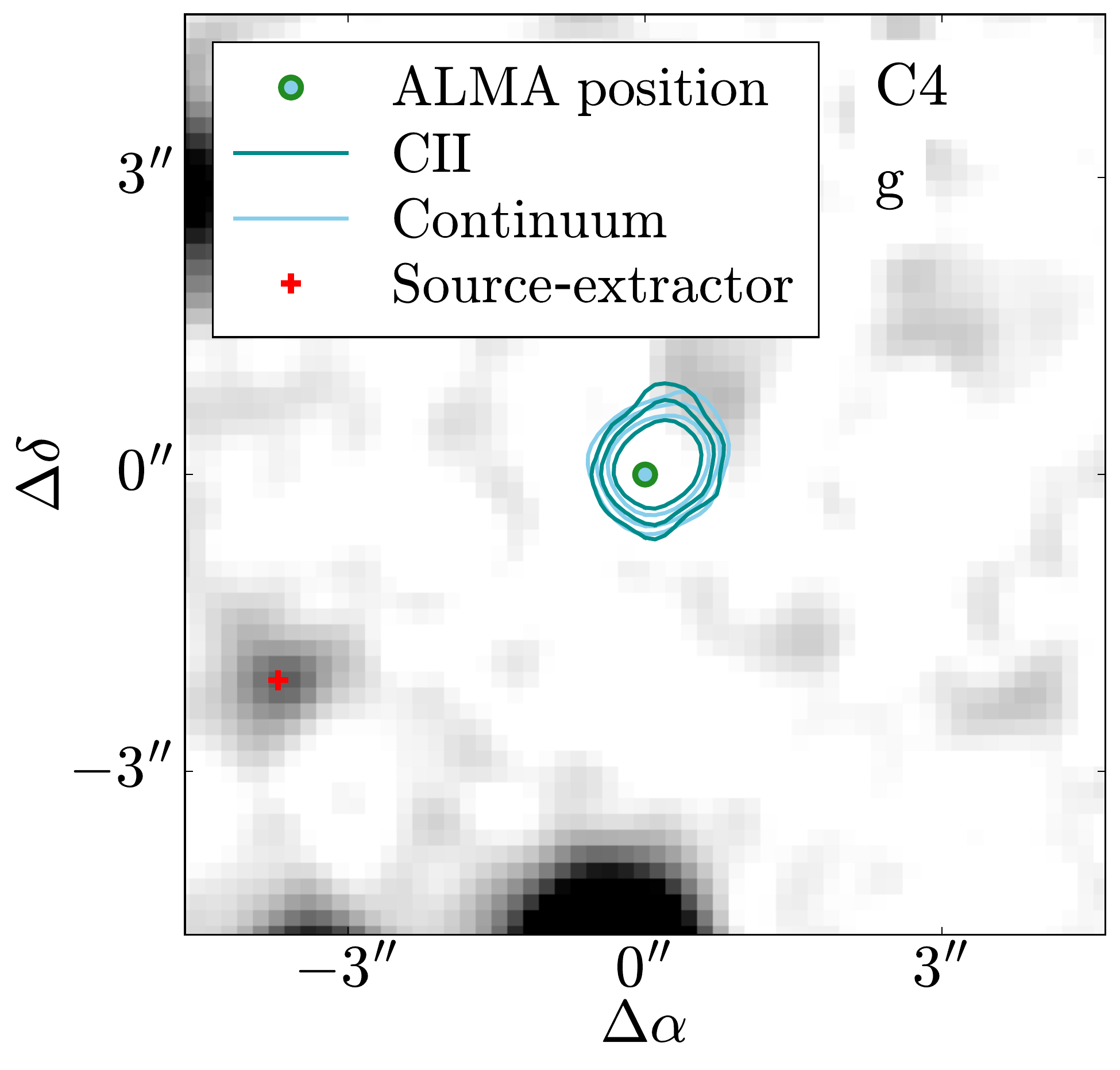}
\includegraphics[width=0.24\textwidth]{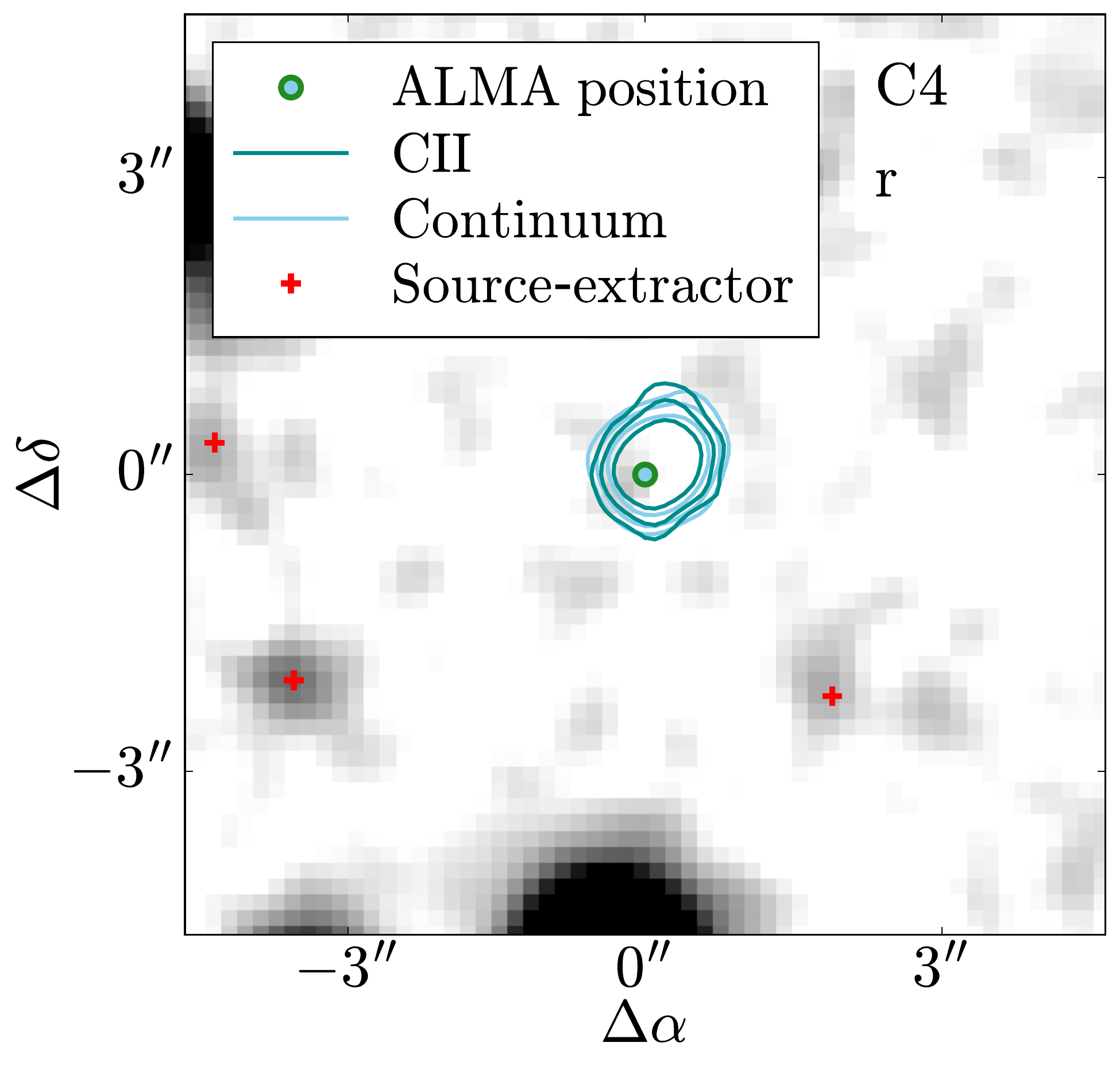}
\includegraphics[width=0.24\textwidth]{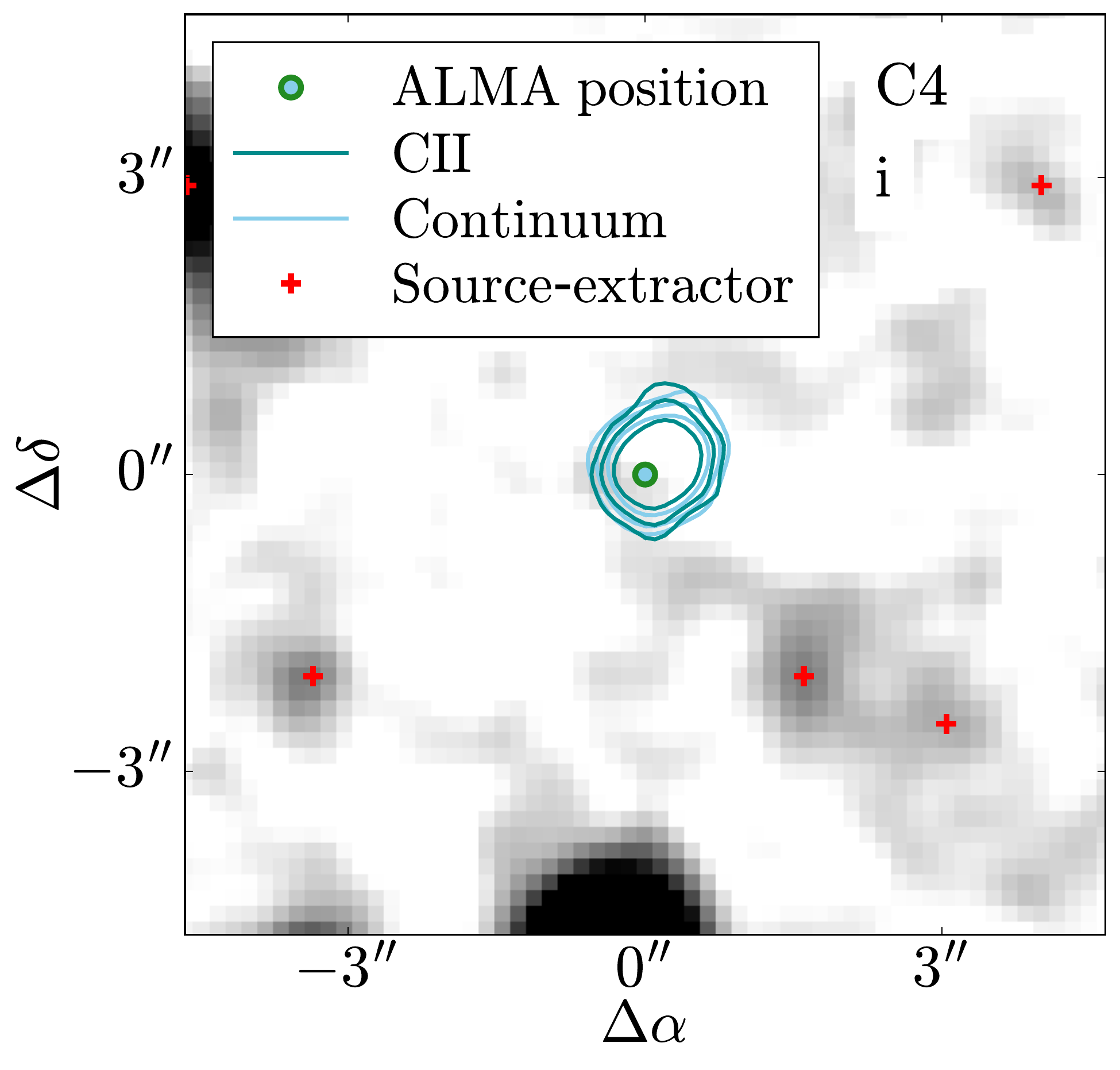}
\includegraphics[width=0.24\textwidth]{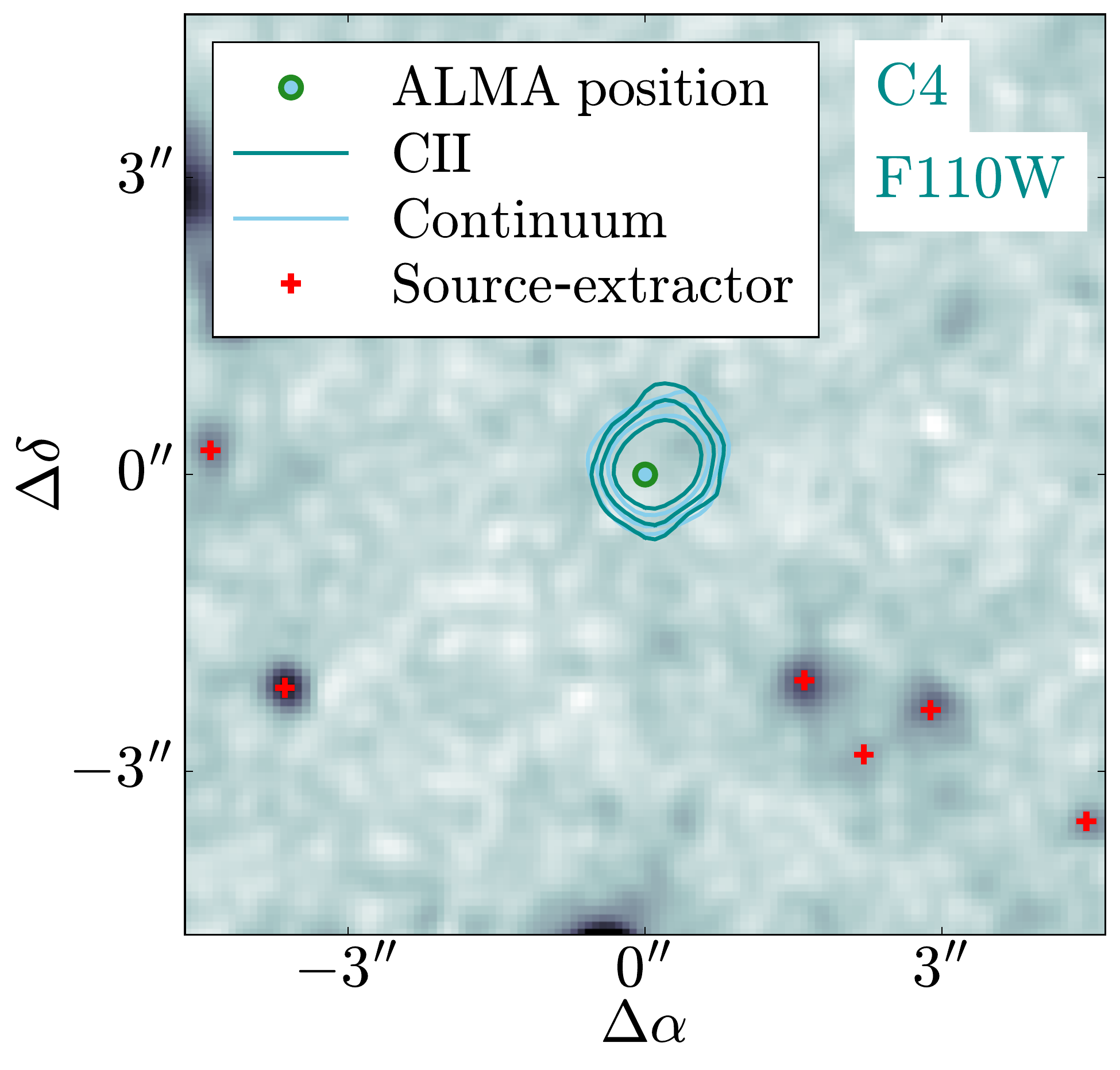}
\includegraphics[width=0.24\textwidth]{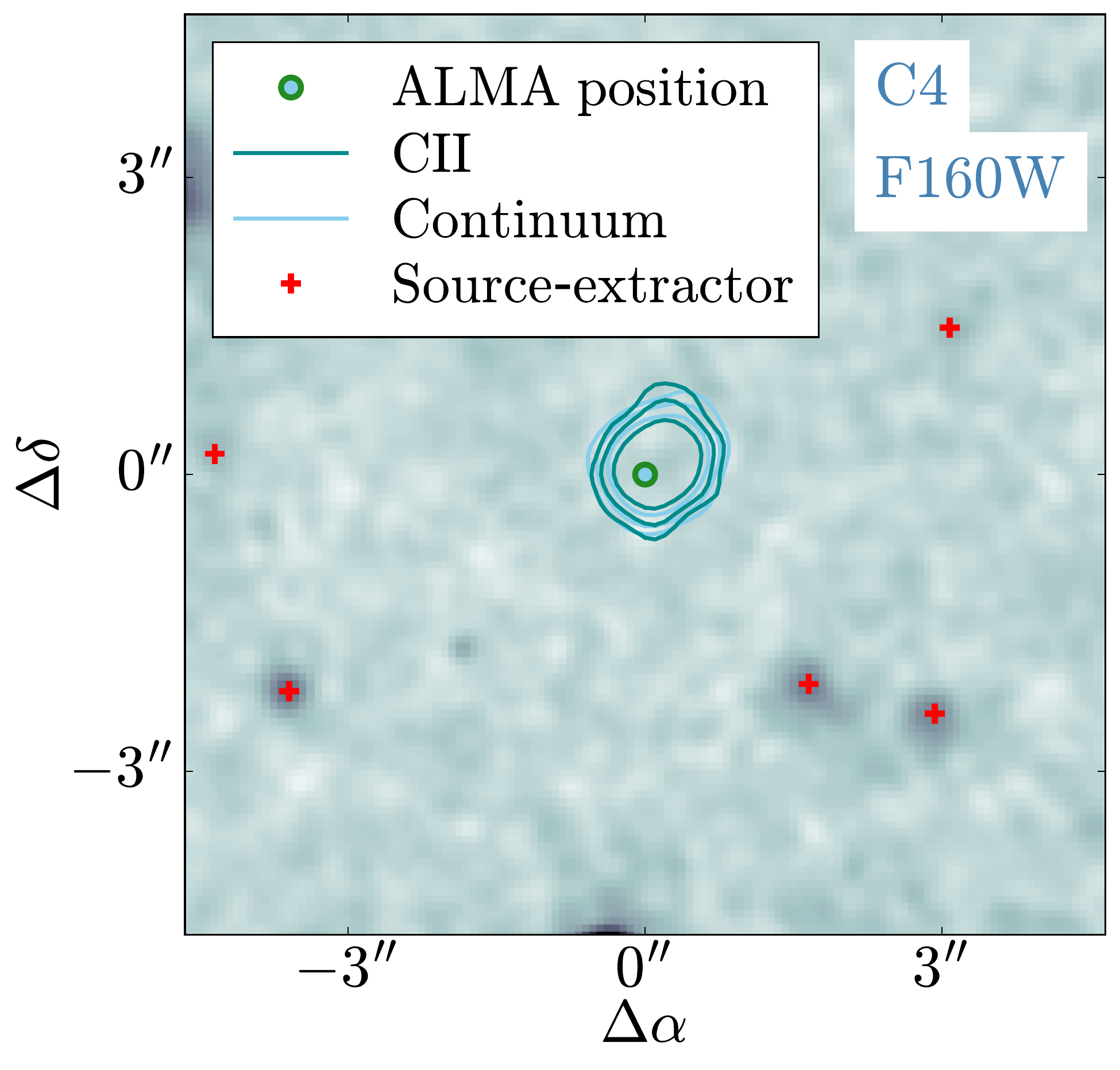}
\includegraphics[width=0.248\textwidth]{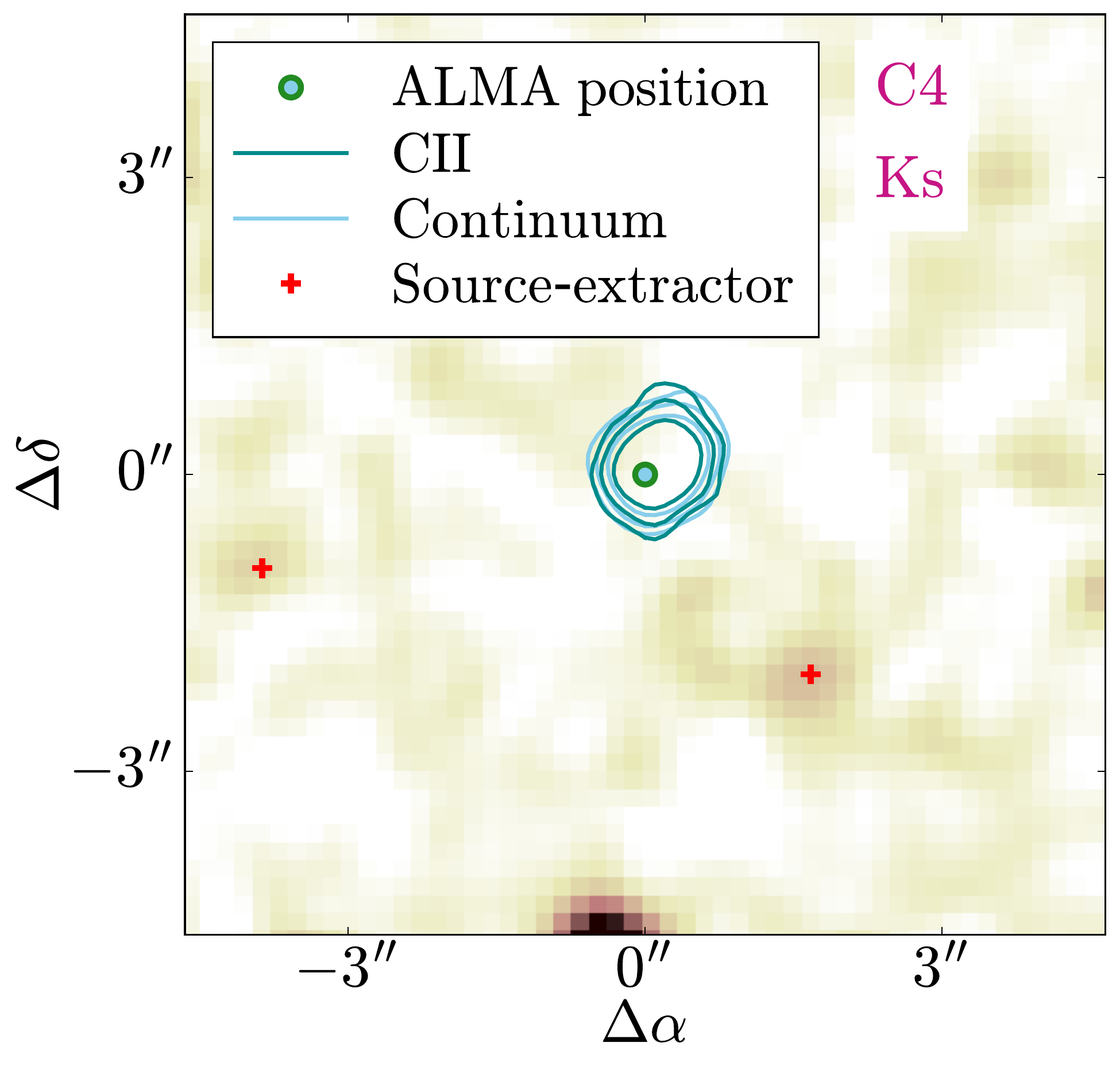}
\includegraphics[width=0.249\textwidth]{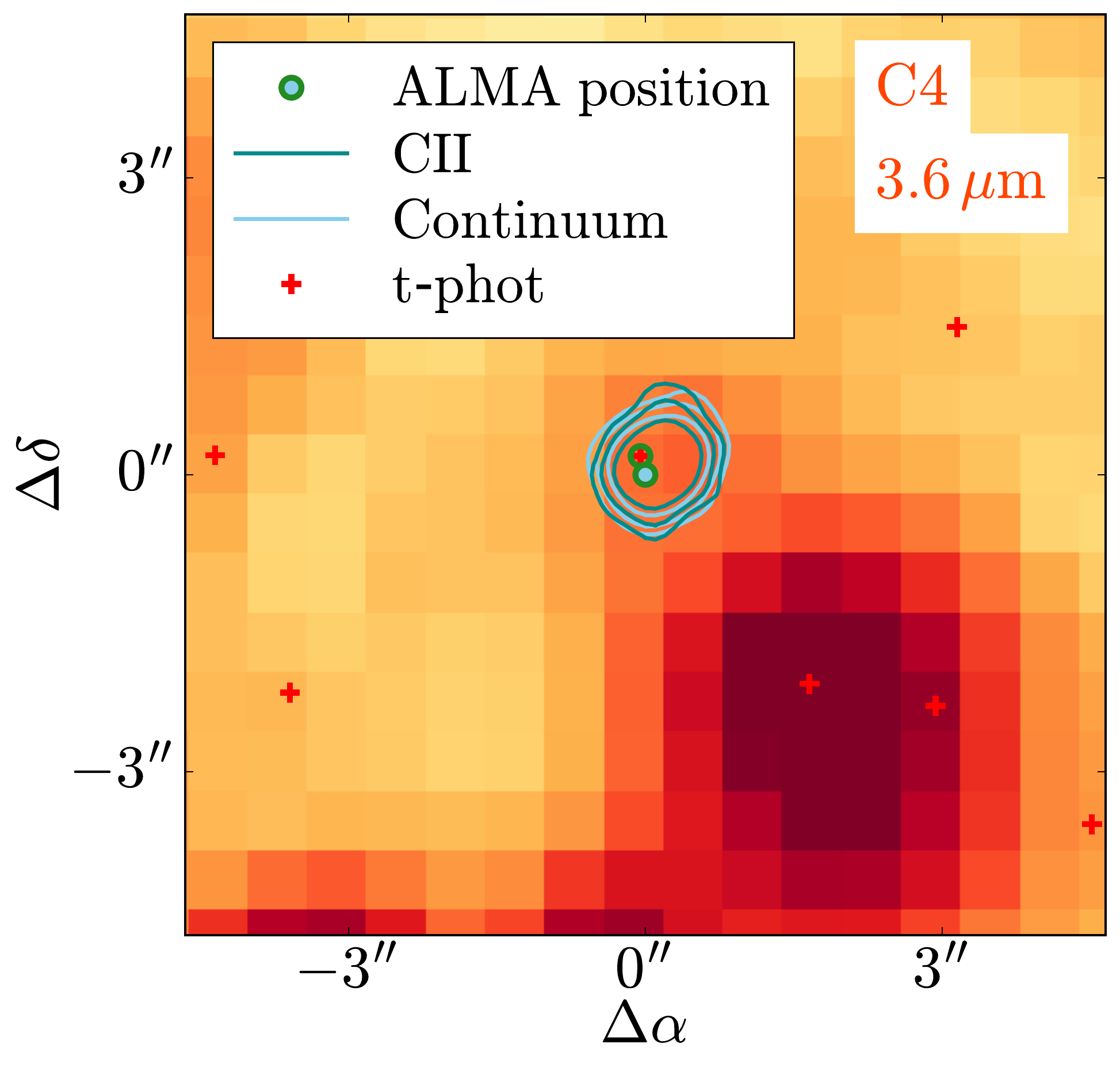}
\includegraphics[width=0.249\textwidth]{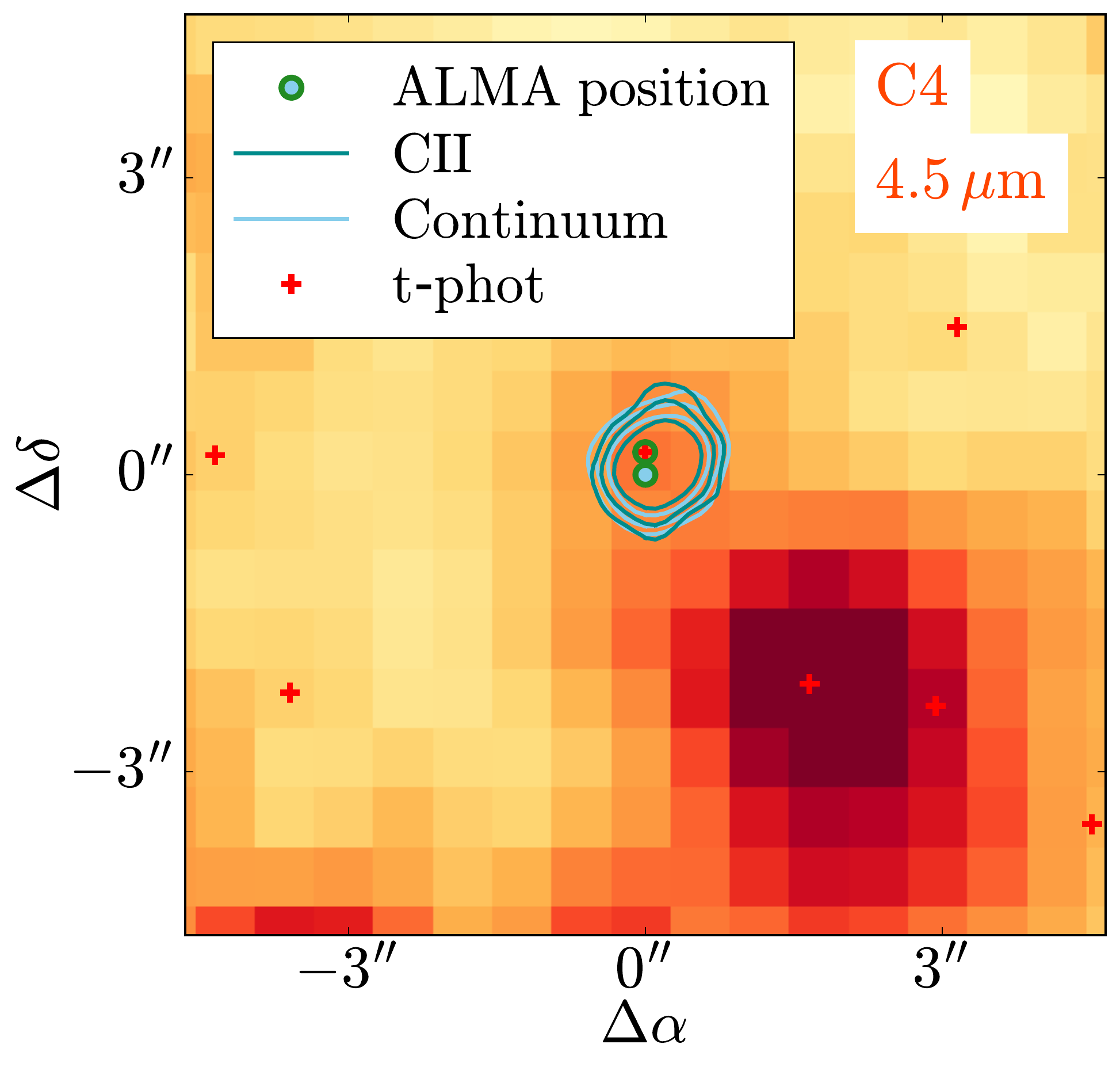}
\end{framed}
\end{subfigure}
\begin{subfigure}{0.85\textwidth}
\begin{framed}
\includegraphics[width=0.24\textwidth]{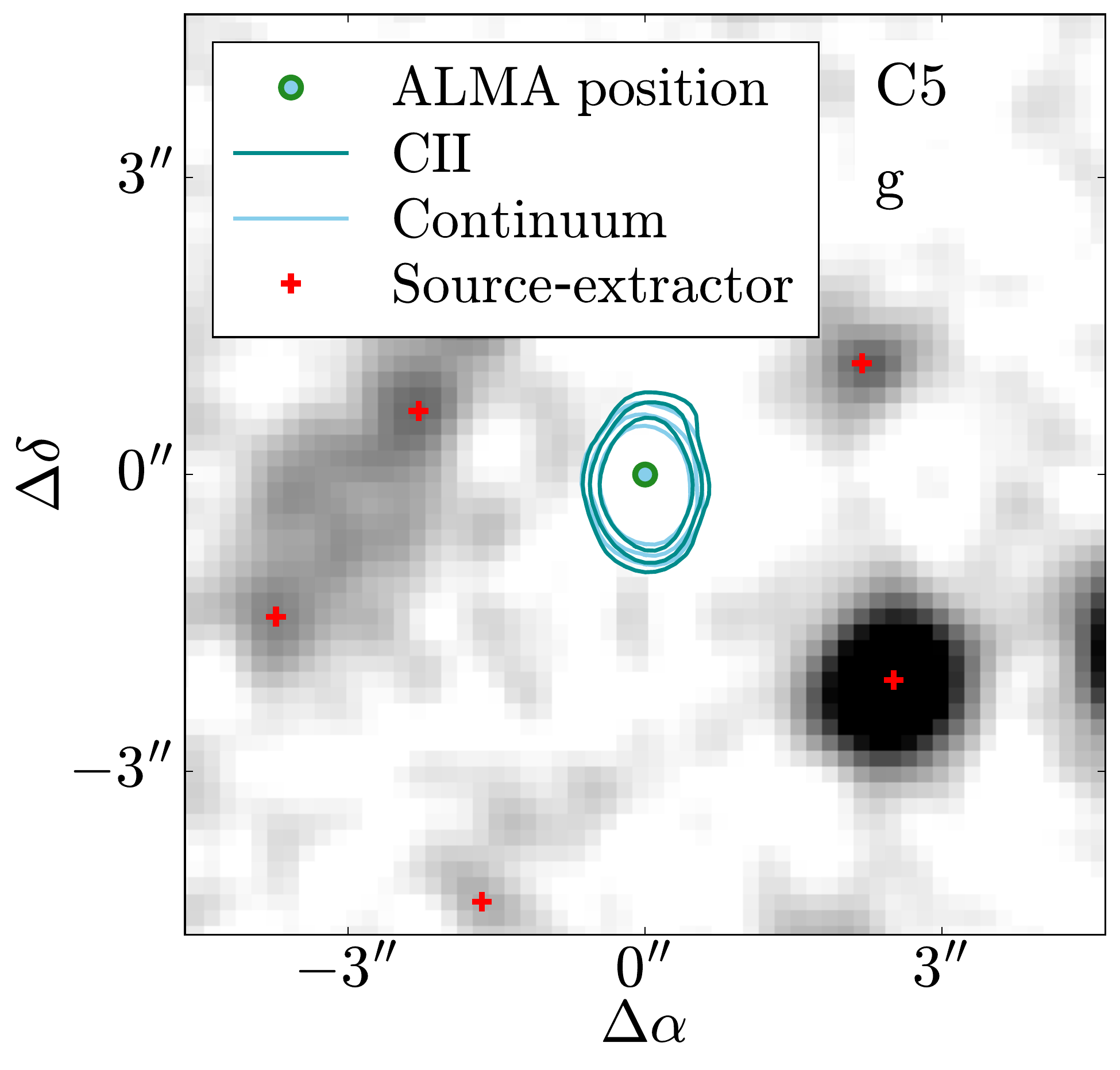}
\includegraphics[width=0.24\textwidth]{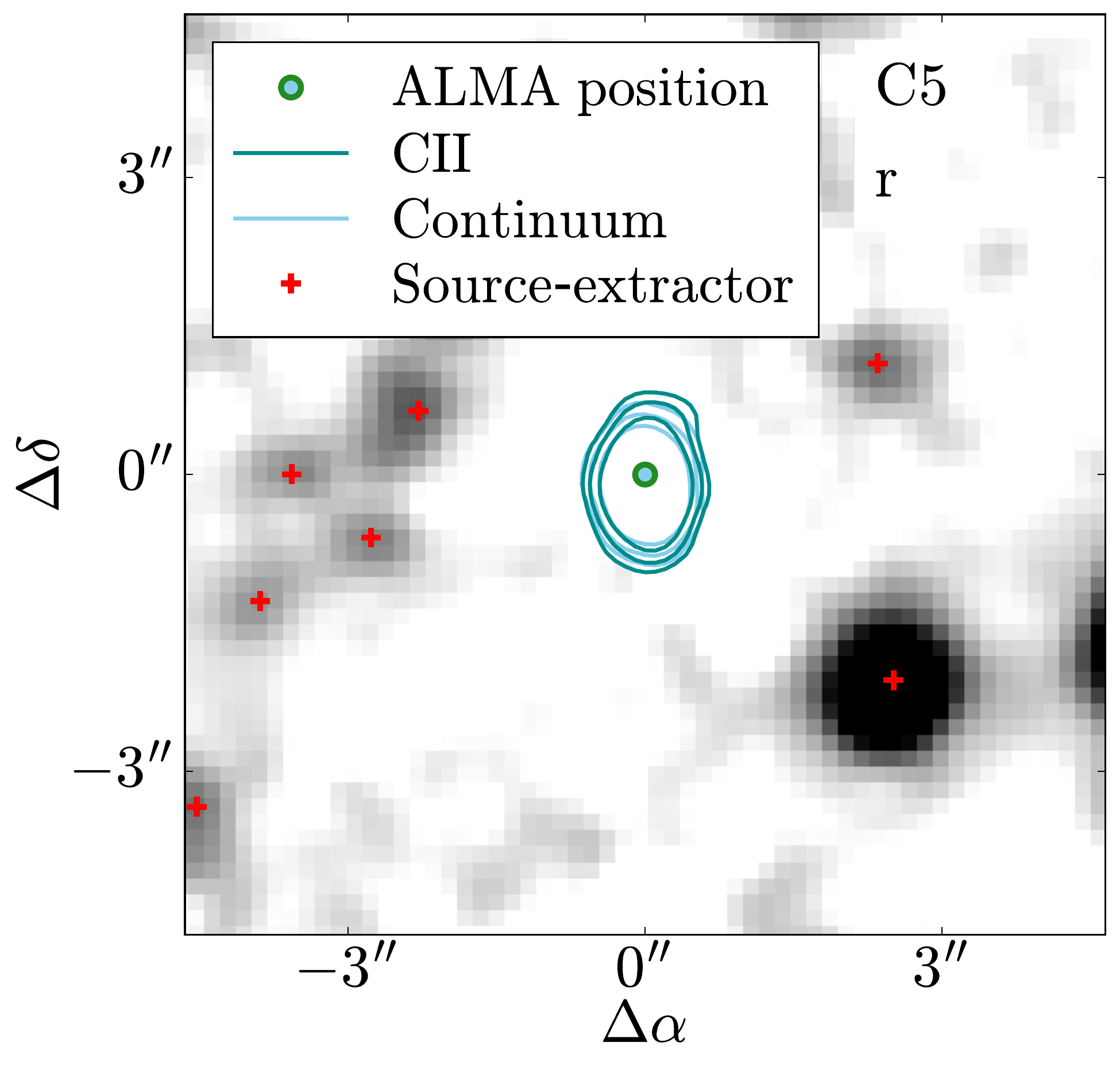}
\includegraphics[width=0.24\textwidth]{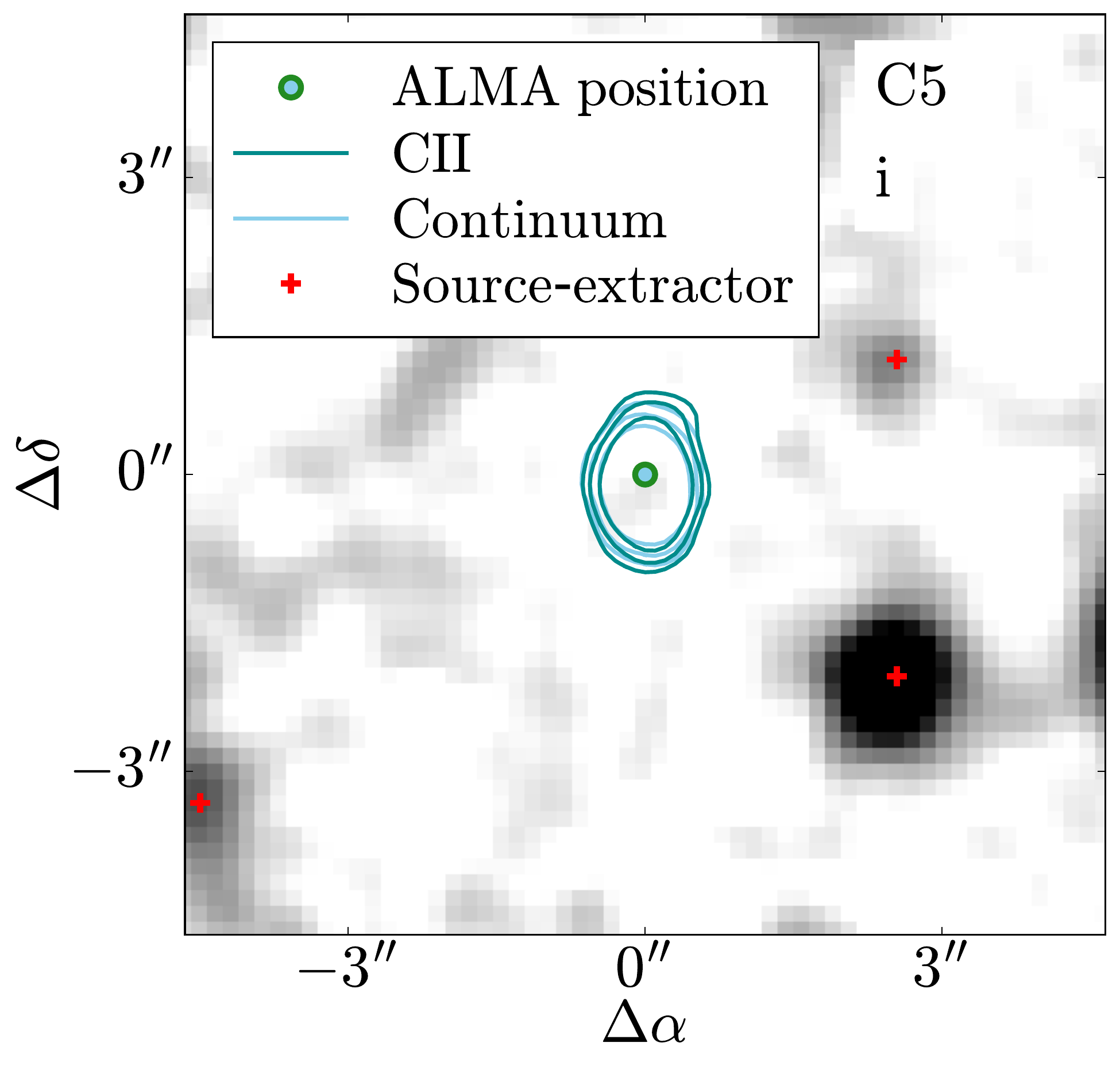}
\includegraphics[width=0.24\textwidth]{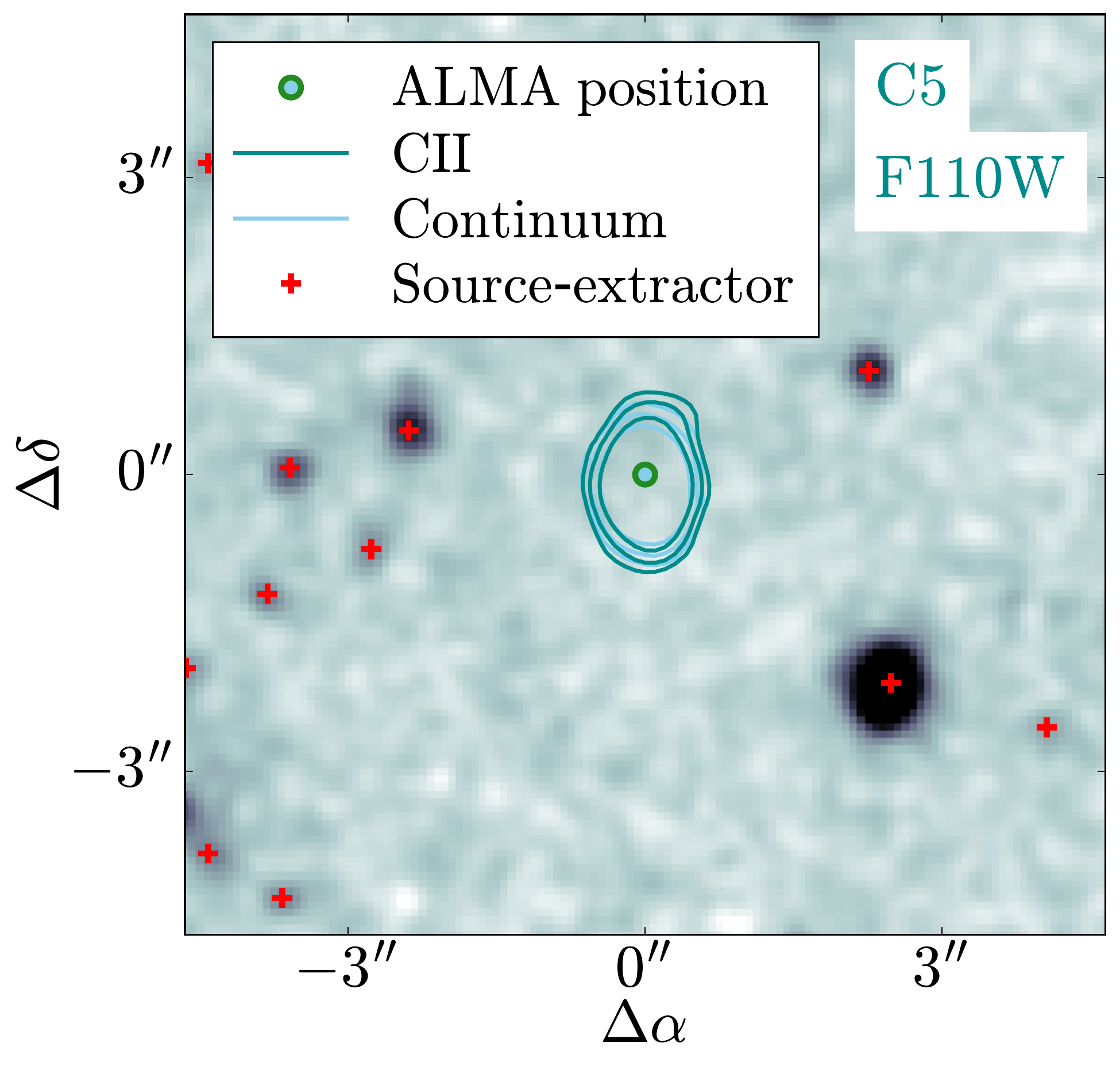}
\includegraphics[width=0.24\textwidth]{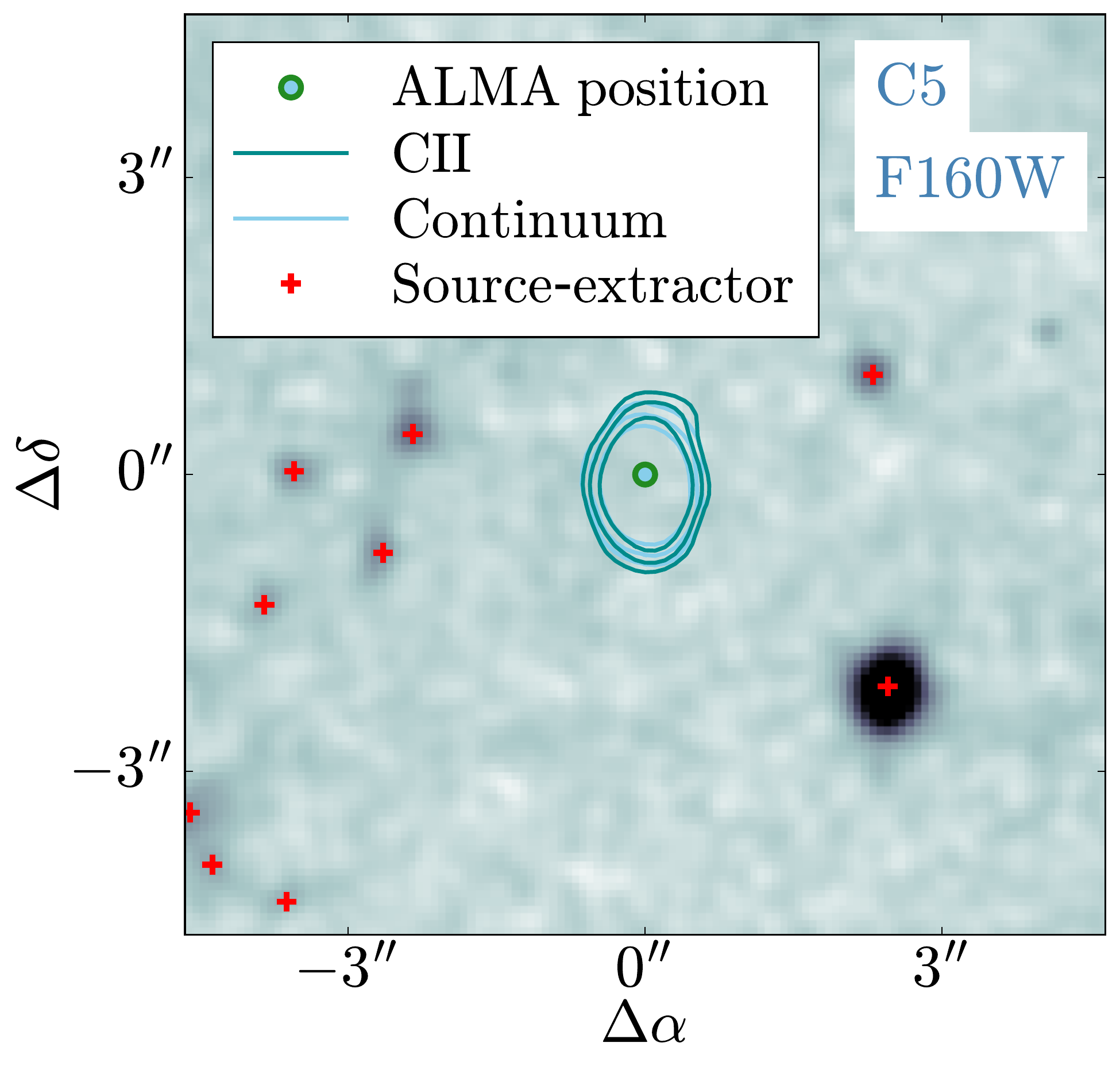}
\includegraphics[width=0.248\textwidth]{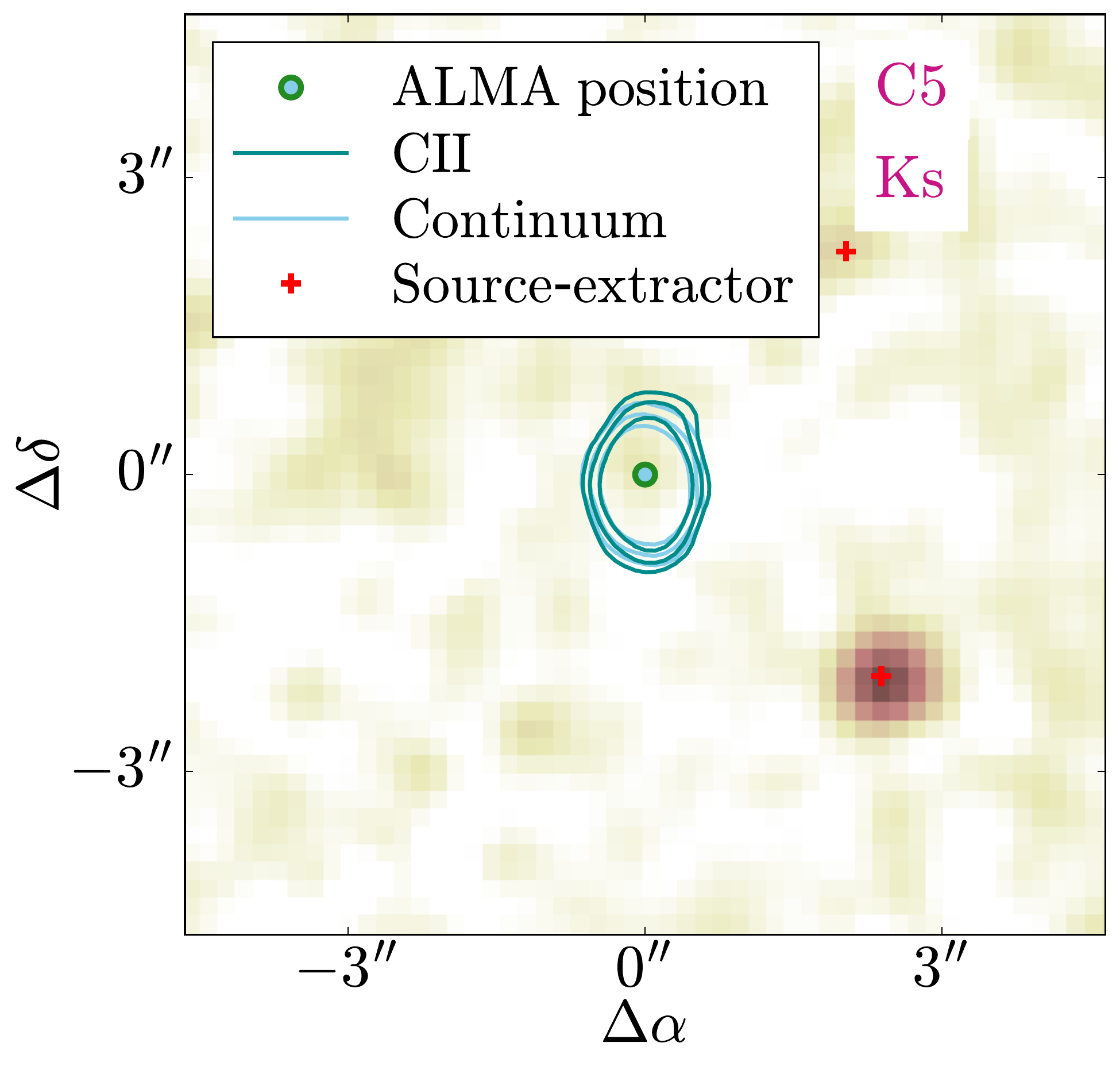}
\includegraphics[width=0.249\textwidth]{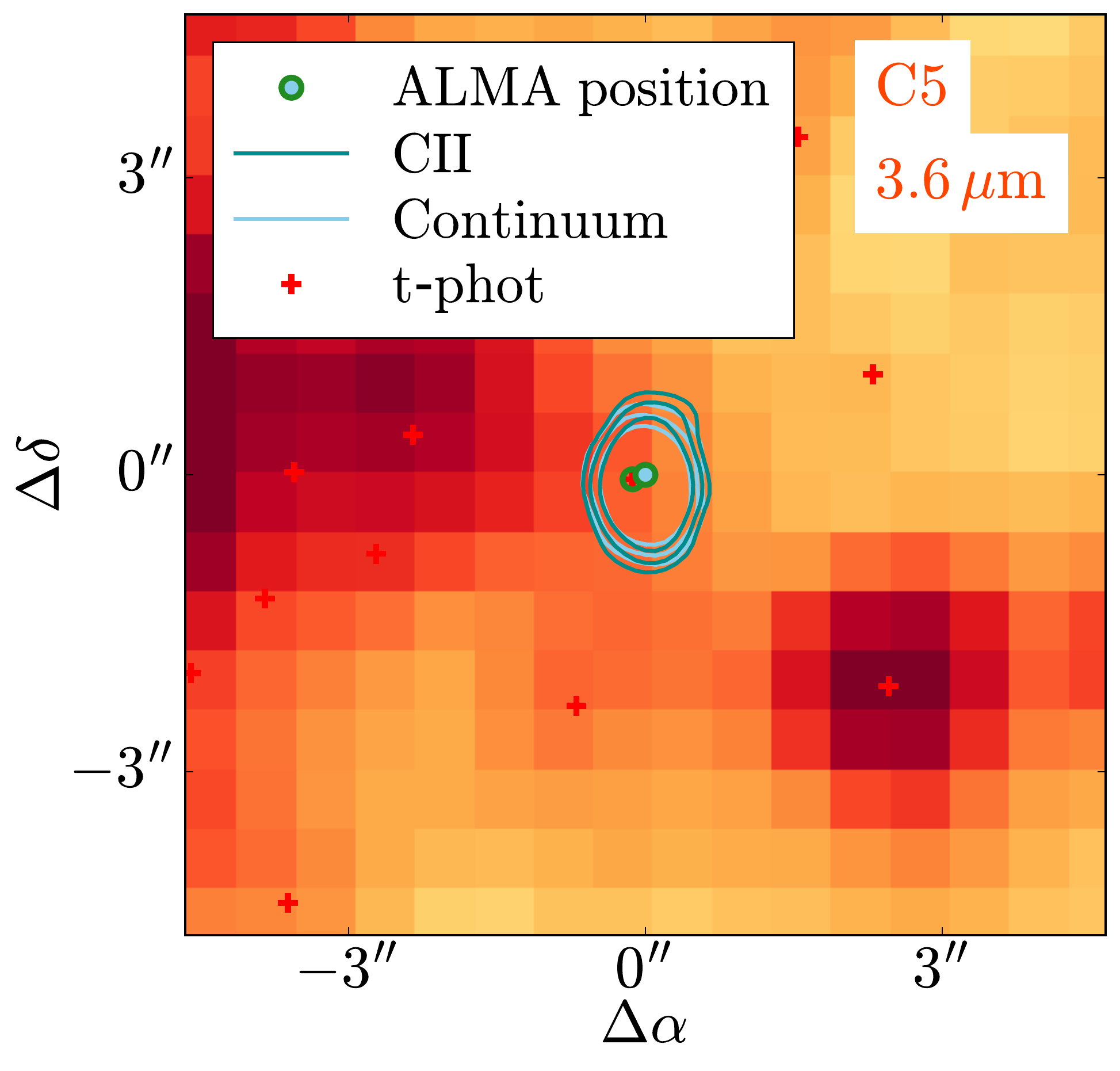}
\includegraphics[width=0.249\textwidth]{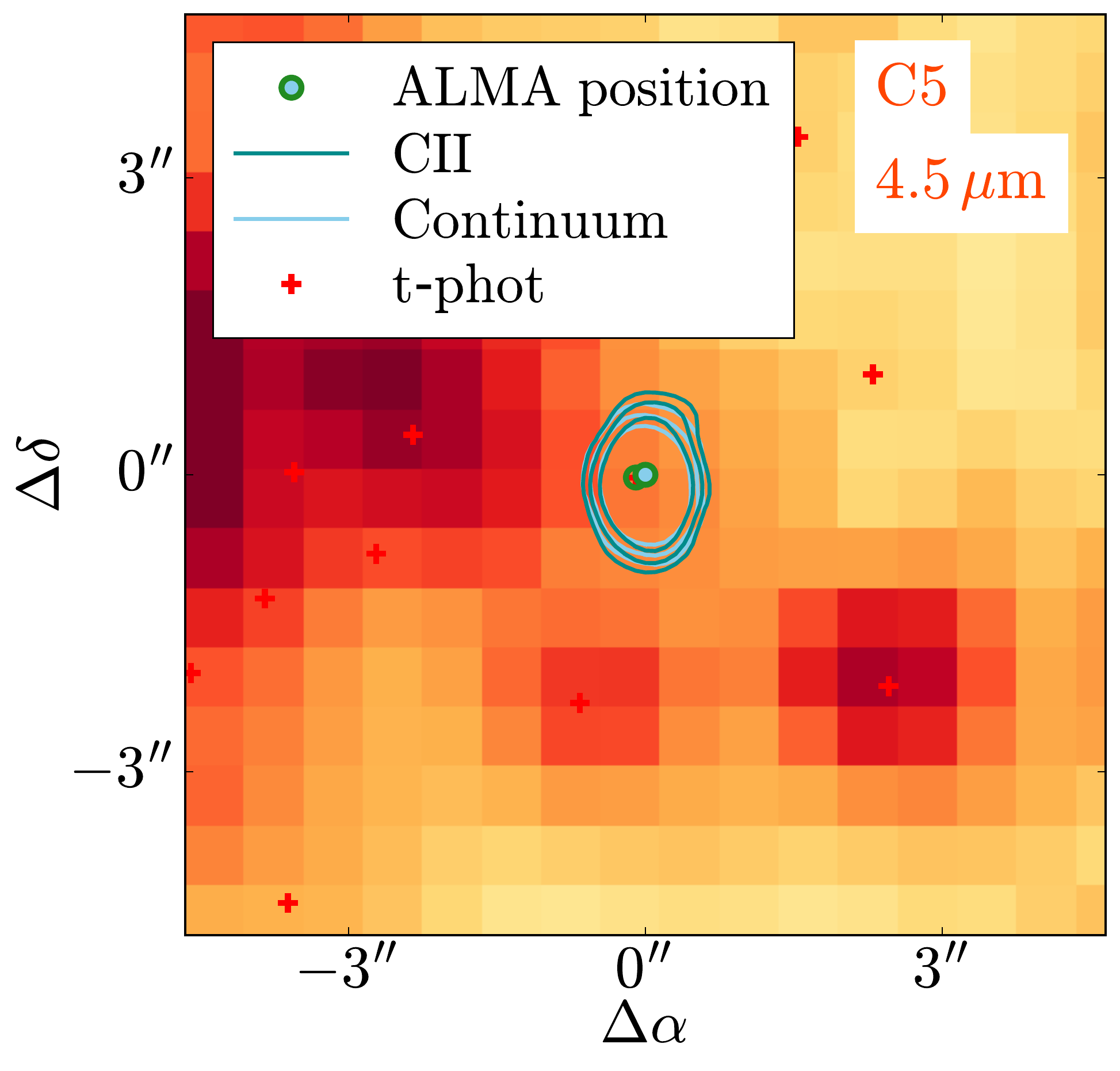}
\end{framed}
\end{subfigure}
\begin{subfigure}{0.85\textwidth}
\begin{framed}
\includegraphics[width=0.24\textwidth]{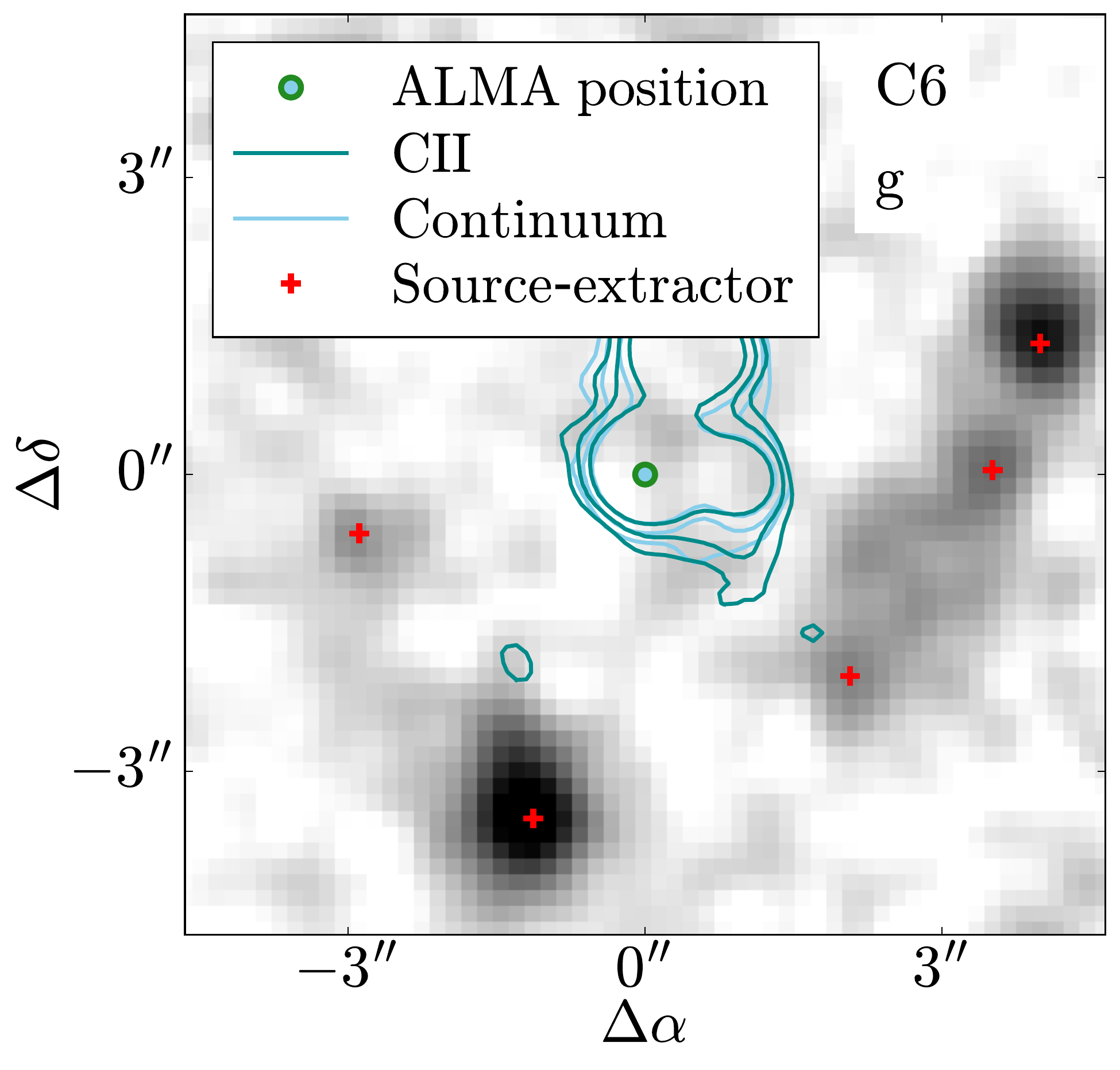}
\includegraphics[width=0.24\textwidth]{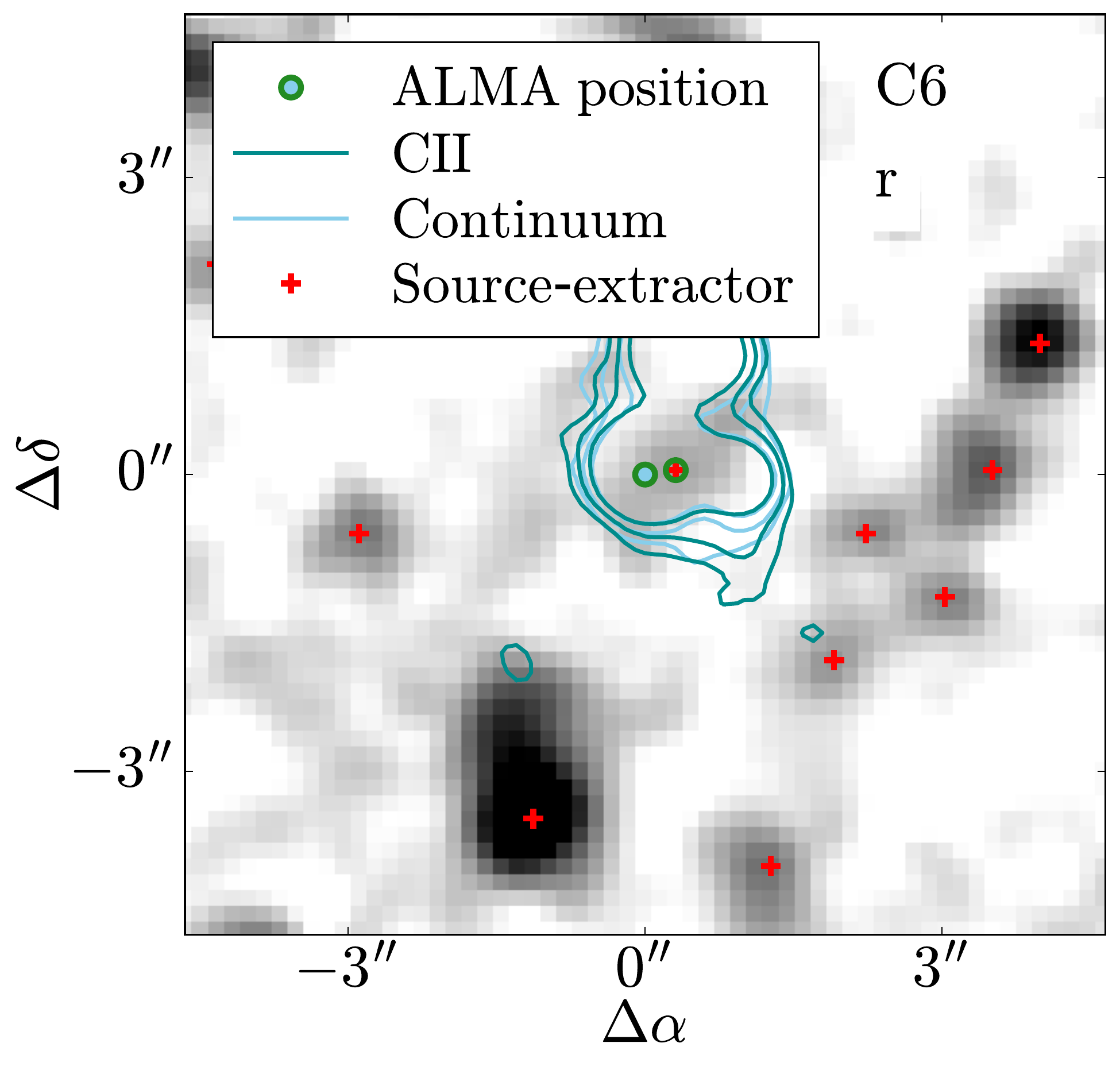}
\includegraphics[width=0.24\textwidth]{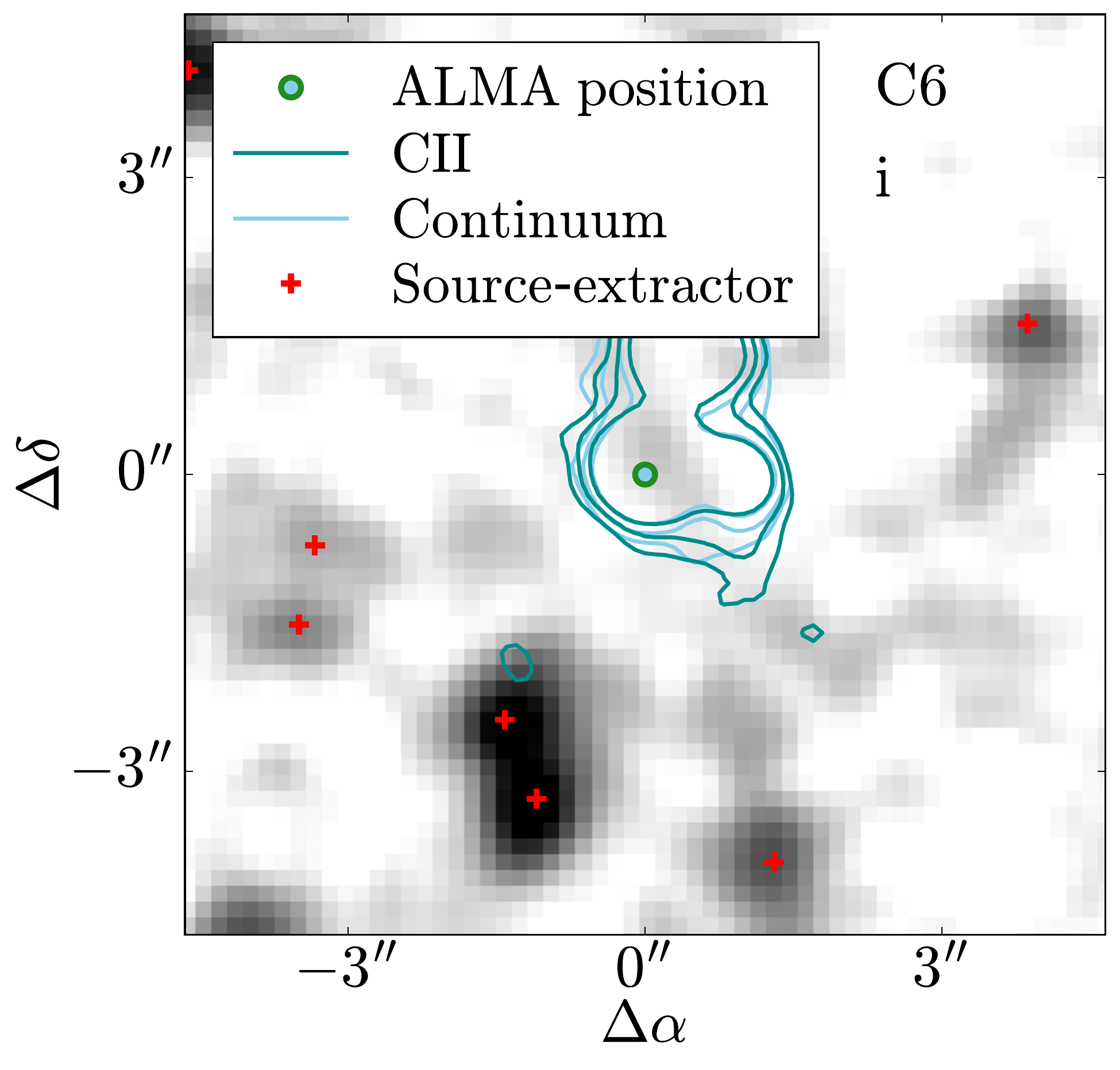}
\includegraphics[width=0.24\textwidth]{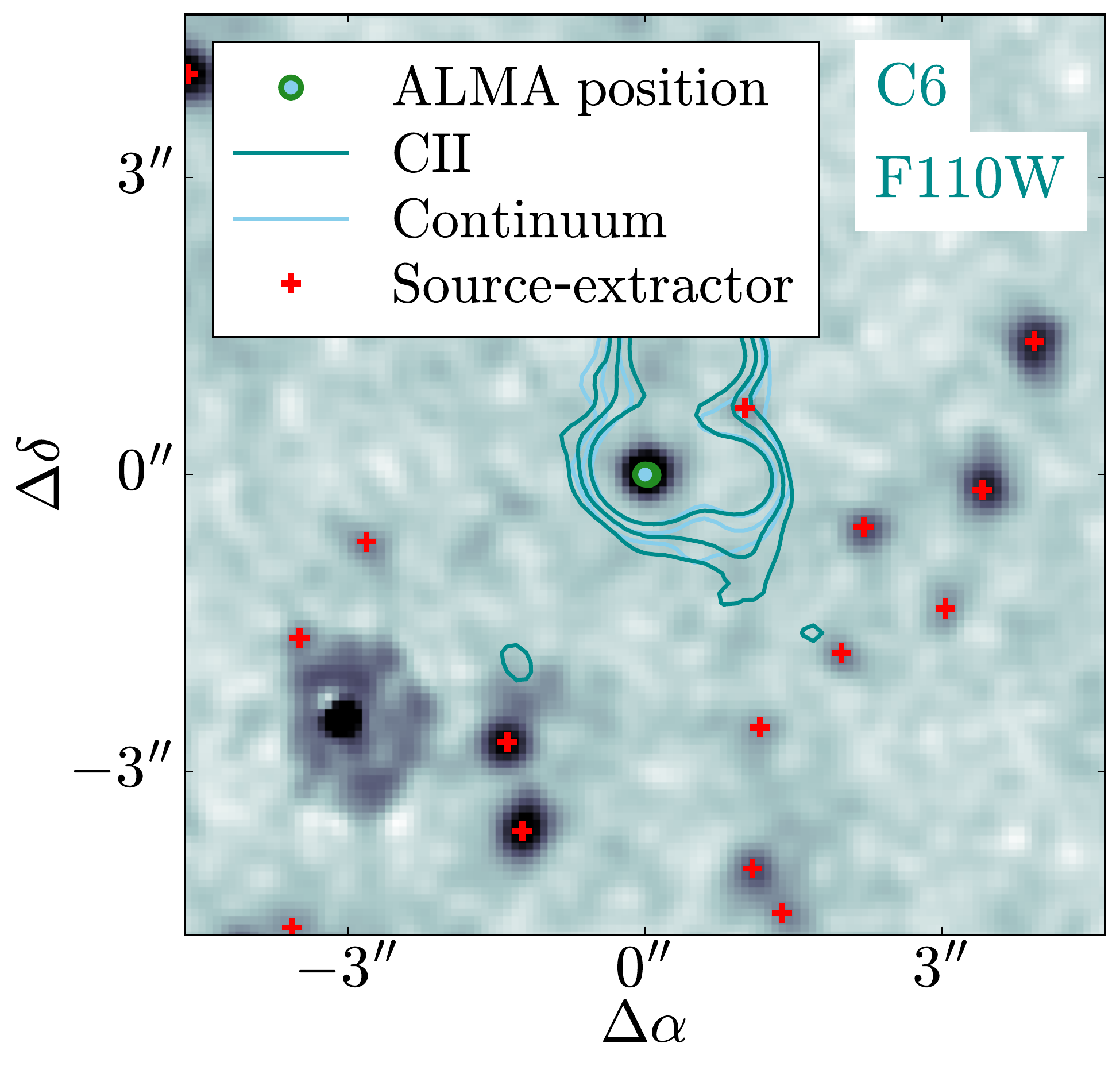}
\includegraphics[width=0.24\textwidth]{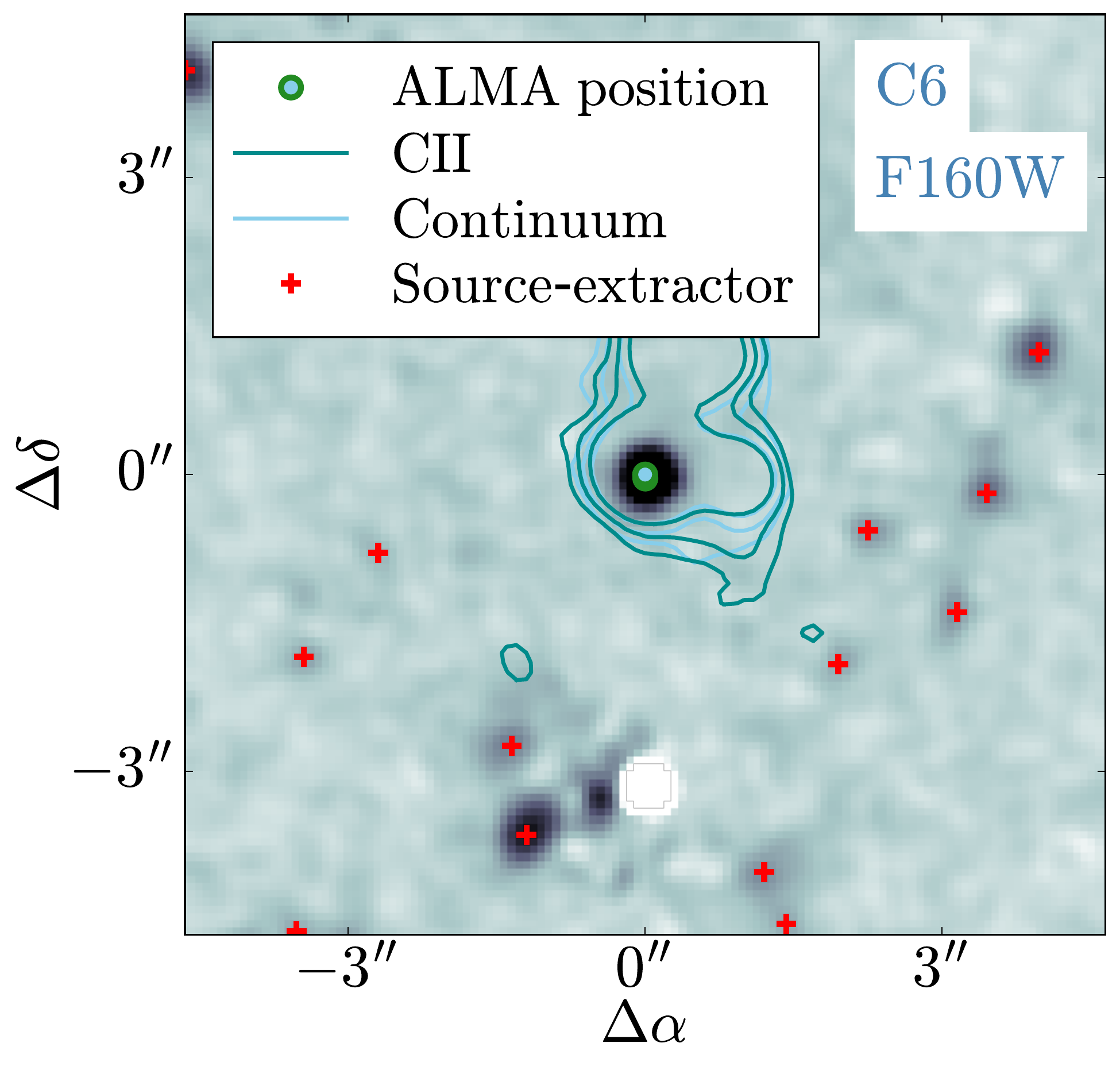}
\includegraphics[width=0.248\textwidth]{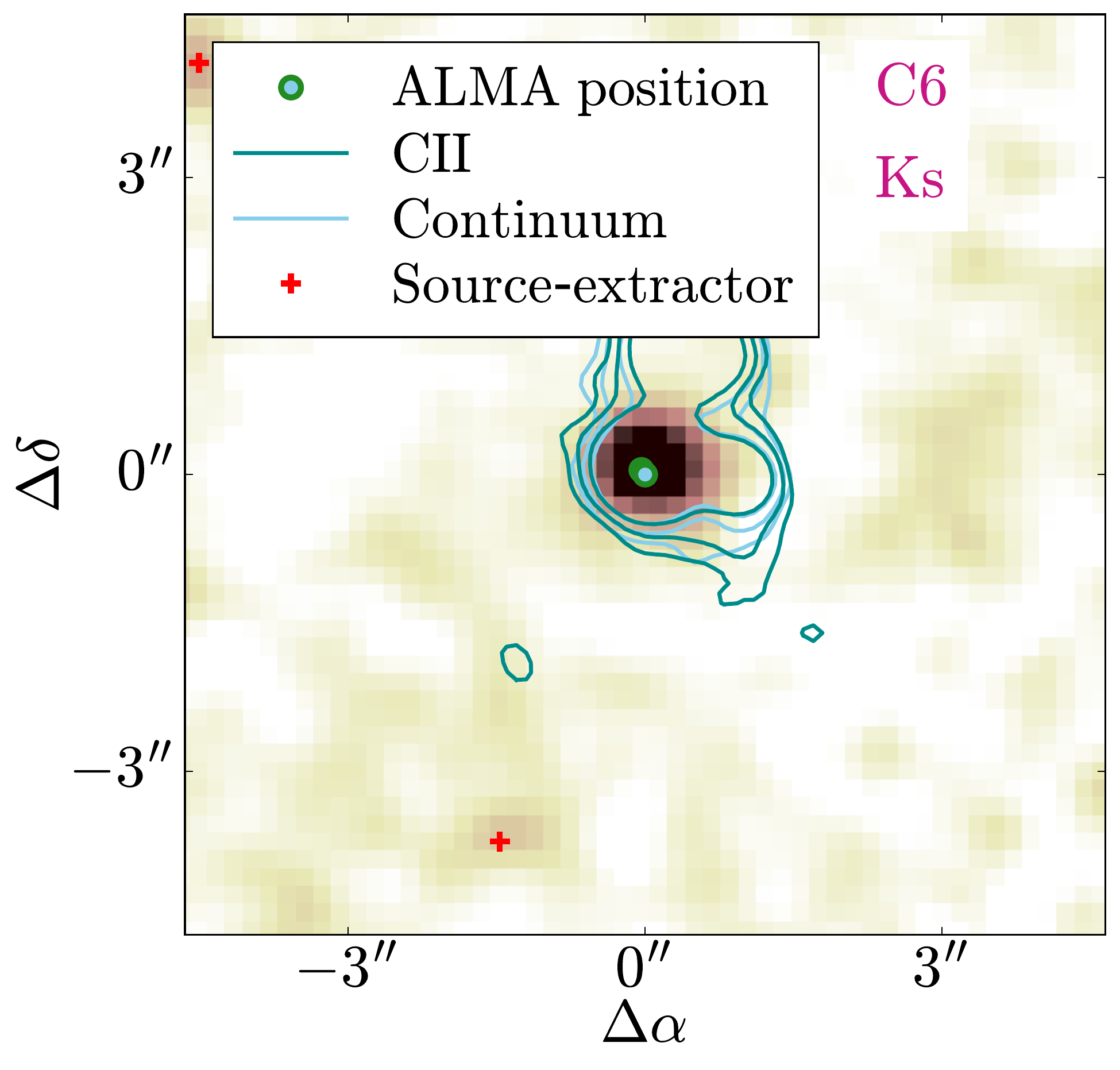}
\includegraphics[width=0.249\textwidth]{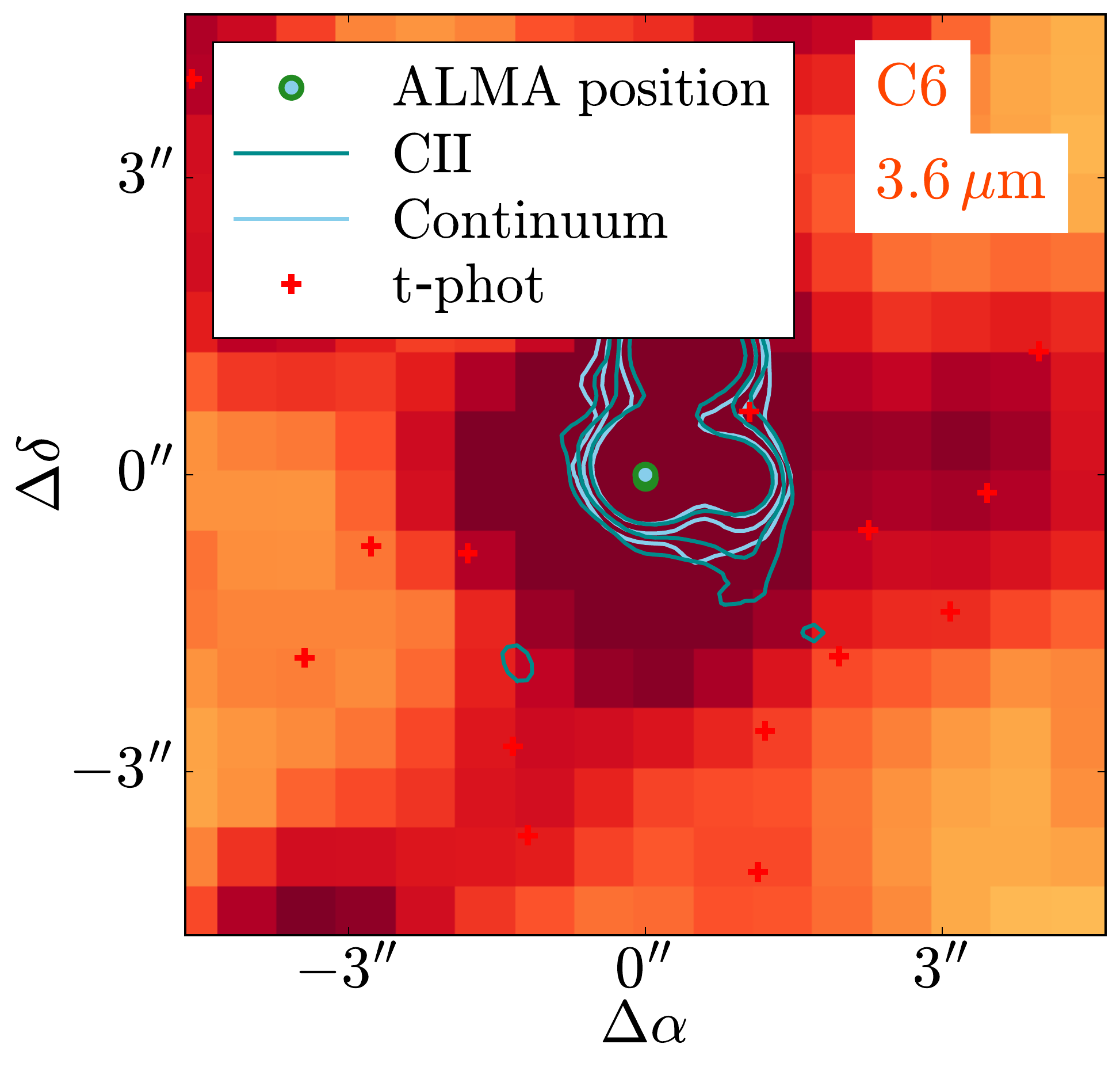}
\includegraphics[width=0.249\textwidth]{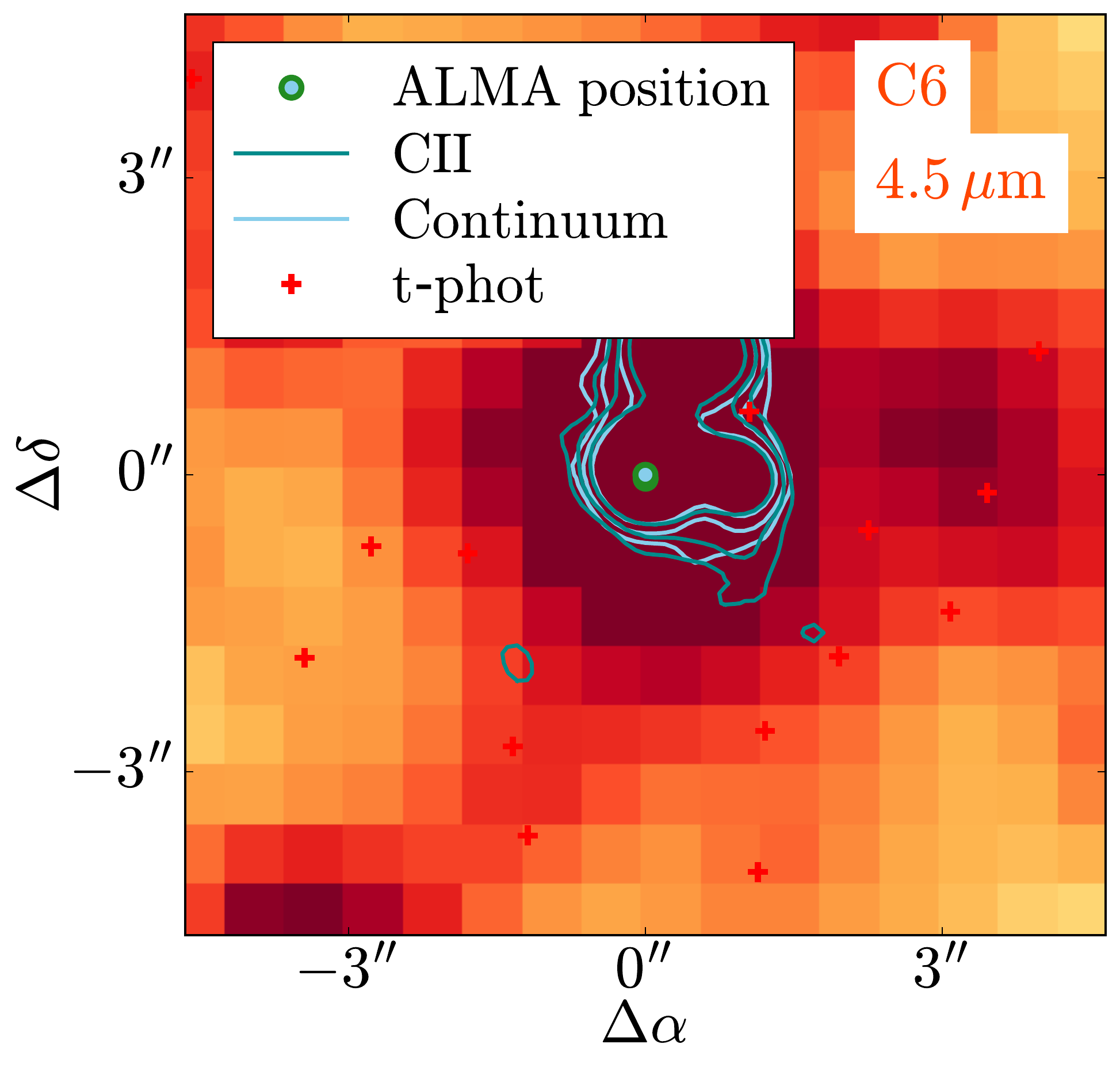}
\end{framed}
\end{subfigure}
\caption{}
\end{figure*}
\renewcommand{\thefigure}{\arabic{figure}}

\renewcommand{\thefigure}{B\arabic{figure} (Cont.)}
\addtocounter{figure}{-1}
\begin{figure*}
\begin{subfigure}{0.85\textwidth}
\begin{framed}
\includegraphics[width=0.24\textwidth]{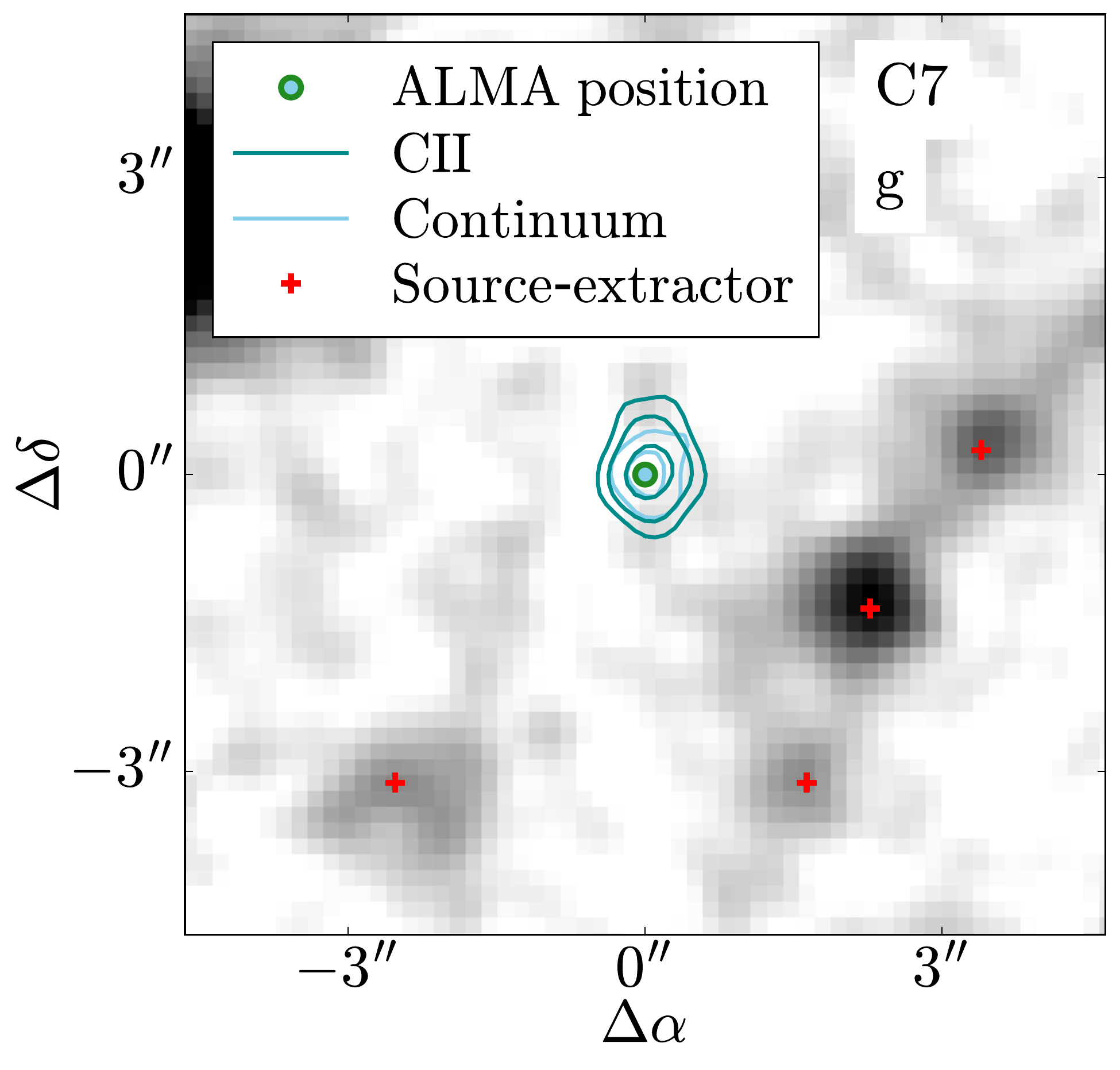}
\includegraphics[width=0.24\textwidth]{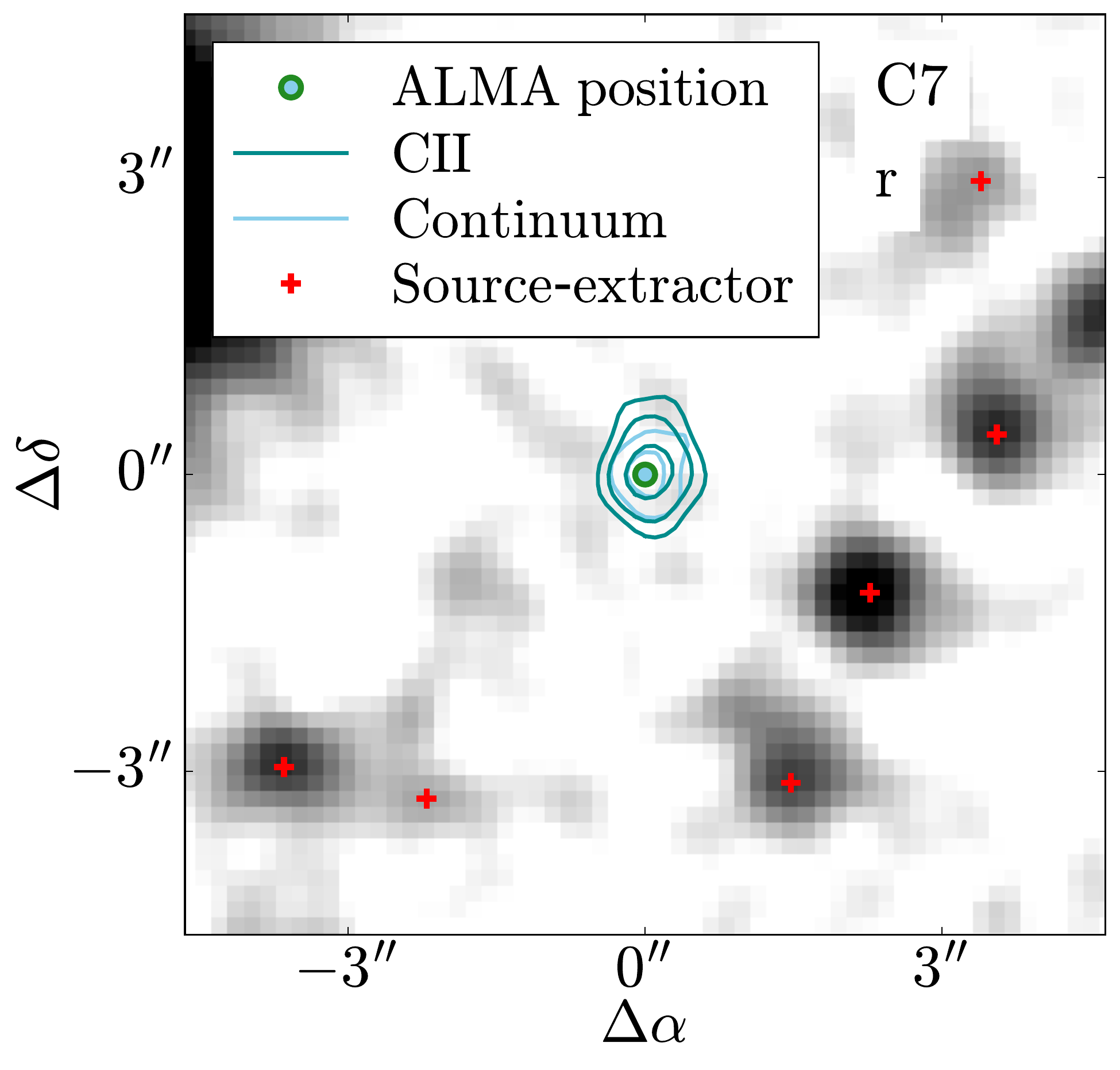}
\includegraphics[width=0.24\textwidth]{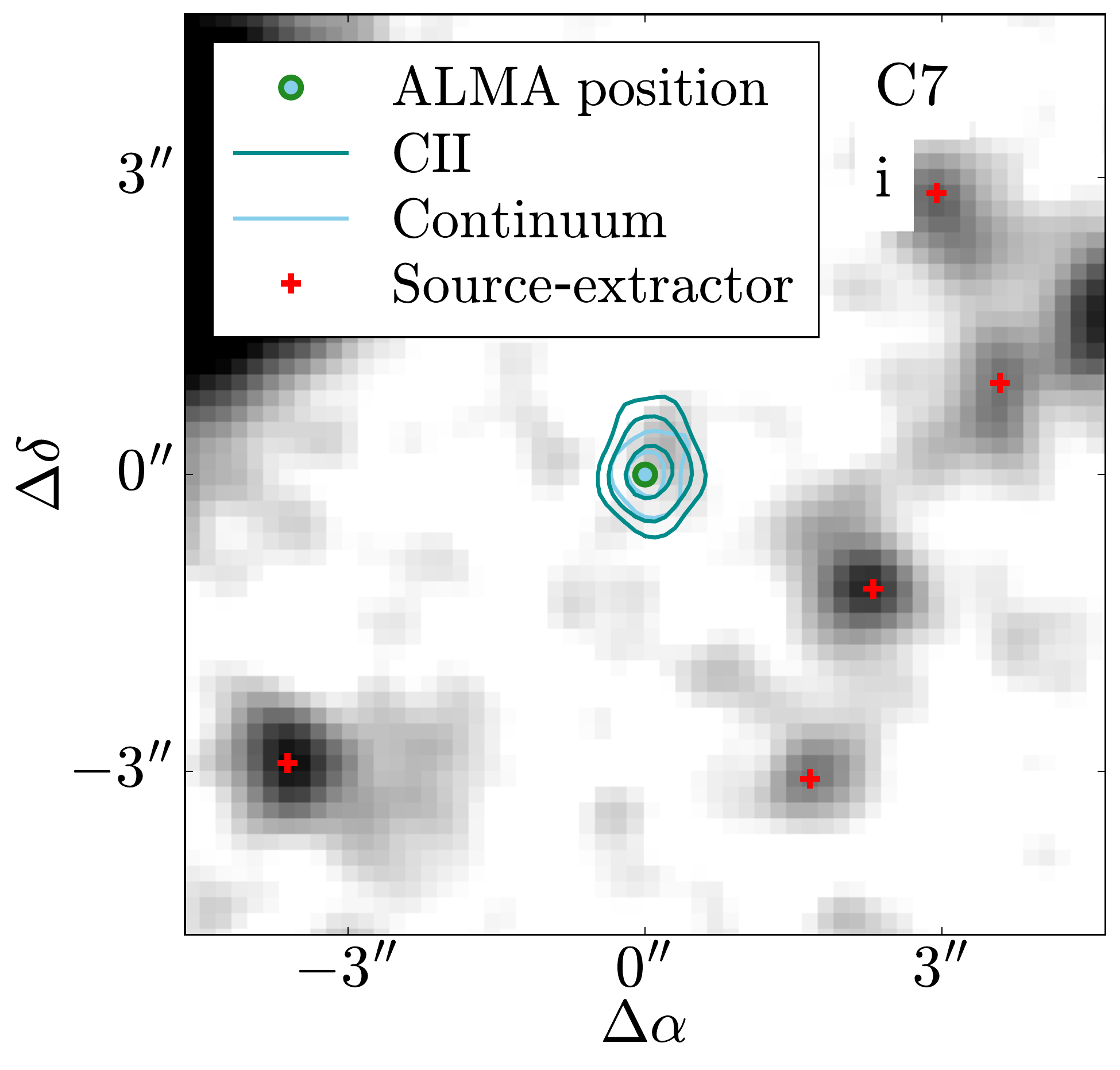}
\includegraphics[width=0.24\textwidth]{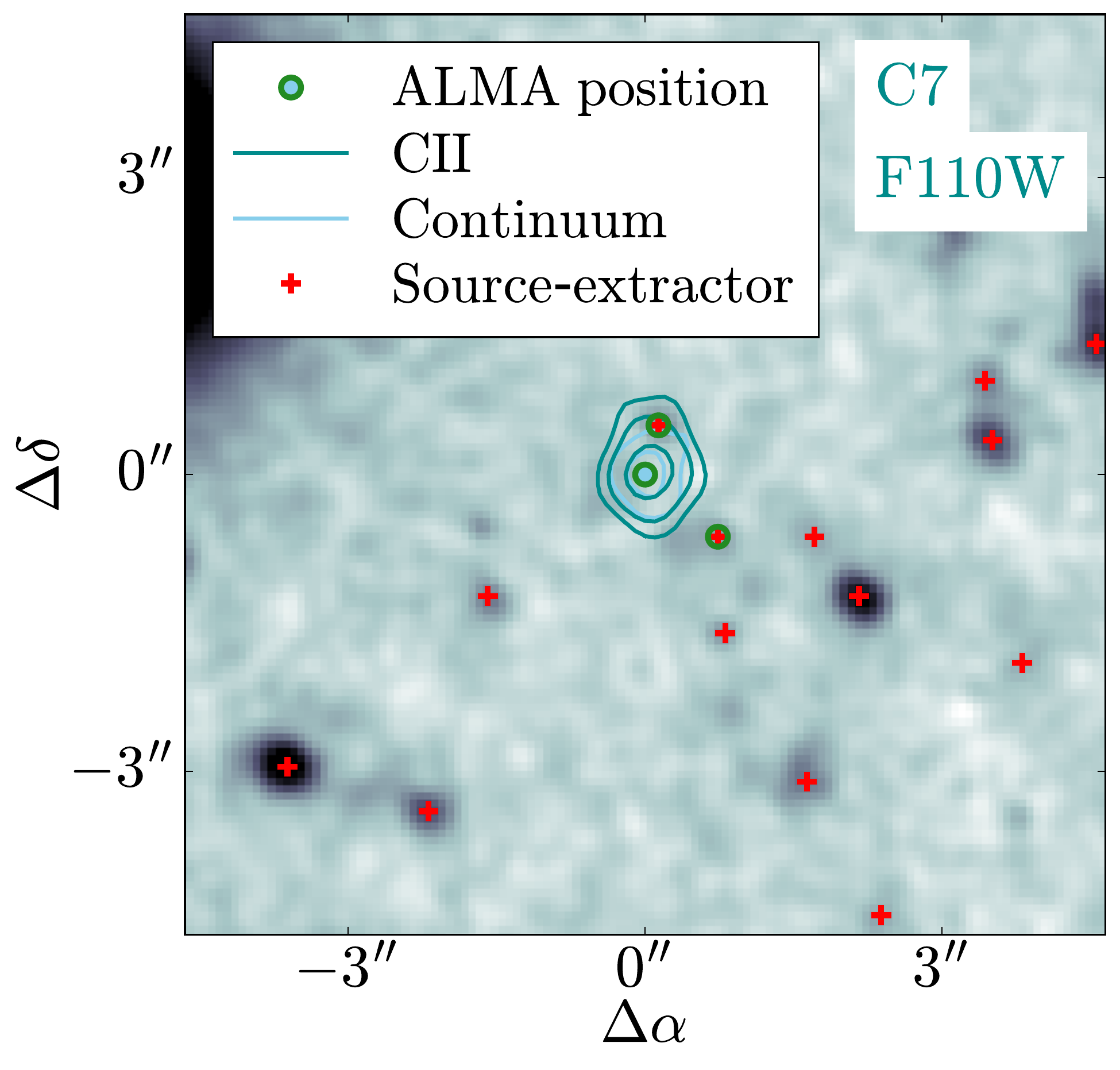}
\includegraphics[width=0.24\textwidth]{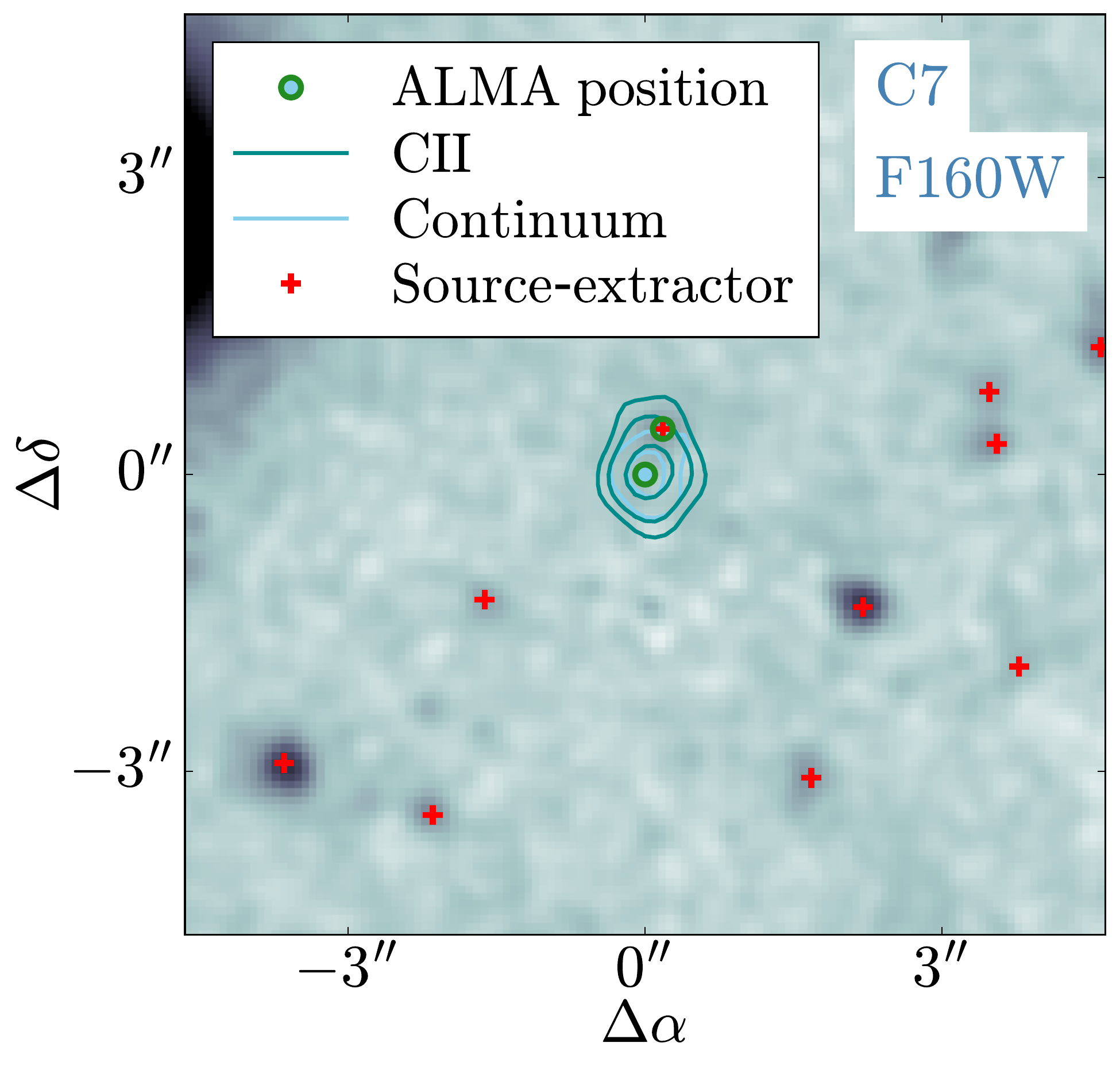}
\includegraphics[width=0.248\textwidth]{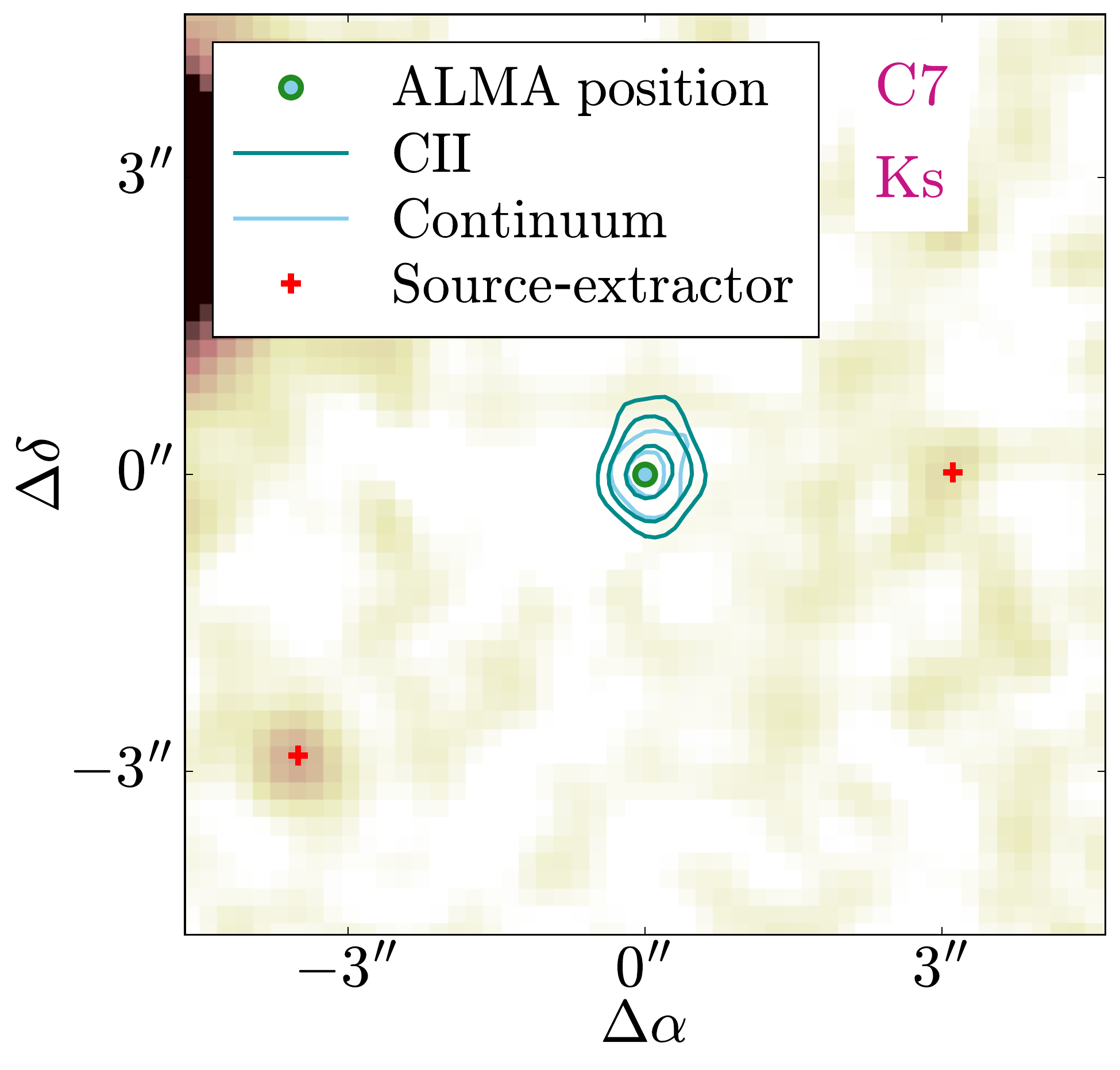}
\includegraphics[width=0.249\textwidth]{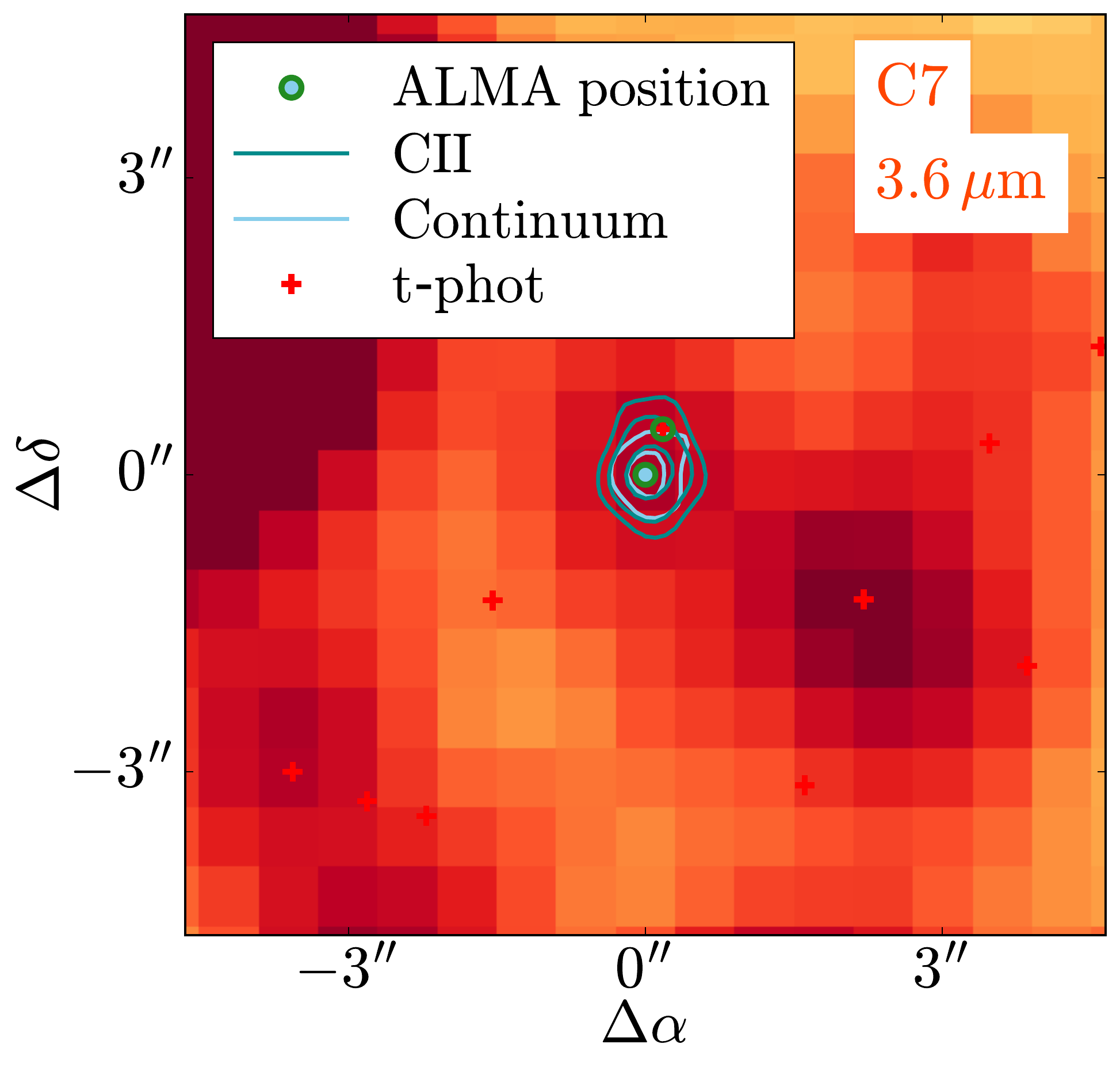}
\includegraphics[width=0.249\textwidth]{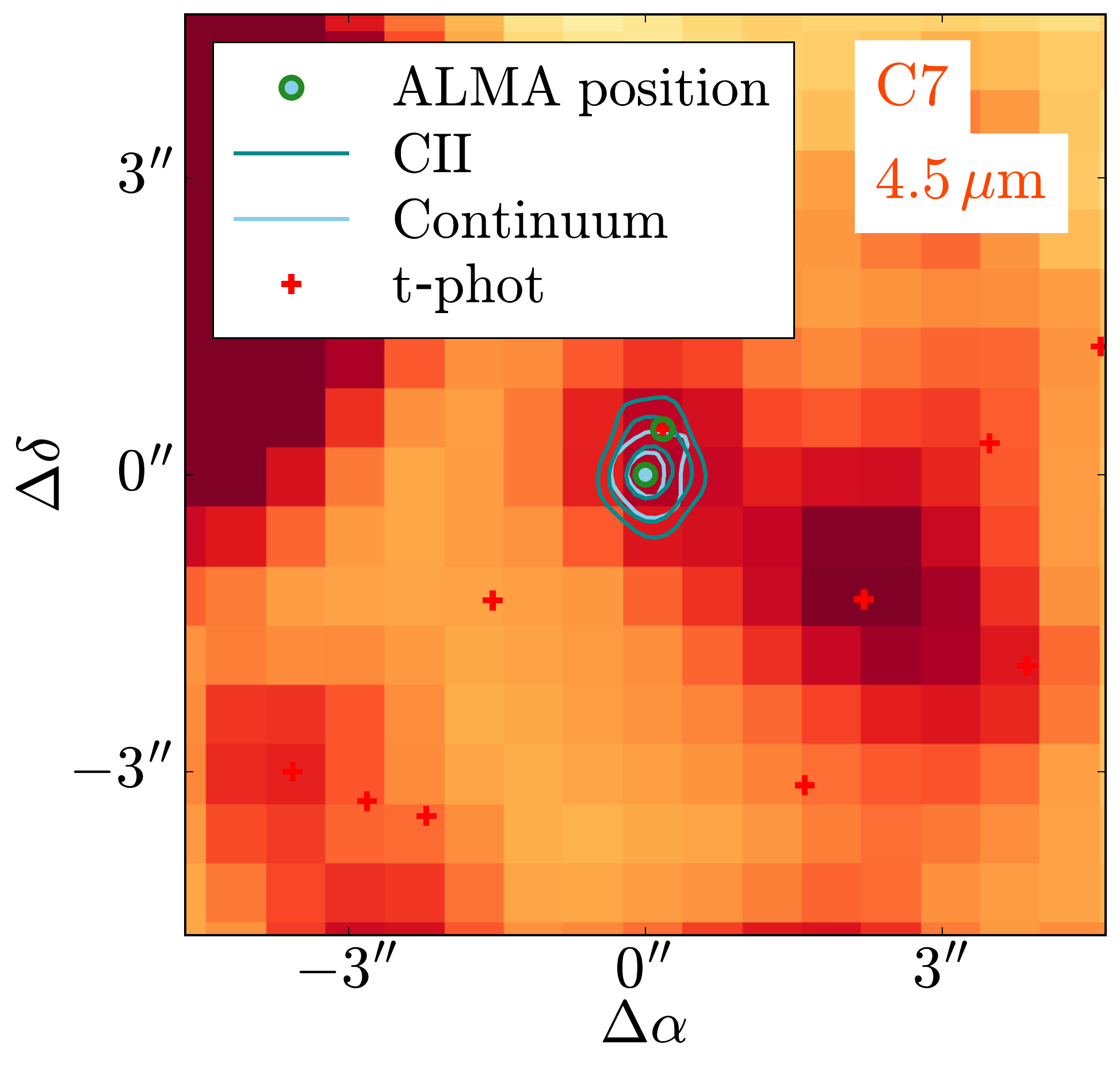}
\end{framed}
\end{subfigure}
\begin{subfigure}{0.85\textwidth}
\begin{framed}
\includegraphics[width=0.24\textwidth]{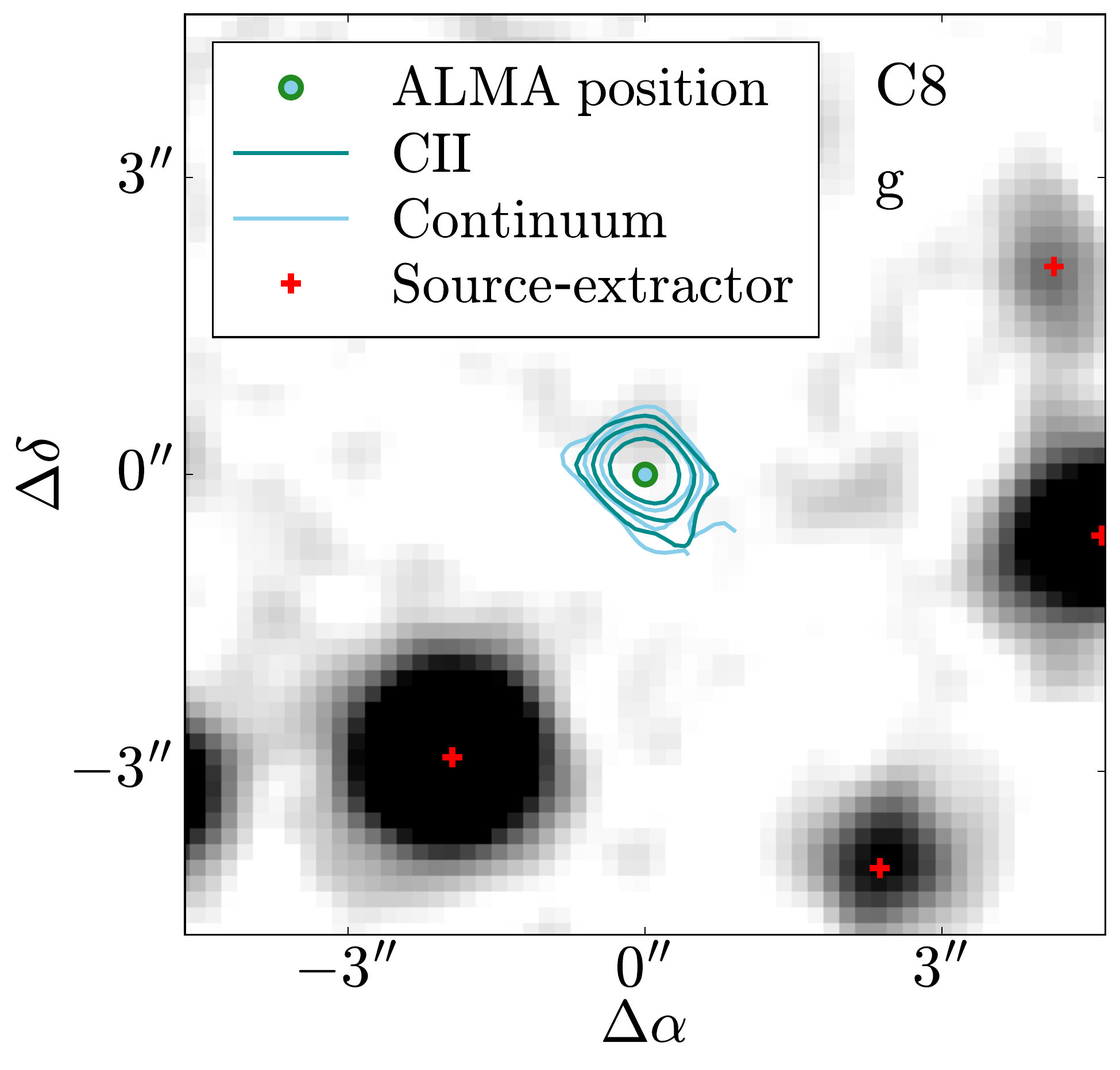}
\includegraphics[width=0.24\textwidth]{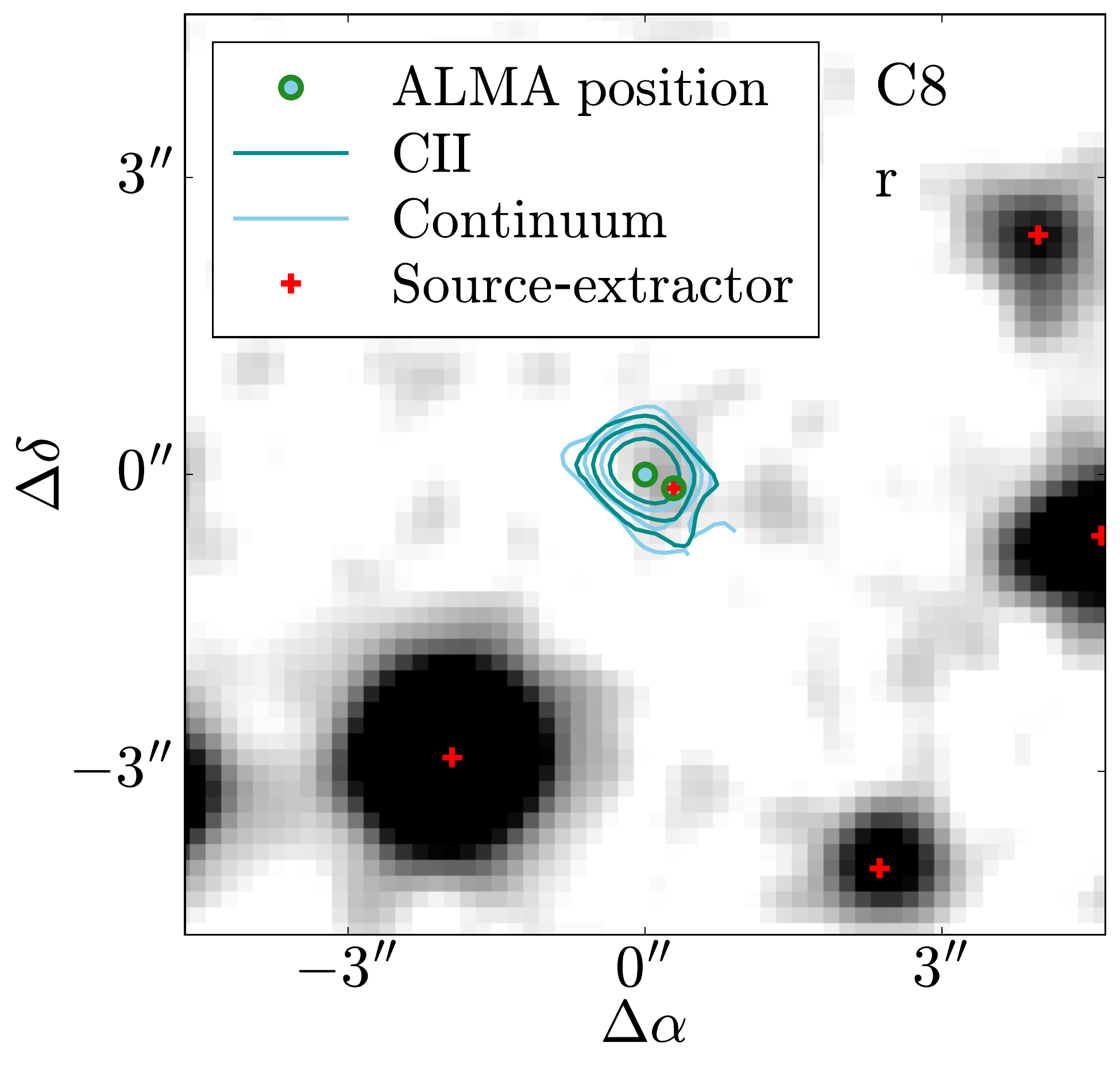}
\includegraphics[width=0.24\textwidth]{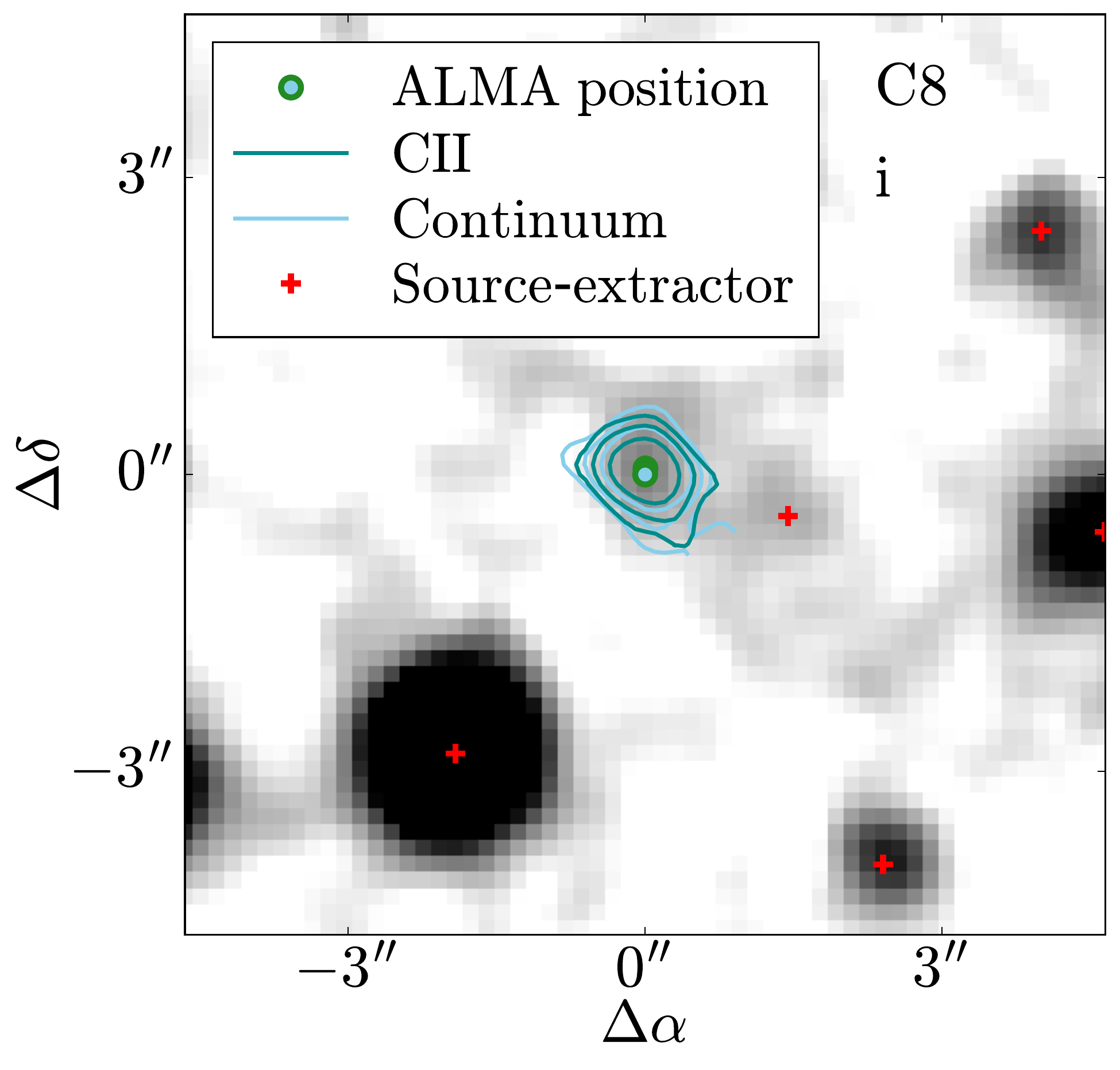}
\includegraphics[width=0.24\textwidth]{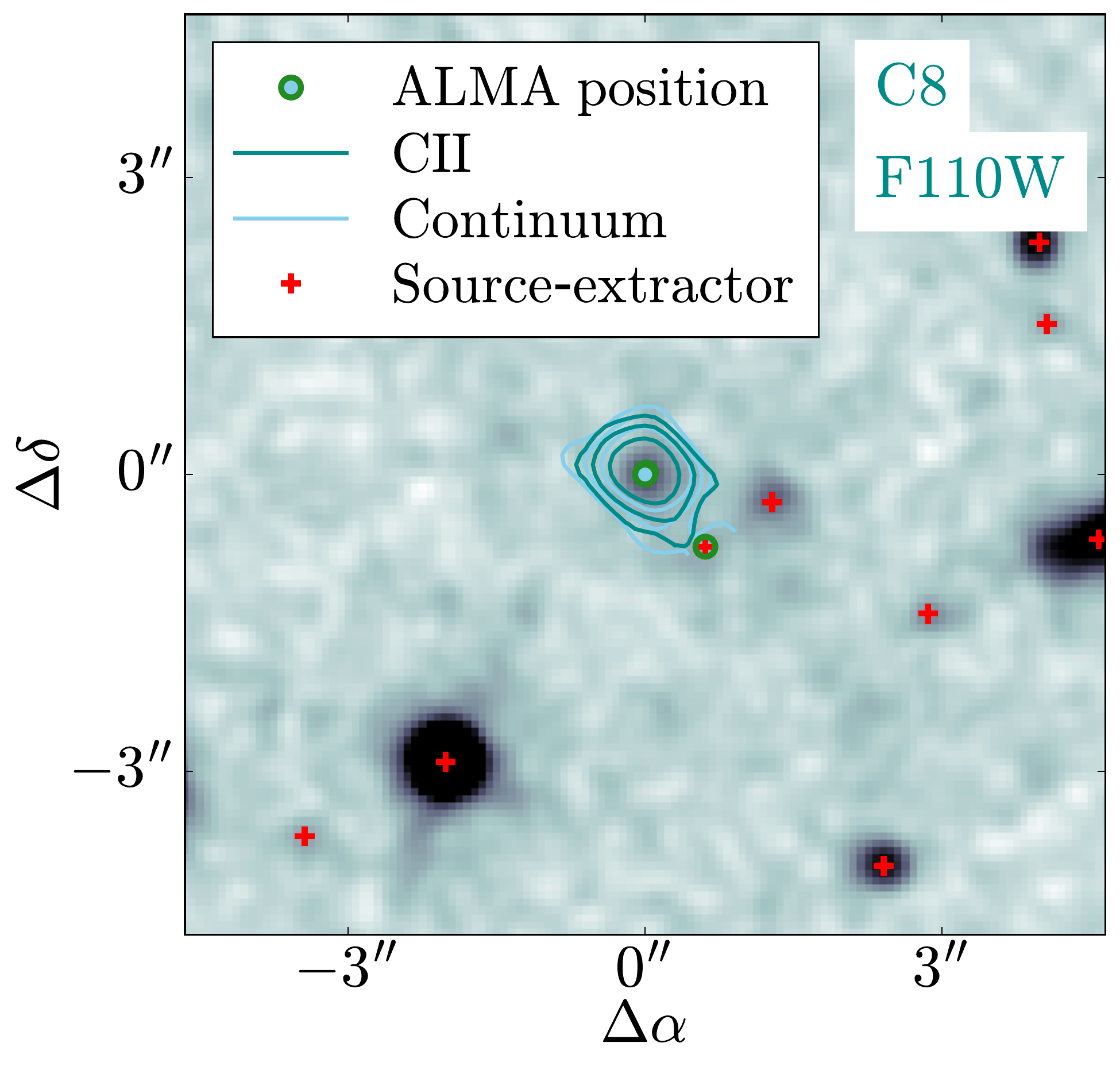}
\includegraphics[width=0.24\textwidth]{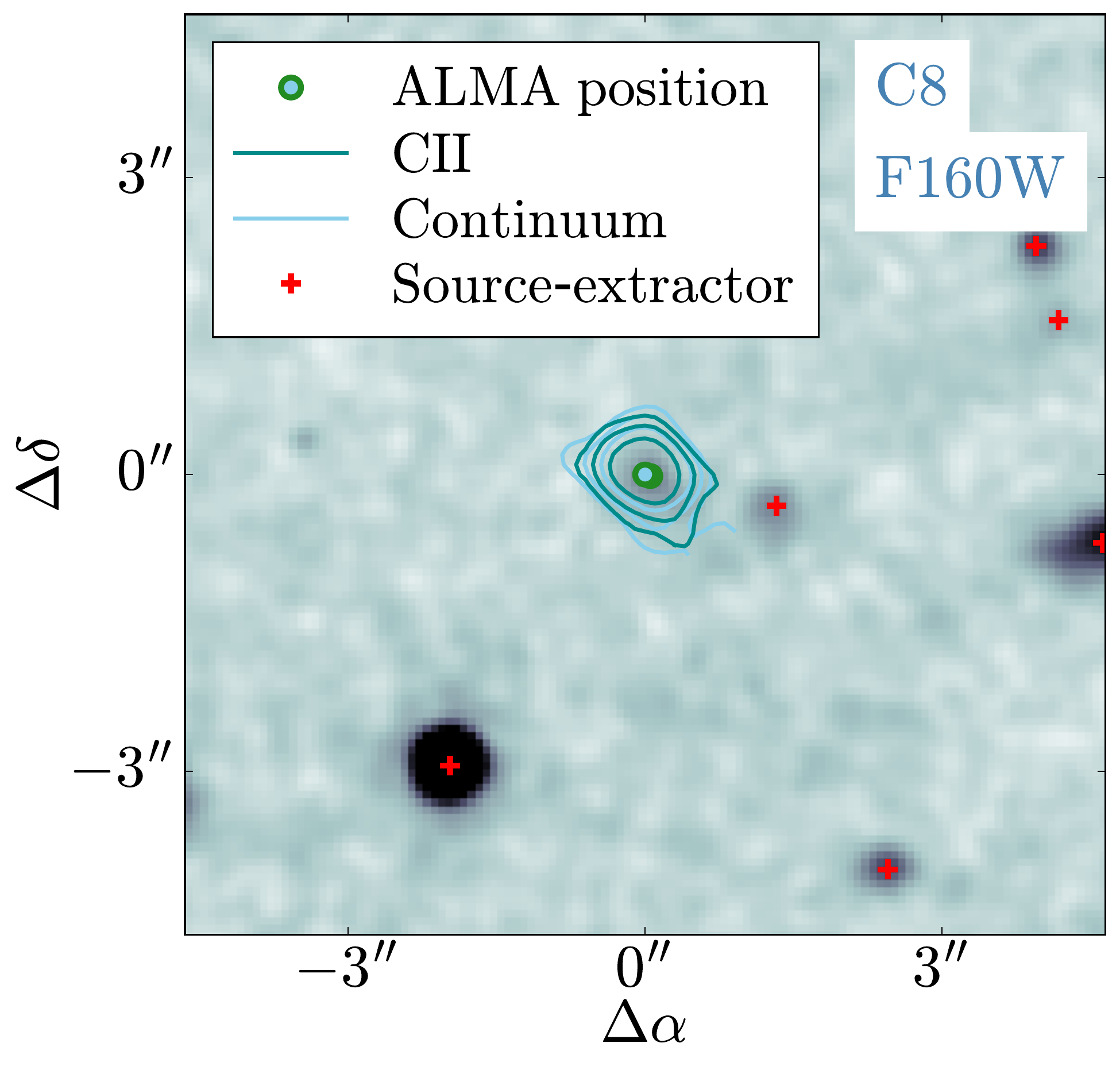}
\includegraphics[width=0.248\textwidth]{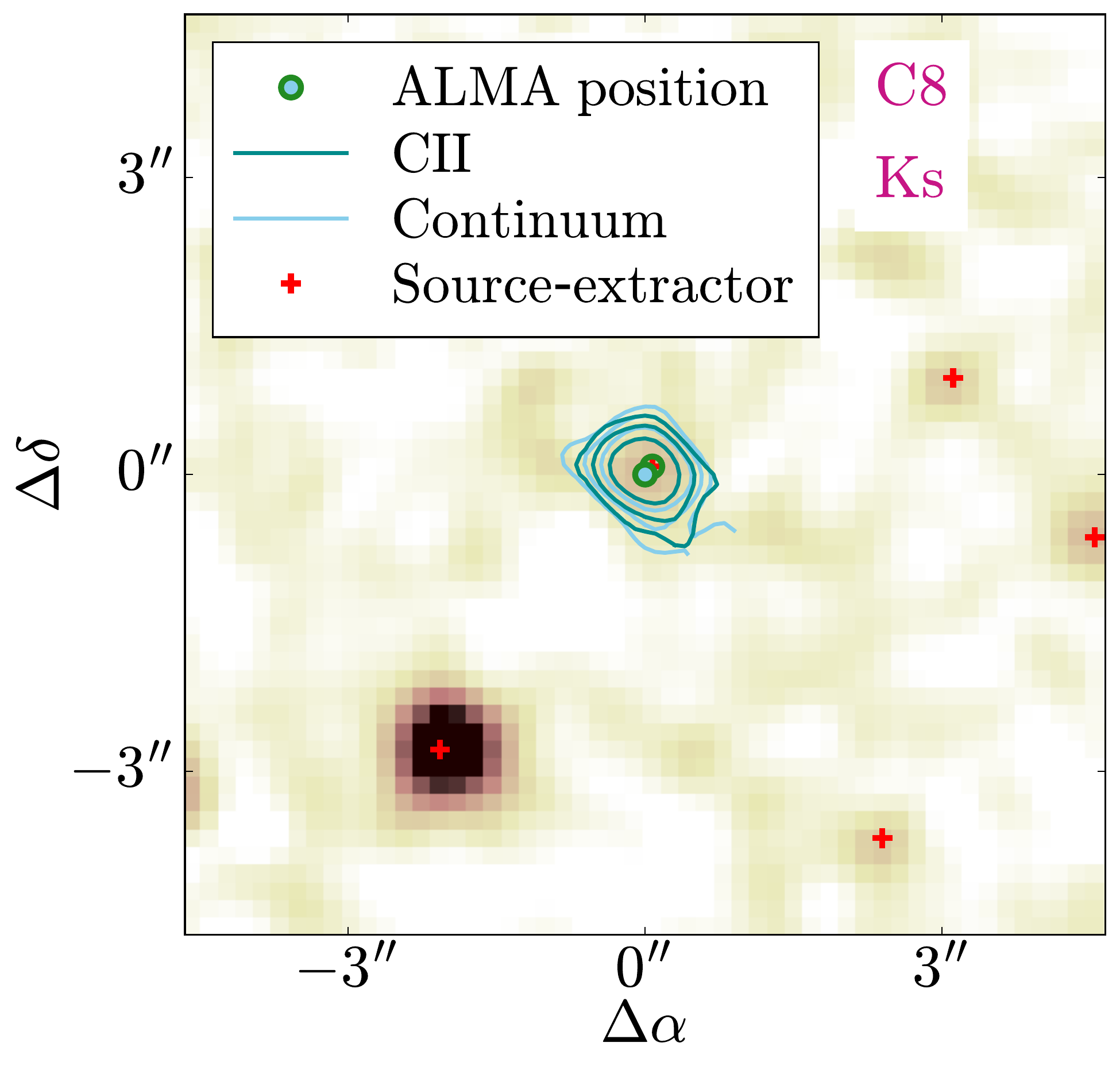}
\includegraphics[width=0.249\textwidth]{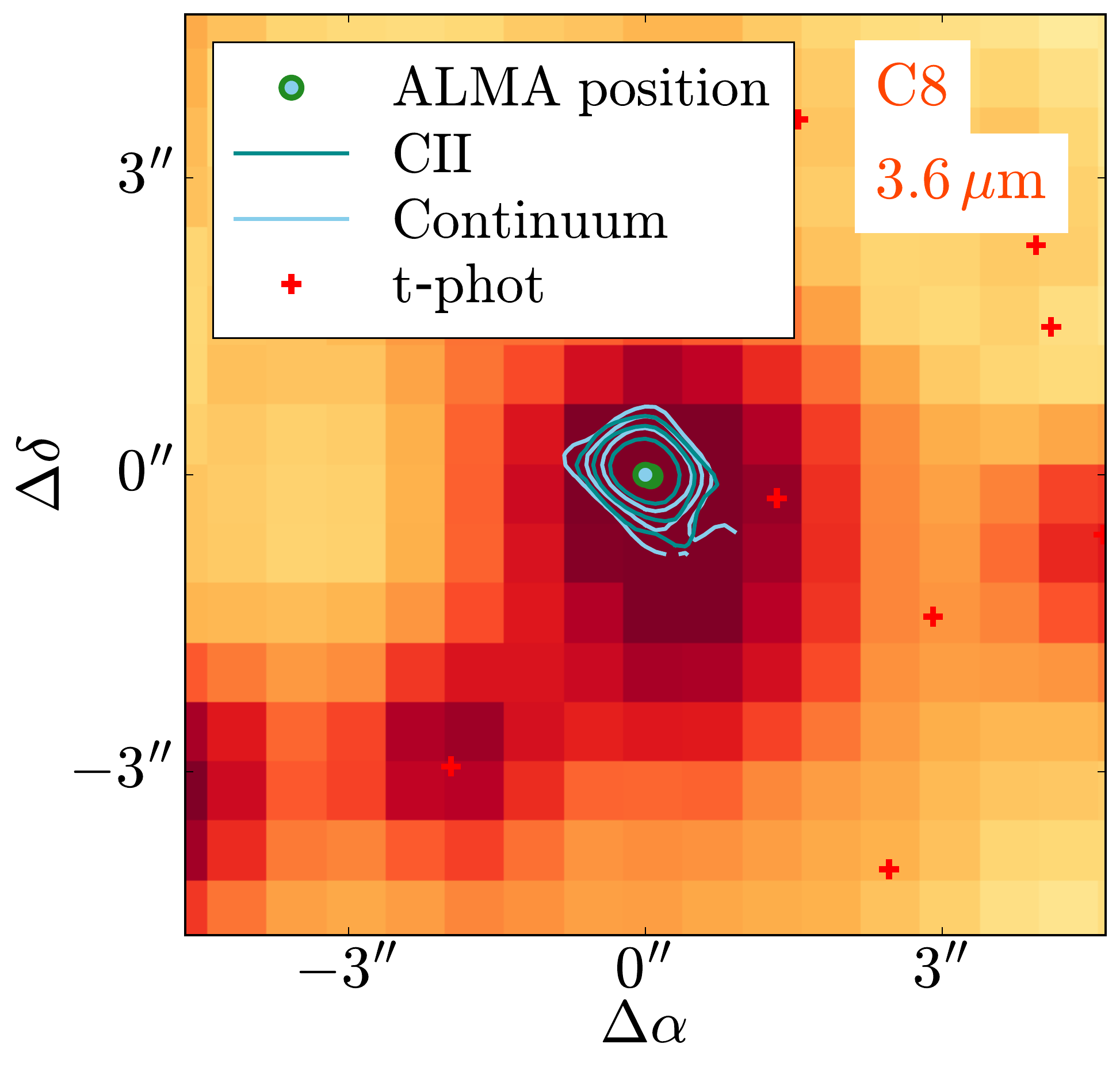}
\includegraphics[width=0.249\textwidth]{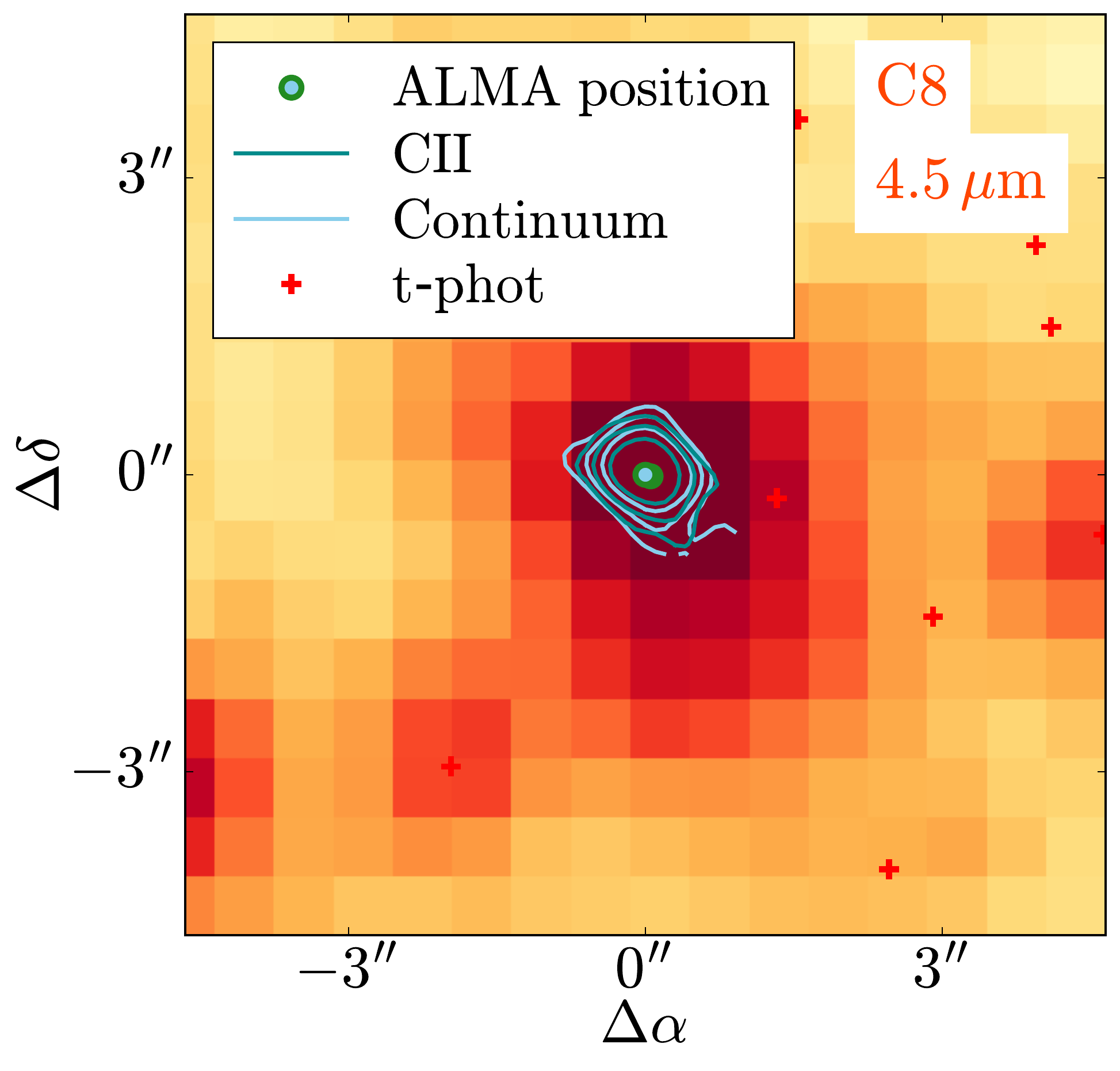}
\end{framed}
\end{subfigure}
\begin{subfigure}{0.85\textwidth}
\begin{framed}
\includegraphics[width=0.24\textwidth]{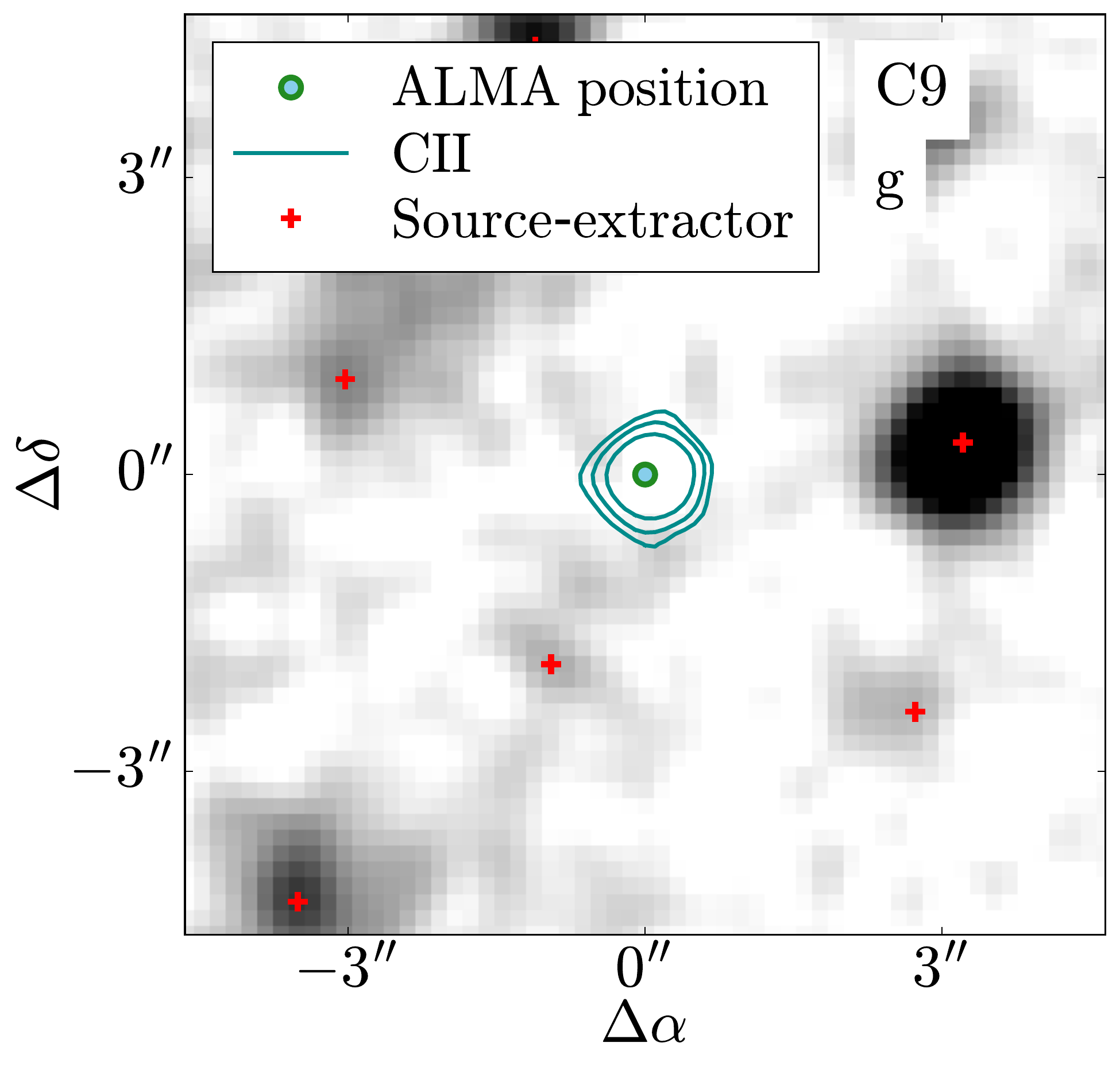}
\includegraphics[width=0.24\textwidth]{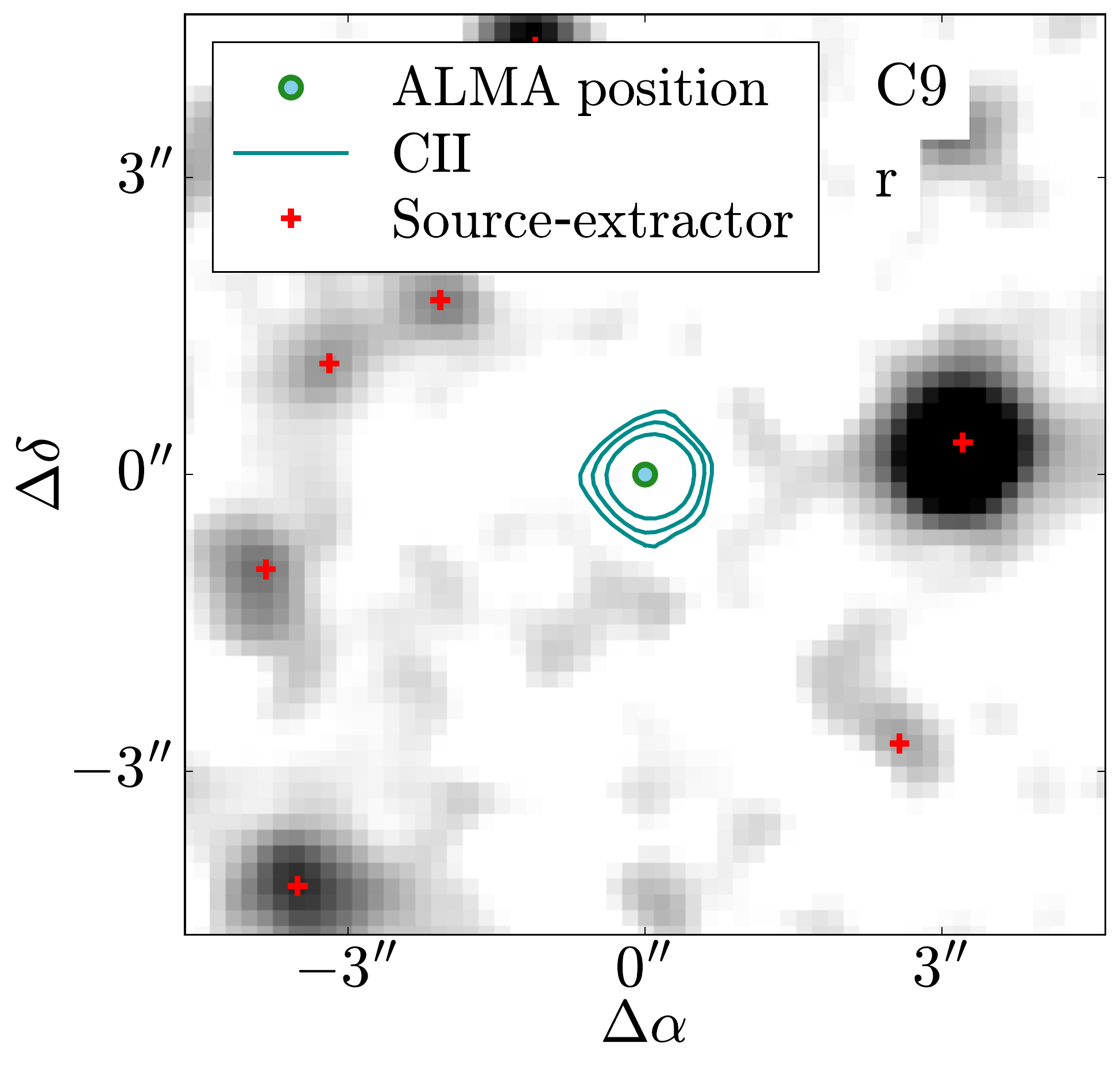}
\includegraphics[width=0.24\textwidth]{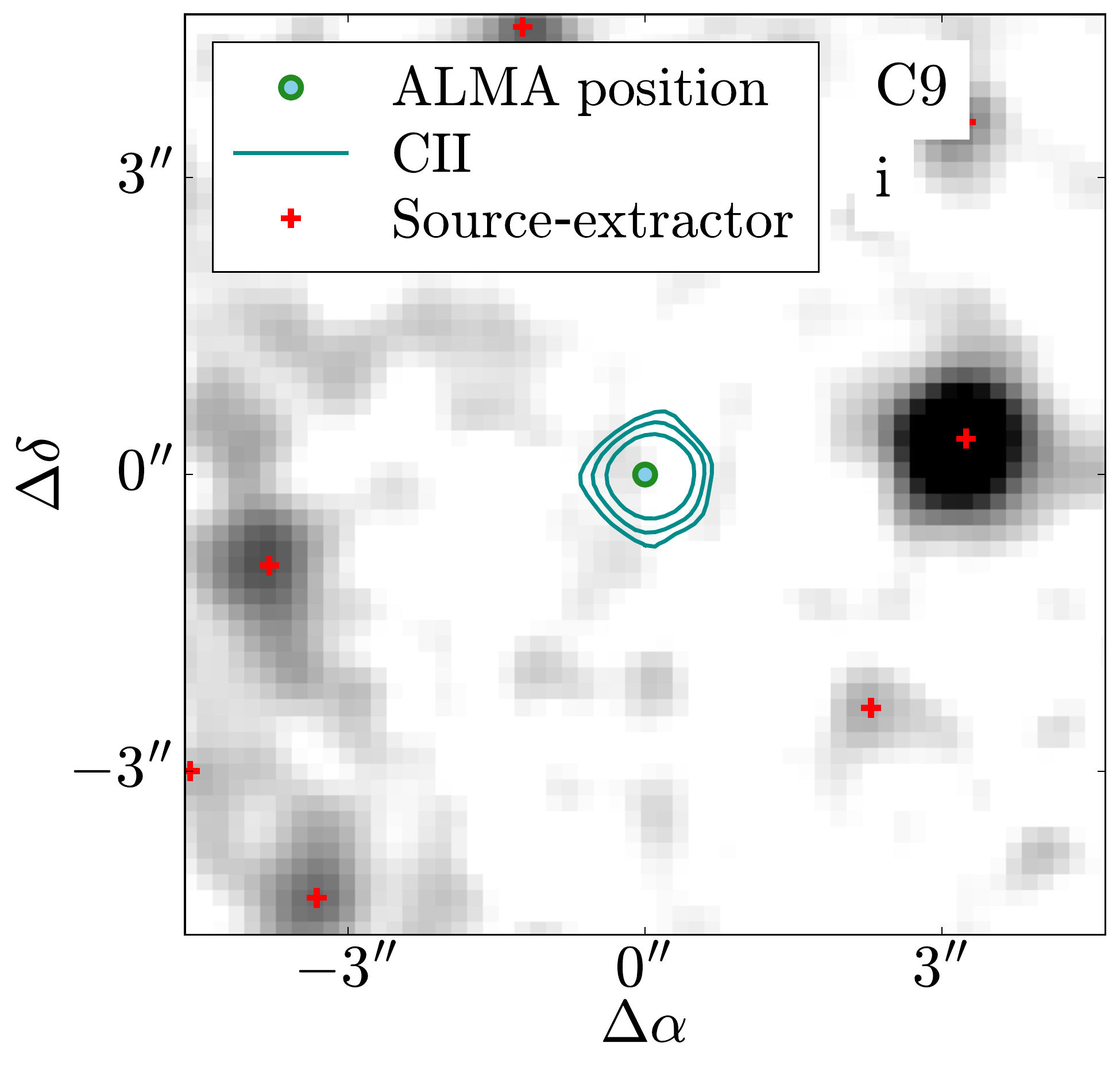}
\includegraphics[width=0.24\textwidth]{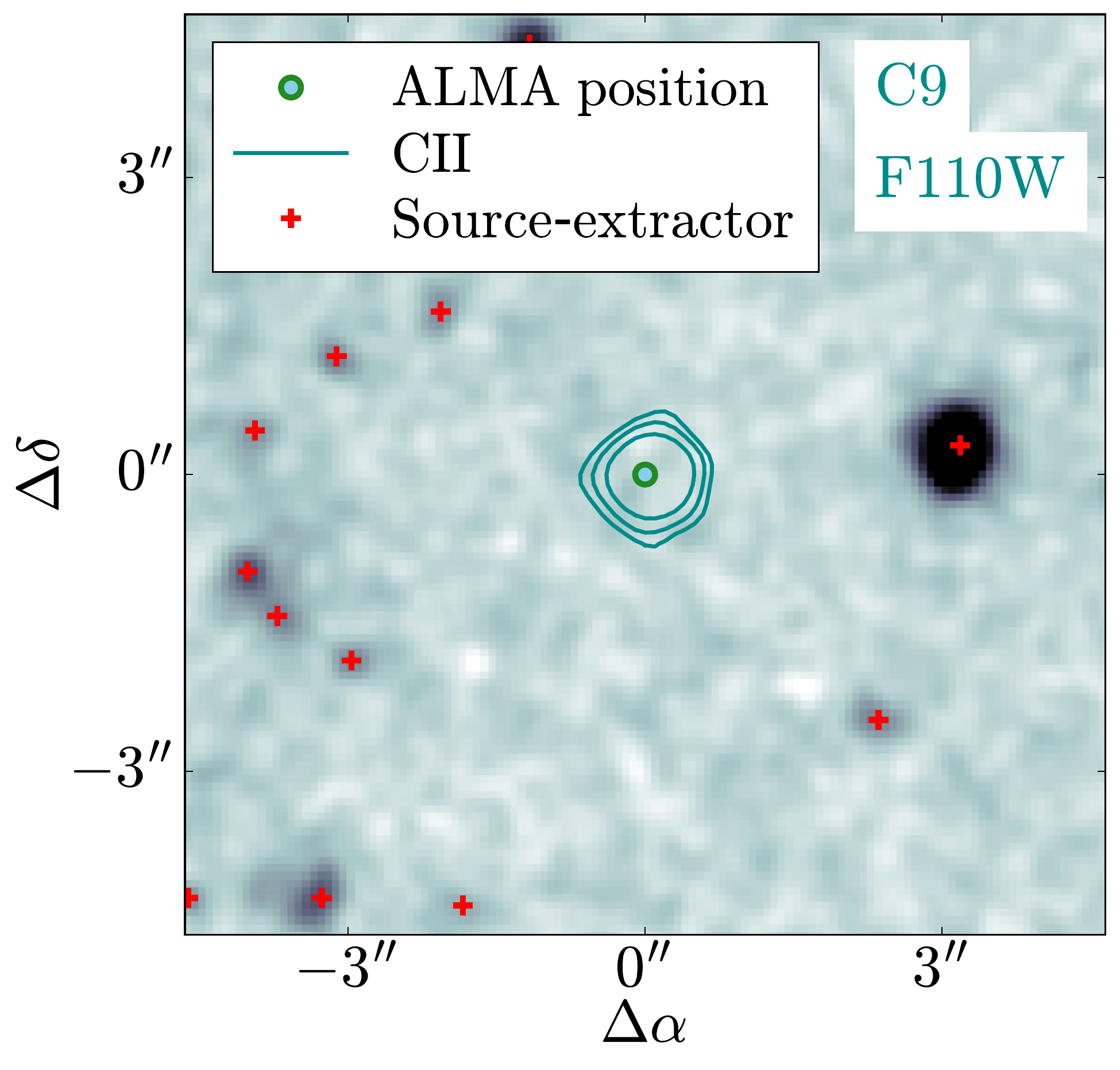}
\includegraphics[width=0.24\textwidth]{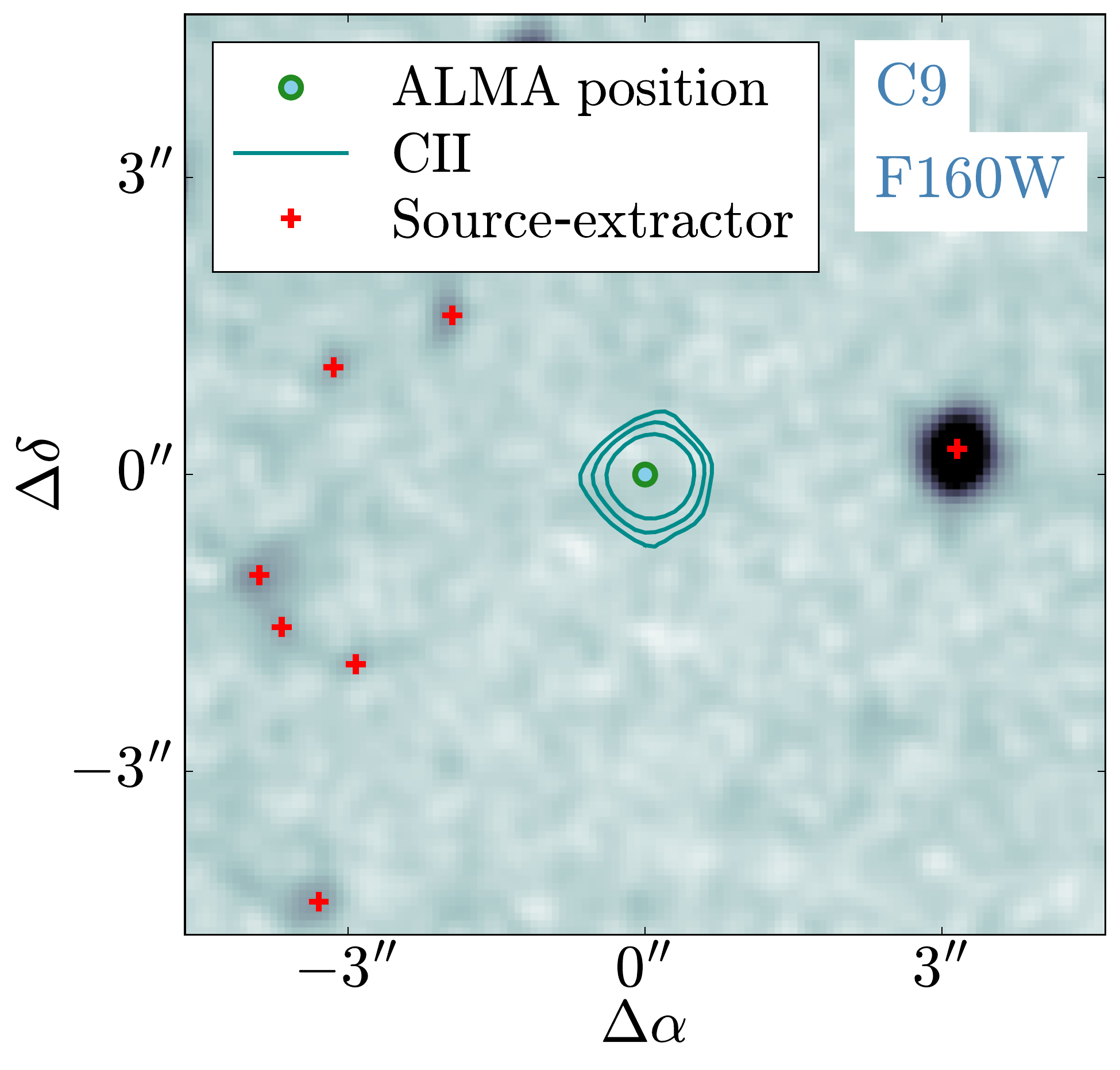}
\includegraphics[width=0.248\textwidth]{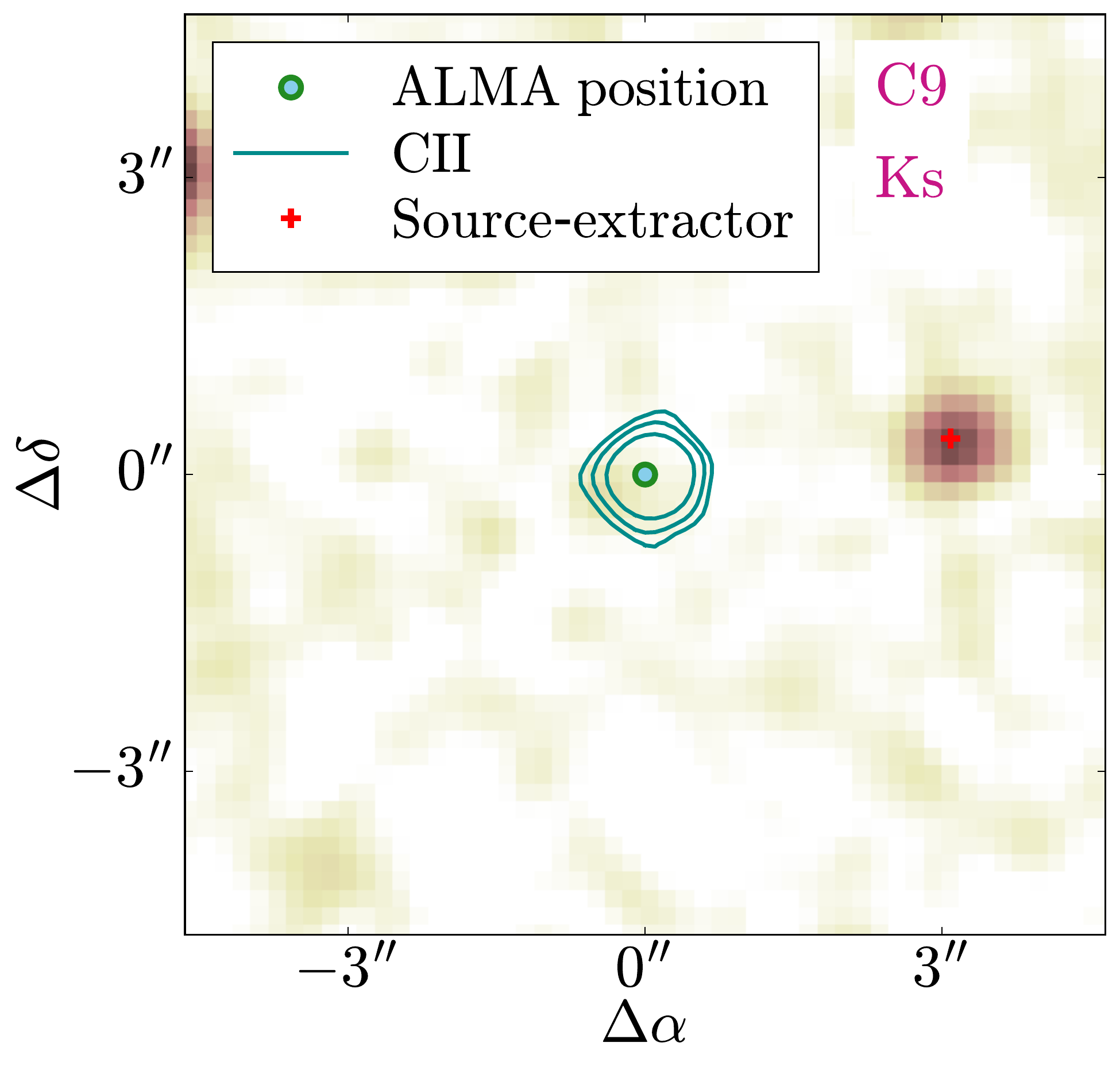}
\includegraphics[width=0.249\textwidth]{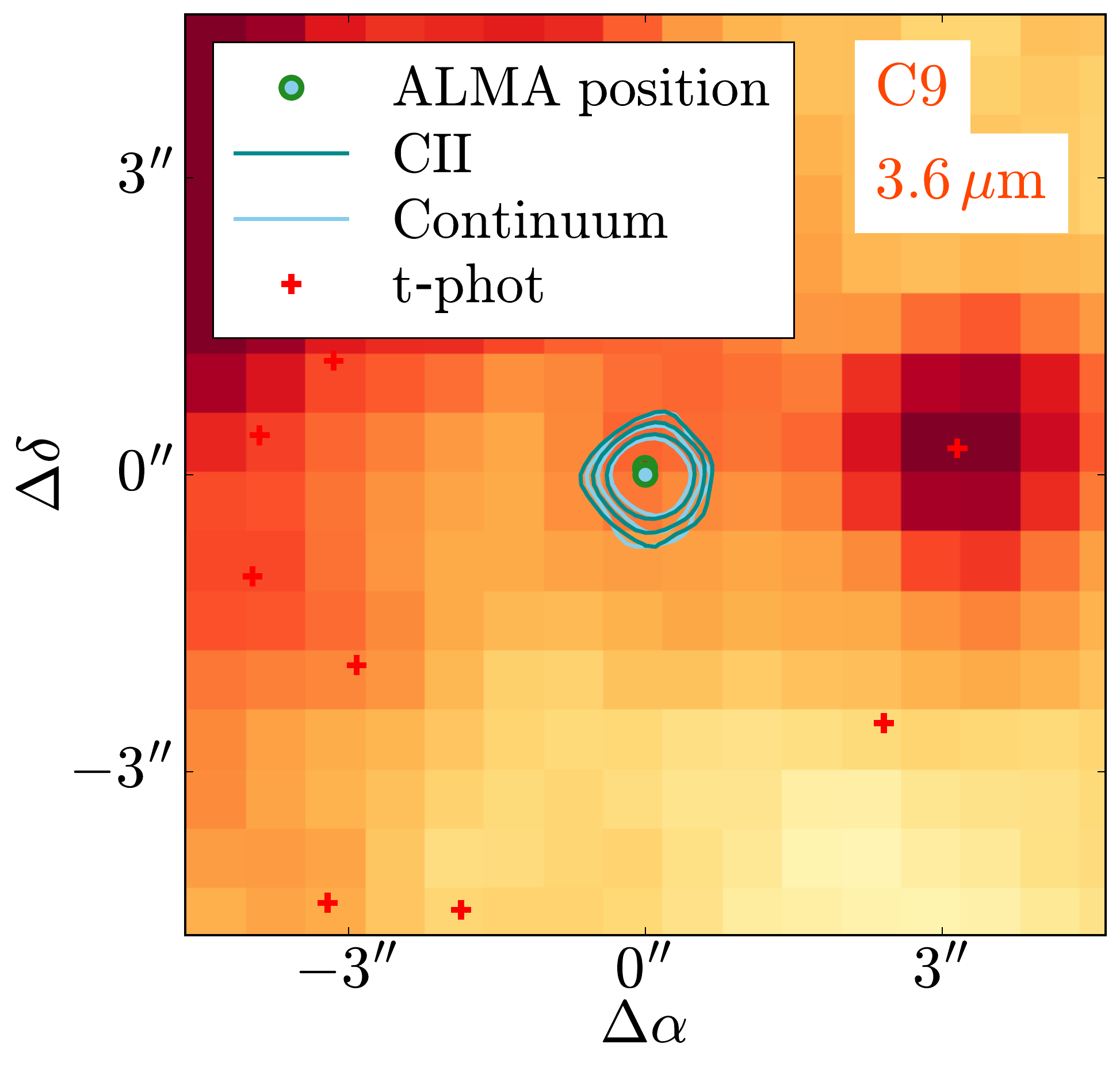}
\includegraphics[width=0.249\textwidth]{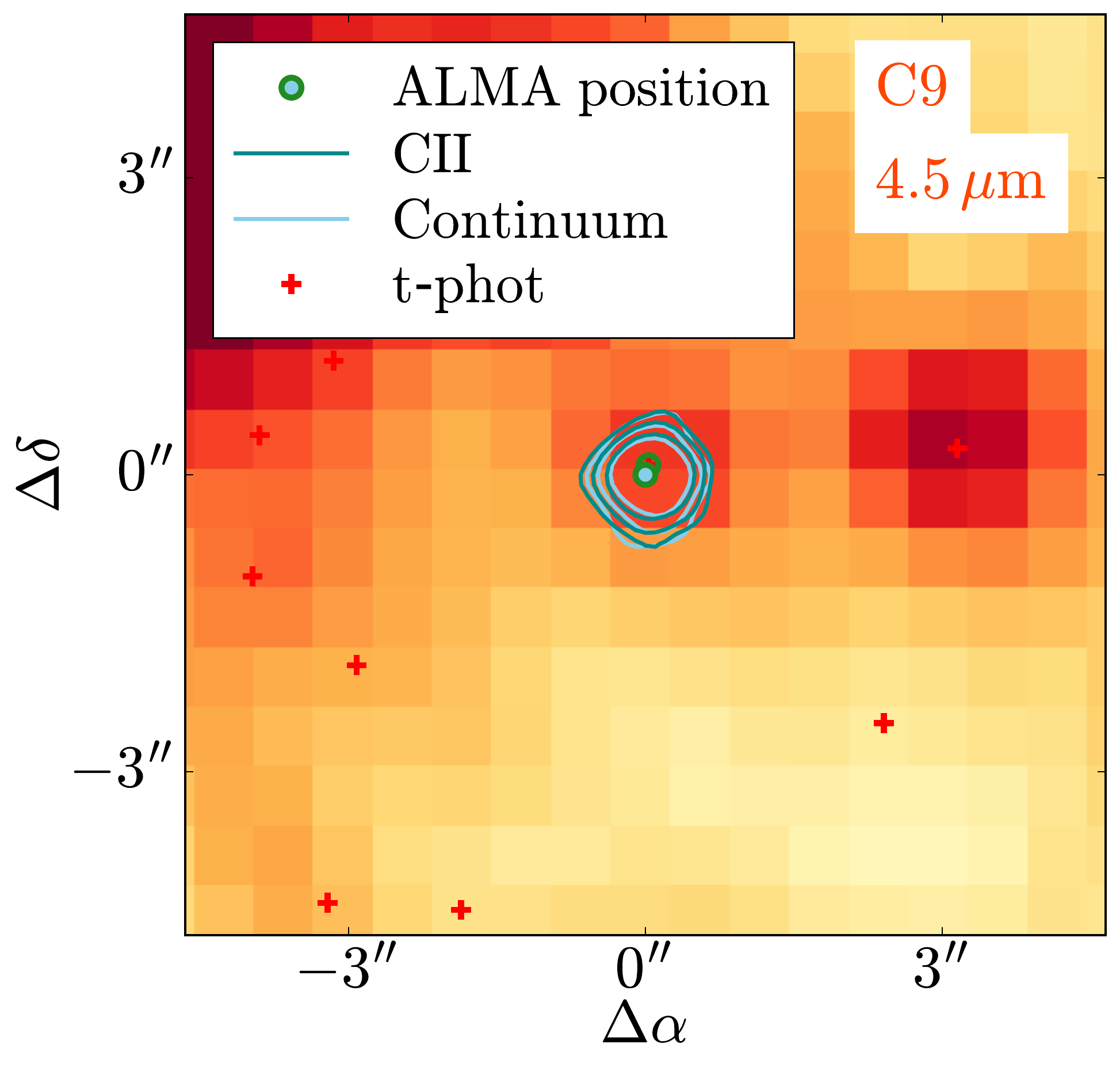}
\end{framed}
\end{subfigure}
\caption{}
\end{figure*}
\renewcommand{\thefigure}{\arabic{figure}}

\renewcommand{\thefigure}{B\arabic{figure} (Cont.)}
\addtocounter{figure}{-1}
\begin{figure*}
\begin{subfigure}{0.85\textwidth}
\begin{framed}
\includegraphics[width=0.24\textwidth]{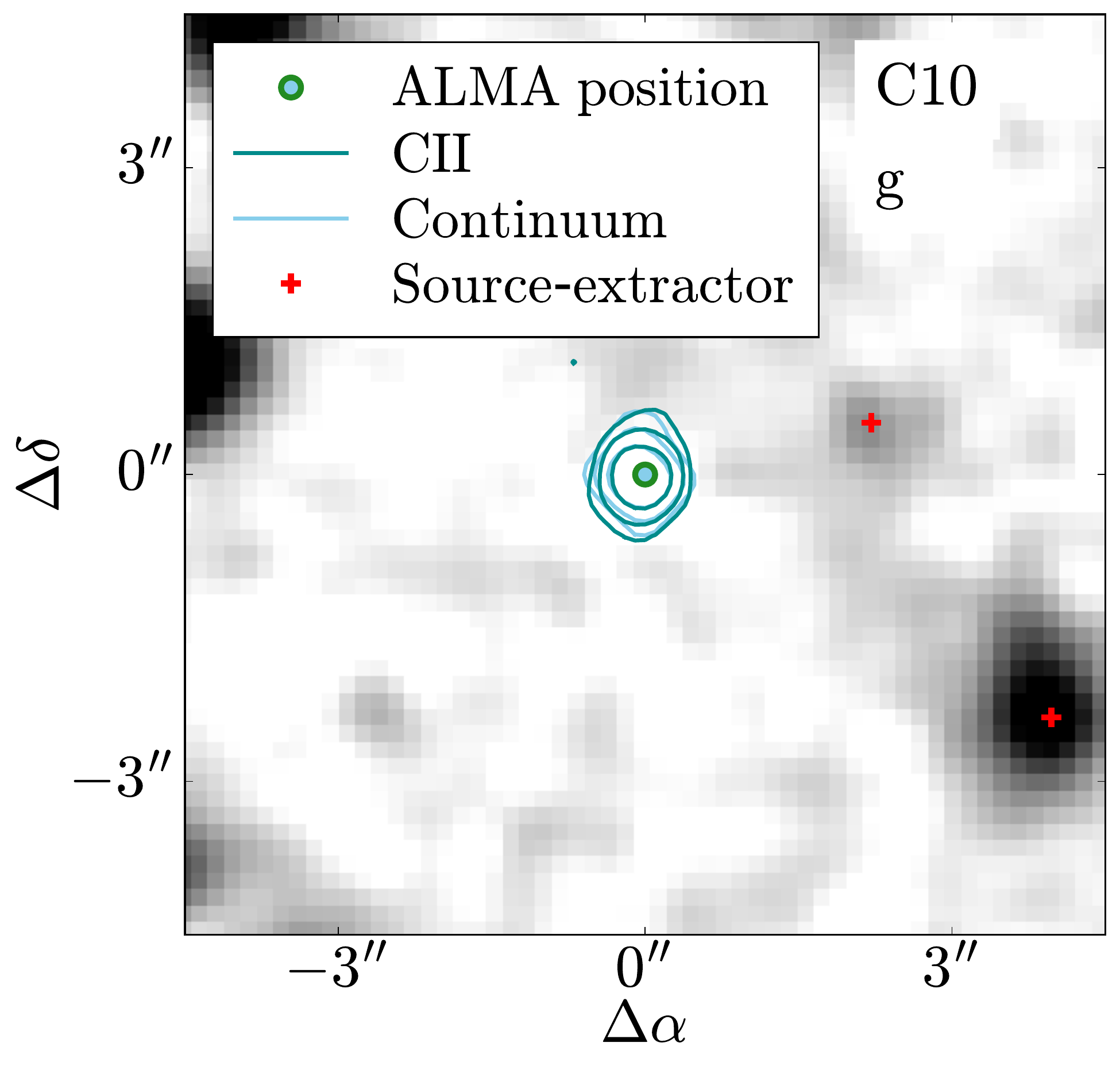}
\includegraphics[width=0.24\textwidth]{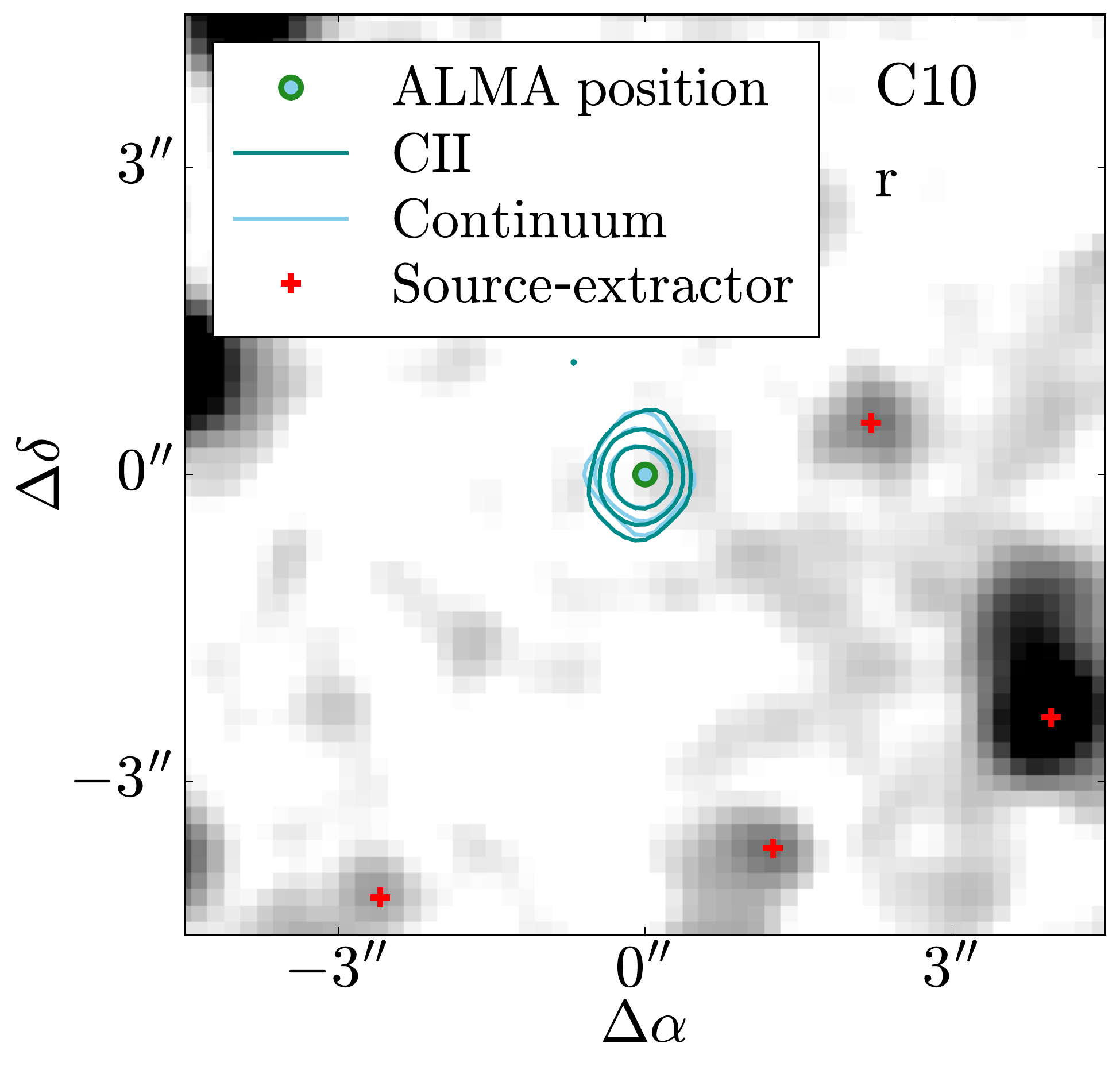}
\includegraphics[width=0.24\textwidth]{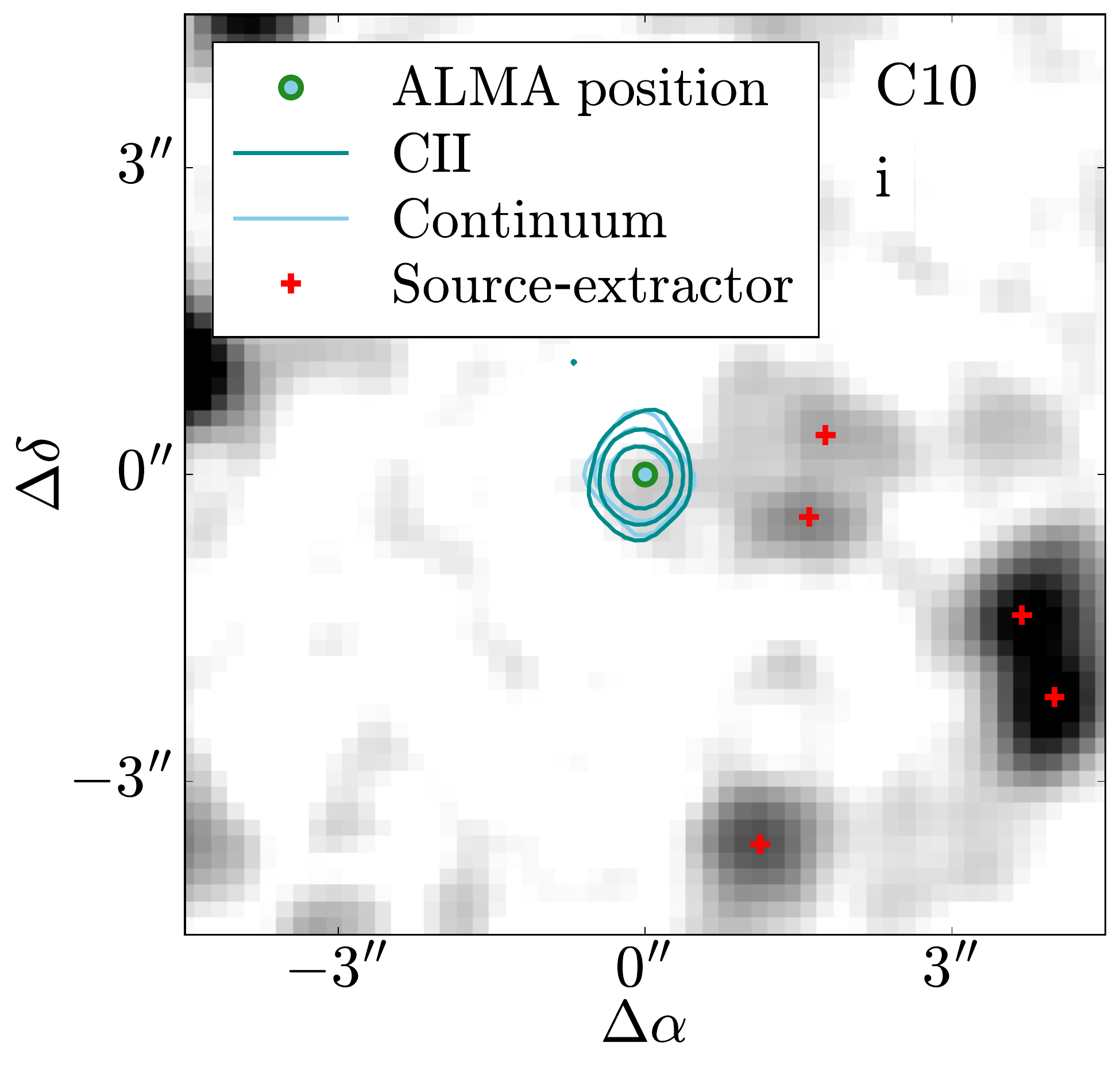}
\includegraphics[width=0.24\textwidth]{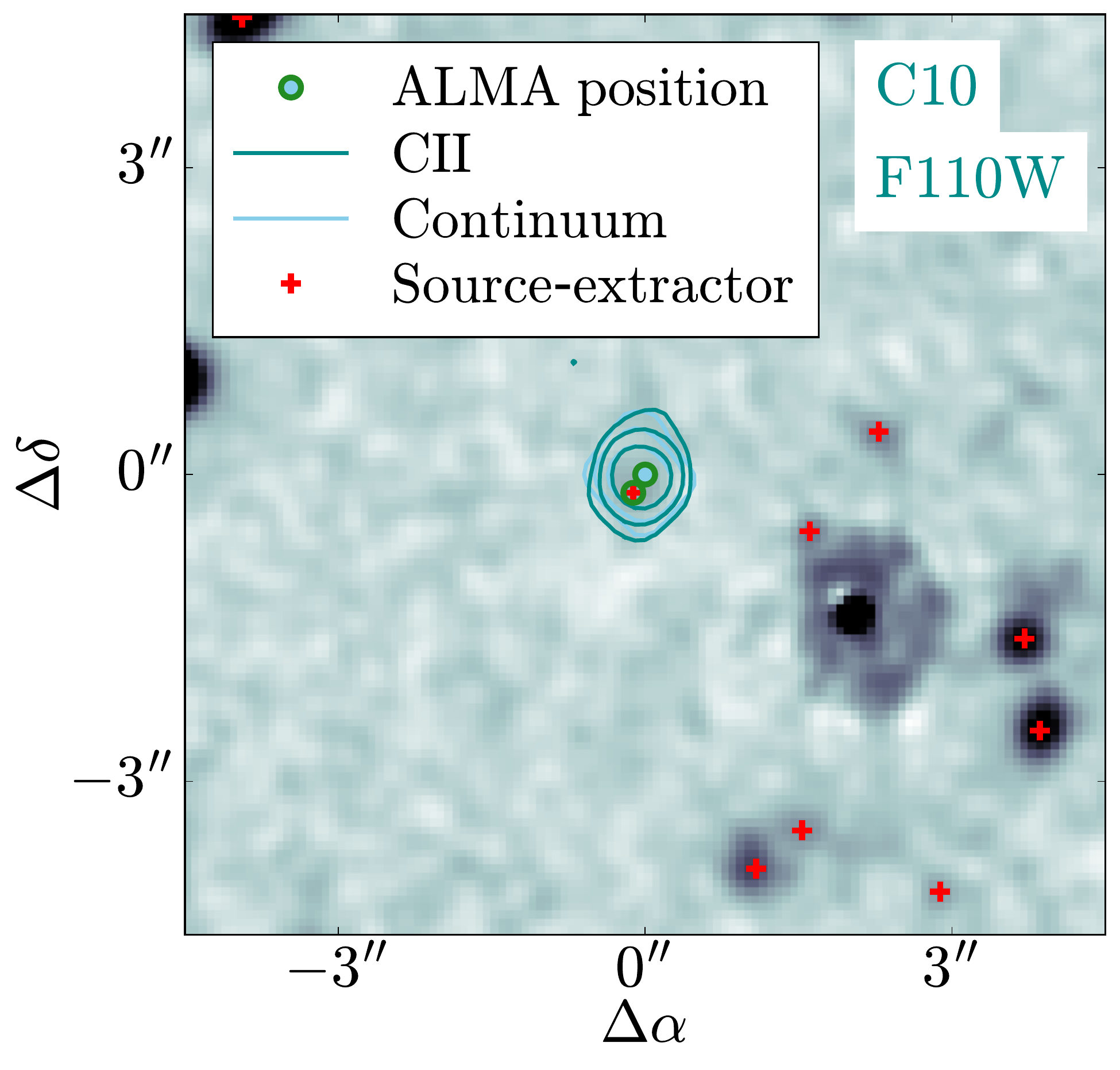}
\includegraphics[width=0.24\textwidth]{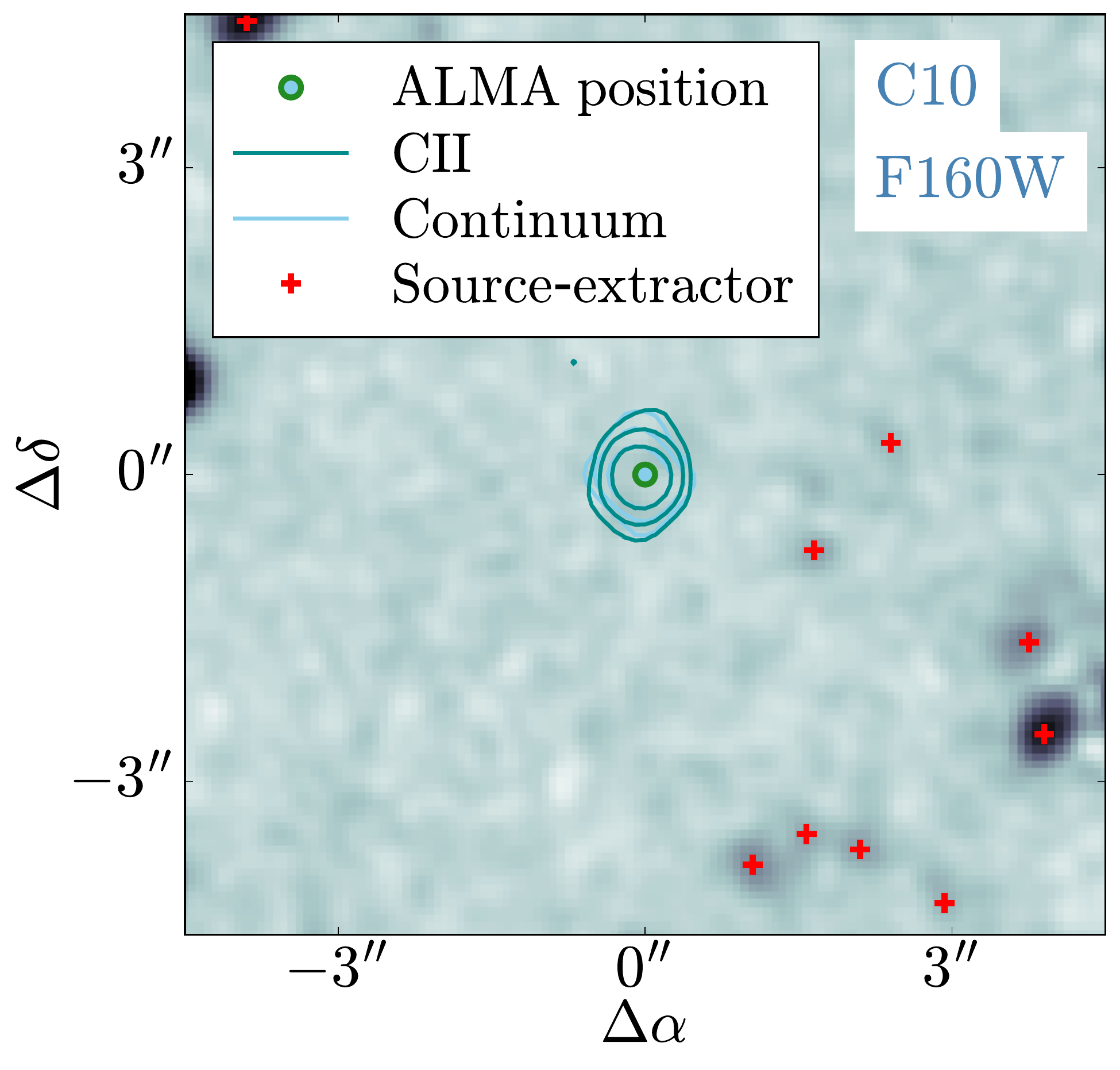}
\includegraphics[width=0.248\textwidth]{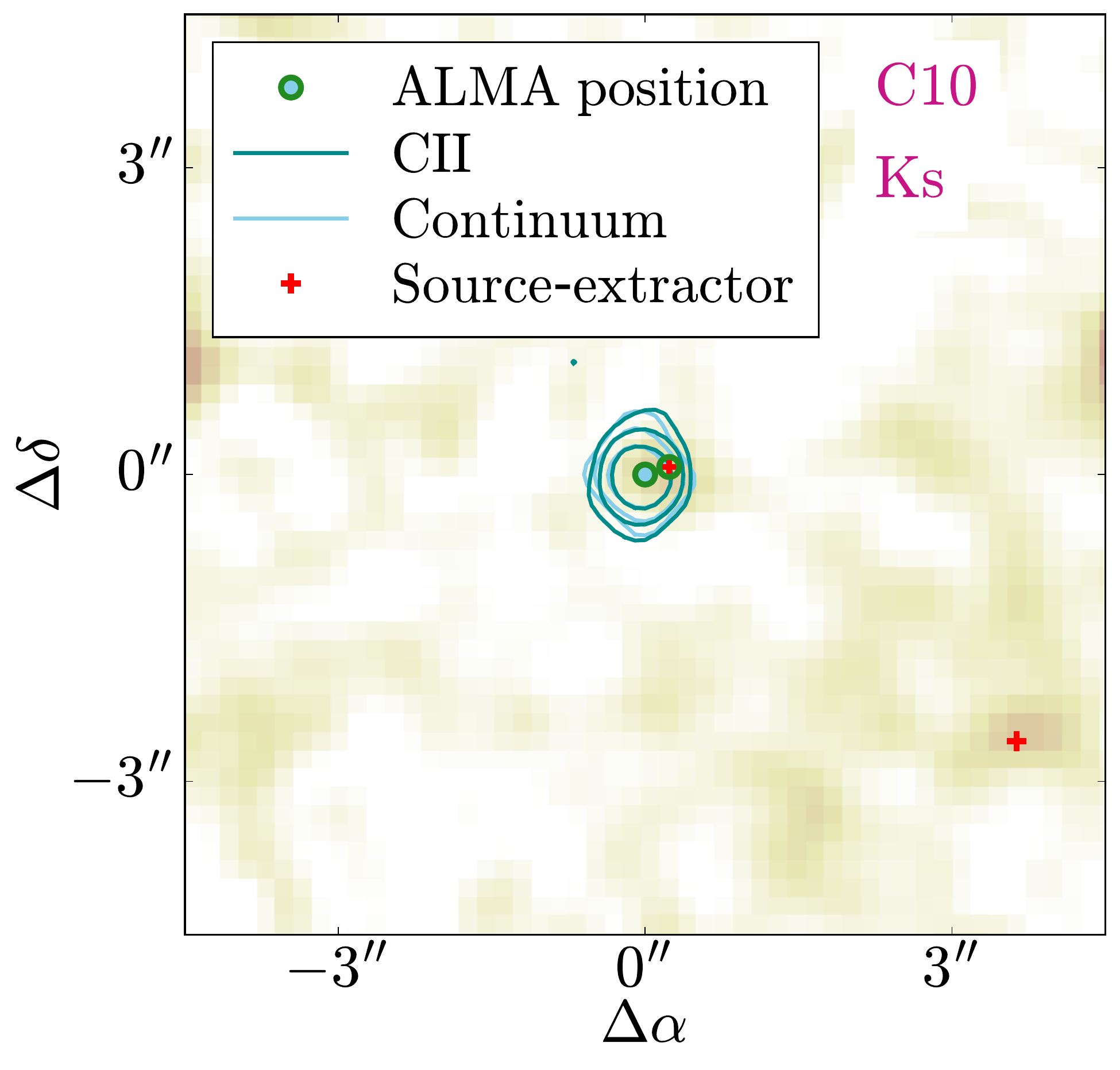}
\includegraphics[width=0.249\textwidth]{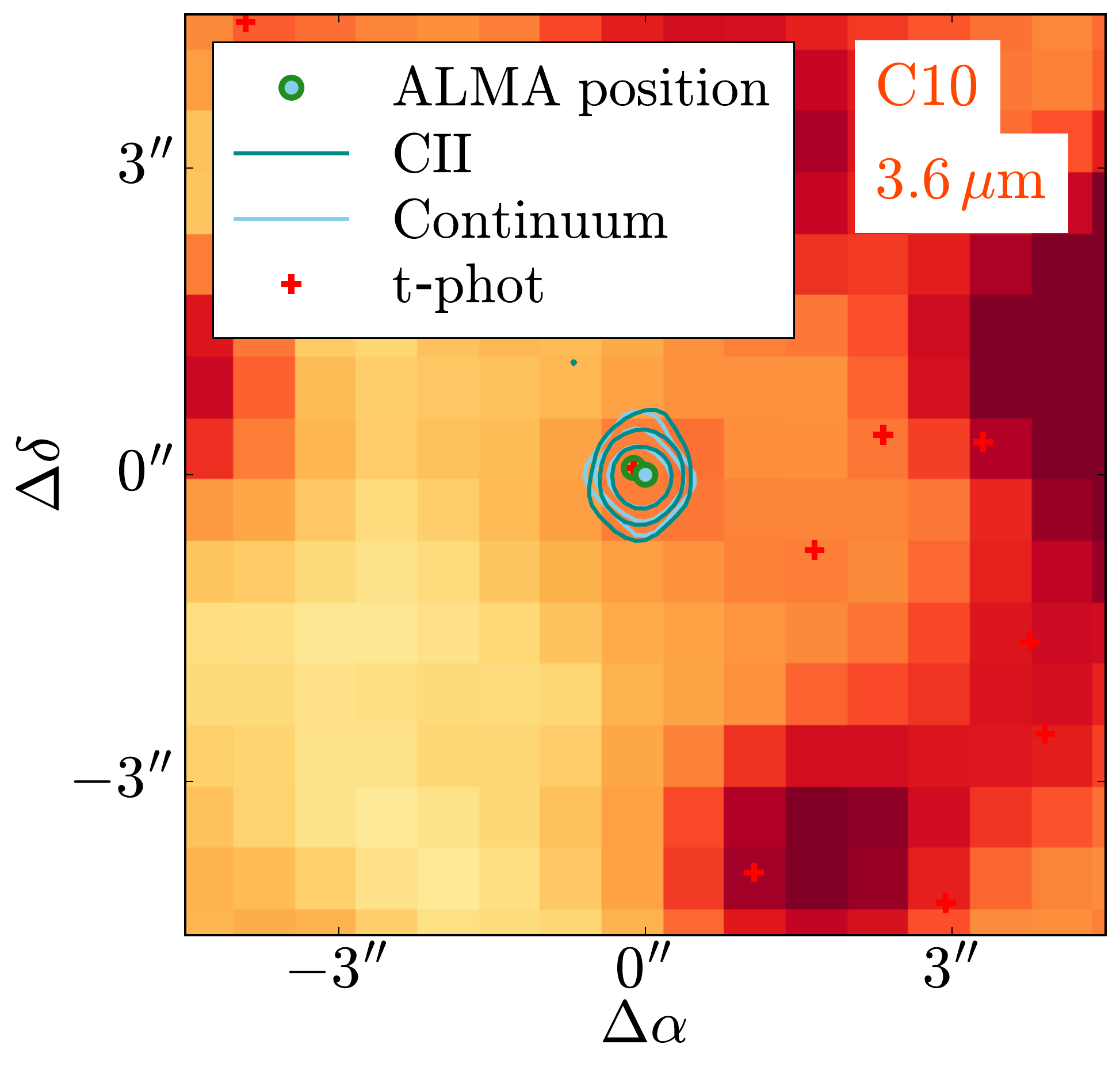}
\includegraphics[width=0.249\textwidth]{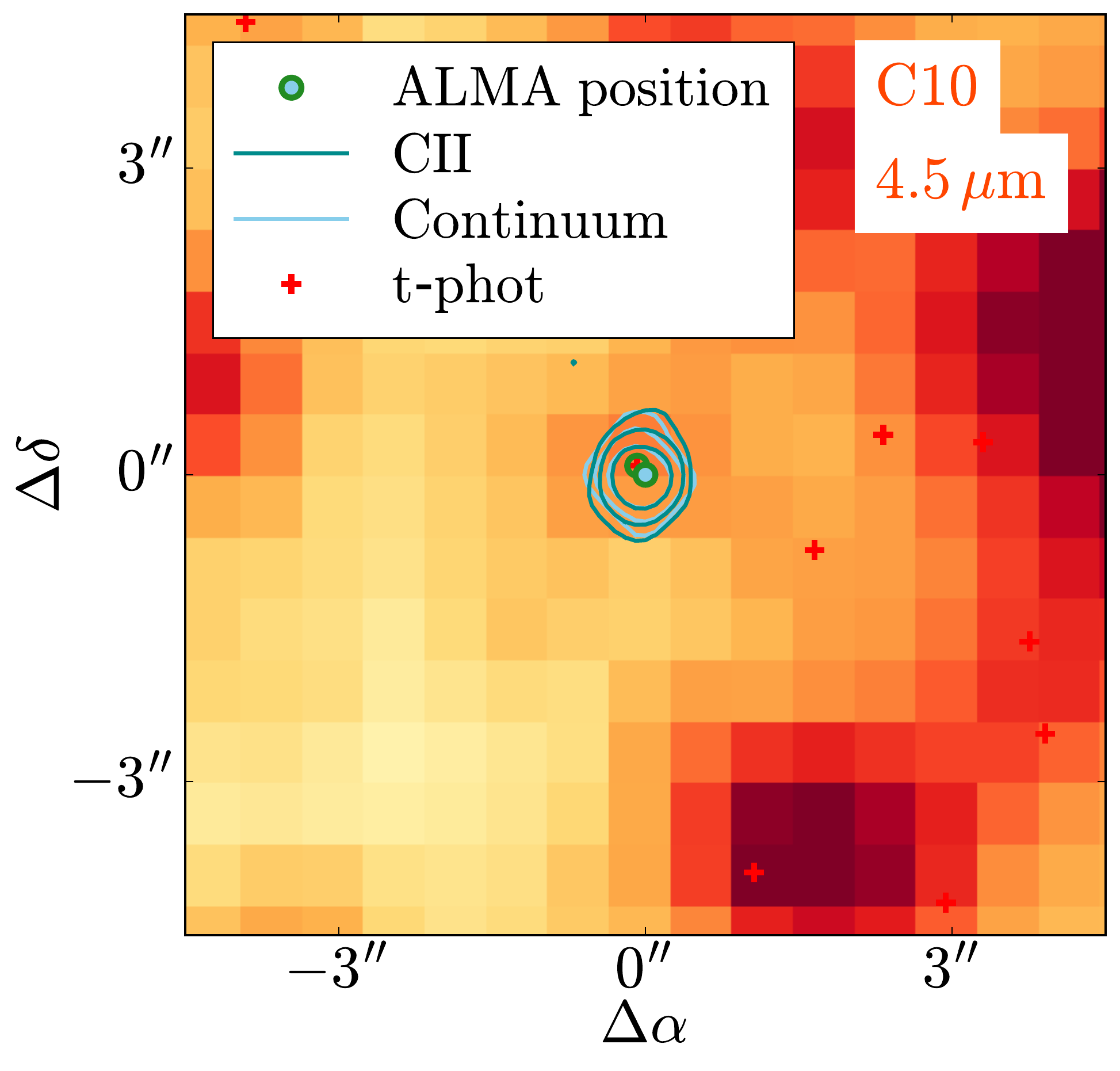}
\end{framed}
\end{subfigure}
\begin{subfigure}{0.85\textwidth}
\begin{framed}
\includegraphics[width=0.24\textwidth]{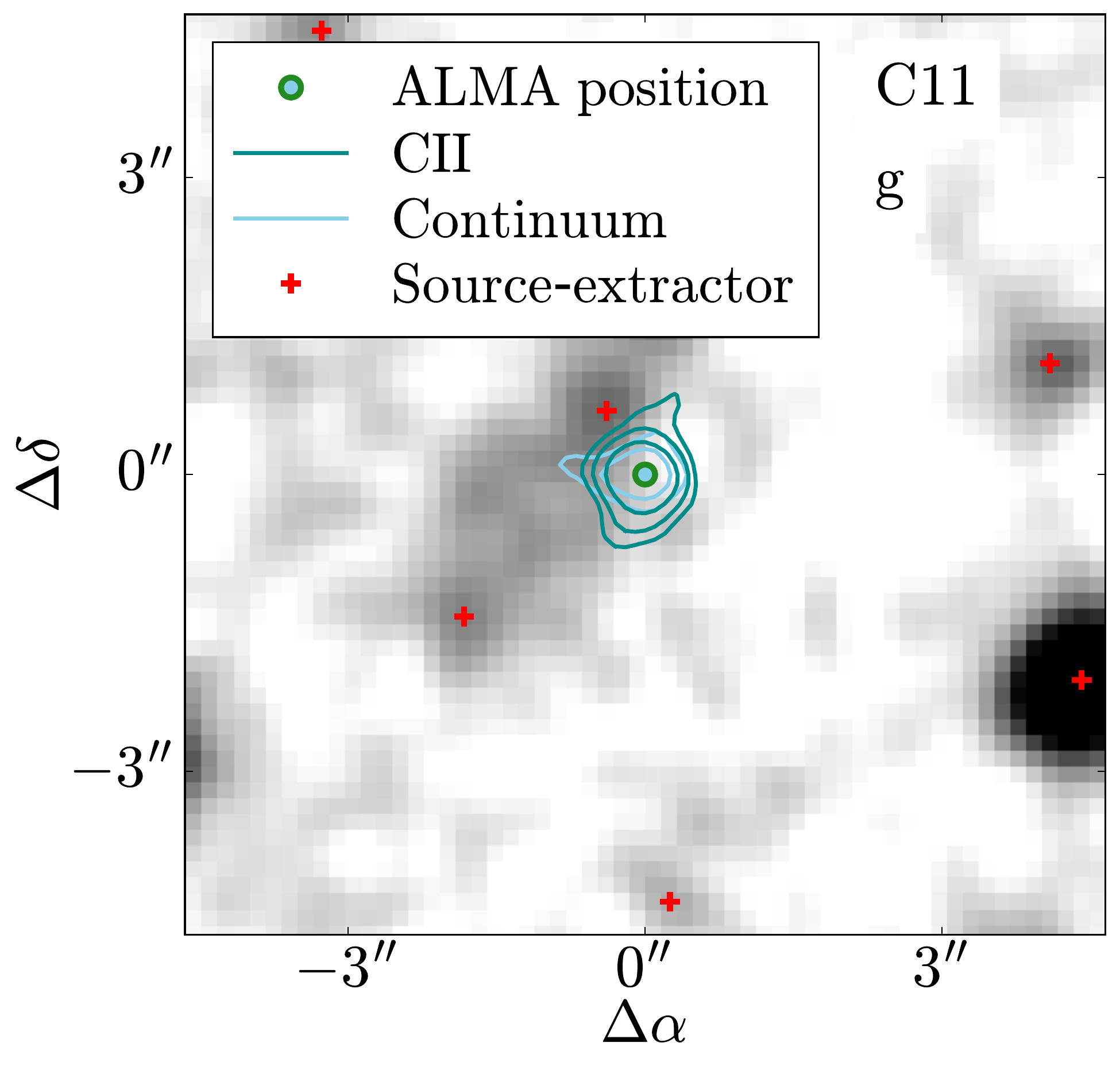}
\includegraphics[width=0.24\textwidth]{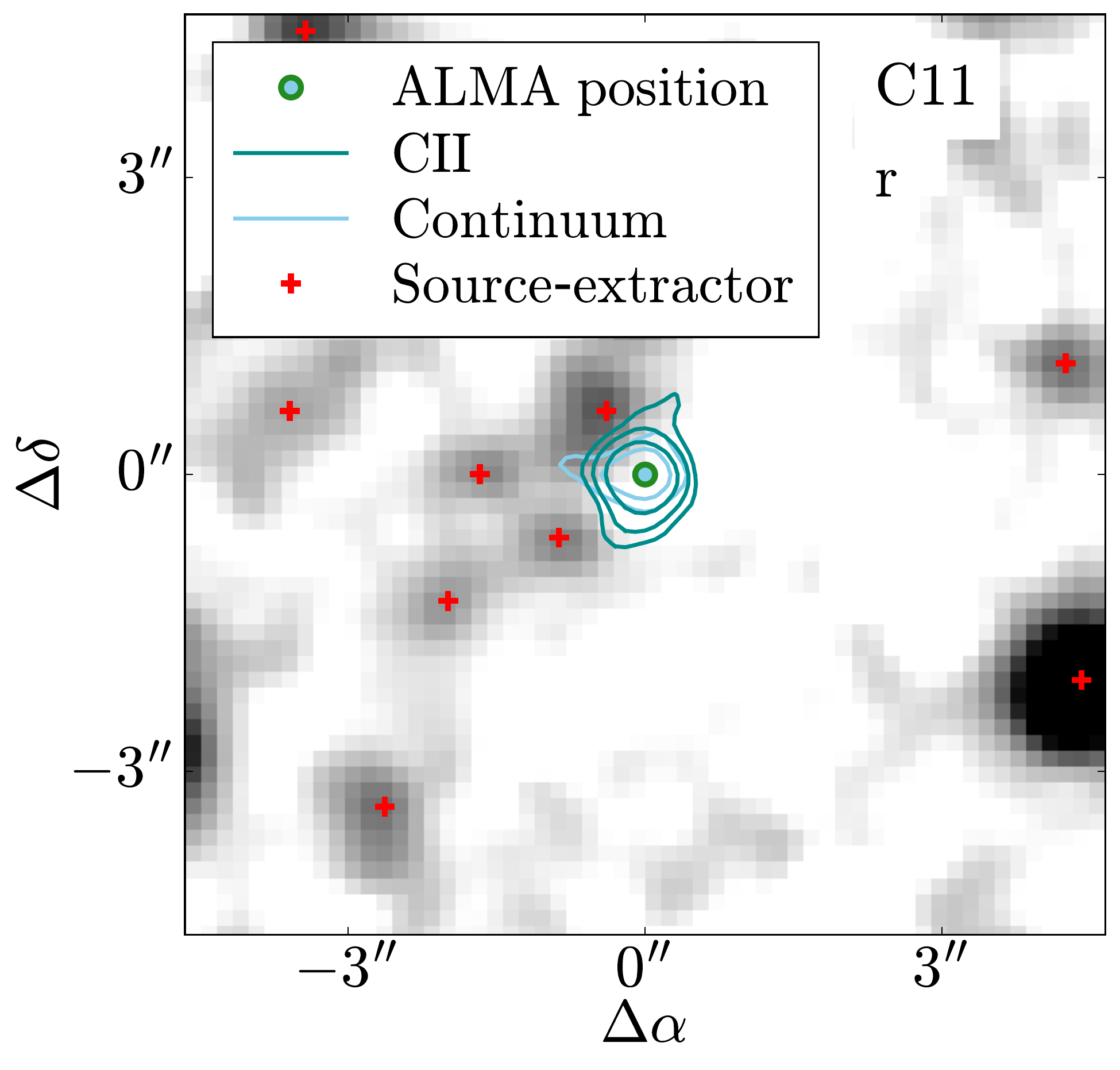}
\includegraphics[width=0.24\textwidth]{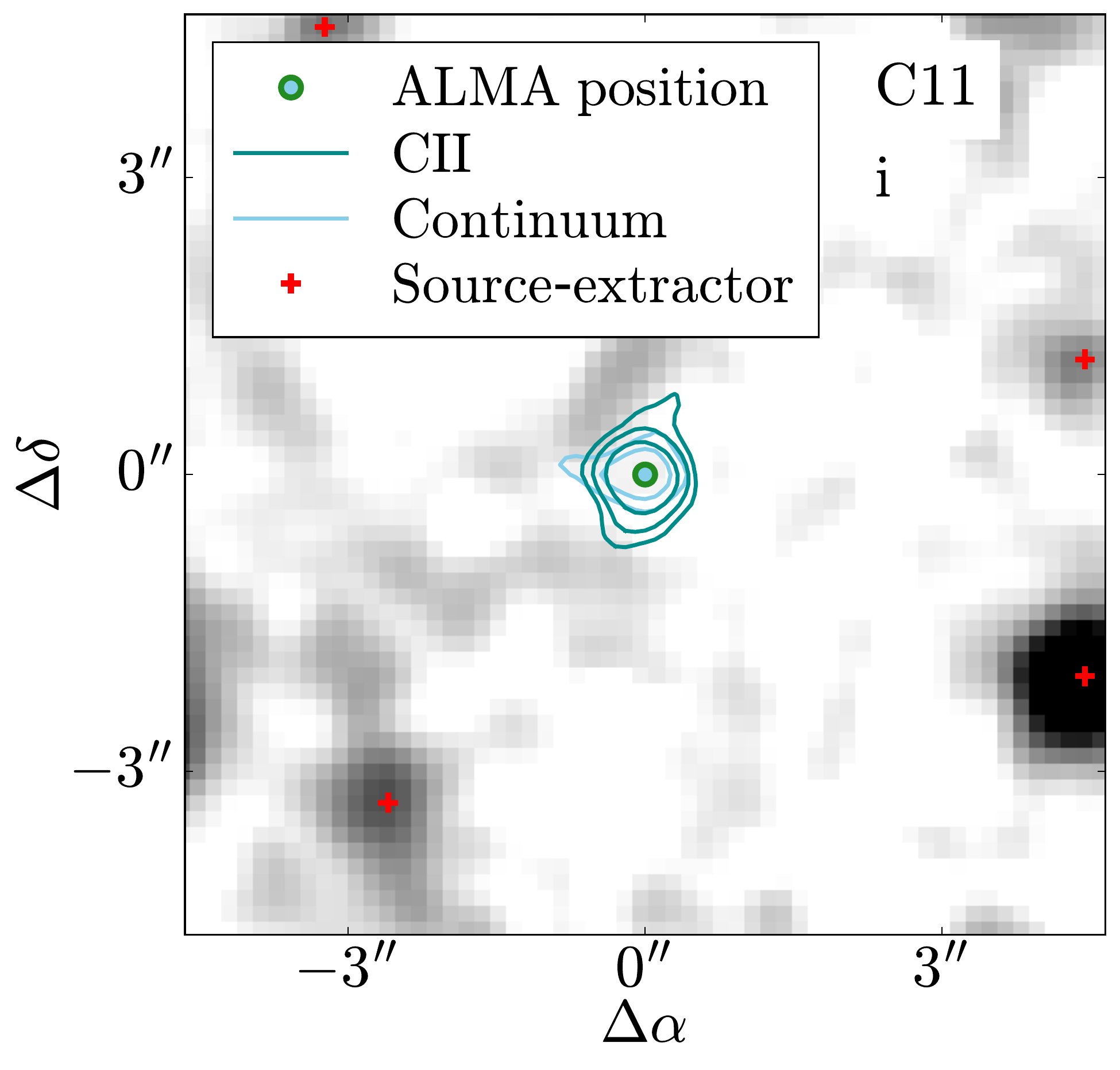}
\includegraphics[width=0.24\textwidth]{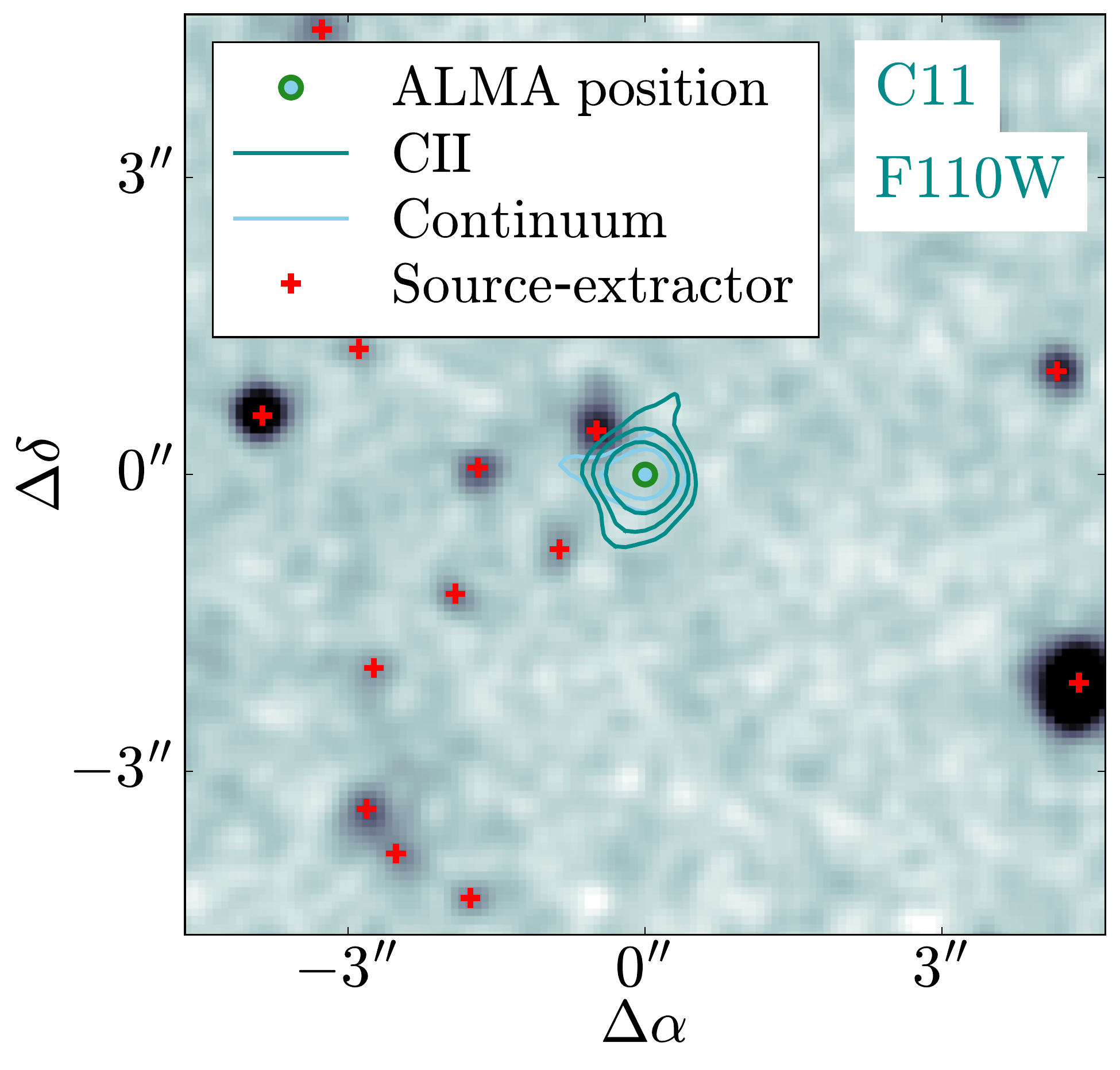}
\includegraphics[width=0.24\textwidth]{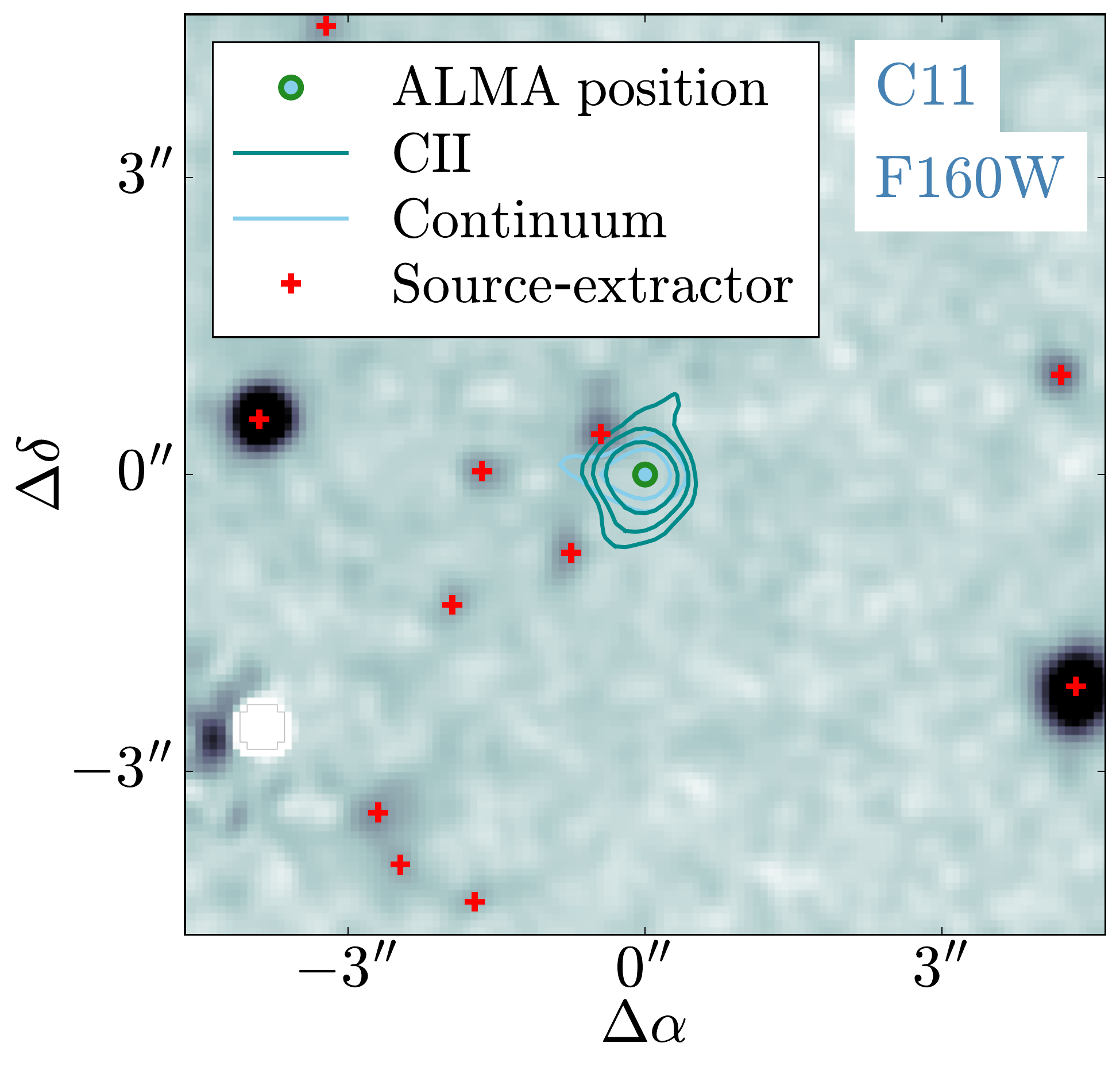}
\includegraphics[width=0.248\textwidth]{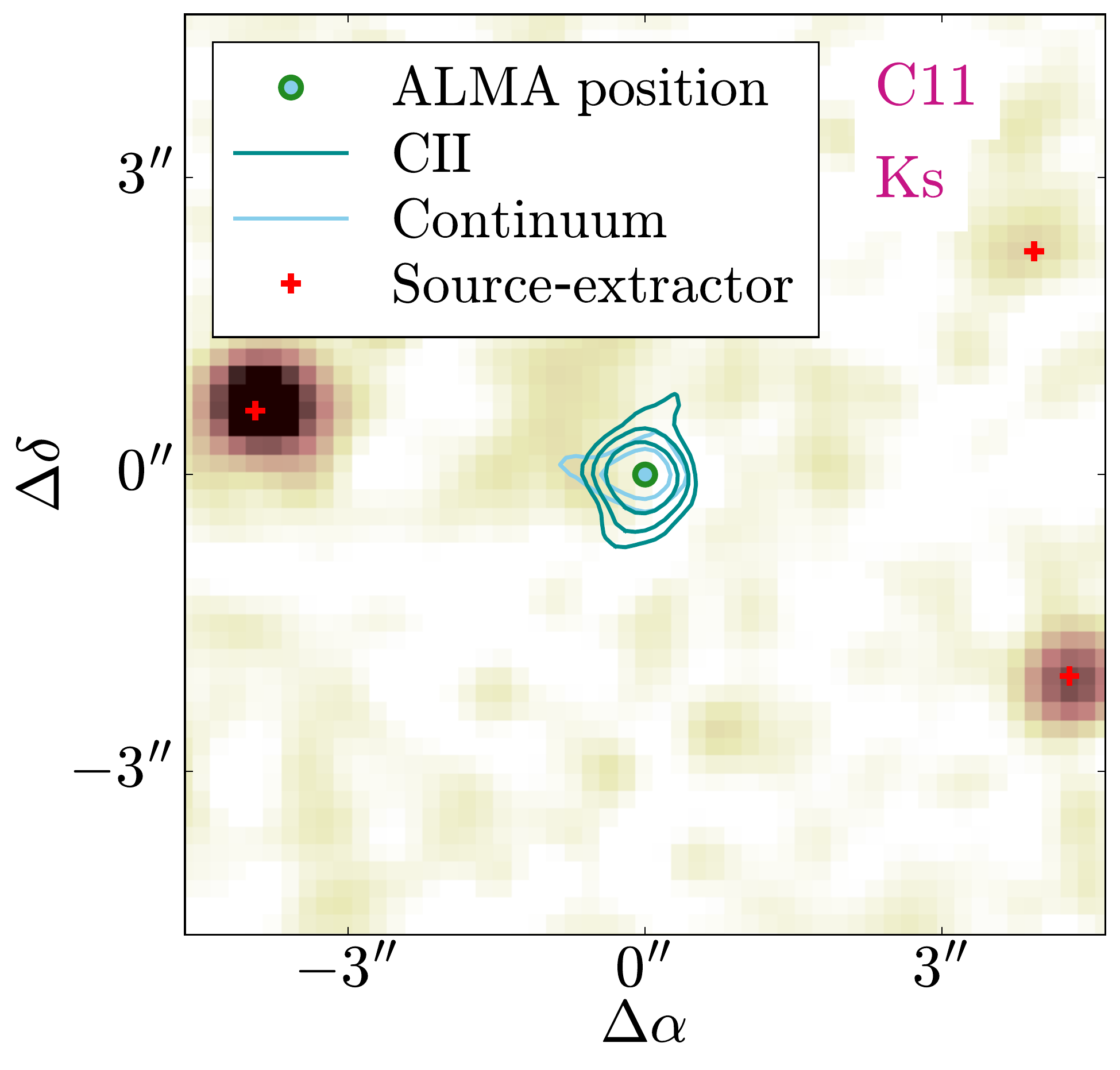}
\includegraphics[width=0.249\textwidth]{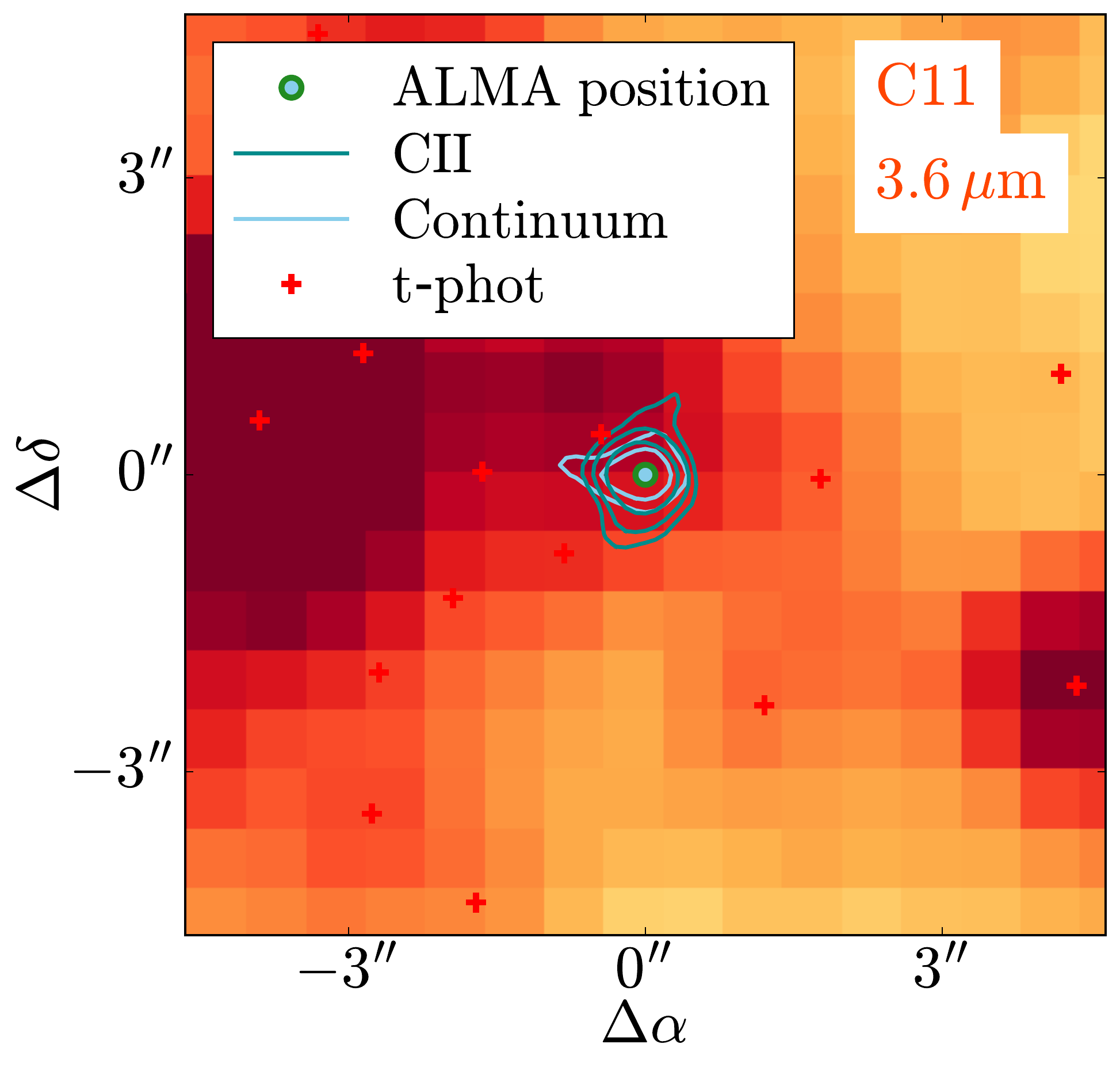}
\includegraphics[width=0.249\textwidth]{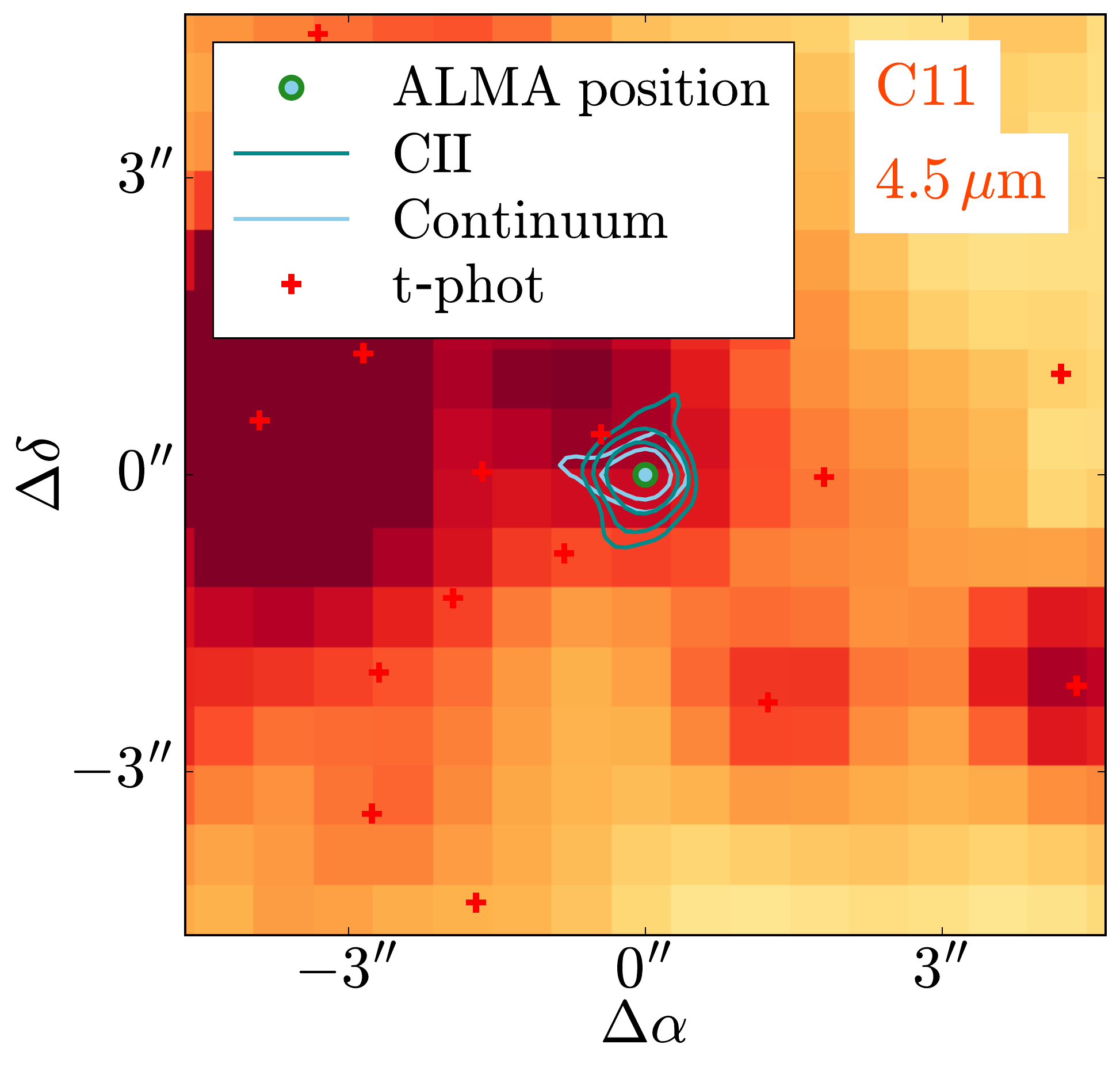}
\end{framed}
\end{subfigure}
\begin{subfigure}{0.85\textwidth}
\begin{framed}
\includegraphics[width=0.24\textwidth]{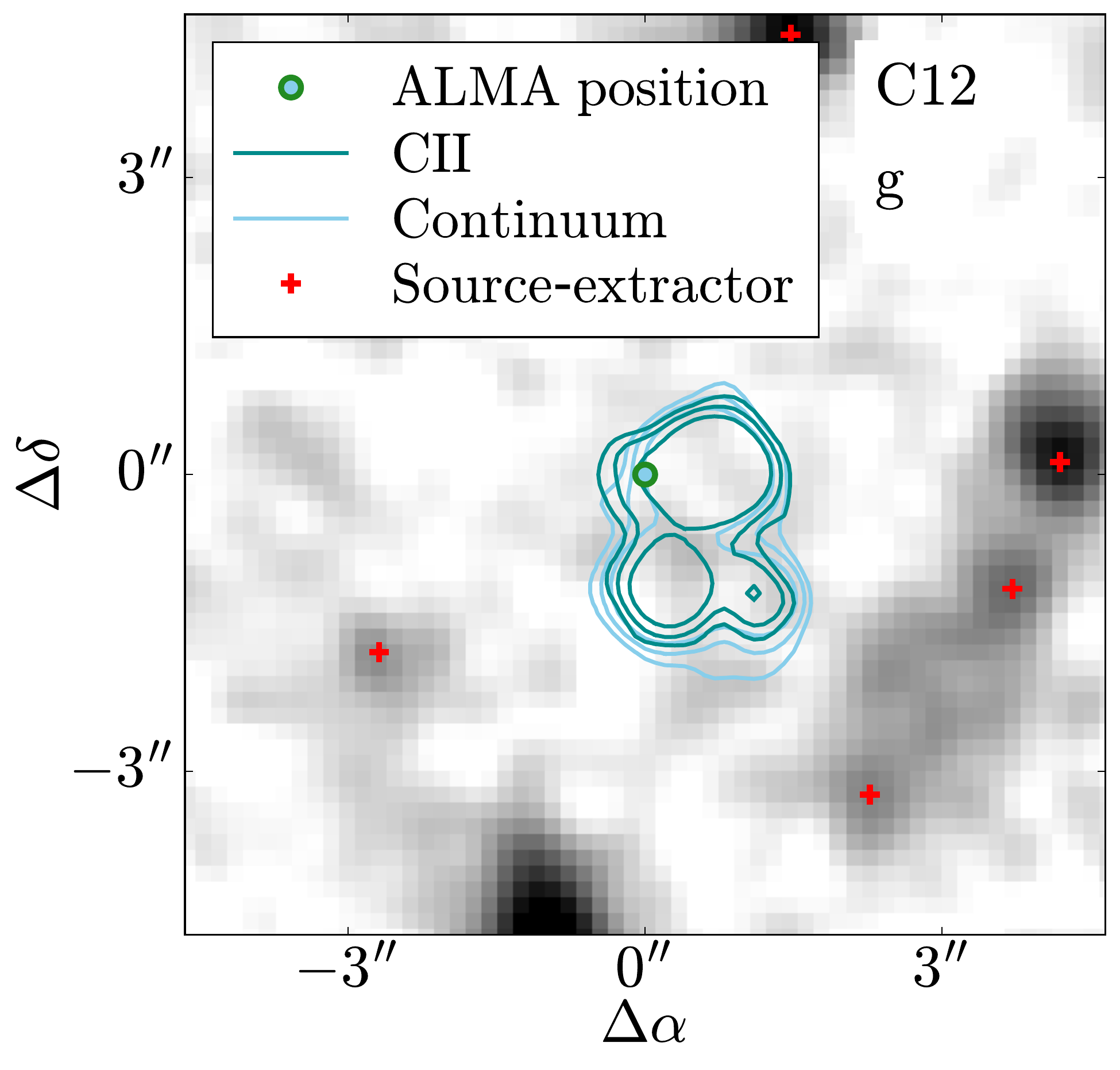}
\includegraphics[width=0.24\textwidth]{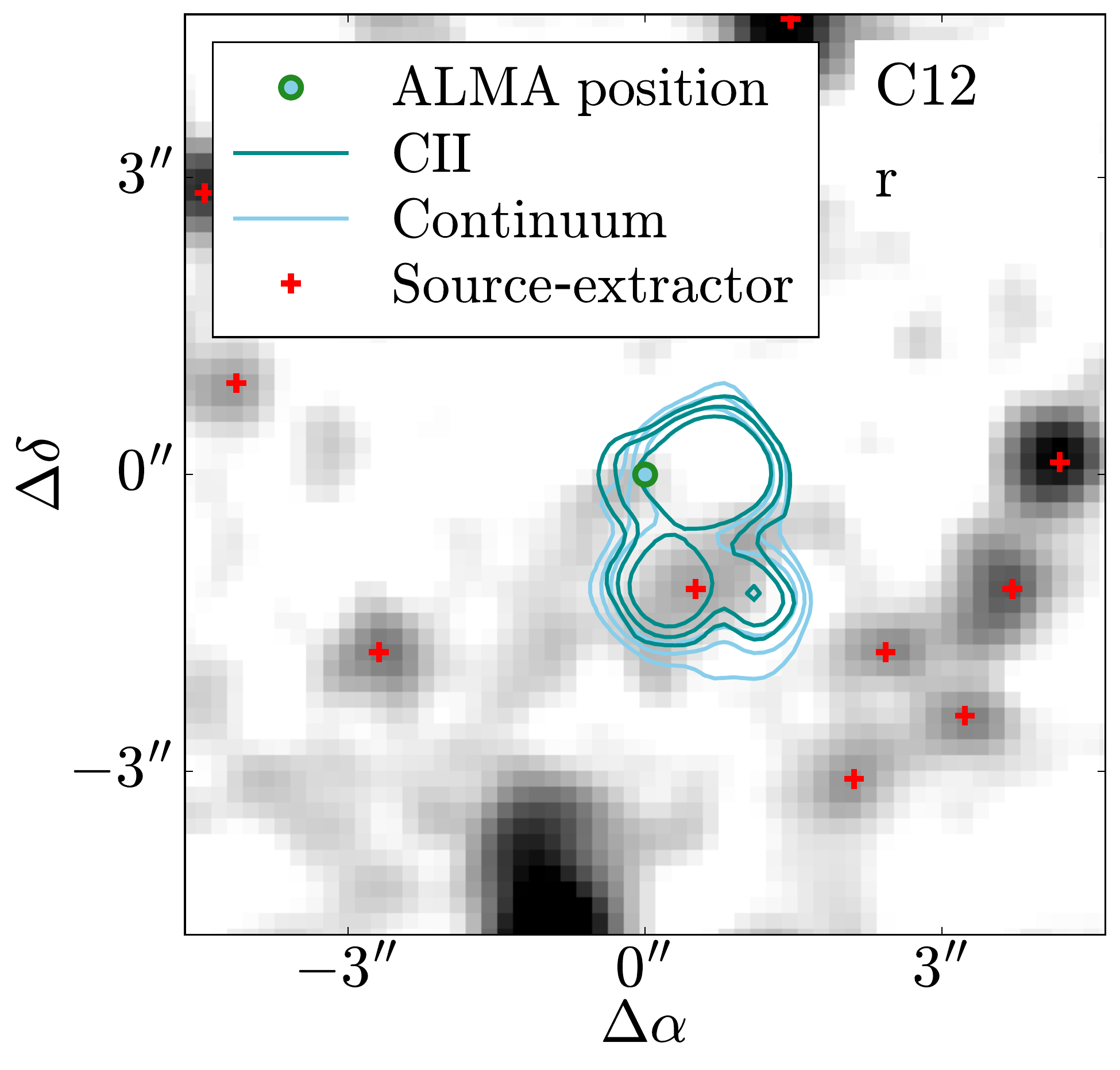}
\includegraphics[width=0.24\textwidth]{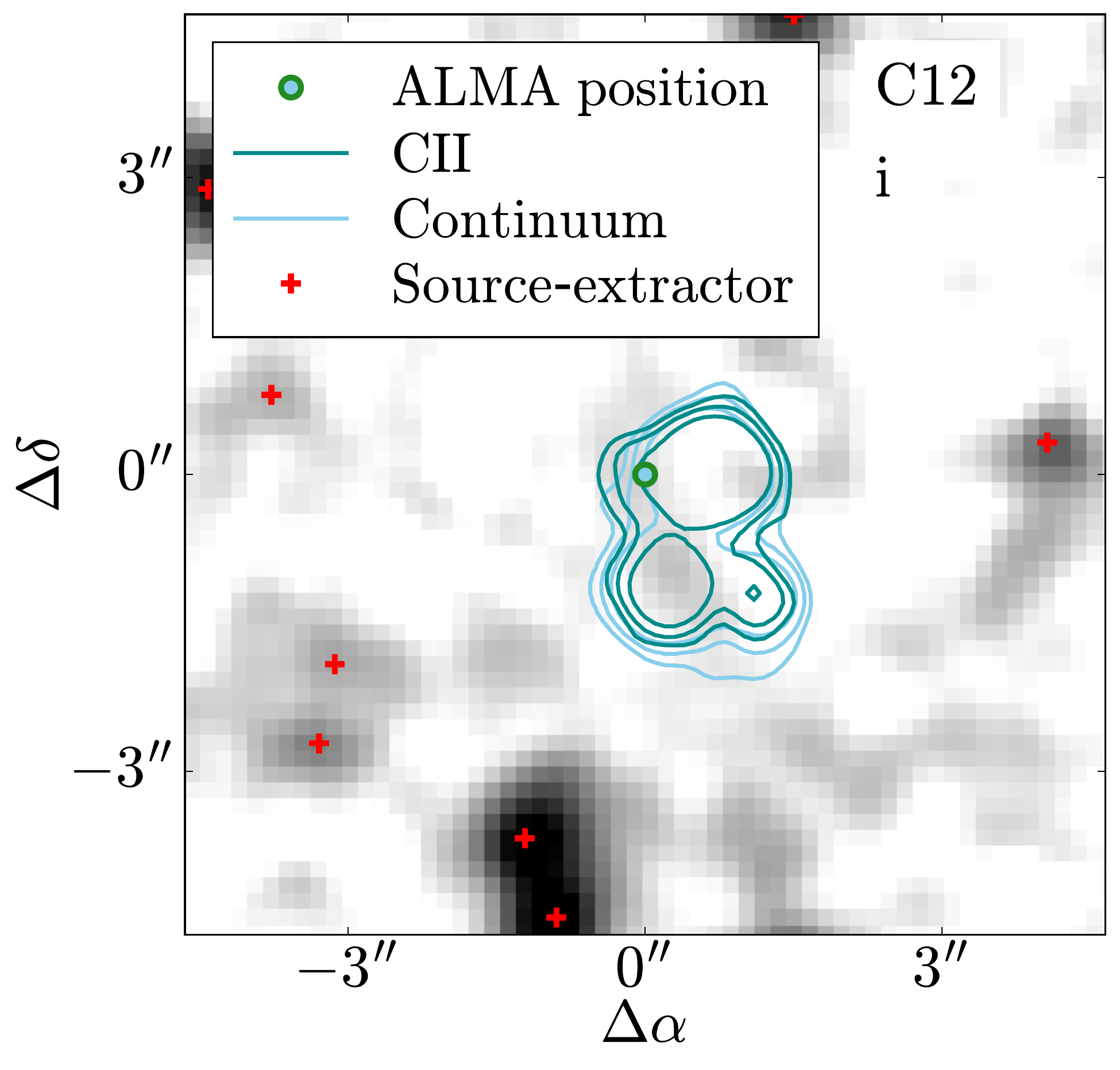}
\includegraphics[width=0.24\textwidth]{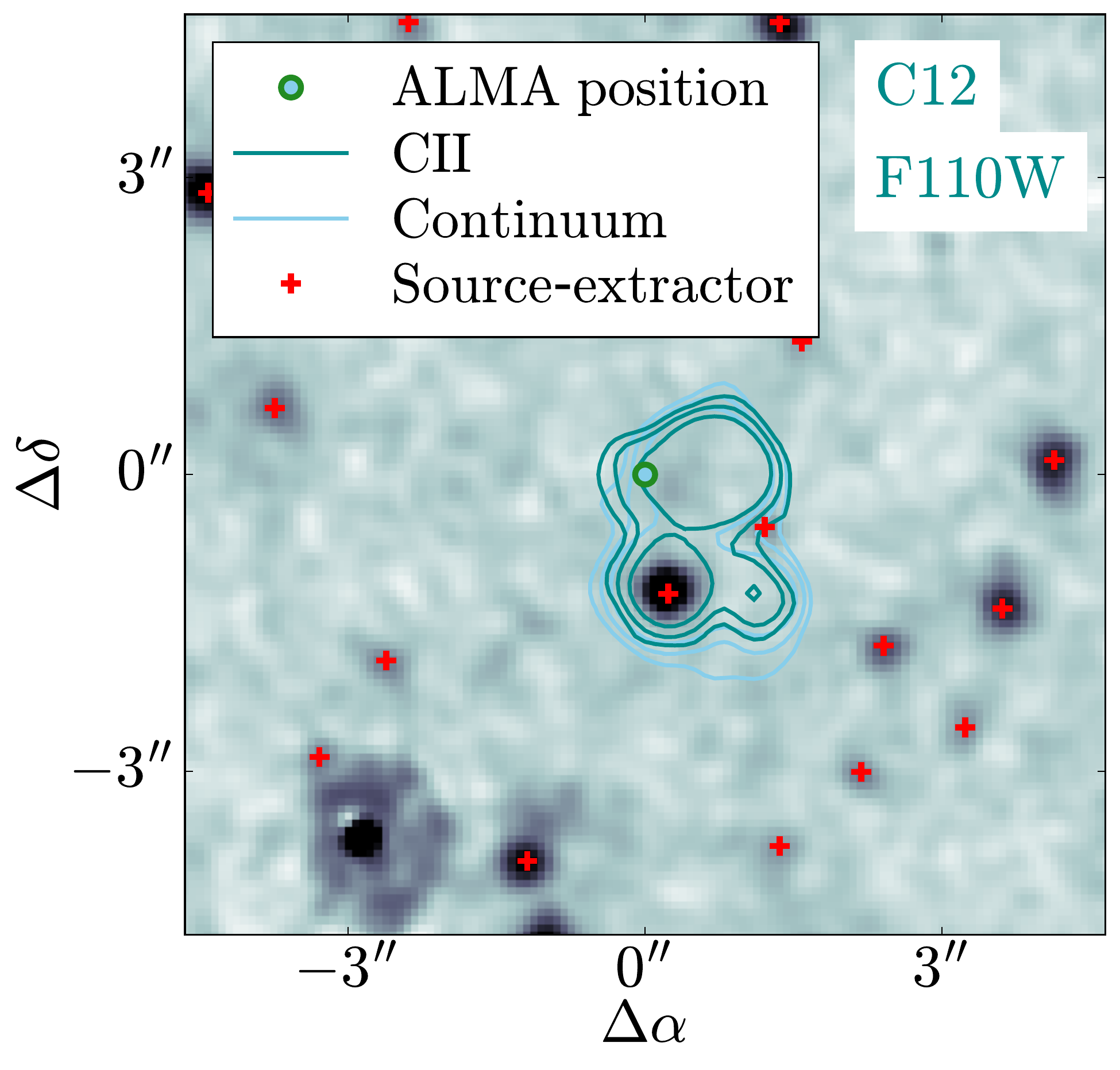}
\includegraphics[width=0.24\textwidth]{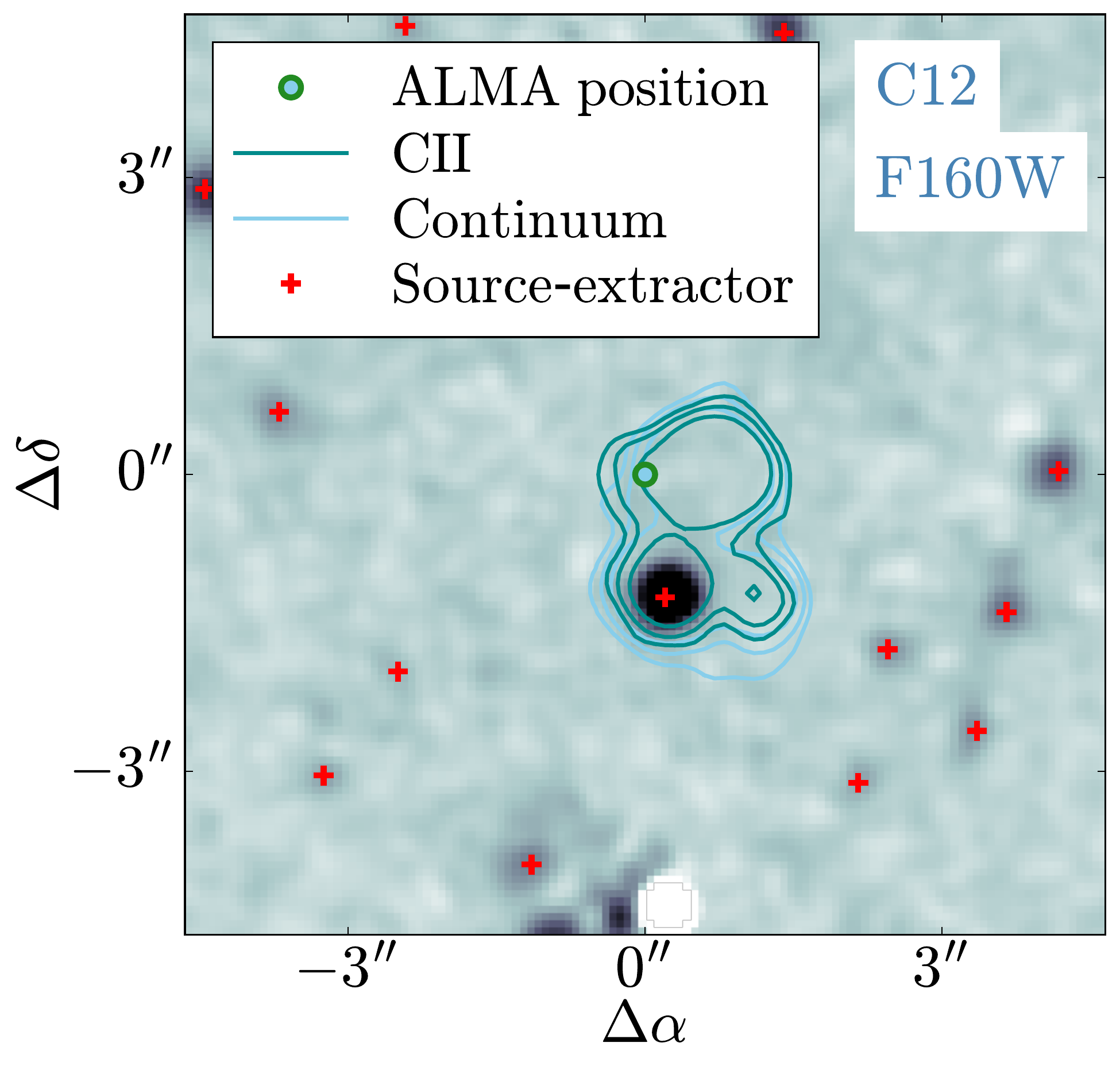}
\includegraphics[width=0.248\textwidth]{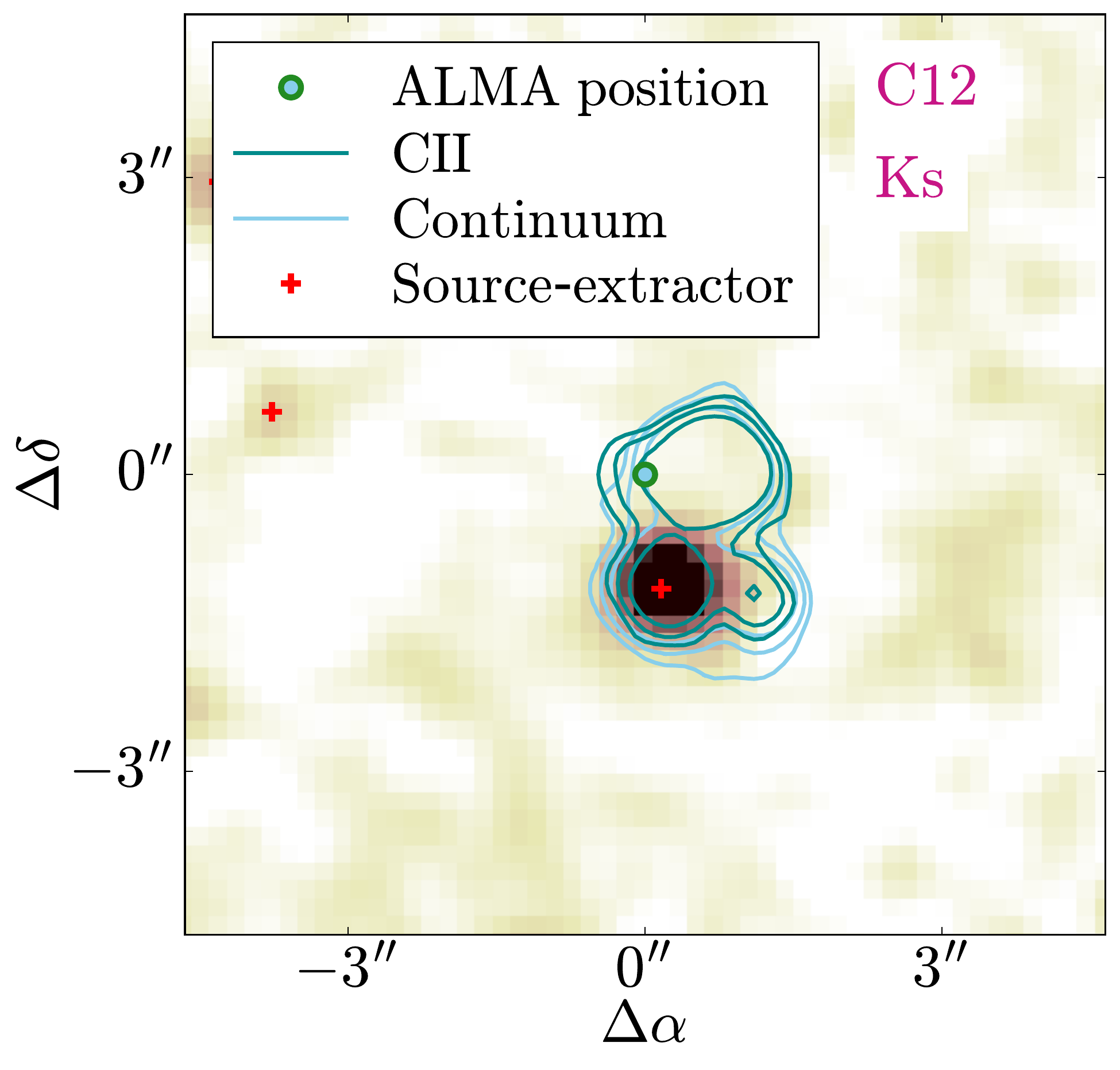}
\includegraphics[width=0.249\textwidth]{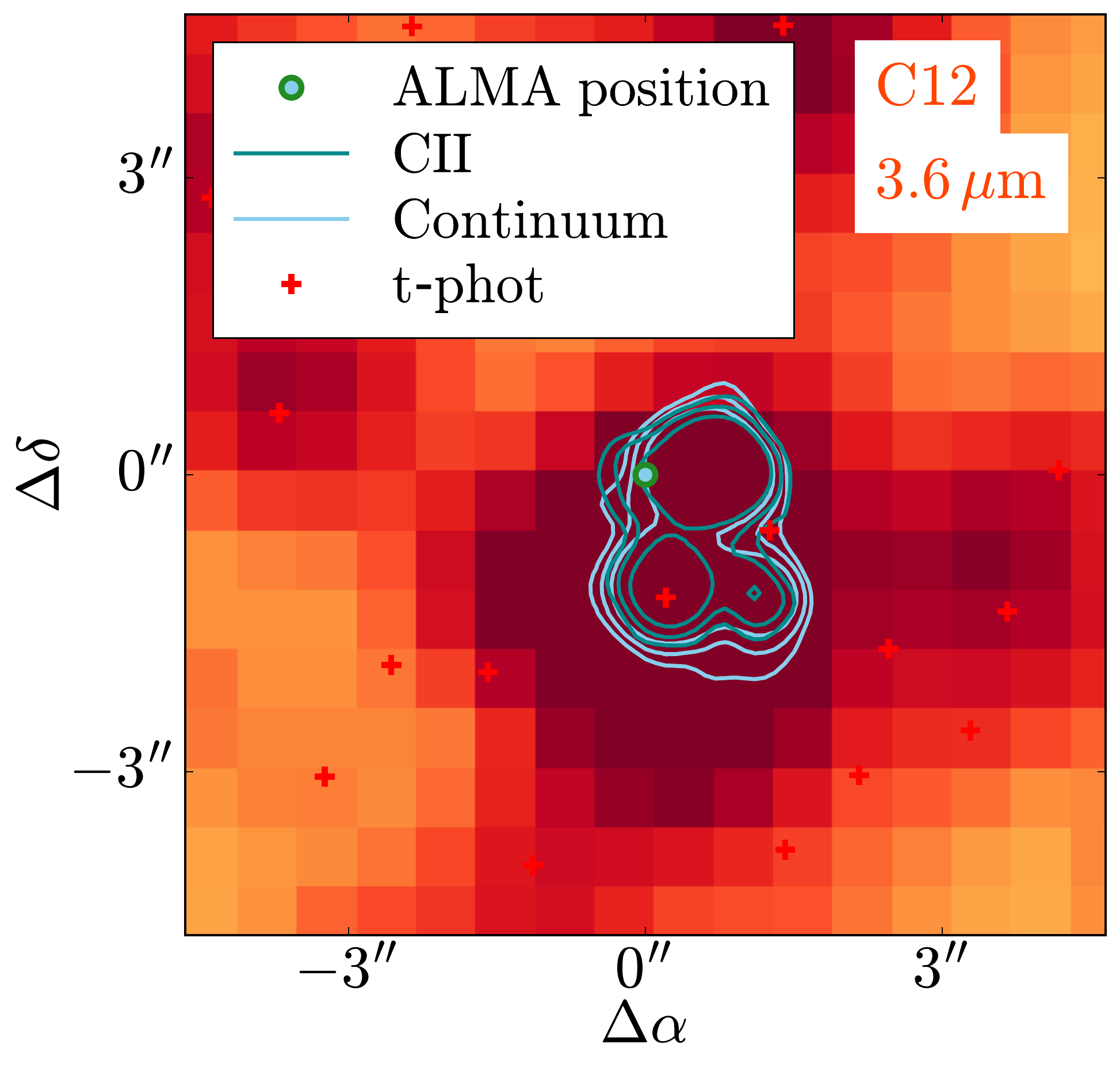}
\includegraphics[width=0.249\textwidth]{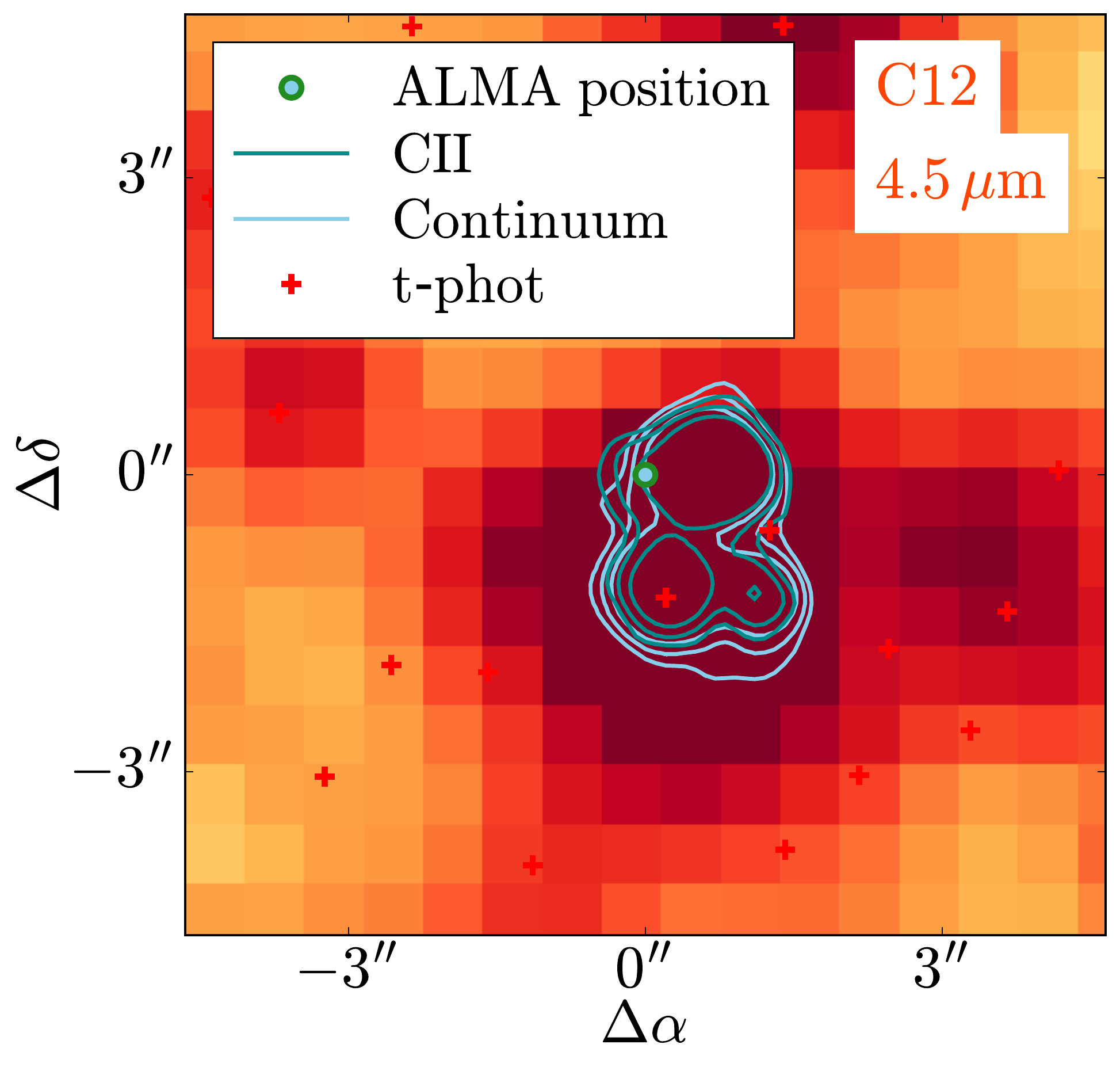}
\end{framed}
\end{subfigure}
\caption{}
\end{figure*}
\renewcommand{\thefigure}{\arabic{figure}}

\renewcommand{\thefigure}{B\arabic{figure} (Cont.)}
\addtocounter{figure}{-1}
\begin{figure*}
\begin{subfigure}{0.85\textwidth}
\begin{framed}
\includegraphics[width=0.24\textwidth]{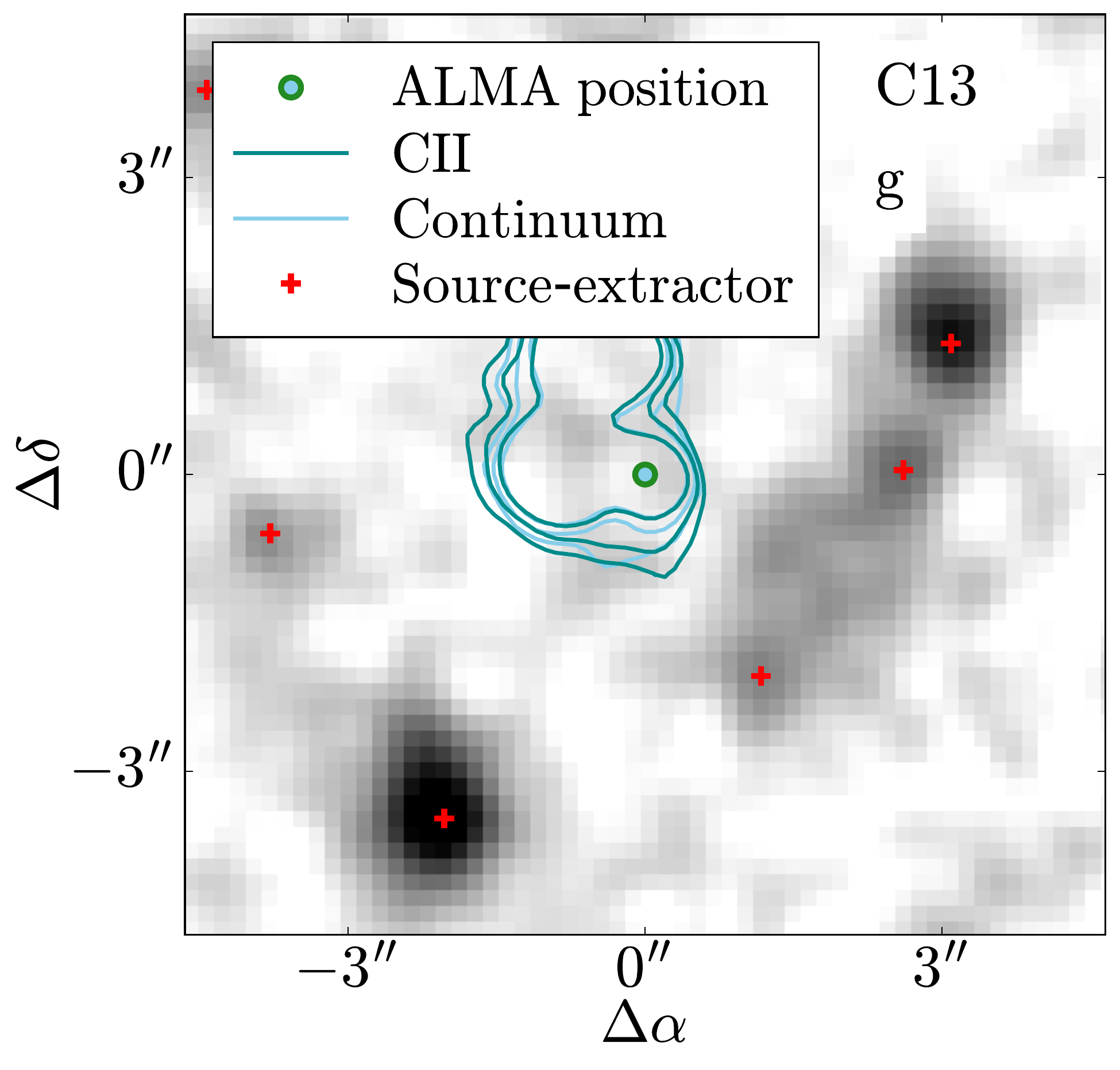}
\includegraphics[width=0.24\textwidth]{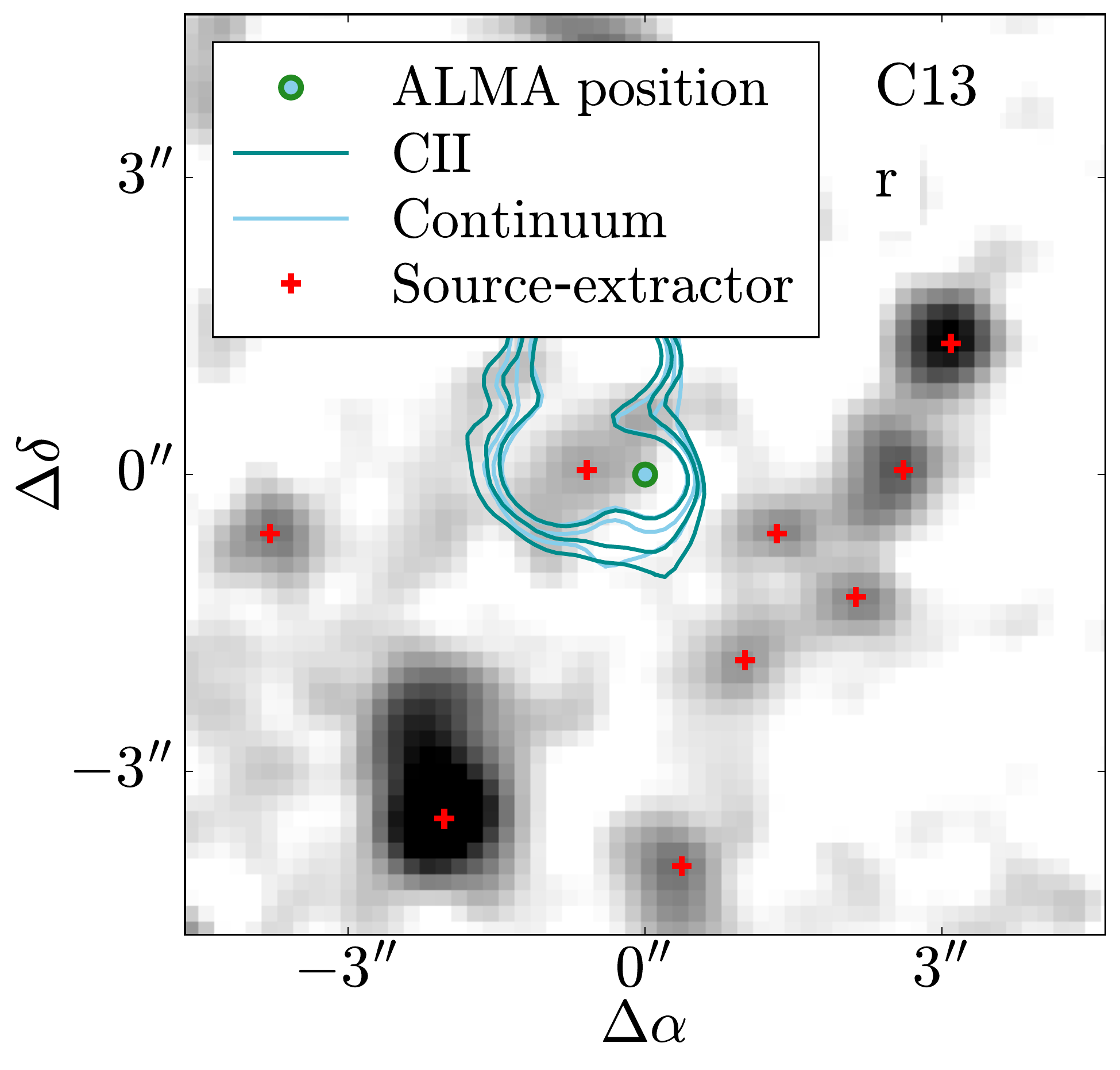}
\includegraphics[width=0.24\textwidth]{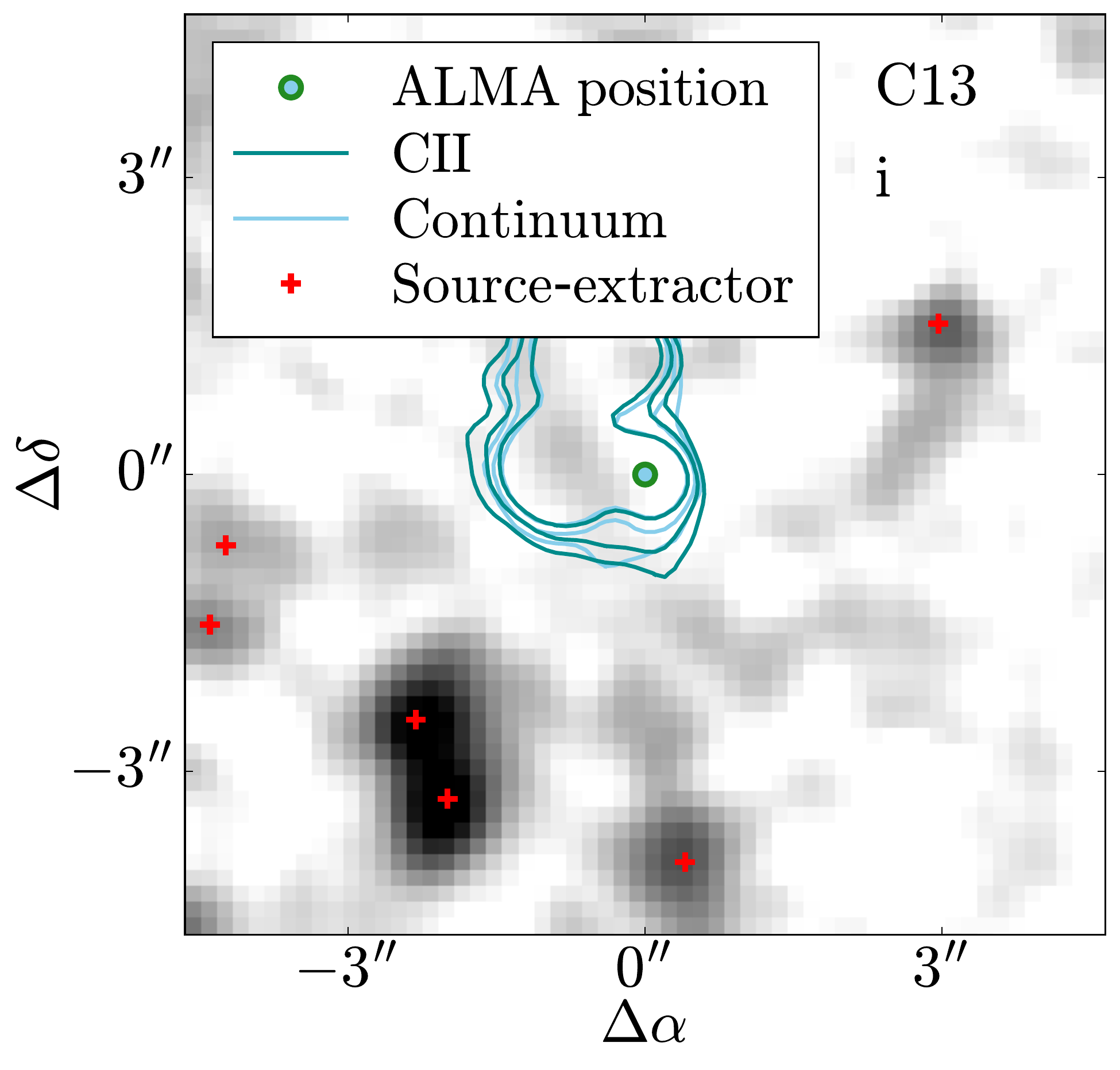}
\includegraphics[width=0.24\textwidth]{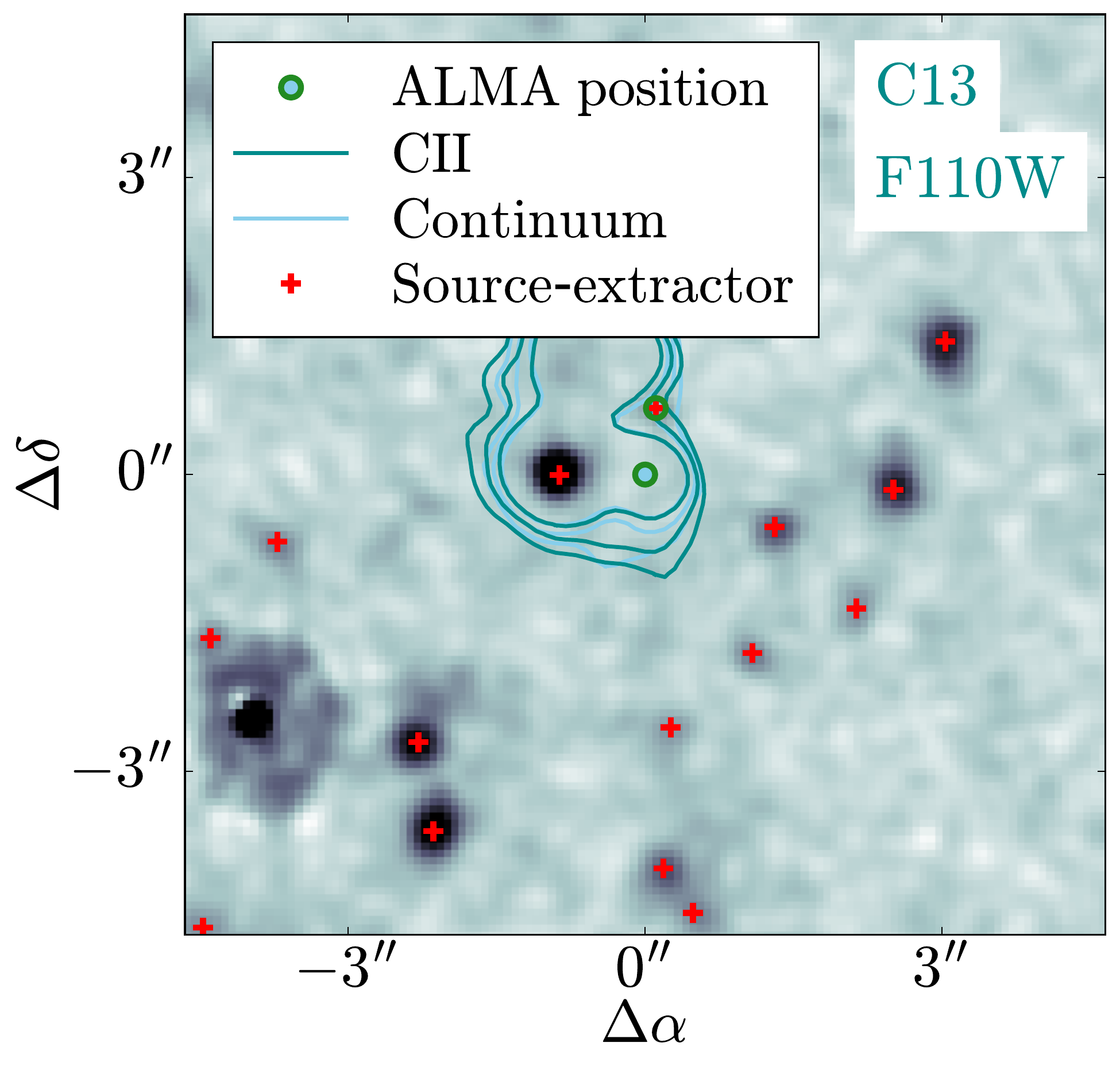}
\includegraphics[width=0.24\textwidth]{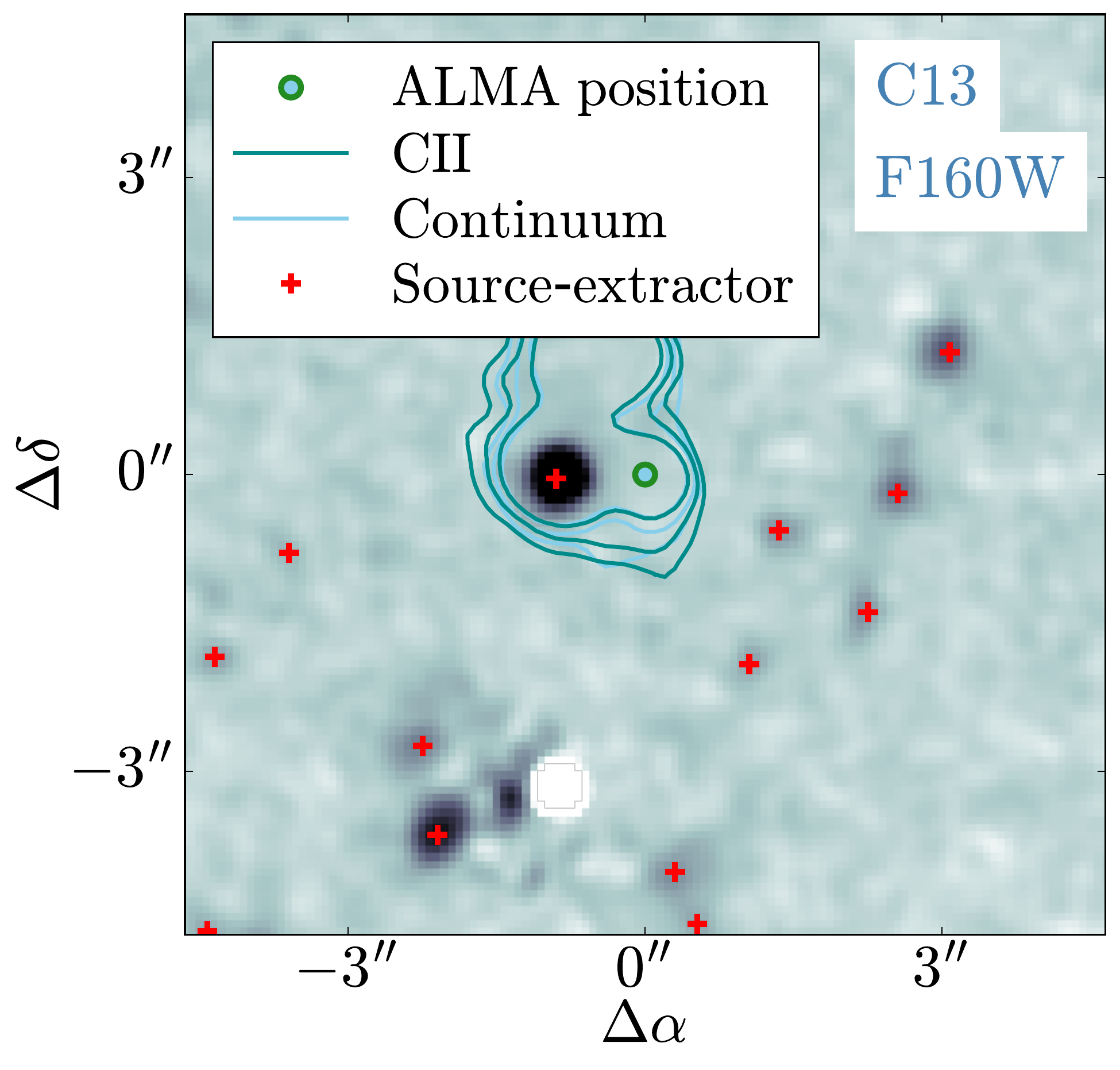}
\includegraphics[width=0.248\textwidth]{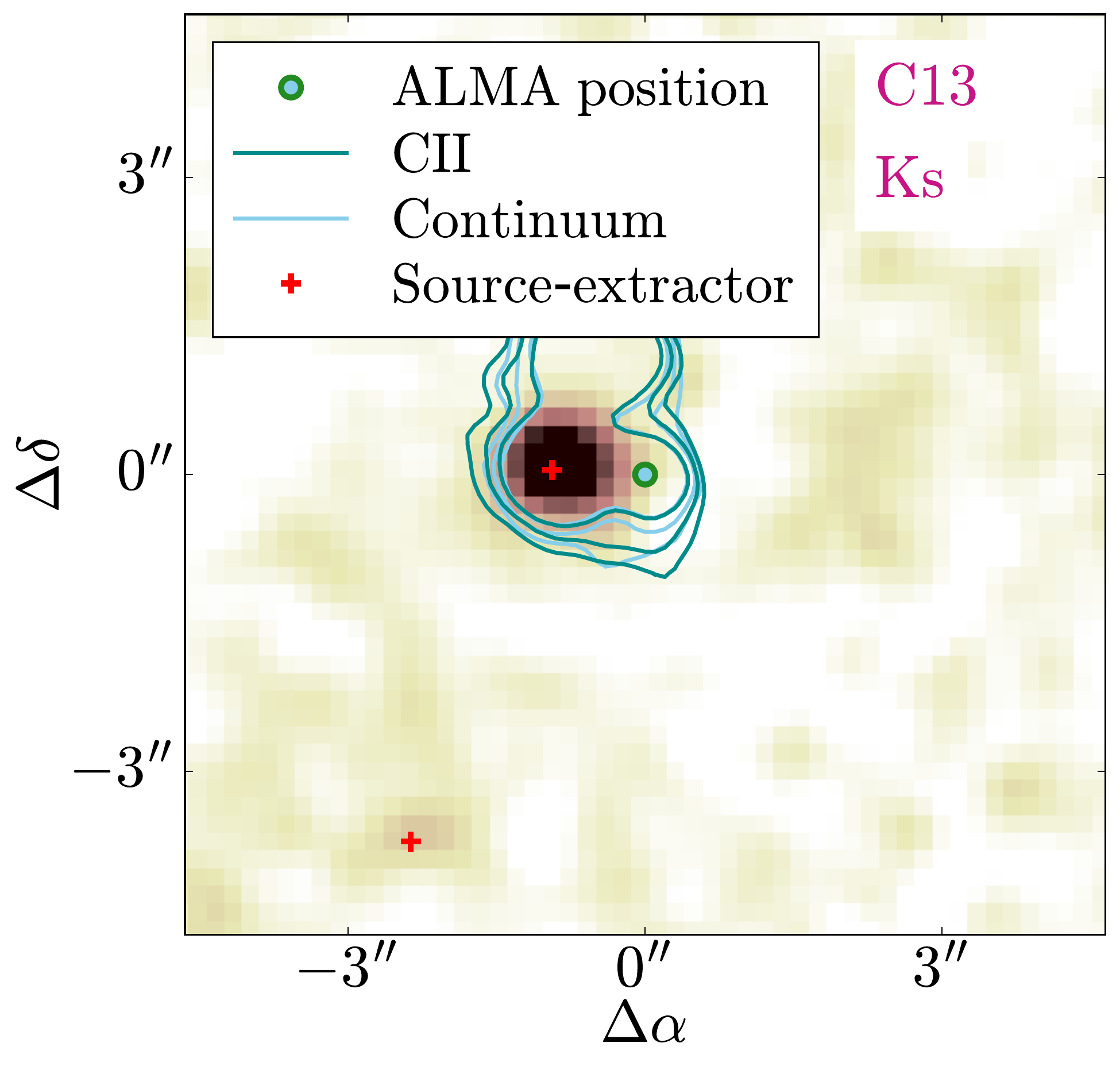}
\includegraphics[width=0.249\textwidth]{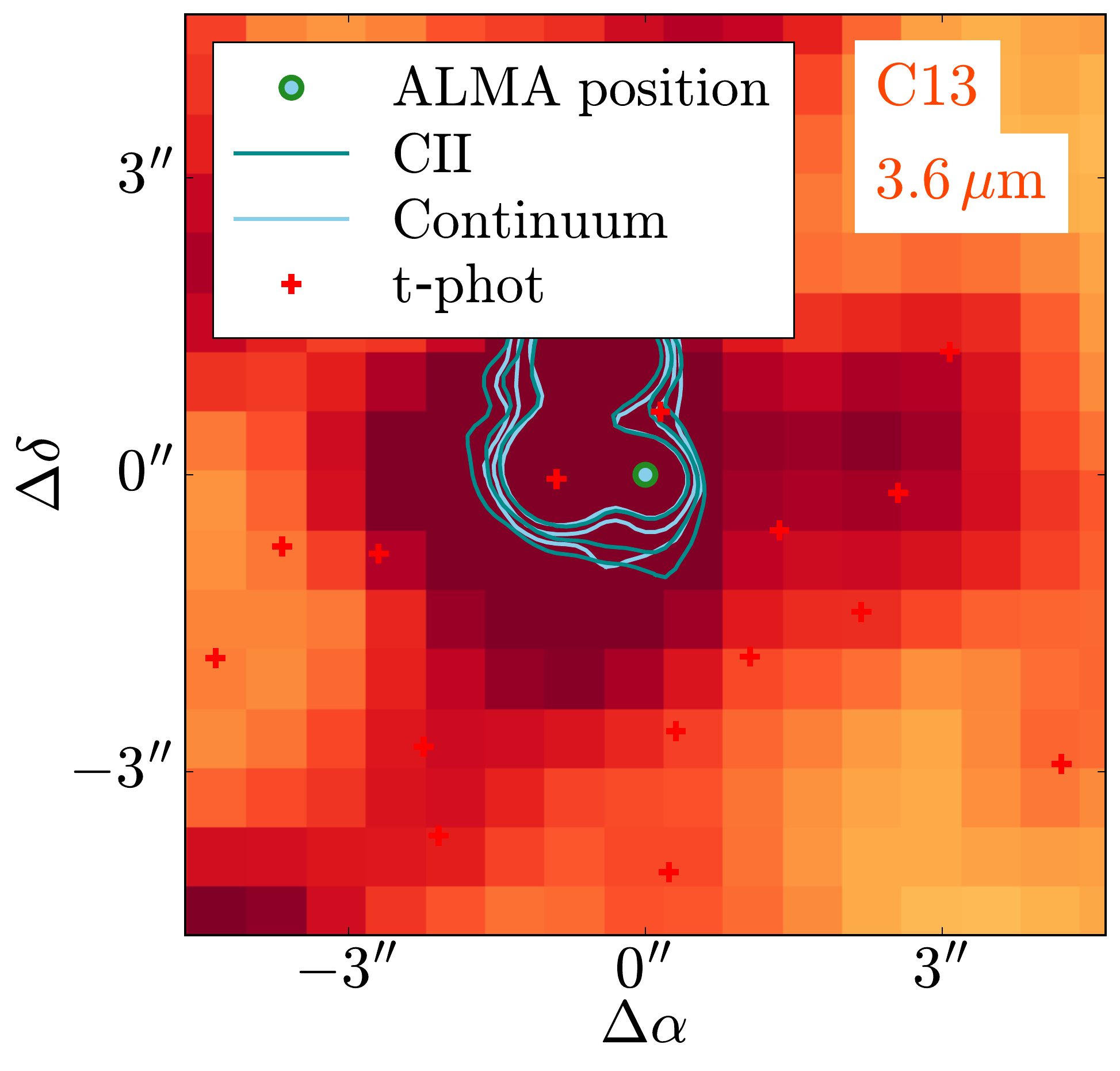}
\includegraphics[width=0.249\textwidth]{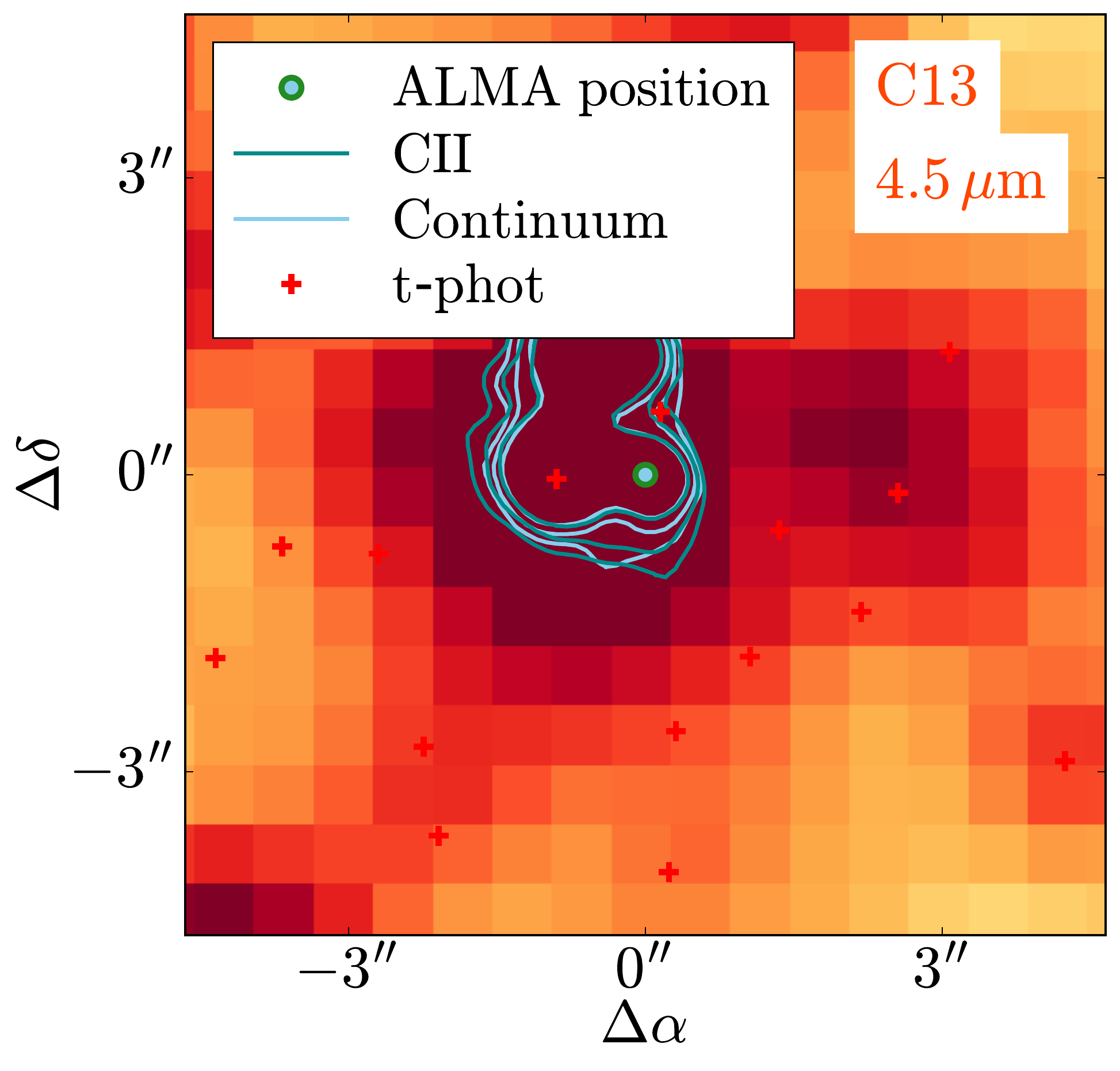}
\end{framed}
\end{subfigure}
\begin{subfigure}{0.85\textwidth}
\begin{framed}
\includegraphics[width=0.24\textwidth]{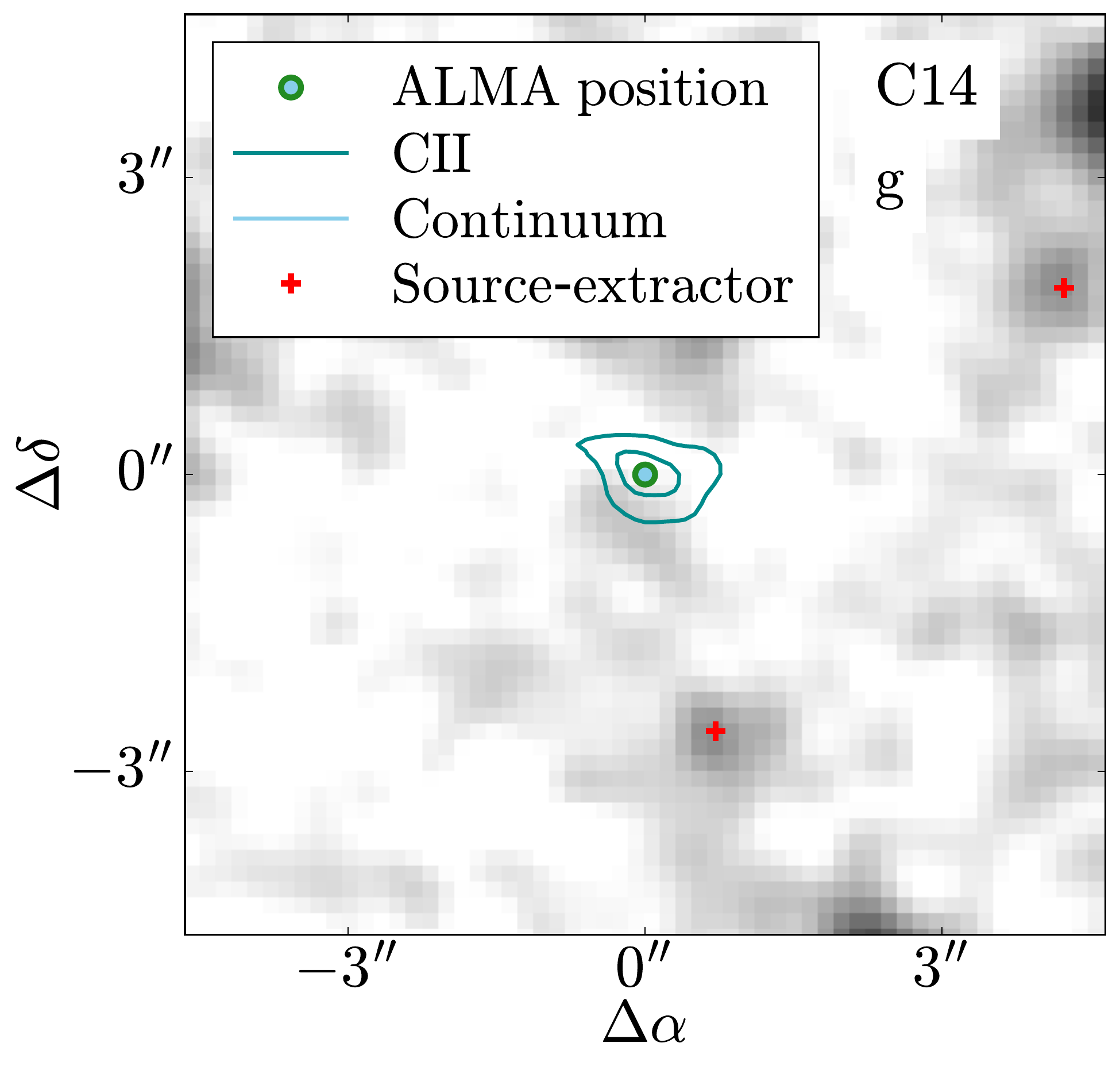}
\includegraphics[width=0.24\textwidth]{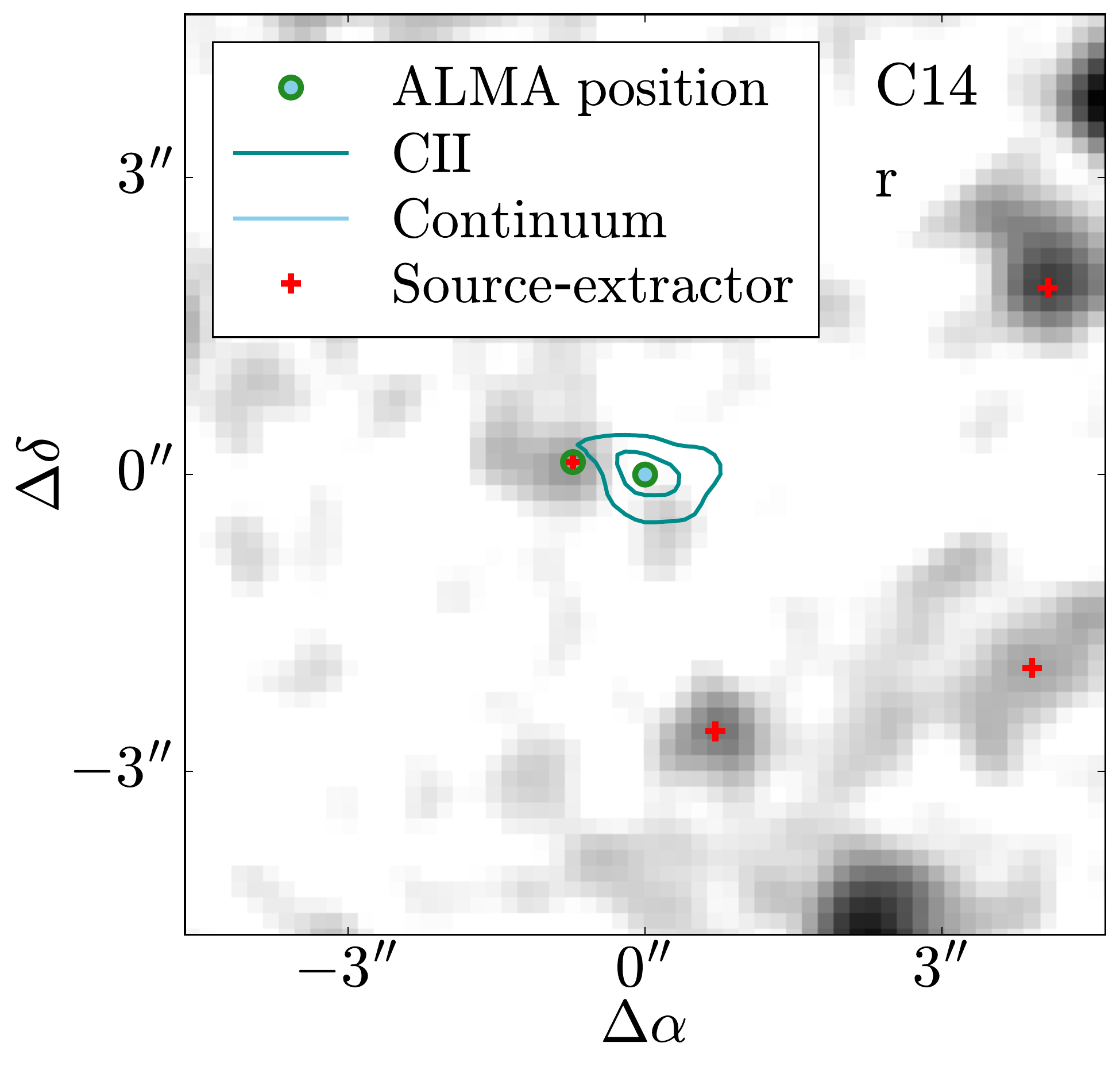}
\includegraphics[width=0.24\textwidth]{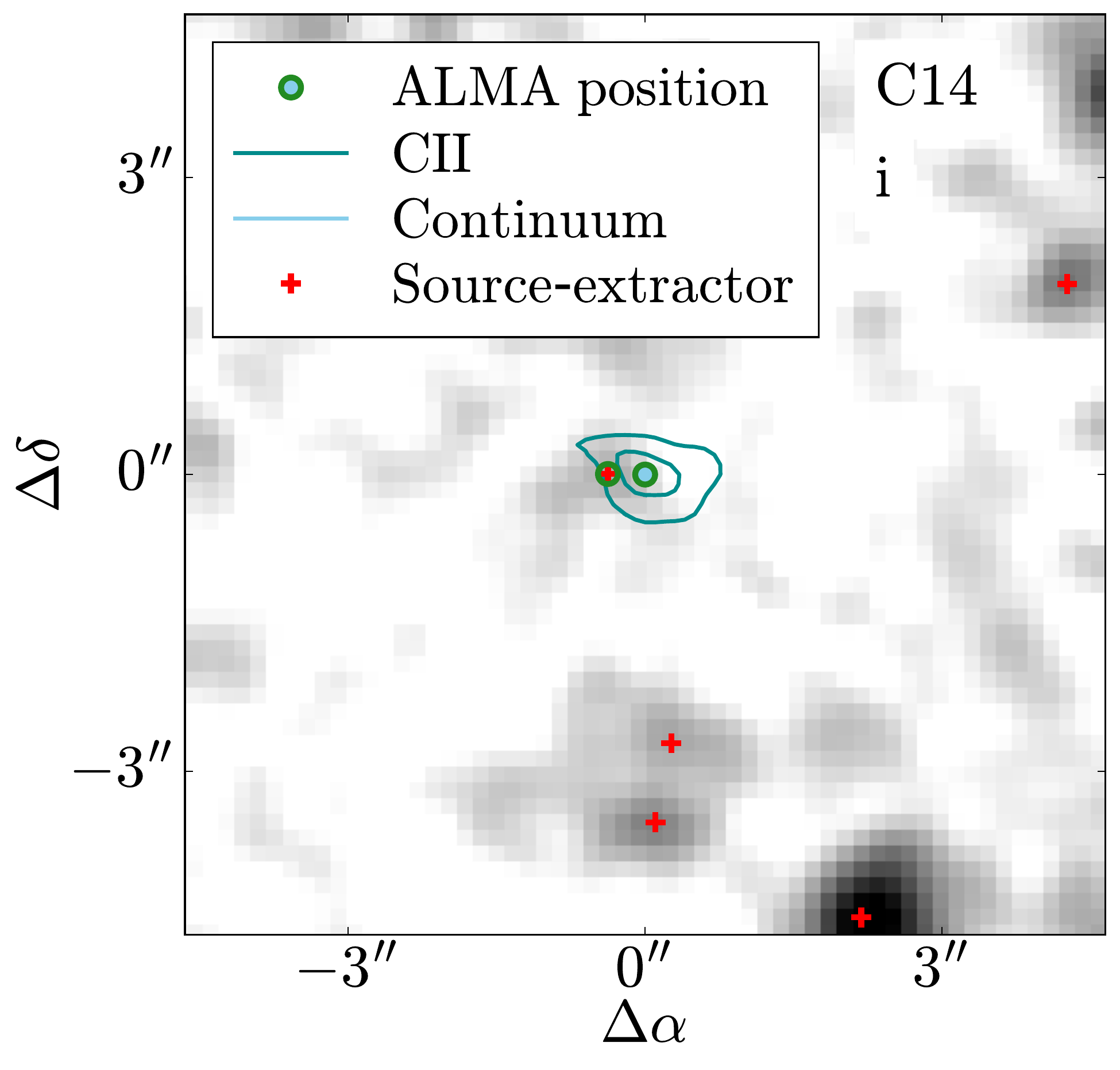}
\includegraphics[width=0.24\textwidth]{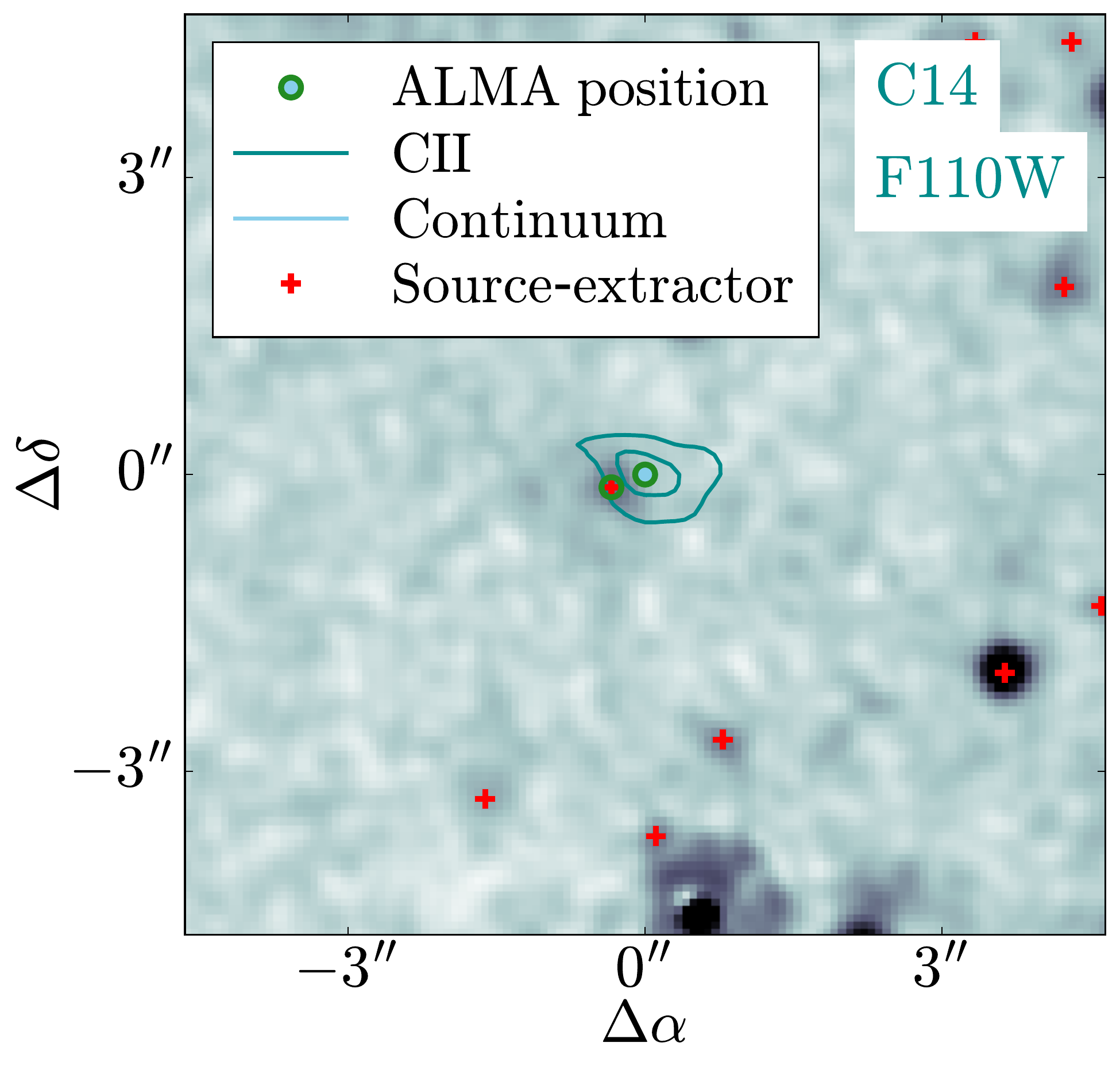}
\includegraphics[width=0.24\textwidth]{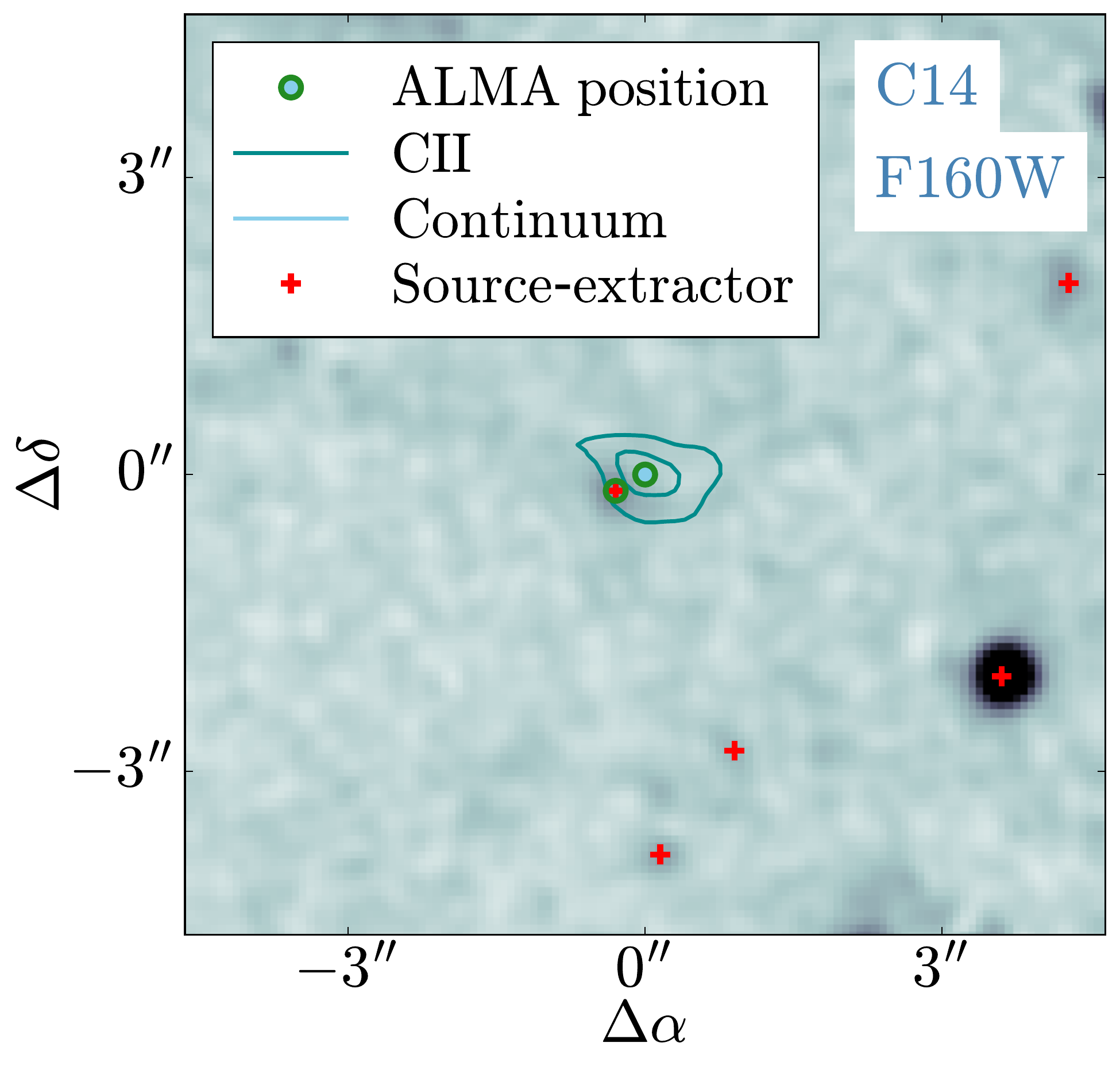}
\includegraphics[width=0.248\textwidth]{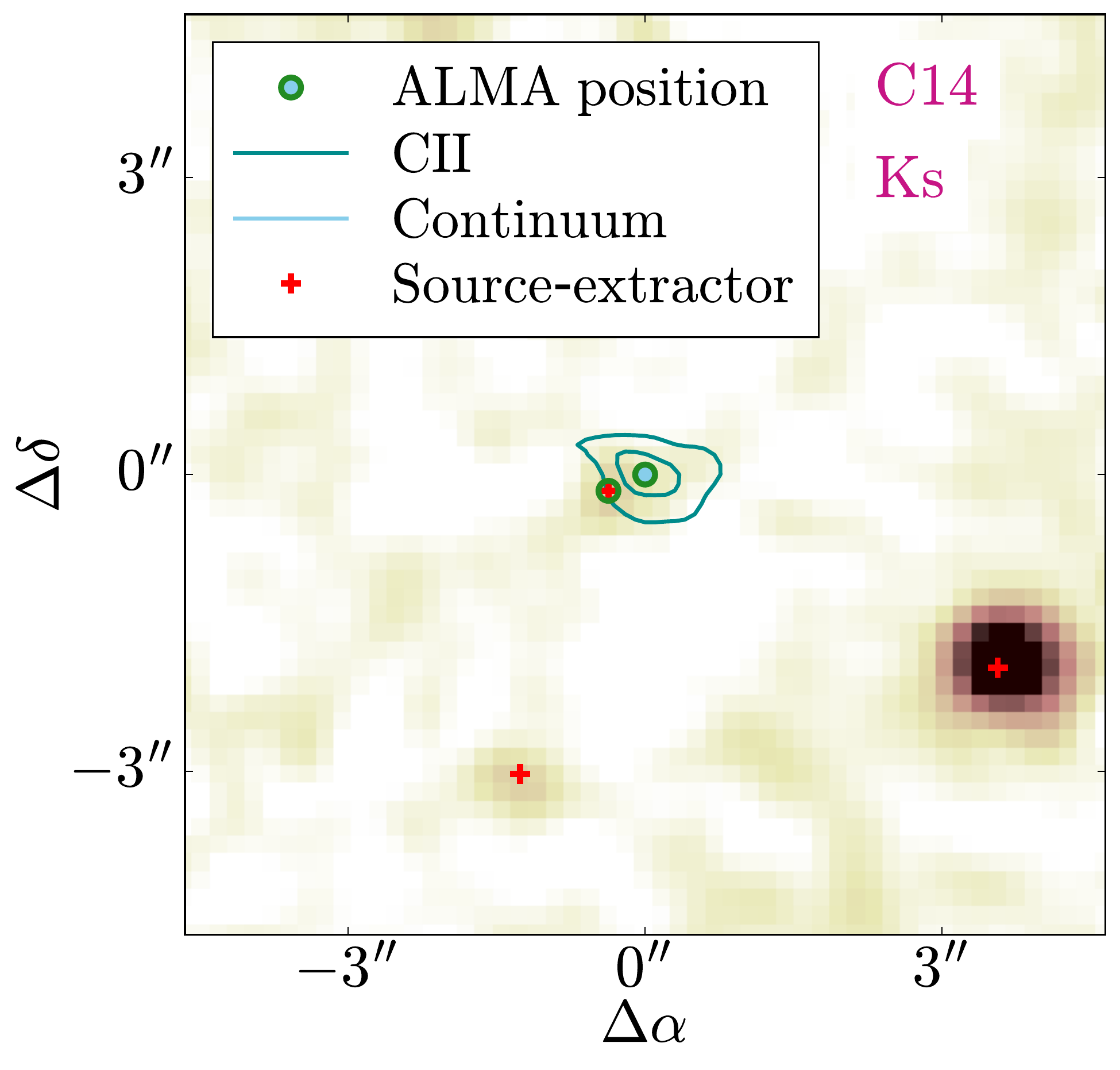}
\includegraphics[width=0.249\textwidth]{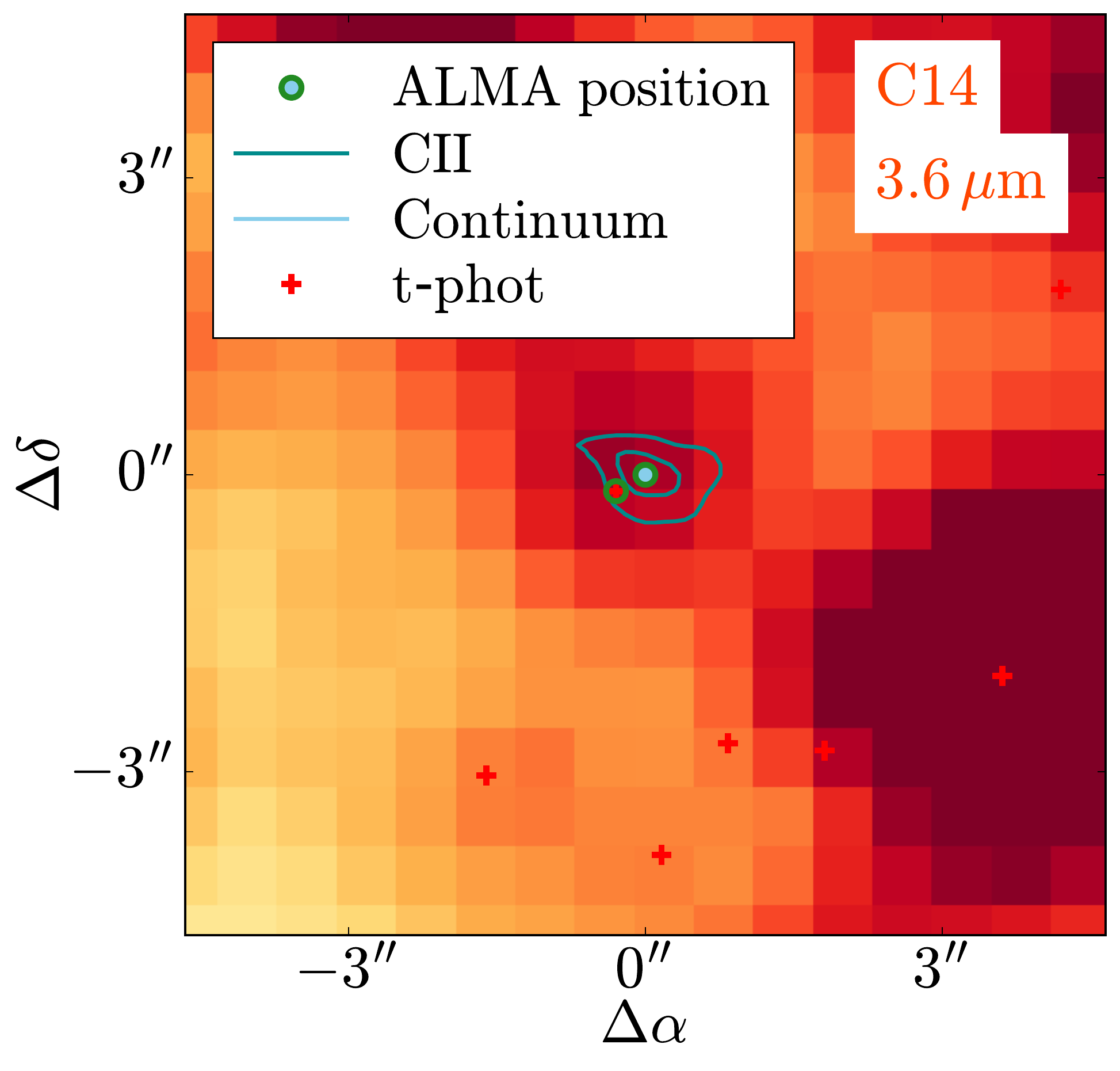}
\includegraphics[width=0.249\textwidth]{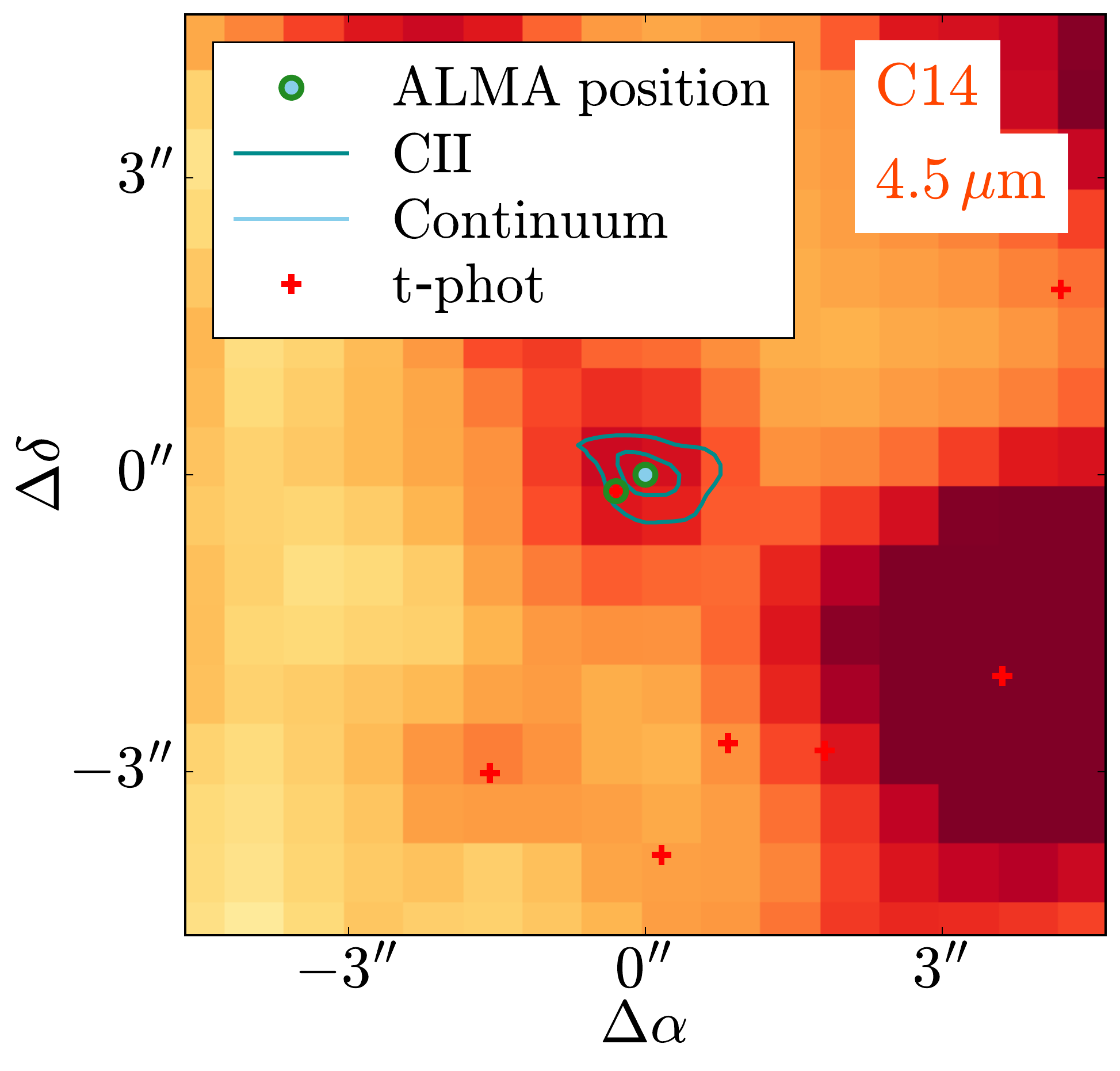}
\end{framed}
\end{subfigure}
\begin{subfigure}{0.85\textwidth}
\begin{framed}
\includegraphics[width=0.24\textwidth]{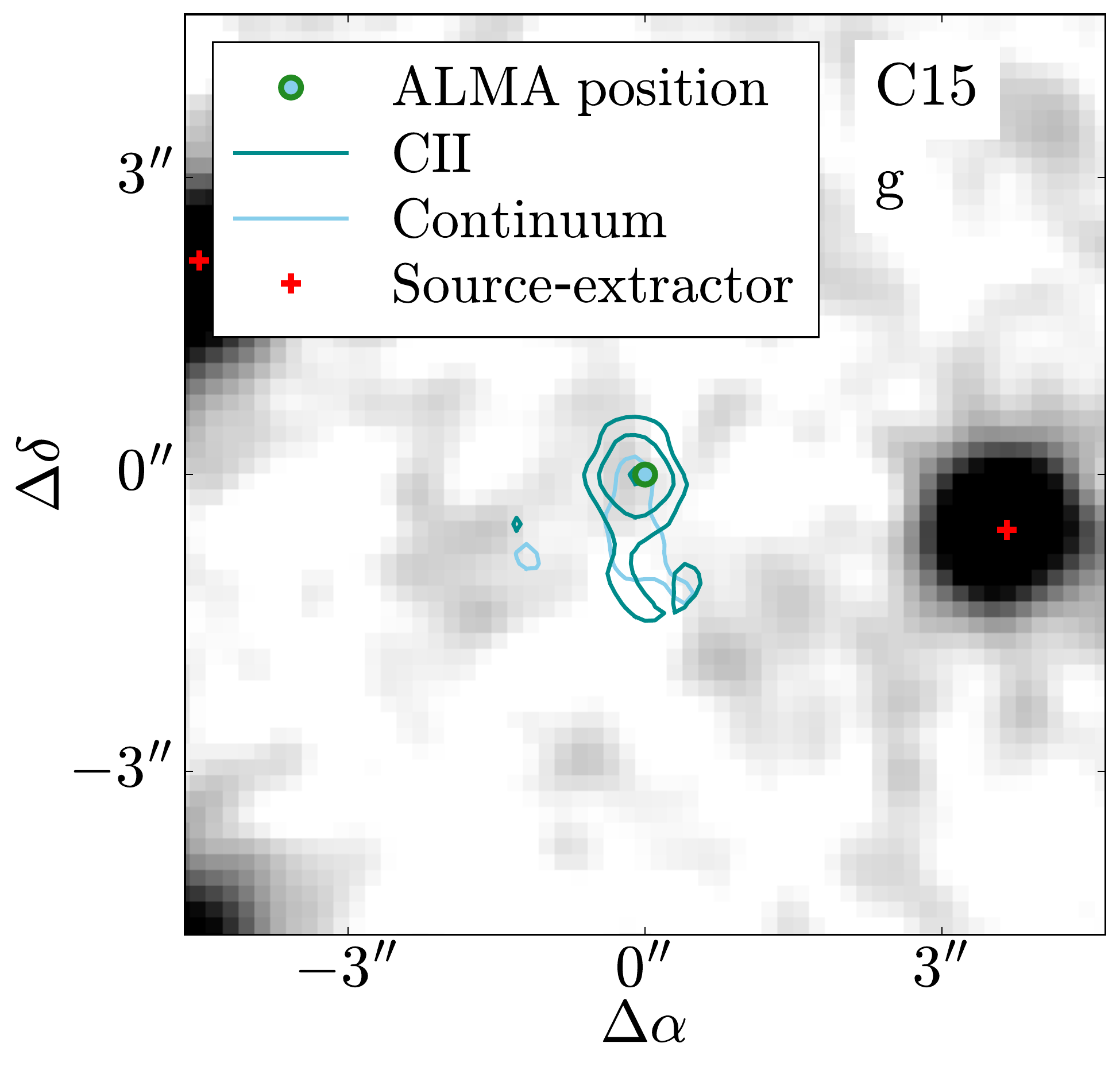}
\includegraphics[width=0.24\textwidth]{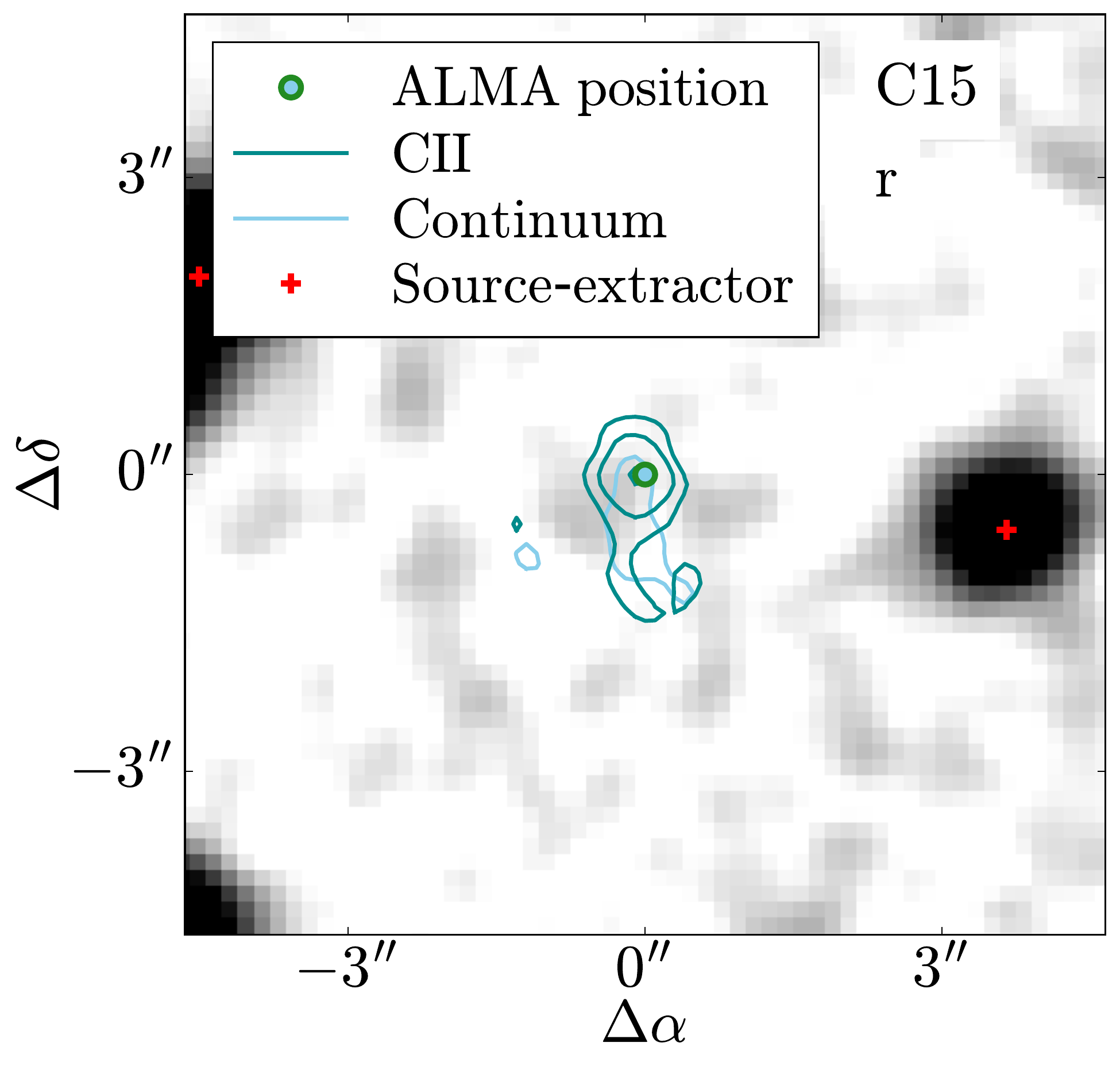}
\includegraphics[width=0.24\textwidth]{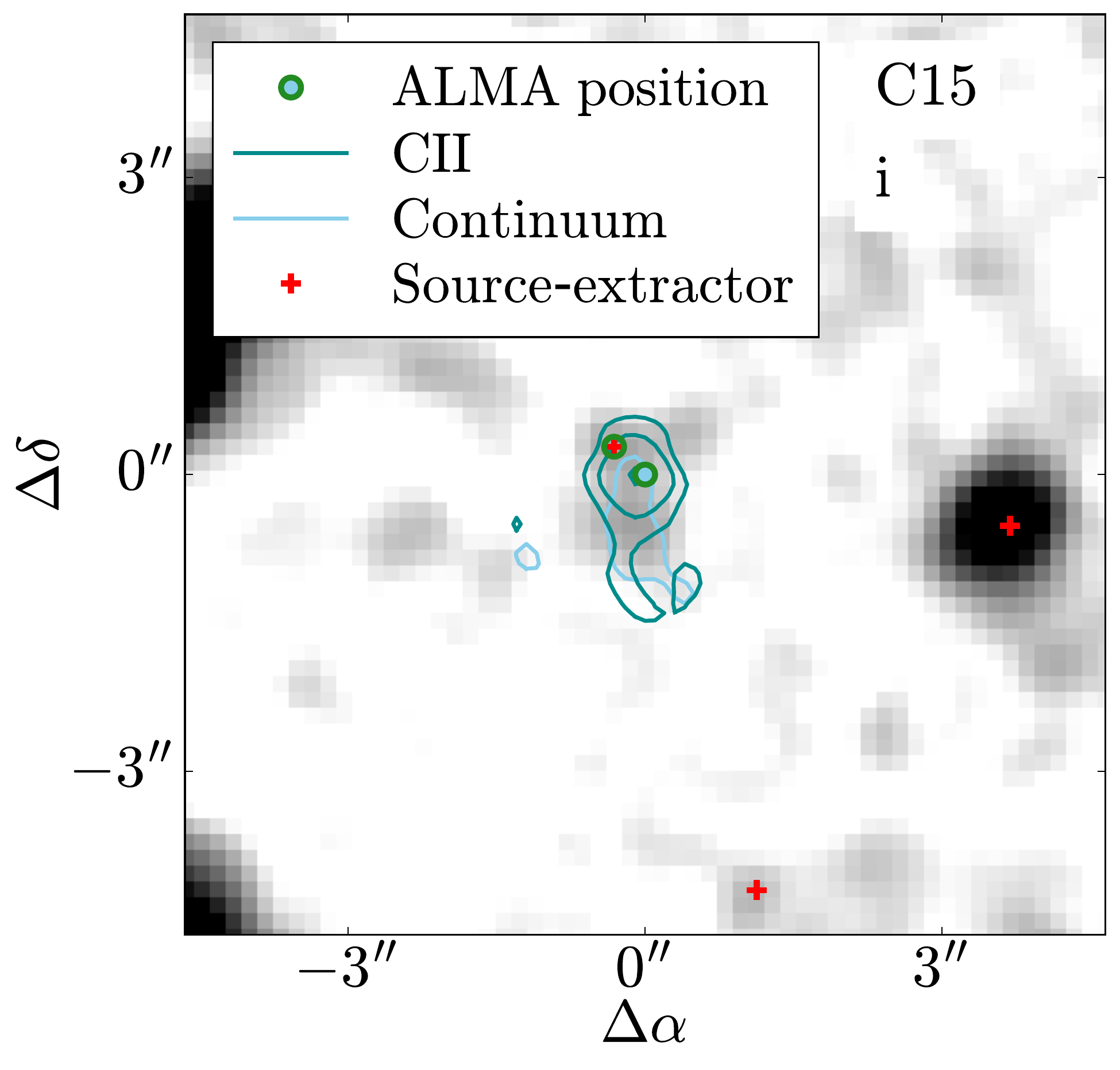}
\includegraphics[width=0.24\textwidth]{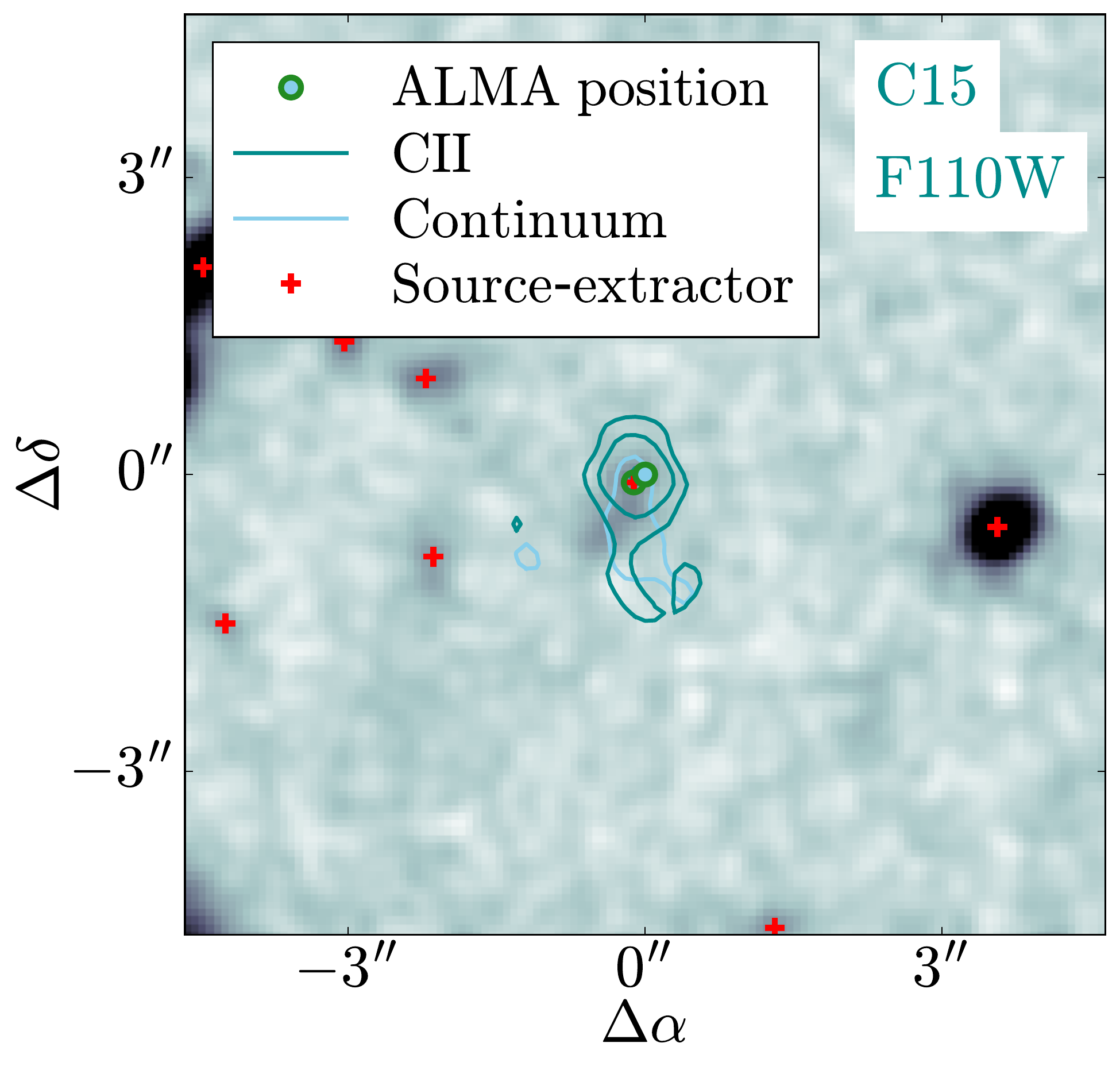}
\includegraphics[width=0.24\textwidth]{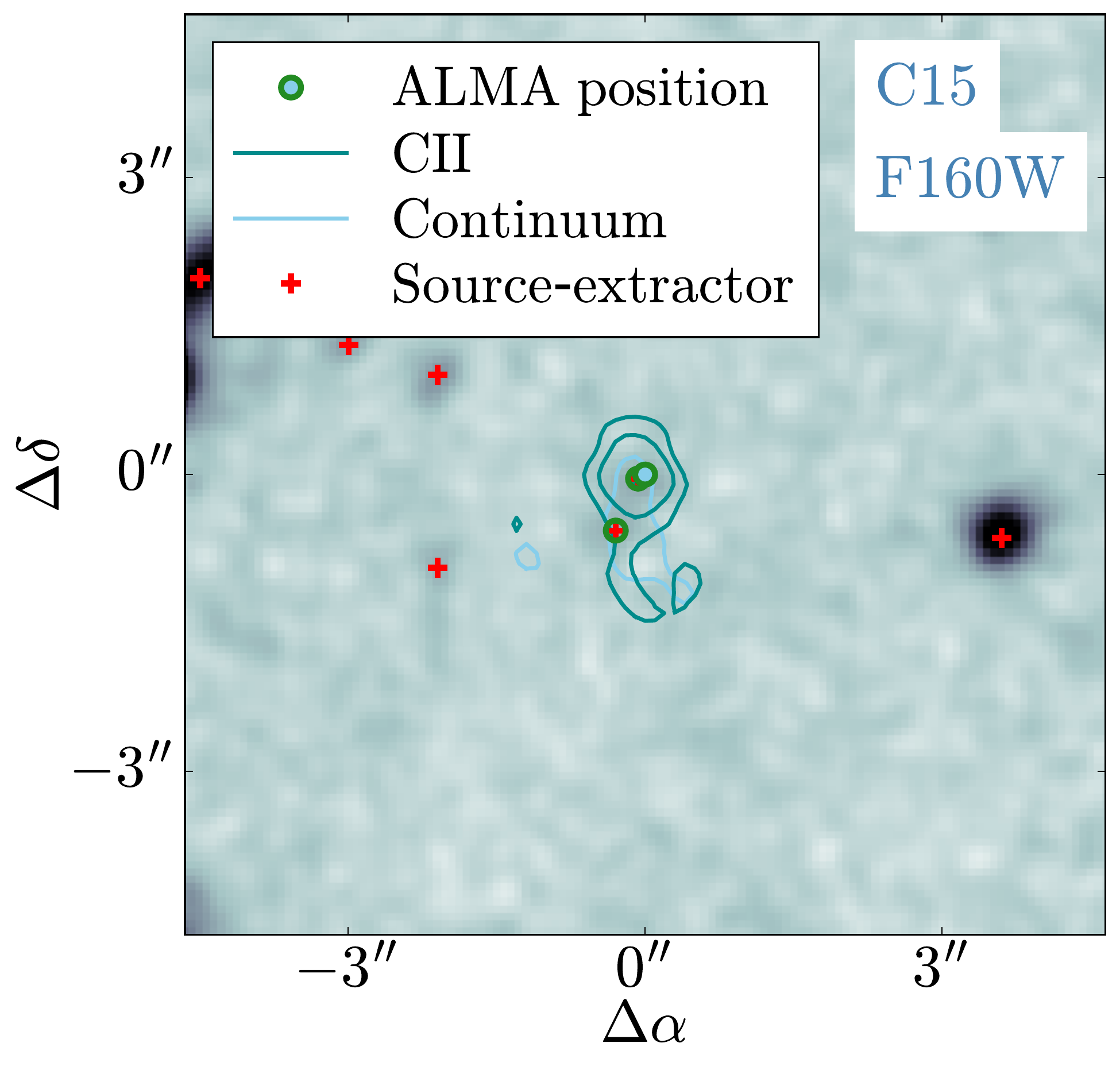}
\includegraphics[width=0.248\textwidth]{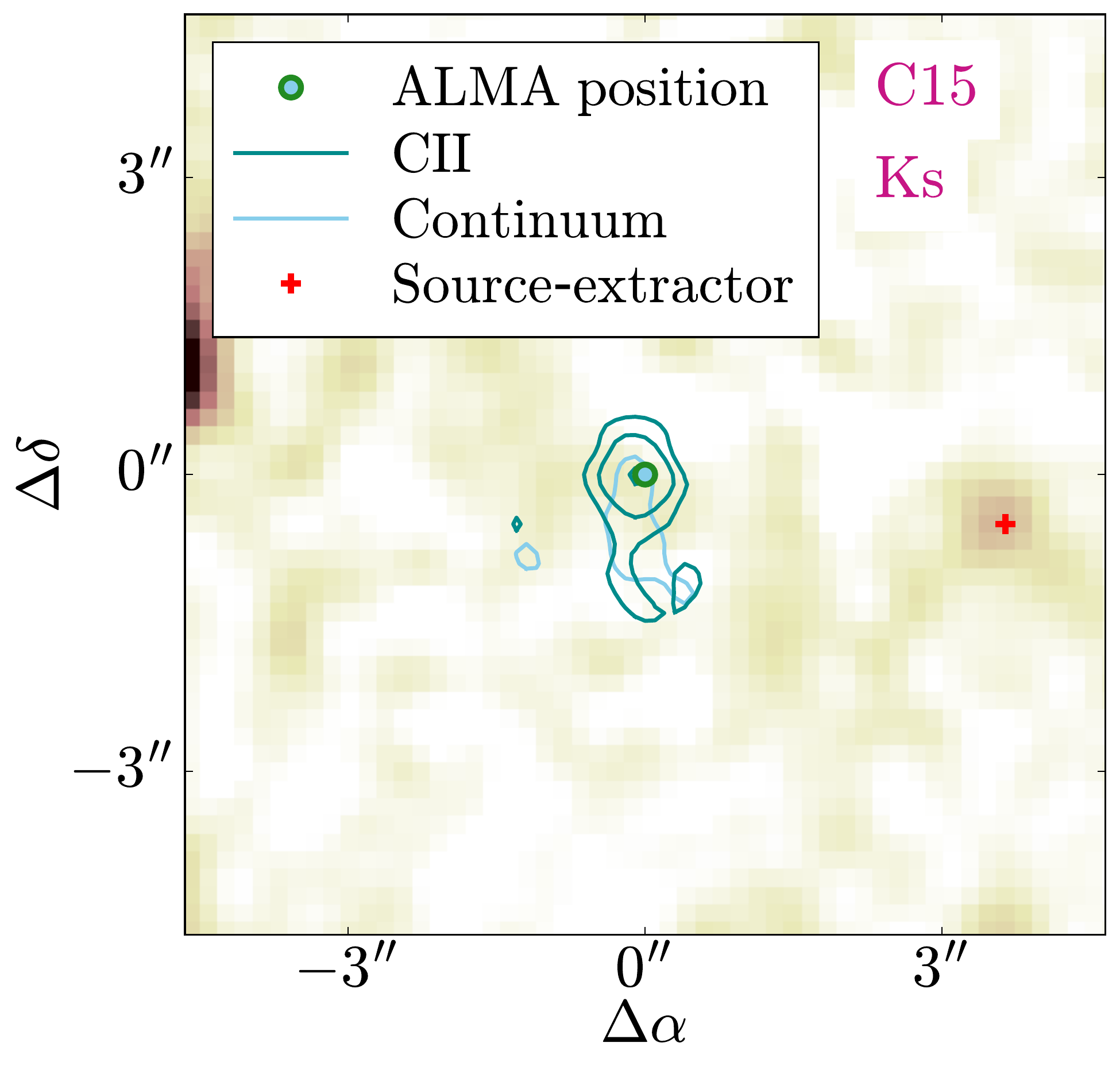}
\includegraphics[width=0.249\textwidth]{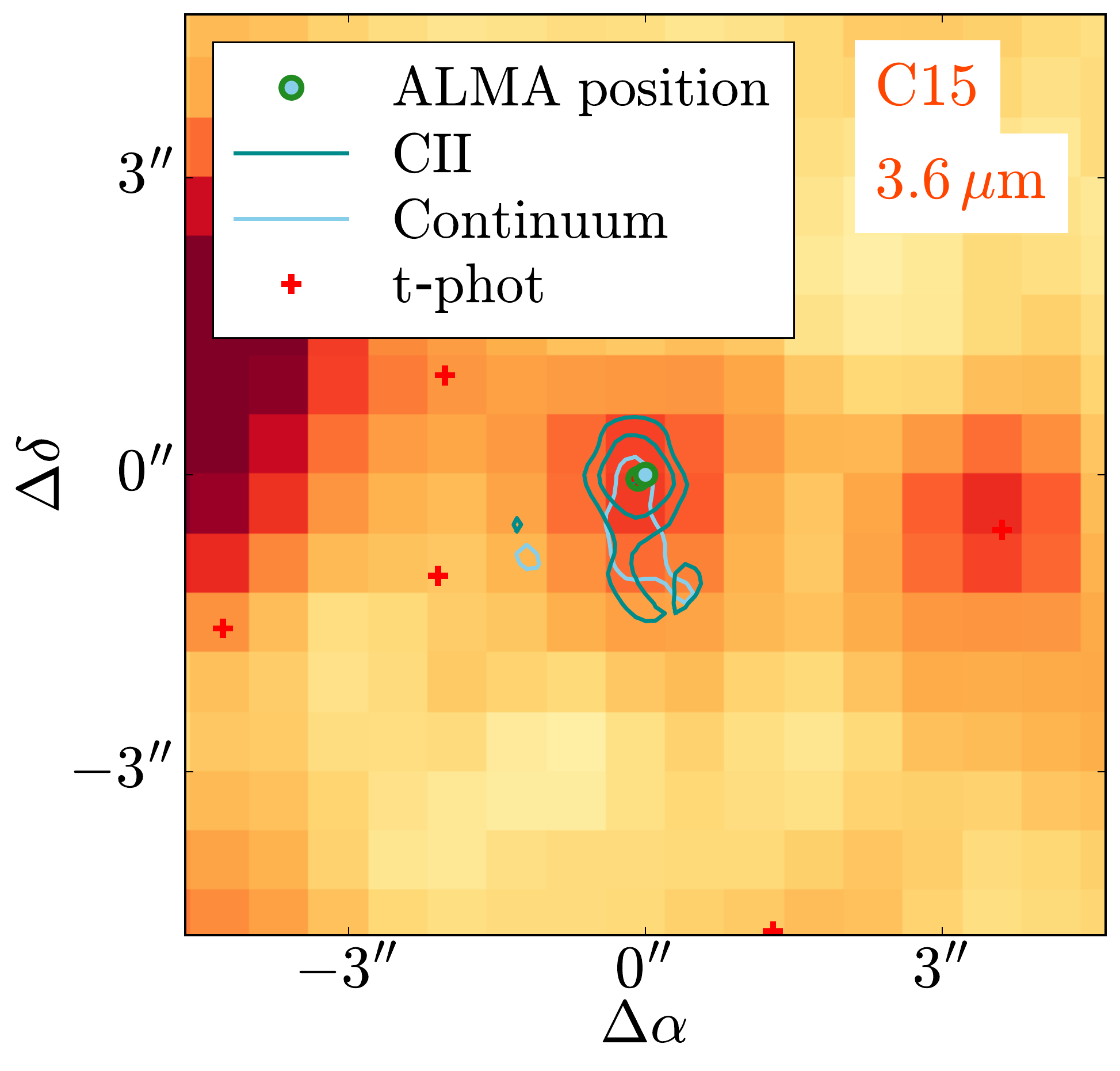}
\includegraphics[width=0.249\textwidth]{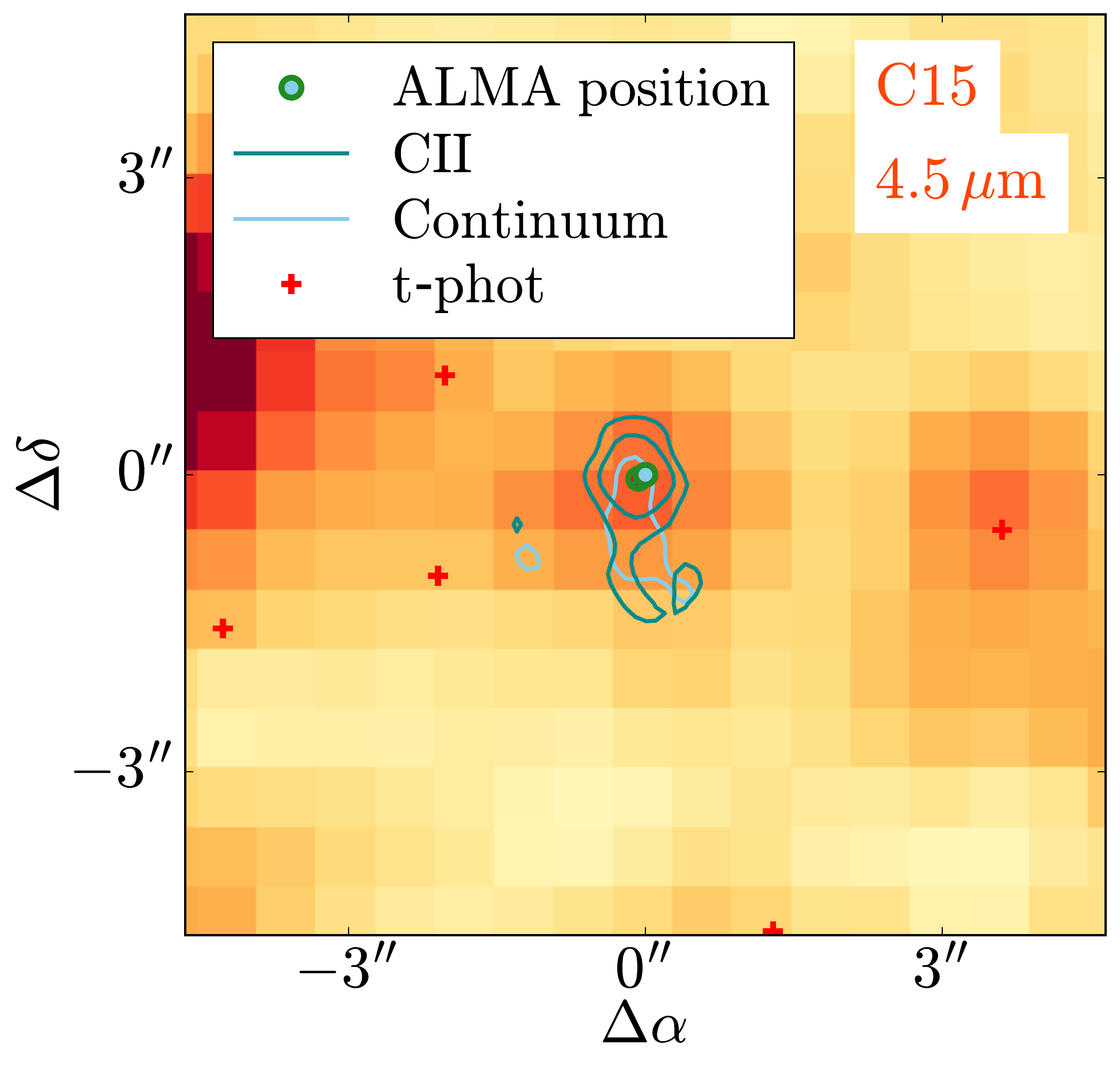}
\end{framed}
\end{subfigure}
\caption{}
\end{figure*}
\renewcommand{\thefigure}{\arabic{figure}}

\renewcommand{\thefigure}{B\arabic{figure} (Cont.)}
\addtocounter{figure}{-1}
\begin{figure*}
\begin{subfigure}{0.85\textwidth}
\begin{framed}
\includegraphics[width=0.24\textwidth]{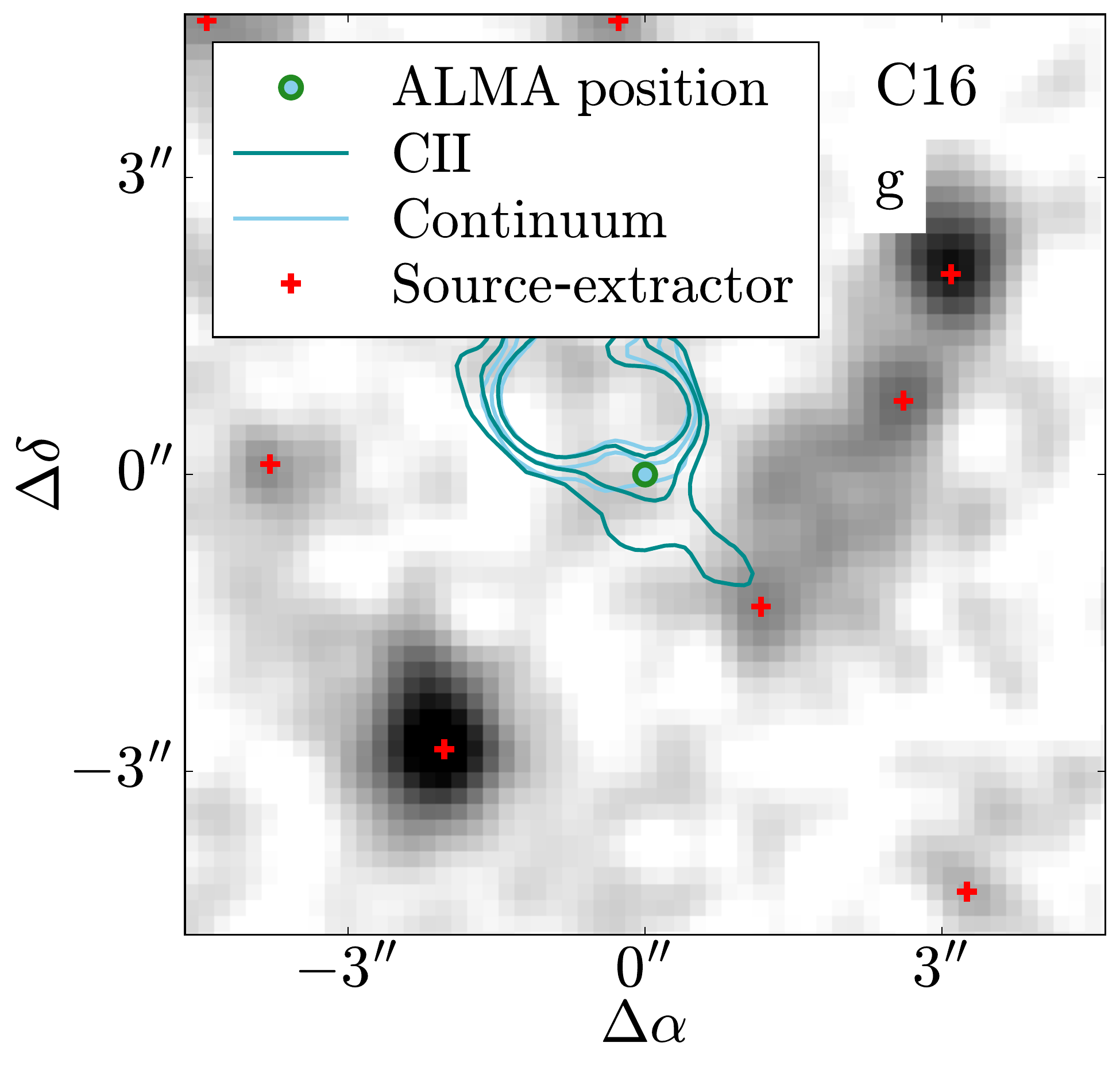}
\includegraphics[width=0.24\textwidth]{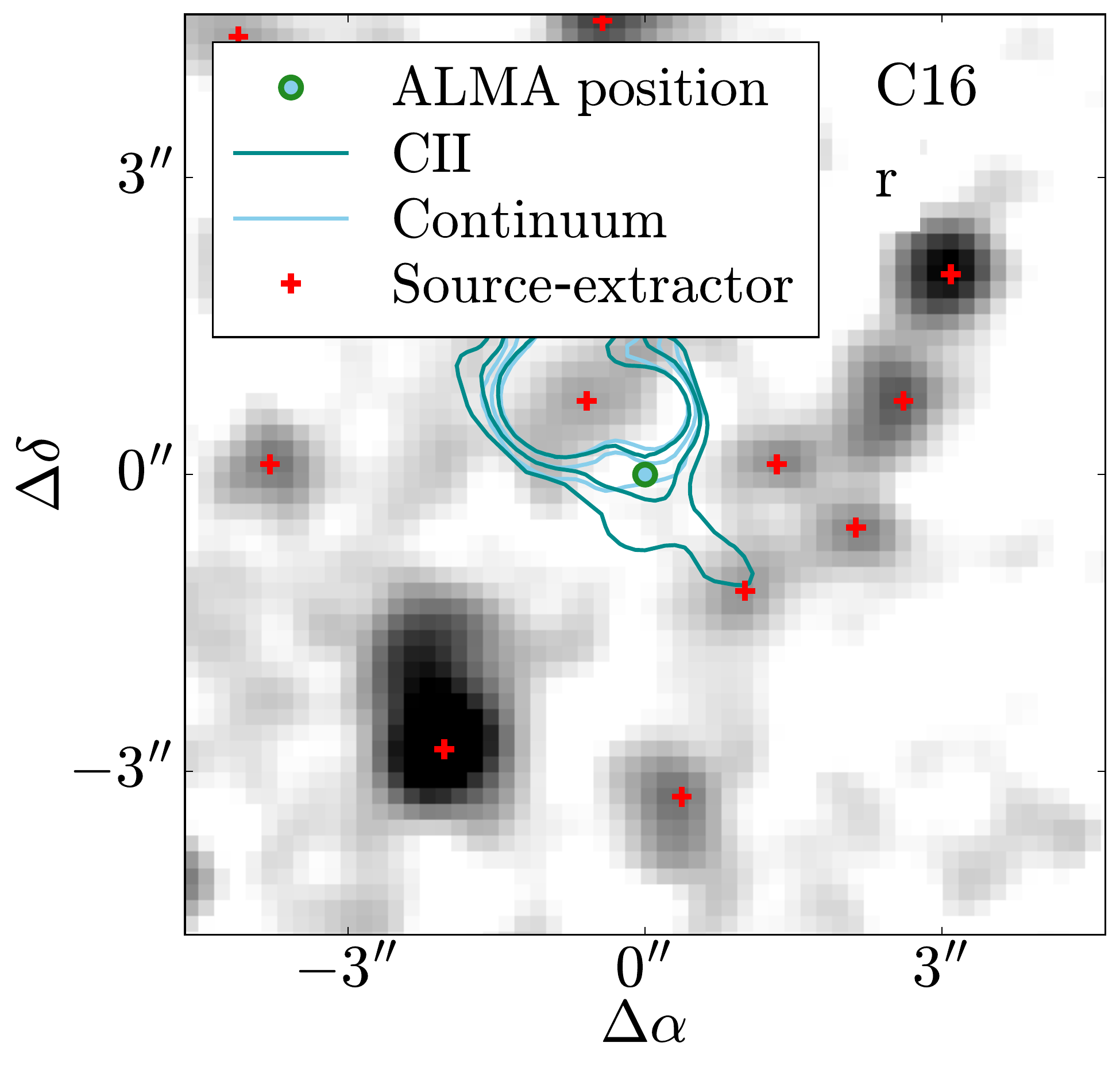}
\includegraphics[width=0.24\textwidth]{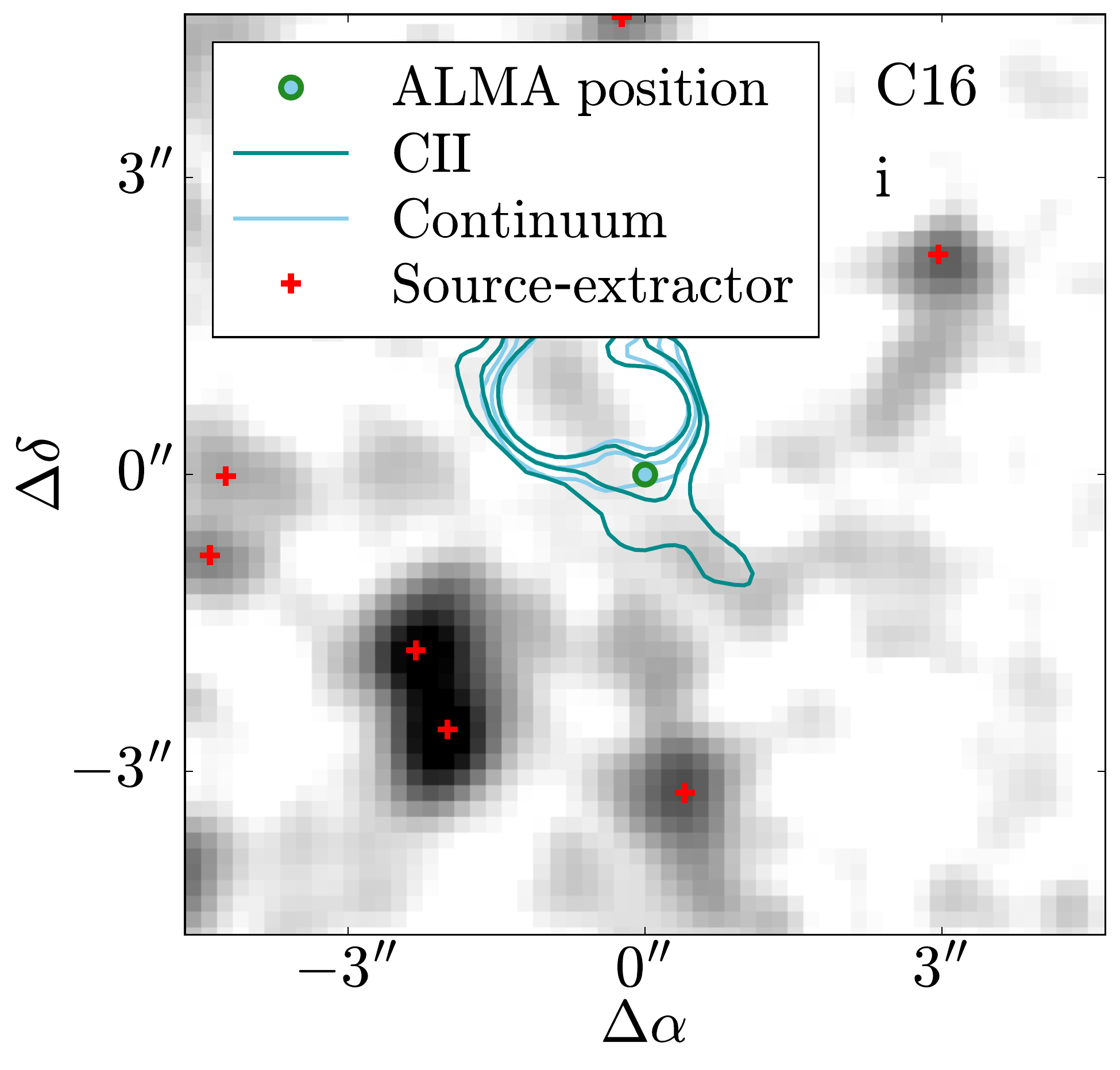}
\includegraphics[width=0.24\textwidth]{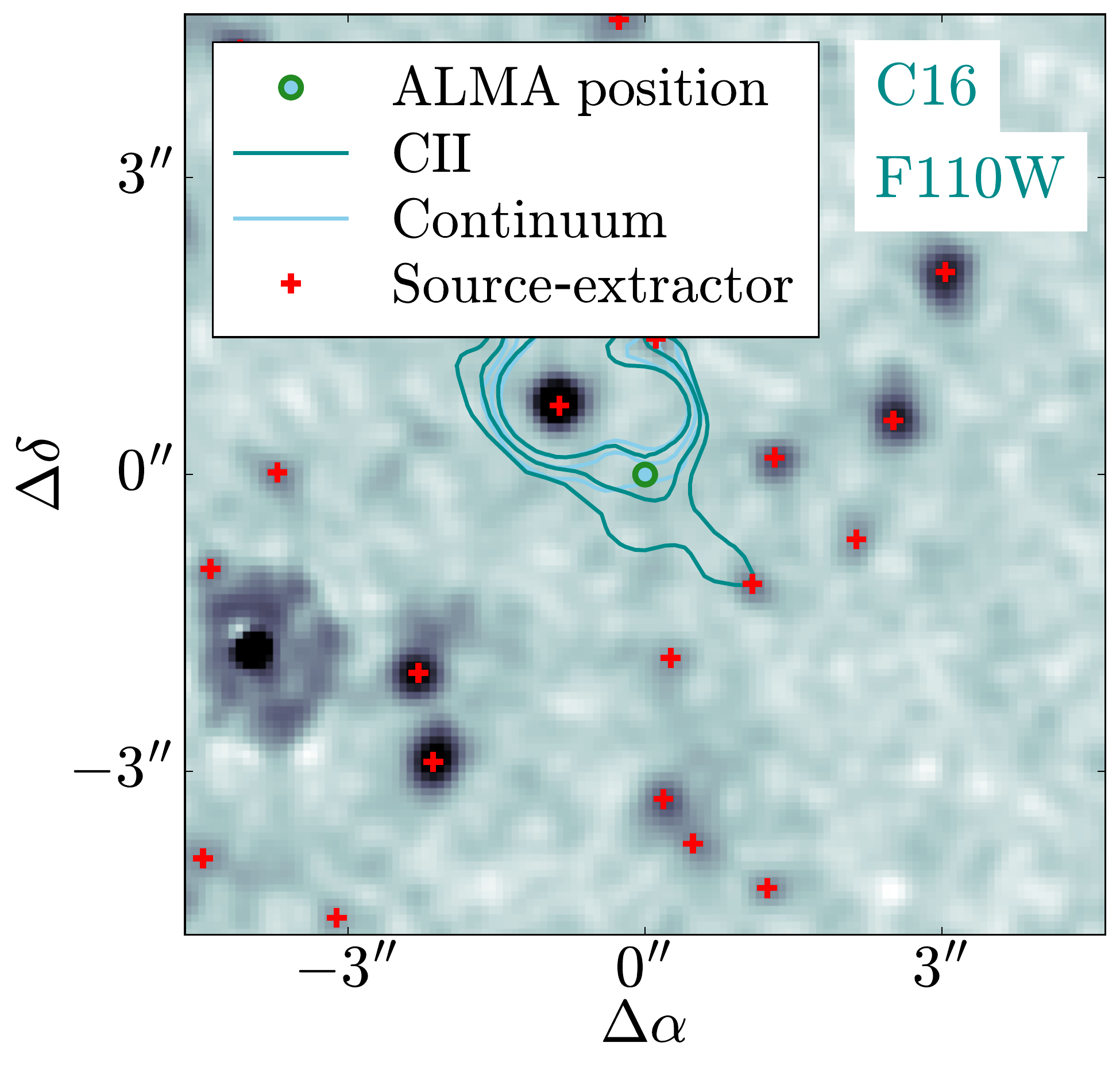}
\includegraphics[width=0.24\textwidth]{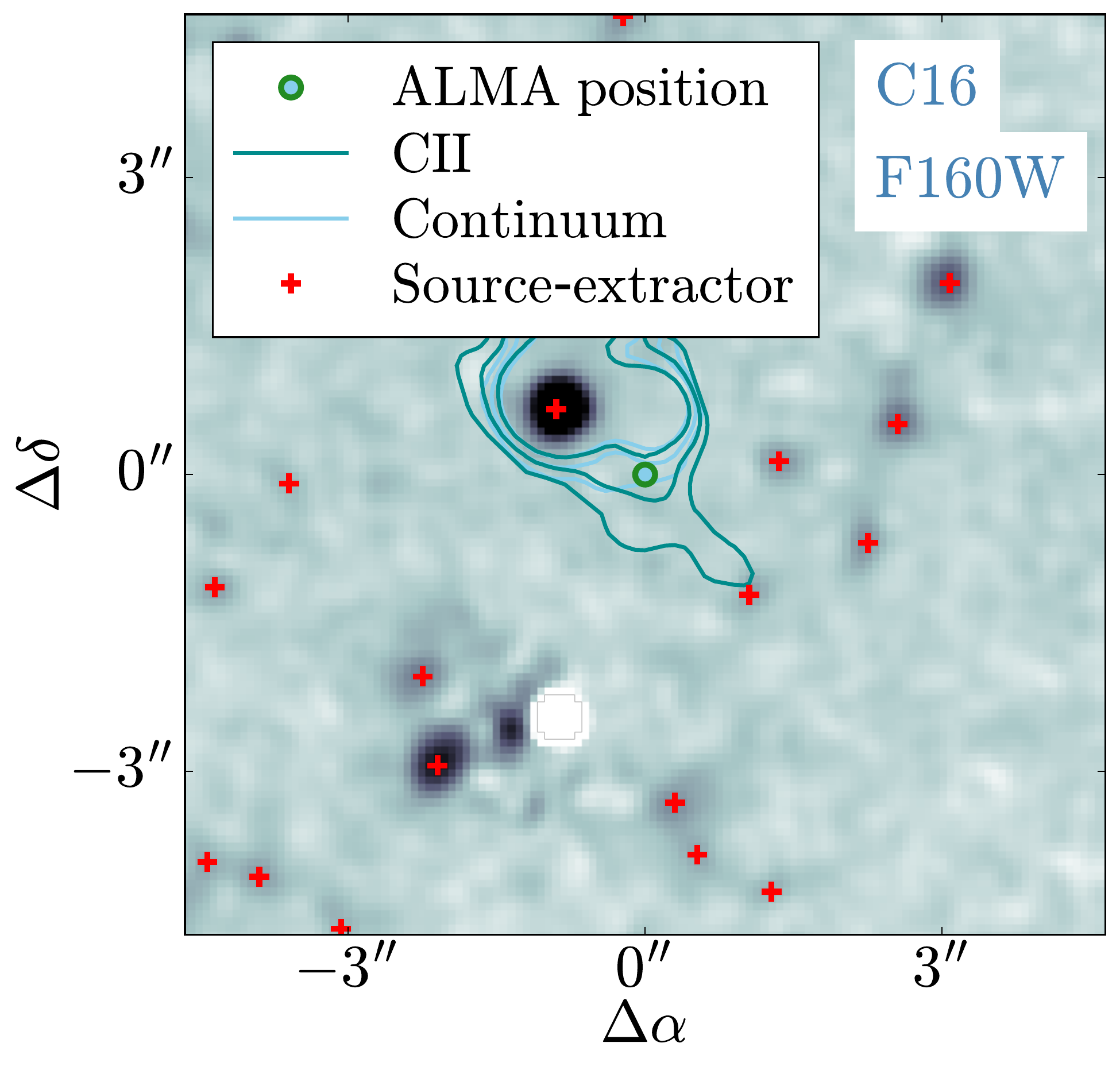}
\includegraphics[width=0.248\textwidth]{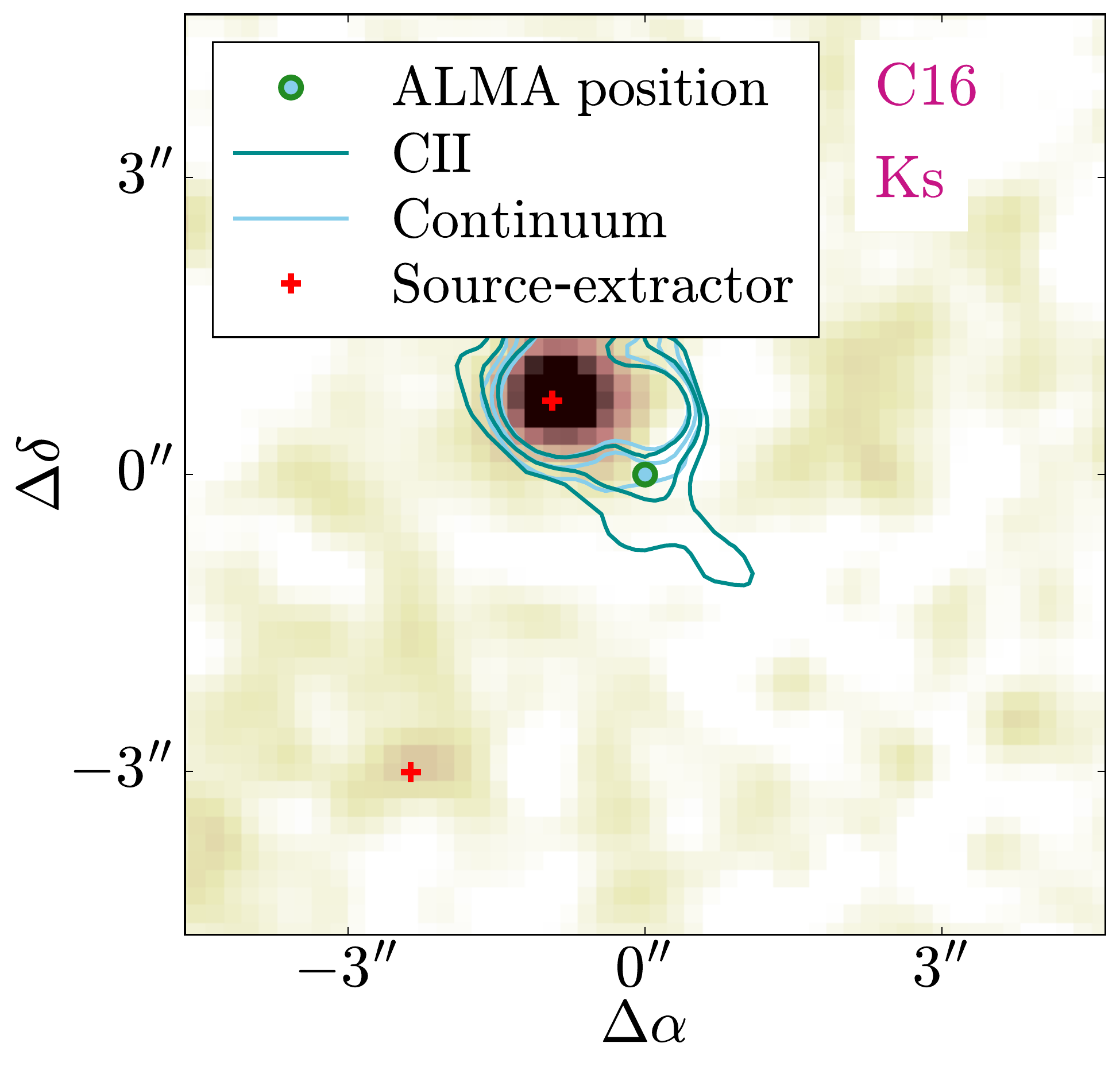}
\includegraphics[width=0.249\textwidth]{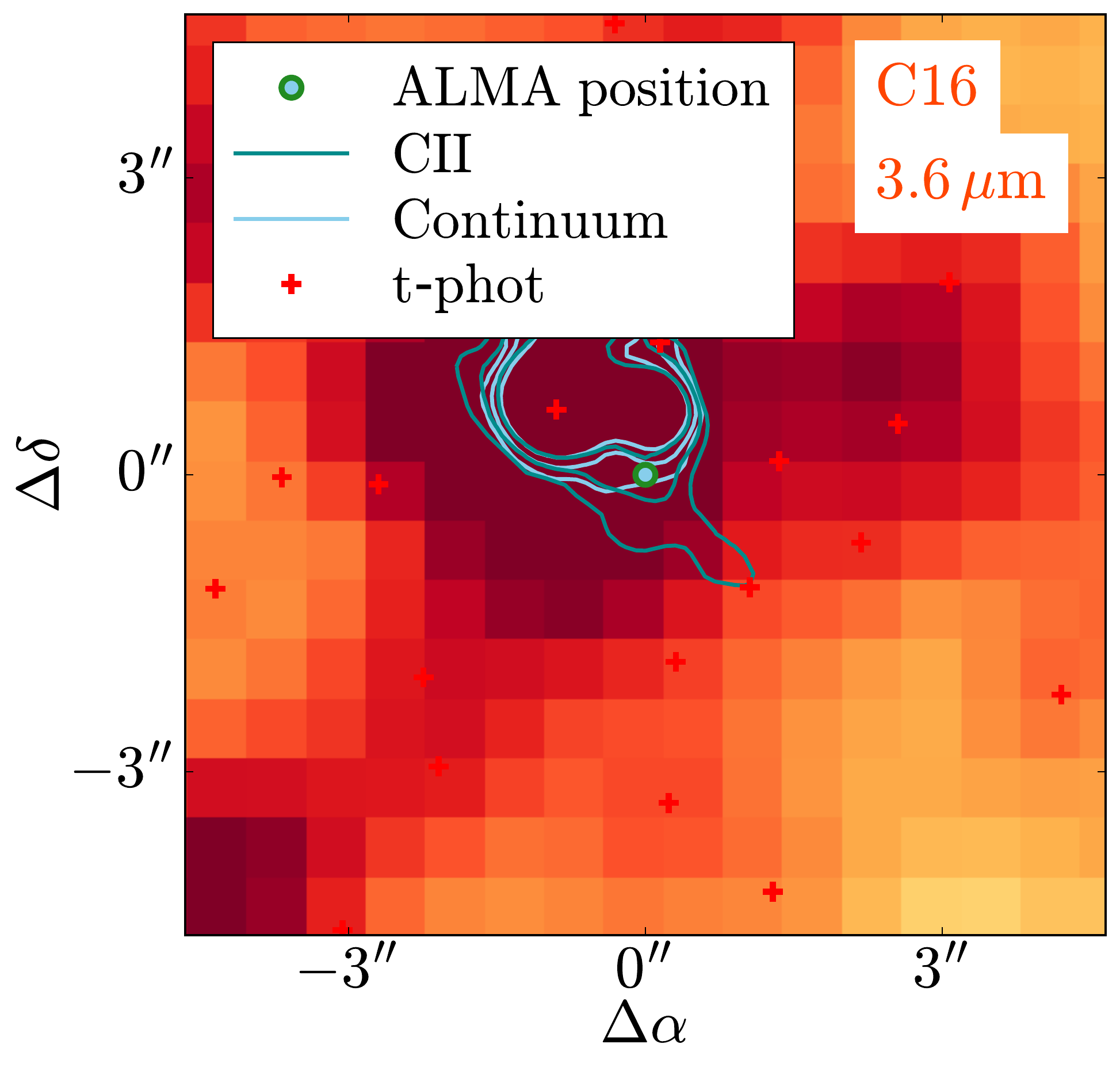}
\includegraphics[width=0.249\textwidth]{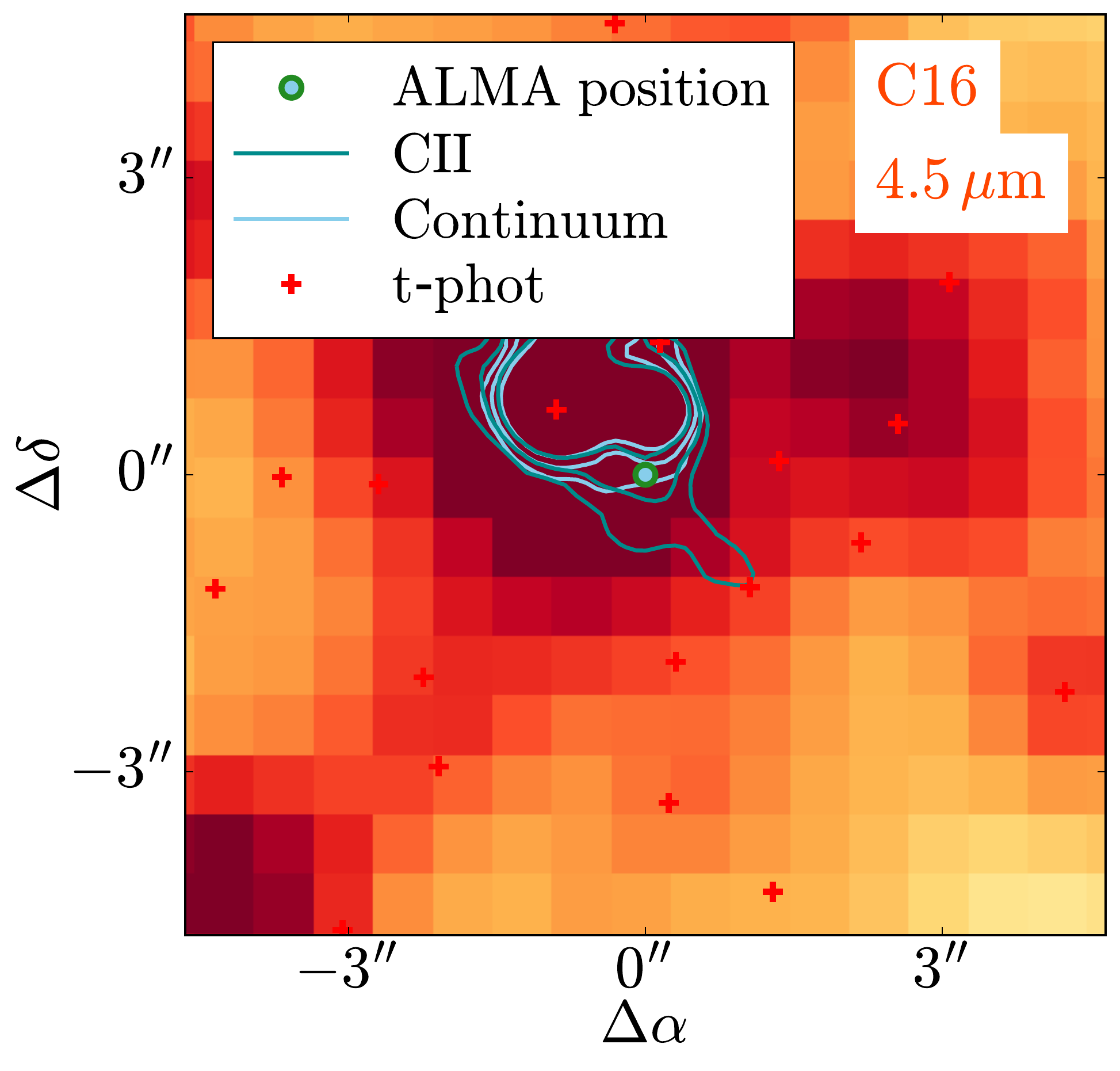}
\end{framed}
\end{subfigure}
\begin{subfigure}{0.85\textwidth}
\begin{framed}
\includegraphics[width=0.24\textwidth]{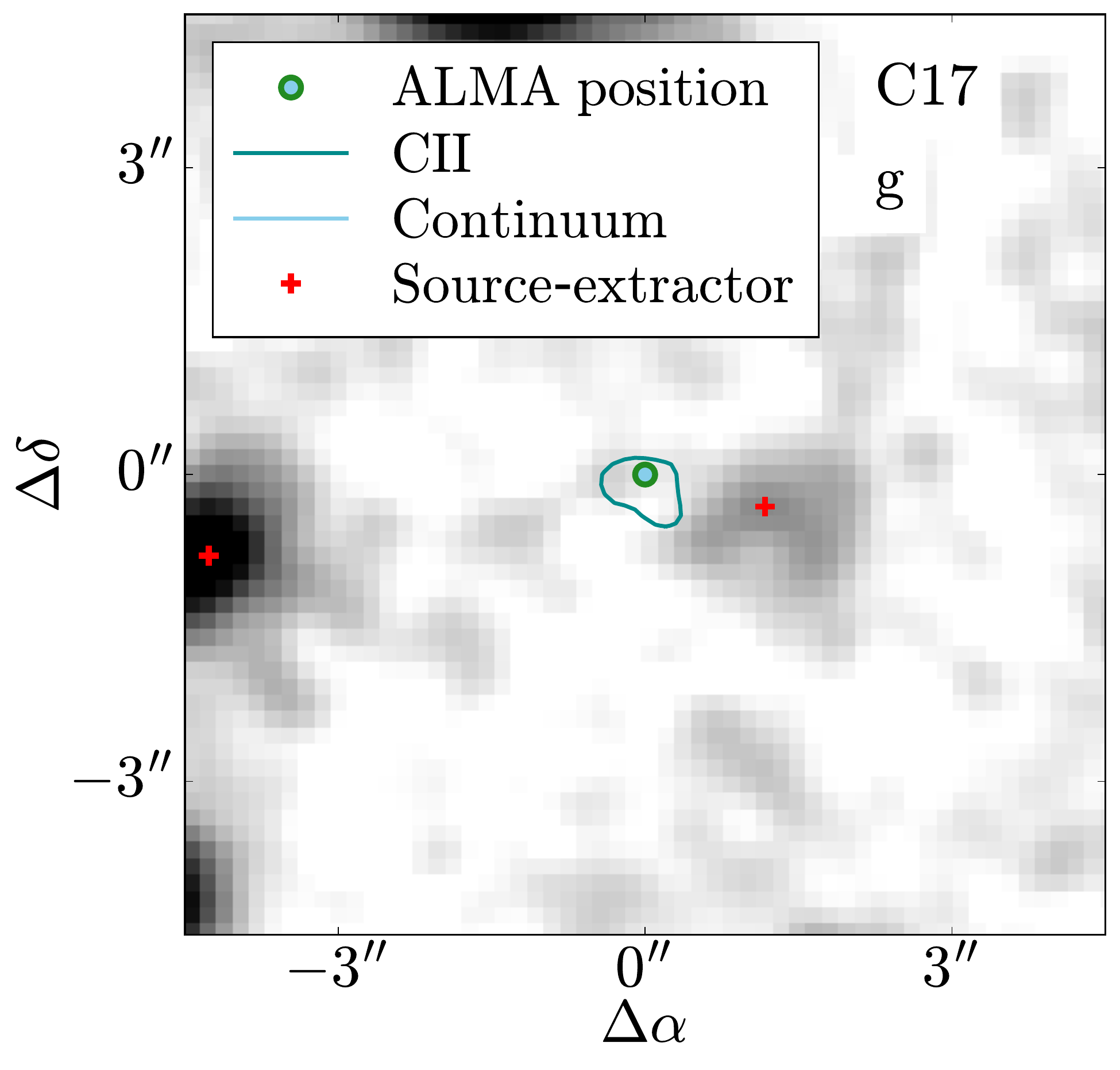}
\includegraphics[width=0.24\textwidth]{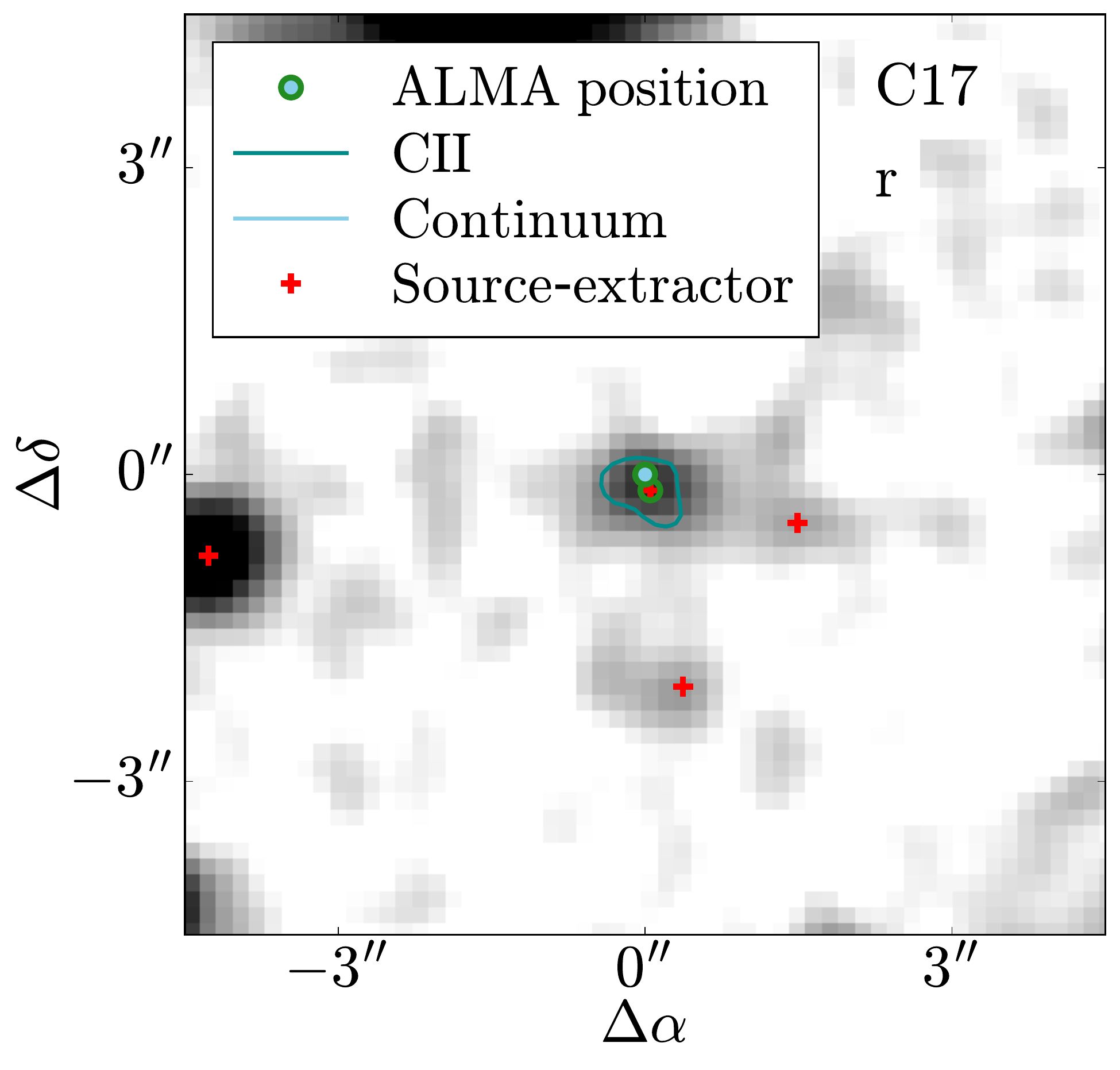}
\includegraphics[width=0.24\textwidth]{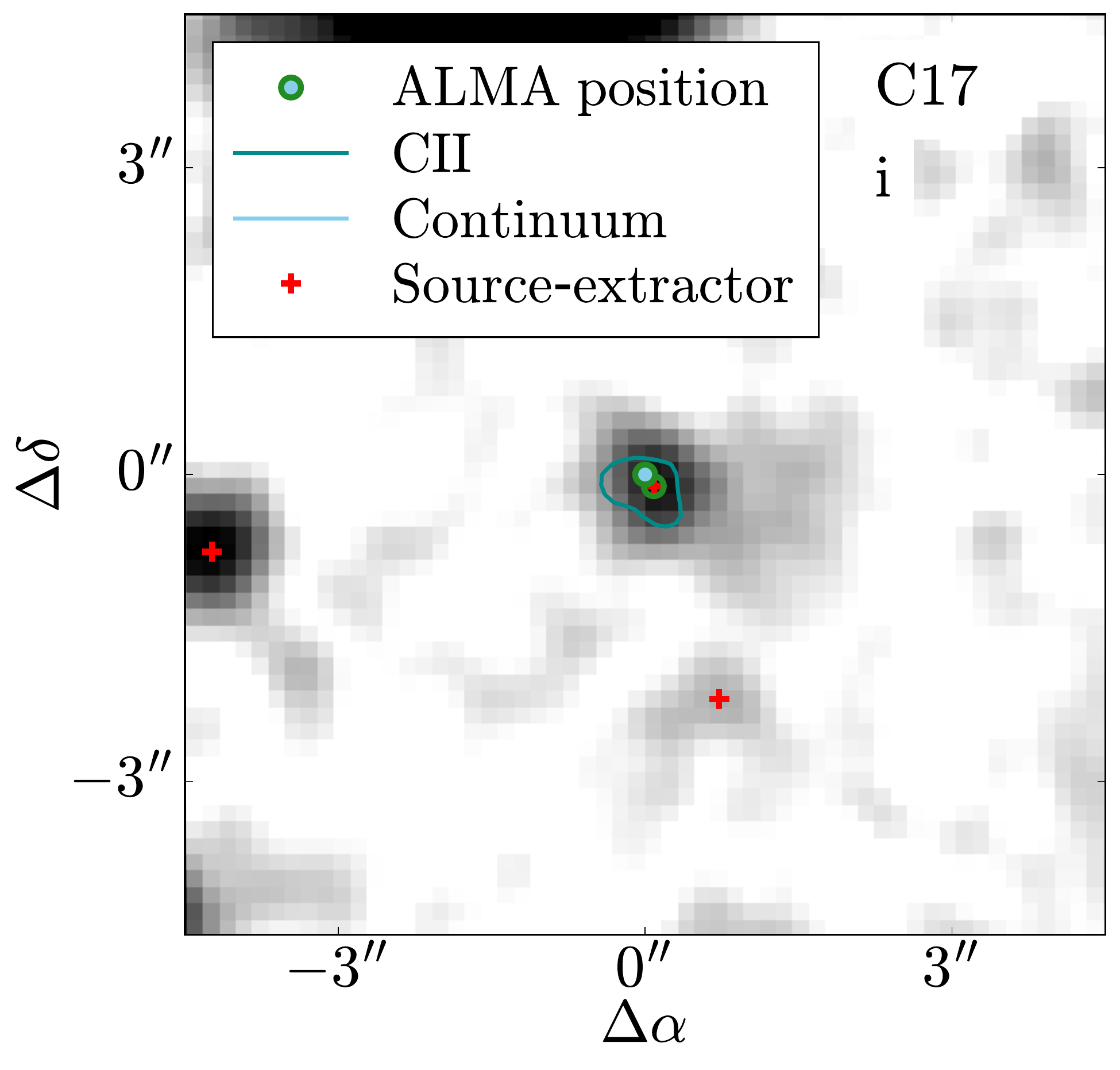}
\includegraphics[width=0.24\textwidth]{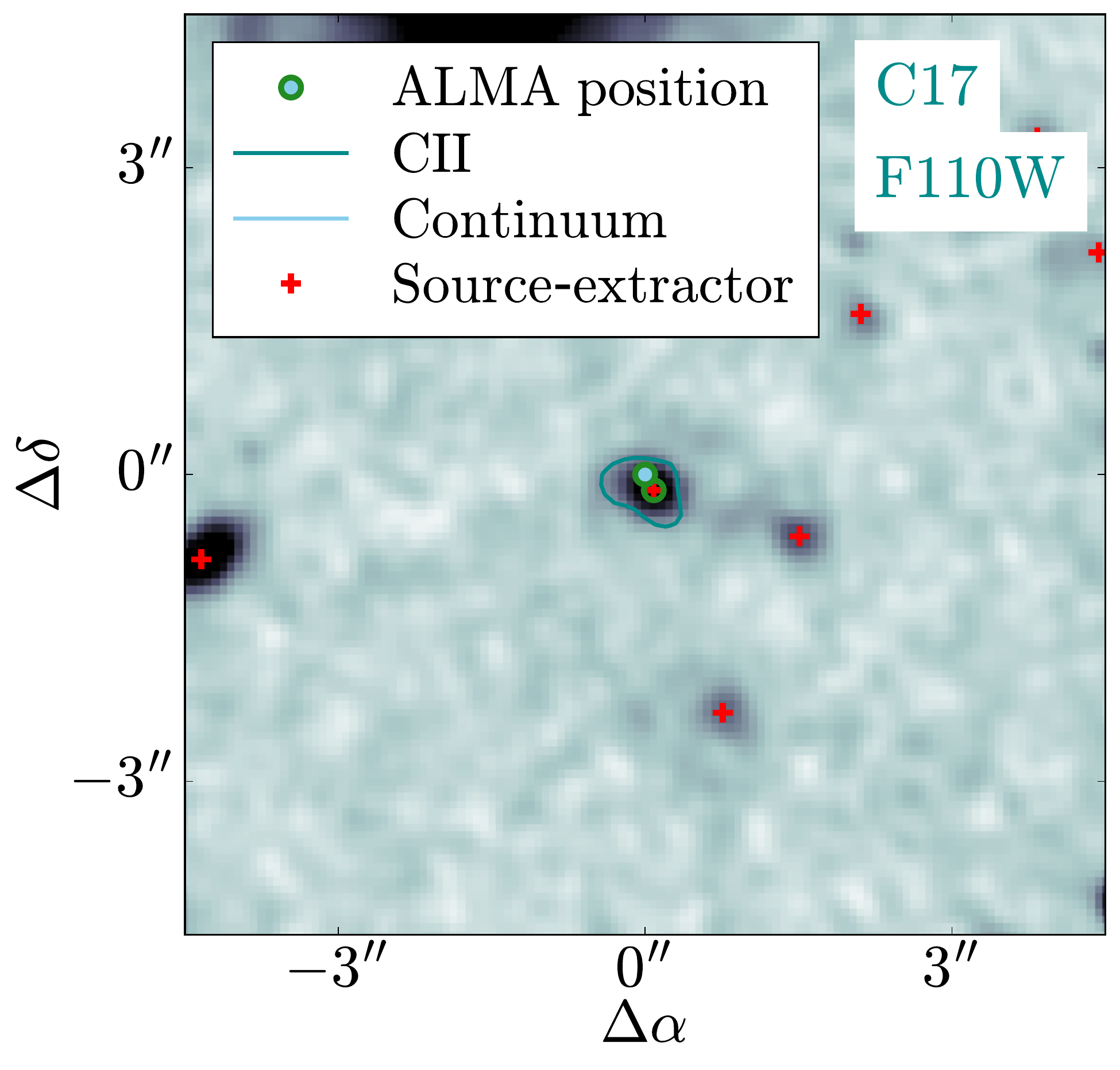}
\includegraphics[width=0.24\textwidth]{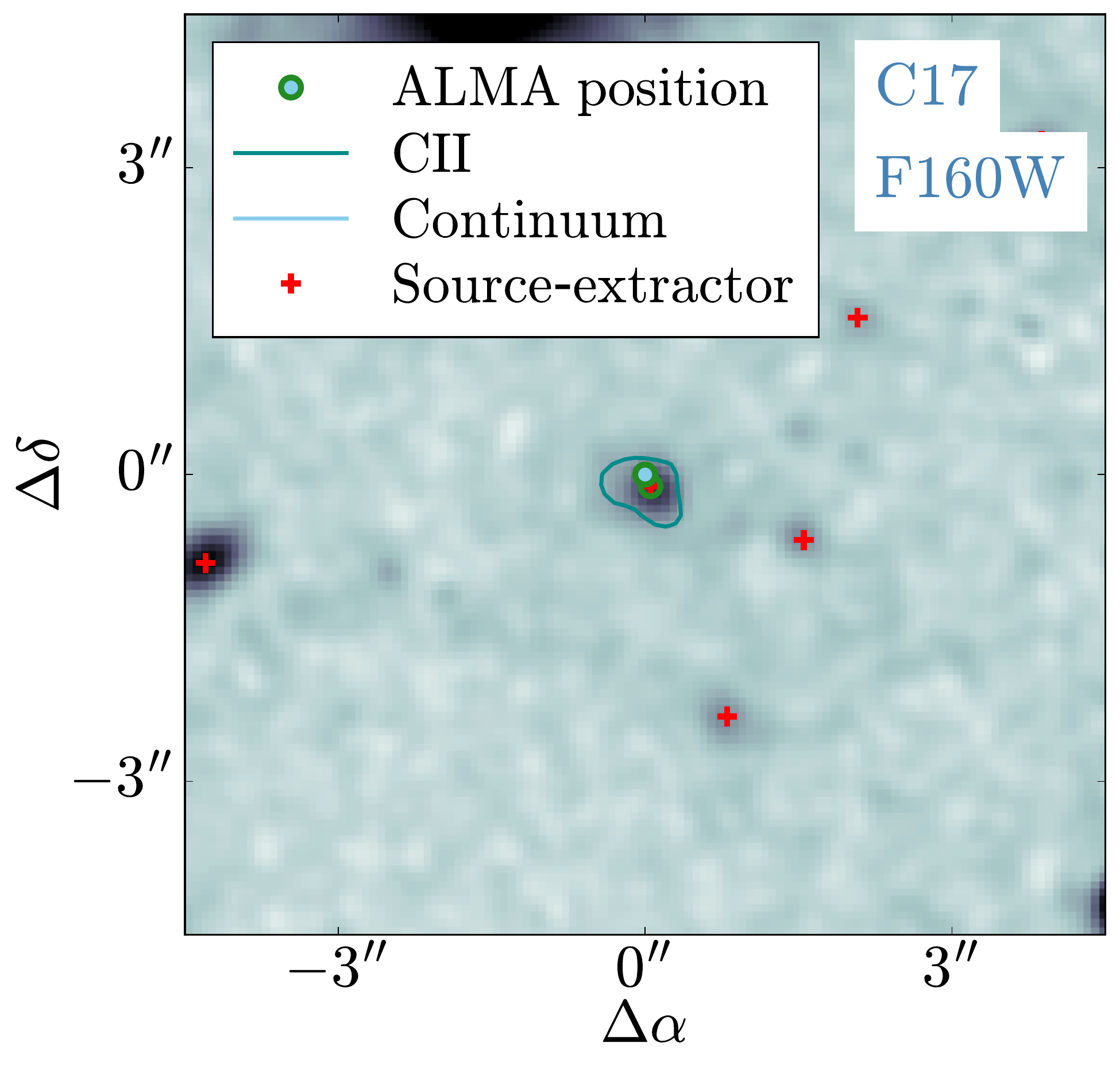}
\includegraphics[width=0.248\textwidth]{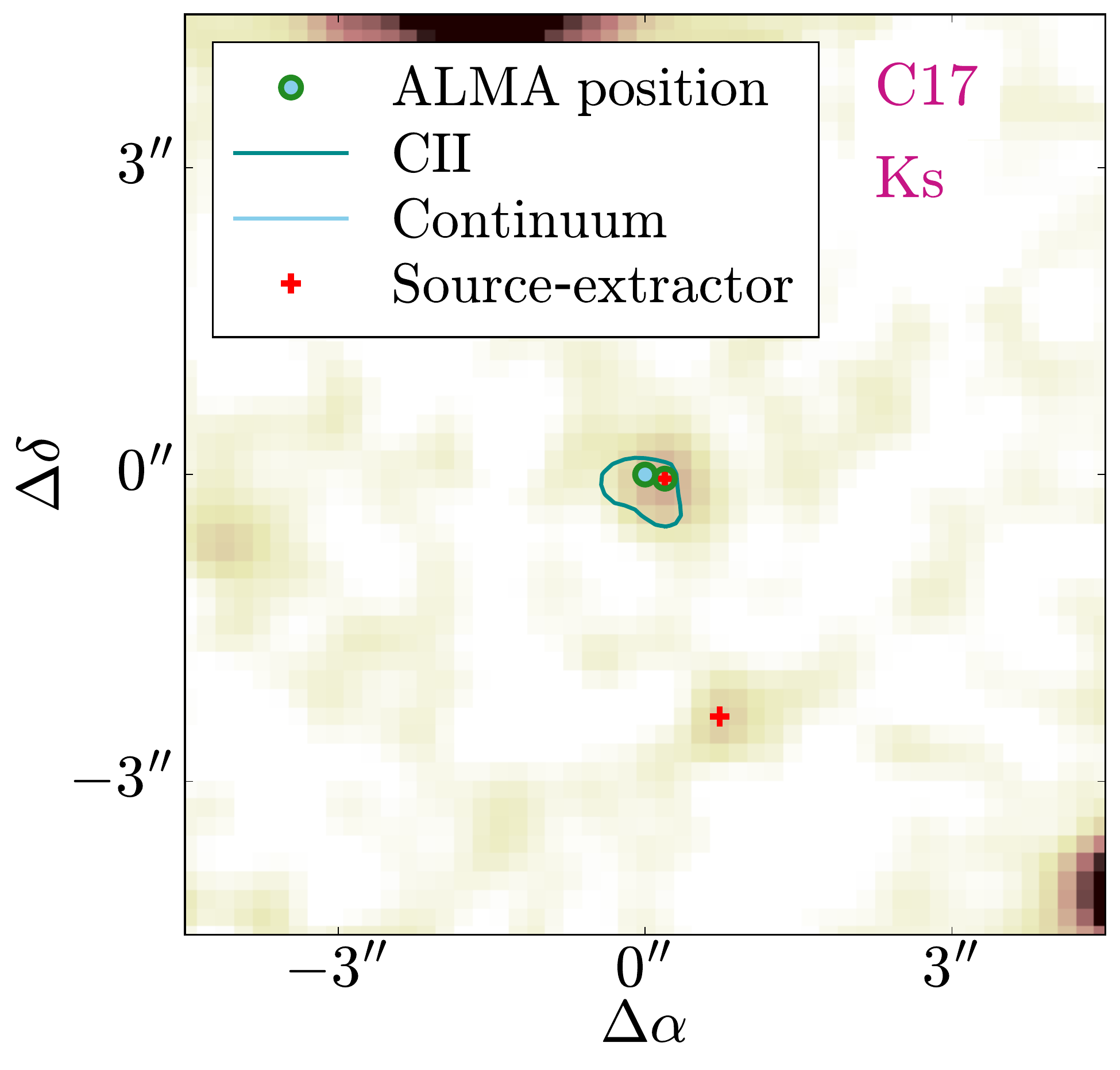}
\includegraphics[width=0.249\textwidth]{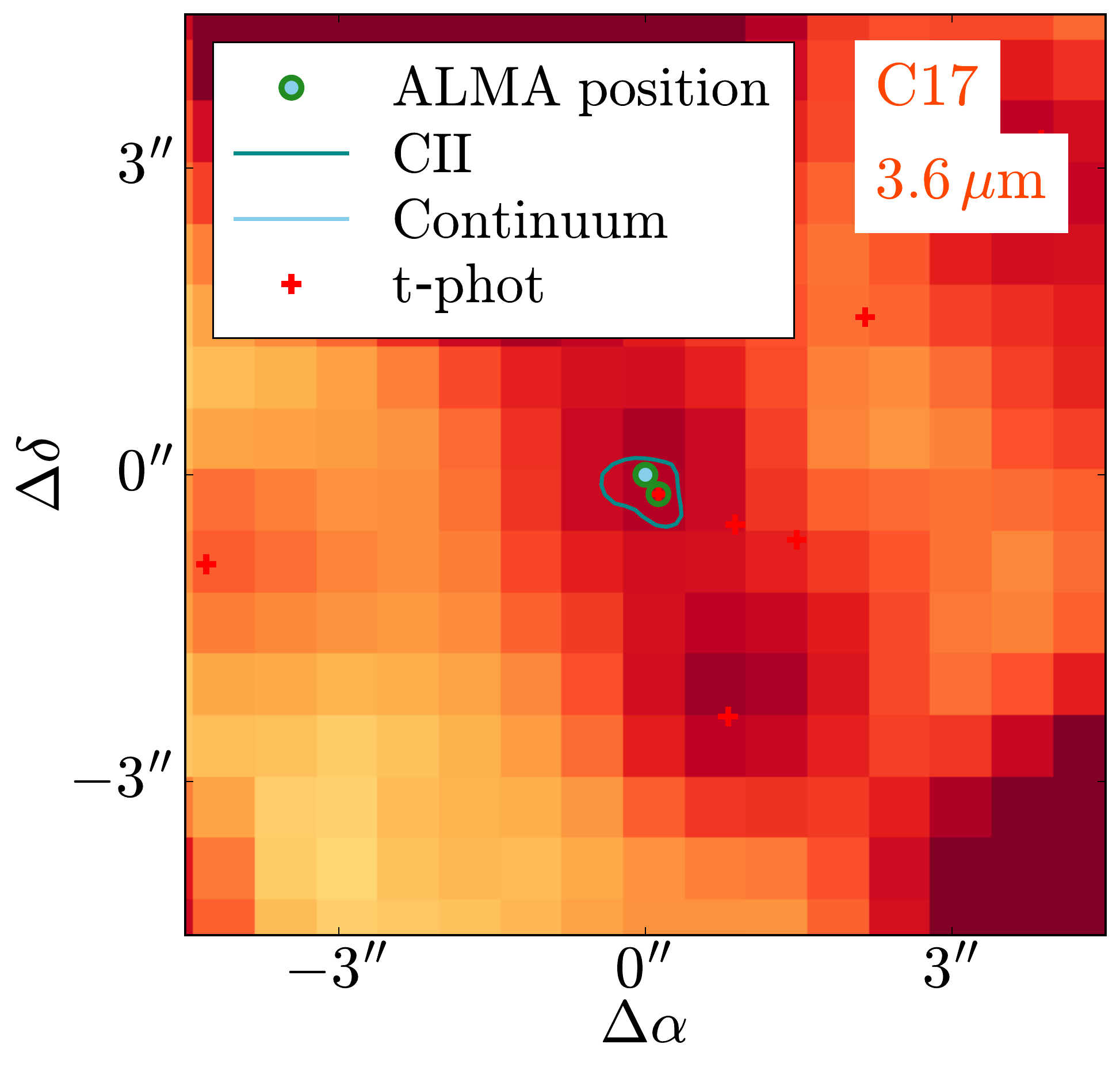}
\includegraphics[width=0.249\textwidth]{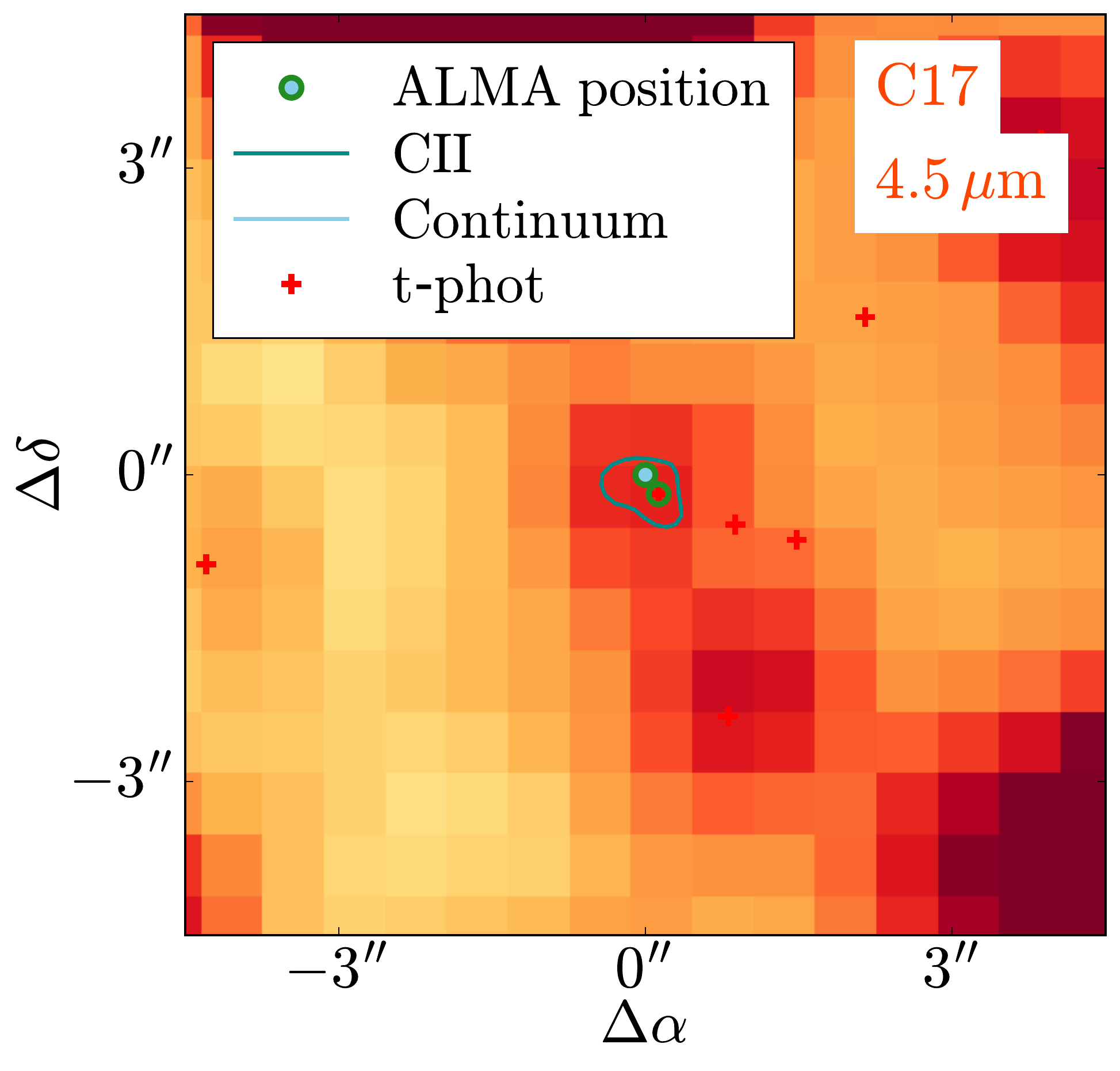}
\end{framed}
\end{subfigure}
\begin{subfigure}{0.85\textwidth}
\begin{framed}
\includegraphics[width=0.24\textwidth]{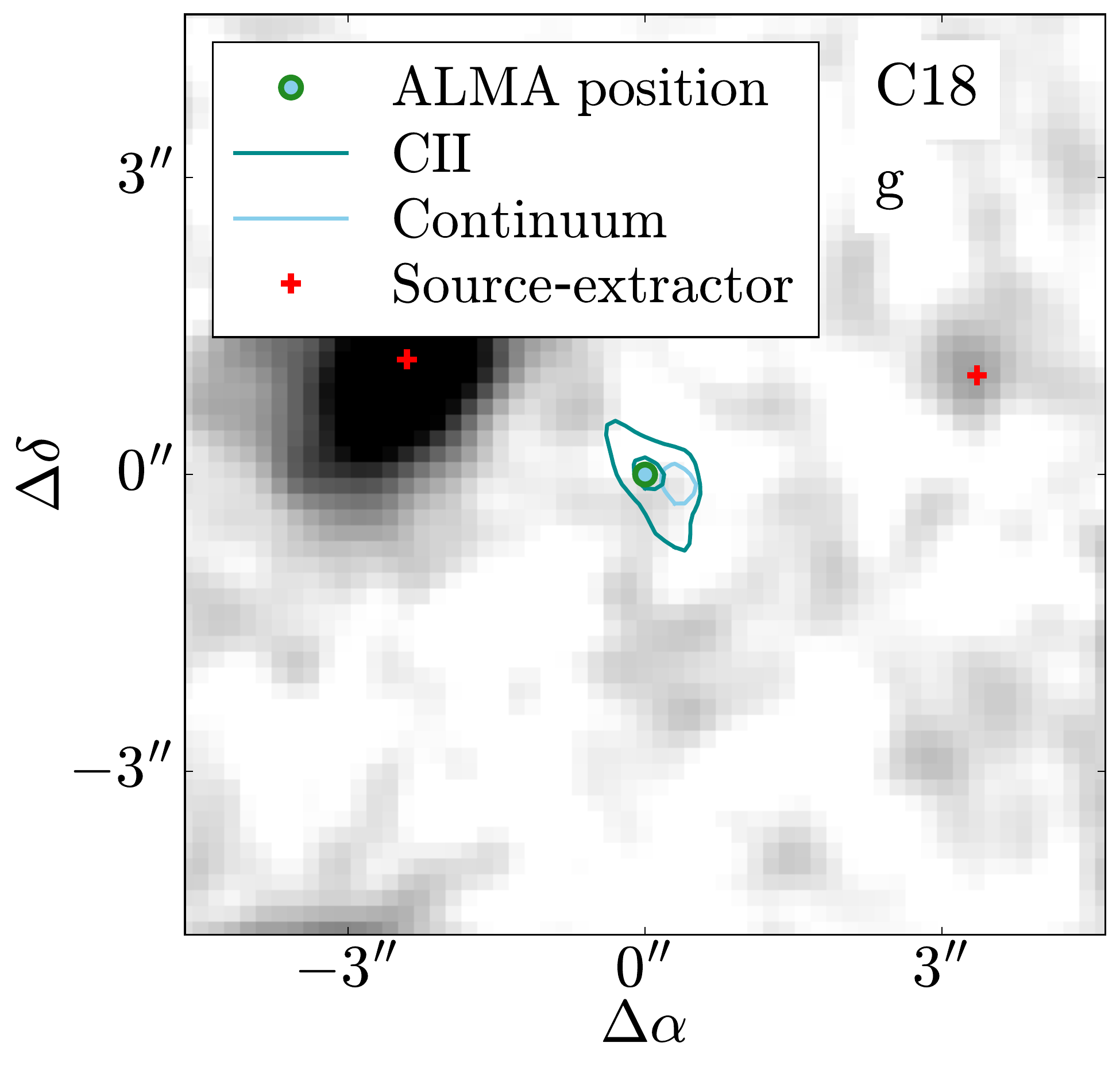}
\includegraphics[width=0.24\textwidth]{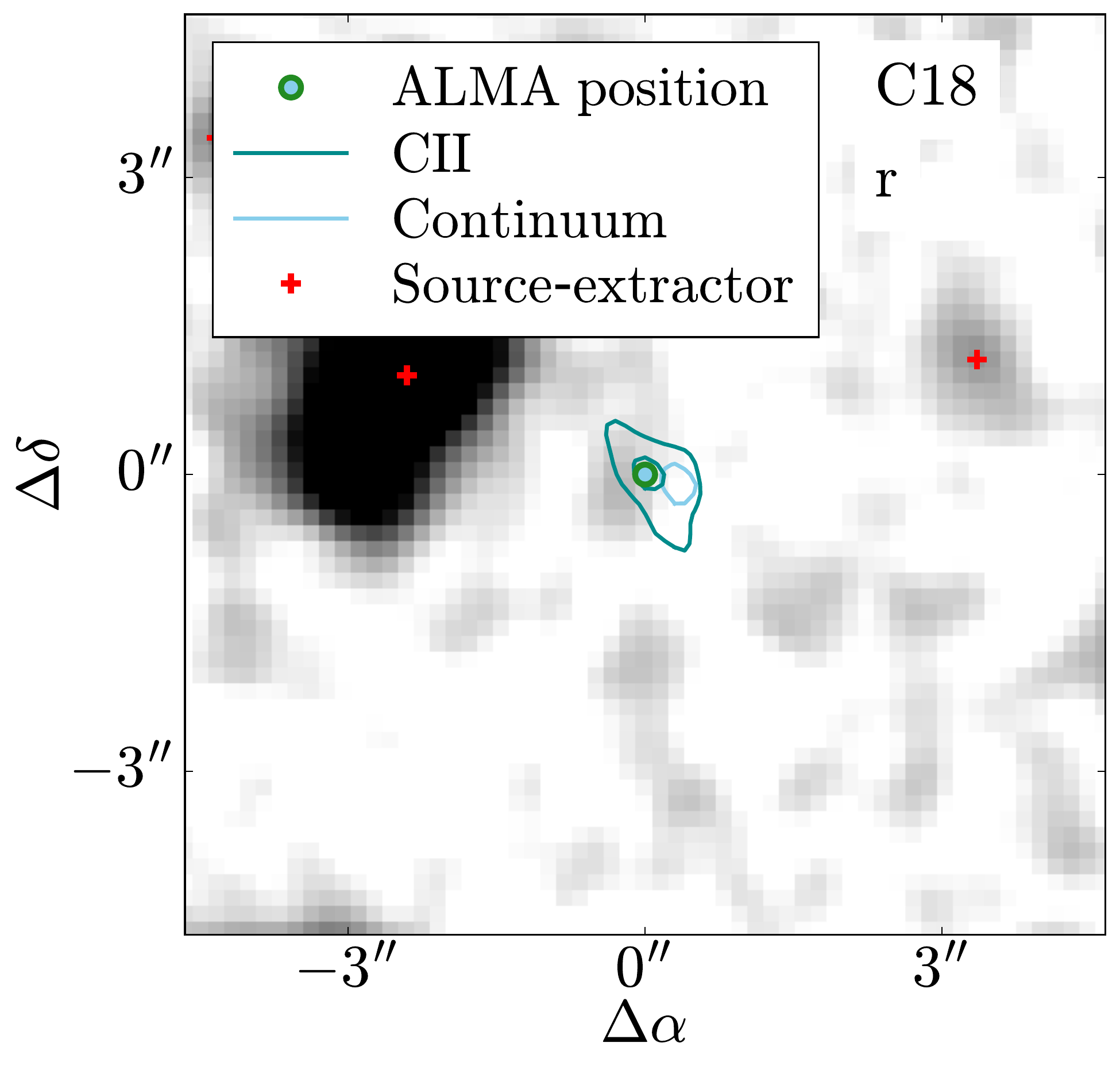}
\includegraphics[width=0.24\textwidth]{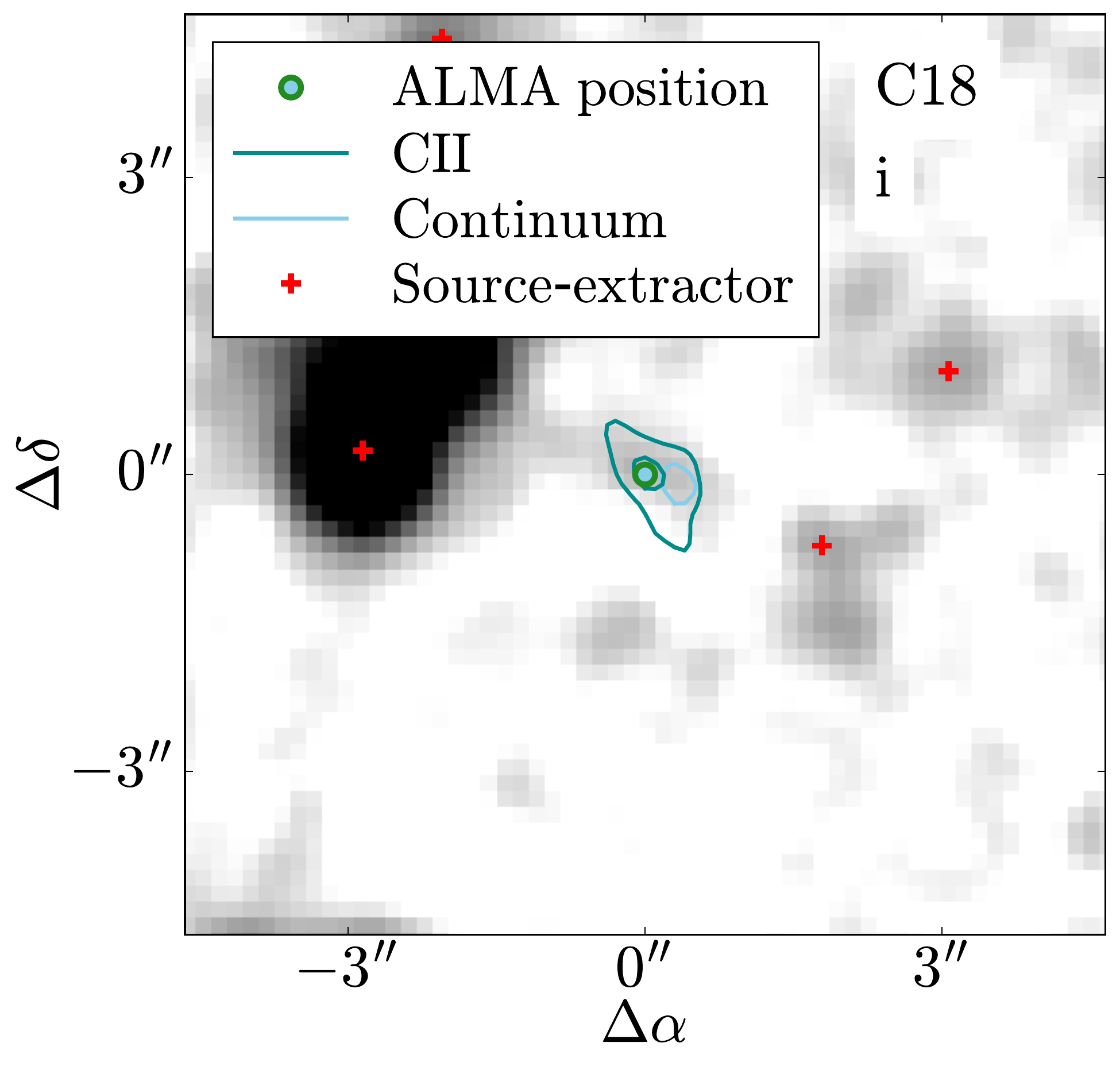}
\includegraphics[width=0.24\textwidth]{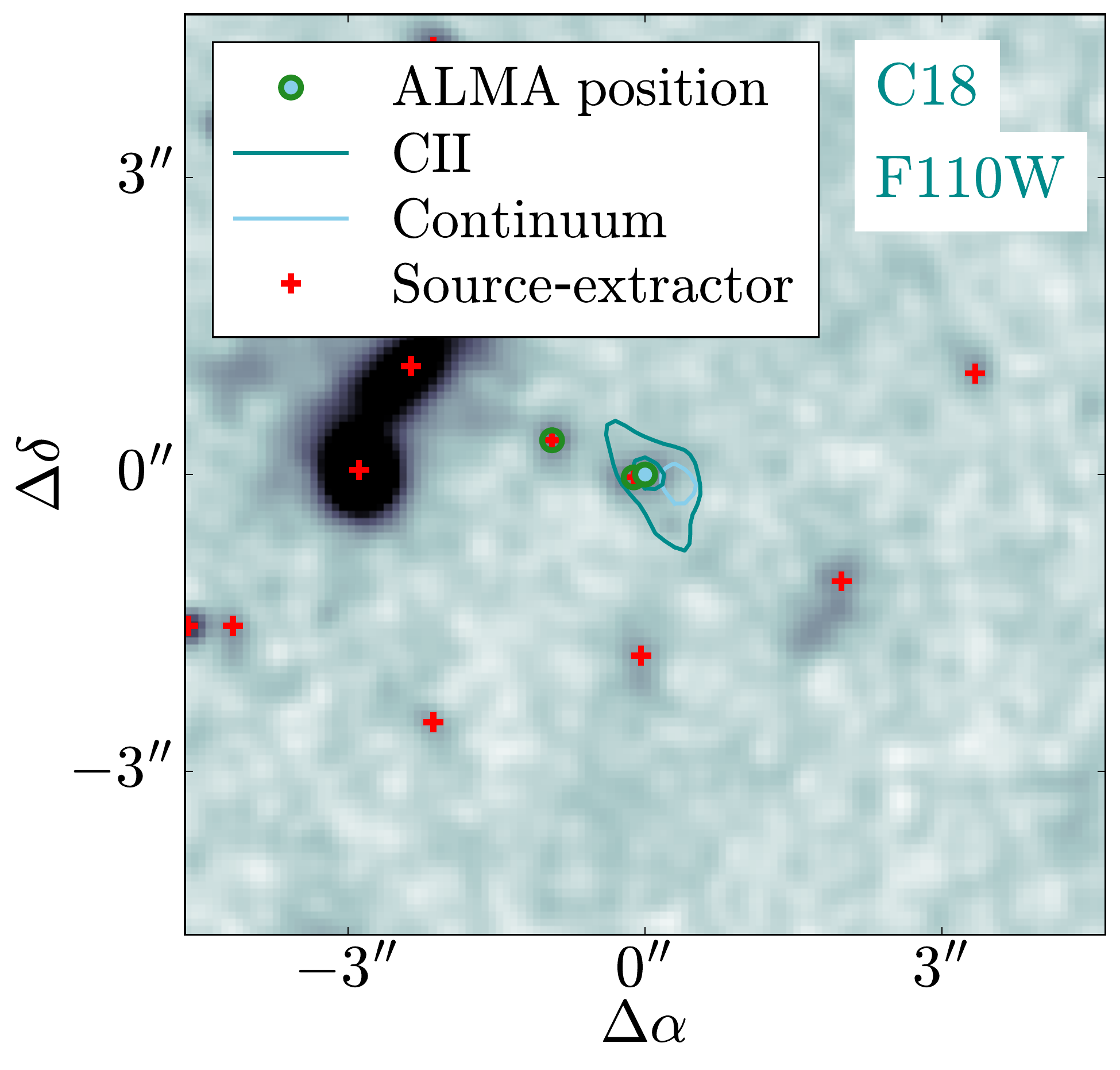}
\includegraphics[width=0.24\textwidth]{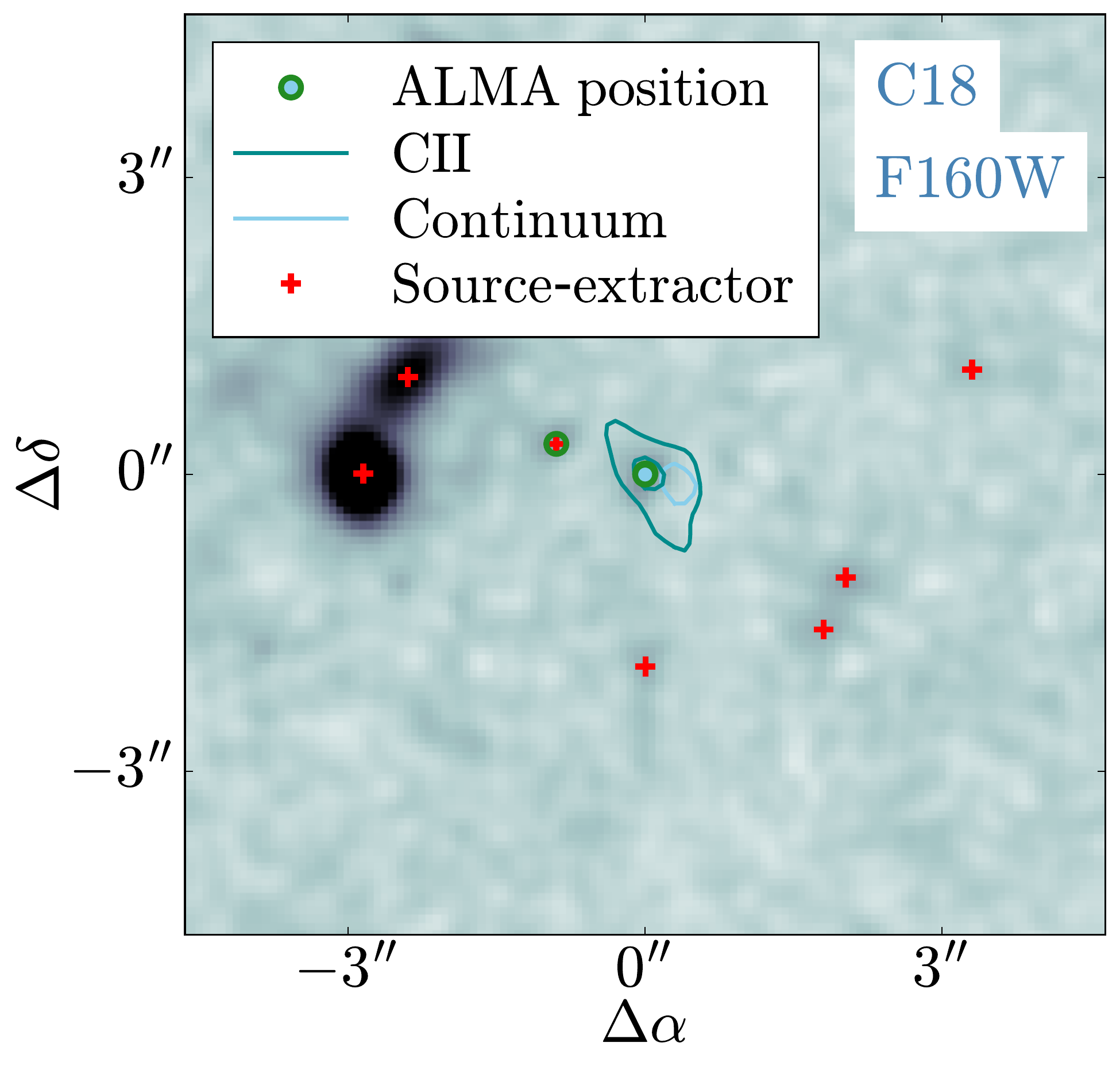}
\includegraphics[width=0.248\textwidth]{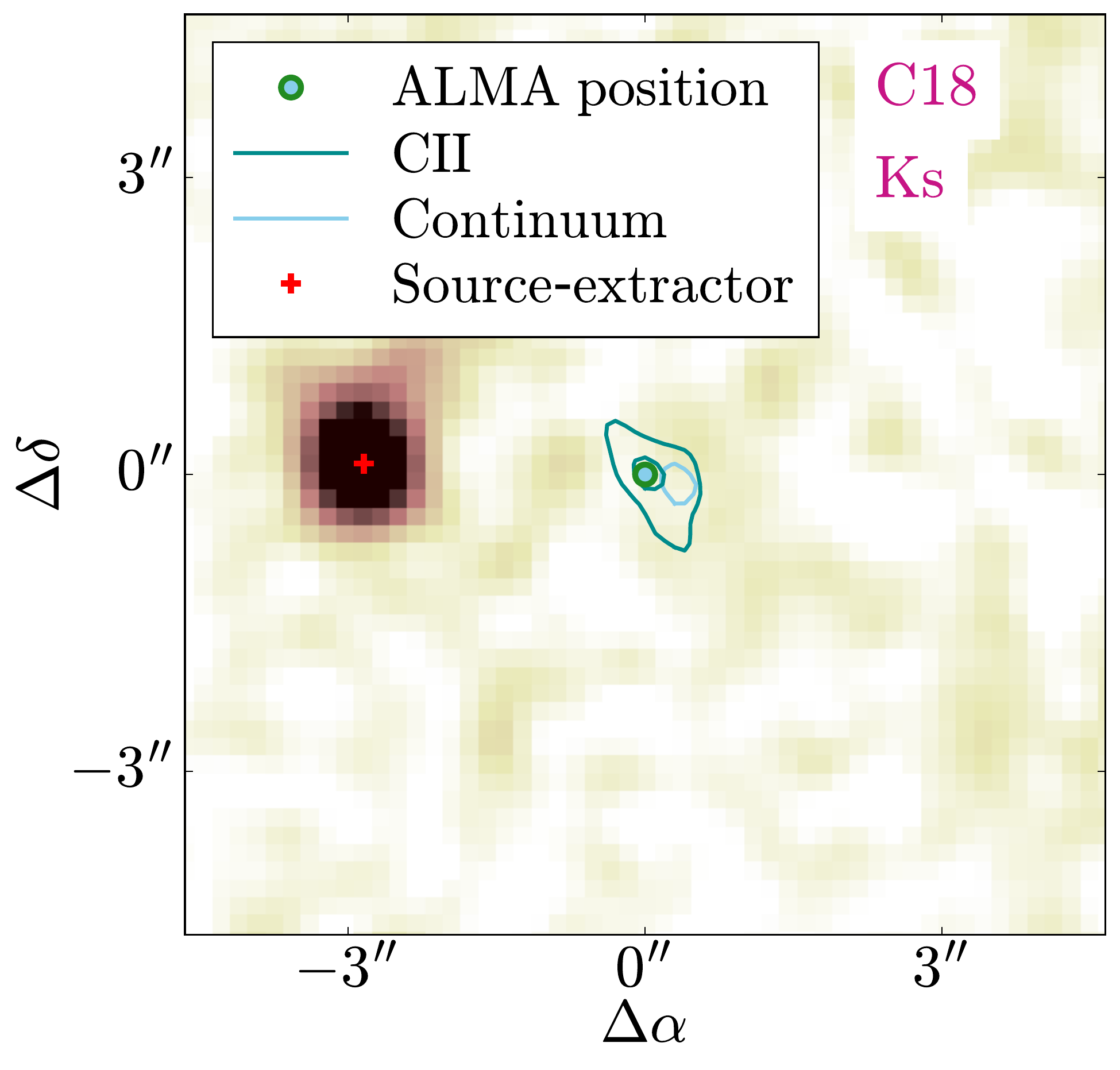}
\includegraphics[width=0.249\textwidth]{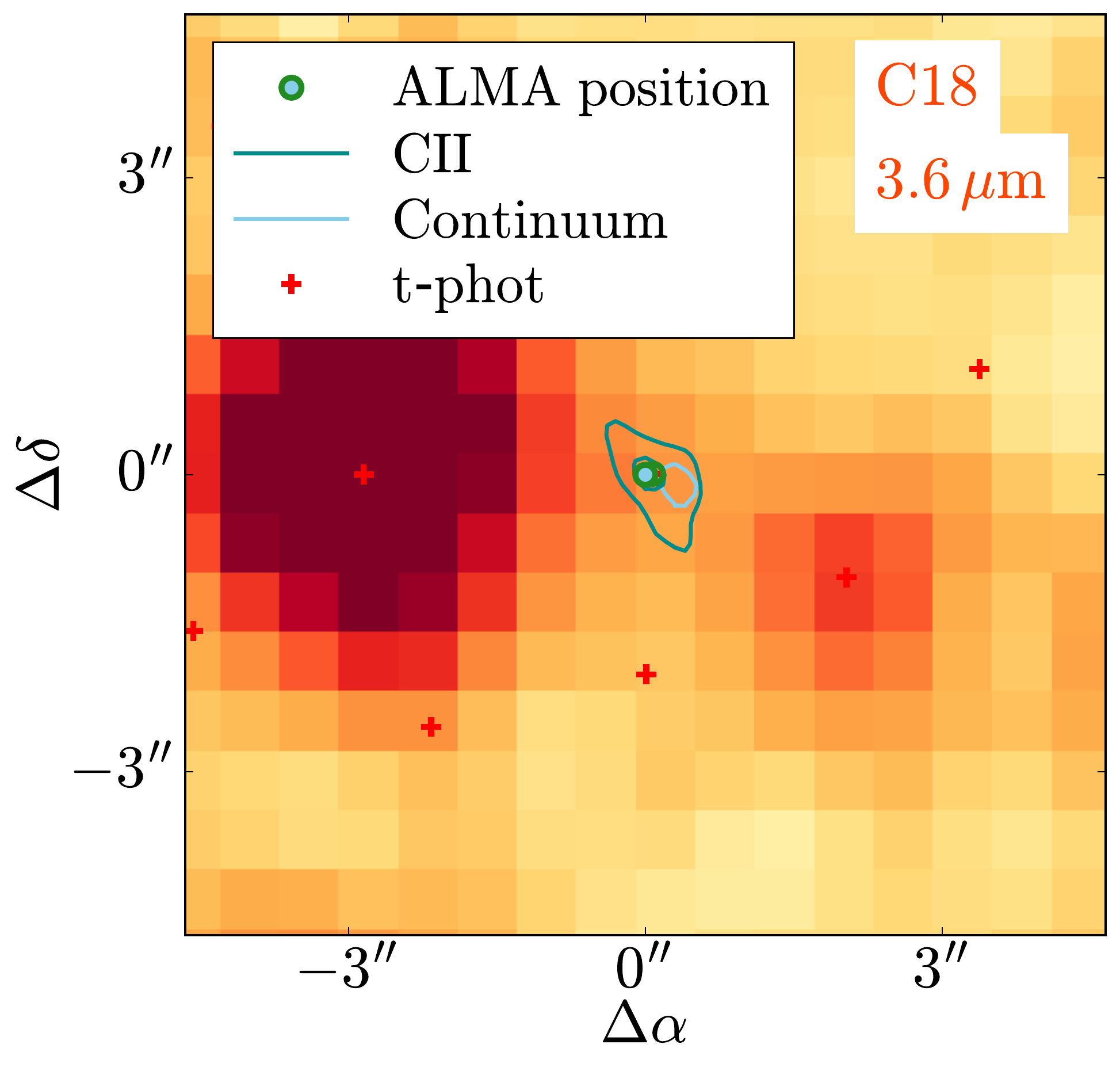}
\includegraphics[width=0.249\textwidth]{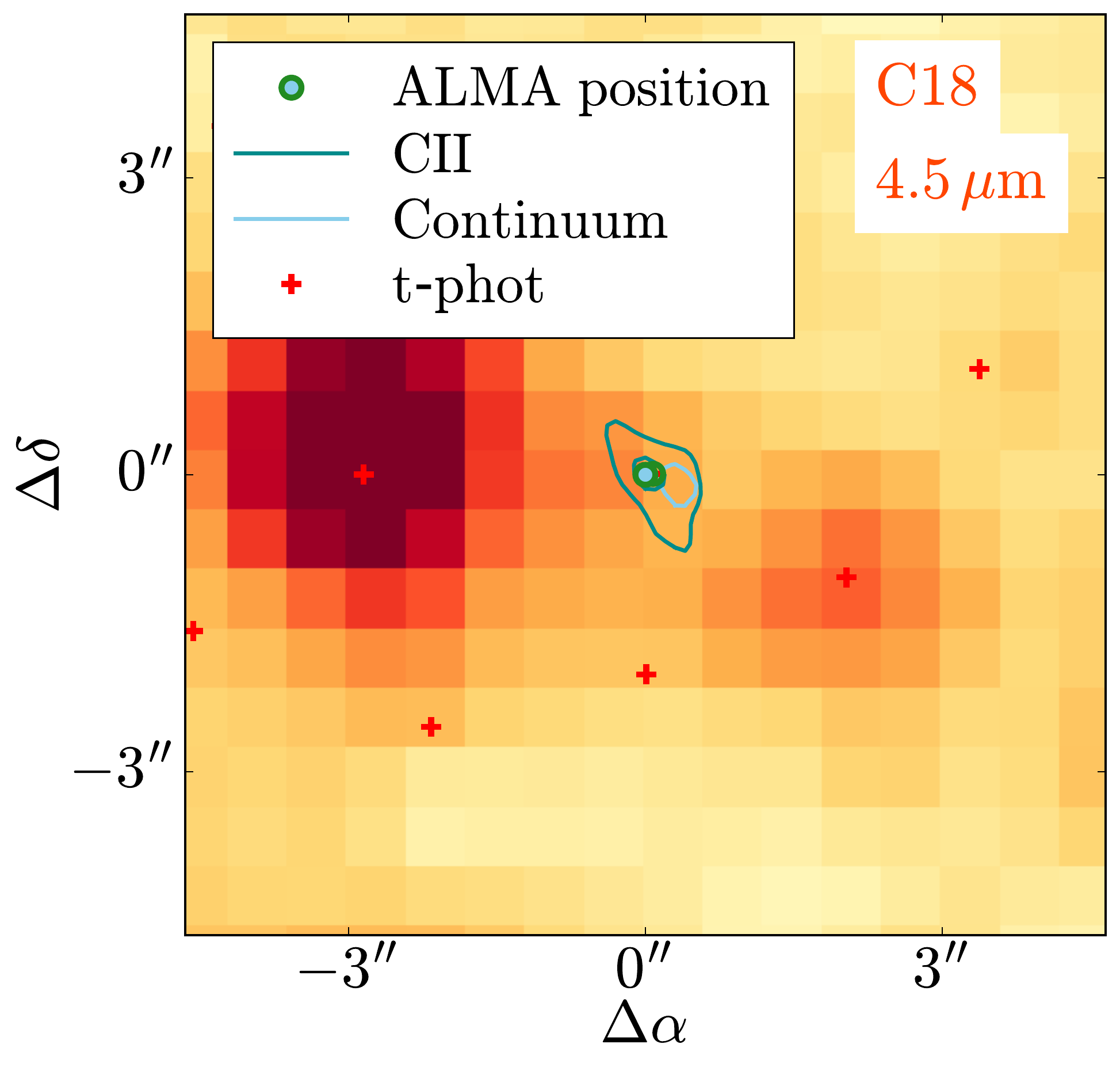}
\end{framed}
\end{subfigure}
\caption{}
\end{figure*}
\renewcommand{\thefigure}{\arabic{figure}}

\renewcommand{\thefigure}{B\arabic{figure} (Cont.)}
\addtocounter{figure}{-1}
\begin{figure*}
\begin{subfigure}{0.85\textwidth}
\begin{framed}
\includegraphics[width=0.24\textwidth]{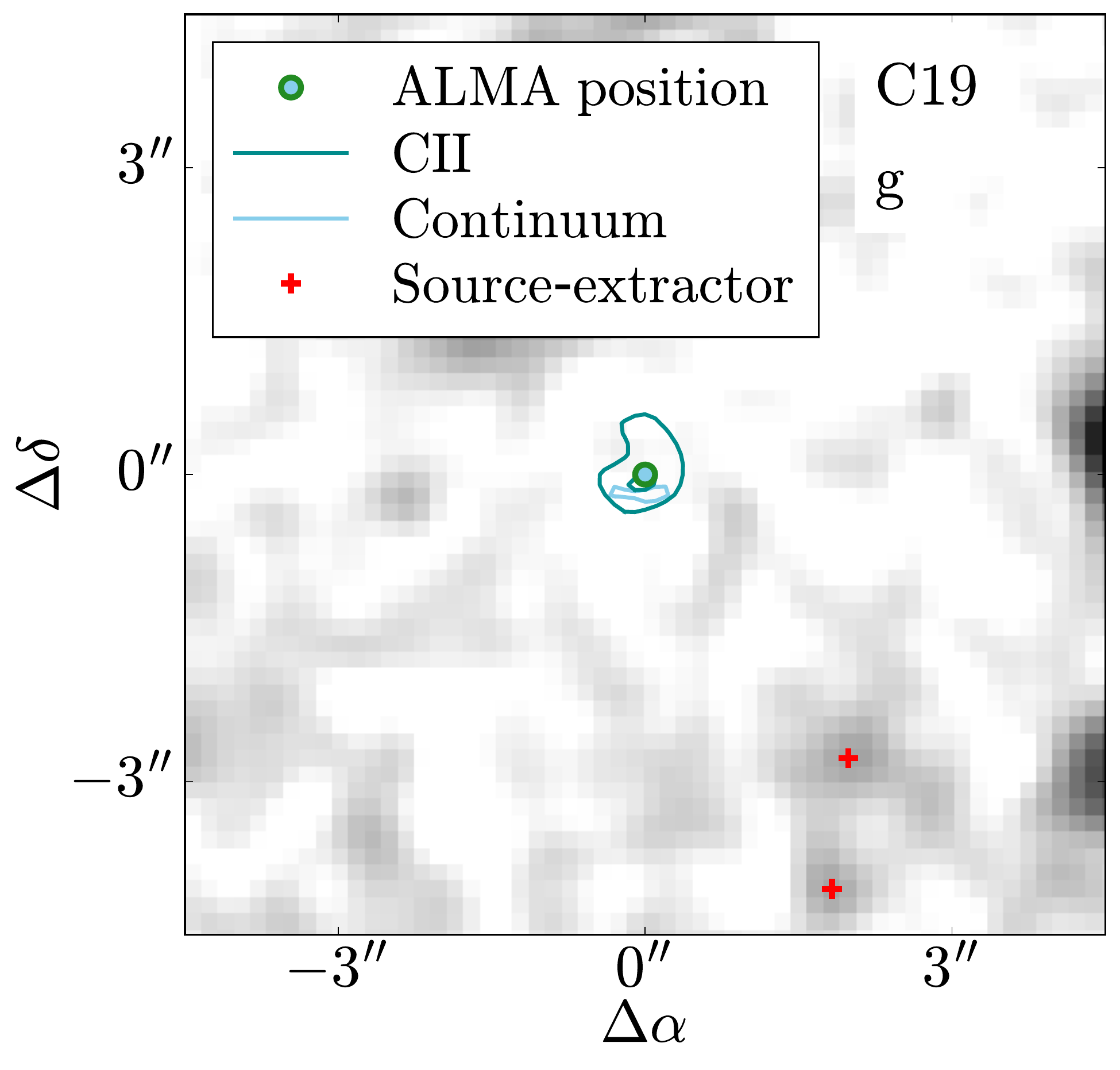}
\includegraphics[width=0.24\textwidth]{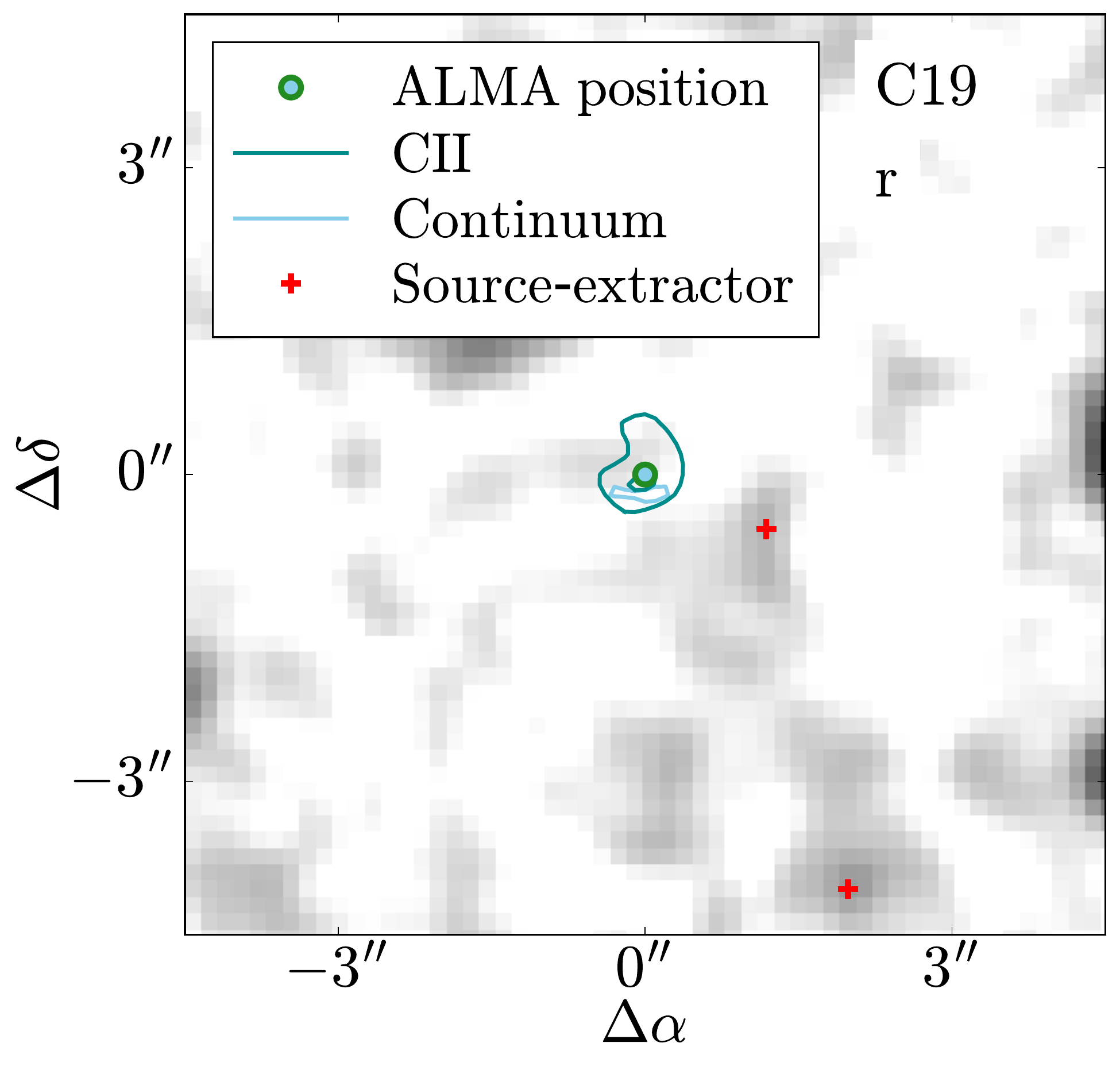}
\includegraphics[width=0.24\textwidth]{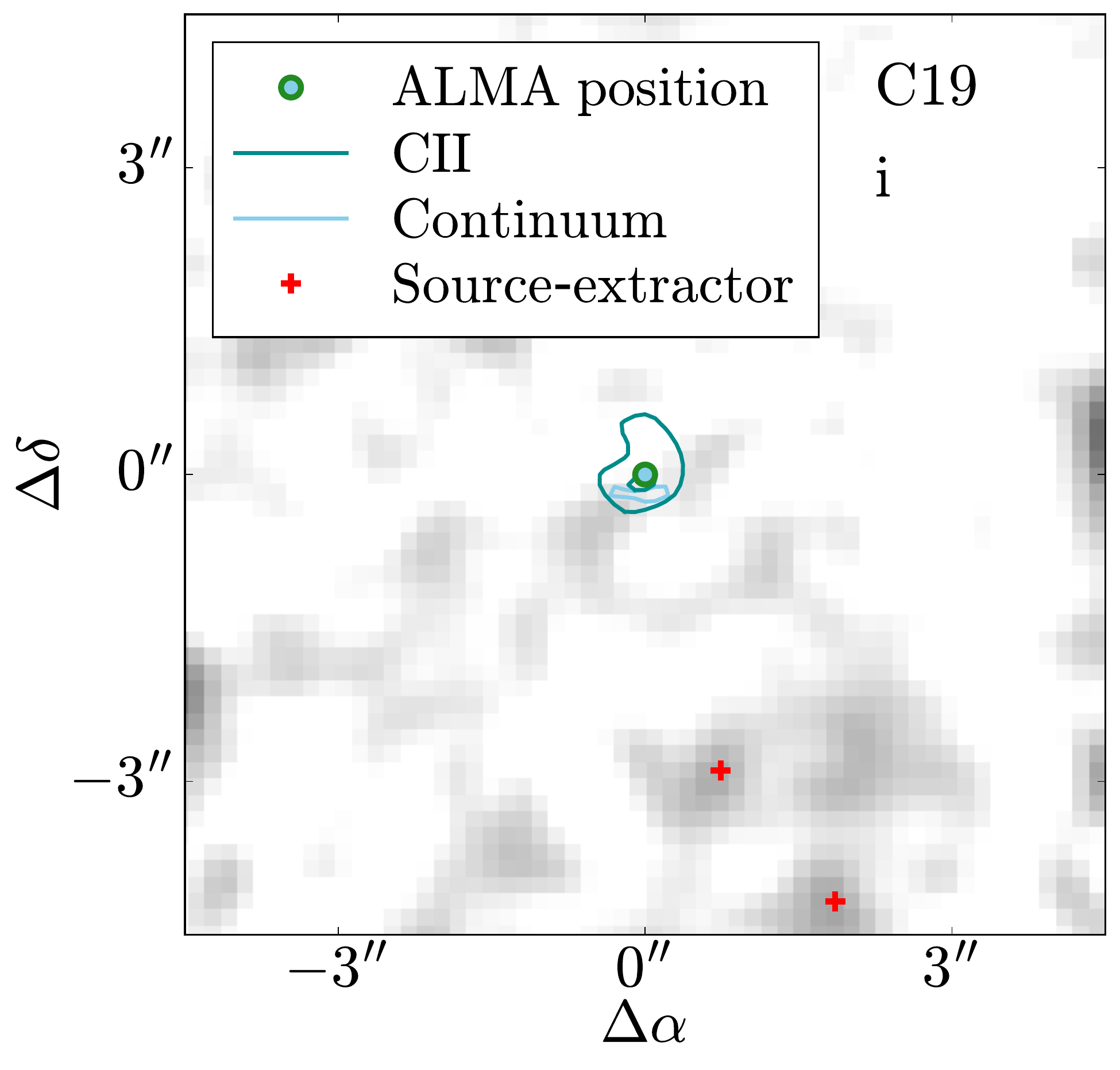}
\includegraphics[width=0.24\textwidth]{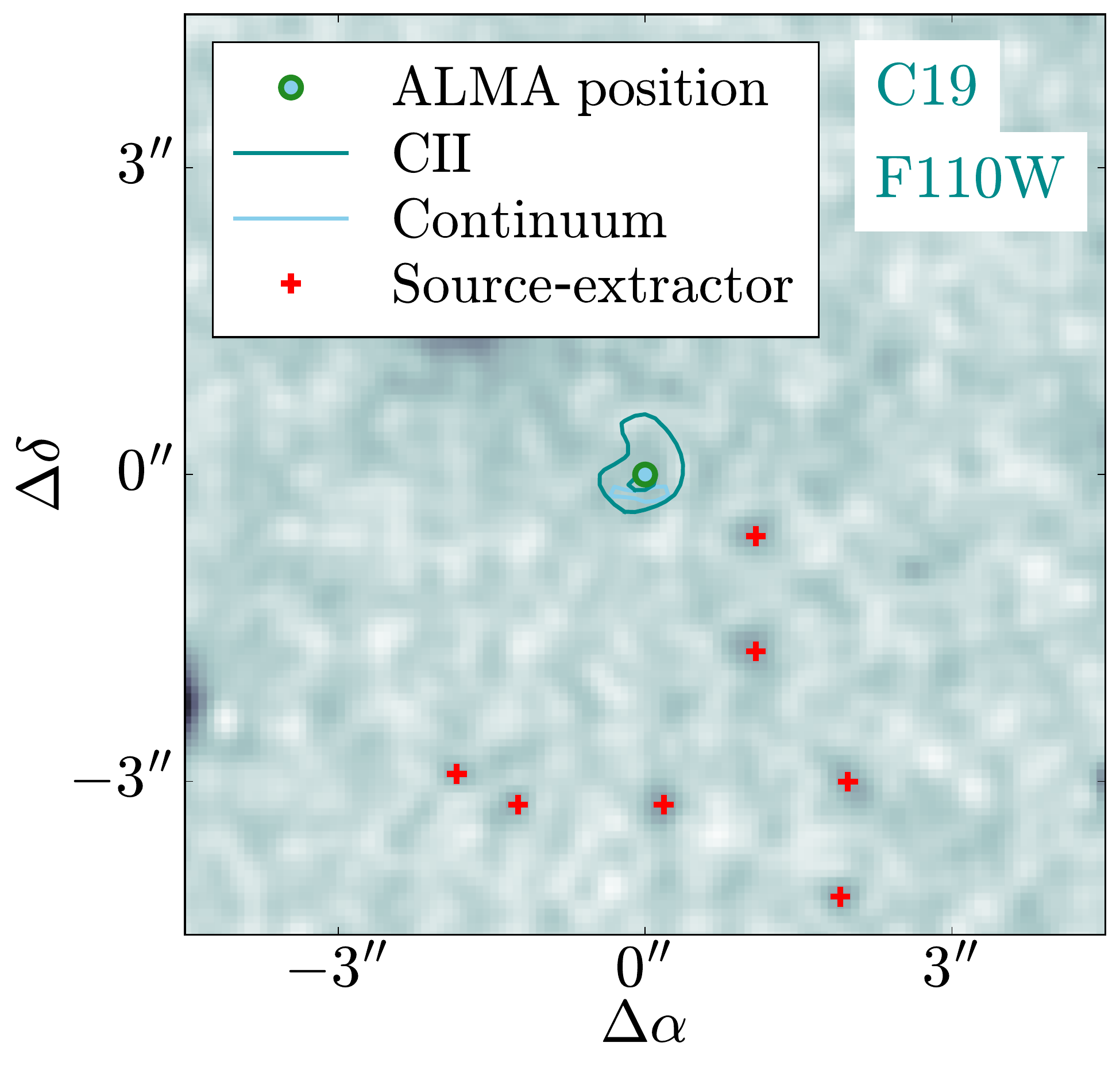}
\includegraphics[width=0.24\textwidth]{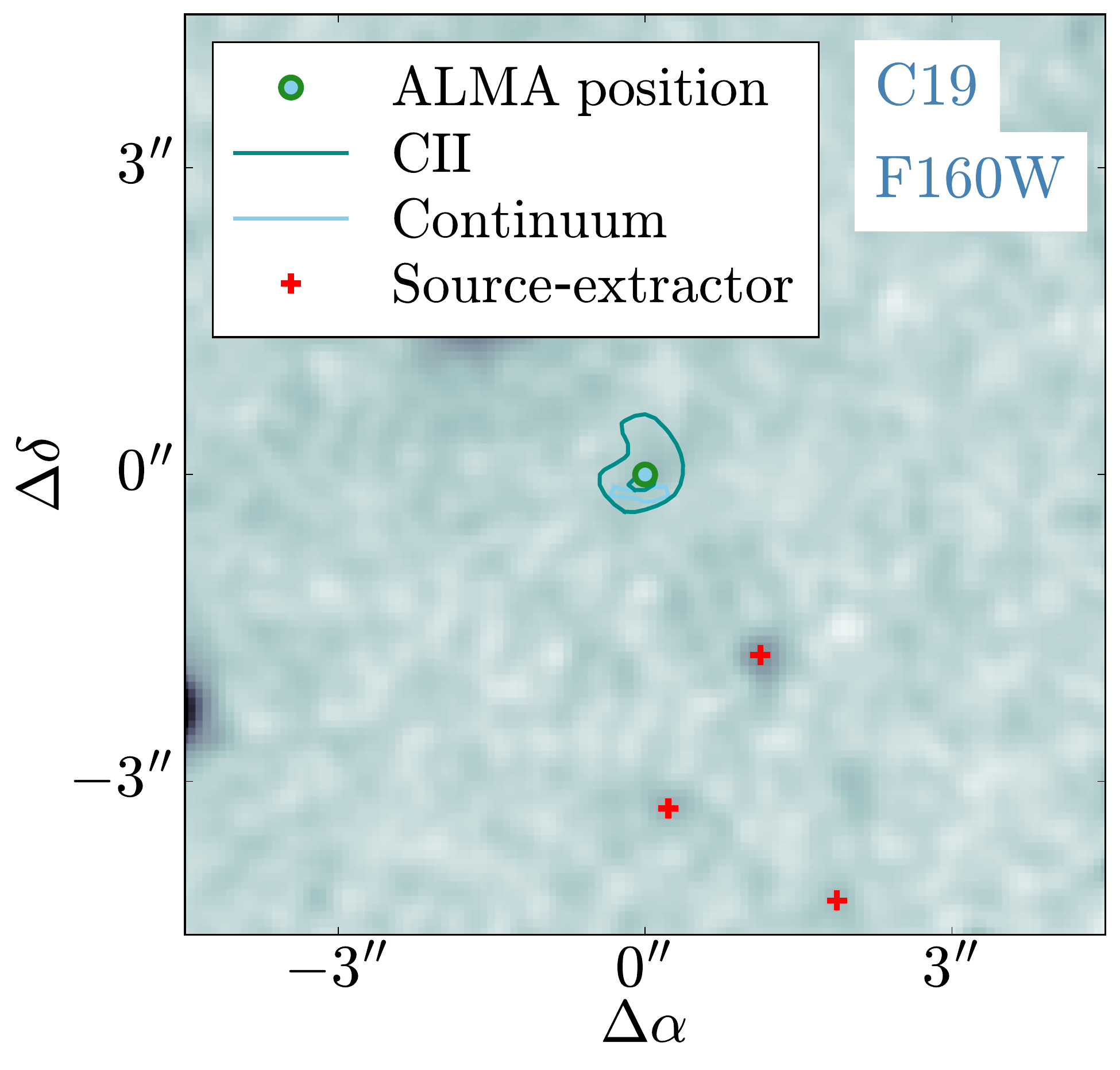}
\includegraphics[width=0.248\textwidth]{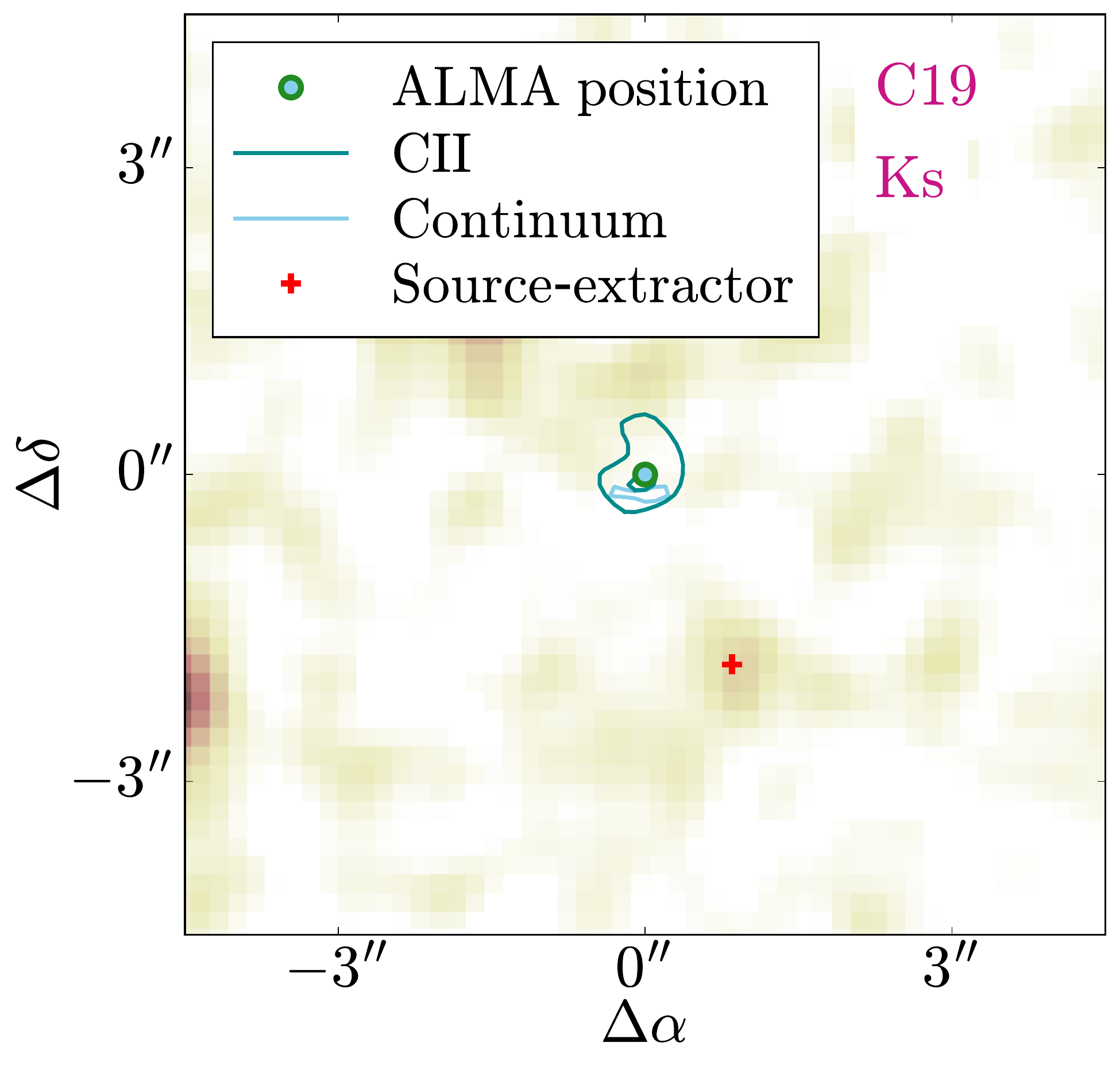}
\includegraphics[width=0.249\textwidth]{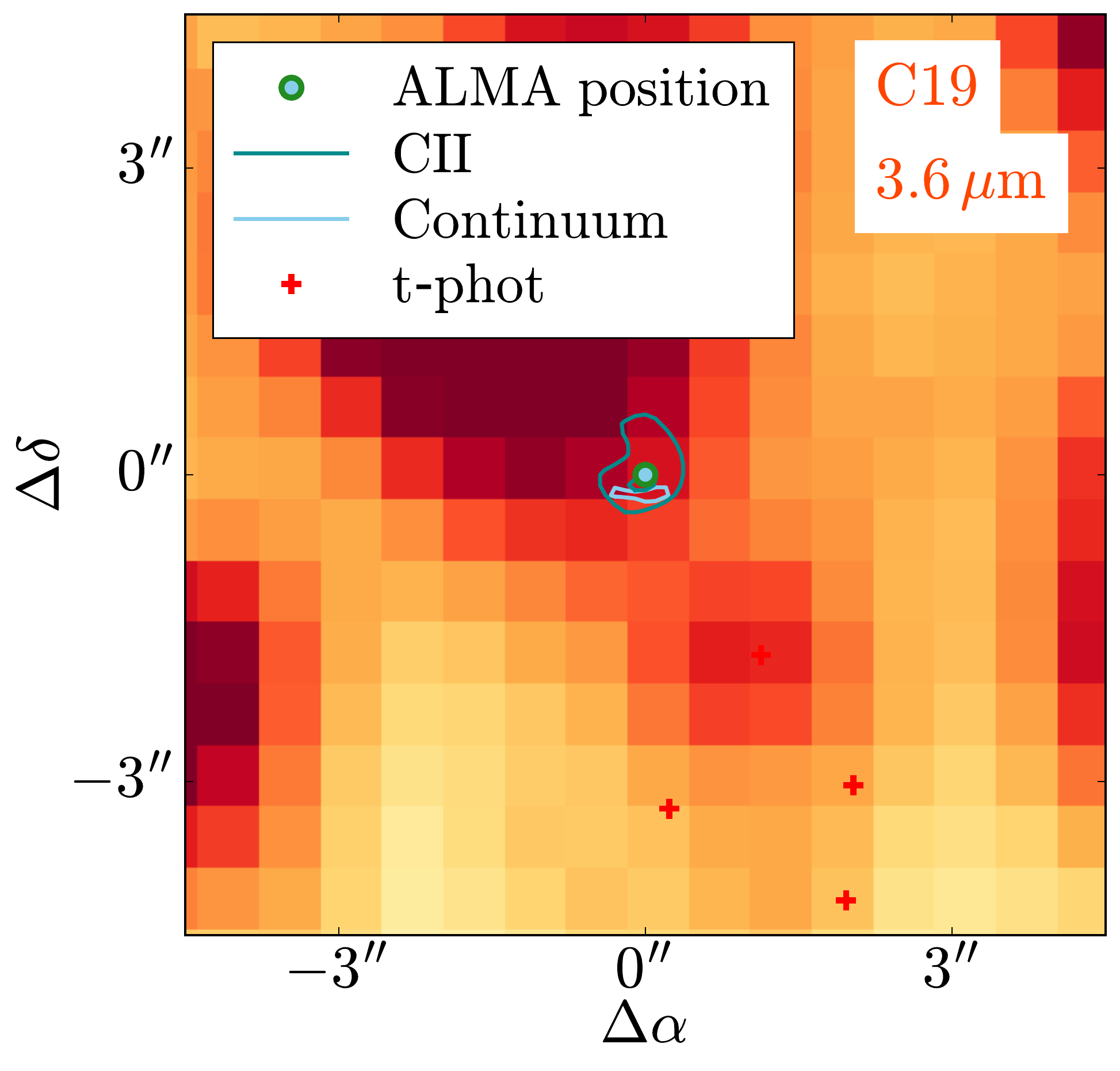}
\includegraphics[width=0.249\textwidth]{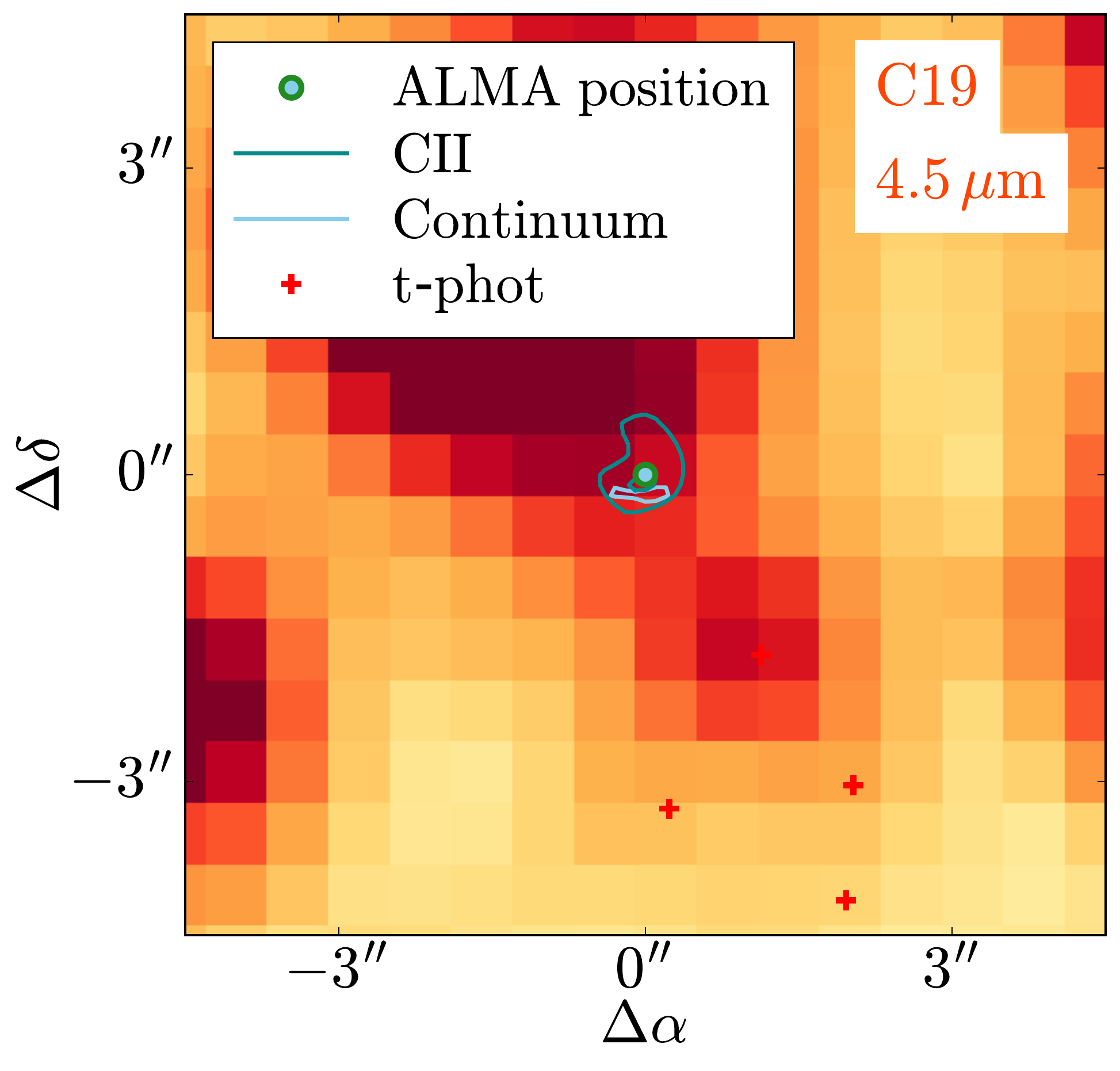}
\end{framed}
\end{subfigure}
\begin{subfigure}{0.85\textwidth}
\begin{framed}
\includegraphics[width=0.24\textwidth]{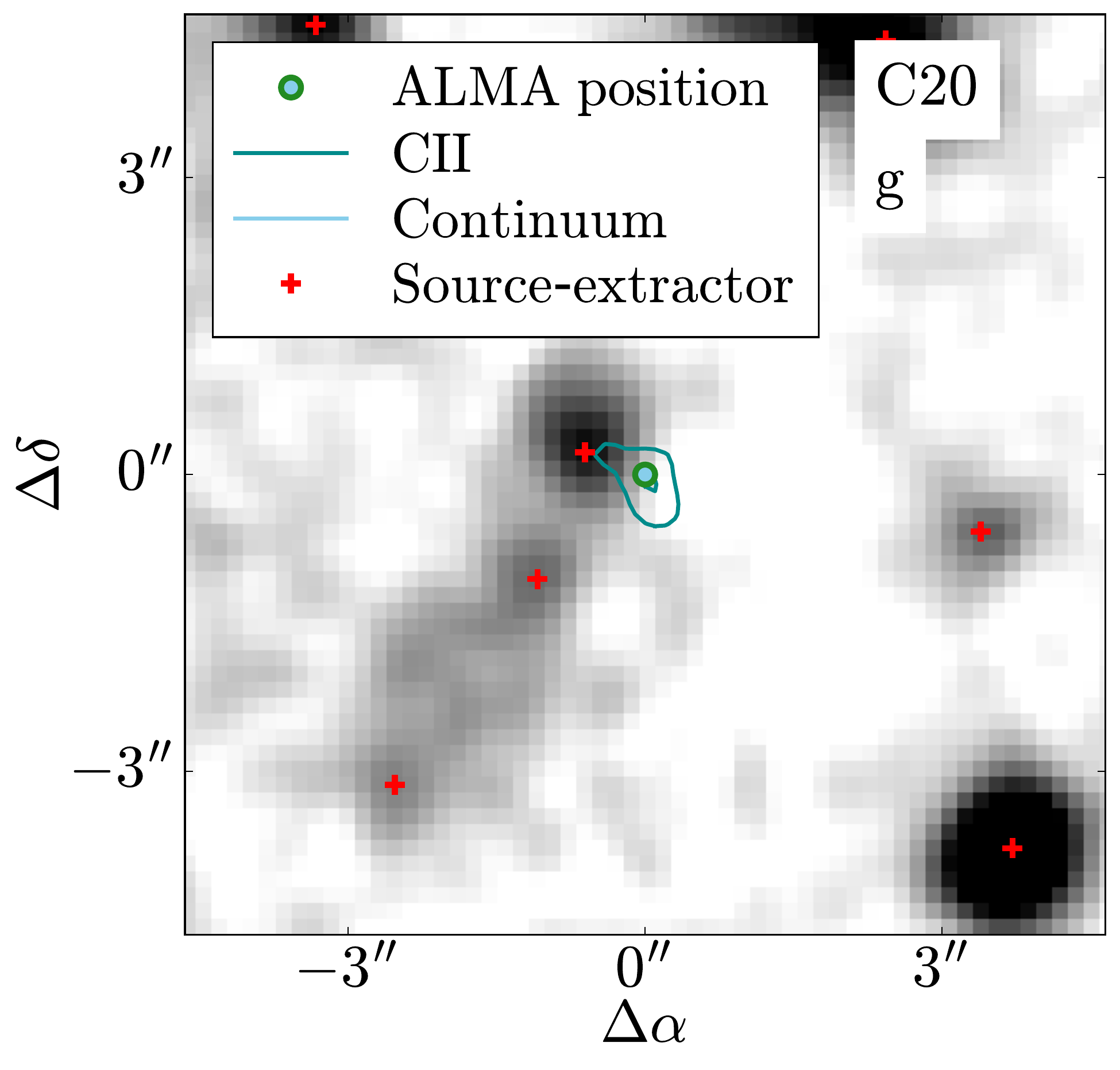}
\includegraphics[width=0.24\textwidth]{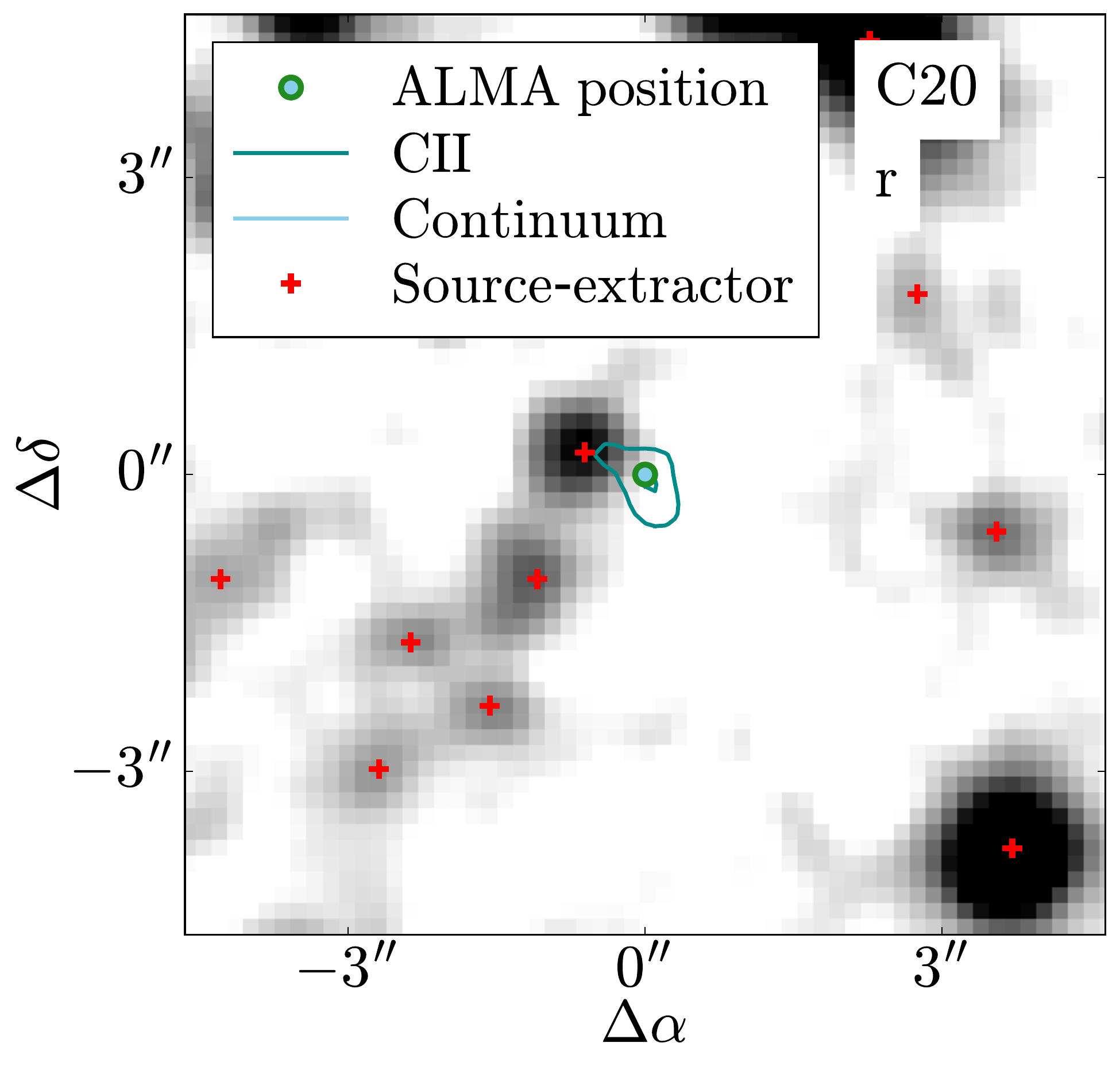}
\includegraphics[width=0.24\textwidth]{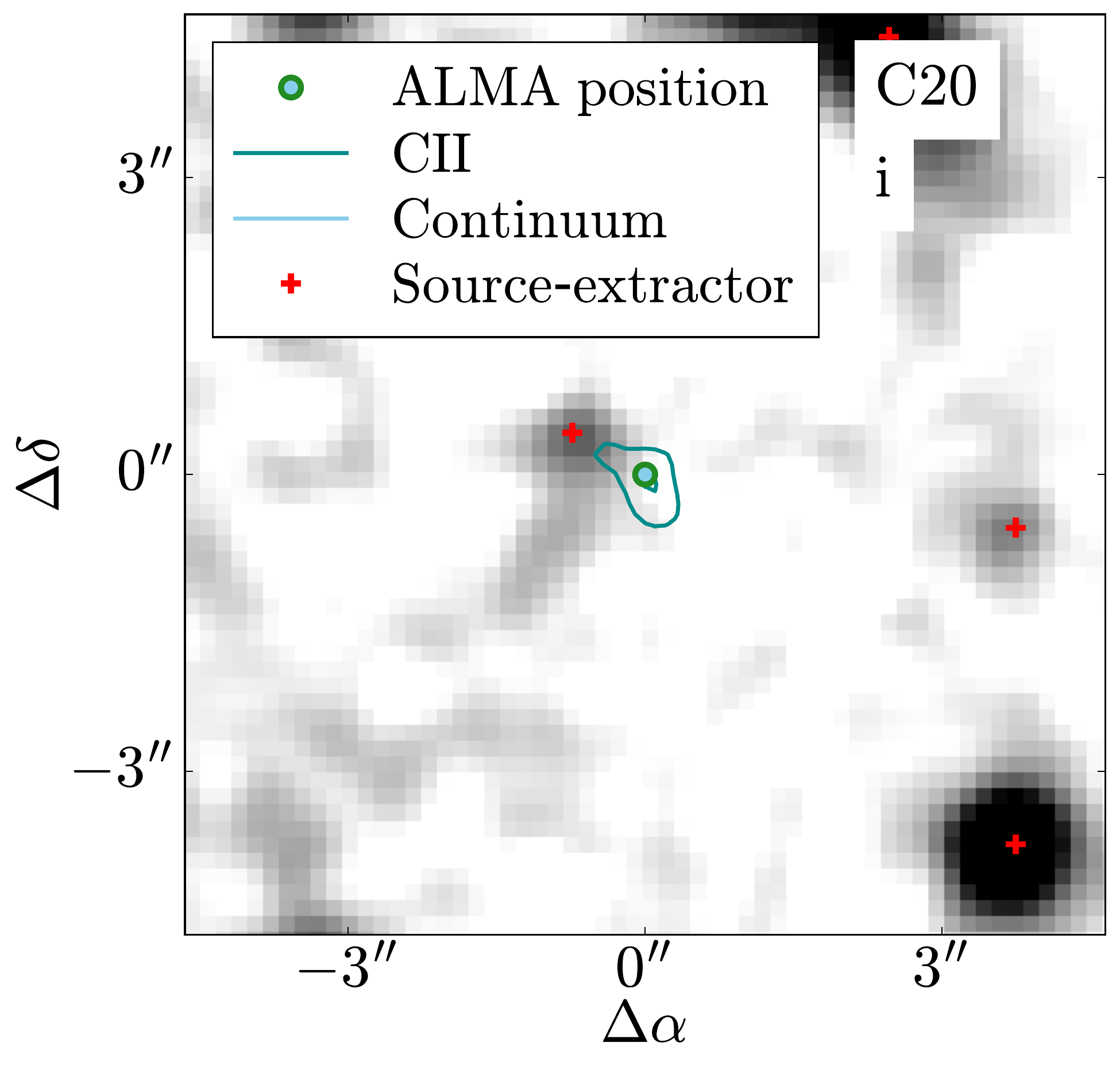}
\includegraphics[width=0.24\textwidth]{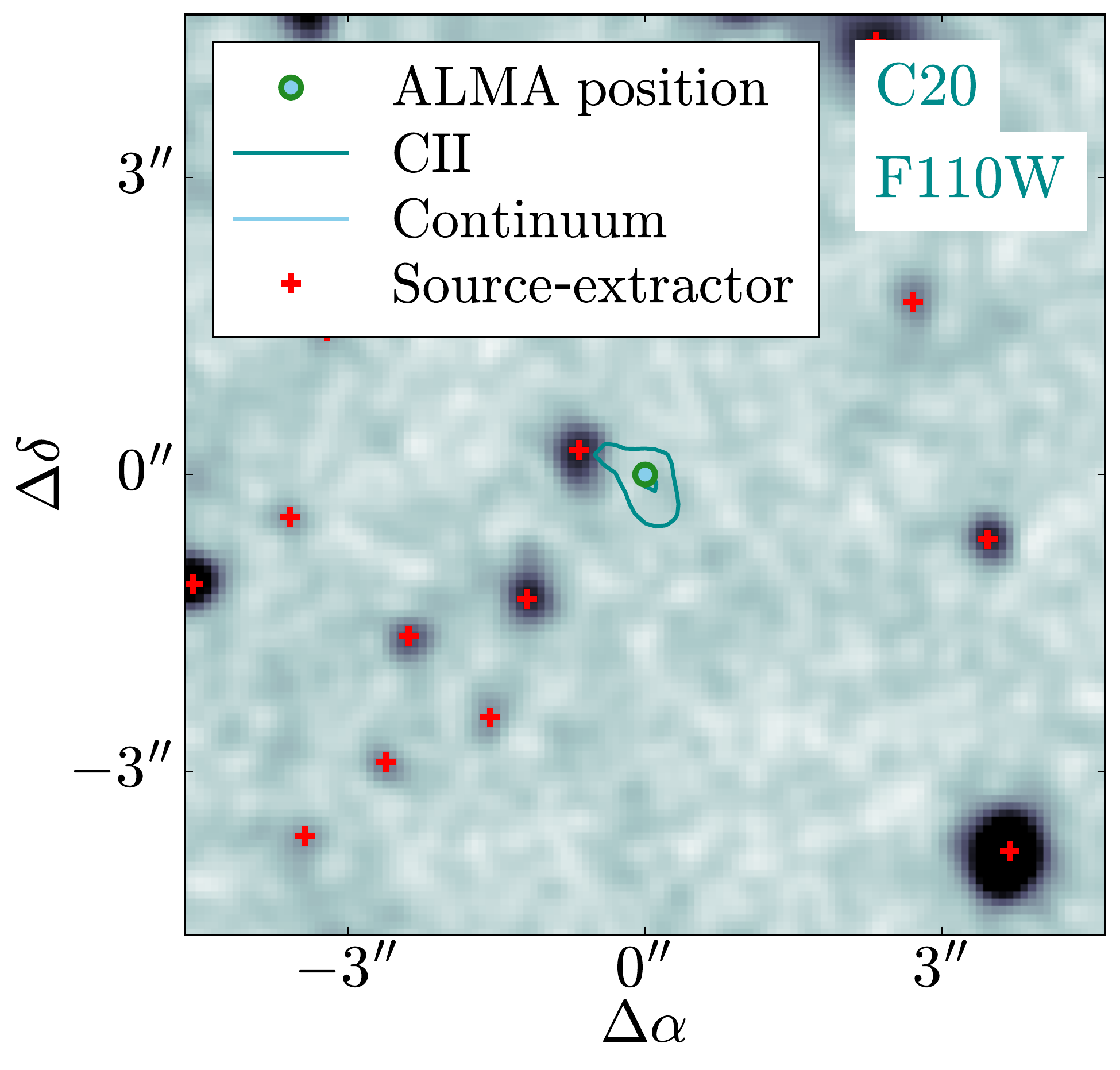}
\includegraphics[width=0.24\textwidth]{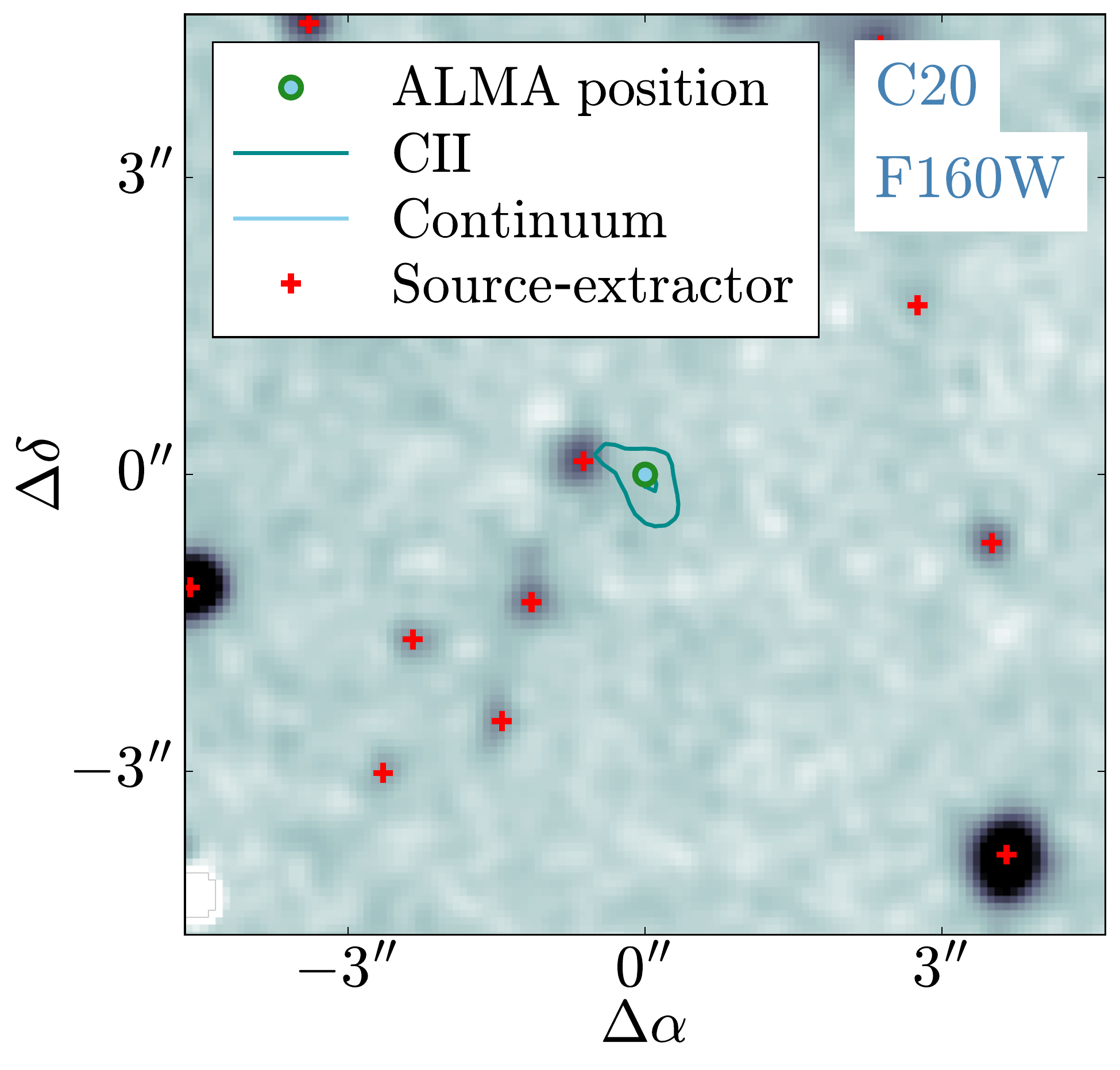}
\includegraphics[width=0.248\textwidth]{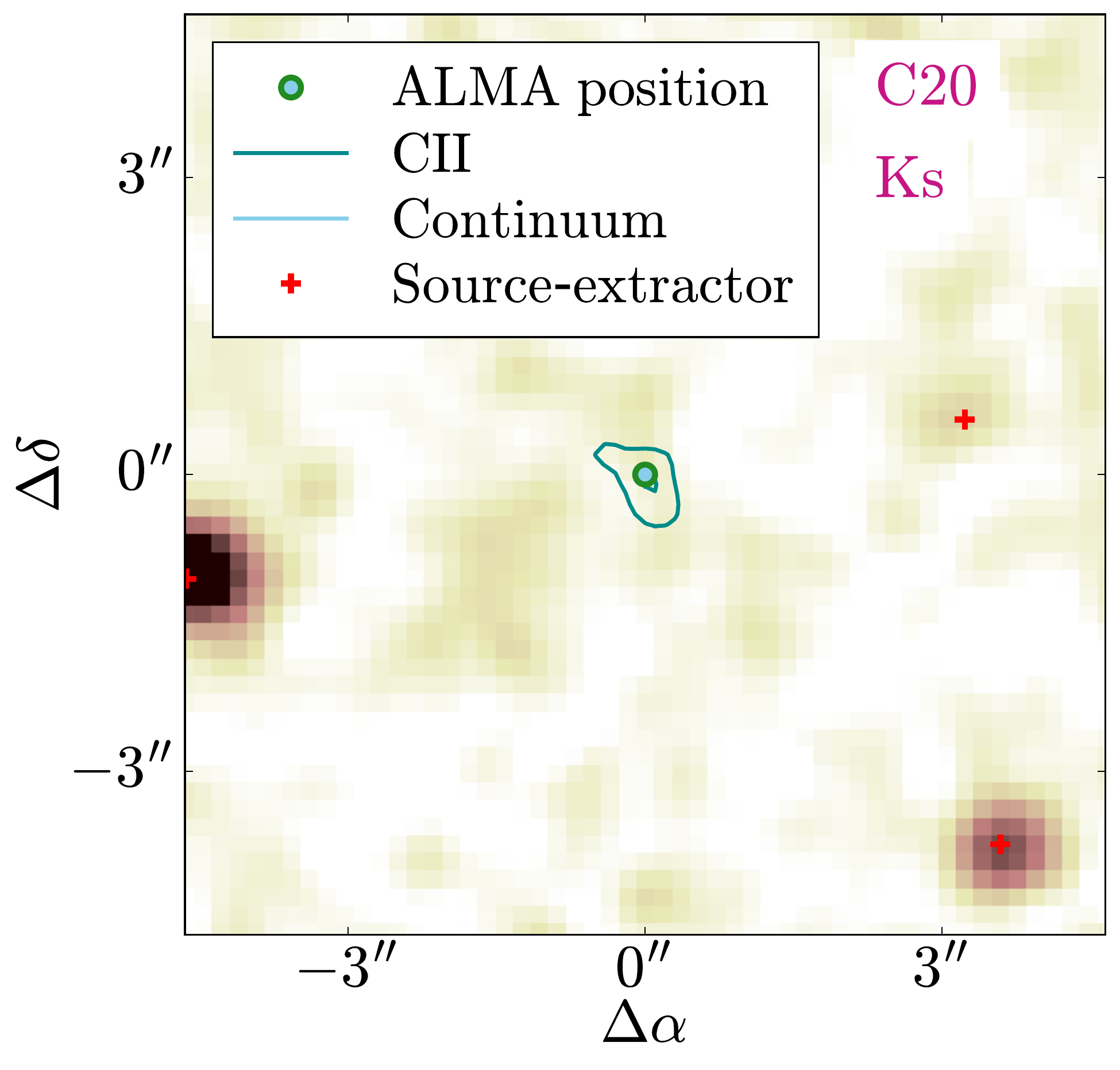}
\includegraphics[width=0.249\textwidth]{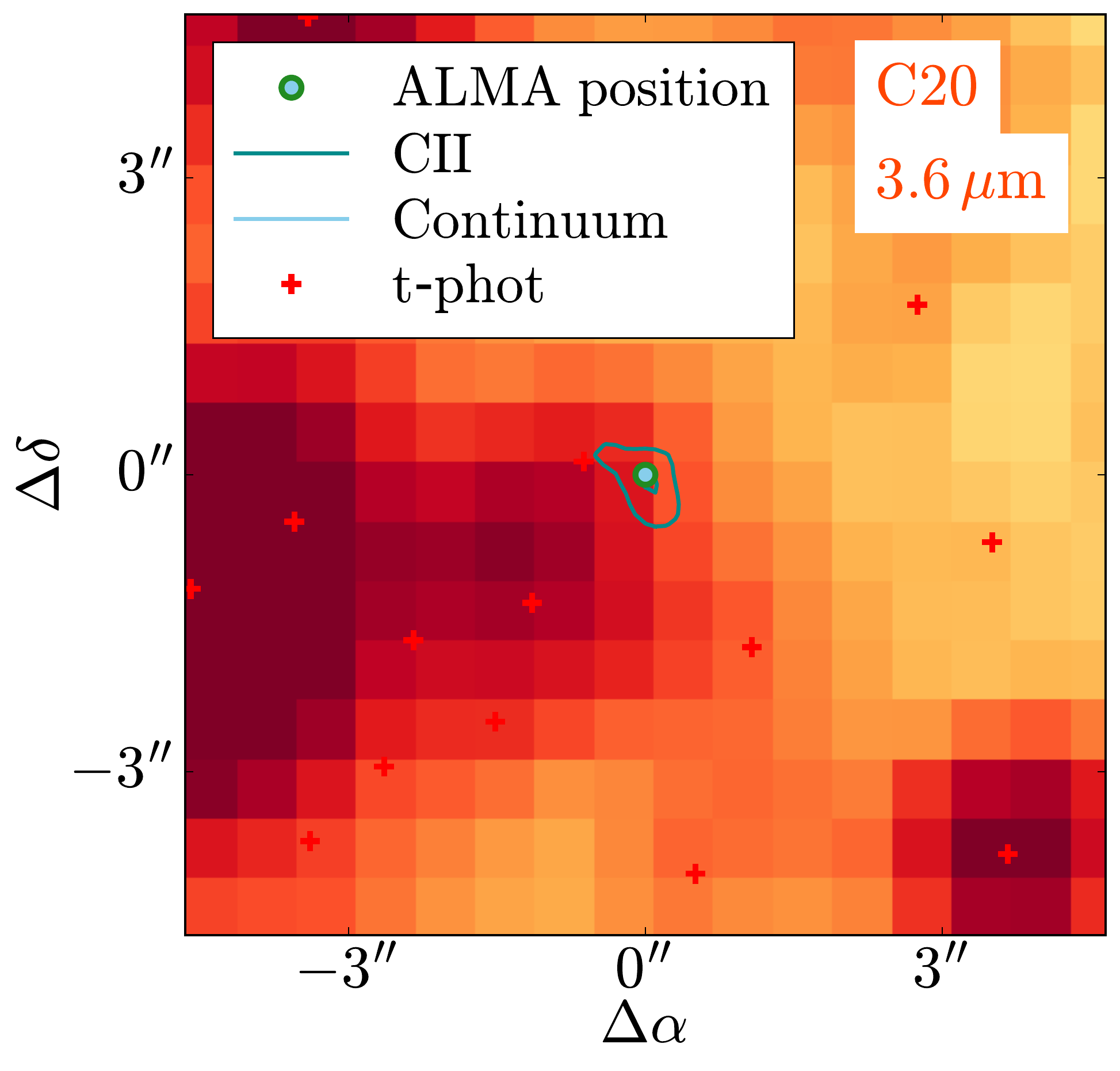}
\includegraphics[width=0.249\textwidth]{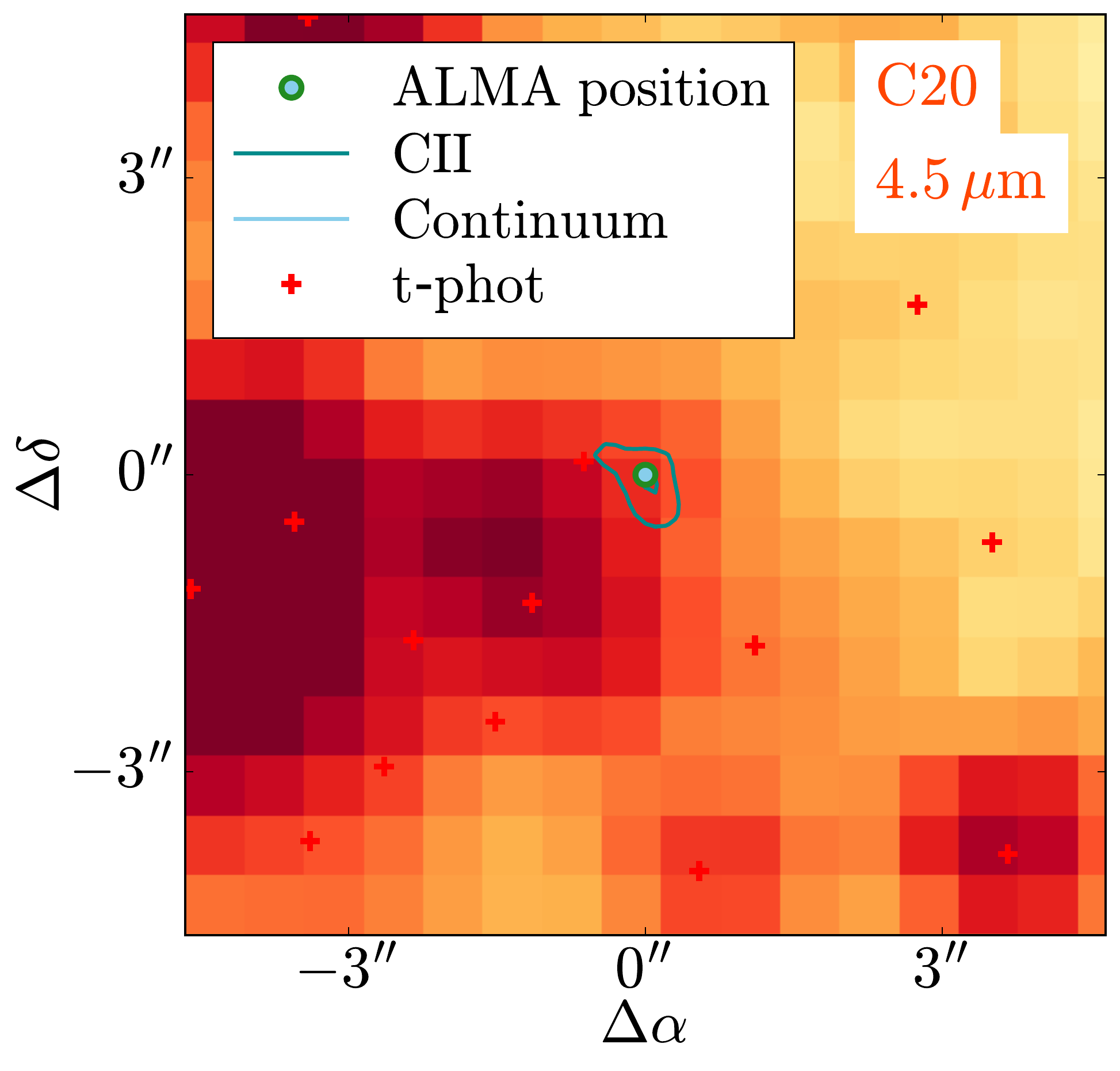}
\end{framed}
\end{subfigure}
\begin{subfigure}{0.85\textwidth}
\begin{framed}
\includegraphics[width=0.24\textwidth]{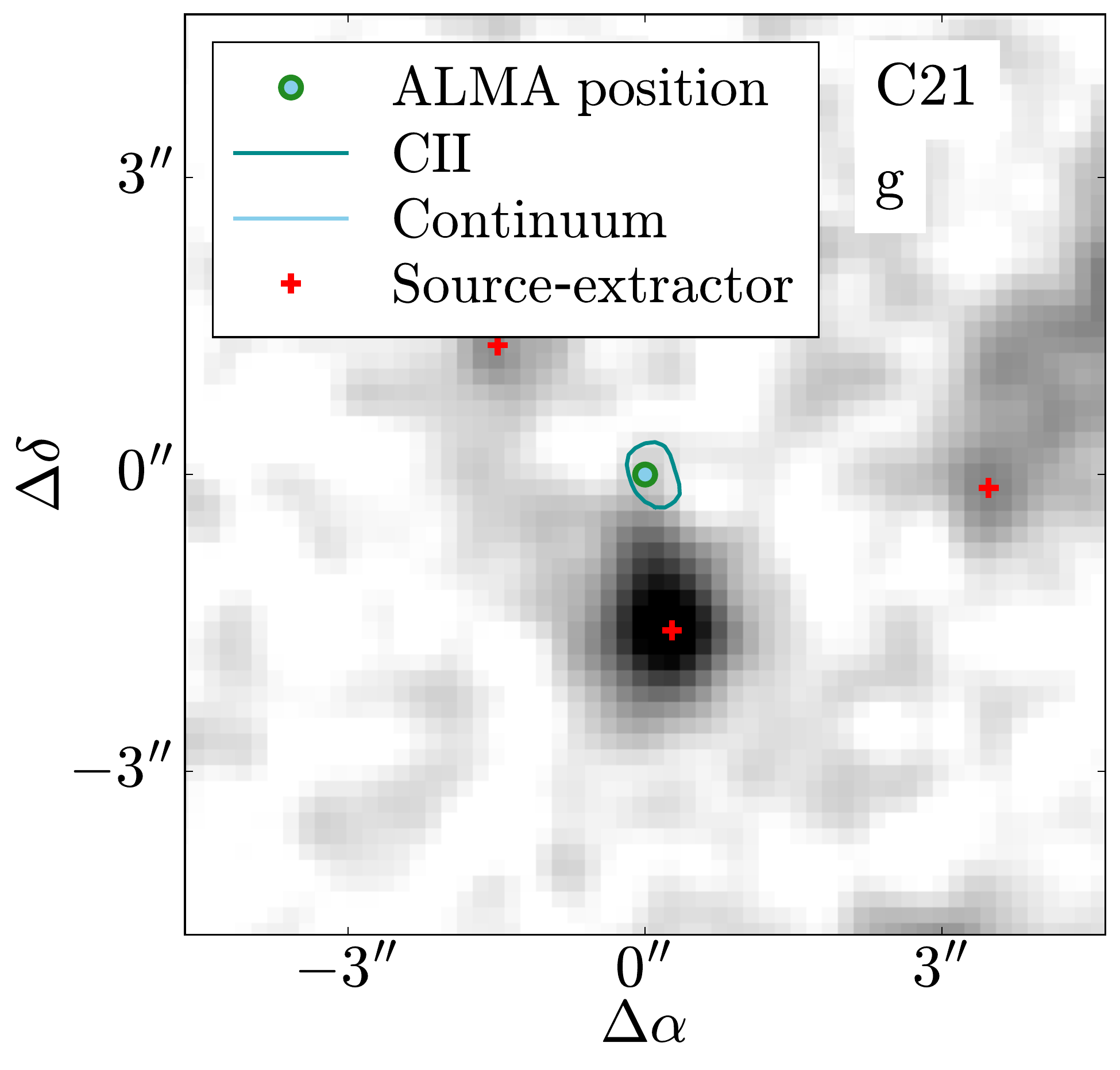}
\includegraphics[width=0.24\textwidth]{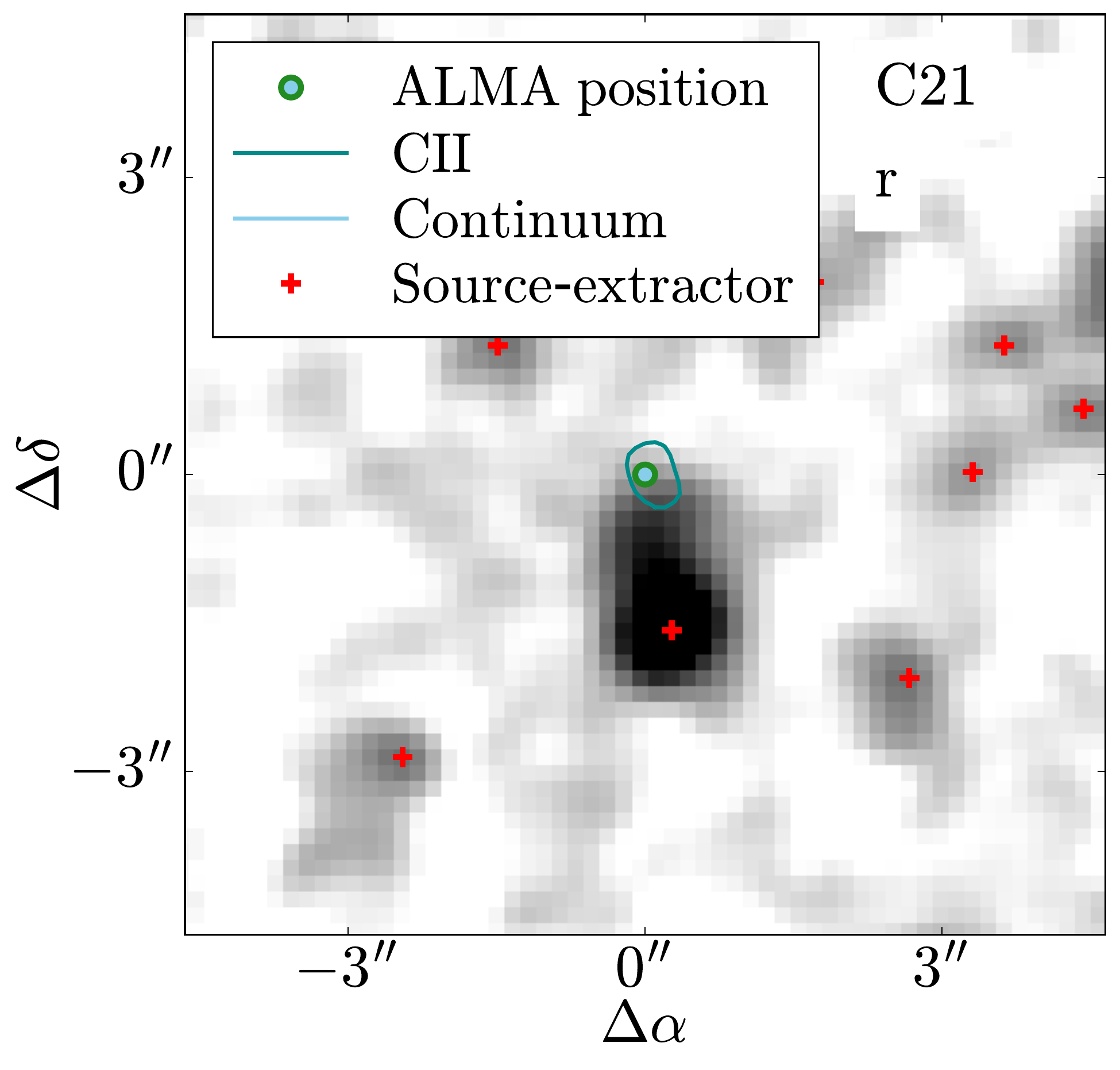}
\includegraphics[width=0.24\textwidth]{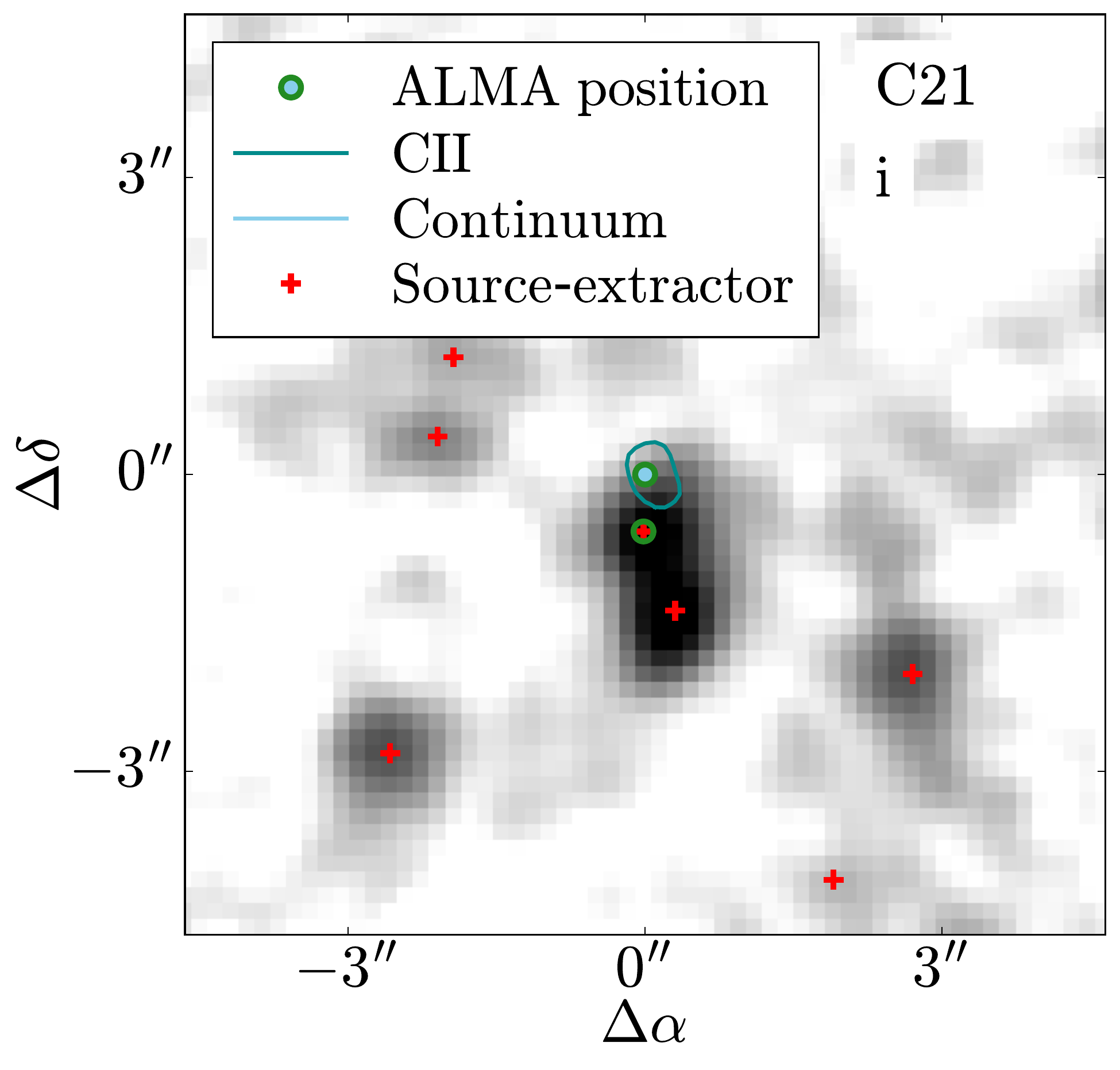}
\includegraphics[width=0.24\textwidth]{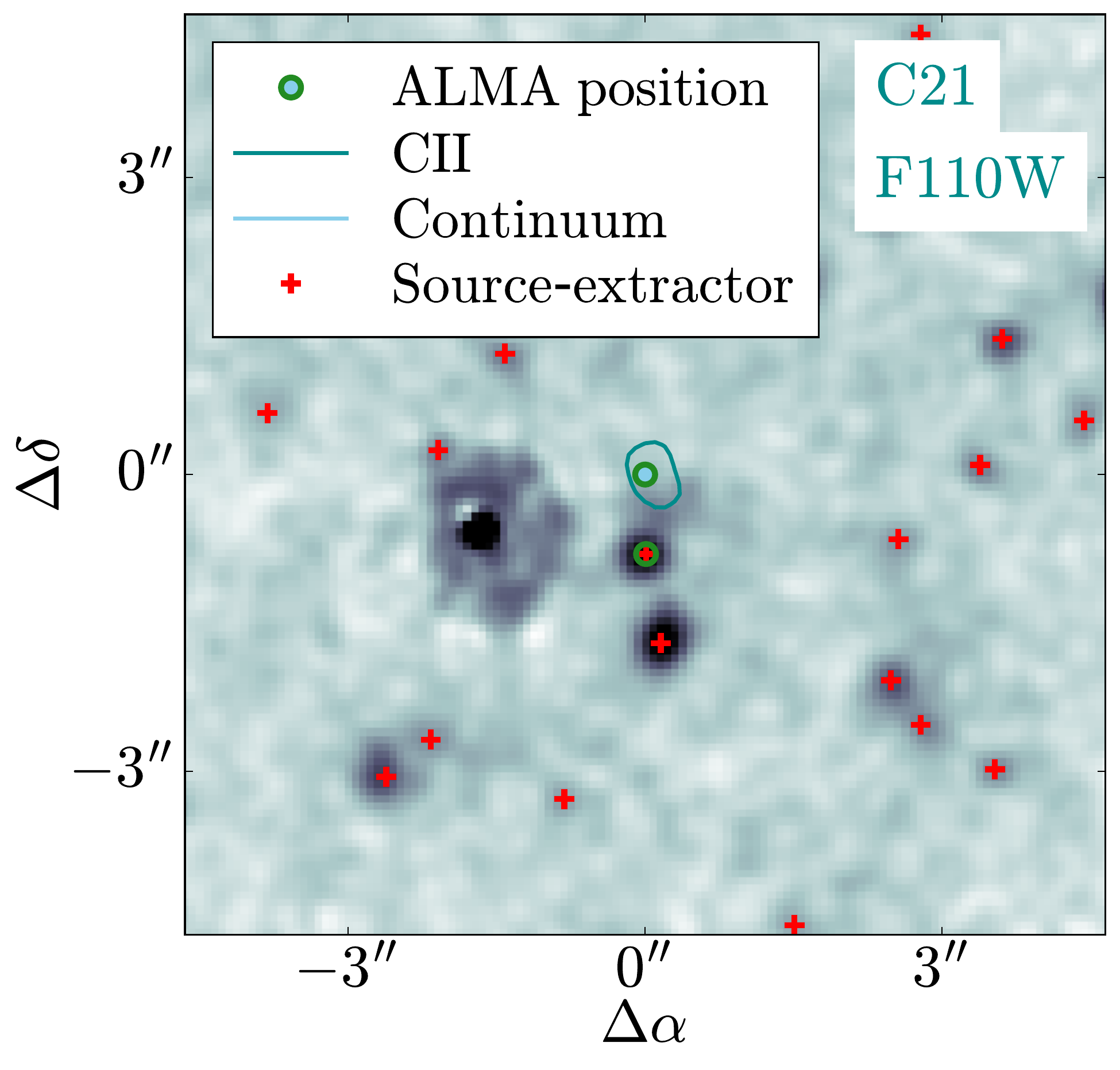}
\includegraphics[width=0.24\textwidth]{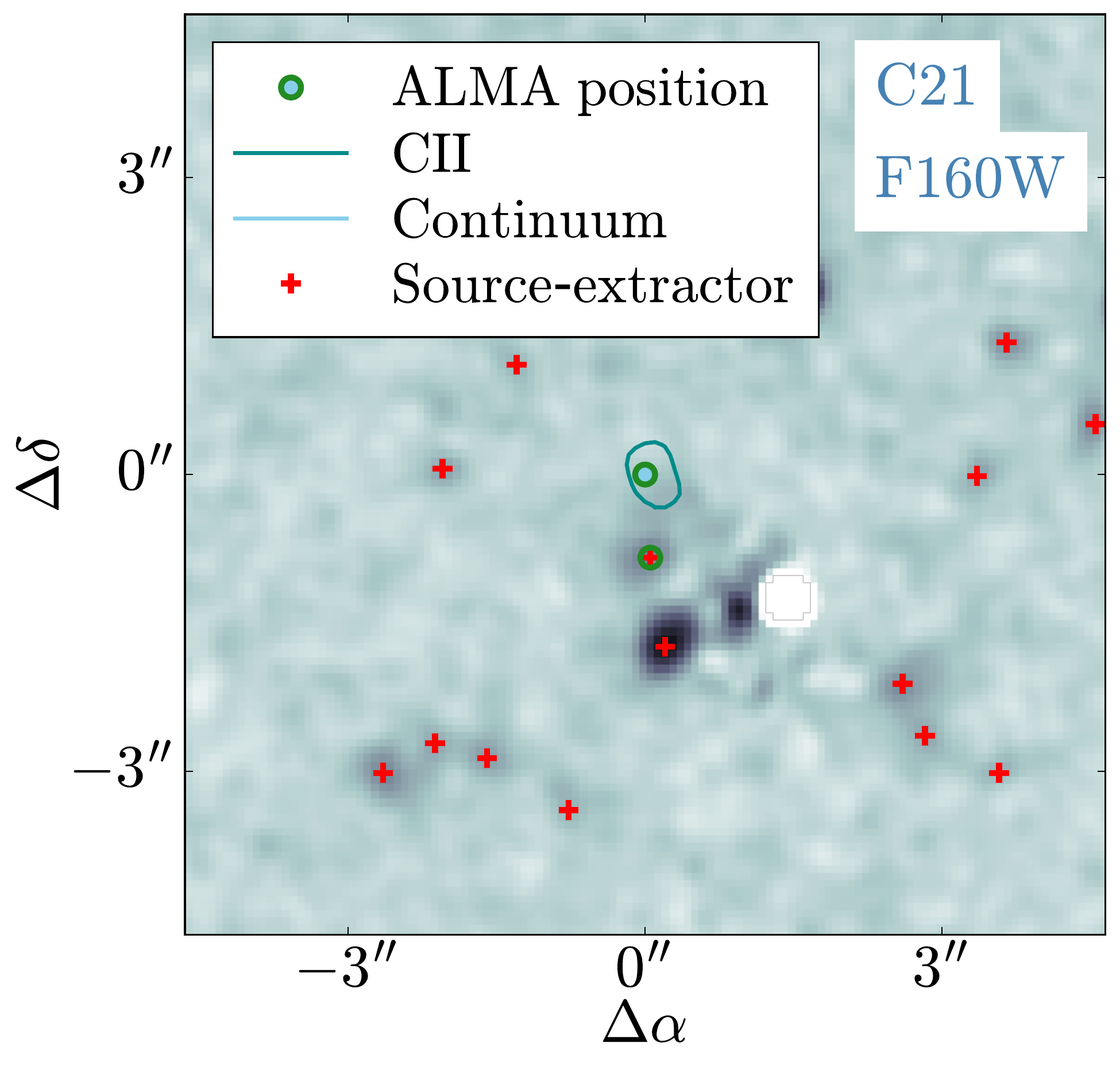}
\includegraphics[width=0.248\textwidth]{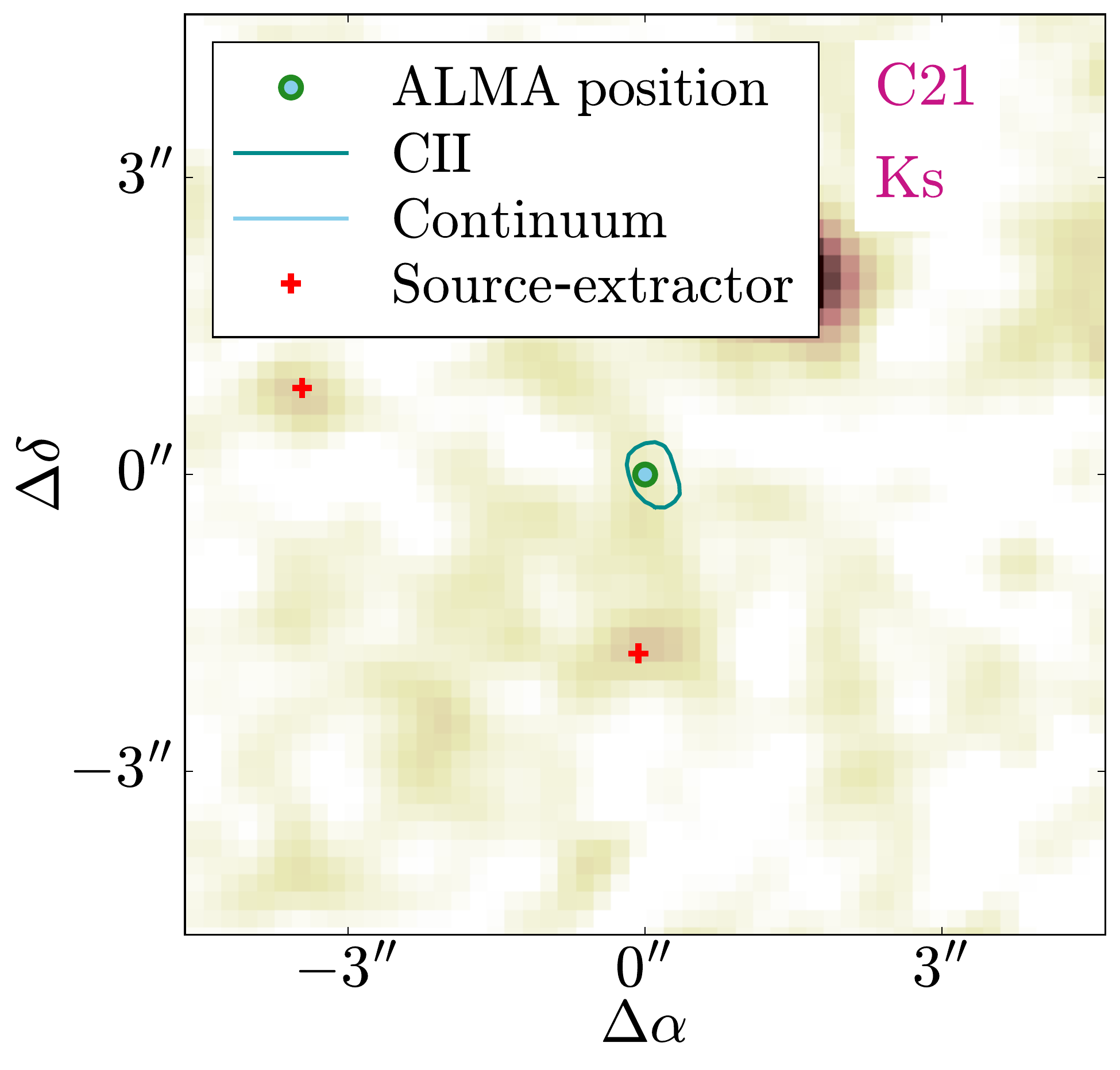}
\includegraphics[width=0.249\textwidth]{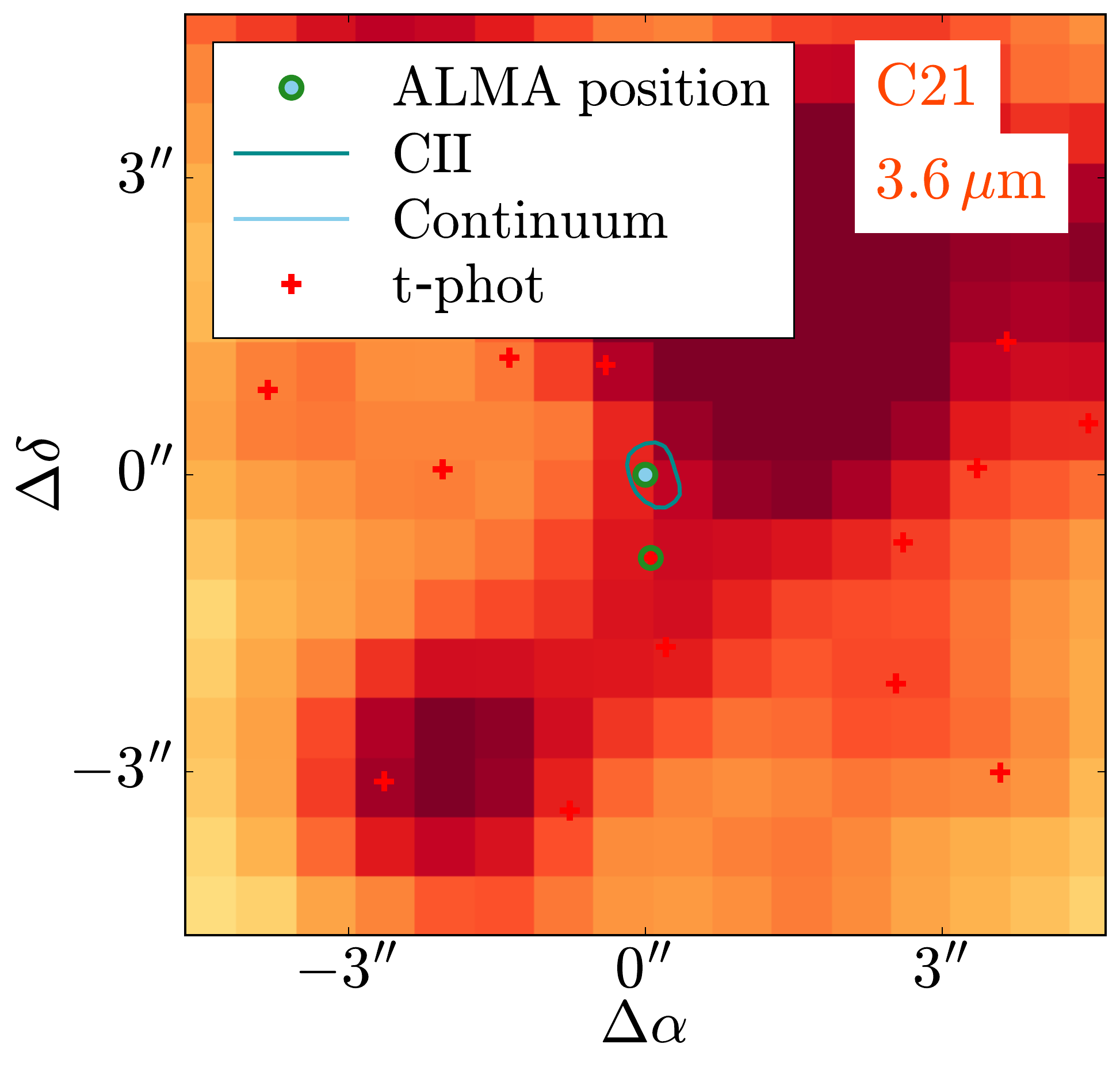}
\includegraphics[width=0.249\textwidth]{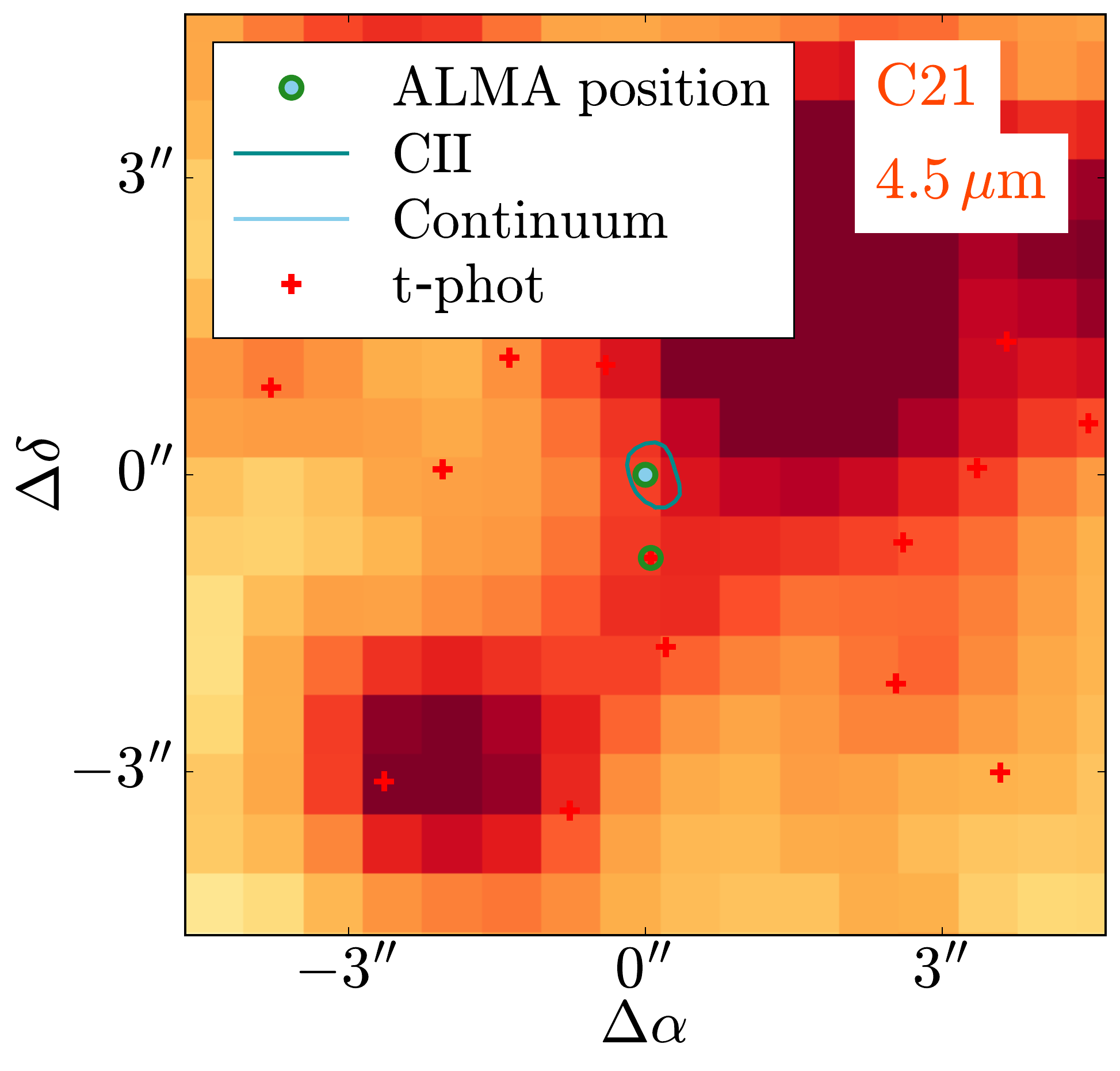}
\end{framed}
\end{subfigure}
\caption{}
\end{figure*}
\renewcommand{\thefigure}{\arabic{figure}}

\renewcommand{\thefigure}{B\arabic{figure} (Cont.)}
\addtocounter{figure}{-1}
\begin{figure*}
\begin{subfigure}{0.85\textwidth}
\begin{framed}
\includegraphics[width=0.24\textwidth]{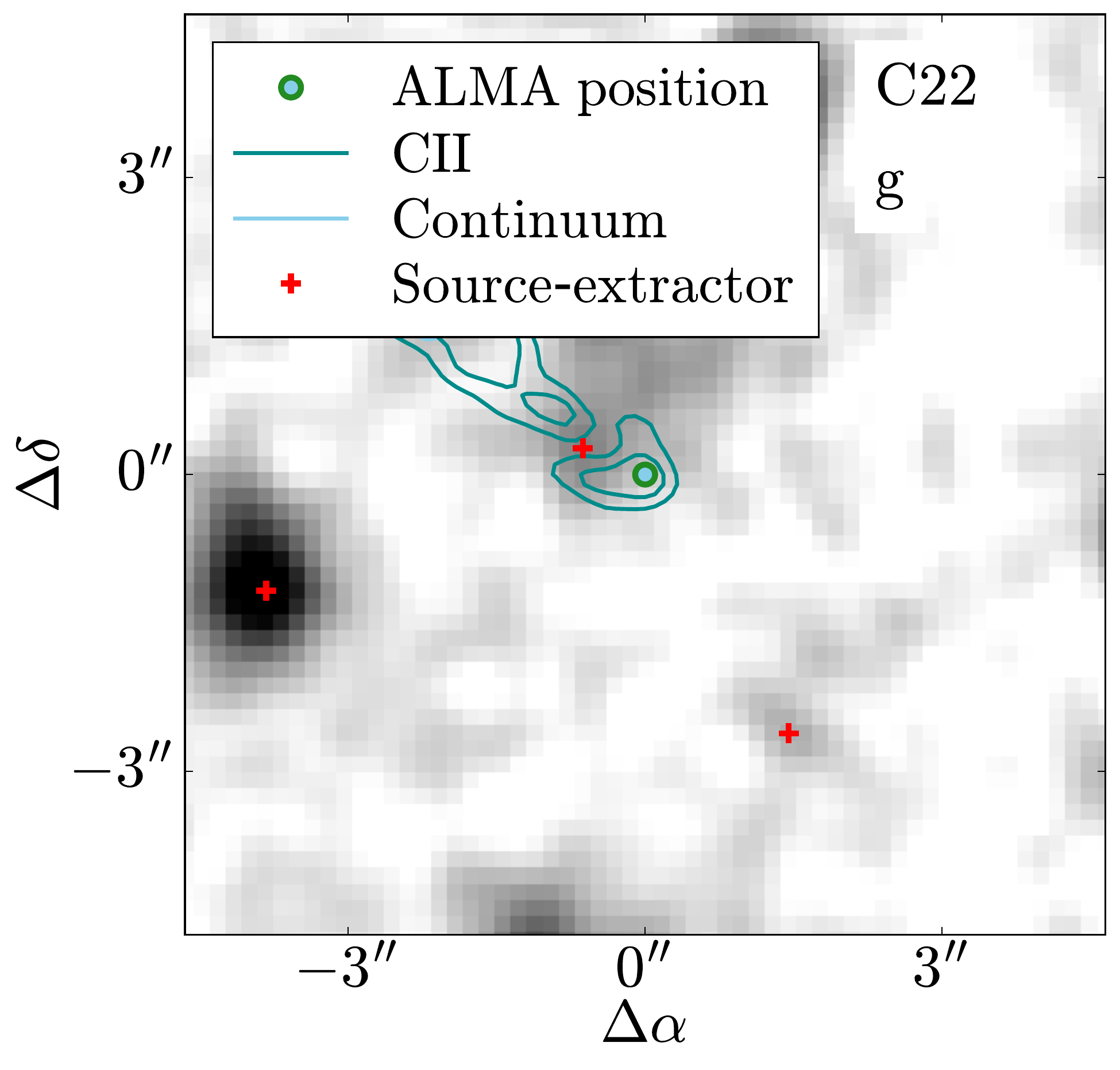}
\includegraphics[width=0.24\textwidth]{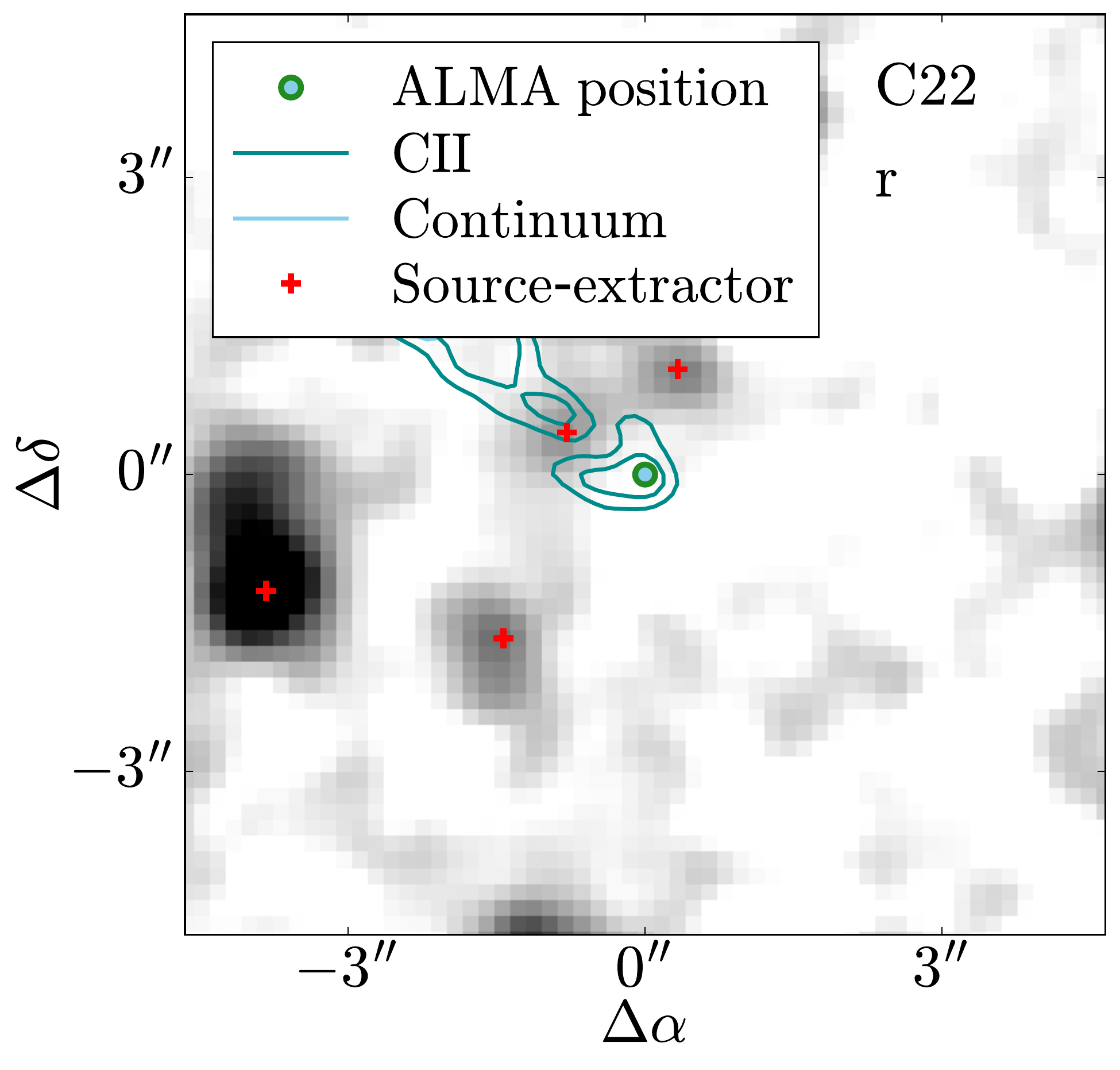}
\includegraphics[width=0.24\textwidth]{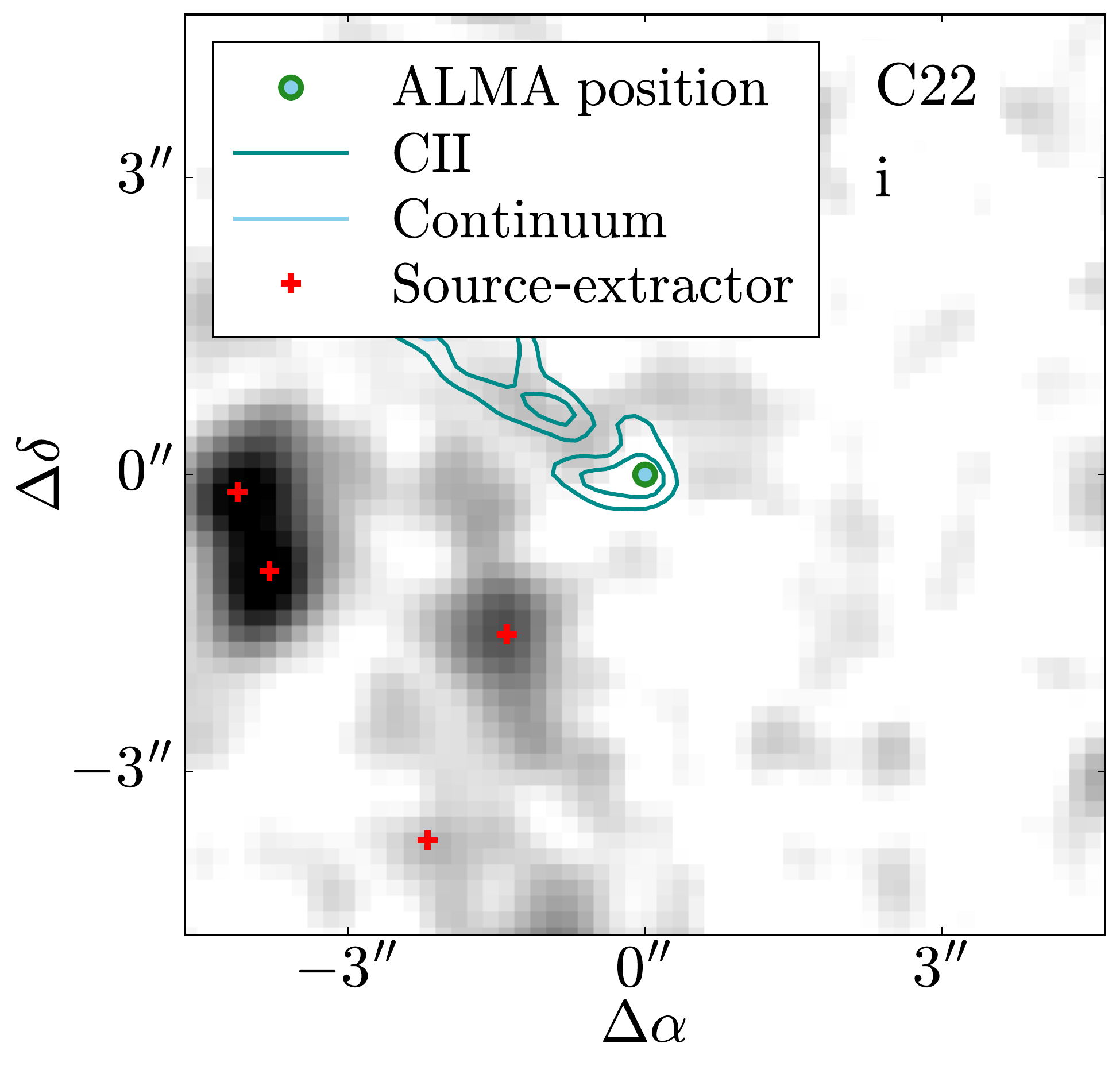}
\includegraphics[width=0.24\textwidth]{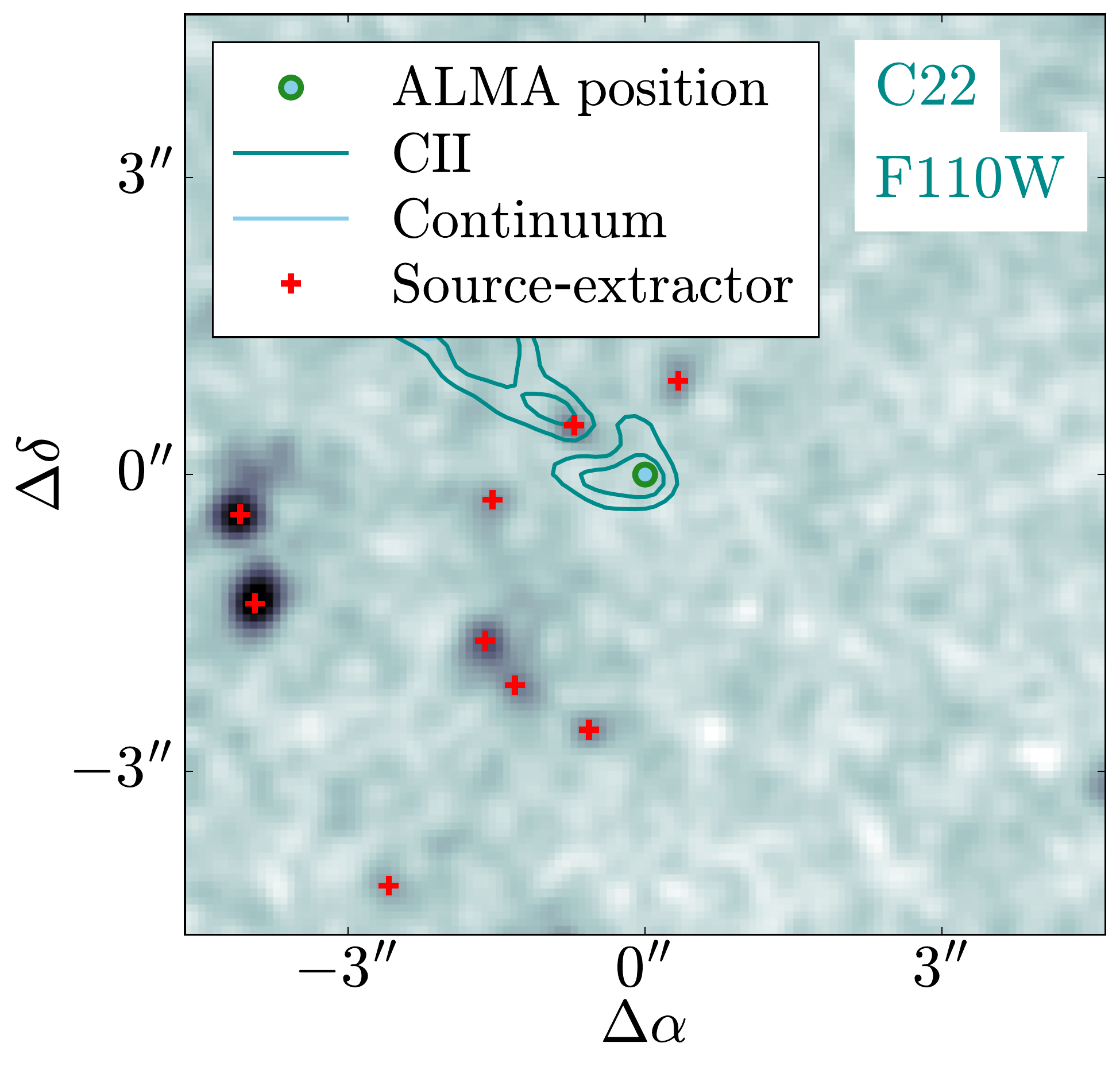}
\includegraphics[width=0.24\textwidth]{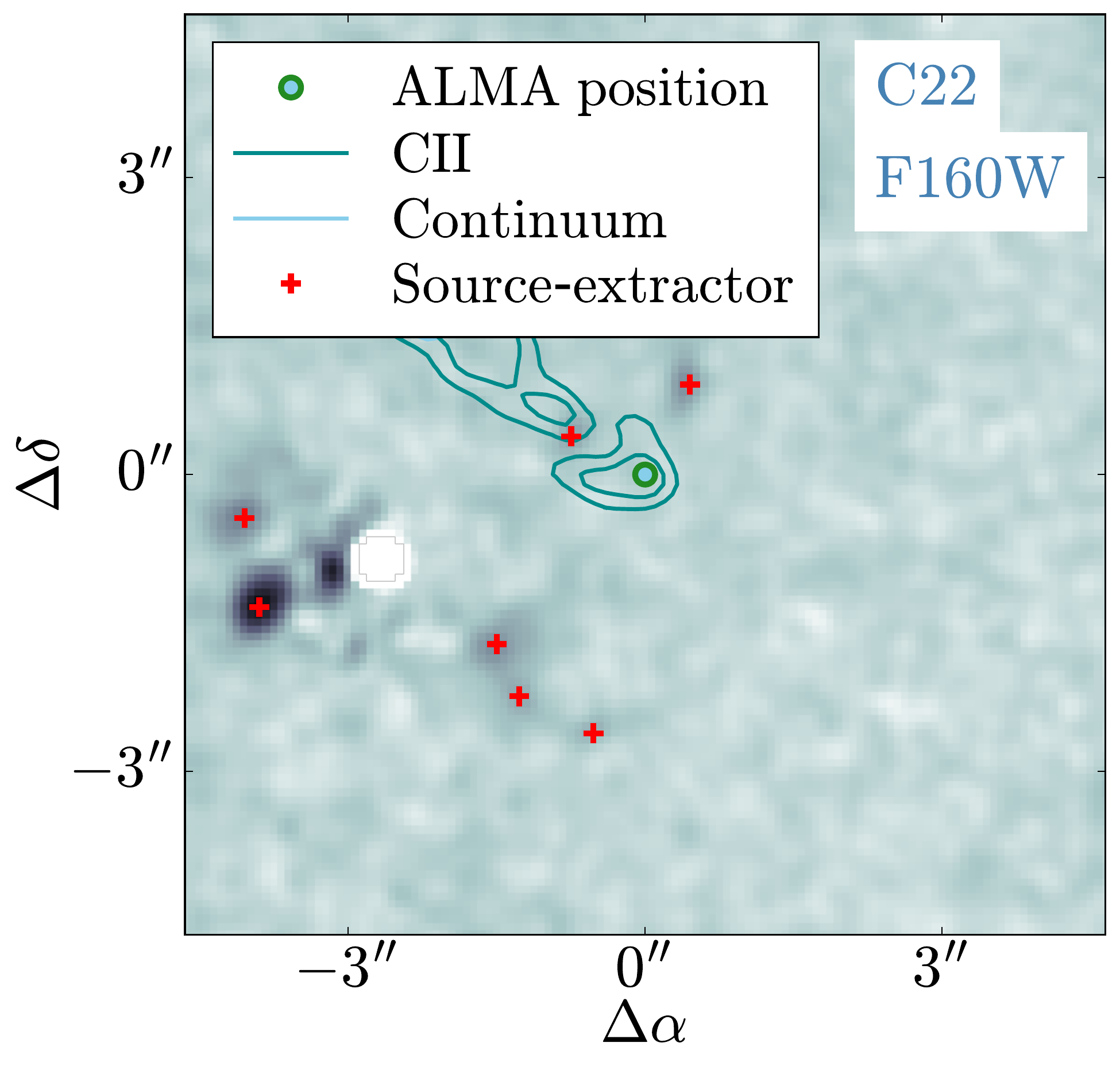}
\includegraphics[width=0.248\textwidth]{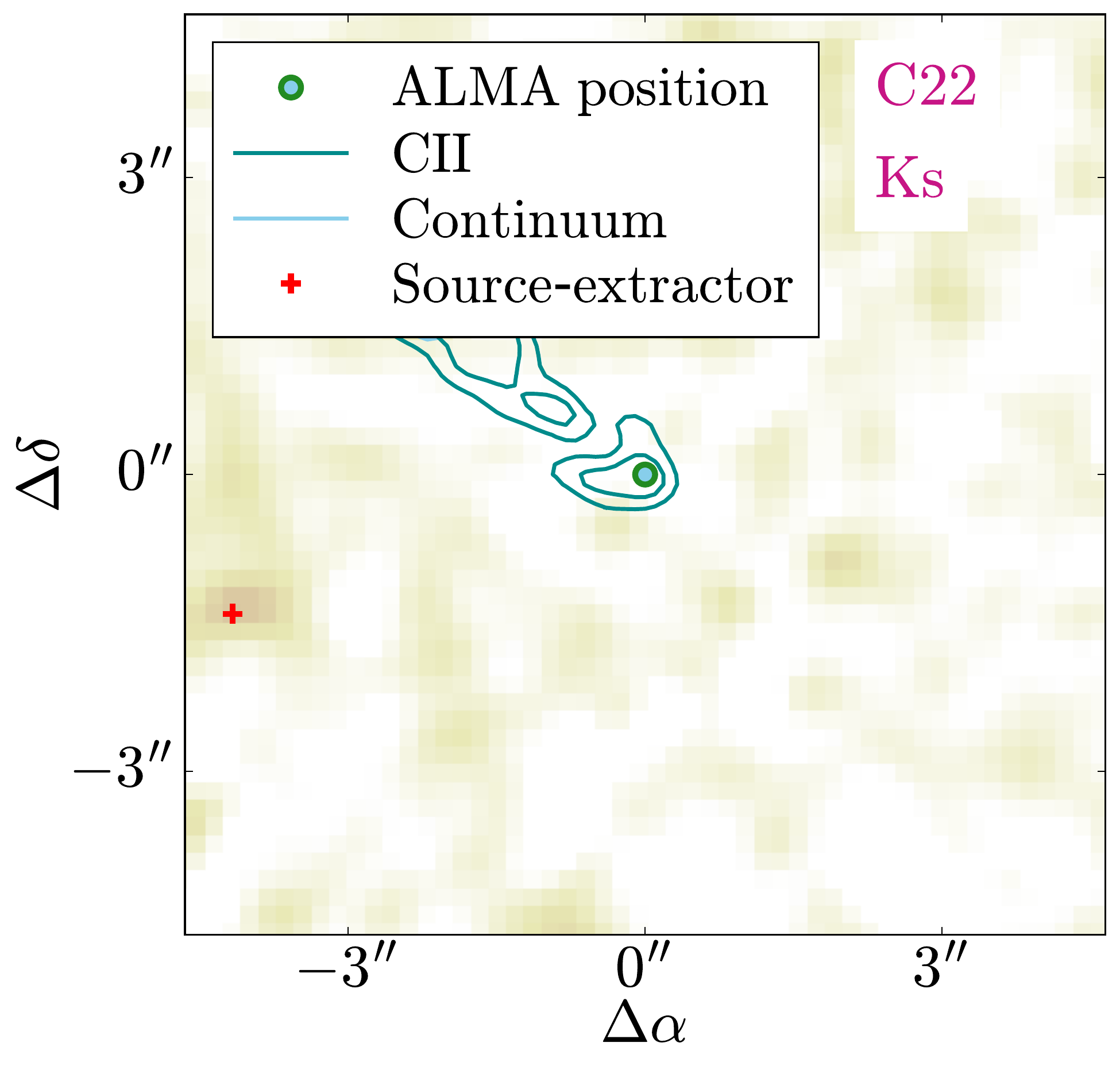}
\includegraphics[width=0.249\textwidth]{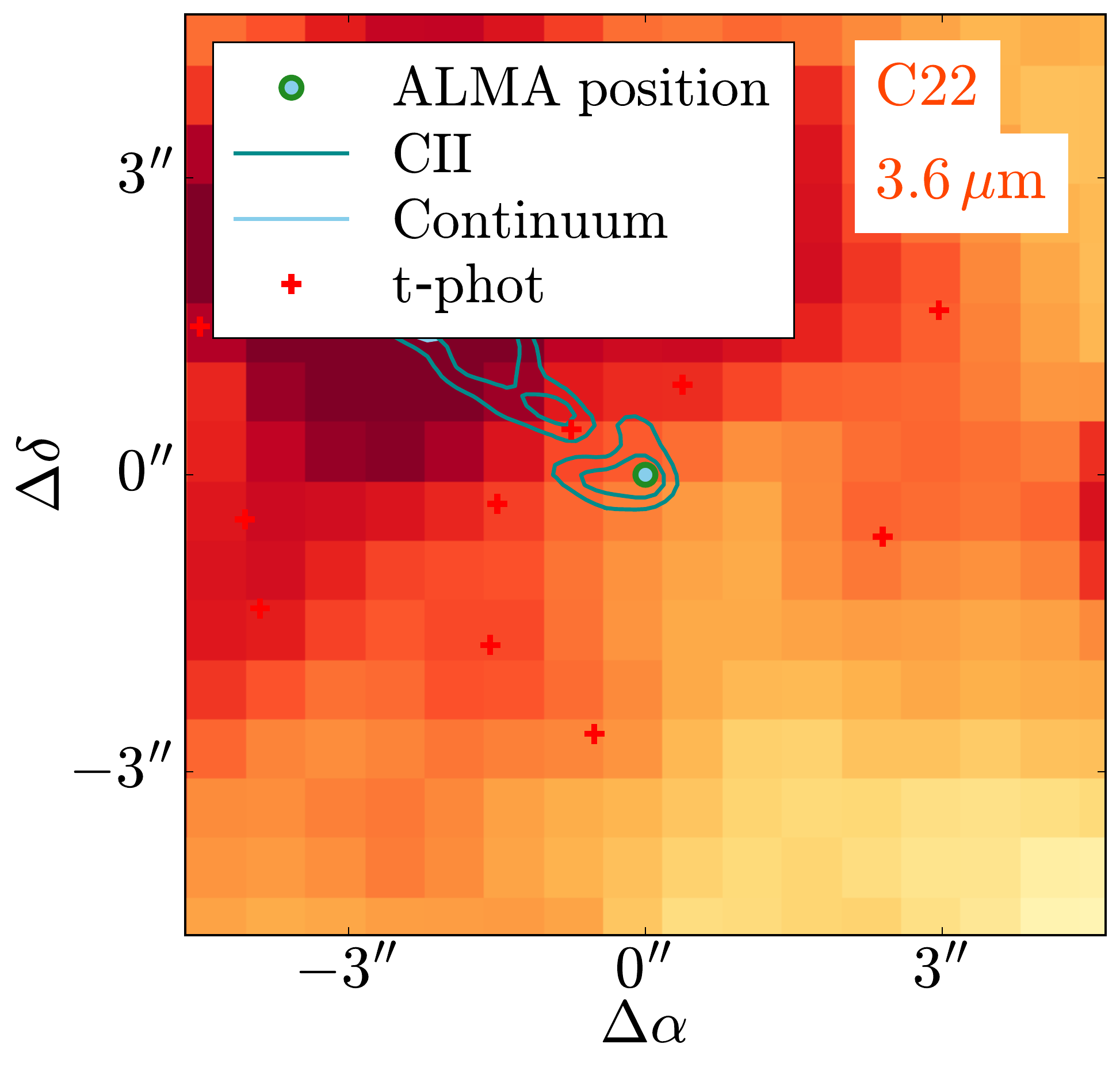}
\includegraphics[width=0.249\textwidth]{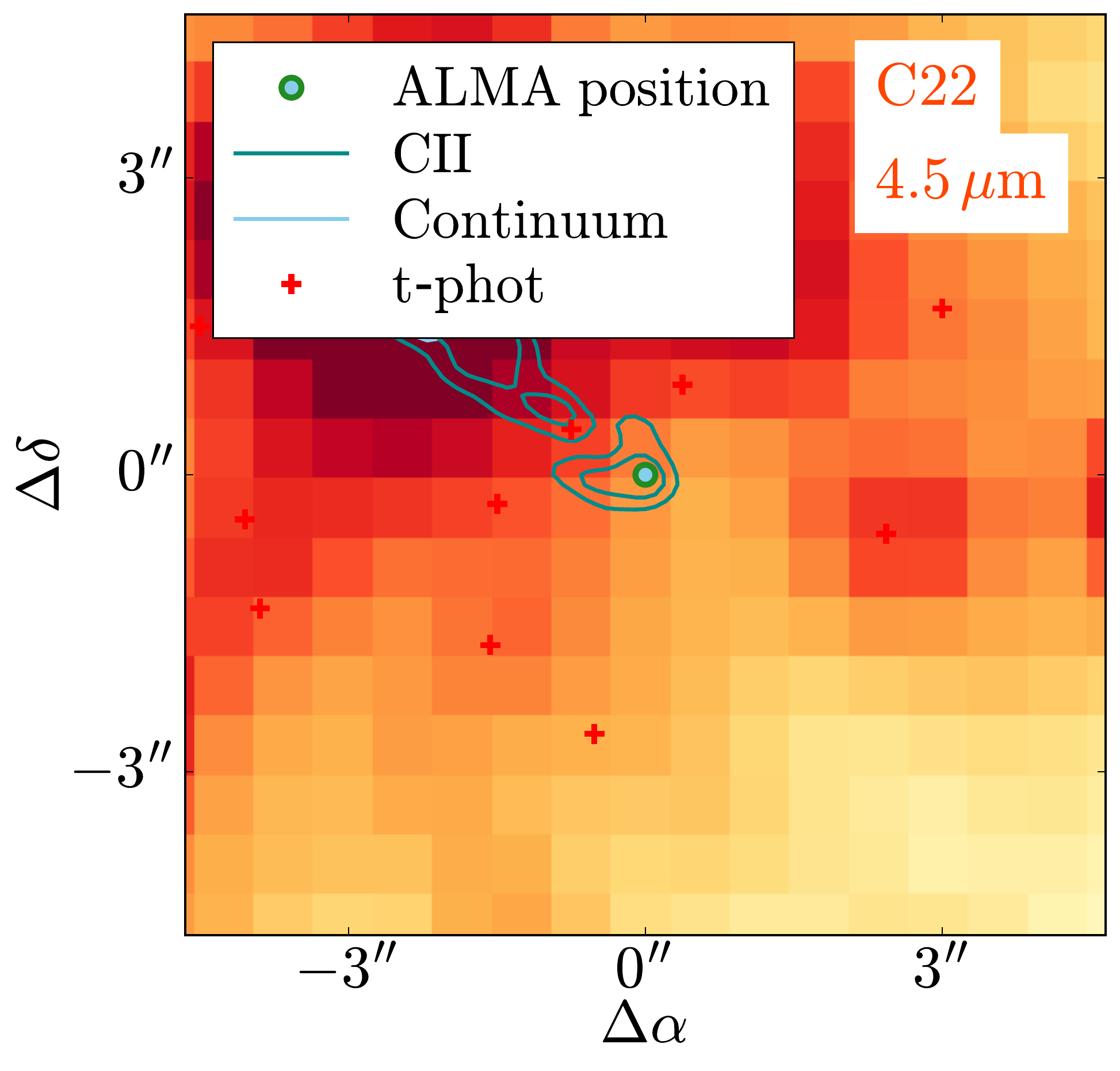}
\end{framed}
\end{subfigure}
\begin{subfigure}{0.85\textwidth}
\begin{framed}
\includegraphics[width=0.24\textwidth]{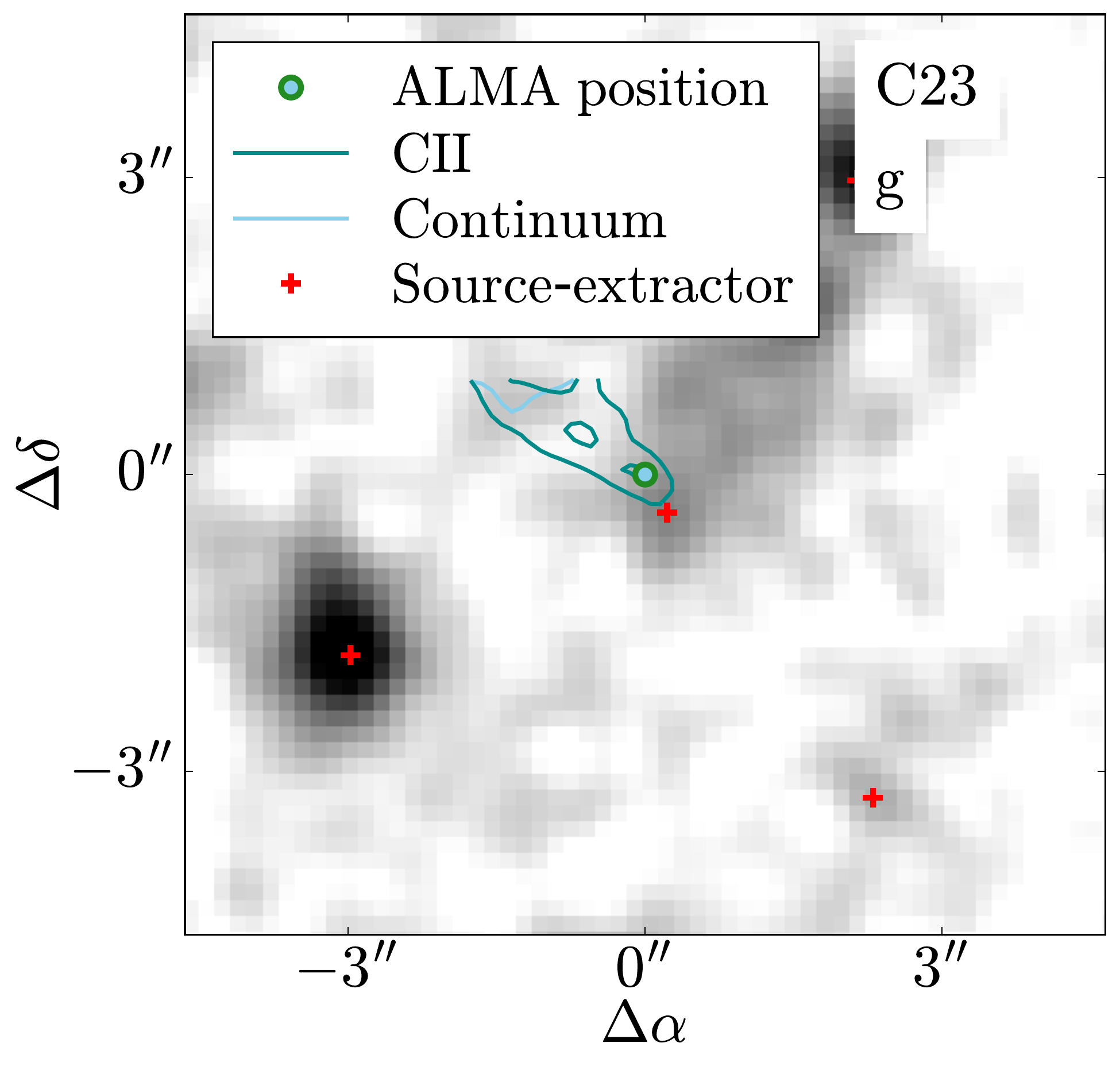}
\includegraphics[width=0.24\textwidth]{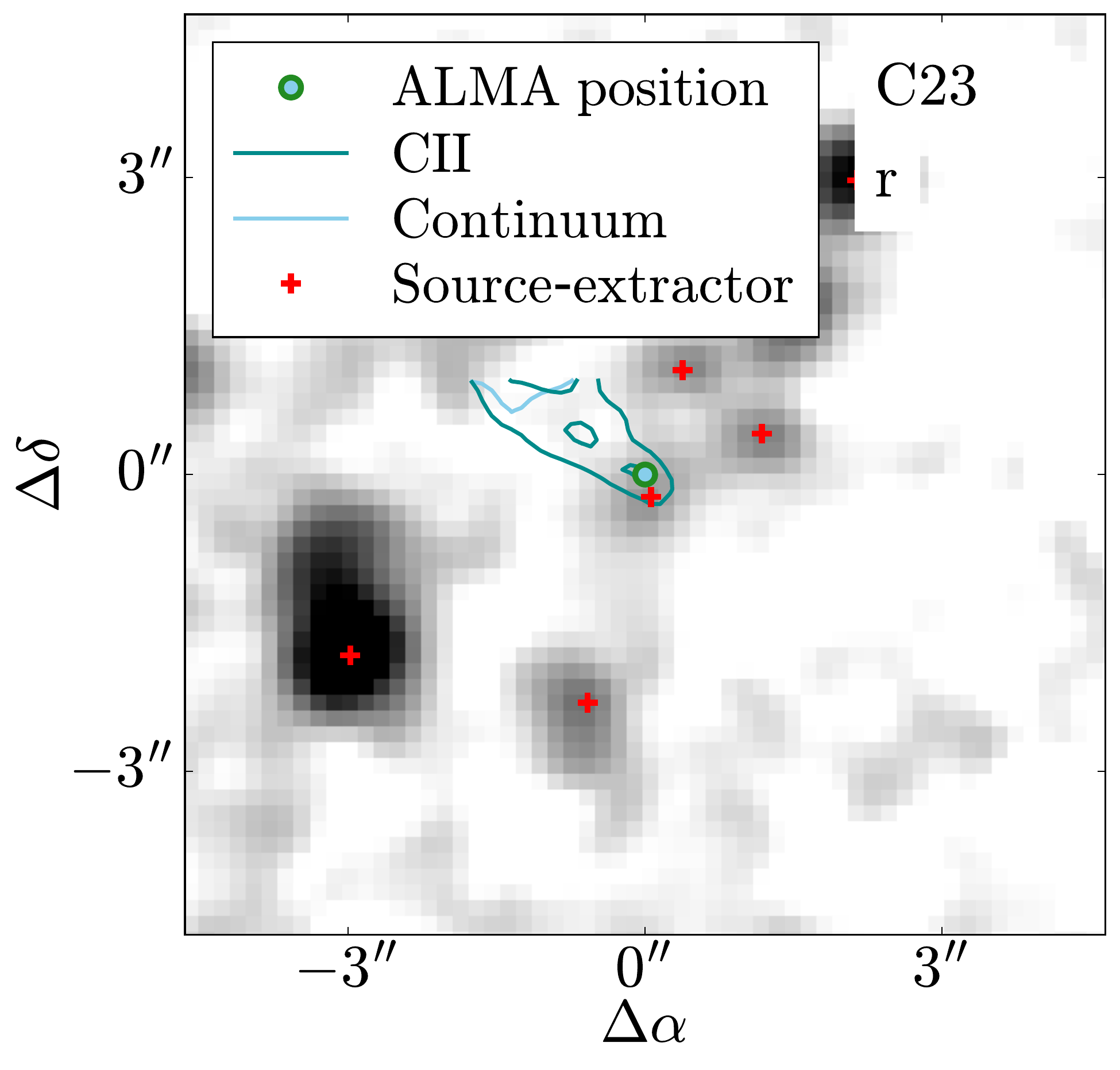}
\includegraphics[width=0.24\textwidth]{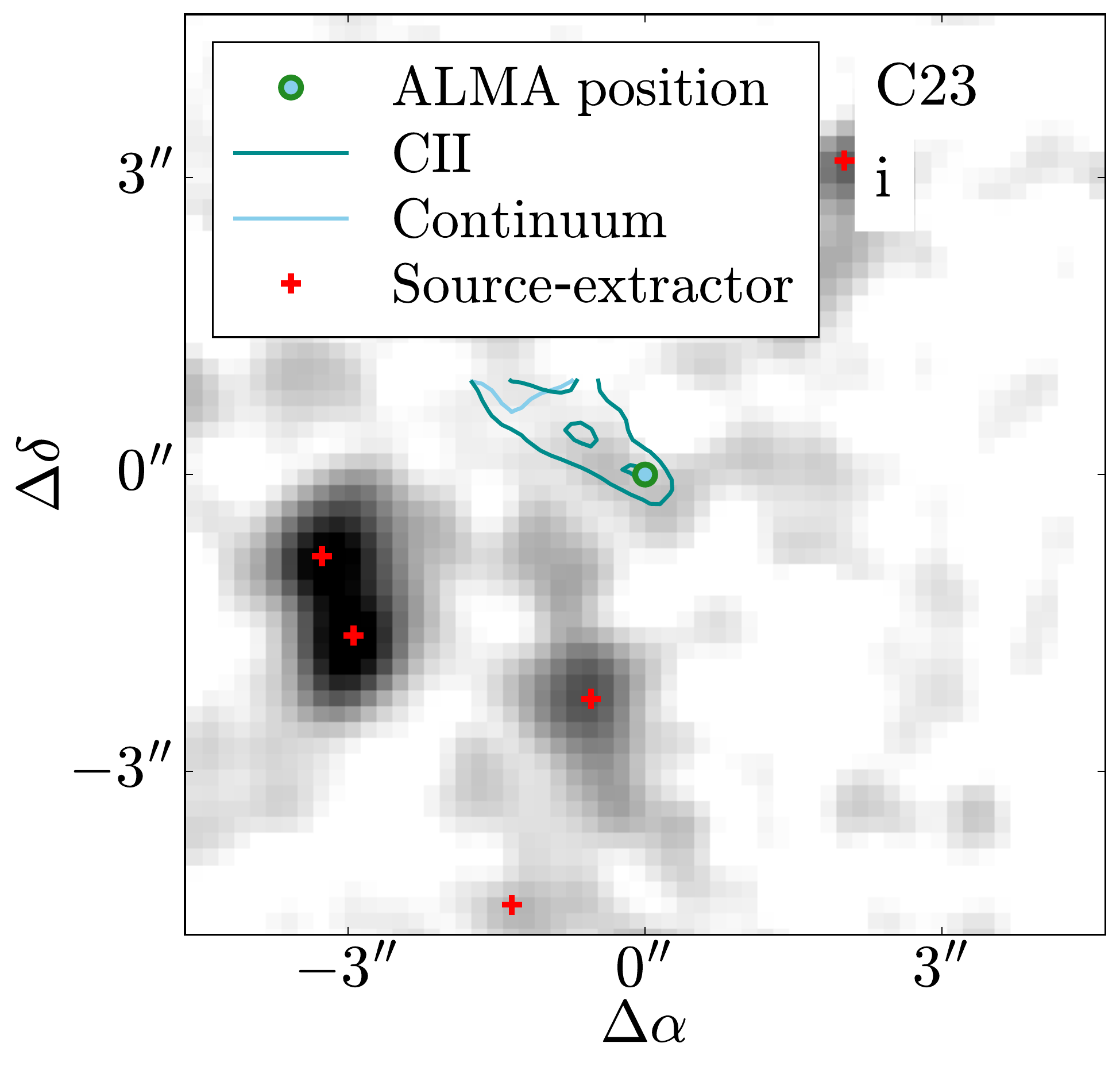}
\includegraphics[width=0.24\textwidth]{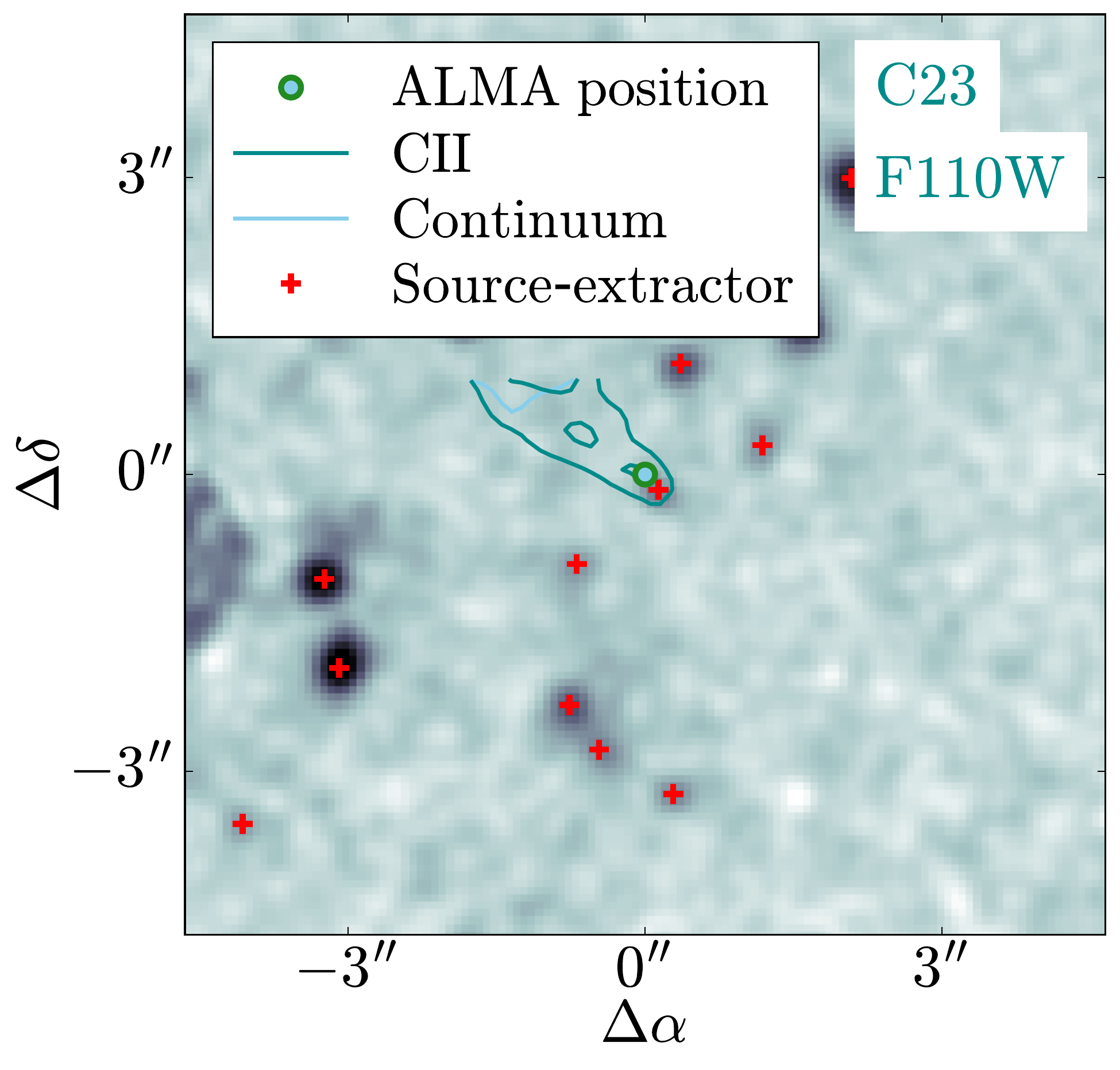}
\includegraphics[width=0.24\textwidth]{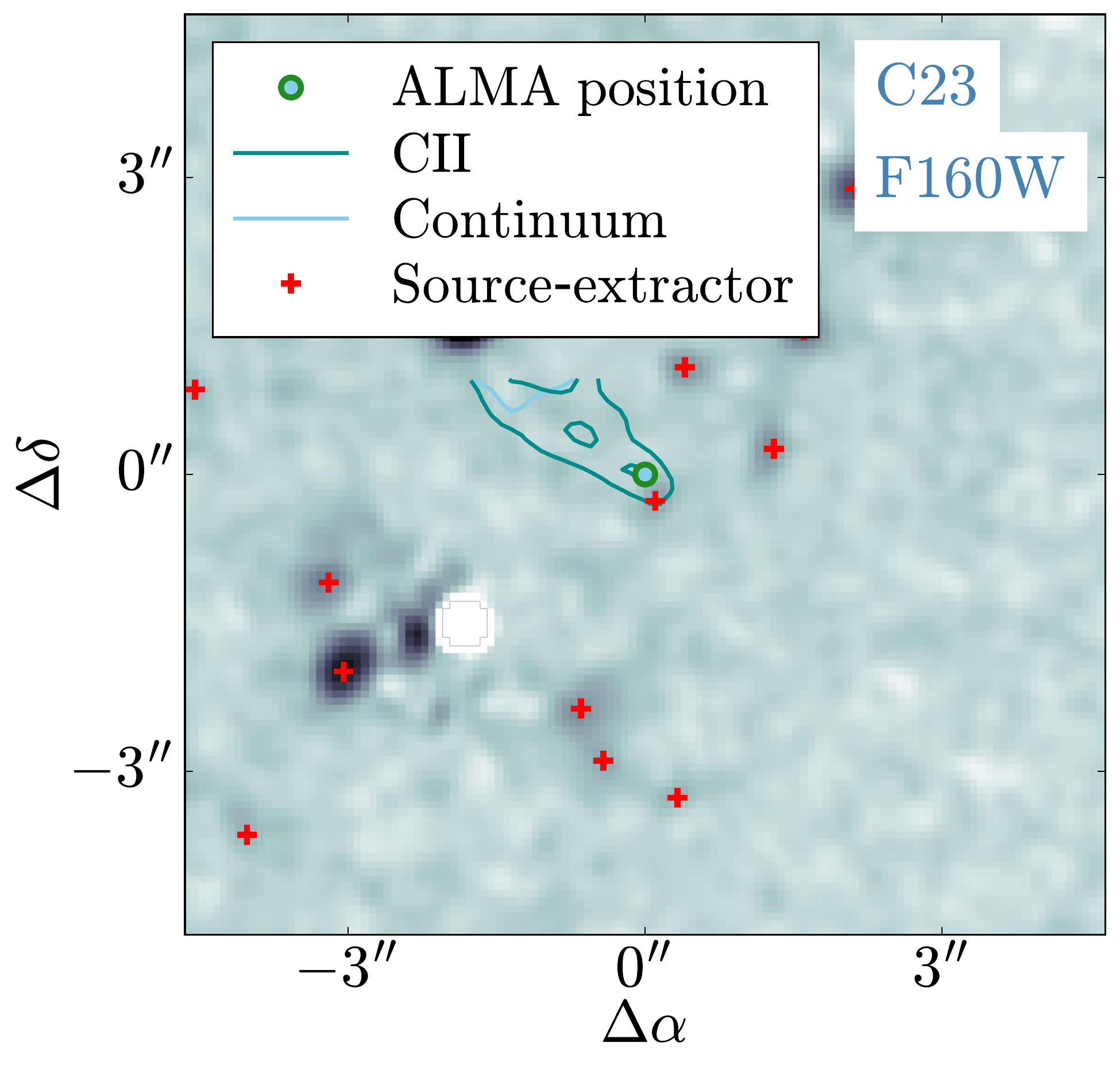}
\includegraphics[width=0.248\textwidth]{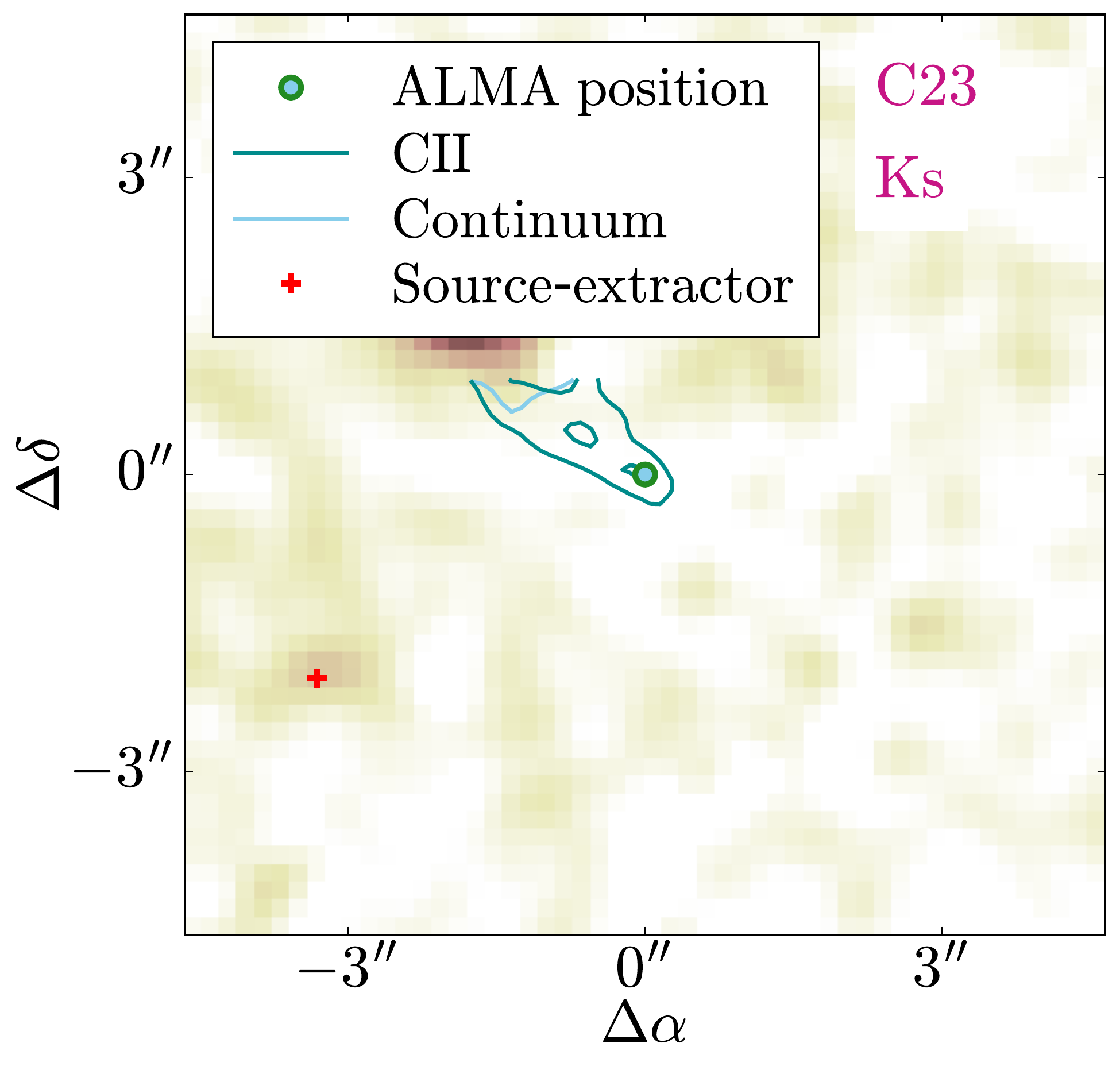}
\includegraphics[width=0.249\textwidth]{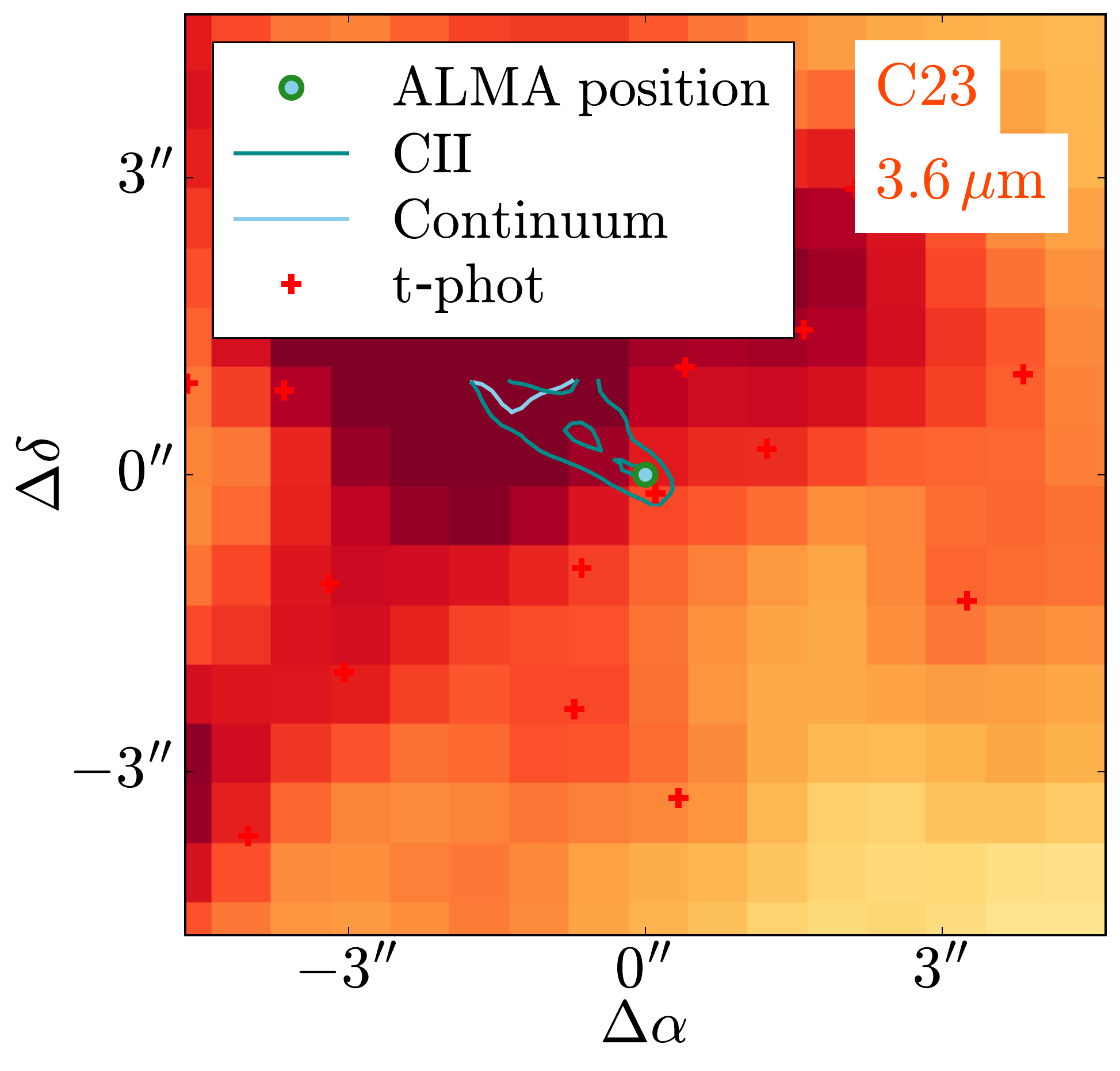}
\includegraphics[width=0.249\textwidth]{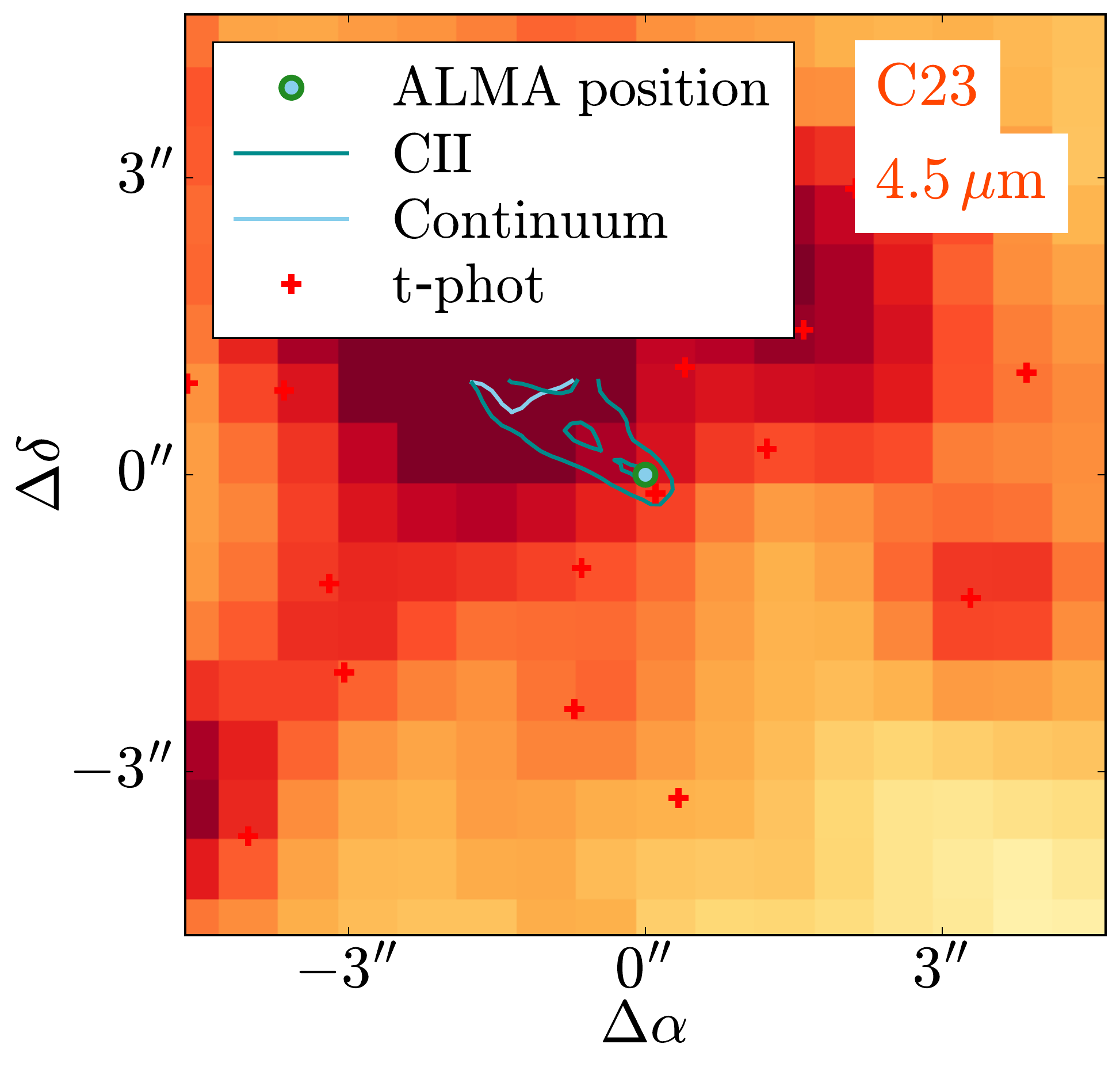}
\end{framed}
\end{subfigure}
\begin{subfigure}{0.85\textwidth}
\begin{framed}
\includegraphics[width=0.24\textwidth]{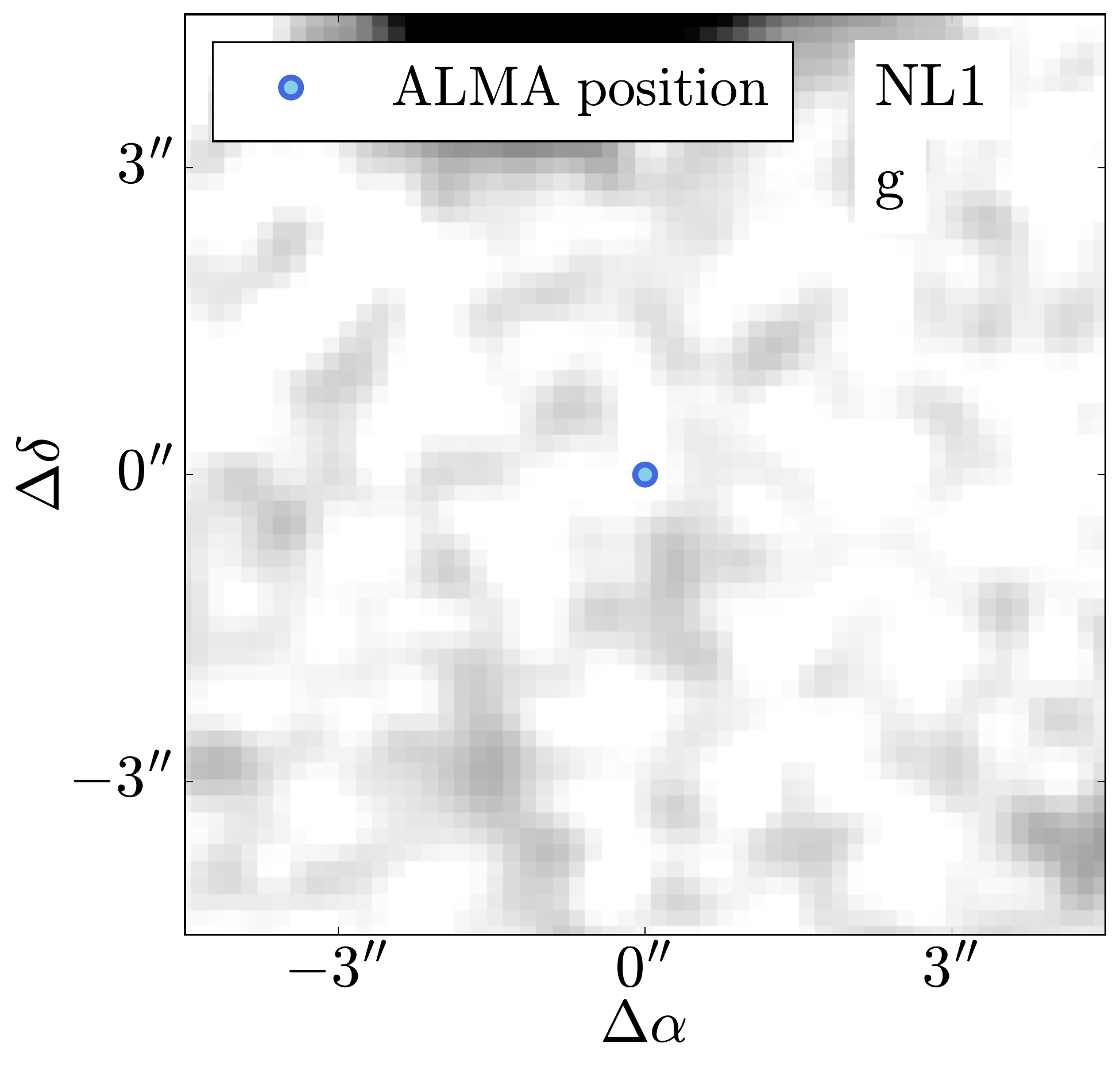}
\includegraphics[width=0.24\textwidth]{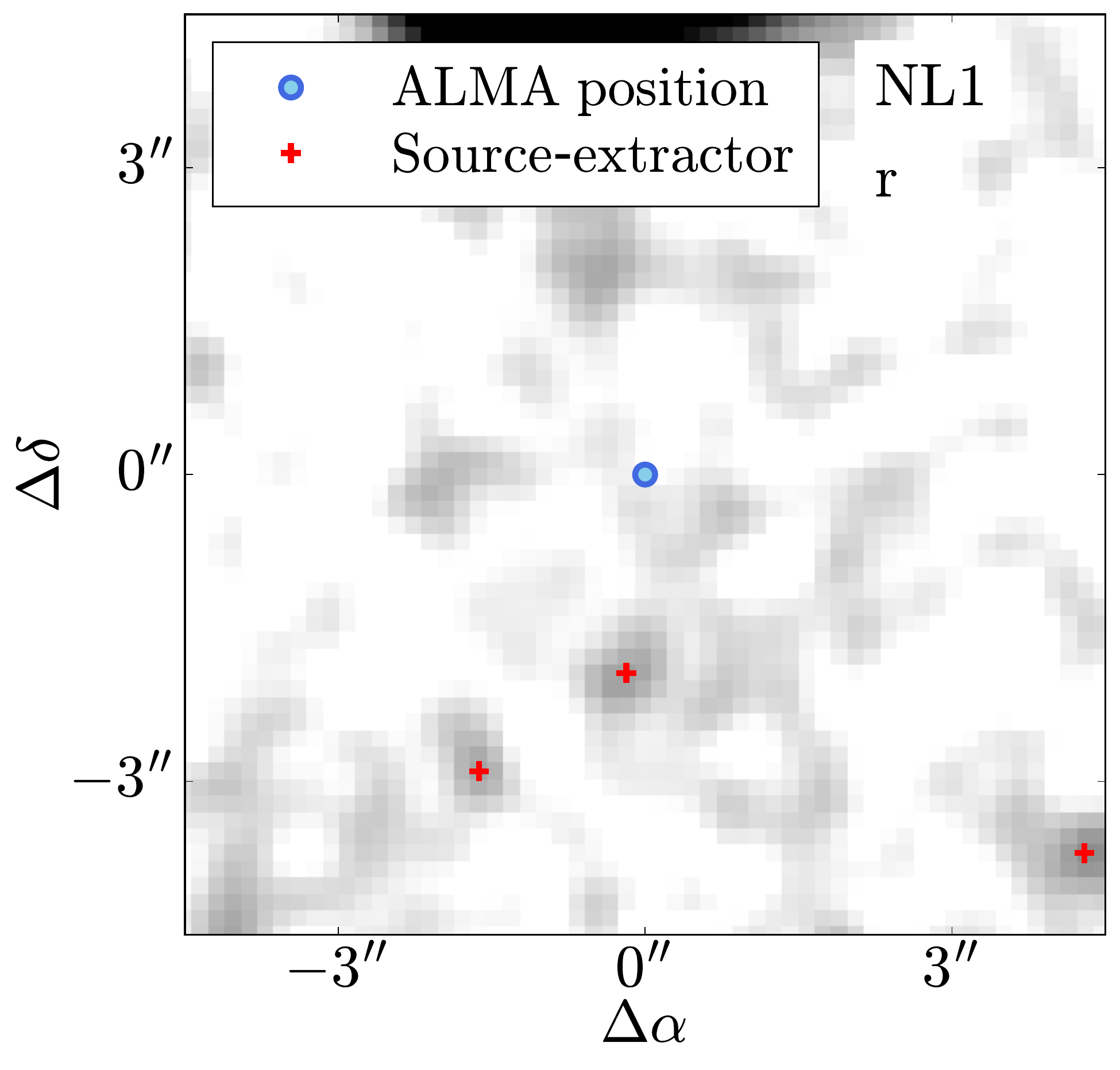}
\includegraphics[width=0.24\textwidth]{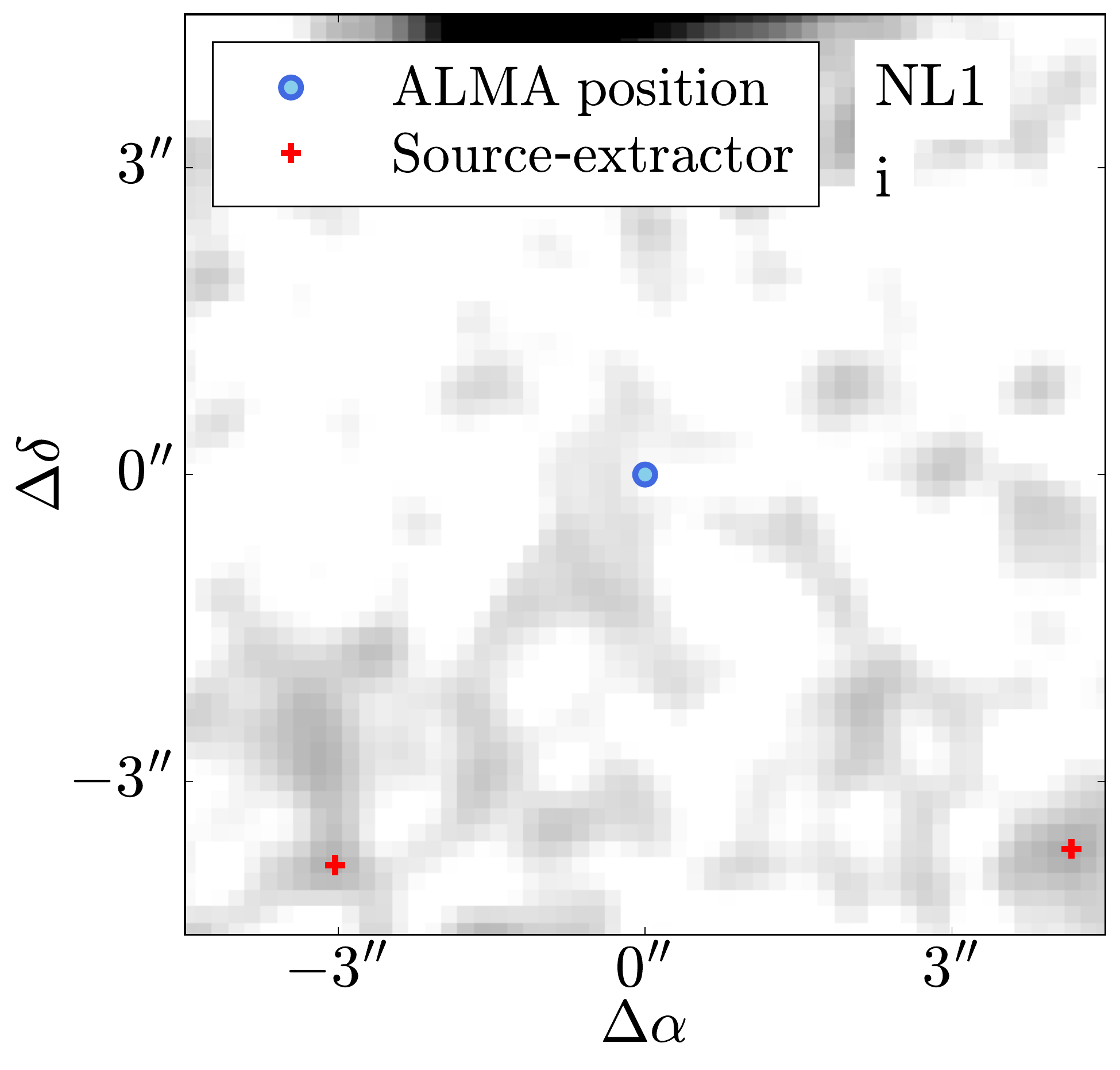}
\includegraphics[width=0.24\textwidth]{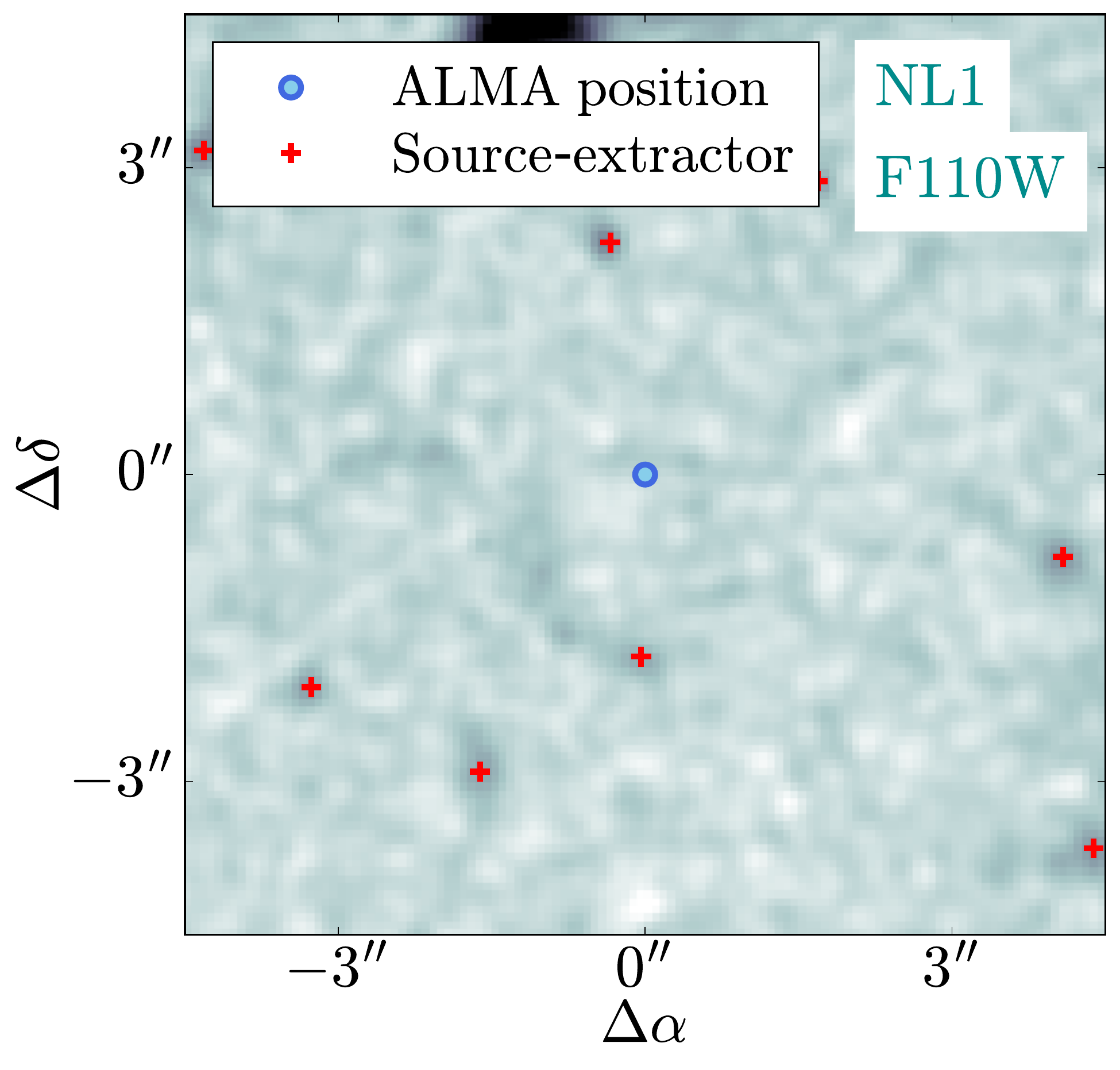}
\includegraphics[width=0.24\textwidth]{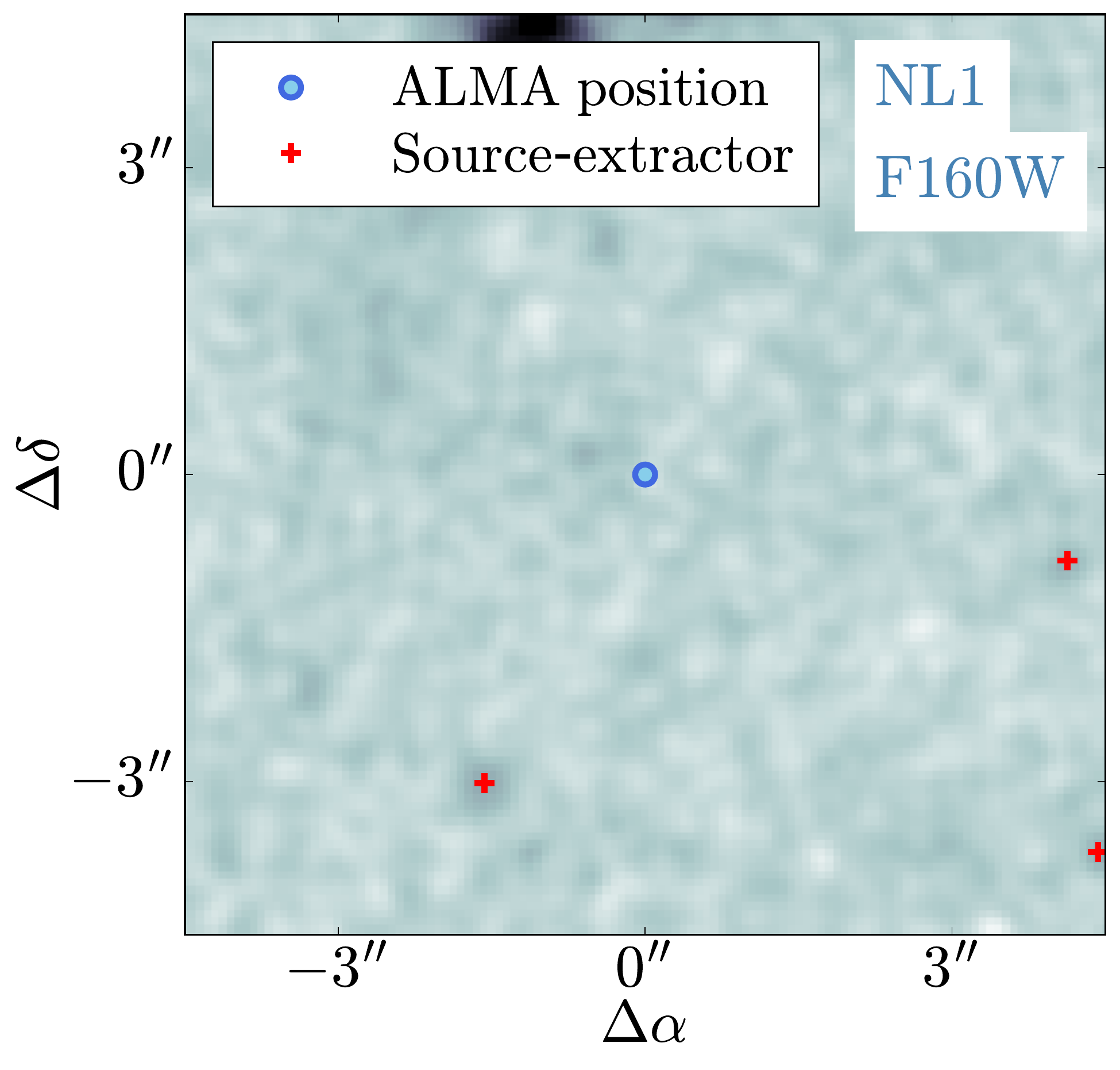}
\includegraphics[width=0.248\textwidth]{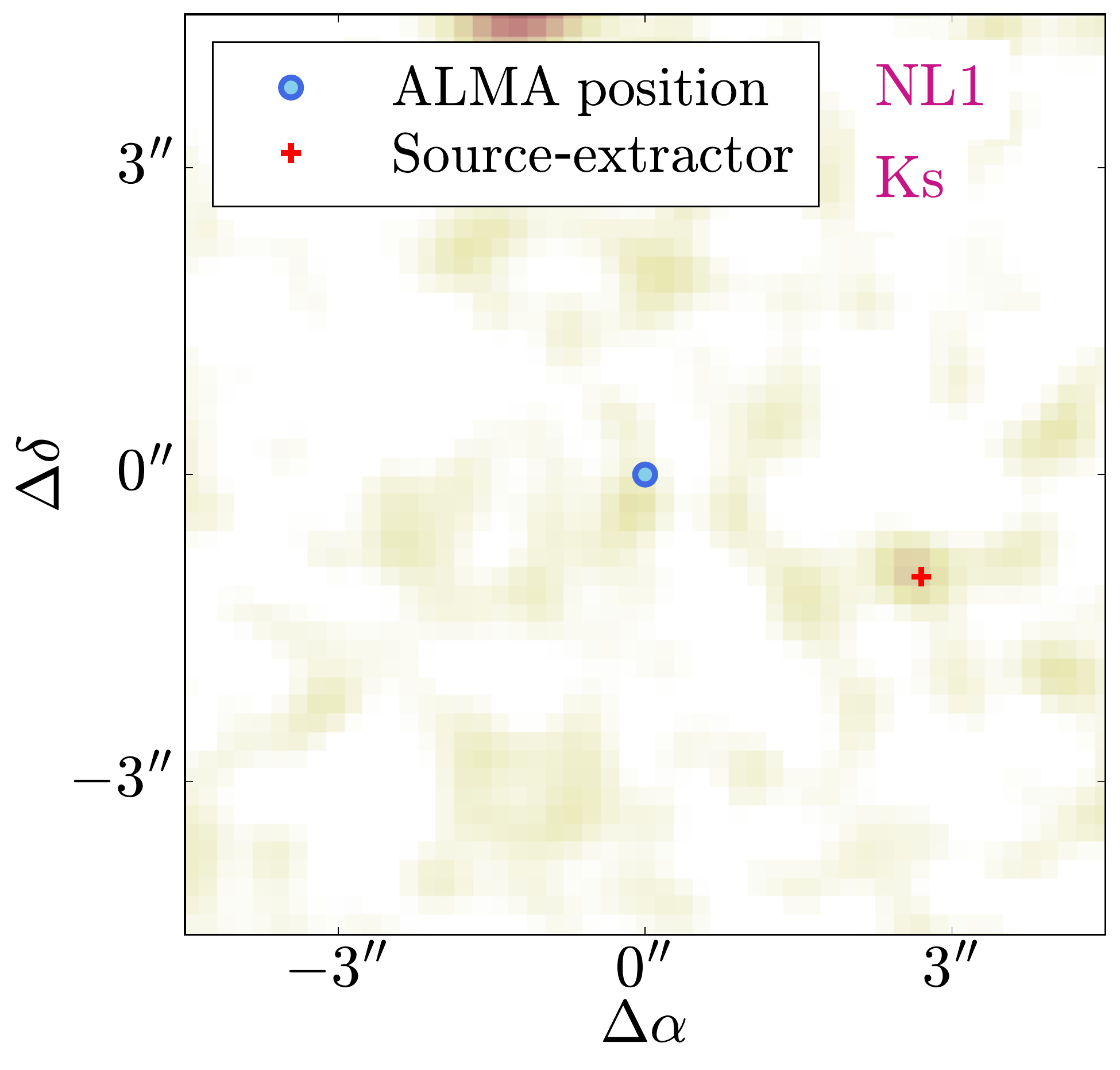}
\includegraphics[width=0.249\textwidth]{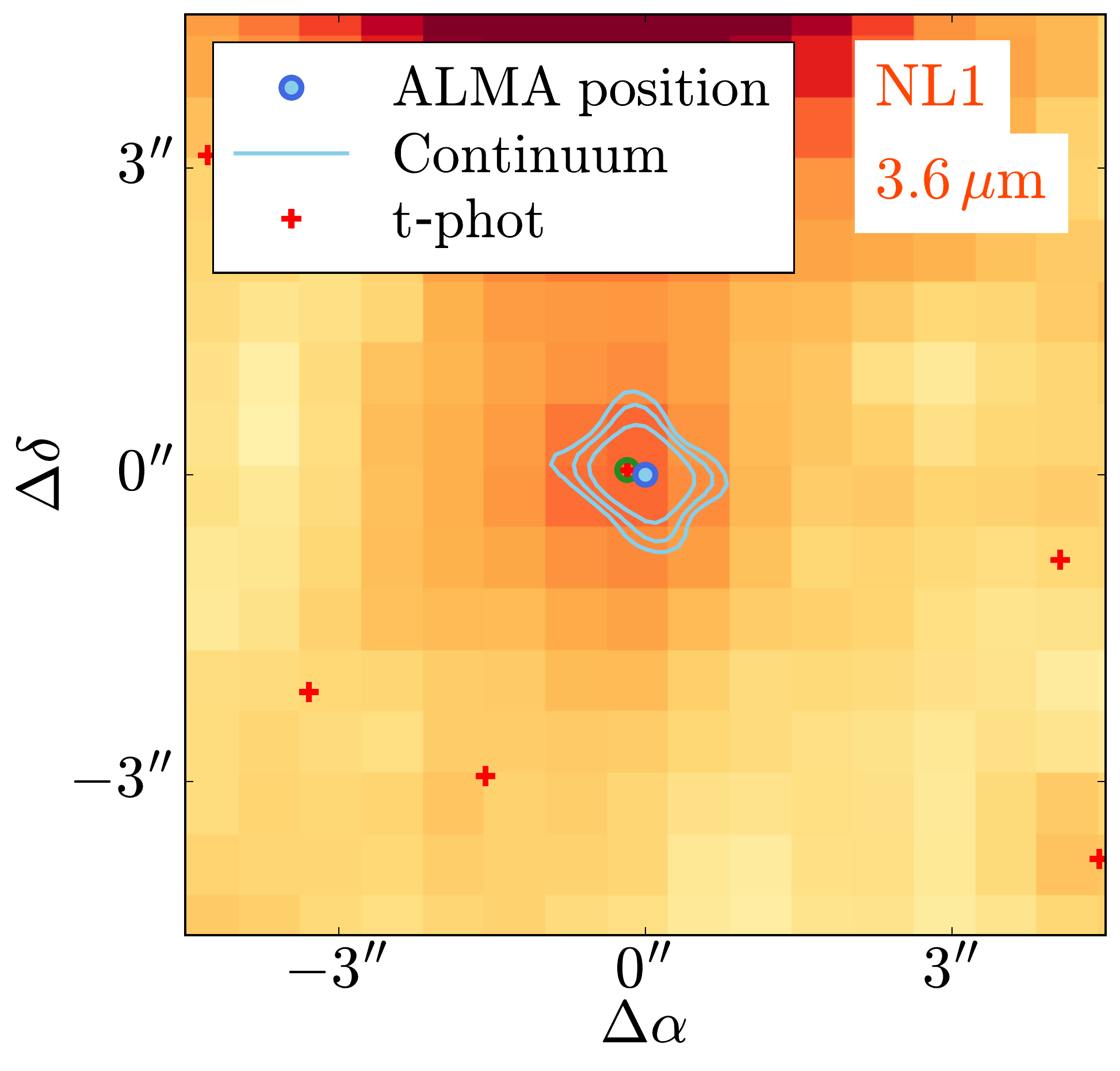}
\includegraphics[width=0.249\textwidth]{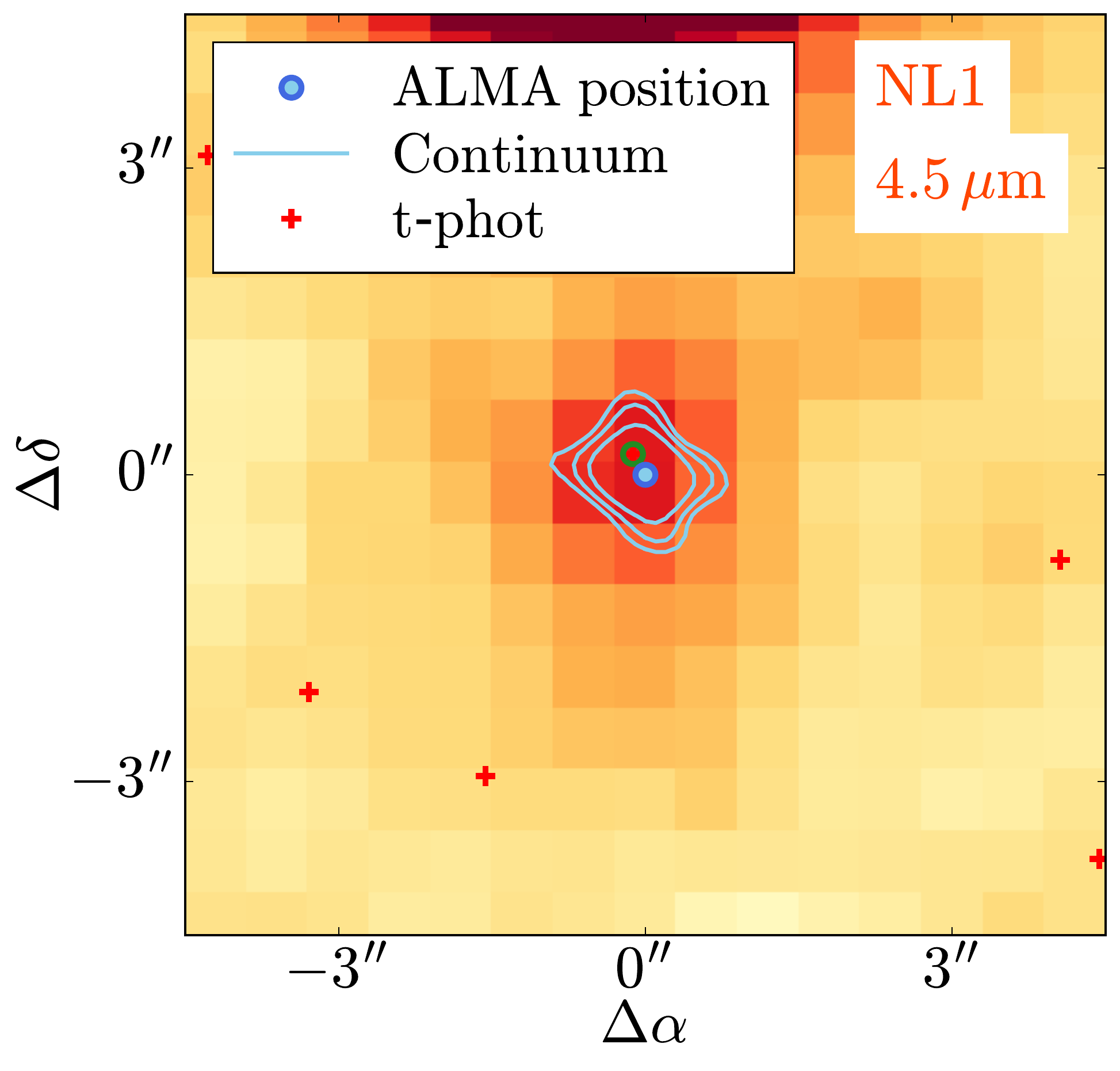}
\end{framed}
\end{subfigure}
\caption{}
\end{figure*}
\renewcommand{\thefigure}{\arabic{figure}}

\renewcommand{\thefigure}{B\arabic{figure} (Cont.)}
\addtocounter{figure}{-1}
\begin{figure*}
\begin{subfigure}{0.85\textwidth}
\begin{framed}
\includegraphics[width=0.24\textwidth]{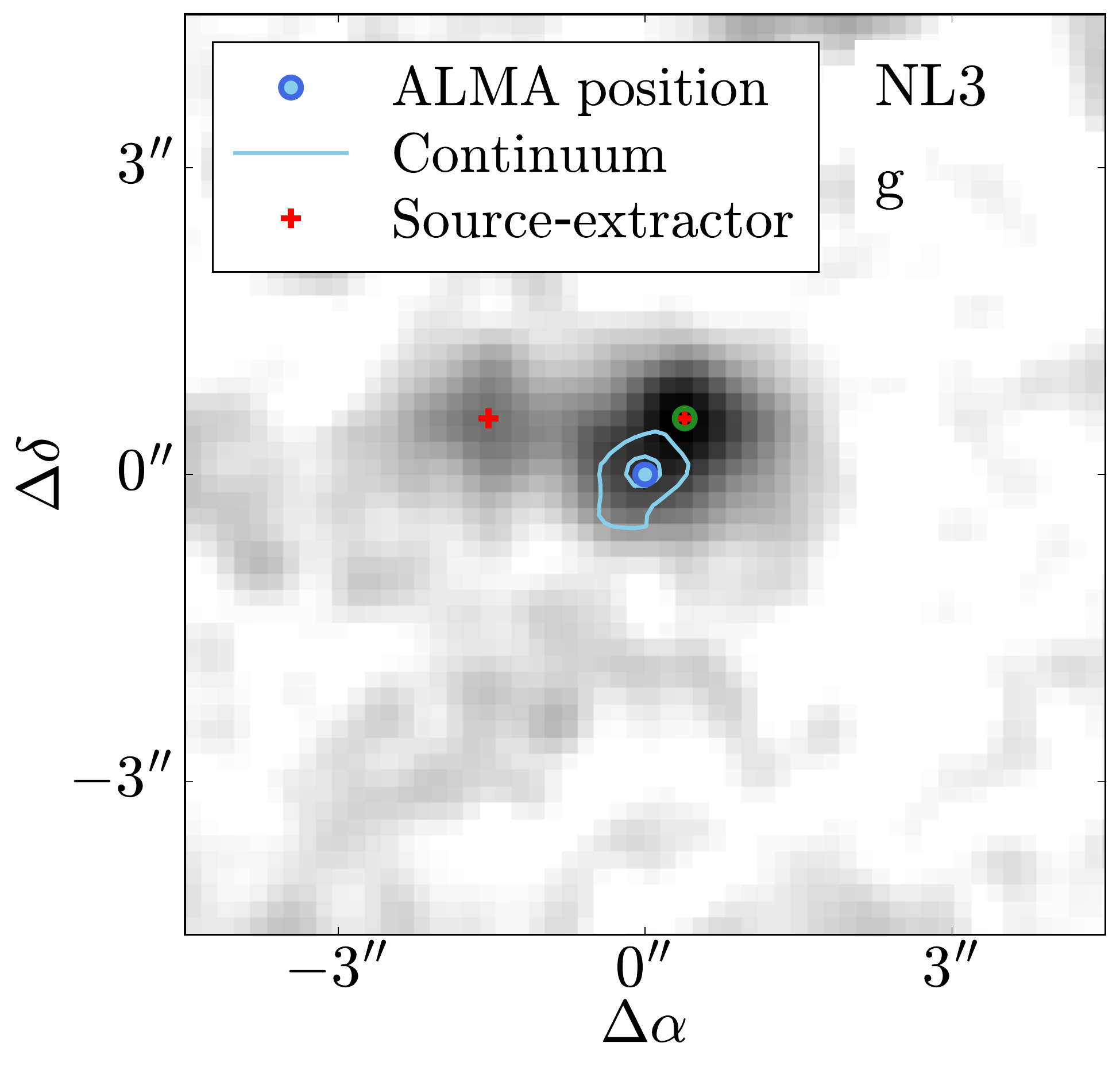}
\includegraphics[width=0.24\textwidth]{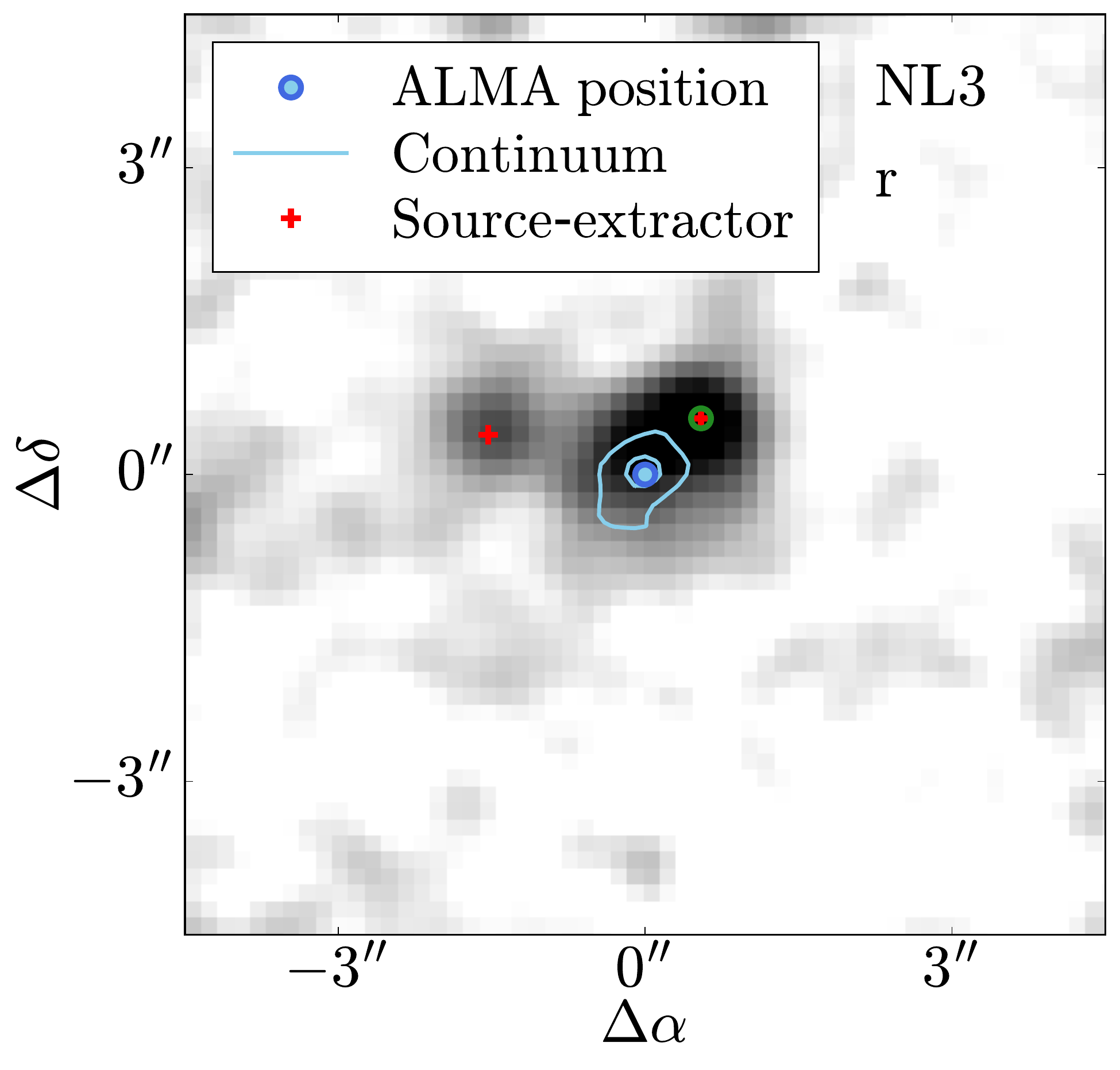}
\includegraphics[width=0.24\textwidth]{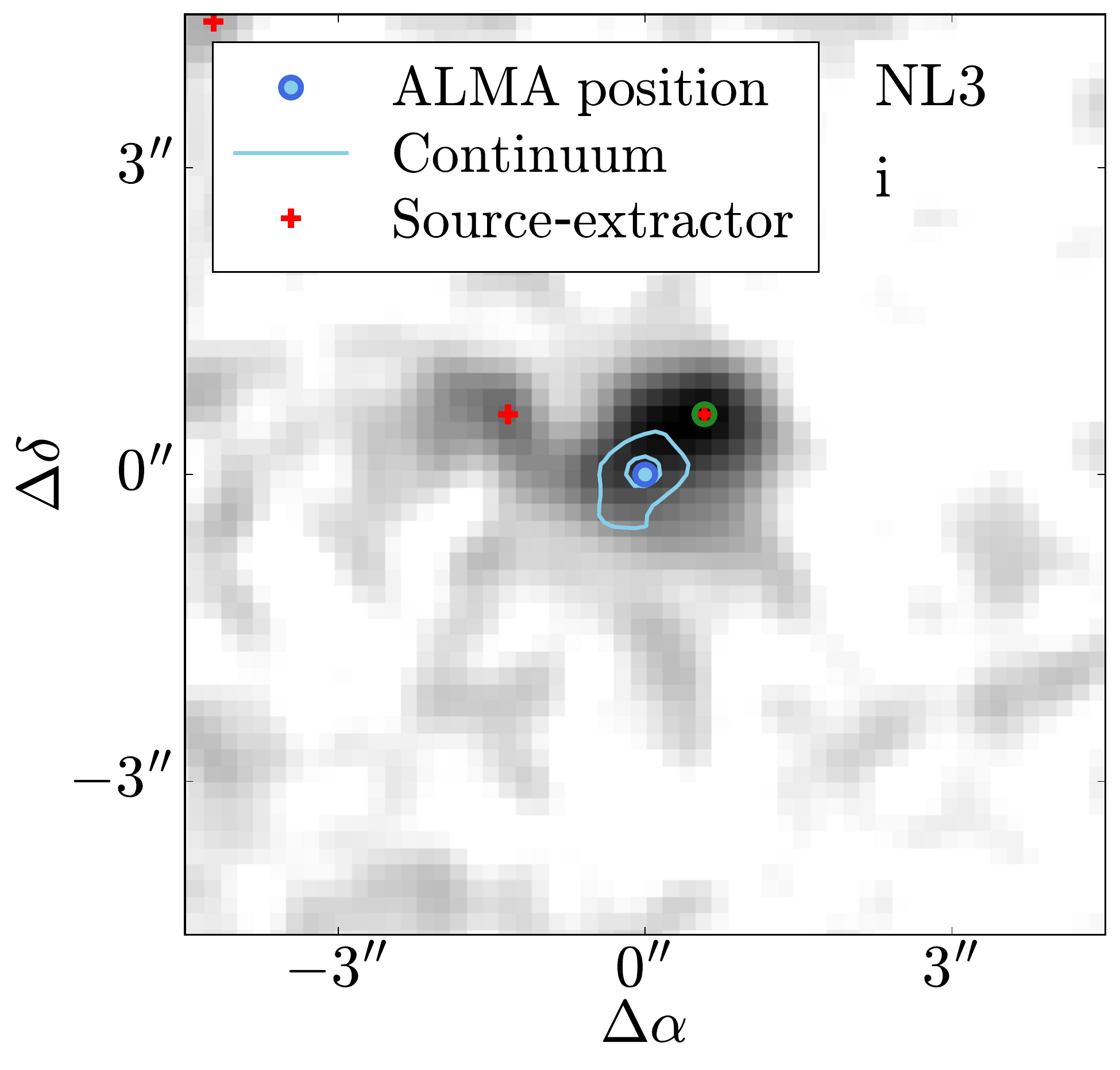}
\includegraphics[width=0.24\textwidth]{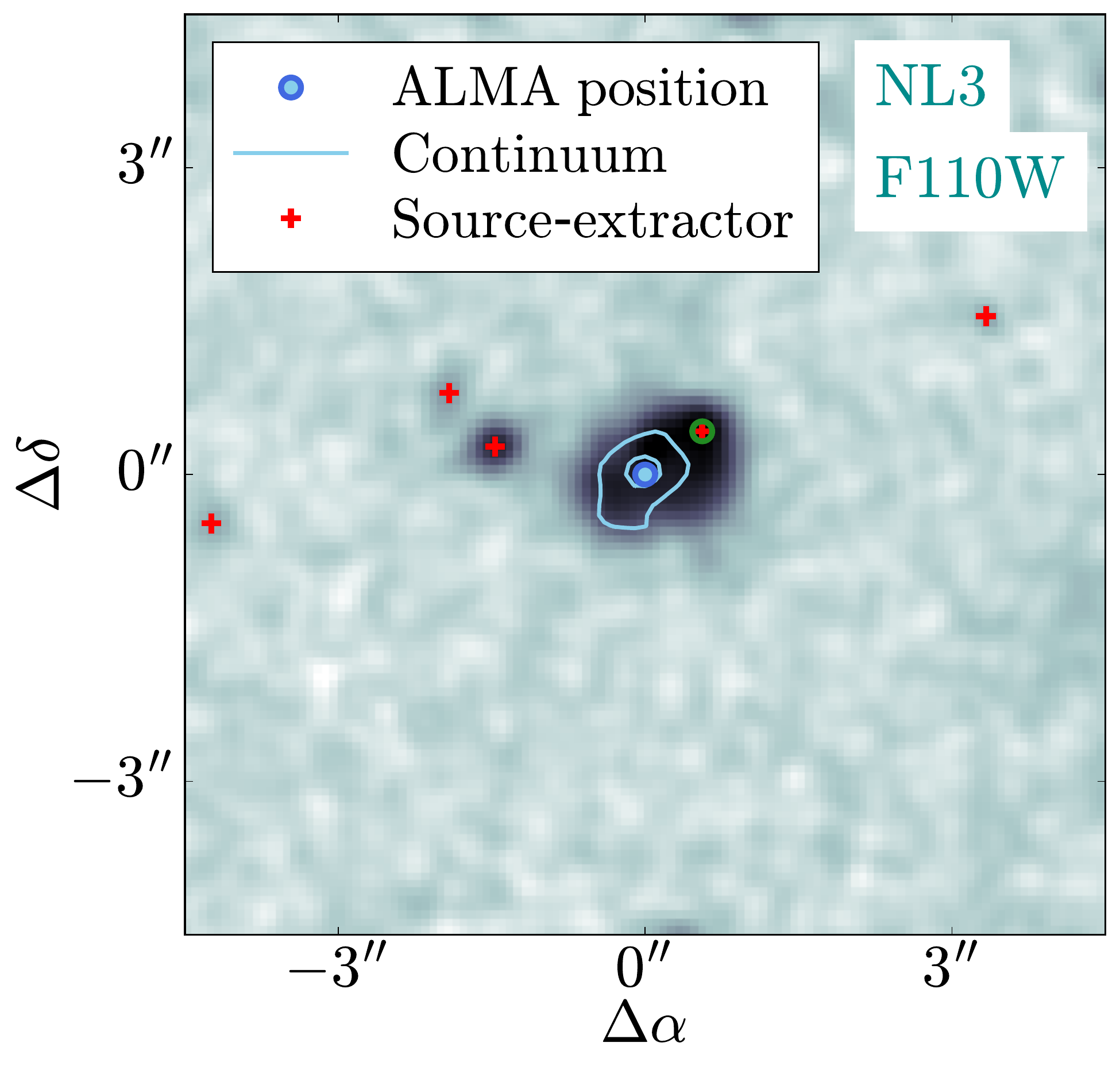}
\includegraphics[width=0.24\textwidth]{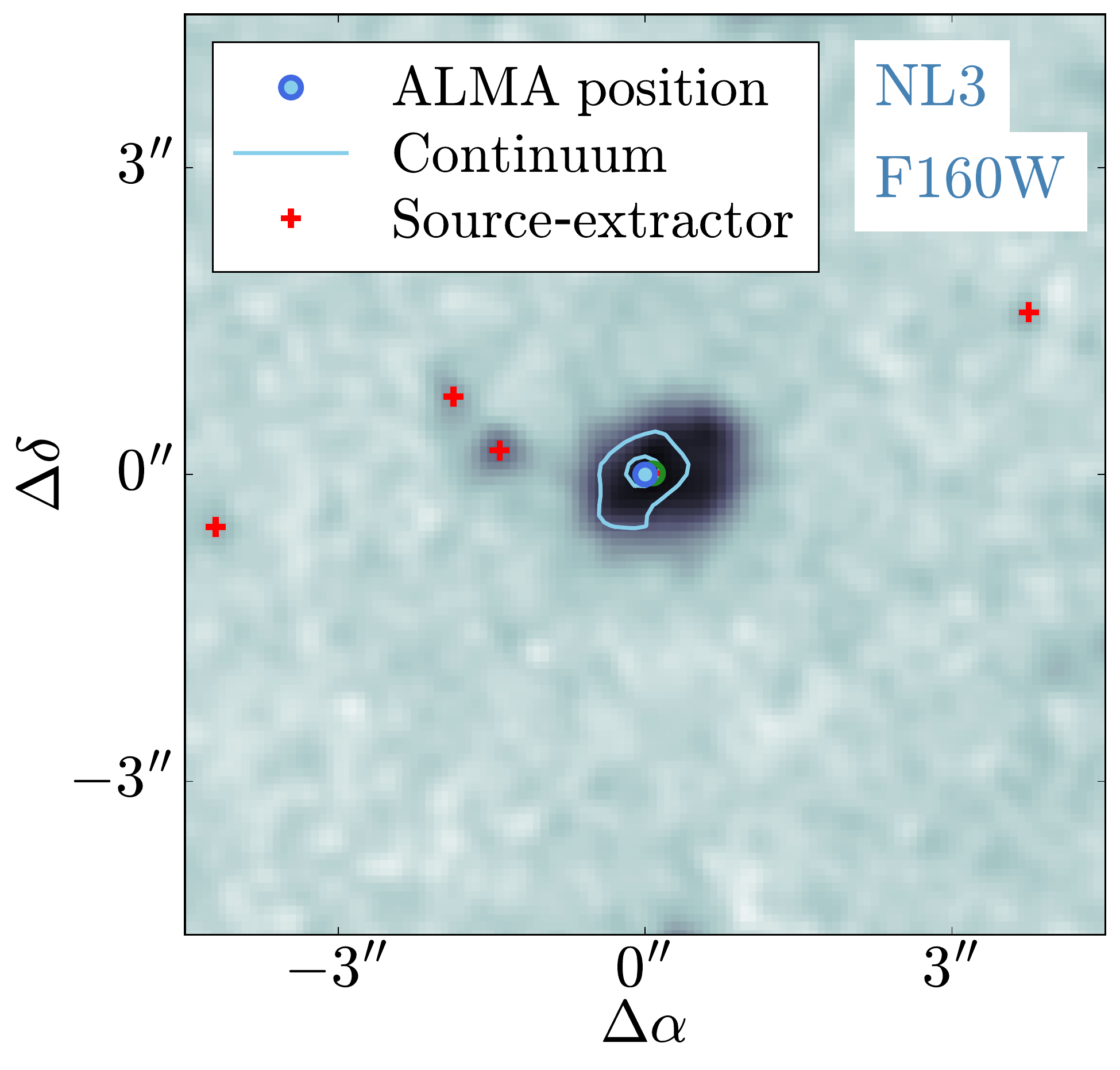}
\includegraphics[width=0.248\textwidth]{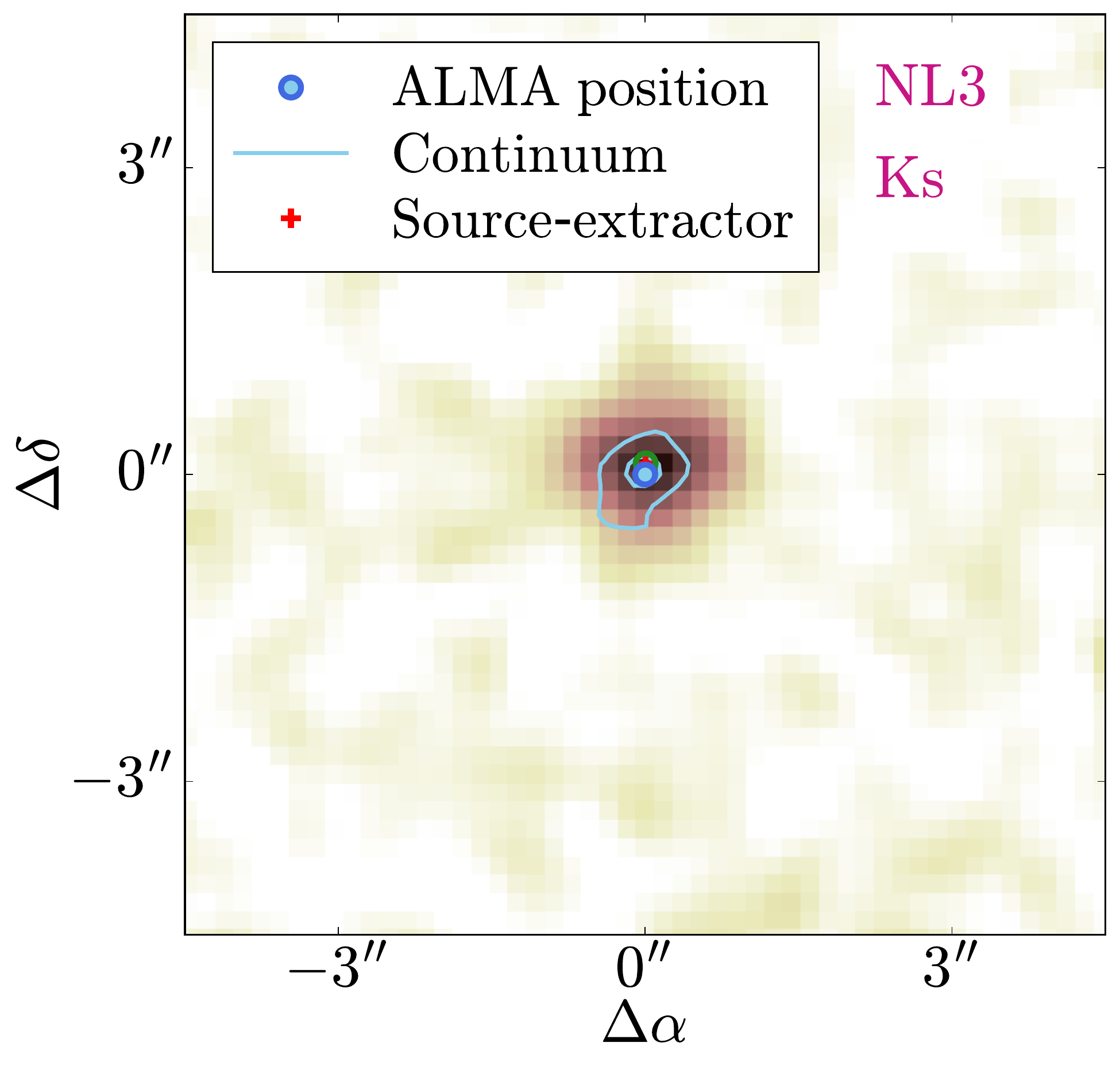}
\includegraphics[width=0.249\textwidth]{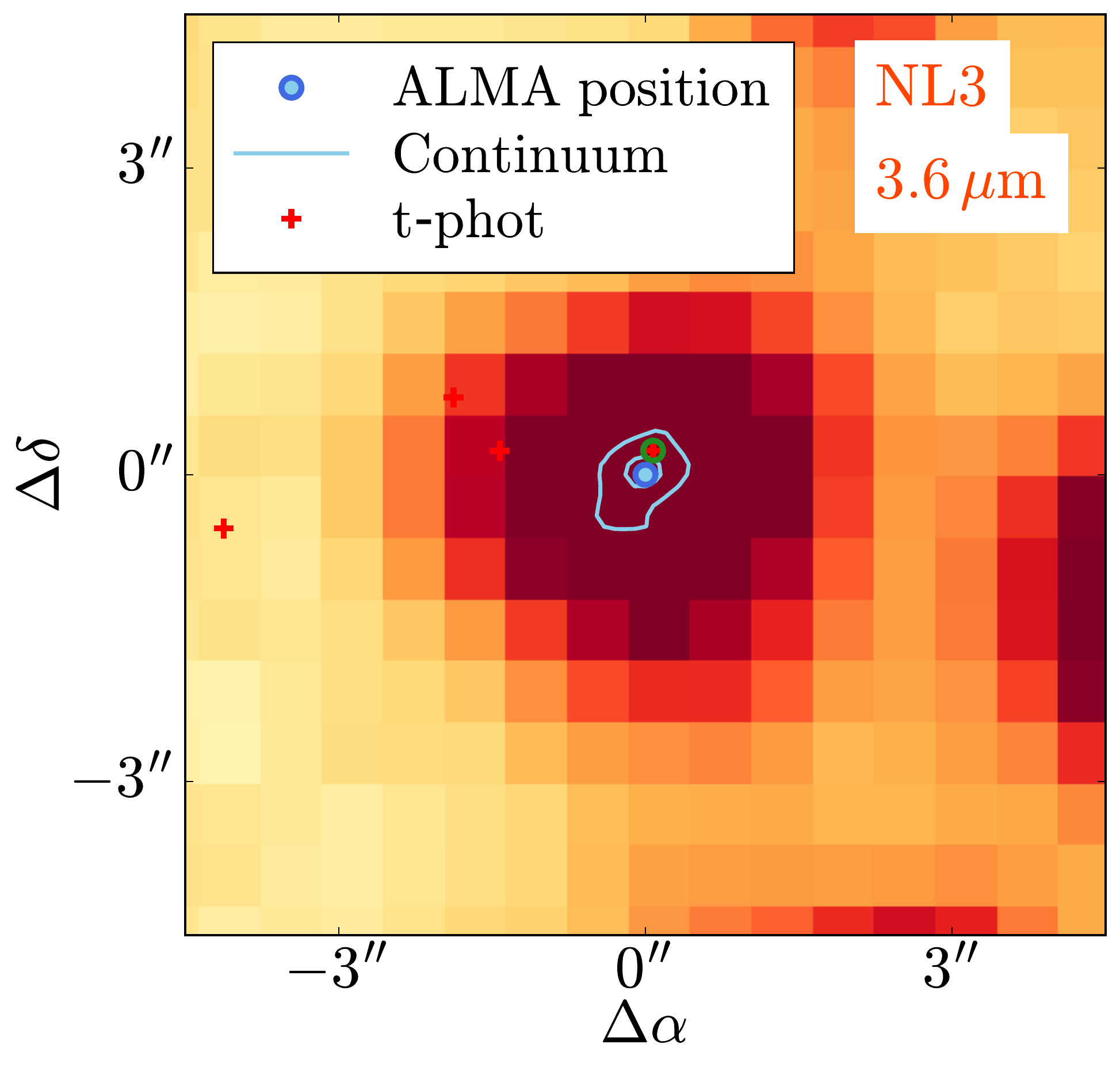}
\includegraphics[width=0.249\textwidth]{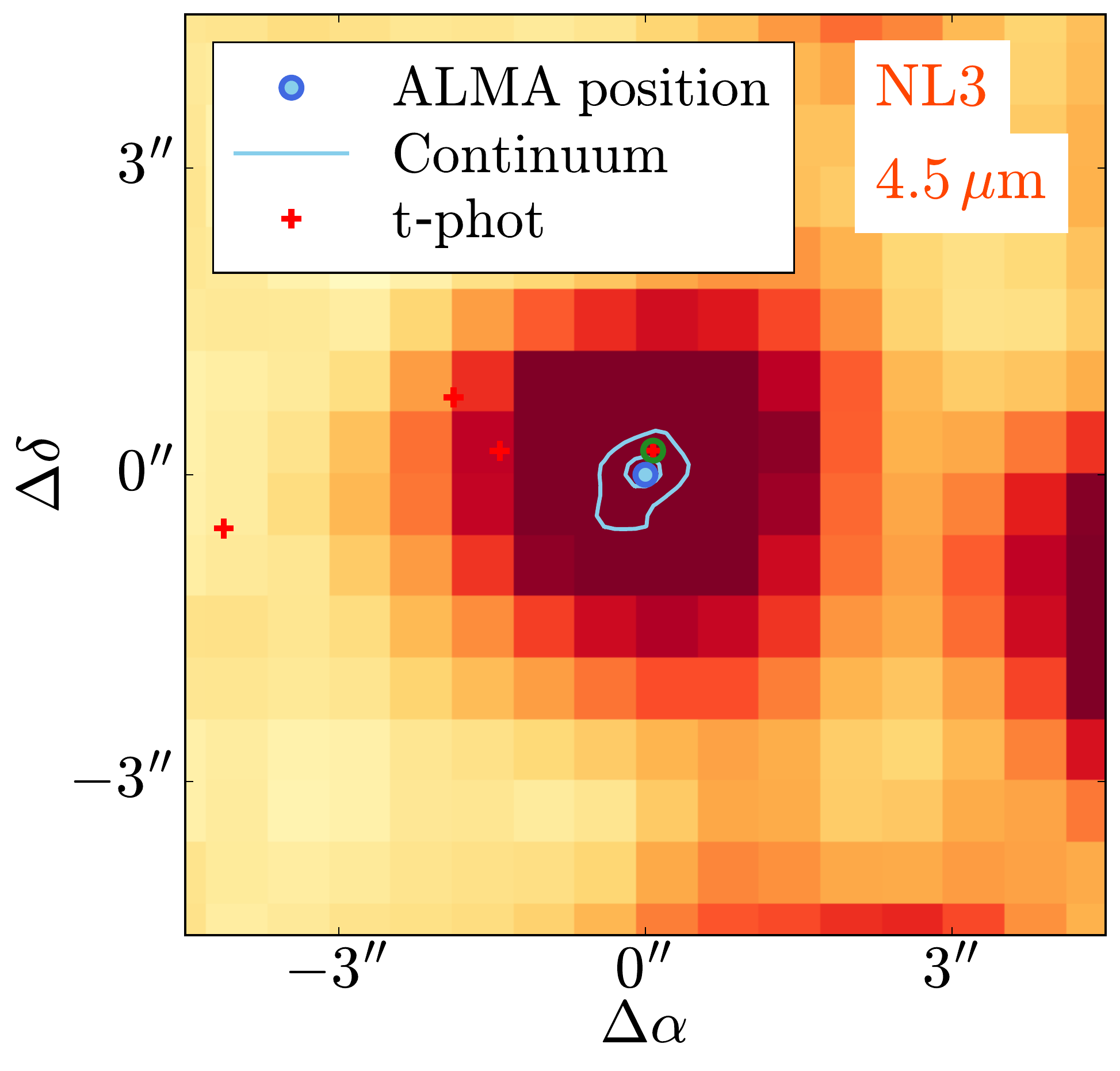}
\end{framed}
\end{subfigure}
\begin{subfigure}{0.85\textwidth}
\begin{framed}
\includegraphics[width=0.24\textwidth]{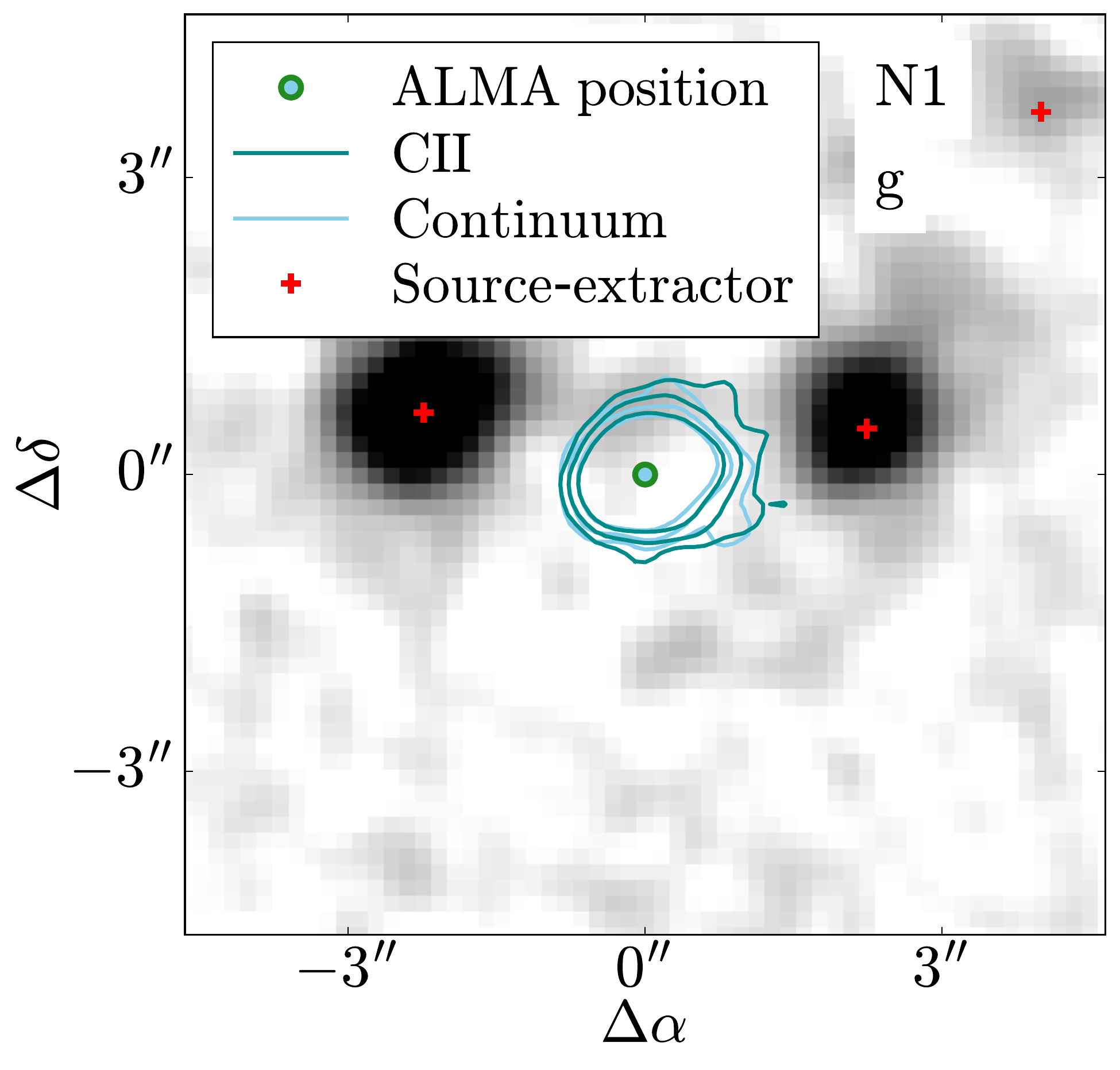}
\includegraphics[width=0.24\textwidth]{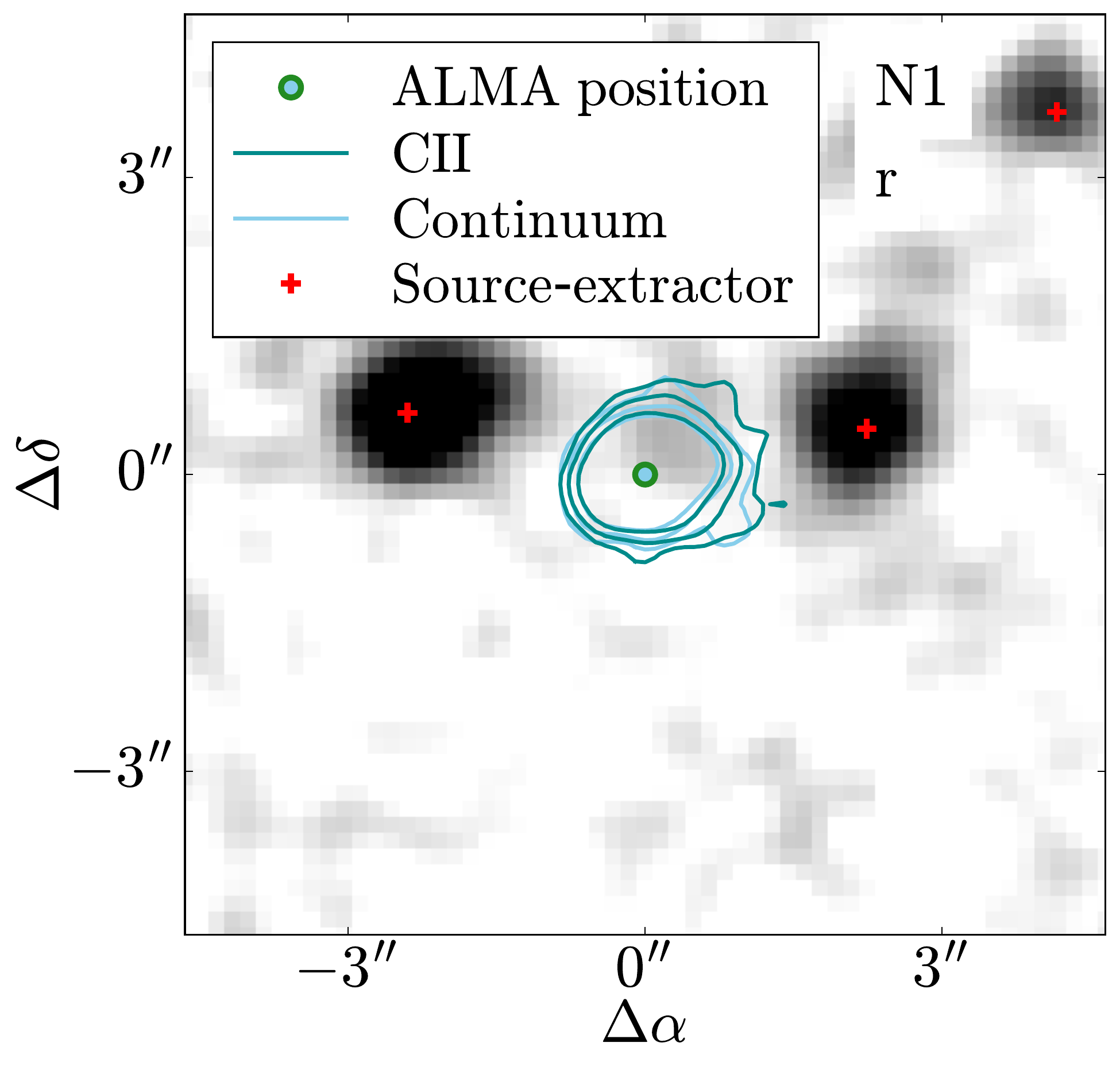}
\includegraphics[width=0.24\textwidth]{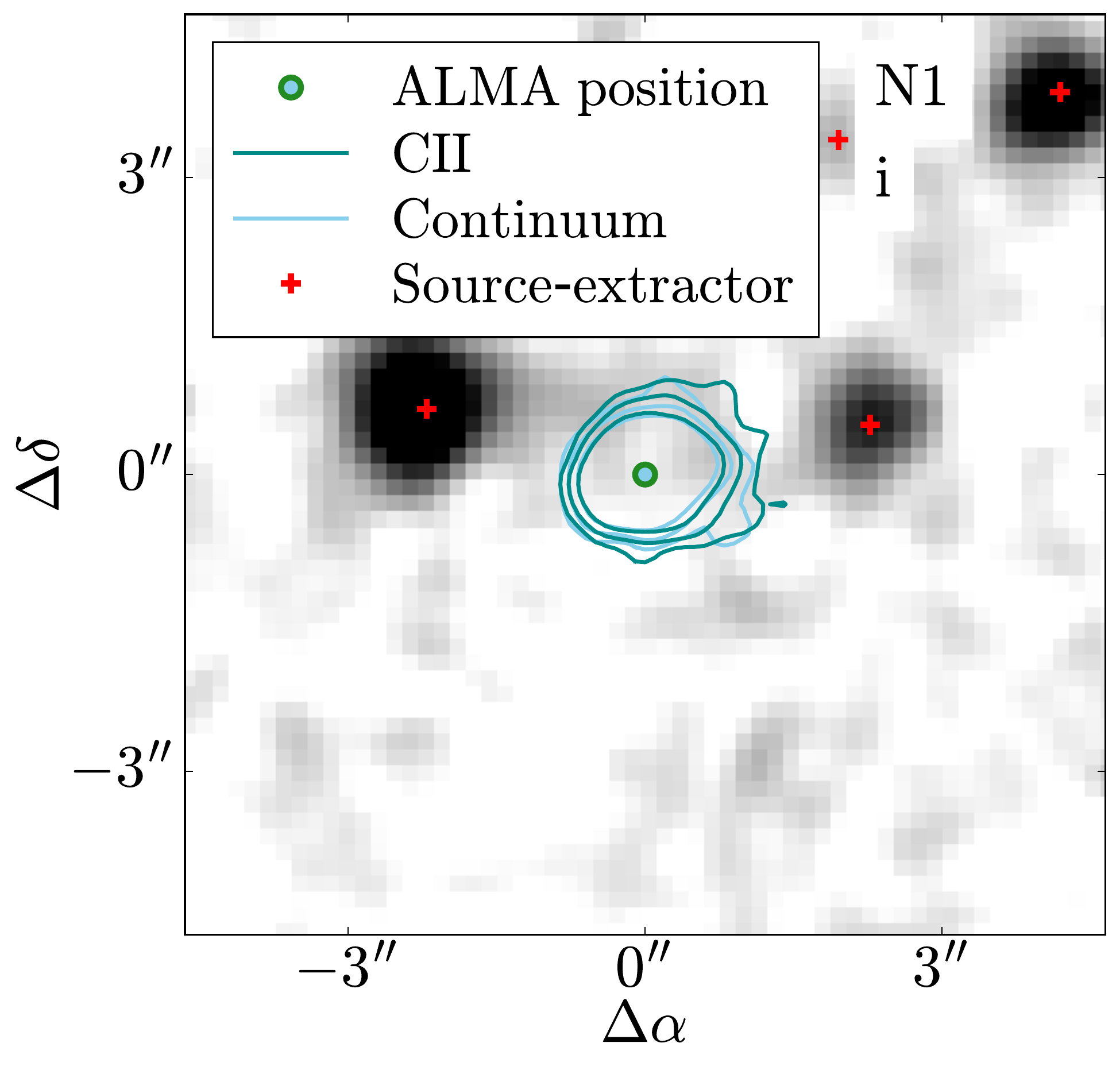}
\includegraphics[width=0.24\textwidth]{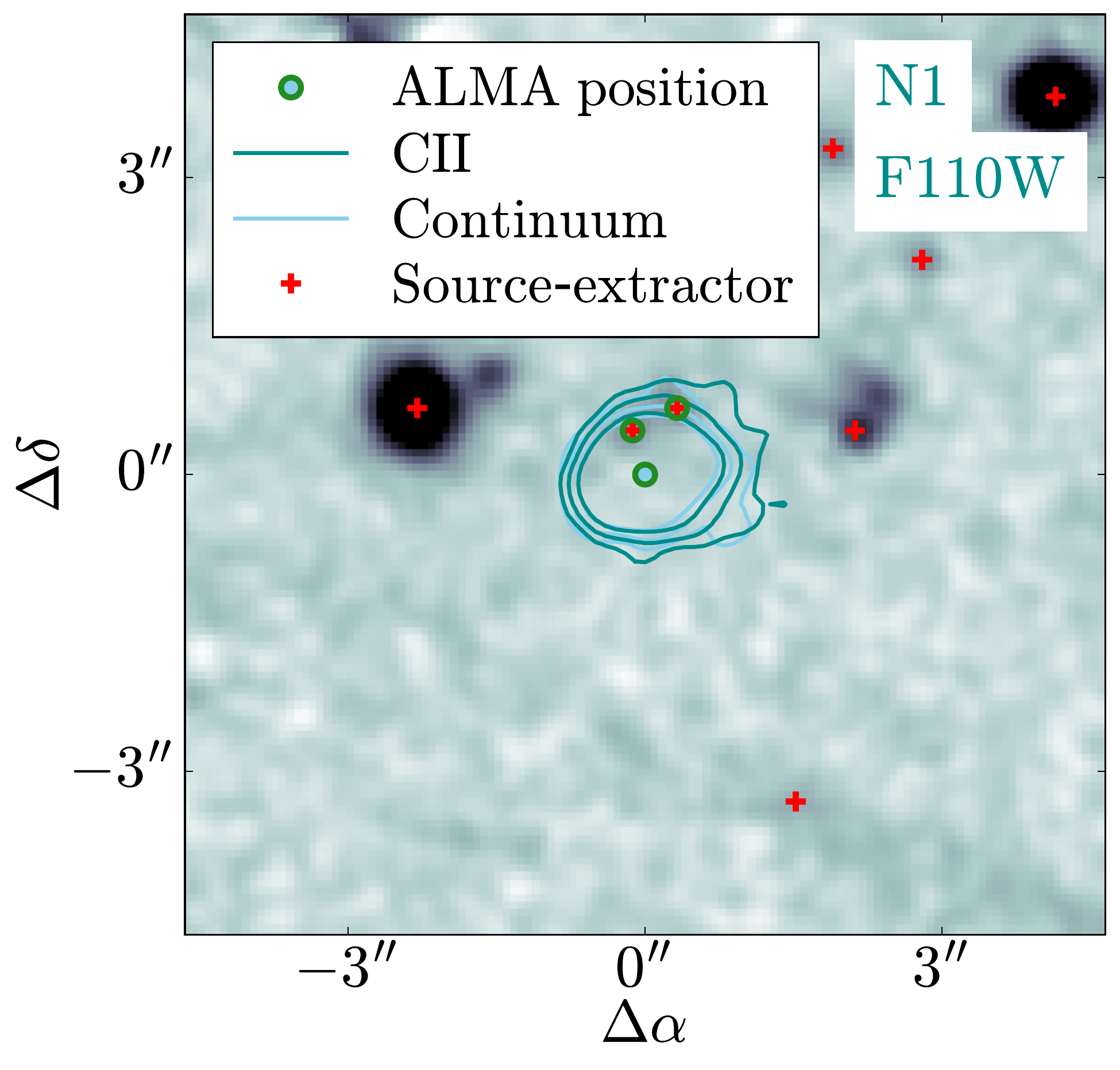}
\includegraphics[width=0.24\textwidth]{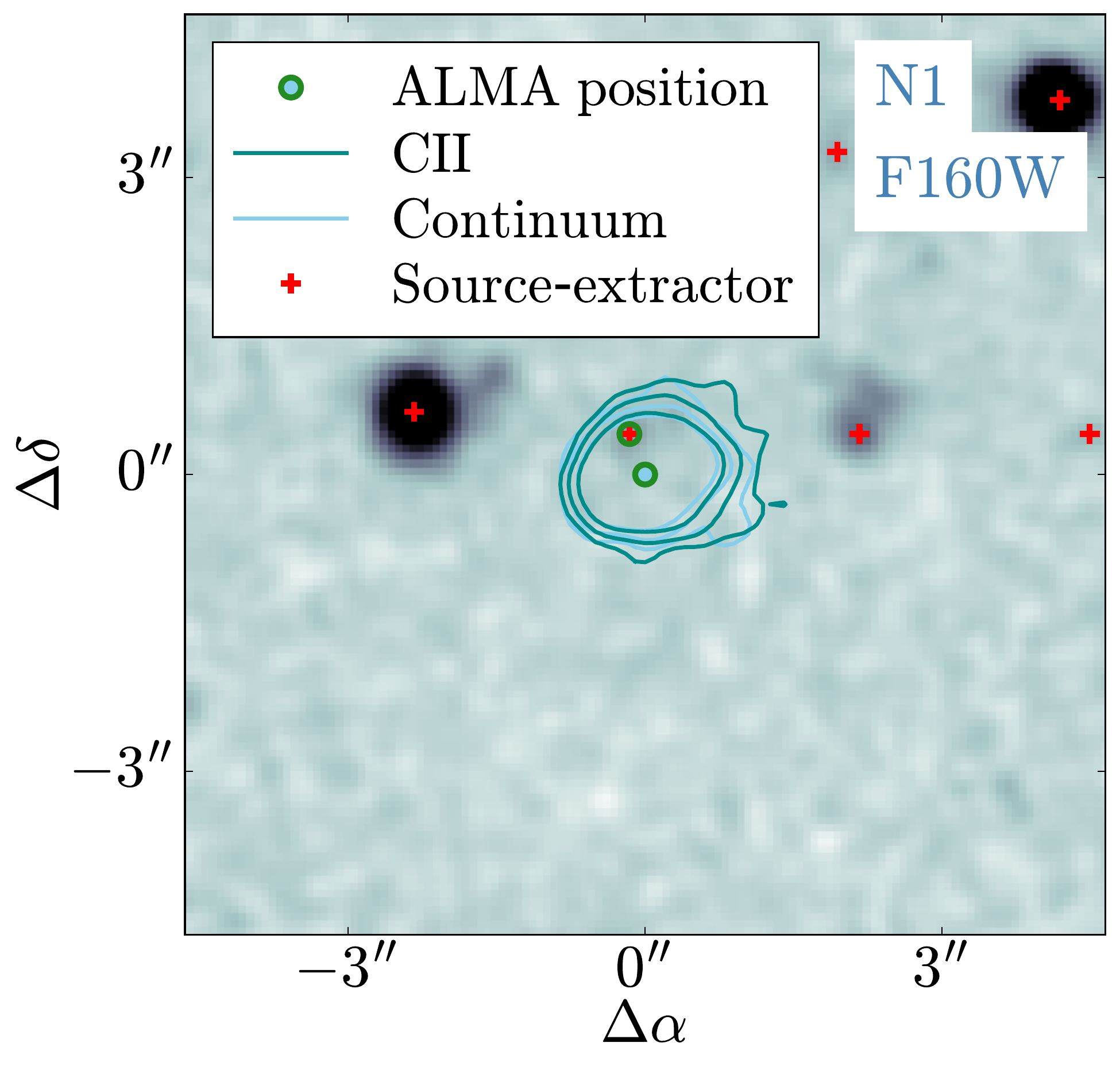}
\includegraphics[width=0.248\textwidth]{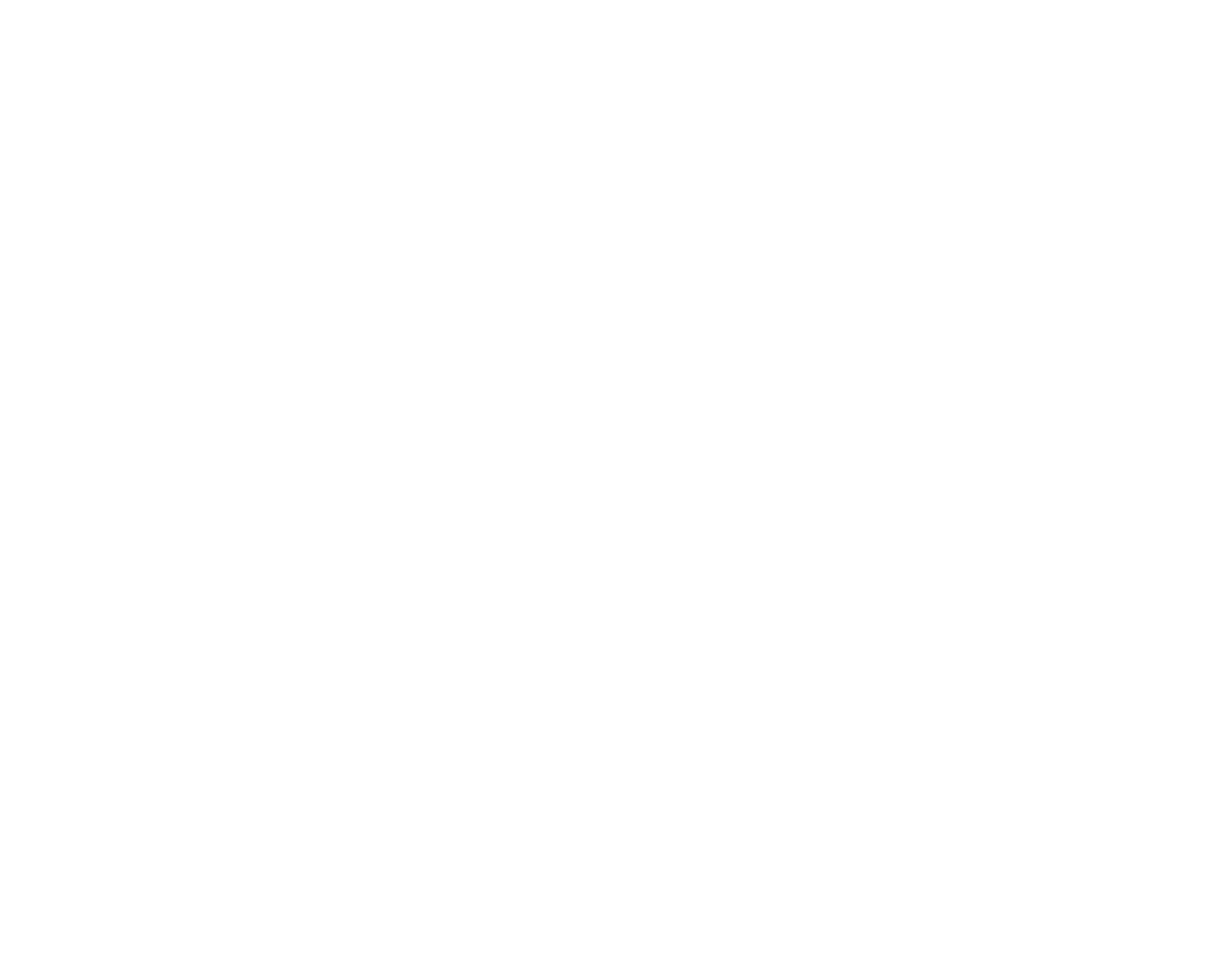}
\includegraphics[width=0.249\textwidth]{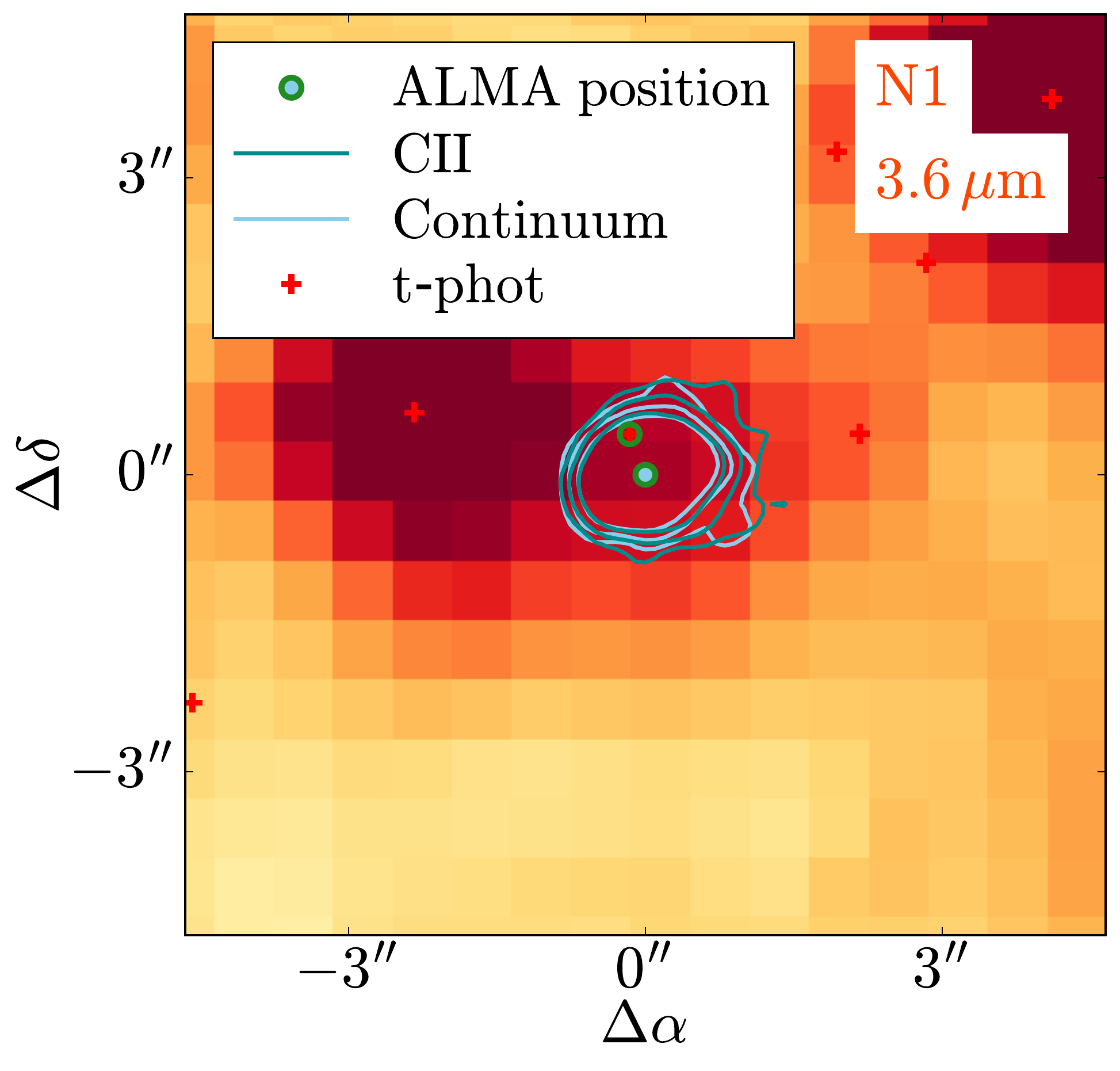}
\includegraphics[width=0.249\textwidth]{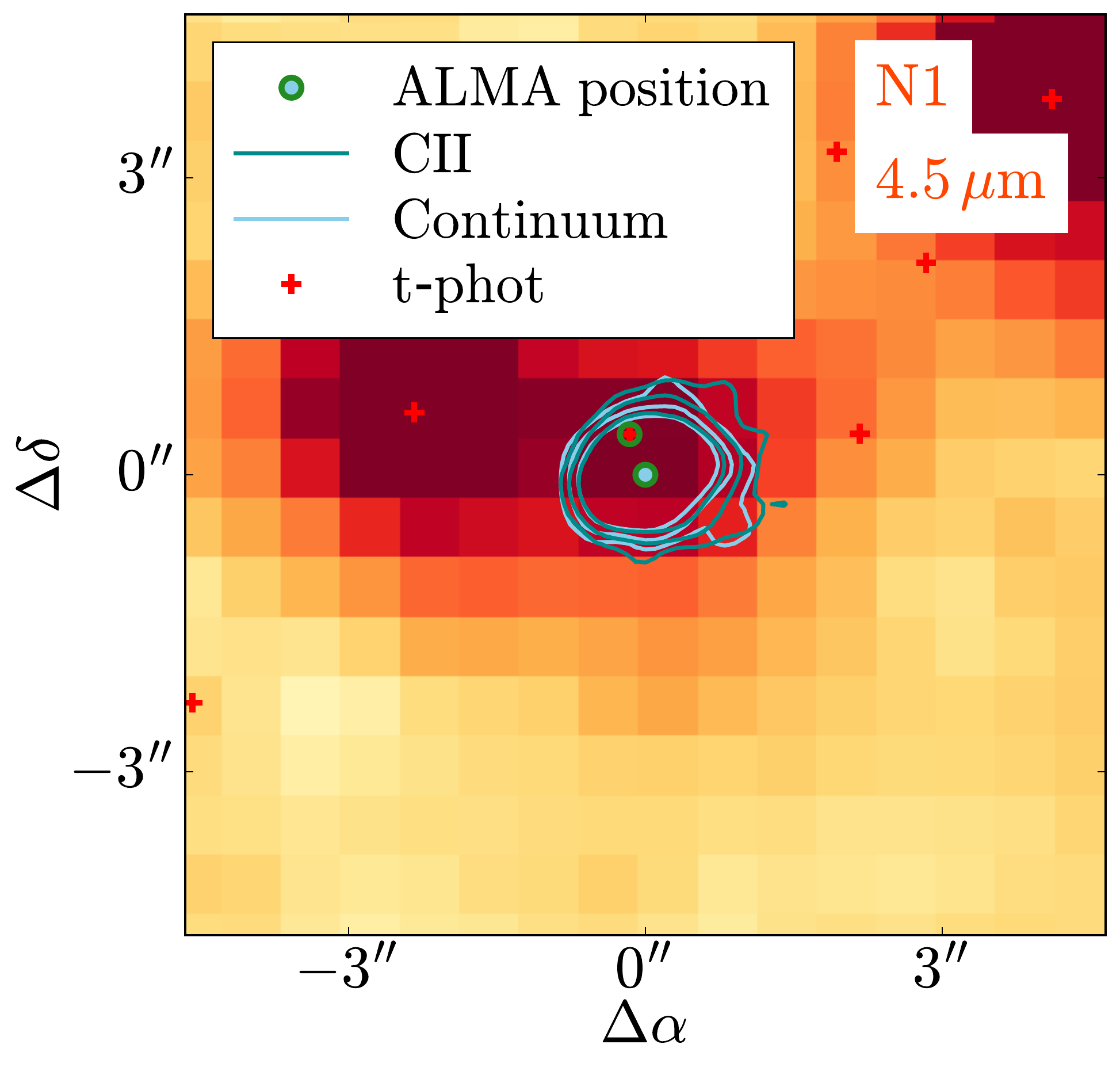}
\end{framed}
\end{subfigure}
\begin{subfigure}{0.85\textwidth}
\begin{framed}
\includegraphics[width=0.24\textwidth]{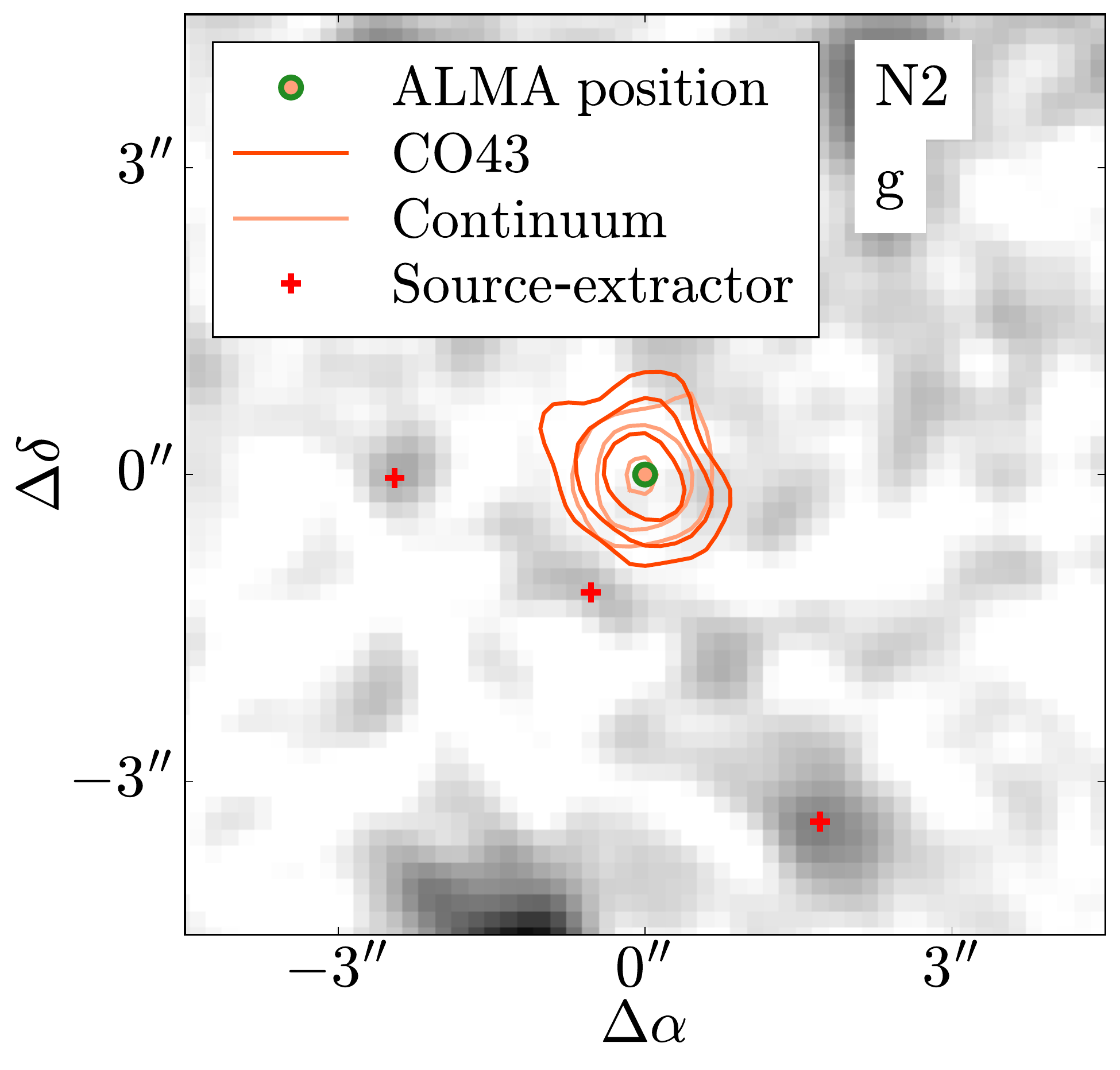}
\includegraphics[width=0.24\textwidth]{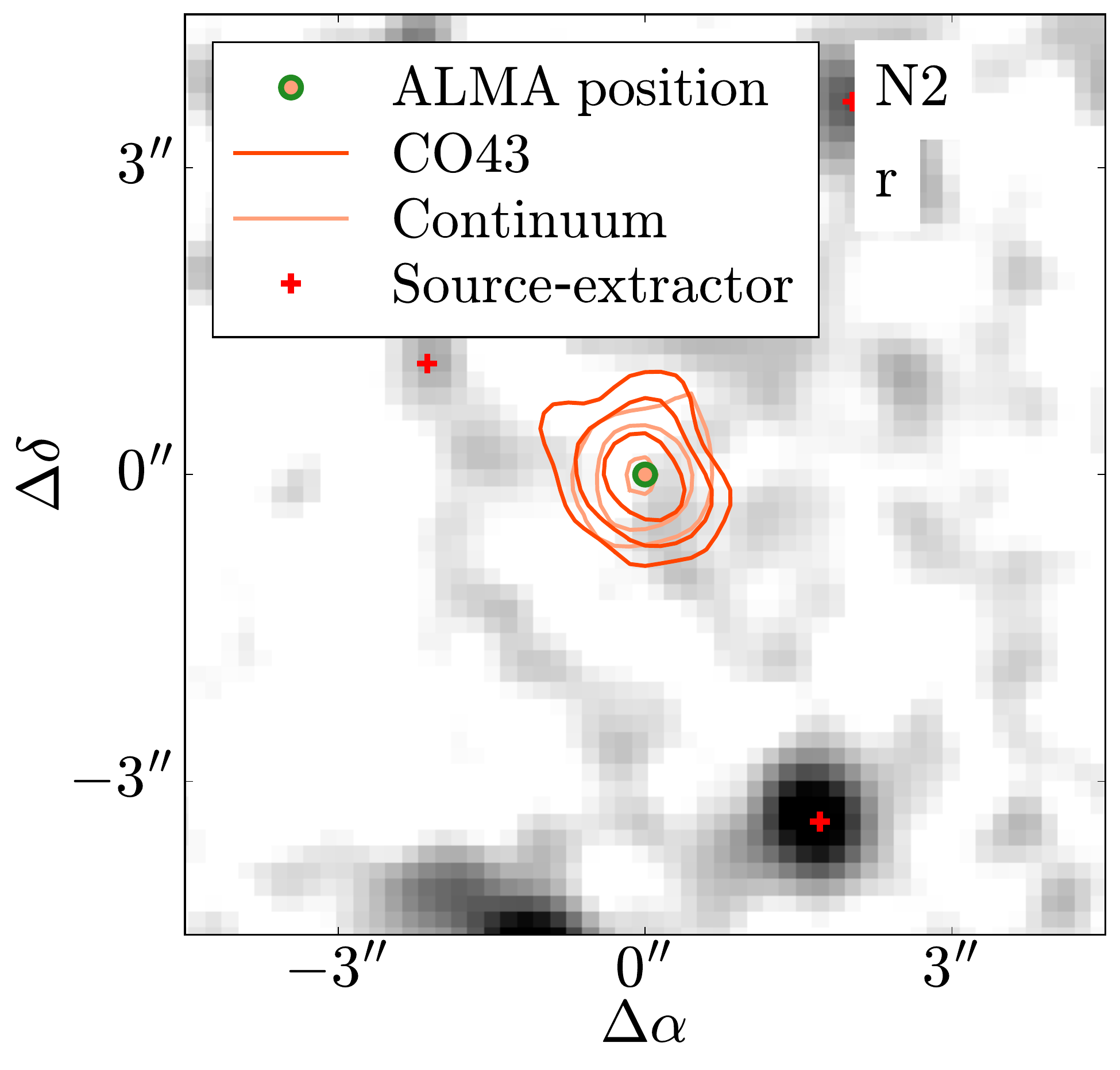}
\includegraphics[width=0.24\textwidth]{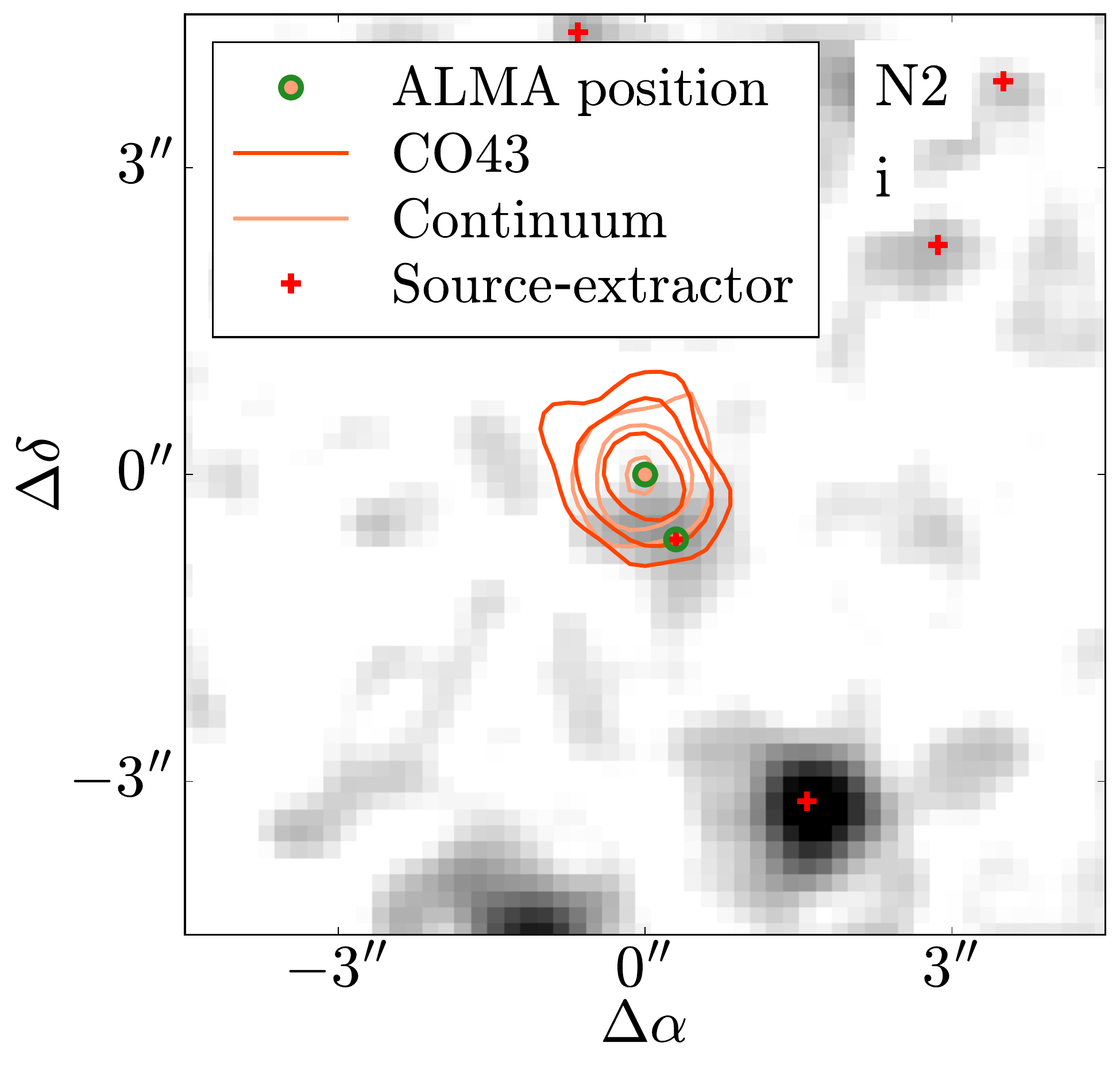}
\includegraphics[width=0.24\textwidth]{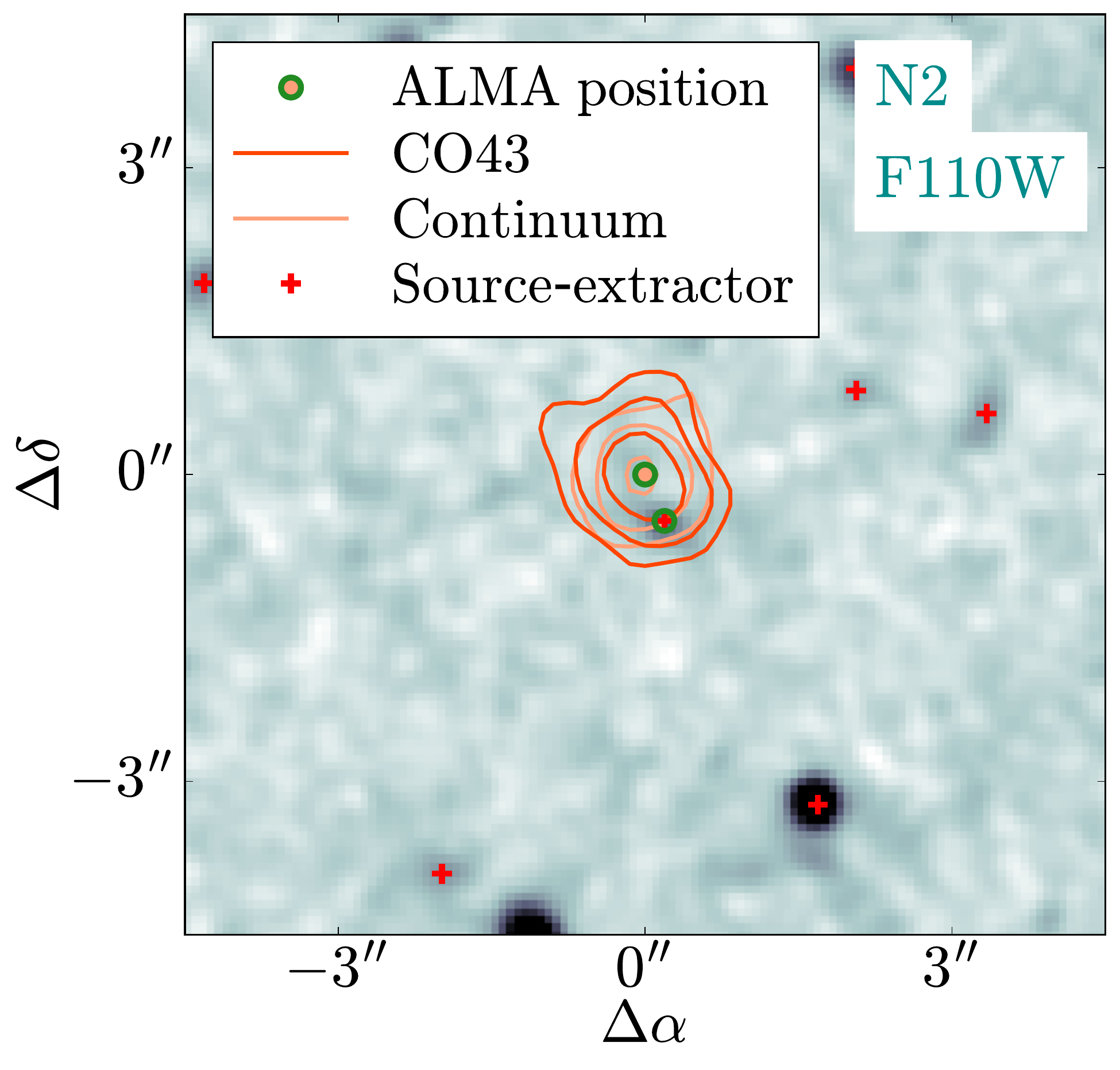}
\includegraphics[width=0.24\textwidth]{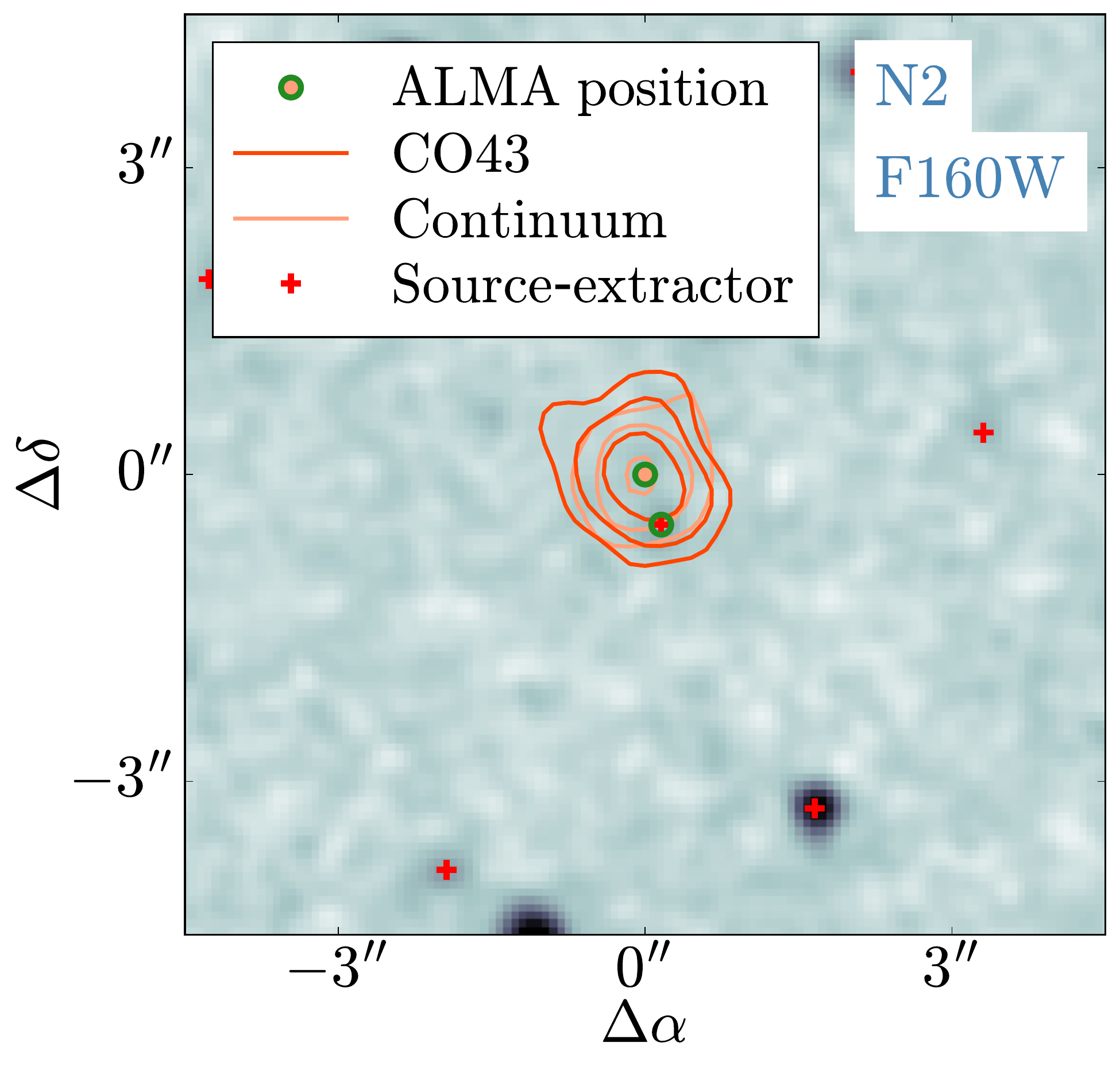}
\includegraphics[width=0.248\textwidth]{Ks/blank.pdf}
\includegraphics[width=0.249\textwidth]{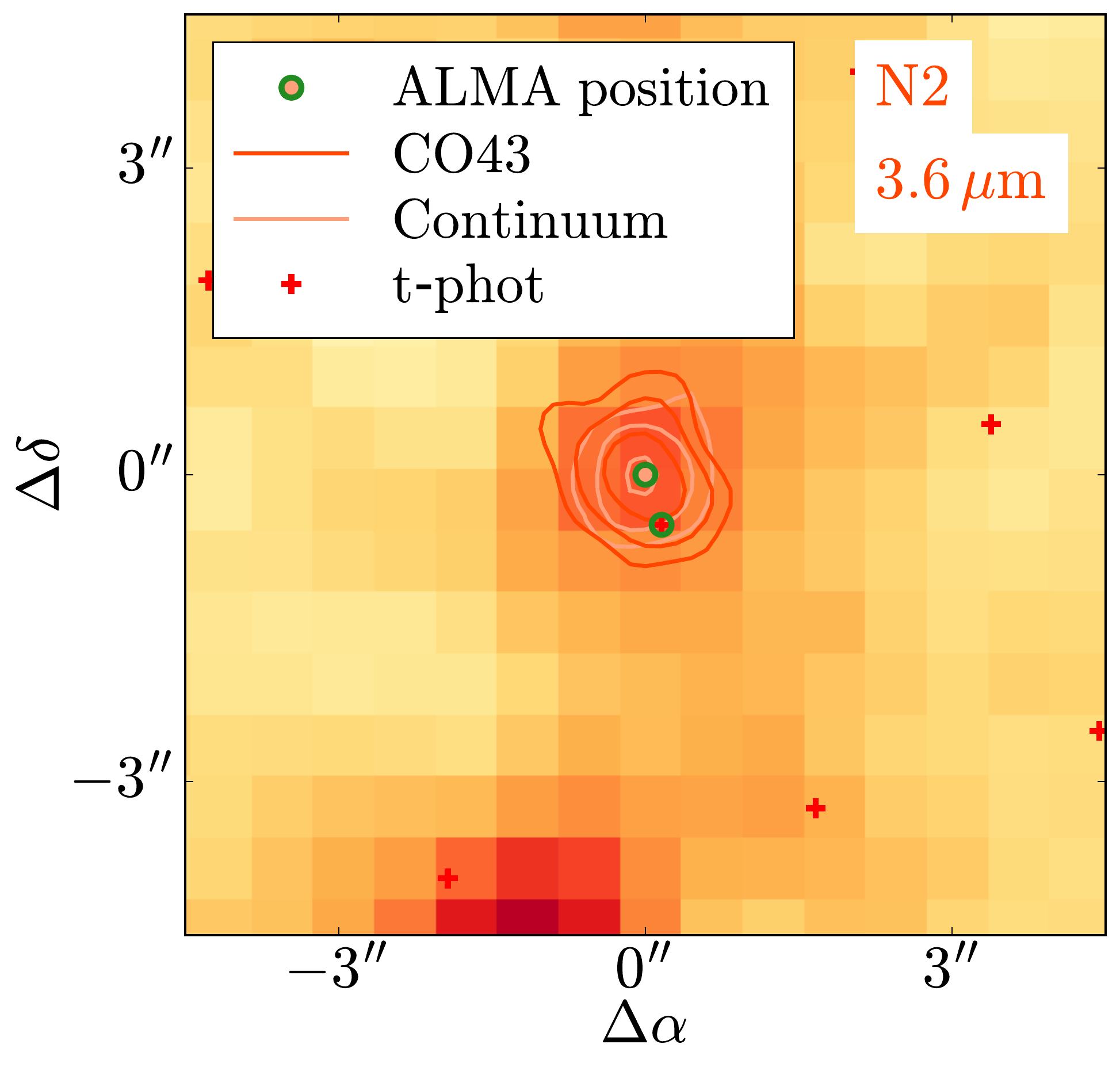}
\includegraphics[width=0.249\textwidth]{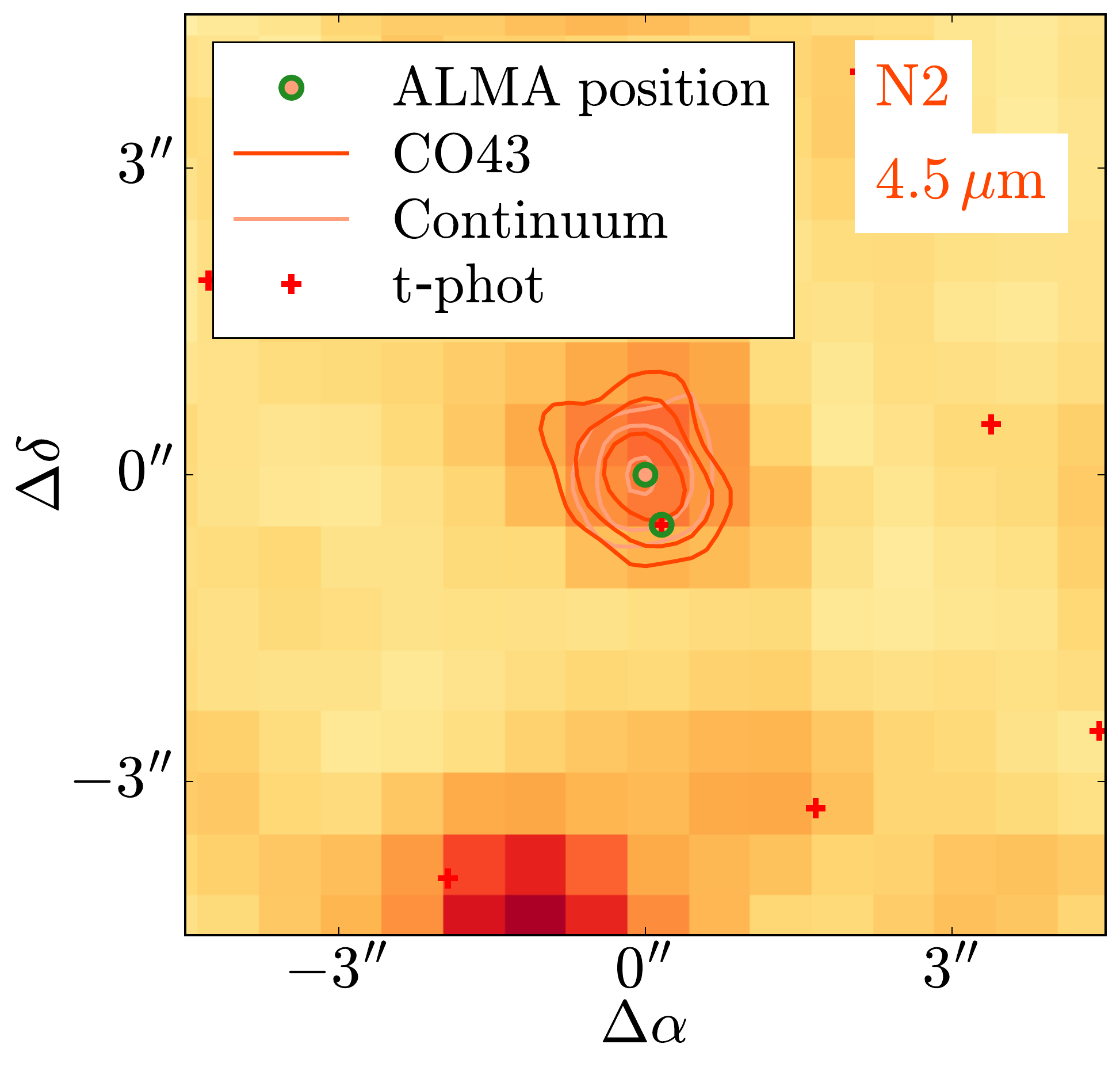}
\end{framed}
\end{subfigure}
\caption{}
\end{figure*}
\renewcommand{\thefigure}{\arabic{figure}}

\renewcommand{\thefigure}{B\arabic{figure} (Cont.)}
\addtocounter{figure}{-1}
\begin{figure*}
\begin{subfigure}{0.85\textwidth}
\begin{framed}
\includegraphics[width=0.24\textwidth]{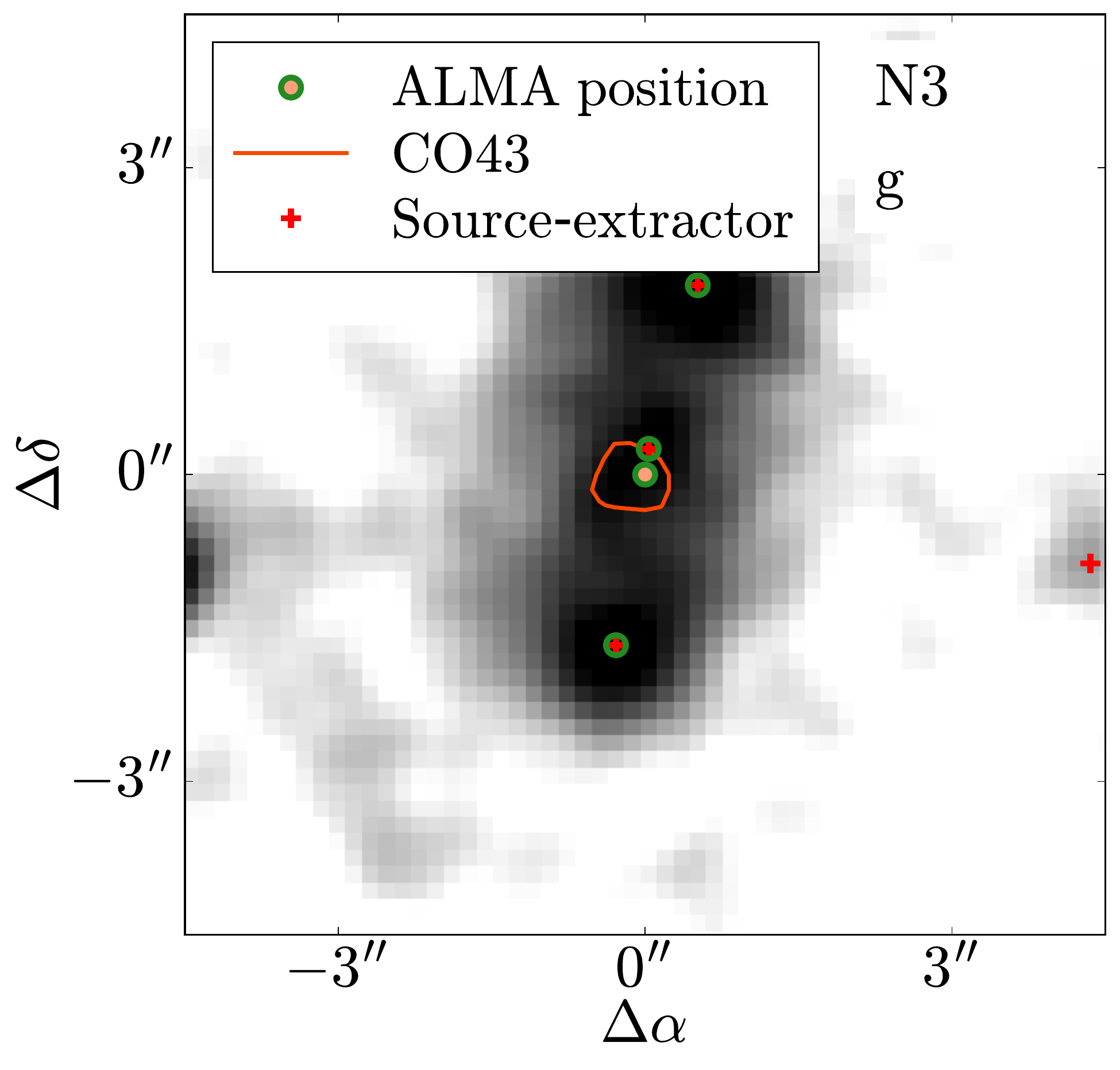}
\includegraphics[width=0.24\textwidth]{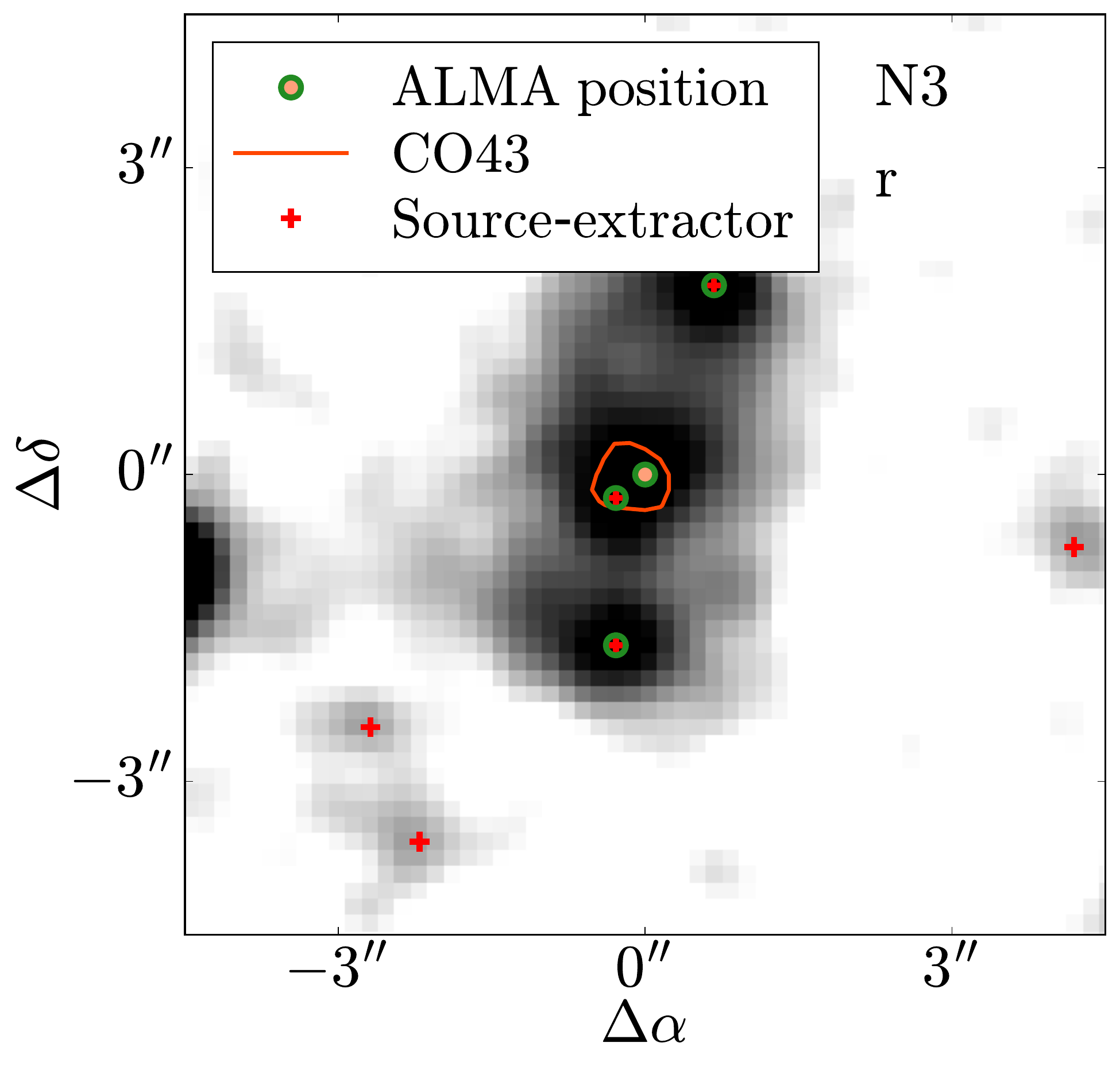}
\includegraphics[width=0.24\textwidth]{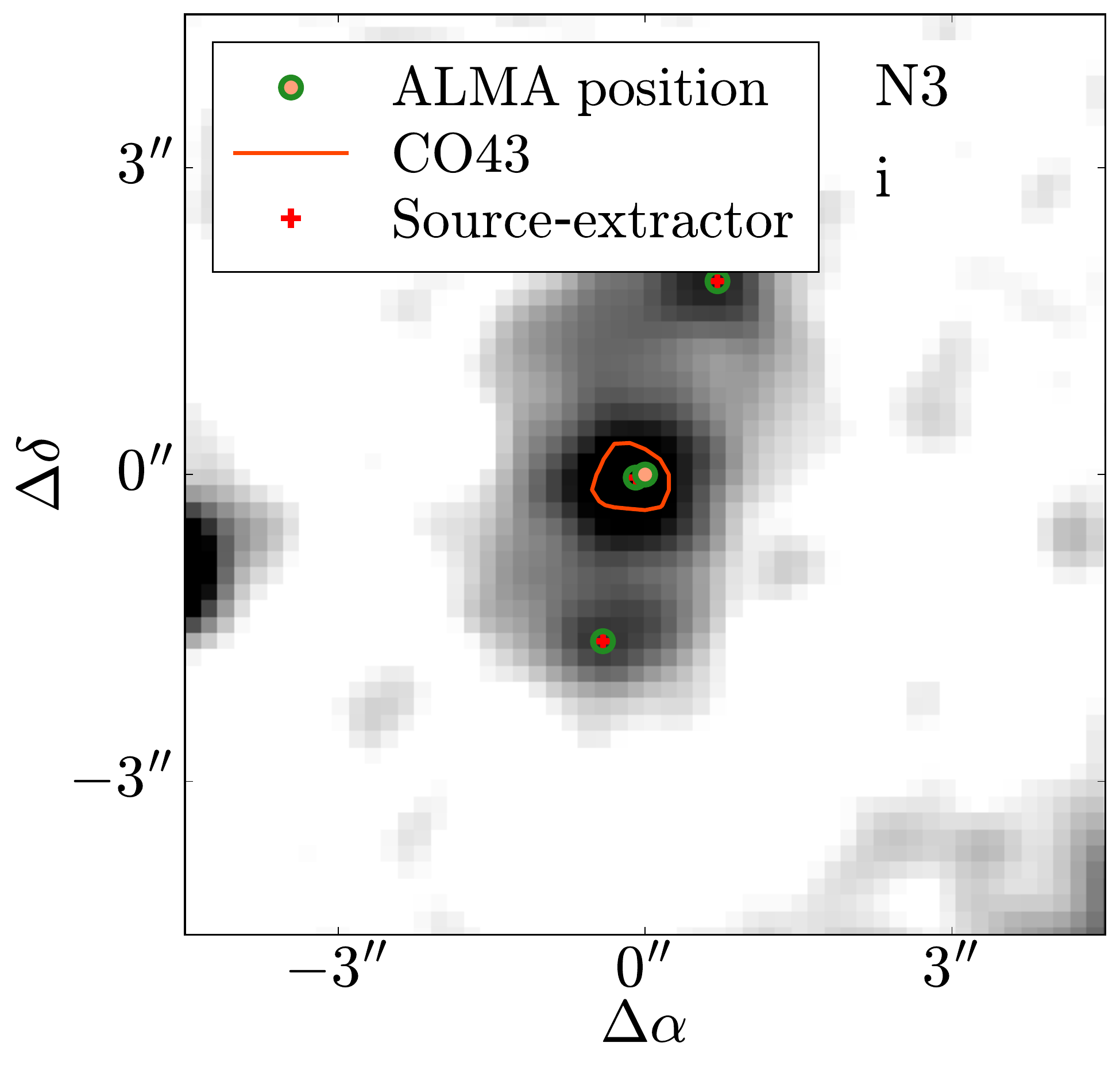}
\includegraphics[width=0.24\textwidth]{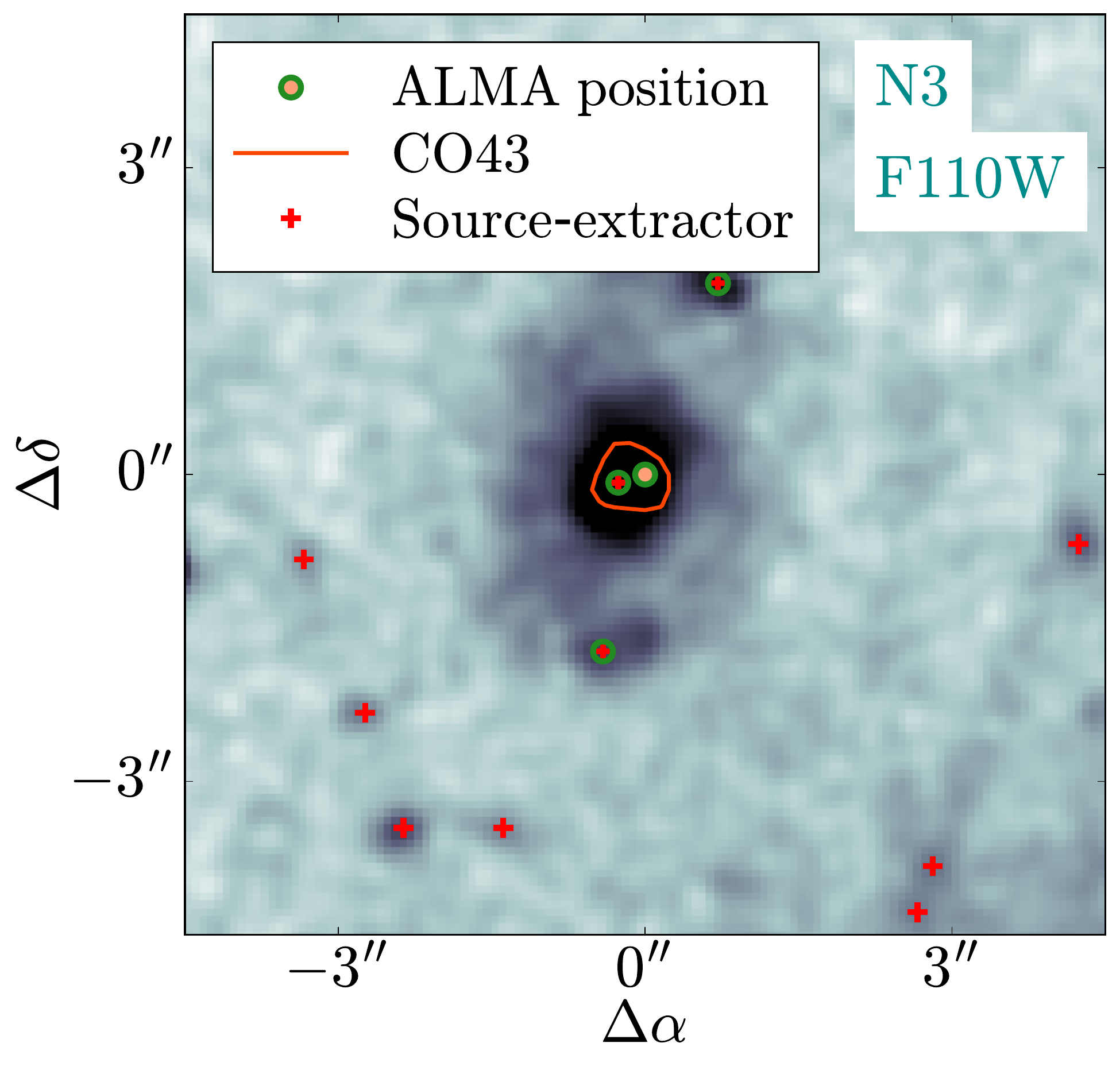}
\includegraphics[width=0.24\textwidth]{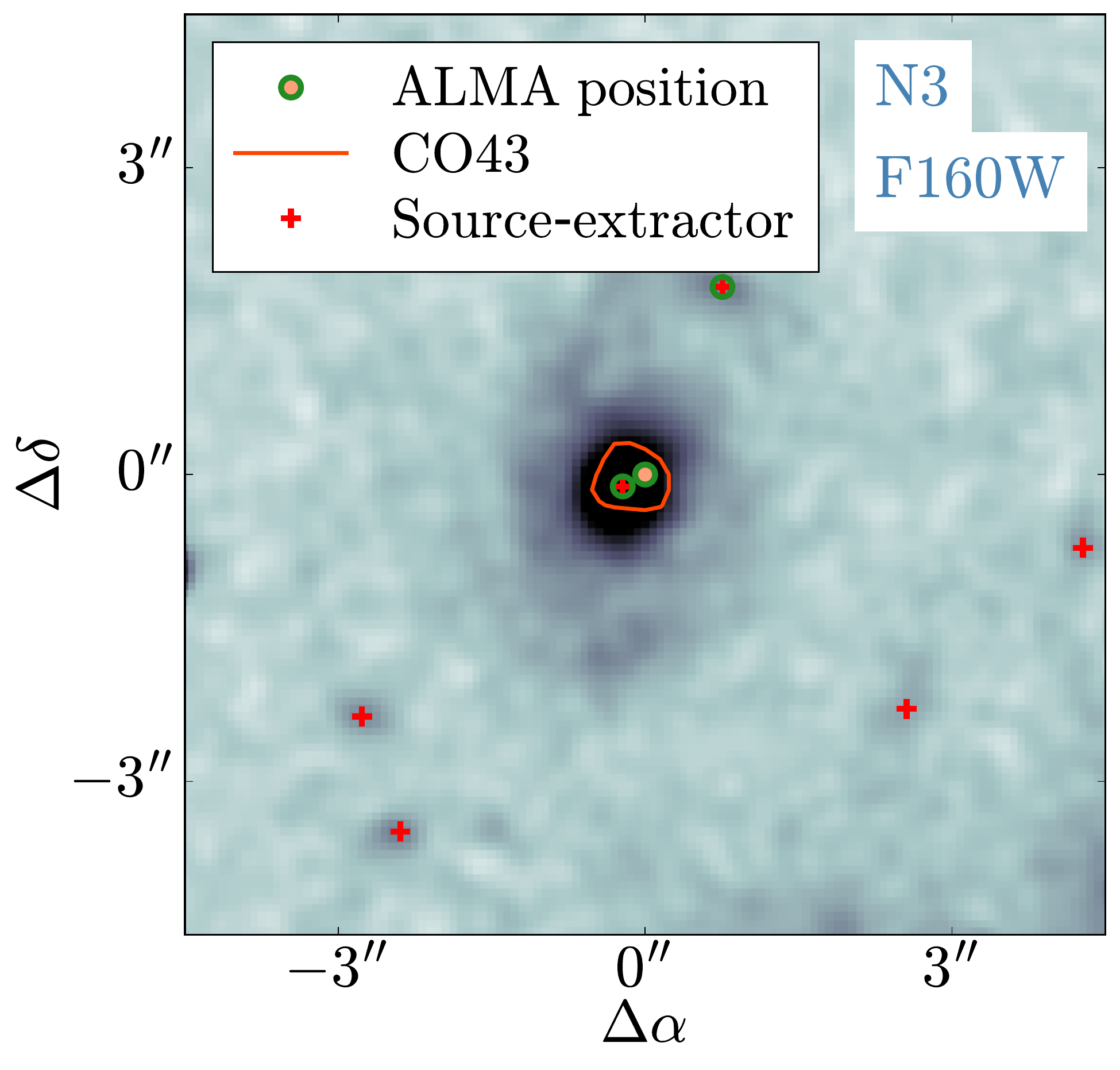}
\includegraphics[width=0.248\textwidth]{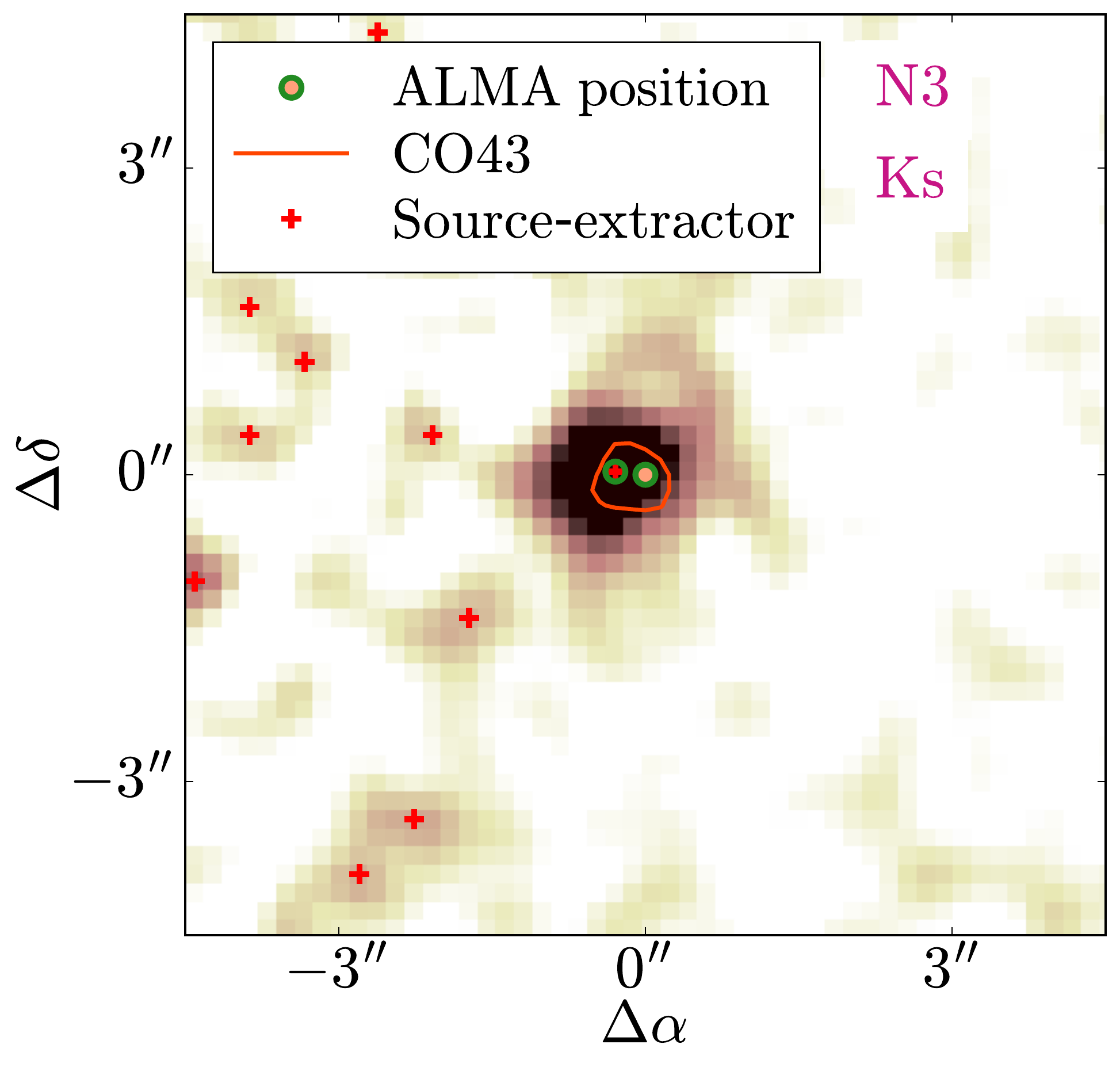}
\includegraphics[width=0.249\textwidth]{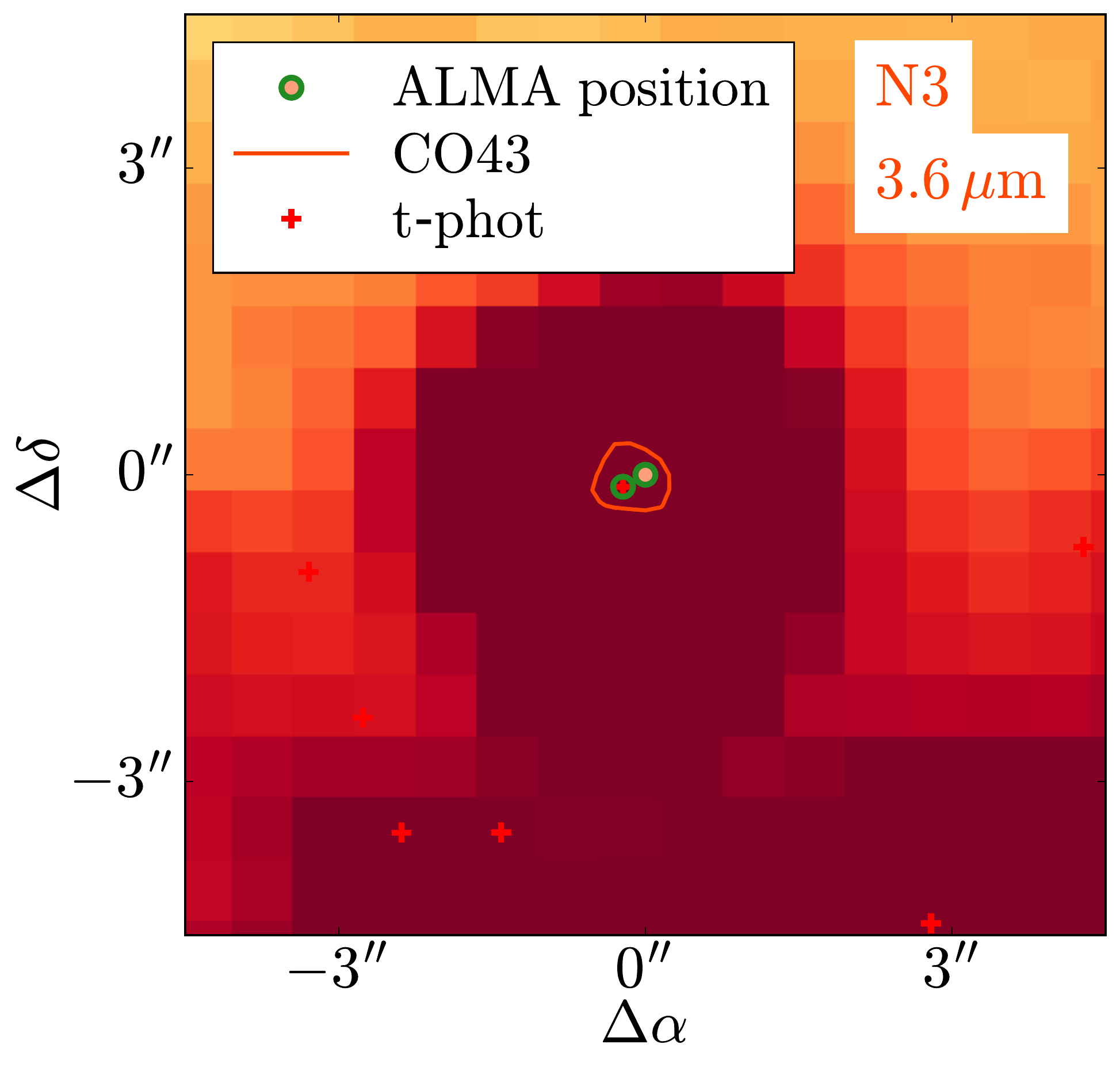}
\includegraphics[width=0.249\textwidth]{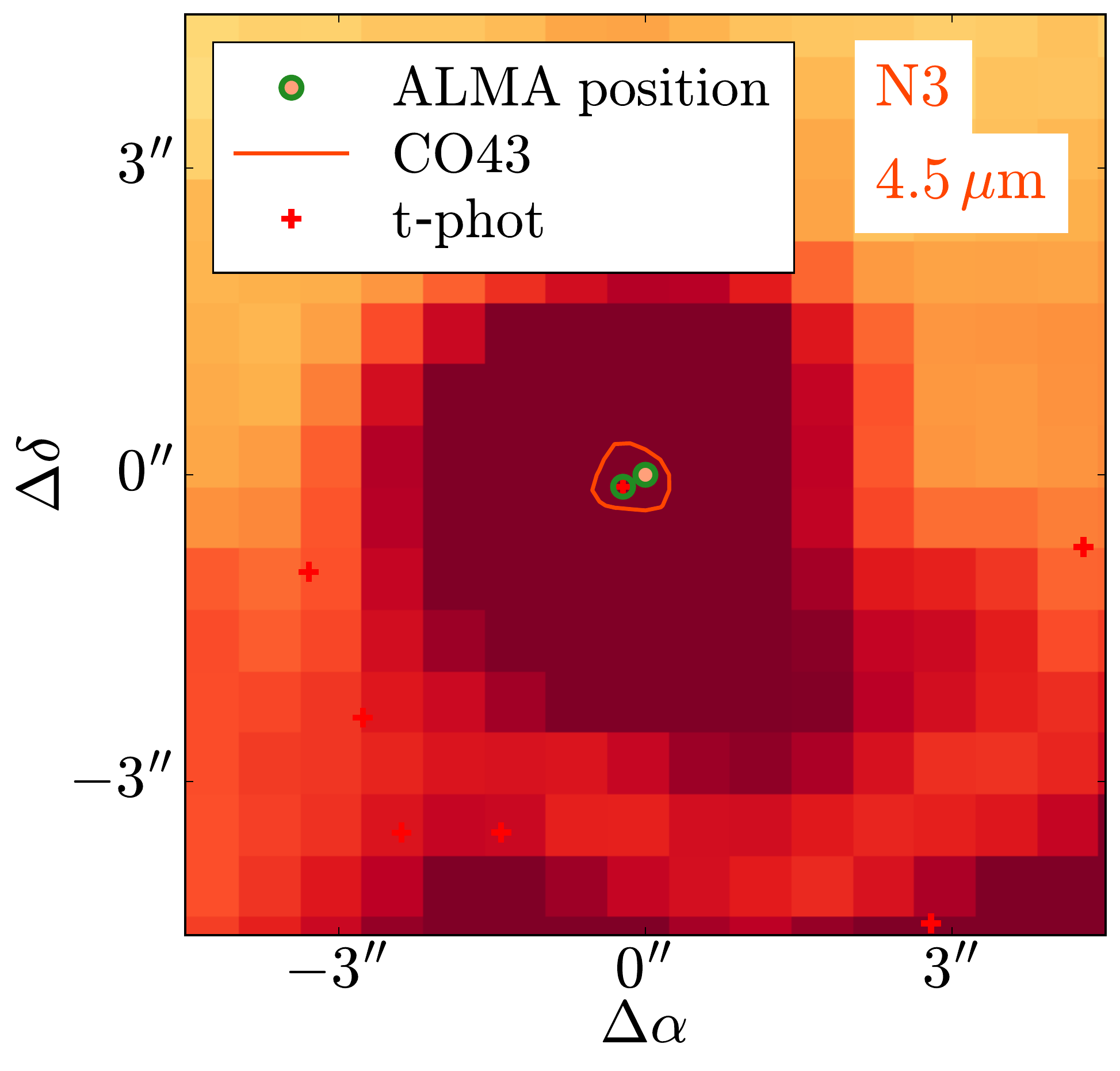}
\end{framed}
\end{subfigure}
\begin{subfigure}{0.85\textwidth}
\begin{framed}
\includegraphics[width=0.24\textwidth]{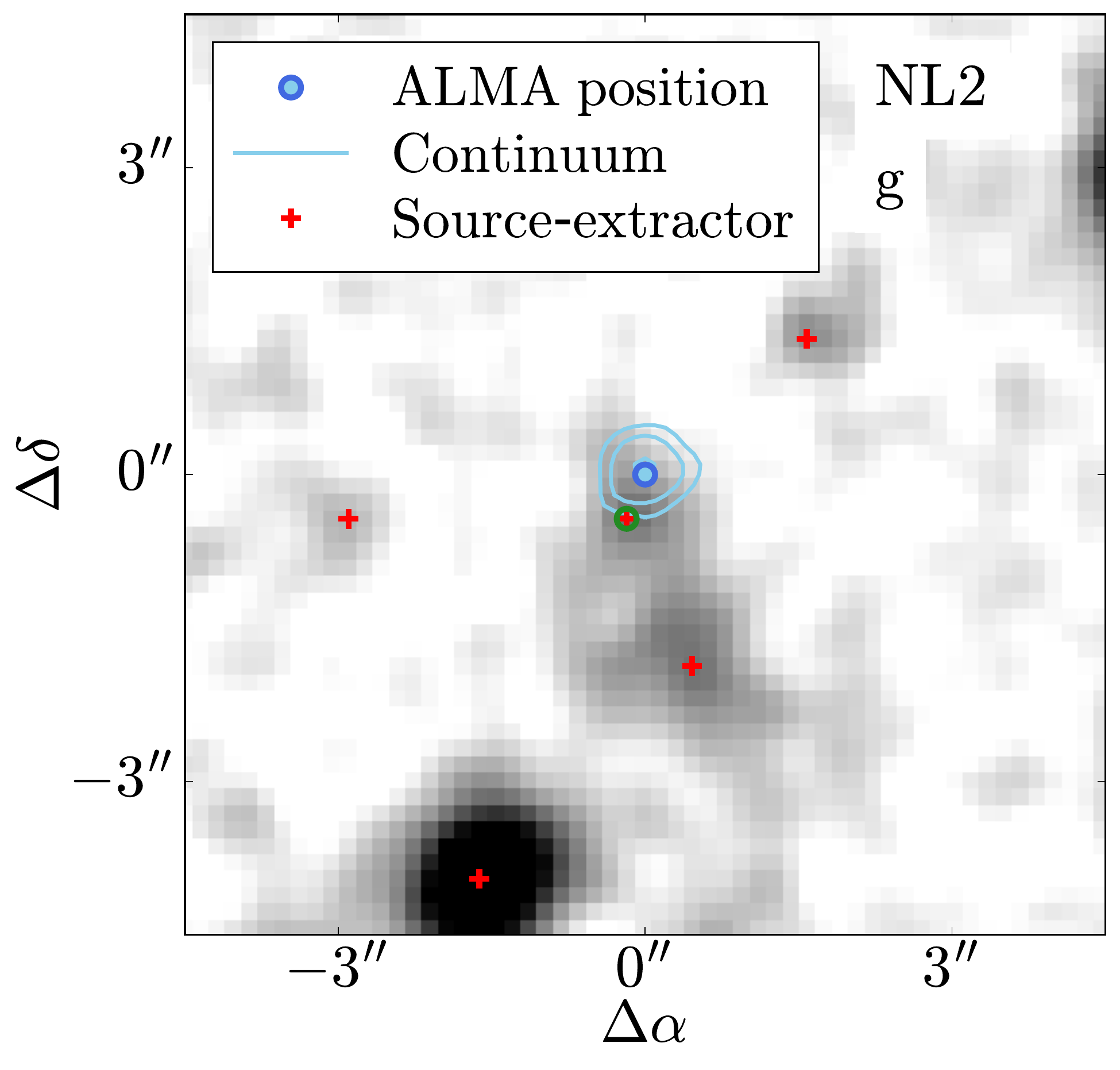}
\includegraphics[width=0.24\textwidth]{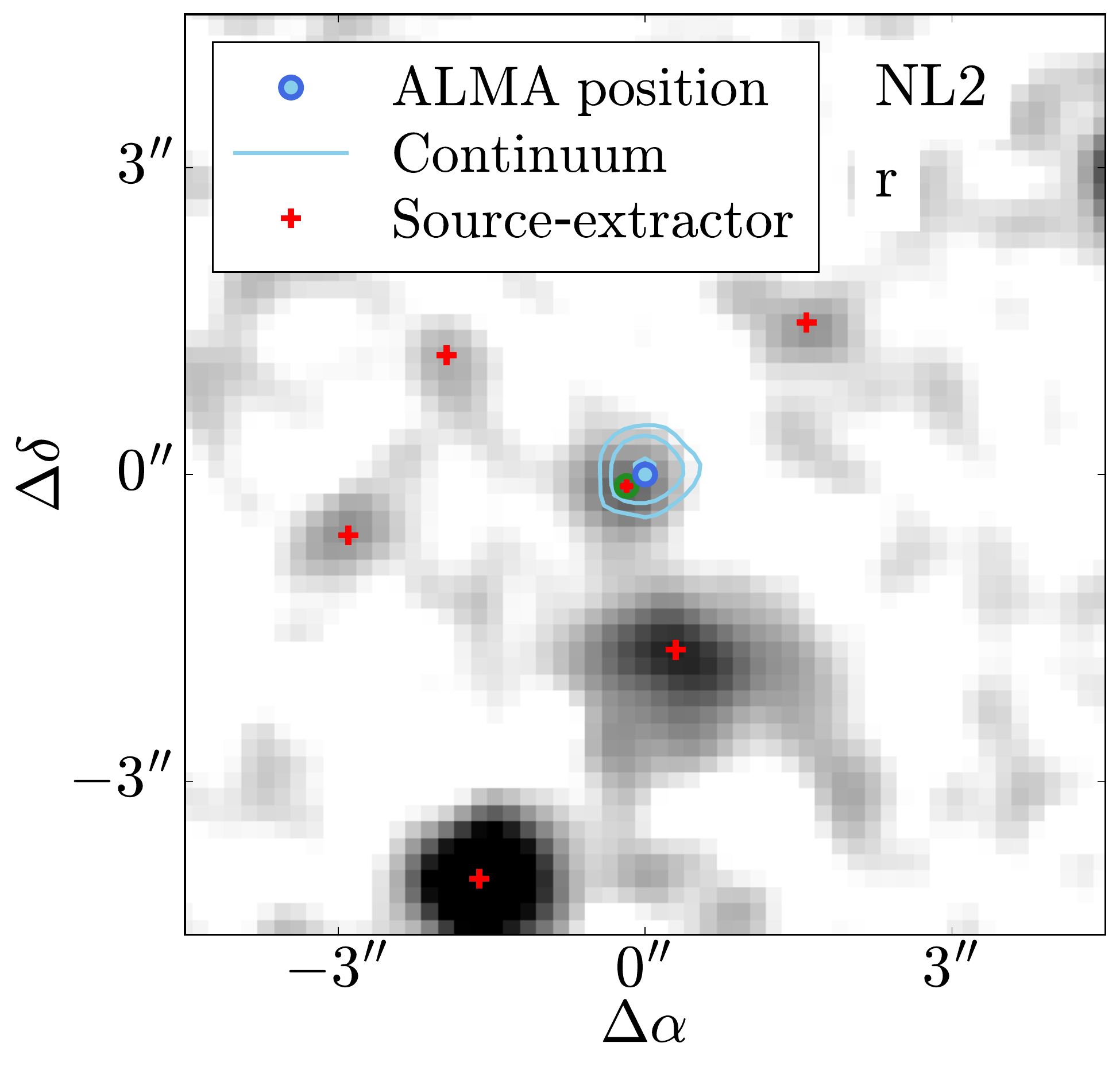}
\includegraphics[width=0.24\textwidth]{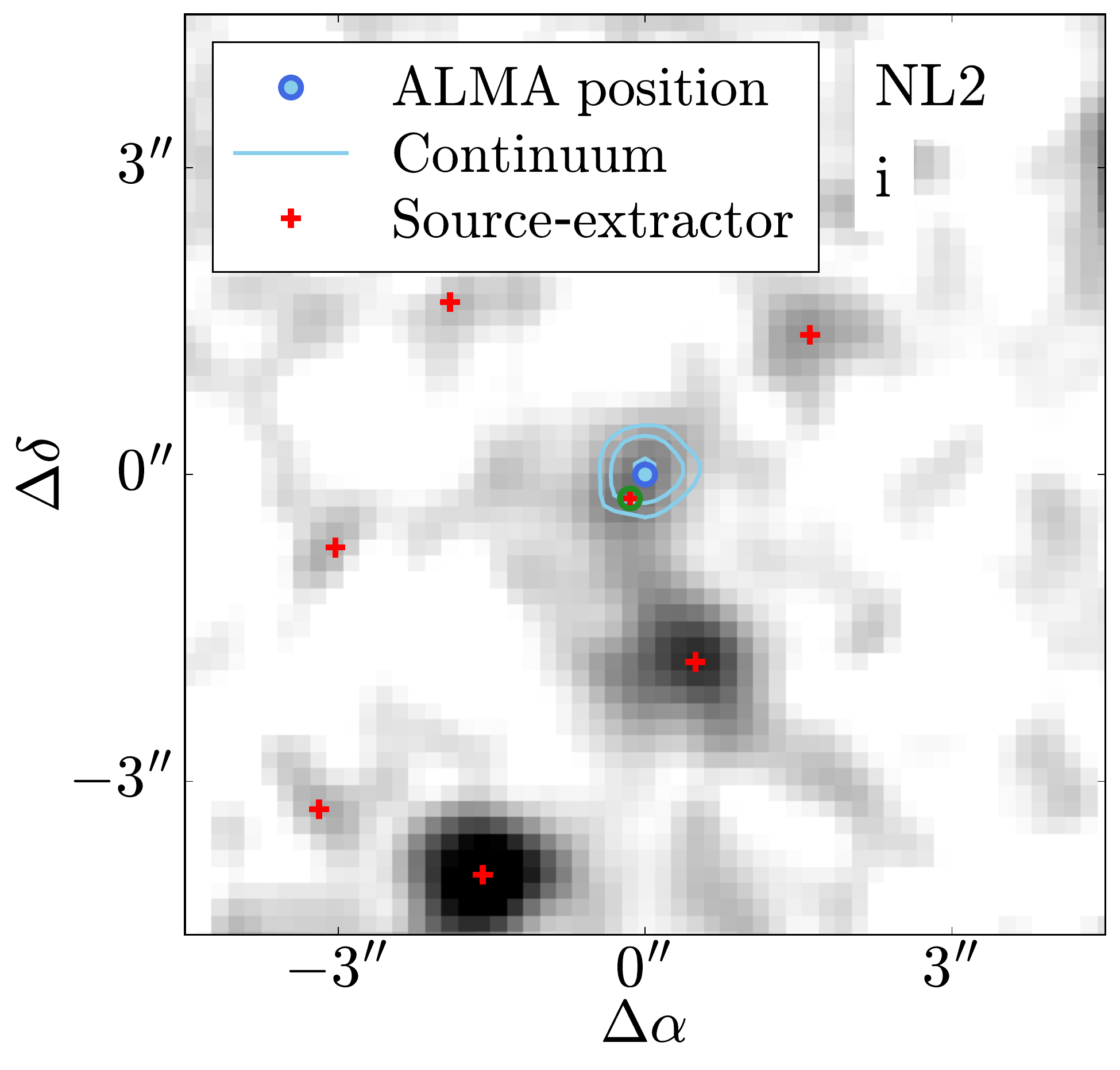}
\includegraphics[width=0.24\textwidth]{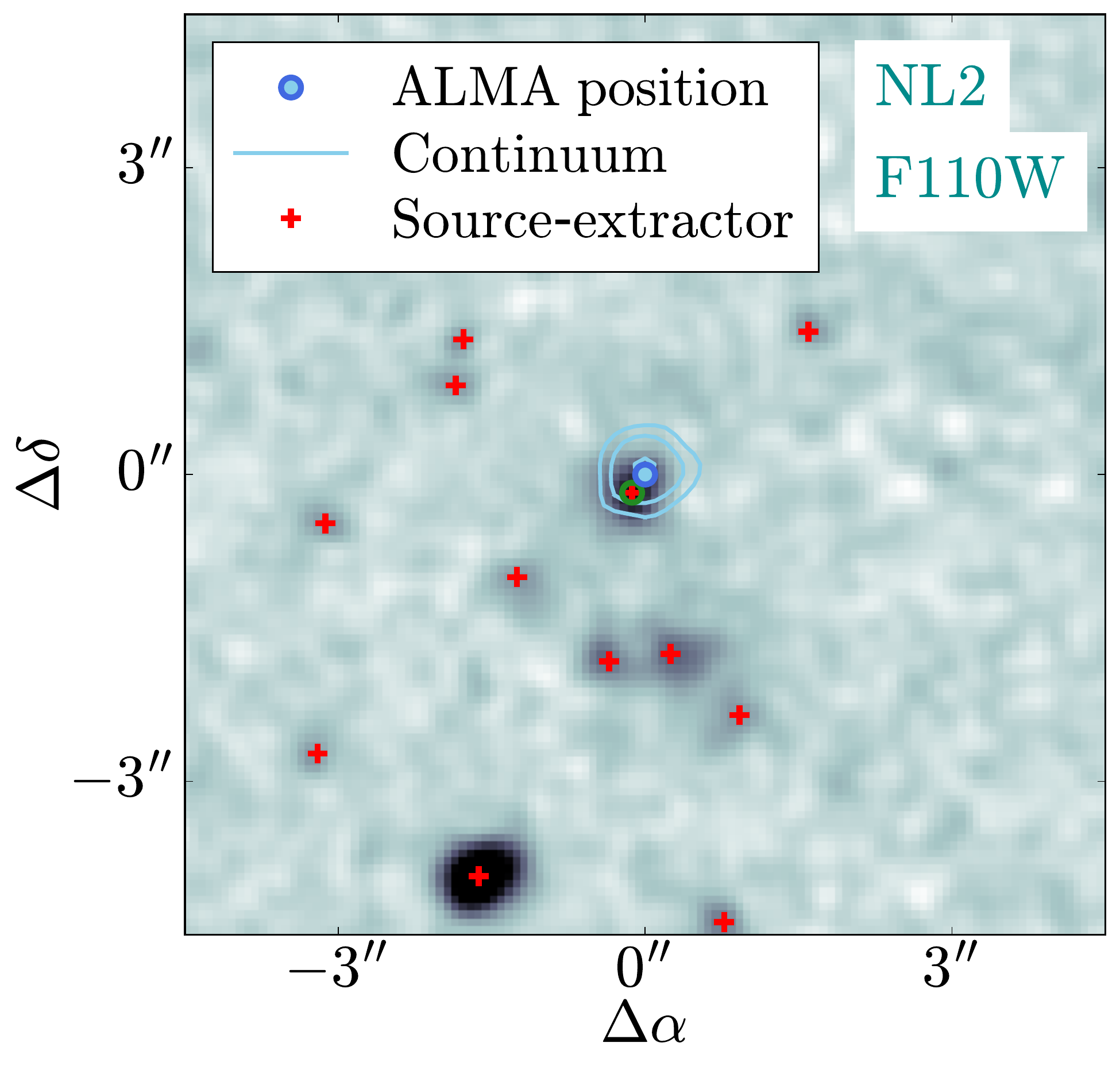}
\includegraphics[width=0.24\textwidth]{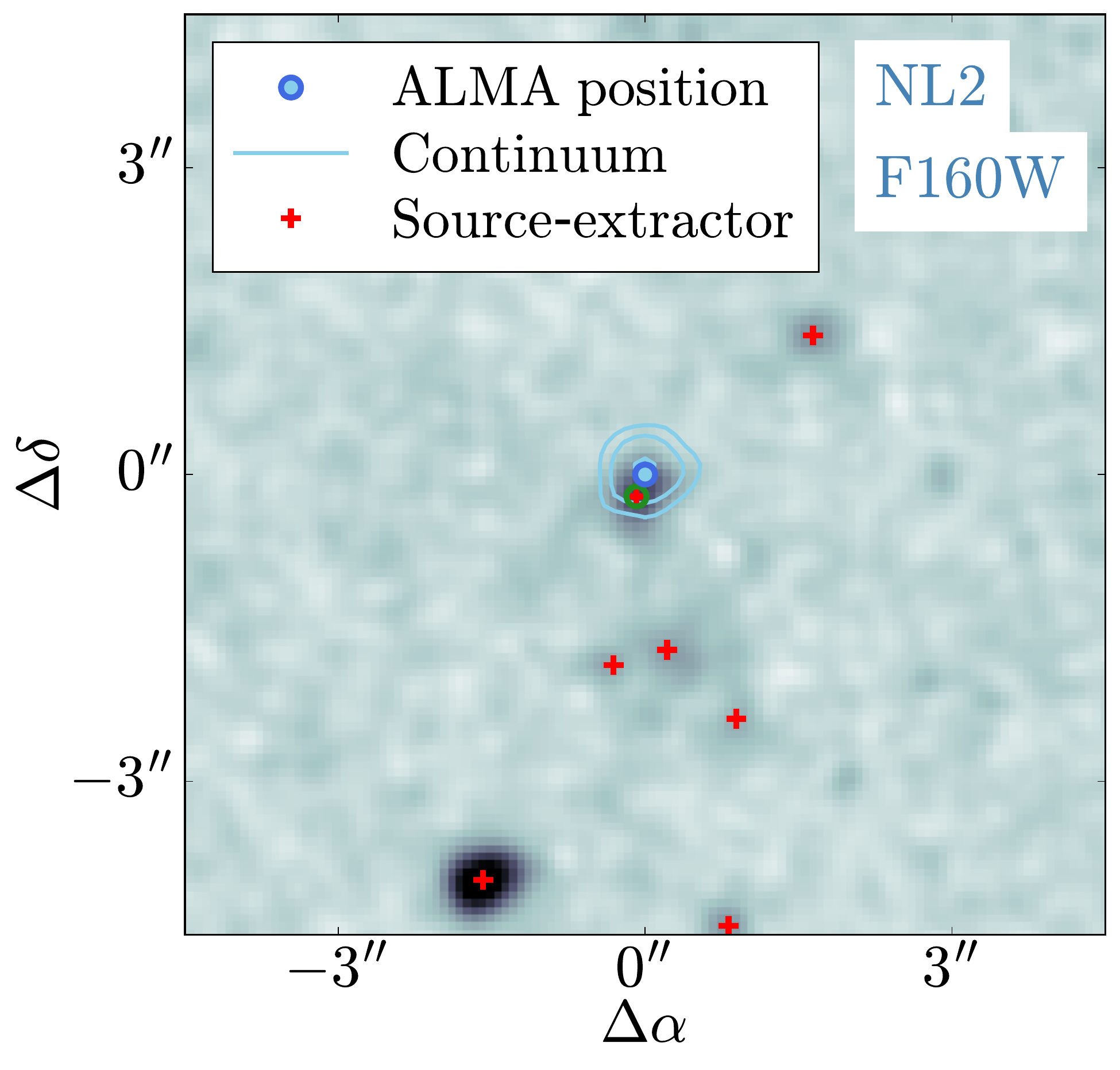}
\includegraphics[width=0.248\textwidth]{Ks/blank.pdf}
\includegraphics[width=0.249\textwidth]{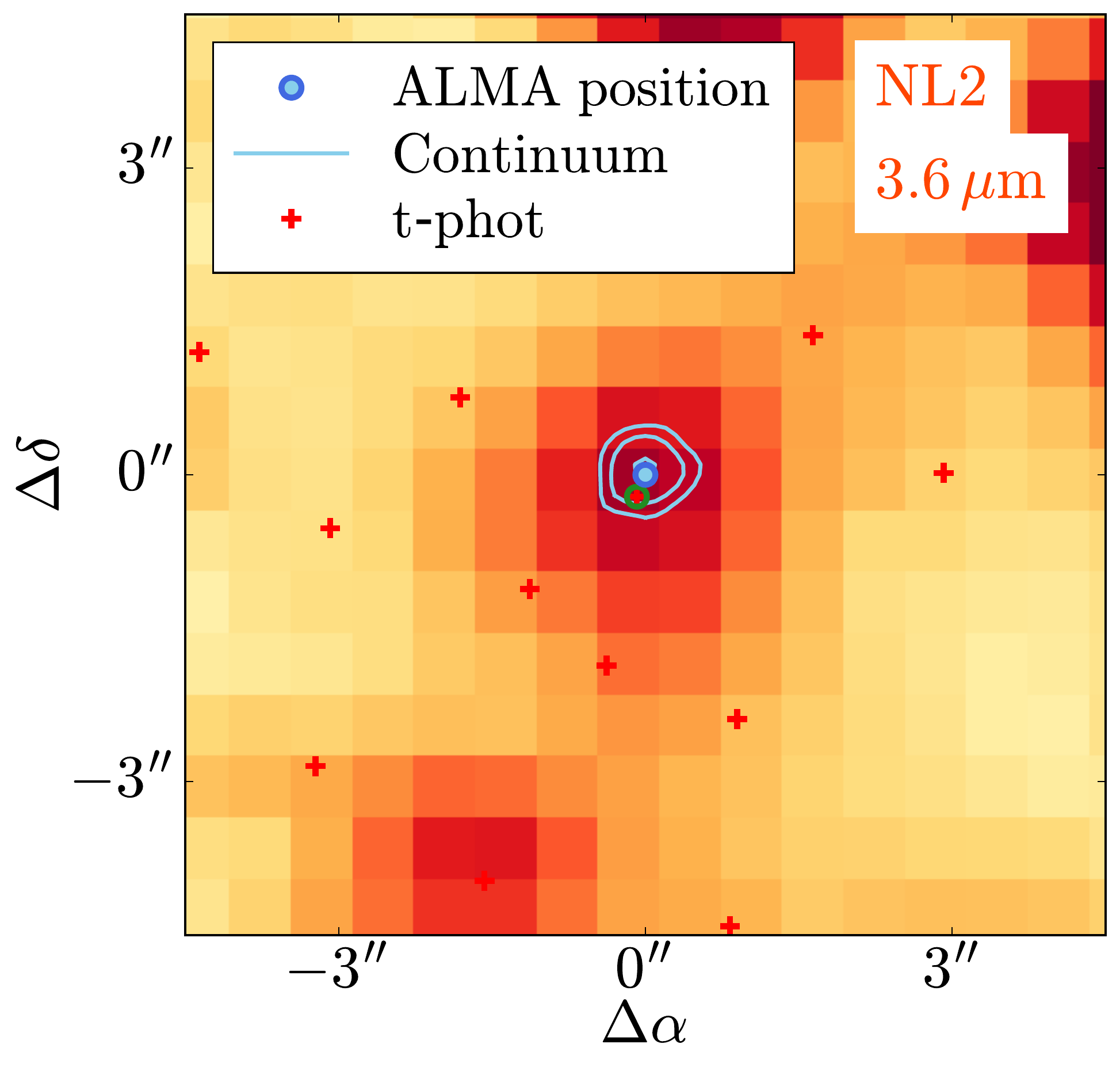}
\includegraphics[width=0.249\textwidth]{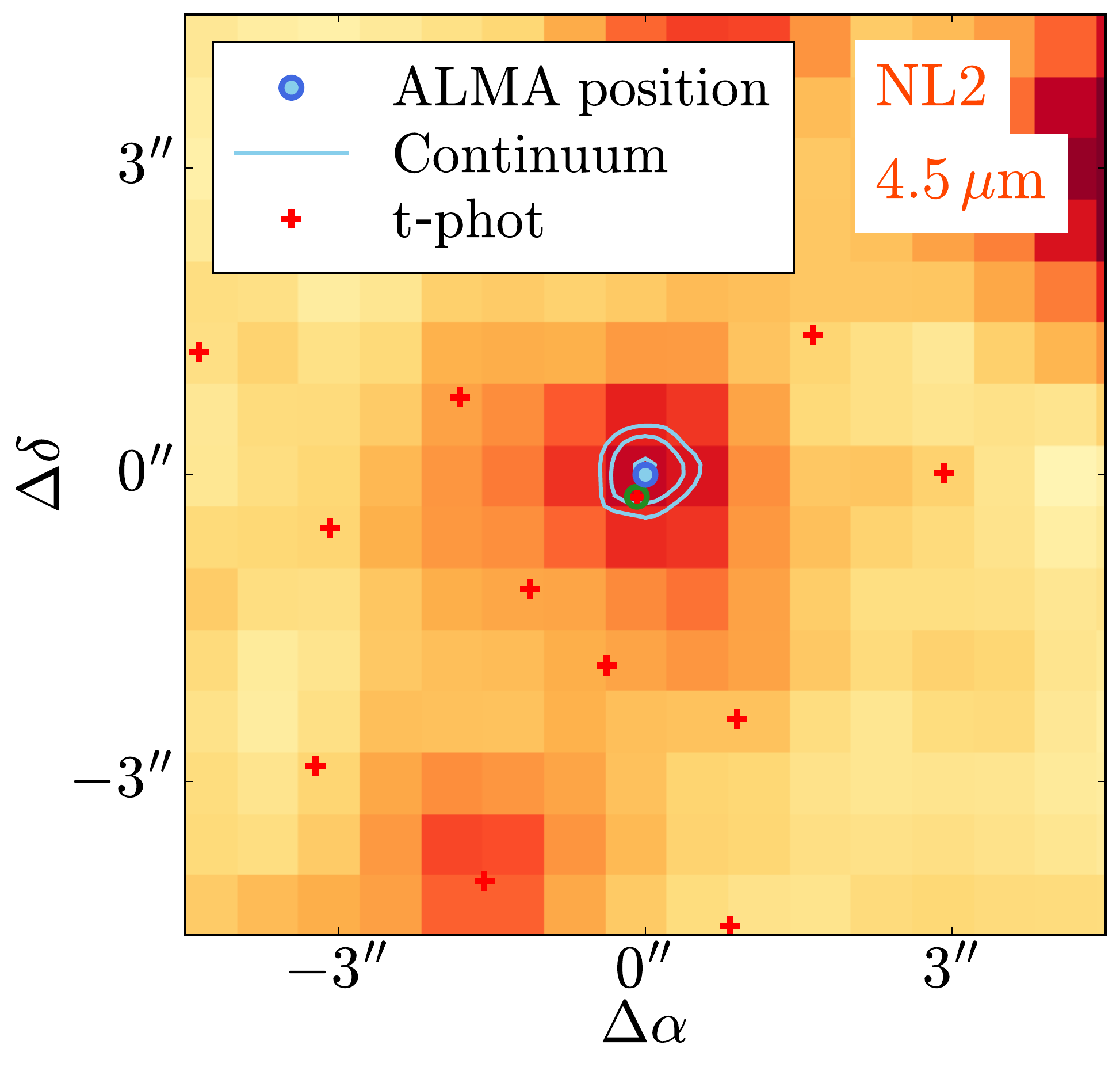}
\end{framed}
\end{subfigure}
\begin{subfigure}{0.85\textwidth}
\begin{framed}
\includegraphics[width=0.24\textwidth]{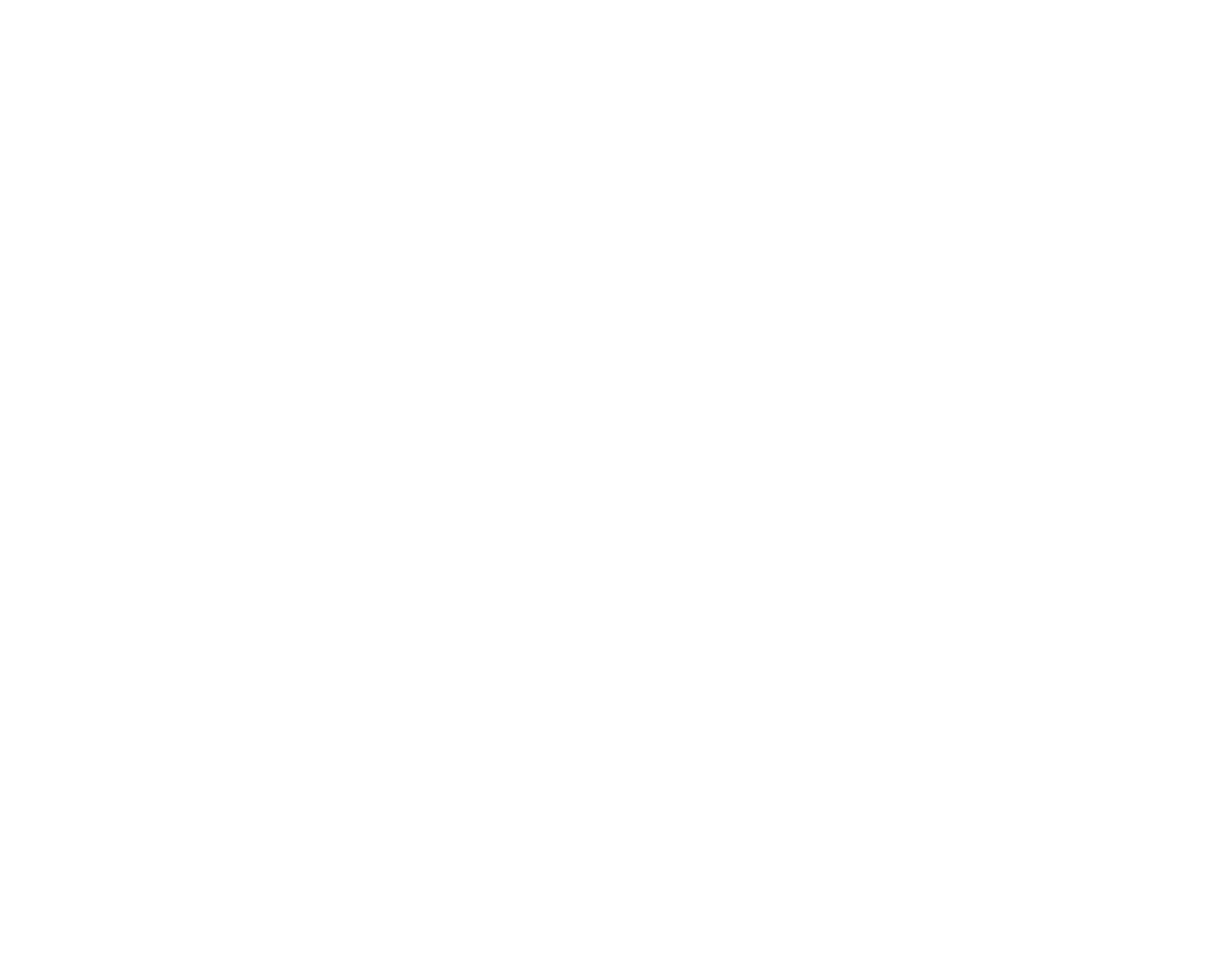}
\includegraphics[width=0.24\textwidth]{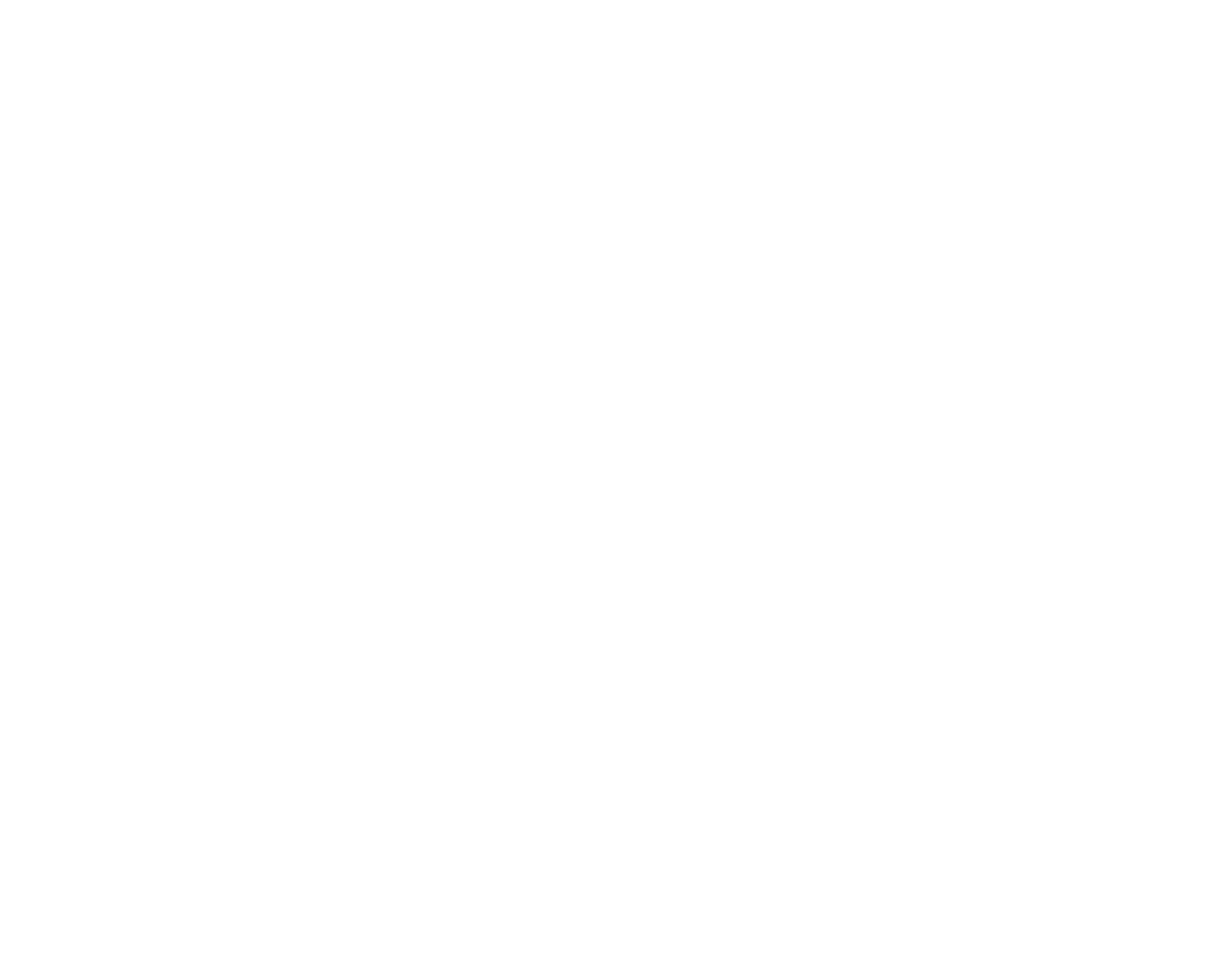}
\includegraphics[width=0.24\textwidth]{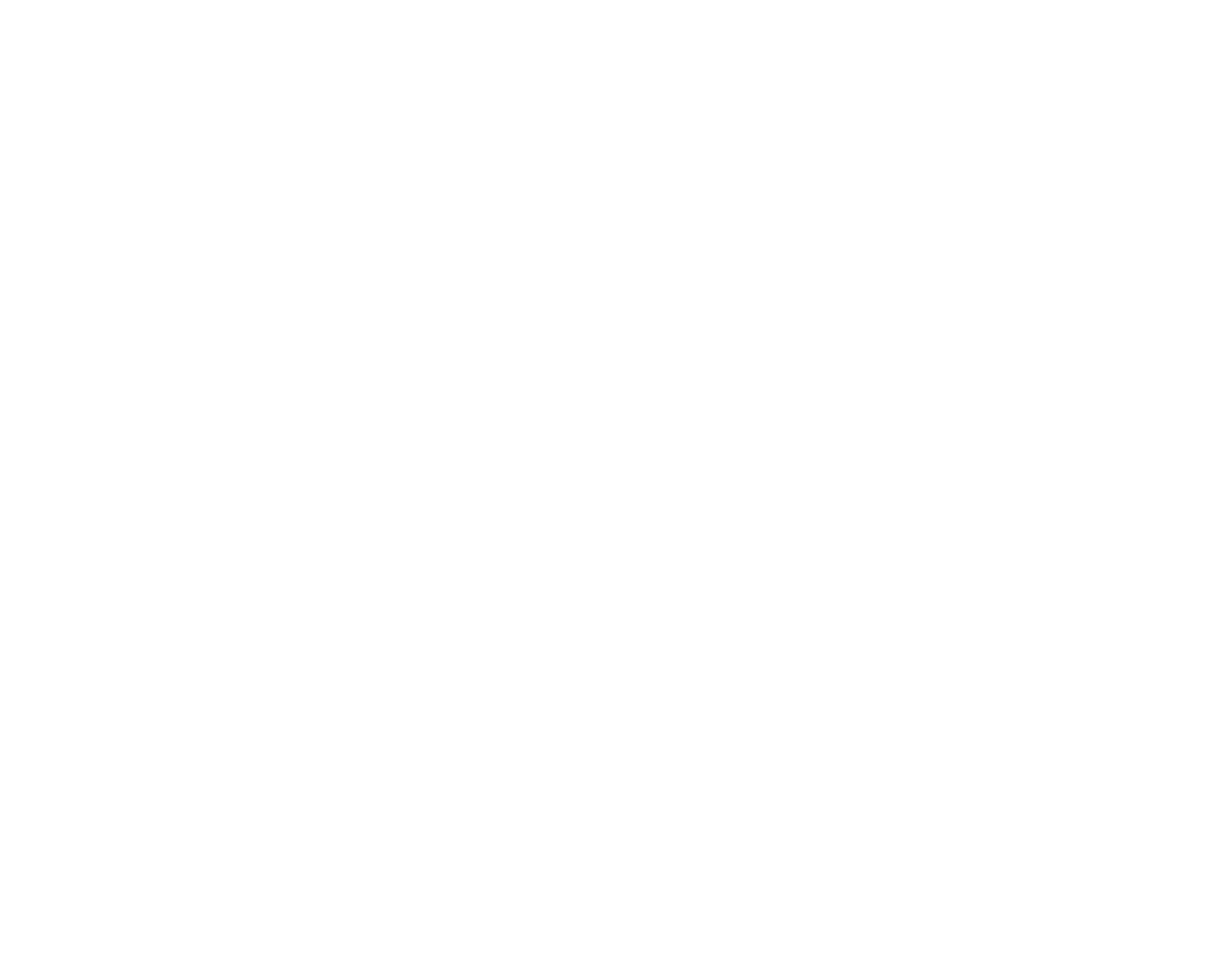}
\includegraphics[width=0.24\textwidth]{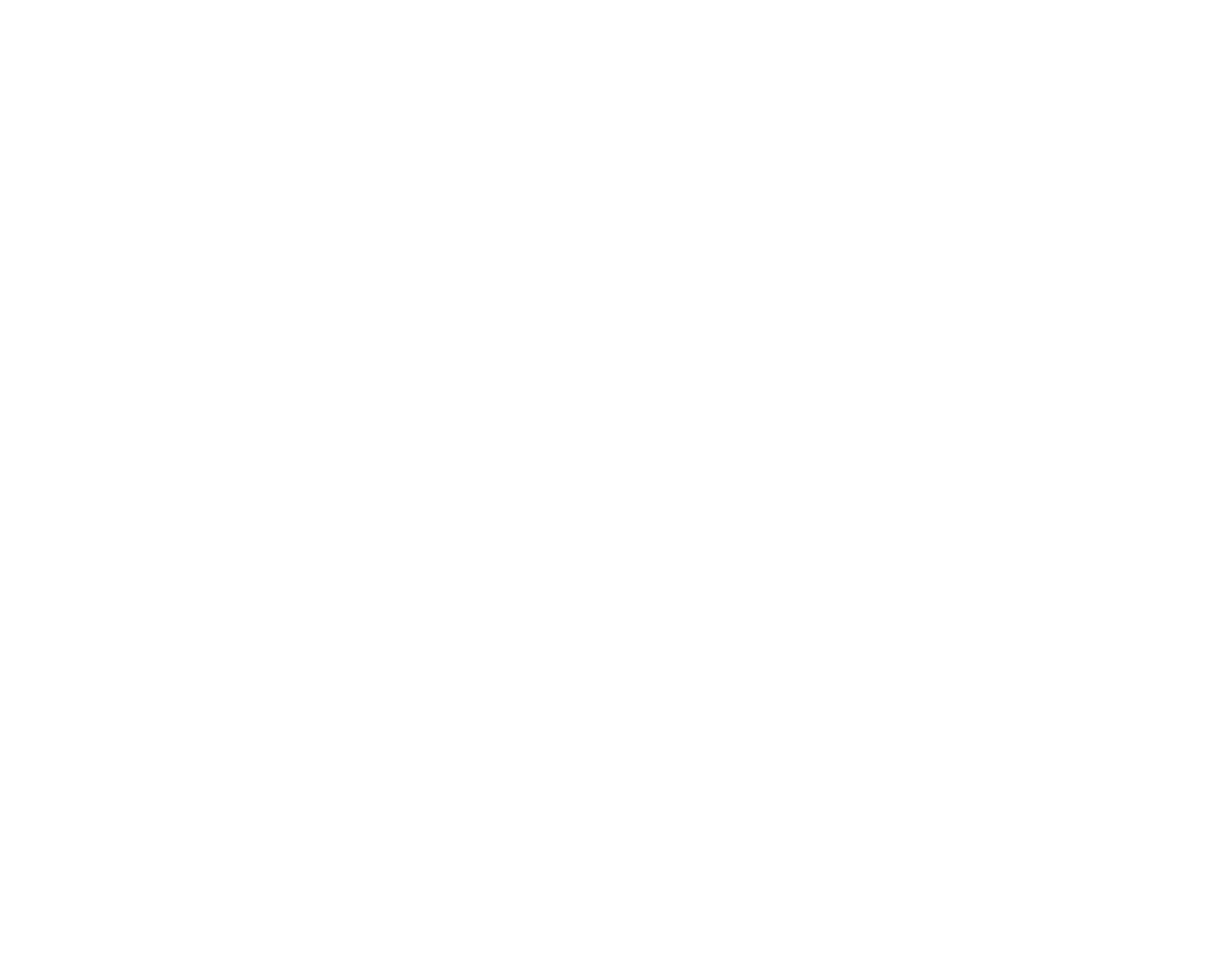}
\includegraphics[width=0.24\textwidth]{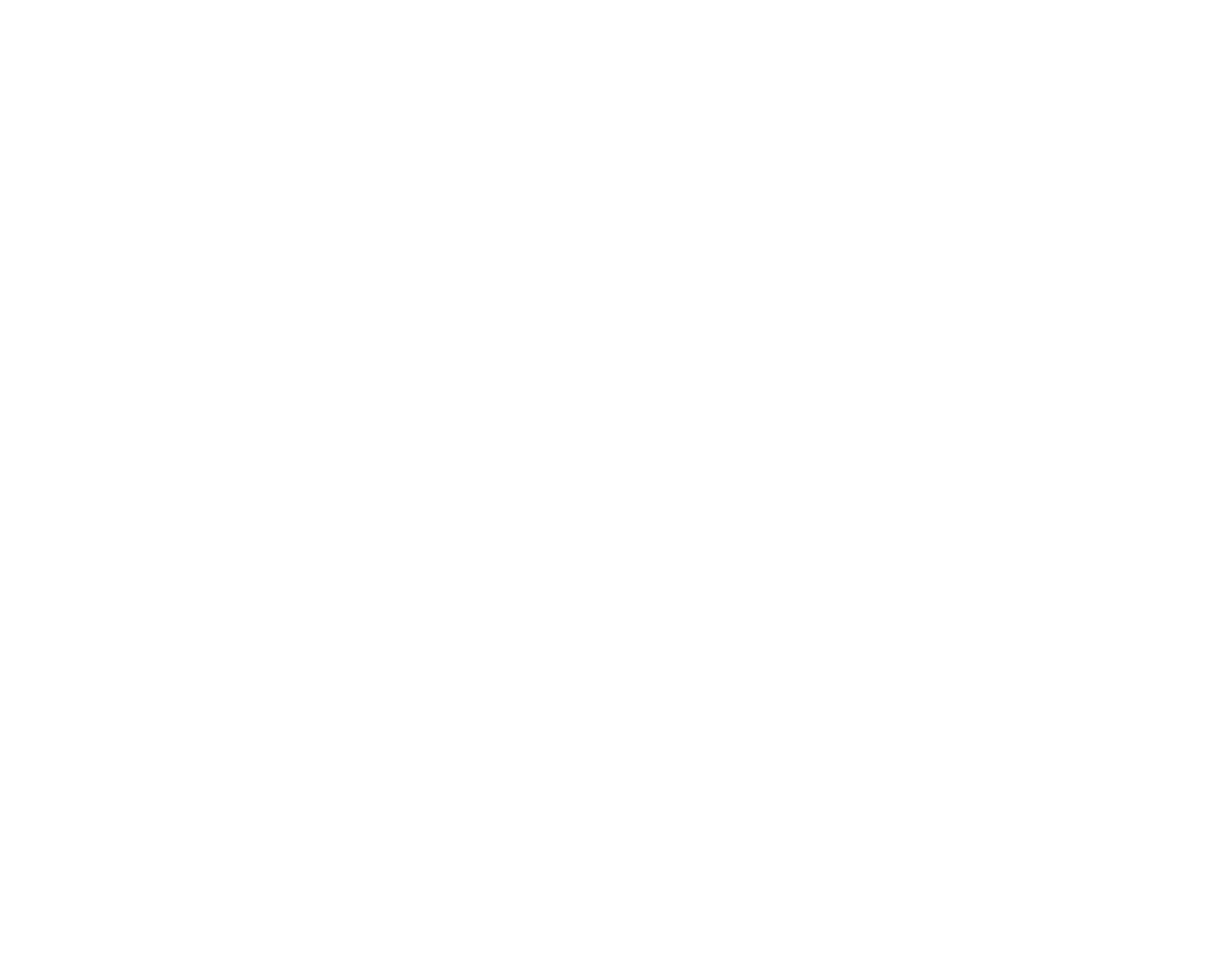}
\includegraphics[width=0.248\textwidth]{Ks/blank.pdf}
\includegraphics[width=0.249\textwidth]{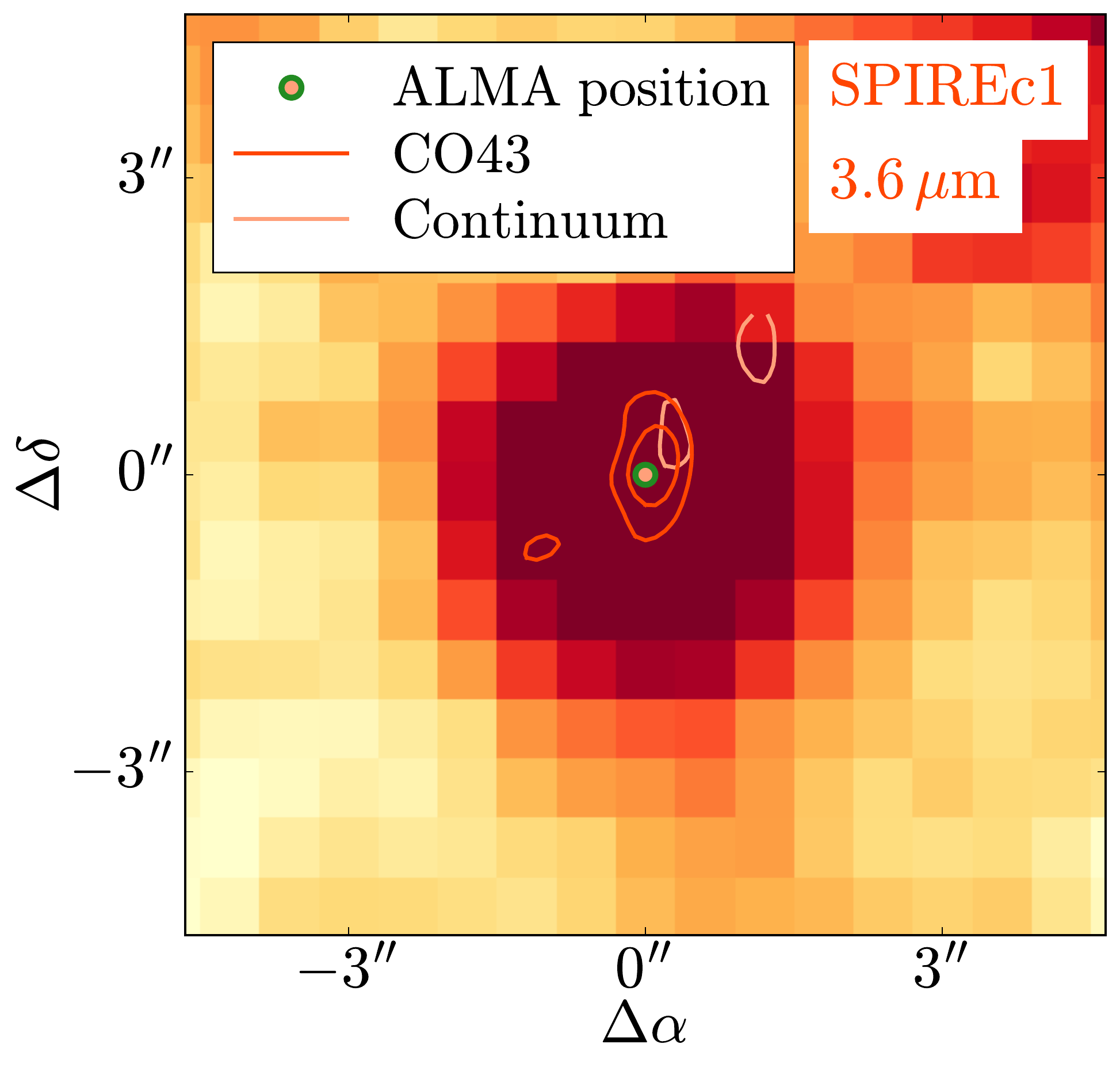}
\includegraphics[width=0.249\textwidth]{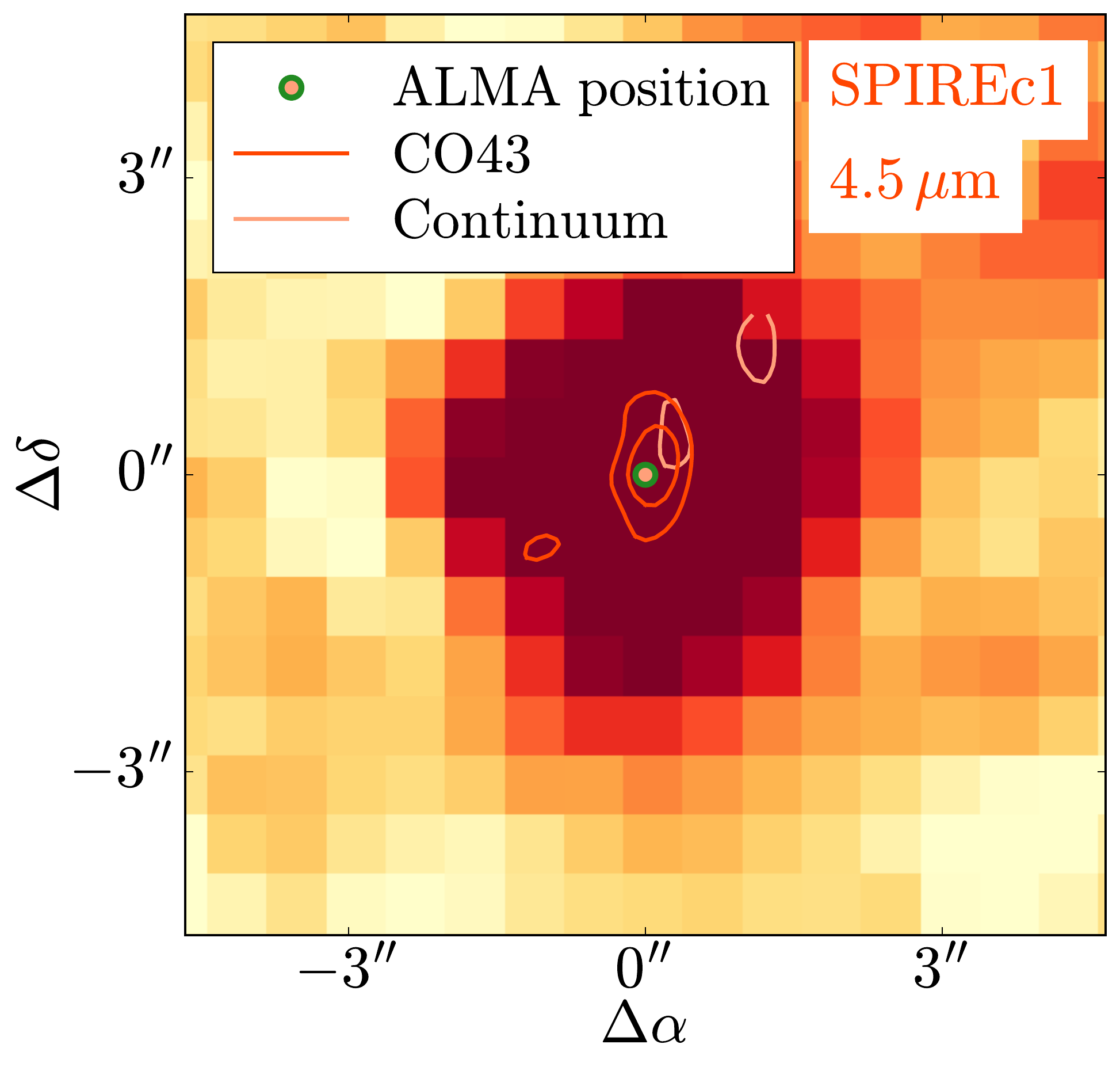}
\end{framed}
\end{subfigure}
\caption{}
\end{figure*}
\renewcommand{\thefigure}{\arabic{figure}}

\renewcommand{\thefigure}{B\arabic{figure} (Cont.)}
\addtocounter{figure}{-1}
\begin{figure*}
\begin{subfigure}{0.85\textwidth}
\begin{framed}
\includegraphics[width=0.24\textwidth]{g/blank.pdf}
\includegraphics[width=0.24\textwidth]{r/blank.pdf}
\includegraphics[width=0.24\textwidth]{i/blank.pdf}
\includegraphics[width=0.24\textwidth]{F110W/blank.pdf}
\includegraphics[width=0.24\textwidth]{F160W/blank.pdf}
\includegraphics[width=0.248\textwidth]{Ks/blank.pdf}
\includegraphics[width=0.249\textwidth]{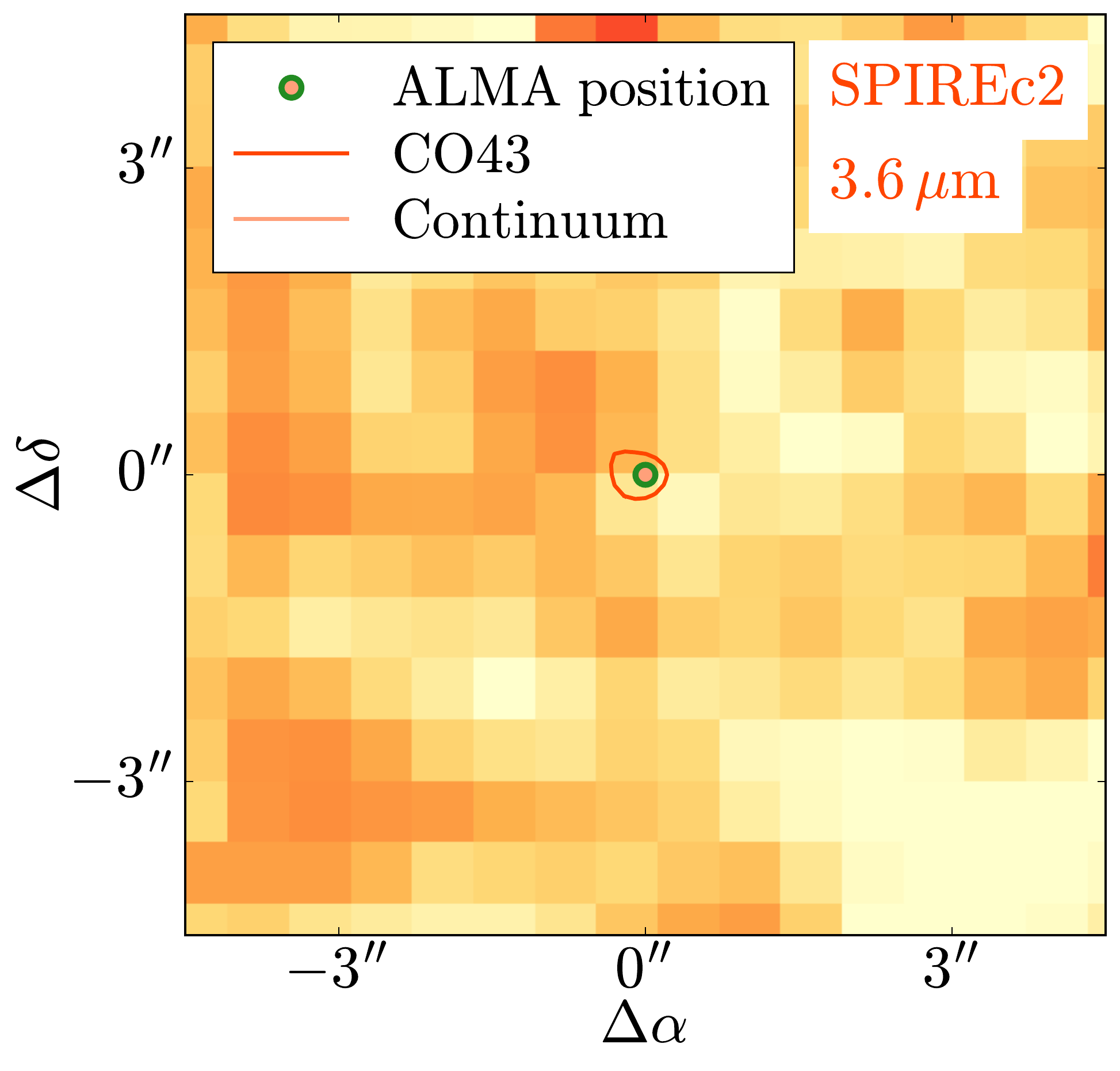}
\includegraphics[width=0.249\textwidth]{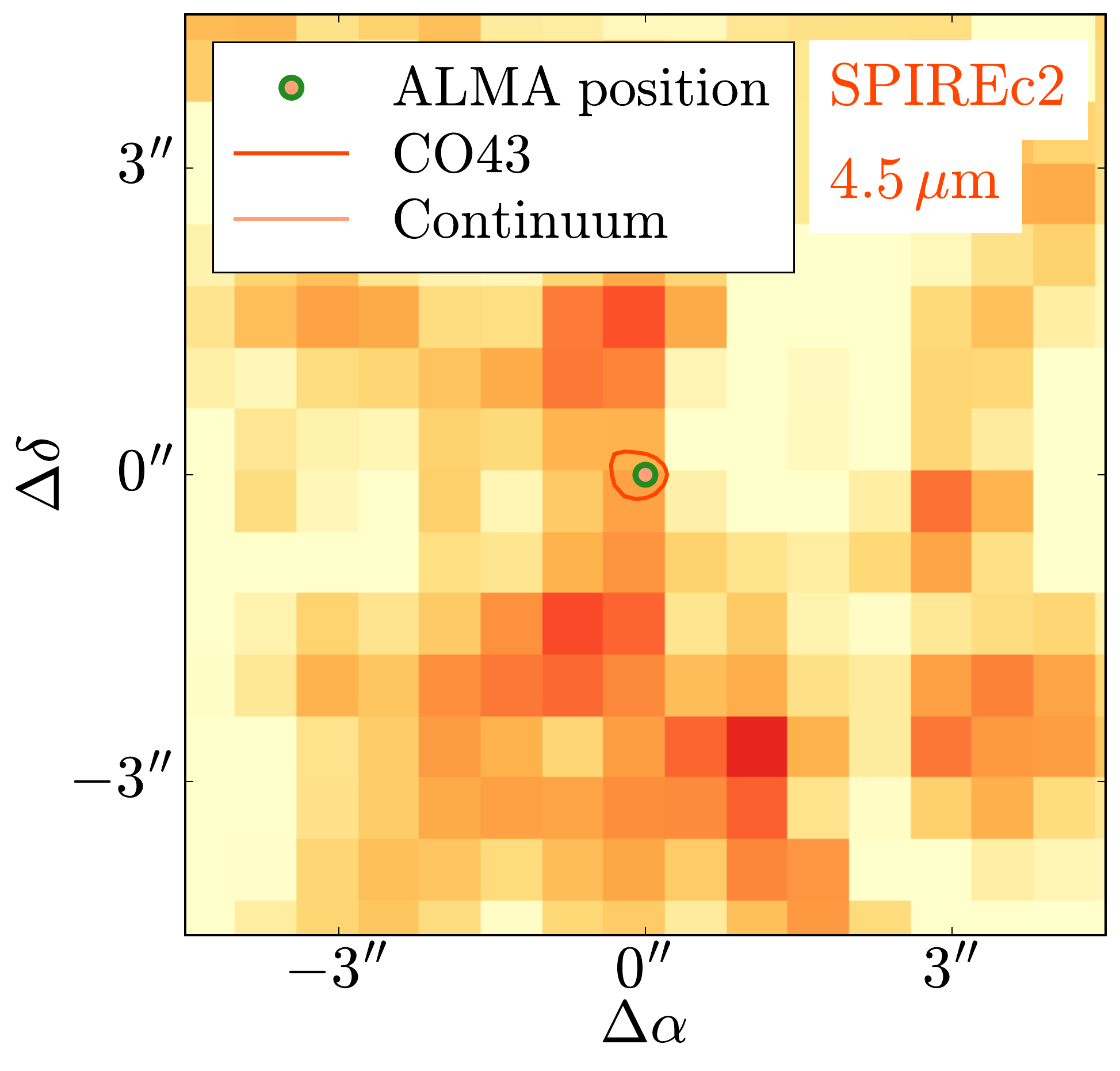}
\end{framed}
\end{subfigure}
\begin{subfigure}{0.85\textwidth}
\begin{framed}
\includegraphics[width=0.24\textwidth]{g/blank.pdf}
\includegraphics[width=0.24\textwidth]{r/blank.pdf}
\includegraphics[width=0.24\textwidth]{i/blank.pdf}
\includegraphics[width=0.24\textwidth]{F110W/blank.pdf}
\includegraphics[width=0.24\textwidth]{F160W/blank.pdf}
\includegraphics[width=0.248\textwidth]{Ks/blank.pdf}
\includegraphics[width=0.249\textwidth]{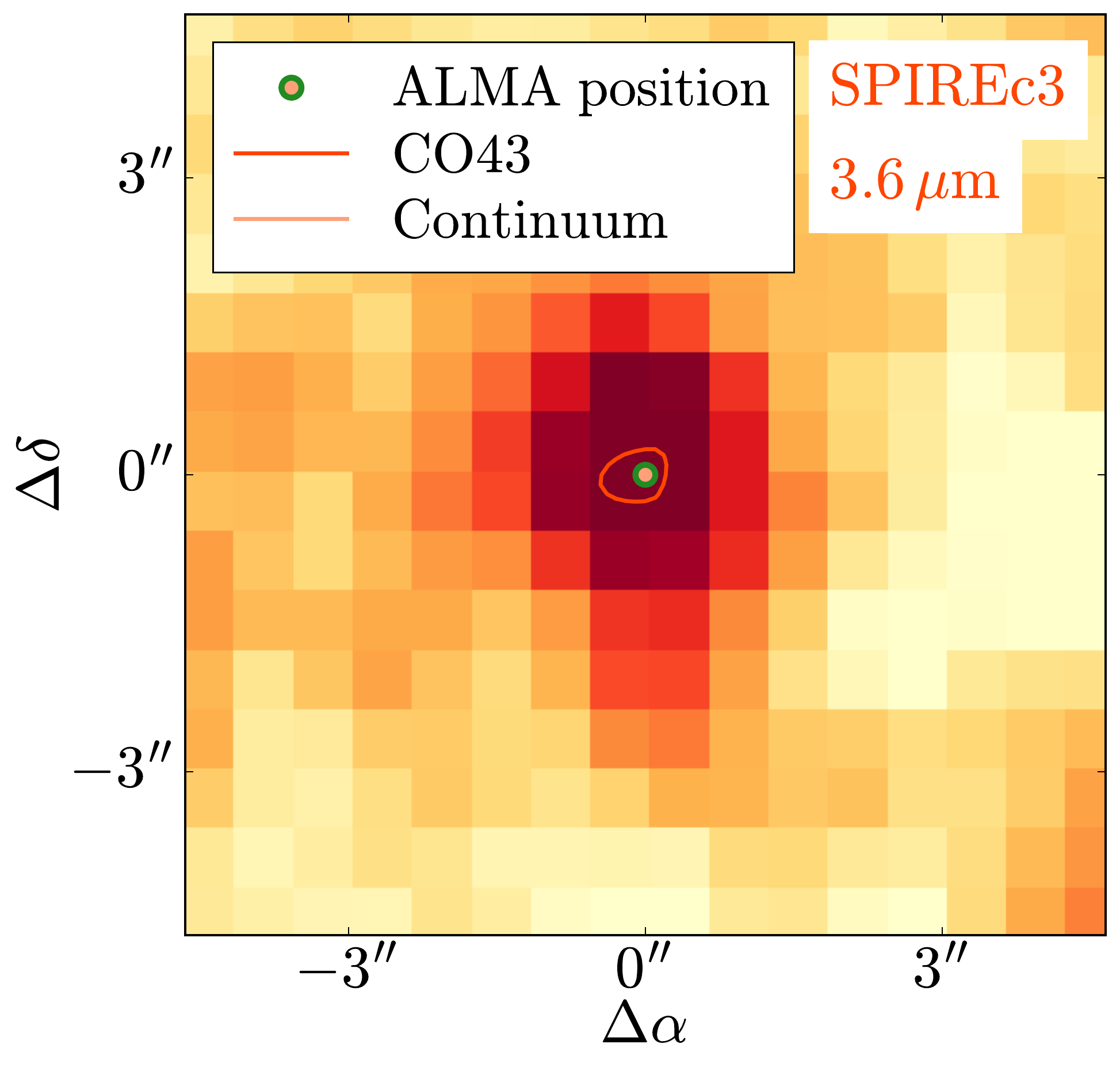}
\includegraphics[width=0.249\textwidth]{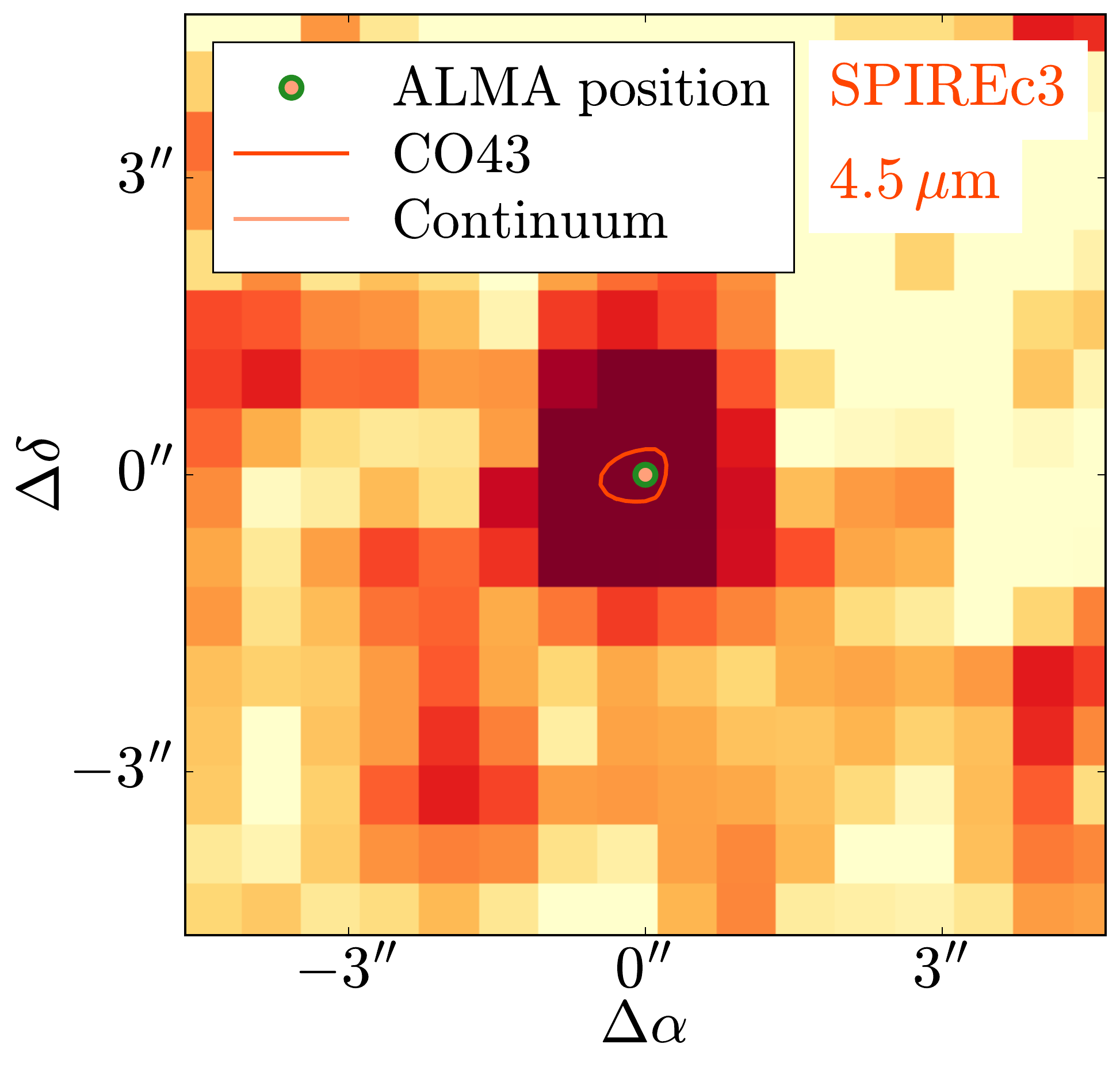}
\end{framed}
\end{subfigure}
\begin{subfigure}{0.85\textwidth}
\begin{framed}
\includegraphics[width=0.24\textwidth]{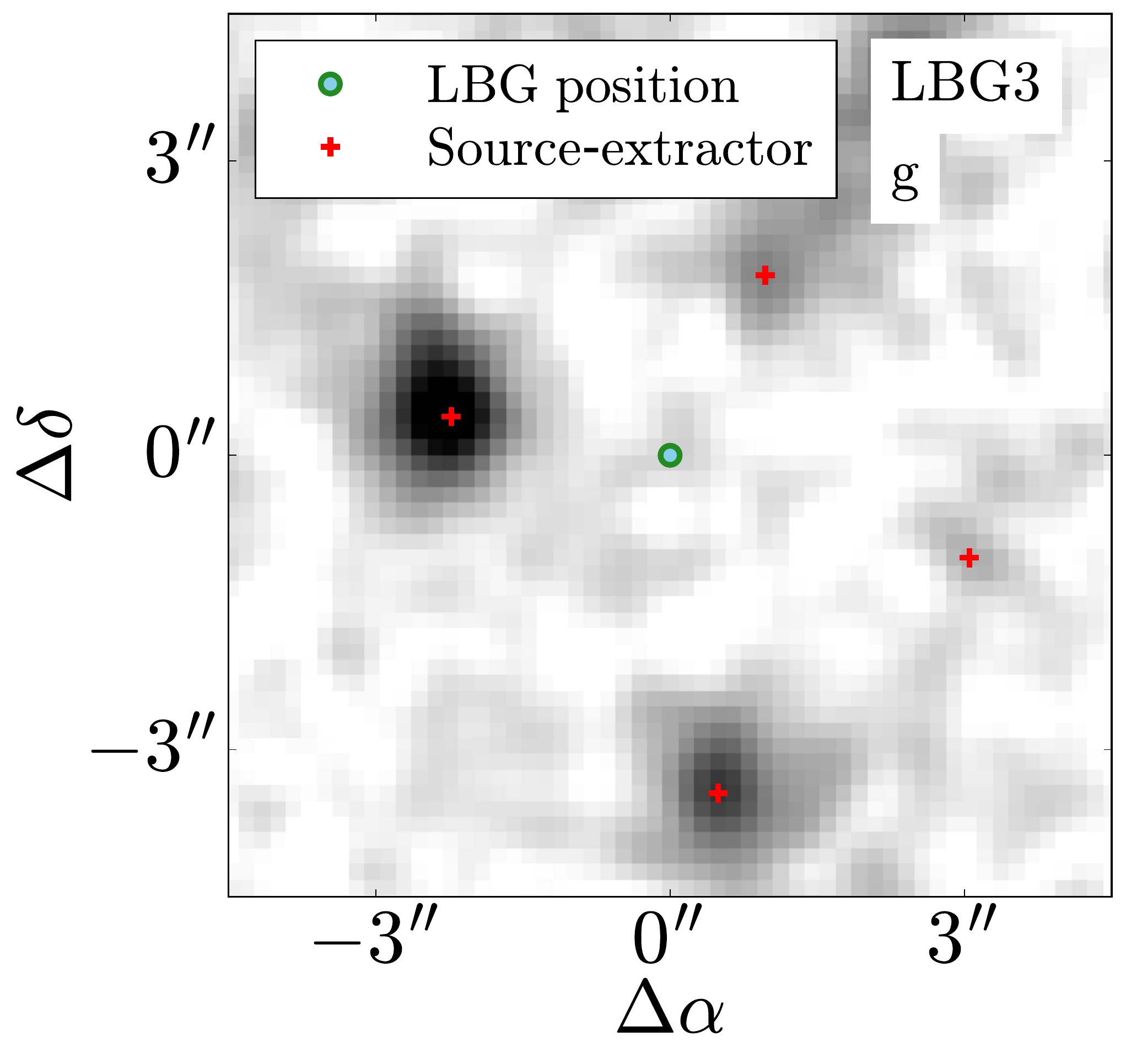}
\includegraphics[width=0.24\textwidth]{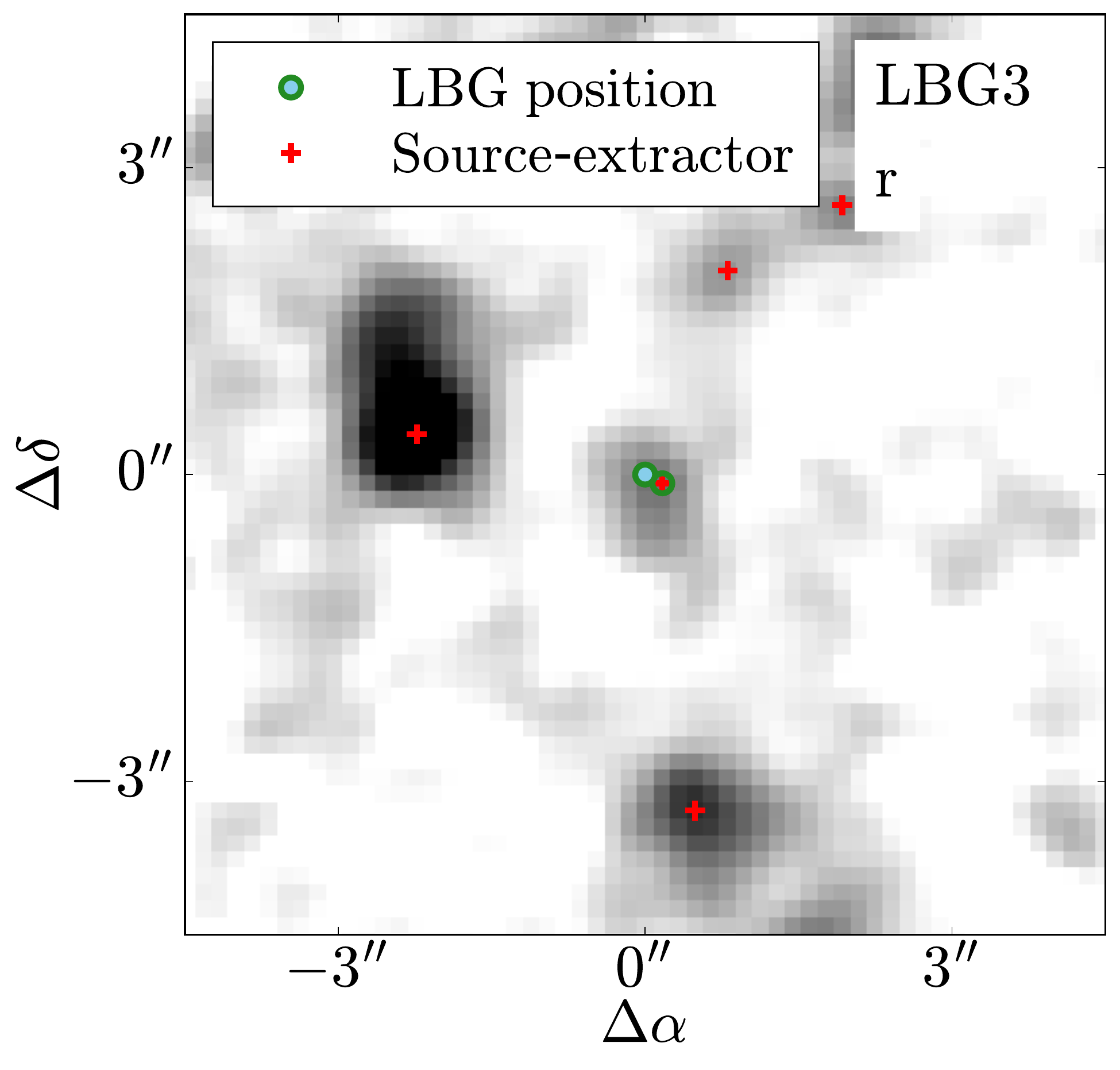}
\includegraphics[width=0.24\textwidth]{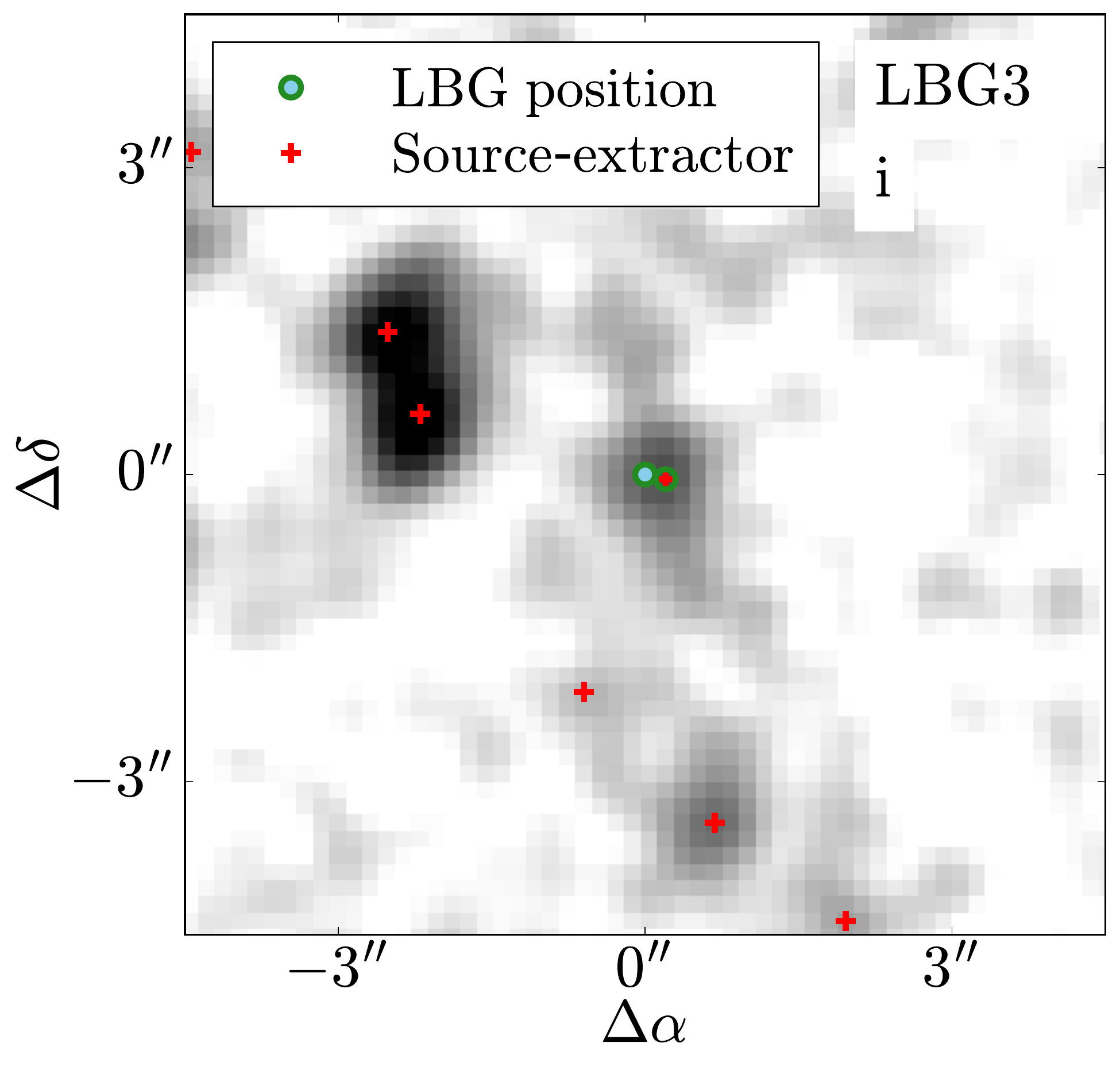}
\includegraphics[width=0.24\textwidth]{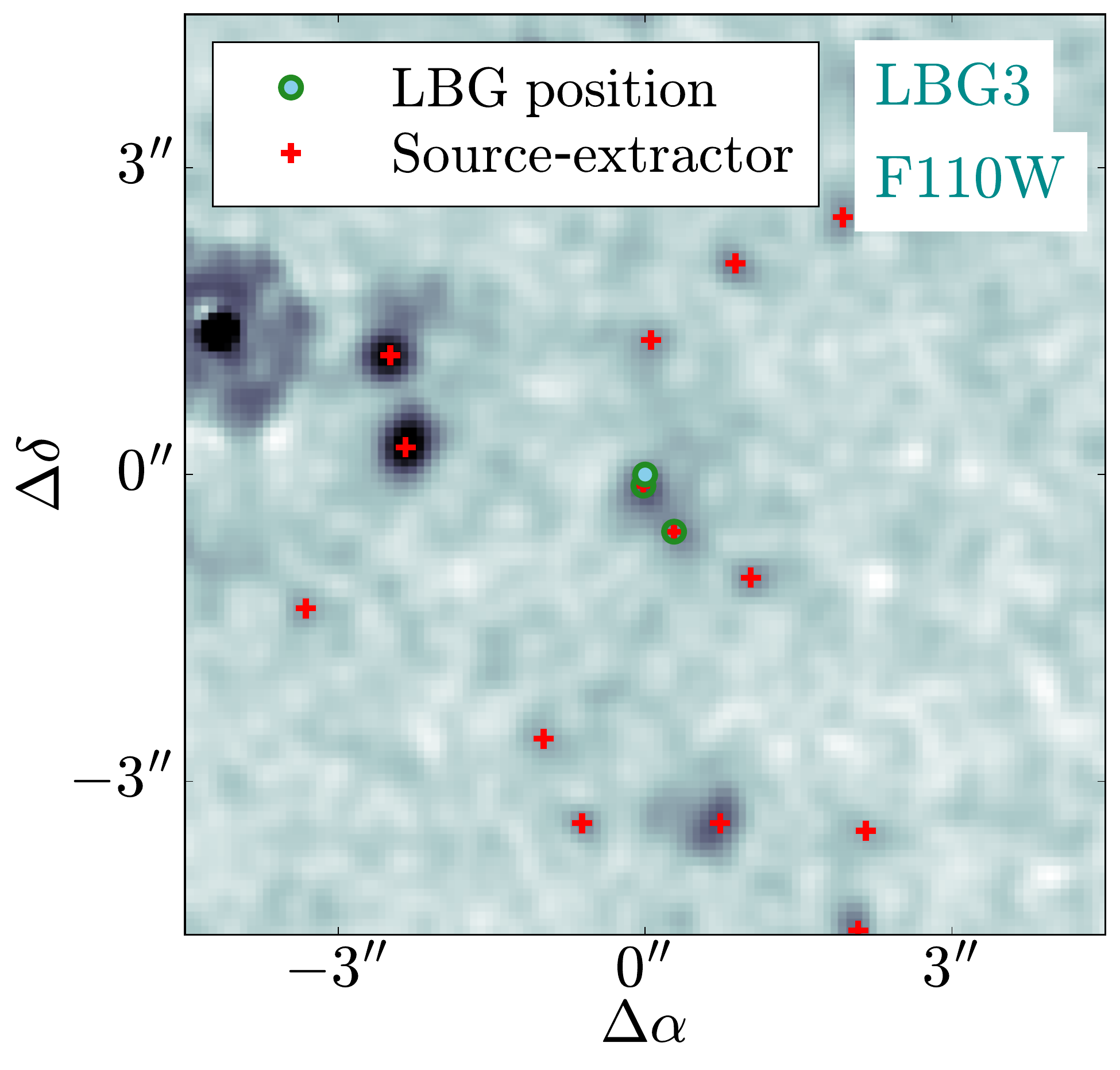}
\includegraphics[width=0.24\textwidth]{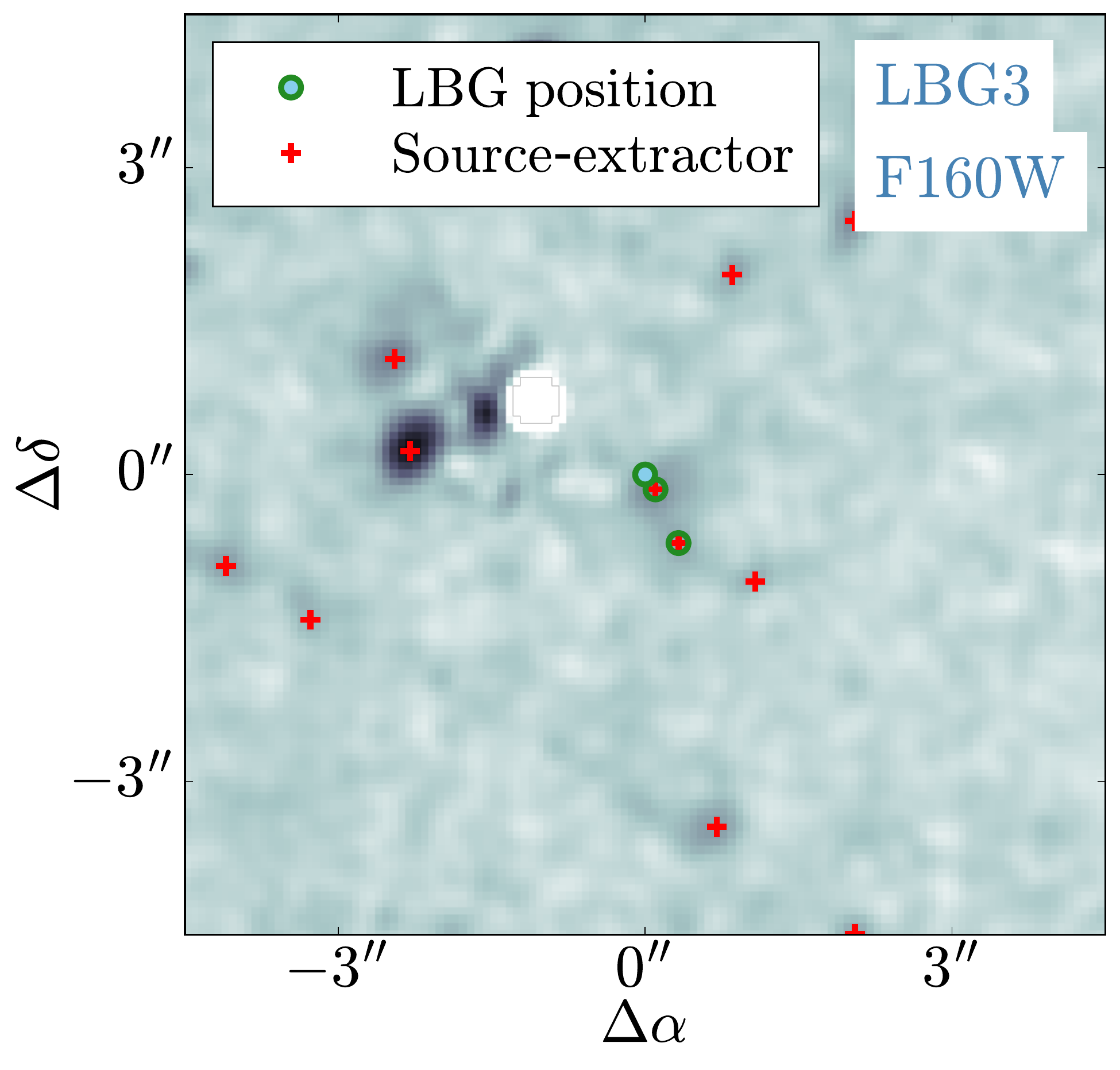}
\includegraphics[width=0.248\textwidth]{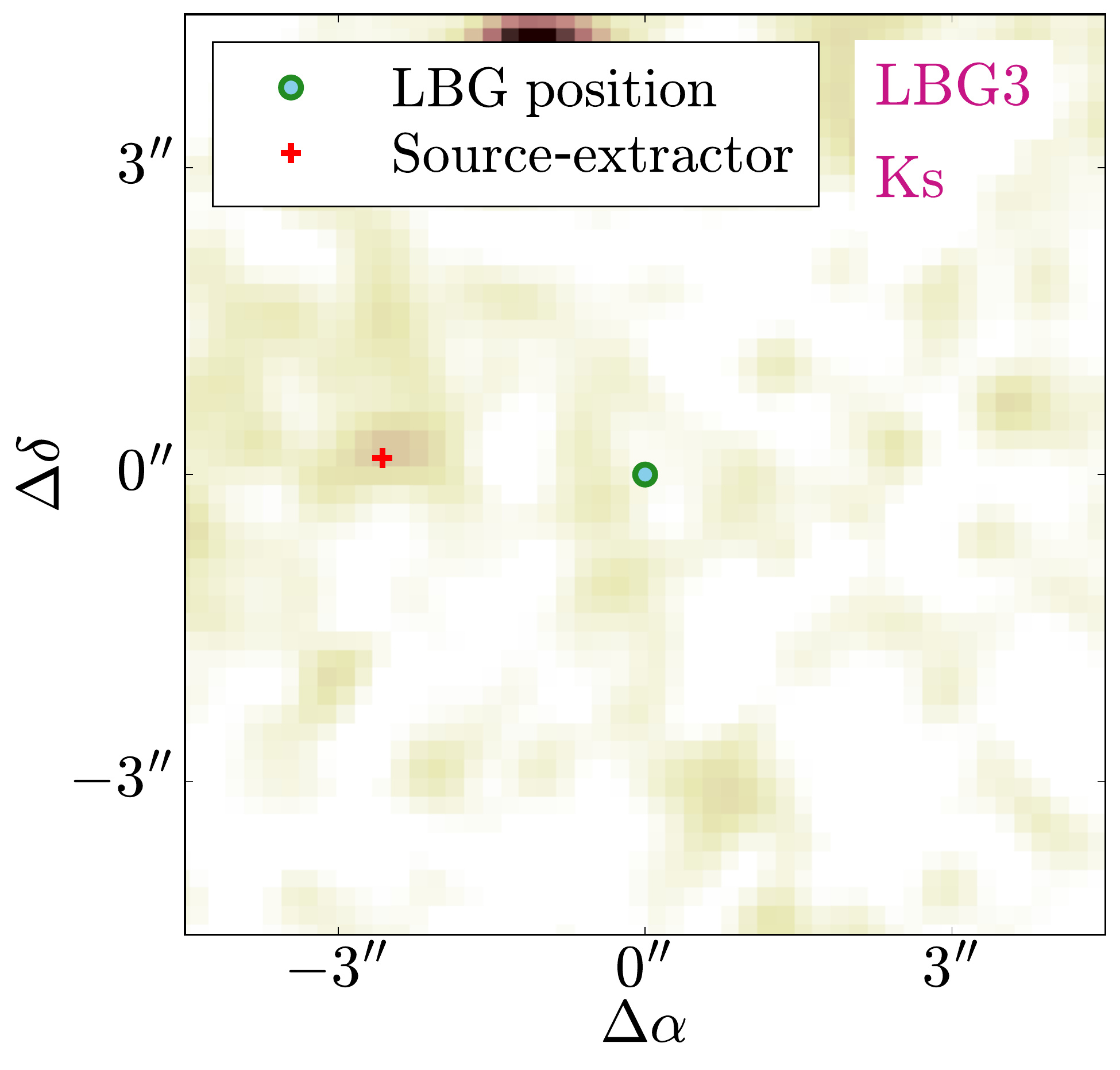}
\includegraphics[width=0.249\textwidth]{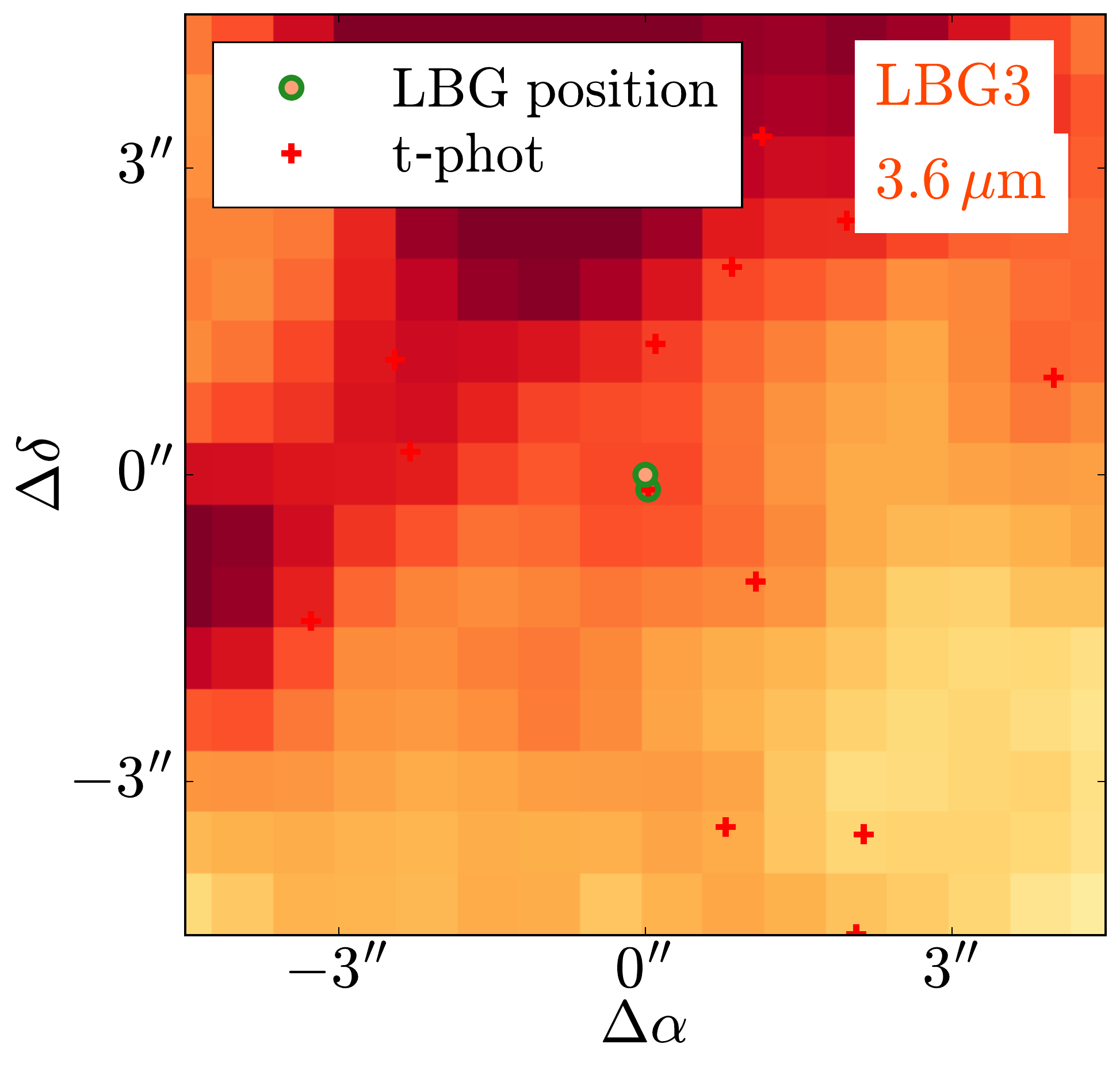}
\includegraphics[width=0.249\textwidth]{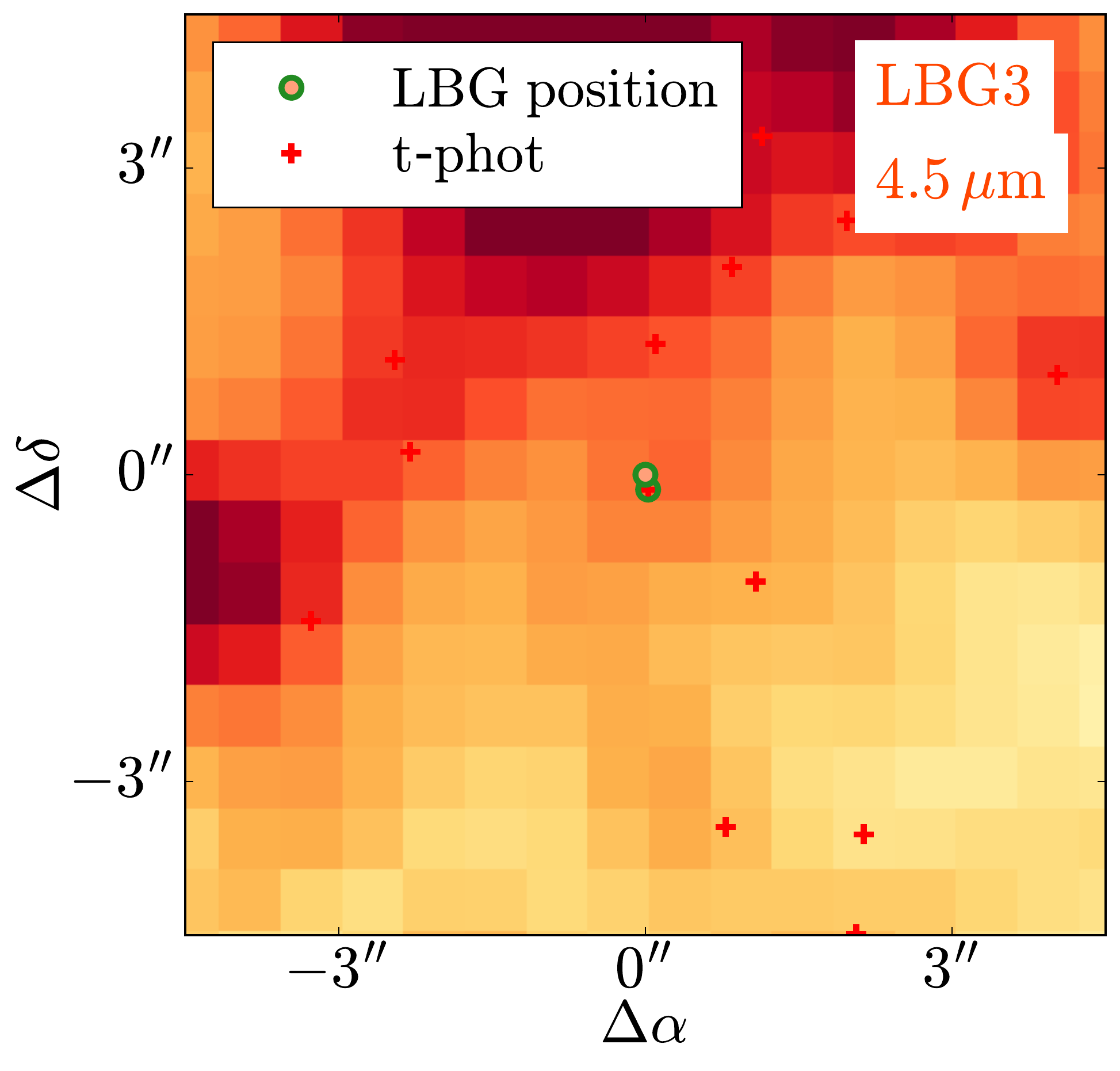}
\end{framed}
\end{subfigure}
\caption{}
\end{figure*}
\renewcommand{\thefigure}{\arabic{figure}}

\renewcommand{\thefigure}{B\arabic{figure} (Cont.)}
\addtocounter{figure}{-1}
\begin{figure*}
\begin{subfigure}{0.85\textwidth}
\begin{framed}
\includegraphics[width=0.24\textwidth]{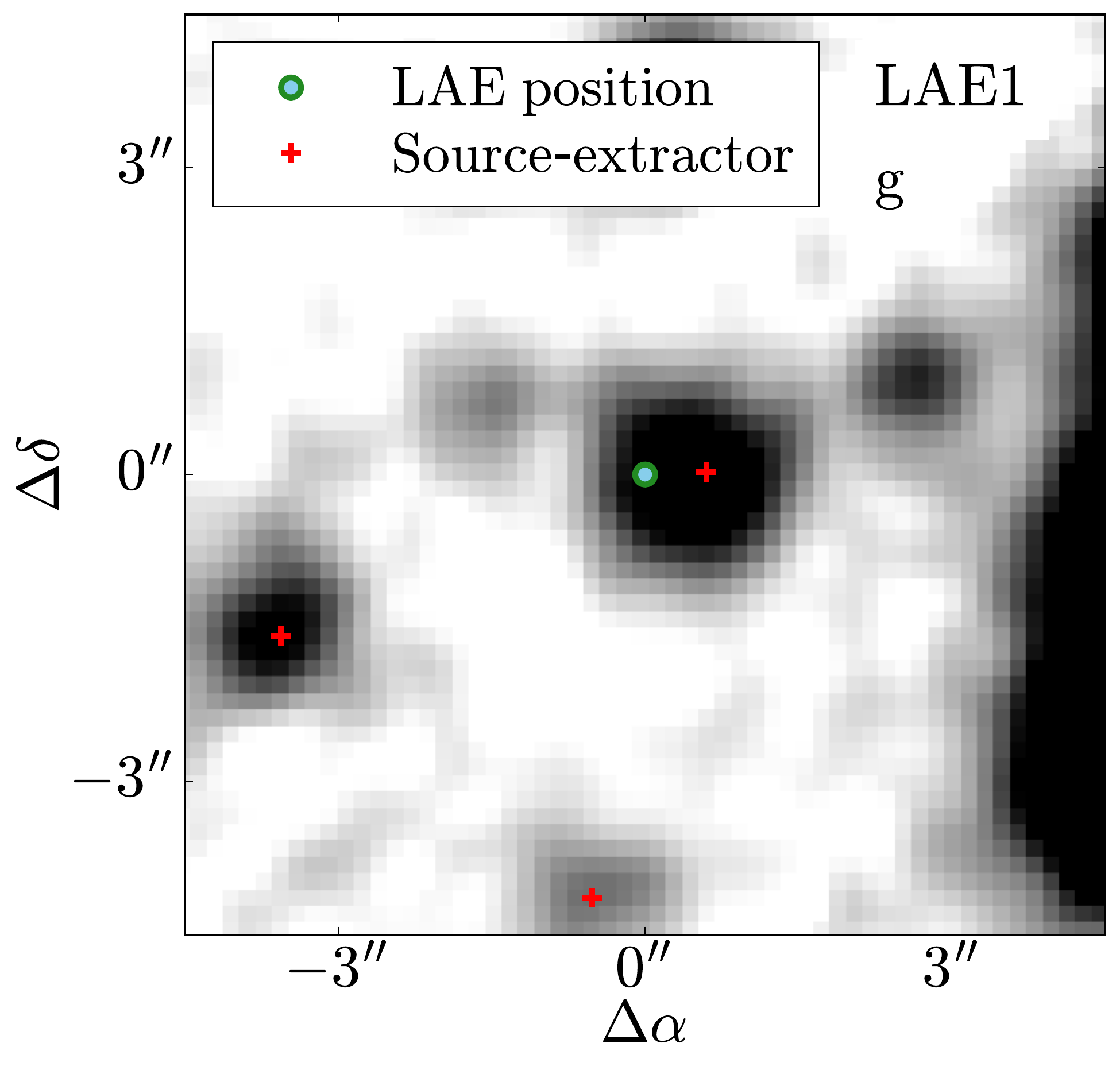}
\includegraphics[width=0.24\textwidth]{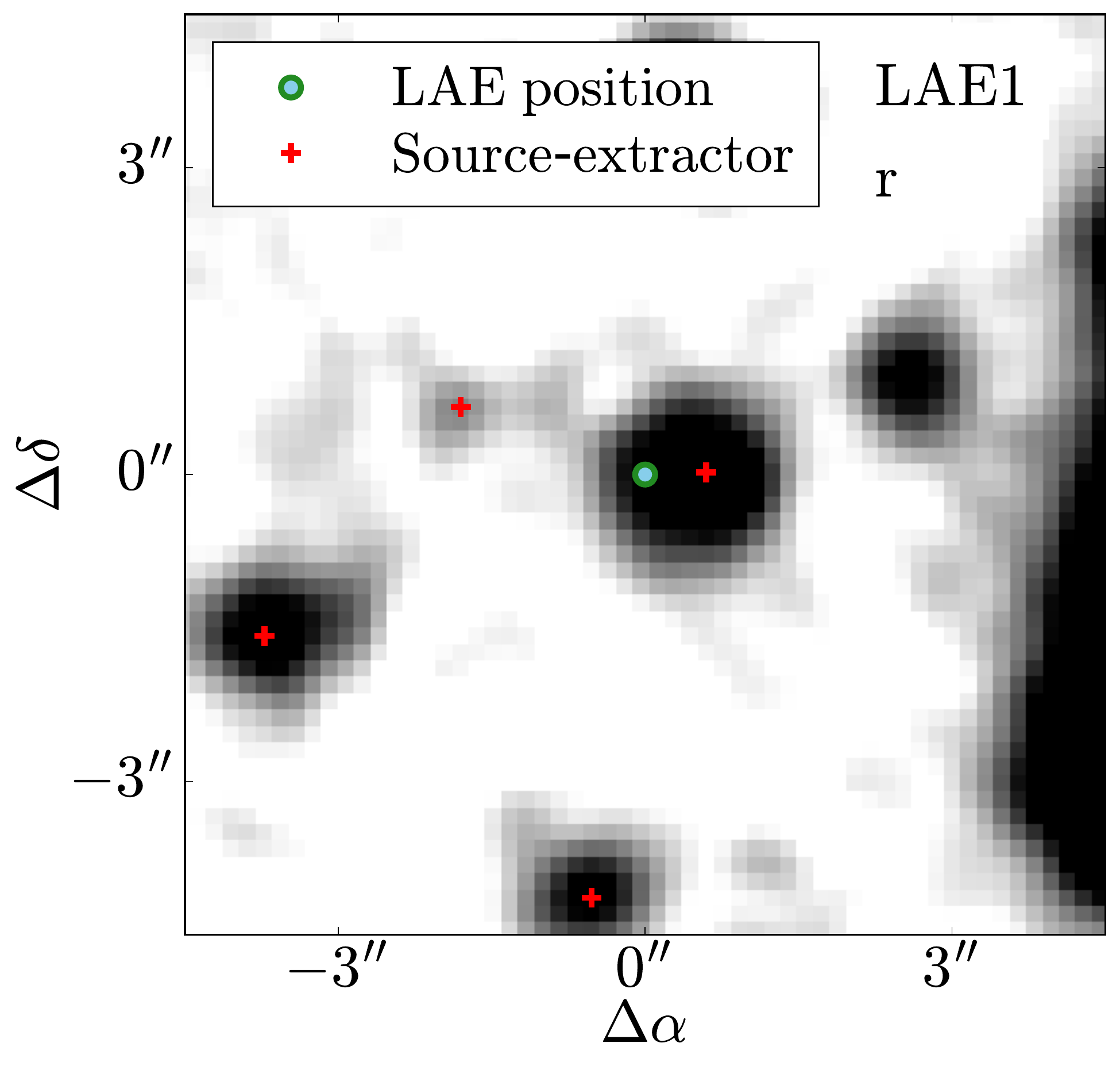}
\includegraphics[width=0.24\textwidth]{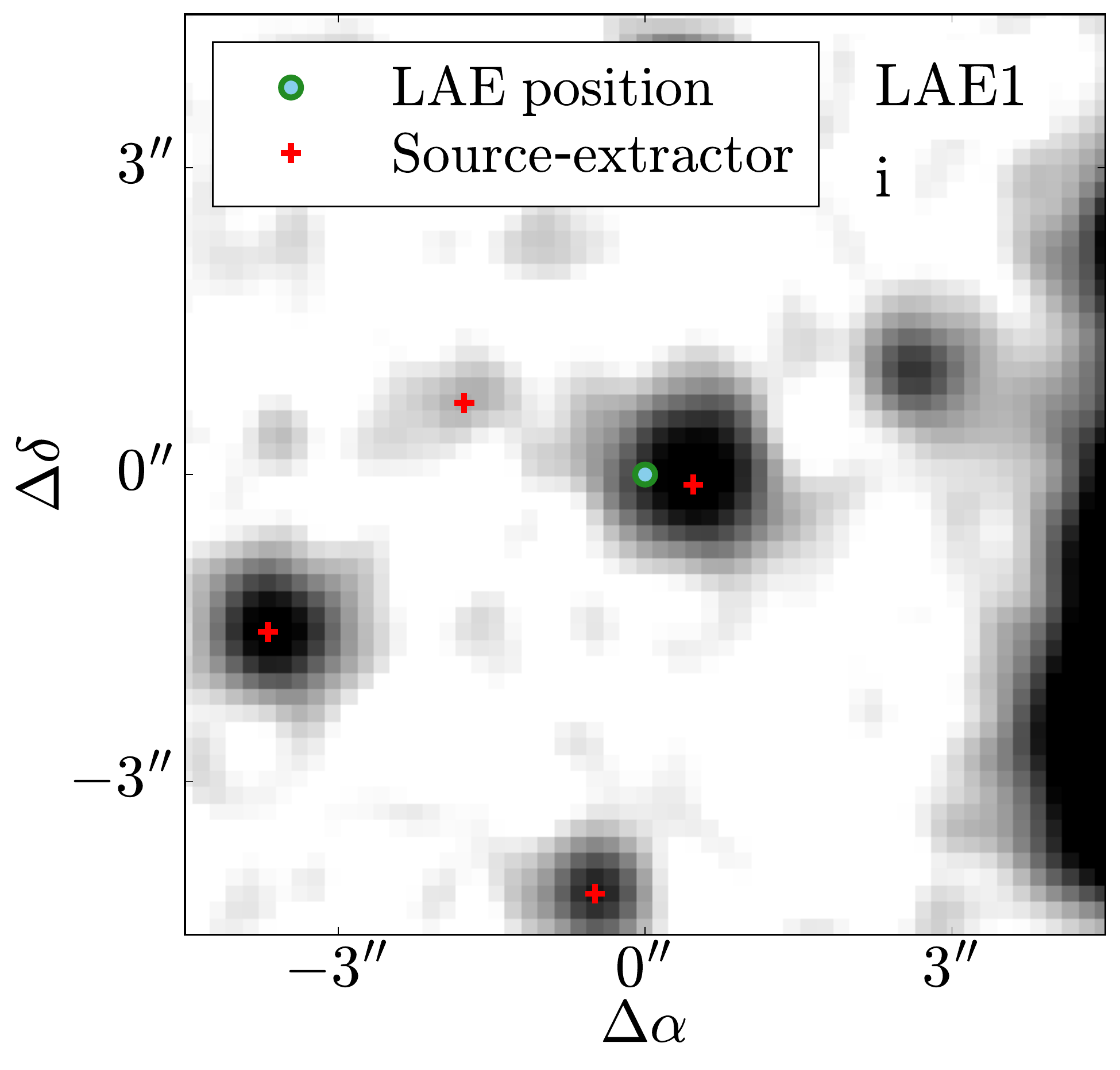}
\includegraphics[width=0.24\textwidth]{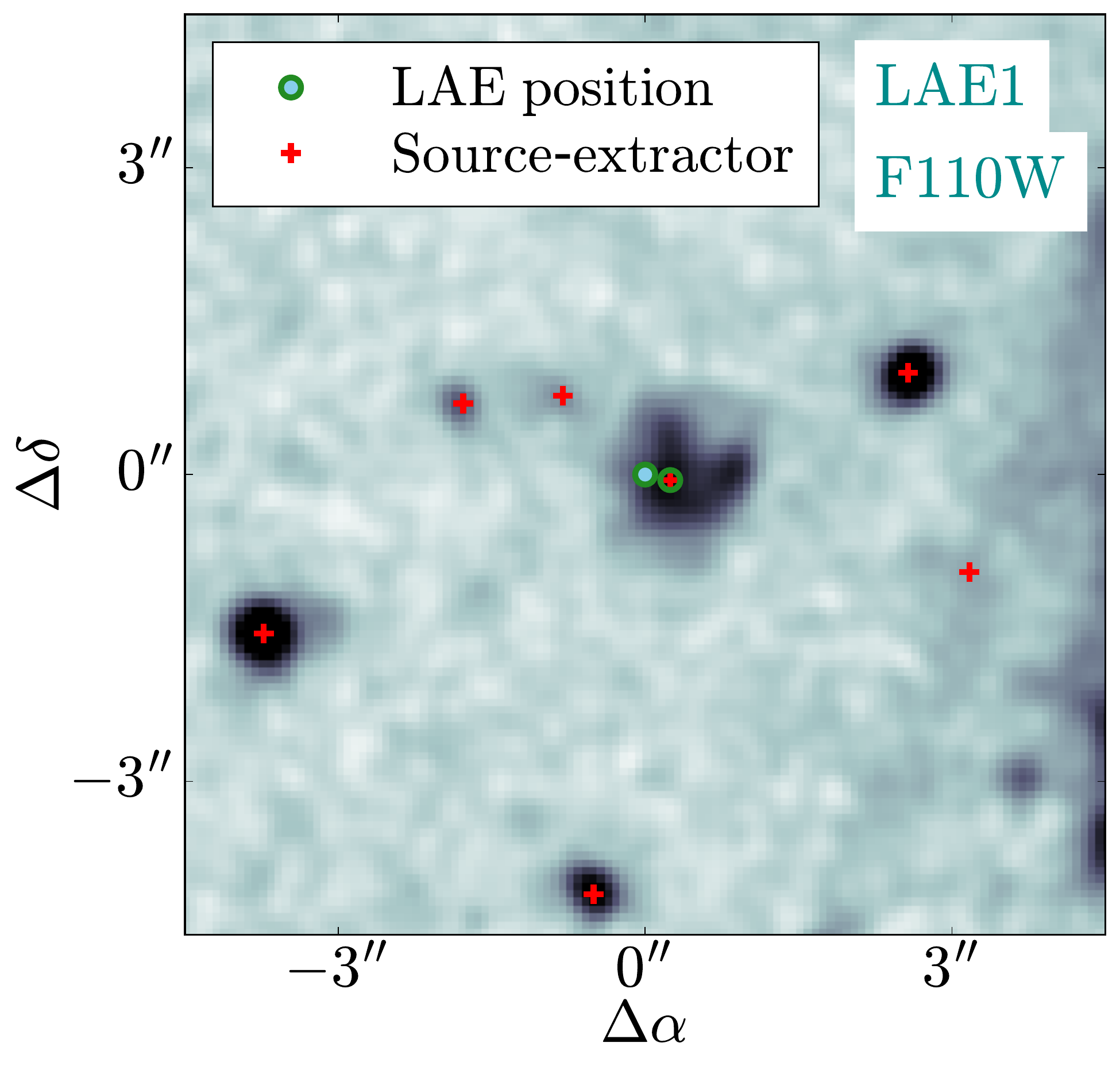}
\includegraphics[width=0.24\textwidth]{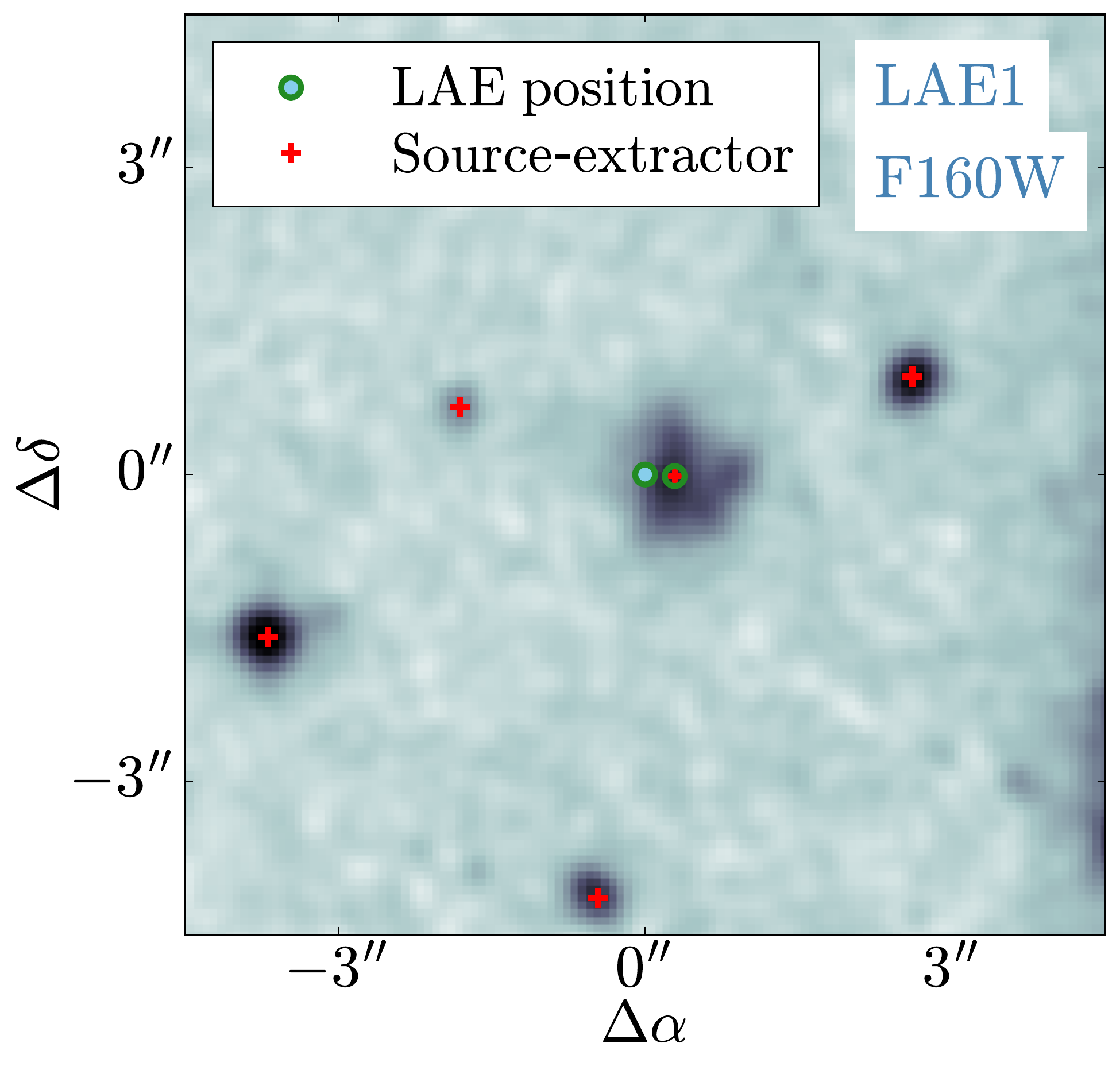}
\includegraphics[width=0.248\textwidth]{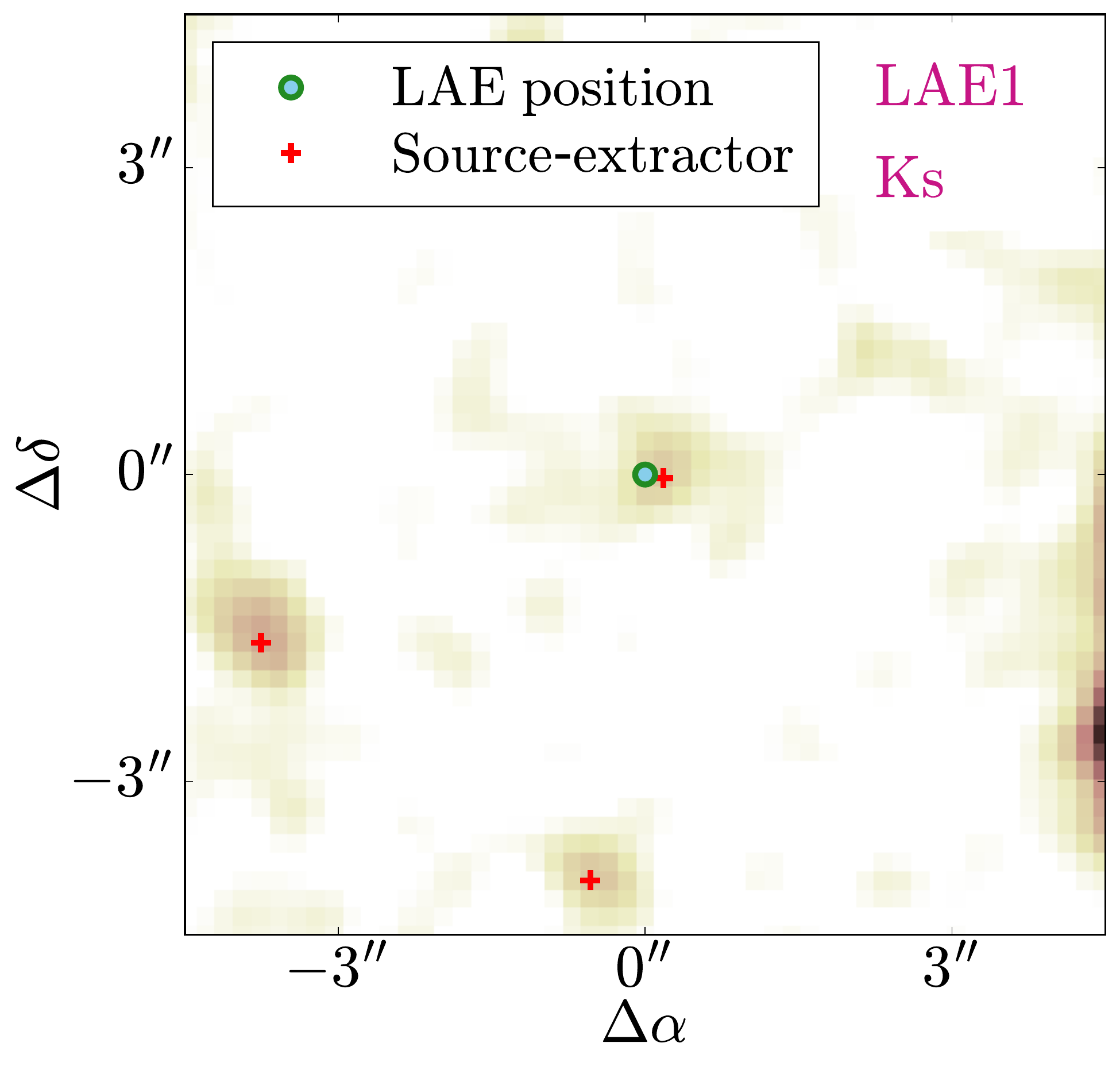}
\includegraphics[width=0.249\textwidth]{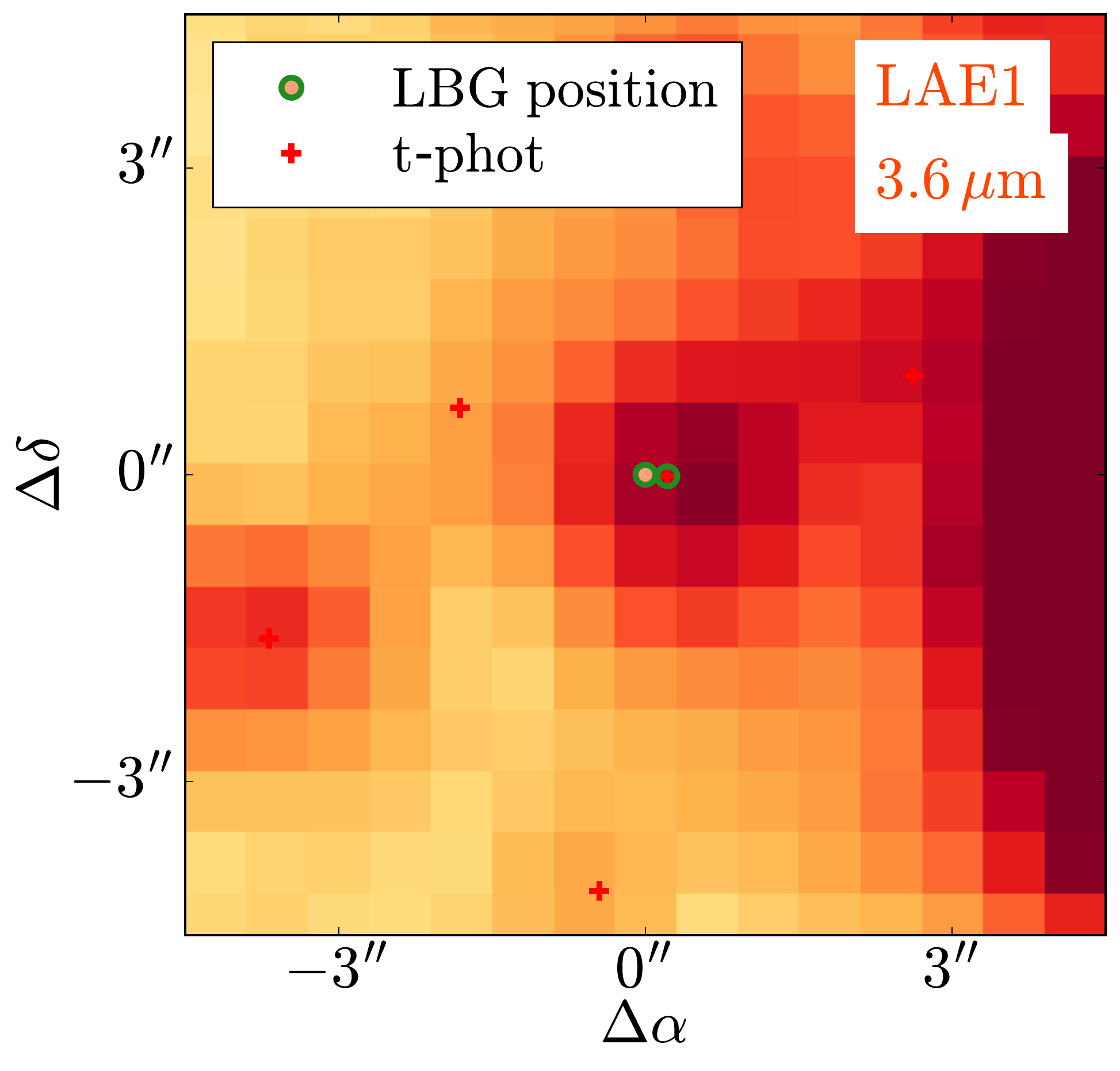}
\includegraphics[width=0.249\textwidth]{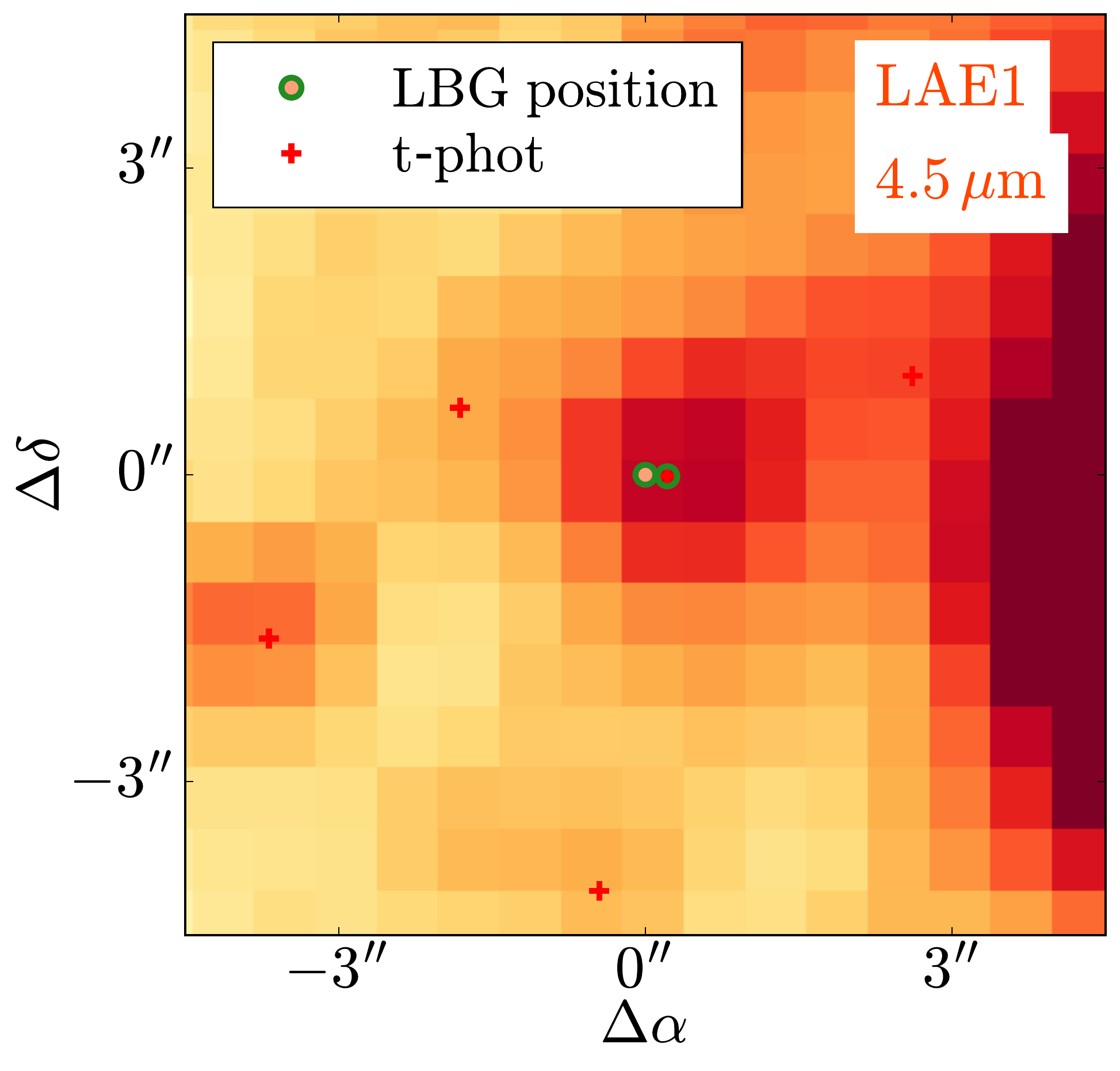}
\end{framed}
\end{subfigure}
\begin{subfigure}{0.85\textwidth}
\begin{framed}
\includegraphics[width=0.24\textwidth]{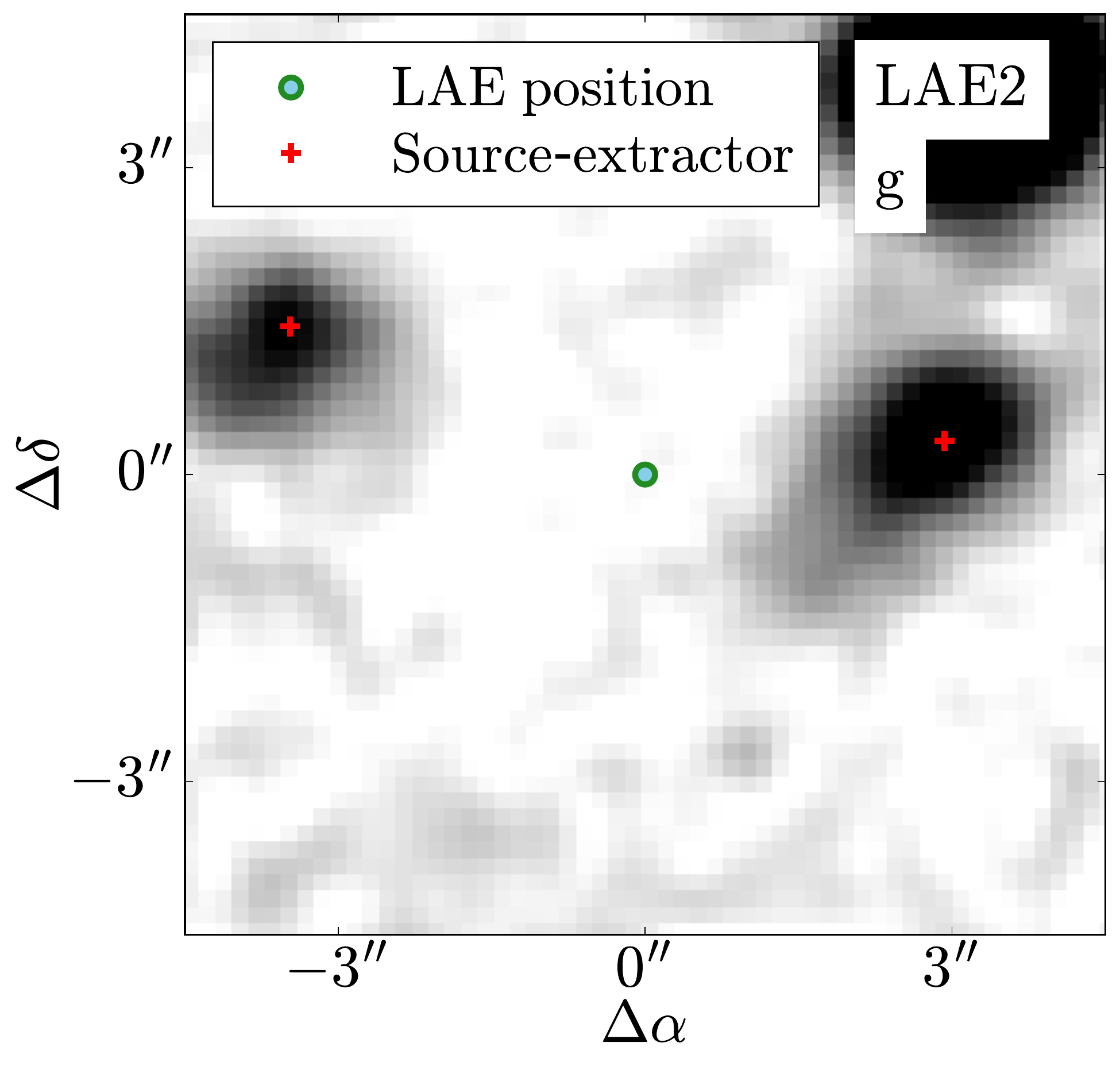}
\includegraphics[width=0.24\textwidth]{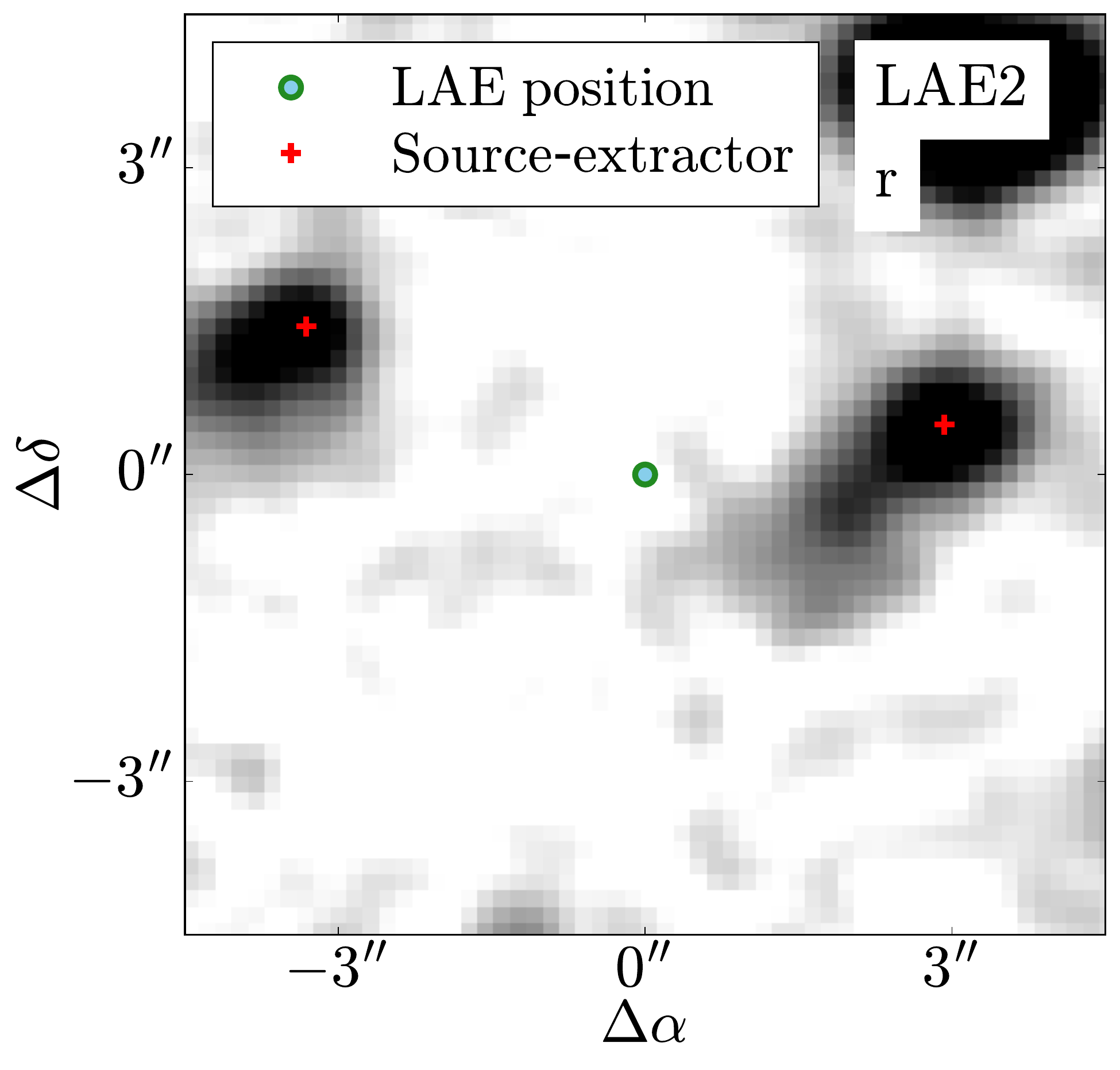}
\includegraphics[width=0.24\textwidth]{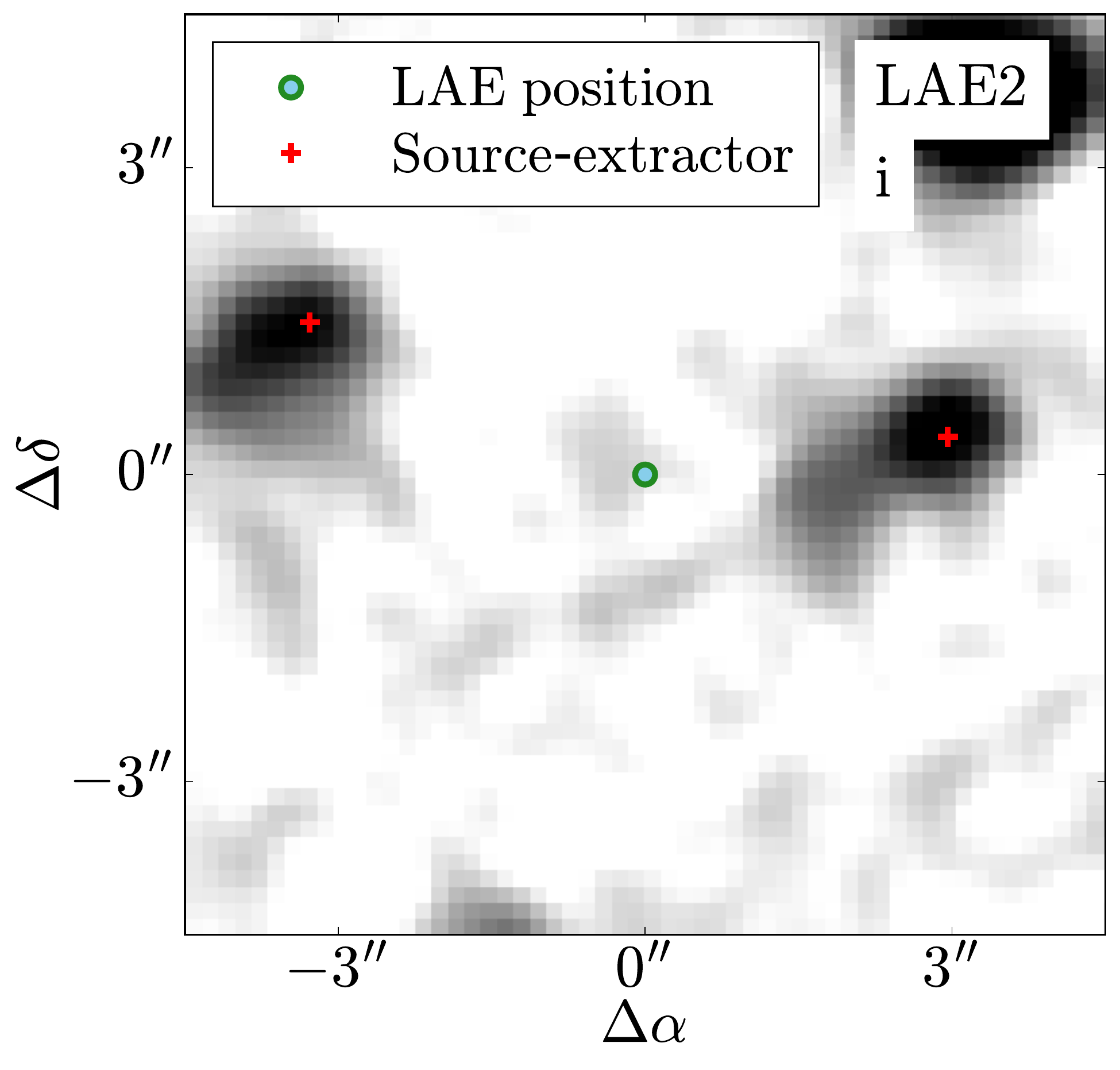}
\includegraphics[width=0.24\textwidth]{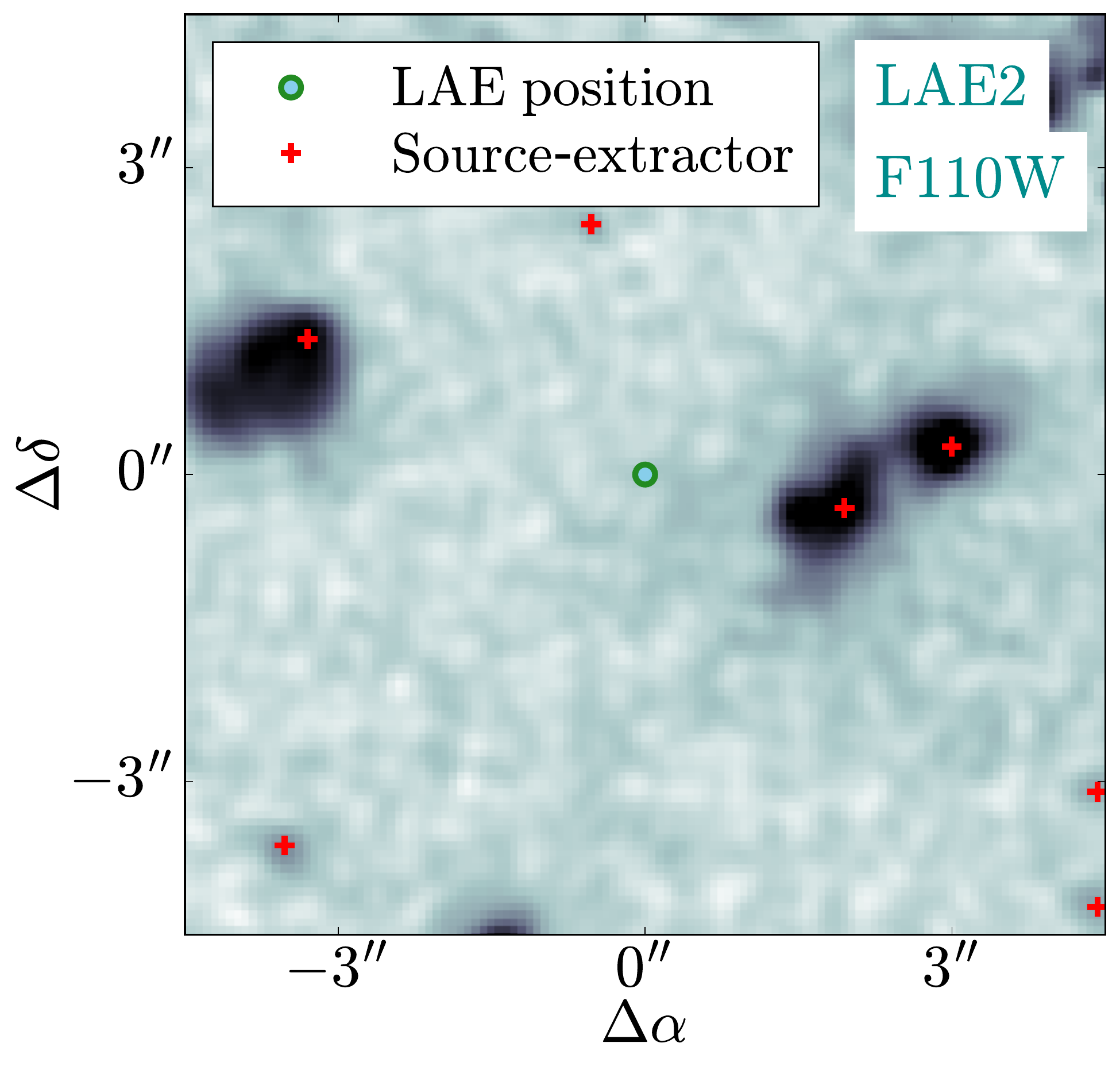}
\includegraphics[width=0.24\textwidth]{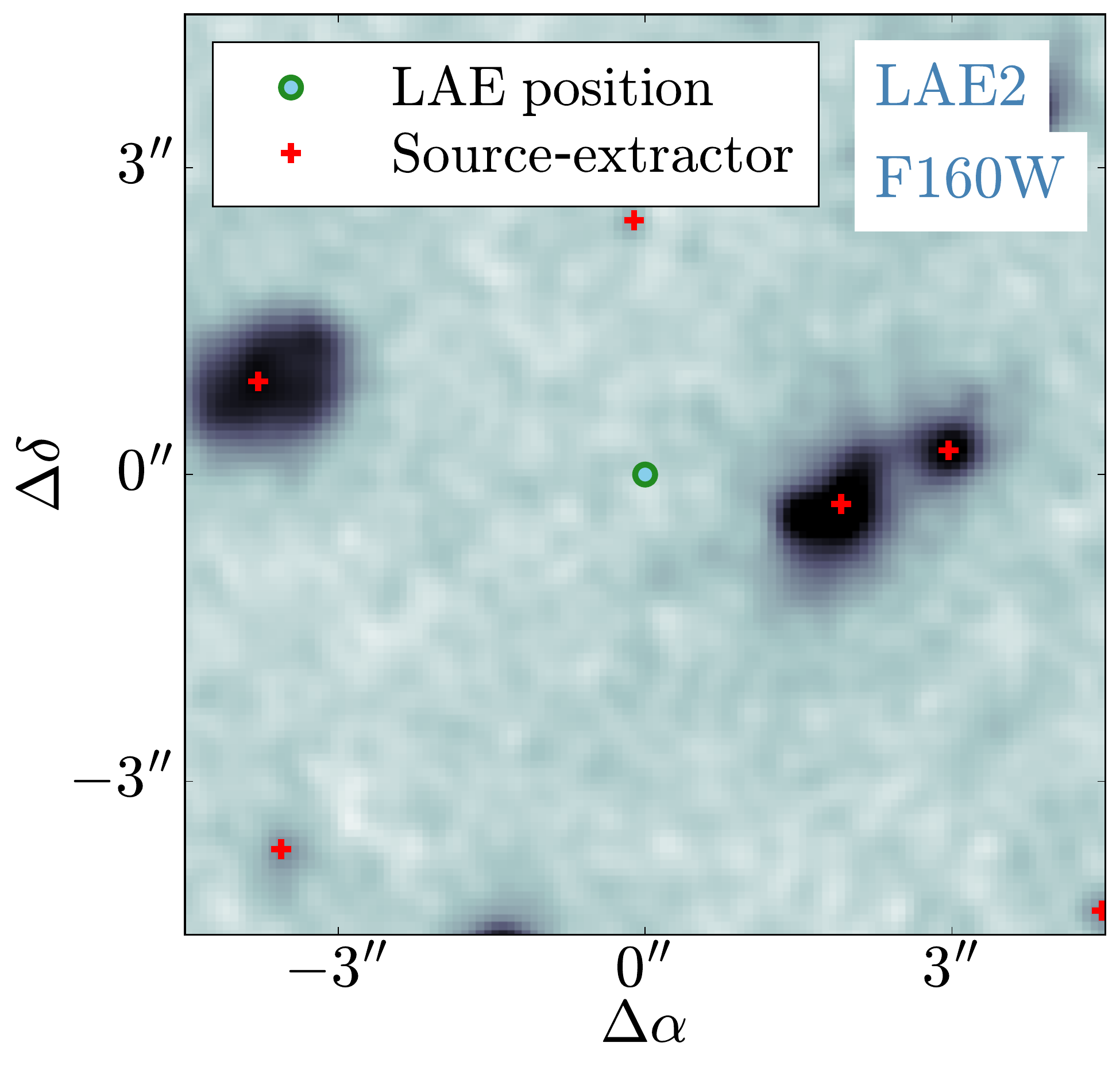}
\includegraphics[width=0.248\textwidth]{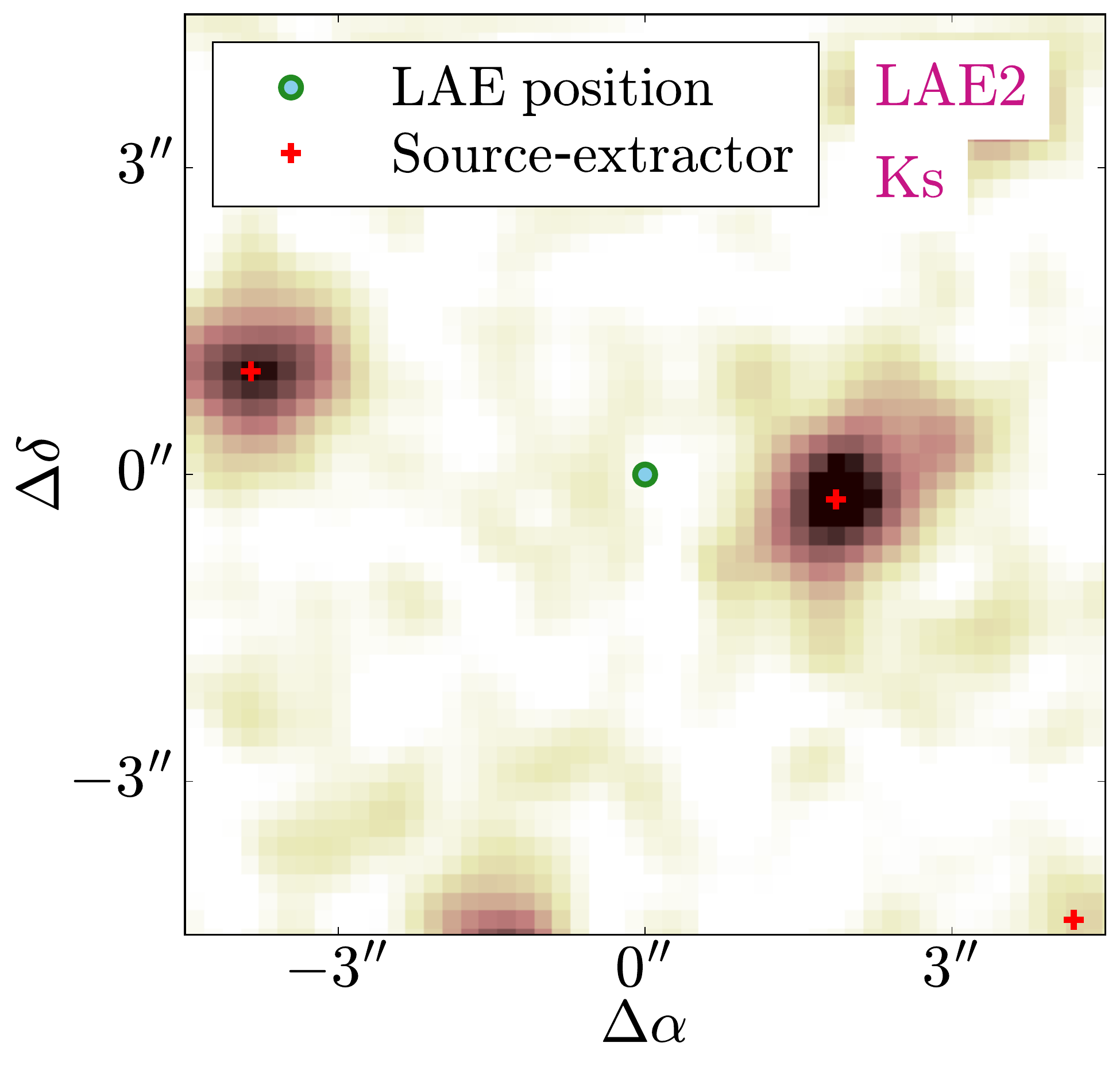}
\includegraphics[width=0.249\textwidth]{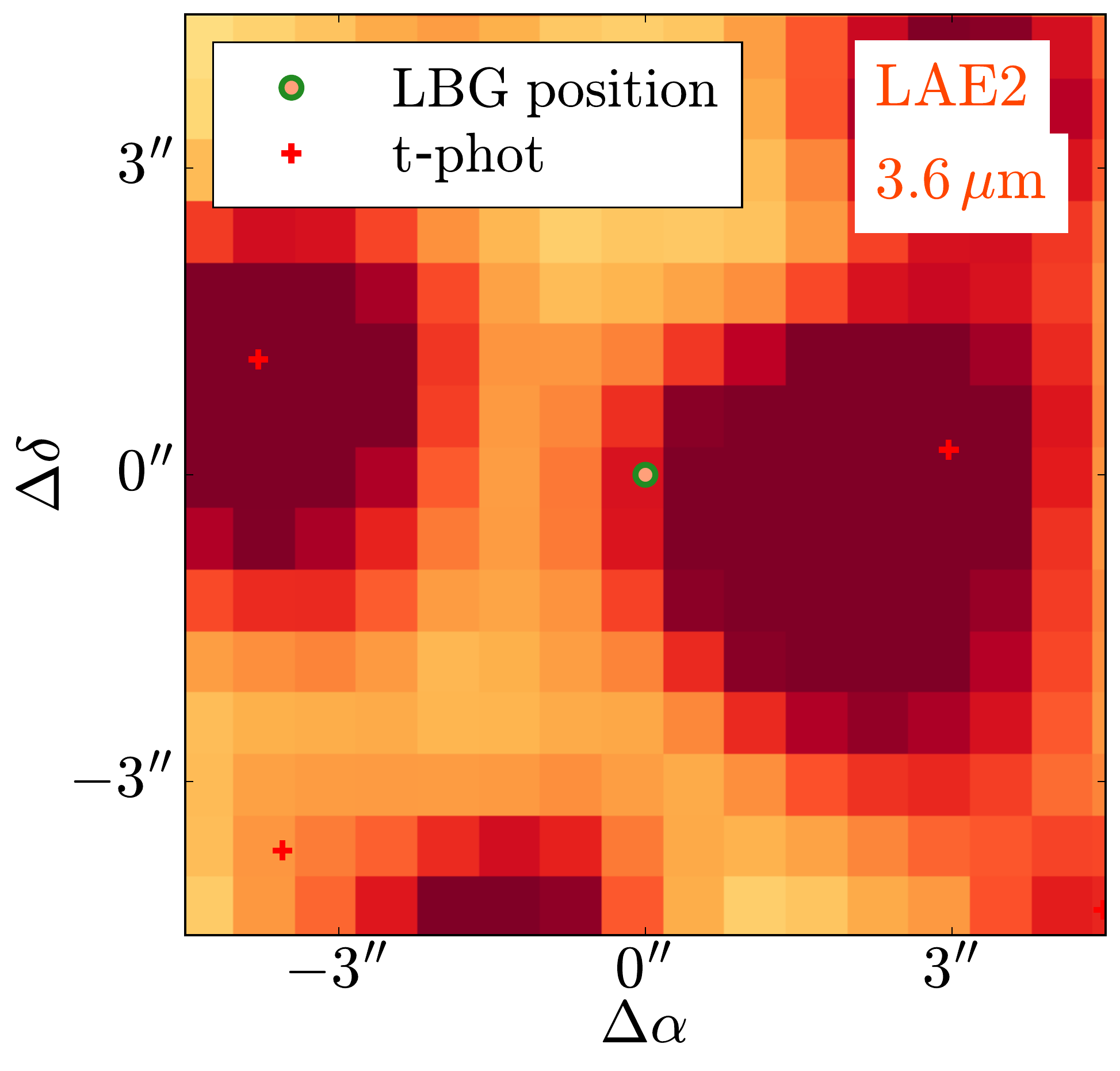}
\includegraphics[width=0.249\textwidth]{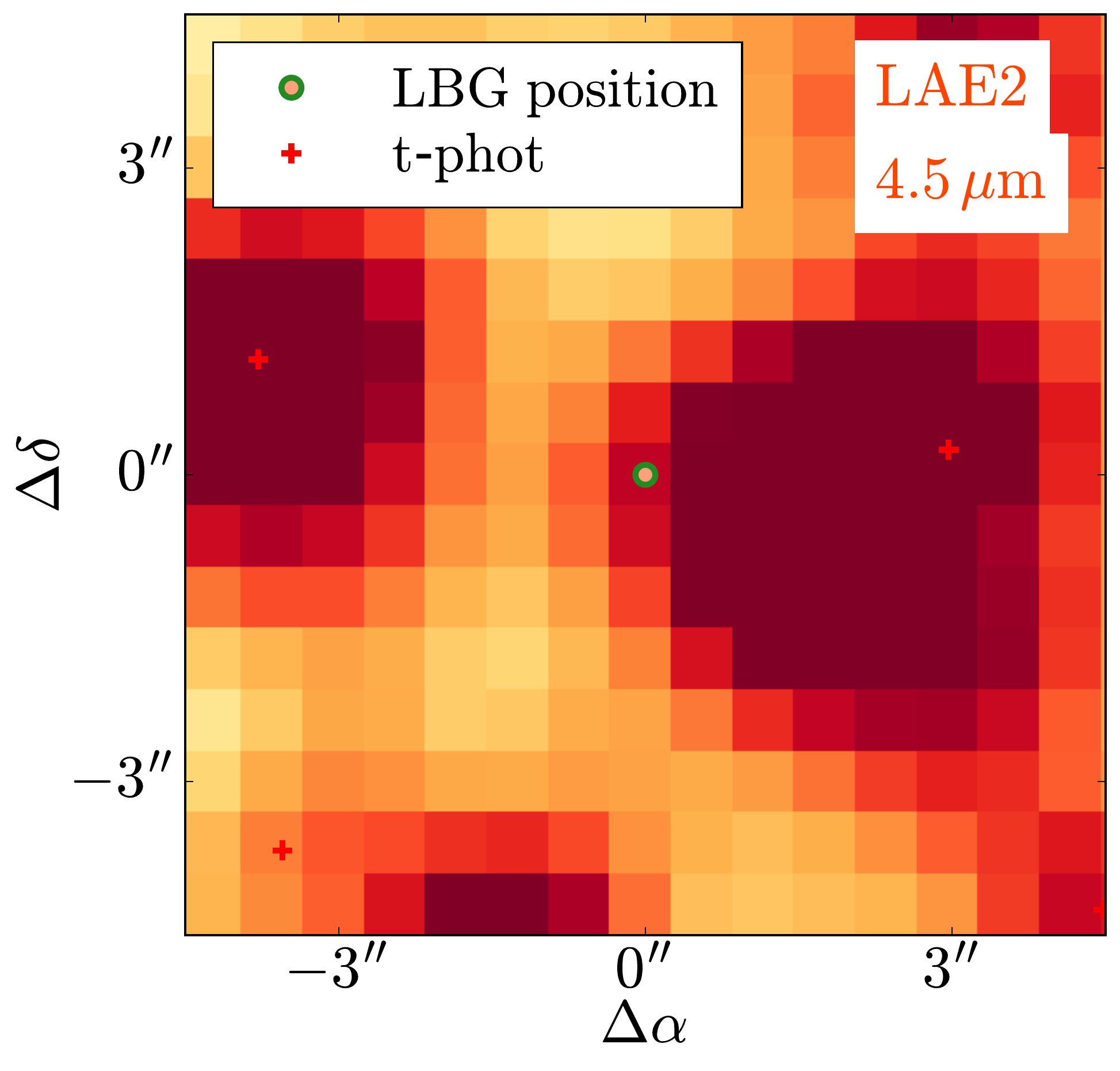}
\end{framed}
\end{subfigure}
\begin{subfigure}{0.85\textwidth}
\begin{framed}
\includegraphics[width=0.24\textwidth]{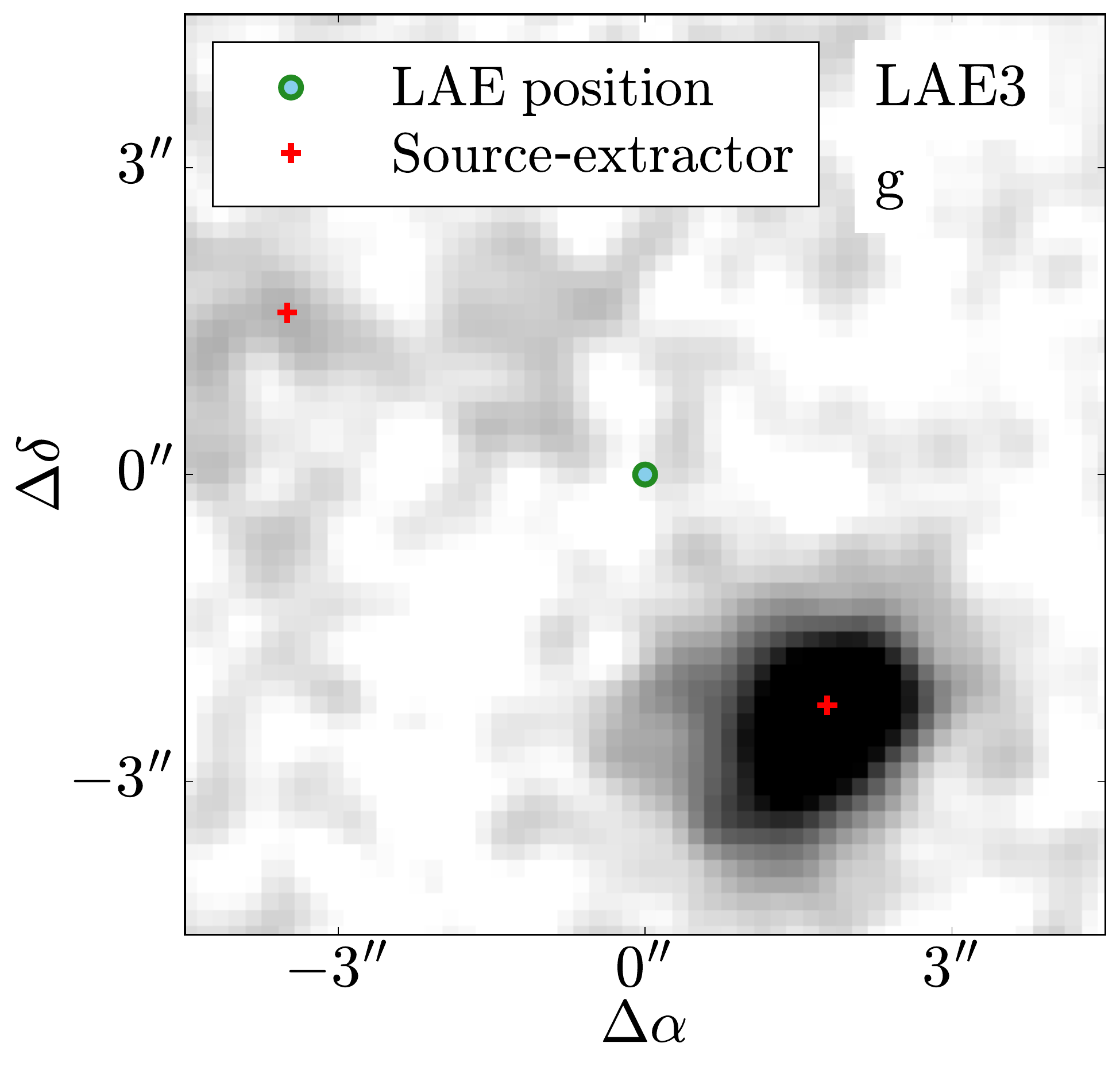}
\includegraphics[width=0.24\textwidth]{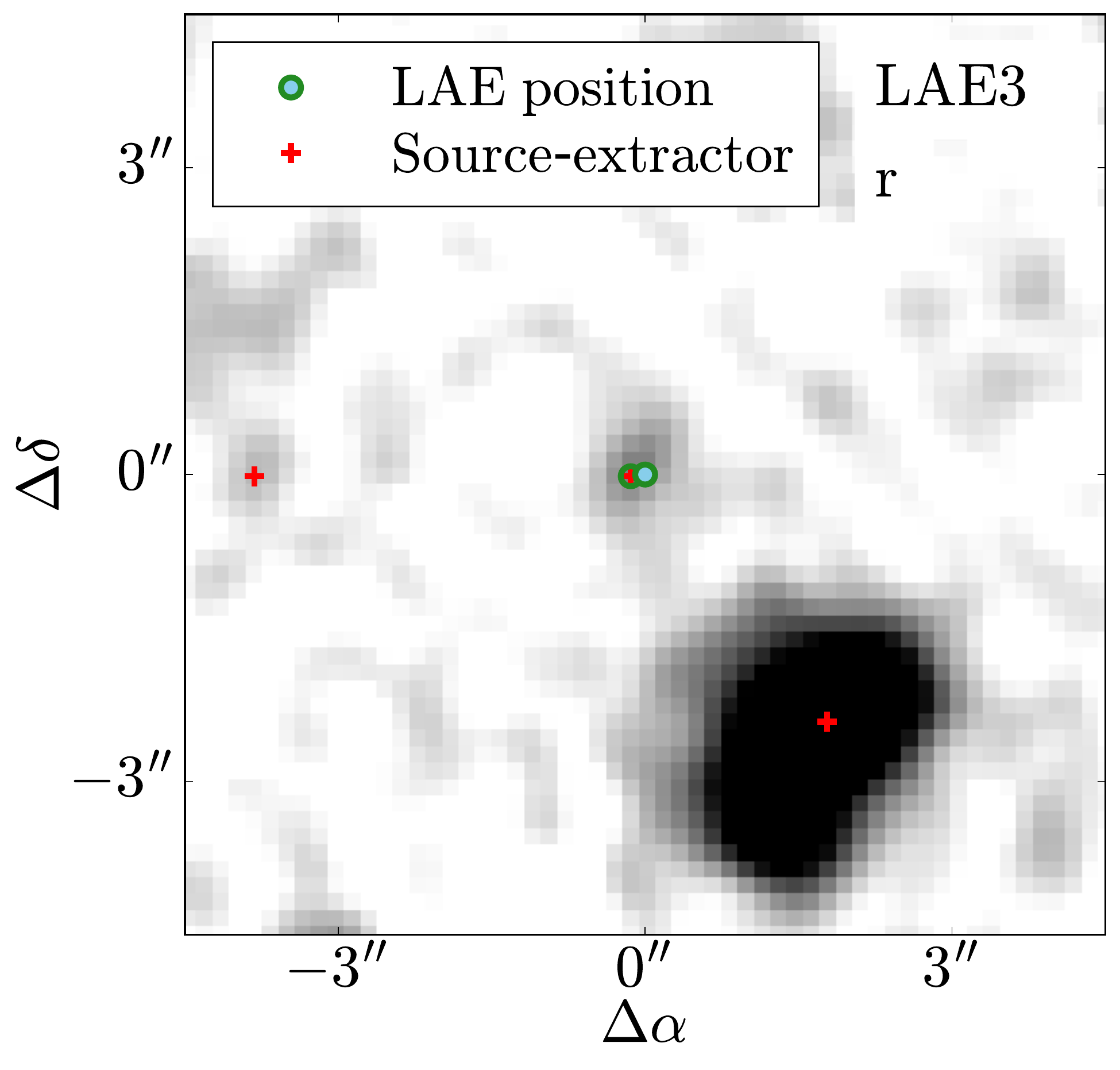}
\includegraphics[width=0.24\textwidth]{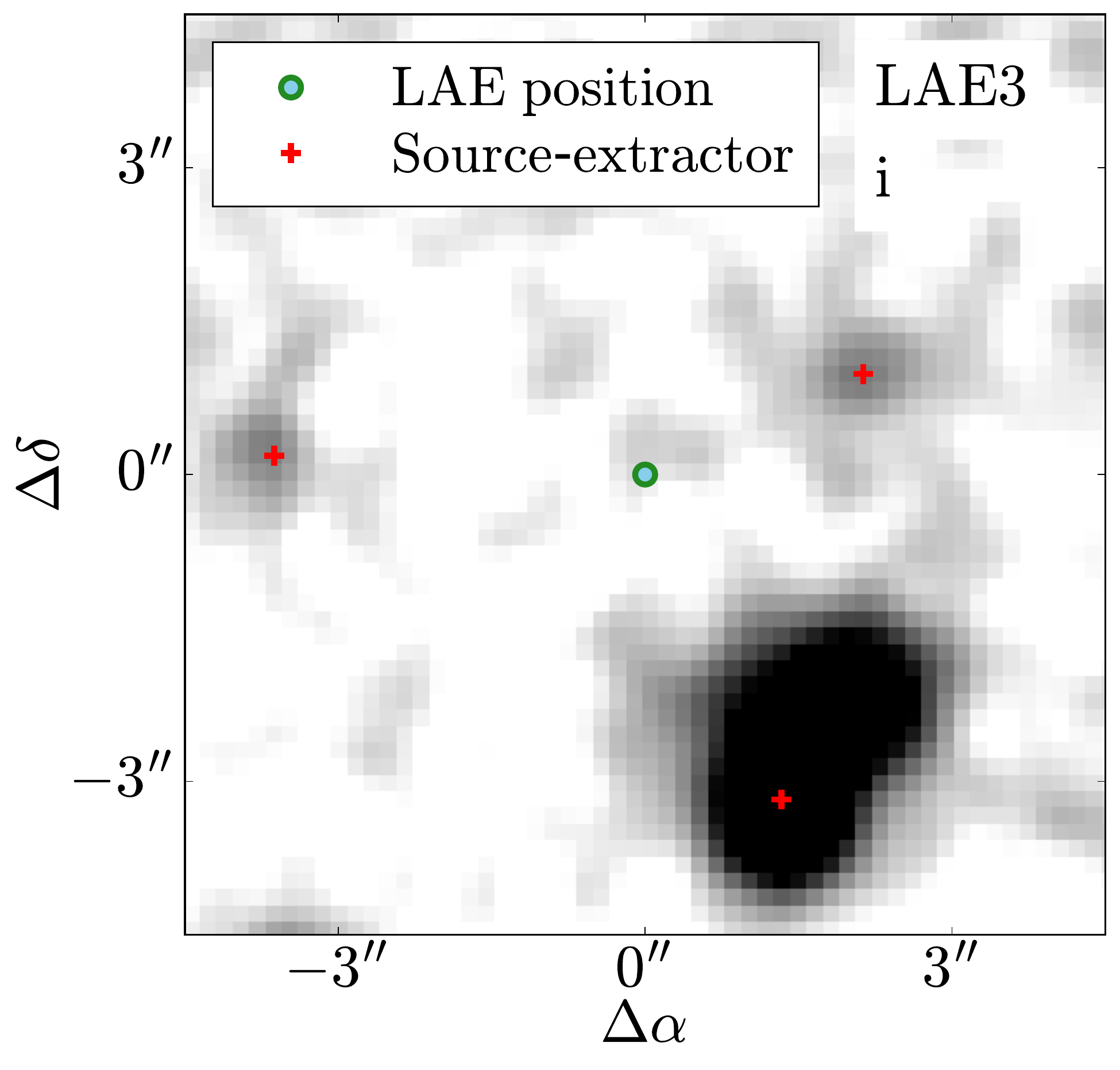}
\includegraphics[width=0.24\textwidth]{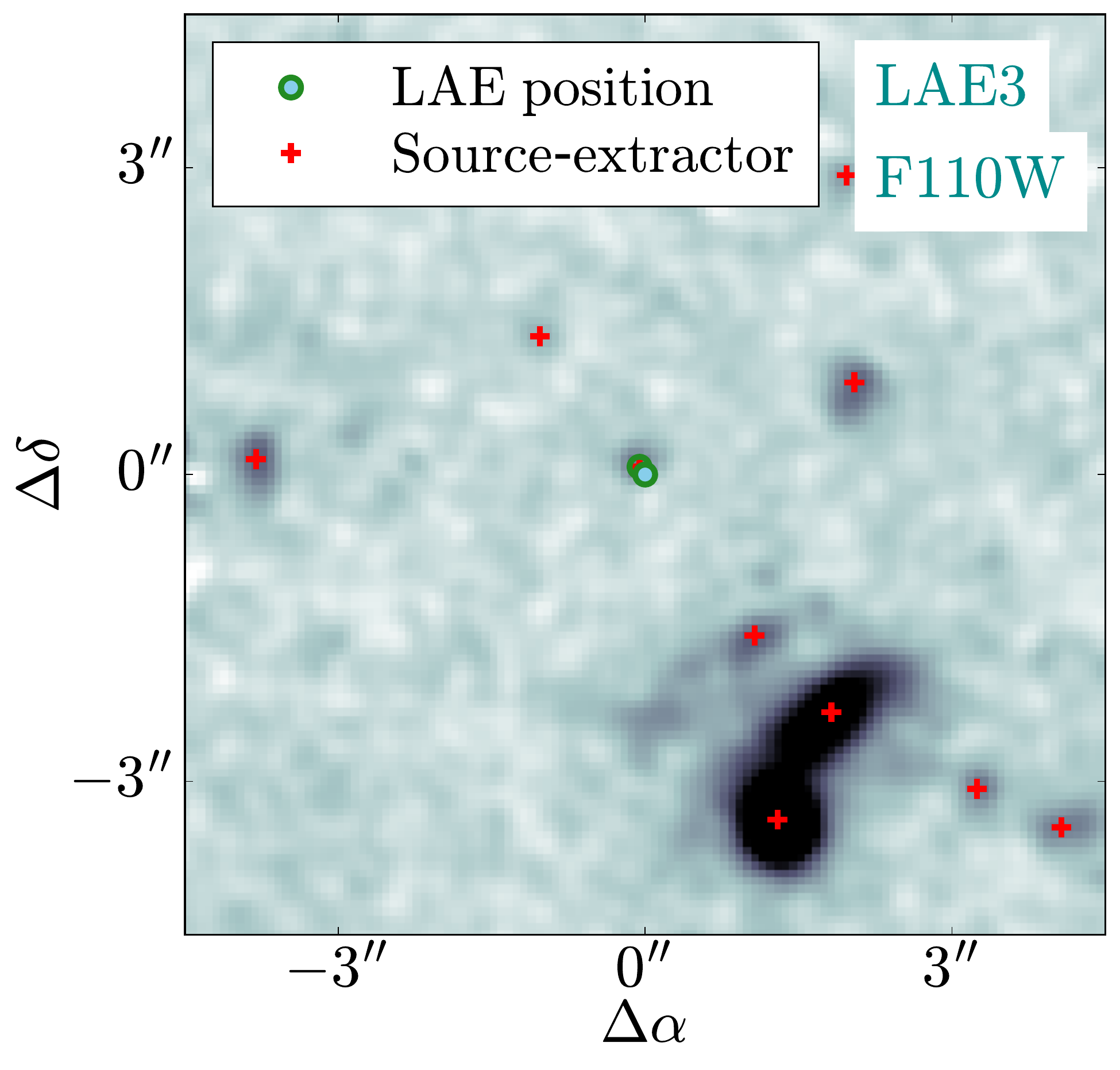}
\includegraphics[width=0.24\textwidth]{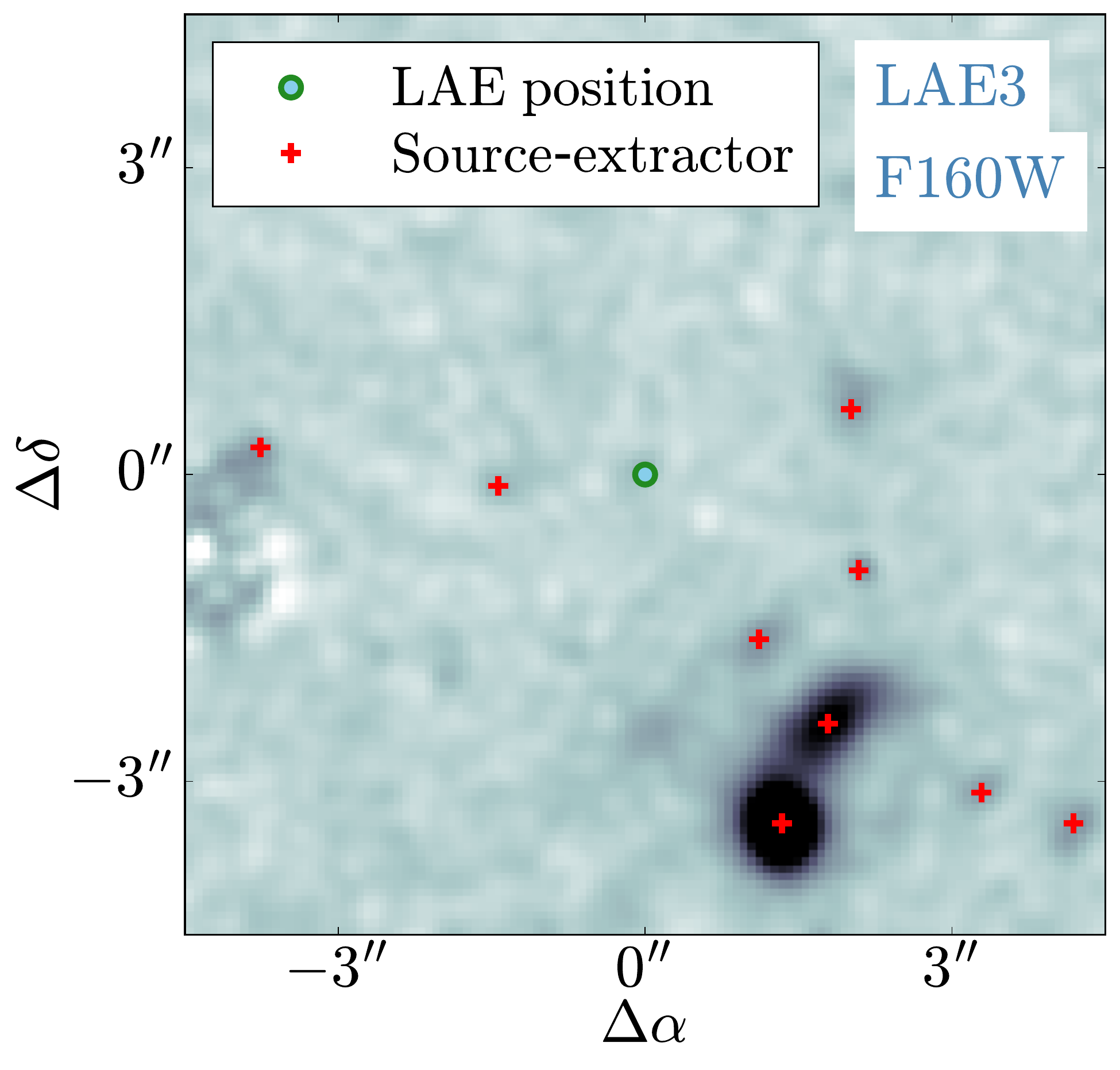}
\includegraphics[width=0.248\textwidth]{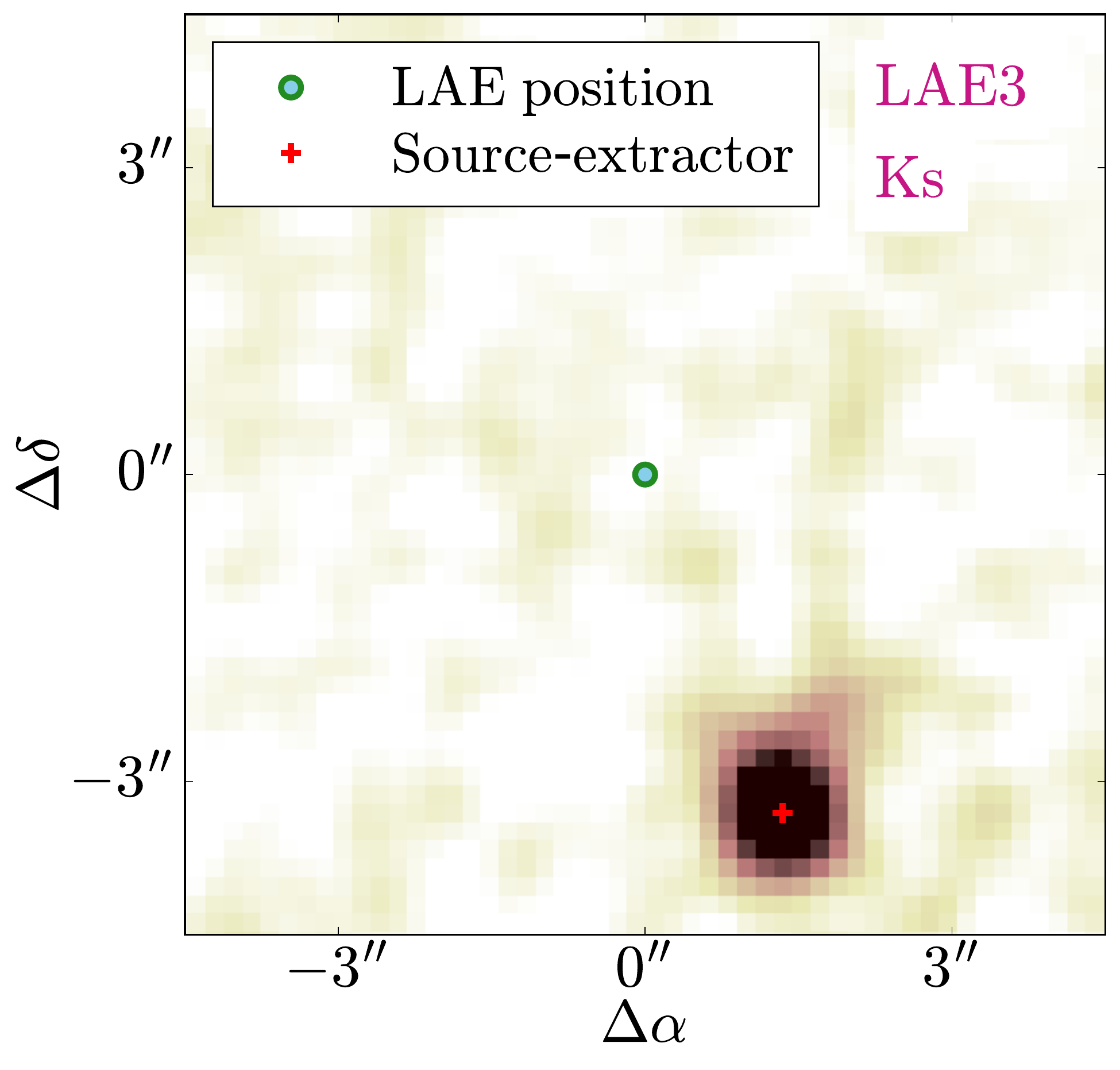}
\includegraphics[width=0.249\textwidth]{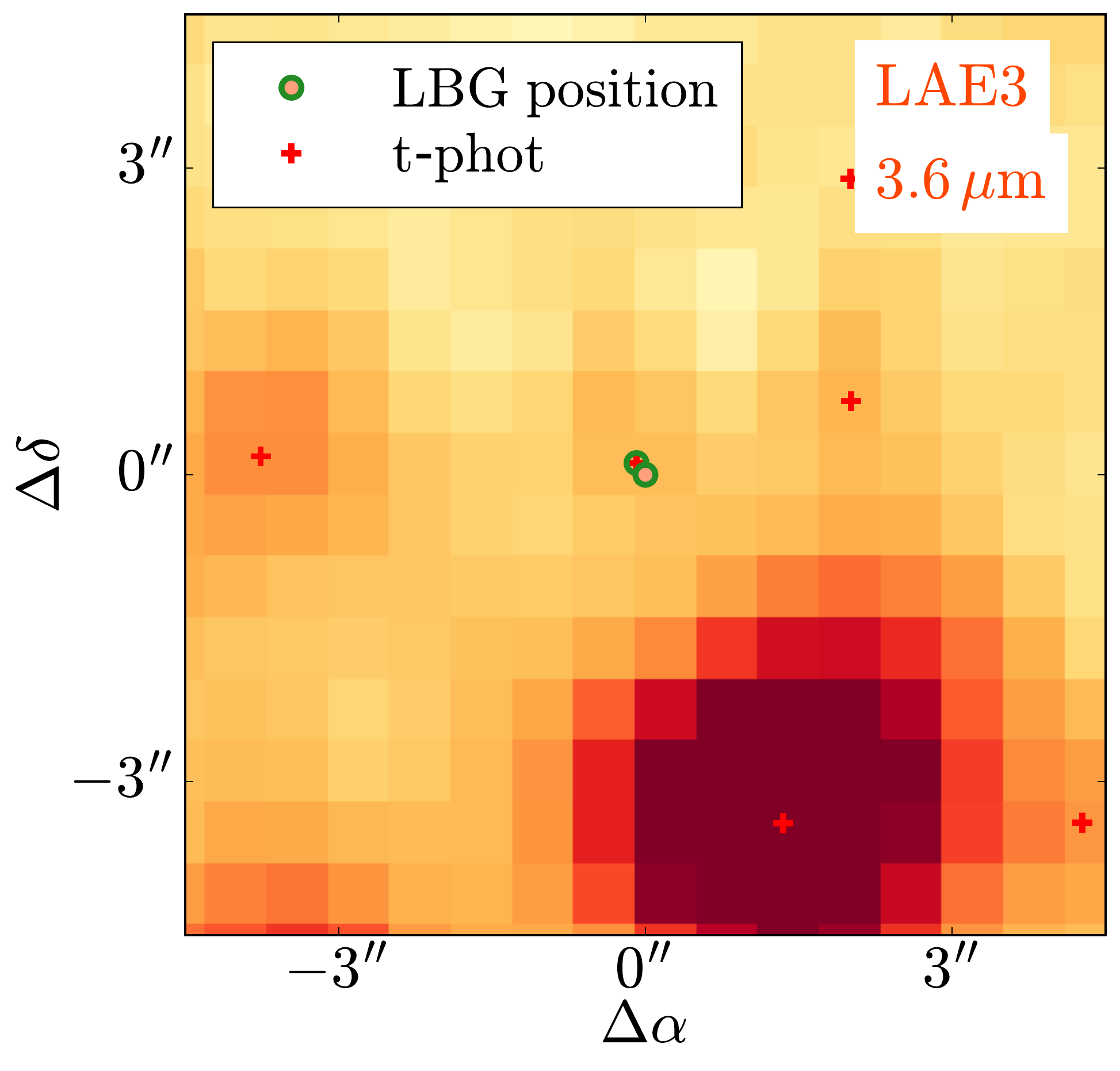}
\includegraphics[width=0.249\textwidth]{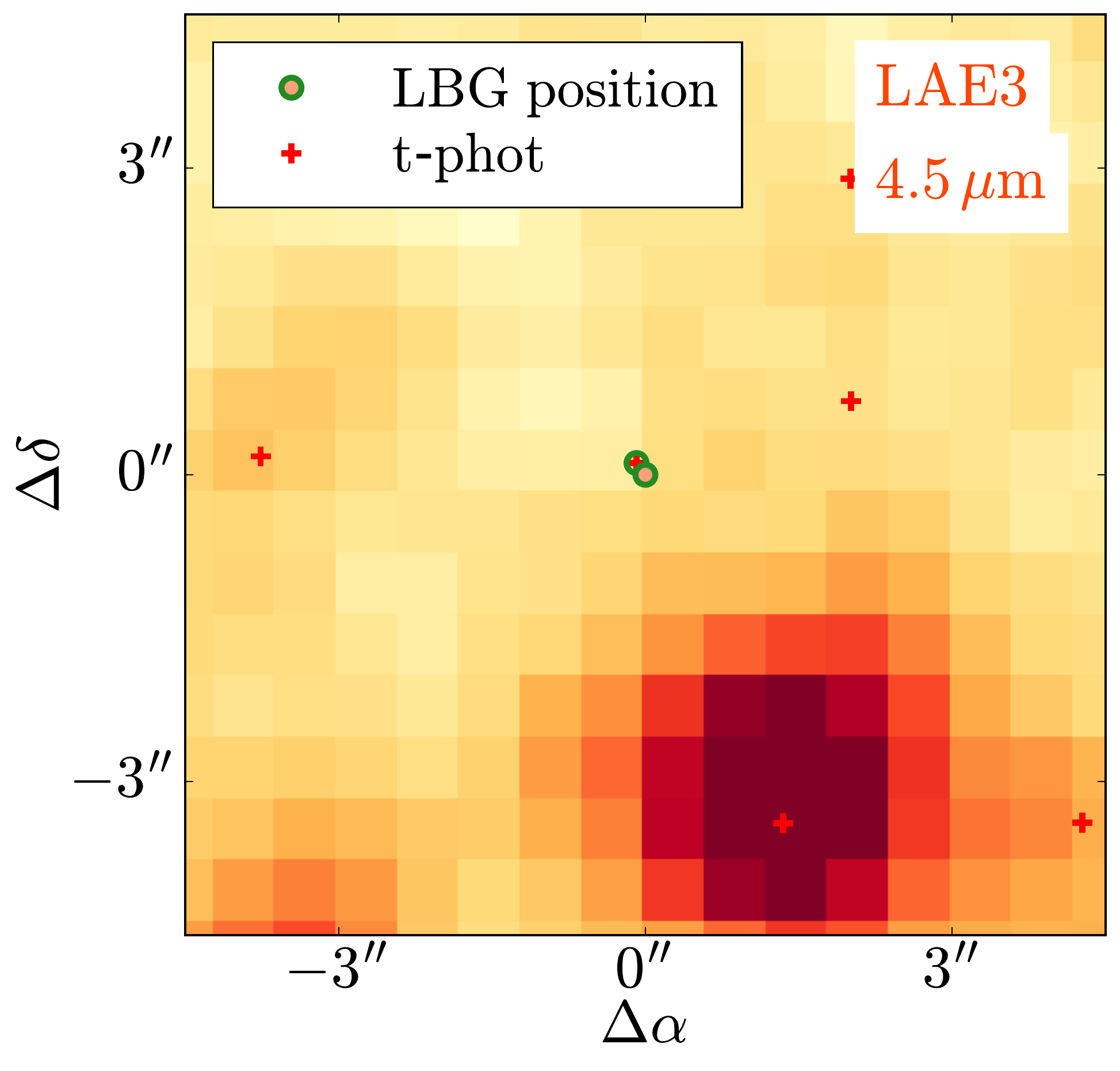}
\end{framed}
\end{subfigure}
\caption{}
\end{figure*}
\renewcommand{\thefigure}{\arabic{figure}}

\renewcommand{\thefigure}{B\arabic{figure} (Cont.)}
\addtocounter{figure}{-1}
\begin{figure*}
\begin{subfigure}{0.85\textwidth}
\begin{framed}
\includegraphics[width=0.24\textwidth]{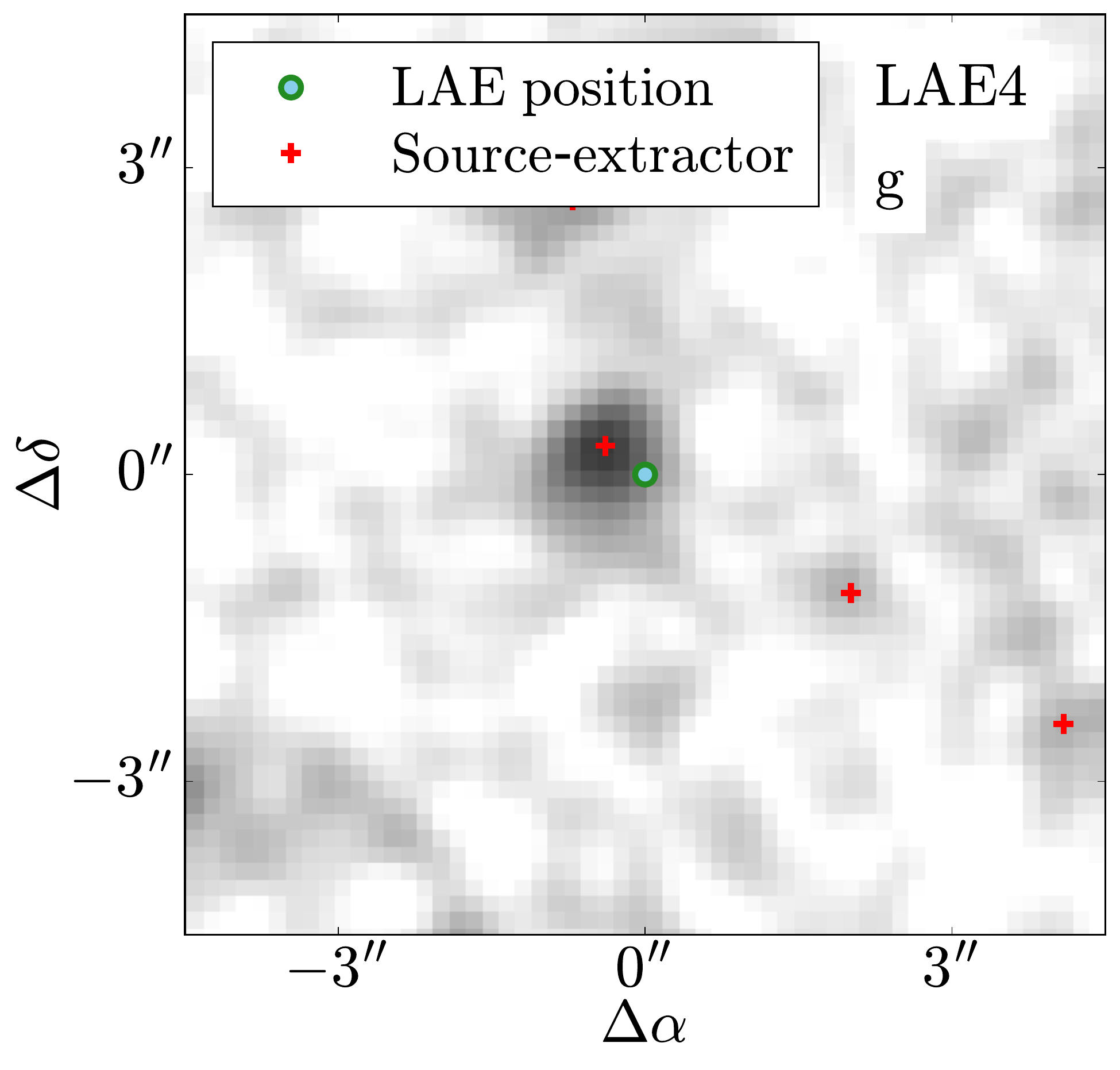}
\includegraphics[width=0.24\textwidth]{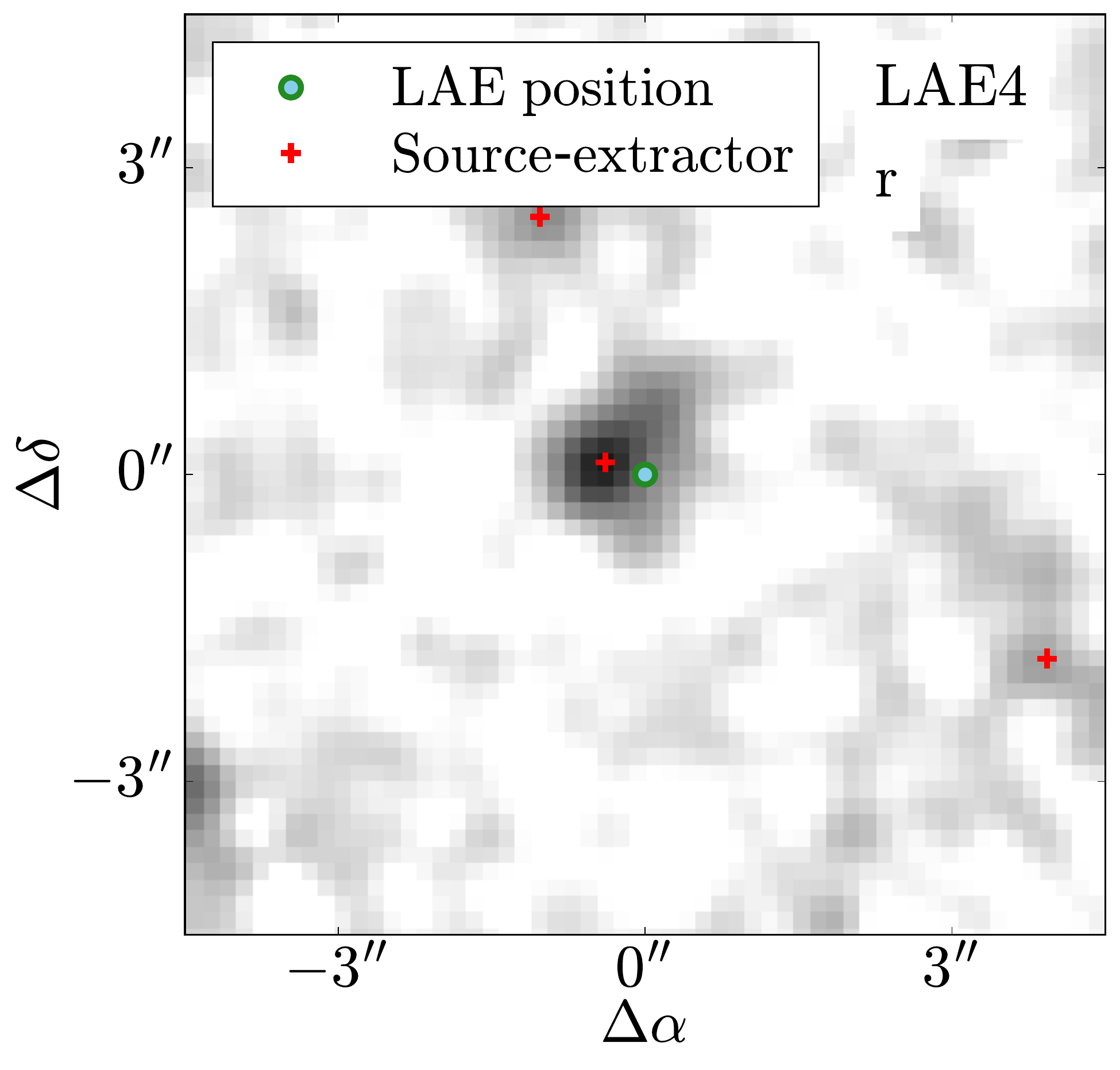}
\includegraphics[width=0.24\textwidth]{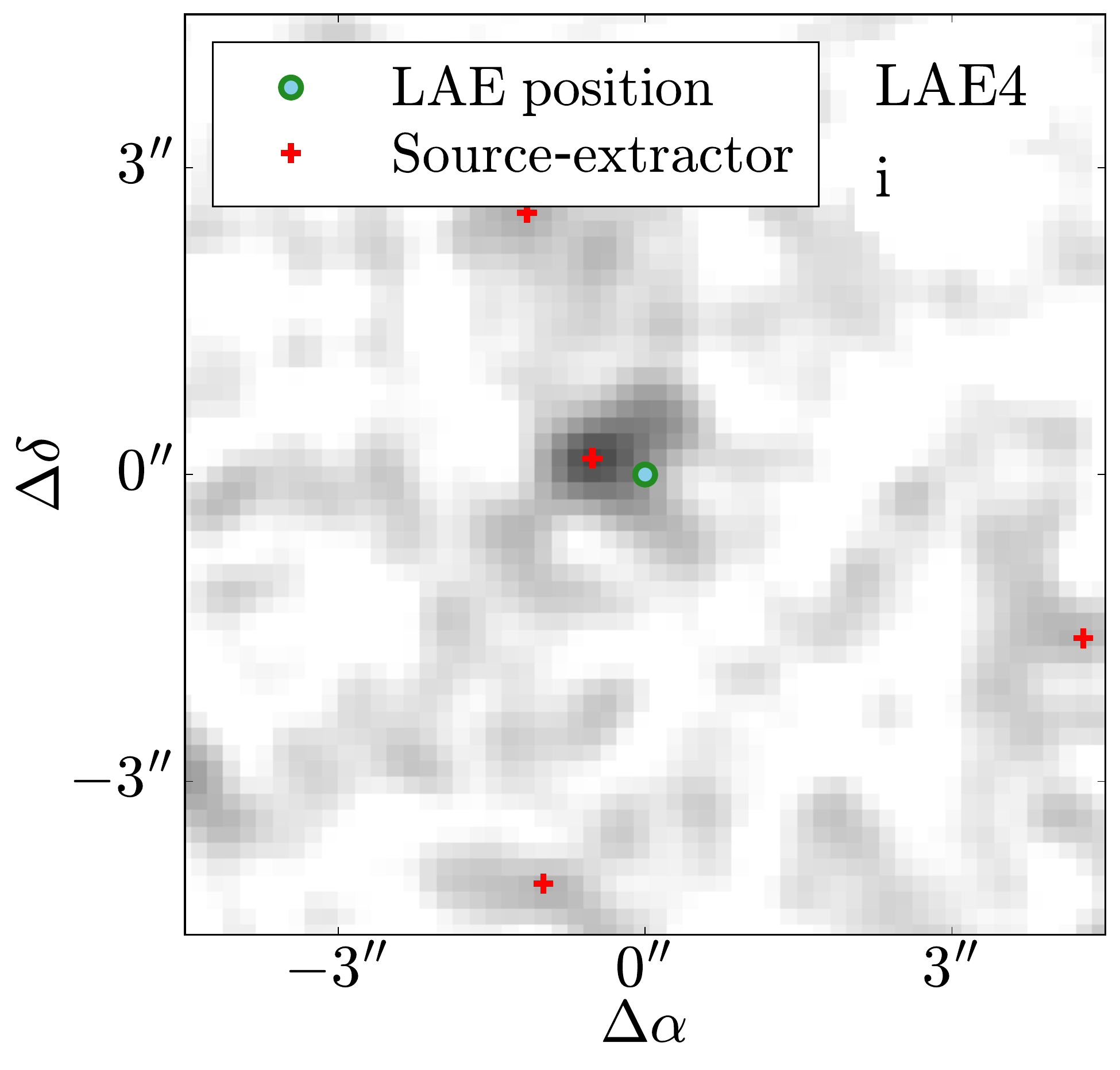}
\includegraphics[width=0.24\textwidth]{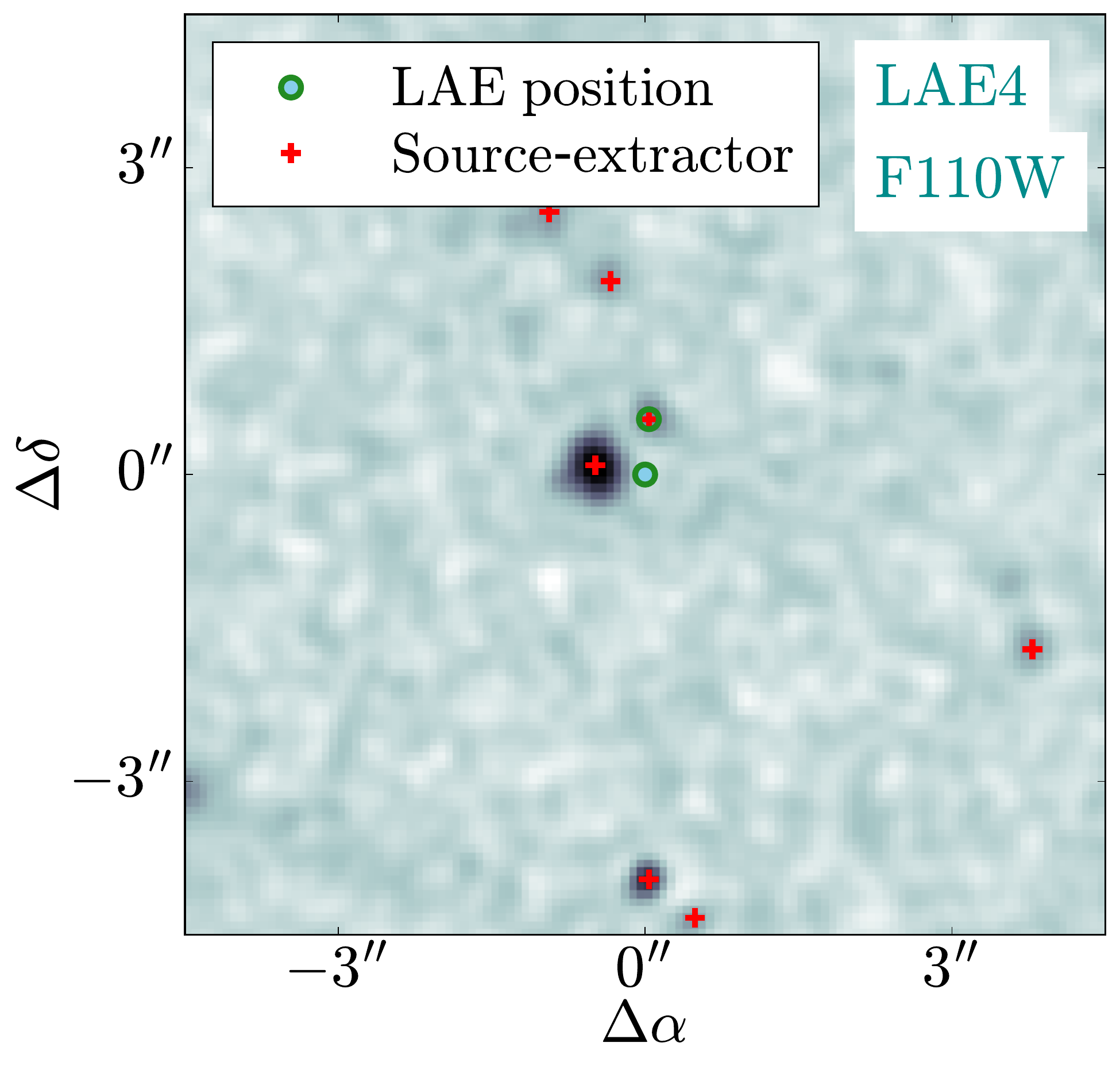}
\includegraphics[width=0.24\textwidth]{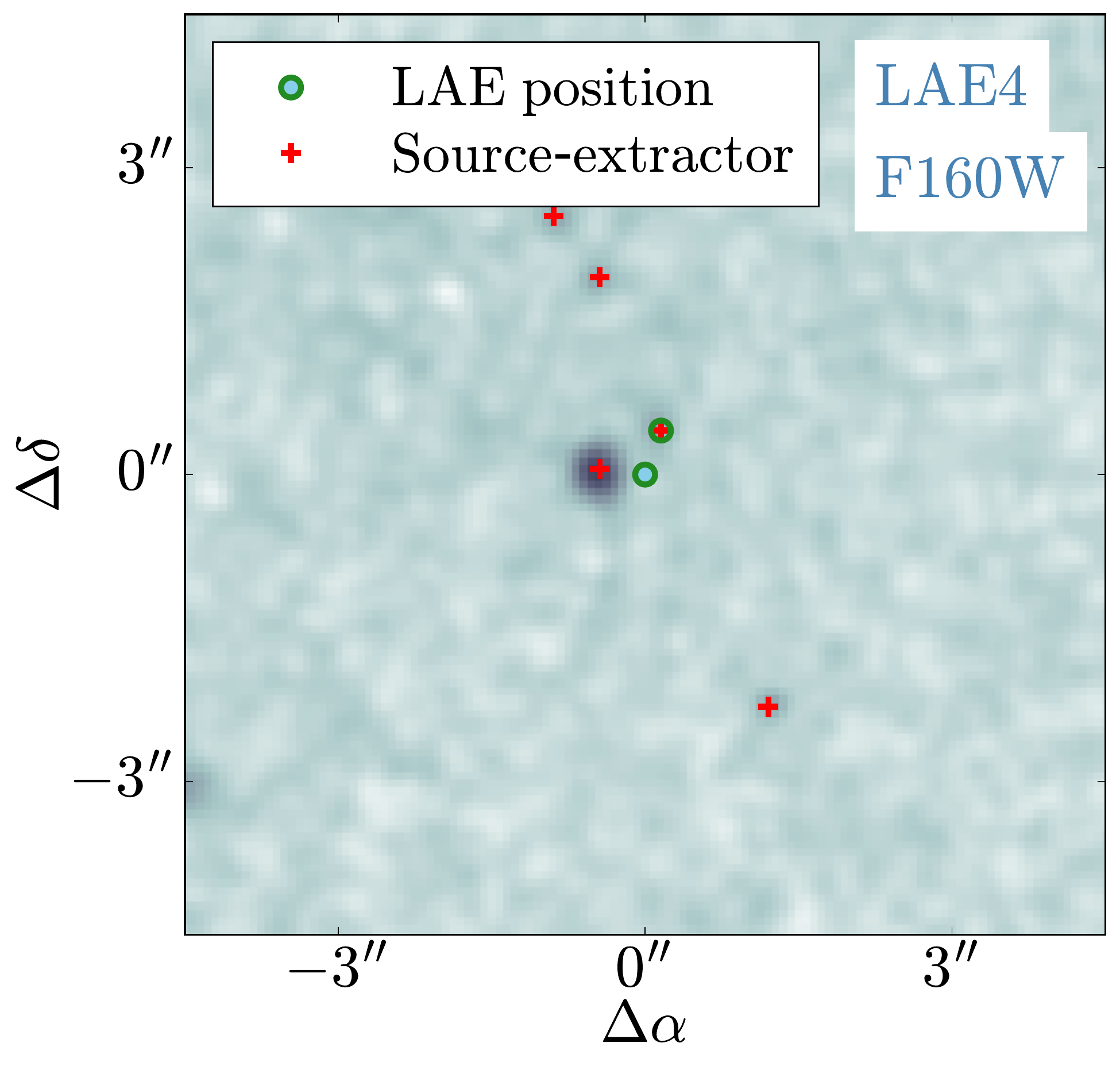}
\includegraphics[width=0.248\textwidth]{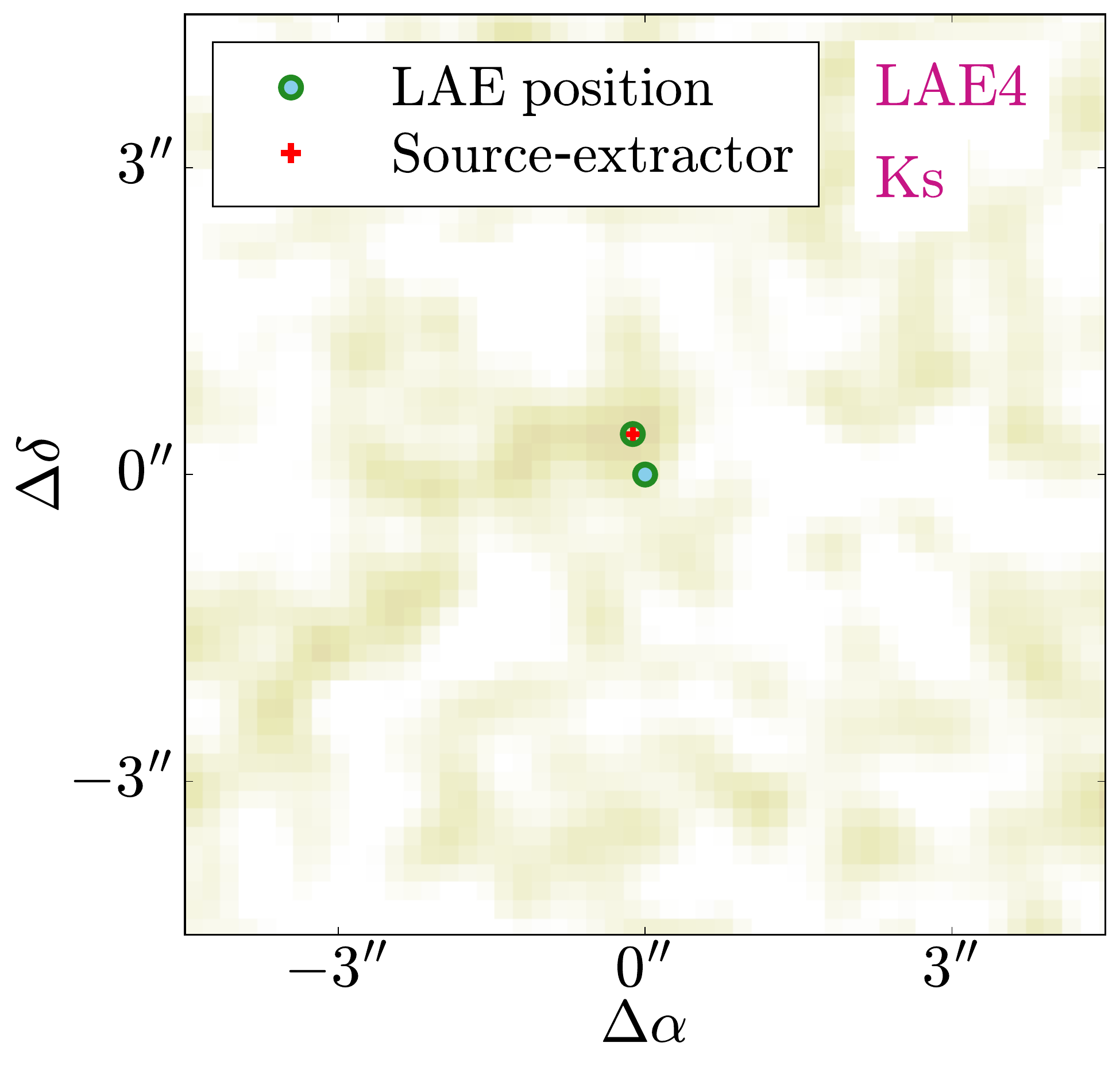}
\includegraphics[width=0.249\textwidth]{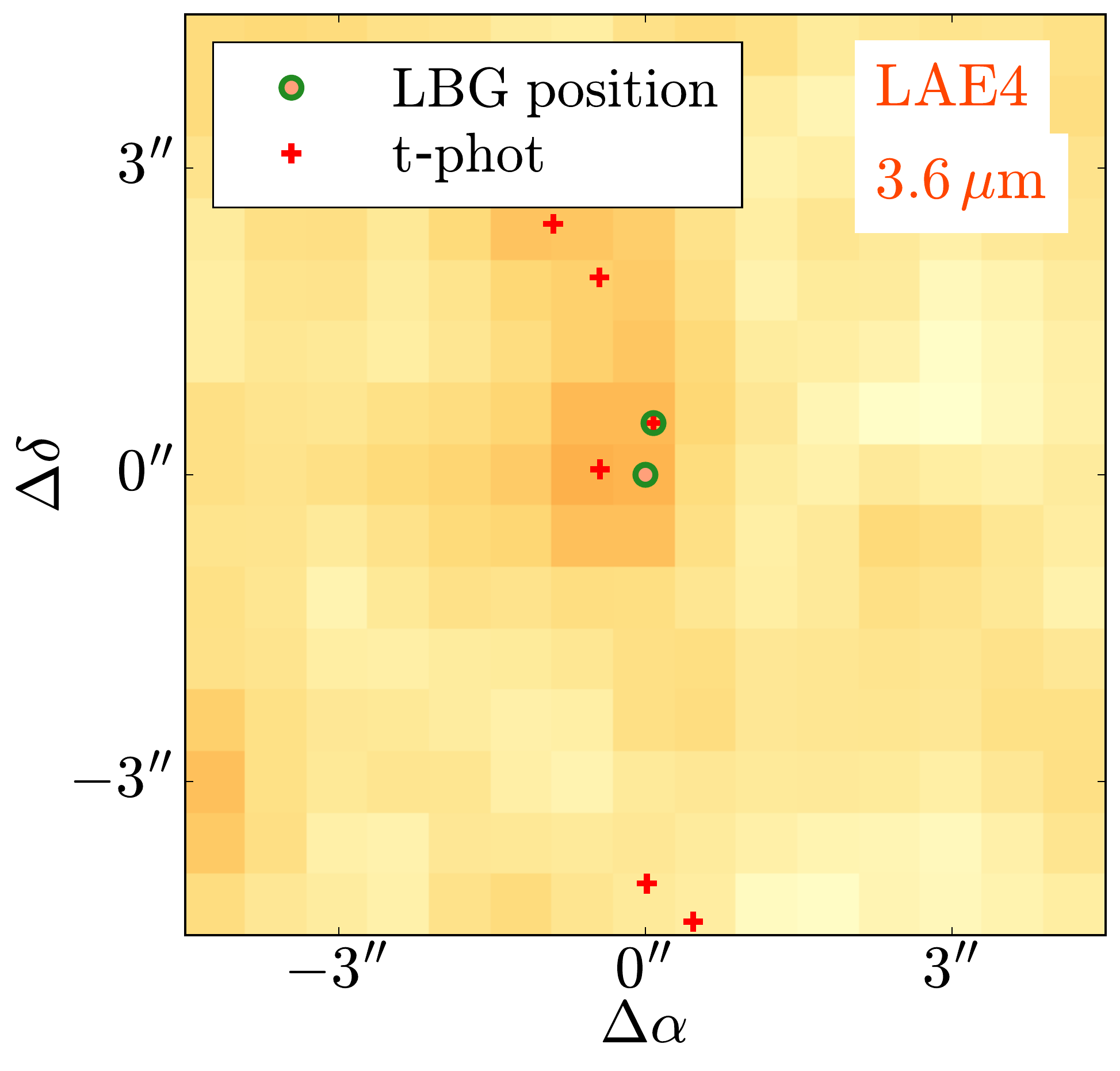}
\includegraphics[width=0.249\textwidth]{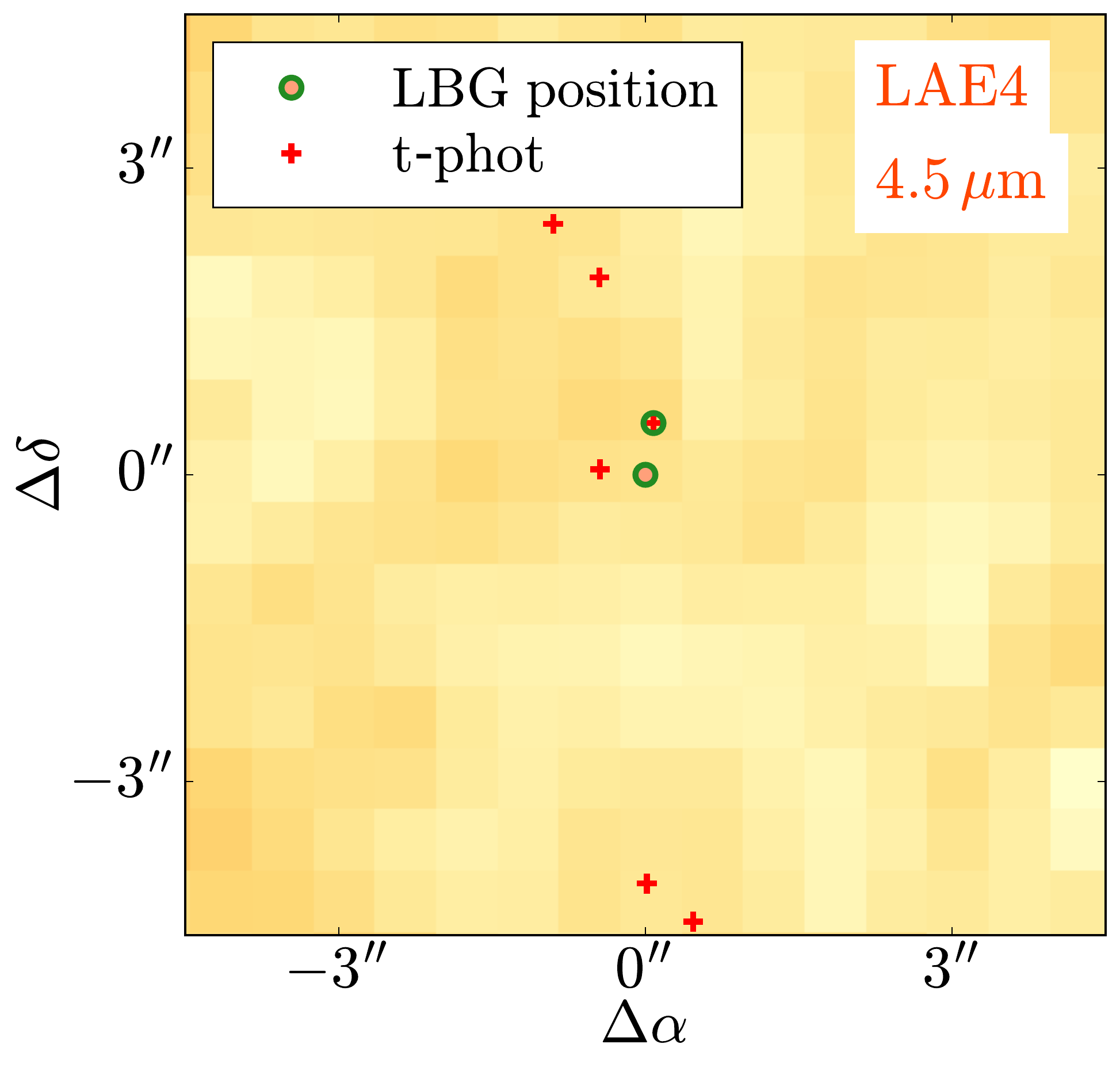}
\end{framed}
\end{subfigure}
\begin{subfigure}{0.85\textwidth}
\begin{framed}
\includegraphics[width=0.24\textwidth]{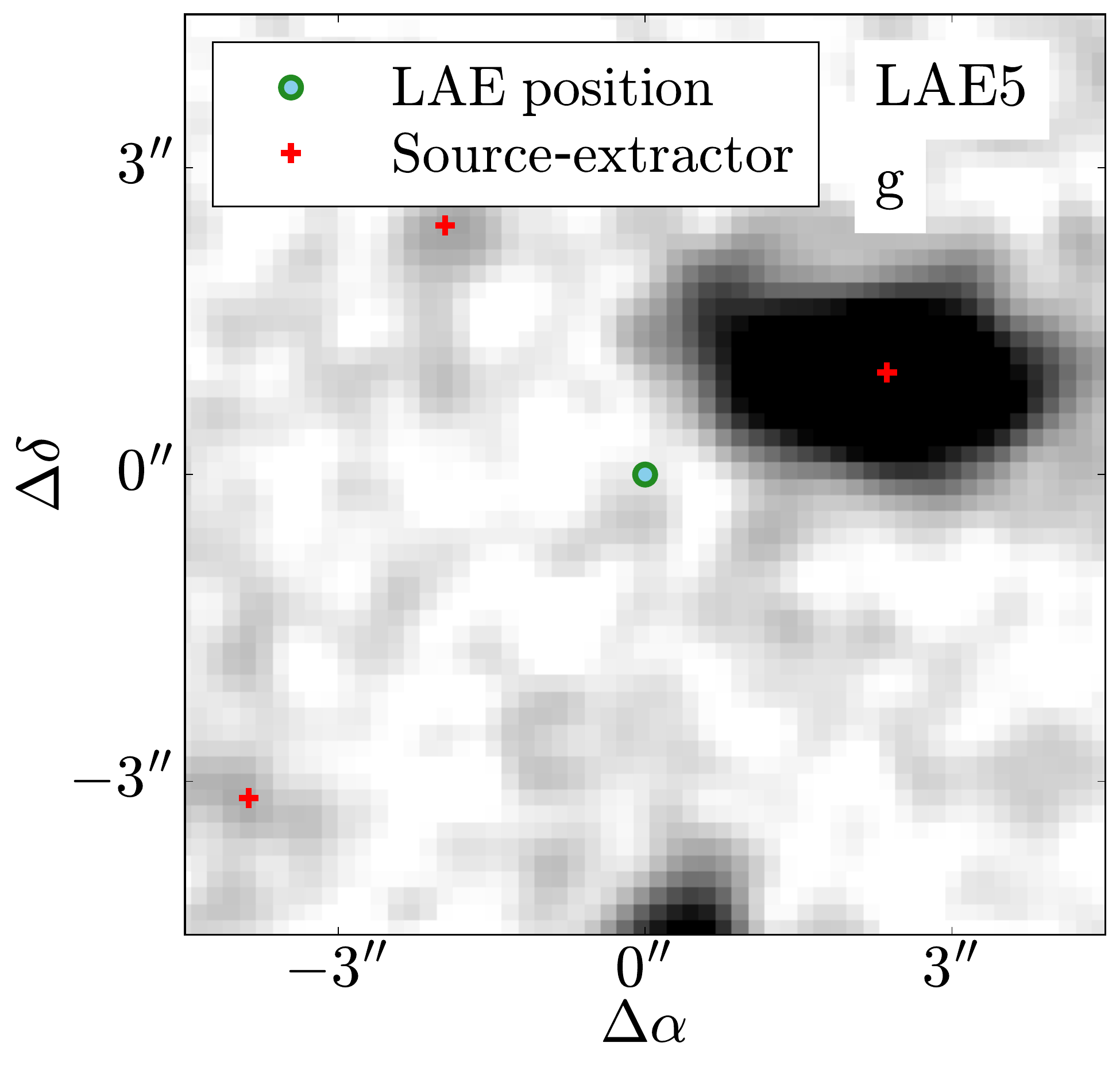}
\includegraphics[width=0.24\textwidth]{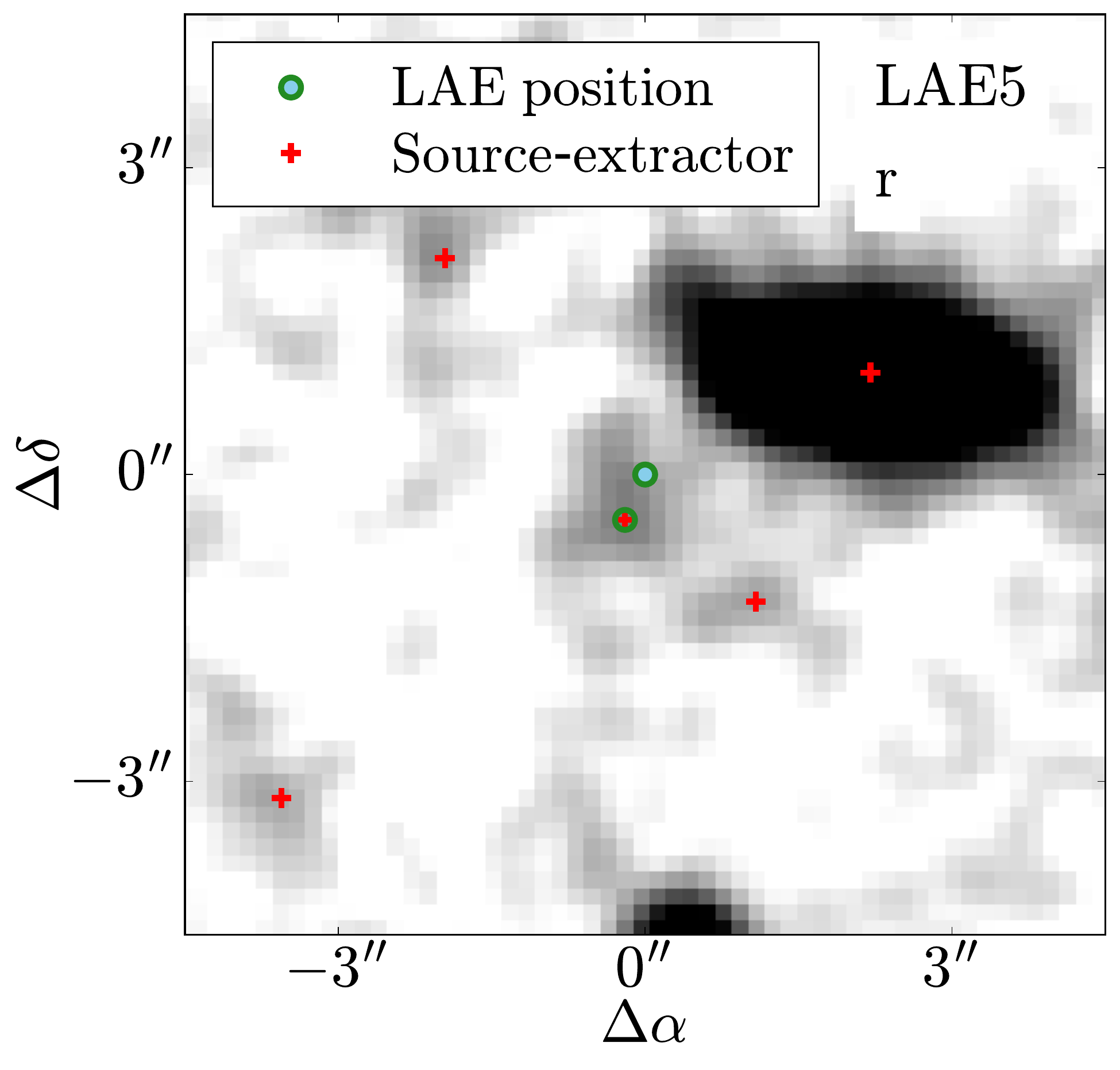}
\includegraphics[width=0.24\textwidth]{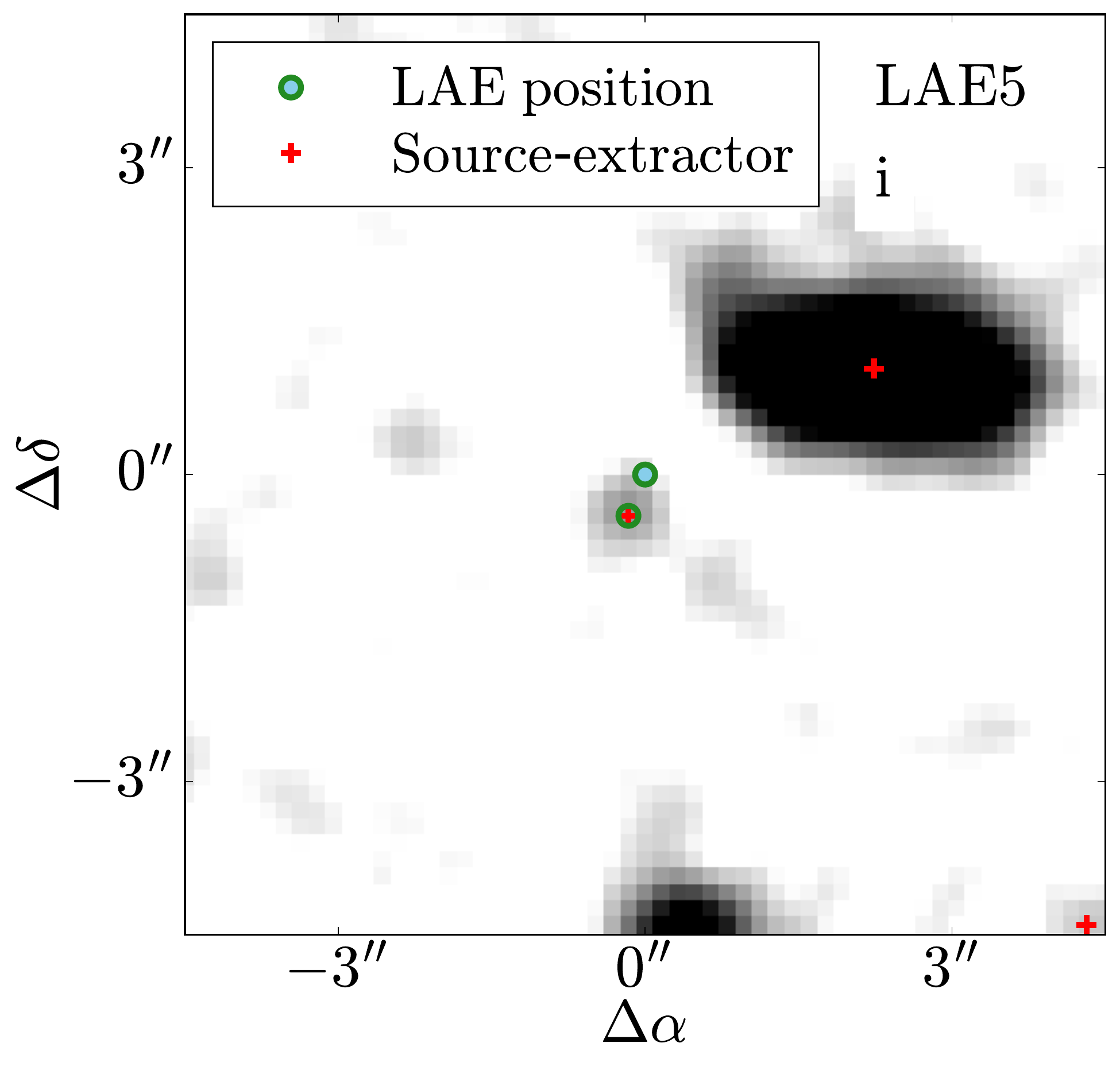}
\includegraphics[width=0.24\textwidth]{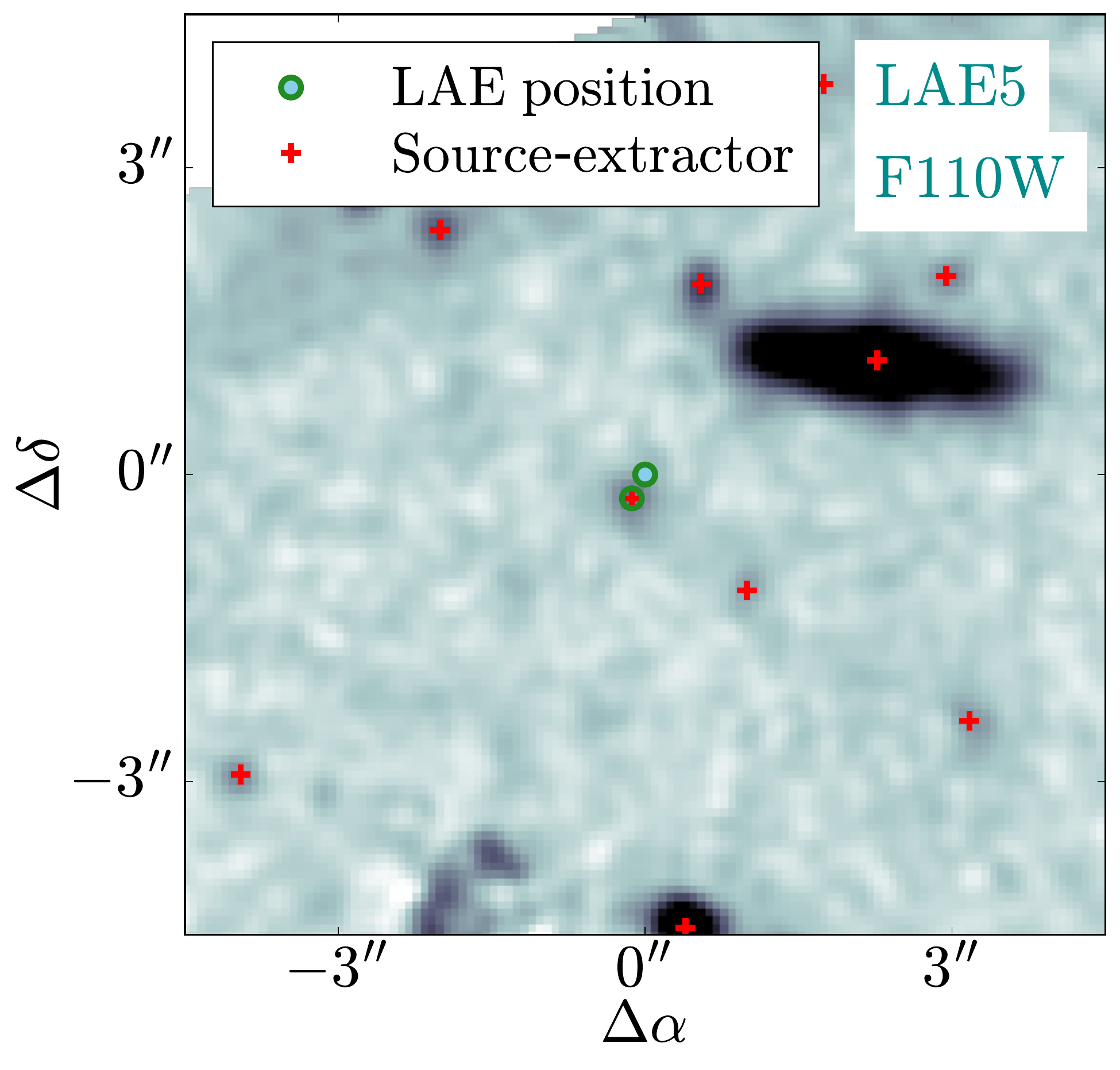}
\includegraphics[width=0.24\textwidth]{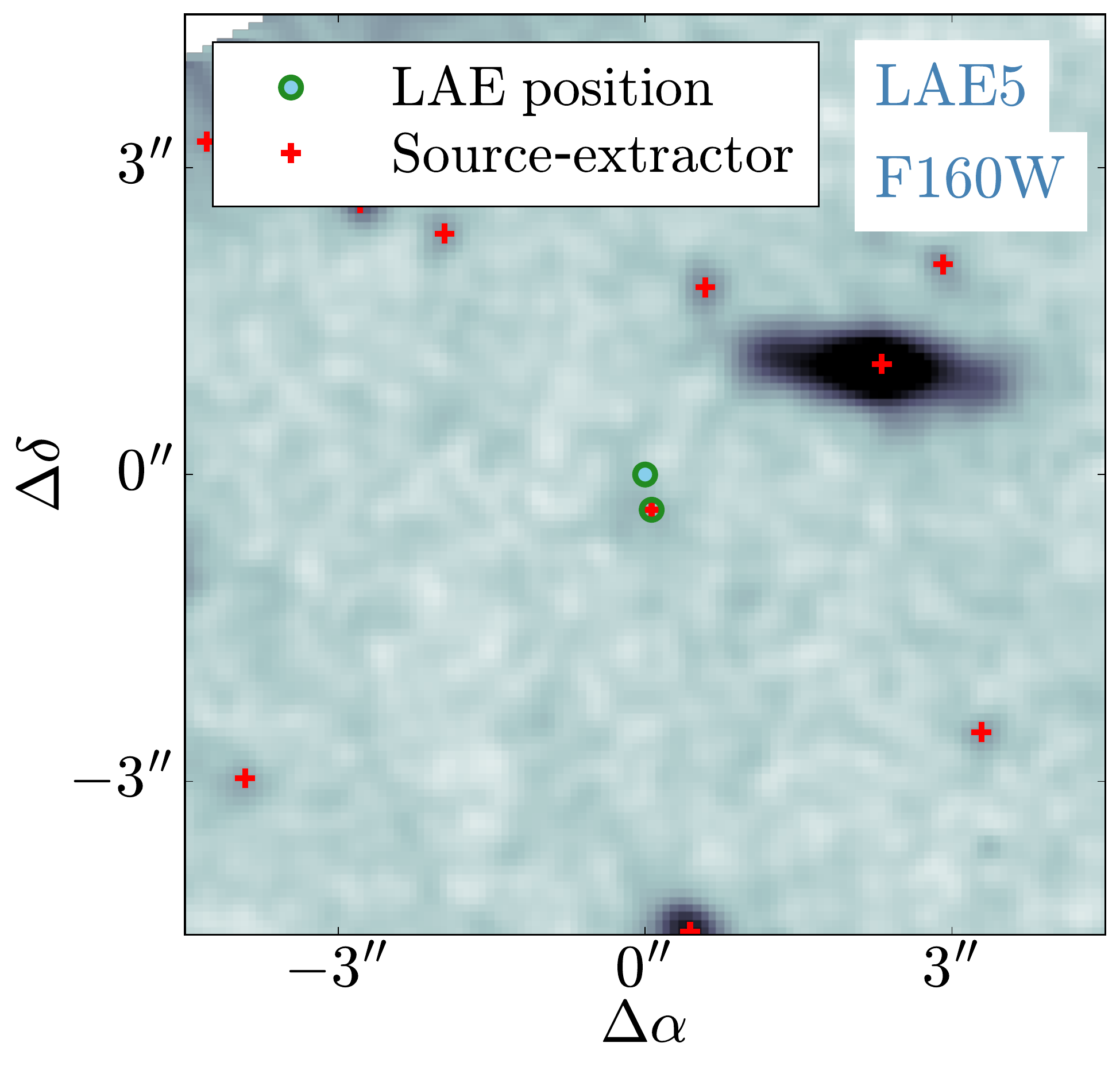}
\includegraphics[width=0.248\textwidth]{Ks/blank.pdf}
\includegraphics[width=0.249\textwidth]{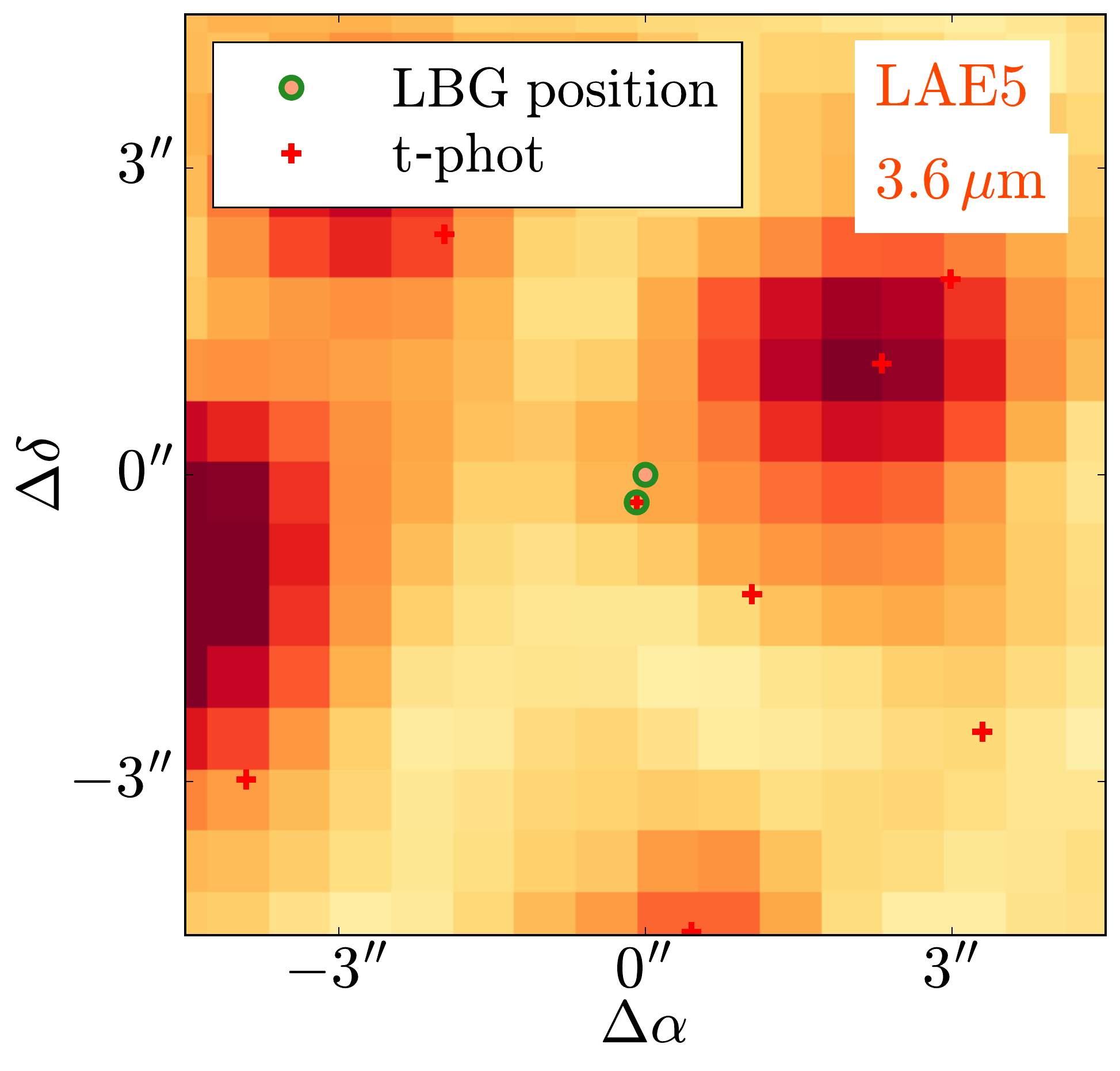}
\includegraphics[width=0.249\textwidth]{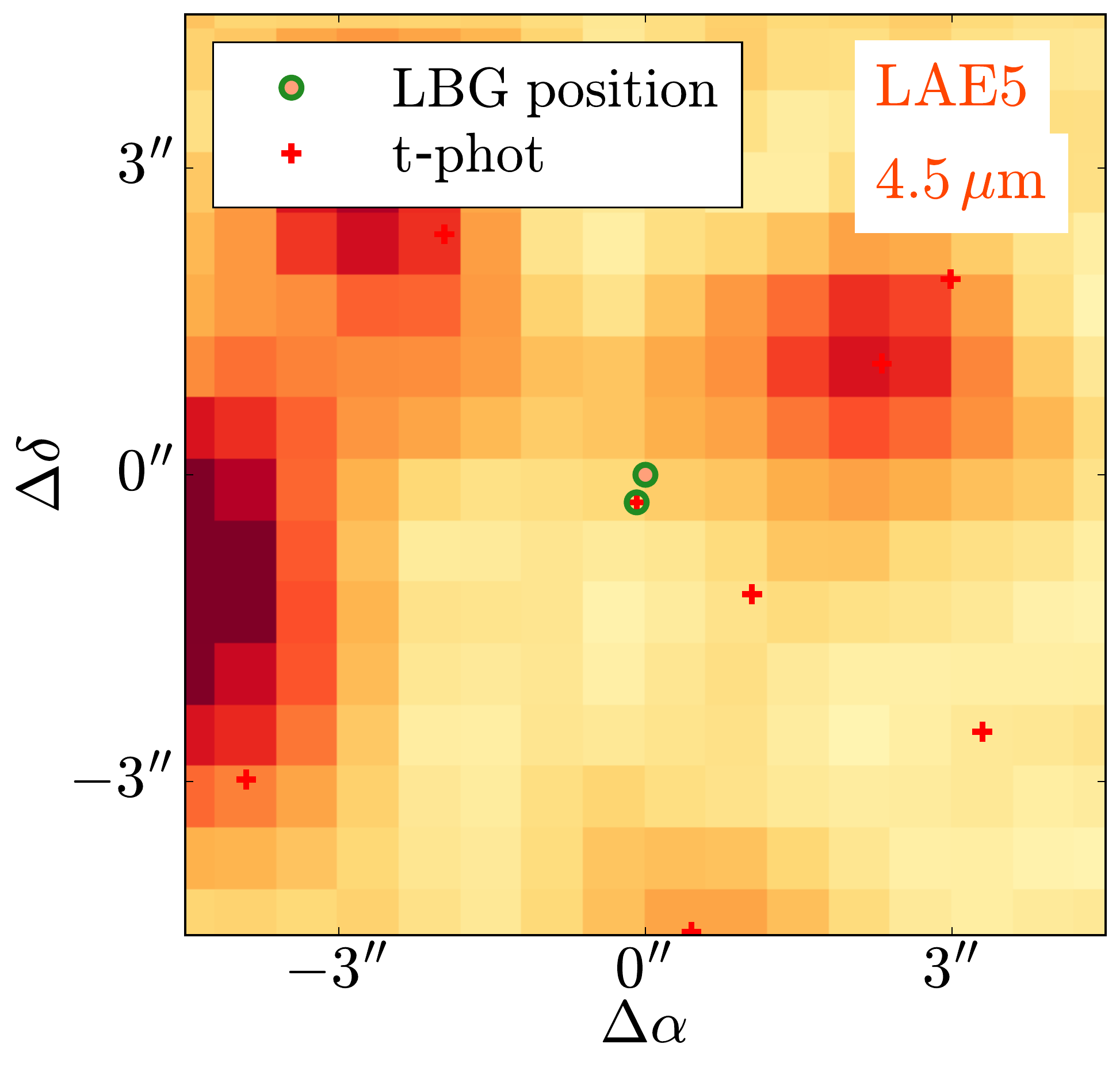}
\end{framed}
\end{subfigure}
\begin{subfigure}{0.85\textwidth}
\begin{framed}
\includegraphics[width=0.24\textwidth]{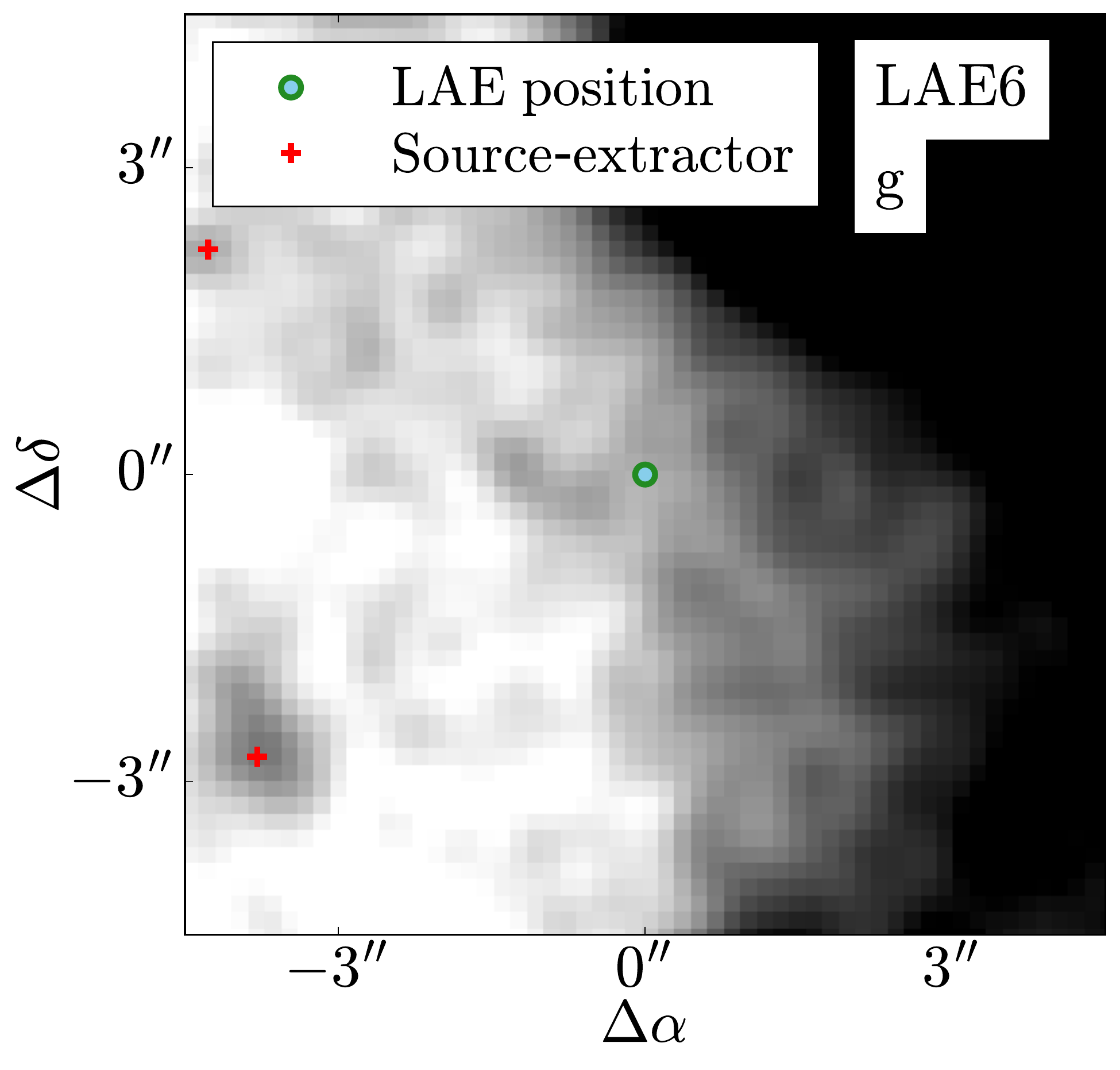}
\includegraphics[width=0.24\textwidth]{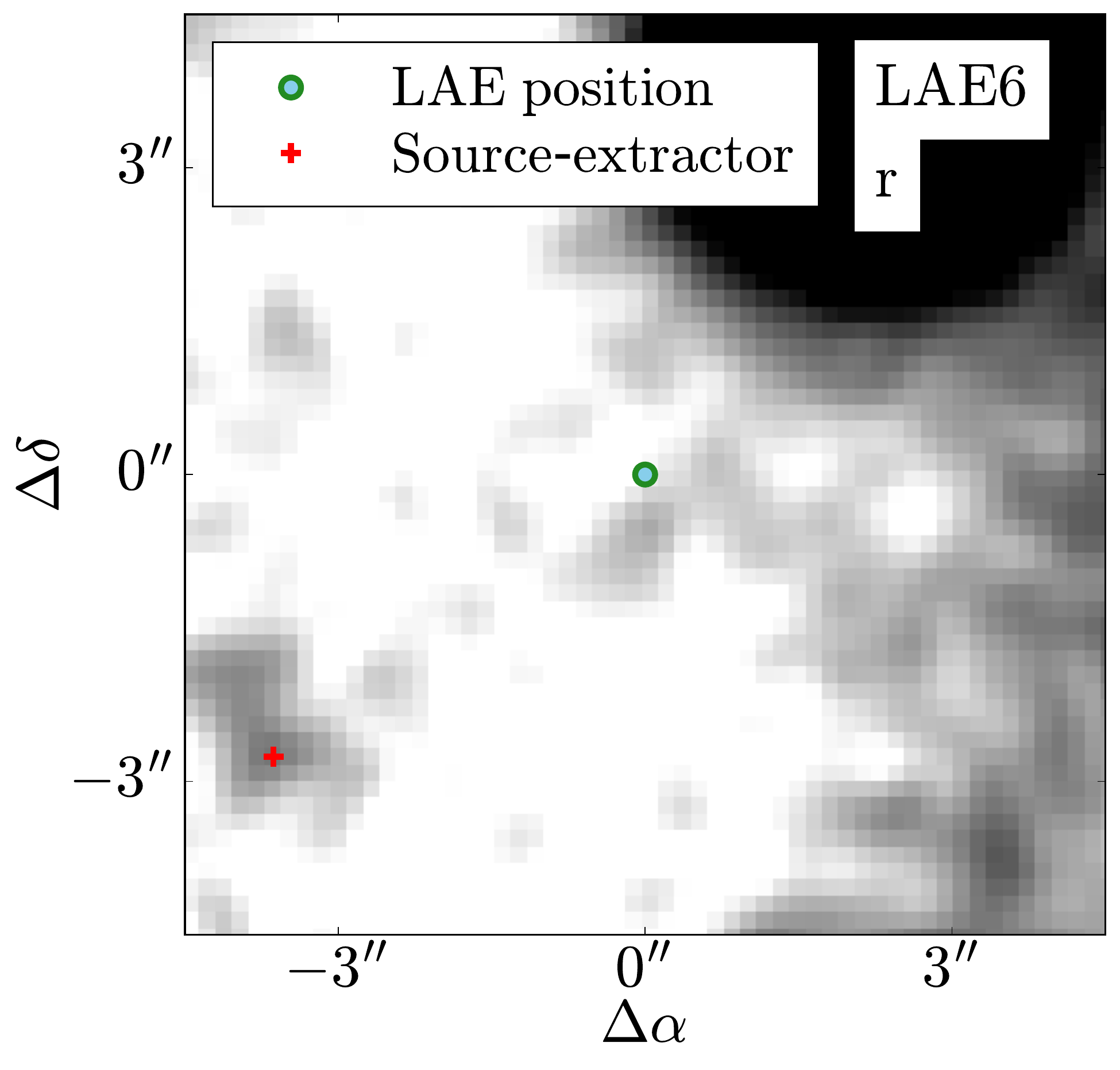}
\includegraphics[width=0.24\textwidth]{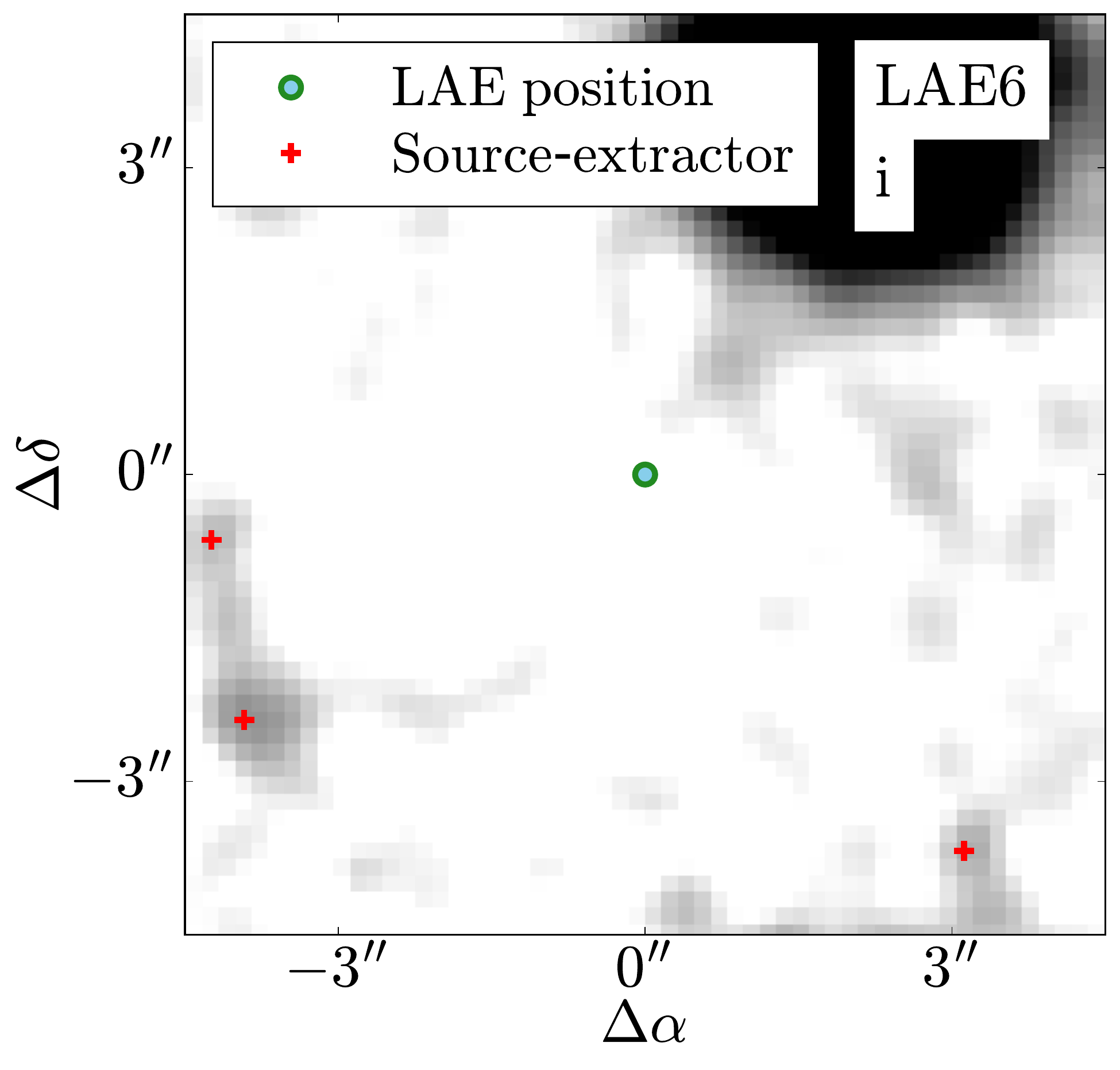}
\includegraphics[width=0.24\textwidth]{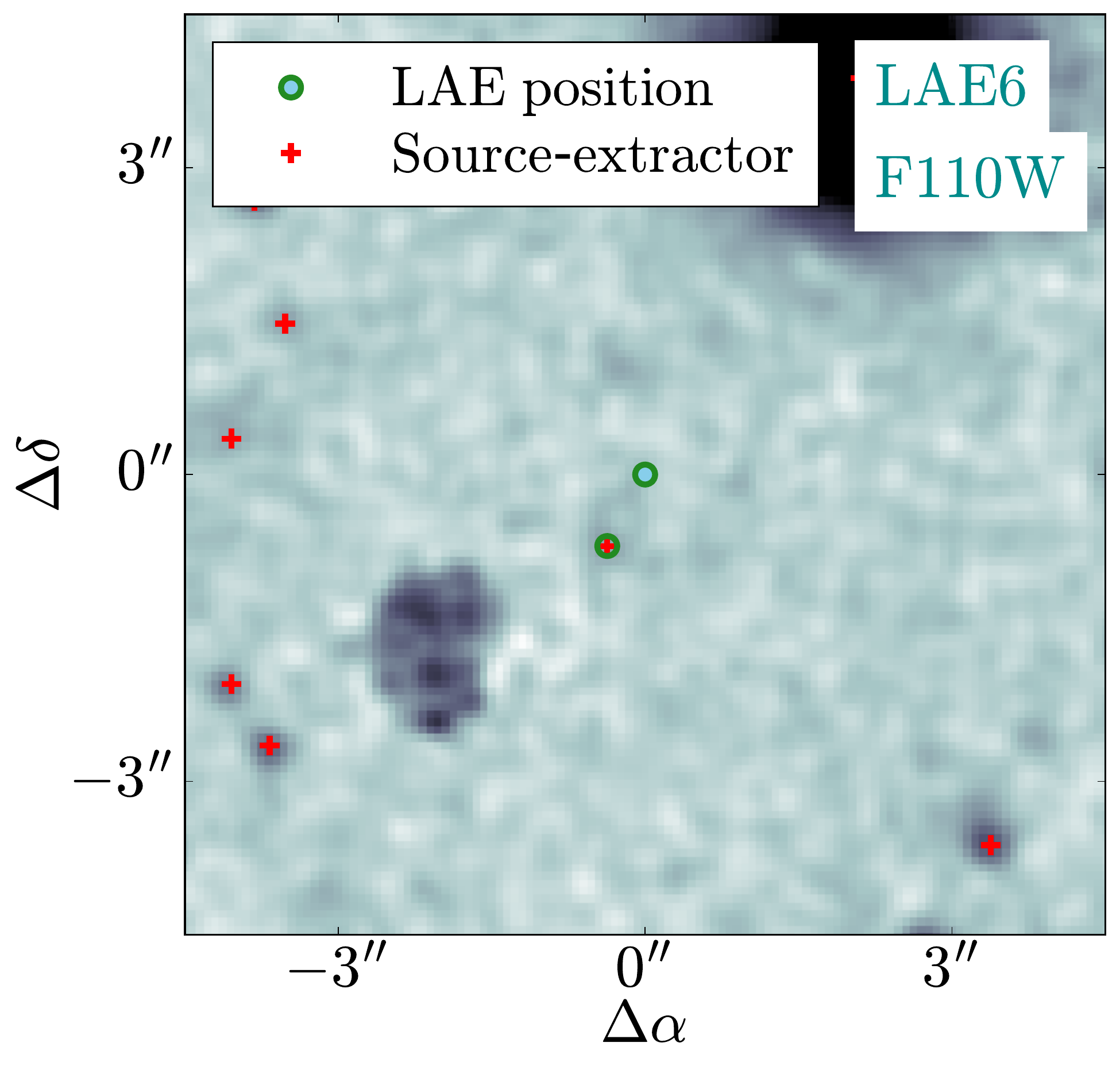}
\includegraphics[width=0.24\textwidth]{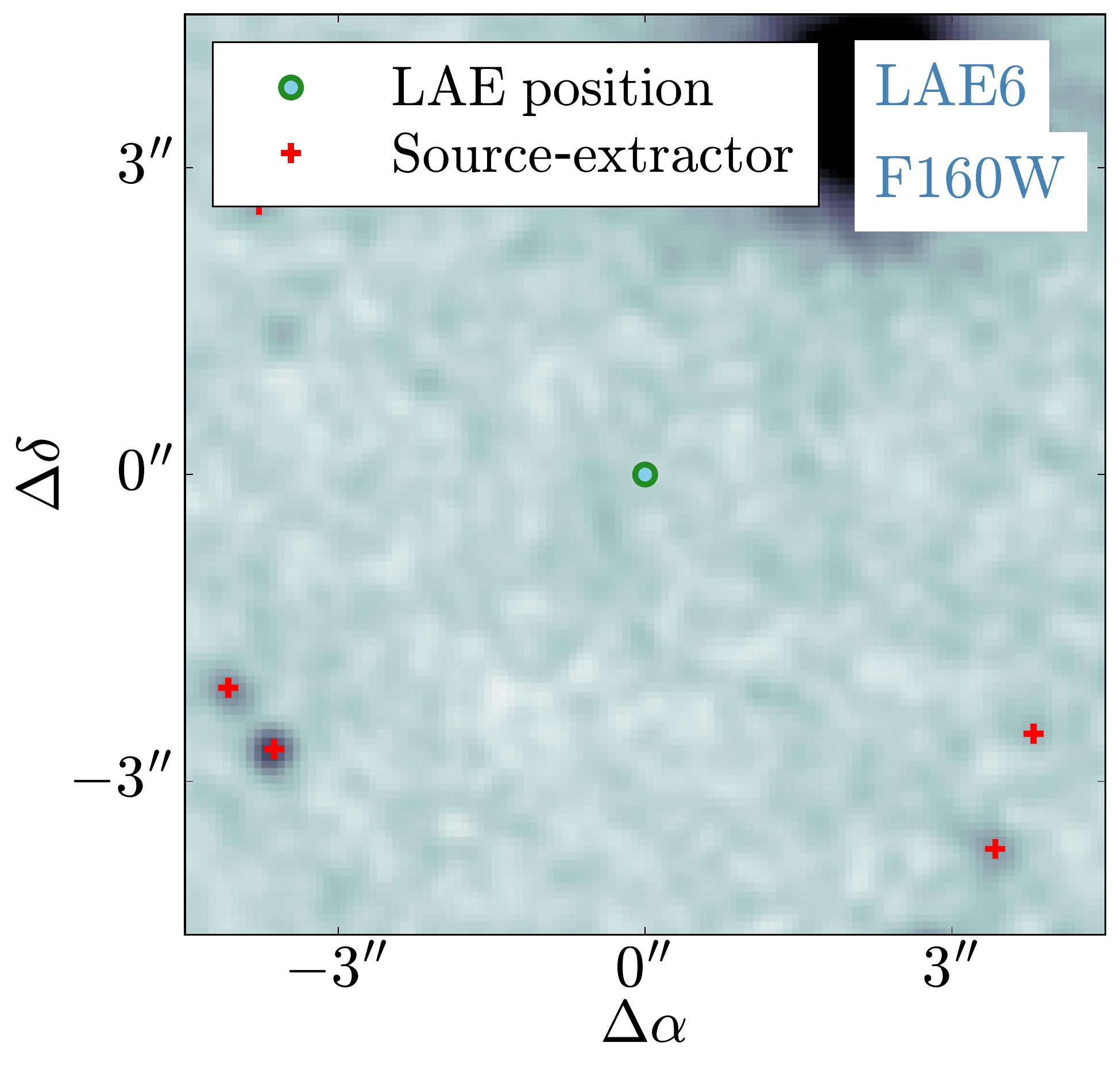}
\includegraphics[width=0.248\textwidth]{Ks/blank.pdf}
\includegraphics[width=0.249\textwidth]{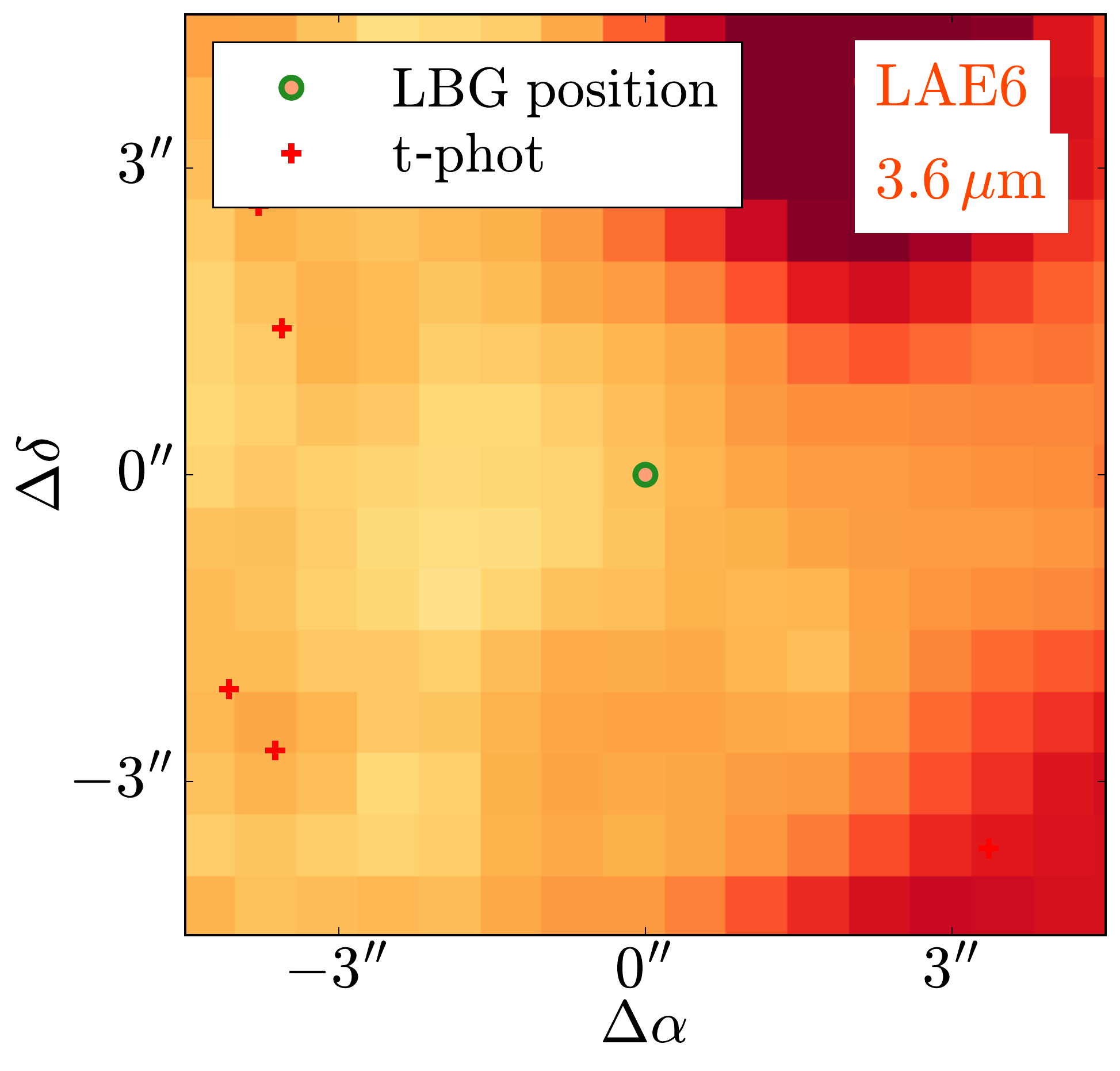}
\includegraphics[width=0.249\textwidth]{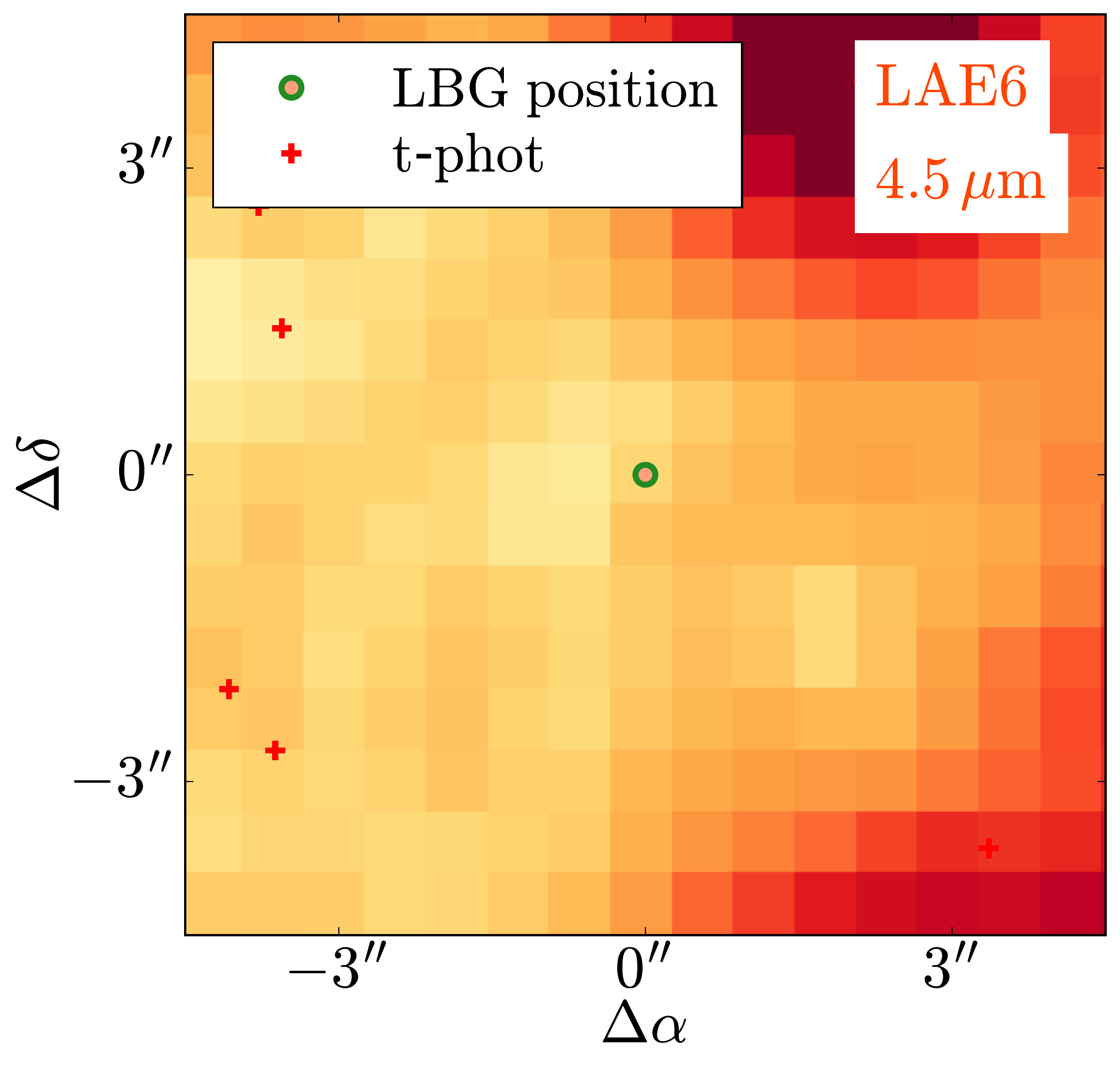}
\end{framed}
\end{subfigure}
\caption{}
\end{figure*}
\renewcommand{\thefigure}{\arabic{figure}}

\renewcommand{\thefigure}{B\arabic{figure} (Cont.)}
\addtocounter{figure}{-1}
\begin{figure*}
\begin{subfigure}{0.85\textwidth}
\begin{framed}
\includegraphics[width=0.24\textwidth]{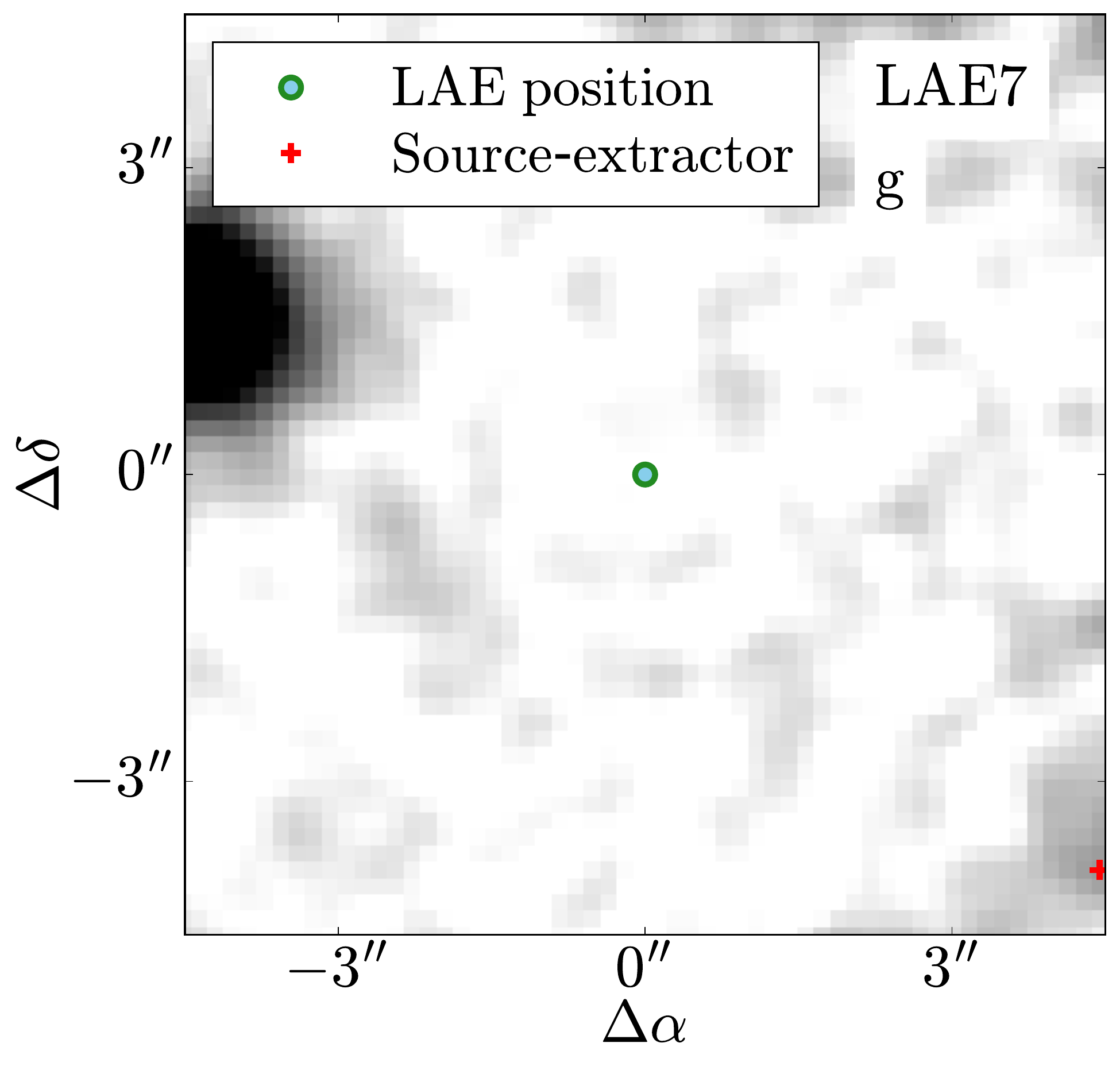}
\includegraphics[width=0.24\textwidth]{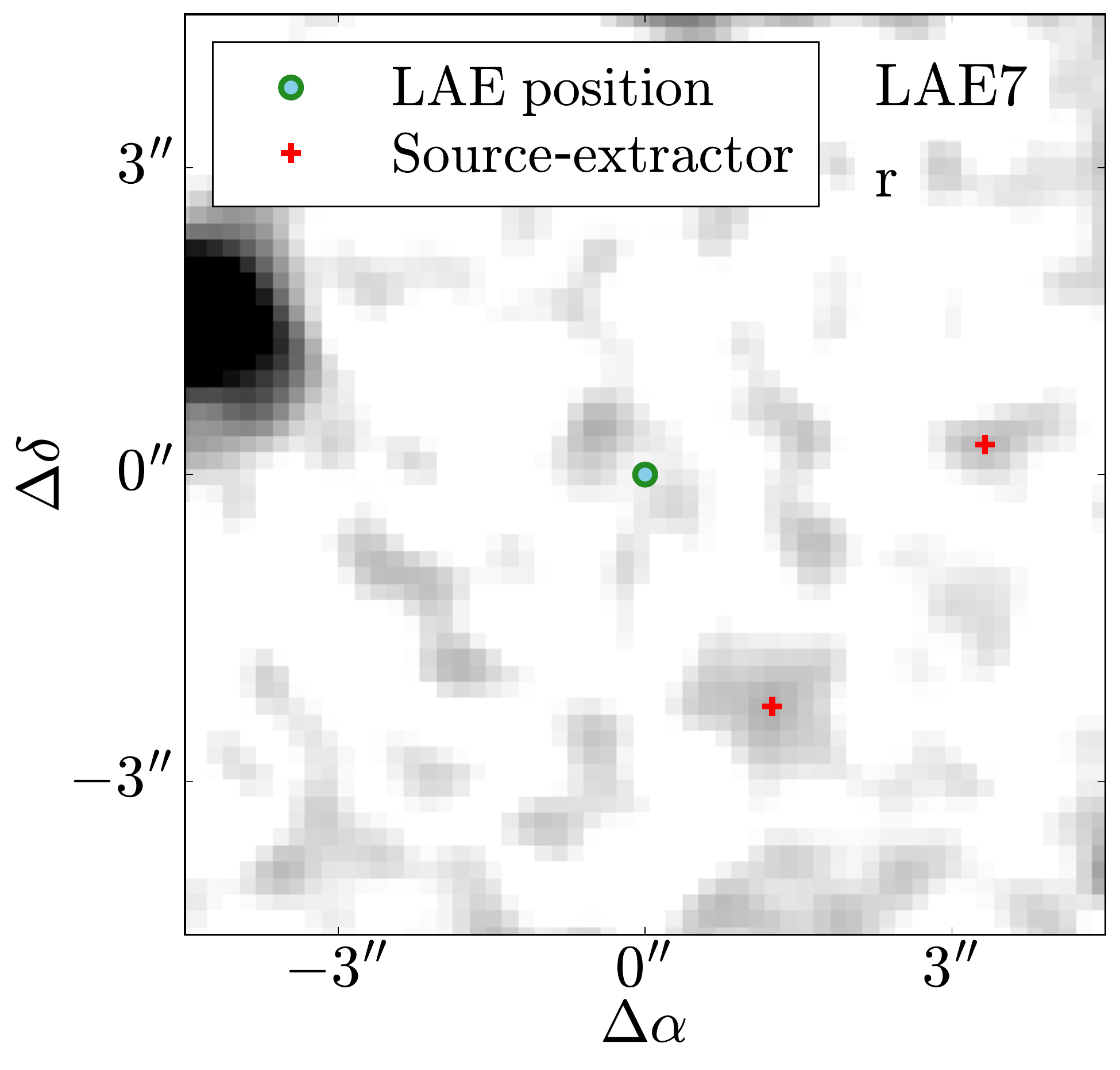}
\includegraphics[width=0.24\textwidth]{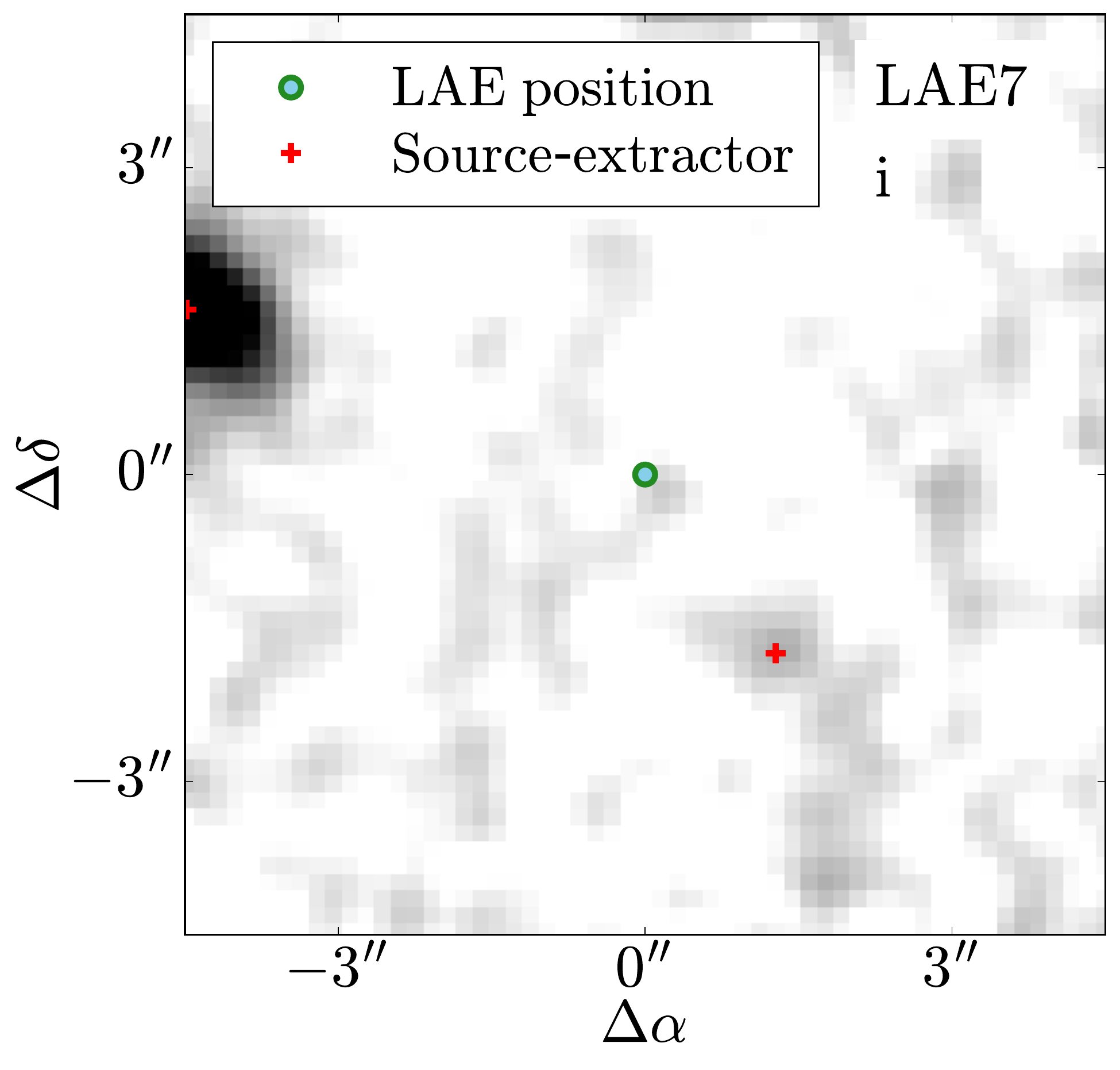}
\includegraphics[width=0.24\textwidth]{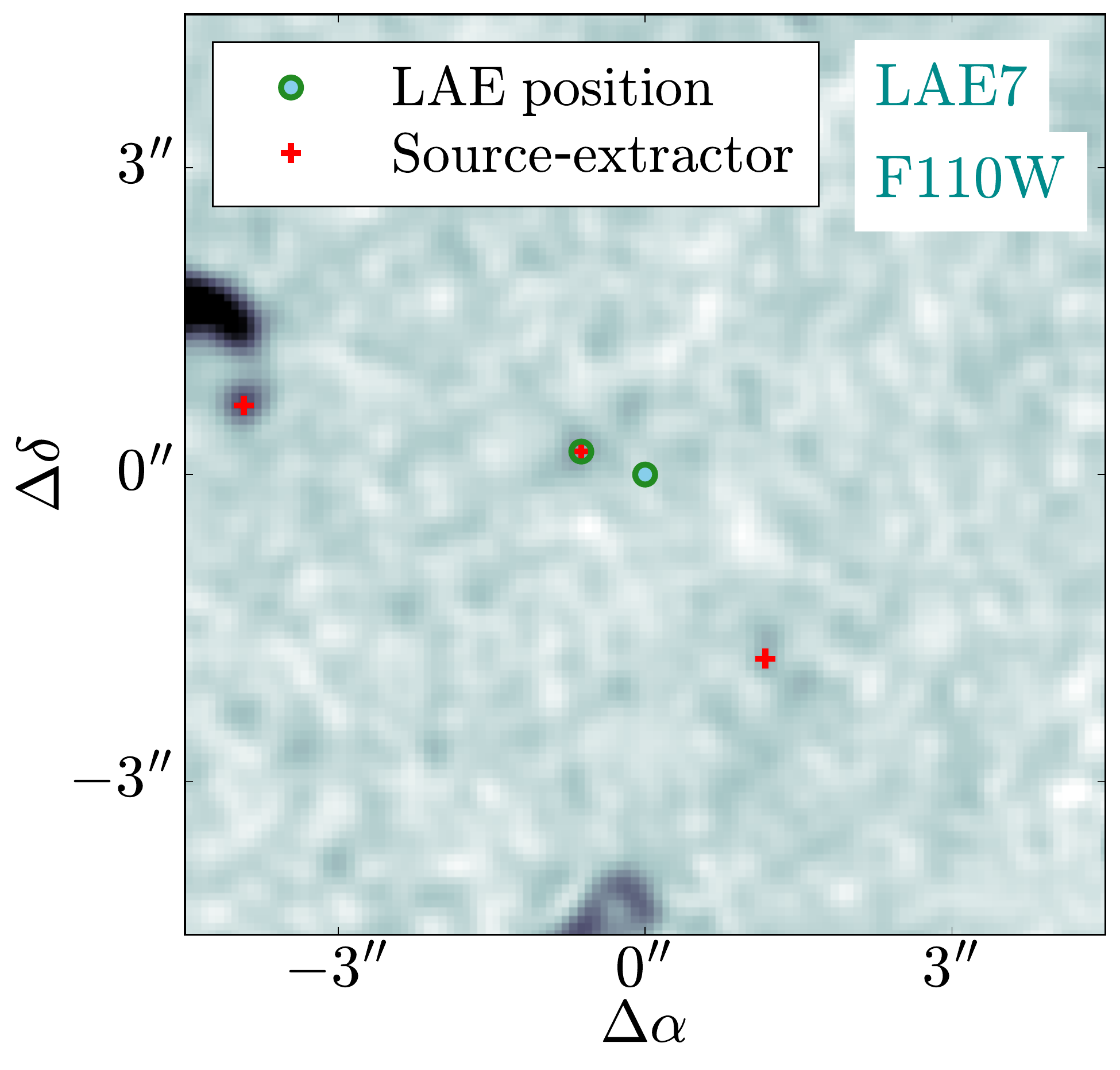}
\includegraphics[width=0.24\textwidth]{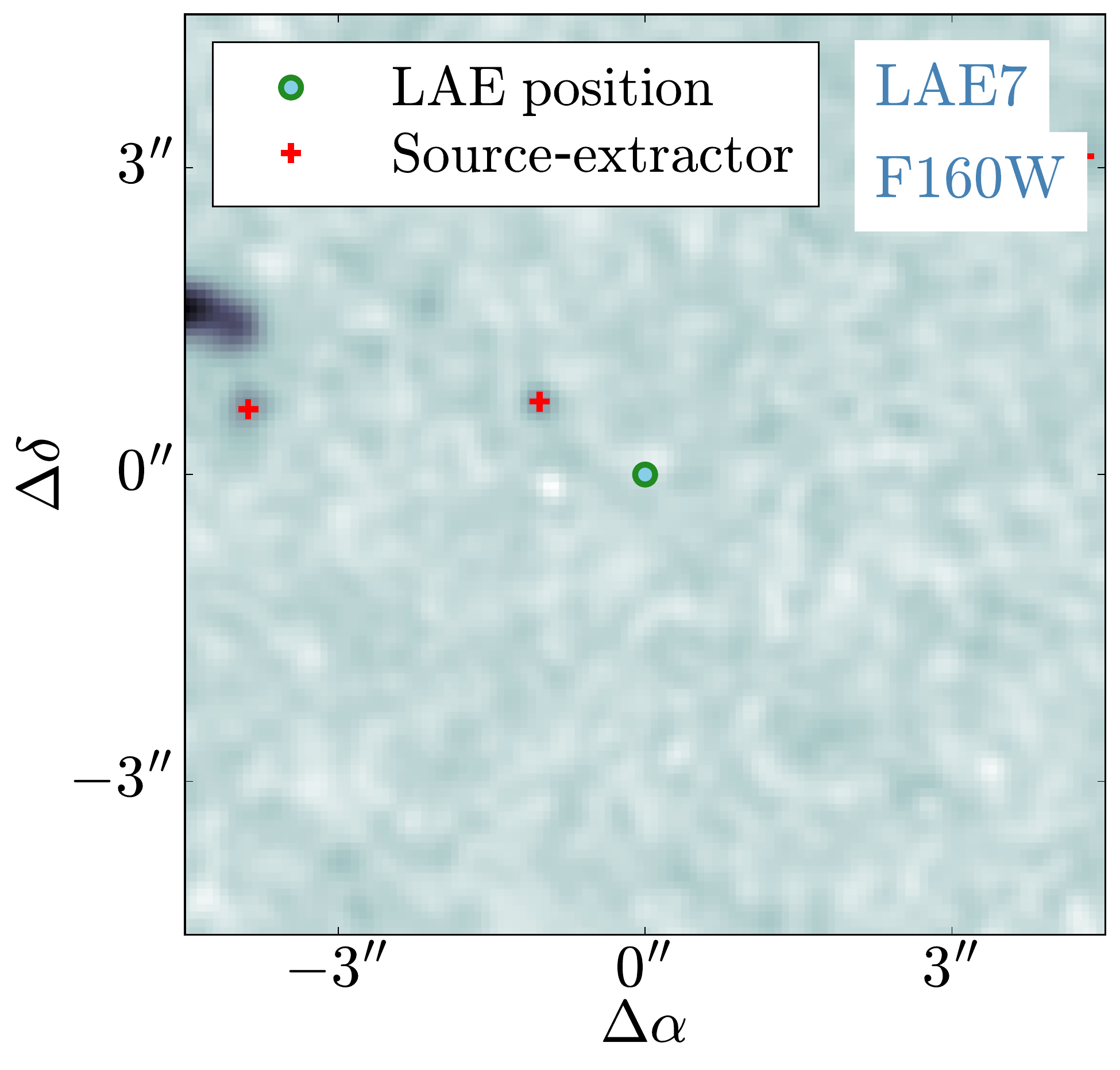}
\includegraphics[width=0.248\textwidth]{Ks/blank.pdf}
\includegraphics[width=0.249\textwidth]{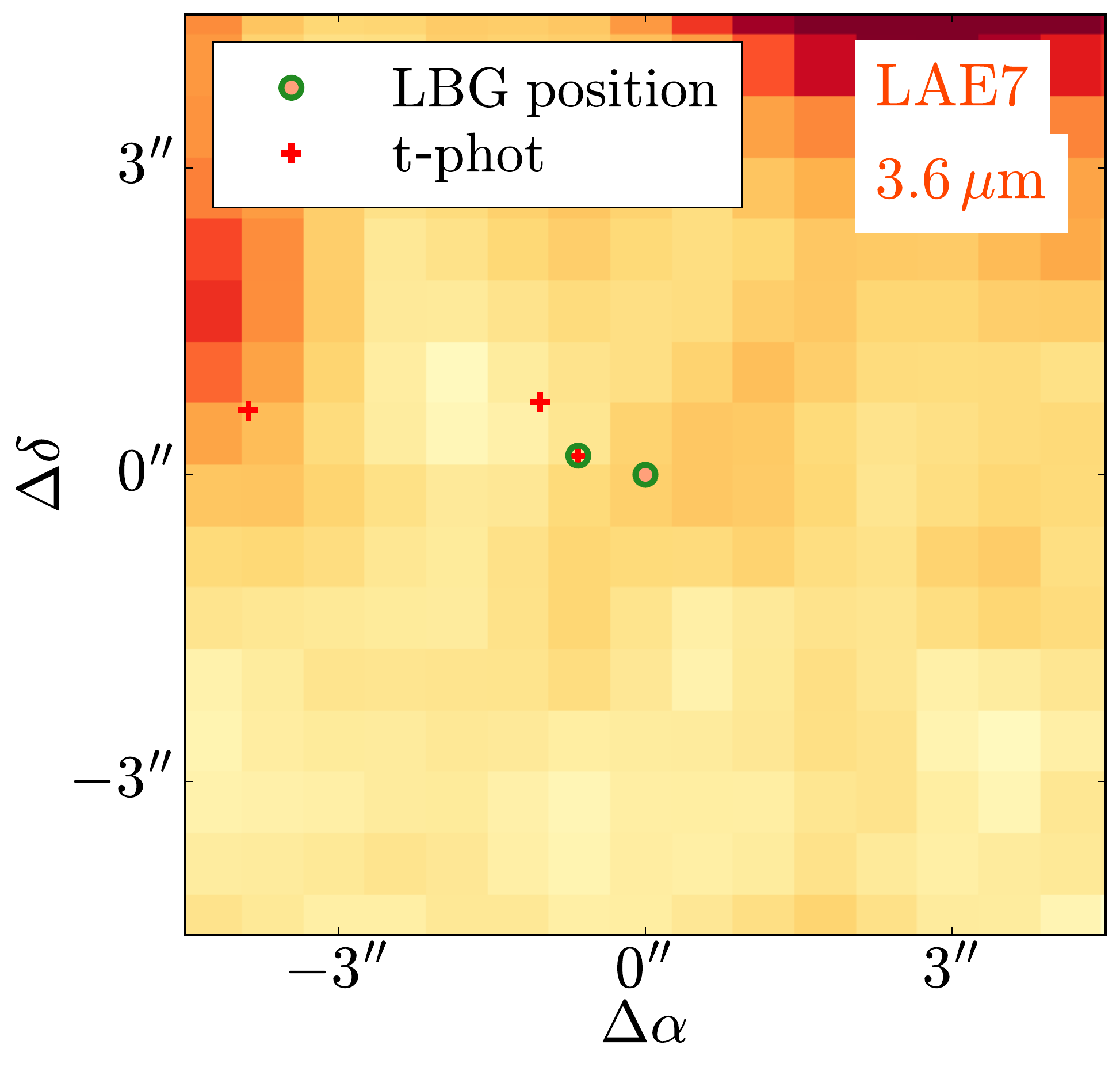}
\includegraphics[width=0.249\textwidth]{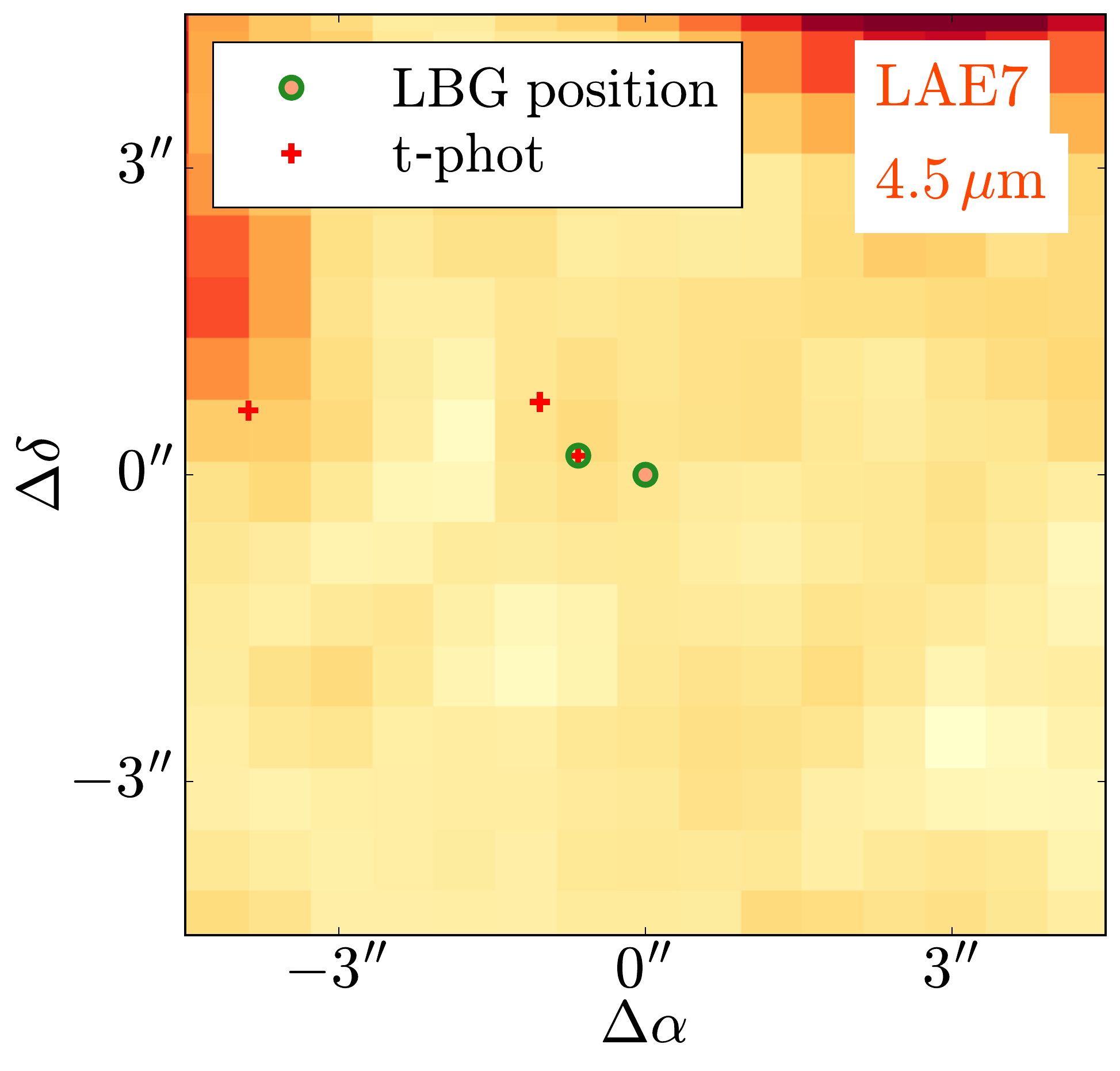}
\end{framed}
\end{subfigure}
\begin{subfigure}{0.85\textwidth}
\begin{framed}
\includegraphics[width=0.24\textwidth]{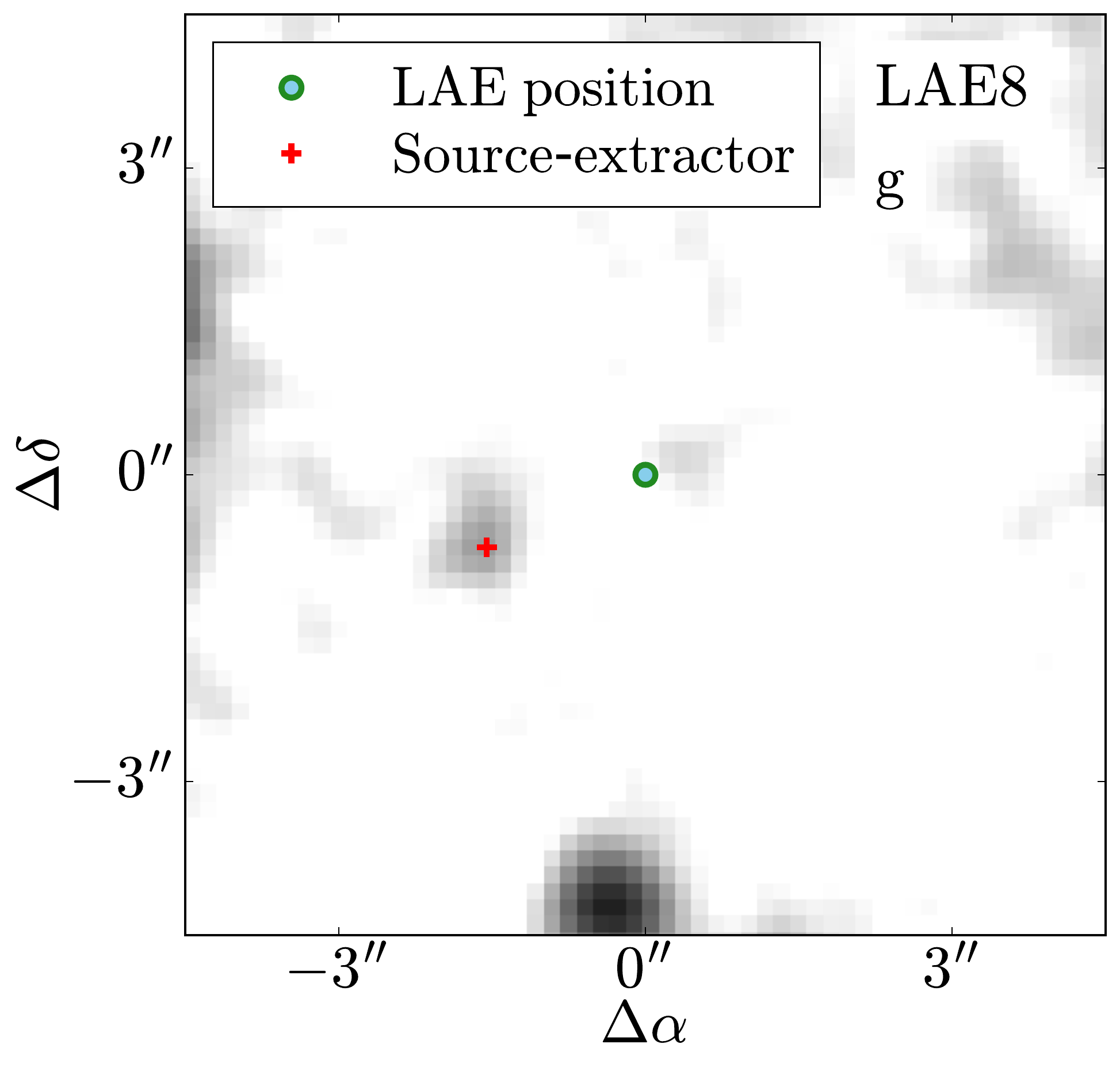}
\includegraphics[width=0.24\textwidth]{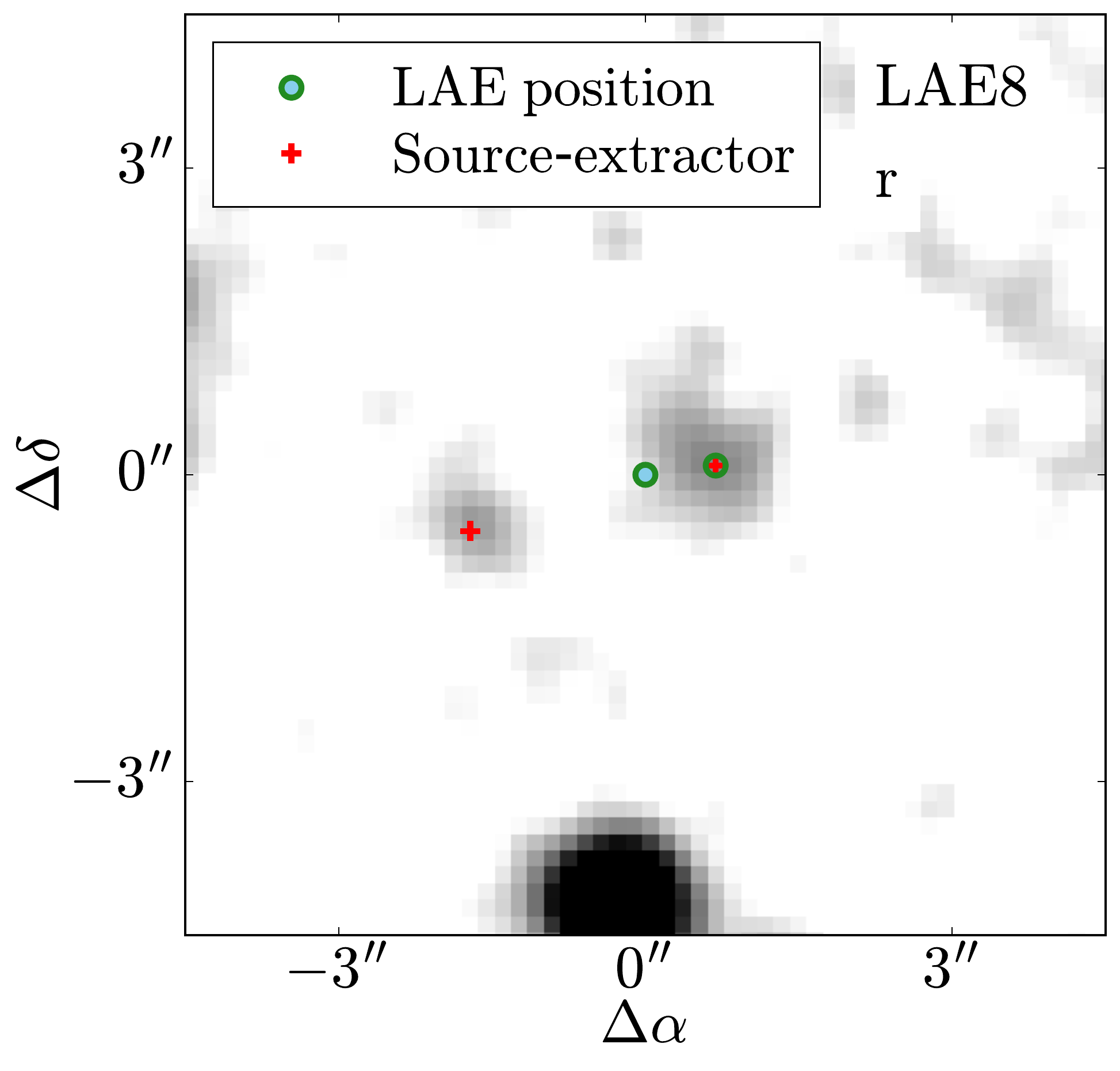}
\includegraphics[width=0.24\textwidth]{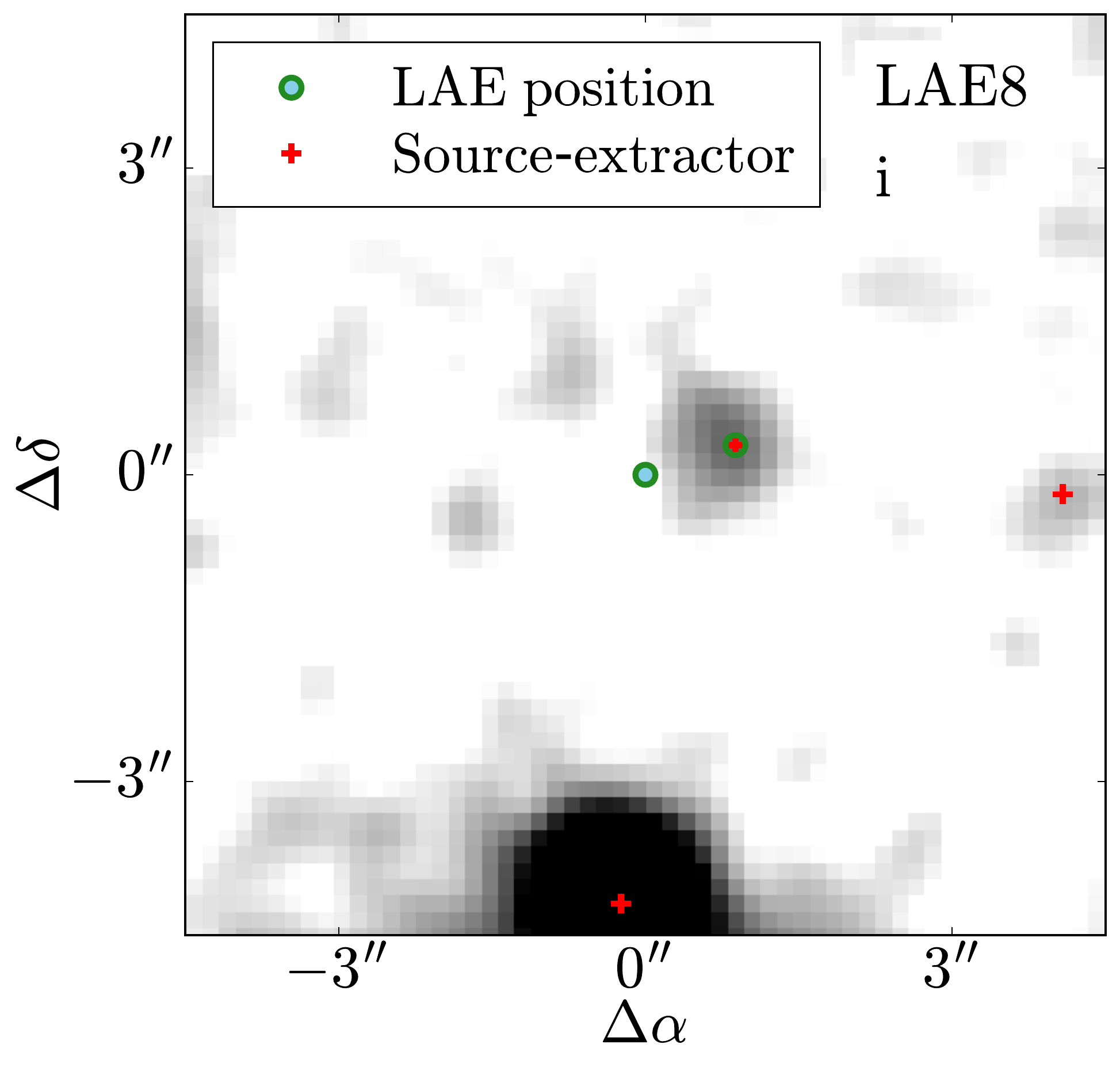}
\includegraphics[width=0.24\textwidth]{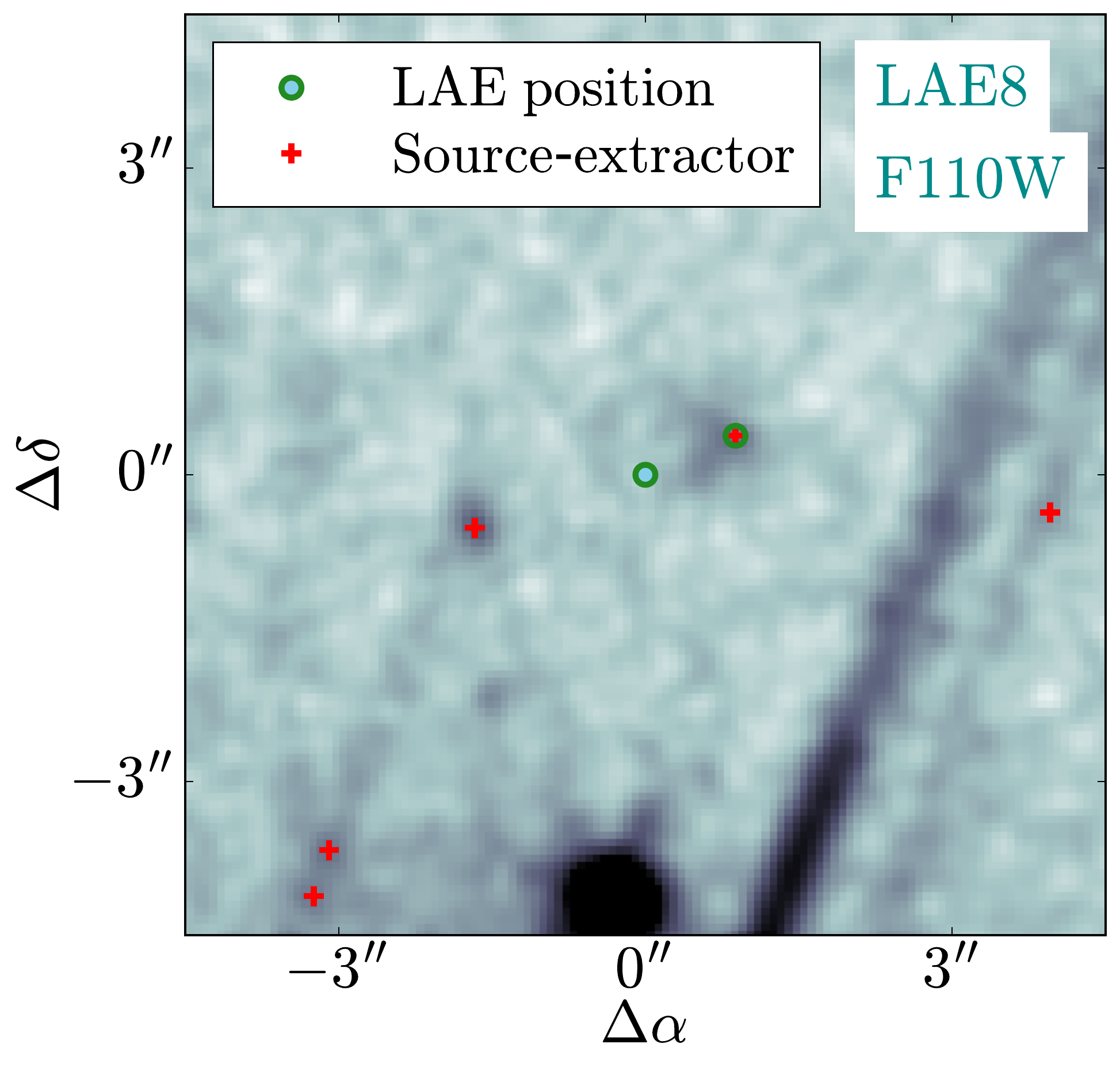}
\includegraphics[width=0.24\textwidth]{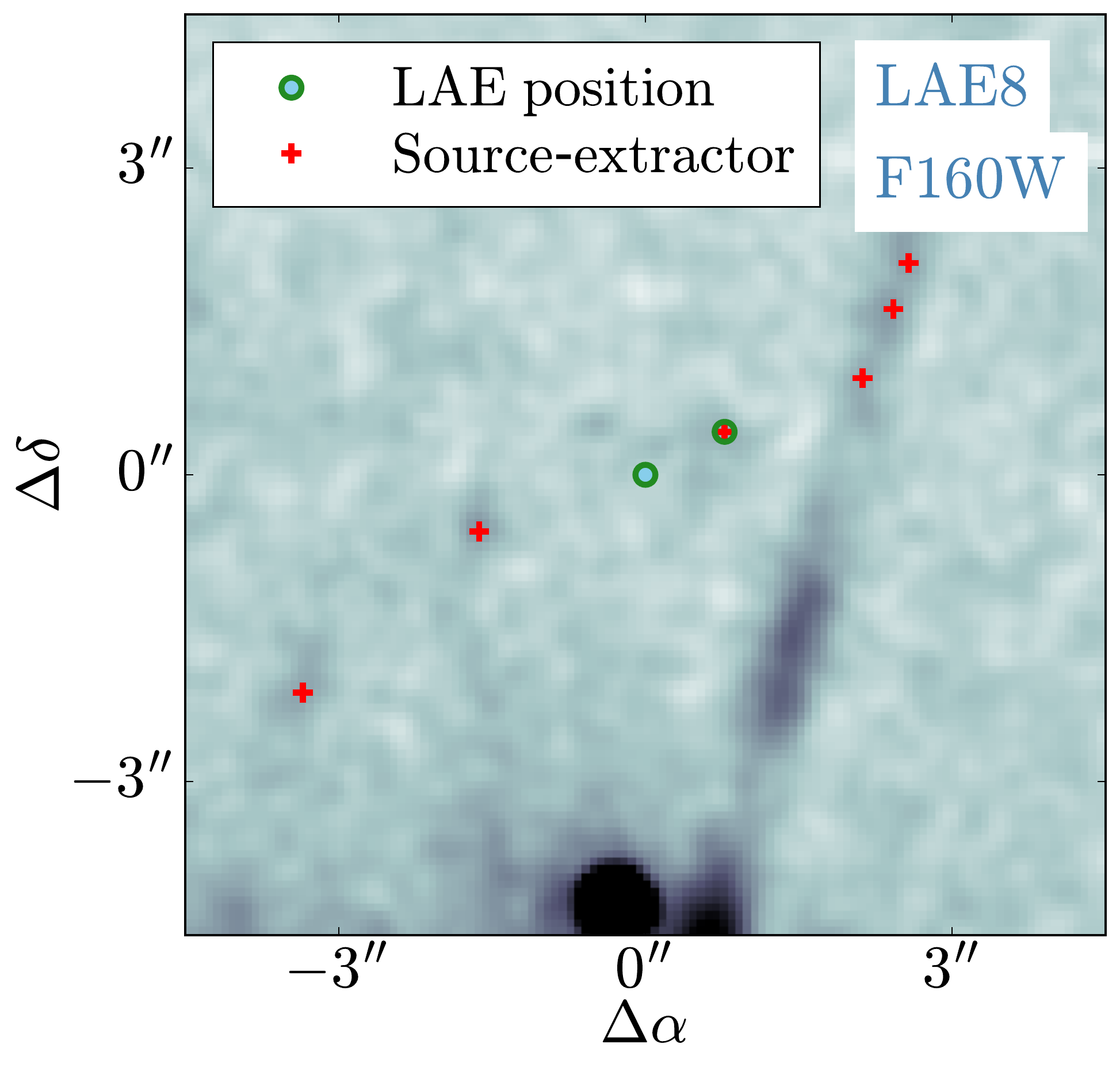}
\includegraphics[width=0.248\textwidth]{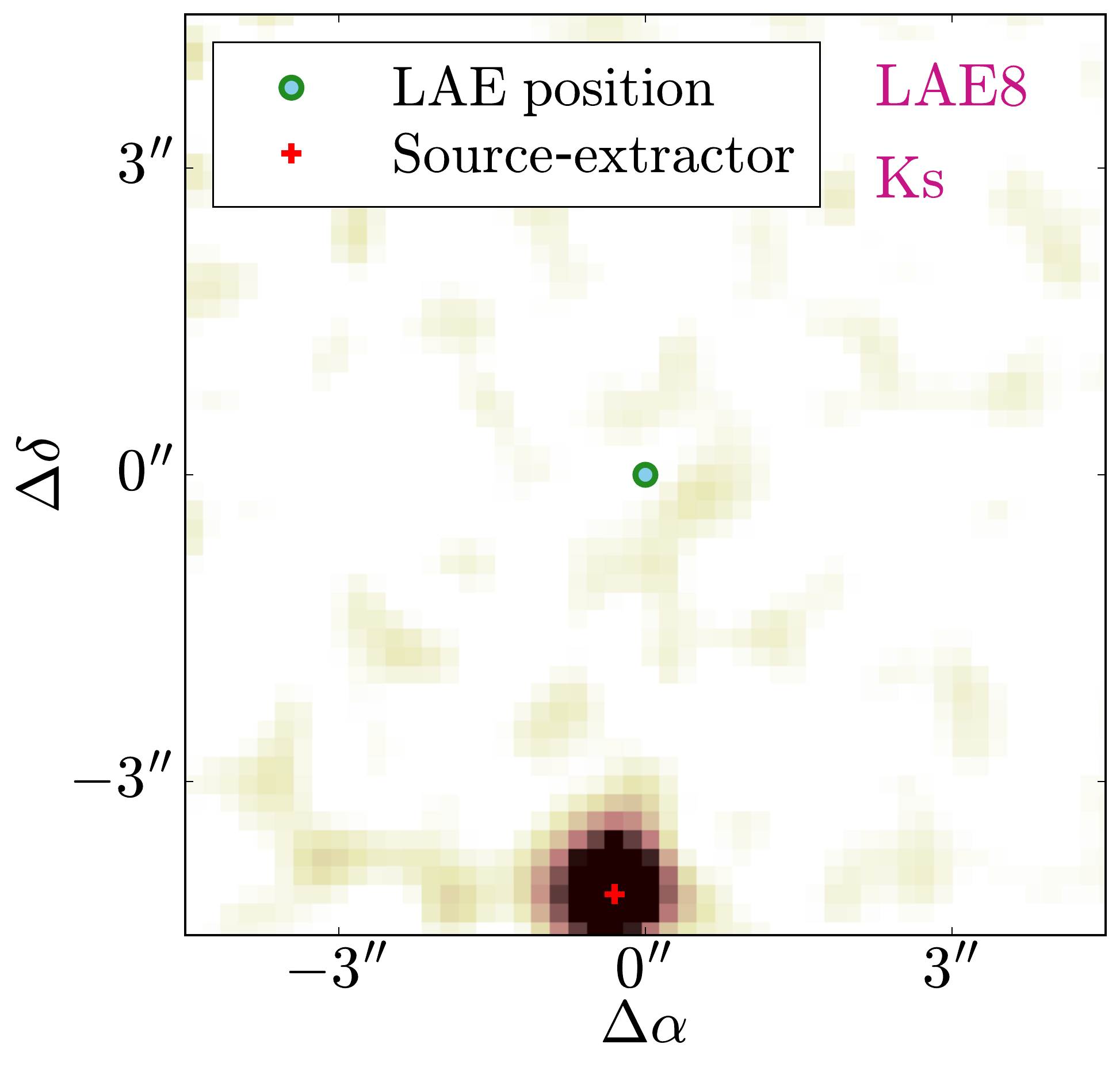}
\includegraphics[width=0.249\textwidth]{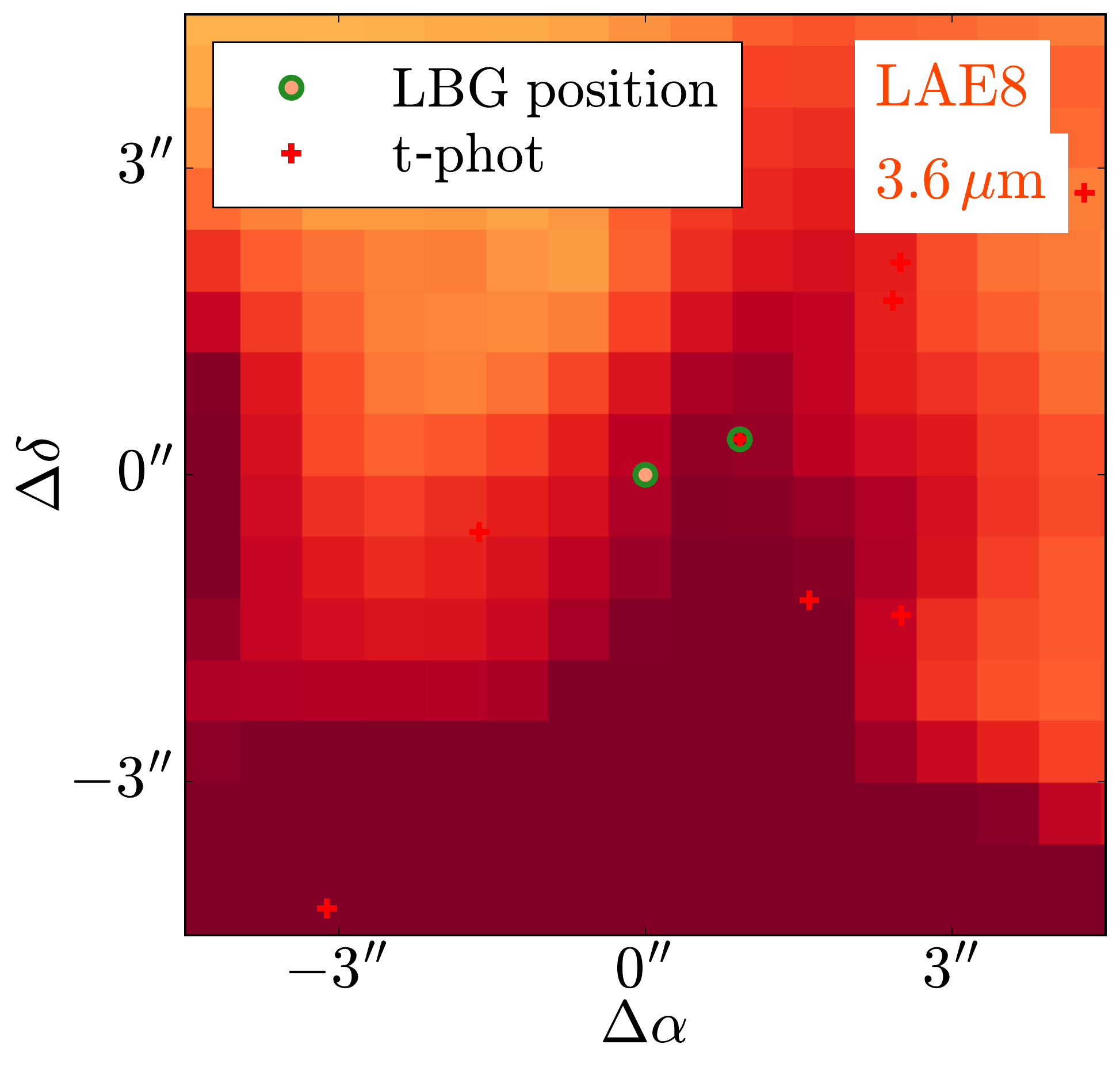}
\includegraphics[width=0.249\textwidth]{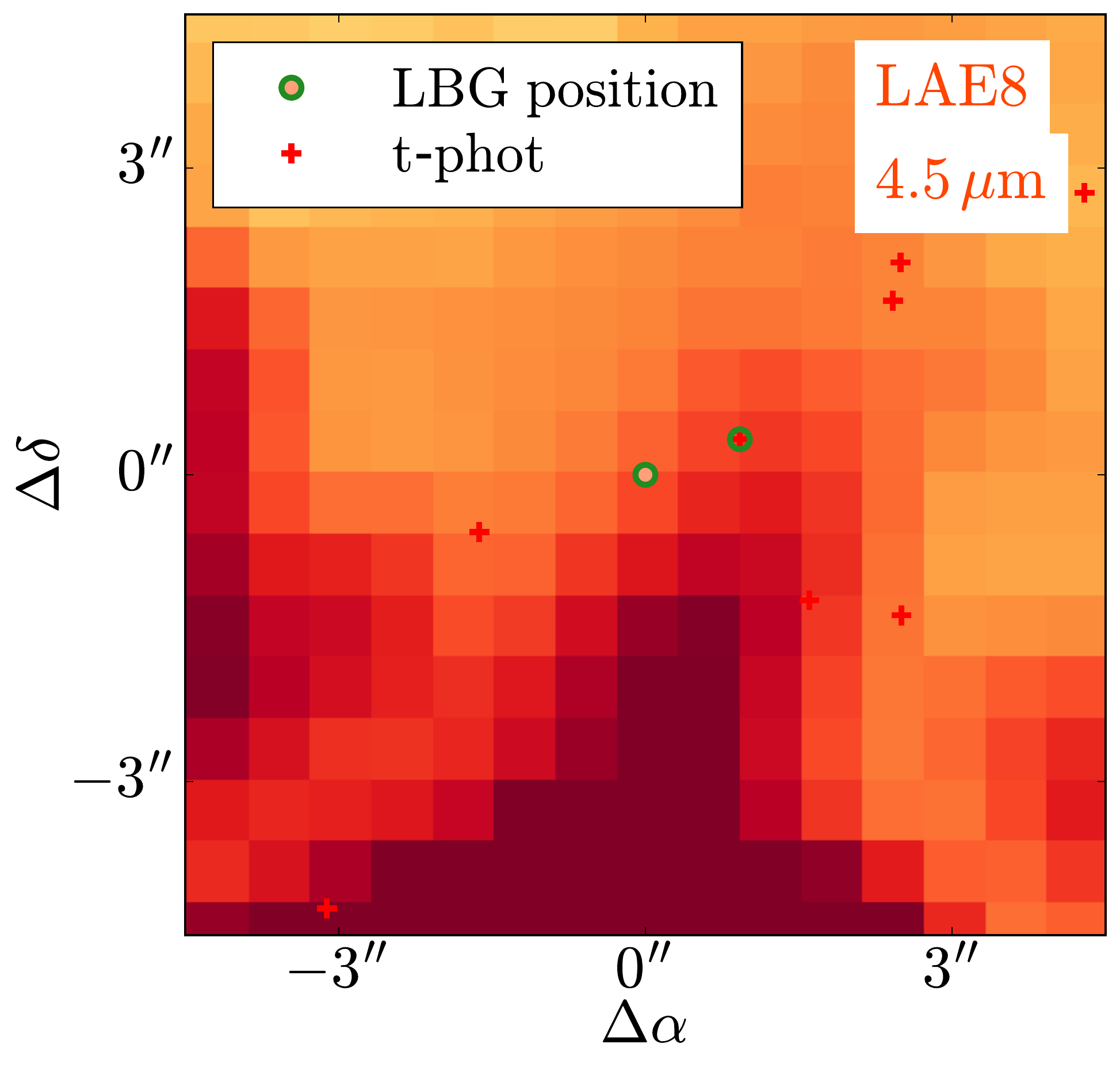}
\end{framed}
\end{subfigure}
\caption{}
\end{figure*}
\renewcommand{\thefigure}{C\arabic{figure}}

\newpage

\section{Best fit SEDs}
\label{appendix2}

\begin{figure*}
\includegraphics[width=0.49\textwidth]{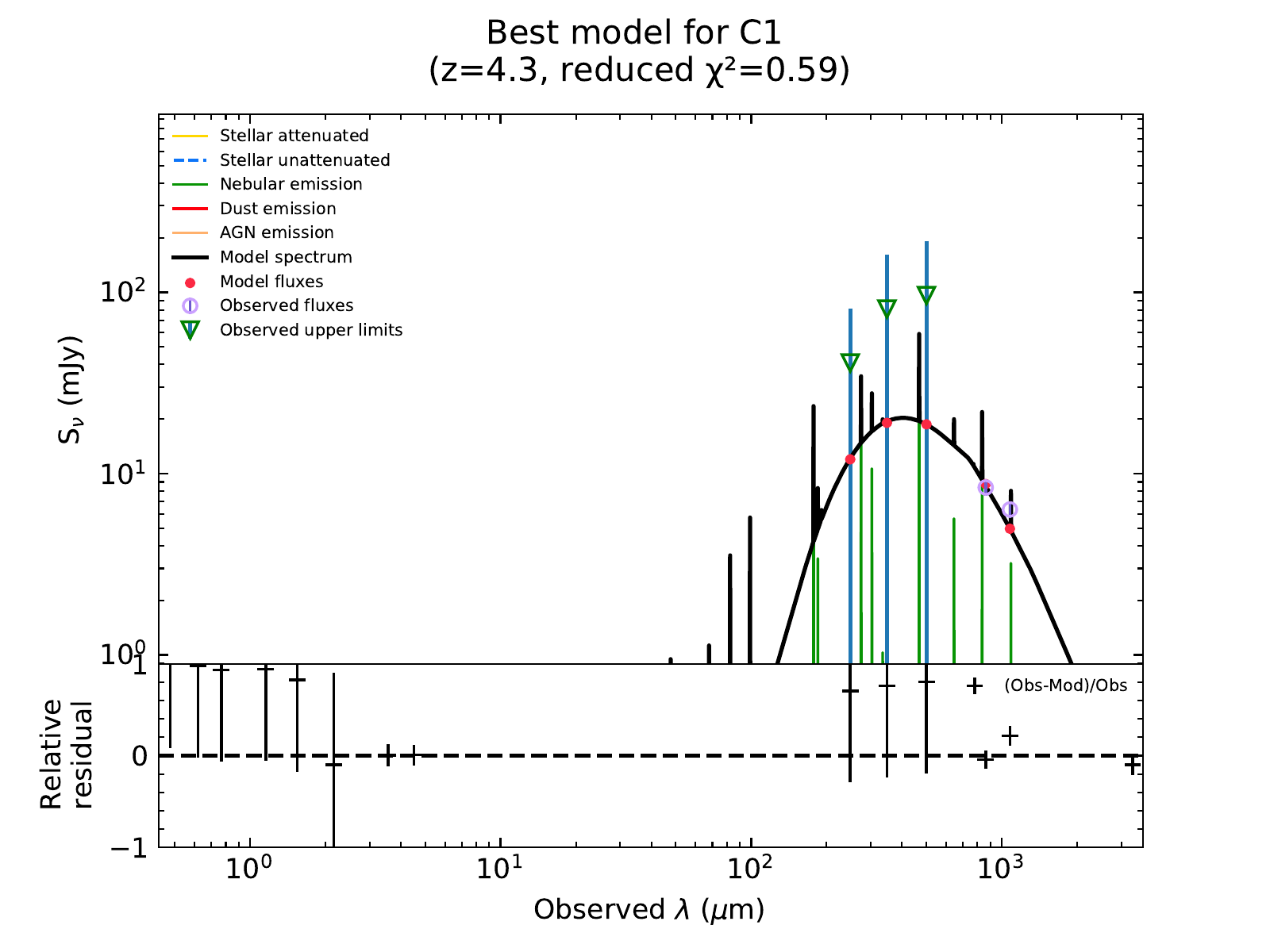}
\includegraphics[width=0.49\textwidth]{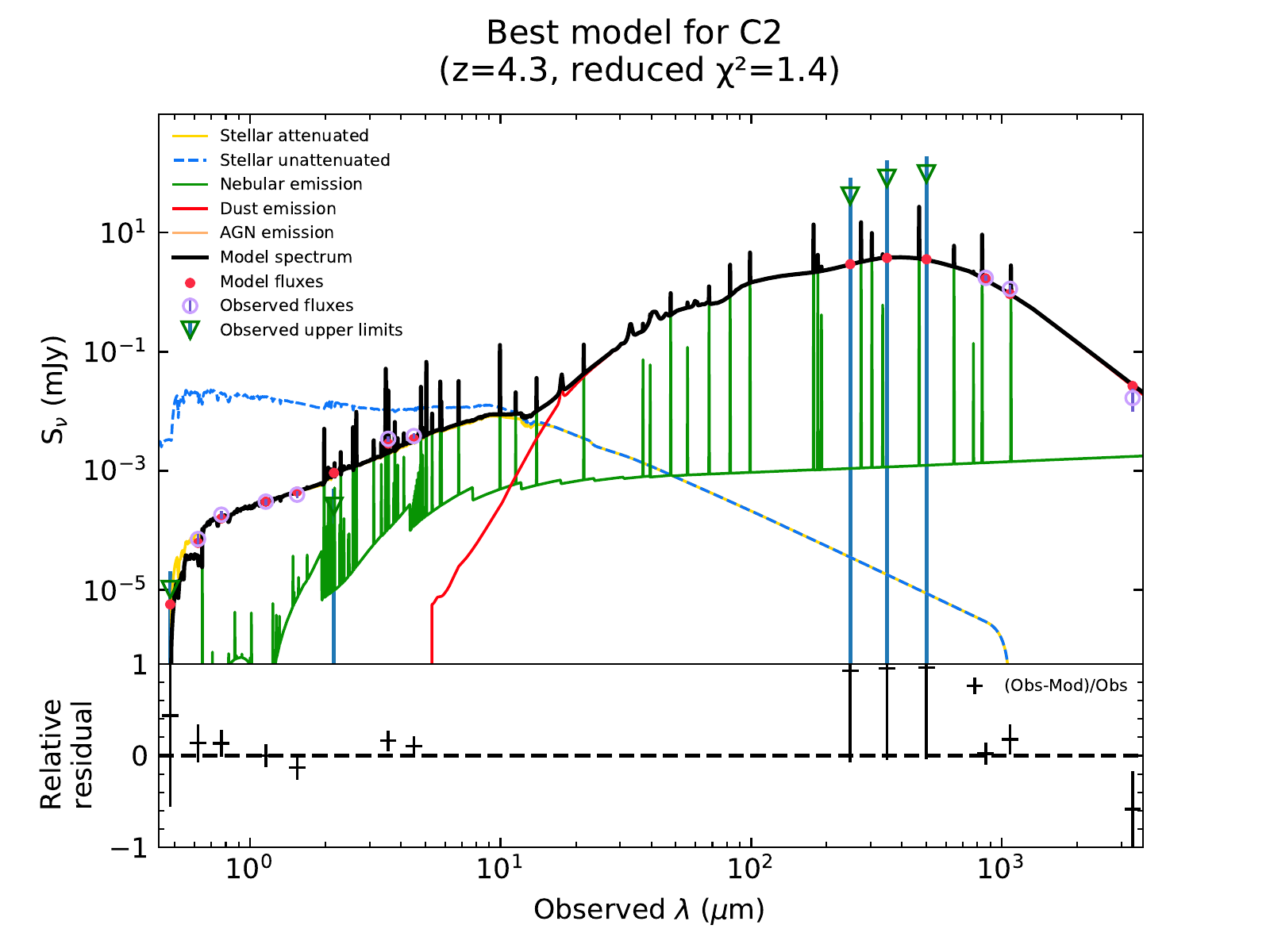}
\includegraphics[width=0.49\textwidth]{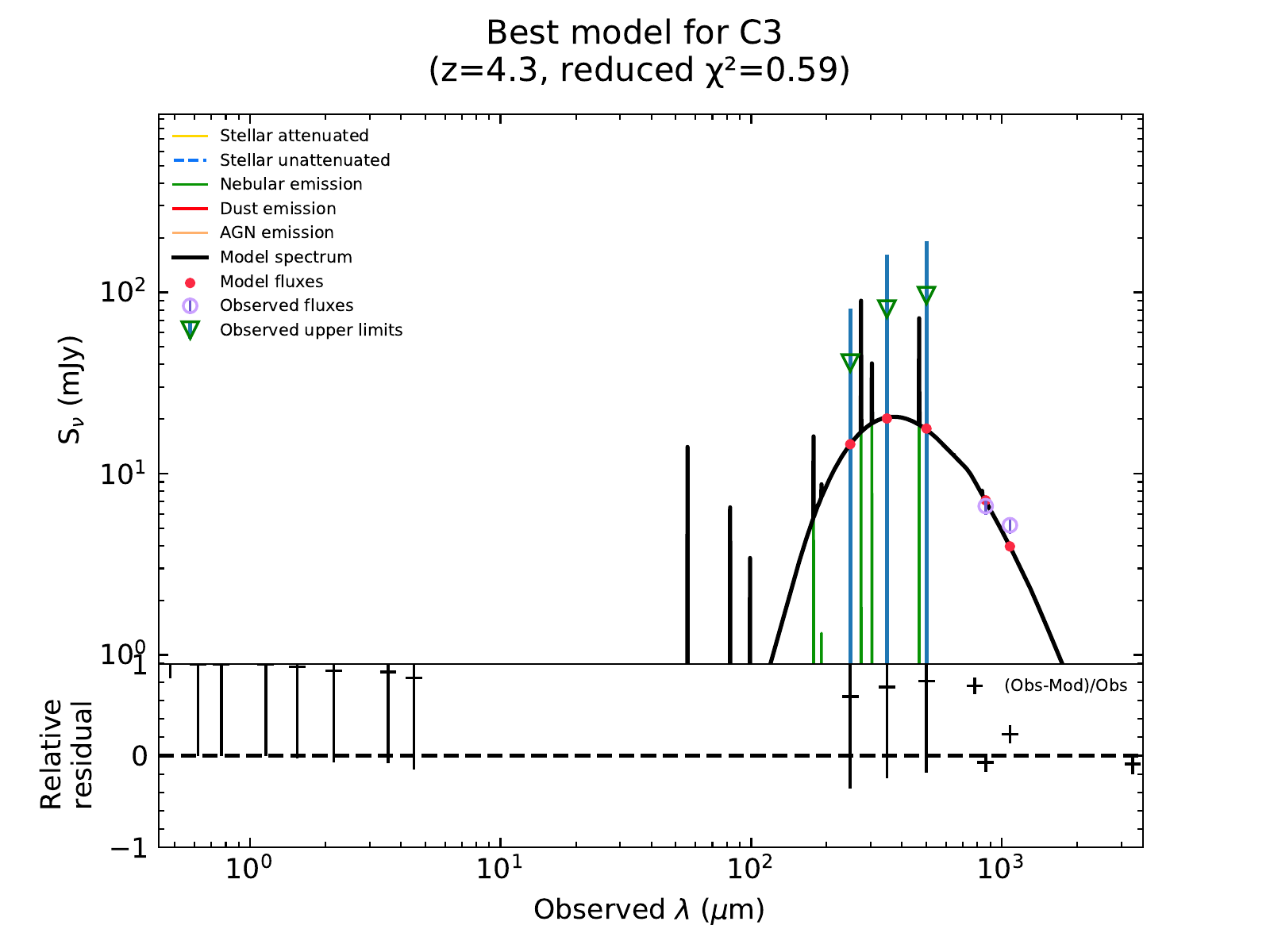}
\includegraphics[width=0.49\textwidth]{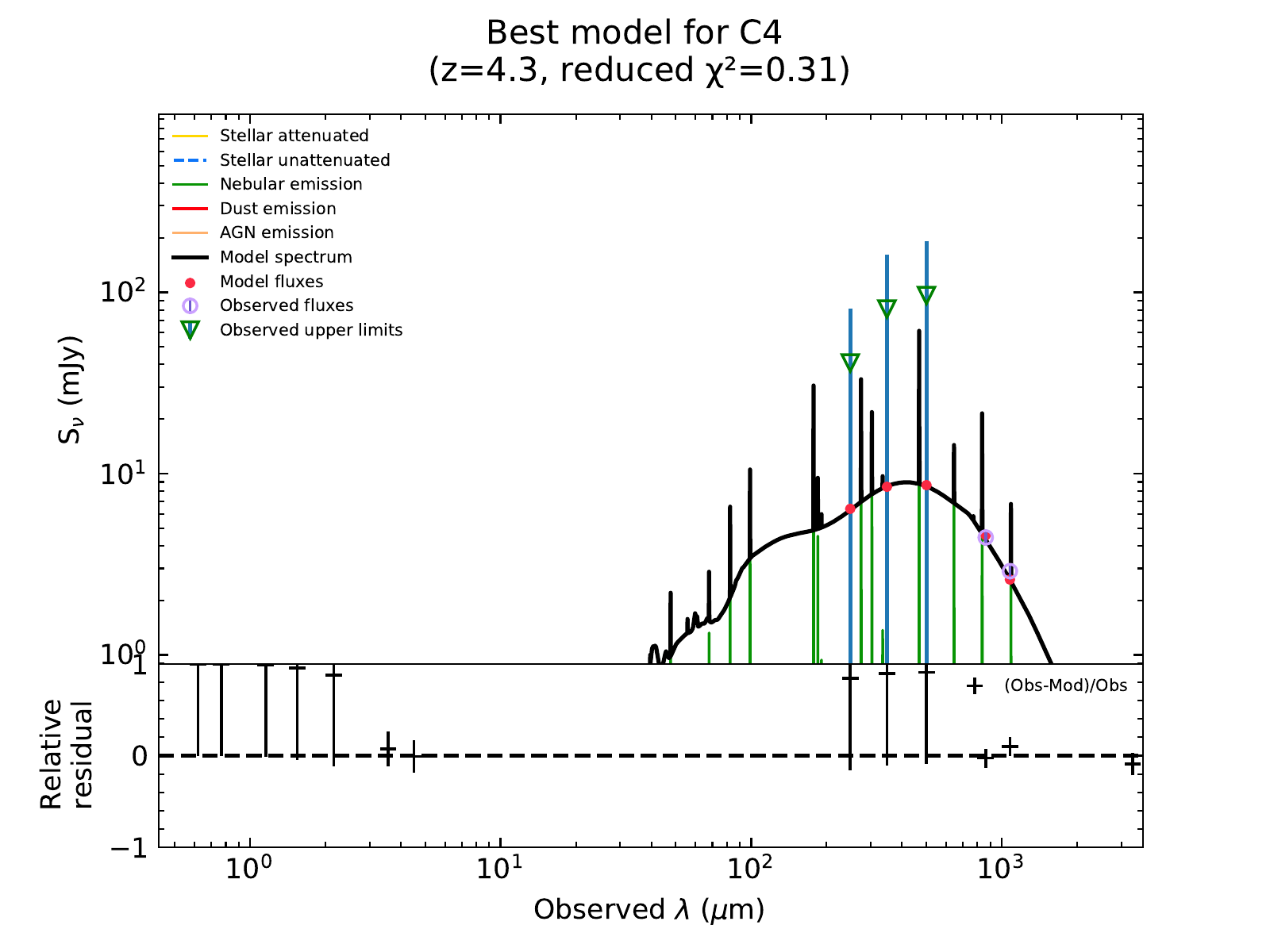}
\includegraphics[width=0.49\textwidth]{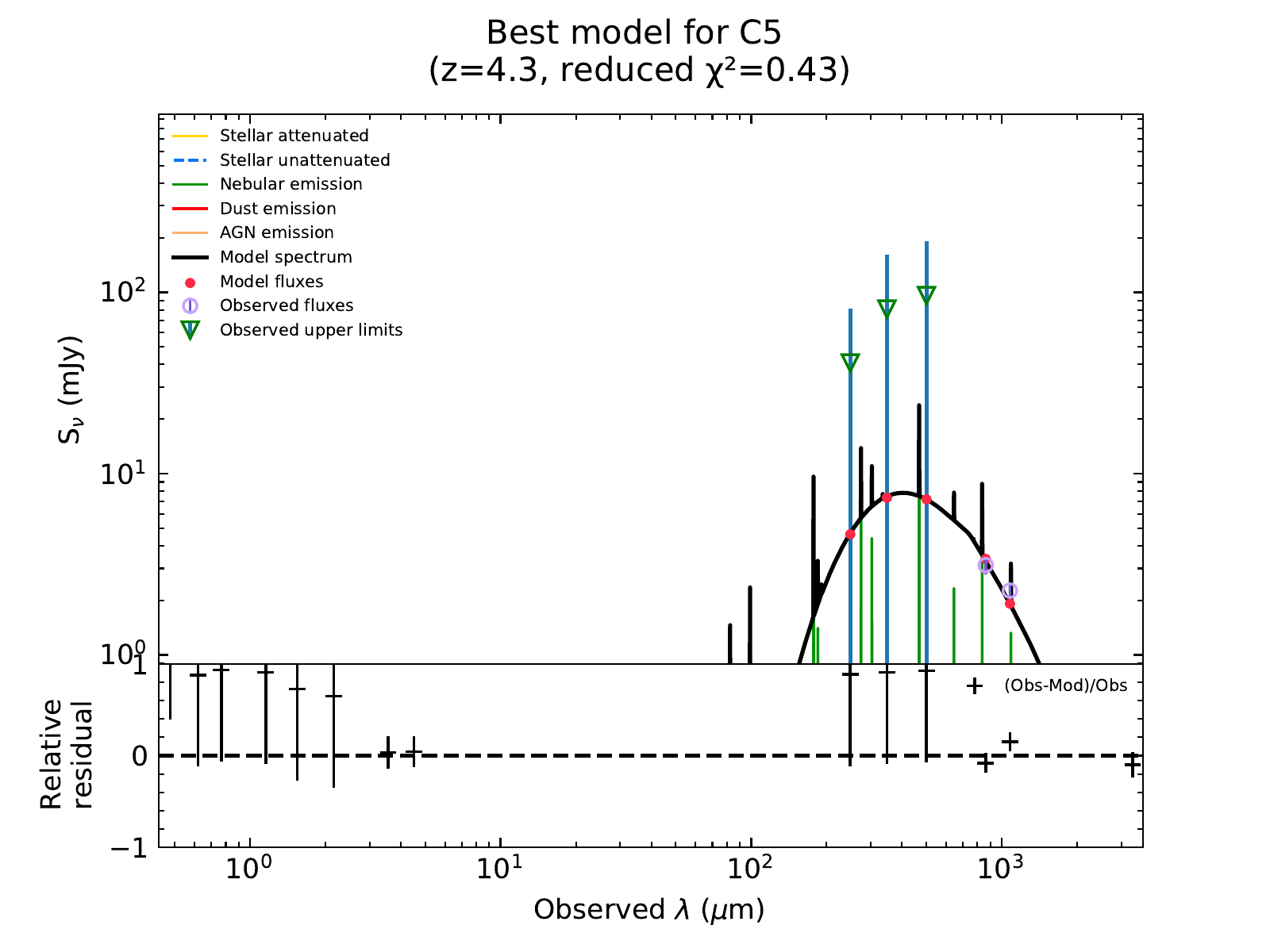}
\includegraphics[width=0.49\textwidth]{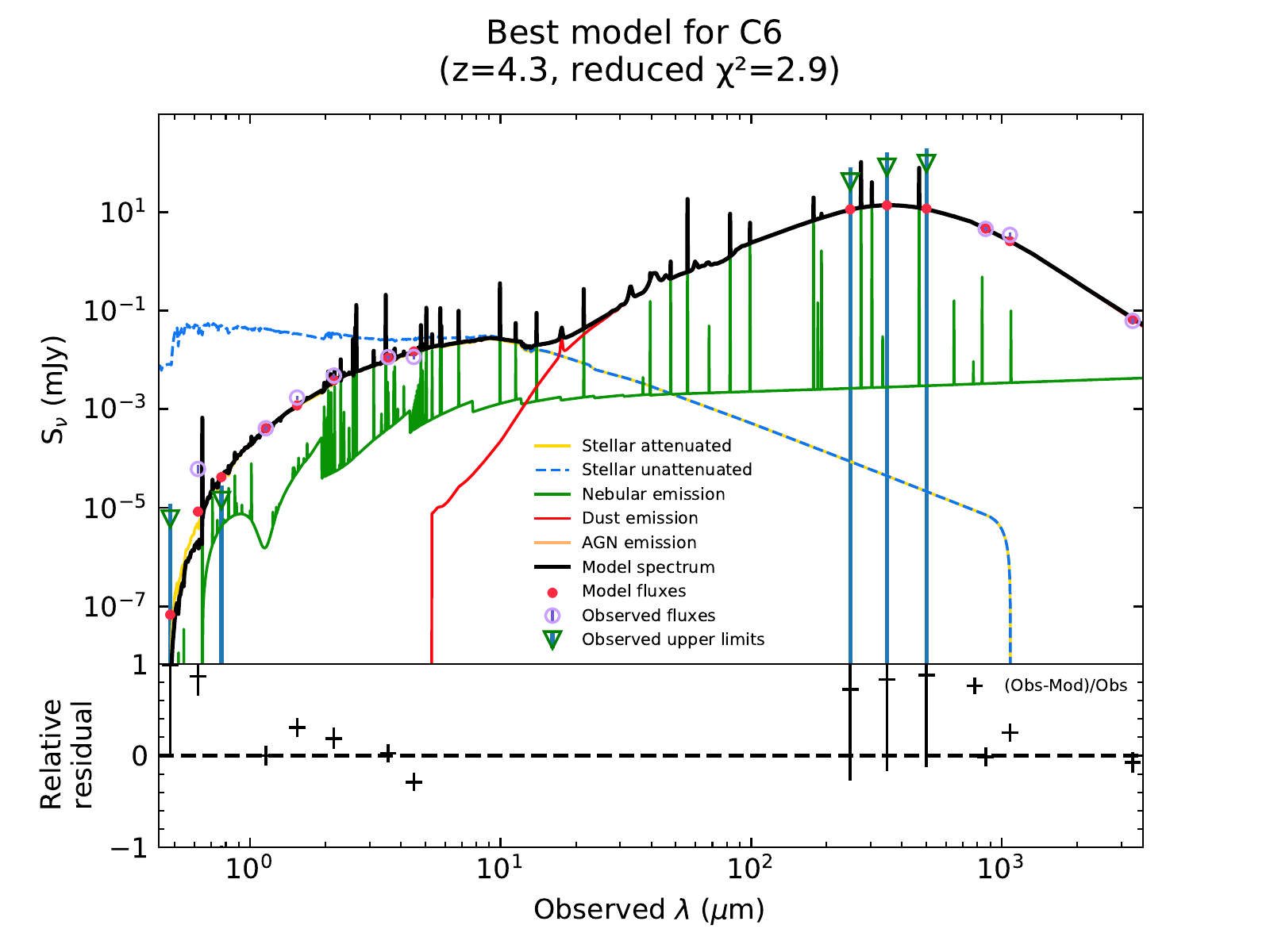}
\caption{Best-fit SEDs from {\tt CIGALE} using the photometry in Table \ref{table:cont} and in \citet{hill2020}. Detected flux densities are shown as circles and flux density upper limits are shown as downward-pointing arrows, while flux densities predicted from the SED model are shown as red circles. The total best-fit SED is the black solid line, plotted along with contributions from stellar emission (unattenuated as the blue dashed line, attenuated as the yellow solid line), dust emission (red solid line), AGN emission (orange solid line), and nebular emission containing various strong emission lines (green solid line). Residuals are plotted at the bottom of each panel.}
\label{seds}
\end{figure*}

In Fig.~\ref{seds} we show best-fit SEDs for all galaxies in our sample, obtained from {\tt CIGALE} using the photometry in Table \ref{table:cont} and in \citet{hill2020}. Detected flux densities are shown as circles, and flux density upper limits are shown as downward-pointing arrows. In these fits we allowed the stellar mass, dust extinction, age, and star-formation timescale to vary. The best-fit parameters are provided in Table \ref{table:cigale}. For galaxies C12, C16, C20, C22, C23, LAE1, and LAE2, only upper limits are available for their photometry across all wavelengths (both submm and optical/infrared), so we do not attempt to fit SEDs or derive stellar masses.

\renewcommand{\thefigure}{C\arabic{figure} (Cont.)}
\addtocounter{figure}{-1}
\begin{figure*}
\includegraphics[width=0.49\textwidth]{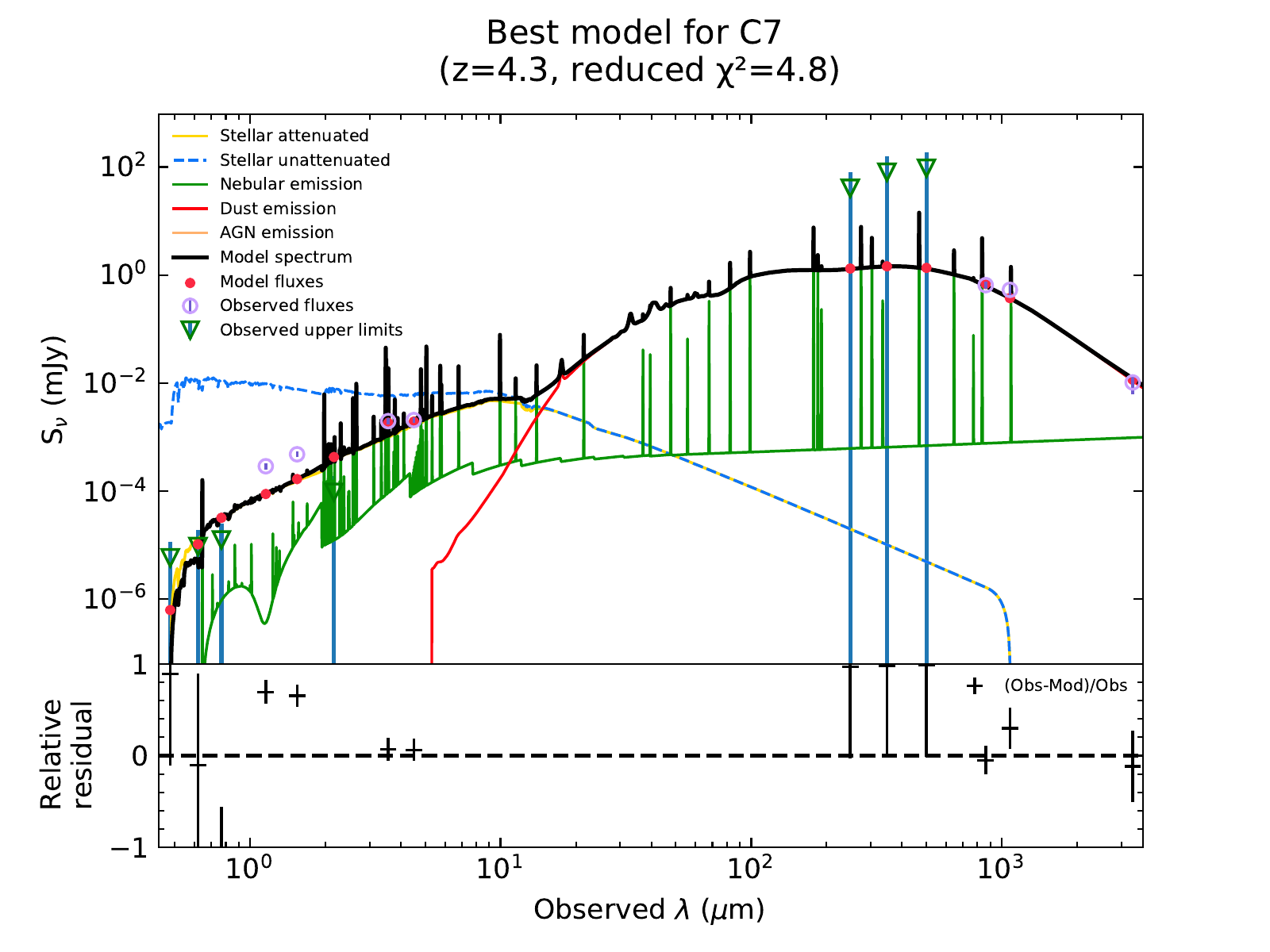}
\includegraphics[width=0.49\textwidth]{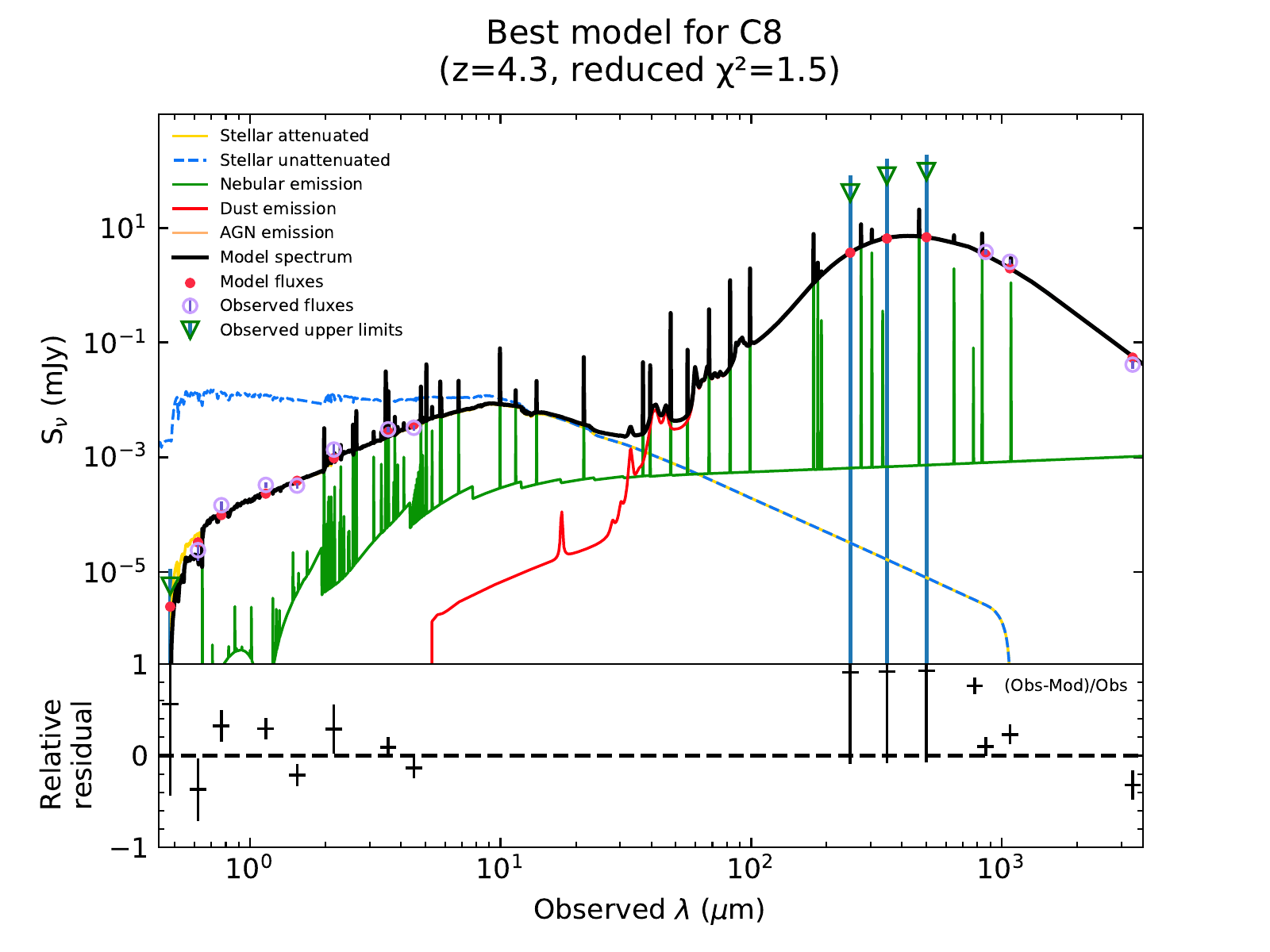}
\includegraphics[width=0.49\textwidth]{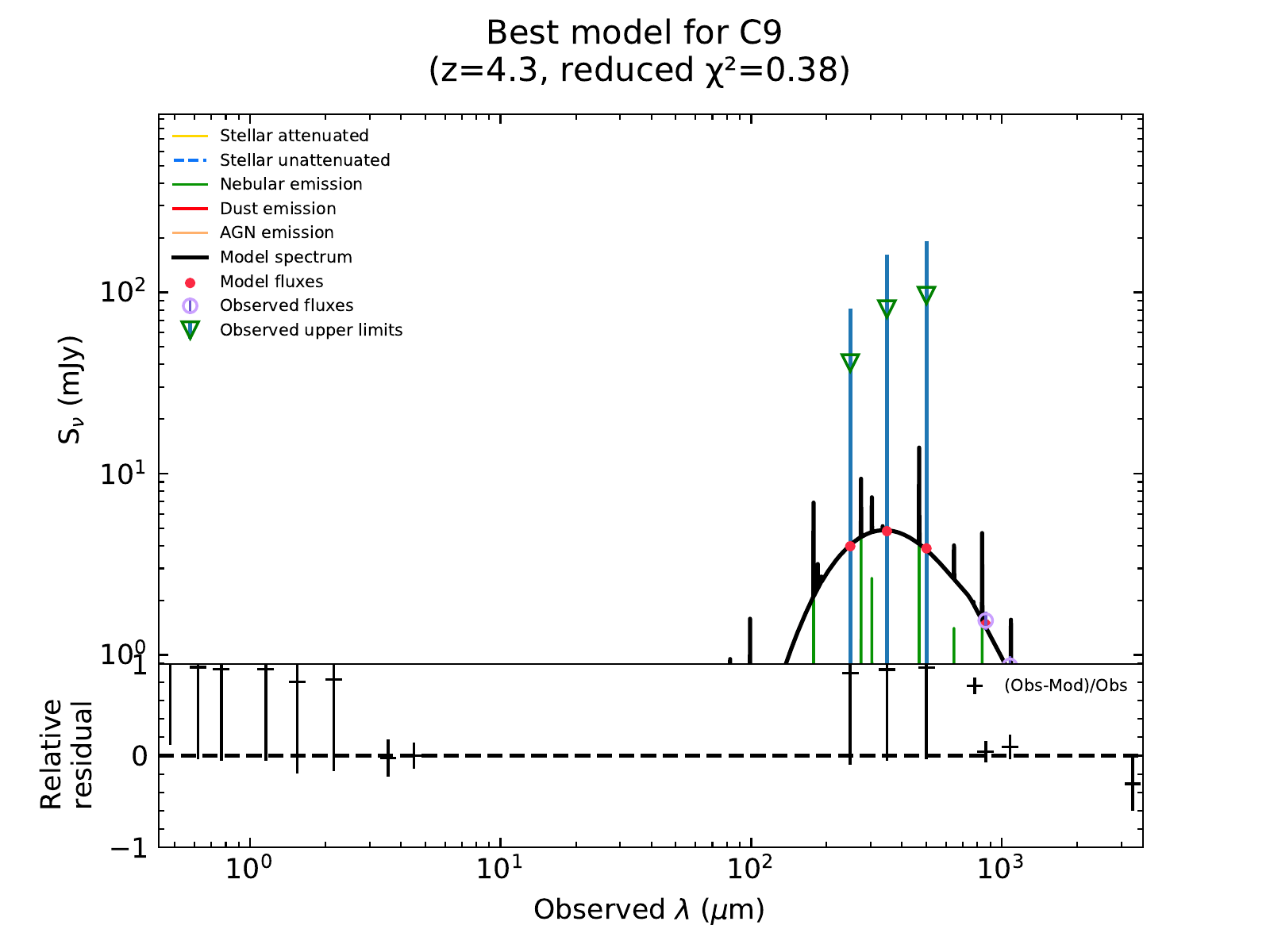}
\includegraphics[width=0.49\textwidth]{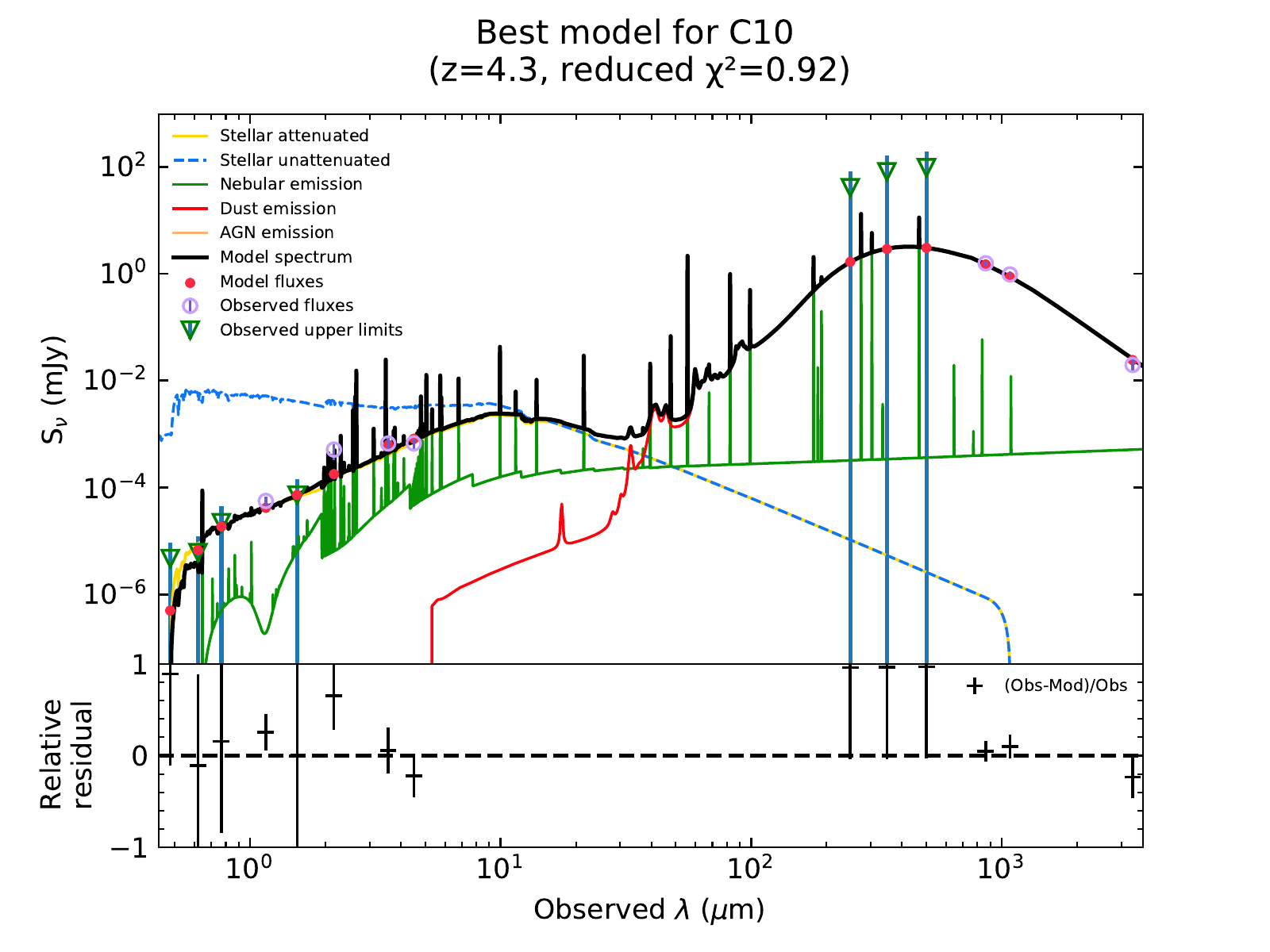}
\includegraphics[width=0.49\textwidth]{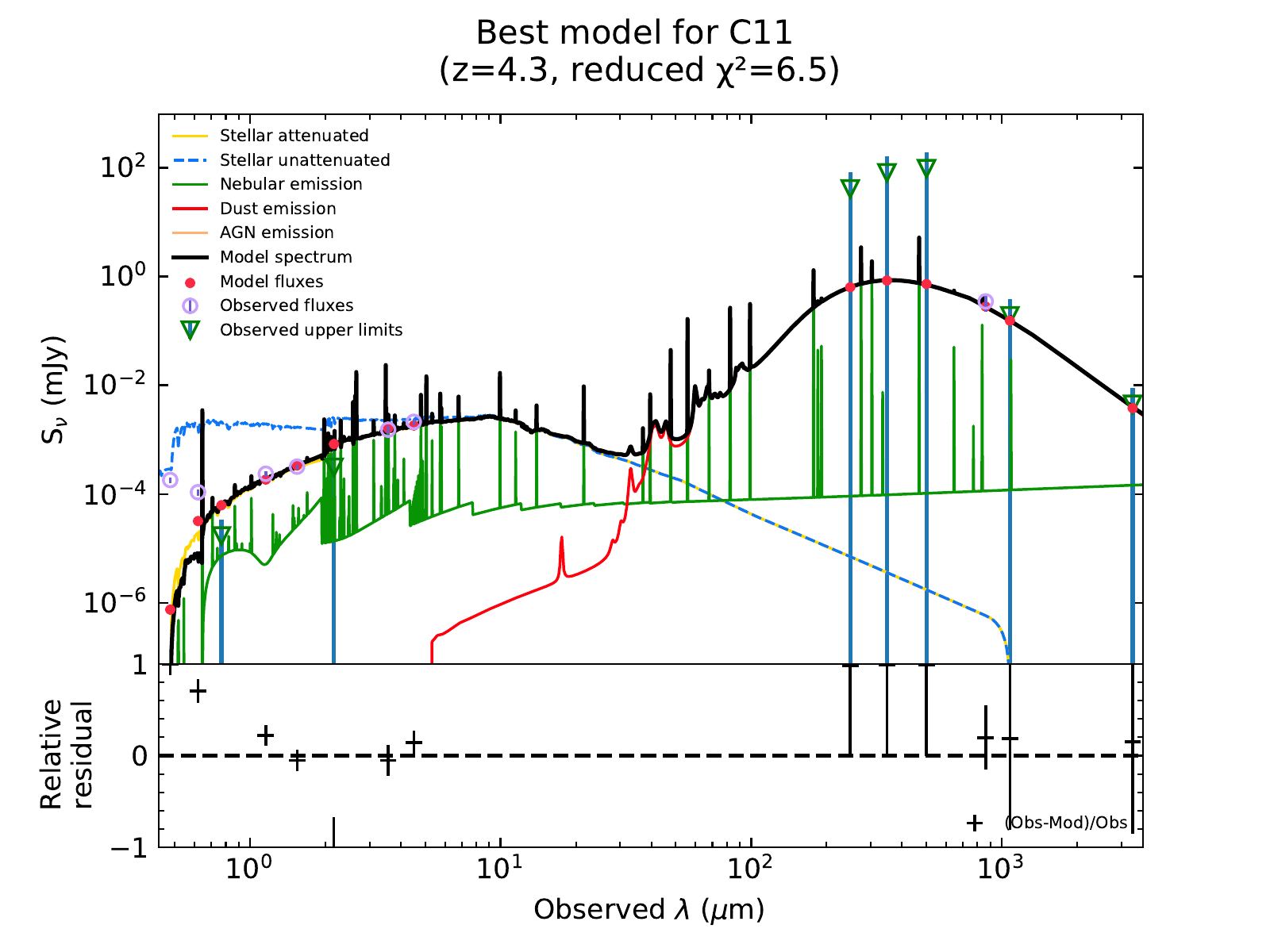}
\includegraphics[width=0.49\textwidth]{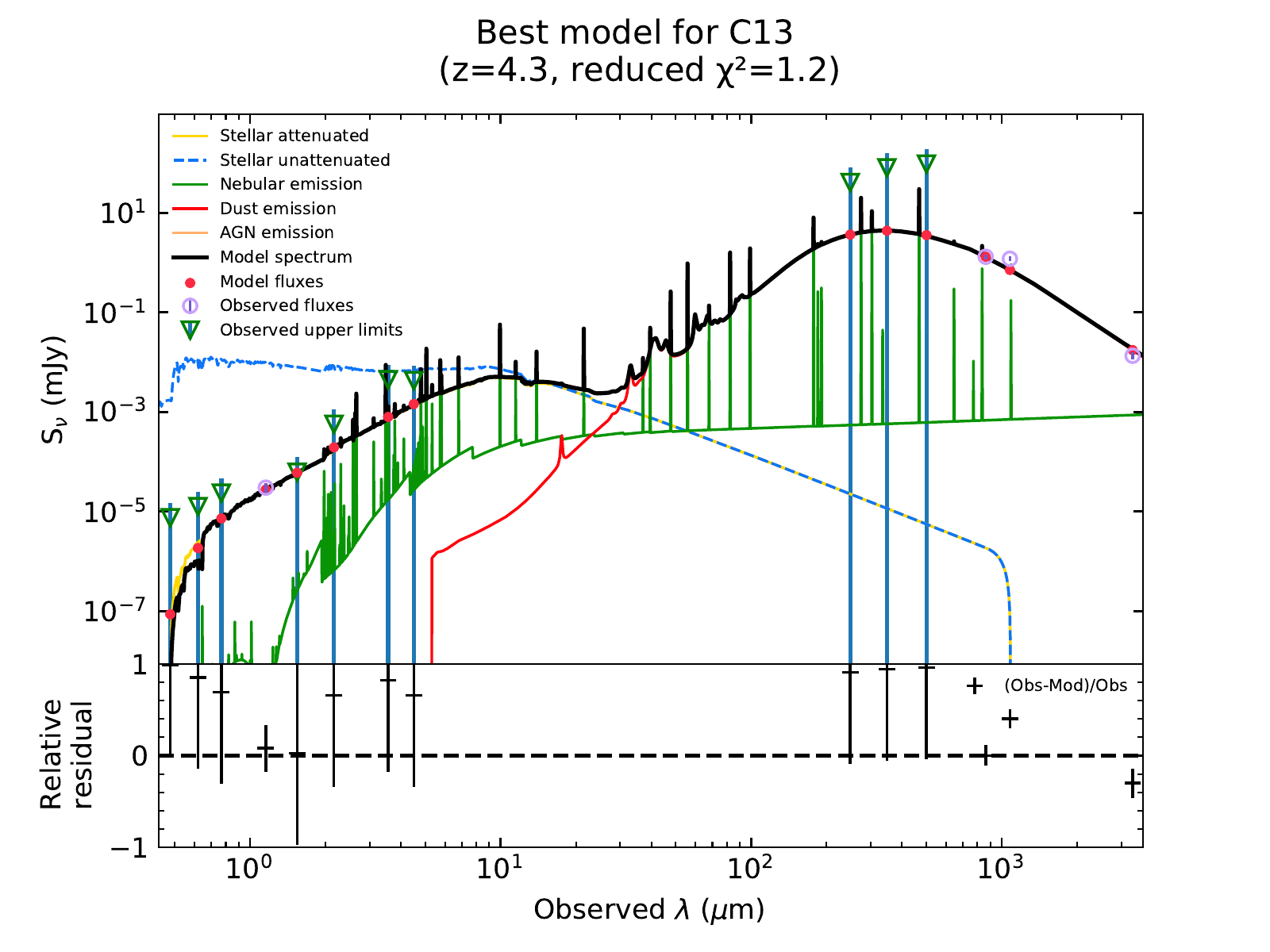}
\caption{}
\end{figure*}
\renewcommand{\thefigure}{\arabic{figure}}

\renewcommand{\thefigure}{C\arabic{figure} (Cont.)}
\addtocounter{figure}{-1}
\begin{figure*}
\includegraphics[width=0.49\textwidth]{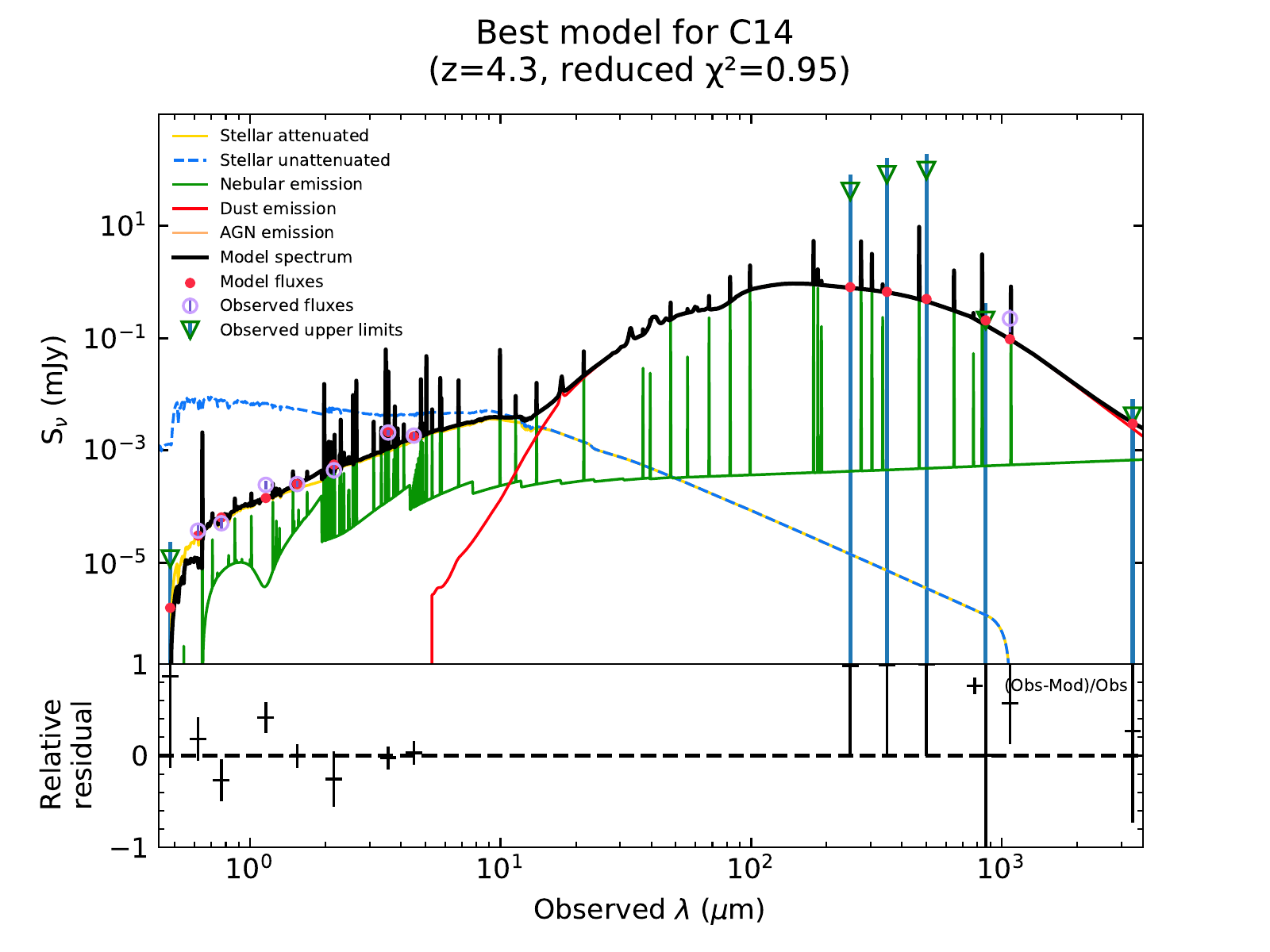}
\includegraphics[width=0.49\textwidth]{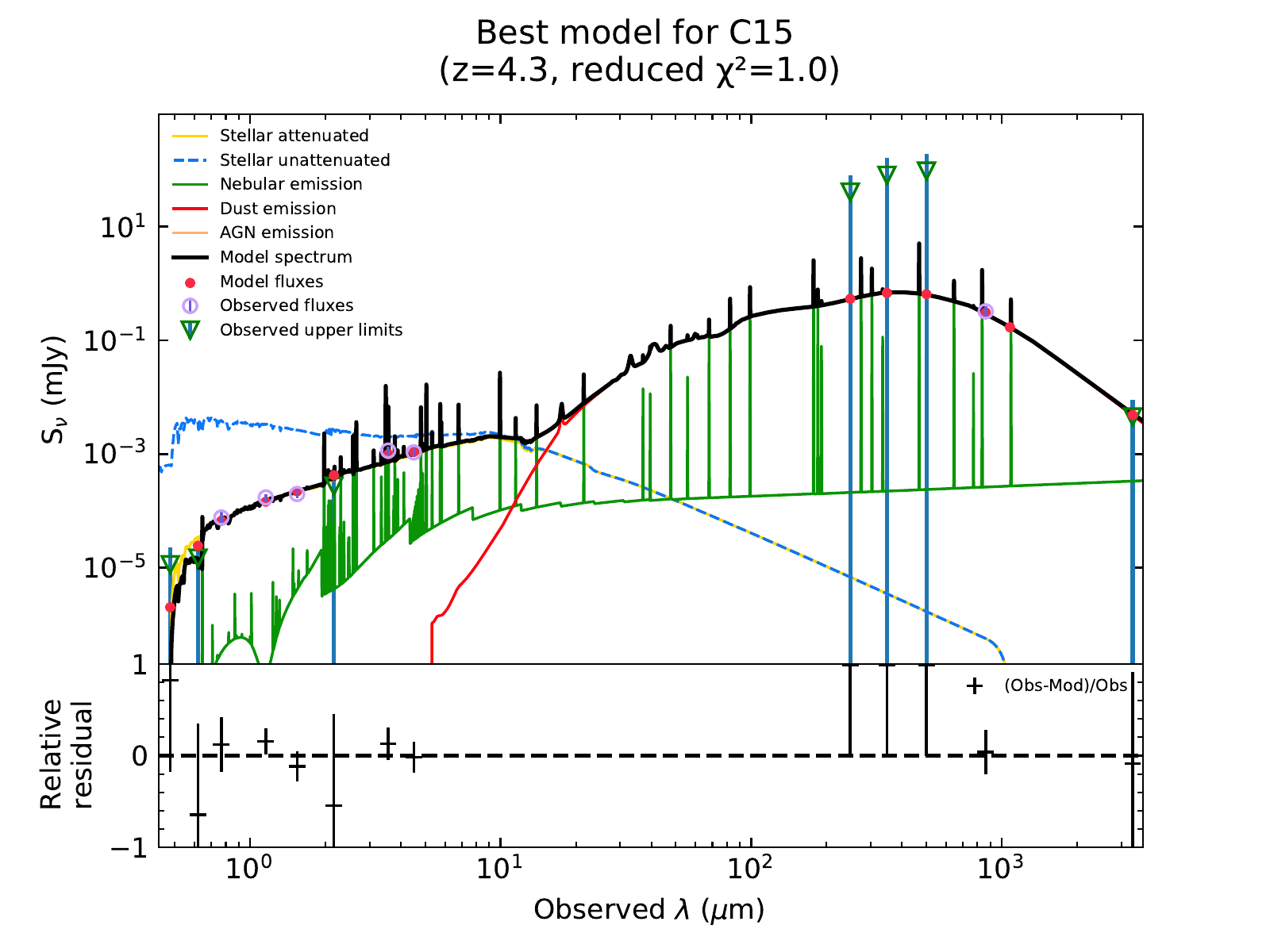}
\includegraphics[width=0.49\textwidth]{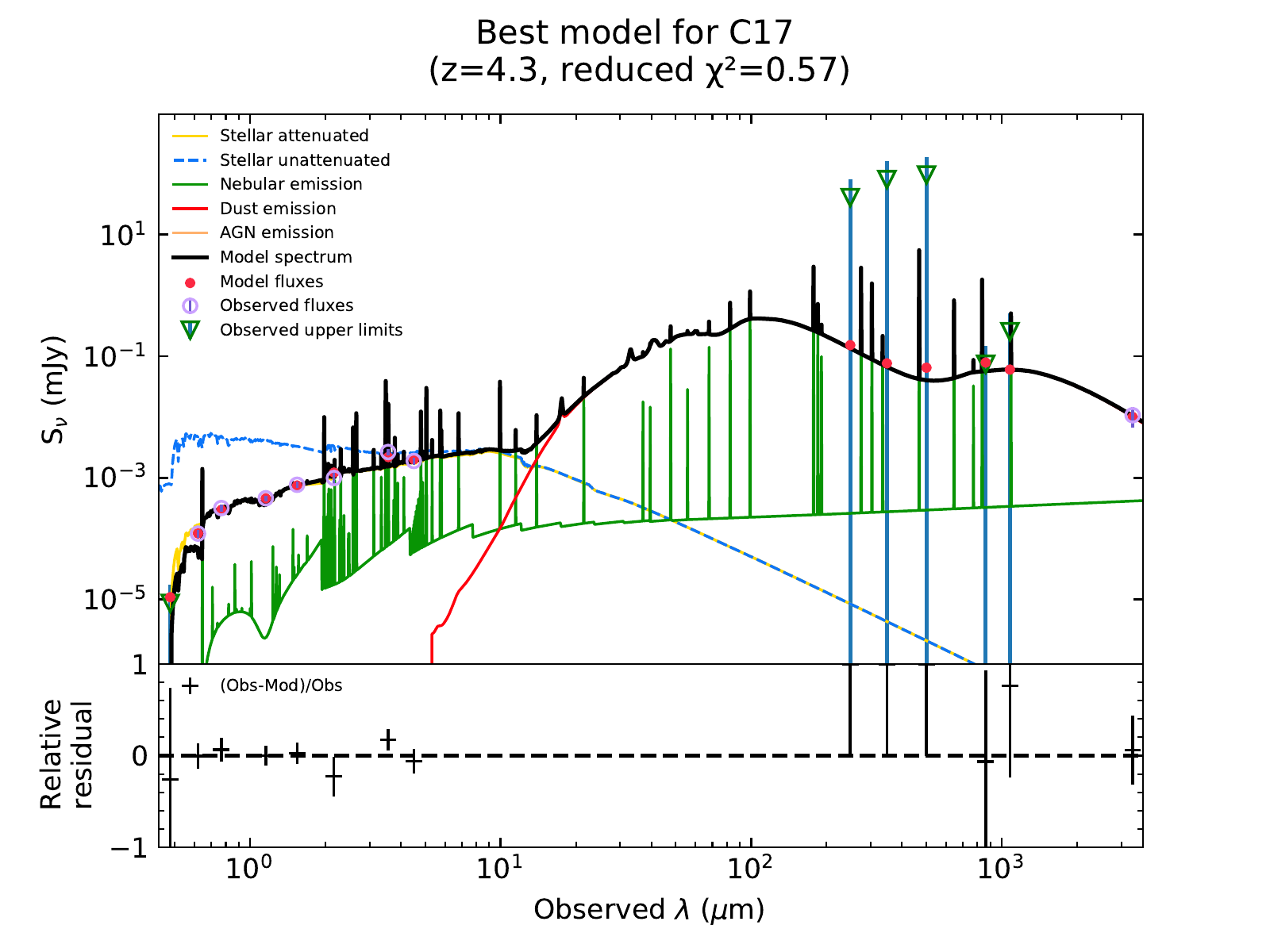}
\includegraphics[width=0.49\textwidth]{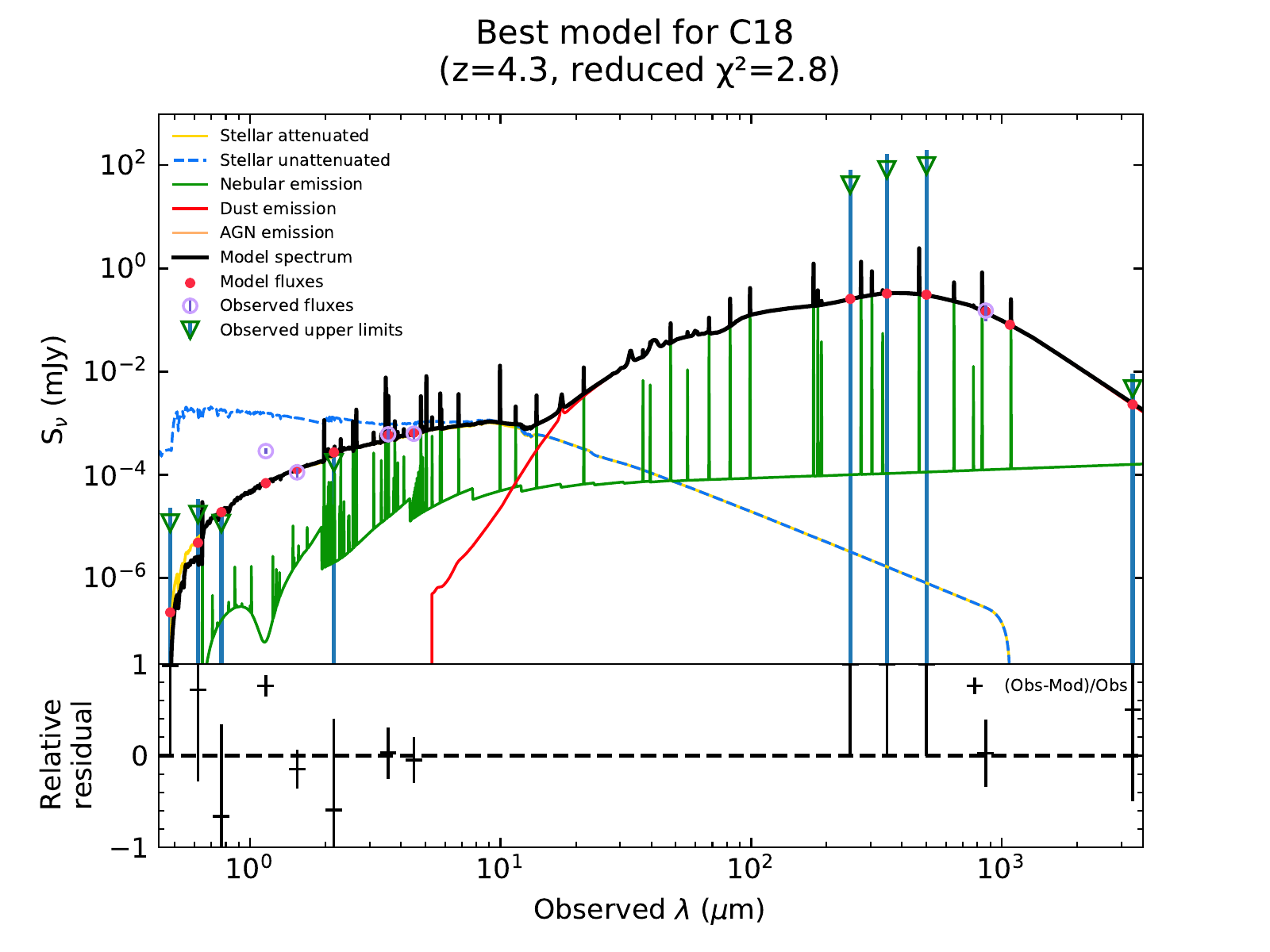}
\includegraphics[width=0.49\textwidth]{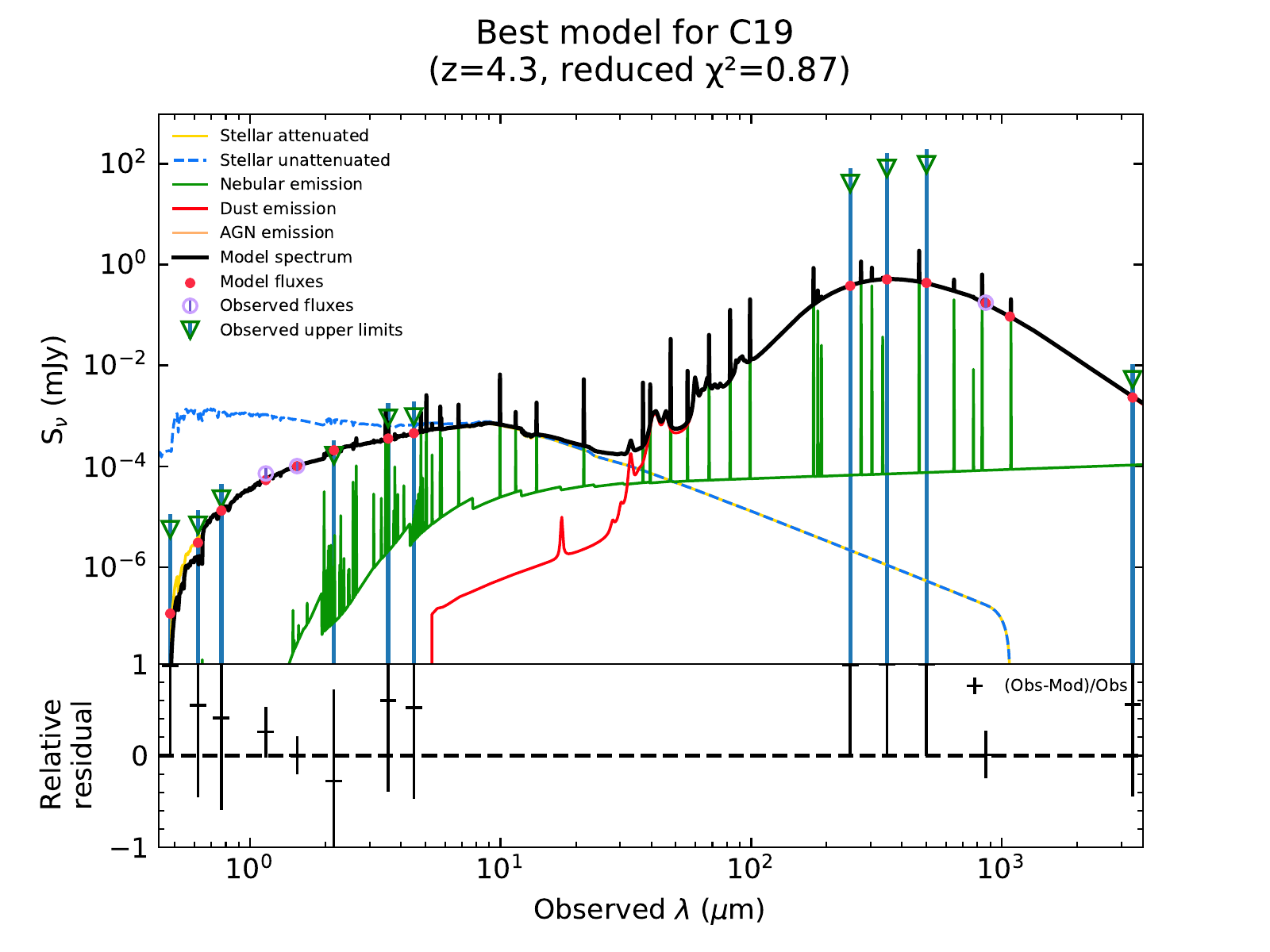}
\includegraphics[width=0.49\textwidth]{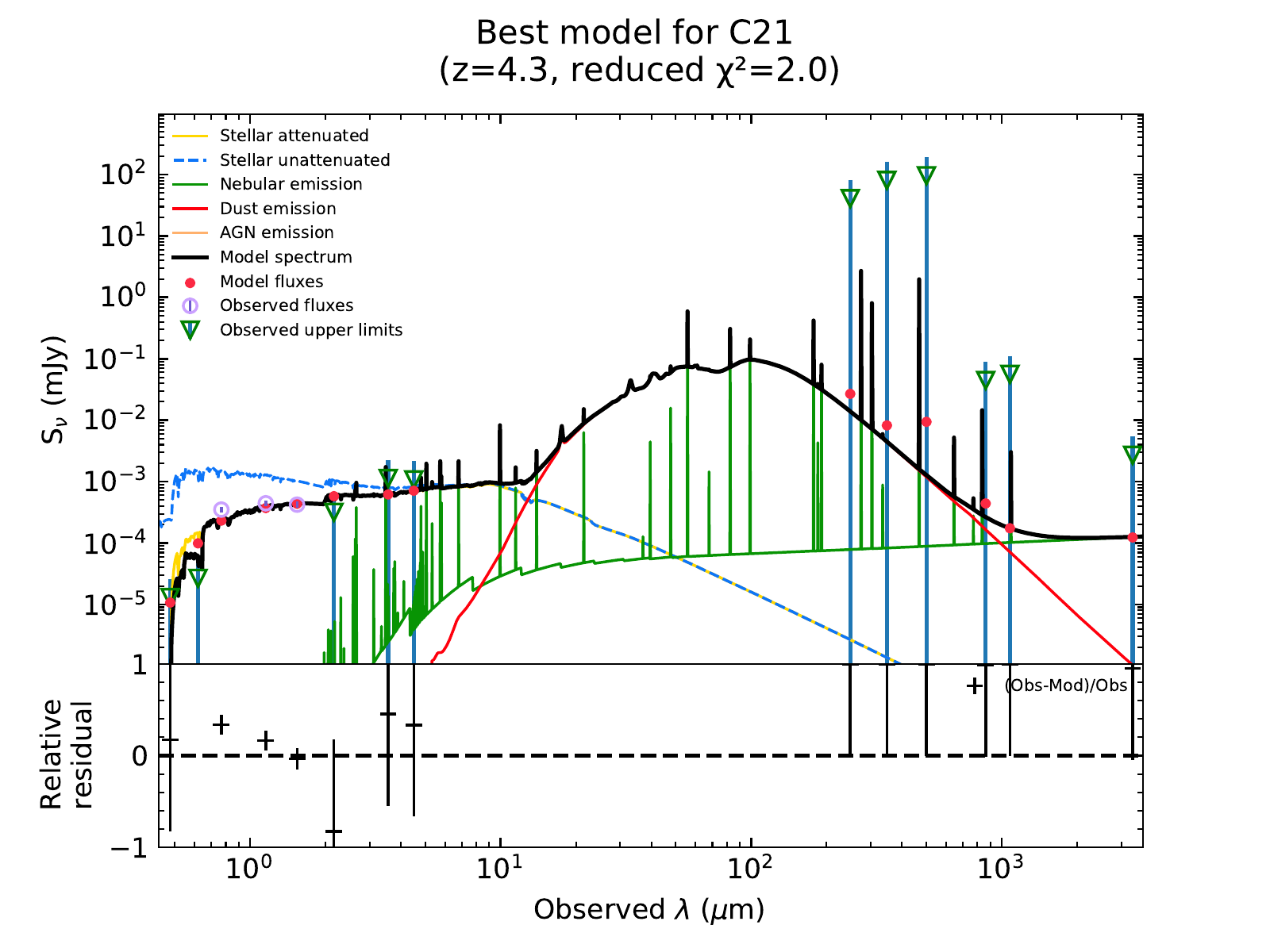}
\caption{}
\end{figure*}
\renewcommand{\thefigure}{\arabic{figure}}

\renewcommand{\thefigure}{C\arabic{figure} (Cont.)}
\addtocounter{figure}{-1}
\begin{figure*}
\includegraphics[width=0.49\textwidth]{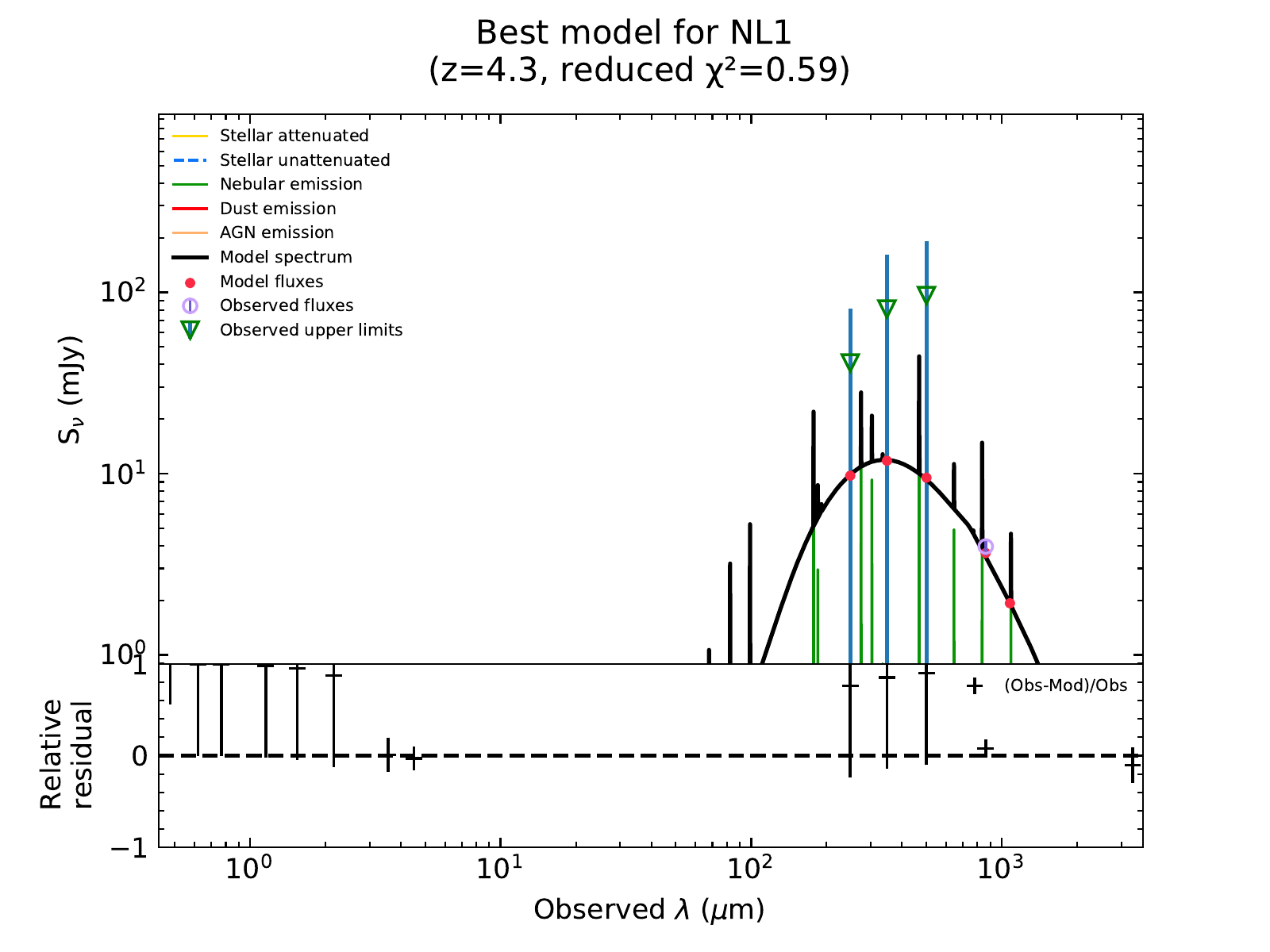}
\includegraphics[width=0.49\textwidth]{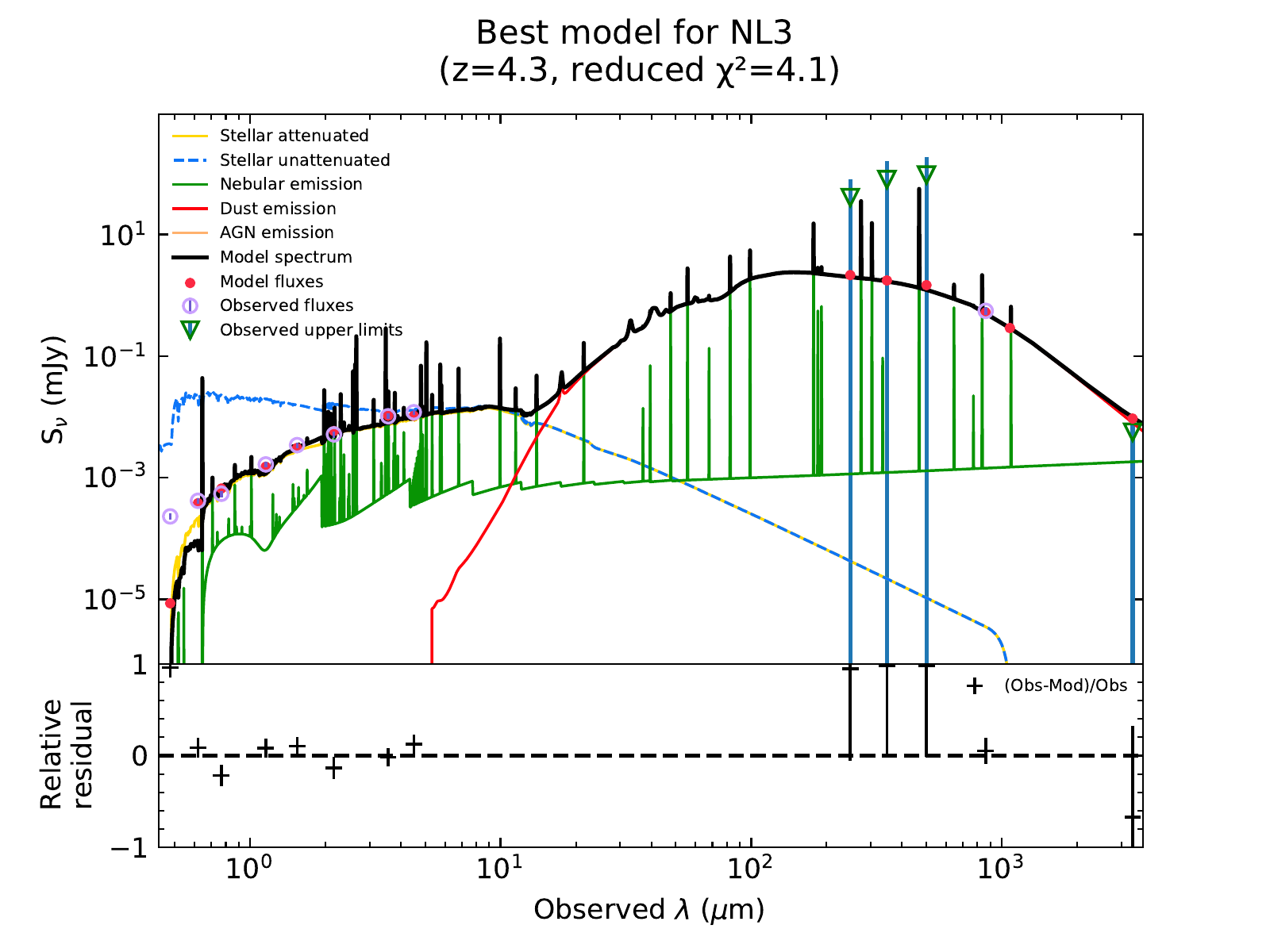}
\includegraphics[width=0.49\textwidth]{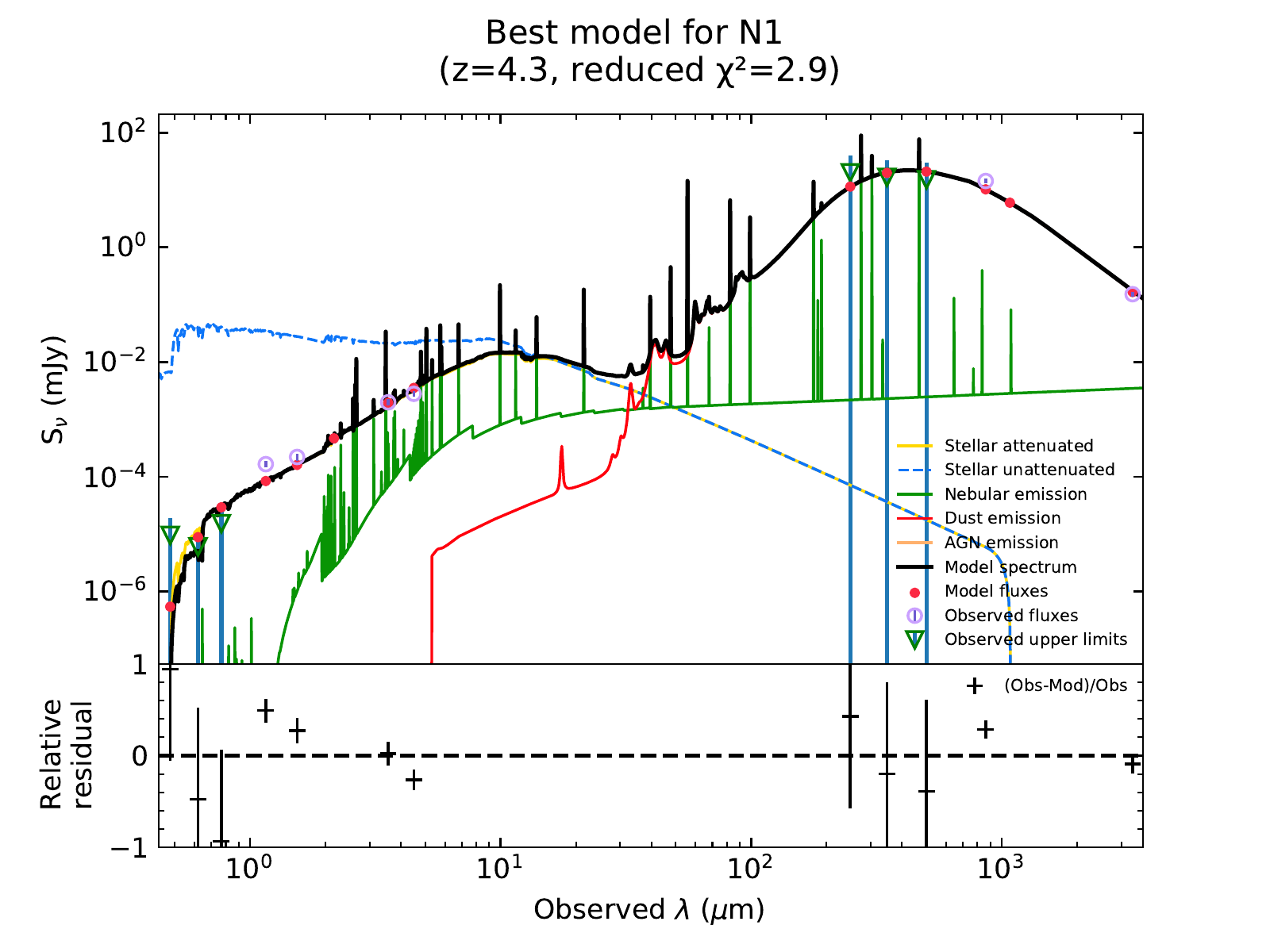}
\includegraphics[width=0.49\textwidth]{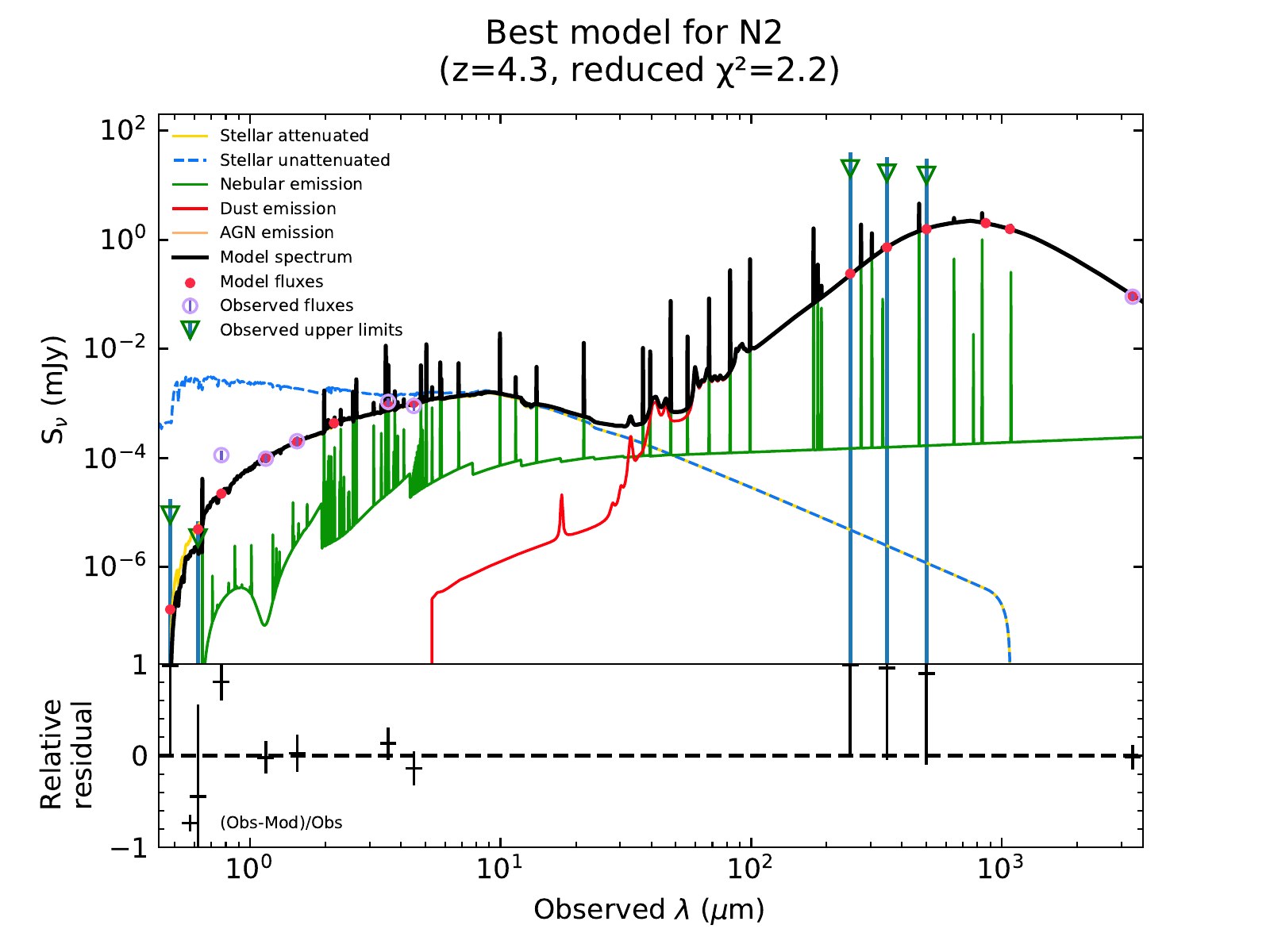}
\includegraphics[width=0.49\textwidth]{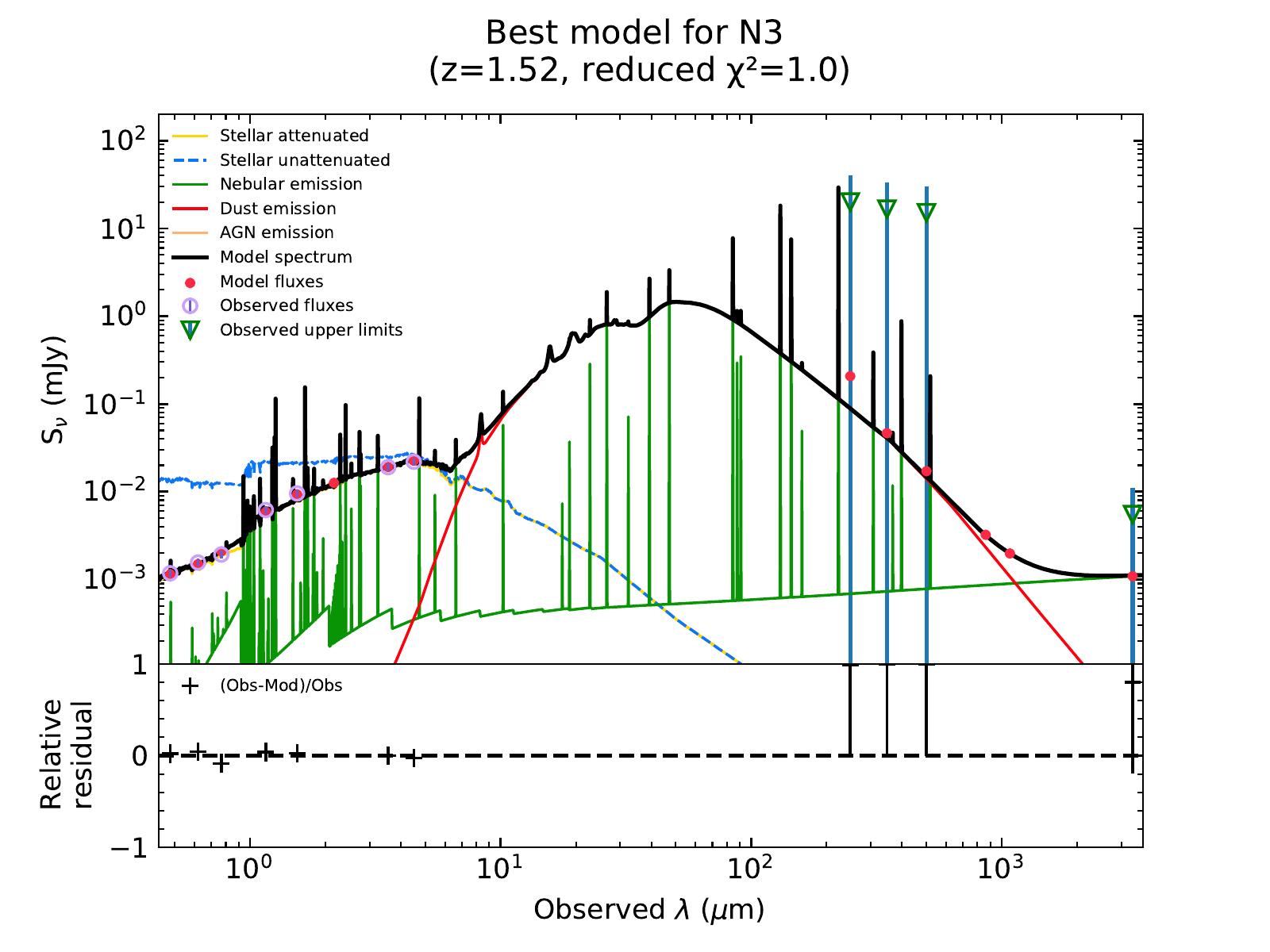}
\includegraphics[width=0.49\textwidth]{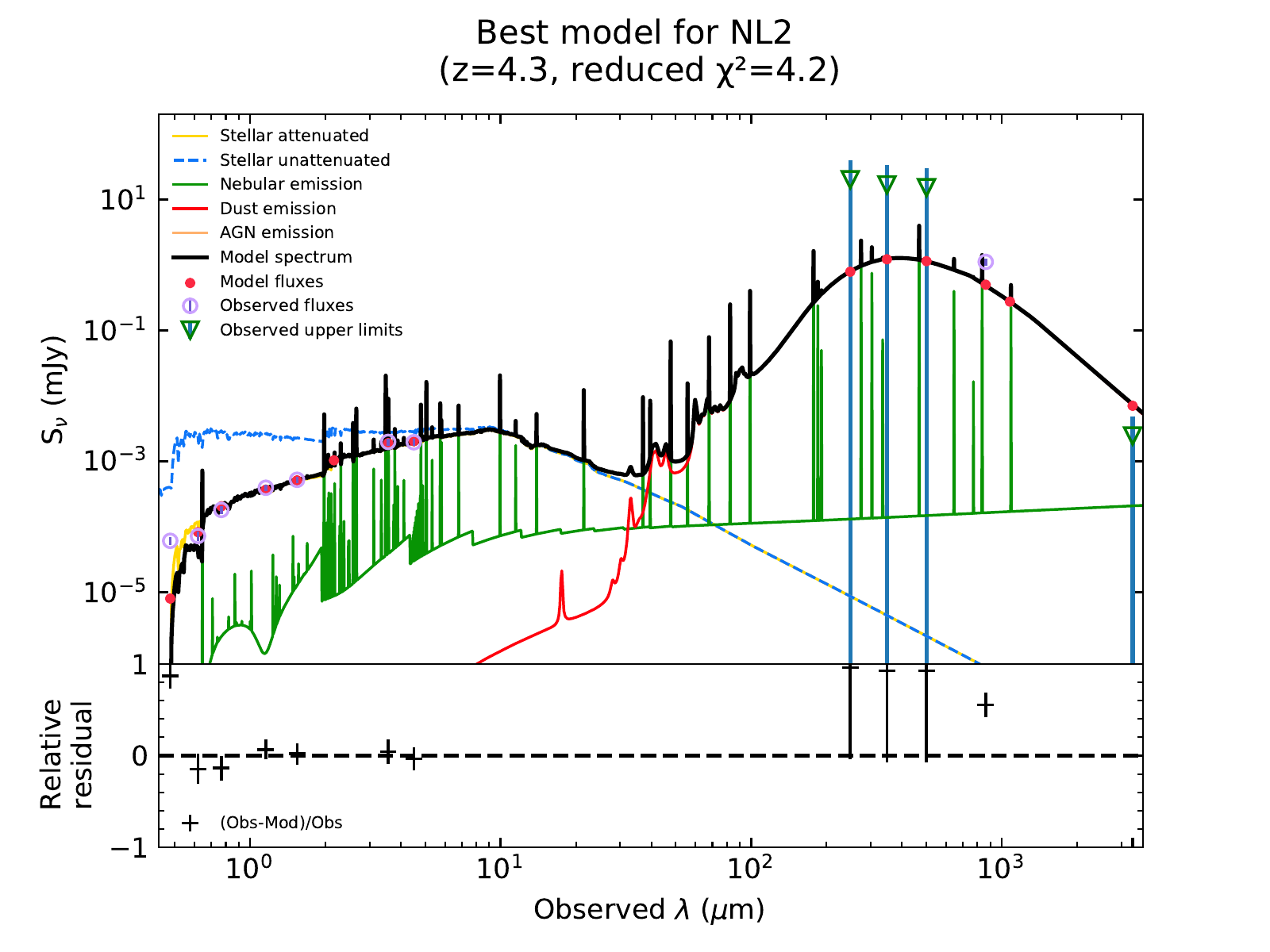}
\caption{}
\end{figure*}
\renewcommand{\thefigure}{\arabic{figure}}

\renewcommand{\thefigure}{C\arabic{figure} (Cont.)}
\addtocounter{figure}{-1}
\begin{figure*}
\includegraphics[width=0.49\textwidth]{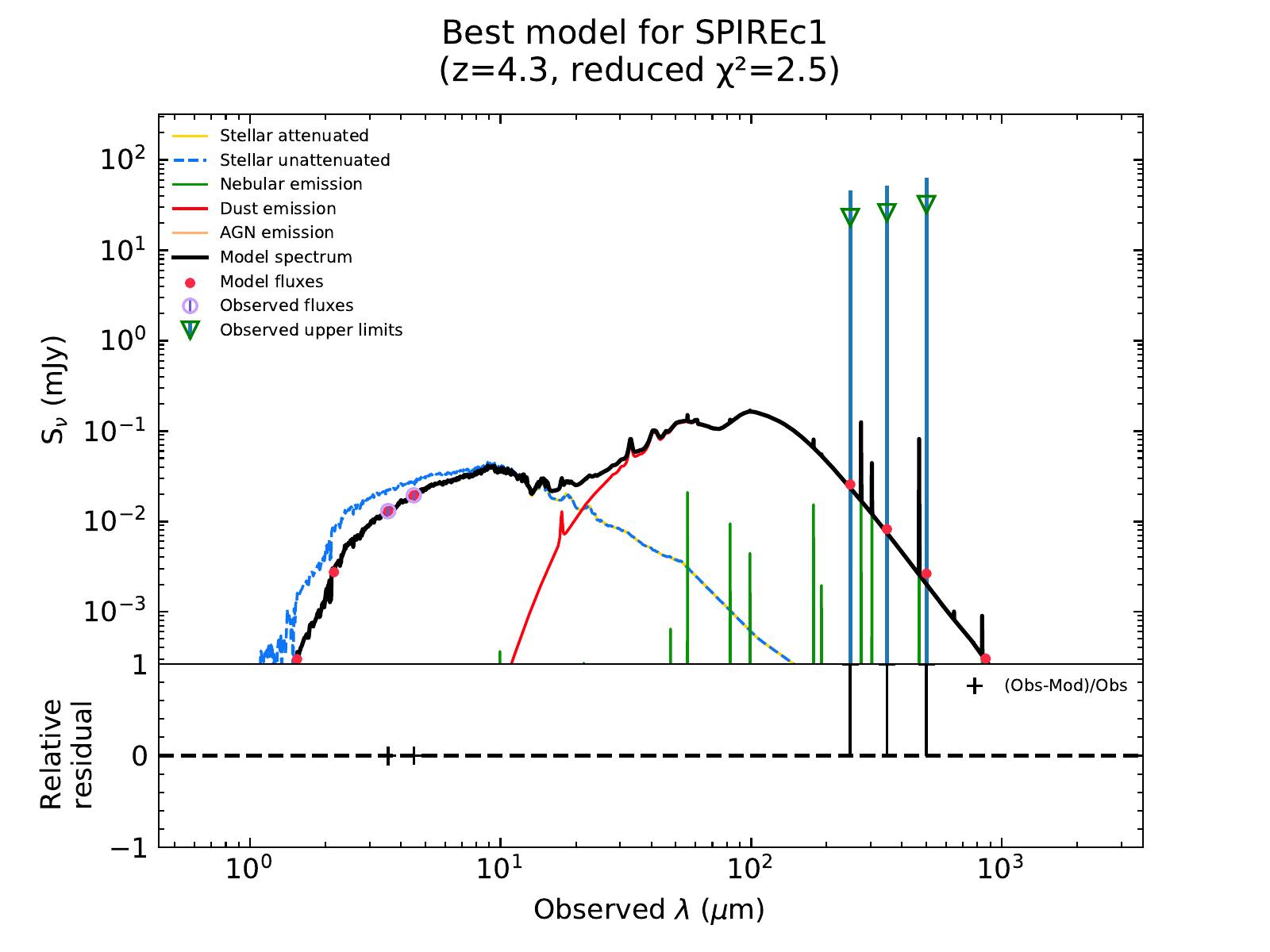}
\includegraphics[width=0.49\textwidth]{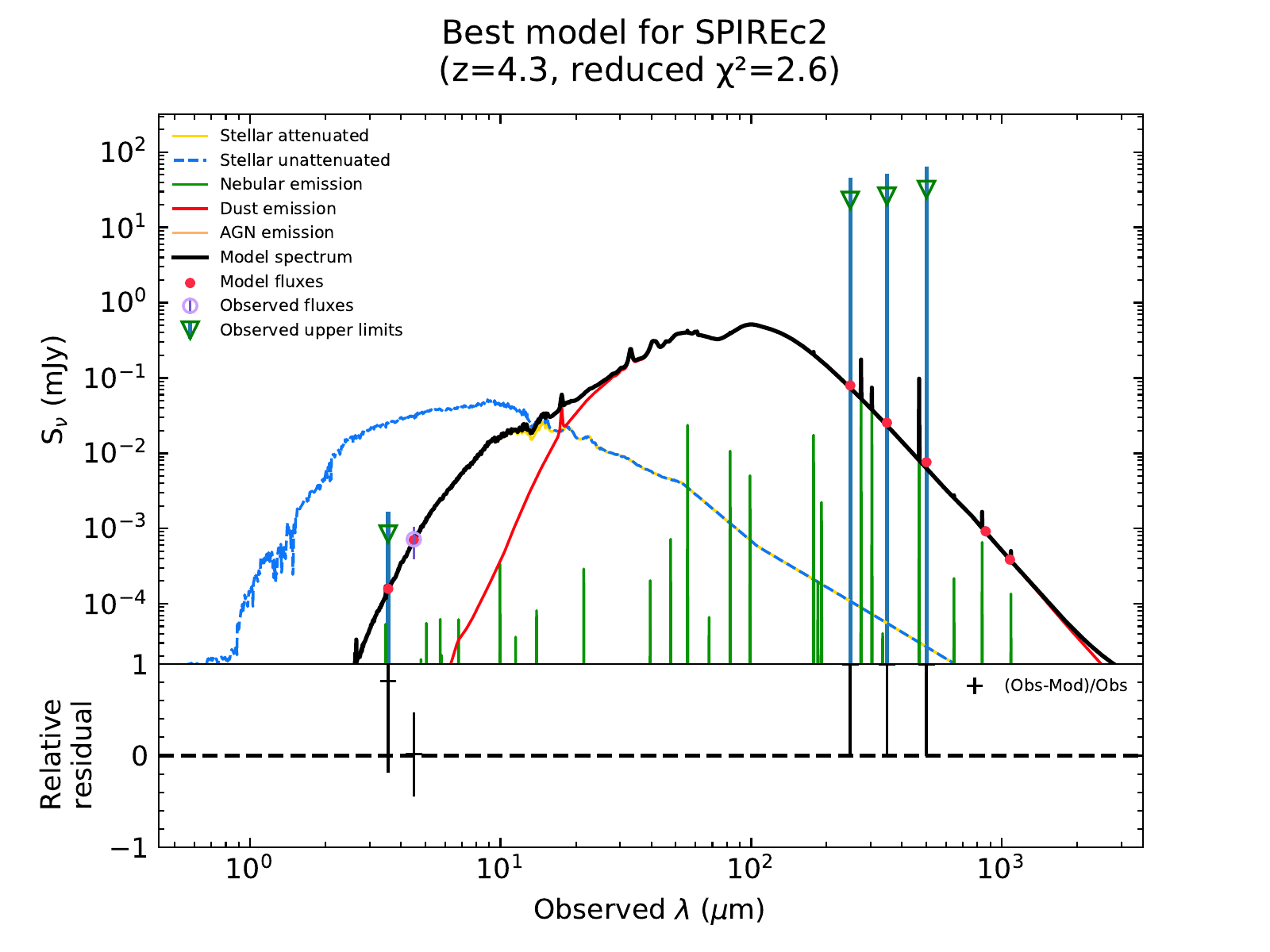}
\includegraphics[width=0.49\textwidth]{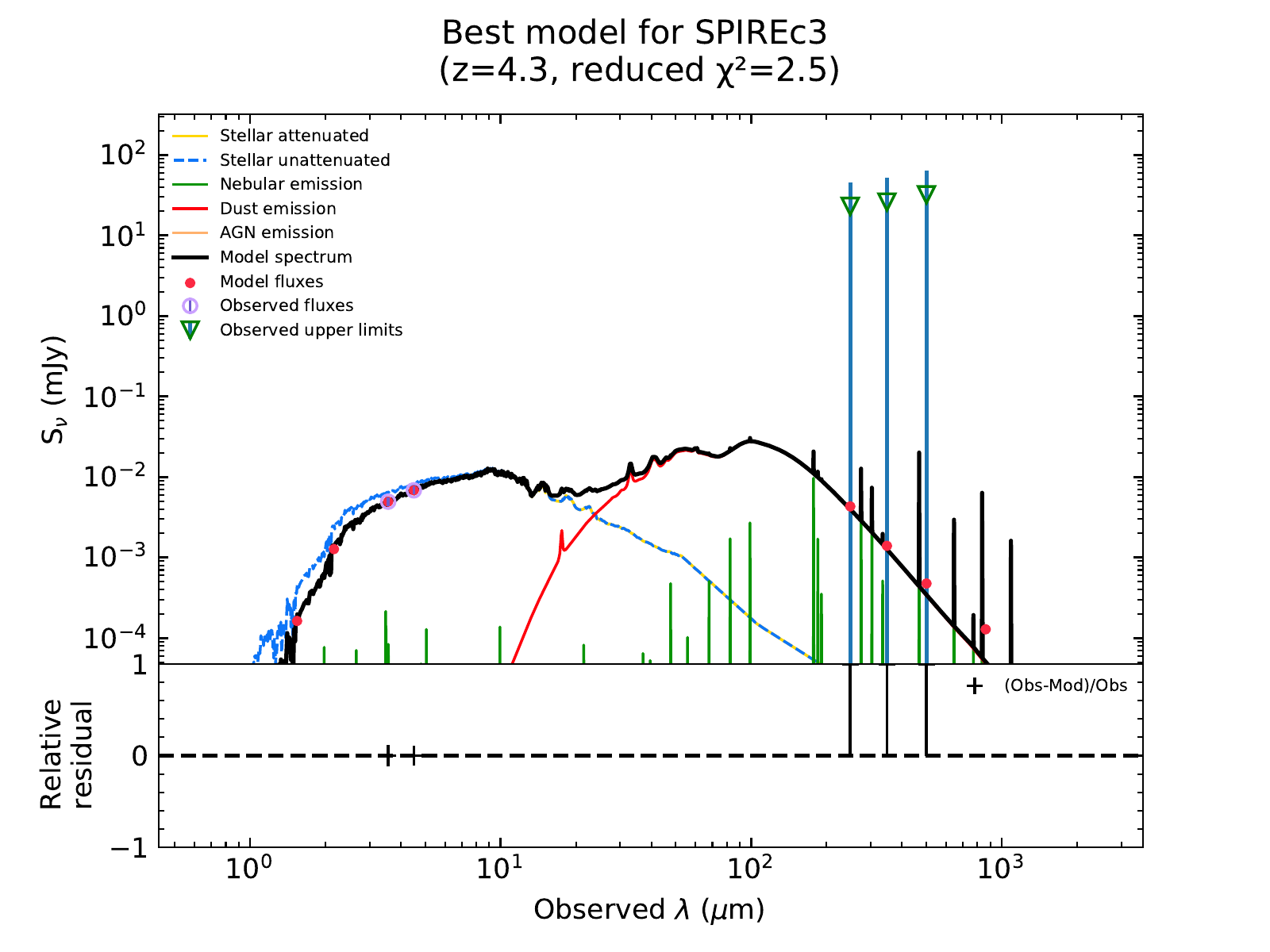}
\includegraphics[width=0.49\textwidth]{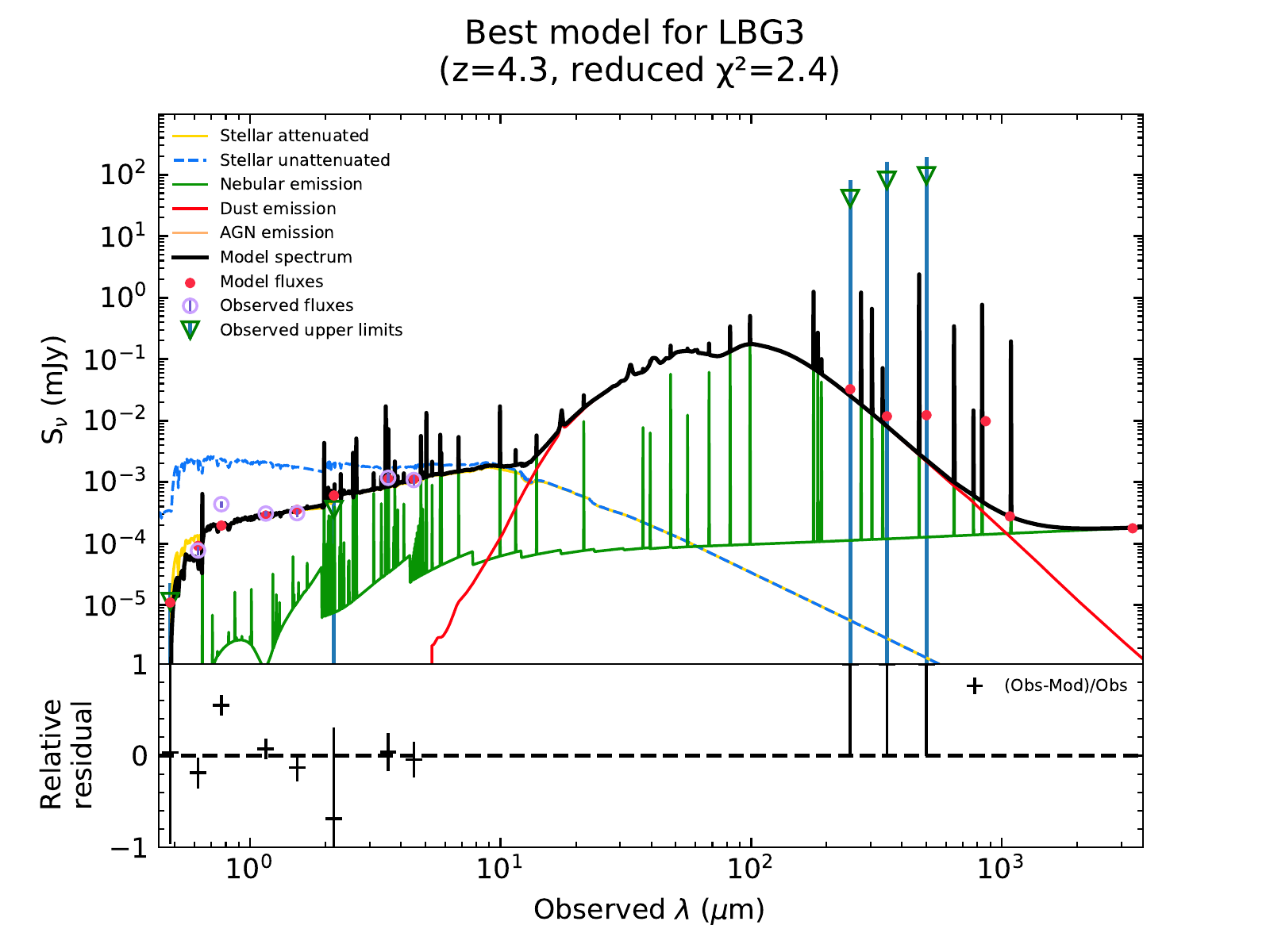}
\includegraphics[width=0.49\textwidth]{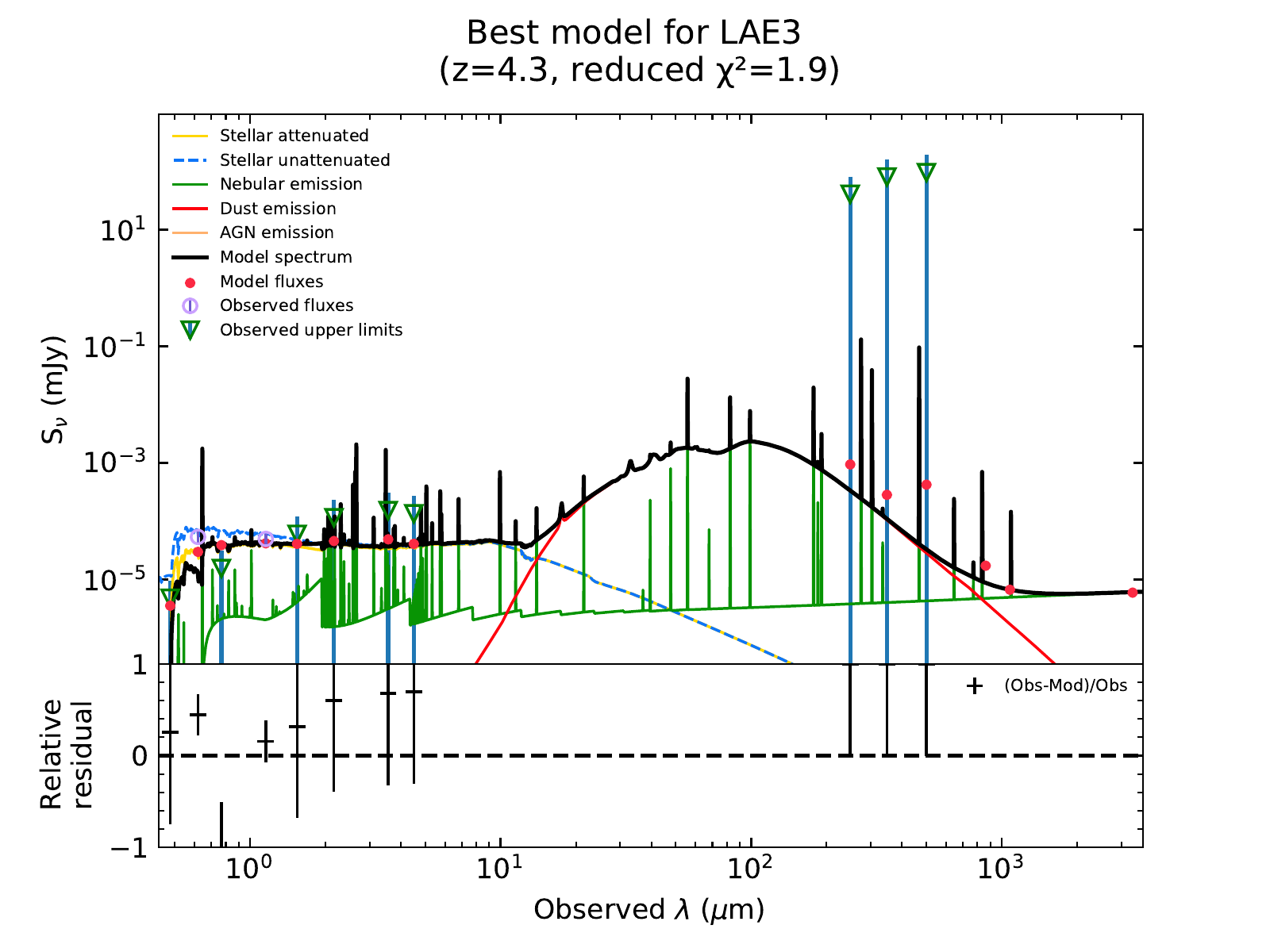}
\includegraphics[width=0.49\textwidth]{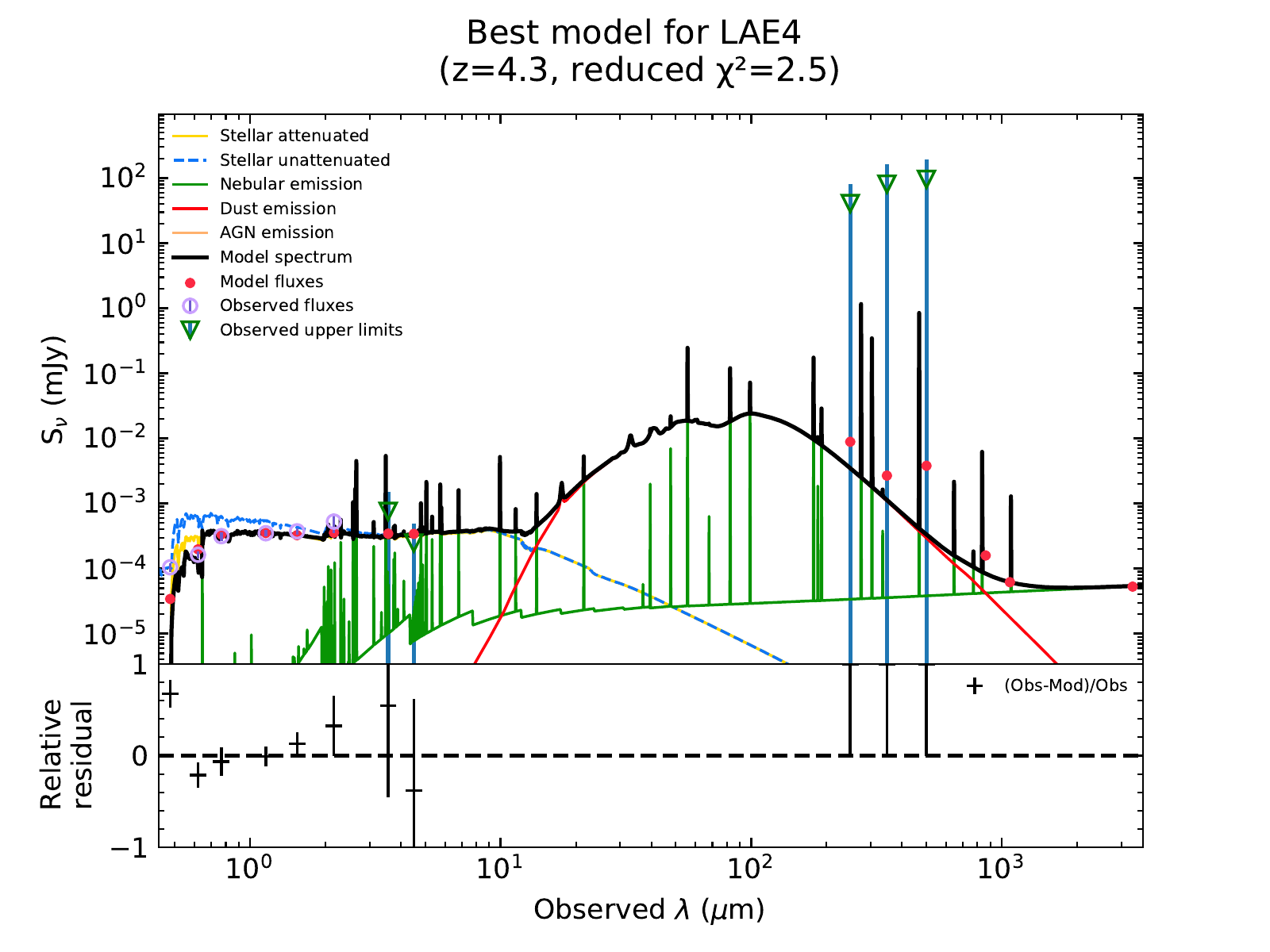}
\caption{}
\end{figure*}
\renewcommand{\thefigure}{\arabic{figure}}

\renewcommand{\thefigure}{C\arabic{figure} (Cont.)}
\addtocounter{figure}{-1}
\begin{figure*}
\includegraphics[width=0.49\textwidth]{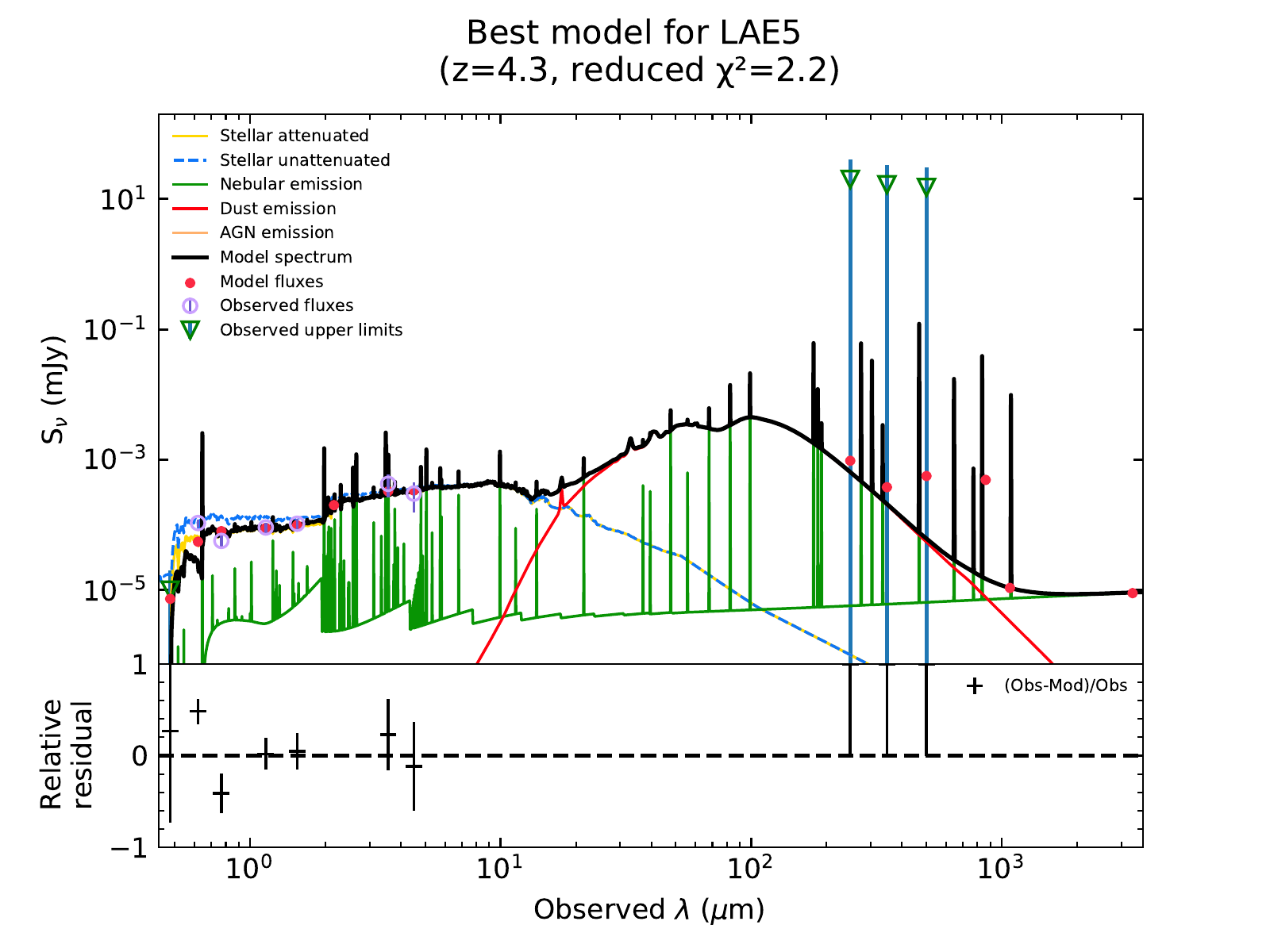}
\includegraphics[width=0.49\textwidth]{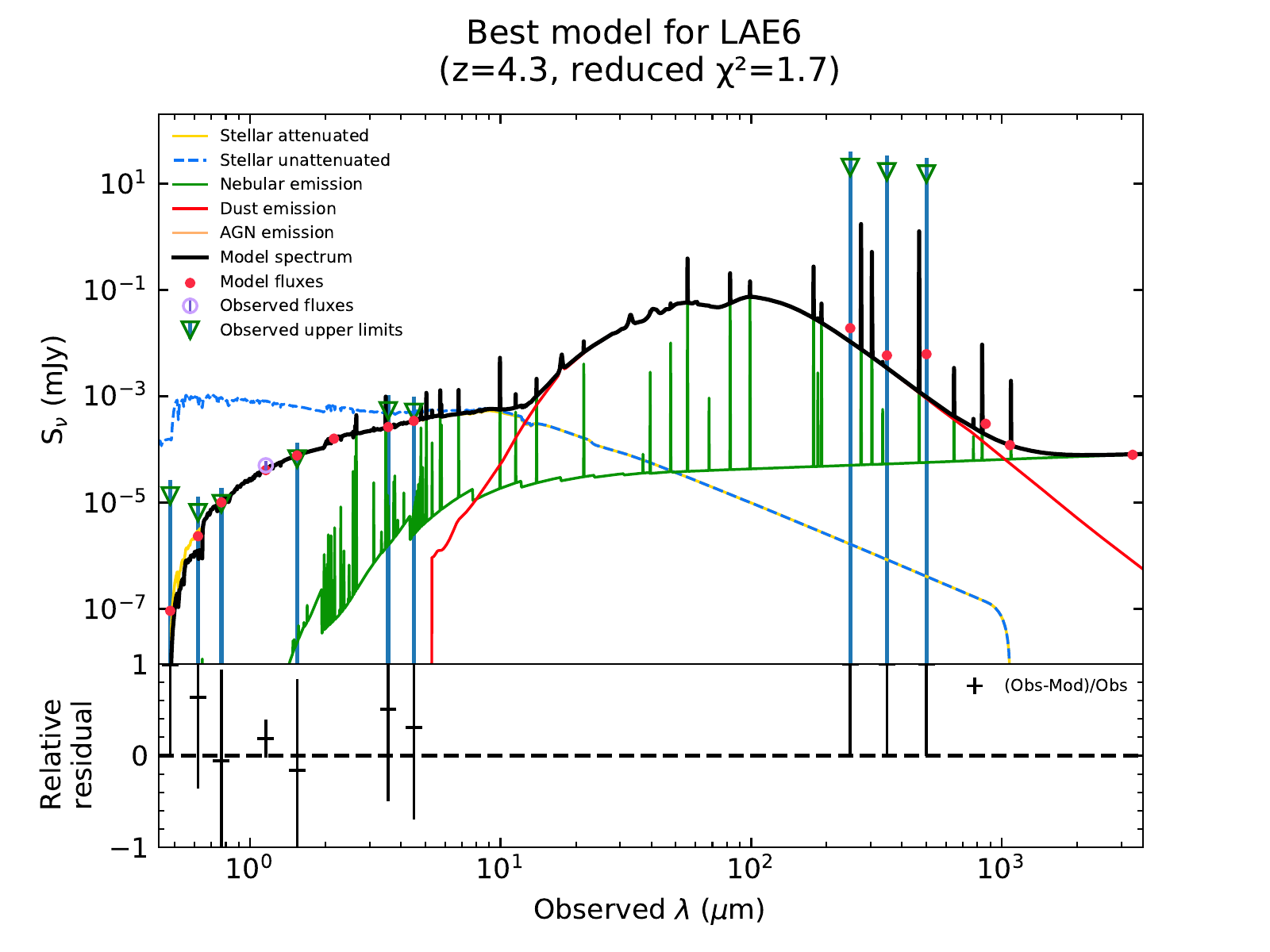}
\includegraphics[width=0.49\textwidth]{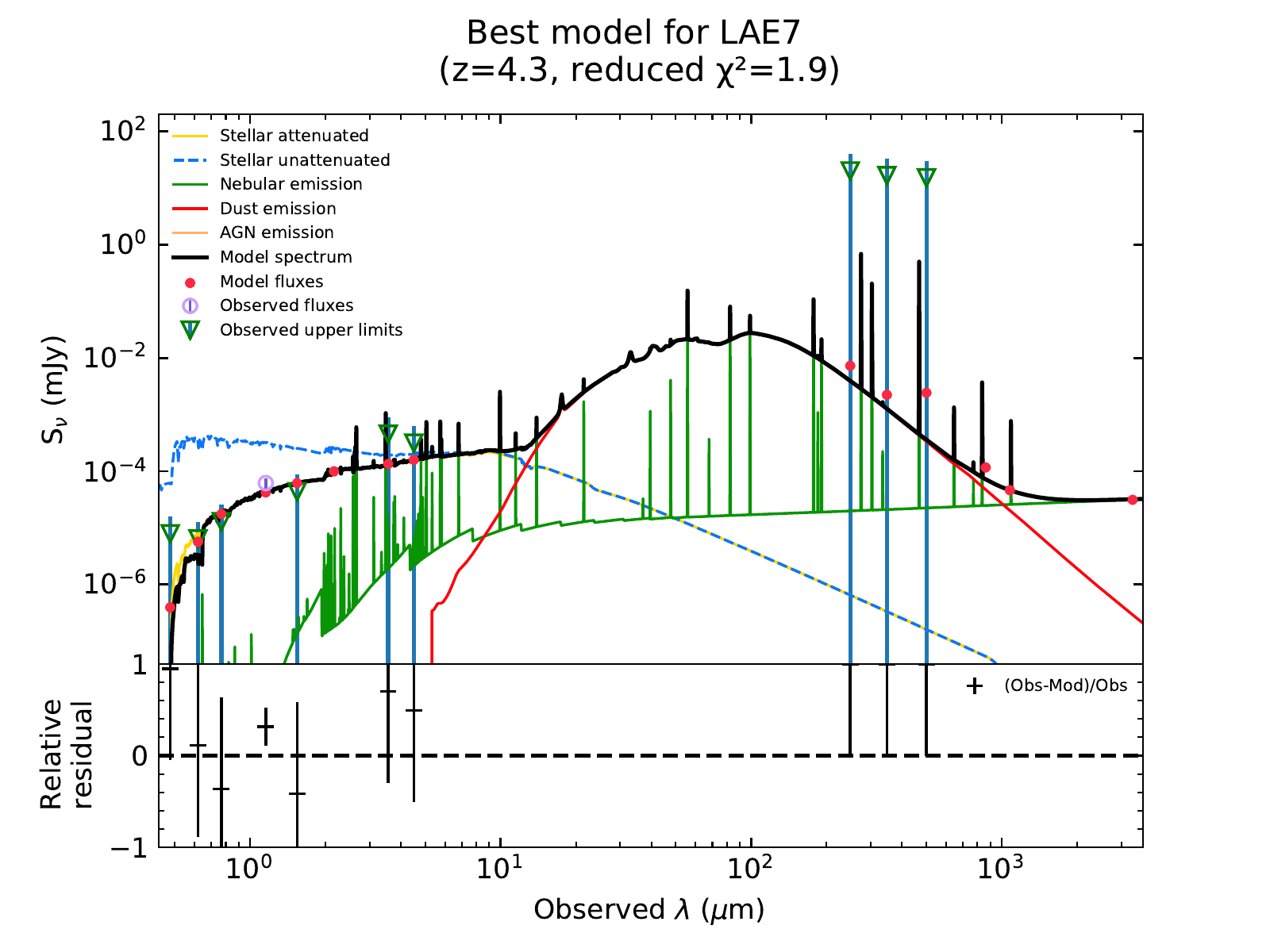}
\caption{}
\end{figure*}
\renewcommand{\thefigure}{D\arabic{figure}}

\newpage

\section{Optical profile fits}
\label{appendix3}

\begin{figure*}
\begin{subfigure}{0.45\textwidth}
\begin{framed}
\includegraphics[width=\textwidth]{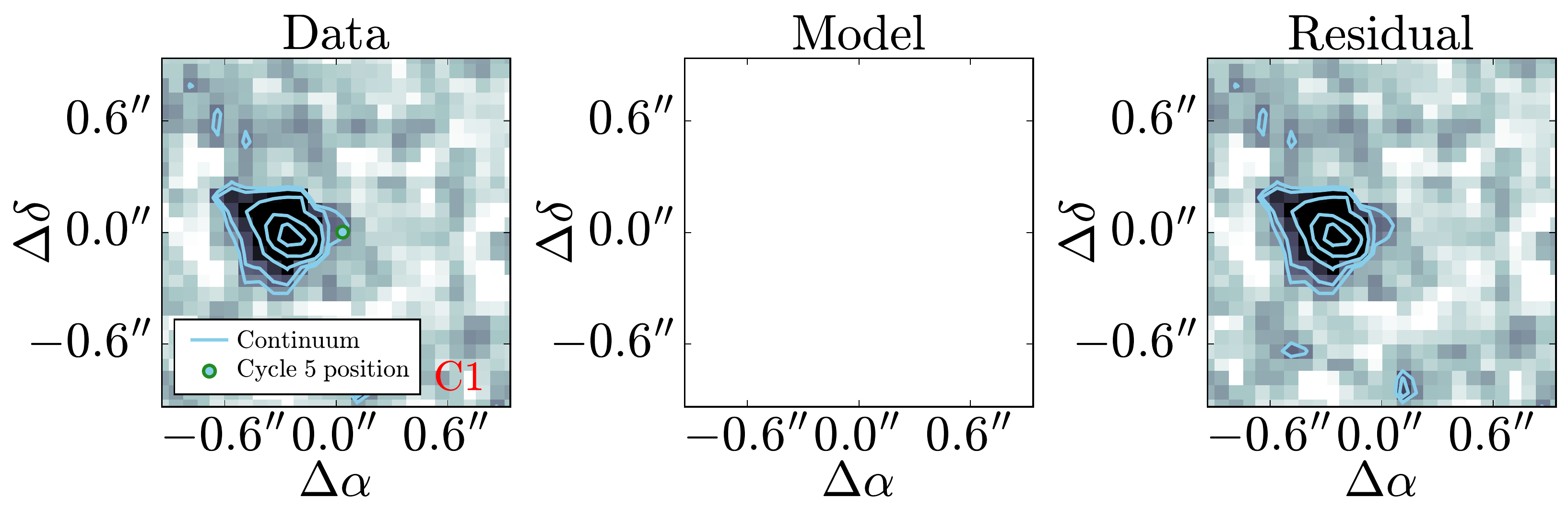}
\end{framed}
\end{subfigure}
\begin{subfigure}{0.45\textwidth}
\begin{framed}
\includegraphics[width=\textwidth]{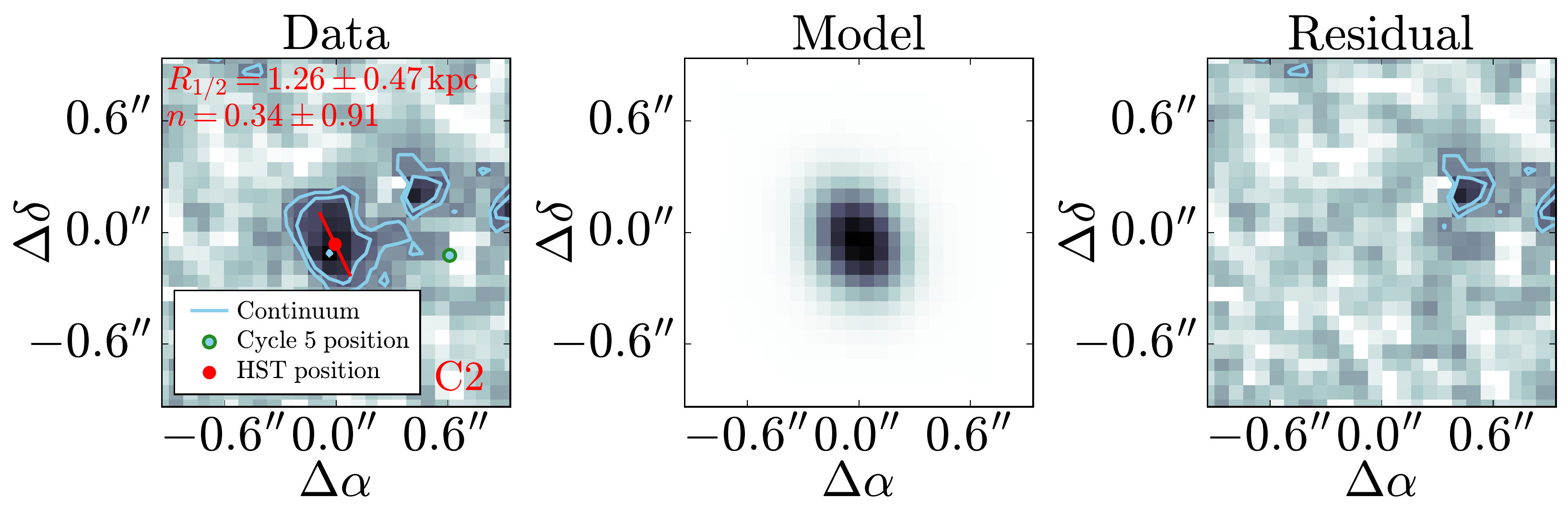}
\end{framed}
\end{subfigure}
\begin{subfigure}{0.45\textwidth}
\begin{framed}
\includegraphics[width=\textwidth]{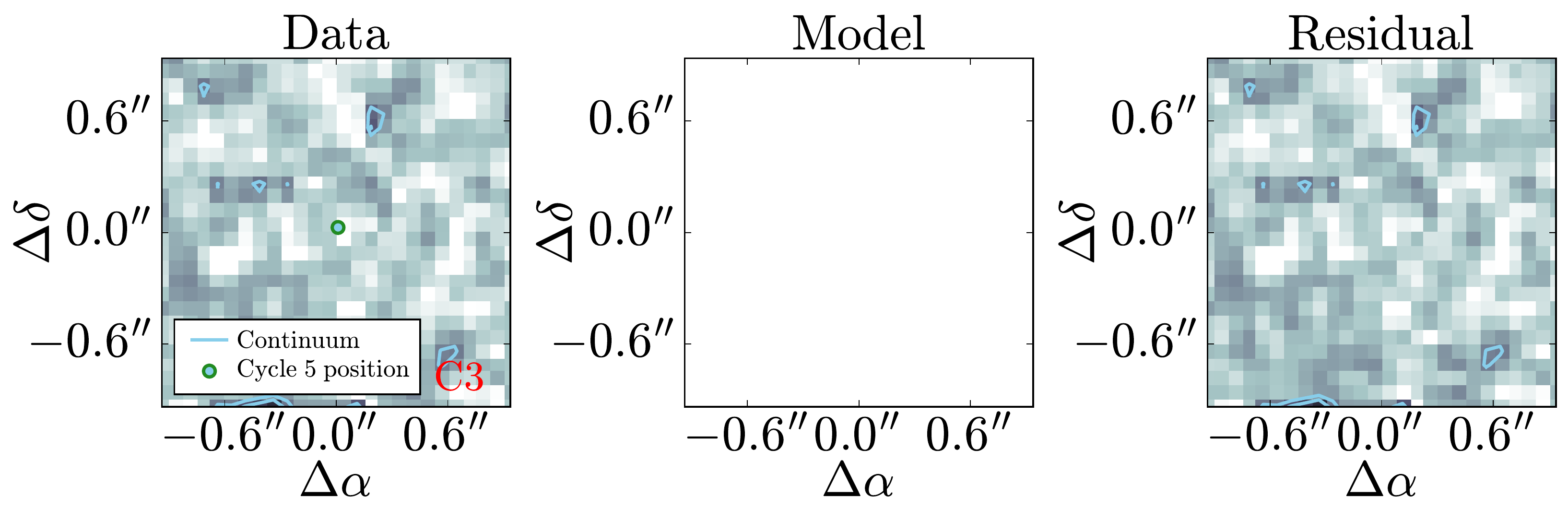}
\end{framed}
\end{subfigure}
\begin{subfigure}{0.45\textwidth}
\begin{framed}
\includegraphics[width=\textwidth]{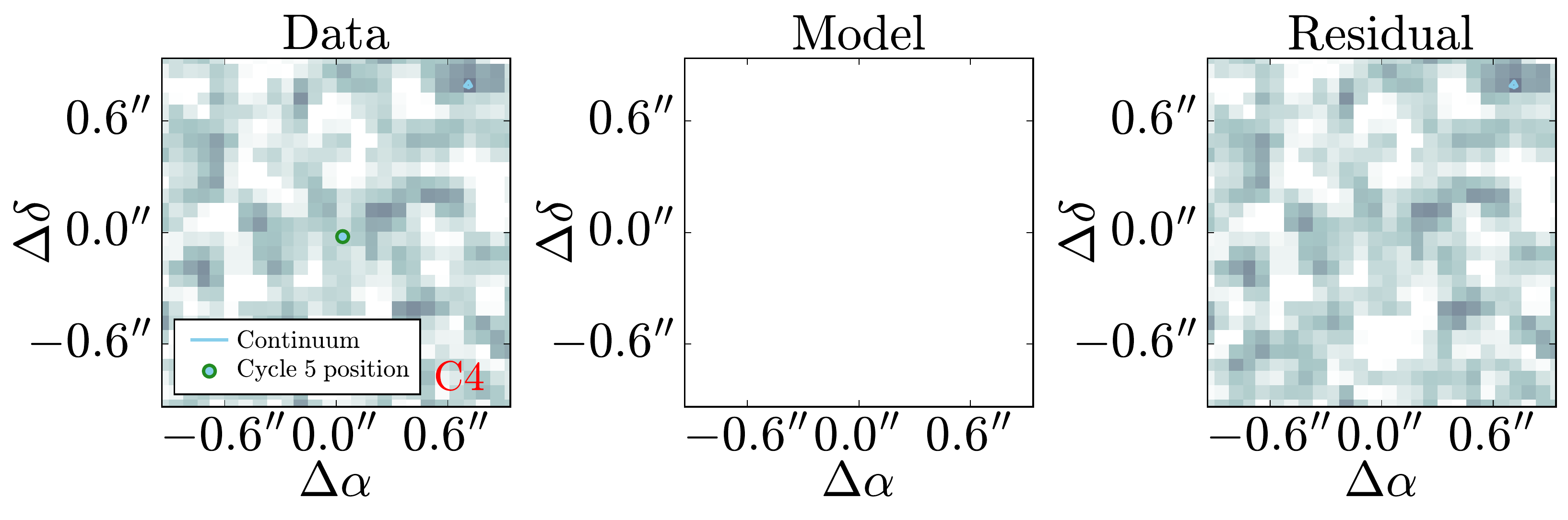}
\end{framed}
\end{subfigure}
\begin{subfigure}{0.45\textwidth}
\begin{framed}
\includegraphics[width=\textwidth]{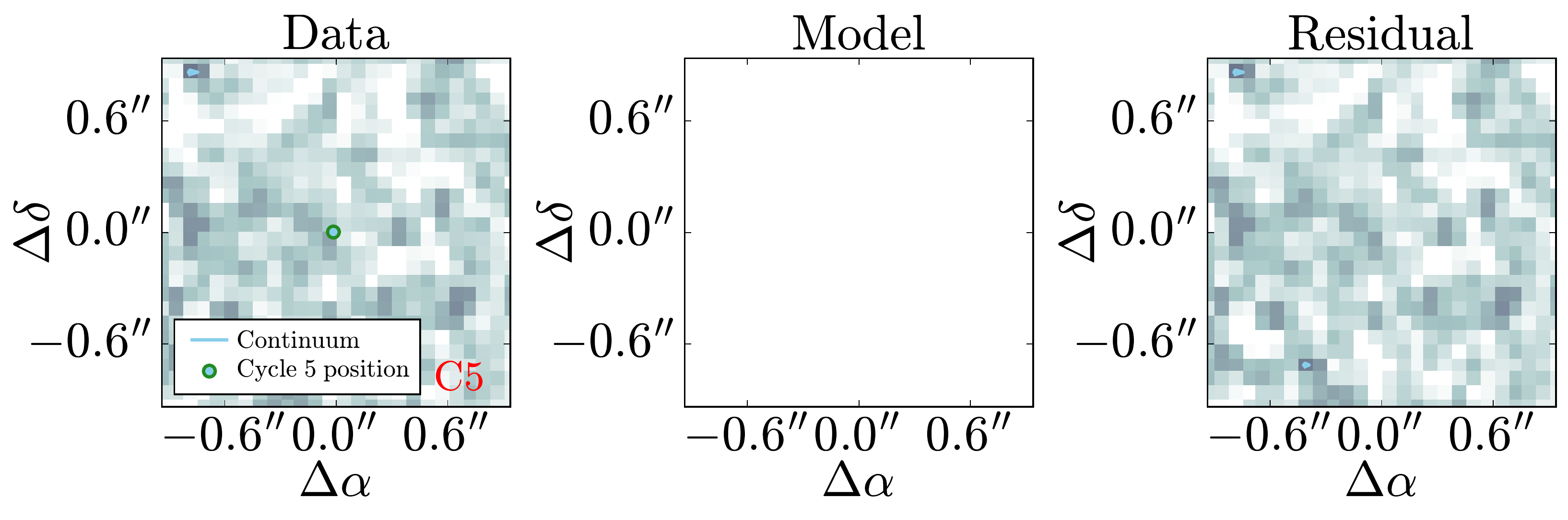}
\end{framed}
\end{subfigure}
\begin{subfigure}{0.45\textwidth}
\begin{framed}
\includegraphics[width=\textwidth]{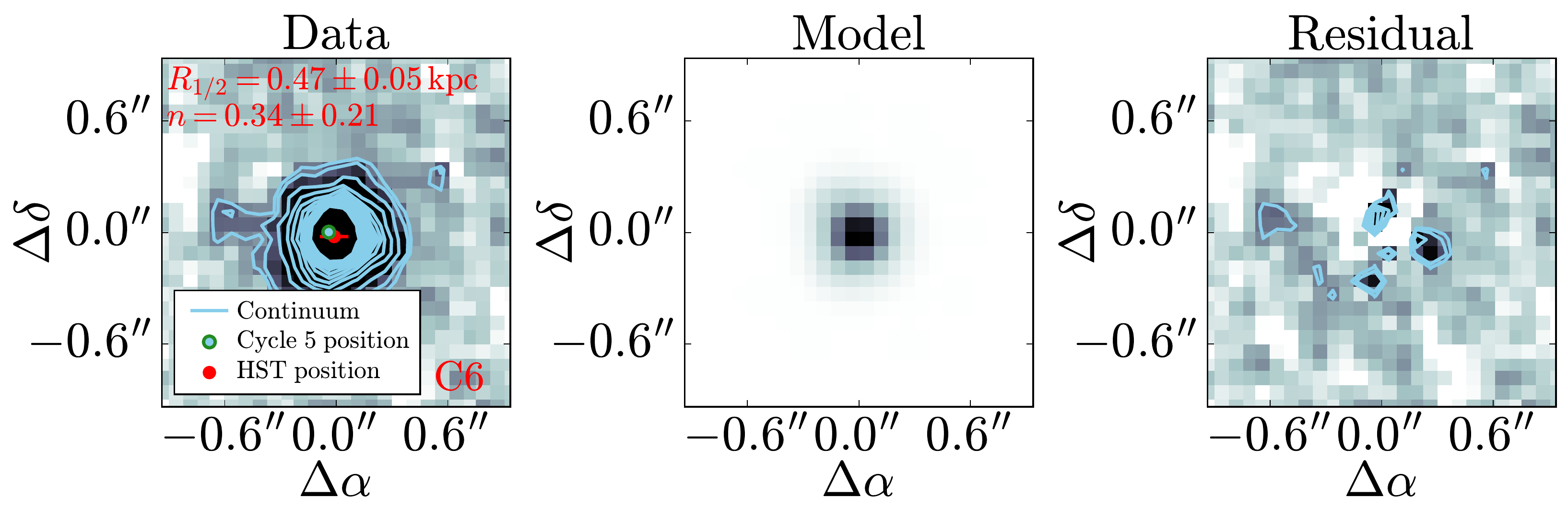}
\end{framed}
\end{subfigure}
\begin{subfigure}{0.45\textwidth}
\begin{framed}
\includegraphics[width=\textwidth]{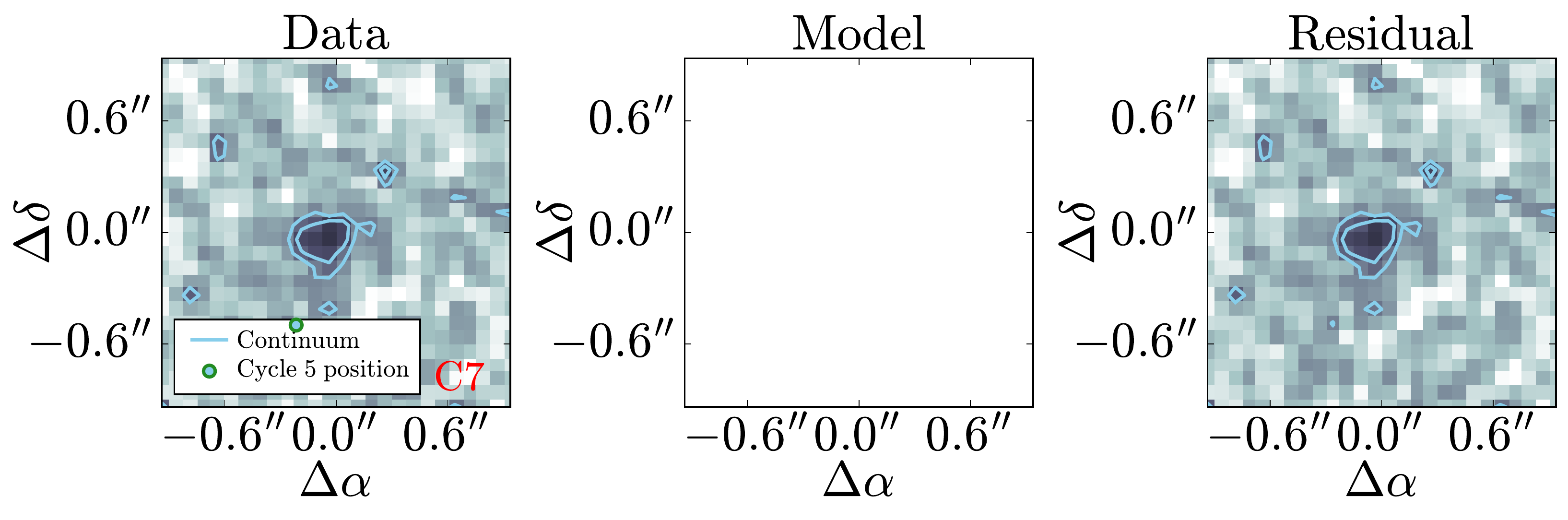}
\end{framed}
\end{subfigure}
\begin{subfigure}{0.45\textwidth}
\begin{framed}
\includegraphics[width=\textwidth]{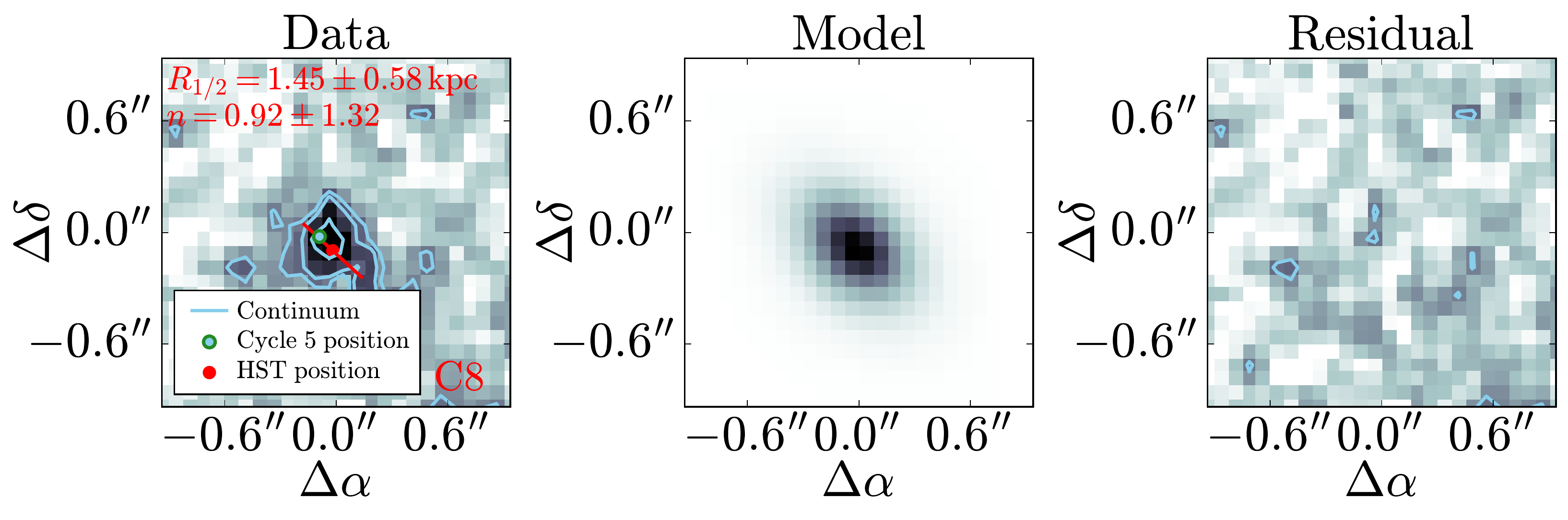}
\end{framed}
\end{subfigure}
\begin{subfigure}{0.45\textwidth}
\begin{framed}
\includegraphics[width=\textwidth]{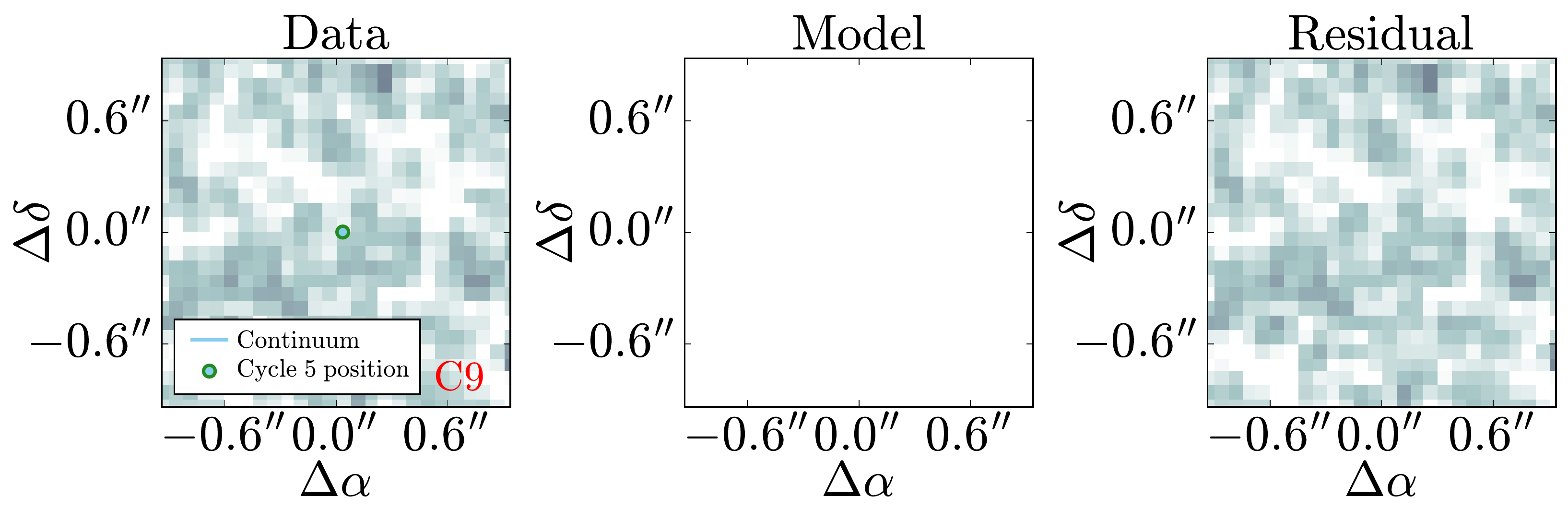}
\end{framed}
\end{subfigure}
\begin{subfigure}{0.45\textwidth}
\begin{framed}
\includegraphics[width=\textwidth]{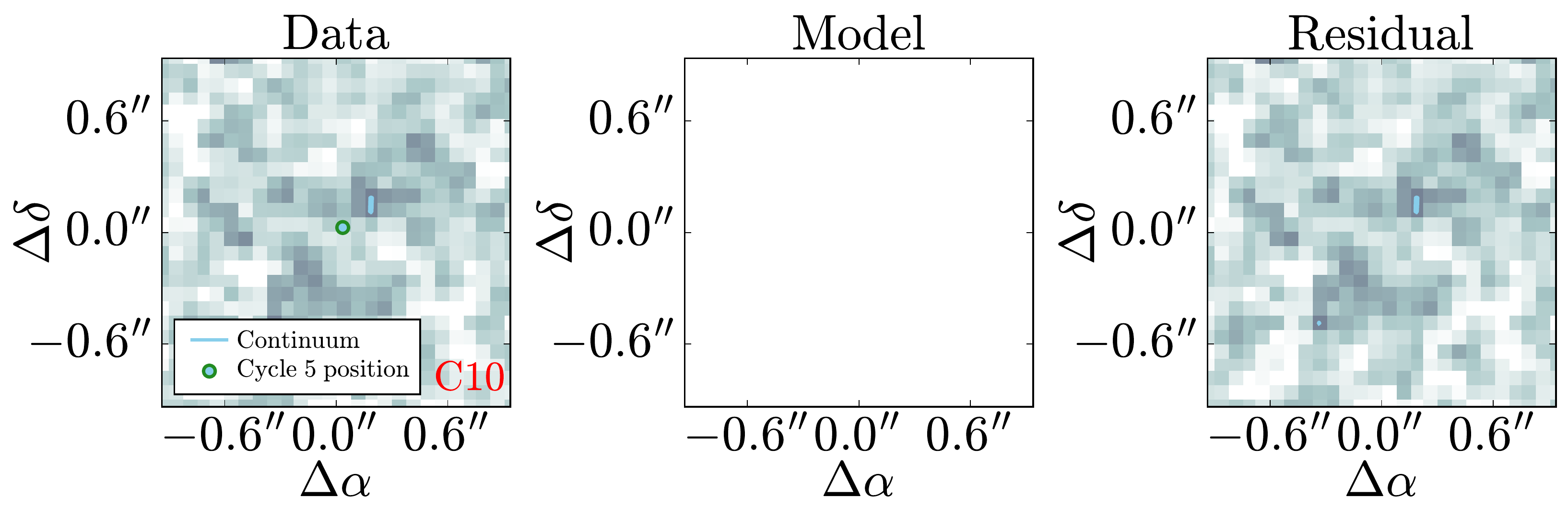}
\end{framed}
\end{subfigure}
\begin{subfigure}{0.45\textwidth}
\begin{framed}
\includegraphics[width=\textwidth]{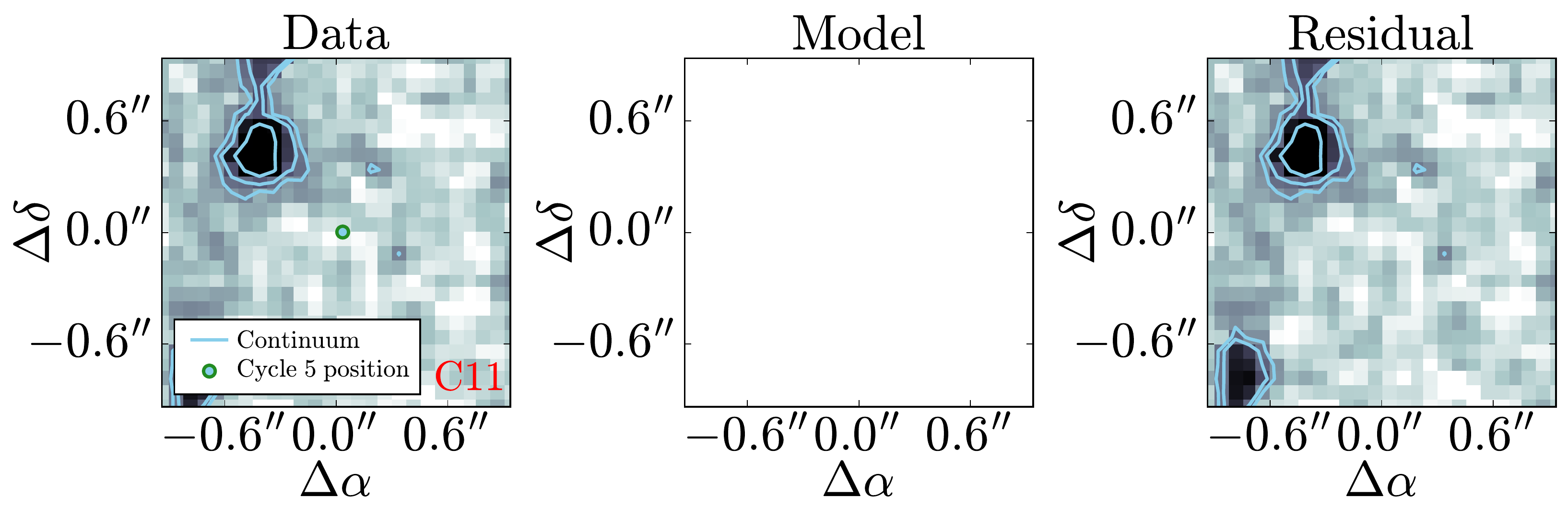}
\end{framed}
\end{subfigure}
\begin{subfigure}{0.45\textwidth}
\begin{framed}
\includegraphics[width=\textwidth]{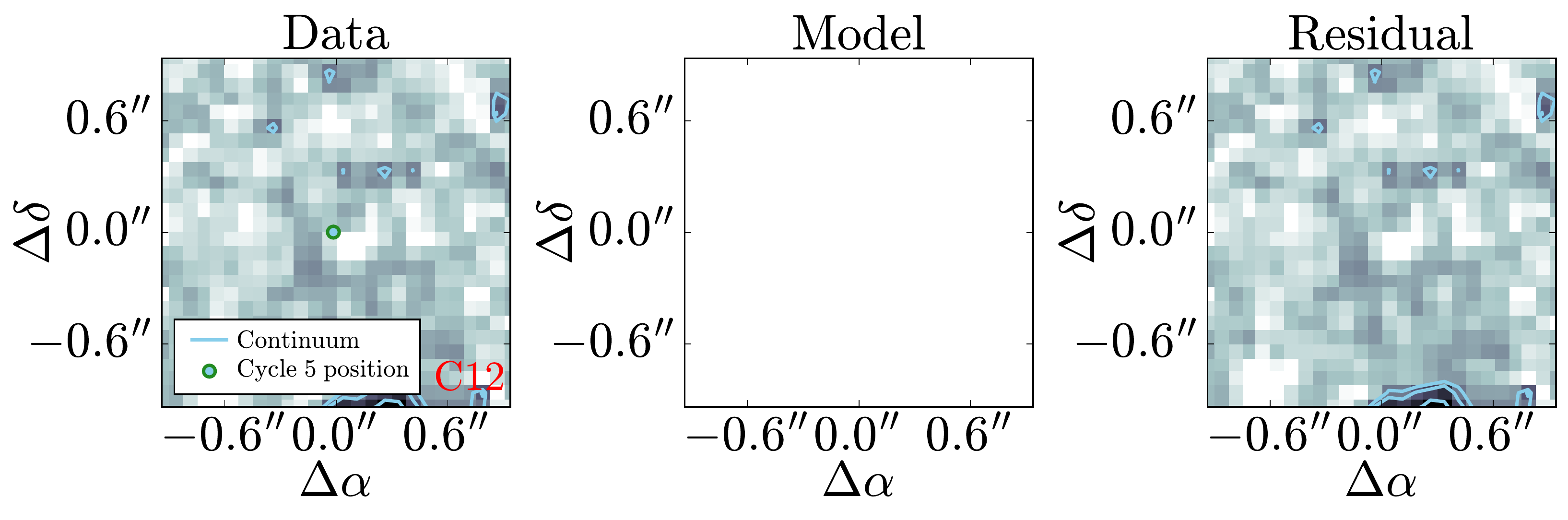}
\end{framed}
\end{subfigure}
\begin{subfigure}{0.45\textwidth}
\begin{framed}
\includegraphics[width=\textwidth]{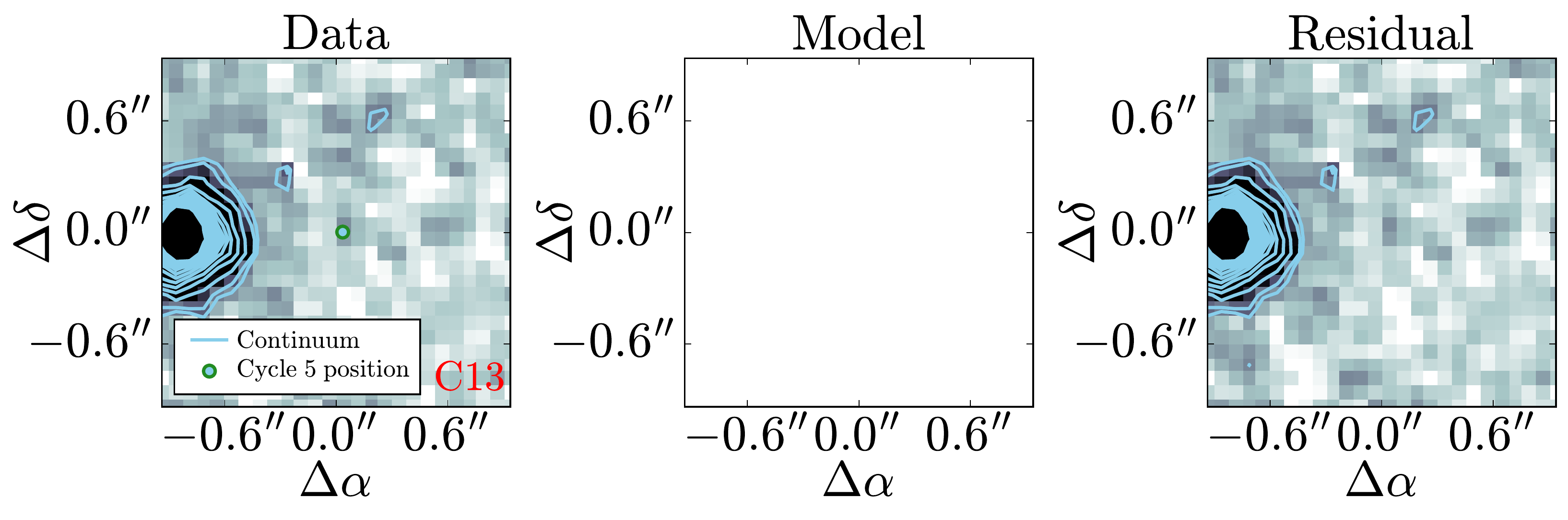}
\end{framed}
\end{subfigure}
\begin{subfigure}{0.45\textwidth}
\begin{framed}
\includegraphics[width=\textwidth]{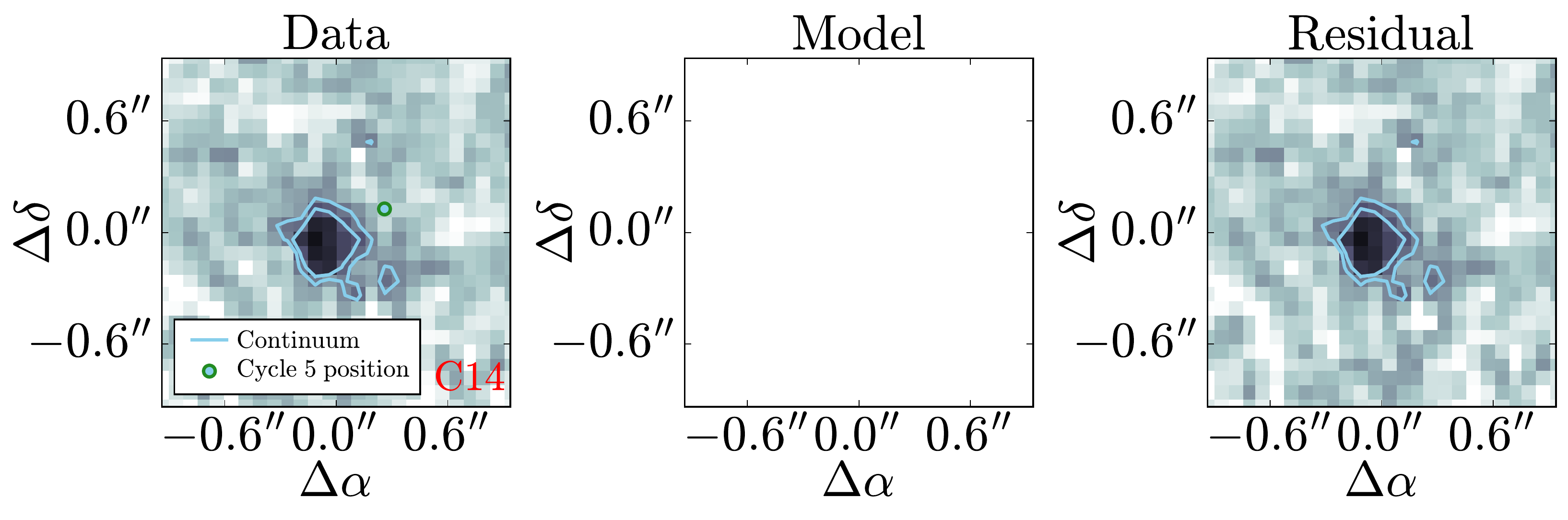}
\end{framed}
\end{subfigure}
\caption{{\it Left:} Rest-frame optical (observed-frame 1.5\,$\mu$m) images from our {\it HST\/} F160W imaging. Contours are 2 and 3$\sigma$, then increase in steps of 3$\sigma$. The blue points are positions found in our lower resolution Cycle 5 data, and red points are the centres of the S{\'e}rsic profiles fit to the optical image. The red bars show the lengths of the half-light diamtres determined from the best-fitting S{\'e}rsic profiles, and best-fitting half-light radii and S{\'e}rsic indices are shown in the top left. {\it Middle:} Best-fitting model S{\'e}rsic profiles. Sources without enough pixels above 5$\sigma$ were not fitted, and for these cases we leave this panel blank. {\it Right:} Residuals from the S{\'e}rsic profile fits.}
\label{fits}
\end{figure*}

In Fig.~\ref{fits} we provide rest-frame optical (observed-frame 1.5\,$\mu$m) cut-outs of our sources obtained from our {\it HST\/} F160W imaging. S{\'e}rsic profiles were fit to all sources detected in the images above 3$\sigma$, including convolution with the beam, and half-light radii were estimated from the fits.

The left panels show these stacked images with 2 and 3$\sigma$ contours, then increasing in steps of 3$\sigma$, with positions found in the Cycle 5 data shown as blue points and positions found from the S{\'e}rsic profiles shown as red points. The red bars indicate the sizes of the half-light radii resulting from the S{\'e}rsic profiles, with best-fitting half-light radii and S{\'e}rsic indices shown in the top left. For sources NL2, NL3, and N3 we provide the best-fitting half-light radii in units of arcseconds (as the redshifts have not been confirmed), otherwise the best-fitting half-light radii are in units of kiloparsecs. The middle panels show our S{\'e}rsic profile models, and the right panels show the residuals. For sources below 5$\sigma$, where we did not attempt to fit S{\'e}rsic profiles, we leave the middle panel blank.

\renewcommand{\thefigure}{D\arabic{figure} (Cont.)}
\addtocounter{figure}{-1}
\begin{figure*}
\begin{subfigure}{0.45\textwidth}
\begin{framed}
\includegraphics[width=\textwidth]{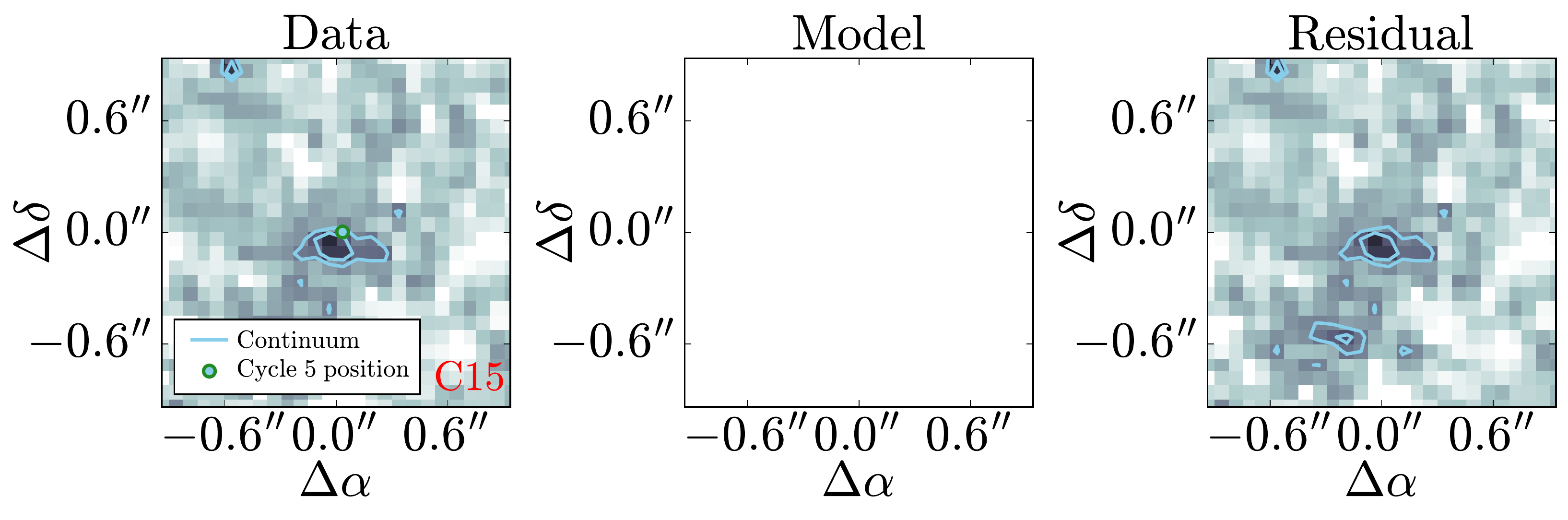}
\end{framed}
\end{subfigure}
\begin{subfigure}{0.45\textwidth}
\begin{framed}
\includegraphics[width=\textwidth]{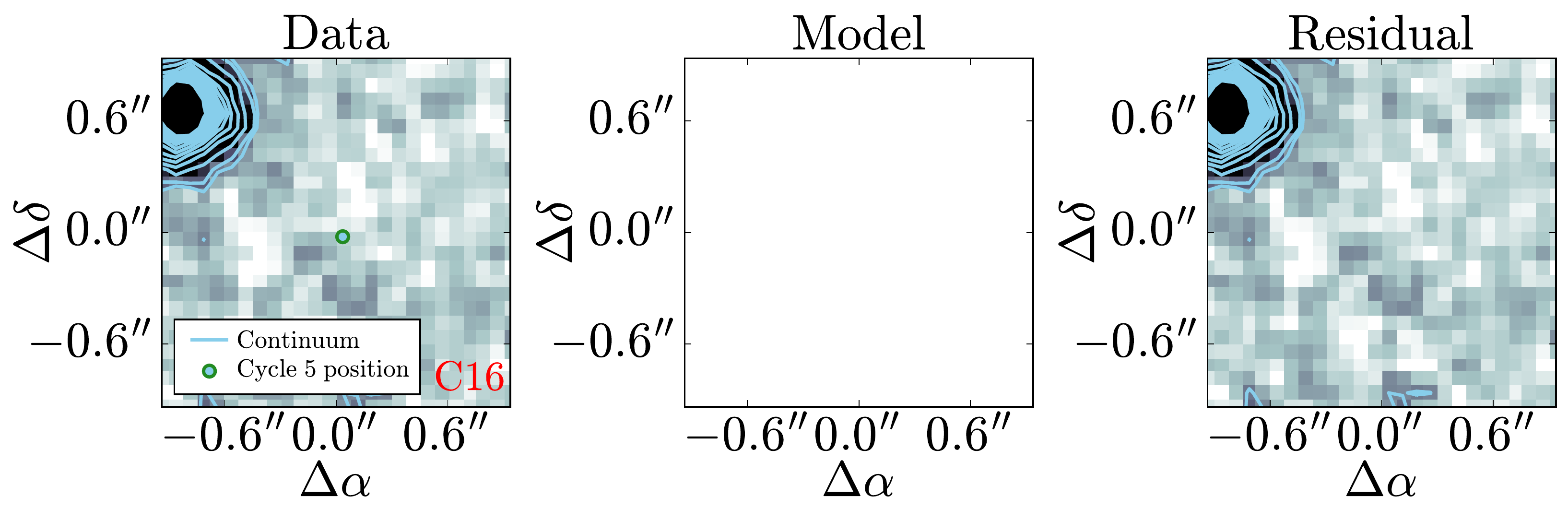}
\end{framed}
\end{subfigure}
\begin{subfigure}{0.45\textwidth}
\begin{framed}
\includegraphics[width=\textwidth]{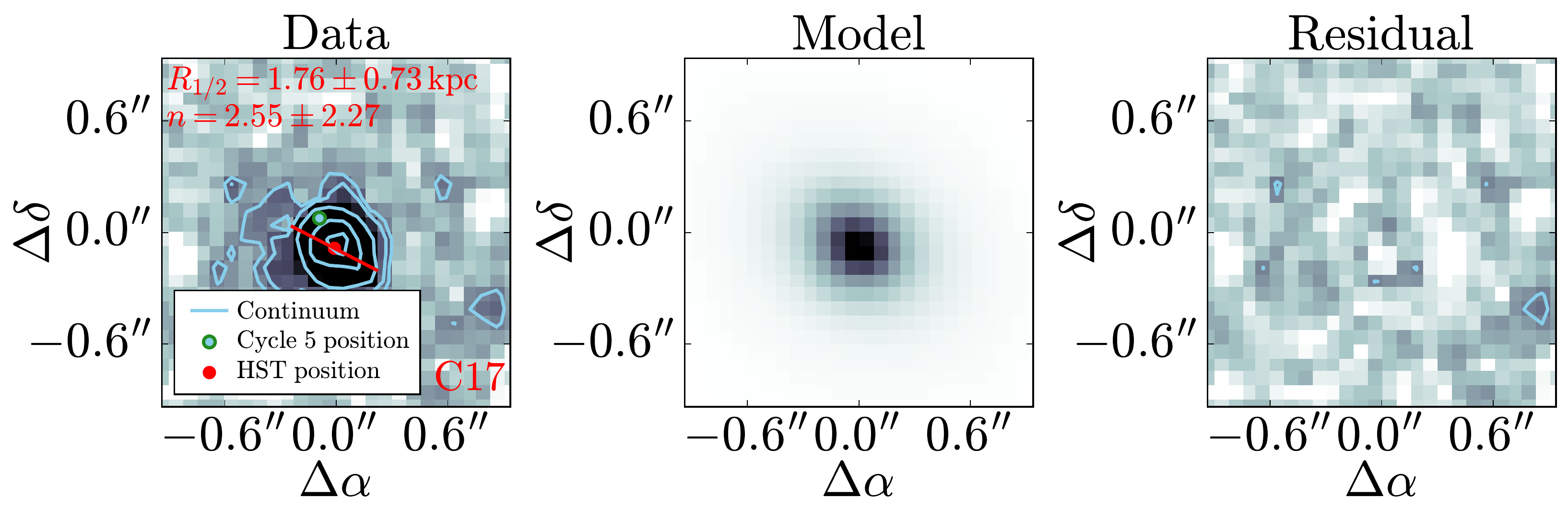}
\end{framed}
\end{subfigure}
\begin{subfigure}{0.45\textwidth}
\begin{framed}
\includegraphics[width=\textwidth]{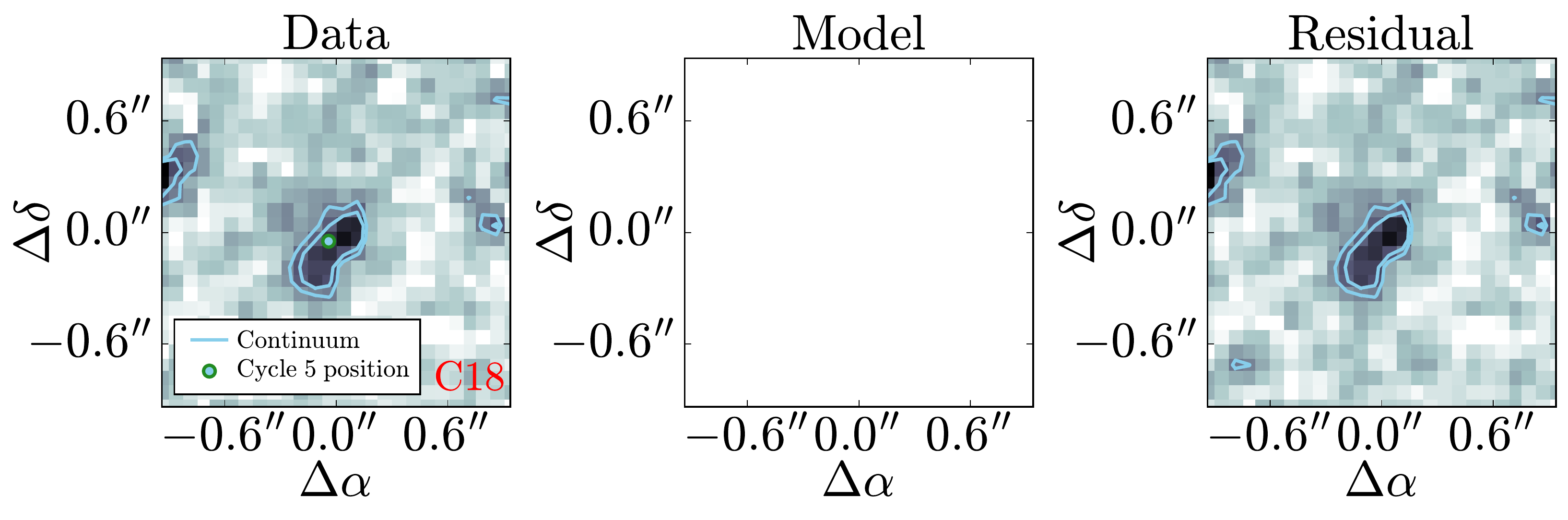}
\end{framed}
\end{subfigure}
\begin{subfigure}{0.45\textwidth}
\begin{framed}
\includegraphics[width=\textwidth]{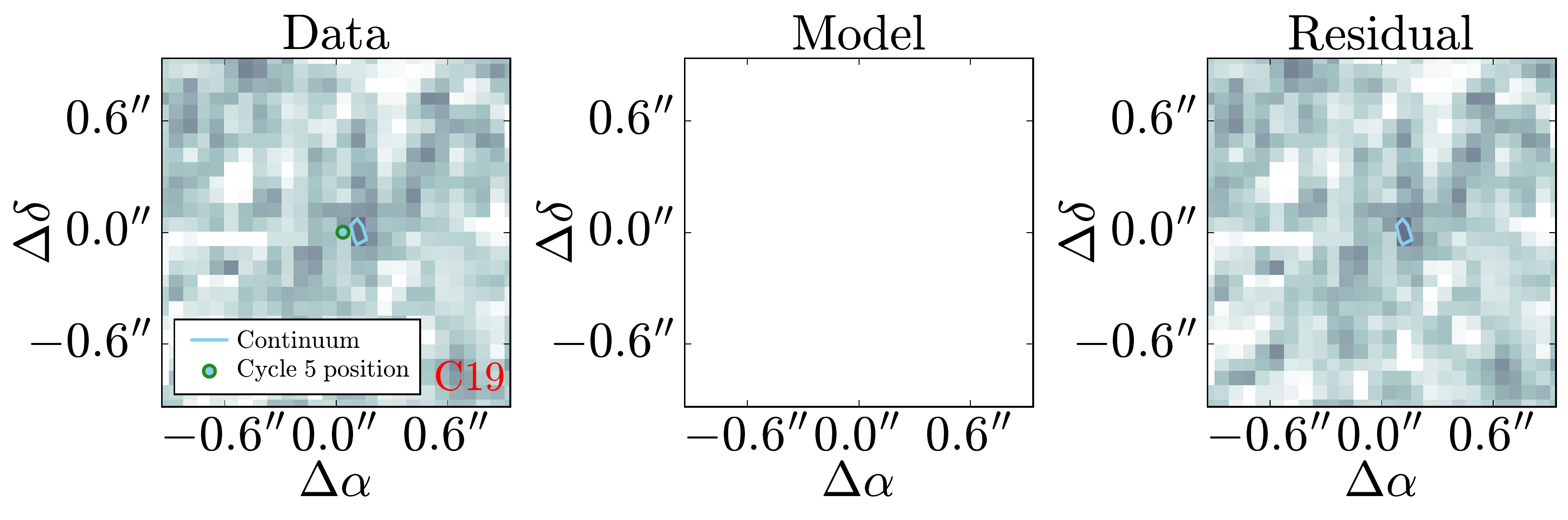}
\end{framed}
\end{subfigure}
\begin{subfigure}{0.45\textwidth}
\begin{framed}
\includegraphics[width=\textwidth]{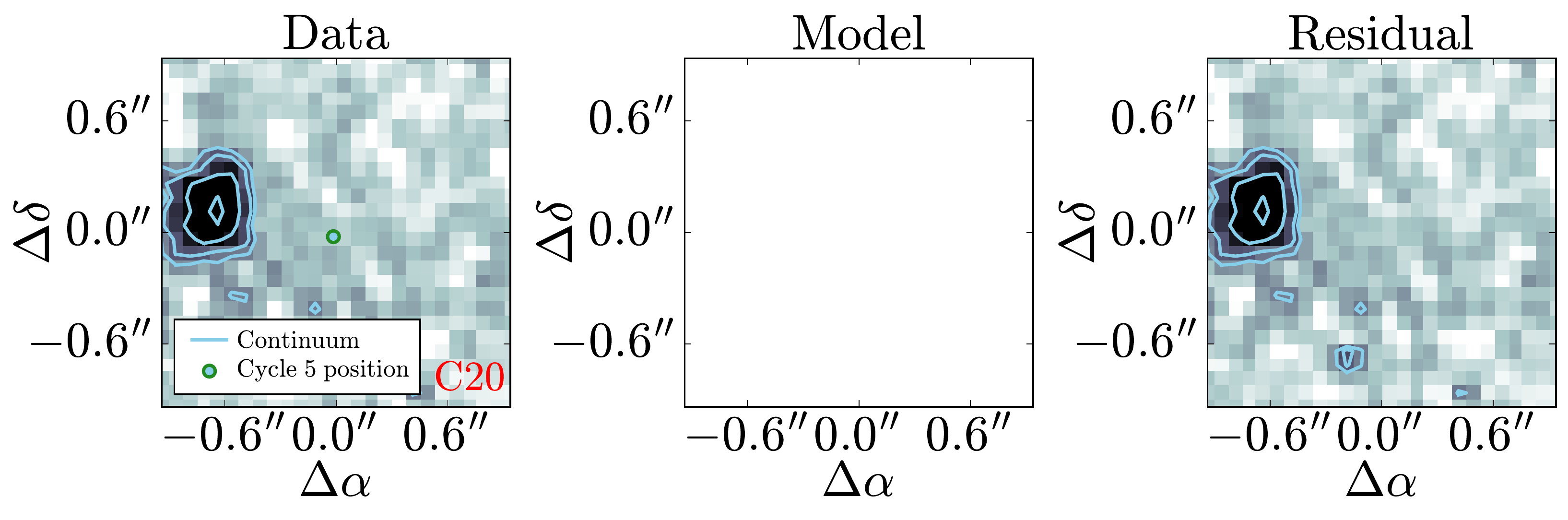}
\end{framed}
\end{subfigure}
\begin{subfigure}{0.45\textwidth}
\begin{framed}
\includegraphics[width=\textwidth]{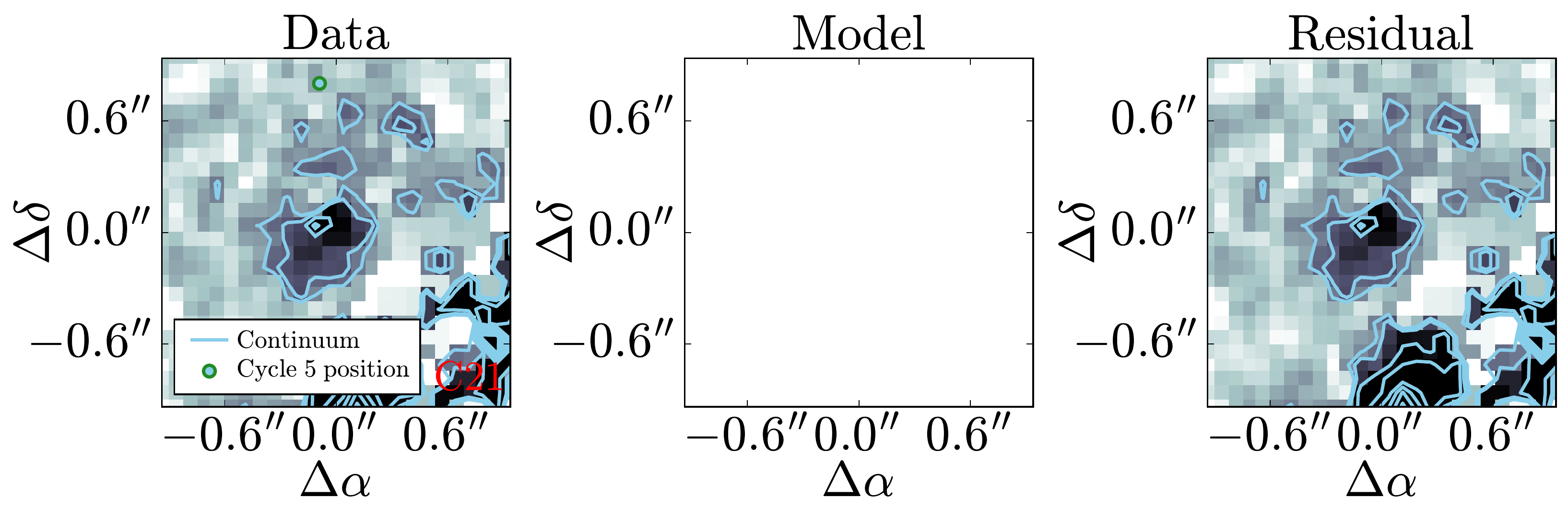}
\end{framed}
\end{subfigure}
\begin{subfigure}{0.45\textwidth}
\begin{framed}
\includegraphics[width=\textwidth]{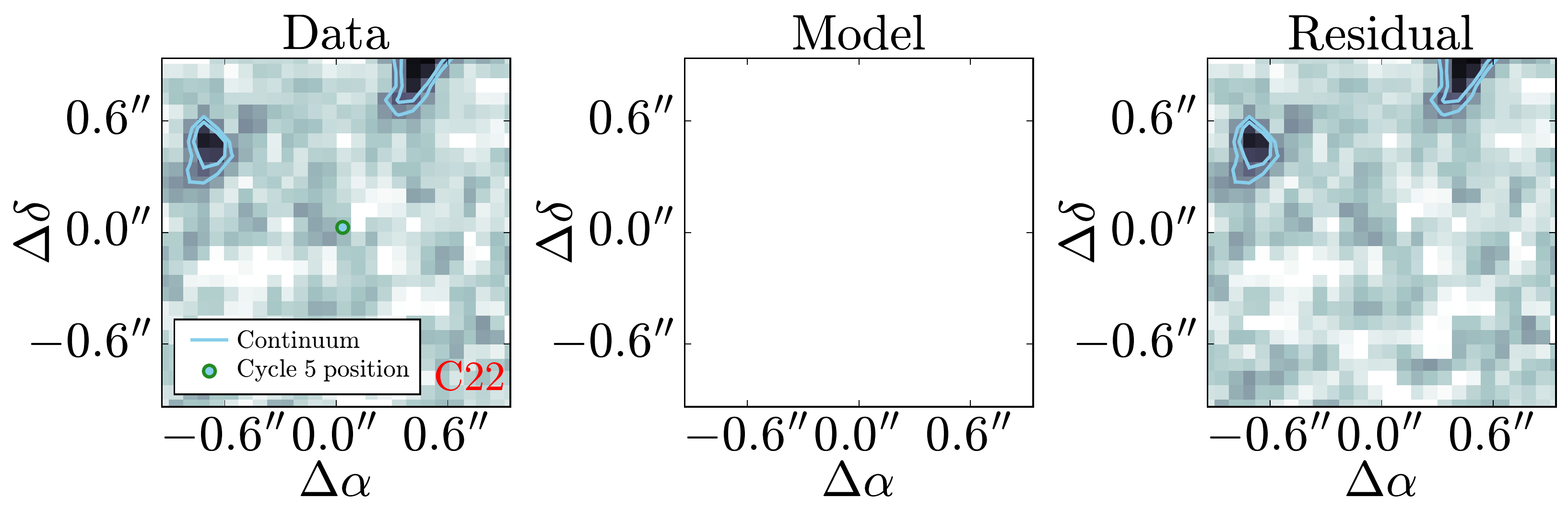}
\end{framed}
\end{subfigure}
\begin{subfigure}{0.45\textwidth}
\begin{framed}
\includegraphics[width=\textwidth]{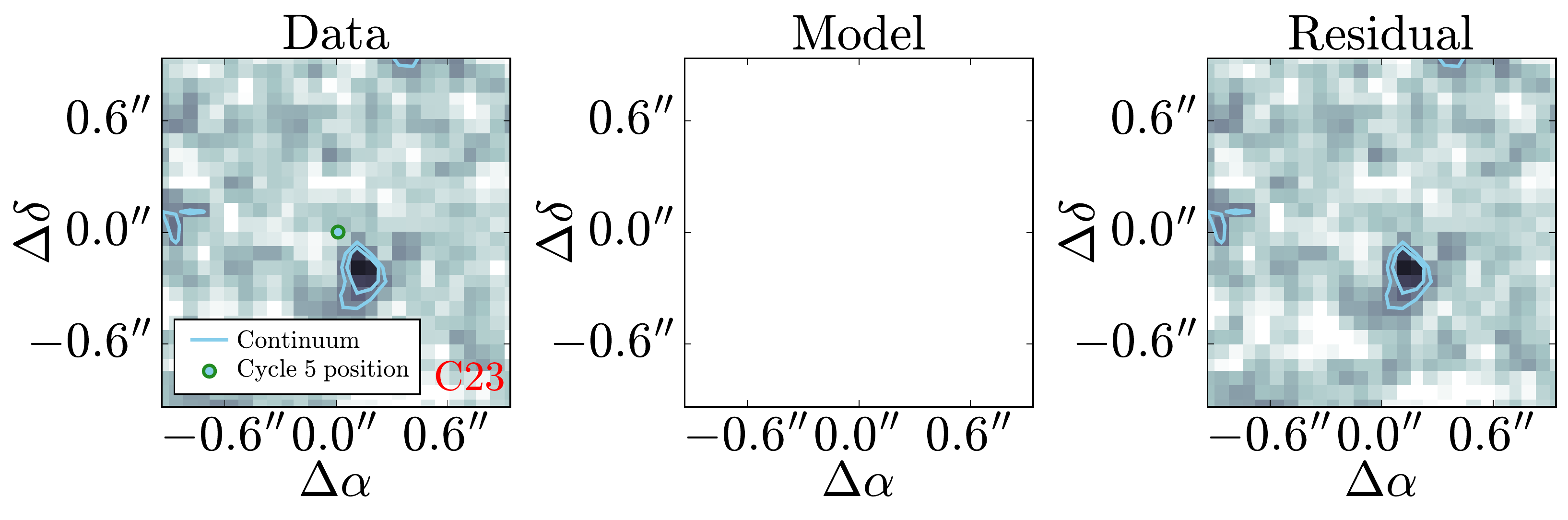}
\end{framed}
\end{subfigure}
\begin{subfigure}{0.45\textwidth}
\begin{framed}
\includegraphics[width=\textwidth]{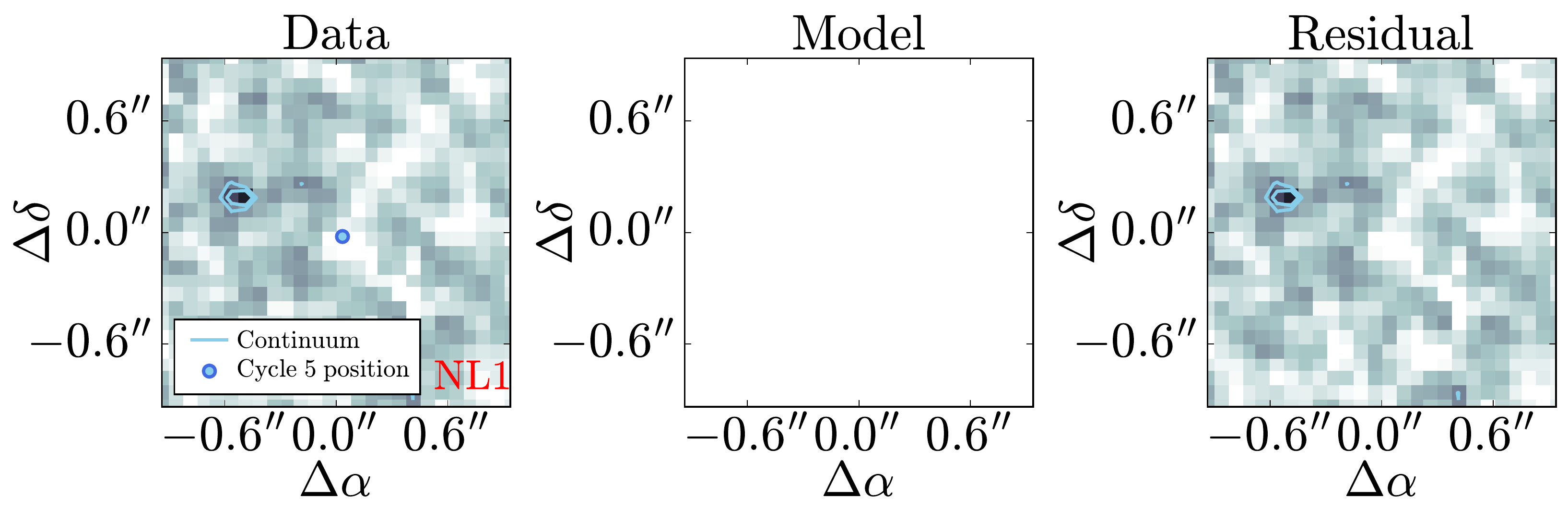}
\end{framed}
\end{subfigure}
\begin{subfigure}{0.45\textwidth}
\begin{framed}
\includegraphics[width=\textwidth]{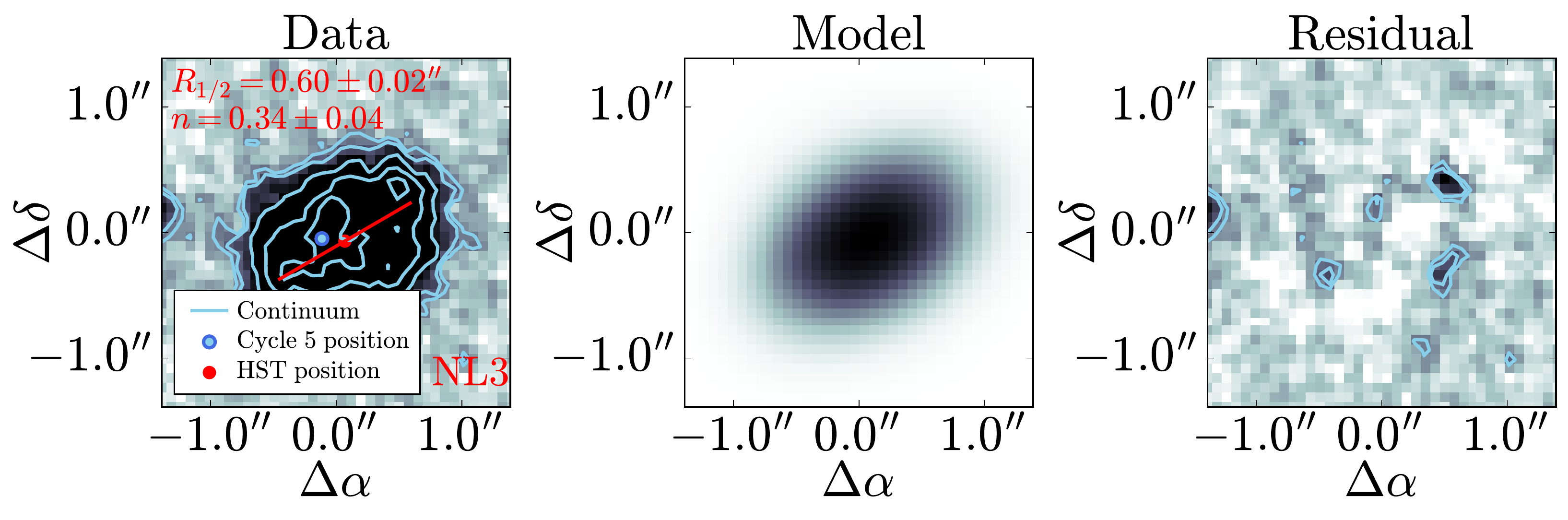}
\end{framed}
\end{subfigure}
\begin{subfigure}{0.45\textwidth}
\begin{framed}
\includegraphics[width=\textwidth]{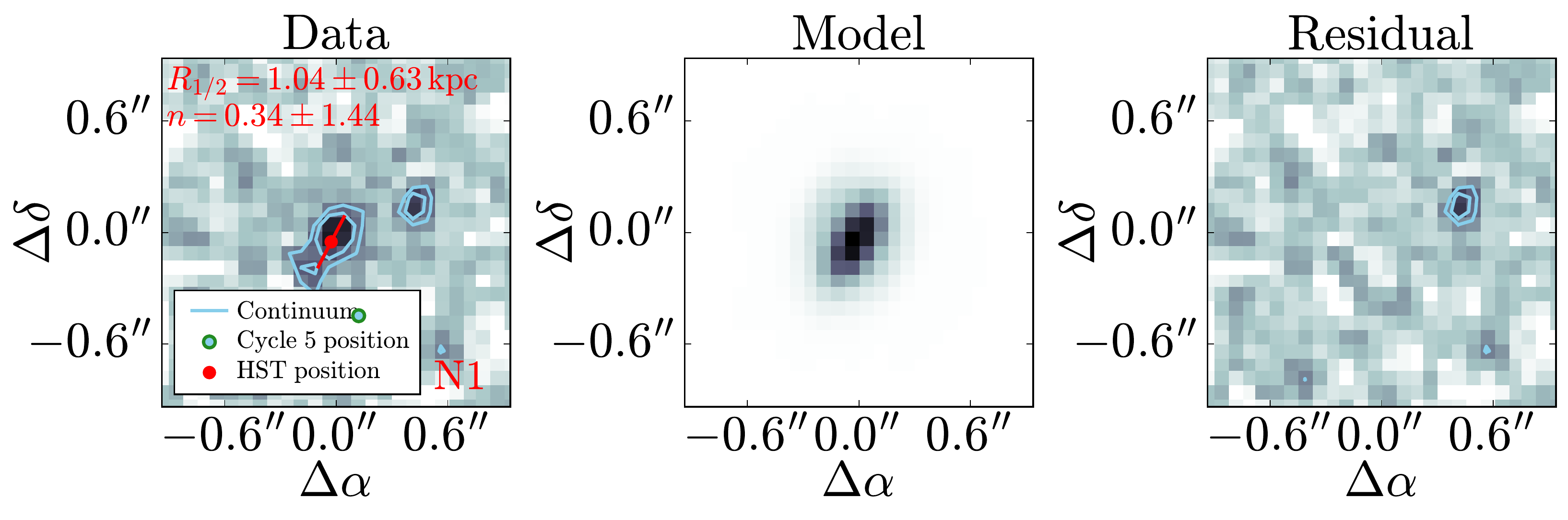}
\end{framed}
\end{subfigure}
\begin{subfigure}{0.45\textwidth}
\begin{framed}
\includegraphics[width=\textwidth]{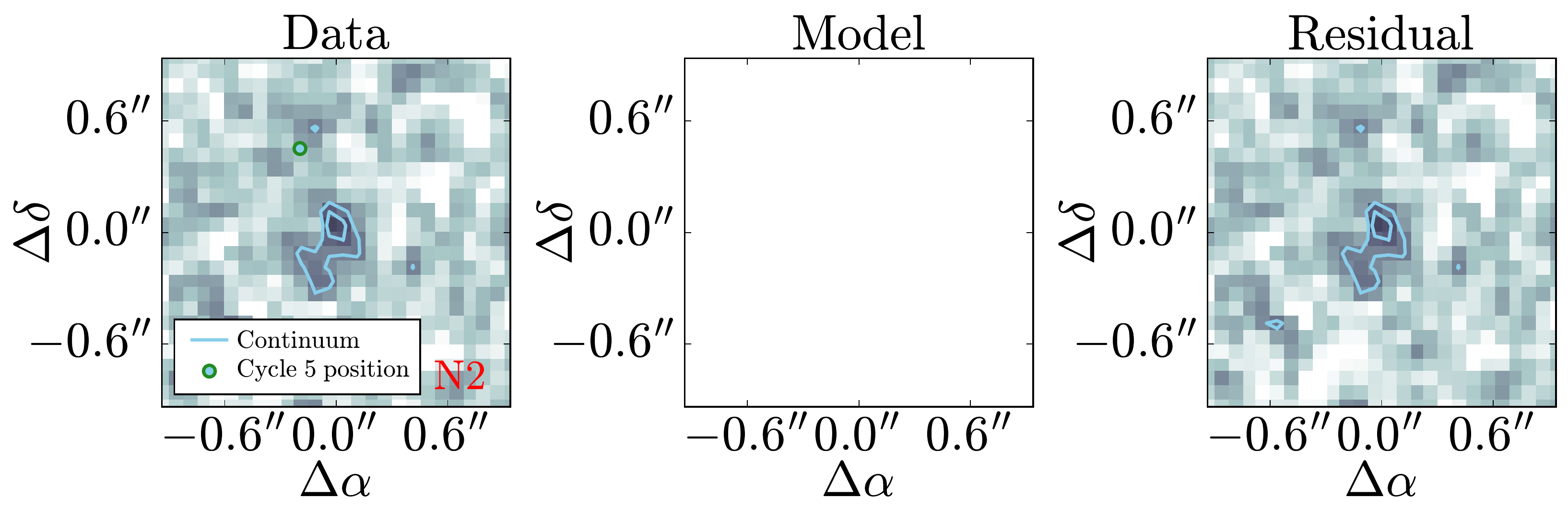}
\end{framed}
\end{subfigure}
\begin{subfigure}{0.45\textwidth}
\begin{framed}
\includegraphics[width=\textwidth]{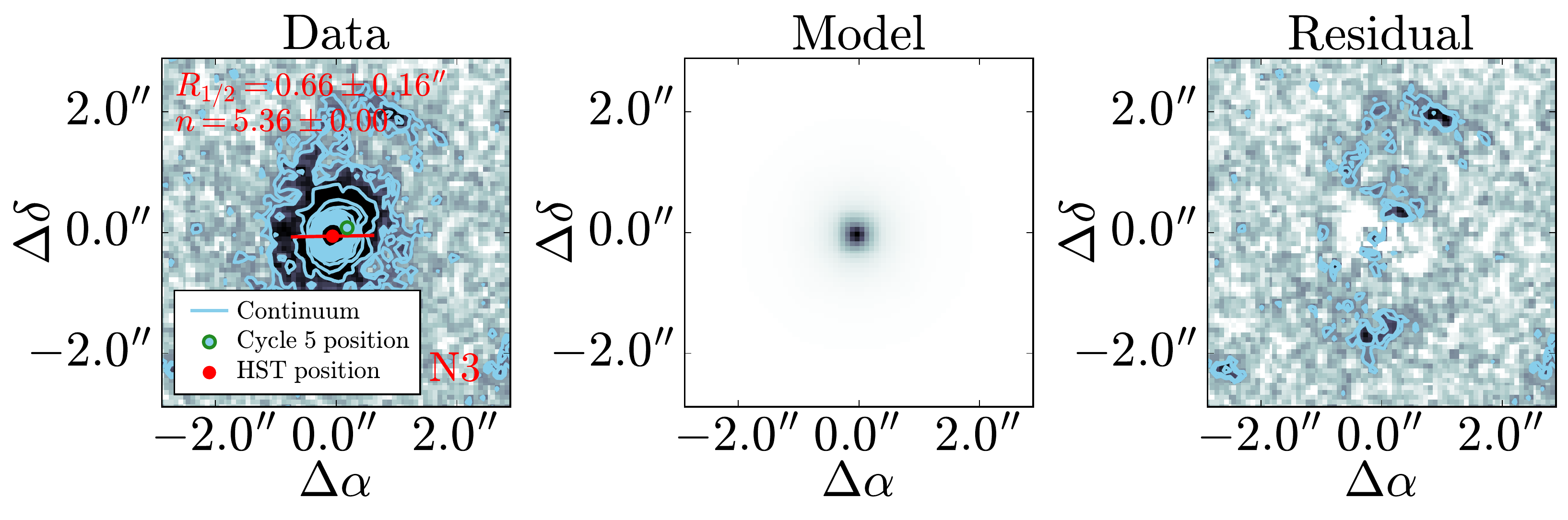}
\end{framed}
\end{subfigure}
\caption{}
\end{figure*}
\renewcommand{\thefigure}{\arabic{figure}}

\renewcommand{\thefigure}{D\arabic{figure} (Cont.)}
\addtocounter{figure}{-1}
\begin{figure*}
\begin{subfigure}{0.45\textwidth}
\begin{framed}
\includegraphics[width=\textwidth]{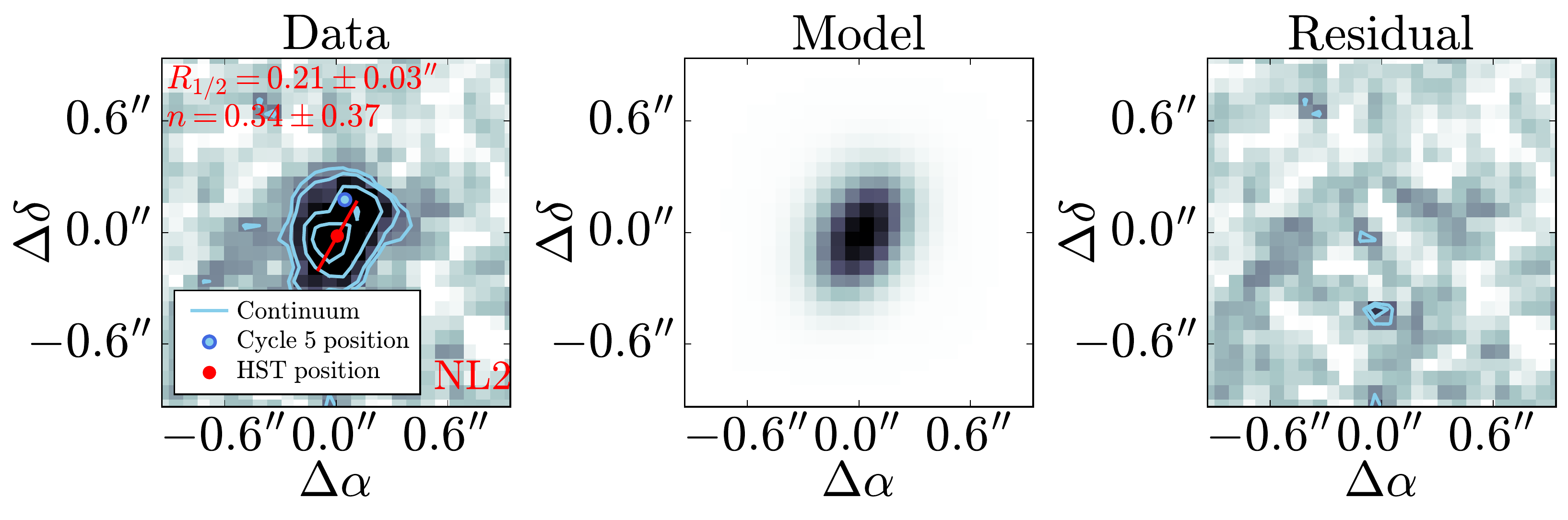}
\end{framed}
\end{subfigure}
\begin{subfigure}{0.45\textwidth}
\begin{framed}
\includegraphics[width=\textwidth]{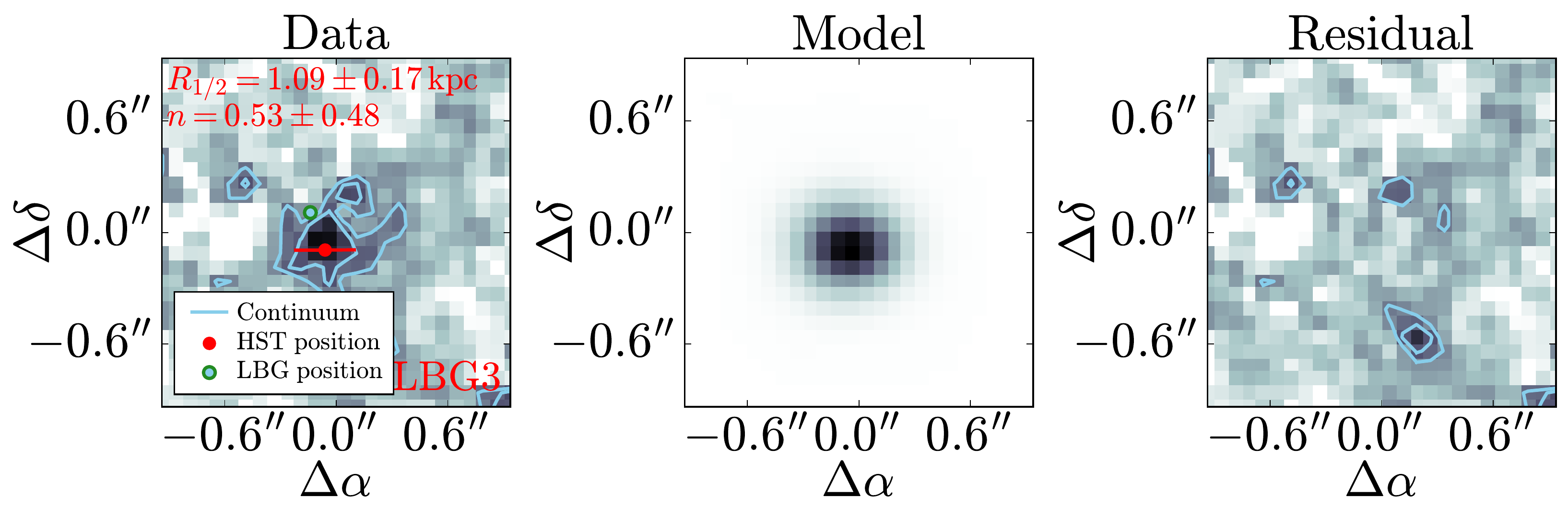}
\end{framed}
\end{subfigure}
\begin{subfigure}{0.45\textwidth}
\begin{framed}
\includegraphics[width=\textwidth]{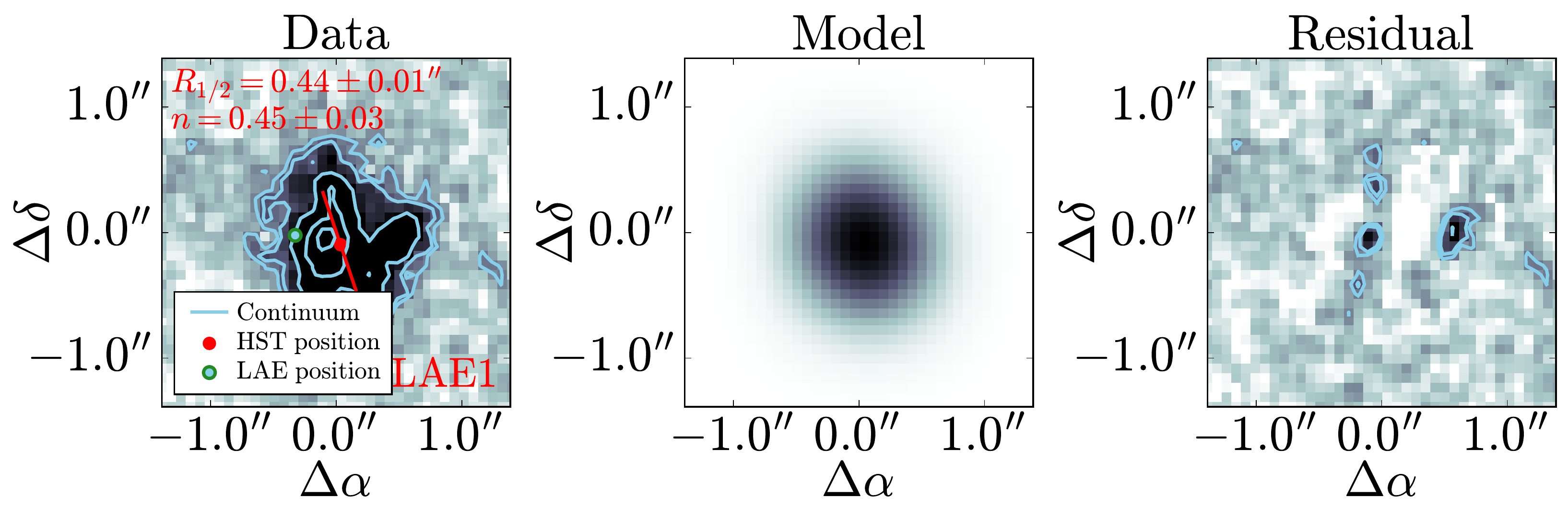}
\end{framed}
\end{subfigure}
\begin{subfigure}{0.45\textwidth}
\begin{framed}
\includegraphics[width=\textwidth]{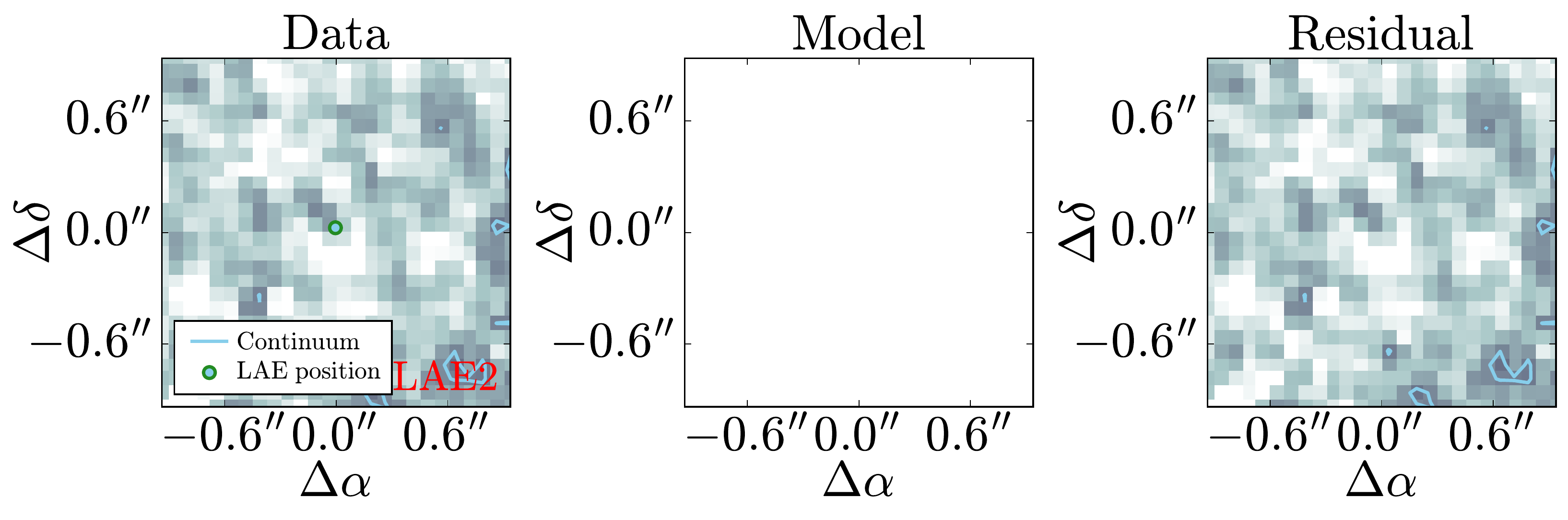}
\end{framed}
\end{subfigure}
\begin{subfigure}{0.45\textwidth}
\begin{framed}
\includegraphics[width=\textwidth]{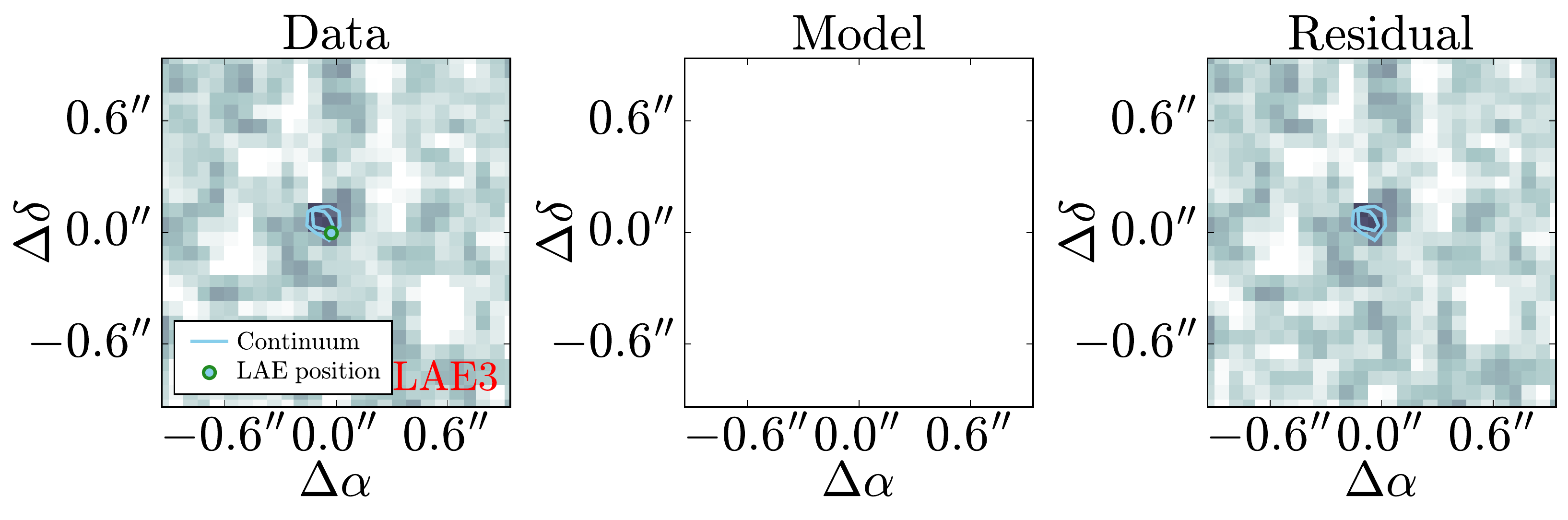}
\end{framed}
\end{subfigure}
\begin{subfigure}{0.45\textwidth}
\begin{framed}
\includegraphics[width=\textwidth]{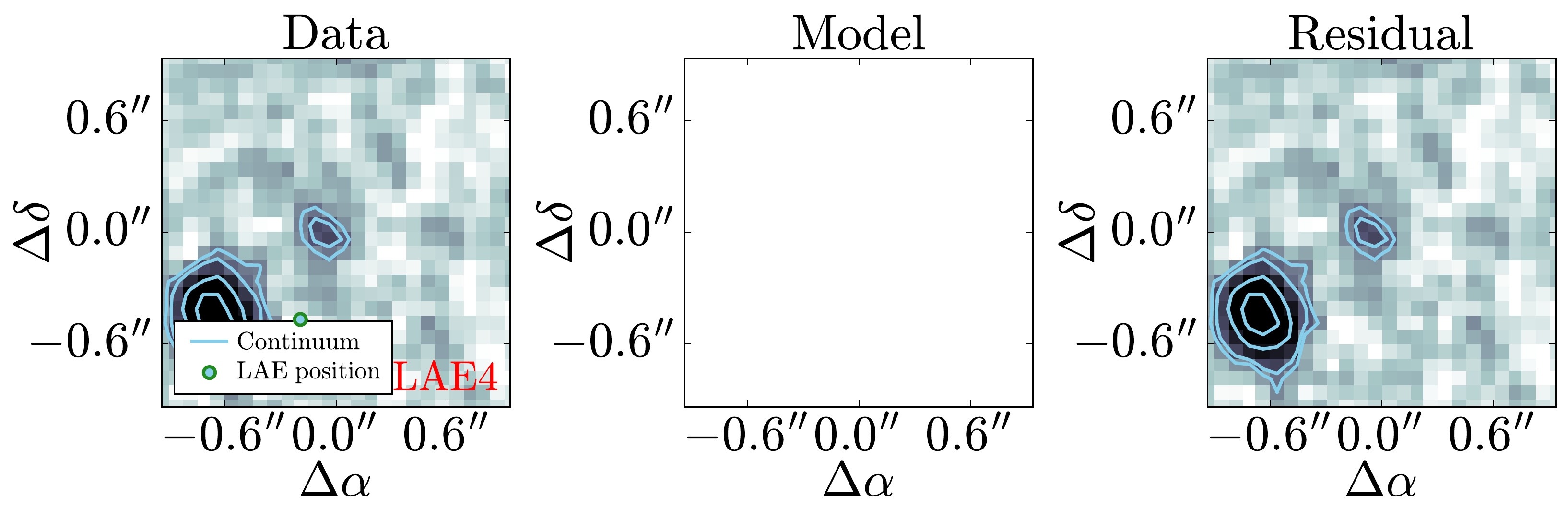}
\end{framed}
\end{subfigure}
\begin{subfigure}{0.45\textwidth}
\begin{framed}
\includegraphics[width=\textwidth]{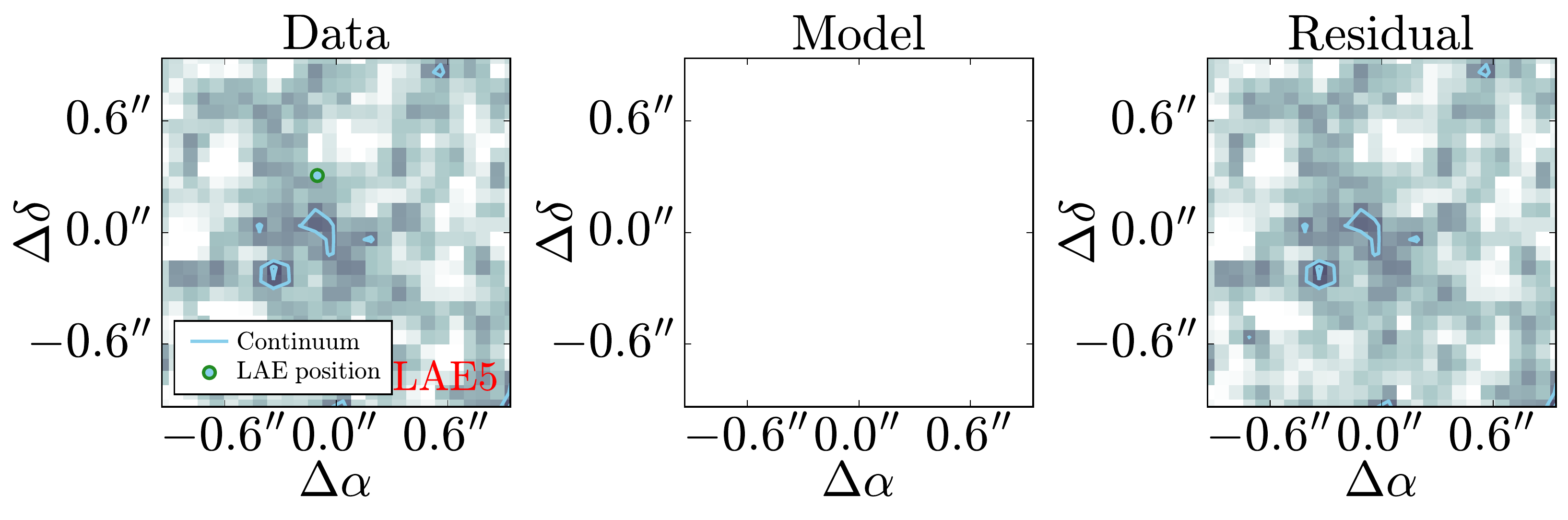}
\end{framed}
\end{subfigure}
\begin{subfigure}{0.45\textwidth}
\begin{framed}
\includegraphics[width=\textwidth]{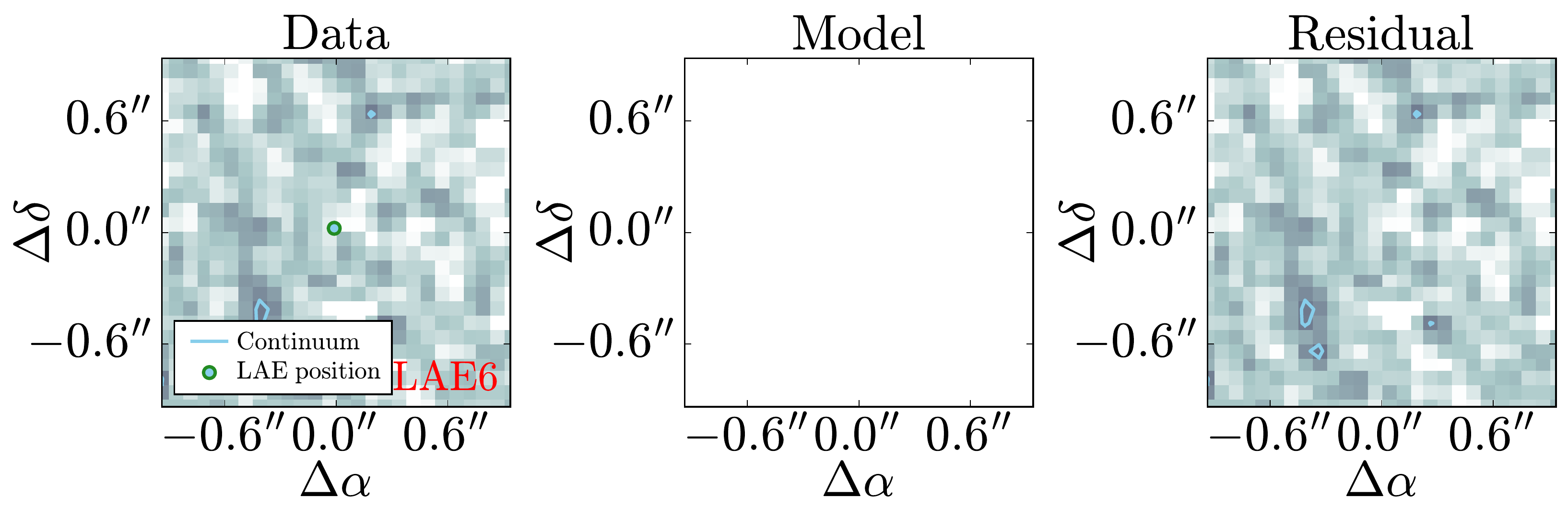}
\end{framed}
\end{subfigure}
\begin{subfigure}{0.45\textwidth}
\begin{framed}
\includegraphics[width=\textwidth]{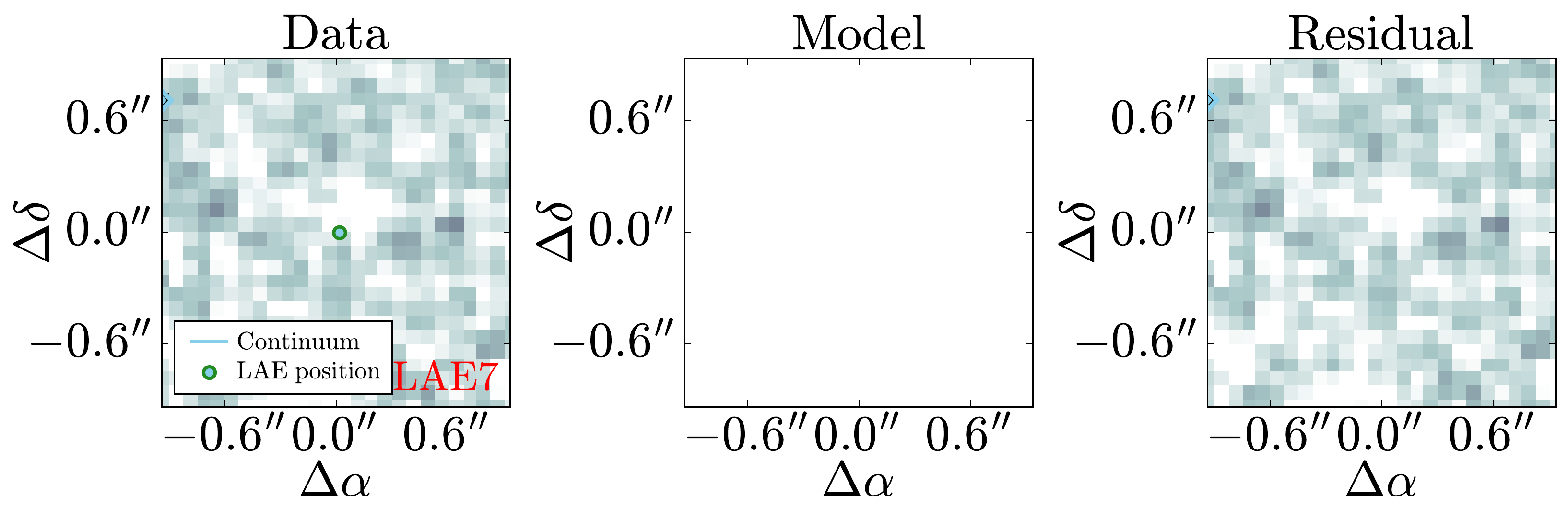}
\end{framed}
\end{subfigure}
\begin{subfigure}{0.45\textwidth}
\begin{framed}
\includegraphics[width=\textwidth]{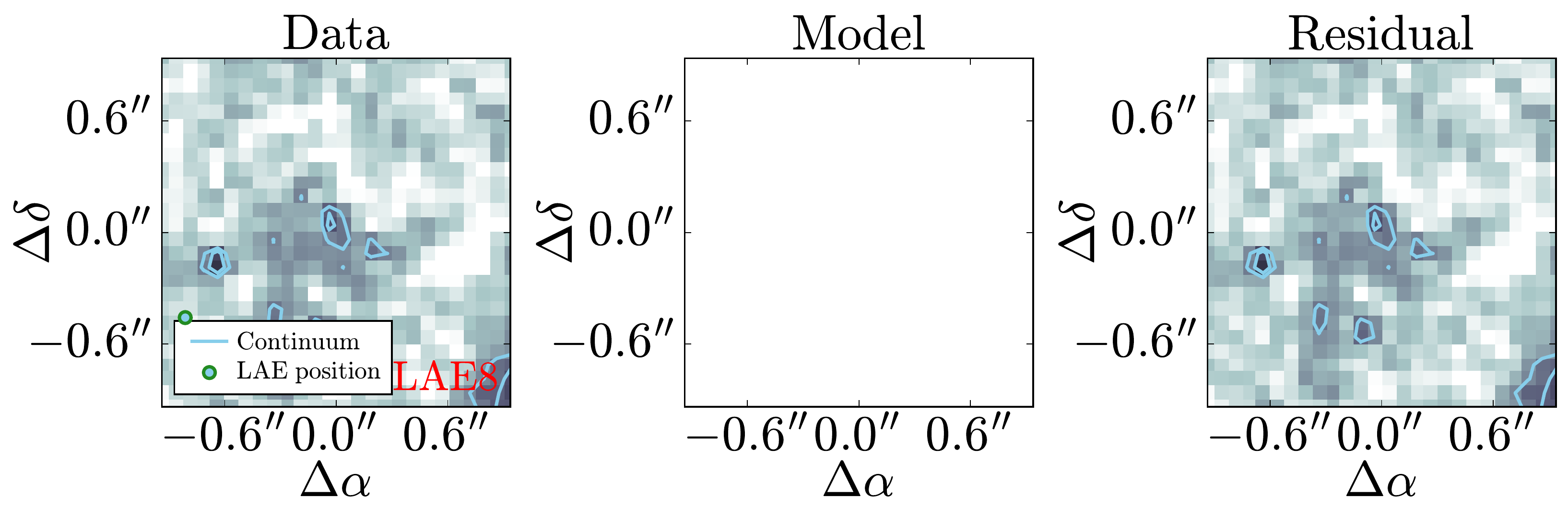}
\end{framed}
\end{subfigure}
\caption{}
\end{figure*}
\renewcommand{\thefigure}{E\arabic{figure}}

\newpage

\section{[C{\sc ii}] profile fits}
\label{appendix4}

\begin{figure*}
\begin{subfigure}{0.45\textwidth}
\begin{framed}
\includegraphics[width=\textwidth]{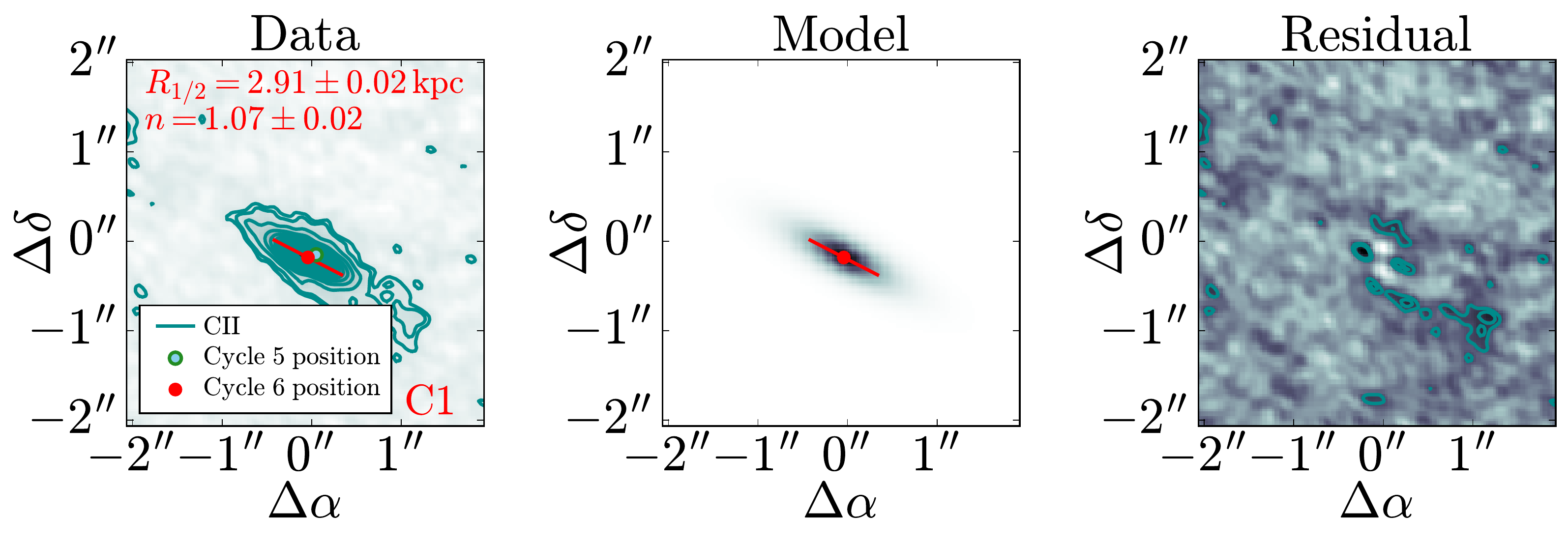}
\end{framed}
\end{subfigure}
\begin{subfigure}{0.45\textwidth}
\begin{framed}
\includegraphics[width=\textwidth]{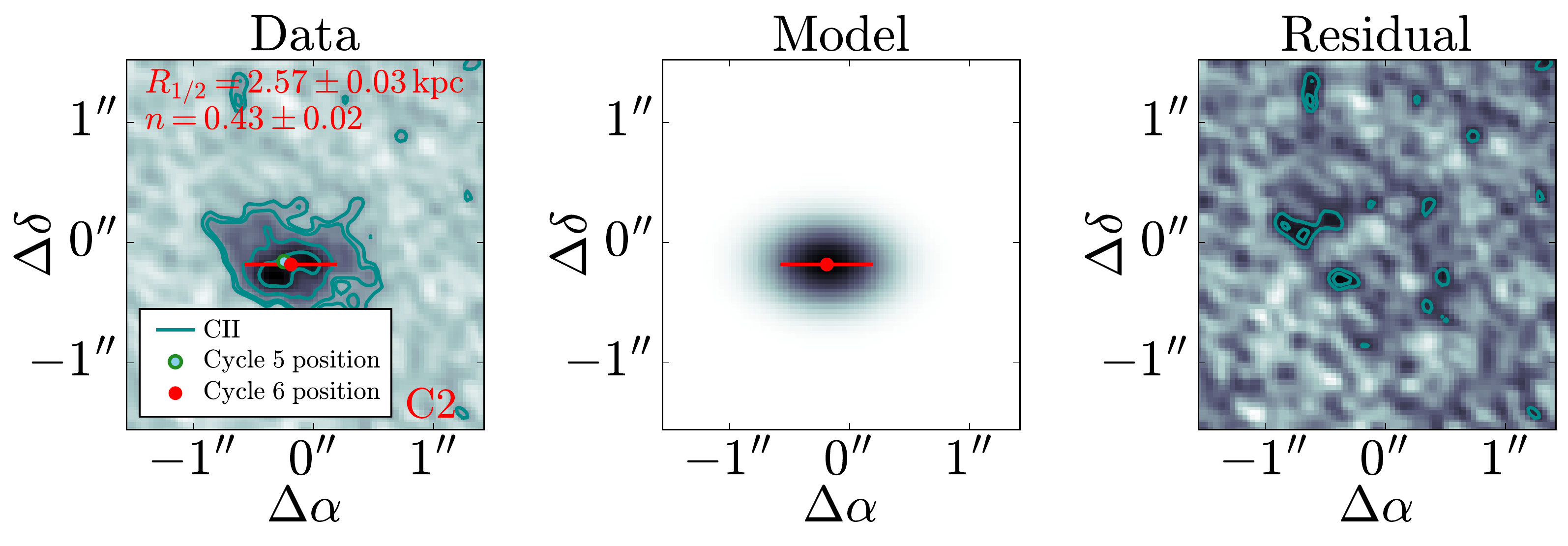}
\end{framed}
\end{subfigure}
\begin{subfigure}{0.45\textwidth}
\begin{framed}
\includegraphics[width=\textwidth]{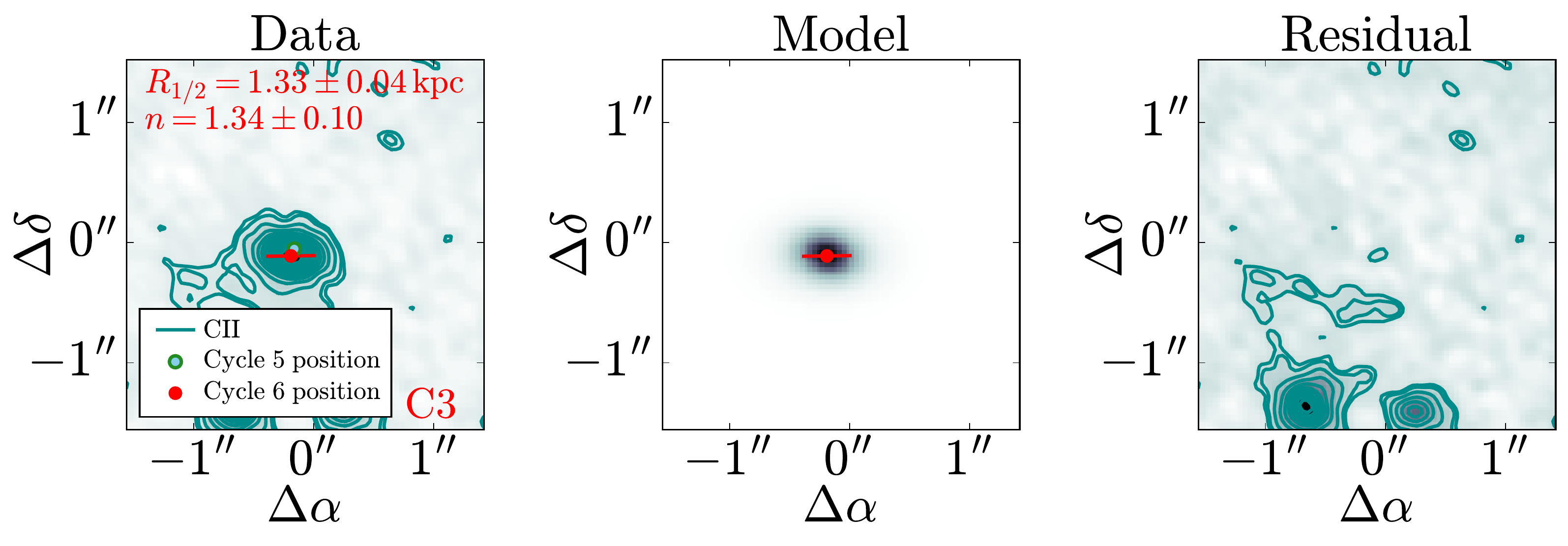}
\end{framed}
\end{subfigure}
\begin{subfigure}{0.45\textwidth}
\begin{framed}
\includegraphics[width=\textwidth]{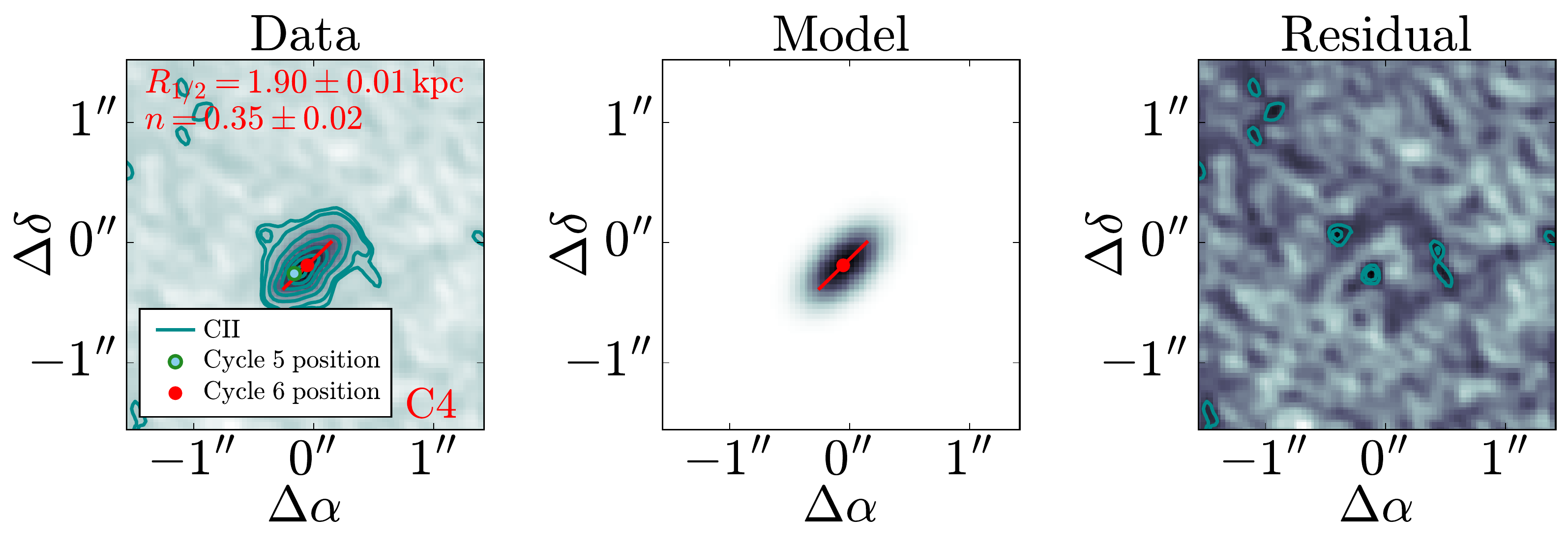}
\end{framed}
\end{subfigure}
\begin{subfigure}{0.45\textwidth}
\begin{framed}
\includegraphics[width=\textwidth]{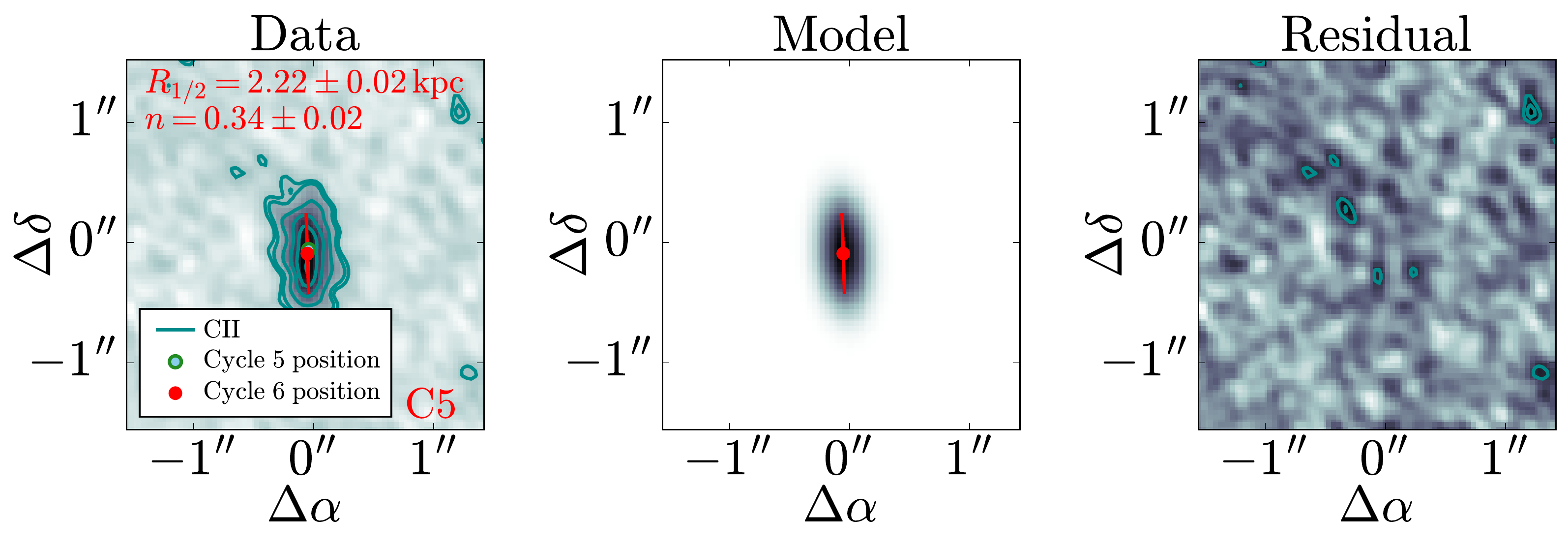}
\end{framed}
\end{subfigure}
\begin{subfigure}{0.45\textwidth}
\begin{framed}
\includegraphics[width=\textwidth]{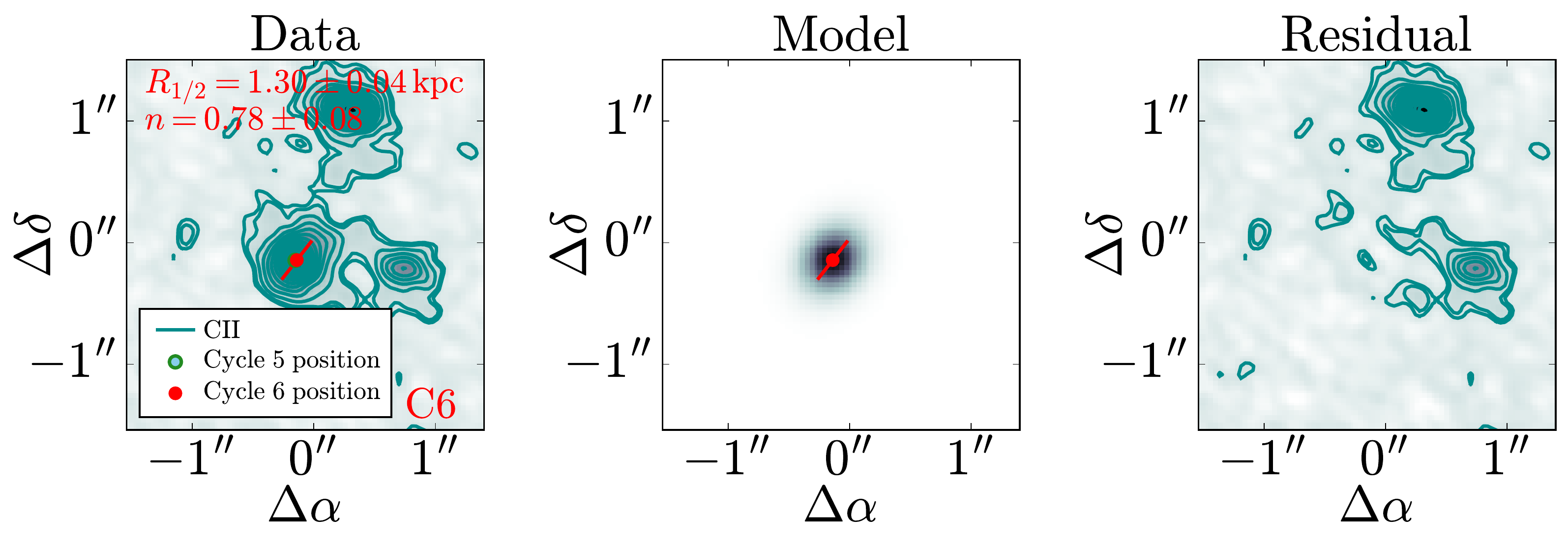}
\end{framed}
\end{subfigure}
\begin{subfigure}{0.45\textwidth}
\begin{framed}
\includegraphics[width=\textwidth]{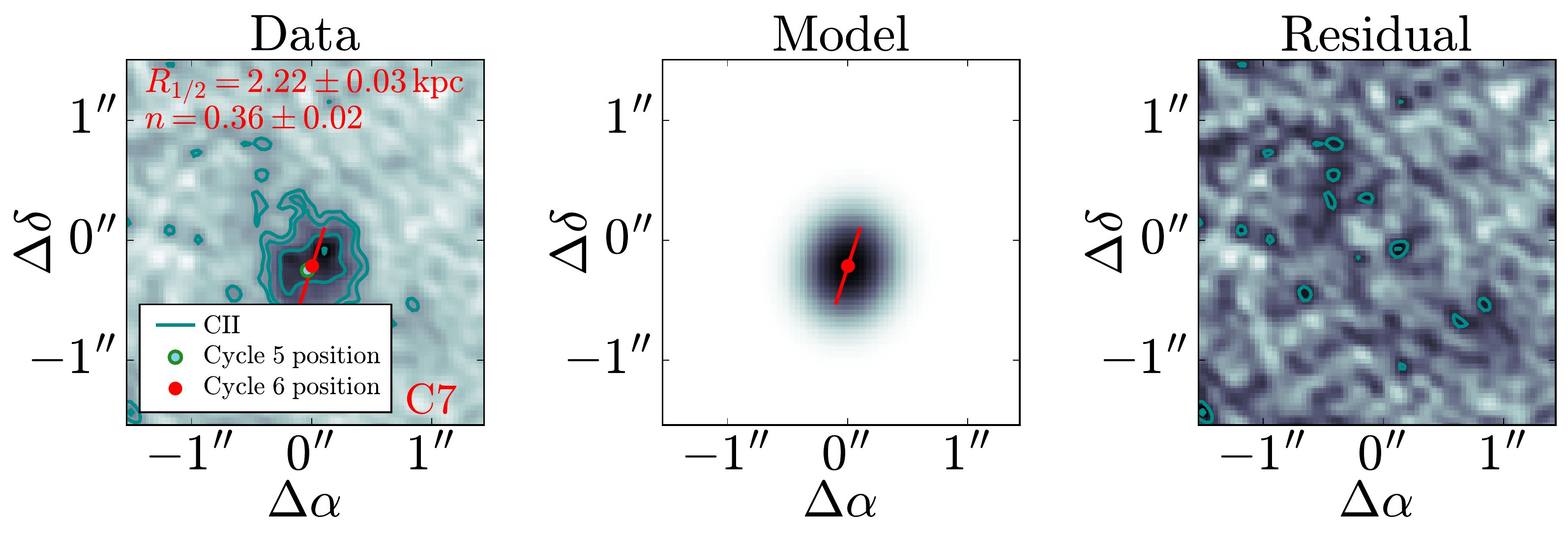}
\end{framed}
\end{subfigure}
\begin{subfigure}{0.45\textwidth}
\begin{framed}
\includegraphics[width=\textwidth]{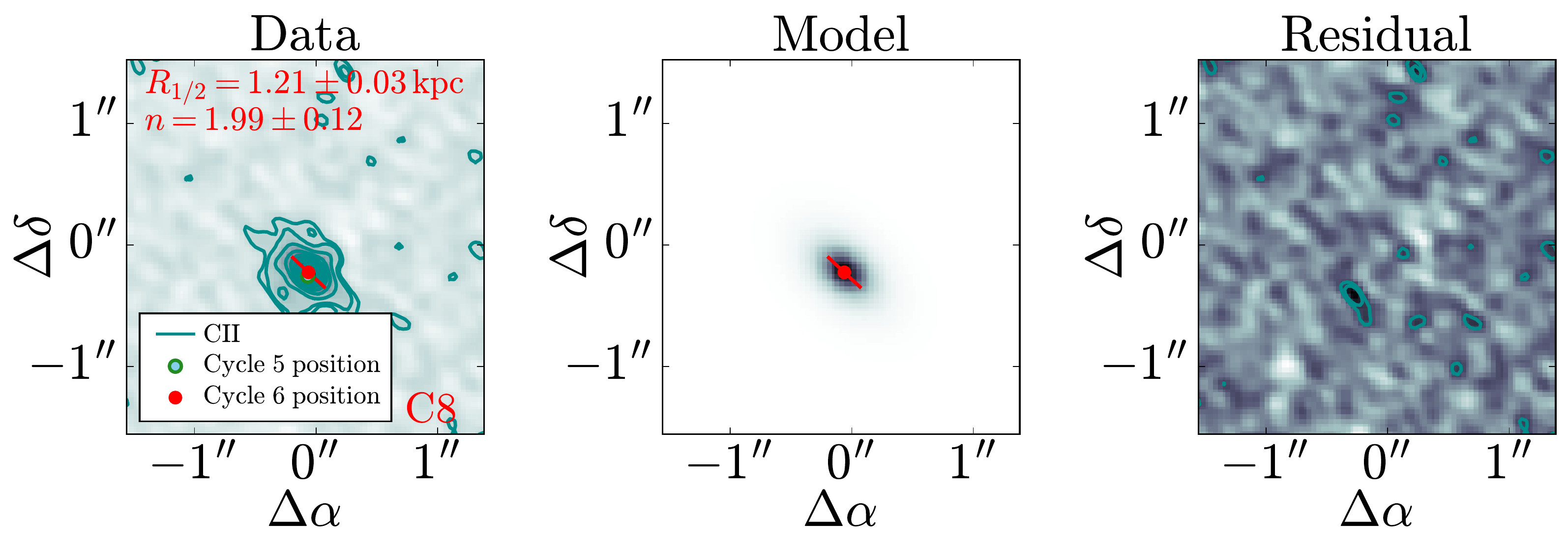}
\end{framed}
\end{subfigure}
\begin{subfigure}{0.45\textwidth}
\begin{framed}
\includegraphics[width=\textwidth]{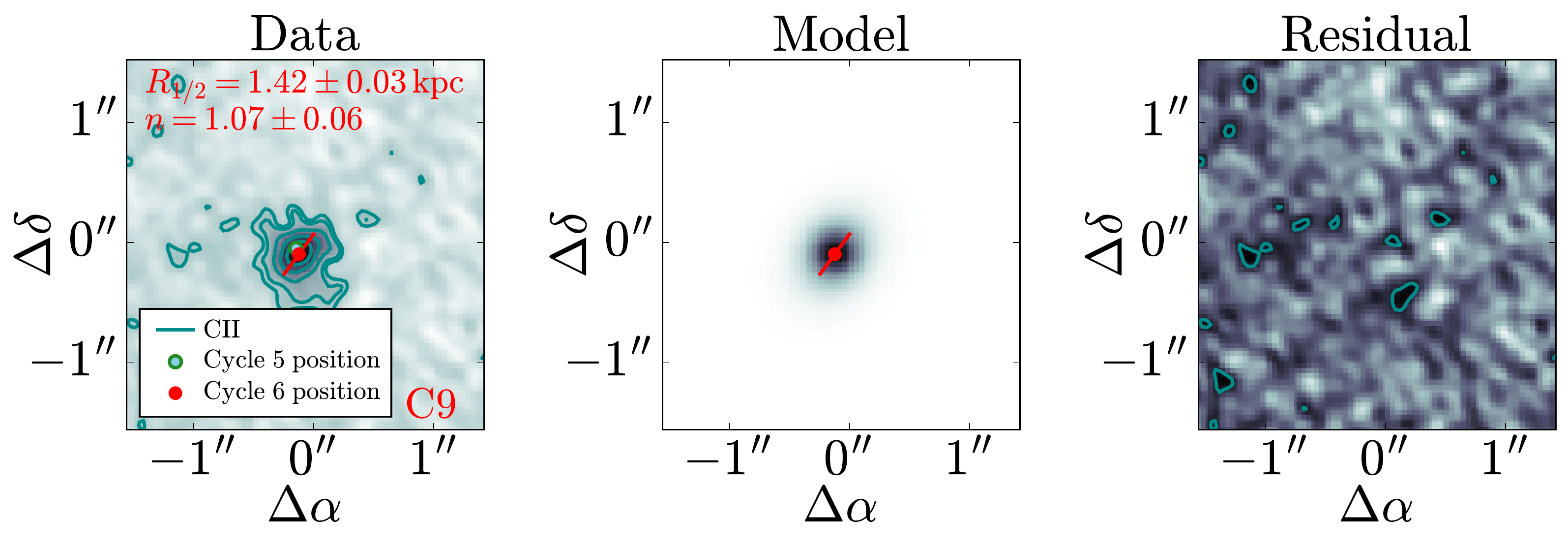}
\end{framed}
\end{subfigure}
\begin{subfigure}{0.45\textwidth}
\begin{framed}
\includegraphics[width=\textwidth]{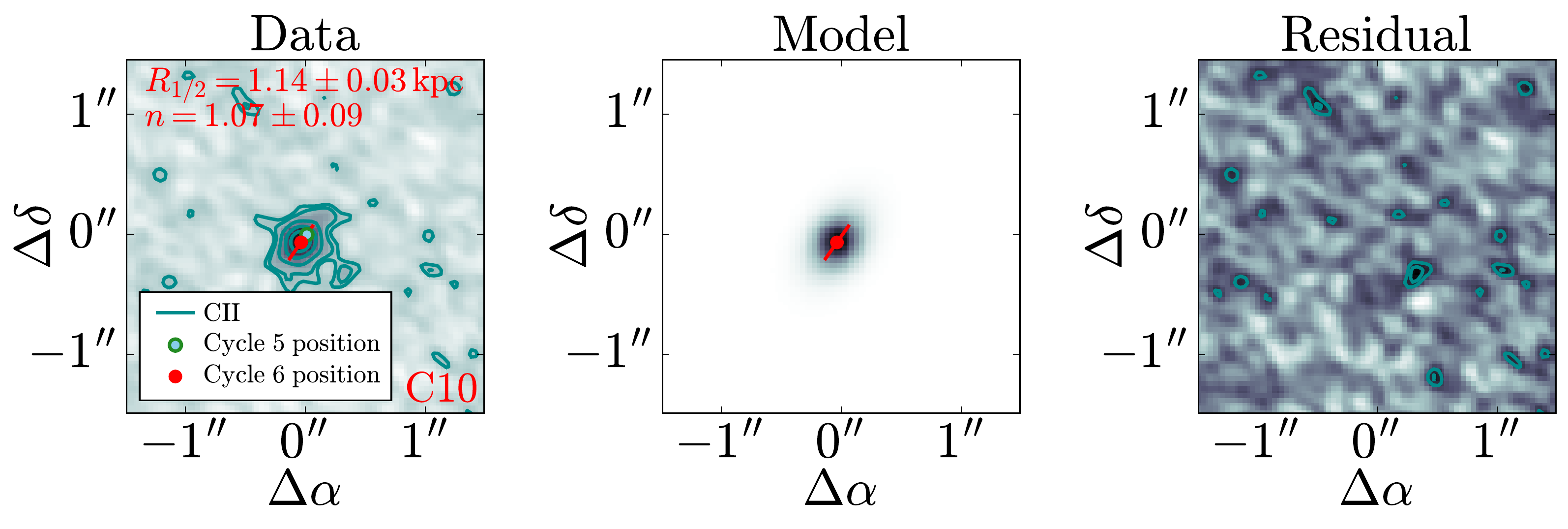}
\end{framed}
\end{subfigure}
\begin{subfigure}{0.45\textwidth}
\begin{framed}
\includegraphics[width=\textwidth]{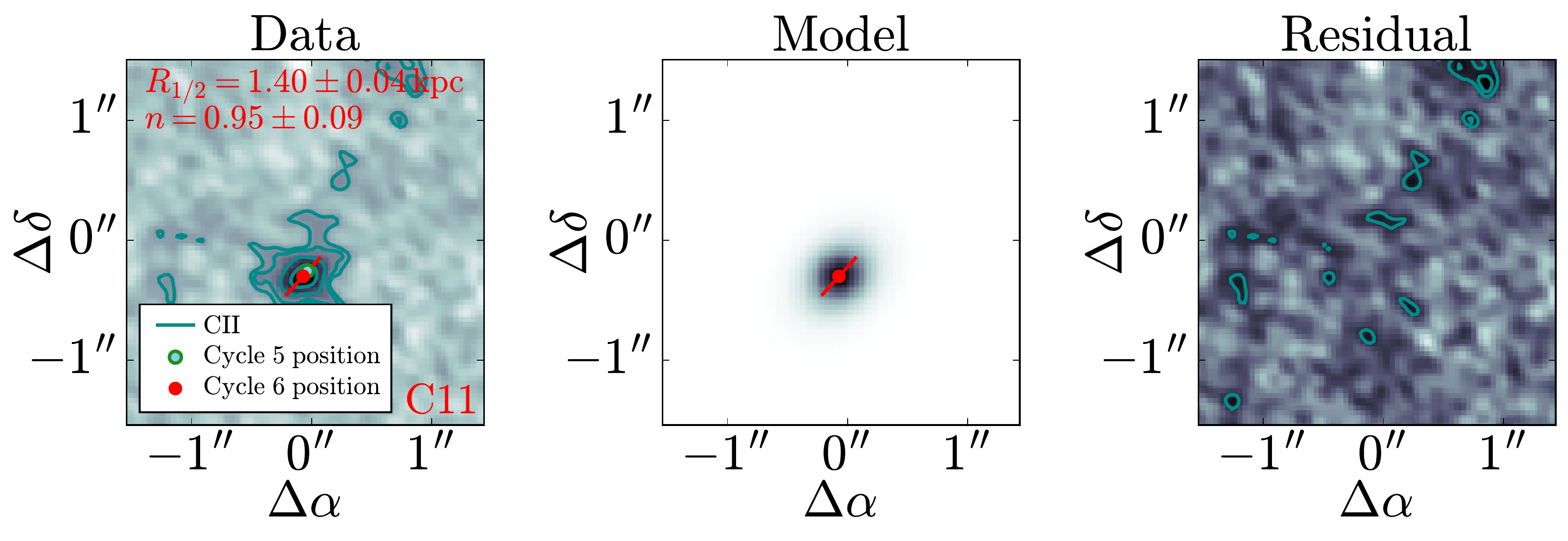}
\end{framed}
\end{subfigure}
\begin{subfigure}{0.45\textwidth}
\begin{framed}
\includegraphics[width=\textwidth]{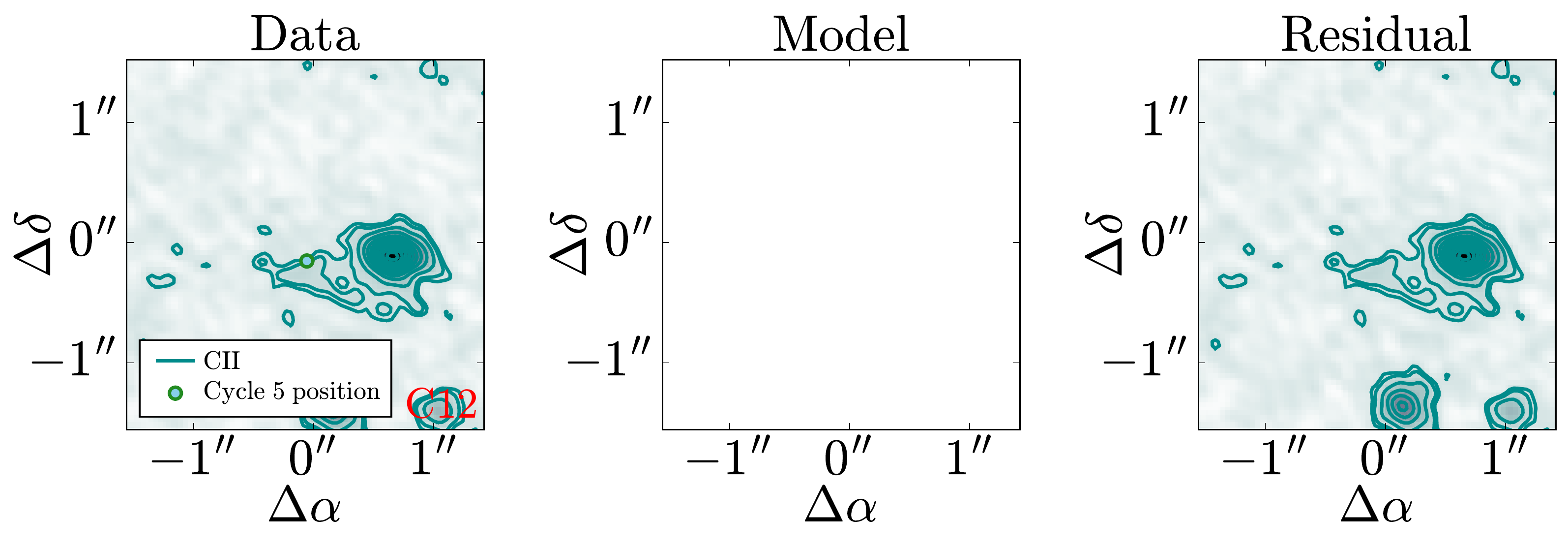}
\end{framed}
\end{subfigure}
\begin{subfigure}{0.45\textwidth}
\begin{framed}
\includegraphics[width=\textwidth]{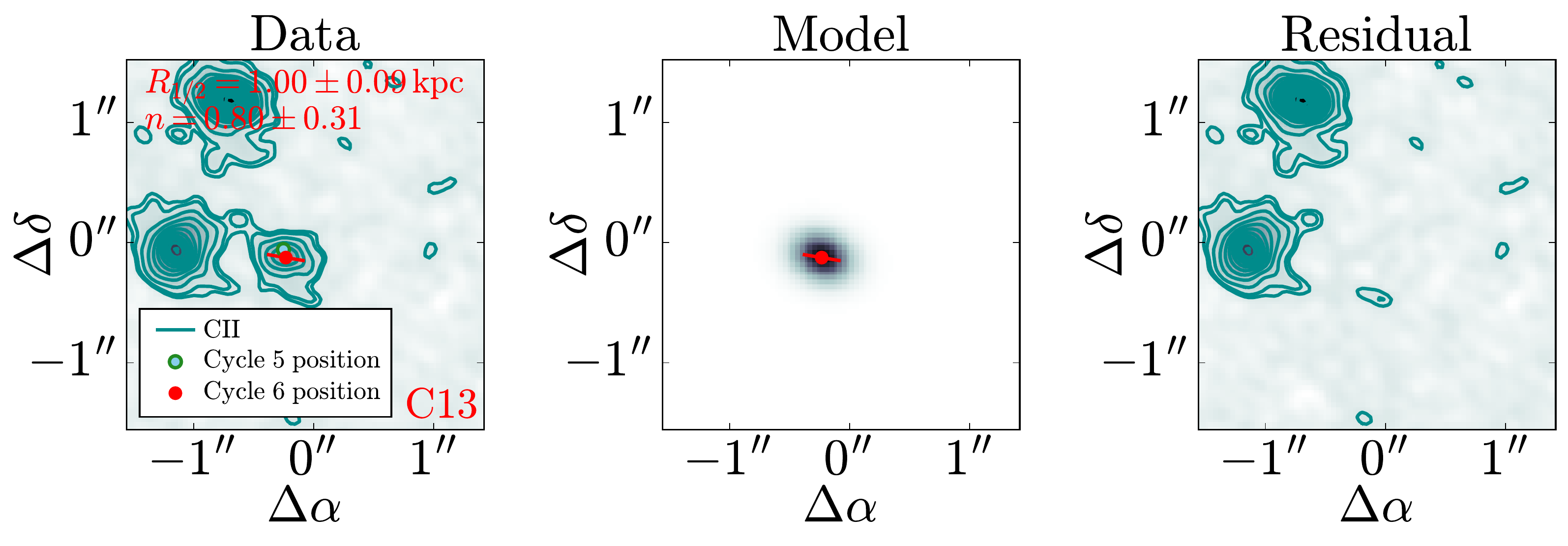}
\end{framed}
\end{subfigure}
\begin{subfigure}{0.45\textwidth}
\begin{framed}
\includegraphics[width=\textwidth]{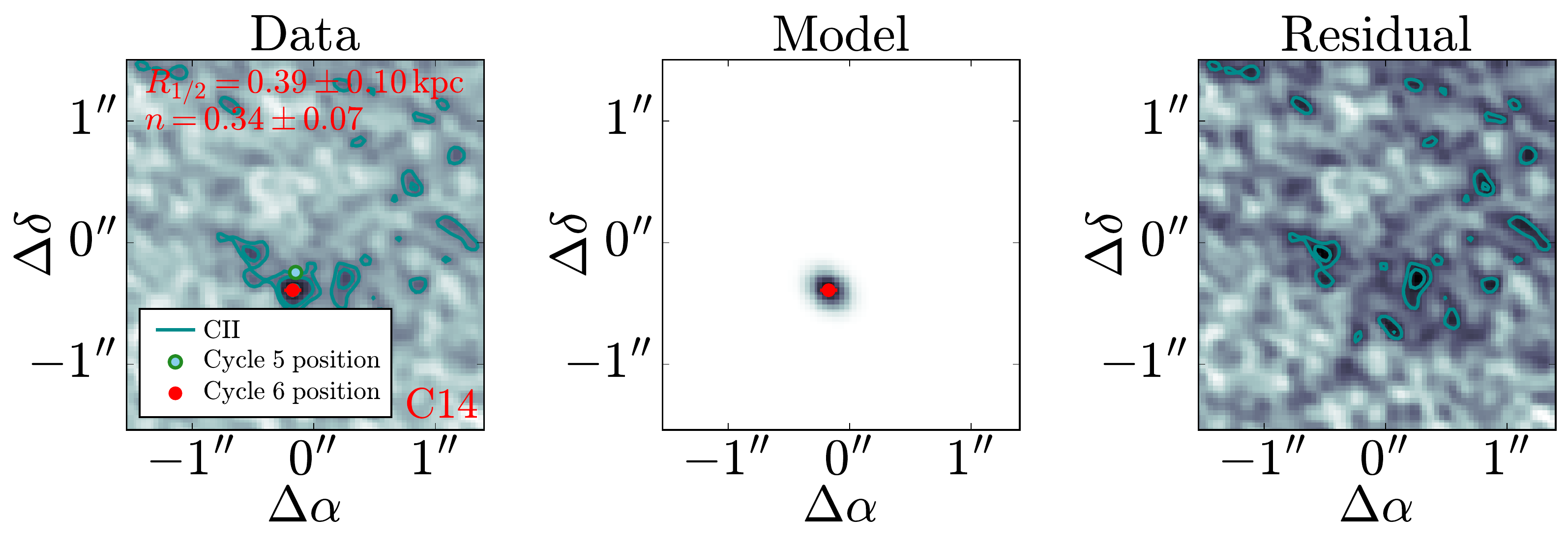}
\end{framed}
\end{subfigure}
\caption{{\it Left:} [C{\sc ii}] images from the high-resolution Band~7 Cycle 6 data. Contours are 2 and 3$\sigma$, then increase in steps of 3$\sigma$. The blue points are positions found in our lower resolution Cycle 5 data, and red points are the centres of the S{\'e}rsic profiles fit to these higher resolution images. The red bars show the lengths of the half-light diametres determined from the best-fitting S{\'e}rsic profiles, while best-fitting half-light radii and S{\'e}rsic indices are shown in the top left. {\it Middle:} Best-fitting model S{\'e}rsic profiles. Sources undetected above 3$\sigma$ were not fitted, and for these cases we leave this panel blank. {\it Right:} Residuals from the S{\'e}rsic profile fits.}
\label{fits_submm}
\end{figure*}

In Fig.~\ref{fits_submm} we provide [C{\sc ii}] cut-outs of our sources obtained from the high-resolution ALMA Band~7 imaging. Line-emission channels were determined from our deeper Cycle 5 data by fitting Gaussian profiles to each spectrum \citep[see][]{hill2020}, and these channels were stacked in the high-resolution data to create moment-0 maps. S{\'e}rsic profiles were fit to all sources with pixels detected in the moment-0 maps above 5 times the local rms, taking into account the instrumental PSF by convolving the model with the best-fit synthesized beam, and half-light radii were estimated from the fits.

The left panels show these stacked images with 2 and 3$\sigma$ countours, then increasing in steps of 3$\sigma$, with positions found in the Cycle 5 data shown as blue points and positions found from the S{\'e}rsic profiles shown as red points. The red bars indicated the sizes of the half-light radii resulting from the S{\'e}rsic profiles, and best-fitting half-light radii in units of kiloparsecs and S{\'e}rsic indices are shown in the top left. The middle panels show our S{\'e}rsic profile models, and the right panels show the residuals. For sources below 3$\sigma$, where we did not attempt to fit S{\'e}rsic profiles, we leave the middle panel blank. 

\renewcommand{\thefigure}{E\arabic{figure} (Cont.)}
\addtocounter{figure}{-1}
\begin{figure*}
\begin{subfigure}{0.45\textwidth}
\begin{framed}
\includegraphics[width=\textwidth]{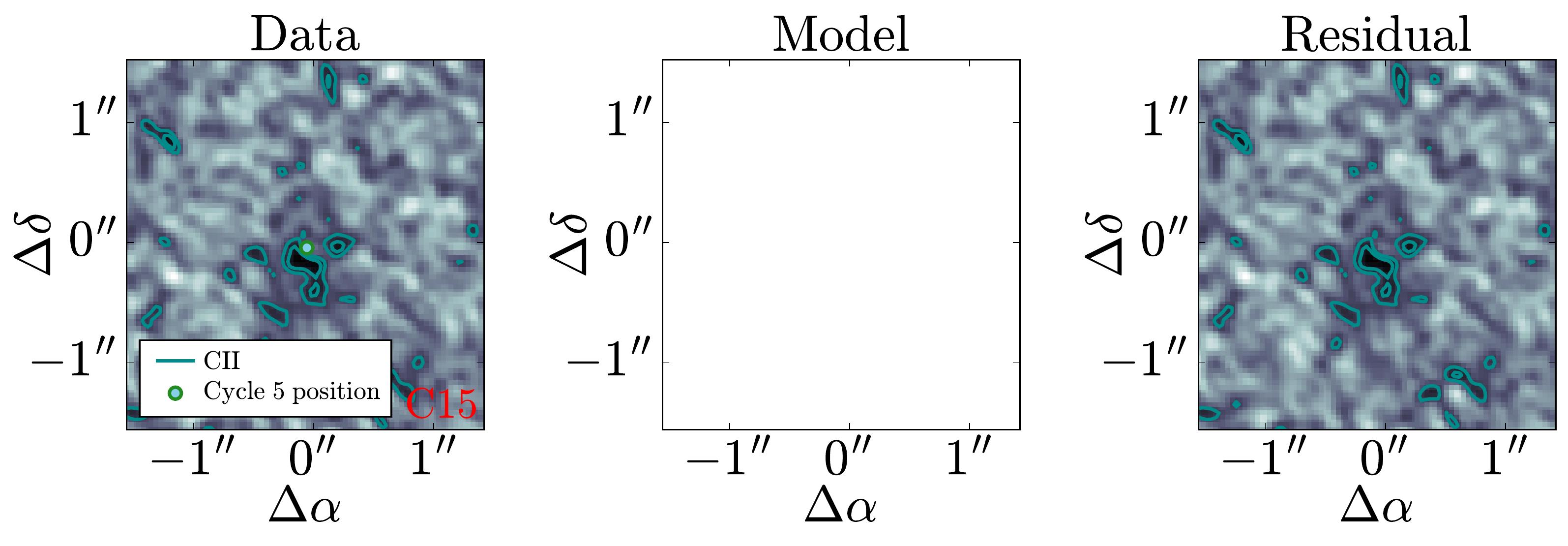}
\end{framed}
\end{subfigure}
\begin{subfigure}{0.45\textwidth}
\begin{framed}
\includegraphics[width=\textwidth]{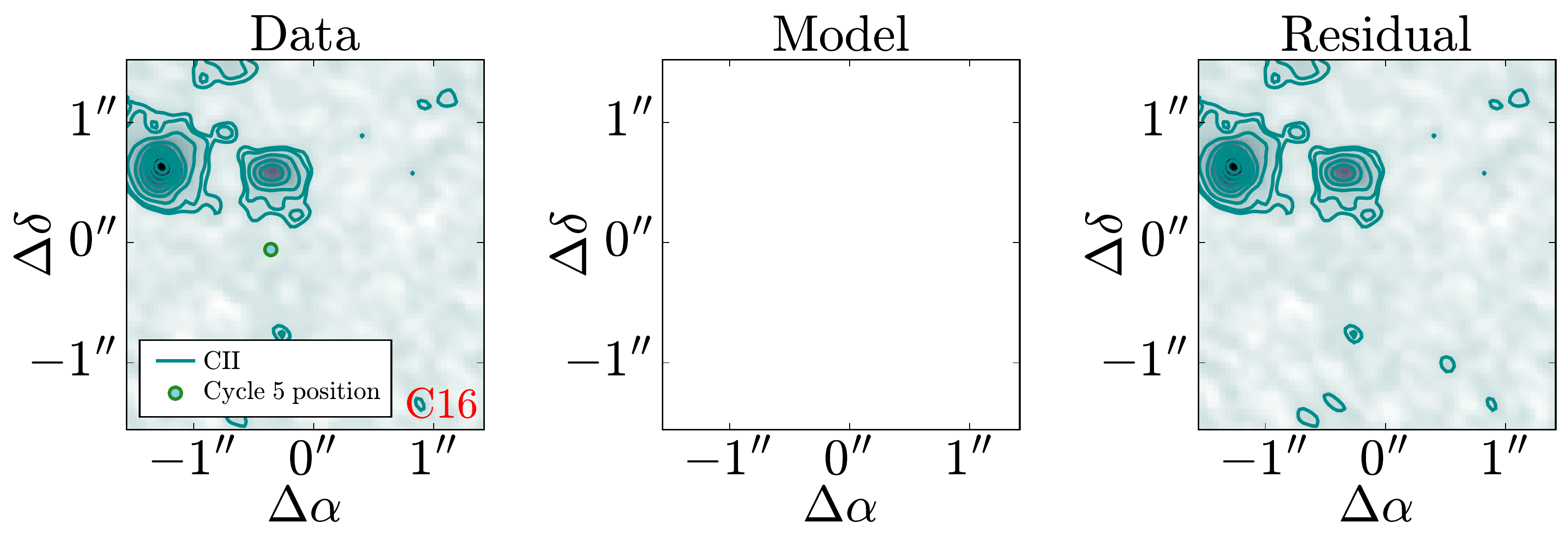}
\end{framed}
\end{subfigure}
\begin{subfigure}{0.45\textwidth}
\begin{framed}
\includegraphics[width=\textwidth]{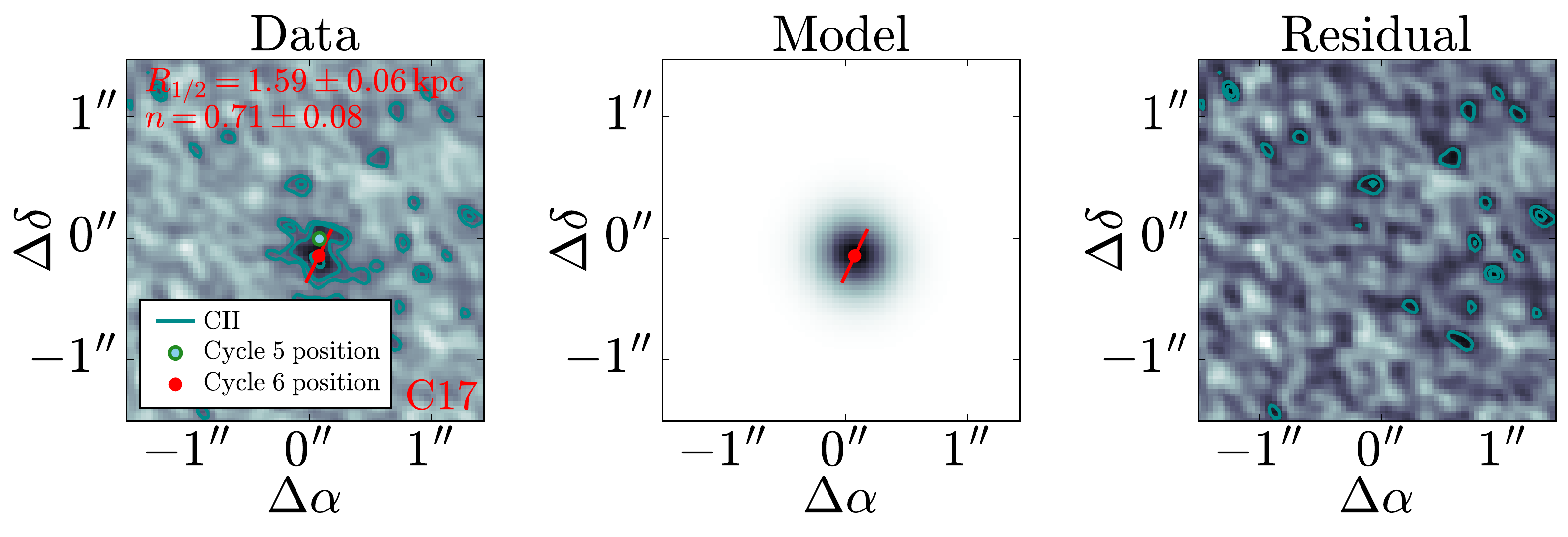}
\end{framed}
\end{subfigure}
\begin{subfigure}{0.45\textwidth}
\begin{framed}
\includegraphics[width=\textwidth]{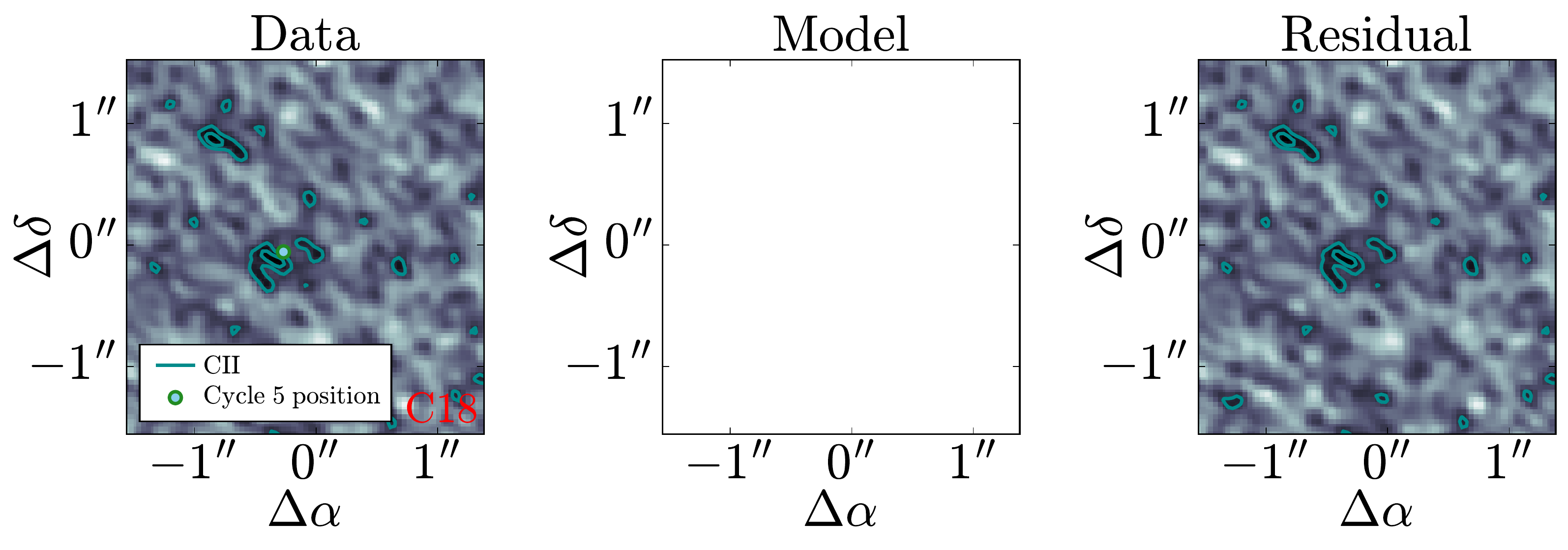}
\end{framed}
\end{subfigure}
\begin{subfigure}{0.45\textwidth}
\begin{framed}
\includegraphics[width=\textwidth]{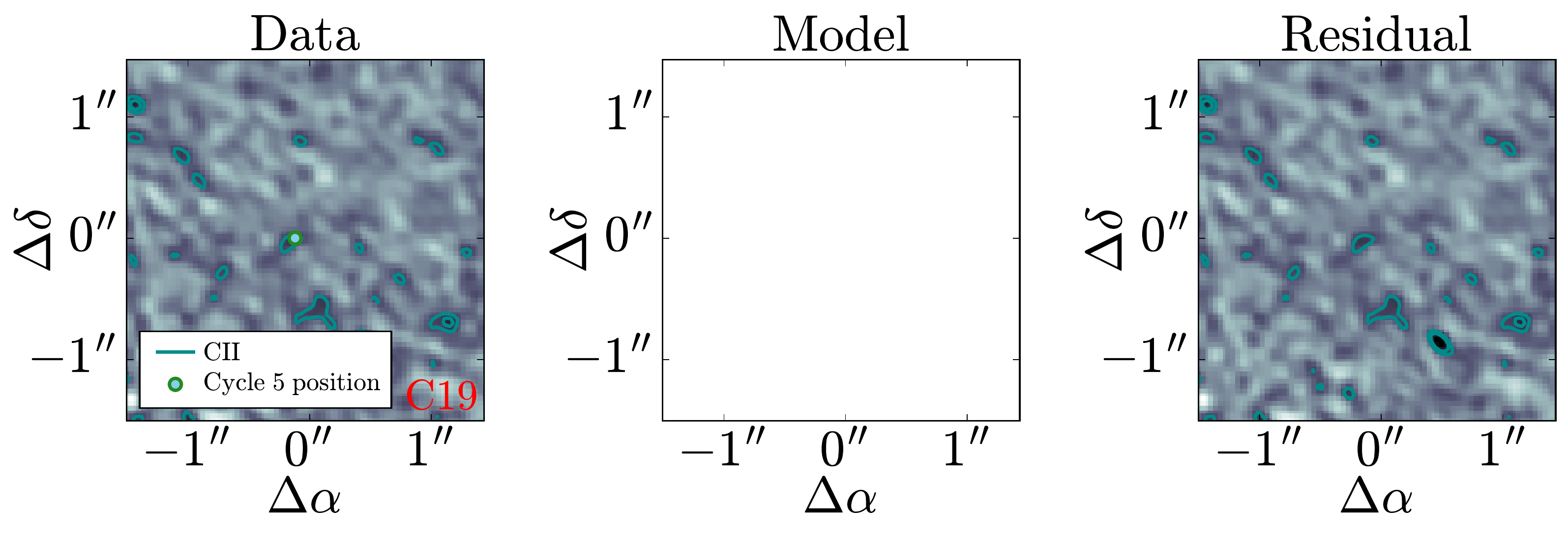}
\end{framed}
\end{subfigure}
\begin{subfigure}{0.45\textwidth}
\begin{framed}
\includegraphics[width=\textwidth]{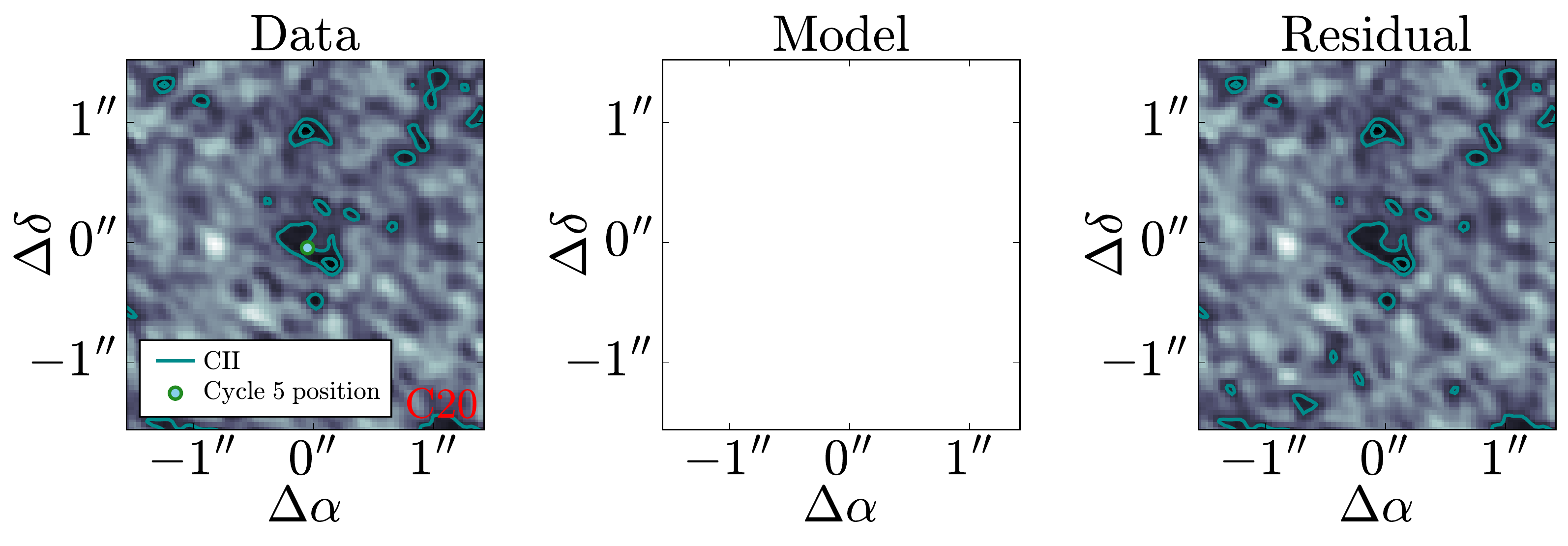}
\end{framed}
\end{subfigure}
\begin{subfigure}{0.45\textwidth}
\begin{framed}
\includegraphics[width=\textwidth]{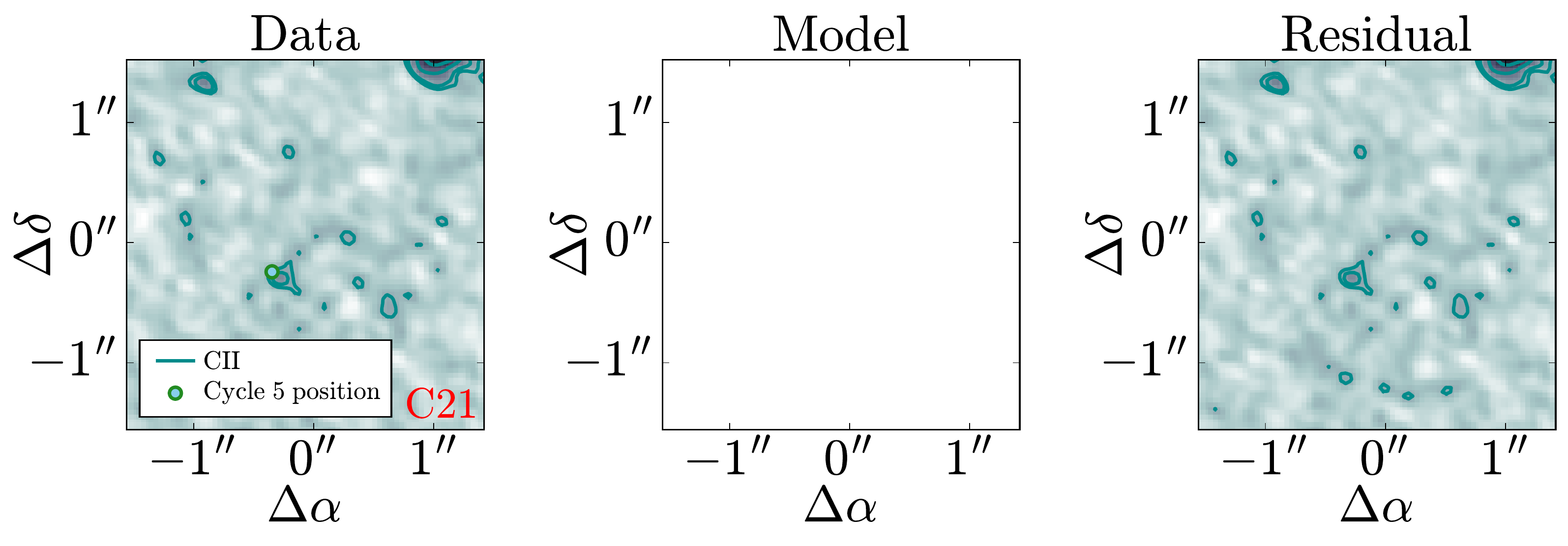}
\end{framed}
\end{subfigure}
\begin{subfigure}{0.45\textwidth}
\begin{framed}
\includegraphics[width=\textwidth]{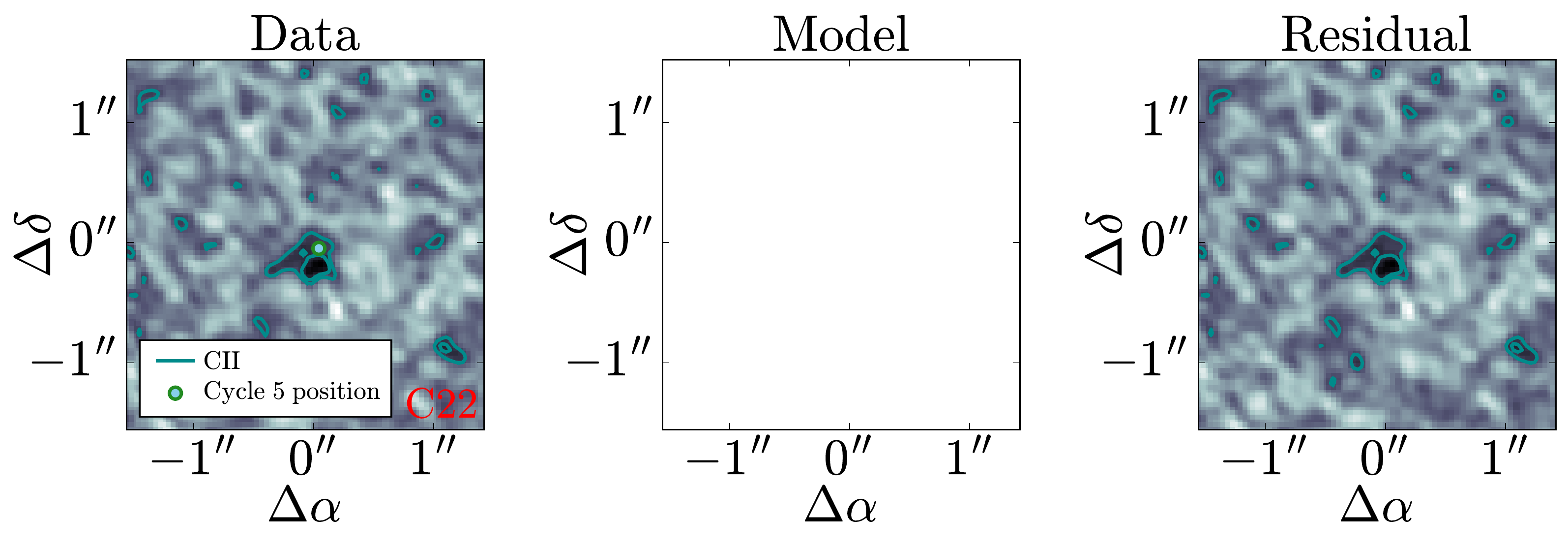}
\end{framed}
\end{subfigure}
\begin{subfigure}{0.45\textwidth}
\begin{framed}
\includegraphics[width=\textwidth]{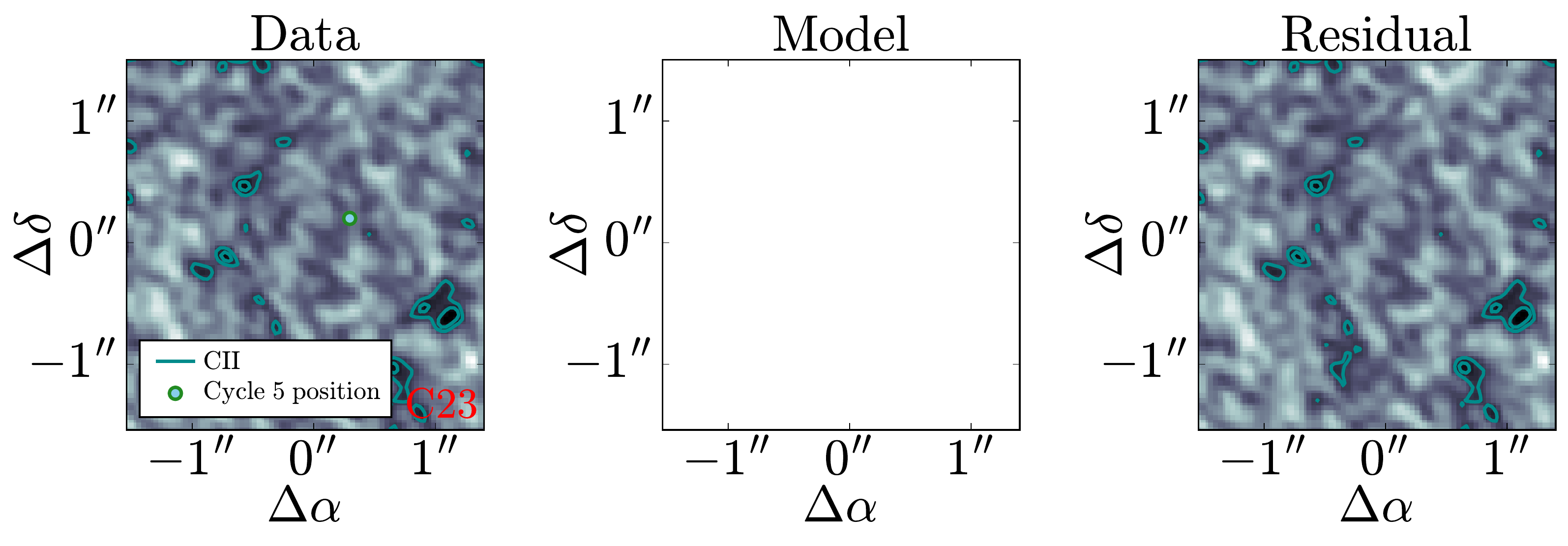}
\end{framed}
\end{subfigure}
\caption{}
\end{figure*}
\renewcommand{\thefigure}{F\arabic{figure}}

\bsp
\label{lastpage}
\end{document}